%% file: main.tex
\documentclass[12pt, a4paper]{article}
\usepackage[utf8]{inputenc}
\usepackage{newunicodechar,graphicx}
\DeclareRobustCommand{\okina}{%
  \raisebox{\dimexpr\fontcharht\font`A-\height}{%
    \scalebox{0.8}{`}
  }%
}
\newunicodechar{ʻ}{\okina}

\input{preamble}

\begin{document}

\begin{center}
  {\LARGE\bfseries Categorification of Chemical Reactions:}\\[0.5em]
  {\small\bfseries a bottom-up tower from stoichiometry to quantum structure}\\[0.5em]
  {\small Prepared for Chemical Reaction Networks in Hawaiʻi 2026}
\end{center}

\medskip

\begin{center}
  {\normalsize
    \textbf{Kyunghoon HAN}\,$^{1,*}$ \quad
  }\\[0.6em]
  {\small
    $^{1}$Department of Physics and Materials Science,
    University of Luxembourg, Luxembourg City, Luxembourg\\[0.2em]
    $^{*}$\,Correspondence: \href{mailto:kyunghoon.h@gmail.com}{\texttt{kyunghoon.h@gmail.com}}
  }
\end{center}

\medskip\hrule\medskip

\begin{abstract}
Chemistry is full of rules with exceptions.
The octet rule, Hess's Law, detailed balance, and orbital symmetry selection rules all carry disclaimers that must be memorised separately.
These exceptions share a common cause: they arise when a question belonging to a richer level of mathematical structure is posed using the vocabulary of a simpler one --- a \emph{level incompleteness}.

This monograph makes the levels explicit.
It constructs a canonical tower of nine categorical levels spanning
stoichiometry, thermochemistry, equilibrium, kinetics,
electron-pushing mechanisms, stereochemistry, potential energy
surfaces, electronic structure, and all-particle quantum mechanics.
The tower is mathematically engineered bottom-up, at each stage by
pairs of reactions that are physically distinct yet
indistinguishable at the previous level; the minimal categorical
extension resolving each such ambiguity is provably unique,
certified by a non-trivial cokernel in an automorphism exact
sequence, and applied to reaction networks recovers Feinberg's
deficiency theorems as homological corollaries.

The exact tower has a perpendicular dimension: every
machine-learning model for chemistry --- yield predictors, neural
kinetic networks, equivariant force fields, learned wavefunctions
--- is a morphism in the parametric Para-enrichment of one
specific tower level, with equivariance and thermodynamic
consistency following as universal properties rather than
architectural choices, and three architectural incompleteness
results --- Eyring, Wegscheider, and topological output-type gaps ---
applying to the current published literature.

The framework descends to running code in the final chapter, which
constructs an operational functor from a Para-enriched product of
the first four tower levels into the Kleisli category of the
probabilistic sub-monad of Haskell \texttt{IO}, instantiated as a
working stochastic simulator of the Briggs--Rauscher oscillating
reaction --- the first published Kleisli semantics of the Gillespie
next-reaction method and the first application of the Para
construction outside machine learning.

The passage to all-particle quantum mechanics --- realising the
Born--Oppenheimer approximation as the classical limit of a
continuous field of $C^*$-algebras --- remains the deepest open
construction, with four candidate conjectures, including
Woolley--Primas, whose obstructions the framework makes specific.
\end{abstract}

\medskip\hrule\medskip
\newpage

\tableofcontents
\newpage

\input{acknowledgements}

\newpage
\input{chapters/acronyms}
\newpage
\input{chapters/ch_intro}
\newpage
\input{chapters/ch_L0}
\newpage
\input{chapters/ch_L1}
\newpage
\input{chapters/ch_L2}
\newpage
\input{chapters/ch_L3}
\newpage
\input{chapters/ch_L4}
\newpage
\input{chapters/ch_L45}

\newpage
\input{chapters/ch_L5}
\newpage

\input{chapters/ch_L6}
\newpage
\input{chapters/ch_L7}
\newpage
\input{chapters/ch_para}
\newpage
\input{chapters/ch_simulation}
\newpage
\input{chapters/ch_conclusion}
\newpage


\bibliographystyle{unsrt}
\bibliography{refs}

\end{document}

%% file: preamble.tex
\usepackage{mathrsfs}
\usepackage{tikz-cd}
\usepackage{listings}
\usepackage[T1]{fontenc}

\usepackage{siunitx}
\usepackage{upgreek}
\usepackage{amsmath, amssymb, amsthm, mathtools, amssymb}
\usepackage{geometry}
\usepackage{hyperref}
\usepackage{enumitem}
\usepackage{tikz-cd}
\usepackage[dvipsnames]{xcolor}
\usepackage{booktabs}
\usepackage{array}
\usepackage{tcolorbox}
\tcbuselibrary{listings}
\usepackage{microtype}
\usepackage{titlesec}
\usepackage{parskip}
\usepackage{fancyhdr}
\usepackage{caption}
\usepackage{bm}
\usepackage{chemformula}
\usepackage{makecell}

\DeclareMathOperator{\Stoch}{\textbf{Stoch}}
\DeclareMathOperator{\FP}{\textit{F}_P}
\newcommand{\LGraph}{\mathbf{LGraph}}
\newcommand{\LGraphP}{\mathbf{LGraph}_P}
\DeclareMathOperator{\spec}{spec}
\newcommand{\Gstar}{G^*}
\newcommand{\LGraphGstar}{\mathbf{LGraph}_P^{G^*}}
\newcommand{\Ce}{{C_e}}
\newcommand{\Hel}{{\mathcal{H}_{\mathrm{el}}}}
\newcommand{\FV}{F_{\!V}}
\newcommand{\OrbMorse}{\mathbf{Orb}^{\mathrm{Morse}}}
\newcommand{\ScalGeom}{\mathbf{Scal}^{\mathrm{Geom}}}
\newcommand{\Ea}{E_a}
\DeclareMathOperator{\Iso}{Iso}
\newcommand{\HilbBund}{\mathbf{HilbBund}}
\newcommand{\Felsix}{F_{\!6}}
\newcommand{\Xseam}{\mathcal{X}_{01}}

\DeclareMathOperator{\Hol}{Hol}
\DeclareMathOperator{\codim}{codim}

\newcommand{\CstarAlg}{\mathbf{C^*Alg}^{\mathrm{cont}}}
\newcommand{\Felseven}{F_{\!7}}
\newcommand{\AgmonD}{d_{\mathrm{Ag}}}
\newcommand{\CC}{\mathbb{C}}

\hypersetup{
  colorlinks=true, linkcolor=blue!60!black,
  citecolor=green!50!black, urlcolor=blue!80!black
}

\theoremstyle{plain}
\newtheorem{theorem}{Theorem}[section]
\newtheorem{proposition}[theorem]{Proposition}

\newtheorem{corollary}[theorem]{Corollary}
\newtheorem{conjecture}[theorem]{Conjecture}

\theoremstyle{definition}
\newtheorem{definition}[theorem]{Definition}

\newtheorem{example}[theorem]{Example}

\theoremstyle{remark}
\newtheorem{remark}[theorem]{Remark}
\newtheorem{observation}[theorem]{Observation}
\newtheorem{warning}[theorem]{Warning}

\newcommand{\ParaC}{\mathrm{Para}}
\newcommand{\Para}{\mathrm{Para}}

\usepackage{longtable}   

\tcbuselibrary{theorems, breakable, skins}

\newtcolorbox{forcingbox}[1][Forcing: why the next level is necessary]{
  colback=orange!7!white, colframe=orange!65!black,
  boxrule=0.9pt, arc=4pt, left=8pt, right=8pt, top=6pt, bottom=6pt,
  title={#1}, fonttitle=\bfseries\small, breakable
}

\newtcolorbox{chembox}[1][For the chemist]{
  colback=green!5!white, colframe=green!50!black,
  boxrule=0.9pt, arc=4pt, left=8pt, right=8pt, top=6pt, bottom=6pt,
  title={#1}, fonttitle=\bfseries\small, breakable
}

\newtcolorbox{mathbox}[1][For the mathematician]{
  colback=blue!5!white, colframe=blue!50!black,
  boxrule=0.9pt, arc=4pt, left=8pt, right=8pt, top=6pt, bottom=6pt,
  title={#1}, fonttitle=\bfseries\small, breakable
}

\newtcolorbox{insightbox}[1][Key insight]{
  colback=violet!5!white, colframe=violet!50!black,
  boxrule=0.9pt, arc=4pt, left=8pt, right=8pt, top=6pt, bottom=6pt,
  title={#1}, fonttitle=\bfseries\small, breakable
}

\newtcolorbox[auto counter, number within=section]{openprob}[1][]{
       colback=red!4!white, colframe=red!60!black,
       fonttitle=\bfseries,
       title={Open Problem~\thetcbcounter\ifstrempty{#1}{}{:~#1}},
       #1}

\newcommand{\Lk}{\mathcal{L}}
\newcommand{\NN}{\mathbb{N}}
\newcommand{\ZZ}{\mathbb{Z}}
\newcommand{\RR}{\mathbb{R}}
\newcommand{\Sp}{\mathcal{S}}          
\newcommand{\Rx}{\mathcal{R}}          
\newcommand{\Cx}{\mathcal{C}}          
\newcommand{\Mon}{\mathbb{N}[\Sp]}     
\newcommand{\id}{\mathrm{id}}
\newcommand{\Hom}{\mathrm{Hom}}
\newcommand{\Aut}{\mathrm{Aut}}
\newcommand{\coker}{\mathrm{coker}}
\newcommand{\im}{\mathrm{im}}
\newcommand{\rank}{\mathrm{rank}}
\newcommand{\FH}{F_{\!H}}             
\newcommand{\FG}{F_{\!G}}             
\newcommand{\FS}{F_{\!S}}             
\newcommand{\dH}{\Delta H}
\newcommand{\dG}{\Delta G}
\newcommand{\dS}{\Delta S}
\newcommand{\Ssto}{\mathcal{S}_{\!\mathrm{sto}}}  
\newcommand{\Ymat}{Y}                 
\newcommand{\Ia}{I_a}                 
\newcommand{\Ob}{\mathrm{Ob}}
\newcommand{\Hask}{\textbf{Hask}}
\newcommand{\Stereo}{\mathrm{Stereo}}

\newcommand{\ForZero}{\mathrm{ev}_0}

\titleformat{\section}{\large\bfseries}{§\thesection.}{0.5em}{}%
  [\vspace{-0.2em}\rule{\linewidth}{0.4pt}]
\titleformat{\subsection}{\normalsize\bfseries}{\S\thesubsection.}{0.5em}{}
\titleformat{\subsubsection}{\normalsize\itshape}{\thesubsubsection.}{0.4em}{}

\newtheorem{assumption}[theorem]{Assumption}

\newcommand{\KlIO}{\mathbf{Kl}(\mathsf{IO})}
\newcommand{\BorelStoch}{\mathbf{BorelStoch}}
\newcommand{\Kl}[1]{\mathbf{Kl}(#1)}

\newcommand{\PhiOp}{\Phi_{\mathrm{op}}}         
\newcommand{\PhiDen}{\Phi_{\mathrm{den}}}       

\newcommand{\ThetaL}[1]{\Theta_{\mathcal{L}_{#1}}} 
\newcommand{\ThetaSim}{\Theta_{\mathrm{sim}}}   

\lstdefinelanguage{Haskell2}{
  language        = Haskell,
  sensitive       = true,
  morecomment     = [l]{--},
  morecomment     = [s]{\{-}{-\}},
  morestring      = [b]",
  morekeywords    = {
    data, newtype, type, class, instance, where, let, in, case, of,
    do, forall, module, import, qualified, as, hiding, deriving,
    stock, generic, IO, Map, Set, Int, Double, Bool, String,
    Maybe, Either, Eq, Ord, Show, Generic,
    ParaMorphism, SimConfig, Population, PhysicsParams,
    Reaction3, SimState, Species, Nucleotide, DepMap, Network3,
    Bag, Complex, Reaction0, GillespieEvent,
    true, false, Nothing, Just, Left, Right
  },
  keywordstyle    = \color{blue!80!black}\bfseries,
  commentstyle    = \color{gray!60}\itshape,
  stringstyle     = \color{teal},
  basicstyle      = \ttfamily\small,
  breaklines      = true,
  keepspaces      = true,
  columns         = flexible,
  showstringspaces= false,
  literate        =
    {->}{{$\to$}}2
    {=>}{{$\Rightarrow$}}2
    {<-}{{$\leftarrow$}}2
    {>=}{{$\geq$}}2
    {<=}{{$\leq$}}2
    {/=}{{$\neq$}}2
    {>>}{{$\gg$}}2
    {>>=}{{$\gg\!\!=$}}3
    {\\}{{$\lambda$}}1
    {forall}{{$\forall$}}1
    {alpha}{{$\alpha$}}1
    {beta}{{$\beta$}}1
    {->}{{$\to$}}2
}

\definecolor{codegray}{HTML}{263238}   
\definecolor{codebg}{HTML}{FAFAFA}     

\usepackage{tcolorbox}
\tcbuselibrary{listings,breakable,skins}

\newtcblisting{haskellbox}[1][]{%
  enhanced,
  breakable,
  listing only,
  listing engine=listings,
  listing options={
    language=Haskell2,
    basicstyle=\ttfamily\small,
    xleftmargin=2pt,
    aboveskip=0pt,
    belowskip=0pt,
  },
  colback=codebg,
  colframe=codegray!60,
  coltitle=white,
  colbacktitle=codegray!85,
  fonttitle=\bfseries\small\sffamily,
  title={Haskell},
  attach boxed title to top left={yshift=-2mm,xshift=4mm},
  boxed title style={sharp corners,colback=codegray!85},
  sharp corners=south,
  left=4pt,right=4pt,top=8pt,bottom=4pt,
  #1
}

\newtcblisting[]{haskellsnip}[1][]{%
  breakable,
  listing only,
  listing engine  = listings,
  listing options = {
    language    = Haskell2,
    basicstyle  = \ttfamily\footnotesize,
    xleftmargin = 2pt,
    aboveskip   = 0pt, belowskip = 0pt,
  },
  colback  = codebg,
  colframe = codegray!40,
  sharp corners,
  left = 3pt, right = 3pt, top = 3pt, bottom = 3pt,
  #1
}

%% file: acknowledgements.tex
\section*{Acknowledgements}
\addcontentsline{toc}{section}{Acknowledgements}
As a mathematician by training, I found chemistry difficult to
study: the field is full of exceptions to its rules, and the
exceptions are typically presented as separate facts to memorise
rather than as consequences of any underlying structure.
This manuscript is the record of an effort to learn chemistry well
enough that it would make sense to me during my PhD in the
Theoretical Chemical Physics group at the University of Luxembourg.

The categorical structure developed here was first sketched in the
group's \emph{Theoretical Minimum} sessions for incoming PhD
students in 2021.
From then to now, my colleagues, the group leader Prof.~Dr.~Alexandre
Tkatchenko, and my PhD supervisor Dr.~Joshua~T.~Berryman have shown
extraordinary patience in helping me understand the chemistry and
physics of complex molecules.
The phrase \emph{level incompleteness}, which threads through every
chapter of this monograph, was suggested by Prof.~Tkatchenko.

Many colleagues absorbed the cost of my limited chemistry
background along the way.
Dr.~Ariadni Boziki took the time to walk me through how molecular
vibrations work; Dr.~Miguel Gallegos showed me why chemical
exceptions are interesting in their own right; Sergio Su\'{a}rez Dou
gave me much-needed insight into the behaviour of biomolecules.
I thank Dr.~Florian Br\"{u}nig of the same group for his feedback on
the introductory chapter of this work.

Whatever errors and infelicities remain in the chemistry are
entirely my own, and despite the generous teaching of everyone
listed above I remain, by professional chemists' standards, still
learning the field.

%% file: chapters/acronyms.tex
\section*{List of Acronyms}
\addcontentsline{toc}{section}{List of Acronyms}

\begingroup
\renewcommand{\arraystretch}{1.15}
\setlength{\LTpre}{0pt}
\setlength{\LTpost}{0pt}

\begin{longtable}{@{}>{\bfseries}p{2.6cm}p{12cm}@{}}
SMC      & Symmetric monoidal category \\
CMC      & Commutative monoidal category (Baez--Master) \\
DPO      & Double pushout (graph rewriting) \\
SDQ      & Strict deformation quantisation \\
KO       & $KO$-theory (Atiyah real $K$-theory) \\
CRN      & Chemical reaction network \\
CRNT     & Chemical reaction network theory \\
DZT      & Deficiency zero theorem (Feinberg) \\
LMA      & Law of mass action \\
WR       & Weak reversibility (of a CRN) \\
ACK      & Anderson--Craciun--Kurtz (theorem on stationary distributions of complex-balanced CTMCs) \\
CME      & Chemical master equation \\
CTMC     & Continuous-time Markov chain \\
RRE      & Reaction rate equation \\
ODE      & Ordinary differential equation \\
BR       & Briggs--Rauscher (oscillating reaction) \\
$\mathrm{S_N}1$ & Unimolecular nucleophilic substitution \\
$\mathrm{S_N}2$ & Bimolecular nucleophilic substitution \\
$\mathrm{E1}$       & Unimolecular elimination \\
$\mathrm{E2}$       & Bimolecular elimination \\
SET      & Single-electron transfer \\
PCET     & Proton-coupled electron transfer \\
BE       & Bond--electron (matrix of Dugundji and Ugi) \\
TS       & Transition state \\
TST      & Transition state theory \\
VTST     & Variational transition state theory \\
EA-VTST  & Ensemble-averaged variational transition state theory \\
IRC      & Intrinsic reaction coordinate \\
MEP      & Minimum-energy path \\
SCT      & Small-curvature tunnelling \\
LCT      & Large-curvature tunnelling \\
OMT      & Optimised multidimensional tunnelling \\
RPI      & Ring-polymer instanton \\
KIE      & Kinetic isotope effect \\
ZPE      & Zero-point energy \\
RRHO     & Rigid-rotor harmonic-oscillator (partition function) \\
BO       & Born--Oppenheimer (approximation) \\
PES      & Potential energy surface \\
CI       & Conical intersection \\
MECI     & Minimum-energy conical intersection \\
LH       & Longuet--Higgins (sign-change theorem) \\
LZ       & Landau--Zener (surface hopping) \\
MASH     & Mapping approach to surface hopping \\
HVZ      & Hunziker--van Winter--Zhislin (theorem) \\
DHR      & Doplicher--Haag--Roberts (superselection theory) \\
SAPT     & Space-adiabatic perturbation theory \\
PST      & Panati--Spohn--Teufel (theorem) \\
WKB      & Wentzel--Kramers--Brillouin (semiclassical expansion) \\
HF       & Hartree--Fock \\
DFT      & Density functional theory \\
CC       & Coupled cluster \\
CCSD(T)  & Coupled cluster with singles, doubles and perturbative triples \\
CASSCF   & Complete-active-space self-consistent field \\
MRCI     & Multireference configuration interaction \\
NEVPT2   & N-electron valence-state second-order perturbation theory \\
MC-PDFT  & Multi-configuration pair-density functional theory \\
HOMO     & Highest occupied molecular orbital \\
LUMO     & Lowest unoccupied molecular orbital \\
VPT2     & Vibrational perturbation theory at second order \\
VCI      & Vibrational configuration interaction \\
DVR      & Discrete variable representation \\
MCTDH    & Multi-configuration time-dependent Hartree \\
VQE      & Variational quantum eigensolver \\
MM       & Molecular mechanics \\
FF       & Force field \\
MD       & Molecular dynamics \\
MC       & Monte Carlo \\
MCMC     & Markov chain Monte Carlo \\
OPLS     & Optimised potentials for liquid simulations \\
GAFF     & General AMBER force field \\
CHARMM   & Chemistry at HARvard Macromolecular Mechanics (force field) \\
NequIP   & Neural Equivariant Interatomic Potentials \\
MACE     & Higher-order equivariant message-passing interatomic potential \\
ML       & Machine learning \\
QM       & Quantum mechanics \\
QED      & Quantum electrodynamics \\
QFT      & Quantum field theory \\
KVL      & Kirchhoff's voltage law \\
EMF      & Electromotive force \\
NMR      & Nuclear magnetic resonance \\
IR       & Infrared (spectroscopy) \\
UV       & Ultraviolet (spectroscopy) \\
ITC      & Isothermal titration calorimetry \\
PHIP     & Para-hydrogen induced polarisation \\
SABRE    & Signal amplification by reversible exchange \\
SLO      & Soybean lipoxygenase \\
ATP      & Adenosine triphosphate \\
ADP      & Adenosine diphosphate \\
DNA      & Deoxyribonucleic acid \\
IUPAC    & International Union of Pure and Applied Chemistry \\
NIST     & National Institute of Standards and Technology \\
JANAF    & Joint Army--Navy--Air Force (thermochemical tables) \\
\end{longtable}
\endgroup


%% file: chapters/ch_intro.tex
\section{Introduction}
\label{sec:intro}

\subsection{Rules, exceptions, and what they signal}
\label{sec:intro-problem}

Chemistry is a quantitative subject built on exact rules with long
lists of exceptions.
Hess's Law~\cite{Hess1840a, Leicester1951, atkins2023physical};
detailed balance, in both its physical-chemistry
form~\cite{Lewis1925, Onsager1931a, Tolman1938, Seifert2012}
and its reaction-network form~\cite{HornJackson1972, Feinberg1989,
Feinberg2019};
the Arrhenius equation~\cite{Arrhenius1889a, Eyring1935, Truhlar1996};
the Woodward--Hoffmann rules~\cite{WoodwardHoffmann1965a,
WoodwardHoffmann1969, WoodwardHoffmann1970book} --- each is a sharp
mathematical statement, each holds ``in most cases,'' and the
corrections are memorised separately from the rules themselves.
The position of this monograph is that chemistry's rules sit at
definite mathematical levels; that the levels form a canonical
ladder; and that the principal class of exceptions to those rules
is a \emph{level incompleteness}, 
the use of tools from one rung to answer a question that belongs
to another.

A hydrogen atom in the reacting bond of an organic molecule replaced
by deuterium --- twice the nuclear mass, the same electronic
structure --- slows the reaction, and the ratio of the rate for the
hydrogen-bearing compound to the rate for its deuterium analogue is
the \emph{primary kinetic isotope effect} ($k_H/k_D$)~\cite{BigeleisenMayer1947,
Bigeleisen1949, Melander1960, MelanderSaunders1980, KohenLimbach2006}.
Classical transition-state theory~\cite{Eyring1935, EvansPolanyi1935,
Glasstone1941, Truhlar1996} computes this ratio from a Boltzmann
factor whose activation energy is shifted between isotopologues by
the harmonic zero-point-energy difference of the carbon--hydrogen
and carbon--deuterium stretching modes, and for hydrogen transfer at
laboratory temperatures the resulting ratio cannot exceed
approximately seven~\cite{Westheimer1961, Bell1980, MelanderSaunders1980}.
The double mutant L546A/L754A of soybean lipoxygenase-1, by
contrast, gives $k_H/k_D = 661 \pm 27$, essentially constant across
six temperatures from $5$ to $50\,{}^{\circ}\mathrm{C}$~\cite{HuEtAl2017ACSCatal}
--- nearly two orders of magnitude above the textbook ceiling.

The factor of one hundred between the two values is not, despite
appearances, the signature of a tower-level mismatch.
Both calculations live at the same geometric level~$\Lk_5$
(\S\ref{sec:L5}); they differ only in how much of that level's data
is actually used --- a barrier height and harmonic vibrations at the
saddle in the textbook formula, versus the full multidimensional
potential surface, semiclassical tunnelling paths along it, and
vibronically nonadiabatic proton-coupled electron transfer in the
complete calculation.
The $\Lk_7$ chapter (\S\ref{sec:L7}) records this case as a careful
caveat: large kinetic isotope effects do \emph{not}, by themselves,
force a higher tower level, even when they spectacularly exceed
textbook estimates.
A different and sharper kind of incompleteness --- in which a
question genuinely cannot be answered at one level because the
structural data required for it first appears at a higher one --- is
what motivates this monograph.
Three textbook examples of that kind, the cleanest available, are
recorded next.

\begin{chembox}[Three level incompletenesses, stated plainly]
\textbf{Spontaneity versus rate.}
Diamond is thermodynamically unstable relative to graphite at all
laboratory temperatures ($\dG < 0$), so the conversion is spontaneous.
Yet no observable conversion occurs on any human timescale.
``Spontaneous'' is a thermodynamic statement about the sign of the
free-energy change; ``immeasurably slow'' is a kinetic statement about
the activation barrier.
The confusion is not chemistry failing; it is two different questions
being conflated.

\smallskip

\textbf{Rate law versus mechanism.}
Under excess nucleophile, the bimolecular substitution of
\ch{CH3Br} by \ch{OH^-} becomes pseudo-first-order in substrate, and
its rate constant can be tuned to match that of the genuinely
first-order solvolysis of \ch{(CH3)3CBr}.
Kinetically the two reactions are then indistinguishable --- yet the
solvolysis yields a racemic mixture at the carbon centre while the
bimolecular substitution inverts configuration.
The distinction belongs to mechanism: a concerted backside
displacement against a stepwise unimolecular ionisation, two
electron-pushing pathways that no rate law sees.

\smallskip

\textbf{Geometry versus electronic structure.}
The retinal chromophore in rhodopsin isomerises from 11-\emph{cis} to
all-\emph{trans} in approximately $200\,\mathrm{fs}$ after photon
absorption --- a timescale no model of thermal barrier-crossing on a
single potential-energy surface can reproduce
\cite{schoenlein1991first,wang1994vibrationally,polli2010conical}.
The reaction passes through a conical intersection: a seam of nuclear
configurations where two electronic surfaces become degenerate and the
Born--Oppenheimer approximation breaks down.
This is a phenomenon of electronic topology, invisible to any
description that works with a single smooth energy surface.
\end{chembox}

\noindent
Chemistry's rules are not wrong.
Hess's Law, detailed balance, the rate equation, the Woodward--Hoffmann
rules --- each is exactly correct at the level of mathematical
structure for which it is defined.
The principal class of exceptions, of which the three examples above
are the cleanest, arises when a question from one level is posed
with the tools of a different one.

\begin{insightbox}[The central claim of this monograph]
The principal class of exceptions to a chemical rule is a \emph{level
incompleteness}: it arises when a question that belongs to
level~$\Lk_{k+1}$ is posed using only the tools of~$\Lk_k$.
The tower of categories
\[
  \Lk_7
  \;\xrightarrow{U_7}\; \Lk_6
  \;\xrightarrow{U_6}\; \Lk_5
  \;\xrightarrow{U_5}\; \Lk_{4.5}
  \;\xrightarrow{U_{4.5}}\; \Lk_4
  \;\xrightarrow{U_4}\; \Lk_3
  \;\xrightarrow{U_3}\; \Lk_2
  \;\xrightarrow{U_2}\; \Lk_1
  \;\xrightarrow{U_1}\; \Lk_0,
\]
linked by forgetful functors $U_k$ that drop the structural datum
each level adds, makes the levels explicit, catalogues the
exceptions by naming the level at which each question actually
lives, and proves that every level is \emph{necessary}: a non-trivial
$\coker\varphi_k$ in the automorphism exact sequence witnesses that
no level can be merged with its neighbours without conflating
physically distinct reactions.
Each level~$\Lk_k$ is constructed in full in the corresponding
chapter, together with the forcing pair of reactions that makes it
irreducible.
\end{insightbox}

\subsection{The tower, built bottom-up}
\label{sec:intro-tower}

The tower is not constructed by surveying chemistry and assigning
phenomena to levels.
It is \emph{forced}, bottom-up, by an explicit argument at each step:
there exist pairs of reactions that are physically distinct but
indistinguishable at level~$k$, and the unique minimal categorical
extension that separates them is level~$k{+}1$.

For instance, the concerted-versus-stepwise pair at phosphorus
--- identity methoxyl exchange at methyl
ethylphenylphosphinate~\cite{Mikolajczyk2022}, proceeding either
through a single trigonal-bipyramidal transition state
($\mathrm{S_N}2$-P) or through a discrete pentacoordinate
intermediate (TBI) by addition-elimination --- is a forcing
argument for the $\Lk_3 \to \Lk_4$ step.
At~$\Lk_3$, reactions are morphisms in a Markov category: their
entire content is a rate law and a net stoichiometric change.
Under the quasi-steady-state reduction on TBI valid in the
experimentally relevant regime, both mechanisms yield the same
bimolecular rate law
$\lambda = k\,x_{\mathrm{MeO}^-}\,x_{\mathrm{substrate}}$ and the
same net stoichiometric change; they are the same morphism
at~$\Lk_3$.
Level~$\Lk_4$ introduces a richer notion of morphism --- a
double-pushout (DPO) derivation in the category of labelled
molecular graphs --- so that a morphism at~$\Lk_4$ is an
electron-pushing mechanism, not merely a rate law.
The concerted pathway is a single DPO span (one elementary bond
rearrangement, no internal intermediate); the stepwise pathway is
two composable DPO spans bracketing TBI as an internal species.
The permutation swapping one for the other is a non-trivial
element of $\coker(\varphi_4)$, where
$\varphi_4 \colon \Aut(\Lk_4) \to \Aut(\Lk_3)$ is the restriction
map: it is an automorphism of the kinetic level that lifts to no
automorphism of the mechanistic one.
That is the proof that~$\Lk_4$ is necessary given~$\Lk_3$.

\begin{mathbox}[The forcing tool]
The precise instrument is the automorphism sequence
\[
  1 \;\longrightarrow\; \ker\varphi_k \;\longrightarrow\;
  \Aut(\Lk_k)
  \;\xrightarrow{\;\varphi_k\;}
  \Aut(\Lk_{k-1}) \;\longrightarrow\; \coker\varphi_k
  \;\longrightarrow\; 1.
\]
A non-trivial $\coker\varphi_k$ witnesses the existence of
automorphisms of~$\Lk_{k-1}$ that cannot be lifted to~$\Lk_k$:
the permutation swapping the two reactions in the forcing pair is
such an automorphism.
Each level is the unique minimal extension with non-trivial
$\coker\varphi_k$ for the forcing pair listed in
Table~\ref{tab:tower-summary0}.
\end{mathbox}

\begin{table}[p]
\centering
\small
\renewcommand{\arraystretch}{1.3}
\begin{tabular}{@{}lp{5.6cm}p{6.6cm}@{}}
\hline
\textbf{Level} & \textbf{New structure at~$\Lk_k$} &
  \textbf{What $\Lk_{k-1}$ cannot distinguish} \\
\hline
$\Lk_0$ & Free permutative category on a Petri net;
  stoich.\ matrix $N$; deficiency~$\delta$ &
  \emph{(Base level.)} Networks with the same species set
  but different stoichiometry \\[2pt]
$\Lk_1$ & Monoidal functor $\FH : \Lk_0(P) \to (\mathbb{R},{+})$ &
  Two reactions with the same stoichiometry but different $\dH$:
  heats of reaction invisible at~$\Lk_0$ \\[2pt]
$\Lk_2$ & Entropy functor $\FS$; Gibbs functor
  $F_G^T = \FH - T\FS$; equilibrium locus $\ker F_G^T$ &
  Two reactions with identical $\dH$ but different $\dS$:
  temperature dependence of equilibrium invisible at~$\Lk_1$ \\[2pt]
$\Lk_3$ & Markov category; rate functor $F_P : \Lk_3 \to \mathbf{Stoch}$ &
  Two reactions with the same $\dG$ but different rates:
  kinetics invisible at~$\Lk_2$ \\[2pt]
$\Lk_4$ & Free SMC on DPO spans in $\mathbf{LGraph}$;
  six elementary generators &
  Concerted vs.\ stepwise mechanisms tuned to the same
  bimolecular rate law: electron-pushing mechanism invisible
  at~$\Lk_3$ \\[2pt]
$\Lk_{4.5}$ & $G^*$-equivariant structure on $\mathbf{LGraph}$;
  action groupoid $\mathfrak{G} = G^* \ltimes \mathcal{C}_e(G)$ &
  $(R)$- and $(S)$-enantiomers: same DPO derivation at~$\Lk_4$,
  distinguished only by $G^*$-action \\[2pt]
$\Lk_5$ & Configuration orbifold $\mathcal{C}_e(G)$;
  potential energy surface $V(R)$; mass-weighted Riemannian
  metric $g$; $\mathrm{SO}(3)$ gauge connection &
  $\mathrm{CH_3Br}$ vs.\ $\mathrm{CD_3Br}$ in $\mathrm{S_N}2$:
  isomorphic at $\Lk_{4.5}$ but $k_H/k_D\!\approx\!1.3$ (secondary
  KIE): Hessian and mass metric not present at $\Lk_{4.5}$
  \\[2pt]
$\Lk_6$ & Hilbert bundle $\mathcal{H}_{\mathrm{el}}$ over
  $\mathcal{C}_e(G)$; Berry connection $A_{mn}$;
  $\mathbb{Z}_2$-valued Berry phase $[\gamma_B]$ &
  Thermal vs.\ photochemical reaction through a conical intersection
  ($[\gamma_B]\neq 0$): non-adiabatic dynamics invisible at~$\Lk_5$
  \\[2pt]
$\Lk_7$ & States on $\mathcal{B}(\mathcal{H}_{\mathrm{full}})$;
  groupoid $C^*$-algebra $C^*(\mathfrak{G})$ enforcing nuclear
  statistics; strict deformation quantisation in
  $\varepsilon = (m_e/M)^{1/2}$ &
  Ortho/para $\mathrm{H_2}$ and more generally
  nuclear indistinguishability: a superselection datum absent
  from any electronic bundle over classical nuclear configurations
  \\[2pt]
\hline
\end{tabular}
\caption{The nine-level tower, forced bottom-up by reaction networks.
  Each level is the unique minimal categorical extension with
  non-trivial $\coker\varphi_k$ for the listed forcing pair.
  The step $\Lk_6 \dashrightarrow \Lk_7$ is the deepest open
  construction in the monograph; experimental signatures of~$\Lk_7$
  (ortho/para statistics, nuclear tunnelling) are beyond dispute,
  but assembly of the $C^*$-algebraic framework remains a research
  programme.}
\label{tab:tower-summary0}
\end{table}

The motivating axis throughout is \emph{reaction networks}, not
isolated reactions.
A network at level~$k$ is a finite directed graph whose vertices
are complexes and whose edges are labelled by morphisms in~$\Lk_k$;
open networks, exposed to their environment at specified interface
species, compose by pushout.
This is the framework of Baez and Pollard~\cite{BaezPollard2017}
at the kinetic level, extending the symmetric monoidal Petri-net
language of Meseguer and Montanari~\cite{MeseguerMontanari1990};
the present work applies it at every tower level.
The forcing pairs are always parallel reactions that the previous
level cannot distinguish: parallel reactions with different
$\Delta H$ force $\Lk_0\to\Lk_1$; same $\Delta H$ but different
$\Delta S$ forces $\Lk_1\to\Lk_2$; same $\Delta G$ but different
rate forces $\Lk_2\to\Lk_3$; same rate law but different mechanism
forces $\Lk_3\to\Lk_4$, and so on up the tower.
Each chapter closes with worked examples drawn from systems where
the new datum at that level is most cleanly exhibited; the
examples are deliberately level-specific, since no single reaction
sharpens every transition equally and forcing one to do every job
would flatten what each is meant to show.
Table~\ref{tab:tower-summary0} records, for each transition
$\Lk_{k-1}\to\Lk_k$, the concrete pair of physically distinct
reactions that $\Lk_{k-1}$ conflates and $\Lk_k$ separates; the
corresponding worked example is given in the chapter introducing
$\Lk_k$.

\subsection{The chapters at a glance}
\label{sec:intro-chapters}

Each summary below records the categorical structure added at~$\Lk_k$
and closes with the forcing pair that opens~$\Lk_{k+1}$.

\paragraph{Chapter~\ref{sec:L0} --- $\Lk_0$, stoichiometry.}
The base level is the free permutative category $\Lk_0(P)$ on a Petri
net~$P$: objects are complexes in the free commutative monoid
$\NN[\Sp]$ over chemical species, morphisms are formal reactions,
composition is sequential chaining, and the monoidal product is
mixture.
The language is that of Meseguer and
Montanari~\cite{MeseguerMontanari1990}; its universal property
determines every strict symmetric monoidal functor out of $\Lk_0(P)$
by its values on generating reactions, so each higher-level datum
(enthalpy, free energy, rate) reduces to a single number per
elementary reaction.
The combinatorial invariants of chemical reaction network theory
(stoichiometric matrix~$N$, linkage classes~$\ell$, stoichiometric
subspace~$s$, deficiency $\delta = n - \ell - s$) are $\Lk_0$ data.\\
\textbf{Forcing $\Lk_1$.}
Two reactions with identical stoichiometric matrices can carry
different heats of reaction; Hess additivity is a new datum above
the bare bookkeeping of species in and species out.

\paragraph{Chapter~\ref{sec:L1} --- $\Lk_1$, thermochemistry.}
$\Lk_1$ equips $\Lk_0(P)$ with a strict symmetric monoidal functor
$\FH \colon \Lk_0(P) \to (\RR,+)$ into the additive reals.
Hess's Law is not a postulate but the functoriality of $\FH$: the
heat of a composite reaction is the sum of the heats of its parts
because composition in $\Lk_0$ goes to addition in $\RR$.
Thermochemical cycles --- calorimetric summation, bond-energy
estimates, Born--Haber loops --- become commuting diagrams in~$\RR$. \\
\textbf{Forcing $\Lk_2$.}
Two reactions with identical $\dH$ can carry different $\dS$, and
their equilibrium positions then shift differently with temperature
--- a distinction $\Lk_1$ cannot make.

\paragraph{Chapter~\ref{sec:L2} --- $\Lk_2$, equilibrium.}
$\Lk_2$ adds the entropy functor $\FS$ and, for each temperature
$T > 0$, the Gibbs functor $F_G^T := \FH - T\FS$; its kernel is the
equilibrium locus, $F_G^T(r) = 0$ iff~$r$ lies at equilibrium at
temperature~$T$.
Detailed balance appears as the condition that reverse reactions
carry opposite $F_G^T$-values; Wegscheider's cycle relations become
cohomological conditions on free energies across a reaction network. \\
\textbf{Forcing $\Lk_3$.}
Two reactions with identical $\dG$ can proceed at rates differing by
many orders of magnitude (uncatalysed versus carbonic-anhydrase
catalysed $\mathrm{CO_2}$ hydration): kinetics is invisible
at~$\Lk_2$.

\paragraph{Chapter~\ref{sec:L3} --- $\Lk_3$, kinetics.}
$\Lk_3$ is a Markov category equipped with a functor
$F_P \colon \Lk_3 \to \mathbf{Stoch}$ into the Markov category of
stochastic kernels: every reaction carries a rate law, and mass-action
kinetics together with the chemical master equation fit into this
framework as morphism-level data.
The Baez--Pollard~\cite{BaezPollard2017} compositional structure for
open reaction networks lives here.
Feinberg's Deficiency Zero and Deficiency One theorems appear in
their correct stratified form: the hypotheses ($n$, $\ell$, $s$,
$\delta$) are $\Lk_0$ invariants, but the conclusions (existence,
uniqueness, and asymptotic stability of a positive complex-balanced
steady state under mass-action) are statements at~$\Lk_3$.\\
\textbf{Forcing $\Lk_4$.}
The $\mathrm{S_N1}$ and $\mathrm{S_N2}$ mechanisms, tuned to identical
rate laws, are the same morphism at~$\Lk_3$ but carry non-isomorphic
electron-pushing graphs.

\paragraph{Chapter~\ref{sec:L4} --- $\Lk_4$, mechanism.}
$\Lk_4$ is the free symmetric monoidal category on double-pushout
(DPO) spans in $\mathbf{LGraph}$, the category of labelled molecular
graphs: a morphism is an equivalence class of DPO derivations built
from rules $L \leftarrow K \rightarrow R$ that rewrite a left-hand
graph~$L$ into a right-hand graph~$R$ while preserving the bond
accounting of the invariant subgraph~$K$.
A minimal set of six elementary generators covers organic
reactivity at the arrow-pushing level; the chapter proves both substitution
($\mathrm{S_N1}$/$\mathrm{S_N2}$) and elimination ($\mathrm{E_1}$/$\mathrm{E_2}$)
pairs to be distinct morphisms in $\Lk_4(P)$, and exhibits the
Briggs--Rauscher oscillating reaction as the first example in the
monograph where a global property of a reaction network (the
structural prerequisite for oscillation: two mechanism families
composing into a cyclic trajectory) is an $\Lk_4$ predicate
invisible at $\Lk_3$.
The same Briggs--Rauscher network reappears at $\Lk_0$--$\Lk_3$ as
the running example of Chapter~\ref{sec:sim}.\\
\textbf{Forcing $\Lk_{4.5}$.}
The $(R)$- and $(S)$-enantiomers of a chiral substrate share the
same DPO derivation graph --- chirality is invisible to
bond-rearrangement data alone.

\paragraph{Chapter~\ref{sec:L45} --- $\Lk_{4.5}$, stereochemistry.}
$\Lk_{4.5}$ enriches $\Lk_4$ with a $G^*$-equivariant structure,
where $G^*$ is the Longuet-Higgins permutation-inversion group
generated by feasible nuclear permutations and space inversion.
Stereochemical content is carried by the action groupoid
$\mathfrak{G} = G^* \ltimes \mathcal{C}_e(G)$ on the nuclear
configuration space that the next level will construct formally.
Chirality appears as a non-trivial orbit of the $\mathbb{Z}/2$
subgroup of~$G^*$ generated by inversion; Walden inversion at an
$\mathrm{S_N2}$ carbon is a non-trivial element of
$\coker(\varphi_{4.5})$.\\
\textbf{Forcing $\Lk_5$.}
$\mathrm{CH_3Br}$ and $\mathrm{CD_3Br}$ undergoing the same
$\mathrm{S_N}2$ reaction are isomorphic at $\Lk_{4.5}$ (same labelled
graph, same DPO rule, no stereocentre) yet react at measurably
different rates ($k_H/k_D \approx 1.3$, a secondary kinetic
isotope effect): explaining the gap requires Hessians at the
minimum and at the transition state, and therefore a mass-weighted
Riemannian metric on a configuration space --- the geometric data
of $\Lk_5$, absent at every preceding level.

\paragraph{Chapter~\ref{sec:L5} --- $\Lk_5$, the potential energy surface.}
$\Lk_5$ attaches to each molecular graph~$G$ the configuration
orbifold $\mathcal{C}_e(G)$ of nuclear geometries modulo rigid motions
and graph automorphisms, together with a smooth potential-energy
function $V \colon \mathcal{C}_e(G) \to \RR$, the mass-weighted
Riemannian metric $g$, and the $\mathrm{SO}(3)$ gauge connection
inherited from the orbifold quotient.
Transition-state theory, minimum-energy paths, and intrinsic reaction
coordinates all live here as geometric objects on~$\mathcal{C}_e(G)$.
A revised KIE stratification (Bigeleisen--Mayer in the harmonic
limit, semiclassical tunnelling on the minimum-energy path, and
vibronically nonadiabatic PCET for enzymatic systems with kinetic
isotope effects up to $\sim\!700$) places the entire textbook
catalogue of isotope effects inside this single level.\\
\textbf{Forcing $\Lk_6$.}
Photochemical pathways branch at conical intersections --- seams of
nuclear geometry where two electronic surfaces meet --- and a single
smooth $V(R)$ cannot distinguish a thermal reaction (no seam
encircled) from an ultrafast photochemical one ($\sim\!200\,\mathrm{fs}$
in retinal); the canonical molecular witness is the half-integer
pseudorotational quantisation in the $\mathrm{Na_3}$
$2^2 E^\prime$ state, an obstruction class in
$H^1(\mathcal{C}_e(G)\setminus X_{\mathrm{seam}}, \mathbb{Z}/2)$ that
no scalar correction to $V$ can produce.

\paragraph{Chapter~\ref{sec:L6} --- $\Lk_6$, electronic structure.}
$\Lk_6$ replaces the single surface $V(R)$ with a Hilbert bundle
$\mathcal{H}_{\mathrm{el}} \to \mathcal{C}_e(G)$ whose fibre at~$R$
is the electronic Hilbert space, together with the Berry connection
$A_{mn}$ on the adiabatic eigenbundles.
For real molecular Hamiltonians the primary topological invariant is
the $\mathbb{Z}_2$-valued Berry phase
$[\gamma_B] \in H^1(\mathcal{C}_e(G)\setminus X_{\mathrm{seam}},
\mathbb{Z}/2)$, which records the parity of conical-intersection
encirclements of a loop and thereby distinguishes thermal from
photochemical pathways.
Non-adiabatic couplings, avoided crossings, and geometric-phase
effects are all morphism-level structure at~$\Lk_6$.\\
\textbf{Forcing $\Lk_7$.}
The ortho/para statistical split of $\mathrm{H}_2$, and more
generally the superselection structure distinguishing identical
nuclei by spin, is absent from any electronic bundle over a classical
nuclear configuration space; nuclear indistinguishability demands a
quantum treatment of all particles.

\paragraph{Chapter~\ref{sec:L7} --- $\Lk_7$, full quantum structure.}
$\Lk_7$ is the target of a candidate \emph{object-level} functor
$F_7^{\mathrm{obj}} \colon \Lk_6(P) \to \mathbf{C}^*\mathbf{Alg}_{\mathrm{cont}}$
into continuous fields of $C^*$-algebras: states live on
$\mathcal{B}(\mathcal{H}_{\mathrm{full}})$, nuclear statistics are
enforced by a groupoid $C^*$-algebra $C^*(\mathfrak{G})$ for
$\mathfrak{G} = G^* \ltimes \mathcal{C}_e(G)$, and the
Born--Oppenheimer approximation becomes the $\varepsilon \to 0$
fibre of a continuous field $\{A_\varepsilon\}_{\varepsilon \in [0,1]}$
in the parameter $\varepsilon = (m_e/M)^{1/2}$, in the sense of
strict deformation quantisation.
This chapter is different in character from its predecessors: the
physical necessity of~$\Lk_7$ is established beyond doubt, but its
mathematical construction as a complete categorical level is a
research programme rather than a completed theory.
Four constructions (C1--C4) are stated as formal conjectures; the
chapter separates what is proved, what is formal, and what is
genuinely open, and identifies the tools from the existing literature
that wait to be assembled.
The Woolley--Primas problem of recovering molecular identity as an
effective superselection or correlation-sector label is developed
in depth (\S\ref{sec:L7}) as the tower statement of C4.
A careful warning records what large kinetic isotope effects can
and cannot serve as forcing evidence for $\Lk_7$: anomalous primary
KIEs of order $10^2$--$10^3$, although routinely cited as evidence
for ``quantum nuclei,'' are reproducible by semiclassical or
path-integral approximations built from $\Lk_5$-level geometric
data and so do not by themselves force the full $\Lk_7$ structure.

\paragraph{Chapter~\ref{sec:Para} --- the Para enrichment (perpendicular dimension).}
Running perpendicular to the exact tower is a second axis: every
level~$\Lk_k$ admits a Para enrichment
$\Lk_k^{\mathrm{Para}} := \mathrm{Lax\text{-}Alg}_{M_k}(\mathrm{Para}(\Lk_k))$
of parametric $M_k$-equivariant maps in the sense of Gavranovi\'{c}
et al.~\cite{GavRanovic2024CDL}.
The Para enrichment is \emph{not} an additional tower level --- it
runs alongside every level simultaneously --- and every
machine-learning model for chemistry is a morphism in exactly one
$\Lk_k^{\mathrm{Para}}$: yield predictors at~$\Lk_0^{\mathrm{Para}}$,
neural-ODE kinetic surrogates at~$\Lk_3^{\mathrm{Para}}$,
$E(3)$-equivariant force fields at~$\Lk_5^{\mathrm{Para}}$, neural
wavefunctions at~$\Lk_6^{\mathrm{Para}}$.
Equivariance and thermodynamic-consistency constraints are thereby
\emph{theorems} at the relevant level, not architectural choices;
a model violating them is not an object of the corresponding
$\Lk_k^{\mathrm{Para}}$.
The chapter organises around three structural questions an
architecture poses to the tower: whether its equivariance is a
theorem or a design choice, what it can represent independently of
its training data (\emph{categorical completeness} at level~$k$),
and which tower-coherence conditions it enforces.
Three architectural incompleteness results for the current
published literature follow: the Eyring transition-state-theory
coherence gap (no model couples its learned rate law to the
activation barrier of its own potential surface;
$\Lk_3 \leftrightarrow \Lk_5$), the Wegscheider consistency gap
(forward and reverse rate constants unconstrained by the model's
own $\Delta G$; $\Lk_2 \leftrightarrow \Lk_3$), and the
topological output-type gap (a scalar-energy output cannot carry
the Berry-phase invariant $[\gamma_B]$ regardless of training,
body order, or receptive field; $\Lk_5 \to \Lk_6$).
MACE serves as the primary worked example at~$\Lk_5^{\mathrm{Para}}$,
So3krates and SO3LR as a paired comparison of architectural designs
at the same level, and the QIM triply-directed encoder of Fallani
et al.~\cite{Fallani2024QIM} as the running case study at the
$\Lk_5$--$\Lk_6$ boundary.

\paragraph{Chapter~\ref{sec:sim} --- the simulation functor.}
The final chapter delivers a single operational functor
\[
  \Phi_{\mathrm{op}} \colon
    \mathbf{Para}\bigl(\Lk_0 \otimes \Lk_1 \otimes \Lk_2 \otimes \Lk_3\bigr)
    \;\longrightarrow\; \mathbf{Kl}(\mathsf{IO})
\]
into the Kleisli category of the probabilistic sub-monad of Haskell
\texttt{IO}, instantiated as a working stochastic simulator of the
Briggs--Rauscher oscillating reaction in the De~Kepper--Epstein
twelve-channel formulation~\cite{DeKepperEpstein1982}.
Each modelling choice in the source is a morphism in a named tower
category: species populations are elements of the free commutative
monoid at~$\Lk_0$; atom and charge balance are $\Lk_0$ structural
checks on the stoichiometric matrix; Wegscheider's relations on the
paired forward/reverse channels (R3/R3-rev and R4/R4-rev) enforce
$\Lk_1$--$\Lk_2$ thermochemistry; and the scalar rate constants together
with the rate-law saturation forms feed a Markov kernel $F_P$ at~$\Lk_3$,
realised twice from the same network value --- by the Gibson--Bruck
next-reaction method (SSA) and, in the large-volume limit, by the
implicit-midpoint mass-action ODE.
The ionic Process~A versus radical Process~B distinction --- the
canonical $\Lk_3 \to \Lk_4$ forcing pair of the Briggs--Rauscher
system, sharing the same overall stoichiometry and the same bulk
HOI-production rate form --- lives at $\Lk_4$ and is provably
invisible to the simulator by construction: the chemical motivation
for locating the simulation precisely at $\Lk_0$--$\Lk_3$ and no
higher.
Three lax cells of $\Phi_{\mathrm{op}}$ --- $\alpha$ (parameter unit),
$\phi_{\mathrm{SSA}}$ (IO non-commutativity, vanishing in total
variation), and $\phi_{\mathrm{ODE}}$ (Kurtz-limit deviation,
non-trivial in the low-copy quiescent phase) --- are identifiable
line-by-line in the source.
A denotational companion $\Phi_{\mathrm{den}} := F \circ \Phi_{\mathrm{op}}$
marginalises over the pseudorandom seed and lands in the Markov
category $\mathbf{BorelStoch}$, restoring the Markov structure that
$\mathbf{Kl}(\mathsf{IO})$ does not itself support.
To our knowledge this provides both the first published Kleisli
semantics of the Gillespie next-reaction method and the first
application of the Para construction outside machine learning.
A separate $\Lk_4 \to \Lk_3$ parameter projection $\pi_4$ embeds
bond-level mechanism data into scalar rate constants plus rate-law
saturation forms, making the chapter's several network variants
(canonical De~Kepper--Epstein, buffered-$\mathrm{H}^+$, Furrow-like
pooled) different $\Lk_3$ reductions of the same underlying
mechanism, comparable as Markov kernels in $\mathbf{BorelStoch}$.
The simulation results verify three structural predictions of the
framework: the relaxation-oscillator dynamics emerge from both the
SSA samples and the ODE Kurtz limit; the two simulators agree in
the high-copy spike phase and diverge in the low-copy quiescent
phase, making the $\phi$-cell distinction operationally visible;
and the form of the divergence is interpretable through the Kurtz
theorem rather than as numerical accident.
The framework's deliverable is not a validation of the tower but
the discipline of distinguishing, in a working simulator, which
features are essential, which are tunable parameters of~$\Theta_{\mathrm{sim}}$,
and which are scale-dependent fictions of the chosen
representation.

\bigskip

\noindent
Four coloured environments serve mathematical and chemical audiences
simultaneously throughout: green for chemical content, blue for
categorical statements, orange for forcing pairs, and purple for the
conceptual connections between them.

\subsection{A brief guide to categorical language}
\label{sec:cat-language}

This subsection introduces the categorical vocabulary;
readers comfortable with enriched category theory may proceed directly
to Chapter~\ref{sec:L0}.

Category theory provides the mathematical grammar of this programme.
The compact dictionary below is aimed at readers fluent in chemistry
or physics but new to categorical ideas; every term is paired with
its chemical meaning and a concrete example.

\paragraph{Category.}
A \emph{category} $\mathcal{C}$ consists of:
\begin{itemize}
  \item a collection of \emph{objects} --- the states or species pools
        under study;
  \item for each ordered pair of objects $X, Y$, a set $\Hom(X, Y)$ of
        \emph{morphisms} from $X$ to $Y$, i.e.\ all processes or
        transformations that lead from state $X$ to state $Y$;
  \item a \emph{composition law}: if $f\in\Hom(X,Y)$ and
        $g\in\Hom(Y,Z)$, their composite $g\circ f\in\Hom(X,Z)$ is the
        process obtained by performing $f$ first and then $g$;
  \item for each object $X$, an \emph{identity morphism}
        $\id_X\in\Hom(X,X)$ --- the trivial ``do-nothing'' process;
\end{itemize}
satisfying \emph{associativity} $(h\circ g)\circ f = h\circ(g\circ f)$
and \emph{unitality} $\id_Y\circ f = f = f\circ\id_X$~\cite{riehl,MacLane1998}.

\smallskip\noindent
\textit{Chemical meaning.}
Objects are \emph{complexes} --- formal non-negative integer combinations
of chemical species such as $\mathrm{CH_4}+2\,\mathrm{O_2}$.
The term \emph{complex} is standard in chemical reaction network
theory~\cite{Feinberg2019}: it denotes precisely one side of a balanced
reaction equation.
Morphisms are \emph{reactions} or multi-step pathways between complexes.
The identity $\id_\mathbf{u}$ is the ``no reaction'' process on complex
$\mathbf{u}$; composition $r_2\circ r_1$ chains two steps sequentially.

\smallskip\noindent
\textit{A key example: $B\RR$.}
The real line forms a one-object category $B\RR$ with a single object
$*$ and one morphism $a\in\RR$ for each real number, with composition
$b\circ a := a+b$.
Think of $B\RR$ as a single node with one directed arrow per real number:
concatenating two arrows means adding their labels.
A functor $F_H\colon\Lk_0(P)\to B\RR$ assigns to every reaction a real
number (e.g.\ enthalpy $\Delta H$) that is automatically additive along
sequential and parallel steps --- this is the categorical formulation of
Hess's Law, developed fully at $\Lk_1$.

\paragraph{Functor.}
A \emph{functor} $F\colon\mathcal{C}\to\mathcal{D}$ is a
structure-preserving map between categories: it sends each object $X$
to an object $F(X)$ and each morphism $f\colon X\to Y$ to a morphism
$F(f)\colon F(X)\to F(Y)$, respecting composition
($F(g\circ f) = F(g)\circ F(f)$) and identities
($F(\id_X) = \id_{F(X)}$).

\smallskip\noindent
\textit{Chemical meaning.}
A functor from the reaction category into $B\RR$ is a \emph{consistent
numerical scoring} of reactions: it assigns a real number to each
elementary step in a way that the score of any sequential or parallel
combination is the sum of the individual scores.
Enthalpy ($\Delta H$), Gibbs free energy ($\Delta G$), and entropy
change ($\Delta S$) are all consistent scorings of this kind --- each
is a functor from the appropriate tower level into $B\RR$.
The functoriality conditions ($F(g\circ f) = F(g)\circ F(f)$ and
$F(\id) = 0$) are precisely the additivity and trivial-process conditions
these thermodynamic quantities must satisfy.

\paragraph{Monoidal category.}
A \emph{monoidal category} $(\mathcal{C},\otimes,I)$ is a category
equipped with a \emph{tensor product}
$\otimes\colon\mathcal{C}\times\mathcal{C}\to\mathcal{C}$ combining two
objects or morphisms ``in parallel'', a \emph{unit object} $I$
satisfying $I\otimes X\cong X\cong X\otimes I$, and coherent
isomorphisms $(X\otimes Y)\otimes Z\xrightarrow{\sim}X\otimes(Y\otimes
Z)$ (associator) and $I\otimes X\xrightarrow{\sim}X$,
$X\otimes I\xrightarrow{\sim}X$ (unitors).
When the associator and unitors are identity maps, the category is
\emph{strict monoidal}~\cite{MacLane1998}.

\smallskip\noindent
\textit{Chemical meaning.}
$X\otimes Y$ is the \emph{mixture} of $X$ and $Y$:
$\mathrm{CH_4}\otimes 2\,\mathrm{O_2}$ is the mixture
$\mathrm{CH_4}+2\,\mathrm{O_2}$.
The unit object is the empty mixture $\mathbf{0}\in\NN[\Sp]$.
Since addition in $\NN[\Sp]$ is already strictly associative and
$\mathbf{0}+\mathbf{u}=\mathbf{u}$, the associator and unitors are
identity maps: the tensor is \emph{strict} monoidal for chemical
mixtures.

\paragraph{Symmetric monoidal category.}
A monoidal category is \emph{symmetric} if there is a coherent family
of isomorphisms $\sigma_{X,Y}\colon X\otimes Y\xrightarrow{\;\sim\;}
Y\otimes X$.
In chemistry: a solution ``of $A$ and $B$'' is the same system as one
``of $B$ and $A$''; no experiment distinguishes the two descriptions,
so the symmetry isomorphism is physically mandatory.
All symmetric monoidal categories in this chapter are
\emph{permutative}: symmetric \emph{strict} monoidal.
In this manuscript an acronym ``SMC'' is used for symmetric monoidal category and ``permutative'' for symmetric strict monoidal.
Theorem~\ref{thm:UP-L0} holds for any strict SMC codomain; the primary
application uses $B\RR$, which is permutative.

\paragraph{Free construction.}
A \emph{free} object on given generators is the most general object of
its type built from those generators, subject to \emph{no} relations
beyond those forced by the axioms of the type.

\smallskip\noindent
\textit{Example.}
The free group on one generator is $\ZZ$ --- the integers with addition.
Starting from the generator $+1$ and its inverse $-1$, all combinations
by addition are formed; no shortcut is imposed (``$+1$ applied three
times'' is never identified with ``$+2$ applied twice'' unless the group
axioms force it, which they do not).
The result is the smallest group containing the generator, with no
redundant identifications~\cite{MacLane1998}.
The key property is \emph{universality}: once $+1$ is sent to any
element $g$ of a target group $G$, the entire homomorphism $\ZZ\to G$
is uniquely forced ($n\mapsto ng$).

\smallskip\noindent
\textit{Chemical meaning.}
\emph{No equivalences} between distinct reaction labels are imposed beyond
those forced by the permutative axioms.
Two routes $r_1, r_2$ with the same reactant and product complexes are
kept as distinct morphisms at $\Lk_0$: no physical identification ---
same $\Delta H$, same rate, same mechanism --- is made until the tower
level at which that information first becomes available.
Once you fix where each species and each reaction label map in a target
category, the entire strict symmetric monoidal functor out of $\Lk_0(P)$
is uniquely forced (Theorem~\ref{thm:UP-L0}): there is no modelling
choice left.

\paragraph{Commutative diagram.}
A diagram of objects and morphisms \emph{commutes} when every directed
path between the same two objects yields the same composite morphism.
The following example shows what this means in stoichiometric terms.

\begin{chembox}[Commutative diagrams in stoichiometry: nitrogen oxidation]
Consider two routes from $\mathrm{N_2}+2\,\mathrm{O_2}$ to
$2\,\mathrm{NO_2}$, with species ordered
$(\mathrm{N_2},\mathrm{O_2},\mathrm{NO},\mathrm{NO_2})$:
\[
\includegraphics{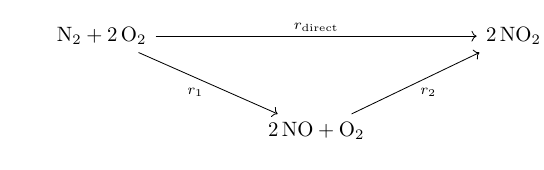}
\]
where $r_1\colon \mathrm{N_2}+\mathrm{O_2}\to 2\,\mathrm{NO}$
and $r_2\colon 2\,\mathrm{NO}+\mathrm{O_2}\to 2\,\mathrm{NO_2}$.
The functor $F_{\Delta\nu}\colon\Lk_0(P)\to(\ZZ^4,+)$ assigns each
reaction its net stoichiometric change vector:
\[
  \Delta\nu(r_1) = \begin{pmatrix}-1\\-1\\+2\\0\end{pmatrix},
  \qquad
  \Delta\nu(r_2) = \begin{pmatrix}0\\-1\\-2\\+2\end{pmatrix},
  \qquad
  \Delta\nu(r_{\mathrm{direct}}) = \begin{pmatrix}-1\\-2\\0\\+2\end{pmatrix}.
\]
The diagram commutes in $(\ZZ^4,+)$ if and only if
$\Delta\nu(r_{\mathrm{direct}}) = \Delta\nu(r_1)+\Delta\nu(r_2)$,
which holds: both sides equal $(-1,-2,0,+2)^\top$.
This is \emph{atom balance} --- the net species change is the same
whether the reaction proceeds directly or via the nitric oxide
intermediate.
Functoriality $F_{\Delta\nu}(r_2\circ r_1)=F_{\Delta\nu}(r_2)+
F_{\Delta\nu}(r_1)$ \emph{is} atom balance: it encodes a column of
the stoichiometric matrix $N$ as commutativity of a triangle.
Any violation would mean atoms are created or destroyed along one route,
i.e.\ $F_{\Delta\nu}$ is not a functor.
\end{chembox}

\paragraph{Petri net.}
\emph{Petri nets} were introduced by Carl Adam Petri in 1962 as a
mathematical framework for concurrent processes~\cite{Petri1962}.
Places hold tokens (molecular populations of species) and transitions
represent reactions, consuming tokens from input places and depositing
tokens into output places according to stoichiometric arc multiplicities.
This is exactly the combinatorial skeleton of stoichiometric
chemistry~\cite{MeseguerMontanari1990,Kock2022}.
The precise definition, a worked chemical example, and the construction
of $\Lk_0(P)$ from a Petri net are given in \S\ref{sec:petri}.

\begin{chembox}[The central gain of the categorical framework]
A category is a precise language for processes that compose sequentially
and run in parallel.
A functor is a consistent numerical scoring of those processes.
Once a thermodynamic or kinetic quantity is identified as a functor from
the reaction category into an appropriate target, its core properties
--- additivity along sequences, additivity over parallel reactions ---
become \emph{theorems} rather than separate postulates.

Each higher tower level adds a specific categorical extension to
the previous level: a new functor, an enrichment of the morphism
category by a symmetry, a geometric decoration, a topological
invariant, or (at the final step) a deformation of the algebraic
structure itself.
The taxonomy of the six extension types that arise across the
present tower is recorded in \S\ref{sec:L7-retrospective};
each is forced at its level by a concrete pair of reactions the
previous level cannot distinguish.
Freeness means no equivalences between reactions are imposed beyond
those forced by the algebra of mixtures; every reaction is kept
distinct unless explicitly identified for a physical reason.

\smallskip\noindent
\emph{Reading the tower as an ML foundation.}
Each level $\Lk_k$ admits a perpendicular Para enrichment
$\Lk_k^\mathrm{Para}$ of parametric equivariant maps
(Chapter~\ref{sec:Para}).
The exact level $\Lk_k$ is recovered as the zero-parameter limit of
$\Lk_k^\mathrm{Para}$; architectures for machine-learning models of
chemistry --- yield predictors, equivariant force fields, learned
wavefunctions --- are lax algebra morphisms in $\Lk_k^\mathrm{Para}$,
with equivariance and thermodynamic-consistency constraints
following from universal properties rather than from architectural
choices.

\smallskip\noindent
In summary: a stoichiometric scheme equipped with sequential
composition (one reaction following another), parallel composition
(reactions in independent mixtures), and order-independence of
mixtures is the defining data of a \emph{symmetric monoidal
category}; assigning numerical chemical content is providing a
\emph{functor}; starting from the most general such structure is
taking the \emph{free} construction.
\end{chembox}

\paragraph{Monad.}
A \emph{monad} on a category $\mathcal{C}$ is an endofunctor
$T\colon\mathcal{C}\to\mathcal{C}$ (a functor whose source and
target are the same category) equipped with two natural
transformations (coherent families of morphisms, one per object):
a \emph{unit} $\eta\colon\id_\mathcal{C}\Rightarrow T$ that injects
each object $X$ into a $T$-decorated version $T(X)$, and a
\emph{multiplication} $\mu\colon T\circ T\Rightarrow T$ that
collapses two layers of decoration into one.
These satisfy associativity ($\mu\circ T\mu=\mu\circ\mu T$) and unit
laws ($\mu\circ T\eta=\mu\circ\eta T=\id_T$)~\cite{MacLane1998,riehl}.

\smallskip\noindent
\textit{Operational meaning.}
A monad packages a kind of ``effect'' or ``extra structure'' that
maps can carry while remaining composable --- randomness, state,
partial information, time, or symmetry.
$T(X)$ is to be read as ``a computation that ultimately yields a
value of type $X$ while carrying the effect of $T$''; $\eta$
produces a trivially-decorated value, and $\mu$ is what makes two
effectful computations compose into a single one.

\smallskip\noindent
\textit{The Kleisli category.}
The natural setting for $T$-effectful computations is the
\emph{Kleisli category} $\mathbf{Kl}(T)$, which has the same
objects as $\mathcal{C}$ but takes morphisms $X\to Y$ to be
$\mathcal{C}$-morphisms $X\to T(Y)$.
The composite of $f\colon X\to T(Y)$ and $g\colon Y\to T(Z)$ is
\[
  g\bullet f \;:=\; \mu_Z\circ T(g)\circ f \;\colon\; X\to T(Z),
\]
lifting $g$ to act on $T(Y)$ via $T$'s functoriality, then
flattening the resulting $T(T(Z))$ via $\mu$.

\smallskip\noindent
\textit{Key example: $\mathbf{Kl}(\mathsf{IO})$.}
Haskell's $\mathsf{IO}$ monad packages every interaction with the
outside world --- file reads, terminal output, random-number
generation --- into the type $\mathsf{IO}(X)$ of computations that
eventually produce an $X$, possibly with side effects on the way.
A morphism in $\mathbf{Kl}(\mathsf{IO})$ from $X$ to $Y$ is then a
function $X\to\mathsf{IO}(Y)$: a procedure that, given an input,
may consume random bits or perform other side effects and yields an
output.
Stochastic simulation of a reaction network is exactly such a
procedure: the operational functor $\Phi_{\mathrm{op}}$ of
Chapter~\ref{sec:sim} lands in the Kleisli category of the
\emph{probabilistic sub-monad} of $\mathsf{IO}$ --- the sub-monad
generated by pseudo-random-number sampling.

%% file: chapters/ch_L0.tex
\section{\texorpdfstring{$\Lk_0$}{Lk0}: The Stoichiometric Level}
\label{sec:L0}

The simplest possible question about a chemical reaction is:
\emph{what goes in and what comes out?}
Before energy, before time, before geometry, before mechanism ---
purely the bookkeeping of species consumed and produced.

Three structural features are present at this barest level of description:

\begin{enumerate}[label=(\alph*)]
  \item \textbf{Steps compose.}
    If reaction $r_1$ produces a species that reaction $r_2$ consumes,
    the two steps together form a new composite reaction $r_2 \circ r_1$.
  \item \textbf{Steps run in parallel.}
    Two reactions that do not interfere with each other can be combined
    into a joint process $r_1 \otimes r_2$, run simultaneously.
  \item \textbf{The ordering of independent species is irrelevant.}
    There is no intrinsic left-to-right order in a mixture;
    $A + B$ and $B + A$ describe the same physical system.
\end{enumerate}

Expressing these structural features in a precise mathematical language
--- rather than leaving them as implicit chemical common sense ---
makes it possible to prove rigorously which properties follow from
which hypotheses and to compare different levels of chemical description
on equal footing.
The language appropriate to features~(a)--(c) is that of
\emph{symmetric monoidal categories}: mathematical structures carrying
exactly the sequential, parallel, and order-irrelevant composition
grammar those features demand.
This language was established by Meseguer and
Montanari~\cite{MeseguerMontanari1990} for Petri nets, and extended
into a compositional framework for open reaction networks by Baez and
Pollard~\cite{BaezPollard2017}.
The exact tower backbone developed in this chapter and the
subsequent ones admits a perpendicular Para-enrichment dimension
(Chapter~\ref{sec:Para}, building on
Gavranovi\'{c} et al.~\cite{GavRanovic2024CDL}) which places
machine-learning models for chemistry~\cite{Akitsu2023,FongSpivak2019}
on an exact mathematical foundation as parametric equivariant maps
over the same tower levels.

A brief categorical glossary, aimed at readers fluent in chemistry
or physics but new to category theory, is provided in \S\ref{sec:cat-language}.

\input{chapters/L0/species}

\input{chapters/L0/petri}

\input{chapters/L0/universal}

\input{chapters/L0/new}

\input{chapters/L0/stoich}

\input{chapters/L0/application}

\input{chapters/L0/forcing}

%% file: chapters/L0/species.tex
\subsection{Species, complexes, and the stoichiometric monoid}
\label{sec:stoichiometric-monoid}

\emph{Stoichiometry} is the branch of chemistry concerned with the
quantitative relationships between species in a chemical reaction:
how many molecules of each substance are consumed, and how many are
produced.
For a mathematician, it is the study of the additive structure of
molecular populations --- which integer combinations of named species
can appear as the left- or right-hand side of a balanced equation,
and how those combinations pool together.
No energy, no dynamics, no geometry: purely the arithmetic of
molecular counts.

The algebraic object capturing this arithmetic is the \emph{free
commutative monoid} on the set of species.

\begin{definition}[Species and the stoichiometric monoid]
\label{def:FCM}
  Fix a finite set $\Sp$ of \emph{chemical species}
  (e.g.\ $\Sp = \{\mathrm{H_2O,\,CH_4,\,O_2,\,CO_2}\}$).
  Species are abstract labels at this level; no geometric, electronic,
  or energetic information is attached.

  The \emph{stoichiometric monoid} is the free commutative monoid on $\Sp$:
  \begin{equation}
    \label{eq:freefreefree}
    \NN[\Sp]
    \;:=\;
    \Bigl\{\,
      \textstyle\sum_{S \in \Sp} n_S \cdot S
      \;\Big|\;
      n_S \in \NN,\;
      \text{all but finitely many } n_S = 0
    \,\Bigr\},
  \end{equation}
  with binary operation
  $\bigl(\sum_S n_S \cdot S\bigr) + \bigl(\sum_S m_S \cdot S\bigr)
   := \sum_S (n_S + m_S) \cdot S$
  and identity element $\mathbf{0}$ (the empty complex, ``vacuum'').
  Elements of $\NN[\Sp]$ are called \emph{complexes}.
\end{definition}

\begin{remark}[Relation to prior mathematical treatments of stoichiometry]
\label{rem:prior-stoich}
The stoichiometric matrix $N \in \ZZ^{|\Sp|\times|\Rx|}$, which encodes
the net change in species counts per reaction, has been the standard
mathematical representation of stoichiometry since the foundational work
of Feinberg~\cite{Feinberg1987,Feinberg2019} and
Horn--Jackson~\cite{HornJackson1972}.
Meseguer--Montanari~\cite{MeseguerMontanari1990} recognise that Petri
net semantics naturally produces free commutative monoids, and
Baez--Master~\cite{BaezMaster2020} exploit this structure for open Petri
nets.
The present treatment differs in making the \emph{universal property} of
$\NN[\Sp]$ the primary organising principle of the stoichiometric level,
rather than a computational convenience: every additive species
observable --- mass, atom count, enthalpy, any additive potential ---
is exactly a function out of $\NN[\Sp]$ induced by the universal
property, and the entire tower of functors in Chapters~3--10 is built
by iterating this observation upward.
The stoichiometric matrix $N$ is recovered as the linear-algebraic
shadow of this structure in \S\ref{sec:N-matrix}.
\end{remark}

The claim that $\NN[\Sp]$ as defined above \emph{is} the free
commutative monoid on $\Sp$ --- not merely a commutative monoid
containing $\Sp$ --- is a theorem whose proof is immediate but worth
making explicit, since the universal property it establishes is the
algebraic engine of everything that follows.

\begin{proposition}[{$\NN[\Sp]$ is the free commutative monoid on $\Sp$}]
\label{prop:FCM-UP}
For any commutative monoid $(M, +, 0)$ and any function $f: \Sp \to M$,
there exists a \emph{unique} monoid homomorphism
$\bar{f}: \NN[\Sp] \to M$ satisfying $\bar{f}(S) = f(S)$ for all
$S \in \Sp$, given concretely by
\[
  \bar{f}\!\Bigl(\sum_{S\in\Sp} n_S \cdot S\Bigr)
  \;=\;
  \sum_{S\in\Sp} n_S \cdot f(S),
\]
where $n\cdot m$ denotes the $n$-fold sum $m+\cdots+m$ in $M$.
\end{proposition}

\begin{proof}
\textit{Existence.}
Define $\bar{f}$ by the displayed formula.
Since only finitely many $n_S$ are non-zero, the sum is finite and
$\bar{f}$ is well-defined.
It is a monoid homomorphism:
\begin{align*}
  \bar{f}(\mathbf{u}+\mathbf{v})
  &= \bar{f}\!\Bigl(\sum_S(n_S+m_S)\cdot S\Bigr)
   = \sum_S(n_S+m_S)\cdot f(S) \\
  &= \sum_S n_S\cdot f(S) + \sum_S m_S\cdot f(S)
   = \bar{f}(\mathbf{u})+\bar{f}(\mathbf{v}),
\end{align*}
and $\bar{f}(\mathbf{0}) = \sum_S 0\cdot f(S) = 0_M$.

\textit{Uniqueness.}
Any monoid homomorphism $g:\NN[\Sp]\to M$ extending $f$ must satisfy,
by the homomorphism property applied $n_S$ times,
$g(n_S\cdot S) = n_S\cdot g(S) = n_S\cdot f(S)$.
Additivity then forces
$g\bigl(\sum_S n_S\cdot S\bigr)
 = \sum_S n_S\cdot f(S) = \bar{f}\bigl(\sum_S n_S\cdot S\bigr)$,
so $g = \bar{f}$.
\end{proof}

\begin{chembox}[What is a complex?]
A complex is what chemists call one ``side'' of a balanced equation.
For $\mathrm{CH_4} + 2\,\mathrm{O_2} \to \mathrm{CO_2} + 2\,\mathrm{H_2O}$,
the reactant complex is $1\cdot\mathrm{CH_4} + 2\cdot\mathrm{O_2}$
and the product complex is $1\cdot\mathrm{CO_2} + 2\cdot\mathrm{H_2O}$,
both elements of $\NN[\Sp]$.

The monoid operation is simply \emph{pooling}:
\[
(1\cdot\mathrm{CH_4} + 2\cdot\mathrm{O_2})
 + (1\cdot\mathrm{CO_2} + 2\cdot\mathrm{H_2O})
 = 1\cdot\mathrm{CH_4} + 2\cdot\mathrm{O_2}
   + 1\cdot\mathrm{CO_2} + 2\cdot\mathrm{H_2O}.
\]
This is the mathematical version of mixing two portions together.
The reaction arrow says: this pool transforms into that pool.
The monoid carries no information about whether or how the transformation
occurs; that belongs to later levels of the tower.

Proposition~\ref{prop:FCM-UP} makes precise something a chemist already
knows: once you decide what numerical value to assign to each individual
species (its mass, its atom count, its standard enthalpy of
formation,~\ldots), the value for any complex is completely forced ---
there is no choice.
The total mass of $1\cdot\mathrm{CH_4}+2\cdot\mathrm{O_2}$ is
$M(\mathrm{CH_4})+2\,M(\mathrm{O_2})$ and nothing else; the universal
property is precisely the theorem that says this.
\end{chembox}

Universal properties are one of the central organising tools of category
theory: MacLane~\cite{MacLane1998} identifies them as the concept that
unifies free constructions, adjunctions, and limits, and
Riehl~\cite{riehl} shows systematically how they eliminate arbitrary
choices from mathematical definitions.
The stoichiometric monoid $\NN[\Sp]$ is the \emph{universal} receptacle
for additive species data: any quantity that is additive over species ---
mass, elemental composition, standard enthalpy of formation --- extends
uniquely to all complexes via Proposition~\ref{prop:FCM-UP}, with no
modelling freedom left.
Moreover, the universal property of $\NN[\Sp]$ is the first in a chain:
the universal property of $\Lk_0(P)$ (Theorem~\ref{thm:UP-L0}) extends
it one step further, from species and complexes to reactions and their
composites, and every subsequent tower level adds one more universal
property governing the next kind of chemical data.

\begin{mathbox}[{Universal property of $\NN[\Sp]$}]
\label{box:UP-NSSp}
Let $(M, +, 0)$ be any commutative monoid --- for instance
$(\RR, +, 0)$, $(\NN^{|\mathsf{E}|}, +, \mathbf{0})$, or
$(\RR_{>0}, \times, 1)$ --- and let $f: \Sp \to M$ be any function
assigning an element of $M$ to each species.
Proposition~\ref{prop:FCM-UP} guarantees two things:
\begin{enumerate}[label=(\roman*)]
  \item \textbf{Existence.} There is a monoid homomorphism
    $\bar{f}: \NN[\Sp] \to M$ extending $f$, given by
    $\bar{f}\!\bigl(\sum_S n_S \cdot S\bigr) = \sum_S n_S \cdot f(S)$.
  \item \textbf{Uniqueness.} $\bar{f}$ is the \emph{only} monoid
    homomorphism $\NN[\Sp] \to M$ that agrees with $f$ on individual
    species.
\end{enumerate}
Concretely: once you decide what value each species receives, the value
of every complex is completely determined --- there is no freedom left.
This universal property is invoked at every level of the tower:
atom-balance maps ($f(S) = $ elemental composition of $S$) extend to
all complexes and reaction stoichiometries (\S\ref{sec:N-matrix});
enthalpy assignments ($f(S) = \Delta H^\circ_f(S)$) extend to Hess
networks ($\Lk_1$, \S3); and rate assignments extend to kinetic
networks ($\Lk_3$, \S5).
\end{mathbox}

\begin{remark}[Rate constants are not species data]
\label{rem:rate-constants}
Rate constants are associated with \emph{reactions}, not with individual
species, and so lie outside the scope of Proposition~\ref{prop:FCM-UP}.
The universal property of $\NN[\Sp]$ applies to species-level
assignments only.
Rate-constant data enters at $\Lk_3$ as part of the kinetic functor
$F_P$, whose domain is the category of reactions $\Lk_0(P)$, not the
monoid of complexes $\NN[\Sp]$; its universal property is
Theorem~\ref{thm:UP-L0}.
\end{remark}

%% file: chapters/L0/petri.tex
\subsection{Petri nets: the categorical input data}
\label{sec:petri}

This subsection gives the precise definition of a Petri net and
establishes the category $\mathbf{Petri}$ of all Petri nets.
The raw data of a stoichiometric scheme is a finite list of species and a
finite list of reaction labels, each label carrying a source complex (reactants) and
a target complex (products) in $\NN^{|\Sp|}$.
This structure is a \emph{Petri net}, introduced by Carl Adam Petri in
1962 as a mathematical framework for concurrent
processes~\cite{Petri1962}, and recognised by Meseguer--Montanari to be
the natural input data for free symmetric monoidal
categories~\cite{MeseguerMontanari1990}.
The stoichiometric category $\Lk_0(P)$ is the free categorical structure
generated by a Petri net $P$: one imposes no relations beyond those
forced by the axioms of a symmetric monoidal category.
The free construction occupies \S\ref{sec:L0-construction}.

\begin{definition}[Petri net~{\cite{Petri1962,MeseguerMontanari1990,Kock2022}}]
\label{def:petri}
A \emph{Petri net} is a quadruple $P = (\Sp, \Rx, s, t)$ where
$\Sp$ is a finite set of \emph{species},
$\Rx$ is a finite set of \emph{reaction labels},
and $s, t: \Rx \to \NN^{|\Sp|}$ assign to each label $r \in \Rx$ its
\emph{source complex} $s(r)$ and \emph{target complex} $t(r)$, with $s(r) \neq t(r)$ in general.
\end{definition}

\begin{chembox}[A Petri net in chemistry: hydrogen combustion]
\label{box:petri-example}
Consider the net stoichiometry of hydrogen combustion:
\[
  2\,\mathrm{H_2} \;+\; \mathrm{O_2}
  \;\xrightarrow{\;r\;}\;
  2\,\mathrm{H_2O}.
\]
As a Petri net: $\Sp = \{\mathrm{H_2},\,\mathrm{O_2},\,\mathrm{H_2O}\}$,
$\Rx = \{r\}$, with
\[
  s(r) = 2\cdot\mathrm{H_2} + 1\cdot\mathrm{O_2} \;\in\; \NN^{|\Sp|},
  \qquad
  t(r) = 2\cdot\mathrm{H_2O} \;\in\; \NN^{|\Sp|}.
\]
In the standard Petri net picture, each species is a \emph{place}
(a pool holding molecular tokens) and each reaction is a
\emph{transition} (consuming tokens from its input places and depositing
tokens into its output places according to the stoichiometric
coefficients).
The arc multiplicities $(2, 1, 2)$ encode stoichiometry; the transition
$r$ has no internal structure beyond its input and output counts.

\smallskip
\noindent\textbf{What the Petri net captures.}
The stoichiometric ratios $(2{:}1{:}2)$, the direction of transformation,
and the connectivity between species and reaction.
Conservation vectors can be computed from the stoichiometric matrix
$N$ alone (see \S\ref{sec:N-matrix}): here, hydrogen-atom balance gives
$w_\mathrm{H} = (2, 0, 2)^\top$, confirming two H atoms in each
$\mathrm{H_2}$ and two H atoms in each $\mathrm{H_2O}$.

\smallskip
\noindent\textbf{What the Petri net deliberately omits.}
\begin{itemize}
  \item \emph{Energy}: the standard enthalpy of combustion
    $\Delta H^\circ = -484\;\mathrm{kJ\,mol^{-1}}$~\cite{Chase1998}
    is not part of $P$; it enters at $\Lk_1$.
  \item \emph{Rate}: the pre-exponential factor and activation energy are
    absent; they enter at $\Lk_3$.
  \item \emph{Mechanism}: whether the reaction proceeds via radical
    intermediates ($\mathrm{H}$, $\mathrm{OH}$, $\mathrm{HO_2}$) or in
    a single elementary step is invisible at $\Lk_0$; mechanism enters
    at $\Lk_4$.
  \item \emph{Geometry}: bond lengths, angles, and three-dimensional
    structure are absent; they enter at $\Lk_5$.
\end{itemize}
The Petri net is the \emph{skeletal} description: the minimal
combinatorial input from which the categorical construction $\Lk_0(P)$
is built. Every additional piece of chemical information corresponds to
additional structure at a higher tower level.
\end{chembox}

\begin{remark}[Reaction labels are named generators]
\label{rem:labels}
$\Rx$ is a set of named generators, not a set of ordered pairs of
complexes.
Two distinct labels $r_1, r_2 \in \Rx$ with $s(r_1) = s(r_2)$ and
$t(r_1) = t(r_2)$ represent two different reactions between the same
complexes --- for instance, two mechanistically distinct routes with the
same net stoichiometry.
The free construction of \S\ref{sec:L0-construction} keeps them as
distinct generators; no physical identification is imposed until a later
level.
\end{remark}

Petri nets assemble into a category: one can map one network into
another by translating species and reaction labels while respecting
stoichiometry, and these translations compose in the obvious way.
This categorical structure on Petri nets is established in
Meseguer--Montanari~\cite{MeseguerMontanari1990} and is needed for two
reasons: it allows chemical networks to be compared and translated
systematically, and it ensures that the construction
$P \mapsto \Lk_0(P)$ is itself a functor between categories rather than
an ad hoc assignment.

\begin{definition}[The category $\mathbf{Petri}$]
\label{def:Petri-cat}
$\mathbf{Petri}$ is the category whose objects are Petri nets and whose
morphisms $f: P \to P'$ are pairs $(f_\Sp, f_\Rx)$ of functions
\[
  f_\Sp: \Sp \to \Sp', \qquad f_\Rx: \Rx \to \Rx',
\]
satisfying the \emph{source--target compatibility conditions}
\[
  s' \circ f_\Rx = \bar{f}_\Sp \circ s,
  \qquad
  t' \circ f_\Rx = \bar{f}_\Sp \circ t,
\]
where $\bar{f}_\Sp: \NN^{|\Sp|} \to \NN^{|\Sp'|}$ is the unique monoid
homomorphism extending $f_\Sp$ (Proposition~\ref{prop:FCM-UP}).
Composition is componentwise; the identity on $P$ is $(\id_\Sp, \id_\Rx)$.
\end{definition}

\begin{remark}[What a Petri morphism does]
\label{rem:Petri-morphism}
A morphism $f: P \to P'$ translates species to species and reaction
labels to reaction labels, in a way that respects stoichiometry: if
$r$ has source complex $\mathbf{u}$ in $P$, then $f_\Rx(r)$ has source
complex $\bar{f}_\Sp(\mathbf{u})$ in $P'$, and similarly for targets.
Non-injectivity is permitted: a non-injective $f_\Sp$ coarsens species
(merging two species into one); a non-injective $f_\Rx$ equates two
distinct reaction labels at the stoichiometric level.
The compatibility conditions are exactly what is needed for
$\bar{f}_\Sp$ --- the unique extension supplied by
Proposition~\ref{prop:FCM-UP} --- to be well-defined on complexes built
from the translated species.
\end{remark}

\subsection{The stoichiometric category \texorpdfstring{$\Lk_0(P)$}{Lk0(P)}}
\label{sec:L0-construction}

The categorical language of \S\ref{sec:cat-language} now assembles into
a single construction.
A \emph{permutative category} is a strict symmetric monoidal category:
associativity and unit hold as strict equalities, while the symmetry
$\sigma_{X,Y}: X \otimes Y \to Y \otimes X$ remains a (possibly
non-trivial) isomorphism between two distinct objects.
The category $\Lk_0(P)$ we construct is a \emph{skeletal} permutative
category: in addition to the strict associativity and unit, the
underlying object monoid $(\NN^{|\Sp|}, +, \mathbf{0})$ is itself
strictly commutative, so $\mathbf{u}+\mathbf{v} = \mathbf{v}+\mathbf{u}$
holds as an equality of objects, not merely as an isomorphism.
This skeletal structure is forced on us by chemistry: a mixture has
no intrinsic ordering, so $A+B$ and $B+A$ must literally name the
same complex.
The non-trivial symmetry $\sigma_{\mathbf{u},\mathbf{v}}:
\mathbf{u}+\mathbf{v} \to \mathbf{u}+\mathbf{v}$ then becomes an
endomorphism encoding the permutation data that a non-skeletal
permutative category would store in distinct objects.

A Petri net $P$ provides the chemical generators of $\Lk_0(P)$: the
species $\Sp$ generate the objects (via $\NN^{|\Sp|}$), and the
reaction labels $\Rx$ become the basic morphisms, each typed by its
source and target complex.
The \emph{free} skeletal permutative category on $P$ is built from
these generators by closing under sequential composition, parallel
tensor, and the structural morphisms (identities and symmetries),
imposing \emph{only} the permutative axioms and no additional chemical
identifications.
This is $\Lk_0(P)$.

\begin{definition}[Stoichiometric category $\Lk_0(P)$]
\label{def:L0}
Let $P = (\Sp, \Rx, s, t)$ be a Petri net.
The \emph{stoichiometric category} $\Lk_0(P)$ is the \emph{free
permutative category} on $P$~\cite{MeseguerMontanari1990,Kock2022}:
the unique (up to strict symmetric monoidal equivalence) permutative
category satisfying the following.
\begin{enumerate}[label=(\roman*)]
  \item \textbf{Objects.}
    The object set is $\NN^{|\Sp|}$, with monoidal product
    $\mathbf{u} \otimes \mathbf{v} := \mathbf{u} + \mathbf{v}$
    and monoidal unit $\mathbf{0}$.
    Since $+$ is commutative in $\NN^{|\Sp|}$, the tensor is strictly
    commutative: $\mathbf{u} \otimes \mathbf{v} = \mathbf{v} \otimes
    \mathbf{u}$ as the same object.
    Associativity and unitality are also strict.

  \item \textbf{Generating morphisms.}
    The morphisms of $\Lk_0(P)$ have two distinct origins.
    \begin{itemize}
      \item \emph{Chemical generators (input data).}
        The reaction generators $r: s(r) \to t(r)$ for each
        $r \in \Rx$, supplied by the Petri net $P$.
        These are the only chemical input.
      \item \emph{Structural morphisms (forced by the skeletal
        permutative structure).}
        The identity morphisms
        $\id_{\mathbf{u}}: \mathbf{u} \to \mathbf{u}$ for each
        $\mathbf{u} \in \NN^{|\Sp|}$, and the symmetry endomorphisms
        $\sigma_{\mathbf{u},\mathbf{v}}: \mathbf{u}+\mathbf{v} \to
        \mathbf{u}+\mathbf{v}$ for each pair
        $\mathbf{u}, \mathbf{v} \in \NN^{|\Sp|}$.
        These are not chemical input; they are forced on $\Lk_0(P)$
        by the requirement that it be a skeletal permutative category
        at all.
    \end{itemize}
    The symmetry $\sigma_{\mathbf{u},\mathbf{v}}$ is an endomorphism ---
    not an isomorphism between two distinct objects --- because
    $\mathbf{u}+\mathbf{v} = \mathbf{v}+\mathbf{u}$ as the same element
    of $\NN^{|\Sp|}$, the skeletal property of item (i) above.

  \item \textbf{All morphisms.}
    Closed under sequential composition $g \circ f$
    (when $\operatorname{cod}(f) = \operatorname{dom}(g)$, which holds in general as chemistry demands it)
    and parallel tensor $f \otimes g$.

  \item \textbf{Relations.}
    Morphisms are equivalence classes under the smallest congruence
    compatible with $\circ$ and $\otimes$ generated by the
    \emph{permutative axioms}:
    identity laws; associativity of $\circ$; interchange
    $(g_1 \otimes g_2) \circ (f_1 \otimes f_2)
     = (g_1 \circ f_1) \otimes (g_2 \circ f_2)$;
    tensor unit; involutivity of $\sigma$; naturality of $\sigma$;
    and the hexagon axiom (all defined in
    Remark~\ref{rem:permutative-axioms} below).
    \emph{No} equation $r_1 \equiv r_2$ is imposed for distinct
    $r_1, r_2 \in \Rx$, even if $s(r_1) = s(r_2)$ and
    $t(r_1) = t(r_2)$.

  \item \textbf{Functoriality.}
    A Petri morphism $f = (f_\Sp, f_\Rx): P \to P'$ induces a strict
    symmetric monoidal functor $\Lk_0(f): \Lk_0(P) \to \Lk_0(P')$,
    acting as $\bar{f}_\Sp$ on objects and by $r \mapsto f_\Rx(r)$
    on generators, making $\Lk_0: \mathbf{Petri} \to \mathbf{PermCat}$
    a functor.
\end{enumerate}
\end{definition}

Here $\mathbf{PermCat}$ denotes the category whose objects are small
\emph{skeletal permutative categories} (strict symmetric monoidal
categories whose object monoid $(\Ob\mathcal{C}, \otimes, I)$ is
itself strictly commutative, matching the structure of $\Lk_0(P)$)
and whose morphisms are strict symmetric monoidal functors.

\begin{remark}[The permutative axioms: involutivity, naturality, hexagon]
\label{rem:permutative-axioms}
The three axioms governing the symmetry morphisms $\sigma$ in
Definition~\ref{def:L0}(iv) are standard in the theory of symmetric
monoidal categories~\cite{MacLane1963,MacLane1998}; we spell them out
explicitly for readers encountering them for the first time.

\begin{itemize}
  \item \textbf{Involutivity.}
    $\sigma_{\mathbf{v},\mathbf{u}} \circ \sigma_{\mathbf{u},\mathbf{v}}
    = \id_{\mathbf{u}+\mathbf{v}}$.
    Swapping the order of $\mathbf{u}$ and $\mathbf{v}$ twice returns to
    the original arrangement.\\
    Chemically: reordering a mixture and then reordering it back leaves
    it unchanged.

  \item \textbf{Naturality.}
    For any morphisms $f: \mathbf{u}\to\mathbf{u}'$ and
    $g: \mathbf{v}\to\mathbf{v}'$,
    \[
      \sigma_{\mathbf{u}',\mathbf{v}'} \circ (f \otimes g)
      \;=\;
      (g \otimes f) \circ \sigma_{\mathbf{u},\mathbf{v}}.
    \]
    Reordering before or after running two reactions in parallel gives
    the same result.\\
    Chemically: it does not matter whether you relabel species before or
    after performing the reactions --- the stoichiometric outcome is
    identical.

  \item \textbf{Hexagon axiom.}
    For all complexes $\mathbf{u}, \mathbf{v}, \mathbf{w}$,
    \[
      \sigma_{\mathbf{u},\,\mathbf{v}+\mathbf{w}}
      \;=\;
      (\id_{\mathbf{v}} \otimes \sigma_{\mathbf{u},\mathbf{w}})
      \circ
      (\sigma_{\mathbf{u},\mathbf{v}} \otimes \id_{\mathbf{w}}).
    \]
    The name comes from the hexagonal commutative diagram the equation
    generates.
    It says that swapping $\mathbf{u}$ past the combined pool
    $\mathbf{v}+\mathbf{w}$ is the same as swapping $\mathbf{u}$ past
    $\mathbf{v}$ first and then past $\mathbf{w}$.\\
    Chemically: reordering one component past a mixture of two others
    can be done in a single step or two sequential steps with the same
    result.
\end{itemize}
In a non-skeletal permutative category, the symmetry
$\sigma_{X,Y}: X \otimes Y \to Y \otimes X$ runs between distinct
objects, and involutivity composes the two arrows to land back at
$X \otimes Y$.
In the skeletal $\Lk_0(P)$, where $\mathbf{u}+\mathbf{v} =
\mathbf{v}+\mathbf{u}$ as the same object, $\sigma_{\mathbf{u},\mathbf{v}}$
becomes an endomorphism and involutivity says it is its own inverse:
$\sigma_{\mathbf{u},\mathbf{v}}$ generates a $\ZZ/2$-action on
$\Hom(\mathbf{u}+\mathbf{v}, \mathbf{u}+\mathbf{v})$ rather than
collapsing to the identity.
\end{remark}

\begin{remark}[The symmetry morphisms and coherence]
\label{rem:sigma-role}
In a non-skeletal symmetric monoidal category, the symmetry
$\sigma_{A,B}: A \otimes B \xrightarrow{\;\sim\;} B \otimes A$
is an isomorphism between two \emph{distinct} objects $A\otimes B$
and $B\otimes A$.
In $\Lk_0(P)$, because $\NN^{|\Sp|}$ is strictly commutative,
$\mathbf{u}+\mathbf{v}$ and $\mathbf{v}+\mathbf{u}$ are literally the
same object, so $\sigma_{\mathbf{u},\mathbf{v}}$ is an endomorphism
--- an arrow from a complex back to itself --- rather than an
isomorphism between two different complexes.

Despite living in $\Hom(\mathbf{u}+\mathbf{v},\,\mathbf{u}+\mathbf{v})$,
$\sigma_{\mathbf{u},\mathbf{v}}$ is \emph{not} the identity morphism
$\id_{\mathbf{u}+\mathbf{v}}$: it is a non-trivial endomorphism, the
involutive generator of a $\ZZ/2$-action on the hom-space.
This distinction matters for functors: a strict symmetric monoidal
functor $F: \Lk_0(P) \to \mathcal{C}$ into a target satisfying the
hypothesis of Theorem~\ref{thm:UP-L0} (object monoid strictly
commutative) must send $\sigma_{\mathbf{u},\mathbf{v}}$ to the
symmetry $\sigma^{\mathcal{C}}_{F(\mathbf{u}),F(\mathbf{v})}$, which
is itself an endomorphism of $F(\mathbf{u})\otimes F(\mathbf{v})$.
That target endomorphism may be trivial (as in the Baez--Master CMC,
Remark~\ref{rem:BM-comparison}) or non-trivial; $\Lk_0(P)$ maps
coherently into both kinds of skeletal permutative target.
\end{remark}

\begin{remark}[No additional chemical relations]
\label{rem:no-extra-relations}
The only relations in $\Lk_0(P)$ are the permutative axioms of
Definition~\ref{def:L0}(iv).
No equation $r_1 \equiv r_2$ is ever imposed between distinct reaction
labels: two routes sharing the same source and target complexes remain
distinct morphisms, kept apart by their names alone.
Any physical identification --- same enthalpy change, same mechanism,
same rate law --- must be imposed explicitly at a higher level of the
tower, where the relevant structure is available to make such an
identification meaningful.
In this sense $\Lk_0(P)$ is the \emph{most general} stoichiometric
category consistent with the Petri net $P$: it retains every distinction
that stoichiometry alone cannot collapse.
\end{remark}

\begin{example}[$\Lk_0$ of a three-reaction network]
\label{ex:L0-small-network}
Let $\Sp = \{A, B, C\}$ and $r_1: A \to B$,
$r_2: B \to A$, $r_3: A{+}B \to C$.

\noindent\textbf{Sequential composition.}
$r_2 \circ r_1: A \to A$ is well-typed ($t(r_1) = s(r_2) = B$) and is
\emph{not} equal to $\id_A$: no permutative axiom identifies a composite
of two generating reactions with an identity.

\noindent\textbf{Type enforcement.}
The composite $r_3 \circ (r_1 \otimes \id_B)$ is ill-typed:
$t(r_1 \otimes \id_B) = B + B = 2B \neq A + B = s(r_3)$.
The category refuses to compose steps whose stoichiometries do not match.

\noindent\textbf{Distinct labels.}
Adding a label $r_1': A \to B$ (a mechanistically distinct
isomerisation) produces $r_1 \neq r_1'$ as morphisms $A \to B$.
No permutative axiom identifies them; any physical identification
(same $\Delta H$, same rate) must wait for $\Lk_1$ or $\Lk_3$.
\end{example}

%% file: chapters/L0/universal.tex
\subsection{The universal property of \texorpdfstring{$\Lk_0$}{Lk0}}
\label{sec:base-adjunction}

A \emph{strict symmetric monoidal functor} out of $\Lk_0(P)$ is the
mathematical notion of a consistent assignment of chemical content to a
network: it sends species to objects of the target category, reaction
labels to morphisms of the correct type, and is required to preserve
both sequential composition and parallel tensor.
The universal property of $\Lk_0(P)$ says that any such functor is
completely determined by its values on the elementary generators ---
species and individual reaction labels --- with no further consistency
conditions to verify.
This is the reaction-level analogue of Proposition~\ref{prop:FCM-UP}:
just as a monoid homomorphism out of $\NN^{|\Sp|}$ is fixed by its values
on individual species, a symmetric monoidal functor out of $\Lk_0(P)$
is fixed by its values on individual species and individual reactions.

\begin{theorem}[Universal property of $\Lk_0(P)$]
\label{thm:UP-L0}
Let $P = (\Sp, \Rx, s, t)$ be a Petri net and $\mathcal{C}$ a strict
symmetric monoidal category whose object monoid
$(\Ob\mathcal{C}, \otimes, I)$ is strictly commutative --- i.e.\
$X \otimes Y = Y \otimes X$ as the same object for all
$X, Y \in \Ob\mathcal{C}$.
There is a bijection --- which varies consistently as $P$ and
$\mathcal{C}$ vary --- between:
\begin{equation}
\label{eq:UP-L0}
  \left\{\,
  \begin{array}{l}
       \text{strict symmetric } \\
       \text{monoidal functors}\\
       F: \Lk_0(P) \to \mathcal{C}
  \end{array}
  \,\right\}
  \;\cong\;
  \left\{\,(g,h)
  \;\left|\;
  \begin{array}{l}
    g: \Sp \to \Ob(\mathcal{C}),\\[2pt]
    h(r): \bar{g}(s(r)) \to \bar{g}(t(r))\\[2pt]
    h(r) \text{ in } \mathcal{C} \text{ for each } r \in \Rx
  \end{array}
  \right.
  \,\right\},
\end{equation}
where $\bar{g}: \NN^{|\Sp|} \to \Ob(\mathcal{C})$ is the unique monoid
homomorphism extending $g$ (Proposition~\ref{prop:FCM-UP}), given by
$\bar{g}\!\bigl(\sum_S n_S \cdot S\bigr) = \bigotimes_S g(S)^{\otimes n_S}$, 
and $\Ob(\mathcal{C})$ stands for the objects of the category $\mathcal{C}$.
Under the bijection, $F(S) = g(S)$ for each species $S \in \Sp$ and
$F(r) = h(r)$ for each $r \in \Rx$; all remaining values of $F$ are
then forced by strict functoriality and monoidality.
\end{theorem}

\begin{proof}
Given a strict symmetric monoidal functor $F: \Lk_0(P) \to \mathcal{C}$,
define $g(S) := F(S)$ and $h(r) := F(r)$.
Strict monoidality forces $F(\mathbf{u}) = \bar{g}(\mathbf{u})$ for all
$\mathbf{u} \in \NN^{|\Sp|}$ (both sides are monoid homomorphisms
$\NN^{|\Sp|} \to \Ob(\mathcal{C})$ agreeing on generators, so they agree
everywhere by Proposition~\ref{prop:FCM-UP}), and strict functoriality
forces all values on composites and tensors.
Conversely, given any pair $(g, h)$, define $F$ on objects by $\bar{g}$
and on generators by $h$, extending to all morphisms by
$F(f \circ k) := F(f) \circ F(k)$,
$F(f \otimes k) := F(f) \otimes F(k)$, and
$F(\sigma_{\mathbf{u},\mathbf{v}}) :=
\sigma^{\mathcal{C}}_{F(\mathbf{u}),F(\mathbf{v})}$.
This is well-defined on equivalence classes (Definition~\ref{def:L0}(iv))
because each permutative axiom in $\Lk_0(P)$ maps to the corresponding
axiom of $\mathcal{C}$, which holds since $\mathcal{C}$ is a strict
symmetric monoidal category.
The bijection varies consistently with $P$ and $\mathcal{C}$ because
a Petri morphism $f: P \to P'$ precomposes $(g,h)$ by
$(g \circ f_\Sp,\; r \mapsto h(f_\Rx(r)))$, which is exactly how
$\Lk_0(f)$ acts.
\end{proof}

\begin{remark}[Why the strict-commutativity hypothesis on $\mathcal{C}$ is harmless in practice]
\label{rem:strict-comm-target}
The hypothesis that $(\Ob\mathcal{C}, \otimes, I)$ is strictly
commutative is forced by the skeletal nature of $\Lk_0(P)$:
because $\mathbf{u}+\mathbf{v} = \mathbf{v}+\mathbf{u}$ as the same
object in $\Lk_0(P)$, any strict symmetric monoidal functor must
send these to the same object of $\mathcal{C}$, and this is only
guaranteed when the object monoid of $\mathcal{C}$ is itself strictly
commutative.
Every target category used in this monograph satisfies this
condition automatically: $B\RR$ (Corollary~\ref{cor:BR}),
$B\ZZ^{|\mathsf{E}|}$ (Example~\ref{ex:atom-balance}), and more
generally $BM$ for any commutative monoid $M$ are one-object
categories, so strict commutativity on objects holds vacuously.
For target categories with multiple objects and genuinely
non-commutative tensor (where $X\otimes Y$ and $Y\otimes X$ are
distinct objects), one must replace strict symmetric monoidal
functors by \emph{strong} ones, equipped with coherence isomorphisms
$F(\mathbf{u}+\mathbf{v}) \xrightarrow{\sim} F(\mathbf{u}) \otimes F(\mathbf{v})$
that absorb the object-level reordering.
The strong version is not needed in this chapter and is not pursued.
\end{remark}

\begin{corollary}[Additive functors into $B\RR$]
\label{cor:BR}
The real line $\RR$ carries the structure of a one-object category
$B\RR$: there is a single object $*$, and the morphisms from $*$ to
itself are the real numbers, with composition defined by addition
$b \circ a := a + b$.
Every real number is an arrow, and composing two arrows means adding
their labels; the identity morphism on $*$ is the number $0$.
This is a permutative category with trivial symmetry
$\sigma_{*,*} = 0$ (the zero element of $\RR$, since swapping a
one-element tensor with itself leaves the number unchanged).

By Theorem~\ref{thm:UP-L0}, strict symmetric monoidal functors
$F: \Lk_0(P) \to B\RR$ correspond bijectively to arbitrary maps
$h: \Rx \to \RR$ (one real number per reaction label), since the
unique species assignment is $g(S) = *$ for all $S$.
The extension to all morphisms is forced:
\[
  F(f \circ k) = F(f) + F(k),
  \qquad
  F(f \otimes k) = F(f) + F(k),
  \qquad
  F(\id_{\mathbf{u}}) = 0,
  \qquad
  F(\sigma_{\mathbf{u},\mathbf{v}}) = 0.
\]
Such a functor assigns an additive numerical weight to every composite
and parallel process.
For $F$ to model thermodynamic enthalpy as a state function --- i.e.\
for $h(r)$ to depend only on $s(r)$ and $t(r)$, not on the particular
label $r$ --- the additional condition that $h$ factor through an object
potential $\Delta H: \NN^{|\Sp|} \to \RR$ via
$h(r) = \Delta H(t(r)) - \Delta H(s(r))$ must be imposed separately.
This state-function condition is \emph{not} part of the $\Lk_1$
structure, which supplies only the additivity condition
$h: \Rx \to \RR$; it is a further constraint on the particular functor
$F_H$, discussed in \S\ref{sec:L0-forcing}.
\end{corollary}

\begin{remark}[Relation to the Baez--Master free CMC]
\label{rem:BM-comparison}
A \emph{commutative monoidal category} (CMC)~\cite{BaezMaster2020} is a
symmetric monoidal category in which every symmetry morphism is the
identity: $\sigma_{A,B} = \id_{A\otimes B}$.
Baez and Master~\cite{BaezMaster2020} construct the free CMC
$F_{\mathrm{Petri}}(P)$ on a Petri net $P$; it is the quotient of
$\Lk_0(P)$ by the congruence $\sigma_{\mathbf{u},\mathbf{v}} =
\id_{\mathbf{u}+\mathbf{v}}$, and Theorem~\ref{thm:UP-L0} specialises
to their result when the target $\mathcal{C}$ is a CMC.
We use $\Lk_0(P)$ rather than $F_{\mathrm{Petri}}(P)$ as the base of
the tower because the higher levels $\Lk_2$ ($\dagger$-SMC), $\Lk_3$
(Markov category), and $\Lk_4$ (DPO spans in $\mathbf{LGraph}$) all
carry non-trivial symmetry structure: a functor out of a CMC, where all
symmetries are identities, cannot map coherently into a target where they
are not.
\end{remark}

\begin{chembox}[What the universal property means for chemists]
Any consistent additive assignment of chemical content to species and to
individual reaction labels extends uniquely to all composites and parallel
combinations.
The extension is forced by the requirement of being a strict symmetric
monoidal functor; it is not a modelling choice.

\smallskip\noindent\textit{Examples.}
Assigning enthalpies $h(r) \in \RR$ to generators and demanding
functoriality gives Hess's Law (additivity over sequences and parallel
reactions) as a consequence of Corollary~\ref{cor:BR}: not a separate
postulate but a theorem.
Assigning per-species atom counts and demanding functoriality
(Example~\ref{ex:atom-balance}) extends each generator's atom
imbalance additively along all sequential and parallel composites,
giving atom conservation as a theorem whenever every generator is
balanced.
Rate constants at $\Lk_3$ require different machinery: rates are not
path-additive, so they enter via a kinetic functor into a Markov
category, not via the additive universal property of $B\RR$.

\smallskip\noindent\textit{What is not forced.}
The state-function condition --- that $h(r)$ depends only on the
complexes $s(r)$ and $t(r)$, not on the particular reaction label $r$
--- is an additional physical constraint, not a consequence of
functoriality alone.
Two distinct reaction labels $r_1, r_2$ with the same source and target
can carry different enthalpies $h(r_1) \neq h(r_2)$, and the functor
axioms do not prohibit this.
The distinction between functorial additivity (supplied by $\Lk_1$)
and the state-function condition (a separate physical constraint
on the particular functor $F_H$) is developed in
\S\ref{sec:L0-forcing}.
\end{chembox}

%% file: chapters/L0/new.tex
\subsection{The automorphism exact sequence and the forcing principle}
\label{sec:aut-exact}

Theorem~\ref{thm:UP-L0} establishes $\Lk_0(P)$ as the free skeletal
permutative category on the Petri net $P$: once species are assigned to objects of
a target category and reaction labels to morphisms of the correct type,
the extension to all composites and tensors is uniquely determined.
This universal property now gives us a precise handle on the
automorphisms of $\Lk_0(P)$ itself, and --- more importantly --- on
what they cannot see.

As previewed in \S\ref{sec:intro-tower}, the tower
$\Lk_0\hookrightarrow\Lk_1\hookrightarrow\cdots\hookrightarrow\Lk_7$
is built bottom-up: at each step there exist pairs of reactions that
are physically distinct yet indistinguishable at the previous level.
The tool that makes ``indistinguishable'' mathematically precise is
the \emph{automorphism sequence}, an adaptation to the categorical
setting of a classical algebraic device.
For a homomorphism $\varphi: G \to H$ of groups, the kernel
$\ker\varphi$ measures what $\varphi$ collapses, and the coset space
$H/\im\varphi$ measures what lies outside its image~\cite{riehl,mclarty2007saunders}.
Applied to the restriction map between the automorphism groups of
adjacent tower levels, the kernel records new symmetries that level
$k$ adds beyond level $k-1$, and the coset space records symmetries
of $\Lk_{k-1}$ that fail to lift to $\Lk_k$ --- precisely the
distinctions a given level cannot express, and that the next level
must add.
We do not assume the image of the restriction map to be a normal
subgroup, so the coset space is in general only a pointed set, not a
group; this is sufficient for the forcing diagnostic developed below.

\paragraph{Presentation-preserving automorphisms and the restriction map.}
A \emph{presentation-preserving automorphism} of $\Lk_k$ is a strict
symmetric monoidal autoequivalence $\phi:\Lk_k\to\Lk_k$ that preserves
the chemical input data --- species set $\Sp$ and reaction labels
$\Rx$ --- as subsets of objects and morphisms respectively.
We write $\Aut_{\mathrm{Petri}}(\Lk_k)$ for the group of all such
automorphisms under composition.
The presentation-preserving qualifier is essential: an arbitrary
categorical autoequivalence might rearrange composites and tensor
products in ways unrelated to any Petri-net relabelling, and the
results below would not apply to it.

If some $\phi \in \Aut_{\mathrm{Petri}}(\Lk_k)$ swaps two reactions
$r_1, r_2 \in \Rx$, then the two are indistinguishable by every
$\Lk_k$-level observable: any functor out of $\Lk_k$ must commute
with $\phi$ and therefore cannot separate them.
The converse fails in general --- two reactions can agree under every
$\Lk_k$-functor without there being a global presentation-preserving
automorphism realising the swap --- so a swapping automorphism is a
\emph{sufficient diagnostic} for indistinguishability, not a
characterisation.
This is exactly what the forcing argument needs: a swap that fails
to lift to $\Lk_{k+1}$ exhibits a real distinction $\Lk_k$ misses.

Each tower inclusion $\iota_k:\Lk_{k-1}\hookrightarrow\Lk_k$ induces a
\emph{restriction homomorphism}
\[
  \varphi_k \;:\; \Aut_{\mathrm{Petri}}(\Lk_k) \;\longrightarrow\;
  \Aut_{\mathrm{Petri}}(\Lk_{k-1}),
  \qquad
  \phi \;\longmapsto\; \phi\!\restriction_{\Lk_{k-1}},
\]
sending each presentation-preserving automorphism of the richer
level to its restriction to the simpler one.
Not every element of $\Aut_{\mathrm{Petri}}(\Lk_{k-1})$ arises this
way: those that do not are symmetries present at level $k-1$ that
level $k$ \emph{breaks}, precisely because the extra structure at
level $k$ distinguishes what level $k-1$ could not.

\begin{forcingbox}[The automorphism sequence and the forcing principle]
\label{box:aut-exact}
For each tower step, the restriction map $\varphi_k$ yields an exact
sequence of pointed sets
\begin{equation}
\label{eq:aut-exact}
  1
  \;\longrightarrow\;
  \ker\varphi_k
  \;\longrightarrow\;
  \Aut_{\mathrm{Petri}}(\Lk_k)
  \;\xrightarrow{\;\varphi_k\;}
  \Aut_{\mathrm{Petri}}(\Lk_{k-1})
  \;\longrightarrow\;
  \coker\varphi_k
  \;\longrightarrow\;
  1,
\end{equation}
where $\coker\varphi_k := \Aut_{\mathrm{Petri}}(\Lk_{k-1})/\im\varphi_k$
is the set of left cosets, with basepoint the trivial coset.
The first three terms carry group structure (kernel of a group
homomorphism, the two automorphism groups themselves), but
$\coker\varphi_k$ is in general only a pointed set: we do not assume
$\im\varphi_k$ to be a normal subgroup of
$\Aut_{\mathrm{Petri}}(\Lk_{k-1})$.
Exactness at each term means the image of the incoming map equals
the preimage of the basepoint of the outgoing one.
The three terms carry distinct chemical content.
\begin{itemize}
  \item $\ker\varphi_k$: presentation-preserving automorphisms of
    $\Lk_k$ restricting to the identity on $\Lk_{k-1}$ ---
    \emph{new symmetries first visible at level $k$}.
    Example: time-reversal $\dagger$ (sending each reaction to its
    reverse) is an automorphism of $\Lk_2$ invisible at $\Lk_1$,
    where no reverse reaction exists.
  \item $\coker\varphi_k$: cosets of presentation-preserving
    automorphisms of $\Lk_{k-1}$ modulo those that lift to $\Lk_k$.
    A non-trivial coset is represented by an automorphism of
    $\Lk_{k-1}$ that fails to lift --- a symmetry that $\Lk_k$
    breaks.
    A non-trivial cokernel is the precise statement that $\Lk_{k-1}$
    conflates objects (reactions, complexes, or configurations) that
    $\Lk_k$ separates.
  \item $\varphi_k$ itself: records how much symmetry survives when
    the extra structure of level $k$ is forgotten.
\end{itemize}

\medskip
\noindent\textbf{Forcing principle (diagnostic direction).}
If $\coker\varphi_k$ is non-trivial, then $\Lk_{k-1}$ admits a genuine
symmetry that $\Lk_k$ breaks: $\Lk_{k-1}$ cannot distinguish reactions
that $\Lk_k$ separates, so the extension is necessary --- i.e., no
reformulation within $\Lk_{k-1}$ can capture the distinction.

\medskip
\noindent\textbf{Forcing principle (constructive direction).}
At each observed non-trivial $\coker\varphi_k$, an explicit minimal
extension $\Lk_{k-1}\to\Lk_k$ that breaks the cokernel class is
exhibited --- adding exactly the new functor, enrichment, or
deformation required and no more.
That this extension is the \emph{unique} minimal extension breaking
the cokernel class is claimed level by level on construction, not as
a general theorem about categorical extensions; the full taxonomy of
the six extension types that occur across the present tower is given
in \S\ref{sec:L7-retrospective}.
\end{forcingbox}

\begin{chembox}[What the exact sequence means for a chemist]
An automorphism is a relabelling of the mathematical structure that
leaves every observable at that level unchanged.
If a relabelling swaps two reactions $r_1$ and $r_2$, then no
measurement available at level $k-1$ can tell them apart.

The cokernel $\coker\varphi_k$ collects exactly those relabellings of
$\Lk_{k-1}$ that swap reactions which are genuinely distinct in
physical reality.
A non-trivial cokernel means $\Lk_{k-1}$ is \emph{blind} to a real
distinction: it lacks the vocabulary to express it.
Adding the minimal structure needed to break that spurious symmetry
yields $\Lk_k$, and nothing more.

This is the categorical formalisation of a familiar experience:
every time a chemical rule has an exception, the exception reveals
that a finer level of description was implicitly being assumed.
The automorphism exact sequence turns this intuition into a working
mathematical diagnostic.
\end{chembox}

\begin{table}[p]
\centering
\small
\setlength{\tabcolsep}{5pt}
\renewcommand{\arraystretch}{1.45}
\begin{tabular}{@{}
  >{\raggedright}p{1.0cm}
  >{\raggedright}p{3.3cm}
  >{\raggedright}p{4.3cm}
  >{\raggedright\arraybackslash}p{4.1cm}
  @{}}
\toprule
\textbf{Level}
  & \textbf{$\Aut_{\mathrm{Petri}}(\Lk_k)$ (surviving symmetry)}
  & \textbf{Forcing pair}
  & \textbf{$\coker\varphi_k$ class} \\
\midrule

$\Lk_0$
  & $\mathrm{Sym}(\Sp)_{\!P}$ and label permutations within
    source-target fibres
  & --- (base level)
  & --- \\

$\Lk_1$
  & Isoenthalpic permutations
  & Two reactions with same source/target, different $\Delta H$
  & $\Delta H$-rescaling swap \\

$\Lk_2$
  & Isothermodynamic permutations with $\dagger$
  & Reactions with same $\Delta H$ but different $\Delta S$;
    equilibrium shifts with $T$ differently
  & $T$-dependent $\Delta G$-rescaling \\

$\Lk_3$
  & Rate-preserving permutations
  & Reactions with same $\Delta G$ but different rate;
    kinetics invisible at $\Lk_2$
  & Rate-rescaling \\

$\Lk_4$
  & $\Aut(G)$ on typed molecular graphs
  & Concerted vs.\ stepwise substitution at phosphorus,
    same $\Lk_3$ rate under steady-state on TBI;
    mechanism invisible at $\Lk_3$
  & Mechanism relabelling \\

$\Lk_{4.5}$
  & Full permutation-inversion group $G^*$
  & $(R)$- vs.\ $(S)$-enantiomers with same DPO;
    stereochemistry invisible at $\Lk_4$
  & $G^*$ sectors \\

$\Lk_5$
  & Isometries of $\Ce(G)$ preserving $V$
  & Distinct activation barriers with same $G^*$;
    PES geometry invisible at $\Lk_{4.5}$
  & PES deformation \\

$\Lk_6$
  & $\mathbb{Z}_2$ gauge on the real eigenbundle
  & CI-crossing vs.\ adiabatic reactions with same $V$
  & Berry $\mathbb{Z}_2$-holonomy $[\gamma_B]$ \\

$\Lk_7$
  & $S_N\times S_M$ particle-exchange
  & H vs.\ D; ortho vs.\ para $\mathrm{H_2}$;
    identity of chemical species
  & Isotope, identical-particle exchange \\

\bottomrule
\end{tabular}
\caption{\textit{Roadmap.} Surviving symmetries, forcing pairs, and
cokernel classes across the tower; full development of each
transition occupies Chapters~3--10.
Each row records the presentation-preserving automorphism group
$\Aut_{\mathrm{Petri}}(\Lk_k)$, the concrete reaction pair
distinguishing $\Lk_k$ from $\Lk_{k-1}$, and the
$\coker\varphi_k$ class (a coset in $\Aut_{\mathrm{Petri}}(\Lk_{k-1})$)
broken by the extension.
The retrospective table of \S\ref{sec:L7-retrospective} records the
extension type of each step and serves as the chapter's reference
card.}
\label{tab:aut-table}
\end{table}

\paragraph{The sequence at $k = 0$: identifying the cokernel.}
At the base level, $\Lk_0(P)$ is the free skeletal permutative
category on $P$ (Theorem~\ref{thm:UP-L0}), so its
presentation-preserving automorphisms are determined by their action
on the two families of chemical generators: species $\Sp$ (generating
the object monoid via Proposition~\ref{prop:FCM-UP}) and reaction
labels $\Rx$ (the morphism generators).

\begin{itemize}
  \item \emph{Species permutations.}
    A permutation $\pi\in\mathrm{Sym}(\Sp)$ that \emph{preserves the
    Petri net} --- i.e., whose induced action on $\NN^{|\Sp|}$ sends
    $\{(s(r), t(r)) : r \in \Rx\}$ back to itself --- extends
    (via Proposition~\ref{prop:FCM-UP} on objects and the universal
    property on morphisms) to an element of
    $\Aut_{\mathrm{Petri}}(\Lk_0(P))$.
    We write $\mathrm{Sym}(\Sp)_{\!P}$ for this stabiliser subgroup.

  \item \emph{Label permutations within fibres.}
    For any pair
    $(\mathbf{u},\mathbf{v})\in\NN^{|\Sp|}\times \NN^{|\Sp|}$,
    a permutation of the fibre
    $(s,t)^{-1}(\mathbf{u},\mathbf{v}) = \{r \in \Rx : s(r) =
    \mathbf{u},\; t(r) = \mathbf{v}\}$ --- a relabelling that sends
    each reaction to another with identical source and target ---
    extends to an element of $\Aut_{\mathrm{Petri}}(\Lk_0(P))$.
\end{itemize}

The second family suffices for the $k=0$ forcing argument.
Let $r_1, r_2\in\Rx$ be two distinct labels with
$s(r_1) = s(r_2) = \mathbf{u}$ and $t(r_1) = t(r_2) = \mathbf{v}$,
carrying distinct enthalpies $\Delta H_1\neq\Delta H_2$
(an instance exhibited concretely in
\S\ref{sec:L0-forcing} for the
$\mathrm{N_2}+\mathrm{O_2}\to 2\,\mathrm{NO}$ system).
The swap $r_1\leftrightarrow r_2$ extends to an element of
$\Aut_{\mathrm{Petri}}(\Lk_0(P))$: $\Lk_0(P)$ carries no numerical
data to distinguish the two labels, so no information at this level
is disturbed by the swap.

It is \emph{not} the restriction of any automorphism of $\Lk_1$: at
$\Lk_1$, the functor $F_H:\Lk_0(P)\to B\RR$ assigns
$F_H(r_1)\neq F_H(r_2)$, and any $\Lk_1$-automorphism must commute
with $F_H$, ruling out the swap.
The label swap therefore represents a non-trivial class in
$\coker\varphi_1$,
\begin{equation}
\label{eq:coker-L0}
  [\,r_1\leftrightarrow r_2\,]\;\in\;\coker\varphi_1,
\end{equation}
demonstrating that $\Lk_1$ is necessary.
The minimal extension resolving the conflation is to equip
$\Lk_0(P)$ with a strict symmetric monoidal functor
$F_H:\Lk_0(P)\to B\RR$ assigning a real enthalpy to each reaction
label --- the content of $\Lk_1$, no more and no less.
The full development of this extension, including the distinction
between the additivity condition (which $F_H$ satisfies by
functoriality) and the state-function condition (an additional
physical requirement), is given in \S\ref{sec:L0-forcing}.

\paragraph{The full tower, forced level by level.}
The same argument at every subsequent tower step yields the
automorphism groups and forcing cokernels recorded in
Table~\ref{tab:aut-table}.
The following tour lists the levels in order; full developments
occupy Chapters~3--10.

At $\Lk_1$, the automorphisms are isoenthalpic permutations, and
the cokernel detects reactions sharing $\Delta H$ but differing in
$\Delta S$, which shift their equilibria differently with
temperature and are therefore genuinely distinct at $\Lk_2$.
At $\Lk_2$, adding the $\dagger$-structure (time-reversal,
encoding detailed balance) breaks the remaining isothermodynamic
symmetry; the cokernel witnesses reactions with identical
$\Delta G$ but different rates, forcing $\Lk_3$.
At $\Lk_3$, rate-preserving permutations are the surviving
symmetries, and the cokernel is generated by the swap
$\mathrm{E}2\leftrightarrow\mathrm{S_N}2$: two reactions with
identical empirical rate laws but mechanistically distinct
electron-pushing graphs, forcing $\Lk_4$.
At $\Lk_4$, once full bond-graph structure is present, the residual
symmetry is $\Aut(G)$ on typed molecular graphs; the cokernel
detects the same bond changes realised in different
three-dimensional embeddings, forcing $\Lk_{4.5}$.
At $\Lk_{4.5}$, the full permutation-inversion group $G^*$ governs
enantiomeric and diastereomeric distinctions, including
symmetry-forbidden versus symmetry-allowed pathways under the
Woodward--Hoffmann rules; the cokernel forces $\Lk_5$ by exhibiting
reactions whose transition states occupy geometrically distinct
points on a potential energy surface.
At $\Lk_5$, isometries of the configuration orbifold $\Ce(G)$
preserving the PES $V$ are the automorphisms; the cokernel
distinguishes reactions with the same geometry but different
topological Berry class, separating photochemical from thermal
pathways and forcing $\Lk_6$.
At $\Lk_6$, the real adiabatic eigenbundle admits a
$\mathbb{Z}_2$-valued gauge freedom (the overall sign of the real
eigensection in a simply connected patch), and its cokernel
captures loops encircling the conical-intersection seam on which
the sign cannot be chosen globally; the invariant is the mod-2
Berry holonomy class $[\gamma_B]\in
H^1(\Ce(G)\setminus\Xseam,\mathbb{Z}_2)$, and the cokernel
separates CI-crossing reactions from adiabatic ones, forcing
$\Lk_7$.
Finally, at $\Lk_7$, the symmetry group $S_N\times S_M$ encodes
all-particle exchange statistics (electrons and nuclei
respectively); the cokernel witnesses distinctions invisible at
every earlier level, exhibited by concrete forcing pairs ---
hydrogen versus deuterium (isotope permutation), and ortho- versus
para-$\mathrm{H_2}$ (identical-particle exchange forbidden by Pauli
antisymmetry).
The former produces distinct dynamical algebras through the
$\varepsilon$-dependent Born--Oppenheimer expansion; the latter
splits the compact ideal of $\mathcal{K}(L^2(\Ce(G)))$ into disjoint
nuclear-spin superselection sectors, neither of which an earlier
level can express.

\medskip
\noindent
With the automorphism diagnostic in hand, the remaining combinatorial
content of $\Lk_0$ --- the stoichiometric matrix, Feinberg's
deficiency invariants, the Deficiency Zero Theorem, and open-network
composition --- is the subject of the following subsection.

%% file: chapters/L0/stoich.tex
\subsection{The stoichiometric matrix, conservation laws, and deficiency}
\label{sec:N-matrix}

The previous subsections built $\Lk_0(P)$ as a categorical object.
This subsection extracts its classical linear-algebraic content --- the
stoichiometric matrix $N$, the conservation-law space, and the deficiency
$\delta$ --- and shows precisely which of these arise from the
universal-property machinery already established, and which are purely
presentational invariants of the Petri net $P$.
In every case the derivation uses only Proposition~\ref{prop:FCM-UP},
Theorem~\ref{thm:UP-L0}, and Corollary~\ref{cor:BR}.
The dynamical significance of $\delta$ --- the Deficiency Theorems ---
requires mass-action kinetics and belongs to $\Lk_3$; we state the
theorems here and mark the level boundary throughout.


The notion of a chemical reaction network used throughout the CRNT
literature goes back to Horn, Jackson, and
Feinberg~\cite{Horn1972,HornJackson1972,Feinberg1987,Feinberg2019}.
The following definition adapts their framework to the Petri-net
presentation of $\Lk_0$, retaining named reaction labels so that
mechanistically distinct routes between the same complexes remain
distinct morphisms in $\Lk_0(P)$.

\begin{definition}[Chemical Reaction Network at $\Lk_0$]
\label{def:CRN}
A \emph{chemical reaction network (CRN)} at $\Lk_0$ is a quintuple
$(\Sp, \Cx, \Rx, s, t)$ where $\Cx \subset \NN^{|\Sp|}$ is a finite set
of \emph{complexes}, $\Rx$ is a finite set of \emph{reaction labels},
and $s, t: \Rx \to \Cx$ assign source and target complexes.
The \emph{underlying Petri net} is $P = (\Sp, \Rx, s, t)$ with $s, t$
viewed as maps into $\NN^{|\Sp|}$, and the stoichiometric category of the
CRN is $\Lk_0(P)$.
The \emph{reaction multigraph} $G(\Rx)$ is the directed multigraph with
vertex set $\Cx$ and one directed edge per element of $\Rx$.
\end{definition}

\begin{remark}[Relation to the classical Feinberg CRN]
\label{rem:feinberg-graph}
The classical CRNT triple~\cite{Horn1972,HornJackson1972,Feinberg1987,
Feinberg2019} $(\Sp, \Cx, \mathcal{E})$ uses a set of directed edges
$\mathcal{E} \subseteq \Cx \times \Cx$, collapsing two reactions with
the same source and target to a single edge.
The label-preserving quintuple is necessary for the Petri-net
presentation: distinct labels must remain distinct morphisms in
$\Lk_0(P)$.
The classical triple is recovered by replacing $\Rx$ with
$\{(s(r), t(r)) : r \in \Rx\}$.
\end{remark}


Every generating morphism $r \in \Rx$ of $\Lk_0(P)$ has a source
complex $s(r) \in \NN^{|\Sp|}$ and a target complex $t(r) \in \NN^{|\Sp|}$.
Their difference $t(r) - s(r)$ lives in the group completion
$\ZZ[\Sp] \cong \ZZ^{|\Sp|}$ of $\NN^{|\Sp|}$.
Collecting these differences as columns, after choosing linear orderings
of $\Sp$ and $\Rx$, gives the stoichiometric matrix.

\begin{definition}[Stoichiometric matrix~{\cite{HornJackson1972,Feinberg1987}}]
\label{def:N}
Let $P = (\Sp, \Rx, s, t)$ be a Petri net.
Fix a linear ordering of $\Sp$ (identifying $\ZZ[\Sp] \cong \ZZ^{|\Sp|}$)
and of $\Rx$.
The \emph{stoichiometric matrix} $N \in \ZZ^{|\Sp| \times |\Rx|}$ has
$(i, r)$-entry
\[
  N_{i,r} \;:=\; (t(r))_i - (s(r))_i,
\]
the net change in species $S_i$ produced by reaction $r$.
Column $r$ of $N$ is the image of $t(r) - s(r)$ in $\ZZ^{|\Sp|}$.
\end{definition}

\begin{remark}[$N$ as a presentational invariant of $\Lk_0(P)$]
\label{rem:N-invariant}
Different orderings of $\Sp$ and $\Rx$ give the same $N$ up to row and
column permutations, so $\rank_\RR(N)$ and $\im_\RR(N)$ are independent
of the ordering.
However, $N$ depends on which morphisms are \emph{generators} (elements
of $\Rx$) as opposed to composites or tensors: it is a
\emph{presentational invariant} of $P$, not a categorical invariant of
$\Lk_0(P)$ up to strict symmetric monoidal equivalence.
Two Petri nets with equivalent stoichiometric categories can have different
stoichiometric matrices.

\noindent\textit{Notational convention.}
Throughout this subsection, $\rho := \rank_\RR(N)$ and
$\Ssto := \im_\RR(N) \subseteq \RR^{|\Sp|}$ denotes the
\emph{stoichiometric subspace}.
We reserve $s$ for the source map $s: \Rx \to \NN^{|\Sp|}$.
\end{remark}


The universal property of $\NN^{|\Sp|}$ (Proposition~\ref{prop:FCM-UP})
and the universal property of $\Lk_0(P)$ (Theorem~\ref{thm:UP-L0},
Corollary~\ref{cor:BR}) together give a complete characterisation of
conservation laws in terms of functors out of $\Lk_0(P)$.
This is the first place in the monograph where those universal properties
combine to yield a non-trivial chemical result.

\begin{proposition}[Conservation laws as vanishing functors]
\label{prop:conservation-functors}
Let $P = (\Sp, \Rx, s, t)$ be a Petri net with stoichiometric matrix
$N$.
For each $w_0 : \Sp \to \RR$, Proposition~\ref{prop:FCM-UP} provides a
unique monoid homomorphism $\bar{w} : \NN^{|\Sp|} \to \RR$ extending $w_0$.
Define $h_w : \Rx \to \RR$ by
\[
  h_w(r) \;:=\; \bar{w}(t(r)) - \bar{w}(s(r)),
  \qquad r \in \Rx.
\]
By Corollary~\ref{cor:BR}, $h_w$ determines a unique strict symmetric
monoidal functor $F_w : \Lk_0(P) \to B\RR$ with $F_w(r) = h_w(r)$ on
each generator, and $F_w$ extended to all morphisms by
\[
  F_w(g \circ f) = F_w(g) + F_w(f), \quad
  F_w(f \otimes g) = F_w(f) + F_w(g), \quad
  F_w(\id_\mathbf{u}) = 0.
\]
The following are equivalent:
\begin{enumerate}[label=(\roman*)]
  \item $w_0 \in \ker(N^\top)$, i.e.\ $w_0^\top(t(r)-s(r)) = 0$
    for every $r \in \Rx$.
  \item $F_w \equiv 0$ on all morphisms of $\Lk_0(P)$.
  \item $\bar{w}(t(r)) = \bar{w}(s(r))$ for every generating morphism
    $r \in \Rx$.
\end{enumerate}
\end{proposition}

\begin{proof}
$(i) \Leftrightarrow (iii)$: by definition of $\bar{w}$,
\begin{align*}
    \bar{w}(t(r)) - \bar{w}(s(r))
 &= \sum_S (t(r))_S w_S - \sum_S (s(r))_S w_S\\
 &= w_0^\top(t(r) - s(r)),
\end{align*}
so $\bar{w}(t(r)) = \bar{w}(s(r))$ iff $w_0^\top(t(r)-s(r)) = 0$
iff column $r$ of $N^\top w_0 \in \RR^{|\Rx|}$ is zero.
Since this holds for all $r$, it is equivalent to $N^\top w_0 = 0$.

$(iii) \Leftrightarrow (ii)$: $F_w(r) = h_w(r) = \bar{w}(t(r)) -
\bar{w}(s(r))$, so $F_w(r) = 0$ for all $r \in \Rx$ iff (iii).
By Theorem~\ref{thm:UP-L0}, the functor $F_w$ is entirely determined by
its values on generators; if $F_w(r) = 0$ for all $r \in \Rx$, then
$F_w \equiv 0$ on all morphisms by the forced extension
$F_w(g \circ f) = 0 + 0 = 0$ and $F_w(f \otimes g) = 0 + 0 = 0$.
\end{proof}

\begin{insightbox}[Conservation laws as a tower concept]
Proposition~\ref{prop:conservation-functors} gives the categorical
content of conservation laws entirely within $\Lk_0$:

A species potential $w_0: \Sp \to \RR$ is a conservation law if and
only if the functor $F_w: \Lk_0(P) \to B\RR$ it induces (via
Proposition~\ref{prop:FCM-UP} and Corollary~\ref{cor:BR}) assigns zero
to every morphism of $\Lk_0(P)$.

The reason conservation is preserved by \emph{all} compositions and
parallel combinations --- not just the generating reactions --- is
functoriality: $F_w(g \circ f) = F_w(g) + F_w(f) = 0$ and
$F_w(f \otimes g) = 0$ follow automatically once $F_w = 0$ on
generators.
No separate proof is needed; the functor axioms carry it.

This also explains what $\ker(N^\top)$ is from the tower's point of
view: it is the set of species potentials for which the induced
$\Lk_1$-level functor (Corollary~\ref{cor:BR}) happens to be identically
zero.
Conservation is not an additional axiom imposed on $\Lk_0(P)$; it is a
condition on functors \emph{out of} $\Lk_0(P)$, and functoriality does
the work.
\end{insightbox}

\begin{proposition}[Stoichiometric subspaces]
\label{prop:N-subspaces}
Let $P = (\Sp, \Rx, s, t)$ be a Petri net with matrix $N$.
\begin{enumerate}[label=(\roman*)]
  \item The \emph{stoichiometric subspace}
    $\Ssto := \im_\RR(N) \subseteq \RR^{|\Sp|}$ is spanned by the net
    displacement vectors $\{t(r) - s(r) : r \in \Rx\}$ of the
    generating morphisms of $\Lk_0(P)$.
  \item The \emph{conservation-law space} $\ker(N^\top)$ is the
    orthogonal complement of $\Ssto$ and equals, by
    Proposition~\ref{prop:conservation-functors}, the set of species
    potentials $w_0: \Sp \to \RR$ for which the induced functor
    $F_w: \Lk_0(P) \to B\RR$ is identically zero.
  \item The \emph{stoichiometric cycle space}
    $\ker_\RR(N) := \{\bm{\nu} \in \RR^{|\Rx|} \mid N\bm{\nu} = 0\}$
    consists of formal real-linear combinations $\sum_r \nu_r r$ of
    generating morphisms whose combined net displacement
    $\sum_r \nu_r (t(r) - s(r))$ is zero in $\RR^{|\Sp|}$.
    The role of $\ker_\RR(N)$ in the dynamics of $\Lk_0(P)$ requires
    $\Lk_3$ kinetics; at $\Lk_0$ it records which weighted combinations
    of reactions leave the stoichiometric subspace point-wise fixed.
\end{enumerate}
All three subspaces are computable from $N$ alone; no kinetic data is
required.
\end{proposition}

\begin{proof}
(i) is the definition of $\im_\RR(N)$ applied to the column vectors
$t(r) - s(r)$.
(ii) follows from Proposition~\ref{prop:conservation-functors} and the
standard linear-algebraic fact that $\ker(N^\top) = (\im(N))^\perp$.
(iii) is the definition of $\ker_\RR(N)$.
\end{proof}

The following corollary, which requires $\Lk_3$ kinetics, shows why
$\Ssto$ and $\ker(N^\top)$ matter dynamically: the stoichiometric
subspace confines every trajectory, and the conservation laws become
exact first integrals.
The compatibility classes that $\Ssto$ cuts out are, however,
already determined at $\Lk_0$, before any kinetics is specified.

\begin{corollary}[Trajectory confinement; requires $\Lk_3$ kinetics]
\label{cor:trajectory}
Let $\bm{\nu}: \RR^{|\Sp|} \to \RR^{|\Rx|}$ be any kinetic rate
function (\emph{$\Lk_3$ structure}) and let $\mathbf{x}(t)$ satisfy
$\dot{\mathbf{x}} = N\bm{\nu}(\mathbf{x})$.
Then $\mathbf{x}(t) - \mathbf{x}(0) \in \Ssto$ for all $t \geq 0$.
Moreover, for every $w_0 \in \ker(N^\top)$, the quantity
$w_0^\top \mathbf{x}(t)$ is constant along every trajectory:
$\tfrac{d}{dt}(w_0^\top \mathbf{x}) = 0$.
\end{corollary}

\begin{proof}
$\dot{\mathbf{x}} = N\bm{\nu}(\mathbf{x}) \in \Ssto$ pointwise, giving
the first claim.
$\tfrac{d}{dt}(w_0^\top\mathbf{x}) = w_0^\top N \bm{\nu} =
(N^\top w_0)^\top \bm{\nu} = 0$ since $w_0 \in \ker(N^\top)$.
\end{proof}


The four combinatorial invariants below are computable from
$(\Sp, \Cx, \Rx, s, t)$ alone --- from $\Lk_0$ data.
They were introduced by Feinberg, Horn, and Jackson~\cite{Horn1972,
HornJackson1972,Feinberg1987,Feinberg2019} and underpin the Deficiency
Theorems.
Recent work has confirmed their centrality: deficiency zero is generic
in random reaction networks~\cite{AndersonNguyen2022}, and the interplay
between deficiency, weak reversibility, and steady states is an active
frontier~\cite{Boros2019,JoshiKaihnsaNguyenShiu2023}.

\begin{definition}[Feinberg's combinatorial invariants~{\cite{Feinberg1987,Feinberg2019}}]
\label{def:feinberg}
Let $(\Sp, \Cx, \Rx, s, t)$ be a CRN.
\begin{itemize}
  \item $n := |\Cx|$: number of complexes.
  \item $\ell$: number of \emph{linkage classes} (connected components
    of the undirected graph underlying $G(\Rx)$).
  \item $\rho := \rank_\RR(N)$: dimension of $\Ssto$.
  \item $\delta := n - \ell - \rho \geq 0$: the \emph{deficiency}.
\end{itemize}
\end{definition}

\begin{chembox}[A feel for $n$, $\ell$, $\rho$, $\delta$]
$n$ counts the distinct molecular pools appearing as complexes.
$\ell$ counts the reaction islands: groups of complexes linked by some
directed path, with no connection to other groups.
$\rho$ counts the independent directions in concentration space along
which any trajectory can move.
$\delta$ counts the independent ways the complex graph can ``circulate''
while producing zero net species change: complex-level degrees of
freedom invisible to species-level accounting.
A $\delta = 0$ network has no such hidden structure.
\end{chembox}

To give $\delta$ a precise meaning in terms of $\Lk_0(P)$, we
factor the stoichiometric matrix through the complex level.
The inclusion $\iota: \Cx \hookrightarrow \NN^{|\Sp|}$ (each complex $c \in \Cx$
is already an element of $\NN^{|\Sp|}$) is a function from the set $\Cx$
to the commutative monoid $\NN^{|\Sp|}$.
By Proposition~\ref{prop:FCM-UP} applied with generator set $\Cx$ and
target monoid $\NN^{|\Sp|}$, this extends to a unique monoid homomorphism
\[
  \bar{\iota} \;:\; \NN[\Cx] \;\longrightarrow\; \NN^{|\Sp|},
  \qquad c \;\longmapsto\; c \;\text{(viewed as element of } \NN^{|\Sp|}\text{)},
\]
from the free commutative monoid on complexes to the free commutative
monoid on species.
Its matrix representation (after choosing orderings) is
$\Ymat \in \RR^{|\Sp| \times n}$, the \emph{complex composition
matrix}, whose column $c$ is the coordinate vector of complex
$c \in \Cx$ in $\RR^{|\Sp|}$.

Independently, the reaction multigraph $G(\Rx)$ has an
\emph{incidence matrix} $\Ia \in \{-1, 0, 1\}^{n \times |\Rx|}$
defined by: $(\Ia)_{c,r} = +1$ if $c = t(r)$,
$(\Ia)_{c,r} = -1$ if $c = s(r)$, else $0$.
Here $\Ia$ is a purely graph-theoretic object determined by $G(\Rx)$.

\begin{proposition}[Linear algebraic interpretation of deficiency~{\cite{Feinberg1987,Feinberg2019}}]
\label{prop:deficiency}
With $\Ymat$ and $\Ia$ as above:
\begin{enumerate}[label=(\roman*)]
  \item $N = \Ymat\Ia$.
  \item $\rank(\Ia) = n - \ell$.
  \item $\delta = \dim_\RR(\ker(\Ymat) \cap \im(\Ia)) \geq 0$.
  \item $\delta = 0$ if and only if $\Ymat\!\restriction_{\im(\Ia)}$ is
    injective, i.e.\ every nonzero net complex-flow produces a nonzero
    species displacement.
\end{enumerate}
\end{proposition}

\begin{proof}
(i)~Column $r$ of $\Ymat\Ia$ is
$\Ymat(e_{t(r)} - e_{s(r)}) = t(r) - s(r)$ in $\RR^{|\Sp|}$,
which is column $r$ of $N$.
(ii)~Standard: the rank of the incidence matrix of a directed graph
with $n$ vertices and $\ell$ connected components is $n - \ell$.
(iii)~By rank--nullity on $\Ymat\!\restriction_{\im(\Ia)}$:
$\rank(\Ymat\Ia) = (n-\ell) - \dim(\ker(\Ymat) \cap \im(\Ia))$,
so $\rho = (n - \ell) - \dim(\ker(\Ymat) \cap \im(\Ia))$, giving
$\delta = n - \ell - \rho = \dim(\ker(\Ymat) \cap \im(\Ia)) \geq 0$.
(iv)~$\delta = 0$ iff $\ker(\Ymat) \cap \im(\Ia) = \{0\}$ iff
$\Ymat\!\restriction_{\im(\Ia)}$ is injective.
\end{proof}

\begin{insightbox}[Deficiency as a failure of the canonical monoid map]
Proposition~\ref{prop:deficiency} has a direct tower interpretation.

The matrix $\Ymat$ is the real-linear extension of the monoid
homomorphism $\bar{\iota}: \NN[\Cx] \to \NN^{|\Sp|}$ induced by
Proposition~\ref{prop:FCM-UP} from the inclusion $\iota: \Cx \hookrightarrow \NN^{|\Sp|}$.
This map sends any formal combination of complexes to the corresponding
combination of species counts.

The matrix $\Ia$ records the net complex-level flows induced by the
reaction labels: each column $r$ is $e_{t(r)} - e_{s(r)}$ in $\RR^n$,
the net movement from source complex to target complex.

The factorization $N = \Ymat\Ia$ (part (i)) says: the net species
displacement of a reaction factors as
\emph{first, compute the net complex-flow} ($\Ia$),
\emph{then project complexes to species counts} ($\Ymat$ = $\bar{\iota}$).

Deficiency $\delta$ measures exactly the failure of $\bar{\iota}$ to
be injective on the image of the complex-flow matrix: it counts the
dimension of the space of complex-level circulations that produce zero
net species change.
When $\delta = 0$, $\bar{\iota}$ is injective on $\im(\Ia)$:
every complex-graph flow leaves a detectable trace in species space.
The Deficiency Zero Theorem uses this injectivity crucially.
When $\delta > 0$, there are ``hidden'' complex-level circulations that
$\Lk_0(P)$, viewed through the species map $\bar{\iota}$, cannot detect.
\end{insightbox}


Weak reversibility is a property of the presentation $(\Sp, \Cx, \Rx, s, t)$,
not a categorical invariant of $\Lk_0(P)$ up to equivalence
(Remark~\ref{rem:N-invariant}).
It records whether every directed path in the generating graph $G(\Rx)$
can be returned to its starting complex by further generators --- not by
arbitrary composites, but specifically by further elements of $\Rx$.
This is a strictly $\Lk_0$ datum: no kinetics, no thermodynamics,
no reverse-reaction functor.

It is distinct from the $\dagger$-structure at $\Lk_2$.
At $\Lk_2$, the $\dagger$ operator provides an \emph{explicit} reverse
generator $r^\dagger: t(r) \to s(r)$ for each $r \in \Rx$, subject to
the categorical axiom $r^{\dagger\dagger} = r$.
Weak reversibility at $\Lk_0$ requires only that the reverse direction
is achievable by some directed path of generators in $G(\Rx)$ --- a
much weaker condition, expressible purely in terms of the multigraph.
Introduced by Horn and Jackson~\cite{Horn1972,HornJackson1972}, it is
the graph-theoretic hypothesis that, together with $\delta = 0$,
guarantees the conclusions of the Deficiency Zero Theorem.
Boros~\cite{Boros2019} later showed that weak reversibility alone
(without $\delta = 0$) guarantees existence of a positive steady state
in each compatibility class.

\begin{definition}[Weak reversibility~{\cite{Horn1972,HornJackson1972}}]
\label{def:WR}
A CRN $(\Sp, \Cx, \Rx, s, t)$ is \emph{weakly reversible} if for every
directed path $c_0 \to c_1 \to \cdots \to c_k$ in $G(\Rx)$, there
exists a directed path from $c_k$ back to $c_0$ in $G(\Rx)$.
Equivalently, every connected component of $G(\Rx)$ is strongly
connected.
\end{definition}

\begin{remark}[Level stratification and compatibility classes]
\label{rem:feinberg-levels}
The invariants $n, \ell, \rho, \delta$ and weak reversibility are
purely $\Lk_0$ data: computable from $(\Sp, \Cx, \Rx, s, t)$ with no
functor into any target category.

For the Deficiency Zero Theorem below, the relevant partition of
concentration space is by \emph{stoichiometric compatibility classes}:
the class of $\mathbf{x}_0$ is
\[
  \bigl(\mathbf{x}_0 + \Ssto\bigr) \cap \RR^{|\Sp|}_{>0},
\]
the set of positive concentration vectors reachable from $\mathbf{x}_0$
by stoichiometric changes.
In tower language, $\mathbf{x}$ lies in this class if and only if
\[
  \bar{w}(\mathbf{x}) \;=\; \bar{w}(\mathbf{x}_0)
  \quad \text{for every } w_0 \in \ker(N^\top),
\]
where $\bar{w}: \NN^{|\Sp|} \to \RR$ is the unique monoid homomorphism
extending $w_0$ (Proposition~\ref{prop:FCM-UP}), and
$\bar{w}(\mathbf{x}) = w_0^\top\mathbf{x}$ is its value on a
concentration vector.
The partition of $\RR^{|\Sp|}_{>0}$ into stoichiometric compatibility
classes is determined entirely at $\Lk_0$; the Deficiency Zero Theorem
says which class contains a unique stable $\Lk_3$ fixed point.
\end{remark}


\begin{theorem}[Deficiency Zero Theorem~{\cite{Horn1972,Feinberg1987}}]
\label{the:DZT1}
Let $(\Sp, \Cx, \Rx, s, t)$ be a CRN with $\delta = 0$, endowed with
\emph{mass-action kinetics} (\emph{$\Lk_3$ structure}): the rate
function $\bm{\nu}$ takes the specific form
$\nu_r(\mathbf{x}) = k_r \prod_{S \in \Sp} x_S^{(s(r))_S}$
for positive rate constants $k_r > 0$.
\begin{enumerate}[label=(\roman*)]
  \item If not weakly reversible: no positive steady state exists for
    any choice of positive rate constants.
  \item If weakly reversible: for every choice of positive rate
    constants, there is exactly one positive steady state in each stoichiometric
    compatibility class $\bigl(\mathbf{x}_0 + \Ssto\bigr) \cap
    \RR^{|\Sp|}_{>0}$ (Remark~\ref{rem:feinberg-levels});
    it is locally asymptotically
    stable; and no periodic orbits exist in $\RR^{|\Sp|}_{>0}$.
\end{enumerate}

\noindent\textbf{Tower-language breakdown.}
\begin{itemize}
  \item \emph{From $\Lk_0$ alone:} the hypotheses $\delta = 0$ and weak
    reversibility; the stoichiometric subspace $\Ssto$; the conservation
    laws $\ker(N^\top)$ (Proposition~\ref{prop:conservation-functors});
    and the partition of $\RR^{|\Sp|}_{>0}$ into stoichiometric
    compatibility classes (Remark~\ref{rem:feinberg-levels}).
  \item \emph{From $\Lk_3$, specifically:} mass-action kinetics --- the
    parametric form $\nu_r(\mathbf{x}) = k_r \prod_S x_S^{(s(r))_S}$,
    not an arbitrary kinetic rate function.
    The DZT fails for general $\Lk_3$ kinetics.
  \item \emph{The conclusion:} within each $\Lk_0$-determined class,
    the mass-action dynamics has exactly one fixed point and it is
    stable.
    The class is an $\Lk_0$ object; the fixed point and its stability
    are $\Lk_3$ statements.
\end{itemize}
\end{theorem}

\begin{remark}[Deficiency One and beyond~{\cite{Feinberg1988,Feinberg1995}}]
\label{rem:DOT}
Feinberg~\cite{Feinberg1988,Feinberg1995} proved a Deficiency One
Theorem for $\delta = 1$ under additional structural conditions on
linkage classes.
Boros~\cite{Boros2019} removed the $\delta = 0$ hypothesis from the
existence part of Theorem~\ref{the:DZT1}(i): every weakly reversible
mass-action system has a positive steady state in each stoichiometric compatibility class
$\bigl(\mathbf{x}_0 + \Ssto\bigr) \cap \RR^{|\Sp|}_{>0}$,
regardless of deficiency.
In all cases, the hypotheses are $\Lk_0$ data and the conclusions
require mass-action kinetics at $\Lk_3$.
\end{remark}

\begin{remark}[Tower framework and Feinberg's CRNT]
\label{rem:our-vs-feinberg}
The combinatorial backbone is identical to classical CRNT~\cite{Feinberg1987,
Feinberg1988,Feinberg2019}: species, complexes, reaction multigraph,
stoichiometric matrix, and the Deficiency Theorems originate there.
The architectural difference is that classical CRNT does not separate the
stoichiometric layer from the kinetic layer.
The tower makes this explicit: $\Lk_0$ carries the hypotheses and the
compatibility-class partition; $\Lk_3$ carries the kinetics and the
conclusions.
\end{remark}


\begin{example}[Michaelis--Menten enzyme kinetics]
\label{ex:MM}
Network: $\mathrm{E+S}
\underset{r_2}{\overset{r_1}{\rightleftharpoons}} \mathrm{ES}
\overset{r_3}{\to} \mathrm{E+P}$.
Species $\Sp = \{\mathrm{E,S,ES,P}\}$;
complexes $c_1 = \mathrm{E+S}$, $c_2 = \mathrm{ES}$,
$c_3 = \mathrm{E+P}$.
\[
  N = \begin{pmatrix}
    -1 & +1 & +1 \\
    -1 & +1 &  0 \\
    +1 & -1 & -1 \\
     0 &  0 & +1
  \end{pmatrix},
  \qquad
  n = 3,\quad \ell = 1,\quad \rho = 2,\quad \delta = 0.
\]

\noindent\textbf{Conservation laws (Proposition~\ref{prop:conservation-functors}).}
The conservation-law space $\ker(N^\top)$ has dimension $|\Sp| - \rho = 2$.
We find the two vanishing functors by computing:

For $w_1 = (1,0,1,0)^\top$ (enzyme):
$h_{w_1}(r_1) = \bar{w}_1(c_2) - \bar{w}_1(c_1) = 1 - (1+0) = 0$,
$h_{w_1}(r_2) = \bar{w}_1(c_1) - \bar{w}_1(c_2) = 1 - 1 = 0$,
$h_{w_1}(r_3) = \bar{w}_1(c_3) - \bar{w}_1(c_2) = (1+0) - 1 = 0$.
So $F_{w_1} \equiv 0$: enzyme conservation
$[\mathrm{E}]+[\mathrm{ES}] = \mathrm{const}$.

For $w_2 = (0,1,1,1)^\top$ (substrate):
$h_{w_2}(r_1) = \bar{w}_2(c_2) - \bar{w}_2(c_1) = 1 - (0+1) = 0$,
$h_{w_2}(r_2) = 0$ by symmetry,
$h_{w_2}(r_3) = \bar{w}_2(c_3) - \bar{w}_2(c_2) = (0+1) - 1 = 0$.
So $F_{w_2} \equiv 0$: substrate conservation
$[\mathrm{S}]+[\mathrm{ES}]+[\mathrm{P}] = \mathrm{const}$.

Both are $\Lk_0$ facts derivable from $N$ alone.
Not weakly reversible ($r_3$ has no return path: $c_3 \to \cdots \to c_2$
does not exist in $G(\Rx)$).

\noindent\textbf{Compatibility classes (Remark~\ref{rem:feinberg-levels}).}
The stoichiometric compatibility class of $\mathbf{x}_0$ is the
two-dimensional surface $\bigl(\mathbf{x}_0 + \Ssto\bigr) \cap
\RR^4_{>0}$, cut out by fixing $\bar{w}_1(\mathbf{x}) = e_0$ and
$\bar{w}_2(\mathbf{x}) = s_0$ for constants $e_0, s_0 > 0$.

\noindent\emph{Anticipating $\Lk_3$:}
Theorem~\ref{the:DZT1}(i) implies no positive steady state under
mass-action kinetics.
\end{example}

\begin{example}[Minimal weakly reversible network]
\label{ex:AB}
$A \overset{r_1}{\underset{r_2}{\rightleftharpoons}} B$,
$\Sp = \{A,B\}$, $n = 2$, $\ell = 1$, $\rho = 1$, $\delta = 0$.

For $w = (1,1)^\top$:
$h_w(r_1) = \bar{w}(B) - \bar{w}(A) = 1 - 1 = 0$,
$h_w(r_2) = \bar{w}(A) - \bar{w}(B) = 1 - 1 = 0$.
So $F_w \equiv 0$: $[A] + [B] = \mathrm{const}$ is the unique
conservation law.
Weakly reversible: $G(\Rx)$ is strongly connected.
The stoichiometric compatibility class of $\mathbf{x}_0$ is
$\{[A]+[B] = c\} \cap \RR^2_{>0}$ for $c = \bar{w}(\mathbf{x}_0) > 0$,
determined entirely at $\Lk_0$.

\noindent\emph{Anticipating $\Lk_3$:}
Theorem~\ref{the:DZT1}(ii) gives exactly one positive mass-action steady
state in each such class.
\end{example}


A single closed CRN models an isolated system.
The category $\mathbf{Petri}$ (Definition~\ref{def:Petri-cat}) allows
open systems to be built modularly and assembled by pushout.
A \emph{cospan} in a category $\mathbf{C}$ is a diagram
$A \xrightarrow{f} C \xleftarrow{g} B$: two morphisms sharing a common
target.
Two cospans with matching right/left boundaries compose by
\emph{pushout}: the universal construction that identifies the shared
boundary and takes the union of the bulk.
This is how Baez--Pollard~\cite{BaezPollard2017} compose open reaction
networks; we give the $\Lk_0$ version here.

\begin{definition}[Open CRN at $\Lk_0$]
\label{def:open}
Let $\Sigma_I, \Sigma_O \subseteq \Sp$ be finite sets of \emph{input}
and \emph{output interface species}: those shared with the environment.
Regard $D_I = (\Sigma_I, \emptyset, -, -)$ and
$D_O = (\Sigma_O, \emptyset, -, -)$ as discrete Petri nets.
An \emph{open CRN} at $\Lk_0$ is a cospan
\[
  D_I \;\xrightarrow{\;i\;}\; P \;\xleftarrow{\;o\;}\; D_O
\]
in $\mathbf{Petri}$.
Two open CRNs with $\Sigma_O = \Sigma_I'$ compose by pushout in
$\mathbf{Petri}$, identifying the output species of the first with the
input species of the second.
\end{definition}

\begin{proposition}[Stoichiometric structure under pushout]
\label{prop:pushout-N}
Let $P_1 = (\Sp_1, \Cx_1, \Rx_1, s_1, t_1)$ and
$P_2 = (\Sp_2, \Cx_2, \Rx_2, s_2, t_2)$ be CRNs assembled by pushout
in $\mathbf{Petri}$ along a discrete interface
$D = (\Sigma, \emptyset, -, -)$ with injective species inclusions
$f_{i,\Sp}: \Sigma \to \Sp_i$ ($i = 1, 2$), the standard open-CRN
setup of Definition~\ref{def:open}.
Then $P_{12} = P_1 \cup_D P_2$ has species set
$\Sp_{12} = \Sp_1 \cup_\Sigma \Sp_2$ (identified with
$\Sp_1 \cup \Sp_2$ via the inclusions) and reaction set
$\Rx_{12} = \Rx_1 \sqcup \Rx_2$.
\begin{enumerate}[label=(\roman*)]
  \item The stoichiometric matrix of $P_{12}$ is
    \[
      N_{12} \;=\; \bigl[\,N_1^{\mathrm{ext}} \;\big|\; N_2^{\mathrm{ext}}\,\bigr]
      \;\in\; \ZZ^{|\Sp_{12}| \times |\Rx_{12}|},
    \]
    where $N_i^{\mathrm{ext}}$ denotes $N_i$ zero-extended to have rows
    indexed by all of $\Sp_{12}$ (zero rows for species in
    $\Sp_{12} \setminus \Sp_i$).
  \item The stoichiometric subspace satisfies
    $\Ssto{}_{12} = \Ssto{}_1^{\mathrm{ext}} + \Ssto{}_2^{\mathrm{ext}}$,
    the sum of the extended subspaces in $\RR^{|\Sp_{12}|}$.
  \item A species potential $w_0: \Sp_{12} \to \RR$ satisfies
    $w_0 \in \ker(N_{12}^\top)$ --- equivalently, by
    Proposition~\ref{prop:conservation-functors}, the functor
    $F_w: \Lk_0(P_{12}) \to B\RR$ is identically zero --- if and only
    if both of the following hold:
    \[
      w_0\!\restriction_{\Sp_1} \in \ker(N_1^\top),
      \qquad
      w_0\!\restriction_{\Sp_2} \in \ker(N_2^\top),
    \]
    where $w_0\!\restriction_{\Sp_i}: \Sp_i \to \RR$ denotes the
    restriction of the single function $w_0$ to the coordinates
    indexed by $\Sp_i$.
    Note that both restrictions automatically agree on $\Sigma$ since
    they are restrictions of the same $w_0$.
\end{enumerate}
\end{proposition}

\begin{proof}
(i)~The generating morphisms of $\Lk_0(P_{12})$ are $\Rx_1 \sqcup \Rx_2$,
with $s_i(r)$ and $t_i(r)$ in $\NN[\Sp_i] \subset \NN[\Sp_{12}]$.
The $(j, r)$-entry of $N_{12}$ for $r \in \Rx_i$ is
$(t_i(r))_j - (s_i(r))_j$, which is $(N_i^{\mathrm{ext}})_{j,r}$.

(ii)~$\Ssto{}_{12} = \im_\RR(N_{12}) = \im_\RR(N_1^{\mathrm{ext}}) +
\im_\RR(N_2^{\mathrm{ext}}) = \Ssto{}_1^{\mathrm{ext}} +
\Ssto{}_2^{\mathrm{ext}}$.

(iii)~By Proposition~\ref{prop:conservation-functors} applied to $P_{12}$:
$F_w \equiv 0$ on $\Lk_0(P_{12})$ iff $h_w(r) = 0$ for all
$r \in \Rx_{12}$.
For $r \in \Rx_1$: $h_w(r) = \bar{w}(t_1(r)) - \bar{w}(s_1(r))
= (w_0\!\restriction_{\Sp_1})^\top(t_1(r) - s_1(r))$, which is zero for
all $r \in \Rx_1$ iff $w_0\!\restriction_{\Sp_1} \in \ker(N_1^\top)$,
and analogously for $r \in \Rx_2$.
Both conditions involve restrictions of the same $w_0$, which agrees on
$\Sigma$ by definition.
\end{proof}

\begin{remark}[Conservation laws are modular; deficiency is not]
\label{rem:modularity}
Proposition~\ref{prop:pushout-N}(iii) is a positive result: a species
potential $w_0$ on the assembled network is a conservation law if and
only if it is a conservation law for each sub-network independently.
This follows cleanly from Proposition~\ref{prop:conservation-functors}
applied to $P_{12}$, and is an instance of the tower language doing
genuine work.

Deficiency, by contrast, does not behave well under pushout.
Writing $\delta_{12} = n_{12} - \ell_{12} - \rho_{12}$: the number of
complexes satisfies $n_{12} = n_1 + n_2 - |\Cx_1 \cap \Cx_2|$ (shared
complexes are merged); the number of linkage classes satisfies
$\ell_{12} \leq \ell_1 + \ell_2$ with strict inequality when the
interface creates new connections; and $\rho_{12} = \rank(N_{12}) \leq
\rho_1 + \rho_2$ with strict inequality when interface species create
linear dependencies.
Each of these quantities can change in a way not determined by $\delta_1$
and $\delta_2$ alone, so $\delta_{12}$ is not controlled by $\delta_1 +
\delta_2$: two $\delta = 0$ sub-networks can assemble into a
$\delta > 0$ network, and vice versa.
In particular, the DZT conclusion --- unique stable steady state per
compatibility class --- does not compose under pushout.
\end{remark}

\begin{chembox}[Open networks as composable chemical systems]
An open CRN is a chemical system with designated \emph{inlet species}
($\Sigma_I$) and \emph{outlet species} ($\Sigma_O$).
Pushout composition connects two such systems in series.

Proposition~\ref{prop:pushout-N} tells us what $\Lk_0$ knows about
the assembly: the stoichiometric matrix is the column-concatenation of
the two sub-matrices, and a potential is conserved in the assembly if
and only if it is conserved by both sub-networks.
Conservation laws are therefore modular at $\Lk_0$.

What is not modular is deficiency: the assembly can create or destroy
linkage-class connections and stoichiometric dependencies in ways
$\delta_1$ and $\delta_2$ alone cannot predict.
The DZT therefore does not compose, and the kinetic composition
(gray-boxing) that Baez--Pollard~\cite{BaezPollard2017} develop at
$\Lk_3$ must handle this non-modularity directly.
\end{chembox}

%% file: chapters/L0/application.tex
\subsection{\texorpdfstring{$\Lk_0$}{Lk0} as a working tool}
\label{sec:L0-applications}

The previous sections established $\Lk_0(P)$ as a mathematical object.
This section demonstrates it as a working tool.
The three examples below correspond to the three main structural
components introduced above: \S\ref{sec:L0-construction} (morphism
calculus), \S\ref{sec:base-adjunction} (universal property), and
Definition~\ref{def:open} (open composition).
Each derives a non-trivial fact from $\Lk_0$ data alone --- no energy,
no rates, no geometry.

\begin{example}[The catalytic cycle as a single composite morphism]
\label{ex:MM-composite}
Consider the Michaelis--Menten network
\[
  E + S
  \;\underset{r_2}{\overset{r_1}{\rightleftharpoons}}\;
  ES
  \;\overset{r_3}{\longrightarrow}\;
  E + P,
\]
with $\Sp = \{E, S, ES, P\}$.
The three generating morphisms of $\Lk_0(P)$ are
$r_1: E{+}S \to ES$, $r_2: ES \to E{+}S$, and $r_3: ES \to E{+}P$.

\noindent\textbf{The catalytic cycle.}
The composite
\[
  r_3 \circ r_1 \;:\; E{+}S \;\longrightarrow\; E{+}P
\]
is a well-formed morphism in $\Lk_0(P)$, since $t(r_1) = ES = s(r_3)$.
It represents the full catalytic event --- substrate binding followed by
product release --- as a single process in the stoichiometric calculus.
This composite is a distinct morphism $E{+}S \to E{+}P$ from any
hypothetical direct reaction generator $r': E{+}S \to E{+}P$ that one
might add to the Petri net: no permutative axiom identifies a composite
of two generating reactions with a fresh generator, so $r_3 \circ r_1$
and $r'$ remain distinct morphisms in $\Lk_0(P)$.

\noindent\textbf{Running binding and release in parallel.}
The tensor product
\[
  r_1 \otimes r_2
  \;:\;
  (E{+}S) + ES \;\longrightarrow\; ES + (E{+}S)
\]
is a morphism representing simultaneous substrate binding in one part
of the system and enzyme--substrate dissociation in another.
It is not equal to $\id_{E+S+ES}$: the two processes happen at
different molecular sites and are kept as a non-trivial parallel process.

\noindent\textbf{A type error.}
The composite $r_3 \circ r_2$ is \emph{ill-typed}:
$t(r_2) = E{+}S$ while $s(r_3) = ES$, and $E{+}S \neq ES$ in
$\NN^{|\Sp|}$ since $E + S$ (the unbound enzyme--substrate pair) and $ES$
(the enzyme--substrate complex) are distinct species.
$\Lk_0(P)$ refuses to form this composite, encoding at the categorical
level the mechanistic fact that $ES$ must form before it can release
product: product release cannot follow dissociation.
\end{example}


\begin{example}[Atom balance via the imbalance functor]
\label{ex:atom-balance}
Chemical balancedness --- the requirement that every reaction
preserves atom counts --- is not an axiom of $\Lk_0(P)$.
It is expressed by a functor out of $\Lk_0(P)$: every Petri net,
balanced or not, admits an \emph{atom-imbalance functor} that
records the elemental discrepancy of each reaction, and balancedness
is the condition that this functor takes value zero on every generator.
The construction uses two successive applications of the universal property.

\noindent\textbf{Setup.}
Consider the hydrogen combustion reaction
$r: 2\,\mathrm{H_2} + \mathrm{O_2} \to 2\,\mathrm{H_2O}$,
with $\Sp = \{H_2, O_2, H_2O\}$.

\noindent\textbf{Step 1: species-level atom counts
(Universal Property of $\NN^{|\Sp|}$, Proposition~\ref{prop:FCM-UP}).}
Define the \emph{atomic composition map}
\[
  a: \Sp \;\longrightarrow\; \NN^2, \qquad
  a(H_2) = (2,0), \quad
  a(O_2) = (0,2), \quad
  a(H_2O) = (2,1),
\]
where the two coordinates record H and O atom counts.
By Proposition~\ref{prop:FCM-UP}, $a$ extends to a unique monoid
homomorphism $\bar{a}: \NN^{|\Sp|} \to \NN^2$, given by
$\bar{a}\!\bigl(\sum_S n_S \cdot S\bigr) = \sum_S n_S \cdot a(S)$.
Applied to the complexes of this reaction:
\begin{align*}
  \bar{a}(2\,H_2 + O_2) &= 2(2,0) + (0,2) = (4,2), \\
  \bar{a}(2\,H_2O)       &= 2(2,1)          = (4,2).
\end{align*}
The source and target complexes carry the same atom count.

\noindent\textbf{Step 2: the morphism-level extension
(Universal Property of $\Lk_0(P)$, Theorem~\ref{thm:UP-L0}).}
Let $B\ZZ^2$ be the one-object category with morphisms $\ZZ^2$ and
composition given by vector addition; this is a strict symmetric
monoidal category whose object monoid is trivially commutative
(one object), so Theorem~\ref{thm:UP-L0} applies.
Define the generator assignment $(a^{\mathrm{obj}}, h)$ by
\[
  a^{\mathrm{obj}}(S) := * \quad \text{for every } S \in \Sp,
  \qquad
  h(r) := \bar{a}(t(r)) - \bar{a}(s(r)) \in \ZZ^2,
\]
where $*$ is the unique object of $B\ZZ^2$.
The species-level atom counts $\bar{a}(S)$ enter only through the
morphism assignment $h(r)$, which records each reaction's
\emph{atom imbalance}; they are not objects of $B\ZZ^2$, which has
only one.
For the hydrogen combustion reaction,
$h(r) = (4,2) - (4,2) = (0,0)$.
Theorem~\ref{thm:UP-L0} provides a unique strict symmetric monoidal
functor
\[
  F_a: \Lk_0(P) \;\longrightarrow\; B\ZZ^2,
\]
the \emph{atom-imbalance functor} of the assignment $a$.

\noindent\textbf{What the functor asserts.}
Strict functoriality forces, for every morphism $f$ in $\Lk_0(P)$:
\begin{itemize}
  \item $F_a(\id_{\mathbf{u}}) = (0,0)$: the trivial process changes
    no atom counts.
  \item $F_a(g \circ f) = F_a(g) + F_a(f)$: atom changes add along
    sequential steps.
  \item $F_a(f \otimes g) = F_a(f) + F_a(g)$: atom changes add over
    parallel processes.
\end{itemize}
Since $F_a(r) = (0,0)$, every composite and tensor product of $r$ with
itself also maps to $(0,0)$.
Running a cascade of balanced reactions therefore produces zero net atom
change: additivity of atom counts along reaction sequences is a theorem,
not a postulate.

\noindent\textbf{Balancedness as a vanishing condition on $F_a$.}
For a general Petri net $P'$ and atom map
$a: \Sp' \to \NN^{|\mathsf{E}|}$, the imbalance functor
$F_a: \Lk_0(P') \to B\ZZ^{|\mathsf{E}|}$ \emph{always exists}: the
assignment $h(r) := \bar{a}(t(r)) - \bar{a}(s(r))$ is well-defined
for every generator regardless of whether atoms balance, and
Theorem~\ref{thm:UP-L0} extends it uniquely to all morphisms.
Balancedness is the condition that $F_a$ takes value zero on every
generator --- equivalently, by strict functoriality, that $F_a$ is
the zero functor on every morphism of $\Lk_0(P')$.
An unbalanced Petri net is a perfectly valid stoichiometric category
equipped with a non-zero imbalance functor; the values
$F_a(r) \neq 0$ record exactly which reactions fail to conserve atoms
and by how much.
Balancedness is therefore not a condition on the \emph{existence} of
$F_a$, but a vanishing condition on its values --- a condition on a
functor out of $\Lk_0(P')$, not an axiom of $\Lk_0$ itself.
\end{example}


\begin{example}[Open composition: assembling a network from sub-networks]
\label{ex:open-composition}
We decompose the Michaelis--Menten catalytic cycle into two modular
sub-networks and reassemble it via pushout in $\mathbf{Petri}$, in the
sense of Definition~\ref{def:open}.

\noindent\textbf{Sub-network 1 (binding and dissociation).}
$P_1 = (\{E, S, ES\},\; \{r_1, r_2\},\; s_1, t_1)$
with $r_1: E{+}S \to ES$ and $r_2: ES \to E{+}S$.
The \emph{output interface} is $D_O = (\{ES\}, \emptyset, -, -)$,
embedded by $o_1: D_O \to P_1$ sending $ES$ to $ES$.
The species $E$ and $S$ are treated as background, not exposed at the
interface.

\noindent\textbf{Sub-network 2 (catalysis).}
$P_2 = (\{ES, E, P\},\; \{r_3\},\; s_2, t_2)$
with $r_3: ES \to E{+}P$.
The \emph{input interface} is $D_I' = (\{ES\}, \emptyset, -, -)$,
embedded by $i_2: D_I' \to P_2$ sending $ES$ to $ES$.

\noindent\textbf{Pushout in $\mathbf{Petri}$.}
Both interfaces are $D = (\{ES\}, \emptyset, -, -)$, so we form the
pushout
\[
  P_1 \;\cup_{D}\; P_2
\]
in $\mathbf{Petri}$.
The pushout identifies the two copies of $ES$, takes the disjoint union
of the remaining species and all reaction labels, and gives
\[
  P_1 \cup_D P_2
  \;=\;
  (\{E, S, ES, P\},\; \{r_1, r_2, r_3\},\; s, t),
\]
the full Michaelis--Menten network of Example~\ref{ex:MM-composite}.
The free construction $\Lk_0: \mathbf{Petri} \to \mathbf{PermCat}$ is
a left adjoint and therefore preserves pushouts; the pushout in
$\mathbf{Petri}$ maps to the pushout in $\mathbf{PermCat}$,
\[
  \Lk_0(P_1 \cup_D P_2)
  \;\cong\;
  \Lk_0(P_1) \;+_{\Lk_0(D)}\; \Lk_0(P_2),
\]
under the hypotheses of Proposition~\ref{prop:pushout-N} on the
interface (discrete, injective on species, no reaction-label
identifications), all of which hold here.
Any strict symmetric monoidal functor out of the assembled category
into a strictly commutative target --- the atom-imbalance functor
of Example~\ref{ex:atom-balance}, or an enthalpy assignment at $\Lk_1$
--- decomposes into compatible sub-network assignments glued along the
shared interface, by the universal property of the pushout.

\noindent\textbf{What the decomposition gives.}
The composite morphism $r_3 \circ r_1: E{+}S \to E{+}P$ is visible in
the assembled category but not in either sub-category alone: $r_1$
belongs to $\Lk_0(P_1)$ and $r_3$ belongs to $\Lk_0(P_2)$, and only
the pushout assembles them into a composable pair.
This is the categorical content of modular network assembly: the
assembled category contains morphisms not present in any single module,
but these morphisms are entirely determined by the sub-network data
and the interface identification.
\end{example}

\begin{insightbox}[The limits of $\Lk_0$]
The three examples above exhaust what $\Lk_0$ can do: track
stoichiometric type, extend additive assignments from generators to
all processes via the universal property, and assemble networks from
modular parts via pushout.
What $\Lk_0$ cannot do is distinguish two reactions with the same source
and target by any numerical criterion, because $\Lk_0(P)$ carries no
numerical data as part of its structure.
The next section identifies this gap precisely and shows that
extending $\Lk_0$ by exactly one additional piece of data --- a functor
$F_H: \Lk_0(P) \to B\RR$ --- is the minimal and canonical next step.
\end{insightbox}

%% file: chapters/L0/forcing.tex
\subsection{What \texorpdfstring{$\Lk_0$}{Lk0} cannot express: the forcing of \texorpdfstring{$\Lk_1$}{Lk1}}
\label{sec:L0-forcing}

Section~\ref{sec:aut-exact} established, via the automorphism
sequence~\eqref{eq:aut-exact} at $k = 0$, that any two reaction
labels $r_1, r_2 \in \Rx$ with identical source and target complexes
--- whenever $\Lk_0(P)$ contains such a pair --- represent a class
$[r_1 \leftrightarrow r_2] \in \coker\varphi_1$ that $\Lk_0$
cannot break, forcing the existence of $\Lk_1$.
This subsection gives the chemical realisation of that abstract
argument using the nitrogen--oxygen--nitric oxide system, and then
develops the subtler content of $\Lk_1$ itself: the distinction
between the \emph{additivity} condition (satisfied by any strict
symmetric monoidal functor into $B\RR$) and the \emph{state-function}
condition (satisfied only by functors induced by an object potential).
The first is the categorical content of $\Lk_1$; the second is an
additional physical constraint that only some $\Lk_1$-functors
satisfy.

\begin{forcingbox}[The gap at $\Lk_0$: a worked chemical instance]
\label{box:forcing-L0}
\textbf{A single reaction.}
Consider the oxidation of nitrogen,
\[
  r: \mathrm{N_2(g)} + \mathrm{O_2(g)} \;\longrightarrow\;
  2\,\mathrm{NO(g)},
  \qquad \Delta H^\circ = +180.5\;\mathrm{kJ\,mol^{-1}}~\cite{Chase1998}.
\]
$\Lk_0(P)$ contains the generating morphism
$r : \mathbf{u} \to \mathbf{v}$ with $\mathbf{u} = \mathrm{N_2} +
\mathrm{O_2}$ and $\mathbf{v} = 2\,\mathrm{NO}$.
It does not contain the scalar $+180.5$: $\Lk_0(P)$ carries no
real-valued functor as part of its data, so there is no structure
at this level in which the enthalpy can reside.

\medskip
\noindent\textbf{Parallel routes --- the forcing instance.}
In practice, a reaction network frequently contains two distinct
reaction labels $r_1, r_2 \in \Rx$ sharing source and target
complexes:
\[
  s(r_1) = s(r_2) = \mathbf{u},
  \qquad
  t(r_1) = t(r_2) = \mathbf{v},
\]
for example, two experimentally identified pathways with the same
gross stoichiometry, or the same reaction studied under different
catalytic conditions.
Both appear as distinct generating morphisms in $\Lk_0(P)$
(Remark~\ref{rem:labels}); the free construction imposes no
identification on them.

Suppose these two reaction labels carry distinct enthalpies:
$\Delta H_1 \neq \Delta H_2$.
The label swap $r_1 \leftrightarrow r_2$ extends to an element of
$\Aut_{\mathrm{Petri}}(\Lk_0(P))$ --- specifically, a label
permutation within the fibre $(s, t)^{-1}(\mathbf{u}, \mathbf{v})$,
the second family identified in \S\ref{sec:aut-exact}.
It preserves every structure of $\Lk_0(P)$: source and target
complexes are unchanged, and no numerical data exists at $\Lk_0$
to be disturbed.
Yet this swap is not the restriction of any element of
$\Aut_{\mathrm{Petri}}(\Lk_1)$: every presentation-preserving
automorphism at $\Lk_1$ must commute with the enthalpy functor
$F_H : \Lk_0(P) \to B\RR$ whose existence defines $\Lk_1$, and
$F_H(r_1) \neq F_H(r_2)$ prevents the swap from preserving the
$\Lk_1$-structure.
By~\eqref{eq:coker-L0}, the swap represents a non-trivial class in
$\coker\varphi_1$.
The minimal extension resolving this conflation is to equip
$\Lk_0(P)$ with the functor $F_H$ itself: $\Lk_1$, developed
in Chapter~3.
\end{forcingbox}

\begin{remark}[The additivity condition: equipping $\Lk_0(P)$
  with $F_H$]
\label{rem:L1-additivity}
Passing to $\Lk_1$ means equipping $\Lk_0(P)$ with a strict
symmetric monoidal functor
\[
  F_H : \Lk_0(P) \;\longrightarrow\; B\RR.
\]
By Corollary~\ref{cor:BR}, such a functor is uniquely determined by
specifying one real number $F_H(r) \in \RR$ for each generating
reaction $r \in \Rx$; the values on composites and tensors are then
forced:
\[
  F_H(g \circ f) = F_H(g) + F_H(f),
  \qquad
  F_H(f \otimes g) = F_H(f) + F_H(g),
  \qquad
  F_H(\id_{\mathbf{u}}) = 0.
\]
These are the \emph{additivity conditions} on $F_H$.
The first is the additivity form of Hess's Law --- enthalpy changes
add along sequential reaction steps; the second is additivity over
independent parallel processes.
Both follow from functoriality, not from any separate physical
postulate.

In terms of the automorphism sequence, equipping $\Lk_0(P)$ with $F_H$
breaks the spurious label-swap symmetry in $\coker\varphi_1$: once
distinct real values $F_H(r_1) \neq F_H(r_2)$ are assigned, the swap
$r_1 \leftrightarrow r_2$ no longer preserves structure, and the two
reaction labels are numerically distinguishable.

The values $\{F_H(r)\}_{r \in \Rx}$ are free parameters: any
assignment of real numbers to reaction generators extends to a valid
strict symmetric monoidal functor.
This is the \emph{additivity condition} for $\Lk_1$: it adds
exactly one real parameter per reaction label, and no other
structure.
\end{remark}

\begin{remark}[The state-function condition]
\label{rem:L1-state-function}
The physical fact that enthalpy is a \emph{state function} --- that
the enthalpy change of a reaction depends only on its initial and
final complex, not on the reaction label --- is strictly stronger
than the additivity condition of Remark~\ref{rem:L1-additivity}.
It requires $F_H$ to be induced by an \emph{object potential}: a
monoid homomorphism
$h : (\NN^{|\Sp|}, +, \mathbf{0}) \to (\RR, +, 0)$ satisfying
\[
  F_H(r) \;=\; h(t(r)) - h(s(r))
  \qquad \text{for every generating } r \in \Rx.
\]
By the universal property of $\NN^{|\Sp|}$ (Proposition~\ref{prop:FCM-UP}),
any function $h_0 : \Sp \to \RR$ --- one real value per species,
typically the standard enthalpy of formation $\Delta H^\circ_f(S)$
--- extends to a unique such $h$, and then to a valid $F_H$ via
Corollary~\ref{cor:BR}.
When such an $h$ exists, the enthalpy change of any morphism
$f: \mathbf{u} \to \mathbf{v}$ (whether a single generator, a
composite, or a tensor product) equals $h(\mathbf{v}) -
h(\mathbf{u})$, depending only on the source and target and not on
the particular morphism.
Path independence then holds automatically: for any two morphisms
$f, g : \mathbf{u} \to \mathbf{v}$,
$F_H(f) = h(\mathbf{v}) - h(\mathbf{u}) = F_H(g)$.

\medskip
\noindent\textbf{The two conditions are distinct.}
The additivity condition (Remark~\ref{rem:L1-additivity}) provides
a real number per reaction label, allows
$F_H(r_1) \neq F_H(r_2)$ for two labels with identical source and
target, and is the categorical content of $\Lk_1$.
The state-function condition imposes $F_H(r_1) = F_H(r_2)$ whenever
$s(r_1) = s(r_2)$ and $t(r_1) = t(r_2)$, by requiring $h$ to be
well-defined on complexes, but \emph{does not} identify $r_1$ with
$r_2$ as morphisms in the category: the two reaction labels remain
distinct elements of $\Rx$, they simply receive the same enthalpy
value.
The state-function condition is therefore a constraint on the
particular functor $F_H$, not a relation imposed on $\Lk_0(P)$
itself.

An $F_H$ satisfying only the additivity condition but not the
state-function condition is a perfectly valid strict symmetric
monoidal functor, representing a situation in which the same gross
stoichiometry is associated with distinct energetic signatures
across different realisations (e.g., different catalysts, different
solvents, or mechanistically distinct pathways).
Both conditions are of physical interest, and the distinction
between them is central to how $\Lk_1$ and $\Lk_2$ differ.
\end{remark}

\begin{remark}[Bridge to $\Lk_1$]
\label{rem:bridge-L1}
The pair $(\Lk_0(P), F_H)$, where $F_H : \Lk_0(P) \to B\RR$ is a
strict symmetric monoidal functor satisfying the additivity
condition, is the $\Lk_1$-level data associated with $P$.
The complete development of $\Lk_1$ --- its relationship to
equilibrium thermodynamics via the entropy functor $F_S$, the
$\dagger$-structure of $\Lk_2$ encoding detailed balance, and the
Wegscheider cycle conditions for rate constants --- is given in
Chapter~3.
\end{remark}

%% file: chapters/ch_L1.tex
\section{\texorpdfstring{$\Lk_1$}{Lk1}: The Thermochemical Level}
\label{sec:L1}

\input{chapters/L1/l1_forcing}
\input{chapters/L1/l1_define}
\input{chapters/L1/l1_layer1}
\input{chapters/L1/l1_layer2}
\input{chapters/L1/l1_examples}
\input{chapters/L1/l1_weg}
\input{chapters/L1/l1_forcing_out}

%% file: chapters/L1/l1_forcing.tex
\subsection{The forcing of \texorpdfstring{$\Lk_1$}{Lk1}}
\label{sec:L1-forcing-in}

Section~\ref{sec:L0-forcing} exhibited the concrete $\Lk_0 \to
\Lk_1$ forcing instance: two reaction labels
$r_1, r_2 \in \Rx$ sharing source and target complexes (the
$\mathrm{N_2} + \mathrm{O_2} \to 2\,\mathrm{NO}$ system of
Forcing~Box~\ref{box:forcing-L0}), and the label-permutation within
the fibre $(s, t)^{-1}(\mathbf{u}, \mathbf{v})$ representing a
non-trivial class in $\coker\varphi_1$.
This section picks up the structural answer: \emph{what exactly
is the minimal extension $\Lk_1$ that resolves this class?}
After a brief complementary illustration, three physical
observations constrain the answer to a single categorical object:
a strict symmetric monoidal functor
$\FH : \Lk_0(P) \to B\RR$.

\begin{forcingbox}[A complementary forcing instance for $\Lk_1$]
\label{box:forcing-L1-in}
The \S\ref{sec:L0-forcing} forcing instance used two distinct
reaction labels in the same Petri net, distinguishable by having
different enthalpies.
A second, pedagogically minimal manifestation of the same
$\coker\varphi_1$ non-triviality is the following.

Let $P$ be any Petri net, and consider two \emph{thermochemical
systems} built on it:
\[
  \bigl(\Lk_0(P),\, F_H^{(1)}\bigr)
  \qquad\text{and}\qquad
  \bigl(\Lk_0(P),\, F_H^{(2)}\bigr),
\]
with $F_H^{(1)}(r) \neq F_H^{(2)}(r)$ on at least one generating
reaction $r$ --- the minimal instance being a single-reaction
Petri net $P = (\{A, B\}, \{r\}, s, t)$ with
$r : A \to B$, and $F_H^{(1)}(r) = +50$ versus $F_H^{(2)}(r) = -50$
(in $\mathrm{kJ\,mol^{-1}}$, chosen schematically).
Both decorated pairs have the same image under the forgetful
operation $U_1$ (Definition~\ref{def:L1}): dropping the
thermochemical datum yields the same $\Lk_0(P)$ in either case.

At $\Lk_1$, however, the two decorated pairs are genuinely
distinct: in this minimal example
$\Aut_{\mathrm{Petri}}(\Lk_0(P))$ is trivial, so the only candidate
intertwining $\alpha: \Lk_0(P) \to \Lk_0(P)$ is the identity, and
$F_H^{(2)} \circ \id = F_H^{(1)}$ would force
$F_H^{(2)}(r) = F_H^{(1)}(r)$ --- contradicting the choice.
By the automorphism sequence of \S\ref{sec:aut-exact}, this
separation at $\Lk_1$ (invisible at $\Lk_0$) is precisely what
non-triviality of $\coker\varphi_1$ records: the forgetful
operation $U_1$ is not injective on decorated pairs --- the
underlying $\Lk_0(P)$ alone does not determine $\FH$.
Distinguishing the two pairs requires at least the
$\Lk_1$-decoration data.
\end{forcingbox}

\medskip
\noindent
The \S\ref{sec:L0-forcing} fibre-permutation instance and the
complementary instance above exhibit the same underlying fact --- 
$\coker\varphi_1$ is non-trivial --- from two different directions:
the former fixes a single Petri net with two distinct labels and
asks which automorphism of the $\Lk_0$-presentation fails to lift;
the latter fixes a single reaction label and asks which pair of
$\Lk_1$-decorated objects projects to the same $\Lk_0$-object.
Both confirm that $\Lk_0$ cannot distinguish endothermic from
exothermic realisations, and that a real-valued datum on reactions
is the minimal structure needed to break this ambiguity.

\medskip
\noindent\textbf{What structure resolves the ambiguity?}
Before writing any mathematics, three physical observations
constrain the answer.

\begin{enumerate}[label=(\alph*)]
  \item \textbf{Sequential additivity (Hess's Law).}
    If reaction $r_1$ is followed by $r_2$, the total enthalpy
    change is $\dH(r_2 \circ r_1) = \dH(r_1) + \dH(r_2)$.
    The label must be compatible with categorical composition.
  \item \textbf{Parallel additivity.}
    If $r_1$ and $r_2$ proceed independently,
    $\dH(r_1 \otimes r_2) = \dH(r_1) + \dH(r_2)$.
    The label must be compatible with the monoidal product.
  \item \textbf{No spurious energy.}
    $\dH(\id_{\bf u}) = 0$ for every complex $\mathbf{u}$ ---
    unitality.
\end{enumerate}

\noindent
These three constraints specify a strict symmetric monoidal
functor
\[
  \FH : \Lk_0(P) \;\longrightarrow\; B\RR
\]
--- the one-object additive category $B\RR$ introduced in
\S\ref{sec:cat-language}.
The fit between the three physical axioms and this categorical
structure is exact in both directions.

\begin{itemize}
  \item \emph{No weaker structure suffices.}
    A plain (non-monoidal) functor satisfies (a) but misses (b):
    functoriality constrains composition, not the tensor product.
    A lax monoidal functor weakens (b) to
    \[
      \FH(r_1 \otimes r_2)
      \;=\;
      \FH(r_1) + \FH(r_2) + \phi_{r_1, r_2},
    \]
    introducing a coherence term $\phi_{r_1, r_2} \in \RR$ absent
    from any physical enthalpy assignment.
    Dropping the unit condition loses (c).
    Each weakening fails at least one of the three physical axioms.

  \item \emph{No stronger structure is needed.}
    The three axioms fix $\FH$ uniquely on all morphisms given its
    values on generators (Proposition~\ref{prop:FH-unique} in
    \S\ref{sec:L1-def}).
    Any further structure --- requiring $\FH$ to be a monoidal
    equivalence, an adjunction, or a faithful functor --- would
    impose conditions (surjectivity, invertibility, injectivity on
    morphisms) with no physical interpretation for an enthalpy
    assignment.
\end{itemize}

\medskip
\noindent
The extension presented here is the minimal categorical object
resolving the $\coker\varphi_1$ class identified at $\Lk_0$:
a decorator functor adjoined to $\Lk_0(P)$ without modifying its
underlying category.
That this is the \emph{unique} minimal extension resolving the
cokernel is claimed constructively at the present level, not as a
general theorem about categorical extensions; the full taxonomy of
the six extension types that occur across the tower is recorded in
\S\ref{sec:L7-retrospective}.
The next subsection makes the construction precise.

%% file: chapters/L1/l1_define.tex
\subsection{Definition of \texorpdfstring{$\Lk_1(P)$}{Lk1P}}
\label{sec:L1-def}

\begin{definition}[Thermochemical level $\Lk_1(P)$]
\label{def:L1}
  Let $P = (\Sp, \Rx, s, t)$ be a Petri net.
  The \emph{thermochemical level} of $P$ is the pair
  \[
    \Lk_1(P) \;:=\; \bigl(\,\Lk_0(P),\;\FH\,\bigr),
  \]
  where $\FH: \Lk_0(P) \to B\RR$ is a strict symmetric monoidal
  functor from the stoichiometric category $\Lk_0(P)$ into $B\RR$.
  The notation $\Lk_1(P)$ refers to this pair --- a decorated
  structure, not itself a category.
  We write $U_1$ for the \emph{forgetful operation}
  \[
    U_1\bigl(\Lk_0(P), \FH\bigr) \;:=\; \Lk_0(P),
  \]
  which returns the underlying stoichiometric category from the
  decorated pair, dropping $\FH$.
\end{definition}

\begin{mathbox}[Unpacking Definition~\ref{def:L1}]
The domain of $\FH$ is $\Lk_0(P)$, the free skeletal permutative
category already constructed in Section~\ref{sec:L0}.
The pair $(\Lk_0(P), \FH)$ is $\Lk_1(P)$; it is emphatically
\emph{not} the case that $\FH$ maps $\Lk_1(P)$ to $B\RR$ (that
would be circular --- $\Lk_1(P)$ is the decorated pair, not a
category in its own right).

The target $B\RR$ is a one-object category, so its object monoid
$\{\ast\}$ is trivially strictly commutative; this is exactly the
hypothesis on the target required by Theorem~\ref{thm:UP-L0}, and
it is satisfied here without further work.
The action of $\FH$ is therefore entirely determined by its action
on morphisms:
\begin{itemize}
  \item Every object ${\bf u} \in \Lk_0(P)$ maps to the single
    object $\ast$.
  \item Every morphism $r: {\bf u} \to {\bf v}$ in $\Lk_0(P)$ maps
    to a real number $\FH(r) \in \RR$.
\end{itemize}
The ``strict symmetric monoidal'' conditions then read:
\begin{itemize}
  \item \textbf{Functoriality:}
    $\FH(r_2 \circ r_1) = \FH(r_2) + \FH(r_1)$.
  \item \textbf{Monoidality:}
    $\FH(r_1 \otimes r_2) = \FH(r_1) + \FH(r_2)$.
  \item \textbf{Unitality:}
    $\FH(\id_{\bf u}) = 0$ for all ${\bf u}$.
\end{itemize}
The adjective \emph{strict} means that the coherence isomorphisms
of a monoidal functor---the natural transformations
$\phi_{A,B}: \FH(A) \otimes \FH(B) \to \FH(A \otimes B)$
and $\phi_0: I \to \FH(I)$---are required to be
\emph{identity morphisms}, not merely natural isomorphisms
\cite[Ch.~XI, \S1]{MacLane1998}.
In $B\RR$ each $\phi_{A,B}$ is an endomorphism of the unique object
$\ast$, i.e.\ a real number; strictness requires it to equal $0$
(the identity in $B\RR$), which is precisely the
monoidality equation above.
These three equations are the \emph{complete content} of the
thermochemical decoration.
\end{mathbox}

The existence and uniqueness of $\FH$ given its values on generating
reactions follows directly from the universal property of $\Lk_0(P)$.

\begin{proposition}[Existence and uniqueness of $\FH$]
\label{prop:FH-unique}
  Let $P$ be a Petri net with reaction set $\Rx$.
  Given any assignment $\dH_0: \Rx \to \RR$, there is a unique strict
  symmetric monoidal functor
  $\FH: \Lk_0(P) \to B\RR$
  satisfying $\FH(r) = \dH_0(r)$ for every generating reaction
  $r \in \Rx$.
\end{proposition}

\begin{proof}
By Theorem~\ref{thm:UP-L0}, $\Lk_0(P)$ is the free skeletal
permutative category on $P$: for any strict symmetric monoidal
category $\mathcal{C}$ \emph{whose object monoid is strictly
commutative} and any morphism-assignment
$f: \Rx \to \mathrm{Mor}(\mathcal{C})$ compatible with source/target
types, there is a unique strict symmetric monoidal functor
$\bar{f}: \Lk_0(P) \to \mathcal{C}$ extending $f$
\cite[Ch.~XI, \S3]{MacLane1998}.

Take $\mathcal{C} = B\RR$, the one-object category with morphisms
$\RR$ and composition given by addition.
Its object monoid is a singleton, so trivially strictly commutative
--- the hypothesis of Theorem~\ref{thm:UP-L0} is satisfied.
The source/target compatibility condition on $f = \dH_0$ is also
trivial since $B\RR$ has only one object.
The unique extension $\bar{f} = \FH$ satisfies all three conditions
of Definition~\ref{def:L1} by construction.
\end{proof}

\begin{insightbox}[One real number per reaction]
Proposition~\ref{prop:FH-unique} says: the thermochemical content
of a network at $\Lk_1$ is determined by \emph{exactly one real
number per chemical generator}.
The reason is structural: every morphism of $\Lk_0(P)$ is a finite
composite and tensor product of chemical generators $r \in \Rx$
together with structural morphisms (identities and symmetries
$\sigma_{\mathbf{u},\mathbf{v}}$).
Strictness forces $\FH(\id_{\bf u}) = 0$ and
$\FH(\sigma_{\mathbf{u},\mathbf{v}}) = 0$ (the symmetry of $B\RR$
is trivial because $B\RR$ has one object), so for any morphism
$f = r_{i_k} \circ \cdots \circ (r_{i_1} \otimes r_{i_2})
\circ \cdots$:
\[
  \FH(f) \;=\; \FH(r_{i_k}) + \cdots + \FH(r_{i_1}) + \FH(r_{i_2})
  + \cdots
\]
$\FH(f)$ is a sum of the chemical-generator values $\dH_0(r_i)$,
with multiplicities determined by how many times each generator
appears in $f$.
The values $\{\dH_0(r)\}_{r \in \Rx}$ therefore determine $\FH$
completely and uniquely; no further measurement is needed.
\end{insightbox}

%% file: chapters/L1/l1_layer1.tex
\subsection{Layer~1: the generic additive functor}
\label{sec:L1-layer1}

The functor $\FH$ of Definition~\ref{def:L1} as stated ---
one free real parameter per generating reaction --- constitutes
\emph{Layer~1} of the thermochemical structure.
The following theorem encodes Hess's Law and its monoidal
counterpart as the two defining axioms of $\FH$.
The physical content of Hess's Law has been understood since 1840
\cite{Hess1840}; its formulation as the functoriality condition of a
strict monoidal functor into $B\RR$ is, to the authors' knowledge,
original to this work.

\begin{theorem}[Hess's Law and parallel additivity --- Layer~1]
\label{thm:hess}
  Let $\FH: \Lk_0(P) \to B\RR$ be the thermochemical functor
  (Layer~1 data).
  \begin{enumerate}[label=(\roman*)]
    \item \textup{(Sequential additivity / Hess's Law)}
      For composable reactions
      $r_1: {\bf u} \to {\bf v}$, $r_2: {\bf v} \to {\bf w}$:
      \[
        \FH(r_2 \circ r_1) \;=\; \FH(r_1) + \FH(r_2).
      \]
    \item \textup{(Parallel additivity)}
      For any reactions $r_1, r_2$:
      \[
        \FH(r_1 \otimes r_2) \;=\; \FH(r_1) + \FH(r_2).
      \]
    \item \textup{(No spurious energy)}
      $\FH(\id_{\bf u}) = 0$ for every complex ${\bf u}$.
  \end{enumerate}
\end{theorem}

\begin{proof}
Parts (i)--(iii) are the functoriality, monoidality, and unitality
axioms of $\FH$ respectively.
They hold by definition of a strict symmetric monoidal functor;
the content of the theorem is that such an $\FH$ exists and is
unique given values on generators, as established in
Proposition~\ref{prop:FH-unique}.
\end{proof}

\begin{remark}[Two postulates, one equation]
Traditional thermochemistry textbooks state Hess's Law (part (i))
and parallel additivity (part (ii)) as two separate empirical
postulates \cite{atkins2023physical}.
Categorically, both are instances of the single requirement that
$\FH$ is a monoidal functor: sequential additivity is functoriality,
parallel additivity is monoidality.
The classical textbook presentation reflects the same mathematical
structure, stated in the language of experiments rather than of
categories.

Physically, the full \emph{path-independence} form of Hess's Law
--- the enthalpy of a process depends only on initial and final
complexes, regardless of route --- is classically derived from the
first law together with the identification of enthalpy as a state
function \cite{AtkinsDeP2014}.
Theorem~\ref{thm:hess} captures the additive content of Hess's
Law (along sequential composites and across parallel processes)
directly from functoriality, without invoking the state-function
identification.
The path-independence form proper requires the Layer~2 coboundary
condition $\FH = \delta^0 h$ developed in \S\ref{sec:L1-layer2}.
\end{remark}

\medskip
\noindent\textbf{What Layer~1 does not enforce: the cycle condition.}
Theorem~\ref{thm:hess} guarantees that $\FH$ is additive along
any composable sequence of reactions and across any parallel
combination.
A natural further expectation is that traversing a closed loop of
reactions should accumulate zero net enthalpy --- the physical
content of ``enthalpy is a state function.''
This is \emph{not} a consequence of Layer~1 alone; it requires
an additional constraint.
The following warning, drawn from a real carbon-combustion cycle,
makes this precise.

\begin{warning}[Cycle non-closure is consistent with Layer~1]
\label{warn:cycle}
Consider the carbon--oxygen system
$\Sp = \{\mathrm{C(s)},\, \mathrm{O_2(g)},\, \mathrm{CO(g)},\,
\mathrm{CO_2(g)}\}$
with three generating reactions forming a directed cycle:
\begin{align*}
  r_1 &:\;
    \mathrm{C(s)} + \tfrac{1}{2}\mathrm{O_2(g)} \to \mathrm{CO(g)},
  \\
  r_2 &:\;
    \mathrm{CO(g)} + \tfrac{1}{2}\mathrm{O_2(g)} \to \mathrm{CO_2(g)},
  \\
  r_3 &:\;
    \mathrm{CO_2(g)} \to \mathrm{C(s)} + \mathrm{O_2(g)}.
\end{align*}
The thermochemically correct values are
$\FH(r_1) = -110.5$, $\FH(r_2) = -283.0$,
$\FH(r_3) = +393.5$ (all in kJ/mol) \cite{cox1989codata}.
Composing $r_2$ with $r_1$ requires a spectator
$\tfrac{1}{2}\mathrm{O_2}$ to bridge the source--target gap, so
the directed cycle from $\mathrm{C(s)} + \mathrm{O_2(g)}$ back to
itself is
$r_3 \circ r_2 \circ \bigl(r_1 \otimes \id_{\tfrac{1}{2}\mathrm{O_2}}\bigr)$.
Functoriality, monoidality, and $\FH(\id) = 0$ give a cycle sum
of exactly zero:
\[
  \FH\!\Bigl(r_3 \circ r_2 \circ \bigl(r_1 \otimes
  \id_{\tfrac{1}{2}\mathrm{O_2}}\bigr)\Bigr)
  \;=\; \FH(r_1) + \FH(r_2) + \FH(r_3)
  \;=\; 0.
\]
Now suppose, however, that one assigns the hypothetically inconsistent
value $\FH(r_3) = +380.0\;\mathrm{kJ/mol}$ (as might arise from a
calibration error or an inconsistent reference state).
Functoriality (Layer~1) then gives:
\[
  \FH\!\Bigl(r_3 \circ r_2 \circ \bigl(r_1 \otimes
  \id_{\tfrac{1}{2}\mathrm{O_2}}\bigr)\Bigr)
  \;=\; \FH(r_1) + \FH(r_2) + \FH(r_3)
  \;\neq\; 0.
\]

This assignment satisfies all three axioms of Layer~1 --- Hess's
Law for composable morphisms, parallel additivity, and no spurious
energy on identity morphisms --- yet the directed cycle does not
sum to zero.
Layer~1 is blind to this inconsistency.
The cycle condition $\sum_i \FH(r_i) = 0$ is an independent
constraint; it is \emph{not} derivable from the functor axioms alone.
\end{warning}

\begin{mathbox}[Why the free category is the right domain]
In $\Lk_0(P)$, every type-compatible composite of chemical
generators yields a distinct morphism.
The only equalities that hold are those forced by the axioms of a
skeletal permutative category (associativity, unit laws,
naturality and involutivity of the symmetry endomorphisms
$\sigma_{\mathbf{u},\mathbf{v}}$) --- \emph{never} any equation
among chemical generators.
In particular,
$r_3 \circ r_2 \circ \bigl(r_1 \otimes \id_{\tfrac{1}{2}\mathrm{O_2}}\bigr)
\neq \id_{\mathrm{C(s)} + \mathrm{O_2(g)}}$ in the free category,
even though the composite's source and target both equal
$\mathrm{C(s)} + \mathrm{O_2(g)}$: the free category has no way to
know that going around the cycle ``should'' be trivial.
The physical interpretation is immediate: a thermochemical
assignment that satisfies Layer~1 but not Layer~2 is the
mathematical model of a calorimetric dataset corrupted by
inconsistent reference states or measurement errors.
Layer~2, introduced in Section~\ref{sec:L1-layer2}, is precisely
the condition that eliminates all such inconsistencies.
\end{mathbox}

%% file: chapters/L1/l1_layer2.tex
\subsection{Layer~2: the state-function condition}
\label{sec:L1-layer2}

Warning~\ref{warn:cycle} shows that Layer~1 alone permits
thermochemical cycles to accumulate non-zero enthalpy, in conflict
with the first law of thermodynamics.
The culprit is that Layer~1 assigns a real number to each
\emph{reaction} independently; it imposes no relationship between
the enthalpy of a reaction and any intrinsic property of the
\emph{complexes} it connects.
The first law demands more: enthalpy must be a \emph{state
function}, meaning its value for any process depends only on the
initial and final states, not on the path taken.
Mathematically, this means the enthalpy assignment must be
expressible as a \emph{difference of potentials} defined on the
complexes themselves.
Layer~2 is the additional datum that enforces this.

\begin{definition}[Layer~2: state-function / coboundary condition]
\label{def:L1-layer2}
A thermochemical functor $\FH: \Lk_0(P) \to B\RR$ satisfies the
\emph{state-function condition} (Layer~2) if there exists a monoid
homomorphism
\[
  h \;:\; \bigl(\Mon,+,\mathbf{0}\bigr)
    \longrightarrow
    \bigl(\RR,+,0\bigr)
\]
such that, for every generating reaction $r:{\bf u}\to{\bf v}$,
\[
  \FH(r) \;=\; h({\bf v}) - h({\bf u}).
\]
We write this condition as $\FH=\delta^0 h$ and call $h$ an
\emph{object potential} (or \emph{additive potential}) for $\FH$.
\end{definition}

The notation $\delta^0$ and the term \emph{object potential} require
some comments.
The map $h$ assigns a real number (a ``height'' or ``potential'')
to each complex in $\Mon$, and the identity
$\FH(r)=h({\bf v})-h({\bf u})$ says that the enthalpy of a generating
reaction is the difference of the endpoint potentials.
This is the discrete analogue of a conservative field on a graph:
individual reactions are directed edges, while sequences of reactions
are directed paths.
For an exact edge-labeling, the total enthalpy along a directed path
depends only on its endpoints, and hence vanishes on directed cycles
\cite{frankel2004geometry,soardi2006potential}.

The symbol $\delta^0$ is borrowed from graph cohomology.

\begin{mathbox}[Coboundary notation]
To explain the notation precisely, let
\[
  G_P=(V_P,E_P)
\]
be the directed reaction graph of the network, where
\[
  V_P:=s(\Rx)\cup t(\Rx)\subseteq \Mon,
  \qquad
  E_P:=\Rx,
\]
and $s,t:\Rx\to\Mon$ are the source and target maps.
For the reaction graph $G_P$, the standard graph-cochain spaces are
\[
  \widetilde C^0(G_P;\RR)
  := \{\phi:V_P\to\RR\},
  \qquad
  \widetilde C^1(G_P;\RR)
  := \{\omega:\Rx\to\RR\}.
\]
Thus a $0$-cochain is a real-valued function on vertices and a
$1$-cochain is a real-valued function on directed edges
\cite{lim2020hodge}.
The standard degree-zero graph coboundary is
\[
  \widetilde\delta^0:
  \widetilde C^0(G_P;\RR)\to \widetilde C^1(G_P;\RR),
  \qquad
  (\widetilde\delta^0\phi)(r)
  := \phi\bigl(t(r)\bigr)-\phi\bigl(s(r)\bigr).
\]
This is the graph-theoretic analogue of taking the gradient of a
scalar potential \cite{lim2020hodge}.

In the present thermochemical setting we impose the stronger
requirement that the potential extend additively to all complexes.
Accordingly, we define
\[
  C^0 := \Hom_{\mathbf{Mon}}(\Mon,\RR),
  \qquad
  C^1 := \Hom_{\mathbf{Set}}(\Rx,\RR).
\]
Restriction to the vertex set gives a map
\[
  \mathrm{res}: C^0 \to \widetilde C^0(G_P;\RR),
  \qquad
  h \mapsto h|_{V_P},
\]
and we define
\[
  \delta^0 : C^0 \to C^1
\]
by the same endpoint-difference formula,
\[
  (\delta^0 h)(r)
  \;:=\;
  h\bigl(t(r)\bigr)-h\bigl(s(r)\bigr).
\]
Thus $\delta^0$ is the usual graph coboundary, restricted to the
subspace of additive potentials on complexes.

Since $\mathbb N[\Sp]$ is the free commutative monoid on the
species set $\Sp$, every monoid homomorphism $h:\Mon\to\RR$ is uniquely
determined by its values on species:
\[
  h\!\left(\sum_{S\in\Sp} n_S S\right)
  \;=\;
  \sum_{S\in\Sp} n_S\,h(S).
\]
Hence an additive object potential has at most $|\Sp|$ free parameters,
one for each species.
\end{mathbox}

The state-function condition admits two equivalent algebraic
formulations, together with two immediate consequences.
The equivalence below is the additive-monoid form of Hess's law:
reaction enthalpies arise from species potentials exactly when they are
linear in the net stoichiometric change.
Path-independence and cycle-vanishing then follow by telescoping
\cite{frankel2004geometry,soardi2006potential}.

\begin{proposition}[Equivalent algebraic formulations of Layer~2]
\label{prop:L1-equiv}
Let $\FH:\Lk_0(P)\to B\RR$ be a Layer~1 functor.
The following are equivalent:
\begin{enumerate}[label=(\roman*)]
  \item \textup{(Coboundary)} There exists an object potential
    $h\in C^0$ such that
    \[
      \FH=\delta^0 h.
    \]

  \item \textup{(Species-potential form)} There exist real numbers
    $\eta_S\in\RR$ for each species $S\in\Sp$ such that, for every
    generating reaction $r:{\bf u}\to{\bf v}$,
    \[
      \FH(r)
      \;=\;
      \sum_{S\in\Sp} N_{S,r}\,\eta_S,
    \]
    where
    \[
      N_{S,r}
      := \nu_S^{({\bf v})}-\nu_S^{({\bf u})}
    \]
    is the net stoichiometric coefficient of species $S$ in reaction
    $r$.
\end{enumerate}

Moreover, either of these equivalent conditions implies:

\begin{enumerate}[label=(\roman*),resume]
  \item \textup{(Path independence)} For any two directed paths of
    generating reactions from ${\bf u}$ to ${\bf v}$,
    \[
      p:r_1,\dots,r_m:{\bf u}\leadsto{\bf v},
      \qquad
      p':r'_1,\dots,r'_n:{\bf u}\leadsto{\bf v},
    \]
    one has
    \[
      \sum_{i=1}^m \FH(r_i)
      \;=\;
      \sum_{j=1}^n \FH(r'_j).
    \]

  \item \textup{(Directed-cycle condition)} For every directed closed
    loop of generating reactions
    \[
      r_1:{\bf u}_0\to{\bf u}_1,\;
      r_2:{\bf u}_1\to{\bf u}_2,\;\ldots,\;
      r_k:{\bf u}_{k-1}\to{\bf u}_0,
    \]
    one has
    \[
      \sum_{i=1}^k \FH(r_i)=0.
    \]
\end{enumerate}
\end{proposition}

\begin{proof}
\textbf{(i) $\Rightarrow$ (ii).}
Set $\eta_S:=h(S)$ for each $S\in\Sp$.
Since $h$ is a monoid homomorphism,
\[
  h({\bf u})=\sum_{S\in\Sp}\nu_S^{({\bf u})}\eta_S,
  \qquad
  h({\bf v})=\sum_{S\in\Sp}\nu_S^{({\bf v})}\eta_S.
\]
Therefore
\begin{align*}
  \FH(r)
  &= h({\bf v})-h({\bf u}) = \sum_{S\in\Sp}
     \bigl(\nu_S^{({\bf v})}-\nu_S^{({\bf u})}\bigr)\eta_S \\
  &= \sum_{S\in\Sp} N_{S,r}\eta_S.
\end{align*}

\textbf{(ii) $\Rightarrow$ (i).}
Define
\[
  h\!\left(\sum_{S\in\Sp} n_S S\right)
  := \sum_{S\in\Sp} n_S\,\eta_S.
\]
Since $\Mon=\mathbb N[\Sp]$ is free commutative on $\Sp$, this defines
a monoid homomorphism $h:\Mon\to\RR$.
For any generating reaction $r:{\bf u}\to{\bf v}$,
\[
  h({\bf v})-h({\bf u})
  = \sum_{S\in\Sp}
    \bigl(\nu_S^{({\bf v})}-\nu_S^{({\bf u})}\bigr)\eta_S
  = \FH(r),
\]
so $\FH=\delta^0 h$.

\textbf{(i) $\Rightarrow$ (iii).}
If
\[
  p:r_1,\dots,r_m:{\bf u}\leadsto{\bf v},
\]
then
\[
  \sum_{i=1}^m \FH(r_i)
  = \sum_{i=1}^m \bigl[h({\bf u}_i)-h({\bf u}_{i-1})\bigr]
  = h({\bf v})-h({\bf u}),
\]
and the same formula holds for any other path $p'$ from ${\bf u}$ to
${\bf v}$.

\textbf{(iii) $\Rightarrow$ (iv).}
Apply path independence to a closed path and the trivial path at the
same basepoint.
\end{proof}

\begin{remark}
\label{rem:directed-cycle}
For arbitrary edge-labelings on a directed graph, vanishing on
directed cycles characterises exactness with respect to an arbitrary
vertex-potential.
In the present paper, however, Layer~2 is stronger: the potential must
extend additively to a monoid homomorphism $h:\Mon\to\RR$.
Accordingly, the directed-cycle condition is a consequence of
Layer~2, but not by itself an equivalent replacement for it.
\end{remark}

\begin{chembox}[Why the state-function condition is the first law]
The condition $\FH = \delta^0 h$ encodes the first law of
thermodynamics in a single equation.
Saying $\FH(r) = h({\bf v}) - h({\bf u})$ means enthalpy depends
only on the initial and final complexes, not on which specific
reaction or sequence of reactions connects them.
Going around a closed reaction cycle yields zero net enthalpy
because the potential heights $h$ return to their starting values.
Layer~1 alone is the theory of an \emph{additive} reaction label;
Layer~2 additionally demands that this label derives from an
absolute assignment of values to the objects (complexes) of
the network.
\end{chembox}

\medskip
\noindent\textbf{Parameter count.}

\begin{observation}[Parameter reduction at Layer~2]
\label{obs:params}
The state-function condition reduces the number of free parameters
required to specify a thermochemical network, exchanging
reaction-level freedom for a smaller species-level one.
\begin{itemize}
  \item \textbf{Layer~1}: $|\Rx|$ free parameters (one real number
    per generating reaction).
  \item \textbf{Layer~2}: $|\Sp|$ species values $\{h(S)\}_{S \in \Sp}$
    specify $h$, hence $\FH = \delta^0 h$; the physically
    meaningful content of $h$ is its image under $\delta^0$, which
    has gauge kernel
    \[
      \ker(\delta^0) \cap \Hom_{\mathbf{Mon}}(\Mon, \RR)
      \;\cong\;
      \ker(N^\top) \;\subseteq\; \RR^{|\Sp|}.
    \]
    where $N \in \ZZ^{|\Sp| \times |\Rx|}$ is the stoichiometric
    matrix.
    Equivalently, two object potentials $h, h'$ differ only by an
    irrelevant gauge iff $h - h'$ is a conservation law
    (Proposition~\ref{prop:conservation-functors}).
  \item \textbf{Net dimension}: the reaction-enthalpy parameter
    space $\im(\delta^0) \subseteq \RR^{|\Rx|}$ has dimension
    $\rho := \rank_\RR(N)$, the dimension of the stoichiometric
    subspace (\S\ref{sec:N-matrix}); the gauge kernel is
    $|\Sp| - \rho$ dimensional.
\end{itemize}
The reduction from $|\Rx|$ to $\rho$ parameters proceeds in two
stages, exposing the role of deficiency.
Graph-cycle conditions on the reaction graph alone --- equivalent
to requiring an arbitrary vertex potential, not necessarily
species-additive --- reduce $\RR^{|\Rx|}$ to $\im(I_a^\top)$ of
dimension $n - \ell$, where $n$ is the number of complexes and
$\ell$ the number of linkage classes.
Imposing the further \emph{species-additive} structure of Layer~2
(the potential $h$ must be a monoid homomorphism on $\Mon$, not
merely a function on complexes) reduces this to $\im(N^\top) =
\im(\delta^0)$ of dimension $\rho$.
The difference $(n - \ell) - \rho =: \delta \geq 0$ is the
\emph{deficiency} of the network (\S\ref{sec:N-matrix}); $\rho =
n - \ell$ holds precisely when $\delta = 0$.
The total constraint count $|\Rx| - \rho$ therefore decomposes as
$\gamma + \delta$, where $\gamma = |\Rx| - (n - \ell)$ counts the
graph-cycle conditions.
\end{observation}

\begin{chembox}[Counting calorimetric experiments]
A thermochemical network with $|\Rx|$ generating reactions and
stoichiometric rank $\rho$ requires exactly $\rho$ independent
calorimetric measurements to fully characterise all reaction
enthalpies; the remaining $|\Rx| - \rho$ values are determined by
the cycle conditions of Proposition~\ref{prop:L1-equiv}(iv).
For deficiency-zero networks, $\rho = n - \ell$, recovering the
familiar ``$n - \ell$ measurements'' rule of thumb (where $n$ is
the number of complexes and $\ell$ the number of linkage classes);
non-zero deficiency requires $\rho < n - \ell$ measurements
relative to the rule of thumb.

For the Born--Haber cycle of Example~\ref{ex:born-haber}, the
cycle constraint relating the direct formation reaction to the
five-step elementary decomposition lets the lattice enthalpy
$\FH(r_5)$ --- which is not accessible to direct calorimetry ---
be extracted from the four measurable elementary steps and the
net formation enthalpy.
\end{chembox}

\subsubsection{Formation enthalpies as the canonical trivialisation}

The object potential $h \in C^0$ is determined by $\FH$ only up to
addition of a conservation-law potential ($\ker(N^\top)$); see
Remark~\ref{rem:gauge} below.
Chemistry resolves this gauge freedom by a universal convention:
the elemental reference species in their standard states are
assigned enthalpy zero.
The resulting \emph{standard molar enthalpies of formation}
$\dH_f^\circ(S)$, compiled in thermochemical databases such as the
NIST-JANAF tables \cite{Chase1998}, give the conventionally
canonical choice of object potential.
This is the IUPAC convention \cite{IUPACGreenBook2007}: it picks a
distinguished representative from the gauge equivalence class of
object potentials and is the universally adopted standard in
thermochemical databases.
We now make this precise.

\begin{definition}[Formation enthalpy and elemental references]
\label{def:formation}
  Let $\mathcal{E} \subset \Sp$ be the set of \emph{elemental
  reference species} in their standard states
  (e.g.\ H$_2$(g), C(graphite), Na(s), O$_2$(g)).
  The \emph{standard molar enthalpy of formation} of species $S \in
  \Sp$ is the real number $\dH_f^\circ(S) \in \RR$ defined as the
  enthalpy change for the reaction that forms one mole of $S$ from
  its constituent elements in their standard reference states.
  By convention, $\dH_f^\circ(E) := 0$ for all $E \in \mathcal{E}$
  \cite{Chase1998,AtkinsDeP2014}.
\end{definition}

\begin{proposition}[Formation enthalpies give the canonical object potential]
\label{prop:formation-section}
  Define $h_f: \Mon \to \RR$ by extending the formation enthalpies
  linearly over complexes:
  \[
    h_f\!\left(\sum_{S \in \Sp} n_S\, S\right)
    \;:=\; \sum_{S \in \Sp} n_S\,\dH_f^\circ(S).
  \]
  Then $h_f$ is a monoid homomorphism (an object potential in the
  sense of Definition~\ref{def:L1-layer2}), and for every generating
  reaction $r: {\bf u} \to {\bf v}$ with stoichiometric coefficients
  $\nu_S^{({\bf u})}$ and $\nu_S^{({\bf v})}$ counting the
  multiplicity of species $S$ in complexes ${\bf u}$ and ${\bf v}$
  respectively:
  \[
    \delta^0 h_f(r)
    \;=\; h_f({\bf v}) - h_f({\bf u})
    \;=\; \sum_{S \in \Sp} N_{S,r}\,\dH_f^\circ(S),
  \]
  where $N_{S,r} := \nu_S^{({\bf v})} - \nu_S^{({\bf u})} \in \ZZ$
  is the net stoichiometric coefficient of species $S$ in reaction
  $r$ (positive for products, negative for reactants).
\end{proposition}

\begin{proof}
\textbf{$h_f$ is a monoid homomorphism.}
Since $\Mon = \NN^{|\Sp|}$ is the free commutative monoid on $\Sp$,
every monoid homomorphism $\Mon \to \RR$ is uniquely determined by
its values on generators (species), and the extension
$h_f(\sum_S n_S S) = \sum_S n_S h_f(S)$ is the unique such
homomorphism with $h_f(S) = \dH_f^\circ(S)$.
The monoid homomorphism axioms hold by linearity:
\[
  h_f(\mathbf{0}) = 0, \qquad
  h_f({\bf u} + {\bf v}) = h_f({\bf u}) + h_f({\bf v}).
\]

\textbf{The formula for $\FH(r)$.}
Direct computation:
\begin{align*}
  h_f({\bf v}) - h_f({\bf u})
  &= \sum_S \nu_S^{({\bf v})}\,\dH_f^\circ(S)
   - \sum_S \nu_S^{({\bf u})}\,\dH_f^\circ(S) \\
  &= \sum_S \bigl(\nu_S^{({\bf v})} - \nu_S^{({\bf u})}\bigr)
     \,\dH_f^\circ(S)
   = \sum_S N_{S,r}\,\dH_f^\circ(S). \qedhere
\end{align*}
\end{proof}

\begin{remark}[Gauge freedom and reference state]
\label{rem:gauge}
The object potential $h_f$ is determined by $\FH$ only up to the
gauge $\ker(N^\top) \subset \RR^{|\Sp|}$: two potentials
$h_f, h'_f \in C^0 = \Hom_{\mathbf{Mon}}(\Mon, \RR)$ yield the same
reaction enthalpies $\FH = \delta^0 h_f = \delta^0 h'_f$ if and
only if their difference $g := h'_f - h_f$ lies in $\ker(N^\top)$,
i.e.\ $\sum_S N_{S,r} g(S) = 0$ for every reaction $r$.
The space $\ker(N^\top)$ is precisely the space of conservation
laws (Proposition~\ref{prop:conservation-functors}); two object
potentials are gauge-equivalent exactly when they differ by a
conservation-law potential.
This is the precise categorical content of the thermodynamic
statement that ``absolute enthalpies do not exist, but enthalpy
differences do'' \cite{AtkinsDeP2014}.
The IUPAC convention $\dH_f^\circ(E) = 0$ for elemental reference
species \cite{IUPACGreenBook2007} is the standard chemical gauge
fixing: for typical chemical networks, where the conservation
laws are spanned by atomic-composition counters, setting the
elemental potentials to zero uniquely determines $h_f$ on the
remaining (non-elemental) species.
\end{remark}

\begin{insightbox}[The formation table \emph{is} the object potential]
A standard thermochemical data table --- one entry $\dH_f^\circ(S)$
per species $S \in \Sp$ --- is precisely the specification of the
monoid homomorphism $h_f: \Mon \to \RR$.
From this single table, Proposition~\ref{prop:formation-section}
computes $\FH(r) = \sum_S N_{S,r}\,\dH_f^\circ(S)$ for every
reaction $r$ in the network, with no further data.
The functor $\FH = \delta^0 h_f$ follows automatically from the
coboundary structure.
The table \emph{is} $h_f$; the Layer~2 condition \emph{is} the
statement that every internally consistent thermochemical dataset
has such a table.
\end{insightbox}

%% file: chapters/L1/l1_examples.tex
\subsection{Worked examples}
\label{sec:L1-examples}

The three examples below are standard exercises from undergraduate
thermochemistry curricula \cite{atkins2023physical,Chase1998}.
Every introductory chemistry student learns to add enthalpies along
reaction pathways, to close thermochemical cycles, and to use
tabulated formation enthalpies to predict unmeasurable quantities.
The categorical language of $\Lk_1$ reveals the common mathematical
skeleton behind all three: Hess's Law is functoriality, cycle closure
is the coboundary condition, and a thermochemical data table is an
object potential.
Nothing new is computed; what is new is the identification of the
precise categorical structure that makes the computations valid.

\subsubsection{Hess's Law as parallel morphisms: carbon combustion}

\begin{example}[Two paths to CO$_2$]
\label{ex:hess-carbon}
  Species
  $\Sp = \{\mathrm{C(graphite),\, O_2(g),\, CO(g),\, CO_2(g)}\}$;
  three generating reactions, with reaction enthalpies from
  \cite{Chase1998}:
  \begin{align*}
    r_1 &:\; \mathrm{C(graphite)} + \mathrm{O_2(g)} \to \mathrm{CO_2(g)},
    && \FH(r_1) = -393.5\;\mathrm{kJ\,mol^{-1}}, \\
    r_2 &:\; \mathrm{C(graphite)} + \tfrac{1}{2}\mathrm{O_2(g)} \to \mathrm{CO(g)},
    && \FH(r_2) = -110.5\;\mathrm{kJ\,mol^{-1}}, \\
    r_3 &:\; \mathrm{CO(g)} + \tfrac{1}{2}\mathrm{O_2(g)} \to \mathrm{CO_2(g)},
    && \FH(r_3) = -283.0\;\mathrm{kJ\,mol^{-1}}.
  \end{align*}

  \begin{remark}[Fractional stoichiometric coefficients]
  \label{rem:frac-stoich}
  The coefficient $\tfrac{1}{2}$ in $r_2$ and $r_3$ appears to
  conflict with the definition of $\Lk_0(P)$ as a category of
  complexes in $\NN^{|\Sp|}$ (free commutative monoid on species,
  with non-negative integer multiplicities).
  There is no genuine conflict: the coefficient $\tfrac{1}{2}$
  is chemical shorthand for a scaled version of the reaction.
  One may always clear denominators — writing
  $2r_2: 2\,\mathrm{C(s)} + \mathrm{O_2(g)} \to 2\,\mathrm{CO(g)}$
  and
  $2r_3: 2\,\mathrm{CO(g)} + \mathrm{O_2(g)} \to 2\,\mathrm{CO_2(g)}$
  --- to obtain integer-coefficient reactions in $\NN^{|\Sp|}$, with
  $\FH(2r_k) = 2\FH(r_k)$ by monoidality.
  Fractional coefficients never arise as transcendental or
  irrational numbers: stoichiometry is always rational, and rational
  coefficients always lift to integer coefficients by scaling by the
  least common denominator of all fractions appearing in the network.
  Throughout this section we retain conventional chemical notation.
  \end{remark}

  \textbf{At $\Lk_0$.}
  The complex $\mathrm{C(s)} + \mathrm{O_2(g)}$
  (where $+$ denotes the tensor product $\otimes$ in $\Lk_0(P)$,
  per the chemical-notation convention) is the source of $r_1$.
  To reach $\mathrm{CO_2(g)}$ via $r_2$ then $r_3$, the spectator
  $\tfrac{1}{2}\mathrm{O_2(g)}$ must be carried along until $r_3$
  consumes it: starting from $\mathrm{C} + \mathrm{O_2}
  = \mathrm{C} + \tfrac{1}{2}\mathrm{O_2} +
  \tfrac{1}{2}\mathrm{O_2}$, the two-step path is the composite
  $r_3 \circ \bigl(r_2 \otimes \id_{\tfrac{1}{2}\mathrm{O_2}}\bigr)$.
  Its source is $\mathrm{C} + \mathrm{O_2}$ and its target is
  $\mathrm{CO_2}$, matching $r_1$.
  The direct $r_1$ and the two-step
  $r_3 \circ \bigl(r_2 \otimes \id_{\tfrac{1}{2}\mathrm{O_2}}\bigr)$
  are therefore \emph{distinct parallel morphisms} in $\Lk_0(P)$
  with the same source and target.

  \textbf{At $\Lk_1$, Layer~1.}
  Functoriality, monoidality, and unitality
  (Theorem~\ref{thm:hess}) give
  $\FH\bigl(r_2 \otimes \id_{\tfrac{1}{2}\mathrm{O_2}}\bigr)
  = \FH(r_2) + 0 = \FH(r_2)$, hence:
  \[
    \FH\!\Bigl(r_3 \circ \bigl(r_2 \otimes
      \id_{\tfrac{1}{2}\mathrm{O_2}}\bigr)\Bigr)
    \;=\; \FH(r_2) + \FH(r_3)
    \;=\; -110.5 + (-283.0) \;=\; -393.5\;\mathrm{kJ\,mol^{-1}}
    \;=\; \FH(r_1).
  \]
  The two paths agree because the data are thermodynamically
  consistent.
  Whether this equality is \emph{forced} (a structural consequence)
  or coincidental is the Layer~2 question.

  \textbf{At $\Lk_1$, Layer~2.}
  The equality $\FH(r_1) = \FH(r_3 \circ r_2)$ is path independence
  (Proposition~\ref{prop:L1-equiv}(iii)): the two parallel morphisms
  must have the same $\FH$ value whenever the coboundary condition
  holds.
  Using standard formation enthalpies
  $\dH_f^\circ(\mathrm{CO_2}) = -393.5$,
  $\dH_f^\circ(\mathrm{CO}) = -110.5$,
  $\dH_f^\circ(\mathrm{C}) = \dH_f^\circ(\mathrm{O_2}) = 0$
  (all in $\mathrm{kJ\,mol^{-1}}$, elemental references at zero
  \cite{Chase1998}):
  \[
    h_f(\mathrm{CO_2}) - h_f(\mathrm{C} \otimes \mathrm{O_2})
    = -393.5 - (0 + 0) = -393.5\;\mathrm{kJ\,mol^{-1}}.
  \]
  The commutative diagram below displays both levels simultaneously.
  \[
    \includegraphics{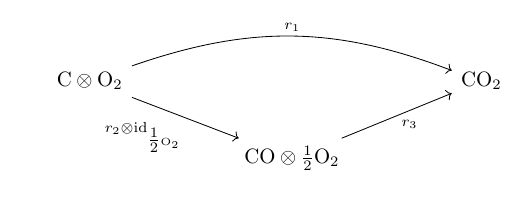}
  \]
  At $\Lk_0$, the diagram does \emph{not} commute: $r_1$ and
  $r_3 \circ \bigl(r_2 \otimes \id_{\tfrac{1}{2}\mathrm{O_2}}\bigr)$
  are distinct parallel morphisms in the free SMC.
  The Layer~2 coboundary condition $\FH = \delta^0 h_f$ forces the
  equality $\FH(r_1) = \FH\bigl(r_3 \circ (r_2 \otimes \id)\bigr)$
  in $B\RR$, making the diagram commute after applying $\FH$.
  This image-commutativity in $B\RR$ is the categorical content of
  Hess's Law.
\end{example}

\subsubsection{Born--Haber cycle: Layer~2 uniquely determines the unmeasurable step}

\begin{example}[Born--Haber cycle for sodium chloride]
\label{ex:born-haber}
  The formation reaction
  $r_{\mathrm{form}}: \mathrm{Na(s)} + \tfrac{1}{2}\mathrm{Cl_2(g)}
  \to \mathrm{NaCl(s)}$
  decomposes into five elementary generators (all values at
  298~K \cite{AtkinsDeP2014,Chase1998}):
  \begin{align*}
    r_1 &:\; \mathrm{Na(s)} \to \mathrm{Na(g)},
    && \FH = +108\;\mathrm{kJ\,mol^{-1}} &\quad\text{(sublimation)},\\
    r_2 &:\; \mathrm{Na(g)} \to \mathrm{Na^+(g)} + e^-,
    && \FH = +496\;\mathrm{kJ\,mol^{-1}} &\quad\text{(ionisation)},\\
    r_3 &:\; \tfrac{1}{2}\mathrm{Cl_2(g)} \to \mathrm{Cl(g)},
    && \FH = +121\;\mathrm{kJ\,mol^{-1}} &\quad\text{(bond dissociation)},\\
    r_4 &:\; \mathrm{Cl(g)} + e^- \to \mathrm{Cl^-(g)},
    && \FH = -349\;\mathrm{kJ\,mol^{-1}} &\quad\text{(electron affinity)},\\
    r_5 &:\; \mathrm{Na^+(g)} + \mathrm{Cl^-(g)} \to \mathrm{NaCl(s)},
    && \FH = \;?\; &\quad\text{(lattice enthalpy)}.
  \end{align*}

  The lattice enthalpy $\FH(r_5)$ --- the enthalpy of converting
  the ionic crystal into infinitely separated gas-phase ions ---
  is not directly accessible by calorimetry.
  The experimental obstacle is fundamental: there is no way to carry
  out the process $\mathrm{NaCl(s)} \to \mathrm{Na^+(g)} +
  \mathrm{Cl^-(g)}$ in a single calorimetric step, because this
  requires vaporising and fully ionising the crystal into isolated
  gas-phase ions with no counter-ions in the vicinity
  \cite{AtkinsDeP2014}.
  What \emph{can} be measured calorimetrically are the four steps
  $r_1$--$r_4$ and the net formation enthalpy $\FH(r_{\mathrm{form}})
  = -411\;\mathrm{kJ\,mol^{-1}}$ \cite{Chase1998}.
  The Born--Haber cycle \cite{BornHaber1919} was introduced precisely
  to extract the lattice enthalpy from these measurable quantities;
  the Layer~2 coboundary condition is its mathematical content.

  \textbf{$\Lk_0$ content.}
  The composite $r_4 \circ r_3$ is ill-typed: $r_3$ produces
  $\mathrm{Cl(g)}$ but $r_4$ requires
  $\mathrm{Cl(g)} + e^-$, and the electron is generated by $r_2$,
  not by $r_3$.
  The two reduction steps must therefore be sequenced so that $r_2$
  fires before $r_4$:
  \[
    r_{\mathrm{form}}'
    \;:=\;
    r_5 \circ
    \bigl(\id_{\mathrm{Na^+(g)}} \otimes r_4\bigr) \circ
    \bigl(r_2 \otimes \id_{\mathrm{Cl(g)}}\bigr) \circ
    \bigl(r_1 \otimes r_3\bigr).
  \]
  Both $r_{\mathrm{form}}$ and the composite $r_{\mathrm{form}}'$
  have source $\mathrm{Na(s)} \otimes \tfrac{1}{2}\mathrm{Cl_2(g)}$
  and target $\mathrm{NaCl(s)}$; they are \emph{distinct parallel
  morphisms} in $\Lk_0(P)$, not an equality of morphisms in the free
  SMC.

  \textbf{$\Lk_1$, Layer~2 (coboundary).}
  Since $r_{\mathrm{form}}$ and $r_{\mathrm{form}}'$ are parallel
  morphisms in $\Lk_0(P)$, path independence forces:
  \[
    \FH(r_{\mathrm{form}}) \;=\; \FH(r_{\mathrm{form}}')
    \;=\;
    \FH(r_1) + \FH(r_2) + \FH(r_3) + \FH(r_4) + \FH(r_5),
  \]
  where the second equality uses functoriality, monoidality, and
  $\FH(\id) = 0$ to collapse the spectator identities.
  
  Inserting $\FH(r_{\mathrm{form}}) = -411\;\mathrm{kJ\,mol^{-1}}$:
  \[
    -411 = \FH(r_5) + 496 + 108 + (-349) + 121
    \;\implies\;
    \FH(r_5) = -787\;\mathrm{kJ\,mol^{-1}}.
  \]
  The Layer~2 coboundary condition uniquely determines the lattice
  enthalpy from the four measurable steps; this is the universal
  property statement: the object potential $h_f$ assigns heights to
  all complexes, and closing the cycle is structurally forced.

  \[
    \includegraphics{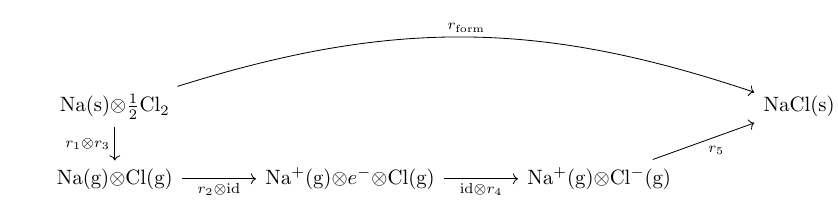}
  \]

  \textbf{Layer~1 vs Layer~2.}
  Layer~1 (functoriality, monoidality, unitality) gives the
  additivity of $\FH$ along the composite $r_{\mathrm{form}}'$:
  $\FH(r_{\mathrm{form}}') = \sum_{i=1}^5 \FH(r_i)$.
  That $r_{\mathrm{form}}'$ and $r_{\mathrm{form}}$ are parallel
  morphisms in $\Lk_0(P)$ is a purely $\Lk_0$ fact (matching
  source--target pairs).
  What Layer~2 adds is that parallel morphisms have equal images
  under $\FH$: $\FH(r_{\mathrm{form}}) = \FH(r_{\mathrm{form}}')$.
  Together these give the numerical determination of the lattice
  enthalpy.
\end{example}

\subsubsection{Organic combustion network: cycle consistency}

\begin{example}[Combustion network: testing Layer~2 consistency]
\label{ex:hess-network}
  The previous two examples used Layer~2 to \emph{predict} an unknown
  quantity.
  This example uses it in the opposite direction: to \emph{check}
  whether a set of experimentally measured enthalpies is internally
  consistent, i.e.\ whether a single object potential $h_f$ exists
  that accounts for all of them.
  This is exactly the test a calorimetrist applies when verifying a
  new thermochemical dataset.

  Consider the alkane combustion network with
  \[\Sp = \{\mathrm{CH_4(g),\, C_2H_6(g),\, H_2(g),\,
  O_2(g),\, CO_2(g),\, H_2O(g)}\}\]
  and four generating reactions (all at 298~K, H$_2$O(g);
  values from \cite{Chase1998}):
  \begin{align*}
    r_1 &:\; \mathrm{CH_4} + 2\,\mathrm{O_2}
             \to \mathrm{CO_2} + 2\,\mathrm{H_2O},
    && \FH = -802.3\;\mathrm{kJ\,mol^{-1}}, \\
    r_2 &:\; \mathrm{C_2H_6} + \tfrac{7}{2}\mathrm{O_2}
             \to 2\,\mathrm{CO_2} + 3\,\mathrm{H_2O},
    && \FH = -1428.5\;\mathrm{kJ\,mol^{-1}}, \\
    r_3 &:\; 2\,\mathrm{CH_4}
             \to \mathrm{C_2H_6} + \mathrm{H_2},
    && \FH = +65.7\;\mathrm{kJ\,mol^{-1}}, \\
    r_4 &:\; \mathrm{H_2} + \tfrac{1}{2}\mathrm{O_2}
             \to \mathrm{H_2O},
    && \FH = -241.8\;\mathrm{kJ\,mol^{-1}}.
  \end{align*}
  Here $r_3$ is a formal dehydrogenative coupling step (no oxygen
  participates) and $r_4$ is hydrogen combustion.

  \textbf{$\Lk_0$ parallel morphisms.}
  At the stoichiometric level, $r_1 \otimes r_1$ and the three-step
  composite formed from $r_3$, $r_4$, and $r_2$ (with spectator
  $\mathrm{O_2}$ molecules carried along to balance the oxidant
  budget at each step) are \emph{distinct parallel morphisms} in
  $\Lk_0(P)$ from $2\,\mathrm{CH_4} + 4\,\mathrm{O_2}$ to
  $2\,\mathrm{CO_2} + 4\,\mathrm{H_2O}$, not an equality of morphisms
  in the free SMC.
  Verbally: two simultaneous methane combustions reach the same
  source--target pair as the indirect path (i) couple two methanes
  to ethane and $\mathrm{H_2}$, (ii) burn the $\mathrm{H_2}$,
  (iii) burn the ethane.

  \textbf{$\Lk_1$, Layer~1 check.}
  Functoriality, monoidality, and $\FH(\id) = 0$ collapse the
  spectator identities, so each path's enthalpy is the sum of the
  $\FH$ values on its non-trivial generators.
  Using the values above:
  \begin{align*}
    \FH(r_3) + \FH(r_4) + \FH(r_2)
    &\;=\; 65.7 + (-241.8) + (-1428.5)
     \;=\; -1604.6\;\mathrm{kJ\,mol^{-1}}, \\
    2\,\FH(r_1)
    &\;=\; 2 \times (-802.3)
     \;=\; -1604.6\;\mathrm{kJ\,mol^{-1}}.
  \end{align*}
  The two sums agree exactly because all four values are drawn from
  a single internally-consistent compilation \cite{Chase1998}.
  A genuine cross-source consistency test --- comparing
  measurements from independent laboratories with independent
  reference states --- typically reveals discrepancies of
  $1$--$3\;\mathrm{kJ\,mol^{-1}}$ per measurement, the calorimetric
  noise floor against which Layer~2 violations would be detectable.

  \textbf{Layer~2 interpretation.}
  Near-zero cycle sum across measurements from independent sources
  means the data lie in (or very close to) $\im(\delta^0)$: they
  are coboundaries of a common object potential $h_f$.
  This is stronger than a Layer~1 check on a single compilation,
  which is internally consistent by construction; cross-source
  consistency actively tests whether the four experiments --- run
  in different laboratories with independent reference states ---
  admit a unified thermochemical description.
  A significant non-closure would signal either measurement error
  or use of inconsistent reference states, a violation of the
  Layer~2 coboundary condition detectable without knowing which
  individual measurement is wrong.
  This is the calorimetric analogue of a Kirchhoff consistency
  check in an electrical network
  (Section~\ref{sec:wegscheider-L1}).
\end{example}

%% file: chapters/L1/l1_weg.tex
\subsection{Thermochemical networks and the Wegscheider pre-condition}
\label{sec:wegscheider-L1}

In 1901, Rudolf Wegscheider showed that in a reversible mass-action 
reaction network obeying detailed balance, the rate constants — equivalently 
the elementary-step equilibrium constants — cannot be chosen independently; 
stoichiometric cycle relations impose multiplicative constraints, 
now called the  \emph{Wegscheider conditions} \cite{Wegscheider1901}.
These conditions are the cornerstone of thermodynamic consistency in
chemical kinetics and underpin everything from enzyme network
analysis to the theory of detailed balance
\cite{Onsager1931, Feinberg1989}.

The constraints enter the tower in two stages.
At $\Lk_2$, the full Wegscheider conditions arise as the coboundary
condition for the Gibbs functor $F_G^T = \FH - T F_S$, restricting
the allowed equilibrium constants via $\ker F_G^T$ (see
Section~\ref{sec:L2}).
At the present level $\Lk_1$, the same mathematical structure
appears in a purely enthalpic form: any thermochemical network with
directed cycles must satisfy \emph{additive} cycle conditions on the
enthalpy assignment.
This is the \emph{thermochemical pre-condition} --- the
enthalpic-additive analogue of the full Wegscheider conditions,
visible already at $\Lk_1$ before entropy and free energy enter at
$\Lk_2$.
Isolating it here makes the tower structure explicit: the
Wegscheider conditions are not a kinetic accident but a consequence
of the coboundary structure that is already forced at the thermochemical level.

The following definition places this classical notion on the
categorical footing provided by $\Lk_0(P)$.
The underlying idea --- that a thermochemical network is a reaction
graph decorated by an enthalpy assignment --- goes back to Hess
\cite{Hess1840} and was made explicit in the cycle-condition language
by Wegscheider \cite{Wegscheider1901}; the formulation below is
new in casting it as a pair $(P, h)$ where $h$ is a monoid
homomorphism into $\RR$ and the enthalpy functor is the induced
coboundary $\FH = \delta^0 h$.

\begin{definition}[State-function thermochemical network]
\label{def:thermo-network}
  A \emph{state-function thermochemical network} (a Layer~2
  thermochemical network) is a pair $(P, h)$ where $P$ is a Petri
  net and $h: \Mon \to \RR$ is a monoid homomorphism (an object
  potential in the sense of Definition~\ref{def:L1-layer2}).
  The induced functor $\FH = \delta^0 h: \Lk_0(P) \to B\RR$
  automatically satisfies both Layer~1 (functoriality) and Layer~2
  (the coboundary condition), and is uniquely determined by $h$.
  This is the Layer~2 specialisation of the generic
  $\Lk_1$-decoration of Definition~\ref{def:L1}: a generic
  $\Lk_1$-functor $\FH$ is required only to be strict symmetric
  monoidal, while a state-function thermochemical network requires
  $\FH$ to factor through an object potential.
\end{definition}

Thermochemical networks in this sense pervade chemistry.
The Born--Haber cycles of ionic compounds (Example~\ref{ex:born-haber}),
the Hess cycles for combustion enthalpies
(Examples~\ref{ex:hess-carbon}, \ref{ex:hess-network}), and the
thermodynamic cycles for ATP hydrolysis coupled to biosynthetic
reactions \cite{Alberty2003} are all instances: in each case, the
physically meaningful quantity is the object potential $h$, and the
measurable reaction enthalpies are its coboundary.
The parameter count $\rho = \rank N$ of
Observation~\ref{obs:params} is the number of independent
calorimetric experiments needed to determine $h$ up to the
$\ker(N^\top)$ gauge (equivalently, up to the choice of reference
state); for deficiency-zero networks this reduces to $n - \ell$.

\begin{proposition}[Kirchhoff's cycle condition as a theorem]
\label{prop:cycle}
  Let $(P, h)$ be a thermochemical network at $\Lk_1$ and let
  $r_1, \ldots, r_k$ form a directed closed loop in the CRN graph
  $G(\Rx)$.
  Then $\displaystyle\sum_{i=1}^k \FH(r_i) = 0$.
\end{proposition}

\begin{proof}
Immediate from Proposition~\ref{prop:L1-equiv}(i)$\Rightarrow$(iv):
the coboundary form of $\FH$ implies the directed-cycle condition.
\end{proof}

To the author's knowledge, the connection between this proposition and
Kirchhoff's Voltage Law has not previously been stated as a theorem
in the chemical reaction network literature, though the analogy
between thermochemical cycles and electrical circuits has been used
informally since the early network thermodynamics of Oster,
Perelson, and Katchalsky \cite{OsterPerelsonKatchalsky1973}.
The categorical proof makes the connection exact.

\begin{chembox}[Kirchhoff's Voltage Law in disguise]
In an electrical circuit, Kirchhoff's Voltage Law (KVL) states that
the algebraic sum of voltage drops around any closed loop of
circuit branches is zero \cite{Kirchhoff1845}: in a conservative
(electrostatic) network, voltage drops are differences of an
electric scalar potential, so traversing a directed cycle returns
to the starting potential.
Formally: for any directed cycle of circuit branches, the sum of
the voltage drops equals zero.

Proposition~\ref{prop:cycle} and KVL are the same mathematical
statement in different physical guises.
In both cases, a functor (the voltage assignment in circuits; the
enthalpy functor $\FH$ in thermochemistry) maps into the one-object
additive category $B\RR$ and must assign zero to every closed loop.
The reason is identical: both the voltage in a circuit and the
enthalpy of a reaction are coboundaries of a scalar potential
(electric potential $\phi$ in circuits; the object potential $h$
in thermochemistry).
The Born--Haber cycle, Hess's Law for formation enthalpies, and the
ATP--ADP free energy cycle in cellular metabolism
\cite{Alberty2003} are all chemical instances of the same
categorical fact.

The non-coincidence is the point: both laws derive from the
coboundary structure $\FH = \delta^0 h$ (or $\Delta\phi =
\delta^0 \phi$ in circuits), and the cycle condition
$\sum_{r \in \gamma} \FH(r) = 0$ is the boundary--coboundary
duality
\[
  \sum_{r \in \gamma} (\delta^0 h)(r)
  \;=\; \langle h, \partial\gamma \rangle
\]
applied to any directed closed cycle $\gamma$, for which the
boundary $\partial\gamma$ vanishes (each complex appears equally
often as source and as target).
\end{chembox}

\begin{observation}[The number of independent cycle conditions]
\label{obs:cycle-rank}
  The \emph{cycle rank} (first Betti number) of the CRN graph is
  $\gamma = |\Rx| - (n - \ell)$, where $n$ is the number of
  distinct complexes and $\ell$ is the number of linkage classes.
  The $\gamma$ independent directed-cycle conditions reduce the
  Layer~1 parameter space $\RR^{|\Rx|}$ to $\im(I_a^\top)$ of
  dimension $n - \ell$ --- the image of arbitrary, not necessarily
  species-additive, vertex potentials on the complex graph.
  Imposing the further species-additive structure of Layer~2
  reduces this to $\im(N^\top) = \im(\delta^0)$ of dimension
  $\rho = (n - \ell) - \delta$, where $\delta \geq 0$ is the
  deficiency.
  This is Observation~\ref{obs:params} restated in terms of the
  topology of the reaction graph: $\gamma$ graph-cycle conditions
  followed by $\delta$ deficiency conditions, totalling $|\Rx| -
  \rho$ constraints overall.
\end{observation}

%% file: chapters/L1/l1_forcing_out.tex
\subsection{What \texorpdfstring{$\Lk_1$}{Lk1} cannot express: forcing of \texorpdfstring{$\Lk_2$}{Lk2}}
\label{sec:L1-forcing-out}

Layer~1 assigns a single real number $\FH(r)$ to each generating
reaction and propagates it additively.
The physical question it cannot answer is: \emph{how does a reaction
network respond to changes in temperature?}
The temperature dependence of equilibrium is governed by the Gibbs
free energy $\dG = \dH - T\dS$; two reactions with identical $\dH$
but different $\dS$ accumulate different free energies at every
temperature, yet are indistinguishable by $\FH$ alone.
The following makes this gap precise.

\begin{forcingbox}[Forcing empirical result for $\Lk_2$: enthalpy-indistinguishable, entropy-separated]
Let $P$ be a \emph{coarse-grained} Petri net with species
\[
  \Sp=\{\mathrm{H},\,\mathrm{I},\,\mathrm{HI}\},
\]
where $\mathrm{H}$ denotes Hsp90, $\mathrm{I}$ denotes the
aryl-dihydroxyphenyl-thiadiazole inhibitor class, and
$\mathrm{HI}$ the bound complex.
Consider two generating reactions
\[
  r_E,\, r_F:\ \mathrm{H}\otimes\mathrm{I}\longrightarrow\mathrm{HI},
\]
representing, after coarse-graining, the binding of ICPD26 and
ICPD34.
Chemically these are distinct ligands; they become parallel
morphisms only because the label $\mathrm{I}$ forgets substituent
identity.

Kazlauskas et al.\ report that the modification
$\mathrm{ICPD26}\to\mathrm{ICPD34}$ weakens binding by about
$3.1\;\mathrm{kJ\,mol^{-1}}$, while the binding enthalpies remain
essentially unchanged within the experimental uncertainty of the
isothermal titration calorimetry (ITC) analysis; the observed change
in binding free energy is therefore attributed to entropy
\cite{kazlauskas2012thermodynamics}.

This exposes a genuine limitation of the $\Lk_1$ description.
At the coarse-grained stoichiometric level, the two binding events
have the same source, the same target, and the same enthalpy label
to experimental resolution:
\[
  \FH(r_E)\approx \FH(r_F).
\]
Yet they are not thermodynamically equivalent, since their binding
free energies differ measurably.
Thus $\FH$ alone does not contain enough information to separate
them.

\textbf{Indistinguishability at $\Lk_1$.}
If $\Lk_1$ records only stoichiometry together with the enthalpy
assignment, then $r_E$ and $r_F$ are indistinguishable at that
level: every invariant visible to $\Lk_1$ treats them as the same
reaction.

\textbf{Why $\Lk_1$ is not enough.}
The Hsp90 example therefore forces an extension of the theory.
There exists thermodynamic information relevant to binding that is
invisible to $\Lk_1$ but necessary to explain the experimentally
observed difference in affinity.
In other words, once one reaches this point, an enthalpy-only level of the tower
is no longer adequate: the tower must be refined by adjoining a
further thermodynamic datum.

\textbf{Methodological caution.}
This formulation must be stated at the level of
\emph{intrinsic} thermodynamic parameters rather than raw observed
heats.
For protein--ligand binding, calorimetric enthalpies can be
shifted by linked protonation and buffer effects, so comparison of
reaction labels should be made only after such corrections
\cite{krishnamurthy2007thermodynamic,kazokaite2021experimental}.
\end{forcingbox}

This forces $\Lk_2$, which must add:
\begin{enumerate}[label=(\roman*)]
  \item A second strict symmetric monoidal functor
    $F_S: \Lk_0(P) \to B\RR$ recording standard entropy changes
    $\dS^\circ_r$ per generating reaction (one additional real
    number per reaction, by the same universal property as $\FH$);
    the level $\Lk_2(P)$ is then the triple $(\Lk_0(P), \FH, F_S)$
    --- a second decoration on the same underlying stoichiometric
    category, exactly parallel to how $\FH$ decorates $\Lk_0(P)$ in
    Definition~\ref{def:L1}.
  \item A \emph{dagger structure} $({\cdot})^\dagger$ assigning to
    each reaction $r: {\bf u} \to {\bf v}$ a formal reverse
    $r^\dagger: {\bf v} \to {\bf u}$, satisfying
    $\FH(r^\dagger) = -\FH(r)$ and $F_S(r^\dagger) = -F_S(r)$,
    encoding microscopic reversibility \cite{Onsager1931}.
  \item A temperature-parametric \emph{Gibbs functor}
    $F_G^T := \FH - T\cdot F_S$ for each $T > 0$, with
    $\ker F_G^T$ characterising the reactions whose standard
    Gibbs free energy vanishes at temperature $T$ ---
    equivalently, those with $K_r(T) = 1$.
    Detailed balance in the kinetic sense is a stronger condition,
    requiring rate data and equilibrium concentrations beyond the
    thermodynamic content of $\Lk_2$, and is deferred to $\Lk_3$.
\end{enumerate}

The Wegscheider conditions --- constraints among equilibrium
constants in a directed cycle --- become the coboundary condition
for $F_G^T$ at $\Lk_2$: Proposition~\ref{prop:cycle} applied to
$F_G^T$ in place of $\FH$.
The full $\Lk_2$ condition $\sum_{r \in \gamma} F_G^T(r) = 0$ at
every $T > 0$ decomposes via $F_G^T = \FH - T\cdot F_S$ into two
$T$-independent vanishing conditions: the enthalpy cycle condition
$\sum_{r \in \gamma} \FH(r) = 0$ (the thermochemical
pre-condition of Section~\ref{sec:wegscheider-L1}) and the entropy
cycle condition $\sum_{r \in \gamma} F_S(r) = 0$ (the additional
content of $\Lk_2$).
The $\Lk_1$ pre-condition is therefore one of two independent
constraints required by Wegscheider, not a singular limit.

%% file: chapters/ch_L2.tex
\section{\texorpdfstring{$\Lk_2$}{Lk2}: The Equilibrium Level}
\label{sec:L2}

\input{chapters/L2/l2_forcing}
\input{chapters/L2/l2_def}
\input{chapters/L2/l2_gibbs}
\input{chapters/L2/l2_layer}
\input{chapters/L2/l2_thm}
\input{chapters/L2/l2_wegscheider}
\input{chapters/L2/l2_examples}
\input{chapters/L2/l2_nextforcing}

%% file: chapters/L2/l2_forcing.tex
\subsection{Forcing the extension: what \texorpdfstring{$\Lk_1$}{L1}
  cannot express}
\label{sec:L2-forcing-in}

Section~\ref{sec:L1-forcing-out} identified the gap at $\Lk_1$:
two reactions with equal enthalpy changes but different entropy
changes respond to temperature in completely opposite ways, yet are
indistinguishable by $\FH$ alone.
Before making this precise categorically, we establish the chemical
reality that such pairs are not exotic but \emph{generic} in
complex reaction networks like a biochemical system.

\begin{forcingbox}[Abstract forcing pair for $\Lk_2$]
\label{box:forcing-L2-in-abstract}
The mathematical content of the gap at $\Lk_1$ is a pair of
generating reactions
\[
  r_1,\; r_2 \;:\; {\bf u} \to {\bf v}
\]
in some Petri net $P$, with the same enthalpy but different
entropies:
\[
  \FH(r_1) = \FH(r_2),
  \qquad
  \FS(r_1) \neq \FS(r_2).
\]
The label swap $\sigma : r_1 \leftrightarrow r_2$ preserves the
$\Lk_1$-decoration ($\FH \circ \sigma = \FH$ on chemical
generators) but does not preserve any decoration recording entropy,
since $\FS \circ \sigma \neq \FS$ on $\{r_1, r_2\}$.
By the automorphism sequence of \S\ref{sec:aut-exact}, $\sigma$
represents a non-trivial coset in $\coker \varphi_2$: the
forgetful operation $U_2$ (Definition~\ref{def:L2}) is not
injective on decorated triples --- the underlying
$\Lk_1$-decoration $(\Lk_0(P), \FH)$ alone does not determine
$\FS$.
$\Lk_1$ conflates the two reactions; $\Lk_2$ separates them, since
their Gibbs free energies
$F_G^T(r_i) = \FH(r_i) - T\,\FS(r_i)$ differ at every $T > 0$.

The chembox below establishes that such pairs are generic in real
chemistry --- a feature of biochemical reaction networks under
coarse-graining, not a mathematical contrivance.
\end{forcingbox}

\begin{chembox}[Binding thermodynamics: $\dH$ and $\dS$ need not covary]
Isothermal titration calorimetry (ITC) studies of
protein--ligand binding show that the enthalpic and entropic
contributions to binding can vary separately across a ligand series.
This is a structural feature of molecular recognition:
$\dH$ and $\dS$ report on different physical aspects of binding and
there is no general law forcing them to change together
\cite{ChoderamMobley2013}.

The binding enthalpy $\dH$ is often associated with the quality of
direct intermolecular interactions, such as hydrogen bonding,
electrostatics, van der Waals contacts, and the enthalpic cost of
desolvating polar groups.
The binding entropy $\dS$ reflects changes in solvent organisation and
molecular freedom, including favorable release of ordered water and
unfavorable losses of ligand and protein conformational freedom upon
complex formation.
Because these contributions arise from different physical mechanisms,
modifying one does not in general determine the other
\cite{freire2008enthalpy,ladbury2010adding,Klebe2015}.

A useful illustration is provided by Biela et al.'s ITC and
crystallographic study of thrombin inhibitors with systematically
varied P3 substituents \cite{BielaKlebe2012}.
These ligands probe the same hydrophobic S3/4 pocket and therefore
belong to the same stoichiometric reaction class at $\Lk_0$:
Inhibitor $+$ Thrombin $\to$ Complex.
Across the series, increasing hydrophobic bulk produced an overall
affinity enhancement of about 40-fold that was attributed mainly to a
more favorable entropy term, consistent with stepwise disruption of
the pocket water network \cite{BielaKlebe2012}.
At the same time, the enthalpy changed much less strongly than the
entropy, so the dominant thermodynamic trend in the series was
entropy-driven rather than enthalpy-driven.

For broader context, thermodynamic profiling of marketed HIV-1
protease inhibitors shows a wide spread of binding enthalpies.
In Freire's analysis, Indinavir is enthalpically unfavourable
($\dH \approx +7.6\;\mathrm{kJ\,mol^{-1}}$),
whereas Darunavir is strongly enthalpy-driven
($\dH \approx -53.1\;\mathrm{kJ\,mol^{-1}}$).
Across this inhibitor class, the progression from earlier to later,
more potent compounds is accompanied by increasingly favourable
binding enthalpies \cite{freire2008enthalpy}.
\end{chembox}

\noindent
What minimal additional structure allows the tower to distinguish
the two reactions of the forcing pair?
The same three physical observations that forced $\FH$ in
Section~\ref{sec:L1-forcing-in} apply to entropy:

\begin{enumerate}[label=(\alph*)]
  \item \textbf{Sequential additivity of standard reaction entropy.}
    $\dS(r_2 \circ r_1) = \dS(r_1) + \dS(r_2)$:
    the total entropy change of a sequence of reactions is the sum
    of each step's contribution.
  \item \textbf{Parallel additivity.}
    $\dS(r_1 \otimes r_2) = \dS(r_1) + \dS(r_2)$:
    independent reactions contribute independently to the total
    entropy change.
  \item \textbf{No spurious entropy.}
    $\dS(\id_{\bf u}) = 0$ for every complex ${\bf u}$:
    the identity process produces no entropy change.
\end{enumerate}

\noindent
These are the same three constraints that characterised $\FH$ in
Section~\ref{sec:L1-forcing-in}, with enthalpy replaced by entropy.
They uniquely identify a strict symmetric monoidal functor
$\FS : \Lk_0(P) \to B\RR$.
The universal property of $\Lk_0(P)$ (Theorem~\ref{thm:UP-L0})
guarantees existence and uniqueness of $\FS$ given one real number
per generating reaction --- the same theorem applied a second time,
with $\dS^\circ_r$ in place of $\dH^\circ_r$.
No new categorical machinery is required.

%% file: chapters/L2/l2_def.tex
\subsection{Definition of \texorpdfstring{$\Lk_2(P)$}{L2(P)}}
\label{sec:L2-def}

\begin{definition}[Equilibrium level $\Lk_2(P)$]
\label{def:L2}
  Let $P$ be a Petri net with reaction set $\Rx$.
  The \emph{equilibrium level} of $P$ is the triple
  \[
    \Lk_2(P) \;:=\; \bigl(\,\Lk_0(P),\;\FH,\;\FS\,\bigr)
    \;=\; \bigl(\,\Lk_1(P),\;\FS\,\bigr),
  \]
  where
  \begin{itemize}
    \item $\FH : \Lk_0(P) \to B\RR$ is the thermochemical functor
      from $\Lk_1$ (Definition~\ref{def:L1}), and
    \item $\FS : \Lk_0(P) \to B\RR$ is a strict symmetric monoidal
      functor assigning to each generating reaction $r \in \Rx$
      its standard molar entropy change $\FS(r) = \dS^\circ_r \in \RR$.
  \end{itemize}
  As with $\Lk_1(P)$ (Definition~\ref{def:L1}), the notation
  $\Lk_2(P)$ refers to this decorated triple --- a structure, not
  itself a category.
  We write $U_2$ for the \emph{forgetful operation}
  \[
    U_2\bigl(\Lk_0(P), \FH, \FS\bigr) \;:=\; \bigl(\Lk_0(P), \FH\bigr),
  \]
  which drops $\FS$ and returns the underlying $\Lk_1$-decorated
  pair.
\end{definition}

\begin{mathbox}[The tower so far: same underlying category throughout]
The domain of both $\FH$ and $\FS$ is $\Lk_0(P)$, the free skeletal
permutative category of Section~\ref{sec:L0}.
The underlying category does not change at this step: $\Lk_0(P)$
retains the same objects (complexes), morphisms (reaction paths),
and compositional structure.
What changes is the amount of functor data decorating it.
The tower so far is:
\[
  \underbrace{\Lk_0(P)}_{\text{free skeletal permutative}}
  \;\hookrightarrow\;
  \underbrace{(\Lk_0(P),\,\FH)}_{\Lk_1(P)}
  \;\hookrightarrow\;
  \underbrace{(\Lk_0(P),\,\FH,\,\FS)}_{\Lk_2(P)}.
\]
Each arrow adds one strict SMC functor into $B\RR$, uniquely
determined by one real number per generating reaction.
The universal property of $\Lk_0(P)$ (Theorem~\ref{thm:UP-L0}) is
invoked once at each step; $B\RR$ is a one-object category, so
its object monoid is trivially strictly commutative and the
hypothesis of Theorem~\ref{thm:UP-L0} is satisfied for both
applications.
\end{mathbox}

\begin{proposition}[Existence and uniqueness of $\FS$]
\label{prop:FS-unique}
  Given any assignment $\dS_0 : \Rx \to \RR$, there is a unique
  strict symmetric monoidal functor
  $\FS : \Lk_0(P) \to B\RR$
  satisfying $\FS(r) = \dS_0(r)$ for every generator $r \in \Rx$.
\end{proposition}

\begin{proof}
Identical to Proposition~\ref{prop:FH-unique}: $B\RR$ is a
one-object category, so its object monoid is trivially strictly
commutative, and Theorem~\ref{thm:UP-L0} provides a unique strict
symmetric monoidal functor extending the generator assignment
$\dS_0$.
\end{proof}

\begin{theorem}[Universal property of $\Lk_2(P)$]
\label{thm:UP-L2}
  Let $P$ be a Petri net with reaction set $\Rx$, and let
  $\Delta H_0, \Delta S_0 : \Rx \to \RR$ be any two assignments
  of real numbers to generators.
  There exists a unique $\Lk_2$-structure on $\Lk_0(P)$, namely the
  unique pair of strict SMC functors $(\FH, \FS)$ satisfying
  \[
    \FH(r) = \Delta H_0(r)
    \qquad\text{and}\qquad
    \FS(r) = \Delta S_0(r)
    \qquad\text{for every } r \in \Rx.
  \]
  Equivalently, $\Lk_2(P)$ is the universal $\RR^2$-decorated strict
  SMC over $P$: any pair of strict SMC functors
  $(\Phi, \Psi) : \Lk_0(P) \to B\RR$ is uniquely and freely
  determined by its values on the generating reactions.
\end{theorem}

\begin{proof}
Apply Theorem~\ref{thm:UP-L0} twice independently: once with the
generator assignment $\Delta H_0$ to produce $\FH$ (this is
Proposition~\ref{prop:FH-unique}), and once with $\Delta S_0$ to
produce $\FS$ (this is Proposition~\ref{prop:FS-unique}).
Each application produces a unique extension; the two applications
do not interact, since the generator data $\Delta H_0$ and
$\Delta S_0$ are independent real-valued assignments to the same
set $\Rx$.
\end{proof}

\begin{remark}[Relation to the $\dagger$-SMC literature]
\label{rem:dagger-smc}
The tower table in Section~\ref{sec:intro} compresses $\Lk_2$ as
a \emph{$\dagger$-symmetric monoidal category} ($\dagger$-SMC),
with $r^\dagger$ the reverse reaction and equilibrium locus
$\ker F_G^T$.
Definition~\ref{def:L2} is the precise version of that shorthand:
$\Lk_2(P) = (\Lk_0(P), \FH, \FS)$ is the free skeletal permutative
category $\Lk_0(P)$ equipped with two real-valued decorating functors.
No additional categorical structure is assumed \cite{Selinger2007, Selinger2011, AbramskyCoecke2008}.

The $\dagger$-SMC structure is available as \emph{supplementary}
data, on $\Lk_2(\bar{P})$ (the equilibrium level of the reversible
closure $\bar{P}$, constructed in Definition~\ref{def:reversible-petri}), but it is not constitutive
of $\Lk_2(P)$ for a general Petri net.
The anti-symmetry axioms of Definition~\ref{def:anti-sym} encode
the $\dagger$-structure as linear constraints on the functor pair
$(\FH, \FS)$, without modifying the underlying category.
In particular: detailed balance (Proposition~\ref{prop:detailed-balance})
and the Wegscheider conditions
(Proposition~\ref{prop:wegscheider1}) are theorems about
$\Lk_2(\bar{P})$, conditional on reversibility --- not theorems
about $\Lk_2(P)$ in general.
\end{remark}

\subsubsection{Reversible Petri nets and the dagger}
\label{sec:dagger}

Physical chemical reactions are, in principle, reversible:
thermodynamics does not forbid the transformation in either direction.
For ordinary chemical systems governed by classical or
non-relativistic quantum dynamics, the principle of microscopic
reversibility is a consequence of the time-reversal symmetry of the
underlying equations, so each elementary reaction step has a
corresponding reverse step \cite{BarronBuckingham2001,KrupkaKaplanLaidler1966,Tolman1925}. 
The thermodynamic data $(\FH,\FS)$
therefore determines not whether the reverse process exists, but how
the equilibrium is biased toward reactants or products
\cite{Onsager1931,AtkinsDeP2014}.
This has a direct modelling consequence: a faithful Petri net for a
system at or near equilibrium should contain both $r$ and $r^\dagger$
for every reaction $r$ it models.

The following definition formalises this and introduces the
\emph{reversible closure} --- the smallest reversible Petri net
containing a given one.
The term \emph{closure} is used in its standard algebraic sense: just
as the algebraic closure of a field adjoins all missing roots and
nothing more, the reversible closure of a Petri net freely adjoins,
for each generator $r$, the reverse generator $r^\dagger$ that is
missing, and nothing more.
The result is characterised by a universal property: it is the initial
reversible Petri net equipped with an embedding of $P$.
Petri nets as freely generated categorical / monoidal structures go
back to Meseguer--Montanari \cite{MeseguerMontanari1990}, while open
Petri nets and reaction networks are treated categorically by
Baez--Pollard and Baez--Master
\cite{BaezPollard2017,BaezMaster2020}.  Reversibility via adjoining
reverse transitions is also standard in the reversible-Petri-net
literature \cite{BarylskaEtAl2018,MelgrattiMezzinaUlidowski2020}.
In the present paper, we call the resulting initial reversible
completion of $P$ its reversible closure.

\begin{definition}[Reversible Petri net and reversible closure]
\label{def:reversible-petri}
  A Petri net $P$ is \emph{reversible} if for every generator
  $r : {\bf u} \to {\bf v}$ in $\Rx$, its reverse
  $r^\dagger : {\bf v} \to {\bf u}$ is also a generator in $\Rx$,
  with $(r^\dagger)^\dagger = r$.
  The \emph{reversible closure} $\bar{P}$ of any Petri net $P$ is the
  reversible Petri net obtained by freely adjoining, for each
  $r \in \Rx$, a new generator $r^\dagger : {\bf v} \to {\bf u}$
  subject to $(r^\dagger)^\dagger = r$.
  It is the initial reversible Petri net equipped with a Petri net
  embedding $\iota : P \hookrightarrow \bar{P}$.
\end{definition}

\begin{mathbox}[Dagger structures and opposite categories]
For any category $\mathcal{C}$, the \emph{opposite category}
$\mathcal{C}^{\mathrm{op}}$ has the same objects as $\mathcal{C}$
but every morphism $f : X \to Y$ in $\mathcal{C}$ becomes a
morphism $f^{\mathrm{op}} : Y \to X$ in $\mathcal{C}^{\mathrm{op}}$,
with composition reversed:
$g^{\mathrm{op}} \circ f^{\mathrm{op}} = (f \circ g)^{\mathrm{op}}$
\cite{Awodey2010,Leinster2014}.
This is a purely formal construction; it does not equip
$\mathcal{C}$ with any extra structure, only relates $\mathcal{C}$
to a ``mirror'' category with the same objects.

A \emph{dagger structure} on $\mathcal{C}$ is something stronger:
an identity-on-objects contravariant involution
\[
  (-)^\dagger \;:\; \mathcal{C}^{\mathrm{op}}
    \;\longrightarrow\; \mathcal{C},
  \qquad
  (f : X \to Y) \;\longmapsto\; (f^\dagger : Y \to X),
\]
satisfying
\[
  (f^\dagger)^\dagger = f,
  \qquad
  \id^\dagger_X = \id_X,
  \qquad
  (g \circ f)^\dagger = f^\dagger \circ g^\dagger.
\]
For a reversible Petri net (Definition~\ref{def:reversible-petri}),
the operation $r \mapsto r^\dagger$ is first defined on chemical
generators and then extended contravariantly to composite
morphisms, giving $\Lk_0(\bar P)$ the structure of a
dagger-symmetric-monoidal category \cite{Selinger2007}.
The opposite category $\Lk_0(\bar P)^{\mathrm{op}}$ exists as a
formal mirror for any $\bar P$; what makes $\Lk_0(\bar P)$ a
\emph{dagger}-SMC is the additional datum that the dagger operation
is identity on objects --- i.e., a canonical identification of
source and target of each reversed morphism with the original
target and source.
\end{mathbox}

\begin{definition}[Anti-symmetry axioms]
\label{def:anti-sym}
  For a reversible Petri net, the functor pair $(\FH, \FS)$ is
  required to satisfy, for every generator $r \in \Rx$:
  \[
    \FH(r^\dagger) = -\FH(r),
    \qquad
    \FS(r^\dagger) = -\FS(r).
  \]
\end{definition}

\begin{remark}[Anti-symmetry is automatic under Layer~2]
\label{rem:antisym-automatic}
  Definition~\ref{def:anti-sym} is stated as a Layer~1 axiom --- an
  additional constraint on the functor pair $(\FH, \FS)$ beyond the
  SMC functor axioms.
  Under Layer~2 for both functors
  (Definition~\ref{def:L1-layer2} for $\FH$, Definition~\ref{def:FS-layer2}
  below for $\FS$), anti-symmetry is not an additional constraint but a
  \emph{theorem}: if $\FH = \delta^0 h_f$ and $\FS = \delta^0 h_S$ on
  the reversible closure $\bar{P}$, then for every generator
  $r : {\bf u} \to {\bf v}$,
  \[
    \FH(r^\dagger) = h_f({\bf u}) - h_f({\bf v}) = -\FH(r),
    \qquad
    \FS(r^\dagger) = h_S({\bf u}) - h_S({\bf v}) = -\FS(r),
  \]
  by applying Proposition~\ref{prop:L1-equiv} separately to each functor.
  Anti-symmetry is thus a Layer~1 axiom --- required to state detailed
  balance without invoking Layer~2 structure --- that becomes a Layer~2
  theorem.
\end{remark}

\begin{chembox}[Anti-symmetry: state functions force sign flips]
Both conditions are direct consequences of the state-function character
of enthalpy and entropy --- among the most firmly established facts in
chemical thermodynamics
\cite{AtkinsDeP2014, Kondepudi2014}.

\textbf{Enthalpy anti-symmetry.}
Since enthalpy $H$ is a state function, its change along any path
from complex ${\bf u}$ to complex ${\bf v}$ depends only on the
endpoints: $\FH(r) = H({\bf v}) - H({\bf u})$ for
$r : {\bf u} \to {\bf v}$ (Layer~2/coboundary form,
Proposition~\ref{prop:formation-section}).
Reversing the reaction gives
\[
  \FH(r^\dagger) = H({\bf u}) - H({\bf v}) = -\FH(r).
\]
This is Hess's Law applied to a single step: if the forward
reaction releases $\dH$ kJ/mol, the reverse absorbs exactly
$|\dH|$ kJ/mol, and vice versa \cite{Hess1840, AtkinsDeP2014}.

\textbf{Entropy anti-symmetry.}
Entropy $S$ is likewise a state function (a consequence of the
second and third laws \cite{AtkinsDeP2014}).
By the same coboundary argument:
$\FS(r^\dagger) = S({\bf u}) - S({\bf v}) = -\FS(r)$.
At the molecular level this reflects \emph{microscopic
reversibility} \cite{Onsager1931}: at equilibrium, the forward and
reverse fluxes through every elementary step are equal, so any
state-function difference associated with the step must reverse
sign upon reversal of the step.
A process that increases the system's entropy by $|\FS(r)|$ in the
forward direction decreases it by the same amount in the reverse.

\textbf{What this is, and what it is not.}
The sign-flip refers to the \emph{reaction entropy change}
$\Delta S^\circ_r$ --- a state-function difference that is
positive, negative, or zero depending on the reaction.
It must not be confused with the \emph{entropy production rate}
$\sigma \geq 0$, a non-negative quantity in nonequilibrium
thermodynamics that quantifies irreversibility and belongs to the
kinetic/open-system level $\Lk_3$, not to the standard-state
$\Lk_2$ decoration.
Reaction entropy changes flip sign upon reversal; entropy
production does not.

Both axioms are imposed as constraints on the functor pair
$(\FH, \FS)$, not on the category $\Lk_0(P)$ itself.
The underlying category remains the free skeletal permutative
category.
\end{chembox}

\begin{remark}[Anti-symmetry halves the parameter count, matching thermochemical tables]
\label{rem:antisym-params}
  For a reversible Petri net with $|\Rx| = 2m$ generators
  (forward/reverse pairs), anti-symmetry reduces the free parameter
  space of $(\FH, \FS)$ from $4m$ to $2m$ real numbers: $m$ values
  of $\dH^\circ$ and $m$ values of $\dS^\circ$, one per forward
  reaction --- precisely the entries of a standard thermochemical
  table \cite{NIST_WebBook, AtkinsDeP2014}.

  This matters for two distinct reasons.

  First, it is \emph{canonical}: the reduced parameter count matches
  exactly the format in which experimental thermochemical data is
  tabulated and used.
  Standard references \cite{NIST_WebBook} list one $\dH_f^\circ$ and
  one $S^\circ$ per species, not separate entries for forward and
  reverse reactions.
  The anti-symmetry axiom is not an extra mathematical assumption but
  the categorical encoding of an empirical convention that is itself
  forced by energy conservation and the state-function property.

  Second, it is \emph{consistency-enforcing}: without anti-symmetry,
  assigning independent values to $r$ and $r^\dagger$ would permit
  $\FH(r) + \FH(r^\dagger) \neq 0$, violating energy conservation and
  making the thermodynamic model internally inconsistent.
  Anti-symmetry is therefore the minimal algebraic condition that
  keeps the decorating functors in agreement with the physical content
  they are supposed to represent.
\end{remark}

%% file: chapters/L2/l2_gibbs.tex
\subsection{The Gibbs functor and the equilibrium locus}
\label{sec:L2-gibbs}

With both $\FH$ and $\FS$ available on $\Lk_0(P)$, the Gibbs free
energy functor is an immediate derived object: no new axiom, no new
structure, only a linear combination of the two functors already in
hand.

\begin{chembox}[Why $\Delta G = \Delta H - T\Delta S$ is the equilibrium criterion]
At constant temperature and constant pressure, the second law of
thermodynamics requires that any spontaneous process decreases the
Gibbs free energy of the system: $dG \leq 0$, with equality at
equilibrium \cite{AtkinsDeP2014, Kondepudi2014}.
For a reaction $r$ proceeding by an infinitesimal extent $d\xi$:
\[
  dG = \Delta G_r \, d\xi,
  \qquad
  \Delta G_r = \Delta G^\circ_r + RT \ln Q,
\]
where $Q$ is the reaction quotient.
At standard conditions ($Q = 1$) this reduces to $\Delta G_r = \Delta G^\circ_r$.
The conditions become:
\begin{itemize}
  \item $\Delta G^\circ_r < 0$: forward reaction spontaneous at standard
    conditions (product-favoured);
  \item $\Delta G^\circ_r > 0$: reverse reaction spontaneous (reactant-favoured);
  \item $\Delta G^\circ_r = 0$: standard equilibrium, $K_{\mathrm{eq}} = 1$.
\end{itemize}
The general equilibrium condition $\Delta G_r = 0$ gives
$\Delta G^\circ_r = -RT \ln K_{\mathrm{eq}}$, connecting the functor
value $F_G^T(r) = \Delta G^\circ_r$ directly to the equilibrium
constant via exponentiation.

From statistical mechanics, $\Delta G^\circ_r = -k_B T \ln(Z_\mathrm{prod}/Z_\mathrm{react})$,
the logarithmic ratio of partition functions of products and reactants
\cite{AtkinsDeP2014}.
The functor $F_G^T$ therefore encodes, in a single real number per
generating reaction, the entire statistical mechanical content of the
reaction's thermodynamic favourability at temperature $T$.
\end{chembox}

\begin{remark}[Precedents for the Gibbs functor]
\label{rem:gibbs-precedent}
To the author's knowledge, the treatment of $\Delta G^\circ$ as a strict SMC
functor $F_G^T : \Lk_0(P) \to B\RR$ is new to this manuscript.
The closest precursor is Baez--Pollard \cite{BaezPollard2017}, who
use functor-language for entropy production and the composition of
open reaction networks, but do not single out the Gibbs functor as a
primary categorical object.
The axiomatic treatment of thermodynamic state functions as
homomorphisms (a structure close to our functor language) appears in
Lieb--Yngvason \cite{LiebYngvason1999}.
The compositional property $F_G^T(r_2 \circ r_1) = F_G^T(r_2) +
F_G^T(r_1)$ — proved as Proposition~\ref{prop:gibbs-functor} below —
is precisely Hess's Law for Gibbs free energy, here elevated from an
empirical observation to a structural consequence of the functor axioms.
\end{remark}
\begin{definition}[Gibbs functor]
\label{def:gibbs}
  For each temperature $T > 0$, the \emph{Gibbs functor at
  temperature $T$} is the strict symmetric monoidal functor
  \[
    F_G^T \;:=\; \FH \;-\; T \cdot \FS
    \;:\; \Lk_0(P) \;\longrightarrow\; B\RR.
  \]
  The \emph{Gibbs family} is the map
  $(0,\infty) \to \{\text{strict SMC functors }\Lk_0(P)\to B\RR\}$,
  $T \mapsto F_G^T$.
  We write $F_G^T$ uniformly throughout; $F_G$ alone is reserved
  only for fixed-$T$ contexts where the temperature has been
  specified explicitly.
\end{definition}

\begin{remark}[Equivalent characterisations of $\Lk_2(P)$: the role of the Gibbs functor]
\label{rem:L2-via-gibbs}
Whether $F_G^T$ alone suffices to characterise $\Lk_2$ depends on
\emph{how many} temperatures are used; the answer is precise.

\textbf{A single Gibbs functor is insufficient from $\Lk_0$.}
Given only $F_G^{T_0} = \FH - T_0 \cdot \FS$ for one temperature
$T_0$, one recovers only the single linear combination
$\FH(r) - T_0\,\FS(r)$ per generator $r$: the individual values
$\FH(r)$ and $\FS(r)$ cannot be separated.
A single Gibbs functor thus gives a strictly weaker structure than
$(\FH, \FS)$.

\textbf{The Gibbs family is equivalent to $(\FH, \FS)$.}
The Gibbs family $\{F_G^T\}_{T > 0}$ assigns to each generator $r$
the affine function of temperature
\[
  T \;\mapsto\; F_G^T(r) \;=\; \underbrace{\FH(r)}_{\text{intercept at }T=0}
  \;-\; T \cdot \underbrace{\FS(r)}_{\text{negative slope}}.
\]
Since an affine function is uniquely determined by its intercept and
slope, the values
\[
  \FH(r) = \lim_{T \to 0^+} F_G^T(r),
  \qquad
  \FS(r) = -\frac{d}{dT} F_G^T(r)
\]
are recovered at every generator $r \in \Rx$.
By Theorem~\ref{thm:UP-L0} these pointwise assignments extend
uniquely to strict SMC functors $\Lk_0(P) \to B\RR$;
the Gibbs family and the pair $(\FH, \FS)$ therefore encode
\emph{identical} information.
$\Lk_2(P)$ may equivalently be defined as $\Lk_0(P)$ equipped with
the Gibbs family $T \mapsto F_G^T$.

The recovery formulas above are an algebraic statement about the
chosen affine approximation, not a physical assertion about the
$T \to 0$ limit of standard Gibbs free energies.
Treating $\FH, \FS$ as temperature-independent makes
$T \mapsto F_G^T(r)$ literally affine; a physically faithful
extrapolation to $T = 0$ would also need heat-capacity data
$C_p(S, T)$ and any phase transitions in the relevant interval ---
content beyond the present $\Lk_2$ decoration (cf.\ the integrated
van~'t~Hoff form in \S\ref{sec:L2-thm}).

\textbf{From $\Lk_1$, one extra temperature suffices.}
If $\FH$ is already known (from $\Lk_1$), then a single
$F_G^{T_0}$ at any $T_0 > 0$ determines
$\FS(r) = (\FH(r) - F_G^{T_0}(r))/T_0$ for every generator.
The extension $\Lk_1 \to \Lk_2$ therefore requires exactly one
additional real number per generator --- the value of $\dG^\circ$ at
one reference temperature --- which is precisely the format of a
standard thermochemical table entry.
\end{remark}

\begin{proposition}[$F_G^T$ is a strict SMC functor for each $T > 0$]
\label{prop:gibbs-functor}
  For every $T > 0$, the map $F_G^T = \FH - T\cdot\FS$ is a strict
  symmetric monoidal functor $\Lk_0(P) \to B\RR$.
\end{proposition}

\begin{proof}
Let $r_1 : {\bf u} \to {\bf v}$ and $r_2 : {\bf v} \to {\bf w}$.
\textbf{Composition:}
\begin{align*}
  F_G^T(r_2 \circ r_1)
  &= \FH(r_2 \circ r_1) - T\,\FS(r_2 \circ r_1) \\
  &= \bigl[\FH(r_2) + \FH(r_1)\bigr] - T\bigl[\FS(r_2) + \FS(r_1)\bigr] \\
  &= F_G^T(r_2) + F_G^T(r_1).
\end{align*}
\textbf{Monoidality:}
\begin{align*}
  F_G^T(r_1 \otimes r_2)
  &= \FH(r_1 \otimes r_2) - T\,\FS(r_1 \otimes r_2) \\
  &= \bigl[\FH(r_1) + \FH(r_2)\bigr] - T\bigl[\FS(r_1) + \FS(r_2)\bigr] \\
  &= F_G^T(r_1) + F_G^T(r_2).
\end{align*}
\textbf{Unitality:} $F_G^T(\id_{\bf u}) = \FH(\id_{\bf u})
- T\,\FS(\id_{\bf u}) = 0 - T \cdot 0 = 0$.
\end{proof}

\begin{mathbox}[A linear combination of functors is a functor]
The morphism-sets $\mathrm{Mor}(\Lk_0(P), B\RR) := \{\text{strict SMC functors }
\Lk_0(P) \to B\RR\}$ form a real vector space under pointwise
addition and scalar multiplication (since $B\RR$ is an abelian group
object in strict SMC categories).
The Gibbs family $T \mapsto \FH - T\cdot\FS$ is a \emph{straight line}
in this space, parametrized by temperature.
The functor axioms are inherited by every element of the family.
\end{mathbox}

\begin{definition}[Standard Gibbs-zero locus]
\label{def:equil-locus}
  The \emph{standard Gibbs-zero locus at temperature $T$} (or
  \emph{$K = 1$ locus}) is
  \[
    Z_G(T) \;:=\; \ker\!\bigl(F_G^T\bigr) \;:=\;
    \bigl\{\,r \in \mathrm{Mor}\!\bigl(\Lk_0(P)\bigr) \;\big|\;
    F_G^T(r) = 0\,\bigr\}.
  \]
  A generating reaction $r \in \Rx$ lies in $Z_G(T)$ iff
  $\FH(r) = T\,\FS(r)$, equivalently iff its standard equilibrium
  constant satisfies $K_r(T) = 1$.
  This is the locus where the standard Gibbs free energy change
  vanishes; it is \emph{not} the same as the system being at
  equilibrium at a given concentration vector $x$, which requires
  $\Delta G_r(x, T) = \Delta G_r^\circ(T) + RT \ln Q_r(x) = 0$.
\end{definition}

\begin{observation}[Temperature dependence of the Gibbs-zero locus]
\label{obs:T-dependence}
  For a generating reaction $r$ with $\FS(r) \neq 0$, define the
  \emph{crossover temperature}
  \[
    T^*(r) \;:=\; \frac{\FH(r)}{\FS(r)}.
  \]
  The behaviour of $F_G^T(r)$ as a function of $T > 0$ splits into
  cases by the signs of $\FH(r)$ and $\FS(r)$:
  \begin{itemize}
    \item \textbf{Same signs ($T^*(r) > 0$):} $r$ enters the
      Gibbs-zero locus $Z_G(T^*(r))$.
      For $T < T^*(r)$, $F_G^T(r)$ has the same sign as $\FH(r)$
      (enthalpy-controlled regime);
      for $T > T^*(r)$, it has the opposite sign
      (entropy-controlled regime).
    \item \textbf{Opposite signs ($T^*(r) \leq 0$):} the crossover
      temperature is unphysical.
      $F_G^T(r)$ has constant sign for all $T > 0$: if
      $\FH(r) < 0$ and $\FS(r) > 0$, the reaction is
      product-favoured under standard conditions at every
      temperature; if $\FH(r) > 0$ and $\FS(r) < 0$, it is
      reactant-favoured under standard conditions at every
      temperature.
  \end{itemize}
  If $\FS(r) = 0$ then $F_G^T(r) = \FH(r)$ for all $T$: $r$ never
  enters the Gibbs-zero locus, and the sign of $F_G^T(r)$ is fixed
  by $\FH$ alone.
\end{observation}

\begin{chembox}[What $\Lk_2$ adds over $\Lk_1$]
At $\Lk_1$, reactions are labelled by enthalpy alone:
exothermic, endothermic, or thermoneutral.
At $\Lk_2$ they are additionally classified at each temperature by
the sign of $F_G^T$: product-favoured under standard conditions
($F_G^T < 0$, $K_r(T) > 1$), reactant-favoured under standard
conditions ($F_G^T > 0$, $K_r(T) < 1$), or in the standard
Gibbs-zero locus ($F_G^T = 0$, $K_r(T) = 1$).
The single most important quantity in chemical thermodynamics ---
the equilibrium constant $K_{\mathrm{eq}}(T)$ --- lives precisely
here, via $K_r(T) = \exp(-F_G^T(r)/RT)$.

``Reactant-favoured under standard conditions'' is not the same as
``forbidden'': a reaction with $F_G^T(r) > 0$ may proceed forward
under non-standard concentrations (when $RT \ln Q_r < -\Delta
G^\circ_r$), by coupling to driven processes, or kinetically.
The classification above is a $\Lk_2$ standard-state statement, not
a universal verdict on the reaction.
\end{chembox}

\begin{remark}[Notes on the term locus and the Gibbs-zero condition]
\label{rem:locus}
The term \emph{locus} (Latin: \emph{place}; plural \emph{loci}) is
standard mathematical vocabulary for the set of all objects
satisfying a given condition: the zero locus of a function $f$ is
$\{x : f(x) = 0\}$.
It is the same concept as \emph{zero set}, \emph{vanishing locus},
or \emph{level set at zero}, and appears throughout algebra and
geometry \cite{Hartshorne1977}.

In the CRNT literature, the analogous object is the
\emph{positive steady-state variety} or \emph{equilibrium ideal}:
the algebraic set of concentration vectors $x \in \RR_{>0}^n$ at
which the ODE $dx/dt = 0$
\cite{CraciunEtAl2009, GrossHill2013, Dickenstein2016}.
Connections between this variety and algebraic geometry underlie
the study of multistationarity and toric geometry in reaction
networks.

The standard Gibbs-zero locus $Z_G(T) = \ker(F_G^T)$ is a different
but related object: it lives in \emph{morphism space} (reaction
space), not concentration space.
It is the set of generating reactions $r$ for which the standard
Gibbs free energy change vanishes at temperature $T$, equivalently
the set of reactions with $K_{\mathrm{eq}}(r, T) = 1$.
This is the \emph{thermodynamic} layer of the equilibrium
structure --- a standard-state statement about reaction labels,
not about concentrations.
The kinetic layer --- which concentration vectors are actually
reached, and at what rates --- belongs to $\Lk_3$.
The algebraic geometry of $Z_G(T)$ as a function of $T$ is
precisely the content of Observation~\ref{obs:T-dependence} and the
van~'t~Hoff equation (Proposition~\ref{prop:vant-hoff}).
\end{remark}

\subsubsection{Reversal symmetry of the standard Gibbs-zero locus}

\emph{Detailed balance} is a central principle of chemical kinetics
and non-equilibrium thermodynamics.
Chemically, it states that at thermodynamic equilibrium, every
elementary reaction step is individually balanced by its reverse:
the forward and reverse \emph{fluxes} through each step are equal,
not merely the net flux around each cycle
\cite{Onsager1931, Feinberg1989}.
For a single reversible step $r : A \rightleftharpoons B$ with
forward rate constant $k_+$ and reverse rate constant $k_-$,
detailed balance at an equilibrium concentration vector requires
\[
  k_+[A]_{\mathrm{eq}} \;=\; k_-[B]_{\mathrm{eq}},
\]
which gives $K_{\mathrm{eq}} = k_+/k_-$.
This is a kinetic flux equality, requiring rate constants and
equilibrium concentrations --- data that lives at $\Lk_3$, not
$\Lk_2$.

What $\Lk_2$ supplies, and the proposition below makes precise, is
the \emph{thermodynamic precondition} for detailed balance: when a
reaction lies in the standard Gibbs-zero locus
$Z_G(T) = \ker F_G^T$, so does its reverse, and the standard
equilibrium constants of the two satisfy
$K_{r^\dagger}(T) = 1/K_r(T)$.
In networks this thermodynamic precondition extends to a cycle
condition --- the \emph{Wegscheider conditions}
\cite{Wegscheider1901}, that the product of standard equilibrium
constants around every directed cycle equals one
(Proposition~\ref{prop:wegscheider1}).
The full kinetic content of detailed balance --- equality of
forward and reverse fluxes at given concentrations --- is deferred
to $\Lk_3$, where it appears as a consistency condition between
the rate constants and the $\Lk_2$ equilibrium constants.

\begin{proposition}[Reversal symmetry of the standard Gibbs-zero locus]
\label{prop:detailed-balance}
  Let $P$ be a reversible Petri net satisfying
  Definition~\ref{def:anti-sym}.
  For any generator $r \in \Rx$ and temperature $T > 0$:
  \[
    F_G^T(r) = 0 \;\iff\; F_G^T(r^\dagger) = 0.
  \]
  Equivalently, the standard Gibbs-zero locus $Z_G(T)$ is closed
  under the dagger involution: $r \in Z_G(T)$ iff
  $r^\dagger \in Z_G(T)$.
\end{proposition}

\begin{proof}
The anti-symmetry axioms give
$F_G^T(r^\dagger) = \FH(r^\dagger) - T\,\FS(r^\dagger)
= -\FH(r) - T(-\FS(r)) = -F_G^T(r)$.
So $F_G^T(r^\dagger) = -F_G^T(r)$, and the claim is immediate.
\end{proof}

\begin{remark}[Scope and categorical status of reversal symmetry]
\label{rem:detailed-balance-scope}
  Proposition~\ref{prop:detailed-balance} holds for reversible Petri
  nets $P$ satisfying the anti-symmetry axioms
  (Definition~\ref{def:anti-sym}); equivalently, it is a theorem
  about $\Lk_2(\bar{P})$, the equilibrium level of the reversible
  closure of $P$.
  For a general Petri net without the dagger structure, the
  anti-symmetry axioms do not apply and the proposition is silent
  (see Remark~\ref{rem:dagger-smc}).

  Within this scope, reversal symmetry of the Gibbs-zero locus is a
  \emph{theorem} --- a consequence of anti-symmetry alone --- not
  an additional postulate on the kinetics or the rate constants.
  The familiar relation
  \[
    K_{\mathrm{eq}}(r^\dagger,\, T) \;=\; \frac{1}{K_{\mathrm{eq}}(r,\, T)}
  \]
  follows immediately by exponentiating
  $F_G^T(r^\dagger) = -F_G^T(r)$.
  This is the thermodynamic content recovered at $\Lk_2$.
  The full kinetic detailed-balance condition --- $K_{\mathrm{eq}}
  = k_+/k_-$ for mass-action kinetics, with equality of forward
  and reverse fluxes at equilibrium concentrations --- belongs to
  $\Lk_3$ and is established there as a consistency condition
  between the levels.
\end{remark}

%% file: chapters/L2/l2_layer.tex
\subsection{Layer~1 and Layer~2 for \texorpdfstring{$\FS$}{FS},
  the third law, and standard chemical potential}
\label{sec:L2-layer}

The Layer~1/Layer~2 split introduced for $\FH$ in
Sections~\ref{sec:L1-layer1} and~\ref{sec:L1-layer2} applies to
$\FS$ with the same categorical structure but one physically
decisive difference at Layer~2.

At \textbf{Layer~1}, the structure is identical for both functors:
$\FS$ is any strict SMC functor $\FS : \Lk_0(P) \to B\RR$,
with no constraint beyond the functor axioms (and anti-symmetry
$\FS(r^\dagger) = -\FS(r)$ for reversible nets,
Definition~\ref{def:anti-sym}).
The Layer~1 parameter space for $\FS$ has $|\Rx|$ free real numbers,
reduced to $|\Rx|/2$ by anti-symmetry for reversible Petri nets.
This is entirely parallel to $\FH$: one real number per generating
reaction, freely chosen, with no further constraint.

At \textbf{Layer~2}, both $\FH$ and $\FS$ are required to satisfy a
state-function condition: each must arise as the coboundary
$\delta^0 h$ of a species-level potential $h$.
The categorical machinery is identical in both cases: the
species-level potential is determined by the reaction data
$F = \delta^0 h$ only up to the gauge $\ker(N^\top) \subset
\RR^{|\Sp|}$ (Remark~\ref{rem:gauge}), the space of conservation
laws viewed as species potentials.
What differs between $\FH$ and $\FS$ is the physical convention
used to fix a distinguished representative within this gauge
equivalence class: for $\FH$, a chosen convention (IUPAC:
$\dH_f^\circ(E) = 0$ for elemental references); for $\FS$, an
external physical principle (the third law: $S(X, 0\,\mathrm{K}) =
0$ for perfect crystalline ground states).
This asymmetry --- between conventional and principled gauge
fixing --- is the subject of Section~\ref{sec:L2-thirdlaw} below.

\subsubsection{Layer~2 for \texorpdfstring{$\FS$}{FS}: the entropy
  state-function condition}
\label{sec:L2-FS-layer2}

Layer~2 requires $\FS$ to satisfy the entropy state-function
condition: the entropy change of a reaction must be expressible as
the difference of species-level entropy values, exactly as the
enthalpy change at Layer~2 for $\FH$ is the difference of formation
enthalpies.

\begin{definition}[Layer~2 for $\FS$: entropy state-function condition]
\label{def:FS-layer2}
  $\FS : \Lk_0(P) \to B\RR$ satisfies \emph{Layer~2} if there
  exists a monoid homomorphism
  $h_S : (\Mon, +, \mathbf{0}) \to (\RR, +, 0)$
  such that, for every generator $r : {\bf u} \to {\bf v}$:
  \[
    \FS(r) \;=\; \delta^0 h_S(r) \;:=\; h_S({\bf v}) - h_S({\bf u}).
  \]
\end{definition}

Since $\Mon = \NN^{|\Sp|}$ is the free commutative monoid, $h_S$ is
uniquely determined by its values on species:
$h_S\!\bigl(\sum_S n_S S\bigr) = \sum_S n_S h_S(S)$.

\begin{proposition}[Standard entropies give the canonical potential
  for $\FS$]
\label{prop:FS-formation}
  Define $h_S(S) := S^\circ(S)$, the standard molar entropy of
  species $S$, and extend by linearity.
  Then $h_S$ is a monoid homomorphism and, for every generator
  $r : {\bf u} \to {\bf v}$:
  \[
    \FS(r) \;=\; h_S({\bf v}) - h_S({\bf u})
    \;=\; \sum_{S \in \Sp} N_{S,r}\,S^\circ(S).
  \]
\end{proposition}

\begin{proof}
Identical to Proposition~\ref{prop:formation-section} with $h_f$
replaced by $h_S$ and $\dH_f^\circ$ replaced by $S^\circ$.
The universal property of $\Mon$ supplies uniqueness.
\end{proof}

\subsubsection{The third law: an absolute reference for \texorpdfstring{$h_S$}{hS}}
\label{sec:L2-thirdlaw}

This is the sole asymmetry between $\FH$ and $\FS$ at Layer~2, and
it has no categorical source: it is a fact about the physical world.
The discussion below is needed because the asymmetry has practical
consequences for what thermochemical data can and cannot be compared
across systems, and because it determines the structure of the
Layer~2 parameter space for $\FS$.

\begin{chembox}[Gauge fixing for $h_f$ vs $h_S$: convention vs universal principle]
The Layer~2 coboundary condition $F = \delta^0 h$ does not fix
$h$ uniquely: categorically, $h$ is determined by $F = \delta^0 h$
only up to the gauge $\ker(N^\top) \subset \RR^{|\Sp|}$
(Remark~\ref{rem:gauge}), the space of conservation laws viewed as
species potentials.
This gauge applies equally to $h_f$ and $h_S$ --- the categorical
machinery does not distinguish them.
What differs is the physical convention used to fix a distinguished
representative.

\textbf{Gauge fixing for $h_f$: a chosen convention.}
The potential $h_f(S) = \dH_f^\circ(S)$ is the standard enthalpy of
formation of $S$ from its constituent elements in their standard
states.
The choice of elemental reference set $\mathcal{E}$ together with
the convention $\dH_f^\circ(E) = 0$ for $E \in \mathcal{E}$ selects
one representative within the $\ker(N^\top)$ equivalence class
\cite{IUPACGreenBook2007}.
A different choice of references (or different convention values
for the elementals) would shift $h_f$ by a conservation-law
potential $g \in \ker(N^\top)$, leaving
$\FH(r) = \delta^0 h_f(r)$ invariant.
There is no physical law that selects a canonical $\mathcal{E}$:
``absolute enthalpies do not exist'' \cite{AtkinsDeP2014}, only
enthalpy differences.
In practice, formation enthalpies from tables using different
elemental reference conventions cannot be compared without
conversion, whereas reaction enthalpies are gauge-invariant.

\textbf{Gauge fixing for $h_S$: a universal physical principle.}
The third law of thermodynamics --- formulated by Nernst and given
its modern form by Planck \cite{Nernst1906, Planck1911} --- states
that the entropy of every perfect crystalline substance vanishes at
absolute zero:
\[
  S(X,\; 0\,\mathrm{K}) = 0
  \qquad \text{for all pure substances } X,
\]
under the assumption of a unique, perfectly ordered ground state.
Species with residual entropy (such as CO, ice $\mathrm{I_h}$, and
amorphous glasses frozen into multiple equivalent configurations)
are the standard practical caveat \cite{AtkinsDeP2014}, but this
does not affect the structural argument: for every species in a
standard thermochemical compilation, a canonical basepoint is
fixed.
The standard molar entropy at any reference temperature
$T_\mathrm{ref} > 0$ is then computed by integrating the heat
capacity from absolute zero:
\[
  S^\circ(X,\; T_\mathrm{ref})
  \;=\; \int_0^{T_\mathrm{ref}} \frac{C_p(X,\, T')}{T'}\, dT',
\]
with additive corrections for any phase transitions
\cite{AtkinsDeP2014}.
The third law thus selects the canonical species representative
$h_S(S) = S^\circ(S, T_\mathrm{ref})$ within the $\ker(N^\top)$
gauge equivalence class --- not by suppressing the categorical
gauge, but by importing an external physical principle that
specifies absolute species entropies.

\textbf{The asymmetry, precisely stated.}
Categorically, both $h_f$ and $h_S$ live in
$C^0 = \Hom_{\mathbf{Mon}}(\Mon, \RR)$ with gauge $\ker(N^\top)$:
only the coboundaries $\FH = \delta^0 h_f$ and
$\FS = \delta^0 h_S$ are physically meaningful.
The asymmetry is in the \emph{kind} of physical input that fixes a
representative: $h_f$ is fixed by a chosen convention (IUPAC
elemental references), while $h_S$ is fixed by a universal physical
principle (the third law).
The third law is external input to the $\Lk_2$ categorical
structure --- a physical datum that has no categorical source.
\end{chembox}

\begin{observation}[The third law as canonical normalisation of $h_S$]
\label{obs:third-law}
  Categorically, the species-level potential $h_S$ is determined by
  $\FS = \delta^0 h_S$ only up to the $\ker(N^\top)$ gauge, exactly
  as for $h_f$.
  The third law of thermodynamics fixes a canonical representative
  within this gauge equivalence class: for any reference temperature
  $T_\mathrm{ref} > 0$,
  \[
    h_S(S)
    \;=\; S^\circ(S,\; T_\mathrm{ref})
    \;=\; \int_0^{T_\mathrm{ref}} \frac{C_p(S,\, T')}{T'}\, dT'
    \;>\; 0
    \qquad \text{for all } S.
  \]
  This is the unique representative consistent with the third-law
  normalisation $S(X, 0\,\mathrm{K}) = 0$ for perfect crystalline
  ground states.
  For $h_f$, the analogous fixing comes from a chosen convention
  (IUPAC: $\dH_f^\circ(E) = 0$ for elemental references), not from
  a universal physical principle.
  The asymmetry between $h_f$ and $h_S$ is therefore at the level
  of \emph{physical conventions used to fix the gauge}, not at the
  level of categorical structure: it is a physical datum (the third
  law) external to the $\Lk_2$ functor structure.
\end{observation}

\subsubsection{Standard chemical potential as a derived object}
\label{sec:L2-chempot}

With Layer~2 satisfied for both $\FH$ and $\FS$, the standard
chemical potential emerges as a canonical derived object that
combines both species-level potentials into a single
temperature-dependent quantity.
It is the height function of the Gibbs functor $\FG^T$, playing the
same organisational role that $h_f$ plays for $\FH$ and $h_S$ plays
for $\FS$.

\begin{chembox}[The standard chemical potential: for chemists and
  mathematicians]
The \emph{standard chemical potential} $\mu^\circ(S, T)$ of species
$S$ at temperature $T$ is the molar Gibbs free energy of $S$ under
standard conditions.
It was introduced by Gibbs \cite{Gibbs1875} as the fundamental
thermodynamic quantity governing the direction of chemical change
and the position of equilibrium.

\textbf{For mathematicians.}
$\mu^\circ(S, T)$ is simply the pointwise combination
$h_f(S) - T h_S(S)$ of the two Layer~2 potentials at a common
species $S$.
It is the height function whose coboundary gives $\FG^T$: just as
$h_f$ encodes the ``enthalpy height'' of each species and $h_S$
encodes the ``entropy height'', $\mu^\circ$ encodes the ``free
energy height'' at temperature $T$.
The drop $h_{\mu^\circ}({\bf v}, T) - h_{\mu^\circ}({\bf u}, T)$
from reactants to products is exactly $\FG^T(r)$: the Gibbs free
energy of the reaction.

\textbf{For chemists.}
The \emph{Gibbs fundamental relation} at constant $T$ and $P$
\cite{AtkinsDeP2014, Kondepudi2014} states
\[
  dG \;=\; \sum_{S \in \Sp} \mu_S\, dn_S,
  \qquad
  \mu_S \;=\;
  \Bigl(\frac{\partial G}{\partial n_S}\Bigr)_{T,P,n_{j\neq S}},
\]
so the chemical potential $\mu_S$ is the partial molar Gibbs free
energy of $S$.
At standard conditions, $\mu_S = \mu^\circ(S, T)$, and the standard
Gibbs free energy of a reaction is the signed sum of chemical
potentials over all species.
\end{chembox}

\begin{definition}[Standard chemical potential]
\label{def:chem-pot}
  For species $S \in \Sp$ and temperature $T > 0$, the
  \emph{standard chemical potential} is
  \[
    \mu^\circ(S,\, T)
    \;:=\; h_f(S) - T\,h_S(S)
    \;=\; \dH_f^\circ(S) - T\,S^\circ(S).
  \]
  Extend to complexes by linearity:
  $h_{\mu^\circ}({\bf u},\,T)
  := \sum_S \nu_S^{({\bf u})}\,\mu^\circ(S,\,T)$.
\end{definition}

The species-level value $\mu^\circ(S, T) = h_f(S) - T\,h_S(S)$
inherits the $\ker(N^\top)$ gauge from $h_f$: the third-law
contribution $T\,h_S(S)$ is canonically fixed (Edit~2 chembox), but
the formation contribution $h_f(S)$ is not.
Only the coboundary $F_G^T = \delta^0 h_{\mu^\circ}$ is
gauge-invariant, and on balanced reactions (the only kind a
chemically valid Petri net contains) it agrees with the textbook
standard reaction Gibbs free energy
$\dG^\circ_r = \sum_S N_{S,r}\,\mu^\circ_{\mathrm{textbook}}(S, T)$.
On non-balanced reactions, the chapter's $\mu^\circ$ would differ
from the textbook standard chemical potential by a gauge term
involving elemental reference entropies; on balanced reactions
this difference vanishes.

The significance of this definition is that it reduces the
thermochemical data of the entire network to species-level data.
At Layer~1, $F_G^T$ requires one real number per reaction; at
Layer~2, one number per species --- the value $\mu^\circ(S, T)$
--- suffices to compute $F_G^T(r)$ for every reaction
simultaneously.
Proposition~\ref{prop:gibbs-mu} makes this precise.

\begin{proposition}[Gibbs functor via standard chemical potential]
\label{prop:gibbs-mu}
  Under Layer~2 for both $\FH$ and $\FS$, the Gibbs functor is
  the coboundary of $h_{\mu^\circ}$:
  for every generator $r : {\bf u} \to {\bf v}$,
  \[
    \FG^T(r)
    \;=\; h_{\mu^\circ}({\bf v},\,T) - h_{\mu^\circ}({\bf u},\,T)
    \;=\; \sum_{S \in \Sp} N_{S,r}\,\mu^\circ(S,\,T).
  \]
\end{proposition}

\begin{proof}
\begin{align*}
  \FG^T(r)
  &= \FH(r) - T\,\FS(r) \\
  &= \bigl[h_f({\bf v}) - h_f({\bf u})\bigr]
     - T\bigl[h_S({\bf v}) - h_S({\bf u})\bigr] \\
  &= \bigl[h_f({\bf v}) - T\,h_S({\bf v})\bigr]
     - \bigl[h_f({\bf u}) - T\,h_S({\bf u})\bigr] \\
  &= h_{\mu^\circ}({\bf v},\,T)
     - h_{\mu^\circ}({\bf u},\,T). \qedhere
\end{align*}
\end{proof}

\begin{insightbox}[The chemical potential as the height function
  of $\Lk_2$]
Proposition~\ref{prop:gibbs-mu} states that, under Layer~2 for both
decorating functors, the Gibbs functor $\FG^T$ is itself a coboundary:
$\FG^T = \delta^0 h_{\mu^\circ}$.
The standard chemical potential $\mu^\circ(S,T)$ is the height
function at $\Lk_2$, playing the same role that $h_f$ plays for
$\FH$ and $h_S$ plays for $\FS$.

This is the precise categorical content of the chemist's formula
\cite{Gibbs1875, AtkinsDeP2014}:
\[
  \dG^\circ_{\mathrm{rxn}} \;=\;
  \sum_{\mathrm{products}} \nu_i\,\mu^\circ_i
  \;-\; \sum_{\mathrm{reactants}} \nu_j\,\mu^\circ_j.
\]
The coboundary structure $\FG^T = \delta^0 h_{\mu^\circ}$ is not an
approximation: it is the exact content of Layer~2, holding for every
reaction in the network simultaneously.
At Layer~1, one real number per reaction is needed to specify $\FG^T$;
at Layer~2, one number per species ($\mu^\circ(S, T)$) suffices for
the entire network.
This is the maximal compression of thermochemical information
available at $\Lk_2$.
\end{insightbox}

%% file: chapters/L2/l2_thm.tex
\subsection{Thermodynamic theorems at \texorpdfstring{$\Lk_2$}{L2}}
\label{sec:L2-thm}

The Gibbs family $T \mapsto \FG^T = \FH - T\cdot\FS$ suffices to
derive, within the categorical framework of $\Lk_2$, four classical
results of chemical thermodynamics as theorems for reaction networks:
the second law criterion for chemical spontaneity (Proposition~\ref{prop:second-law}), the van~'t~Hoff
equation for the temperature dependence of equilibrium constants, the
Gibbs--Helmholtz equation recovering the reaction enthalpy from
free-energy measurements, and Le~Chatelier's principle for temperature
perturbations.
Each is a statement about morphisms in $\Lk_0(P)$ labelled by the
Gibbs family; none requires data beyond what is already in $\Lk_2$.

\subsubsection{The equilibrium constant}

\begin{chembox}[What is the equilibrium constant and why does it
  matter?]
The equilibrium constant $K_{\mathrm{eq}}(r, T)$ is the central
quantitative tool of chemical thermodynamics \cite{AtkinsDeP2014,
Kondepudi2014}.
It encodes the position of equilibrium for a reaction $r$ at
temperature $T$: the ratio of product to reactant concentrations (or
activities) when the system has reached thermodynamic equilibrium.

For a reaction $r : \sum_j \nu_j^- S_j \to \sum_i \nu_i^+ S_i$ under
the law of mass action:
\[
  K_{\mathrm{eq}}(r, T)
  \;=\;
  \frac{\prod_i a_i^{\nu_i^+}}{\prod_j a_j^{\nu_j^-}}
  \Bigg|_{\text{equilibrium}},
\]
where $a_S$ is the thermodynamic activity of species $S$.
A large $K_{\mathrm{eq}} \gg 1$ means the reaction is
product-favoured; $K_{\mathrm{eq}} \ll 1$ means reactant-favoured;
$K_{\mathrm{eq}} = 1$ means the system is at standard equilibrium,
i.e., the reaction lies in the equilibrium locus
$\ker(F_G^T)$ (Definition~\ref{def:equil-locus}).

The connection to the Gibbs functor is the fundamental relation
\cite{Gibbs1875, AtkinsDeP2014}:
\[
  \Delta G^\circ_r \;=\; -RT\ln K_{\mathrm{eq}},
\]
or equivalently $K_{\mathrm{eq}} = \exp(-\Delta G^\circ_r / RT)$.
In our framework, $\Delta G^\circ_r = F_G^T(r)$, so the equilibrium
constant is the exponential of the (negative, rescaled) Gibbs functor
value.
The functor $F_G^T$ thus encodes all equilibrium position information
in a single real number per generating reaction, from which
$K_{\mathrm{eq}}$ is recovered by exponentiation.
\end{chembox}

\begin{definition}[Equilibrium constant]
\label{def:K}
  For a generating reaction $r \in \Rx$ and temperature $T > 0$,
  the \emph{standard equilibrium constant} is
  \[
    K_{\mathrm{eq}}(r,\,T)
    \;:=\;
    \exp\!\left(\frac{-F_G^T(r)}{RT}\right),
  \]
  where $R = 8.314\;\mathrm{J\,mol^{-1}K^{-1}}$ is the gas constant.
\end{definition}

\subsubsection{The second law for chemical reactions}

The second law of thermodynamics, at constant temperature and
pressure, takes the following form for individual reaction steps.

\begin{proposition}[Sign of $F_G^T$ at standard conditions]
\label{prop:second-law}
  Let $r \in \Rx$ be a generating reaction and $T > 0$.
  \begin{enumerate}[label=(\roman*)]
    \item $F_G^T(r) < 0$: $r$ is product-favoured under standard
      conditions, $K_r(T) > 1$.
    \item $F_G^T(r) > 0$: $r$ is reactant-favoured under standard
      conditions, $K_r(T) < 1$; the reverse $r^\dagger$ is
      product-favoured.
      The forward direction can still proceed at non-standard
      concentrations (when $RT \ln Q_r < -F_G^T(r)$), by coupling
      to driven processes, or kinetically.
    \item $F_G^T(r) = 0$: $r$ lies in the standard Gibbs-zero locus
      $Z_G(T) = \ker(F_G^T)$, equivalently $K_{\mathrm{eq}}(r, T) =
      1$.
  \end{enumerate}
  The Gibbs-zero locus $Z_G(T)$ partitions the generating reactions
  at temperature $T$ into these three classes by the sign of
  $F_G^T$.
\end{proposition}

\begin{proof}
At constant temperature and constant pressure, the second law
requires $dG \leq 0$ for any spontaneous process
\cite{AtkinsDeP2014, Kondepudi2014}.
For a reaction proceeding by extent $d\xi > 0$,
$dG = \Delta G_r(x, T)\,d\xi$, where
$\Delta G_r(x, T) = F_G^T(r) + RT\ln Q_r(x)$ depends on
concentrations through the reaction quotient $Q_r$.
At standard conditions ($Q_r = 1$), $\Delta G_r$ reduces to
$F_G^T(r)$, and the three cases follow from
$K_{\mathrm{eq}} = \exp(-F_G^T(r)/RT)$: $F_G^T < 0$ gives
$K_{\mathrm{eq}} > 1$ (product-favoured under standard conditions);
$F_G^T > 0$ gives $K_{\mathrm{eq}} < 1$ (reactant-favoured under
standard conditions); $F_G^T = 0$ gives $K_{\mathrm{eq}} = 1$ ($r$
in the Gibbs-zero locus).
\end{proof}

\begin{remark}[The second law as a level-stratified statement]
At $\Lk_0$: no thermodynamic quantity is defined; spontaneity
cannot be formulated.
At $\Lk_1$ with $\FH$ only: exothermic reactions ($\FH < 0$) might
seem spontaneous, but this is only true when $T\FS(r) \ll \FH(r)$;
the general criterion requires $\FS$.
At $\Lk_2$: Proposition~\ref{prop:second-law} is a theorem, and the
enthalpy-only intuition is recovered as the low-temperature limit
of the exact criterion $F_G^T(r) = \FH(r) - T\FS(r) < 0$.
\end{remark}

\subsubsection{The van~'t~Hoff equation}

\begin{chembox}[The van~'t~Hoff equation: history and use]
The van~'t~Hoff equation describes how the equilibrium constant of a
reaction changes with temperature.
It was derived by Jacobus Henricus van~'t~Hoff in 1884 from the
Clausius--Clapeyron equation and Hess's Law
\cite{vantHoff1884, AtkinsDeP2014}, and is one of the most
practically important equations in chemical thermodynamics.

Its primary use is measurement: plotting $\ln K_{\mathrm{eq}}$ against
$1/T$ (the \emph{van~'t~Hoff plot}) gives a straight line with slope
$-\FH(r)/R$ and intercept $\FS(r)/R$ when $\FH$ and $\FS$ are
temperature-independent.
This allows $\FH(r)$ and $\FS(r)$ to be extracted from
equilibrium constant measurements at multiple temperatures, without
calorimetry.
The van~'t~Hoff plot is the experimental foundation for the Layer~1
parameter space: it is the direct measurement procedure that assigns
one real number ($F_G^T$, at any chosen reference temperature) to each reaction.

In our framework, the equation is not a separate physical input but
a two-line consequence of differentiating the affine function
$T \mapsto \ln K_{\mathrm{eq}}(r, T) = -F_G^T(r)/RT$ with respect
to $T$.
The approximation that $\FH(r)$ and $\FS(r)$ are
temperature-independent corresponds to a $T$-independent reading of
the Layer~2 potentials --- i.e., using fixed $h_f(S, T_\mathrm{ref})$
and $h_S(S, T_\mathrm{ref})$ at a reference temperature and ignoring
heat-capacity corrections.
Fully temperature-dependent Layer~2 data --- available for both
$h_f(S, T)$ and $h_S(S, T)$ in the references like JANAF tables \cite{Chase1998} ---
gives the integrated form of the van~'t~Hoff equation accounting
for $C_p(T)$.
\end{chembox}

\begin{proposition}[Van~'t~Hoff equation]
\label{prop:vant-hoff}
  Treating $\FH(r)$ and $\FS(r)$ as temperature-independent
  (the standard approximation for $\dH^\circ$ and $\dS^\circ$
  \cite{AtkinsDeP2014}):
  \[
    \frac{d\ln K_{\mathrm{eq}}(r,\,T)}{dT}
    \;=\; \frac{\FH(r)}{RT^2}.
  \]
\end{proposition}

\begin{proof}
Expand $\ln K_{\mathrm{eq}}$ using Definition~\ref{def:K} and the
Gibbs functor:
\[
  \ln K_{\mathrm{eq}}(r,\,T)
  = \frac{-F_G^T(r)}{RT}
  = \frac{-\FH(r) + T\,\FS(r)}{RT}
  = \frac{-\FH(r)}{RT} + \frac{\FS(r)}{R}.
\]
Differentiating with respect to $T$ (both $\FH(r)$ and $\FS(r)$
constant by assumption):
\[
  \frac{d\ln K_{\mathrm{eq}}}{dT}
  = \frac{\FH(r)}{RT^2} + 0
  = \frac{\FH(r)}{RT^2}. \qedhere
\]
\end{proof}

\begin{remark}[The van~'t~Hoff equation as a level-stratified
  statement]
At $\Lk_0$: neither $K_{\mathrm{eq}}$ nor $T$ is part of the structure; the
equation cannot be formulated.
At $\Lk_1$ with $\FH$ only: $K_{\mathrm{eq}}$ could be defined via a partially
specified $F_G^T$, but its $T$-derivative would require $\FS$, which
is absent.
At $\Lk_2$: the equation is a two-line computation from
$F_G^T = \FH - T\cdot\FS$.
It is the statement $\partial(-F_G^T/RT)/\partial T = \FH/RT^2$,
an identity in the definition of the Gibbs family.
The van~'t~Hoff plot (slope $= -\FH/R$, intercept $= \FS/R$) is the
geometric picture of the Gibbs family as a straight line in the
space of strict SMC functors $\Lk_0(P) \to B\RR$
(Mathbox, Section~\ref{sec:L2-gibbs}).
\end{remark}

\subsubsection{The Gibbs--Helmholtz equation}

\begin{chembox}[The Gibbs--Helmholtz equation: physical meaning and
  practical use]
The Gibbs--Helmholtz equation \cite{Gibbs1875, Helmholtz1882,
AtkinsDeP2014} is the relation between the temperature dependence of
the Gibbs free energy and the enthalpy of a process.
Physically, it answers the following question: if one measures
$\Delta G^\circ_r = F_G^T(r)$ at two temperatures $T_1$ and $T_2$
without calorimetry, can one recover $\FH(r)$?

The answer is yes: $\FH(r)$ is the \emph{slope} of the function
$F_G^T(r)/T$ plotted against $1/T$.
Under the constant-$\FH, \FS$ approximation,
$F_G^T(r)/T = \FH(r)\cdot(1/T) - \FS(r)$, so the plot is a straight
line with slope $\FH(r)$ and intercept $-\FS(r)$ --- a second route
to the Layer~1 parameter $\FH(r)$ that does not require measuring
heat directly.
This is important in practice whenever calorimetry is difficult (e.g.,
for reactions in solution or in biological systems), where $\Delta G$
can be measured via $K_{\mathrm{eq}}$ but $\Delta H$ is harder to
access \cite{AtkinsDeP2014, freire2008enthalpy}.
\end{chembox}

\begin{proposition}[Gibbs--Helmholtz equation]
\label{prop:gibbs-helmholtz}
  For any generator $r$ and temperature $T > 0$:
  \[
    \FH(r)
    \;=\;
    -T^2\,\frac{\partial}{\partial T}\!\left[\frac{F_G^T(r)}{T}\right].
  \]
\end{proposition}

\begin{proof}
\[
  \frac{\partial}{\partial T}\!\left[\frac{F_G^T(r)}{T}\right]
  = \frac{\partial}{\partial T}\!\left[\frac{\FH(r)}{T} - \FS(r)\right]
  = \frac{-\FH(r)}{T^2}.
\]
Multiplying by $-T^2$ gives the result.
\end{proof}

\begin{chembox}[Gibbs--Helmholtz in practice and in the tower]
The Gibbs--Helmholtz equation extracts $\FH(r)$ from measurements
of $F_G^T(r)$ at two temperatures, without direct calorimetry
\cite{AtkinsDeP2014}.
Categorically, $\FH$ and $\FS$ are the ``intercept'' and
``negative slope'' of the affine function
$T \mapsto F_G^T(r) = \FH(r) - T\,\FS(r)$.
The Gibbs--Helmholtz equation recovers $\FH(r)$ from a single
derivative,
$\FH(r) = -T^2\,\partial(F_G^T/T)/\partial T$, and this is also
connected to Remark~\ref{rem:L2-via-gibbs}: the affine structure of
$T \mapsto F_G^T(r)$ means that two temperature measurements
suffice to determine both $\FH(r)$ and $\FS(r)$ individually,
providing a complete $\Lk_2$ characterisation from $\Lk_1$ data at
two temperatures.
\end{chembox}

\subsubsection{Le~Chatelier's principle}

\begin{observation}[Le~Chatelier's principle, under the constant-$\FH$ approximation]
\label{obs:le-chatelier}
  Under the temperature-independent-enthalpy approximation
  inherited from Proposition~\ref{prop:vant-hoff}, for any generator
  $r$:
  \begin{itemize}
    \item $\FH(r) > 0$ (endothermic): $d\ln K/dT > 0$, so $K$
      increases with $T$.
      Heating shifts equilibrium toward products.
    \item $\FH(r) < 0$ (exothermic): $d\ln K/dT < 0$, so $K$
      decreases with $T$.
      Heating shifts equilibrium toward reactants.
    \item $\FH(r) = 0$ (thermoneutral): $K$ is independent of $T$.
  \end{itemize}
  This is Le~Chatelier's principle \cite{LeChatelier1884} for
  temperature perturbations: a system at equilibrium responds to
  heating by shifting in the endothermic direction.
  At $\Lk_2$ it is a theorem in the same approximation that supports
  the van~'t~Hoff equation; if heat-capacity effects change the sign
  of $\dH^\circ$ over the temperature interval considered, the
  directional statement must be re-evaluated using the integrated
  van~'t~Hoff form.
\end{observation}

%% file: chapters/L2/l2_wegscheider.tex
\subsection{Wegscheider conditions at \texorpdfstring{$\Lk_2$}{L2}}
\label{sec:L2-wegscheider}

\begin{chembox}[Wegscheider conditions in the CRNT literature]
The Wegscheider conditions were introduced by Rudolf Wegscheider in
1901 \cite{Wegscheider1901} as necessary constraints on the
equilibrium constants of a reversible reaction network with cycles.
For a directed cycle $r_1, \ldots, r_k$ in the reaction graph, the
conditions state:
\[
  \prod_{i=1}^k K_{\mathrm{eq}}(r_i,\, T) \;=\; 1,
  \qquad\text{equivalently,}\qquad
  \sum_{i=1}^k \ln K_{\mathrm{eq}}(r_i,\, T) \;=\; 0.
\]
Their physical content is that thermodynamics forbids perpetual
motion: no network of reactions can have a cycle in which every step
is spontaneous, since that would allow work to be extracted from a
system at uniform temperature and pressure.

In the CRNT literature, the Wegscheider conditions appear as linear
constraints on the vector of $\ln K_{\mathrm{eq}}$ values.
Horn and Jackson \cite{HornJackson1972} showed that they are
equivalent to the existence of a thermodynamic potential (a species-
level free energy function) whose coboundary gives the reaction free
energies --- precisely the Layer~2 condition $\FG^T = \delta^0
h_{\mu^\circ}$ of Proposition~\ref{prop:gibbs-mu}.
Feinberg \cite{Feinberg1989} used the conditions as a necessary
criterion for detailed balance in mass-action systems.
M\"{u}ller and Regensburger \cite{MuellerRegensburger2012} gave
a systematic treatment of the conditions in the context of
generalised mass-action kinetics.

What is new in Proposition~\ref{prop:wegscheider1} below is not the
conditions themselves but their derivation: they follow directly and
necessarily from Layer~2 applied independently to $\FH$ and $\FS$,
without invoking kinetics or rate constants.
In the standard CRNT treatment, the conditions on $\dH^\circ$ and
$\dS^\circ$ are stated separately from the multiplicative condition
on $K_{\mathrm{eq}}$ values.
The proof below shows they are the same condition at different levels
of the tower.
\end{chembox}

\noindent
Before stating the proposition, we fix notation for the reaction
graph.

\begin{definition}[Reaction graph $G(\Rx)$ and directed cycle]
\label{def:reaction-graph}
  The \emph{reaction graph} $G(\Rx)$ of a Petri net $P$ is the
  directed graph whose vertices are the complexes
  ${\bf u} \in \NN^{|\Sp|}$ and whose directed edges are the generating
  reactions $r : {\bf u} \to {\bf v}$ in $\Rx$.
  A \emph{directed cycle} in $G(\Rx)$ of length $k$ is a sequence
  of generators $r_1, \ldots, r_k \in \Rx$ such that the target of
  each $r_i$ equals the source of $r_{i+1}$ (indices modulo $k$):
  \[
    r_1 : {\bf u}_1 \to {\bf u}_2,\quad
    r_2 : {\bf u}_2 \to {\bf u}_3,\quad
    \ldots,\quad
    r_k : {\bf u}_k \to {\bf u}_1.
  \]
  Such a cycle determines a composite morphism
  $r_k \circ \cdots \circ r_1 : {\bf u}_1 \to {\bf u}_1$ in
  $\Lk_0(P)$ whose source and target coincide.
  This composite need not equal $\id_{{\bf u}_1}$ --- in the free
  skeletal permutative category $\Lk_0(P)$, distinct sequences of 
  chemical generators can yield distinct morphisms with the same source and target.
  The cycle conditions that follow are the statement that this
  composite maps to $0$ under $\FH$, $\FS$, and $\FG$, despite not
  being the identity morphism.
\end{definition}

\noindent
For a Petri net satisfying Layer~2 for $\FH$ (equivalently, for any
coboundary functor), the cycle condition $\sum_i \FH(r_i) = 0$ is
an immediate consequence of the coboundary property
$\FH(r) = h_f({\bf v}) - h_f({\bf u})$: the sum telescopes around
the cycle.
This is Kirchhoff's cycle law, or equivalently Hess's Law for
networks \cite{Hess1840}.
The same argument applies to $\FS$ under Layer~2 for $\FS$.
Proposition~\ref{prop:wegscheider1} combines both.

\begin{proposition}[Wegscheider conditions]
\label{prop:wegscheider1}
  Let $(P,\, h_f,\, h_S)$ be a thermochemical network at $\Lk_2$
  with Layer~2 satisfied for both $\FH$ and $\FS$.
  Let $r_1, \ldots, r_k$ form a directed cycle in $G(\Rx)$.
  Then for every $T > 0$:
  \begin{equation}
    \label{eq:wegscheider-add}
    \sum_{i=1}^k F_G^T(r_i) \;=\; 0,
    \qquad\text{equivalently,}\qquad
    \prod_{i=1}^k K_{\mathrm{eq}}(r_i,\, T) \;=\; 1.
  \end{equation}
\end{proposition}

\begin{proof}
\textbf{Additive form.}
By Proposition~\ref{prop:cycle} (Kirchhoff's cycle condition applied
to $\FH$ at $\Lk_1$, a consequence of $\FH = \delta^0 h_f$):
$\sum_{i=1}^k \FH(r_i) = 0$.
The same proposition applied to $\FS$ (since Definition~\ref{def:FS-layer2}
imposes the same coboundary condition $\FS = \delta^0 h_S$):
$\sum_{i=1}^k \FS(r_i) = 0$.
Therefore, for all $T > 0$:
\[
  \sum_{i=1}^k F_G^T(r_i)
  = \sum_{i=1}^k \bigl[\FH(r_i) - T\,\FS(r_i)\bigr]
  = 0 - T \cdot 0 \;=\; 0.
\]
\textbf{Multiplicative form.}
$\displaystyle\prod_{i=1}^k K_{\mathrm{eq}}(r_i,\,T)
= \exp\!\Bigl(-\tfrac{1}{RT}\sum_i F_G^T(r_i)\Bigr)
= \exp(0) = 1$.
\end{proof}

\begin{mathbox}[Why universality in $T$ forces two independent
  conditions]
The Wegscheider conditions in the CRNT literature are sometimes
stated only at a single reference temperature.
The categorical structure at $\Lk_2$ shows why this is insufficient
for full thermodynamic consistency.

Suppose the Layer~2 cycle condition were imposed on $F_G^{T_0}$ only at a single temperature $T_0$.
The condition $\sum_i F_G^{T_0}(r_i) = 0$ would hold at $T_0$ but
fail at other temperatures: a model satisfying it would be
thermodynamically inconsistent outside a narrow temperature range.
Requiring the cycle condition for \emph{all} $T > 0$:
\[
  \sum_i \FH(r_i) - T\sum_i \FS(r_i) = 0
  \quad\forall\,T > 0.
\]
Since this is an affine function of $T$ that vanishes identically,
both coefficients must vanish independently:
\[
  \sum_i \FH(r_i) = 0
  \qquad\text{and}\qquad
  \sum_i \FS(r_i) = 0.
\]
These are the independent coboundary conditions for $\FH$ and $\FS$.
The Wegscheider conditions are exactly their joint content, and
full temperature-universality requires both.
This is also the content of Remark~\ref{rem:L2-via-gibbs}: the
Gibbs family encodes $(\FH, \FS)$ independently, and consistency
at all temperatures forces both to satisfy the cycle condition
separately.
\end{mathbox}

\begin{observation}[Degrees of freedom at $\Lk_2$ and the deficiency hierarchy]
\label{obs:dof-L2}
  The Layer~2 cycle conditions inherit the three-tier hierarchy of
  Observation~\ref{obs:cycle-rank}, applied independently to $\FH$,
  $\FS$, and the Gibbs family:
  \[
    \RR^{|\Rx|}
    \;\supseteq\;
    \im(I_a^\top)
    \;\supseteq\;
    \im(N^\top),
    \qquad
    |\Rx| \;\geq\; n - \ell \;\geq\; \rho.
  \]
  Imposing graph-cycle conditions on the reaction graph alone
  reduces $\RR^{|\Rx|}$ to $\im(I_a^\top)$ of dimension $n - \ell$
  (arbitrary complex potentials, not necessarily species-additive);
  imposing the further species-additive structure of Layer~2
  reduces this to $\im(N^\top)$ of dimension $\rho$.
  The gap $\delta := (n - \ell) - \rho \geq 0$ is the deficiency
  of the network; for deficiency-zero networks $\rho = n - \ell$.
  The \emph{intrinsic} reaction-level parameter count of $\Lk_2$
  thermodynamics is therefore $2\rho$: $\rho$ each for $\FH$ and
  $\FS$ via $\im(N^\top) = \im(\delta^0)$.

  In the species-table representation used in standard
  thermochemistry, this content is encoded by:
  \begin{itemize}
    \item $|\Sp|$ formation enthalpies $\dH_f^\circ(S)$,
      determined up to the $\ker(N^\top)$ gauge --- the IUPAC
      convention $\dH_f^\circ(E) = 0$ for elemental references
      provides the standard fixing.
    \item $|\Sp|$ standard entropies $S^\circ(S)$, also with the
      $\ker(N^\top)$ gauge categorically --- the third law fixes
      the canonical representative
      (Section~\ref{sec:L2-thirdlaw}).
  \end{itemize}
  These $2|\Sp|$ entries determine all reaction enthalpies,
  entropies, Gibbs free energies, and equilibrium constants at
  every temperature; the Wegscheider conditions are then
  automatically satisfied.
  The species-table representation has $2(|\Sp| - \rho)$ gauge
  degrees of freedom relative to the intrinsic $2\rho$
  reaction-level parameters.
\end{observation}

%% file: chapters/L2/l2_examples.tex
\subsection{Worked examples at \texorpdfstring{$\Lk_2$}{L2}}
\label{sec:L2-examples}

All thermochemical data in this section are taken from the
NIST Chemistry WebBook \cite{NIST_WebBook} and the NIST-JANAF
Thermochemical Tables \cite{Chase1998} at the standard reference
temperature $T_\mathrm{ref} = 298.15\,\mathrm{K}$ and pressure
$p^\circ = 1\,\mathrm{bar}$, unless otherwise noted.

\subsubsection{Thermal dissociation of dinitrogen tetroxide}

\begin{example}[$\mathrm{N_2O_4}(g) \rightleftharpoons 2\,\mathrm{NO_2}(g)$]
\label{ex:N2O4}
Species: $\Sp = \{\mathrm{N_2O_4(g)},\, \mathrm{NO_2(g)}\}$.
Reversible Petri net $\bar{P}$ with generators:
\[
  r_1 : \mathrm{N_2O_4(g)} \to 2\,\mathrm{NO_2(g)},
  \qquad
  r_1^\dagger : 2\,\mathrm{NO_2(g)} \to \mathrm{N_2O_4(g)}.
\]

\textbf{$\Lk_1$ data.}
Standard formation enthalpies:
$\dH_f^\circ(\mathrm{NO_2}) = +33.2\;\mathrm{kJ\,mol^{-1}}$,
$\dH_f^\circ(\mathrm{N_2O_4}) = +9.16\;\mathrm{kJ\,mol^{-1}}$.
\[
  \FH(r_1) = 2(33.2) - 9.16 = +57.2\;\mathrm{kJ\,mol^{-1}}
  \quad\text{(endothermic)}.
\]

\textbf{$\Lk_2$ data.}
Standard molar entropies at 298.15~K:
$S^\circ(\mathrm{NO_2}) = 240.1\;\mathrm{J\,mol^{-1}K^{-1}}$,
$S^\circ(\mathrm{N_2O_4}) = 304.4\;\mathrm{J\,mol^{-1}K^{-1}}$.
\[
  \FS(r_1) = 2(240.1) - 304.4 = +175.8\;\mathrm{J\,mol^{-1}K^{-1}}.
\]

\textbf{Anti-symmetry check (Definition~\ref{def:anti-sym}).}
$\FH(r_1^\dagger) = -57.2\;\mathrm{kJ\,mol^{-1}}$ (exothermic),
$\FS(r_1^\dagger) = -175.8\;\mathrm{J\,mol^{-1}K^{-1}}$ (entropy decrease).
Both signs flip as expected.

\textbf{Gibbs functor at selected temperatures.}
\begin{align*}
  F_G^{298}(r_1)
    &= 57.2 - 298 \times 0.1758 = +4.85\;\mathrm{kJ\,mol^{-1}} > 0
    &\Rightarrow\; &\mathrm{N_2O_4}\text{ favoured},\\
  F_G^{400}(r_1)
    &= 57.2 - 400 \times 0.1758 = -13.1\;\mathrm{kJ\,mol^{-1}} < 0
    &\Rightarrow\; &\mathrm{NO_2}\text{ favoured}.
\end{align*}
Equilibrium constants (Definition~\ref{def:K}):
\[
  K(298) = \exp\!\!\left(\frac{-4850}{8.314\times 298}\right) \approx 0.14,
  \qquad
  K(400) = \exp\!\!\left(\frac{-(-13100)}{8.314\times 400}\right) \approx 51.
\]

\textbf{Crossover temperature (Observation~\ref{obs:T-dependence}).}
\[
  T^*(r_1) = \frac{\FH(r_1)}{\FS(r_1)}
  = \frac{57200}{175.8} \approx 325\;\mathrm{K}.
\]
Below 325~K: $\mathrm{N_2O_4}$ is thermodynamically favoured ($K < 1$).
Above 325~K: $\mathrm{NO_2}$ is favoured ($K > 1$).
At exactly 325~K: $K = 1$.

\textbf{Van~'t~Hoff (Proposition~\ref{prop:vant-hoff}).}
\[
  \frac{d\ln K}{dT} = \frac{+57200}{8.314\,T^2} > 0
  \quad\text{for all }T > 0\text{:}
\]
$K$ is strictly increasing with temperature.
Le~Chatelier (Observation~\ref{obs:le-chatelier}): heating shifts
equilibrium toward $\mathrm{NO_2}$, consistent with the forward
reaction being endothermic.

\textbf{Layer~2 check.}
$\FH(r_1) = \delta^0 h_f(r_1)$ uses formation enthalpies, and $\FS(r_1) = \delta^0 h_S(r_1)$ uses absolute standard entropies (third-law normalised) from the same data used above. \\
By Proposition~\ref{prop:gibbs-mu}, $\FG^T(r_1) = \delta^0 h_{\mu^\circ}(r_1)$ can be obtained immediately. 

\textbf{Reversal symmetry of the Gibbs-zero locus
(Proposition~\ref{prop:detailed-balance}).}
$F_G^T(r_1^\dagger) = -F_G^T(r_1)$, so
$r_1 \in Z_G(T) = \ker(F_G^T)$ iff $r_1^\dagger \in Z_G(T)$.
At $T = 325\;\mathrm{K}$ both lie in the Gibbs-zero locus
simultaneously, with
$K_{\mathrm{eq}}(r_1, 325) = K_{\mathrm{eq}}(r_1^\dagger, 325) = 1$.
\end{example}

\subsubsection{An \texorpdfstring{$\mathrm{S_N2}$}{SN2} reaction \texorpdfstring{$\mathrm{CH_3Cl} + \mathrm{OH^-} \to
\mathrm{CH_3OH} + \mathrm{Cl^-}$}{CH3Cl+OH- to CH3OH+Cl-}: \texorpdfstring{$dG^\circ$}{dGo} and
  \texorpdfstring{$K_{\mathrm{eq}}$}{Keq} (schematic)}

\begin{example}[$\mathrm{S_N2}$ at $\Lk_2$ --- schematic]
\label{ex:SN2-L2}
Consider the reaction: $\mathrm{CH_3Cl} + \mathrm{OH^-} \to
\mathrm{CH_3OH} + \mathrm{Cl^-}$ in aqueous solution.
The data below are presented as a \emph{schematic illustration} of
the $\Lk_2$ apparatus on a familiar reaction, using
order-of-magnitude estimates rather than rigorous aqueous
reference-state thermodynamics; aqueous ionic standard-state
conventions are subtle, and the high-temperature extrapolation
appearing later should not be taken literally.

\textbf{$\Lk_1$ data.}
$\FH(r) = -75.0\;\mathrm{kJ\,mol^{-1}}$ (exothermic; from
Section~\ref{sec:L1-examples}, using standard enthalpies of
formation in aqueous solution).

\textbf{$\Lk_2$ data \cite{AtkinsDeP2014}.}
Entropy change:
$\FS(r) \approx -90\;\mathrm{J\,mol^{-1}K^{-1}}$
(order-of-magnitude estimate consistent with reported ion--molecule
$\mathrm{S_N2}$ reaction entropies in water; the precise value is
solvent- and ionic-strength-dependent \cite{AbrahamEtAl1988}).
Two solvated ions react to give one neutral molecule and one ion;
the net ordering of the solvation shell around the charged species
and the reduction in the number of solute particles decrease the
solution entropy, an effect well-documented for ionic substitution
reactions in polar solvents.

\textbf{Gibbs functor at 298~K.}
\[
  F_G^{298}(r) = -75.0 - 298(-0.090) = -75.0 + 26.8
               = -48.2\;\mathrm{kJ\,mol^{-1}}.
\]
The reaction lies outside the Gibbs-zero locus
$Z_G(298) = \ker(F_G^{298})$: it is strongly product-favoured under
standard conditions.
\[
  K_{\mathrm{eq}}(r,\, 298) =
  \exp\!\!\left(\frac{48200}{8.314\times 298}\right)
  \approx 2.8\times 10^8.
\]

\textbf{Temperature dependence.}
$\FS(r) = -0.090\;\mathrm{kJ\,mol^{-1}K^{-1}} < 0$:
\[
  \frac{d\ln K}{dT} = \frac{-75000}{RT^2} < 0.
\]
$K$ decreases with temperature (Le~Chatelier under the constant-$\FH$
approximation, Observation~\ref{obs:le-chatelier}: exothermic,
heating disfavours products).
The crossover formula gives
\[
  T^*(r) = \frac{-75000}{-90} \approx 833\;\mathrm{K},
\]
but this extrapolation lies well outside the regime of the aqueous
standard-state thermodynamics from which $\dH^\circ$ and $\dS^\circ$
were taken: water's normal boiling point is $373\;\mathrm{K}$, and
above the critical point ($\approx 647\;\mathrm{K}$) liquid-water
reference states are not defined.
The $T^*$ value should therefore be read as the formal
extrapolation of constant-$\dH^\circ, \dS^\circ$ data, not as a
physical prediction.
Within the validity range of the aqueous standard state, the model
predicts that the reaction becomes reactant-favoured under standard
conditions at sufficiently high $T$; nothing more.

\textbf{What $\Lk_1$ gives vs.\ what $\Lk_2$ adds.}

\medskip
\begin{center}
\renewcommand{\arraystretch}{1.4}
\begin{tabular}{@{}l l p{6.5cm}@{}}
  \hline
  \textbf{Level} & \textbf{Data} & \textbf{Conclusion (at 298~K, aqueous)} \\
  \hline
  $\Lk_1$ & $\FH = -75.0\;\mathrm{kJ\,mol^{-1}}$
           & Reaction is exothermic. \\[4pt]
  $\Lk_2$ & $\FS = -90\;\mathrm{J\,mol^{-1}K^{-1}}$
           & $K_{\mathrm{eq}}(298) \approx 2.8\times 10^8$;
             strongly product-favoured under standard conditions.
             $d\ln K/dT < 0$: equilibrium shifts toward reactants
             with heating, within the validity range of the aqueous
             standard state. \\
  \hline
\end{tabular}
\end{center}

\textbf{What $\Lk_2$ cannot express.}
Both $K_{\mathrm{eq}}$ and $\FG^T$ say nothing about how fast
equilibrium is reached.
The rate constant
$k \approx 4\times 10^{-3}\;\mathrm{M^{-1}s^{-1}}$ at 298~K
\cite{OlmsteadBrauman1977} is invisible at $\Lk_2$: it enters at
$\Lk_3$.
\end{example}

%% file: chapters/L2/l2_nextforcing.tex
\subsection{What \texorpdfstring{$\Lk_2$}{L2} cannot express: forcing of \texorpdfstring{$\Lk_3$}{L3}}
\label{sec:L2-forcing-out}

\begin{forcingbox}[Forcing pair for $\Lk_3$: same $\dG^\circ$, different rates]
Consider a coarse-grained reversible Petri net $P$ with species
\[
  \Sp = \{\mathrm{CO_2(aq)},\, \mathrm{H_2O(l)},\,
  \mathrm{H_2CO_3(aq)}\}
\]
--- the catalyst (carbonic anhydrase II) is \emph{not} represented
as a net species --- and \emph{two parallel generating reactions}
for $\mathrm{CO_2}$ hydration:
\[
  r_A,\; r_B \;:\; \mathrm{CO_2(aq)} + \mathrm{H_2O(l)}
      \;\longrightarrow\; \mathrm{H_2CO_3(aq)},
\]
the uncatalysed pathway and the enzyme-catalysed pathway.
Both pathways coexist in biological systems (cells express
carbonic anhydrase wherever fast $\mathrm{CO_2}$/bicarbonate
interconversion is needed); at this coarse-grained level they have
the same source and target complex and are genuinely distinct
elementary reactions in the Petri net, distinguished by their
transition states.
At a finer-grained level where the enzyme is tracked as a species,
$r_B$ would expand into a sequence of enzyme-binding, turnover, and
product-release steps; the two-pathway forcing pair above is
meaningful precisely because we have chosen the coarse-grained
framing in which the catalyst is not a net species.

\medskip
\noindent
\textbf{Thermochemical data coincide.}
Both reactions have the same stoichiometry and therefore identical
$\Lk_2$ data \cite{NIST_WebBook, Chase1998}:
\begin{align*}
  \FH(r_A) = \FH(r_B) &= -2.9\;\mathrm{kJ\,mol^{-1}}, \\
  \FS(r_A) = \FS(r_B) &= -98\;\mathrm{J\,mol^{-1}K^{-1}},
\end{align*}
giving
\[
  F_G^{298}(r_A) \;=\; F_G^{298}(r_B)
  \;=\; -2900 - 298 \times (-98)
  \;\approx\; +26.3\;\mathrm{kJ\,mol^{-1}},
\]
and $K_{\mathrm{eq}}(298) \approx 2.4\times 10^{-5}$ identically.
Their $\dH^\circ$, $\dS^\circ$, $\dG^\circ(T)$, and
$K_{\mathrm{eq}}(T)$ agree for every $T > 0$: as parallel morphisms
in $\Lk_2(P)$ they are \emph{indistinguishable}.

\medskip
\noindent
\textbf{Rates differ by seven orders of magnitude.}
The pathways have vastly different kinetic behaviour
\cite{Khalifah1971, SilverBuss1992}:
\begin{align*}
  r_A &:\; \text{uncatalysed},
    \quad k_A \approx 3.7\times 10^{-2}\;\mathrm{s^{-1}}
    \quad\text{(half-life $\approx 19$~s at 298~K)}, \\
  r_B &:\; \text{catalysed by carbonic anhydrase II},
    \quad k_B \approx 10^6\;\mathrm{s^{-1}}
    \quad\text{(half-life $\approx 0.7\;\mu$s)}.
\end{align*}
The ratio $k_B/k_A \approx 2.7\times 10^7$: carbonic anhydrase II
is among the most efficient enzymes known, accelerating the
hydration of $\mathrm{CO_2}$ by over seven orders of magnitude
\cite{Lindskog1997}.
Equilibrium position is the same in both cases; the time to reach
it differs enormously.

\medskip
\noindent
\textbf{The swap is an $\Lk_2$-automorphism not lifting to $\Lk_3$.}
The swap $\sigma : r_A \leftrightarrow r_B$ preserves the source
complex, the target complex, and both decorating functor values:
\[
  \FH \circ \sigma = \FH,
  \qquad
  \FS \circ \sigma = \FS.
\]
It is therefore a well-defined automorphism of $\Lk_2(P)$, i.e.\
$\sigma \in \Aut(\Lk_2(P))$.
However, $\sigma$ does not lift to an automorphism of any
structure assigning rate constants to generators, since
$k_A \neq k_B$: $\sigma \notin \im(\varphi_3)$, where
$\varphi_3 : \Aut(\Lk_3(P)) \to \Aut(\Lk_2(P))$ is the forgetful
map in the automorphism sequence (\S\ref{sec:aut-exact}).
Equivalently, $\sigma$ represents a non-trivial element of
$\coker(\varphi_3)$, witnessing that $\Lk_2$ conflates two
kinetically distinct reaction pathways of the same stoichiometry.
\end{forcingbox} 

\noindent
This forces $\Lk_3$, which must add:
\begin{enumerate}[label=(\roman*)]
  \item A \emph{rate decoration} on $\Lk_0(P)$ assigning to each
    chemical generator $r \in \Rx$ either a deterministic rate
    constant $k_r > 0$ (mass-action ODE setting) or a stochastic
    propensity function $\lambda_r : \NN^{|\Sp|} \to \RR_{\geq 0}$
    (chemical-master-equation setting).
    The level $\Lk_3(P)$ is then the quadruple
    $\bigl(\Lk_0(P),\,\FH,\,\FS,\,\{k_r\}_{r \in \Rx}\bigr)$ ---
    the $\Lk_2$ thermodynamic data plus per-generator kinetic data.
    The categorical assembly of these rate data into a functor
    $\Lk_0(P) \to \mathbf{Stoch}$ \cite{Fritz2020, ChoJacobs2019}
    (with stochastic kernels arising as exponentials
    $P_t = e^{t\Omega}$ of the per-reaction generator
    contributions $Q_r$, $\Omega = \sum_r Q_r$) is part of the
    $\Lk_3$ construction proper, deferred to that chapter.

  \item For mass-action kinetics, the per-reaction rate is
    \[
      v_r(c) \;=\; k_r \prod_{i \in \Sp} c_i^{s_i(r)}
      \quad\text{(deterministic, on concentrations $c$),}
    \]
    and the stochastic propensity on copy-number states
    $x \in \NN^{|\Sp|}$ has the falling-factorial form
    \[
      \lambda_r(x)
      \;=\; \kappa_r \prod_{i \in \Sp} (x_i)_{s_i(r)}
      \;=\; \kappa_r \prod_{i \in \Sp}
        \frac{x_i!}{(x_i - s_i(r))!},
    \]
    where $s_i(r)$ is the stoichiometric coefficient of species
    $i$ in the source of $r$, and the volume-scaling between $k_r$
    and $\kappa_r$ follows the standard convention
    \cite{Feinberg1972, AtkinsDeP2014}.

  \item The \emph{chemical master equation}
    $\tfrac{d}{dt} p = \Omega\,p$, with generator
    $\Omega = \sum_r Q_r$ summing per-reaction contributions; the
    probability distribution $p$ over copy-number states evolves
    under the Markov semigroup $e^{t\Omega}$.
    In the deterministic large-copy-number limit, this reduces to
    the mass-action ODE system $\dot c = N\,v(c)$
    \cite{Kurtz1972}, where $N$ is the stoichiometric matrix of
    \S\ref{sec:N-matrix} and $v(c)$ is the vector of
    per-reaction rates from item~(ii).

  \item Trajectory-level theorems --- most notably the
    \emph{Deficiency Zero Theorem} of Feinberg--Horn--Jackson
    \cite{Feinberg1972, HornJackson1972}, which gives global
    asymptotic stability of complex-balanced equilibria in weakly
    reversible deficiency-zero networks under mass-action kinetics,
    and the \emph{Anderson--Craciun--Kurtz} theorem
    \cite{AndersonCraciunKurtz2010} for the stochastic analogue
    (product-form Poisson stationary distribution).
    These results require the rate data of $\Lk_3$ --- the $\Lk_0$
    deficiency hypothesis $\delta = 0$ alone is insufficient to
    guarantee them --- and are stated and proved in the $\Lk_3$
    chapter; they are mentioned here only to indicate what becomes
    accessible once kinetic data is added.
\end{enumerate}

\noindent
The bridge from $\Lk_2$ to $\Lk_3$ is the \emph{kinetic
Wegscheider condition} \cite{Wegscheider1901, Feinberg1989,
HornJackson1972}, derived as follows.
For each reversible elementary pair $r,\,r^\dagger$ in a reversible
mass-action network, detailed balance in the kinetic sense
requires that the ratio of forward and reverse rate constants
equal the standard equilibrium constant:
\[
  \frac{k_r^+}{k_r^-} \;=\; K_{\mathrm{eq}}(r,\, T),
\]
with the appropriate standard-state and activity conventions.
Multiplying this elementary relation around a directed cycle
$r_1, \ldots, r_k$ in the reaction graph gives
\[
  \prod_{i=1}^k \frac{k_{r_i}^+}{k_{r_i}^-}
  \;=\; \prod_{i=1}^k K_{\mathrm{eq}}(r_i,\, T)
  \;=\; 1,
\]
where the final equality is the thermodynamic Wegscheider
condition of Proposition~\ref{prop:wegscheider1}.
This bridge operates at the interface of $\Lk_2$ and $\Lk_3$: the
equilibrium constants on the right are computed from $\Lk_2$
thermochemical data ($\FH$, $\FS$), while the rate constants on
the left are $\Lk_3$ kinetic data.
The $\Lk_2$ Wegscheider conditions constrain the $\Lk_3$ rate
constants without determining them: knowing $K_{\mathrm{eq}}$
fixes the \emph{ratio} $k^+/k^-$ per reversible pair but leaves
their absolute magnitudes free, which is precisely the information
added at $\Lk_3$.
Conversely, rate constants at $\Lk_3$ determine $K_{\mathrm{eq}}(T)$
through this ratio condition, independent of the thermochemical
data used to compute $K$ at $\Lk_2$ --- providing a non-trivial
cross-level consistency check between the two levels.

%% file: chapters/ch_L3.tex
\section{\texorpdfstring{$\Lk_3$}{L3}: The Kinetic Level}
\label{sec:L3}

\input{chapters/L3/l3_forcing}
\input{chapters/L3/l3_def}
\input{chapters/L3/l3_massaction}
\input{chapters/L3/l3_cme}
\input{chapters/L3/l3_layer2}
\input{chapters/L3/l3_dzt}
\input{chapters/L3/l3_baezpollard}
\input{chapters/L3/l3_examples}
\input{chapters/L3/l3_nextforcing}

%% file: chapters/L3/l3_forcing.tex
\subsection{Forcing the extension: what \texorpdfstring{$\Lk_2$}{L2}
  cannot express}
\label{sec:L3-forcing-in}

\begin{chembox}[What $\Lk_3$ adds: rates, time, and stochastic dynamics]
At $\Lk_2$, two physically distinct systems can be indistinguishable:
they share the same enthalpy $\FH(r)$, entropy $\FS(r)$,
free-energy change $F_G^T(r)$, and equilibrium constant
$K_{\mathrm{eq}}(r, T)$ for every $T > 0$.
The single new datum at $\Lk_3$ is a positive rate constant
$k_r \in \RR_{>0}$ per generating reaction.

\medskip
\noindent
\textbf{What this unlocks.}
For the $\mathrm{CO_2}$ hydration network of
Section~\ref{sec:L2-forcing-out}, both the uncatalysed system
($k_A \approx 3.7\times 10^{-2}\;\mathrm{s^{-1}}$,
half-life $\approx 19$~s) and the carbonic anhydrase-catalysed
system ($k_B \approx 10^6\;\mathrm{s^{-1}}$,
half-life $\approx 0.7\;\mu\mathrm{s}$) are now
distinguishable --- their $\Lk_2$ data are identical, their
$\Lk_3$ data differ by over seven orders of magnitude
\cite{Khalifah1971, SilverBuss1992, Lindskog1997}.

\medskip
\noindent
\textbf{Stochastic predictions.}
With rate constants in hand, $\Lk_3$ supports a full probability
distribution over species-count states.
For a weakly reversible network with deficiency $\delta = 0$
(an $\Lk_0$ invariant, Definition~\ref{def:feinberg})
equipped with mass-action rate constants,
the stationary distribution of the chemical master equation is a
product-of-Poissons \cite{AndersonCraciunKurtz2010}:
\[
  \pi(\mathbf{x})
  \;=\; \frac{1}{Z}
  \prod_{s \in \Sp} \frac{(c_s^*)^{x_s}}{x_s!},
\]
where $\mathbf{c}^*$ is the unique positive complex-balanced steady
state guaranteed by the Deficiency Zero Theorem
\cite{Horn1972, Feinberg1987} (Section~\ref{sec:L3-dzt}).
This product form --- and the very notion of a ``time to equilibrium''
--- is invisible at $\Lk_2$: it requires the rate functor $\FP$.
\end{chembox}

\noindent
Section~\ref{sec:L2-forcing-out} established the forcing gap at
$\Lk_2$: the uncatalysed and catalysed $\mathrm{CO_2}$ hydration
systems define two distinct points in the fibre of the forgetful
operation $U_3 : \Lk_3(P) \to \Lk_2(P)$ over a single $\Lk_2$
object --- their $\FH$, $\FS$, and $F_G^T$ values agree for
every $T > 0$, while their rate constants differ by seven orders
of magnitude.
Since both points project to the same $\Lk_2$ decoration, the
label swap $r_A \leftrightarrow r_B$ is a well-defined
automorphism of the $\Lk_2$-decoration that does not lift along
$\varphi_3 : \Aut(\Lk_3(P)) \to \Aut(\Lk_2(P))$:
in the automorphism sequence
\[
  1 \;\to\; \ker\varphi_3
  \;\to\; \Aut(\Lk_3(P))
  \;\xrightarrow{\;\varphi_3\;}
  \;\Aut(\Lk_2(P))
  \;\to\; \coker(\varphi_3) \;\to\; 1,
\]
the swap represents a non-trivial coset in the pointed-set
quotient
$\coker(\varphi_3) = \Aut(\Lk_2(P))/\im(\varphi_3)$
(\S\ref{sec:aut-exact}; the cokernel is a pointed-set quotient,
not a group cokernel).
A non-trivial $\coker(\varphi_3)$ proves $U_3$ is not injective
on decorated quadruples and that the extension to $\Lk_3$ is
\emph{necessary}: $\Lk_2$ conflates kinetically distinct systems
that $\Lk_3$ separates.

\medskip\noindent
\textbf{Why the target is not $B\RR$.}
At $\Lk_1$ and $\Lk_2$, the decorating functors targeted $B\RR$:
enthalpy and entropy are real numbers, sequential composition is
addition, and parallel composition is also addition.
The additive structure of $\RR$ matches the additivity of
state-function differences along reaction paths.

For kinetics, the corresponding additive structure lives on
\emph{Markov-semigroup generators}, not on rate constants
themselves.
The generator contribution of a reaction $r$ is the operator
$\FP(r) = M_{\lambda_r}(R_r - I)$ on observables
$f : \NN^{|\Sp|} \to \RR$
(Definitions~\ref{def:propensity},~\ref{def:L3});
per-reaction generator contributions sum into a single CME
generator $\Omega = \sum_r \FP(r)$, and parallel reactions on
disjoint species combine via Kronecker sum
$\FP(r_1) \otimes I + I \otimes \FP(r_2)$
(Propositions~\ref{prop:FP-functorial},~\ref{prop:FP-monoidal}).
The natural ambient algebra is therefore the cone $\mathfrak{g}_\Sp$
of Markov-semigroup generators on $\NN^\Sp$, viewed as a
one-object strict SMC with sequential composition by addition,
monoidal product by Kronecker sum, and unit the zero generator
(Remark~\ref{rem:stoch-strict}).
Every complex $\mathbf{u} \in \Lk_0(P)$ maps to the unique object
of this one-object category, interpreted as the global state
space $\NN^\Sp$.

Markov \emph{kernels} (probability transition functions) live in
the ambient symmetric monoidal category $\Stoch$
\cite{Fritz2020, ChoJacobs2019}, but enter the picture only via
finite-time evolution: a generator $L \in \mathfrak{g}_\Sp$
exponentiates to a one-parameter Markov semigroup
$\{e^{tL}\}_{t \geq 0}$ of kernels in
$\mathrm{End}_{\Stoch}(\NN^{\Sp})$.
This exponentiation $\exp_t : L \mapsto e^{tL}$ is \emph{not} an
SMC functor (sequential composition fails to commute with
addition off the abelian locus, by Trotter--Kato), and that
failure is the precise categorical content of the
$\Lk_3 \to \Lk_4$ forcing pair
(Section~\ref{sec:L3-forcing-out}).
The shorthand ``$\FP : \Lk_0(P) \to \Stoch$'' that appears in
some places in this chapter abbreviates the composite
$\exp_t \circ \FP$, with $\FP$ landing in $\mathfrak{g}_\Sp$ at
the categorical level.

\medskip\noindent
The same physical logic that forced $\FH$ and $\FS$ at earlier
levels forces $\FP : \Lk_0(P) \to \mathfrak{g}_\Sp$ here, with
three axioms whose content is now non-trivial:
\begin{enumerate}[label=(\alph*)]
  \item \textbf{Sequential composition.}
    For composable reactions
    $r_1 : \mathbf{u} \to \mathbf{v}$,
    $r_2 : \mathbf{v} \to \mathbf{w}$ in $\Lk_0(P)$, the
    composite $r_2 \circ r_1$ maps under $\FP$ to the additive
    sum
    \[
      \FP(r_2 \circ r_1) \;=\; \FP(r_2) + \FP(r_1)
      \quad\text{in } \mathfrak{g}_\Sp
    \]
    (Proposition~\ref{prop:FP-functorial}).
    This is the generator of the CTMC in which both reaction
    channels $r_1$ and $r_2$ are simultaneously available; it is
    \emph{not} an effective single-step rate law for the
    coarse-grained reaction $\mathbf{u} \to \mathbf{w}$ obtained
    by eliminating $\mathbf{v}$.
    Such a coarse-graining requires a quasi-steady-state or
    rapid-equilibrium reduction, which lies outside the strict
    $\Lk_3$ functorial assignment and is recovered only as a
    derived approximation under additional hypotheses on the
    rate constants.
  \item \textbf{Parallel composition.}
    If $r_1$ and $r_2$ act on independent subsystems with
    disjoint species sets $\Sp_1, \Sp_2$, their combined
    generator on the joint state space
    $\NN^{|\Sp_1|} \times \NN^{|\Sp_2|}$ is the Kronecker sum
    \[
      \FP(r_1 \otimes r_2) \;=\;
      \FP(r_1) \otimes I + I \otimes \FP(r_2)
    \]
    (Proposition~\ref{prop:FP-monoidal}), the infinitesimal of
    the kernel tensor product in $\Stoch$.
  \item \textbf{Identity.}
    The ``do nothing'' morphism $\id_\mathbf{u}$ produces no
    state change:
    $\FP(\id_\mathbf{u}) = 0 \in \mathfrak{g}_\Sp$
    (Definition~\ref{def:L3}), the zero generator, which
    exponentiates to the identity Markov kernel
    $e^{0} = \id_{\NN^\Sp} \in \mathrm{End}_{\Stoch}(\NN^\Sp)$.
\end{enumerate}
These are the three axioms of a strict symmetric monoidal
functor, now targeting $\mathfrak{g}_\Sp$ rather than $B\RR$.
The extension is \emph{forced} by the same physical logic; what
changes is the additive monoid in which the new datum lives ---
from $\RR$ to the cone $\mathfrak{g}_\Sp$ of Markov-semigroup
generators on $\NN^\Sp$.

%% file: chapters/L3/l3_def.tex
\subsection{Definition of \texorpdfstring{$\Lk_3(P)$}{L3(P)}
  and the target \texorpdfstring{$\Stoch$}{Stoch}}
\label{sec:L3-def}

\subsubsection{The Markov generator algebra and the ambient \texorpdfstring{$\Stoch$}{Stoch}}

The stochastic description of a chemical reaction network is
classical: given the countable state space of species-count
vectors $\mathbf{x} \in \NN^{|\Sp|}$ --- or a finite closed
irreducible subset of it, when conservation laws bound the total
copy numbers --- the network evolves as a continuous-time Markov
chain (CTMC) whose generator encodes the rates at which each
reaction fires
\cite{AndersonCraciunKurtz2010, Feinberg2019}.
This description underlies a large body of work in the CRNT
literature --- on stationary distributions
\cite{CappellettiJoshi2019, HoesslyWiuf2025},
on product-form Poisson distributions and their extensions
\cite{AndersonSchnoerrYuan2020},
and on stochastic approximation of arbitrary distributions
\cite{CappellettiAndersonWinfree2020} ---
without requiring a categorical language.

Two categorical objects are needed to make the compositional
structure of CTMCs explicit:
\begin{itemize}
  \item the symmetric monoidal category $\Stoch$, in which
    finite-time Markov \emph{kernels} $K(\,\cdot \mid x)$
    compose by Chapman--Kolmogorov integration --- the ambient
    setting in which the time-evolved chain lives;
  \item the cone $\mathfrak{g}_\Sp$ of Markov \emph{generators}
    on $\NN^{|\Sp|}$, viewed as the morphism set of the
    one-object strict SMC $B\mathfrak{g}_\Sp$, with sequential
    composition by addition, monoidal product by Kronecker sum,
    and unit the zero generator --- the actual functorial target
    of the kinetic functor $\FP$.
\end{itemize}
Generators and kernels are linked by exponentiation
$\exp_t : L \mapsto e^{tL}$, which sends a generator
$L \in \mathfrak{g}_\Sp$ to its one-parameter Markov semigroup
$\{e^{tL}\}_{t \geq 0}$ in
$\mathrm{End}_{\Stoch}(\NN^{|\Sp|})$.
This section defines both, and shows that $\FP$ targets
$B\mathfrak{g}_\Sp$ at the categorical level --- with $\Stoch$
entering only as the home of finite-time evolution.

\begin{definition}[$\Stoch$]
\label{def:stoch}
  The category $\Stoch$ has:
  \begin{itemize}
    \item \textbf{Objects}: measurable spaces $(X, \Sigma_X)$.
    \item \textbf{Morphisms}: Markov kernels $K : X \to Y$,
  i.e.\ functions $K : X \times \Sigma_Y \to [0,1]$ such
  that (i) for each $x \in X$, $K(x, \cdot)$ is a
  probability measure on $(Y, \Sigma_Y)$, and (ii) for
  each $A \in \Sigma_Y$, the map $K(\cdot, A) : X \to [0,1]$
  is measurable.
    \item \textbf{Composition}: for $K : X \to Y$ and $L : Y \to Z$,
      $(L \circ K)(\cdot \mid x)
      := \int_Y L(\cdot \mid y)\,K(dy \mid x)$
      (Chapman--Kolmogorov integration).
    \item \textbf{Monoidal product}: $(X,\Sigma_X) \otimes (Y,\Sigma_Y)
      := (X \times Y,\, \Sigma_X \otimes \Sigma_Y)$,
      with $(K_1 \otimes K_2)(\cdot \mid x_1, x_2)
      := K_1(\cdot \mid x_1) \otimes K_2(\cdot \mid x_2)$.
    \item \textbf{Monoidal unit}: the one-point space
      $I = (\{*\}, \{\emptyset,\{*\}\})$.
  \end{itemize}
  $\Stoch$ is a symmetric monoidal category under this structure;
  it is moreover a \emph{Markov category} in the sense of Fritz
  \cite{Fritz2020}, with a canonical copy/discard structure that
  encodes conditional independence
  \cite{Fritz2020, ChoJacobs2019}.
\end{definition}

\begin{definition}[The generator algebra $B\mathfrak{g}_\Sp$]
\label{def:gen-S}
The \emph{Markov-generator algebra} $\mathfrak{g}_\Sp$ is the
positive cone of generators of Markov semigroups on
$\NN^{|\Sp|}$: a (possibly unbounded) operator $L$ on observables
$f : \NN^{|\Sp|} \to \RR$ with off-diagonal-non-negative matrix
entries and rows summing to zero, closable to a CTMC under
standard non-explosion conditions \cite{AndersonKurtz2015}.
The \emph{one-object category} $B\mathfrak{g}_\Sp$ has a single
formal object $\ast$ --- interpreted as the global state space
$\NN^{|\Sp|}$, the carrier on which every complex
$\mathbf{u} \in \Lk_0(P)$ acts under $\FP$ --- and morphism set
$\mathrm{End}(\ast) = \mathfrak{g}_\Sp$.
$B\mathfrak{g}_\Sp$ is a strict symmetric monoidal category:
\begin{itemize}
  \item sequential composition: addition, $L_1 + L_2$;
  \item monoidal product: Kronecker sum,
    $L_1 \otimes I + I \otimes L_2$ (the infinitesimal of the
    kernel tensor product in $\Stoch$);
  \item unit: the zero generator on $\NN^{|\Sp|}$.
\end{itemize}
All associativity, unit, and symmetry equations hold as
equalities, not merely up to isomorphism.
\end{definition}

\begin{mathbox}[$B\mathfrak{g}_\Sp$ versus $B\RR$]
At $\Lk_1$ and $\Lk_2$ the target was $B\RR$: one object,
morphisms $\RR$, sequential composition and monoidal product
both given by addition.
$\FH$ and $\FS$ were essentially monoid homomorphisms
$(\mathrm{Mor}(\Lk_0(P)), \circ, \otimes) \to (\RR, +, +)$,
labelling each reaction by a single real number.

At $\Lk_3$ the target $B\mathfrak{g}_\Sp$ is structurally
parallel --- one object, morphism set additive, with sequential
composition and monoidal product both additive in their
respective senses --- but the additive monoid is enriched from
$\RR$ to the cone $\mathfrak{g}_\Sp$ of Markov-semigroup
generators on $\NN^{|\Sp|}$.
This enrichment has three direct probabilistic consequences:

\begin{itemize}
  \item \textbf{Morphisms encode full stochastic transitions.}
    A generator $L \in \mathfrak{g}_\Sp$ encodes, via
    exponentiation $e^{tL}$, the entire family of finite-time
    transition probabilities
    $P(\mathbf{x}(t) \in \cdot \mid \mathbf{x}(0) = \mathbf{x})$
    of the underlying CTMC \cite{AndersonCraciunKurtz2010}.
    At $\Lk_1$ and $\Lk_2$, a reaction $r$ was labelled by a
    single real number; at $\Lk_3$ the label
    $\FP(r) \in \mathfrak{g}_\Sp$ is an unbounded operator on
    observables that determines, after exponentiation, every
    conditional probability associated with $r$.

  \item \textbf{Sequential composition is operator addition,
    extending real addition.}
    For composable reactions,
    $\FP(r_2 \circ r_1) = \FP(r_2) + \FP(r_1)$ in
    $\mathfrak{g}_\Sp$
    (Proposition~\ref{prop:FP-functorial}).
    This sum is the generator of the CTMC in which both
    reaction channels are simultaneously available; it is
    \emph{not} an effective single-step rate law for the
    coarse-grained reaction $\mathbf{u} \to \mathbf{w}$
    obtained by eliminating the intermediate (which would
    require a quasi-steady-state reduction outside the strict
    $\Lk_3$ functorial assignment).
    The collapse to scalar addition recovers $\Lk_1$/$\Lk_2$
    additivity in the degenerate one-state case.

  \item \textbf{Monoidal product is Kronecker sum, extending
    real addition.}
    For reactions on independent subsystems with disjoint
    species sets,
    $\FP(r_1 \otimes r_2) = \FP(r_1) \otimes I + I \otimes
    \FP(r_2)$ as generator contributions on the joint state
    space (Proposition~\ref{prop:FP-monoidal}).
    This is the infinitesimal of the kernel tensor product in
    $\Stoch$ and encodes probabilistic independence
    \cite{Fritz2020}: each subsystem evolves on its own
    Poisson clock.
    Again, this collapses to real addition in the degenerate
    one-state case.
\end{itemize}

\noindent
The three axioms of a strict SMC functor
$\FP : \Lk_0(P) \to B\mathfrak{g}_\Sp$ have non-trivial
probabilistic content while preserving the structural role of
``additive functor'' that $\FH$ and $\FS$ played at lower
levels.
The Markov-\emph{kernel} interpretation of these data --- the
finite-time transition probabilities --- lives in $\Stoch$ and
is recovered from $\FP(r)$ by exponentiation
$e^{t\,\FP(r)}$, which intertwines addition with kernel
composition only on the abelian locus of $\mathfrak{g}_\Sp$
(the Trotter--Kato obstruction; this non-functoriality of
$\exp_t$ is the categorical content of the $\Lk_3 \to \Lk_4$
forcing pair, Section~\ref{sec:L3-forcing-out}).
\end{mathbox}

\begin{remark}[Why $\exp_t$ is not an SMC functor]
\label{rem:stoch-strict}
The functor $\FP : \Lk_0(P) \to B\mathfrak{g}_\Sp$ takes values
in generators, not in finite-time Markov kernels.
The link to $\Stoch$ proper is the exponential
$\exp_t : L \mapsto e^{tL}$, which sends $L \in \mathfrak{g}_\Sp$
to its one-parameter Markov semigroup of kernels in
$\mathrm{End}_{\Stoch}(\NN^{|\Sp|})$ under standard non-explosion
conditions \cite{AndersonKurtz2015}.
This exponential is \emph{not} a symmetric monoidal functor:
it intertwines addition with kernel composition only on the
abelian locus of $\mathfrak{g}_\Sp$ (the Trotter--Kato / BCH
obstruction).
On a sequential composite $r_2 \circ r_1$ in $\Lk_0(P)$,
$\FP$ records the additive generator
$\FP(r_2) + \FP(r_1) \in \mathfrak{g}_\Sp$ ---
the generator of the CTMC in which both reaction channels are
simultaneously available --- whose finite-time kernel
$e^{t\,(\FP(r_2)+\FP(r_1))}$ differs from
$e^{t\,\FP(r_2)} \circ e^{t\,\FP(r_1)}$ by
Baker--Campbell--Hausdorff commutator terms.
This loss of path-order information at $\Lk_3$ is precisely the
categorical content of the $\Lk_3 \to \Lk_4$ forcing pair
(Section~\ref{sec:L3-forcing-out}); recovering bond-level order
information requires the DPO machinery of $\Lk_4$.
\end{remark}

\subsubsection{The state space is \texorpdfstring{$\Lk_0$}{L0} data}

In classical CRNT, species counts in $\NN^{|\Sp|}$
is read off directly from the reaction network: the
species set $\Sp$ is part of the network definition, and a
state is simply a non-negative integer count for each species.
No kinetic data --- no rate constants, no activation energies
--- are needed to write down $\NN^{|\Sp|}$.
The stoichiometric change vector $\mathbf{n}_r$ of a reaction
$r$ is likewise determined by the network's stoichiometry alone
(the left- and right-hand side complexes), as discussed in Chapter~\ref{sec:L0}.
In the tower language, this means $\NN^{|\Sp|}$ and
$\mathbf{n}_r$ are $\Lk_0$ data: they are visible at the
stoichiometric level before any rate constants are introduced.

\begin{observation}[State space from $\mathcal{L}_0$]
\label{obs:state-space}
The kinetic state space $\mathbb{N}^{|\Sp|}$ is determined
entirely by $\Lk_0$ data:
\begin{itemize}
  \item The species set $\Sp$ is a $\Lk_0$ datum.
  \item A state is a species-count vector
    $\mathbf{x} = (x_s)_{s \in \Sp} \in \NN^{|\Sp|}$.
  \item The stoichiometric change vector of generator
    $r : \mathbf{u} \to \mathbf{v}$
    is $\mathbf{n}_r := \boldsymbol{\nu}^{(\mathbf{v})}
    - \boldsymbol{\nu}^{(\mathbf{u})} \in \ZZ^{|\Sp|}$,
    directly read from the $\Lk_0$ morphism $r$.
\end{itemize}
The single new datum at $\Lk_3$ is a positive rate constant
$k_r \in \RR_{>0}$ per generator.
Everything else required to define $\FP(r)$ is already present
in $\Lk_0(P)$.
\end{observation}

\subsubsection{Definition of \texorpdfstring{$\Lk_3(P)$}{L3P}}

In this chapter, the formulation is given in the Heisenberg
picture: $\FP(r)$ acts on observables
$f : \NN^{|\Sp|} \to \RR$ on the maximal domain on which it is
well-defined, with finitely-supported functions as a canonical
core \cite{AndersonKurtz2015}.

\begin{definition}[Kinetic level $\Lk_3(P)$]
\label{def:L3}
  Let $P$ be a Petri net with species set $\Sp$ and reaction set
  $\Rx$.
  The \emph{kinetic level} $\Lk_3(P)$ is the quadruple
  \[
    \Lk_3(P)
    \;:=\; \bigl(\,\Lk_0(P),\;\FH,\;\FS,\;\FP\,\bigr),
  \]
  where $\FH$, $\FS$ are the thermochemical and entropy functors
  from $\Lk_2(P)$, and
  \[
    \FP : \Lk_0(P) \;\longrightarrow\; B\mathfrak{g}_\Sp
  \]
  is a strict symmetric monoidal functor
  (Definition~\ref{def:gen-S}) satisfying:
  \begin{itemize}
    \item every complex $\mathbf{u} \in \Lk_0(P)$ maps to the
      unique object $\ast$ of $B\mathfrak{g}_\Sp$, interpreted
      as the global state space $\NN^{|\Sp|}$;
    \item every identity morphism $\id_{\mathbf{u}}$ maps to the
      zero generator,
      $\FP(\id_{\mathbf{u}}) = 0 \in \mathfrak{g}_\Sp$;
    \item each chemical generator $r \in \Rx$ maps to the
      \emph{generator contribution}
      \[
        \FP(r) \;:=\; M_{\lambda_r}\,(R_r - I),
      \]
      the standard CME generator contribution of reaction $r$ in
      the Heisenberg picture, acting on observables
      $f : \NN^{|\Sp|} \to \RR$, where:
      \begin{itemize}
        \item $R_r$ is the shift operator
          $(R_r f)(\mathbf{x}) := f(\mathbf{x} + \mathbf{n}_r)$;
        \item $M_{\lambda_r}$ is multiplication by the
          mass-action propensity
          \[
            \lambda_r(\mathbf{x})
            \;:=\; k_r \prod_{s \in \Sp}
            \binom{x_s}{\nu_s^{(\mathbf{u}_r)}}
            \qquad (k_r \in \RR_{>0});
          \]
        \item the binomial coefficient vanishes when
          $x_s < \nu_s^{(\mathbf{u}_r)}$, so $\FP(r)$
          automatically vanishes at boundary states with
          insufficient reactants.
      \end{itemize}
      Explicitly,
      \[
        \bigl(\FP(r) f\bigr)(\mathbf{x})
        \;=\; \lambda_r(\mathbf{x})
          \bigl[f(\mathbf{x} + \mathbf{n}_r) - f(\mathbf{x})\bigr]
        \quad\text{\cite{AndersonCraciunKurtz2010, AndersonKurtz2015}.}
      \]
      Propensities are polynomial in $\mathbf{x}$, so $\FP(r)$ is
      in general unbounded; we take its domain to be functions of
      compact support, on which the assembled generator
      $\sum_{r \in \Rx} \FP(r)$ admits a closure generating a
      Markov semigroup under the standard non-explosion conditions
      \cite{AndersonKurtz2015}.
      The full CME generator
      $\Omega = \overline{\sum_{r \in \Rx} \FP(r)}$
      is assembled in Section~\ref{sec:L3-cme}.
  \end{itemize}
  The \emph{forgetful operation} $U_3$ acts on decorated
  quadruples by
  $(\Lk_0(P), \FH, \FS, \FP) \mapsto (\Lk_0(P), \FH, \FS)$,
  dropping $\FP$.
\end{definition}

In the literature, $k_r$ is in general non-negative.
The case $k_r = 0$ corresponds to a vanishing generator
contribution, $\FP(r) = 0 \in \mathfrak{g}_\Sp$, which
exponentiates to the identity kernel $\id_{\NN^{|\Sp|}} \in \Stoch$:
no reaction takes place.

\begin{remark}[$\FP(r)$ versus a Markov kernel]
\label{rem:FP-generator}
The generator contribution $\FP(r) = M_{\lambda_r}(R_r - I)$
is a Markov generator on $\NN^{|\Sp|}$: it is not itself a
Markov kernel (which must be a probability measure for each
input state), but rather its infinitesimal version.
The associated Markov kernel is the integrated semigroup
$e^{t\,\FP(r)}$, which describes the probability distribution
over states after time $t$ if only reaction $r$ could fire.
Working at the generator level --- standard in the chemical
master equation (CME) literature \cite{AndersonCraciunKurtz2010}
--- allows one to sum contributions linearly:
$\Omega = \sum_r \FP(r)$ is itself a valid CME generator,
whereas Markov kernels compose by Chapman--Kolmogorov
integration, not addition.
\end{remark}

The tower now reads:
\[
  \underbrace{\Lk_0(P)}_{\text{free skeletal permutative category}}
  \;\hookrightarrow\;
  \underbrace{\Lk_1}_{+\,\FH}
  \;\hookrightarrow\;
  \underbrace{\Lk_2}_{+\,\FS}
  \;\hookrightarrow\;
  \underbrace{\Lk_3}_{+\,\FP}.
\]
The first two extensions add functors into $B\RR$ (real numbers
under addition), reflecting the additive nature of standard
reaction enthalpies and entropies.
The third adds a functor into $B\mathfrak{g}_\Sp$, whose
additive structure (sequential = addition, monoidal product =
Kronecker sum) extends the additivity of $\RR$ to the genuinely
probabilistic setting of kinetics: this enrichment of the
target's additive monoid --- from $\RR$ to $\mathfrak{g}_\Sp$
--- is the categorical signature of the passage from
thermodynamics to dynamics.
The underlying free skeletal permutative category $\Lk_0(P)$
remains unchanged throughout.

\begin{proposition}[Existence and uniqueness of $\FP$]
\label{prop:FP-unique}
  Given any assignment $k : \Rx \to \RR_{>0}$ of positive rate
  constants to chemical generators, there is a unique strict
  SMC functor
  $\FP : \Lk_0(P) \to B\mathfrak{g}_\Sp$
  extending $k$ via $r \mapsto M_{\lambda_r}(R_r - I)$.
\end{proposition}

\begin{proof}
By Theorem~\ref{thm:UP-L0}, $\Lk_0(P)$ is the free skeletal
permutative category generated by $P$: given any strict SMC
$\mathcal{C}$ with strictly commutative object monoid, an
assignment of (i) species in $\Sp$ to objects of $\mathcal{C}$
and (ii) chemical generators $r \in \Rx$ to morphisms in
$\mathcal{C}$ with matching sources and targets extends
uniquely to a strict SMC functor $\Lk_0(P) \to \mathcal{C}$.
The target $\mathcal{C} = B\mathfrak{g}_\Sp$
(Definition~\ref{def:gen-S}) is one-object, so its object
monoid is the singleton --- trivially strictly commutative ---
and the species assignment (i) is forced to map every species
to the unique object $\ast$; only the morphism assignment (ii)
carries information.

The proposed assignment
$r \mapsto M_{\lambda_r}(R_r - I) \in \mathfrak{g}_\Sp$ is
well-typed: $\lambda_r \geq 0$ pointwise, the matrix entries of
$M_{\lambda_r}(R_r - I)$ are off-diagonal-non-negative with
zero row sums (the defining structure of a Markov generator),
and the assembled $\sum_r \FP(r)$ admits a closure generating
a Markov semigroup under standard non-explosion conditions
\cite{AndersonCraciunKurtz2010, AndersonKurtz2015}.
Applying the universal property of $\Lk_0(P)$ to this
assignment yields the unique strict SMC functor
$\FP : \Lk_0(P) \to B\mathfrak{g}_\Sp$ extending $k$;
the value on each generator is determined by $\mathbf{n}_r$
(from $\Lk_0$, fixing $R_r$) and $k_r$ (the new datum at
$\Lk_3$).
\end{proof}

\noindent
Proposition~\ref{prop:FP-unique} has the same logical
structure as the existence-and-uniqueness statements for
$\FH$ (Proposition~\ref{prop:FH-unique}) and $\FS$
(Proposition~\ref{prop:FS-unique}): in all three cases, the
universal property of the free skeletal permutative category
$\Lk_0(P)$ guarantees that a single real-valued assignment per
chemical generator extends uniquely to a strict SMC functor on
all of $\Lk_0(P)$.
The only difference is what the real number labels and where it
lands, which is precisely what the insightbox below records.

\begin{insightbox}[One positive real per reaction --- again]
Propositions~\ref{prop:FH-unique}, \ref{prop:FS-unique},
and~\ref{prop:FP-unique} are the same theorem applied three
times to three different targets:
\begin{center}
\renewcommand{\arraystretch}{1.3}
\begin{tabular}{llll}
  \hline
  \textbf{Level} & \textbf{Functor} & \textbf{Target} &
  \textbf{New datum per chemical generator}\\
  \hline
  $\Lk_1$ & $\FH$ & $B\RR$ & $\dH^\circ_r \in \RR$ \\
  $\Lk_2$ & $\FS$ & $B\RR$ & $\dS^\circ_r \in \RR$ \\
  $\Lk_3$ & $\FP$ & $B\mathfrak{g}_\Sp$ & $k_r \in \RR_{>0}$ \\
  \hline
\end{tabular}
\end{center}
The universal property of $\Lk_0(P)$ is the single engine
driving all three extensions.
At $\Lk_3$ the target changes from $B\RR$ to $B\mathfrak{g}_\Sp$
--- from the additive monoid of real numbers to the richer
additive cone of Markov-semigroup generators on $\NN^{|\Sp|}$
--- but the argument is structurally identical.
The positivity constraint $k_r > 0$ (absent for $\dH^\circ_r$
and $\dS^\circ_r$, which can be any real number) reflects the
physical requirement that reaction rates are non-negative:
a non-positive $k_r$ would yield a generator contribution
$M_{\lambda_r}(R_r - I)$ that fails to define a valid Markov
semigroup.
\end{insightbox}

%% file: chapters/L3/l3_massaction.tex
\subsection{Mass-action kinetics: unpacking
  \texorpdfstring{$\FP$}{FP}}
\label{sec:L3-massaction}

The assignment of a rate constant $k_r \in \RR_{>0}$ to each
generating reaction $r$ is standard practice in stochastic
chemical kinetics: it is the starting point for Gillespie's
stochastic simulation algorithm \cite{Gillespie1977}, for the
derivation of the chemical master equation \cite{McQuarrie1967,
Gillespie1992}, and for the CTMC models of chemical reaction
networks that underlie the CRNT stochastic literature
\cite{AndersonKurtz2011, AndersonCraciunKurtz2010}.
The role of this section is to make explicit how $k_r$ enters
the functor $\FP$ and how the resulting propensity function
$\lambda_r(\mathbf{x})$ relates to the stoichiometric data
already present in $\Lk_0(P)$.

\subsubsection{The propensity and generator contribution}

For a generator $r : \mathbf{u} \to \mathbf{v}$ in $\Lk_0(P)$
with stoichiometric vector
$\mathbf{n}_r = \boldsymbol{\nu}^{(\mathbf{v})}
- \boldsymbol{\nu}^{(\mathbf{u})} \in \ZZ^{|\Sp|}$
and rate constant $k_r \in \RR_{>0}$, the kinetic functor
assigns the following data.

\begin{definition}[Mass-action propensity and generator contribution]
\label{def:propensity}
  Fix a system volume $V > 0$ (in units consistent with the
  rate-constant convention below).
  The \emph{stochastic mass-action propensity} of reaction
  $r : \mathbf{u} \to \mathbf{v}$ in state
  $\mathbf{x} \in \NN^{|\Sp|}$ is
  \[
    \lambda_r^V(\mathbf{x})
    \;:=\; \kappa_r\,V^{1 - |\mathbf{u}|}
    \prod_{s \in \Sp}
    \binom{x_s}{\nu_s^{(\mathbf{u})}}
    \qquad (\kappa_r \in \RR_{>0}),
  \]
  where:
  \begin{itemize}
    \item $\nu_s^{(\mathbf{u})}$ is the stoichiometric
      coefficient of species $s$ in the source complex
      $\mathbf{u}$;
    \item $|\mathbf{u}| := \sum_s \nu_s^{(\mathbf{u})}$ is
      the molecularity (total reactant count) of $r$;
    \item $\kappa_r$ is the \emph{stochastic mass-action rate
      constant}, fixed independently of $V$ and chosen so that
      the deterministic-limit rate
      $v_r(\mathbf{c}) = \kappa_r \prod_s c_s^{\nu_s^{(\mathbf{u})}}$
      has the conventional concentration units (Proposition~\ref{prop:LMA});
    \item $\binom{x_s}{\nu_s^{(\mathbf{u})}} = 0$ if
      $x_s < \nu_s^{(\mathbf{u})}$, so $\lambda_r^V$ vanishes
      automatically at boundary states with insufficient
      reactants.
  \end{itemize}
  The volume factor $V^{1-|\mathbf{u}|}$ is the standard
  classical scaling
  \cite{Kurtz1972, AndersonKurtz2015, Gillespie1992}
  that makes the propensity behave correctly in the
  large-volume limit: $V^{-1}\lambda_r^V(V\mathbf{c}) \to v_r(\mathbf{c})$
  pointwise as $V \to \infty$ with $\mathbf{x} = V\mathbf{c}$
  fixed (proof in Section~\ref{sec:L3-cme}).
  Without this scaling, bimolecular and higher reactions have
  no finite macroscopic limit.

  The \emph{shift operator} $R_r$ acts on observables
  $f : \NN^{|\Sp|} \to \RR$ by
  $(R_r f)(\mathbf{x}) := f(\mathbf{x} + \mathbf{n}_r)$.
  The \emph{generator contribution} of $r$ at volume $V$ is the
  operator
  \[
    \FP^V(r) \;:=\; M_{\lambda_r^V}\,(R_r - I),
  \]
  acting on observables $f$ in the Heisenberg picture as
  \[
    \bigl(\FP^V(r)\,f\bigr)(\mathbf{x})
    \;=\; \lambda_r^V(\mathbf{x})
      \bigl[f(\mathbf{x} + \mathbf{n}_r) - f(\mathbf{x})\bigr],
  \]
  the standard CME generator contribution
  \cite{AndersonCraciunKurtz2010, AndersonKurtz2015},
  consistent with Definition~\ref{def:L3}.
  $\FP^V(r)$ is in general unbounded (the propensity is
  polynomial in $\mathbf{x}$); it is well-defined on
  finitely-supported functions, on which
  $\sum_r \FP^V(r)$ closes to a Markov-semigroup generator
  under standard non-explosion conditions
  \cite{AndersonKurtz2015}.
  We write $\FP(r)$ in place of $\FP^V(r)$ when the volume is
  fixed and unambiguous.
\end{definition}

\begin{remark}[Convention: binomial vs.\ falling-factorial propensities]
\label{rem:propensity-conventions}
Two equivalent stochastic propensity conventions appear in the
literature, differing only by where the combinatorial factors
$\prod_s \nu_s^{(\mathbf{u})}!$ are placed:
\begin{itemize}
  \item \textbf{Binomial form} (used in this chapter,
    Definition~\ref{def:propensity}):
    $\lambda_r^V(\mathbf{x}) = \kappa_r\,V^{1-|\mathbf{u}|}
    \prod_s \binom{x_s}{\nu_s^{(\mathbf{u})}}$.
  \item \textbf{Falling-factorial form} (Anderson--Kurtz
    \cite{AndersonKurtz2011}, Feinberg
    \cite{Feinberg2019}):
    $\lambda_r^V(\mathbf{x}) = \kappa_r^\mathrm{ff}\,
    V^{1-|\mathbf{u}|}
    \prod_s (x_s)_{\nu_s^{(\mathbf{u})}}$,
    where $(x_s)_n := x_s(x_s-1)\cdots(x_s-n+1)$.
\end{itemize}
The two are related by
$\kappa_r^\mathrm{ff} = \kappa_r / \prod_s \nu_s^{(\mathbf{u})}!$,
since $(x_s)_\nu = \nu!\,\binom{x_s}{\nu}$.
The deterministic limit $v_r(\mathbf{c}) = \kappa_r \prod_s c_s^{\nu_s^{(\mathbf{u})}}$
(Proposition~\ref{prop:LMA}) is the same in both conventions ---
the combinatorial factors cancel between numerator and the
$1/\nu_s!$ in $\binom{x_s}{\nu_s}$ when one passes to the
deterministic rate.

A third \emph{macroscopic} convention, customary in
physical-chemistry textbooks
\cite{AtkinsDeP2014, Feinberg2019}, absorbs both the volume
scaling and the combinatorial factor into a single
concentration-based rate constant
\[
  k_r^\mathrm{macro}
  \;:=\; \kappa_r
  \;=\; \kappa_r^\mathrm{ff} \prod_s \nu_s^{(\mathbf{u})}!,
\]
so that $v_r(\mathbf{c}) = k_r^\mathrm{macro}
\prod_s c_s^{\nu_s^{(\mathbf{u})}}$ directly.
The kinetic Wegscheider condition takes its familiar macroscopic
form
$k_r^\mathrm{macro}/k_{r^\dagger}^\mathrm{macro} = K_\mathrm{eq}(r,T)$
in this convention; the binomial-form analogue
(Definition~\ref{def:L3-layer2}) carries an explicit factorial
factor $\prod_s \nu_s^{(\mathbf{u})}!/\nu_s^{(\mathbf{v})}!$,
which is a convention artefact, not a thermodynamic correction.
The three conventions agree on every gauge-invariant kinetic
prediction (deterministic rates, equilibrium ratios, mean steady
states); they differ only in where the $\nu!$ factors are
bookkept.
\end{remark}

\begin{mathbox}[What is $\mathcal{L}_0$ data and what is new]
In the propensity $\lambda_r(\mathbf{x})$:
\begin{itemize}
  \item The stoichiometric coefficients
    $\nu_s^{(\mathbf{u})}$: $\Lk_0$ data
    (from the source complex $\mathbf{u}$ of the morphism $r$).
  \item The shift vector $\mathbf{n}_r$: $\Lk_0$ data
    (the stoichiometric change, computed from $r$).
  \item The rate constant $k_r$: the single new datum at $\Lk_3$.
\end{itemize}
The falling factorial
$\binom{x_s}{\nu} = x_s(x_s-1)\cdots(x_s - \nu + 1)/\nu!$
is the exact stochastic mass-action term first written down by
McQuarrie \cite{McQuarrie1967} and given its definitive
physical derivation by Gillespie \cite{Gillespie1977}.
For large copy numbers, $\binom{x_s}{\nu} \approx (x_s/V)^\nu
V^\nu / \nu! \to c_s^\nu V^\nu / \nu!$ as $V \to \infty$ with
$c_s = x_s/V$ fixed, recovering the deterministic power-law
rate $k_r \prod_s c_s^{\nu_s}$ after rescaling by $V$
(Proposition~\ref{prop:LMA}).
\end{mathbox}

\subsubsection{Functoriality unpacked}

Two reactions $r_1$ and $r_2$ can be arranged in sequence in
$\Lk_0(P)$: the target complex of $r_1$ is the source complex
of $r_2$, giving a composite morphism $r_2 \circ r_1$ in
$\Lk_0(P)$.
The functor axiom $\FP(r_2 \circ r_1) = \FP(r_2) \circ \FP(r_1)$
in the target $B\mathfrak{g}_\Sp$ unpacks --- since sequential
composition in $B\mathfrak{g}_\Sp$ is operator addition
(Definition~\ref{def:gen-S}) --- as the additivity of generator
contributions:
\[
  \FP(r_2 \circ r_1) \;=\; \FP(r_2) + \FP(r_1)
  \quad\text{in } \mathfrak{g}_\Sp.
\]
This is a statement about generator-level assembly: the
right-hand side is the generator of the CTMC in which both
reaction channels $r_1$ and $r_2$ are simultaneously available
on the joint state space.
It is \emph{not} an effective single-step rate law for the
coarse-grained reaction $\mathbf{u} \to \mathbf{w}$; eliminating
the intermediate complex $\mathbf{v}$ requires a quasi-steady-state
or rapid-equilibrium reduction outside the strict $\Lk_3$
functorial assignment, recovered only as a derived approximation
under additional hypotheses on the rate constants.
At the level of stoichiometry, the same composition gives net
shift vector
$\mathbf{n}_{r_2 \circ r_1} = \mathbf{n}_{r_1} + \mathbf{n}_{r_2}$,
implicit throughout CRNT wherever net reactions are obtained by
adding elementary steps \cite{Feinberg2019}.

\begin{proposition}[Generator additivity along $\Lk_0$ composition]
\label{prop:FP-functorial}
  For composable chemical generators
  $r_1 : \mathbf{u} \to \mathbf{v}$ and
  $r_2 : \mathbf{v} \to \mathbf{w}$ in $\Lk_0(P)$, the
  generator contributions satisfy
  \[
    \FP(r_2 \circ r_1)
    \;=\; \FP(r_2) + \FP(r_1)
    \quad\text{in } \mathfrak{g}_\Sp,
  \]
  the generator of the CTMC in which both reaction channels are
  simultaneously available; and the stoichiometric change of
  the composite equals the sum of the individual changes,
  $\mathbf{n}_{r_2 \circ r_1} = \mathbf{n}_{r_1} + \mathbf{n}_{r_2}
  \in \ZZ^{|\Sp|}$.
\end{proposition}

\begin{proof}
The composite $r_2 \circ r_1$ is a morphism in $\Lk_0(P)$, not
an additional chemical generator: it carries no rate constant of
its own.
The functor axiom is the statement that $\FP$ extends correctly
from chemical generators to such composites under the universal
property of $\Lk_0(P)$ (Proposition~\ref{prop:FP-unique}).
The target $B\mathfrak{g}_\Sp$ has sequential composition given
by operator addition (Definition~\ref{def:gen-S}; the standard
additivity of CME generator contributions
\cite{AndersonCraciunKurtz2010}), so
\[
  \FP(r_2 \circ r_1)
  \;=\; \FP(r_2) + \FP(r_1)
  \quad\text{in } \mathfrak{g}_\Sp.
\]

For the stoichiometric statement, the composite
$r_2 \circ r_1 : \mathbf{u} \to \mathbf{w}$ has shift vector
\[
  \mathbf{n}_{r_2 \circ r_1}
  \;=\; \boldsymbol{\nu}^{(\mathbf{w})}
      - \boldsymbol{\nu}^{(\mathbf{u})}
  \;=\; \bigl(\boldsymbol{\nu}^{(\mathbf{w})}
      - \boldsymbol{\nu}^{(\mathbf{v})}\bigr)
      + \bigl(\boldsymbol{\nu}^{(\mathbf{v})}
      - \boldsymbol{\nu}^{(\mathbf{u})}\bigr)
  \;=\; \mathbf{n}_{r_2} + \mathbf{n}_{r_1},
\]
the additivity of shift vectors in $\ZZ^{|\Sp|}$.

A note on what this proposition does \emph{not} say.
The shift operators $R_r$ acting on observables factor
multiplicatively along composition,
$R_{r_2 \circ r_1} = R_{r_2}\,R_{r_1}$, while the generator
contributions $\FP(r_i) = M_{\lambda_{r_i}}(R_{r_i} - I)$ add.
The two live in different algebras --- the shift semigroup of
bounded operators on observables and the additive cone
$\mathfrak{g}_\Sp$ of Markov generators --- and $\FP$ uses
each in its proper place: the multiplicative structure of $R_r$
inside the individual $M_{\lambda_r}(R_r - I)$, and the additive
structure of $\mathfrak{g}_\Sp$ when assembling contributions
along a $\Lk_0$ composite.
This additive assembly is the generator of the joint CTMC, not
an effective rate law for the coarse-grained one-step reaction
$\mathbf{u} \to \mathbf{w}$; recovering an effective rate law
requires a separate reduction (quasi-steady state, rapid
equilibrium) that is not part of the strict $\Lk_3$ functorial
assignment.
\end{proof}

\begin{chembox}[What this proposition means physically, and what it does not]
\textbf{What it says.}
Sequential reactions compose correctly at the levels of
stoichiometry and generator-of-joint-CTMC: if $r_1$ consumes
species $A$ and produces $B$, and $r_2$ consumes $B$ and
produces $C$, then the net stoichiometric change of doing both
is the shift from the $A$-count to the $C$-count, and the
generator of the CTMC with both channels available is
$\FP(r_2) + \FP(r_1)$.
This is the additive content of Hess's Law applied to
stoichiometry --- the overall stoichiometric balance is
independent of how it is decomposed into elementary steps.

\textbf{What it does not say.}
The proposition does \emph{not} claim that the effective rate
at which the coarse-grained reaction $A \to C$ proceeds is some
simple function of $\kappa_{r_1}$ and $\kappa_{r_2}$.
The actual time evolution of the joint CTMC depends on both
propensities through the Gillespie waiting times
\cite{Gillespie1977}, and eliminating the intermediate species
$B$ to obtain an effective one-step rate law requires
quasi-steady-state or rapid-equilibrium reduction --- a
chemical-engineering approximation valid only when $B$ is
short-lived, and not part of the strict $\Lk_3$ assignment.
The proposition captures stoichiometric composition and
joint-CTMC assembly; effective coarse-grained rates are a
separate, approximation-dependent question.
\end{chembox}

\subsubsection{Monoidality unpacked}

Two reactions $r_1$ and $r_2$ can also be placed in independent
parallel subsystems with disjoint species sets $\Sp_1$ and
$\Sp_2$.
The monoidal product $r_1 \otimes r_2$ is their simultaneous
presence in the joint system, and the functor axiom
$\FP(r_1 \otimes r_2) = \FP(r_1) \otimes \FP(r_2)$
demands that the joint generator contribution decomposes
correctly.
This is the categorical expression of the familiar CTMC
independence result: two reactions on disjoint species evolve
independently, and their joint generator is the sum of the
individual generators, a fact used constantly when constructing
modular kinetic models \cite{AndersonKurtz2011}.

\begin{proposition}[Monoidality of $\FP$: parallel reactions]
\label{prop:FP-monoidal}
  For generators $r_1$ and $r_2$ in independent subsystems
  with disjoint species sets $\Sp_1$, $\Sp_2$,
  \[
    \FP(r_1 \otimes r_2) \;=\; \FP(r_1) \otimes \FP(r_2)
  \]
  as generator contributions on the joint state space
  $\NN^{|\Sp_1|} \times \NN^{|\Sp_2|}$.
  At the level of the assembled CME generators
  (Definition~\ref{def:CME} in Section~\ref{sec:L3-cme}),
  this corresponds to the Kronecker sum:
  \[
    \Omega_{12} = \Omega_1 \otimes I + I \otimes \Omega_2.
  \]
\end{proposition}

\begin{proof}
Independent reactions act on independent coordinates of the
joint state $(\mathbf{x}_1, \mathbf{x}_2)$.
Reaction $r_1$ fires at propensity
$\lambda_{r_1}^V(\mathbf{x}_1)$ independently of $\mathbf{x}_2$,
and $r_2$ fires at propensity
$\lambda_{r_2}^V(\mathbf{x}_2)$ independently of $\mathbf{x}_1$
(volume scaling per Definition~\ref{def:propensity}).
The generator contribution of $r_1$ on the joint space acts as
$M_{\lambda_{r_1}^V}(R_{r_1} - I) \otimes I$, and that of $r_2$
as $I \otimes M_{\lambda_{r_2}^V}(R_{r_2} - I)$.
The combined reaction $r_1 \otimes r_2$ therefore has generator
contribution
$\bigl(M_{\lambda_{r_1}^V}(R_{r_1} - I) \otimes I\bigr)
+ \bigl(I \otimes M_{\lambda_{r_2}^V}(R_{r_2} - I)\bigr)$ in the
Kronecker-sum form of the joint $\mathfrak{g}_{\Sp_1 \sqcup \Sp_2}$,
and summing over chemical generators in each subsystem gives
$\Omega_{12} = \Omega_1 \otimes I + I \otimes \Omega_2$.
\end{proof}

\begin{remark}[Kronecker sum versus tensor product of generators]
The Kronecker sum $\Omega_{12} = \Omega_1 \otimes I
+ I \otimes \Omega_2$ is the generator of \emph{independent}
parallel evolution: each subsystem runs on its own Poisson
clock, and the two clocks are independent.
The tensor product $\Omega_1 \otimes \Omega_2$ would describe
simultaneous firing of both reactions at every event ---
a different (and physically unnatural) process.
This distinction is the categorical precision behind the
everyday CRNT modelling choice to write
$\dot{\mathbf{c}} = N_1 \mathbf{v}_1 + N_2 \mathbf{v}_2$
for a combined network rather than any product structure
\cite{Feinberg2019}.
\end{remark}

%% file: chapters/L3/l3_cme.tex
\subsection{The Chemical Master Equation}
\label{sec:L3-cme}

The chemical master equation (CME) was introduced by McQuarrie
\cite{McQuarrie1967} as the master equation for stochastic
chemical kinetics and was given its definitive probabilistic
foundation by Gillespie \cite{Gillespie1977, Gillespie1992},
who derived it from first principles of molecular collision
theory.
Within the tower, the CME is not an independent postulate: it
is the object obtained by \emph{assembling} all the generator
contributions $\{\FP(r)\}_{r \in \Rx}$ assigned by the kinetic
functor.
Specifically, $\FP$ assigns an operator
$\FP(r) = M_{\lambda_r}(R_r - I)$ to each generating reaction
$r$ via the universal property of $\Lk_0(P)$ (Proposition~\ref{prop:FP-unique}).
The CME generator $\Omega$ is the sum of all these operators;
it is not a single functor value but a derived object assembled
from the full image of $\FP$.

This assembly step --- summing over all reactions --- is what
distinguishes $\Lk_3$ from $\Lk_1$ and $\Lk_2$:
the functors $\FH(r) \in \RR$ and $\FS(r) \in \RR$ are
direct numerical outputs of the functor on a single morphism,
whereas the CME generator requires summing $\FP(r)$ over all
generators in $\Rx$.
The analogue at $\Lk_1$ would be to form
$\sum_r \FH(r)$ along a closed loop --- but that sum vanishes
identically by the additive content of Hess's Law on cycles
(\S\ref{sec:L1-layer1}), so it produces no new object;
here the sum genuinely produces a new mathematical entity, the
infinitesimal generator of a Markov semigroup.

\begin{insightbox}[$\FP$ versus $\Omega$: assignment versus assembly]
\label{ins:FP-vs-Omega}
The functor $\FP$ assigns an individual generator contribution
$\FP(r) = M_{\lambda_r}(R_r - I)$ to each reaction $r \in \Rx$
via the universal property of $\Lk_0(P)$.
The CME generator
\[
  \Omega \;=\; \sum_{r \in \Rx} \FP(r)
  \;=\; \sum_{r \in \Rx} M_{\lambda_r}(R_r - I)
\]
is assembled from the entire image of $\FP$.
It is not a single morphism value of $\FP$, but a global object.

This two-step structure --- functor assigns, sum assembles ---
has no analogue at $\Lk_1$ or $\Lk_2$, where the individual
values $\FH(r)$ and $\FS(r)$ directly gave the thermochemical
invariants without further assembly.
The CME is the characteristic assembled object of $\Lk_3$.
\end{insightbox}

\begin{definition}[CME generator and the Chemical Master Equation
  {\cite{McQuarrie1967, Gillespie1992}}]
\label{def:CME}
  Fix a system volume $V > 0$.
  The \emph{CME generator} of $\Lk_3(P)$ at volume $V$ is the
  operator on observables $f : \NN^{|\Sp|} \to \RR$
  \[
    \Omega^V(\Lk_3(P))
    \;:=\;
    \sum_{r \in \Rx} \FP^V(r),
  \]
  with $\FP^V(r) = M_{\lambda_r^V}(R_r - I)$ and $\lambda_r^V$
  the volume-scaled mass-action propensity of
  Definition~\ref{def:propensity}.
  The \emph{Chemical Master Equation} of $\Lk_3(P)$ is the
  Kolmogorov forward equation of the CTMC with generator
  $\Omega^V(\Lk_3(P))$:
  \[
    \frac{d}{dt} p(\mathbf{x}, t)
    \;=\;
    \bigl(\Omega^V(\Lk_3(P))^\top\, p\bigr)(\mathbf{x}, t)
    \;=\;
    \sum_{r \in \Rx}
    \Bigl[
      \lambda_r^V(\mathbf{x} - \mathbf{n}_r)\,
      p(\mathbf{x} - \mathbf{n}_r, t)
      \;-\;
      \lambda_r^V(\mathbf{x})\, p(\mathbf{x}, t)
    \Bigr],
  \]
  where $p(\mathbf{x}, t) \geq 0$ is the probability of being in
  state $\mathbf{x} \in \NN^{|\Sp|}$ at time $t \geq 0$.
  We write $\Omega(\Lk_3(P))$ in place of $\Omega^V$ when the
  volume is fixed and unambiguous.
\end{definition}

\begin{remark}[Operator form versus gain--loss form]
\label{rem:CME-notation}
The CME above uses the notation established in
Definition~\ref{def:propensity}: $\lambda_r(\mathbf{x})$ is the
full mass-action propensity (incorporating $\kappa_r$ and the
volume factor), so the gain--loss form does not carry an extra
factor of $\kappa_r$.
The two equivalent forms are related by transposition between
the Heisenberg and Schr\"odinger pictures.
On observables $f$, the Heisenberg-picture generator acts as
\[
  (\Omega f)(\mathbf{x})
  \;=\; \sum_{r \in \Rx} \lambda_r(\mathbf{x})\,
    \bigl[f(\mathbf{x} + \mathbf{n}_r) - f(\mathbf{x})\bigr],
\]
which is the operator form $\Omega = \sum_r M_{\lambda_r}(R_r - I)$.
On probabilities $p$, the dual $\Omega^\top$ acts as
\[
  (\Omega^\top p)(\mathbf{x})
  \;=\; \sum_{r \in \Rx}
    \bigl[\lambda_r(\mathbf{x} - \mathbf{n}_r)\,
    p(\mathbf{x} - \mathbf{n}_r)
    - \lambda_r(\mathbf{x})\,p(\mathbf{x})\bigr],
\]
which is the gain--loss form.
The two are interchanged by the duality
$\langle \Omega f, p\rangle = \langle f, \Omega^\top p\rangle$
of summation against $\NN^{|\Sp|}$; the Kolmogorov forward
equation $\dot p = \Omega^\top p$ uses the second form.
\end{remark}

\begin{chembox}[Reading the CME term by term]
Each summand in the CME has two contributions, familiar from
any textbook treatment of CTMCs \cite{AndersonKurtz2011}:
\begin{itemize}
  \item \textbf{Gain term}:
    $\lambda_r(\mathbf{x} - \mathbf{n}_r)\,
    p(\mathbf{x} - \mathbf{n}_r, t)$ --- probability flowing
    \emph{into} state $\mathbf{x}$ from the predecessor state
    $\mathbf{x} - \mathbf{n}_r$ via reaction $r$.
  \item \textbf{Loss term}:
    $\lambda_r(\mathbf{x})\,p(\mathbf{x}, t)$ --- probability
    flowing \emph{out of} state $\mathbf{x}$ via reaction $r$.
\end{itemize}
The net change in $p(\mathbf{x}, t)$ is the algebraic sum of all
gains and losses across all reactions.
Probability is conserved at the generator level:
\begin{align*}
  \Omega^V \mathbf{1} &\;=\; 0
  \quad\text{(Heisenberg picture, on the constant observable),}
  \\
  \sum_{\mathbf{x}} \bigl(\Omega^{V\,\top}\, p\bigr)(\mathbf{x}, t)
  &\;=\; 0
  \quad\text{(Schr{\"o}dinger picture).}
\end{align*}
The two are the same conservation law expressed on observables
and on probability distributions respectively, related by
duality $\langle \Omega^V f, p\rangle = \langle f,
\Omega^{V\,\top} p\rangle$.
In the tower language, this conservation law is the statement
that $\Omega^V$ is a valid Markov generator on $\NN^{|\Sp|}$
(an element of $\mathfrak{g}_\Sp$): the row-sum-zero condition
on the matrix entries of $\Omega^V$ in the Heisenberg picture
is exactly the infinitesimal form of $\sum_{\mathbf{y}}
K(\mathbf{y} \mid \mathbf{x}) = 1$ for Markov kernels $K$
\cite{Fritz2020}.
\end{chembox}

The CME generator $\Omega(\Lk_3(P))$ inherits the
species-permutation symmetry of $\Lk_0(P)$.
This is the first tower-native statement about $\Omega$ as
an object in its own right, and it is the equivariance
property that will lift to the Para shadow
$\Lk_3^{\mathrm{Para}}$ in Chapter~\ref{sec:Para}: a learned
kinetic model is a valid object of $\Lk_3^{\mathrm{Para}}$ only if it satisfies the analogue of Proposition~\ref{prop:CME-covariance} below.

\begin{proposition}[Species-permutation covariance of $\Omega$]
\label{prop:CME-covariance}
  Let $\sigma \in \Aut(\Lk_0(P))$ be an automorphism induced
  by a species permutation $\sigma : \Sp \to \Sp$, and let
  $\sigma \cdot \Lk_3(P)$ denote the kinetic level structure
  with species relabelled by $\sigma$ and rate constants
  transported along the induced bijection
  $\sigma_* : \Rx \to \Rx$ via
  $(\sigma \cdot \kappa)_{\sigma_*(r)} := \kappa_r.$.
  Let $U_\sigma$ denote the operator on observables induced by
  the coordinate permutation $\mathbf{x} \mapsto \sigma \cdot \mathbf{x}$,
  i.e.\ $(U_\sigma f)(\mathbf{x}) = f(\sigma^{-1} \cdot \mathbf{x})$.
  Then
  \[
    U_\sigma\, \Omega(\Lk_3(P))\, U_\sigma^{-1}
    \;=\;
    \Omega(\sigma \cdot \Lk_3(P)).
  \]
\end{proposition}

\begin{proof}
  It suffices to prove the identity generator-by-generator,
  since $\Omega$ is a linear sum over $\Rx$ and conjugation by
  $U_\sigma$ distributes over the sum.
  Fix $r \in \Rx$; we compute the action of
  $U_\sigma\,\FP(r)\,U_\sigma^{-1}
  = U_\sigma\,M_{\lambda_r}(R_r - I)\,U_\sigma^{-1}$
  on a test function $f$, using $(U_\sigma^{-1} g)(\mathbf{x})
  = g(\sigma \cdot \mathbf{x})$ throughout.

  For the shift operator,
  \[
    (U_\sigma R_r U_\sigma^{-1} f)(\mathbf{x})
    \;=\; (R_r (U_\sigma^{-1} f))(\sigma^{-1} \cdot \mathbf{x})
    \;=\; (U_\sigma^{-1} f)(\sigma^{-1} \cdot \mathbf{x} + \mathbf{n}_r)
    \;=\; f(\mathbf{x} + \sigma \cdot \mathbf{n}_r),
  \]
  so $U_\sigma R_r U_\sigma^{-1} = R_{\sigma_*(r)}$, since
  $\sigma \cdot \mathbf{n}_r = \mathbf{n}_{\sigma_*(r)}$ by
  definition of the induced action $\sigma_*$ on reactions.

  For the multiplication operator,
  \[
    (U_\sigma M_{\lambda_r} U_\sigma^{-1} f)(\mathbf{x})
    \;=\; \lambda_r(\sigma^{-1} \cdot \mathbf{x})\,f(\mathbf{x}).
  \]
  Now 
  \[\lambda_r(\sigma^{-1} \cdot \mathbf{x})
= \kappa_r \prod_{s \in \Sp}
  \binom{x_{\sigma(s)}}{\nu_s^{(\mathbf{u}_r)}}
= \kappa_r \prod_{t \in \Sp}
  \binom{x_t}{\nu_{\sigma^{-1}(t)}^{(\mathbf{u}_r)}}\]
  (reindexing $t = \sigma(s)$).
  By definition of the species relabelling, the source complex
  of $\sigma_*(r)$ has stoichiometric coefficients
  $\nu_t^{(\mathbf{u}_{\sigma_*(r)})}
  = \nu_{\sigma^{-1}(t)}^{(\mathbf{u}_r)}$,
  and the rate constant is
  $(\sigma \cdot \kappa)_{\sigma_*(r)} = \kappa_r$, so
  $\lambda_r(\sigma^{-1} \cdot \mathbf{x})
  = \lambda_{\sigma_*(r)}(\mathbf{x})$ in the relabelled
  network.
  Hence $U_\sigma M_{\lambda_r} U_\sigma^{-1}
  = M_{\lambda_{\sigma_*(r)}}$.

  Combining the two,
  \[
    U_\sigma\,\FP(r)\,U_\sigma^{-1}
    \;=\; M_{\lambda_{\sigma_*(r)}}(R_{\sigma_*(r)} - I)
    \;=\; (\sigma \cdot \FP)(\sigma_*(r)).
  \]
  Summing over $r \in \Rx$ and reindexing by the bijection
  $\sigma_* : \Rx \to \Rx$ gives\\
  $U_\sigma \Omega(\Lk_3(P)) U_\sigma^{-1}
  = \Omega(\sigma \cdot \Lk_3(P))$ as claimed.
\end{proof}

\begin{remark}[Two levels of invariance and why it matters]
\label{rem:CME-covariance-meaning}
  Proposition~\ref{prop:CME-covariance} is a
  \emph{covariance} statement, not an invariance statement:
  $\Omega$ changes under $\sigma$, but changes in exactly the
  way the relabelled $\Lk_3(P)$ prescribes.
  Two special cases are worth naming.
  \begin{itemize}
    \item \textbf{Stabilisers of $\Lk_3(P)$.}
      If $\sigma$ fixes the tuple $(\Lk_0(P), \{k_r\})$ ---
      i.e.\ $\sigma \in \Aut(\Lk_3(P))$ ---
      then $\sigma \cdot \Lk_3(P) = \Lk_3(P)$ and the
      proposition reduces to
      $U_\sigma \Omega U_\sigma^{-1} = \Omega$:
      the CME generator commutes with the stabiliser action.
      This is the statement that symmetry-related states
      evolve identically under the CTMC.
    \item \textbf{Forcing-pair diagnostic.}
      If $\sigma$ lies in $\Aut(\Lk_2(P)) \setminus
      \Aut(\Lk_3(P))$ --- the rate-constant-swapping
      automorphisms of Section~\ref{sec:L3-forcing-in}, which
      witness $\coker(\varphi_3) \neq 1$ ---
      then $\sigma \cdot \Lk_3(P) \neq \Lk_3(P)$ and
      $U_\sigma \Omega U_\sigma^{-1} \neq \Omega$.
      The proposition detects the forcing pair: the CME
      generator itself distinguishes what $\Lk_2$ cannot.
  \end{itemize}
  Viewed as an equivariance condition, the proposition says:
  the assignment $\Lk_3(P) \mapsto \Omega(\Lk_3(P))$ is a
  permutation-equivariant map, where the permutations acts on
  $\Lk_3$-structures by species relabelling and on the space
  of generators by conjugation.
  A parametric kinetic model
  $(\Theta, f_\theta) : \Lk_3(P) \to \mathfrak{g}_\Sp$
  is a morphism in $\Lk_3^{\mathrm{Para}}$ only if it is equivariant
  in this sense: the species-permutation symmetry is not an
  architectural choice but a \emph{defining} property of the
  level, exactly as Gavranovi\'c et al.
  \cite{GavRanovic2024CDL} identifies equivariance with the
  $\mathrm{Lax}$-algebra homomorphism condition.
\end{remark}

\subsubsection{The deterministic limit as a forgetful functor}

The reaction rate equation (RRE) is relevant here for the following reason: it shows that $\Lk_3$ strictly contains $\Lk_0$ as a limiting case.
Concretely, the stoichiometric matrix $N$ of $\Lk_0$ and the rate functor $\FP$ of $\Lk_3$ together determine the RRE via a \emph{large-volume forgetful functor} $\Phi_\infty$; the RRE is not a separate postulate but a derived image of $\FP$.
Rewriting Kurtz's theorem \cite{Kurtz1970, Kurtz1972} in tower language makes this factorisation explicit.

\begin{definition}[Large-volume scaling limit]
\label{def:Phi-infty}
  For each volume $V > 0$, define the \emph{scaling map}
  \[
    \phi_V : \NN^{|\Sp|} \;\longrightarrow\; \RR_{>0}^{|\Sp|},
    \qquad
    \phi_V(\mathbf{x}) \;:=\; \mathbf{x}/V.
  \]
  Let $\{A^V\}_{V > 0}$ be a volume-indexed family of operators
  on observables, with $A^V \in \mathfrak{g}_\Sp$ for each $V$;
  the family is said to be \emph{classically scaled} if the
  limit
  \[
    (\Phi_\infty\{A^V\})(\mathbf{c})
    \;:=\; \lim_{V\to\infty}
    V \cdot \bigl((\phi_V)_* A^V\bigr)\!\bigl[\mathbf{c}\bigr]
    \;\in\; T_{\mathbf{c}}\RR_{>0}^{|\Sp|}
  \]
  exists for every $\mathbf{c} \in \RR_{>0}^{|\Sp|}$ in the
  sense of pointwise convergence on smooth compactly-supported
  test functions, where $(\phi_V)_*$ denotes the pushforward of
  $A^V$ along $\phi_V$.
  The \emph{large-volume scaling limit} $\Phi_\infty$ assigns
  to each classically-scaled family the resulting smooth vector
  field on $\RR_{>0}^{|\Sp|}$.

  $\Phi_\infty$ is not a functor on arbitrary generators: it is
  defined only on volume-indexed families satisfying the
  classical density-dependent scaling
  \cite{Kurtz1972, AndersonKurtz2011}.
  Mass-action families
  $A^V = \Omega^V(\Lk_3(P)) = \sum_{r} \FP^V(r)$
  built from the volume-scaled propensities of
  Definition~\ref{def:propensity} are classically scaled by
  construction; this is the content of
  Proposition~\ref{prop:LMA} below.
\end{definition}

\begin{proposition}[The RRE as the scaling limit of the mass-action CME family
  {\cite{Kurtz1970, Kurtz1972, AndersonKurtz2011}}]
\label{prop:LMA}
  The volume-indexed family of CME generators
  $\{\Omega^V(\Lk_3(P))\}_{V > 0}$, with
  $\Omega^V = \sum_{r \in \Rx} \FP^V(r)$ assembled from the
  volume-scaled mass-action propensities of
  Definition~\ref{def:propensity}, is classically scaled, and
  its large-volume scaling limit (Definition~\ref{def:Phi-infty})
  is the reaction-rate vector field:
  \[
    \Phi_\infty\!\bigl(\{\Omega^V\}\bigr)
    \;=\; N \circ \mathbf{v},
  \]
  where:
  \begin{itemize}
    \item $N \in \ZZ^{|\Sp| \times |\Rx|}$ is the stoichiometric
      matrix, an $\Lk_0$ datum.
    \item $\mathbf{v} : \RR_{>0}^{|\Sp|} \to \RR_{>0}^{|\Rx|}$,
      $v_r(\mathbf{c}) = \kappa_r \prod_{s \in \Sp}
      c_s^{\nu_s^{(\mathbf{u}_r)}}$,
      is the mass-action rate function, with $\kappa_r$ from
      $\FP$ (the stochastic mass-action rate constant of
      Definition~\ref{def:propensity}) and exponents
      $\nu_s^{(\mathbf{u}_r)}$ from $\Lk_0$.
  \end{itemize}
  The autonomous ODE
  $\dot{\mathbf{c}} = (N \circ \mathbf{v})(\mathbf{c})$ is the
  \emph{reaction-rate equation (RRE)}.
  Moreover, the family of stochastic processes governed by the
  CME at volume $V$ converges in probability, uniformly on
  compact time intervals, to the flow of the RRE as
  $V \to \infty$ with $\mathbf{c}(0) = \mathbf{x}^V(0)/V$ fixed
  \cite{Kurtz1970, Kurtz1972}.
\end{proposition}

\begin{proof}
Apply $\Phi_\infty$ to a single volume-indexed generator
contribution $\FP^V(r) = M_{\lambda_r^V}(R_r - I)$.
For smooth test functions $f : \RR_{>0}^{|\Sp|} \to \RR$, the
pushforward along $\phi_V$ acts on $f$ at $\mathbf{c}$ by
applying $\FP^V(r)$ to $f \circ \phi_V$ at
$\mathbf{x} = V\mathbf{c}$:
\[
  \bigl[(\phi_V)_* \FP^V(r) f\bigr](\mathbf{c})
  \;=\; \lambda_r^V(V\mathbf{c})\,
  \Bigl[
    f\!\Bigl(\mathbf{c} + \frac{\mathbf{n}_r}{V}\Bigr)
    - f(\mathbf{c})
  \Bigr].
\]
The volume-scaled propensity at $\mathbf{x} = V\mathbf{c}$ has
the limit
\[
  \frac{\lambda_r^V(V\mathbf{c})}{V}
  \;=\; \frac{1}{V}\,\kappa_r\,V^{1 - |\mathbf{u}_r|}
  \prod_{s \in \Sp} \binom{V c_s}{\nu_s^{(\mathbf{u}_r)}}
  \;\xrightarrow{V\to\infty}\;
  \kappa_r \prod_{s \in \Sp} c_s^{\nu_s^{(\mathbf{u}_r)}}
  \;=\; v_r(\mathbf{c}),
\]
since
$V^{-|\mathbf{u}_r|}\binom{V c_s}{\nu_s^{(\mathbf{u}_r)}}
\to c_s^{\nu_s^{(\mathbf{u}_r)}}/\nu_s^{(\mathbf{u}_r)}!$
componentwise (with the falling-factorial convergence absorbing
the $\prod_s \nu_s^{(\mathbf{u}_r)}!$ in the binomial
denominators) and
$\sum_s \nu_s^{(\mathbf{u}_r)} = |\mathbf{u}_r|$ giving the
correct power of $V$.
This is the content of the volume-scaling factor
$V^{1-|\mathbf{u}_r|}$ built into Definition~\ref{def:propensity}:
without it, bimolecular and higher reactions would have no
finite limit.
Combined with
$V\bigl[f(\mathbf{c} + \mathbf{n}_r/V) - f(\mathbf{c})\bigr]
\to \mathbf{n}_r \cdot \nabla f(\mathbf{c})$,
\[
  V \cdot \bigl[(\phi_V)_* \FP^V(r) f\bigr](\mathbf{c})
  \;=\;
  \frac{\lambda_r^V(V\mathbf{c})}{V}
  \cdot
  V\Bigl[f\!\Bigl(\mathbf{c} + \frac{\mathbf{n}_r}{V}\Bigr)
    - f(\mathbf{c})\Bigr]
  \;\xrightarrow{V\to\infty}\;
  v_r(\mathbf{c})\,
  \bigl(\mathbf{n}_r \cdot \nabla f\bigr)(\mathbf{c}).
\]
Summing over $r \in \Rx$,
\[
  \Phi_\infty\!\bigl(\{\Omega^V\}\bigr)(\mathbf{c})
  \;=\; \sum_{r \in \Rx} v_r(\mathbf{c})\,
    \bigl(\mathbf{n}_r \cdot \nabla\bigr)
  \;=\; \bigl(N \mathbf{v}(\mathbf{c})\bigr) \cdot \nabla,
\]
the directional derivative operator of the vector field
$N \mathbf{v}(\mathbf{c})$.
This is the generator of the deterministic flow
$\dot{\mathbf{c}} = N \mathbf{v}(\mathbf{c})$, hence
$\Phi_\infty(\{\Omega^V\}) = N \circ \mathbf{v}$ as claimed.
The trajectory-level in-probability convergence follows from
\cite{Kurtz1970}.
\end{proof}

\begin{remark}[Kurtz's theorem in the literature]
\label{rem:kurtz}
Kurtz's 1970 paper \cite{Kurtz1970} established the
in-probability convergence; the 1972 follow-up
\cite{Kurtz1972} specialised this to mass-action chemical
kinetics and made the power-law rate form explicit.
Anderson and Kurtz \cite{AndersonKurtz2011} give the
definitive modern treatment in the CRNT setting (the
classical-scaling derivation of the deterministic law of
mass action is the content of their Section~3).
The proof sketch above is the tower-language restatement of
that result: $\Phi_\infty$ extracts the drift of the rescaled
process from the volume-indexed family $\{\Omega^V\}$, and
the drift factors as $N \circ \mathbf{v}$ because $N$ is an
$\Lk_0$ datum (it does not depend on $\kappa_r$) while
$\mathbf{v}$ is an $\Lk_3$ datum (it depends linearly on
$\kappa_r$ and polynomially on $\mathbf{c}$ through the
$\Lk_0$ exponents $\nu_s$).
\end{remark}

\begin{mathbox}[Level stratification of the RRE: a commutative diagram]
The factorisation
$\Phi_\infty(\{\FP^V(r)\})(\mathbf{c})
= v_r(\mathbf{c})\,\mathbf{n}_r \cdot \nabla$, established
generator-by-generator in the proof of
Proposition~\ref{prop:LMA} and assembled into
$\Phi_\infty(\{\Omega^V\}) = N \circ \mathbf{v}$ on the full
CME family, makes the following diagram commute on each
$r \in \Rx$ (read: per fixed-$V$ slice for the upper arrows,
and as a scaling limit on the volume-indexed family for the
lower arrow):
\[
\includegraphics{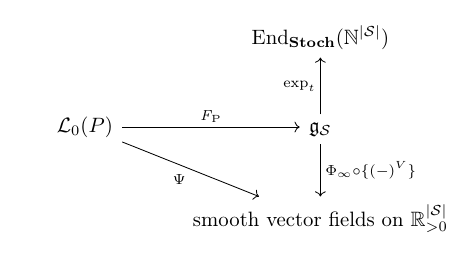}
\]
where the diagonal $\Psi$ sends a chemical generator $r$ to the
vector field $\mathbf{c} \mapsto v_r(\mathbf{c})\,\mathbf{n}_r$
on $\RR_{>0}^{|\Sp|}$, and summation over $r \in \Rx$ gives the
RRE vector field $\mathbf{c} \mapsto N\mathbf{v}(\mathbf{c})$.

The upper arrow $\exp_t : L \mapsto e^{tL}$ recovers the
finite-time Markov kernels in $\Stoch$ from the generator;
this is not an SMC functor (sequential composition fails to
commute with addition off the abelian locus, by Trotter--Kato),
and that failure is the categorical content of the
$\Lk_3 \to \Lk_4$ forcing pair.
The lower arrow combines the volume-scaling
$\FP \mapsto \{\FP^V\}$ with the scaling-limit functional
$\Phi_\infty$ of Definition~\ref{def:Phi-infty}; it is well-defined
only on the mass-action class of generators where the classical
volume scaling applies.

The diagram shows which data live at which level:
\begin{itemize}
  \item $\FP$ (horizontal arrow): full stochastic structure at
    fixed volume, $\Lk_3$ --- the propensity $\lambda_r^V$ and
    the shift $R_r$.
  \item $\exp_t$ (upper arrow): finite-time Markov kernels in
    $\Stoch$, parametrised by $t \geq 0$.
  \item $\Phi_\infty$ (lower arrow): large-volume scaling limit
    on the volume-indexed family $\{\Omega^V\}$, stripping
    stochastic fluctuations down to the deterministic drift.
  \item $\mathbf{n}_r$ (in the diagonal $\Psi$): the
    stoichiometric vector, $\Lk_0$ data.
  \item $v_r(\mathbf{c}) = \kappa_r \prod_{s \in \Sp}
    c_s^{\nu_s^{(\mathbf{u}_r)}}$: $\kappa_r$ is $\Lk_3$ data;
    the source-complex exponents $\nu_s^{(\mathbf{u}_r)}$ are
    $\Lk_0$ data.
\end{itemize}
The RRE is the image of $\FP$ under the scaling limit: a
\emph{derived} object, not a primitive.
The CME is primary at $\Lk_3$; the deterministic mass-action
equation that CRNT practitioners reach for first is, from the
tower's perspective, the large-volume shadow of the stochastic
truth, and the time-$t$ Markov kernels in $\Stoch$ are its
exponentiation.
This ordering --- CME primary, RRE derived --- is the precise
sense in which $\Lk_3$ is a stochastic level.
\end{mathbox}

%% file: chapters/L3/l3_layer2.tex
\subsection{Layer~1 and~2 for \texorpdfstring{$\FP$}{FP}:
  thermodynamic consistency and detailed balance}
\label{sec:L3-layer2}

Every level of the tower so far has had two sub-layers.
Layer~1 is the minimal structure forced by the universal
property of $\Lk_0(P)$: given any assignment of the new
datum to generators, a unique functor extension exists.
Layer~2 is an additional coherence condition that locks
the new functor to the structure already present at lower
levels of the tower.
At $\Lk_1$, Layer~2 was the coboundary condition
$\FH = \delta^0 h_f$ (Proposition~\ref{prop:L1-equiv}),
which forced $\FH$ values around closed loops to sum to zero
--- a condition internal to $\Lk_1$.
At $\Lk_2$, Layer~2 was the Wegscheider cycle condition
(Proposition~\ref{prop:wegscheider1}), requiring $F_G^T$
to vanish on every closed loop in $G(\Rx)$ --- a condition
forced by the $\dagger$-SMC structure of $\Lk_2$.

At $\Lk_3$, the new functor $\FP$ inherits from below: both
$F_G^T$ (an $\Lk_2$ derived quantity) and the $\dagger$-structure
on $\Lk_0(P)$ (which gives every generator $r$ a reverse
$r^\dagger$) are already present in the tower.
The question is: what does it mean for $\FP$ to be
\emph{compatible} with this inherited structure?
The answer, derived below, is that $\FP$ must be
$\dagger$-compatible: the Markov chain generated by
$\Omega = \sum_r \FP(r)$ must satisfy detailed balance
with respect to the thermodynamic equilibrium locus
$\mathcal{E}_T$ from $\Lk_2$.
This compatibility condition is precisely the kinetic
Wegscheider condition \cite{Wegscheider1901, Feinberg1989},
and it is not an independent postulate --- it is forced by
the tower's $\dagger$-structure.

\subsubsection{Layer~1: any positive rate constants form a
  valid CME}

\textbf{Layer~1 for $\FP$}: any assignment
$k : \Rx \to \RR_{>0}$ defines a valid strict SMC functor
$\FP : \Lk_0(P) \to \Stoch$ by Proposition~\ref{prop:FP-unique}.
For a reversible Petri net with $|\Rx| = 2m$ generators
(forward and reverse pairs), there are $2m$ free positive-real
parameters.
The assembled CME generator $\Omega = \sum_{r \in \Rx} \FP(r)$
(Definition~\ref{def:CME}) is a valid Markov generator for
any such assignment: the columns of $\Omega$ sum to zero
and off-diagonal entries are non-negative.

This means $\FP$ at Layer~1 is decoupled from the
thermodynamics of $\Lk_2$: the rate constants
$k_r$ and $k_{r^\dagger}$ can be chosen independently,
with no reference to $\FH$, $\FS$, or $F_G^T$.
The following insightbox shows what goes wrong
categorically when this decoupling persists.

\begin{insightbox}[Layer~1 alone leaves $\Lk_3$ and $\Lk_2$ inter-level constraints unenforced]
\label{ins:L3-layer1}
  Consider a reversible Petri net with a single chemical
  generator pair $(r, r^\dagger)$ for the unimolecular
  interconversion $A \rightleftharpoons B$.
  At Layer~1, choose $\kappa_r = 10^3\;\mathrm{s^{-1}}$ and
  $\kappa_{r^\dagger} = 1\;\mathrm{s^{-1}}$.
  The CME generator is
  \[
    \Omega \;=\; M_{\lambda_r}(R_r - I)
      + M_{\lambda_{r^\dagger}}(R_{r^\dagger} - I),
    \quad
    \lambda_r(\mathbf{x}) = 10^3\, x_A,
    \quad
    \lambda_{r^\dagger}(\mathbf{x}) = x_B,
  \]
  a valid Markov generator assembled from the image of $\FP$
  on $\{r, r^\dagger\}$ (both reactions are unimolecular, so
  the volume factor $V^{1-|\mathbf{u}|} = V^0 = 1$ vanishes
  and the bare $\kappa$ values appear directly in the
  propensities).
  The CTMC is reversible (every two-state birth--death chain
  with positive rates in both directions is), so its
  stationary distribution $\pi^*$ satisfies detailed balance
  \[
    \pi^*(\mathbf{x})\,\lambda_r(\mathbf{x})
    \;=\; \pi^*(\mathbf{x} + \mathbf{n}_r)\,
    \lambda_{r^\dagger}(\mathbf{x} + \mathbf{n}_r).
  \]
  $\pi^*$ is a Poisson product with mean concentrations
  $\mathbf{c}^*$ satisfying
  $c_B^*/c_A^* = \kappa_r/\kappa_{r^\dagger} = 10^3$, the
  \emph{kinetic} equilibrium ratio.

  Now suppose the $\Lk_2$ data give
  $F_G^{298}(r) = +17.1\;\mathrm{kJ\,mol^{-1}}$, so
  $K_\mathrm{eq}(r, 298\,\mathrm{K})
  = \exp\!\bigl(-17100/(8.314 \times 298)\bigr)
  \approx 10^{-3}$ --- the \emph{thermodynamic standard
  equilibrium constant}, which would describe the
  concentration ratio attained by a closed system at
  thermodynamic equilibrium.

  Then $c_B^*/c_A^* = 10^3 \neq K_\mathrm{eq} = 10^{-3}$:
  the kinetic and thermodynamic standard ratios disagree by
  six orders of magnitude, and the Markov chain converges to
  a steady state $\mathbf{c}^* \notin \mathcal{E}_T$ ---
  outside the thermodynamic equilibrium variety defined at
  $\Lk_2$.
  This disagreement is internally consistent at Layer~1: the
  rate constants $\kappa_r, \kappa_{r^\dagger}$ are free
  positive reals, and there is no level-internal axiom
  forcing them to encode the same equilibrium ratio as
  $F_G^T$.
  But for closed-system thermochemistry --- where any
  reversible elementary reaction must reach the same
  equilibrium on both kinetic and thermodynamic accounts ---
  this independence is exactly the property that needs to be
  ruled out.
  Layer~2 supplies the missing inter-level constraint,
  forcing $\FP$ and $F_G^T$ to agree on equilibrium ratios
  via the $\dagger$-compatibility condition derived in the
  next subsection.
\end{insightbox}

\subsubsection{Layer~2: \texorpdfstring{$\dagger$}{+}-compatibility forces detailed balance}

The $\Lk_2$ level already carries a $\dagger$-SMC structure:
every generator $r : \mathbf{u} \to \mathbf{v}$
has a reverse $r^\dagger : \mathbf{v} \to \mathbf{u}$, and the
free-energy functor satisfies $F_G^T(r^\dagger) = -F_G^T(r)$.
The thermodynamic equilibrium variety is
\[
  \mathcal{E}_T \;:=\; \bigl\{\mathbf{c}^* \in \RR_{>0}^{|\Sp|}
    : \textstyle \prod_s (c_s^*)^{n_{r,s}}
    = K_{\mathrm{eq}}(r,T) \text{ for all } r \in \Rx \bigr\},
\]
the locus in concentration space picked out by the
$\Lk_2$ functor $F_G^T$ via the Boltzmann relation
$K_{\mathrm{eq}}(r,T) = \exp(-F_G^T(r)/RT)$.
This concentration-space variety $\mathcal{E}_T$ and the categorical kernel
$\ker F_G^T \subset \mathrm{Mor}(\Lk_0(P))$ are companion
objects: $\ker F_G^T$ records reactions that are
thermodynamically neutral in standard state, while
$\mathcal{E}_T$ records the concentration vectors at which
all reactions in $\Rx$ are simultaneously balanced via the
Boltzmann relation $\prod_s (c_s^*)^{n_{r,s}} = K_\mathrm{eq}(r,T)$.
The latter is the relevant locus for detailed balance of the CME.

For a Poisson product distribution
$\pi^*(\mathbf{x}) = \prod_s e^{-c_s^*}(c_s^*)^{x_s}/x_s!$
with $\mathbf{c}^* \in \mathcal{E}_T$, the pointwise
detailed balance condition on the CTMC generator
$\Omega = \sum_{r \in \Rx} \FP(r)$ reads
\[
  \pi^*(\mathbf{x})\,\lambda_r(\mathbf{x})
  \;=\;
  \pi^*(\mathbf{x} + \mathbf{n}_r)\,
  \lambda_{r^\dagger}(\mathbf{x} + \mathbf{n}_r)
\]
for every state $\mathbf{x}$ and every reversible pair
$(r, r^\dagger)$.
We work in the binomial-form stochastic convention of
Definition~\ref{def:propensity}, with stochastic rate
constants $\kappa_r$; the analogous derivation in
falling-factorial or macroscopic conventions changes only the
combinatorial factor (Remark~\ref{rem:propensity-conventions}).
Rearranged and evaluated on the Poisson ratio
$\pi^*(\mathbf{x}+\mathbf{n}_r)/\pi^*(\mathbf{x})
= \prod_s (c_s^*)^{n_{r,s}} x_s!/(x_s+n_{r,s})!$,
with product-form propensities
$\lambda_r(\mathbf{x}) = \kappa_r \prod_s
\binom{x_s}{\nu_s^{(\mathbf{u})}}$
(Definition~\ref{def:propensity}, fixed-$V$ shorthand), the
state-dependent factorials cancel identically and the
condition collapses to
\[
  \prod_s (c_s^*)^{n_{r,s}}
  \;=\;
  \frac{\kappa_r}{\kappa_{r^\dagger}}
  \prod_s \frac{\nu_s^{(\mathbf{v})}!}{\nu_s^{(\mathbf{u})}!}.
\]
The left-hand side is fixed by $\Lk_2$: the condition
$\mathbf{c}^* \in \mathcal{E}_T$ unpacks as
$\sum_s n_{r,s}[\mu_s^\circ(T) + RT \ln c_s^*] = 0$, giving
\[
  \prod_s (c_s^*)^{n_{r,s}}
  \;=\;
  \exp\!\left(\frac{-F_G^T(r)}{RT}\right)
  \;=\;
  K_\mathrm{eq}(r, T).
\]
Combining the two identities forces
\[
  \frac{\kappa_r}{\kappa_{r^\dagger}}
  \;=\; K_\mathrm{eq}(r, T) \prod_s
    \frac{\nu_s^{(\mathbf{u})}!}{\nu_s^{(\mathbf{v})}!}
  \;=\; \exp\!\left(\frac{-F_G^T(r)}{RT}\right)
    \prod_s \frac{\nu_s^{(\mathbf{u})}!}{\nu_s^{(\mathbf{v})}!}.
\]
This is the content of Layer~2: not a new postulate but
the unique condition that makes $\FP$ compatible with the
$\dagger$-SMC structure of $\Lk_2$ on the thermodynamic
equilibrium locus, in the binomial stochastic convention.

\begin{remark}[The combinatorial factor: where it lives]
\label{rem:combinatorial-factor}
The factor $\prod_s \nu_s^{(\mathbf{u})}!/\nu_s^{(\mathbf{v})}!$
appearing in the kinetic Wegscheider condition above is a
\emph{convention artefact} of the binomial-form propensity, not
a thermodynamic correction
(Remark~\ref{rem:propensity-conventions}).
For first-order kinetics --- every reaction with
$\nu_s^{(\mathbf{u})}, \nu_s^{(\mathbf{v})} \in \{0,1\}$ for
all $s$ --- the factor reduces to $1$ and the condition takes
its familiar form
$\kappa_r/\kappa_{r^\dagger} = K_\mathrm{eq}(r,T)$.
The S$_\mathrm{N}$2 example below is of this kind.
For higher-order reactions like $2A \rightleftharpoons B$
(factor $2$), the factor appears in the binomial-stochastic
Wegscheider statement; in the macroscopic chemical convention
$k_r^\mathrm{macro} = \kappa_r \prod_s \nu_s^{(\mathbf{u})}!$
(Remark~\ref{rem:propensity-conventions}), this absorbs into
the rate-constant definition and the familiar form
$k_r^\mathrm{macro}/k_{r^\dagger}^\mathrm{macro} =
K_\mathrm{eq}(r,T)$ is recovered.
\end{remark}

\begin{definition}[Layer~2 for $\FP$: kinetic Wegscheider
  condition {\cite{Wegscheider1901, HornJackson1972,
  Feinberg1989}}]
\label{def:L3-layer2}
  The functor $\FP$ satisfies \emph{Layer~2} if it is
  $\dagger$-compatible with $\Lk_2$: for every reversible
  chemical generator pair $(r, r^\dagger)$,
  \[
    \frac{\kappa_r}{\kappa_{r^\dagger}}
    \;=\; K_\mathrm{eq}(r,\, T)
      \prod_s \frac{\nu_s^{(\mathbf{u})}!}{\nu_s^{(\mathbf{v})}!}
    \;=\; \exp\!\left(\frac{-F_G^T(r)}{RT}\right)
      \prod_s \frac{\nu_s^{(\mathbf{u})}!}{\nu_s^{(\mathbf{v})}!},
  \]
  with $\kappa_r$ the binomial-form stochastic rate constants
  of Definition~\ref{def:propensity}.
  This is the \emph{kinetic Wegscheider condition}, introduced
  by Wegscheider \cite{Wegscheider1901} and analysed within
  CRNT by Horn and Jackson \cite{HornJackson1972} and Feinberg
  \cite{Feinberg1989} (in the macroscopic concentration form
  $k_r^\mathrm{macro}/k_{r^\dagger}^\mathrm{macro} =
  K_\mathrm{eq}(r,T)$, which absorbs the combinatorial factor;
  see Remarks~\ref{rem:propensity-conventions} and
  ~\ref{rem:combinatorial-factor}).
  It is the algebraic form of the detailed balance condition
  for the assembled CME generator $\Omega$.
\end{definition}

\begin{mathbox}[Layer~2 as $\dagger$-compatibility across levels]
The Layer~2 conditions at each level of the tower share a
common categorical form: they lock the new functor to the
structure present one or two levels below.
\begin{itemize}
  \item \textbf{Layer~2 at $\Lk_1$}: the coboundary condition
    $\FH = \delta^0 h_f$ is a condition \emph{within} $\Lk_1$,
    requiring $\FH$ values to be consistent around closed loops.
  \item \textbf{Layer~2 at $\Lk_2$}: the Wegscheider cycle
    condition $\prod_i K_\mathrm{eq}(r_i, T) = 1$ is a
    condition within $\Lk_2$, forced by the $\dagger$-SMC
    structure linking $\FH$ and $\FS$.
  \item \textbf{Layer~2 at $\Lk_3$}: the kinetic Wegscheider
    condition $k_r/k_{r^\dagger} = K_\mathrm{eq}(r,T)$ is
    the \emph{first inter-level} coherence condition: it
    requires the $\Lk_3$ datum $\FP$ to be compatible with
    the $\Lk_2$ derived quantity $F_G^T$.
\end{itemize}
The categorical formulation makes the pattern visible: each
Layer~2 is a natural transformation condition, requiring the
relevant functor to commute with the $\dagger$-structure
already present in the tower.
At $\Lk_3$ this commutativity is between $\FP$
(the stochastic morphism) and $F_G^T$ (the free-energy
morphism), and it expresses thermodynamic consistency: the
equilibrium is reached at the same concentration ratio
whether approached via kinetics or thermodynamics.
\end{mathbox}

\begin{proposition}[Layer~2 halves the parameter space]
\label{prop:L3-param-reduction}
  For a reversible Petri net with $|\Rx| = 2m$ chemical
  generators (forward and reverse pairs):
  \begin{itemize}
    \item \textbf{Layer~1}: $2m$ free positive reals
      ($\kappa_r$ for each of the $2m$ generators, chosen
      independently).
    \item \textbf{Layer~2}: $m$ free positive reals
      (forward constants $\{\kappa_r\}_{r\;\mathrm{fwd}}$
      only; each reverse constant determined by
      $\kappa_{r^\dagger} = \kappa_r
      \prod_s \nu_s^{(\mathbf{v})}!/\nu_s^{(\mathbf{u})}! /
      K_\mathrm{eq}(r,T)$,
      where $K_\mathrm{eq}(r,T)$ is an $\Lk_2$ datum).
  \end{itemize}
  The reduction from $2m$ to $m$ free parameters is achieved by
  the $m$ kinetic Wegscheider conditions, one per reversible
  pair. The $\Lk_2$ data $\{K_\mathrm{eq}(r,T)\}$ thus
  geometrically constrain the $\Lk_3$ parameter space to an
  $m$-dimensional positive orthant inside the $2m$-dimensional
  Layer~1 parameter space.
\end{proposition}

\subsubsection{Cycle form of the kinetic Wegscheider condition}

The pairwise condition of Definition~\ref{def:L3-layer2}
extends to all closed loops in the reaction graph --- this is
the classical content of Wegscheider's original 1901 result
\cite{Wegscheider1901}, and it is the condition used in
modern CRNT to characterise detailed-balanced networks
\cite{HornJackson1972, Feinberg1989, Feinberg2019}.
In the tower language, a closed loop is a composable sequence
of generators $r_1, \ldots, r_n \in \Rx$ in $\Lk_0(P)$
such that $r_n \circ \cdots \circ r_1 = \mathrm{id}_\mathbf{u}$
for some complex $\mathbf{u}$; equivalently, it is a cycle
in the directed graph $G(\Rx)$.
The cycle form of Layer~2 is then a direct consequence of
applying the functoriality of $F_G^T$ to such a cycle,
combined with Definition~\ref{def:L3-layer2}.

\begin{proposition}[Cycle form of Layer~2
  {\cite{Wegscheider1901, HornJackson1972, Feinberg1989}}]
\label{prop:kinetic-wegscheider}
  Suppose $\FP$ satisfies Layer~2.
  Then for every directed closed loop
  $r_1, \ldots, r_n$ in $G(\Rx)$ (a composable cycle in
  $\Lk_0(P)$):
  \[
    \prod_{i=1}^n \kappa_{r_i}
    \;=\; \prod_{i=1}^n \kappa_{r_i^\dagger}.
  \]
\end{proposition}

\begin{proof}
Apply Definition~\ref{def:L3-layer2} to each pair
$(r_i, r_i^\dagger)$:
\[
  \prod_{i=1}^n \frac{\kappa_{r_i}}{\kappa_{r_i^\dagger}}
  \;=\; \prod_{i=1}^n K_\mathrm{eq}(r_i, T)
  \cdot \prod_{i=1}^n \prod_s
    \frac{\nu_s^{(\mathbf{u}_i)}!}{\nu_s^{(\mathbf{v}_i)}!}.
\]
The combinatorial factor telescopes around the closed loop:
since the source of $r_{i+1}$ is the target of $r_i$, we have
$\nu_s^{(\mathbf{u}_{i+1})} = \nu_s^{(\mathbf{v}_i)}$, and the
closing condition $\mathbf{u}_1 = \mathbf{v}_n$ gives
$\prod_i \nu_s^{(\mathbf{u}_i)}!/\nu_s^{(\mathbf{v}_i)}! = 1$
for each species $s$.
The remaining $K_\mathrm{eq}$ product satisfies
\[
  \prod_{i=1}^n K_\mathrm{eq}(r_i, T)
  \;=\; \exp\!\left(\frac{-1}{RT}\sum_{i=1}^n F_G^T(r_i)\right),
\]
and $\sum_i F_G^T(r_i)$ is the value of the strict SMC functor
$F_G^T : \Lk_0(P) \to B\RR$ on the closed-loop endomorphism
$r_n \circ \cdots \circ r_1$
(Proposition~\ref{prop:gibbs-functor}), which vanishes by the
$\Lk_2$ Wegscheider conditions
(Proposition~\ref{prop:wegscheider1}).
Therefore $\prod_i \kappa_{r_i}/\kappa_{r_i^\dagger} = e^0 \cdot 1 = 1$.
\end{proof}

\begin{remark}[Irreversible cycles are forbidden]
\label{rem:irreversible-cycles}
Proposition~\ref{prop:kinetic-wegscheider} is the precise
categorical reason why irreversible directed cycles
$A_1 \to A_2 \to \cdots \to A_n \to A_1$ are
thermodynamically forbidden in a closed system at equilibrium:
they correspond to a closed loop in $G(\Rx)$ with
$\kappa_{r_i^\dagger} = 0$ for some $i$, which would require
$\prod \kappa_{r_i} = 0$ --- impossible since all
$\kappa_{r_i} \in \RR_{>0}$.
Wegscheider \cite{Wegscheider1901} identified this
as the origin of the conditions bearing his name;
Horn and Jackson \cite{HornJackson1972} established
that detailed balance (i.e.\ Layer~2 for $\FP$) is
equivalent to complex balance plus the cycle conditions;
and Feinberg \cite{Feinberg1989} provided necessary and
sufficient conditions for a mass-action network to satisfy
detailed balance.
In the tower, all of these statements are consequences of
the single requirement that $\FP$ is $\dagger$-compatible
with $\Lk_2$.
\end{remark}

\begin{chembox}[Two routes to $K_\mathrm{eq}$: an example]
The $\Lk_2$--$\Lk_3$ interface provides two independent
computations of the equilibrium constant, and Layer~2 demands
they agree.
We work in the macroscopic chemical convention
$k_r^\mathrm{macro}/k_{r^\dagger}^\mathrm{macro} =
K_\mathrm{eq}(r,T)$
(Remark~\ref{rem:propensity-conventions}); the SN2 reaction is
bimolecular forward and bimolecular reverse, so the
combinatorial factor $\prod_s \nu_s^{(\mathbf{u})}!/
\nu_s^{(\mathbf{v})}! = 1!\cdot 1!/(1!\cdot 1!) = 1$ and the
binomial-form Wegscheider condition reduces to the macroscopic
form for this reaction.
\begin{itemize}
  \item \textbf{Thermodynamic route ($\Lk_2$)}:
    $K_\mathrm{eq} = \exp(-F_G^T(r)/RT)$,
    computed from the standard enthalpy
    $\FH(r) = -75.0\;\mathrm{kJ\,mol^{-1}}$ and entropy
    $\FS(r) = -90\;\mathrm{J\,mol^{-1}K^{-1}}$
    of $\mathrm{CH_3Cl + OH^- \to CH_3OH + Cl^-}$
    \cite{NIST_WebBook}:
    $F_G^{298\,\mathrm{K}}(r) = -48.2\;\mathrm{kJ\,mol^{-1}}$,
    giving $K_\mathrm{eq}(298\,\mathrm{K}) \approx 2.8\times 10^8$.
  \item \textbf{Kinetic route ($\Lk_3$)}:
    $K_\mathrm{eq} = k_r^\mathrm{macro} /
    k_{r^\dagger}^\mathrm{macro}$, where
    $k_r^\mathrm{macro} \approx 6\times 10^{-6}\;
    \mathrm{M^{-1}s^{-1}}$ is the measured forward
    macroscopic rate constant for this nucleophilic
    substitution \cite{AtkinsDeP2014, March1992}.
    Layer~2 then forces the reverse rate constant to be
    $k_{r^\dagger}^\mathrm{macro}
    = k_r^\mathrm{macro} / K_\mathrm{eq}
    \approx 2.1\times 10^{-14}\;\mathrm{M^{-1}s^{-1}}$,
    a value far below the practical detection limit of any
    direct measurement.
\end{itemize}
This illustrates a powerful consequence of Layer~2: the
reverse rate constant need not be measured independently.
It is \emph{determined} by the forward rate constant and the
$\Lk_2$ thermodynamic data, via the $\dagger$-compatibility
condition.
The consistency check
$k_r^\mathrm{macro}/k_{r^\dagger}^\mathrm{macro}
= K_\mathrm{eq}$ is simultaneously an experimental test of
thermodynamic self-consistency and a demonstration that the
tower's inter-level structure has real predictive content:
an $\Lk_3$ measurement ($k_r^\mathrm{macro}$) plus an $\Lk_2$
computation ($K_\mathrm{eq}$) together determine an $\Lk_3$
datum ($k_{r^\dagger}^\mathrm{macro}$) that would otherwise
require an independent experiment.
\end{chembox}

%% file: chapters/L3/l3_dzt.tex
\subsection{The Deficiency Zero Theorem: level stratification}
\label{sec:L3-dzt}

The Deficiency Zero Theorem (DZT) is the central result of
classical CRNT \cite{Horn1972, Feinberg1987}.
Within the tower, it has a specific structural role: its
hypotheses belong entirely to $\Lk_0$, while its conclusion is
an $\Lk_3$ statement holding uniformly across every choice of
$\FP$ over a fixed $\Lk_0$ structure.
This section analyses the $\Lk_0 \to \Lk_3$ transition that
makes such a quantification possible, and relates it to the
forgetful map $U_3 : \Lk_3(P) \to \Lk_0(P)$.

The DZT hypotheses --- weak reversibility and deficiency
$\delta = 0$ --- were stated in
Definition~\ref{def:feinberg} as properties of the Petri net
$P$ alone.
Within the tower, this places them squarely at $\Lk_0$:
the complex count $n$, the linkage-class count $\ell$, the
stoichiometric rank $s$, the deficiency $\delta = n - \ell - s$,
and the weak-reversibility condition on the directed reaction
graph $G(\Rx)$ are all invariants of $\Lk_0(P)$, computable
without reference to rate constants, enthalpies, or entropies.

The conclusion of the DZT --- existence, uniqueness, and
asymptotic stability of a positive steady state --- has no
analogue at $\Lk_0$: there is no notion of dynamics, let alone
a steady state, without the rate functor $\FP$.
The new content at $\Lk_3$ is therefore entirely on the
conclusion side.
Specifically, the DZT makes a statement not about one
particular $\FP$, but about the \emph{entire fiber} of the
forgetful map $U_3 : \Lk_3(P) \to \Lk_0(P)$ over a fixed
$\Lk_0$ structure:
\[
  U_3^{-1}(\Lk_0(P))
  \;\cong\; \mathrm{Map}(\Rx,\, \RR_{>0}),
\]
the space of all positive rate-constant assignments.

\subsubsection{The \texorpdfstring{$\Lk_0 \to \Lk_3$}{L0->L3} jump: forgetful fiber and  the automorphism picture}

The above remarks implies that the DZT makes a statement not about one particular
$\FP$, but about the \emph{entire fiber} of the forgetful
functor $U_3 : \Lk_3(P) \to \Lk_0(P)$ over a fixed $\Lk_0$
structure.

The functor $U_3$ strips the rate constants: its fiber over
$\Lk_0(P)$ is the set of all valid $\FP$ assignments,
\[
  U_3^{-1}(\Lk_0(P))
  \;\cong\; \mathrm{Map}(\Rx,\, \RR_{>0}),
\]
the space of positive rate constants --- a copy of
$\RR_{>0}^{|\Rx|}$ for each Petri net $P$.
The DZT says: for Petri nets $P$ with $\delta = 0$ and weak
reversibility, the large-volume image
$\Phi_\infty(\Omega)$ (Proposition~\ref{prop:LMA}) of
\emph{every} point in this fiber has a unique positive fixed
point in each stoichiometric class.
The following diagram makes the cross-level structure explicit:

\[
\includegraphics{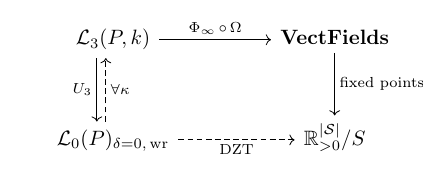}
\]

\noindent
\noindent
Reading the diagram: the upper row sends an $\Lk_3$ structure
$(P, \kappa)$ to the large-volume vector field
$\Phi_\infty(\Omega)$ assembled from the rate functor
$\FP$ (Proposition~\ref{prop:LMA}).
The left column is the forgetful operation $U_3$ extracting
the underlying $\Lk_0$ structure
(Definition~\ref{def:L3}).
The DZT (dashed arrow) says that when $\Lk_0(P)$ satisfies
$\delta = 0$ and weak reversibility, this vector field has a
unique positive fixed point in each stoichiometric class
$\RR_{>0}^{|\Sp|}/S$ --- a conclusion that holds uniformly
across the fiber $U_3^{-1}(\Lk_0(P)) \cong
\mathrm{Map}(\Rx, \RR_{>0})$, i.e.\ for every Layer-1 choice
$\kappa \in \mathrm{Map}(\Rx, \RR_{>0})$.
The jump from $\Lk_0$ data (bottom row) to $\Lk_3$ conclusion
(top row) is therefore tame in this technical sense: the DZT
factors through $U_3$ on the hypothesis side, and the
conclusion is independent of which fiber point $\kappa$ is chosen.

\begin{remark}[The DZT and the automorphism sequence]
\label{rem:DZT-aut}
In the automorphism sequence at $\Lk_3$
\[
  1 \;\to\; \ker\varphi_3
  \;\to\; \Aut(\Lk_3(P))
  \;\xrightarrow{\;\varphi_3\;}
  \Aut(\Lk_2(P))
  \;\to\; \coker(\varphi_3)
  \;\to\; 1
\]
(with $\coker(\varphi_3) = \Aut(\Lk_2(P))/\im(\varphi_3)$ as a
pointed-set quotient, \S\ref{sec:aut-exact}), elements of
$\coker(\varphi_3)$ are the rate-constant swaps that are
automorphisms of $\Lk_2$ (they preserve $\FH$, $\FS$, and
$F_G^T$) but not of $\Lk_3$.
For a general network, such swaps can in principle map a
monostable system to a multistable one, or change the number
of steady states in a stoichiometric class.
The DZT rules this out for $\delta = 0$ weakly reversible
networks: since every rate assignment $\kappa$ gives exactly
one positive steady state per class, no element of
$\coker(\varphi_3)$ can alter the steady-state count.
Equivalently, the ``steady-state section''
$\kappa \mapsto \mathbf{c}^*(\kappa)$ is a well-defined
continuous map on the entire fiber
$\mathrm{Map}(\Rx, \RR_{>0})$, invariant under any
automorphism that permutes rate constants within the fiber.
The DZT is thus, from the tower's perspective, a theorem about
the \emph{tameness} of the $\Lk_0 \to \Lk_3$ jump for special
networks: the $\Lk_3$ structure (any choice of $\FP$) cannot
introduce dynamical complexity that the $\Lk_0$ structure
alone forbids.
\end{remark}

\begin{theorem}[Deficiency Zero Theorem
  {\cite{Horn1972, HornJackson1972, Feinberg1987}}]
\label{thm:DZT}
  Let $\Lk_0(P)$ be weakly reversible with deficiency
  $\delta = 0$ (Definition~\ref{def:feinberg}).
  Then for every Layer-1 rate functor $\FP$
  --- equivalently, for every point $\kappa$ in the fiber
  $U_3^{-1}(\Lk_0(P)) \cong \mathrm{Map}(\Rx, \RR_{>0})$ ---
  the large-volume vector field
  $\Phi_\infty(\Omega) = N\mathbf{v}$
  (Proposition~\ref{prop:LMA}) satisfies:
  \begin{enumerate}[label=(\roman*)]
    \item \textup{(Existence and uniqueness)} In each
      positive stoichiometric compatibility class
      $(\mathbf{c}_0 + S) \cap \RR_{>0}^{|\Sp|}$,
      $\Phi_\infty(\Omega)$ has a unique positive fixed
      point $\mathbf{c}^*(\kappa)$.
    \item \textup{(Asymptotic stability within the positive class)}
      This fixed point is locally asymptotically stable, and
      every trajectory of
      $\dot{\mathbf{c}} = \Phi_\infty(\Omega)$ initiated in
      the open positive stoichiometric compatibility class
      converges to $\mathbf{c}^*(\kappa)$ as $t \to \infty$.
      The function
      $V(\mathbf{c}) = \sum_{s \in \Sp}
      \bigl(c_s \ln(c_s/c_s^*) - c_s + c_s^*\bigr)$
      is a strict Lyapunov function for
      $\Phi_\infty(\Omega)$ on the positive class
      \cite{HornJackson1972}.
      The behaviour of trajectories initiated on the boundary
      $\partial \RR_{\geq 0}^{|\Sp|}$ is not asserted by the
      DZT; that question --- whether boundary trajectories
      also converge to $\mathbf{c}^*$ rather than to a
      boundary equilibrium --- is the persistence content of
      the Global Attractor Conjecture
      \cite{CraciunEtAl2009}, which is
      established for many subclasses but remains open in
      full generality.
    \item \textup{(Complex balance)} At $\mathbf{c}^*$, the
      assembled CME generator $\Omega$ is complex-balanced:
      for each complex $\mathbf{u} \in \Cx$,
      \[
        \sum_{r:\,\mathbf{u}\to\mathbf{v}} \kappa_r\,
        (\mathbf{c}^*)^{\boldsymbol{\nu}^{(\mathbf{u})}}
        \;=\;
        \sum_{r:\,\mathbf{w}\to\mathbf{u}} \kappa_r\,
        (\mathbf{c}^*)^{\boldsymbol{\nu}^{(\mathbf{w})}}.
      \]
  \end{enumerate}
\end{theorem}

\begin{remark}[Reach of the tower]
\label{rem:tower-reach-DZT}
  Theorem~\ref{thm:DZT} is not proved within the $\Lk_0$--$\Lk_3$
  framework of this monograph.
  The existence and uniqueness statement (i) rests on
  Birch's theorem and a convex-analytic argument on
  $\RR_{>0}^{|\Sp|}$ \cite{HornJackson1972, Feinberg1987},
  structures that live transversally to the tower rather than at
  any single level.
  The tower's contribution is different: it fixes the
  \emph{level grammar} of the theorem (hypotheses at $\Lk_0$,
  conclusion quantified uniformly over the $\Lk_3$ fiber
  $U_3^{-1}(\Lk_0(P))$), and it proves the pieces of the
  theorem that do not require this external input.
  The next proposition makes the tower-native content precise.
\end{remark}

\begin{proposition}[Tower-native partial DZT]
\label{prop:tower-DZT}
  Let $\Lk_0(P)$ be weakly reversible with deficiency
  $\delta = 0$, and suppose $\Lk_3(P)$ admits a
  complex-balanced fixed point
  $\mathbf{c}^* \in \RR_{>0}^{|\Sp|}$
  --- that is, a positive concentration at which the
  assembled CME generator $\Omega$ satisfies the
  complex-balance equations of
  Theorem~\ref{thm:DZT}\textup{(iii)}.
  Then:
  \begin{enumerate}[label=(\roman*)]
    \item \textup{(Fixed point)}
      $\mathbf{c}^*$ is a zero of
      $\Phi_\infty(\Omega) = N\mathbf{v}$.
    \item \textup{(Lyapunov descent)}
      The pseudo-Helmholtz function
      $V(\mathbf{c}) = \sum_{s \in \Sp}
      \bigl(c_s \ln(c_s/c_s^*) - c_s + c_s^*\bigr)$
      is a strict Lyapunov function for
      $\Phi_\infty(\Omega)$ on the positive stoichiometric
      class containing $\mathbf{c}^*$.
    \item \textup{(Stochastic lift under Layer~2)}
  Suppose additionally that the Petri net is reversible,
  $\Lk_3(P)$ satisfies Layer~2 (kinetic Wegscheider,
  Definition~\ref{def:L3-layer2}), and $\mathbf{c}^* \in \mathcal{E}_T$.
  Then the Poisson product distribution
  $\pi^*(\mathbf{x}) \propto
  \prod_{s \in \Sp} (c_s^*)^{x_s}/x_s!$
  satisfies detailed balance for the assembled CME generator
  $\Omega$ and is a stationary distribution of the CTMC.
  For first-order kinetics, this additional hypothesis is
  automatic: macroscopic complex balance combined with
  reversibility and Layer~2 implies
  $\mathbf{c}^* \in \mathcal{E}_T$, since the combinatorial
  factor in Layer~2 reduces to $1$
  (Remark~\ref{rem:combinatorial-factor}).
  \end{enumerate}
\end{proposition}

\begin{proof}[Proof of Proposition~\ref{prop:tower-DZT}]
  Throughout, write $\mathbf{v}(\mathbf{c}) \in
  \RR_{>0}^{|\Rx|}$ for the mass-action rate vector with
  components $v_r(\mathbf{c}) = \kappa_r \prod_s
  c_s^{\nu_s^{(\mathbf{u}_r)}}$, so
  $\Phi_\infty(\Omega) = N\mathbf{v}$
  (Proposition~\ref{prop:LMA}).
  For a reaction $r : \mathbf{u}_r \to \mathbf{v}_r$, let
  $y(r) := \mathbf{u}_r \in \Cx$ and
  $y'(r) := \mathbf{v}_r \in \Cx$ denote its source and target
  complexes.

  \smallskip
  \noindent\textbf{(i) Fixed point.}
  The stoichiometric matrix $N$ factors as
  $N = Y \Delta$, where $Y : \RR^{\Cx} \to \RR^{|\Sp|}$ is the
  $\Lk_0$ linear map sending a complex
  $\mathbf{u} = \sum_s \nu_s^{(\mathbf{u})} s$ to its species
  vector $\boldsymbol{\nu}^{(\mathbf{u})}$, and
  $\Delta : \RR^{|\Rx|} \to \RR^{\Cx}$ is the reaction-graph
  incidence map $\Delta(e_r) = e_{y'(r)} - e_{y(r)}$
  (Definition~\ref{def:N}).
  Hence
  \[
    (\Phi_\infty(\Omega))(\mathbf{c}^*)
    \;=\; N\mathbf{v}(\mathbf{c}^*)
    \;=\; Y \,\Delta\, \mathbf{v}(\mathbf{c}^*).
  \]
  It suffices to show $\Delta \mathbf{v}(\mathbf{c}^*) = 0$
  in $\RR^{\Cx}$.
  The $\mathbf{u}$-component of $\Delta \mathbf{v}(\mathbf{c}^*)$
  is
  \[
    \bigl(\Delta \mathbf{v}(\mathbf{c}^*)\bigr)_{\mathbf{u}}
    \;=\;
    \sum_{r:\,y'(r)=\mathbf{u}} v_r(\mathbf{c}^*)
    \;-\;
    \sum_{r:\,y(r)=\mathbf{u}} v_r(\mathbf{c}^*),
  \]
  which is the difference of total rate flowing \emph{into}
  $\mathbf{u}$ and total rate flowing \emph{out of}
  $\mathbf{u}$ under mass-action kinetics at $\mathbf{c}^*$.
  Complex balance at $\mathbf{c}^*$ is precisely the vanishing
  of this difference for every $\mathbf{u} \in \Cx$, so
  $\Delta \mathbf{v}(\mathbf{c}^*) = 0$ and therefore
  $(\Phi_\infty(\Omega))(\mathbf{c}^*) = 0$.

  \smallskip
  \noindent\textbf{(ii) Lyapunov descent.}
  The function
  $V(\mathbf{c}) = \sum_s (c_s \ln(c_s/c_s^*) - c_s + c_s^*)$
  is smooth and strictly convex on $\RR_{>0}^{|\Sp|}$, with
  $\nabla V(\mathbf{c})_s = \ln(c_s / c_s^*)$ and
  $V(\mathbf{c}^*) = 0$, $V(\mathbf{c}) > 0$ for
  $\mathbf{c} \neq \mathbf{c}^*$ in the positive class.
  Along trajectories of $\dot{\mathbf{c}} = N\mathbf{v}$,
  \[
    \dot V(\mathbf{c})
    \;=\; \nabla V(\mathbf{c}) \cdot N \mathbf{v}(\mathbf{c})
    \;=\; \sum_{r \in \Rx}
    v_r(\mathbf{c})\,
    \ln\!\Bigl(\tfrac{\mathbf{c}}{\mathbf{c}^*}\Bigr)^{\mathbf{n}_r},
  \]
  using $\mathbf{n}_r = \boldsymbol{\nu}^{(y'(r))} -
  \boldsymbol{\nu}^{(y(r))}$.
  Write $\xi_{\mathbf{u}} := \ln\bigl((\mathbf{c}/\mathbf{c}^*)
  ^{\boldsymbol{\nu}^{(\mathbf{u})}}\bigr)$ for each complex
  $\mathbf{u}$ and
  $w_r := v_r(\mathbf{c}^*)\,
  e^{\,\xi_{y(r)}} = v_r(\mathbf{c})$
  (the last equality uses
  $v_r(\mathbf{c}) = v_r(\mathbf{c}^*)
  (\mathbf{c}/\mathbf{c}^*)^{\boldsymbol{\nu}^{(y(r))}}$).
  Substituting,
  \[
    \dot V
    \;=\; \sum_r w_r \bigl(\xi_{y'(r)} - \xi_{y(r)}\bigr)
    \;=\; -\,\xi^\top \Delta \mathbf{w}.
  \]
  Complex balance at $\mathbf{c}^*$ is the statement that the
  weighted Laplacian $L$ of the reaction graph (with edge
  weights $v_r(\mathbf{c}^*)$) has $\mathbf{1}$ in its kernel;
  the general inequality
  $\xi^\top \Delta \mathbf{w} \geq 0$
  with equality iff $\xi$ is constant on linkage classes is the
  standard log-sum inequality applied to the incidence structure
  of the reaction graph \cite{HornJackson1972}.
  Hence $\dot V \leq 0$, with equality iff
  $\xi_{y(r)} = \xi_{y'(r)}$ for every $r$, which for weakly
  reversible networks forces $\xi$ constant on each linkage
  class; combined with $\delta = 0$ and the definition of $Y$,
  this pins $\xi$ to zero on the stoichiometric class, hence
  $\mathbf{c} = \mathbf{c}^*$.

  \smallskip
  \noindent\textbf{(iii) Stochastic lift under Layer 2.}
  By Definition~\ref{def:L3-layer2}, reversibility and Layer~2
give $\kappa_r / \kappa_{r^\dagger}
= K_\mathrm{eq}(r, T) \prod_s \nu_s^{(\mathbf{u})}!/\nu_s^{(\mathbf{v})}!
= \exp(-F_G^T(r)/RT) \prod_s \nu_s^{(\mathbf{u})}!/\nu_s^{(\mathbf{v})}!$
for every reversible pair $(r, r^\dagger)$.
For first-order reactions, the combinatorial factor reduces
to $1$ (Remark~\ref{rem:combinatorial-factor}).
  By the Layer-2 detailed-balance derivation of
  Section~\ref{sec:L3-layer2}, the Poisson product
  $\pi^*(\mathbf{x}) \propto
  \prod_s (c_s^*)^{x_s}/x_s!$
  with $\mathbf{c}^* \in \mathcal{E}_T$ satisfies pointwise
  detailed balance for every reversible generator pair:
  \[
    \pi^*(\mathbf{x})\,\lambda_r(\mathbf{x})
    \;=\;
    \pi^*(\mathbf{x} + \mathbf{n}_r)\,
    \lambda_{r^\dagger}(\mathbf{x} + \mathbf{n}_r).
  \]
  Complex balance at $\mathbf{c}^*$ in the reversible setting
  reduces to $\mathbf{c}^* \in \mathcal{E}_T$ pairwise across
  every reversible generator pair, so the hypothesis of
  the proposition is compatible with the Layer-2 condition.
  Summing pointwise detailed balance over $r \in \Rx$
  and using that the Markov kernel gains and losses cancel
  pairwise gives $\Omega^\top \pi^* = 0$; hence $\pi^*$ is
  stationary for the CTMC generated by $\Omega$.
\end{proof}

Proposition~\ref{prop:tower-DZT} captures the largest fragment
of the DZT that the tower proves autonomously: granted a
complex-balanced fixed point, the deterministic fixed-point and
Lyapunov descent follow from $\Lk_0$ structure alone, while the
stochastic stationarity of (iii) is a specifically tower-theoretic
route --- via reversibility and Layer~2 --- to a special case of
Anderson--Craciun--Kurtz (Theorem~\ref{thm:ACK}).
Existence of the complex-balanced $\mathbf{c}^*$ itself remains
outside the tower's reach (Remark~\ref{rem:tower-reach-DZT}).

\begin{mathbox}[Level stratification of the DZT]
\renewcommand{\arraystretch}{1.4}
\begin{center}
\begin{tabular}{p{4.2cm} p{2.8cm} p{5.0cm}}
  \hline
  \textbf{Ingredient} & \textbf{Level} &
  \textbf{Where it lives}\\
  \hline
  $n$, $\ell$, $s$,
  $\delta = n{-}\ell{-}s$ &
  $\Lk_0$ &
  Deficiency formula, Section~\ref{sec:L0}\\[2pt]
  Weak reversibility of $G(\Rx)$ &
  $\Lk_0$ &
  Directed graph of $\Lk_0(P)$\\[2pt]
  Quantifier over $k$ &
  $\Lk_3$ &
  Fiber $U_3^{-1}(\Lk_0(P))$\\[2pt]
  Unique positive fixed
  point $\mathbf{c}^*$ &
  $\Lk_3$ &
  Zeros of $\Phi_\infty(\Omega)$ \\[2pt]
  Asymptotic stability &
  $\Lk_3$ &
  Lyapunov $V$ on positive class \\[2pt]
  Complex balance &
  $\Lk_3$ &
  Assembled CME generator $\Omega$ \\
  \hline
\end{tabular}
\end{center}

\medskip\noindent
The DZT is not formulable at $\Lk_2$: thermodynamics fixes
$K_\mathrm{eq}(r, T) = \prod_s (c_s^*)^{n_{r,s}}$ for each
generator, i.e.\ the equilibrium ratios, but locating the
individual $\mathbf{c}^*$ requires solving
$N\mathbf{v}(\mathbf{c}^*) = \mathbf{0}$ with the full
$\Lk_3$ rate functor $\FP$.
Under Layer~2, the Wegscheider coboundary pins down the rate
ratios $\kappa_r/\kappa_{r^\dagger}$ from $\Lk_2$ data, leaving only an
overall timescale free in $\Lk_3$; the position of
$\mathbf{c}^*$ within the equilibrium locus is then determined.
\end{mathbox}

\subsubsection{The stochastic DZT: Anderson--Craciun--Kurtz}

Theorem~\ref{thm:DZT} and the tower-native fragment of
Proposition~\ref{prop:tower-DZT} are both statements about the
large-volume limit $\Phi_\infty(\Omega)$: they concern the
vector field obtained by sending the stochastic system to its
deterministic mean-field RRE.
Proposition~\ref{prop:tower-DZT}(iii) gave a partial stochastic
counterpart under reversibility and Layer~2.
A natural question within the tower is whether the same
$\Lk_0$ hypotheses --- $\delta = 0$ and weak reversibility,
without a Layer-2 requirement --- enforce a canonical form for
the stationary distribution of the full stochastic system on
the assembled CME generator $\Omega$ directly, before any
large-volume limit.
The answer is yes: the Anderson--Craciun--Kurtz theorem
\cite{AndersonCraciunKurtz2010} establishes that $\Omega$ has a
stationary distribution of product-of-Poissons form, with
Poisson parameters given by the complex-balanced equilibrium
$\mathbf{c}^*(k)$ of Theorem~\ref{thm:DZT}\textup{(iii)}.
This is the stochastic analogue of the DZT: the same $\Lk_0$
hypotheses, now applied to the full $\Lk_3$ object $\Omega$
rather than its large-volume shadow, and --- unlike
Proposition~\ref{prop:tower-DZT}(iii) --- without requiring
reversibility.

\begin{theorem}[Anderson--Craciun--Kurtz
  {\cite{AndersonCraciunKurtz2010}}]
\label{thm:ACK}
  Let $\Lk_3(P)$ admit a positive concentration
  $\mathbf{c}^* \in \RR_{>0}^{|\Sp|}$ satisfying the
  \emph{stochastic complex-balance condition}: for every
  complex $\mathbf{u} \in \Cx$,
  \[
    \sum_{r:\,y(r) = \mathbf{u}}
      \frac{\kappa_r\,(\mathbf{c}^*)^{\boldsymbol{\nu}^{(\mathbf{u})}}}
      {\prod_s \nu_s^{(\mathbf{u})}!}
    \;=\;
    \sum_{r:\,y'(r) = \mathbf{u}}
      \frac{\kappa_r\,(\mathbf{c}^*)^{\boldsymbol{\nu}^{(y(r))}}}
      {\prod_s \nu_s^{(y(r))}!}.
  \]
  Then the assembled CME generator $\Omega$ has a stationary
  distribution on each closed irreducible subset of
  $\NN^{|\Sp|}$ given by the product-of-Poissons form
  \[
    \pi(\mathbf{x})
    \;=\; \frac{1}{Z}
    \prod_{s \in \Sp}
    \frac{(c_s^*)^{x_s}}{x_s!}.
  \]
  In particular, if $\Lk_0(P)$ is weakly reversible with
  deficiency $\delta = 0$, then a positive macroscopic
  complex-balanced $\mathbf{c}^*$ exists by
  Theorem~\ref{thm:DZT}\textup{(iii)} for every Layer-1 rate
  functor $\FP$, and the rescaling
  $\kappa_r \mapsto \kappa_r/\prod_s \nu_s^{(y(r))}!$ converts macroscopic
  complex balance to stochastic complex balance.
  Hence the stationary distribution above exists on each closed
  irreducible subset.
\end{theorem}

\begin{proof}
  The argument is the Anderson--Craciun--Kurtz proof
  \cite{AndersonCraciunKurtz2010} adapted to the binomial-form
  propensity of Definition~\ref{def:propensity}; the
  combinatorial factors $\prod_s \nu_s^{(\mathbf{u})}!$ that the
  falling-factorial or macroscopic conventions absorb into
  rate constants (Remark~\ref{rem:propensity-conventions})
  are tracked explicitly below.
  Fix a state $\mathbf{x} \in \NN^{|\Sp|}$ to directly verify
  $(\Omega^\top \pi)(\mathbf{x}) = 0$.
  For each complex $\mathbf{u} \in \Cx$, define the
  \emph{occupancy function}
  \[
    q_\mathbf{u}(\mathbf{x})
    \;:=\;
    \prod_{s \in \Sp} \binom{x_s}{\nu_s^{(\mathbf{u})}}.
  \]
  With binomial-form mass-action propensities
  $\lambda_r(\mathbf{x}) = \kappa_r \prod_s
  \binom{x_s}{\nu_s^{(y(r))}}$
  (Definition~\ref{def:propensity}), the propensity factors
  through the source complex:
  $\lambda_r(\mathbf{x}) = \kappa_r\, q_{y(r)}(\mathbf{x})$.

  \smallskip
  \noindent\textbf{Poisson--occupancy identity.}
  The Poisson product $\pi(\mathbf{x}) \propto
  \prod_s (c_s^*)^{x_s}/x_s!$ satisfies, for every complex
  $\mathbf{u} \in \Cx$ and every state $\mathbf{x}$ with
  $\mathbf{x} \geq \boldsymbol{\nu}^{(\mathbf{u})}$,
  \[
    q_\mathbf{u}(\mathbf{x})\, \pi(\mathbf{x})
    \;=\;
    \frac{(\mathbf{c}^*)^{\boldsymbol{\nu}^{(\mathbf{u})}}}
      {\prod_s \nu_s^{(\mathbf{u})}!}\,
    \pi\bigl(\mathbf{x} - \boldsymbol{\nu}^{(\mathbf{u})}\bigr).
  \]
  Direct computation: $\binom{x_s}{\nu_s^{(\mathbf{u})}}
  = x_s!/[(x_s - \nu_s^{(\mathbf{u})})!\,\nu_s^{(\mathbf{u})}!]$,
  so $q_\mathbf{u}(\mathbf{x})\,\pi(\mathbf{x})$ has factor
  $\prod_s (c_s^*)^{x_s}/[(x_s - \nu_s^{(\mathbf{u})})!\,
  \nu_s^{(\mathbf{u})}!]$, which factors as
  $\prod_s (c_s^*)^{\nu_s^{(\mathbf{u})}}/\nu_s^{(\mathbf{u})}!$
  times the Poisson factor at the shifted state.

  \smallskip
  \noindent\textbf{Rewriting the master equation.}
  Using $\mathbf{n}_r = \boldsymbol{\nu}^{(y'(r))} -
  \boldsymbol{\nu}^{(y(r))}$, the forward Kolmogorov equation
  at state $\mathbf{x}$ reads
  \[
    (\Omega^\top \pi)(\mathbf{x})
    \;=\;
    \sum_{r \in \Rx}
    \kappa_r\bigl[
      q_{y(r)}(\mathbf{x} - \mathbf{n}_r)\,
      \pi(\mathbf{x} - \mathbf{n}_r)
      \;-\;
      q_{y(r)}(\mathbf{x})\,\pi(\mathbf{x})
    \bigr].
  \]
  Apply the Poisson--occupancy identity to both terms.
  For the loss term,
  \[
    q_{y(r)}(\mathbf{x})\,\pi(\mathbf{x})
    \;=\;
    \frac{(\mathbf{c}^*)^{\boldsymbol{\nu}^{(y(r))}}}
      {\prod_s \nu_s^{(y(r))}!}\,
    \pi\bigl(\mathbf{x} - \boldsymbol{\nu}^{(y(r))}\bigr).
  \]
  For the gain term, $\mathbf{x} - \mathbf{n}_r -
  \boldsymbol{\nu}^{(y(r))} = \mathbf{x} -
  \boldsymbol{\nu}^{(y'(r))}$, so
  \[
    q_{y(r)}(\mathbf{x} - \mathbf{n}_r)\,
    \pi(\mathbf{x} - \mathbf{n}_r)
    \;=\;
    \frac{(\mathbf{c}^*)^{\boldsymbol{\nu}^{(y(r))}}}
      {\prod_s \nu_s^{(y(r))}!}\,
    \pi\bigl(\mathbf{x} - \boldsymbol{\nu}^{(y'(r))}\bigr).
  \]
  Defining the \emph{stochastic rate}
  $\hat v_r(\mathbf{c}^*)
  := \kappa_r (\mathbf{c}^*)^{\boldsymbol{\nu}^{(y(r))}}
    /\prod_s \nu_s^{(y(r))}!$
  and $\tilde\pi_\mathbf{u} := \pi(\mathbf{x} -
  \boldsymbol{\nu}^{(\mathbf{u})})$, the master equation
  collapses to
  \[
    (\Omega^\top \pi)(\mathbf{x})
    \;=\;
    \sum_{r \in \Rx} \hat v_r(\mathbf{c}^*)
    \bigl[\tilde\pi_{y'(r)} - \tilde\pi_{y(r)}\bigr].
  \]

  \smallskip
  \noindent\textbf{Stochastic complex-balance closure.}
  Reindex the sum by the complex hit.  Each reaction $r$
  contributes $+\hat v_r(\mathbf{c}^*)\tilde\pi_{y'(r)}$
  (incoming at $y'(r)$) and
  $-\hat v_r(\mathbf{c}^*)\tilde\pi_{y(r)}$ (outgoing from
  $y(r)$).  Collecting by complex,
  \[
    (\Omega^\top \pi)(\mathbf{x})
    \;=\;
    \sum_{\mathbf{u} \in \Cx}
    \tilde\pi_\mathbf{u}
    \Biggl[
      \sum_{r:\,y'(r) = \mathbf{u}} \hat v_r(\mathbf{c}^*)
      \;-\;
      \sum_{r:\,y(r) = \mathbf{u}} \hat v_r(\mathbf{c}^*)
    \Biggr].
  \]
  The bracket is precisely the stochastic complex-balance defect
  at $\mathbf{u}$, which vanishes by hypothesis for every
  $\mathbf{u} \in \Cx$.  Hence
  $(\Omega^\top \pi)(\mathbf{x}) = 0$ at every state
  $\mathbf{x}$.  The restriction of $\pi$ to any closed
  irreducible subset is a stationary distribution of the
  CTMC generated by $\Omega$ there.
\end{proof}

\begin{mathbox}[Level stratification of the ACK theorem]
The ACK theorem has the same cross-level architecture as the
DZT, but the conclusion now lives at the level of the
assembled CME generator $\Omega$ rather than its
large-volume limit $\Phi_\infty(\Omega)$.
\renewcommand{\arraystretch}{1.4}
\begin{center}
\begin{tabular}{p{4.6cm} p{1.4cm} p{5.4cm}}
  \hline
  \textbf{Ingredient} & \textbf{Level} &
  \textbf{Where it lives}\\
  \hline
  Stochastic complex-balanced $\mathbf{c}^*$ &
  $\Lk_3$ &
  Hypothesis of Theorem~\ref{thm:ACK};
  related to fixed points of $\Phi_\infty(\Omega)$ by the
  rate-constant rescaling
  $\kappa_r \mapsto \kappa_r/\prod_s \nu_s^{(y(r))}!$ \\[2pt]
  Assembled generator $\Omega$ &
  $\Lk_3$ &
  Definition~\ref{def:CME}\\[2pt]
  Product-of-Poissons stationary form on each
  closed irreducible subset &
  $\Lk_3$ &
  Stationary state of $\Omega$ \\[2pt]
  Poisson parameters $c_s^*$ &
  $\Lk_3$ &
  From the hypothesised $\mathbf{c}^*$ \\[2pt]
  \hline
  \multicolumn{3}{l}{\emph{Corollary form (via
    Theorem~\ref{thm:DZT}):}} \\[2pt]
  $\delta = 0$, weak reversibility &
  $\Lk_0$ &
  Definition~\ref{def:feinberg} \\[2pt]
  Quantifier over $\kappa$ &
  $\Lk_3$ &
  Fiber $U_3^{-1}(\Lk_0(P))$ \\
  \hline
\end{tabular}
\end{center}

\medskip\noindent
\textbf{What the product form means, and where it lives.}
The product-of-Poissons formula
$\pi(\mathbf{x}) = (1/Z) \prod_s (c_s^*)^{x_s}/x_s!$
is the stationary distribution of the assembled CME generator
$\Omega$ on each closed irreducible subset
$\Gamma \subseteq \NN^{|\Sp|}$ of the CTMC, after
normalization on $\Gamma$
(Theorem~\ref{thm:ACK}, \cite{AndersonCraciunKurtz2010}).
A reaction network typically has several conservation laws
(elemental balances, charge conservation, total particle
counts within a closed system), so $\Gamma$ is generally a
proper subset of $\NN^{|\Sp|}$ --- a hyperplane fixing the
conserved totals.
\emph{Conditional on $\Gamma$}, the species counts
$(x_s)_{s \in \Sp}$ are not independent: the conservation
constraints introduce deterministic relations among them,
and the conditional distribution
\[
  \pi_\Gamma(\mathbf{x})
  \;=\; \frac{1}{Z_\Gamma}
  \prod_{s \in \Sp} \frac{(c_s^*)^{x_s}}{x_s!}
  \quad (\mathbf{x} \in \Gamma)
\]
is the normalized restriction of an independent product;
species counts within $\Gamma$ are therefore correlated.
For instance, on the closed irreducible class
$\{(x_A, x_B) : x_A + x_B = N\}$ of $A \rightleftharpoons B$,
the conditional distribution is binomial
$\pi_N(x_A) = \binom{N}{x_A}\,
\bigl(c_A^*/(c_A^*+c_B^*)\bigr)^{x_A}
\bigl(c_B^*/(c_A^*+c_B^*)\bigr)^{N-x_A}$,
not a product of independent Poissons.
Independence is recovered only when the network has no
conservation laws and $\NN^{|\Sp|}$ itself is a single closed
irreducible class --- e.g.\ for an open system with
particle injection/decay reactions in every species.

What is genuinely structural, and intrinsically $\Lk_3$, is
that the stationary distribution on every $\Gamma$ is the
\emph{normalized restriction} of an independent product
distribution with parameters $c_s^*$ tied to a single common
complex-balanced point.
The fact has no analogue at $\Lk_2$ (thermodynamics sees only
ratios of concentrations and equilibrium loci, not
distributions over discrete states) or below (no probability
distribution over states is available).
It is accessible only through $\Omega$ and the assembled
$\Lk_3$ structure.
For deficiency-zero weakly reversible networks,
Theorem~\ref{thm:DZT}\textup{(iii)} guarantees a macroscopic
complex-balanced $\mathbf{c}^*$ for every Layer-1 $\FP$;
converting to stochastic complex balance via the rate
rescaling $\kappa_r \mapsto \kappa_r/\prod_s \nu_s^{(y(r))}!$
(Theorem~\ref{thm:ACK}) gives the product-restriction form on
each $\Gamma$.
This stationary structure is therefore forced by $\Lk_0$
topology alone, up to the rescaling.

\medskip\noindent
\textbf{Comparison with the DZT.}
The DZT and ACK form a commuting pair in the tower:
\[
\includegraphics{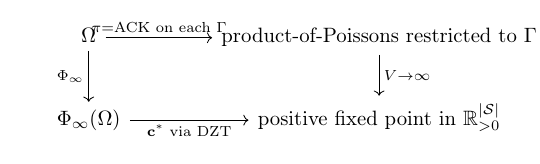}
\]
Given $\delta = 0$ and weak reversibility, $\mathbf{c}^*$ is
the unique positive fixed point of the RRE guaranteed by the
DZT in each positive stoichiometric compatibility class (a
macroscopic complex-balanced point); the rate rescaling
$\kappa_r \mapsto \kappa_r/\prod_s \nu_s^{(y(r))}!$ converts
macroscopic to stochastic complex balance, and ACK then gives
the product-Poisson distribution $\pi$ restricted to each
closed irreducible subset $\Gamma$ with Poisson parameters
$\mathbf{c}^*$.
The large-volume limit $V \to \infty$ with $\mathbf{x}/V \to
\mathbf{c}$ concentrates $\pi$ on the Dirac mass
$\delta_{\mathbf{c}^*}$ within the corresponding stoichiometric
compatibility class --- the RRE fixed point recovered as the
large-volume shadow of the ACK distribution.
The two theorems are the same result viewed at two levels of
the assembled $\Lk_3$ object $\Omega$ and its scaling shadow
$\Phi_\infty(\Omega)$.
\end{mathbox}

%% file: chapters/L3/l3_baezpollard.tex
\subsection{Compositional kinetics: gluing open \texorpdfstring{$\Lk_3$}{L3} networks}
\label{sec:L3-baezpollard}

The tower's $\Lk_3$ assembly is already well-behaved under
two operations on reaction sets: sequential composition of
reactions (Proposition~\ref{prop:FP-functorial}) and disjoint
union of species sets
(Proposition~\ref{prop:FP-monoidal}).
A mild extension of these properties handles a third
operation that is ubiquitous in real chemistry: gluing two
networks at shared boundary species.
Enzymatic modules feeding a metabolic pathway, a receptor
coupled to a downstream signalling cascade, an autocatalytic
network buffered by a reservoir --- all are composites of
smaller networks joined at interface species.
The compositional framework of Baez and Pollard
\cite{BaezPollard2017} makes this gluing rigorous and shows
that the tower's deterministic limit $\Phi_\infty(\Omega)$
and its steady states compose correctly under it.

Informally, an \emph{open $\Lk_3$ network} is a closed
$\Lk_3(P)$ together with two distinguished subsets of $\Sp$
playing the role of inputs and outputs (boundary species
where material may flow in or out of the network), with the
remaining species internal.
Two open networks are glued by identifying the output species
of the first with the input species of the second; shared
species become internal to the composite, and the rate
constants of both networks are retained.
The formal framework uses Fong's \emph{decorated cospans}
\cite{fong2015decorated,BaezPollard2017}: cospans of species sets in
$\mathbf{FinSet}$ decorated by $\Lk_3$ data, with
composition given by pushout of the underlying cospans
combined with functorial composition of decorations.

\begin{theorem}[Compositional kinetics
  {\cite[Thms.~18 and~23]{BaezPollard2017}}]
\label{thm:baez-pollard}
  Gluing of open $\Lk_3$ networks is compatible with both
  the deterministic dynamics and the steady-state boundary
  behaviour.
  Specifically, for composable open networks $\Lk_3(P_1)$
  and $\Lk_3(P_2)$ sharing an interface of boundary species:
  \begin{enumerate}[label=(\roman*)]
    \item \textup{(Gray-boxing)}
      The large-volume vector field
      $\Phi_\infty(\Omega_{12})$ of the composite decomposes
      as the sum of the individual vector fields
      $\Phi_\infty(\Omega_1)$ and $\Phi_\infty(\Omega_2)$,
      extended across the shared interface by restriction
      and summation of reaction velocities.
    \item \textup{(Steady-state composition)}
      The steady-state relation between boundary
      concentrations and boundary flows of the composite is
      the relational composite of the individual
      steady-state relations: if $\mathbf{c}_1^*$ is a
      steady state of $\Lk_3(P_1)$ producing interface
      flow $\mathbf{f}$, and $\mathbf{c}_2^*$ is a steady
      state of $\Lk_3(P_2)$ accepting $\mathbf{f}$ at
      its input, then together they define a steady state
      of the composite.
  \end{enumerate}
\end{theorem}

\begin{proof}[Tower-native content proof]
  Part~(i) follows from the $\Lk_3$ assembly rules directly.
  Let $\Sp_{12} = \Sp_1 \cup_{\Sp_\partial} \Sp_2$ be the
  pushout of species sets along the shared interface
  $\Sp_\partial$, and let $\Rx_{12} = \Rx_1 \sqcup \Rx_2$
  be the disjoint union of reaction sets.
  Each reaction $r \in \Rx_i$ has source complex, target
  complex, and propensity supported on $\Sp_i \subseteq
  \Sp_{12}$; canonically extending $\FP(r)$ to act on
  observables $f : \NN^{|\Sp_{12}|} \to \RR$ by acting
  trivially on species in $\Sp_{12} \setminus \Sp_i$, the
  assembled generator on the combined state space is
  \[
    \Omega_{12} \;=\; \sum_{r \in \Rx_{12}} \FP(r)
    \;=\; \Omega_1 + \Omega_2 \quad
    \text{in } \mathfrak{g}_{\Sp_{12}},
  \]
  where each $\Omega_i$ denotes its canonical extension along
  the inclusion $\Sp_i \hookrightarrow \Sp_{12}$.
  The volume-indexed family $\{\Omega_{12}^V\}$ is
  mass-action by construction, hence classically scaled
  (Definition~\ref{def:Phi-infty}), and decomposes additively
  as $\Omega_{12}^V = \Omega_1^V + \Omega_2^V$ at every $V$.
  Since the scaling limit commutes with finite sums of
  classically-scaled families (pointwise limits of sums
  equal sums of pointwise limits, and the volume-scaling
  exponents $V^{1-|\mathbf{u}_r|}$ are determined per
  reaction independently of which subnetwork it belongs to),
  \[
    \Phi_\infty\!\bigl(\{\Omega_{12}^V\}\bigr)
    \;=\;
    \Phi_\infty\!\bigl(\{\Omega_1^V\}\bigr)
    + \Phi_\infty\!\bigl(\{\Omega_2^V\}\bigr)
  \]
  as vector fields on $\RR_{>0}^{|\Sp_{12}|}$.
  At an interface species $s \in \Sp_\partial$, the
  $s$-component of the composite vector field is the sum of
  the $s$-components of $\Phi_\infty(\{\Omega_1^V\})$ and
  $\Phi_\infty(\{\Omega_2^V\})$ --- the gray-boxing
  ``summation across the shared interface'' of part~(i).

  Part~(ii) reduces to part~(i) by setting
  $\dot{\mathbf{c}} = 0$ at the composite level and tracking
  how the interface flow matches the output of $\Lk_3(P_1)$
  to the input of $\Lk_3(P_2)$; the full argument involves
  semialgebraic projection (Tarski--Seidenberg) for the
  steady-state variety and is external to the tower
  \cite[Thm.~23~proof]{BaezPollard2017}.
\end{proof}

\begin{chembox}[Michaelis--Menten as gray-boxing under quasi-steady-state reduction]
The classical enzyme mechanism
$E + S \rightleftharpoons ES \to E + P$ is an open $\Lk_3$
network with input species $\{S\}$, output species $\{P\}$,
and internal species $\{E, ES\}$.
The Michaelis--Menten rate law
\[
  \text{(product outflow)}
  \;=\; \frac{V_{\max}\,c_S}{K_M + c_S},
  \quad
  V_{\max} = k_\mathrm{cat}\,[E]_\mathrm{tot},
  \quad
  K_M = \frac{k_{-1} + k_\mathrm{cat}}{k_1},
\]
is the steady-state input--output relation produced by
gray-boxing the internal $\{E, ES\}$ dynamics under the
\emph{quasi-steady-state} reduction of Briggs and Haldane
\cite{Briggs1925}: treating $\dot c_{ES} \approx 0$ on the
timescale of substrate variation and using the
conserved-enzyme constraint
$[E]_\mathrm{tot} = c_E + c_{ES}$ to eliminate $c_{ES}$.
This reduction is not part of the strict $\Lk_3$ functorial
assignment --- $\FP$ only assembles the joint CTMC generator
of the four reactions, not an effective rate law for the
coarse-grained $S \to P$ conversion --- but is the standard
chemical-engineering reduction valid when $c_{ES}$ is
short-lived (cf.\ the chembox after
Proposition~\ref{prop:FP-functorial}).
Both $V_{\max}$ and $K_M$ depend on individual Layer-1 rate
constants $k_1, k_{-1}, k_\mathrm{cat}$ carried by $\FP$
(together with the conserved $[E]_\mathrm{tot}$); none is
available from $\Lk_2$ data alone.

For a metabolic pathway built from $n$ enzymatic modules,
each reduced to its Michaelis--Menten input--output relation
by gray-boxing, Theorem~\ref{thm:baez-pollard} says the
pathway's steady-state boundary behaviour is the
\emph{relational composite} of the individual module
relations under the Baez--Pollard cospan structure.
For monotone cascades --- each module's steady-state map
strictly increasing in its input concentration --- the
relational composite is itself functional, and the pathway
inherits a single-valued input--output rate law that
recovers classical pipelined Michaelis--Menten analysis.
For feedback or multistable subnetworks, the relation is
genuinely multi-valued: a single boundary input may
correspond to several internal steady states, and the
relational composition framework is the natural setting in
which to track this --- the cospan structure does not
collapse the multi-valuedness, whereas a functional-composition
framing silently would.
This compositional treatment is the mathematical foundation
of modular metabolic control analysis.
\end{chembox}

%% file: chapters/L3/l3_examples.tex
\subsection{Worked examples at \texorpdfstring{$\Lk_3$}{L3}}
\label{sec:L3-examples}

\subsubsection{Formaldehyde hydration:
  CME, RRE, and Layer~2 consistency}

\begin{example}[$\mathrm{HCHO(aq)} + \mathrm{H_2O}
  \rightleftharpoons \mathrm{CH_2(OH)_2(aq)}$ at $\Lk_3$]
\label{ex:HCHO-L3}
Formaldehyde in dilute aqueous solution hydrates reversibly
to its gem-diol, methanediol:
\[
  \mathrm{HCHO(aq)} + \mathrm{H_2O(l)}
  \;\mathrel{\substack{k_h\\[-0.3ex]\rightleftharpoons\\[-0.3ex]k_d}}\;
  \mathrm{CH_2(OH)_2(aq)}.
\]
Because water is present in large excess at nearly constant
activity, both directions are phenomenologically first-order,
and the network reduces to the unimolecular-unimolecular
isomerisation $\mathrm{HCHO} \rightleftharpoons
\mathrm{CH_2(OH)_2}$ --- the simplest nontrivial $\Lk_3$
structure.
Unlike the $\mathrm{N_2O_4}$ dissociation of
Example~\ref{ex:N2O4}, forward and reverse rate constants
for this reaction have been measured together in a single
kinetic regime (dilute neutral water, 298~K), making it a
clean test of the Layer-2 consistency
(Definition~\ref{def:L3-layer2}).

\textbf{$\Lk_3$ data (298~K, dilute neutral water).}
\[
  k_h = 11\;\mathrm{s^{-1}}
  \quad\text{(hydration)}
  \cite{Winkelman2002, SuttonDownes1972},
  \qquad
  k_d = 5\times 10^{-3}\;\mathrm{s^{-1}}
  \quad\text{(dehydration)}
  \cite{BellEvans1966, Funderburk1978}.
\]

\textbf{State space and stoichiometry (from $\Lk_0$).}
$\Sp = \{\mathrm{HCHO},\, \mathrm{CH_2(OH)_2}\}$;
state $\mathbf{x} = (x_1, x_2) \in \NN^2$.
Two complexes, $\mathbf{u}_1 = \mathrm{HCHO}$ and
$\mathbf{u}_2 = \mathrm{CH_2(OH)_2}$, with stoichiometric
change vectors
$\mathbf{n}_{r_h} = (-1, +1)$ and
$\mathbf{n}_{r_d} = (+1, -1)$.

\textbf{Propensities (Definition~\ref{def:propensity}).}
Both reactions are first-order, so the combinatorial
mass-action propensities reduce to linear functions of
molecule counts:
\[
  \lambda_h(x_1, x_2)
  \;=\; k_h\, \binom{x_1}{1}
  \;=\; k_h\, x_1,
  \qquad
  \lambda_d(x_1, x_2)
  \;=\; k_d\, \binom{x_2}{1}
  \;=\; k_d\, x_2.
\]

\textbf{CME generator (Definition~\ref{def:CME}).}
\[
  \Omega \;=\; M_{\lambda_h}(R_{r_h} - I) + M_{\lambda_d}(R_{r_d} - I),
\]
acting on observables $f : \NN^2 \to \RR$ as
\[
  (\Omega f)(x_1, x_2)
  \;=\; k_h\, x_1\,[f(x_1-1, x_2+1) - f(x_1, x_2)]
  + k_d\, x_2\,[f(x_1+1, x_2-1) - f(x_1, x_2)].
\]

\textbf{Large-volume limit (Proposition~\ref{prop:LMA}).}
$\Phi_\infty(\Omega)$ gives the RRE:
\[
  \dot{c}_1 = -k_h\, c_1 + k_d\, c_2,
  \qquad
  \dot{c}_2 = +k_h\, c_1 - k_d\, c_2.
\]

\textbf{Layer~2 consistency check
  (Definition~\ref{def:L3-layer2}).}
The kinetic equilibrium constant from $\Lk_3$ data:
\[
  K_\mathrm{kin}
  \;:=\; \frac{k_h}{k_d}
  \;=\; \frac{11}{5\times 10^{-3}}
  \;\approx\; 2.2\times 10^3
  \quad\text{(dimensionless)}.
\]
The thermodynamic equilibrium constant from an independent
$^{13}\mathrm{C}$ and $^1\mathrm{H}$ NMR integration of the
two species in aqueous solution at 298~K
\cite{Rivlin2015}:
\[
  K_\mathrm{hyd}(298)
  \;\approx\; 2.0\times 10^3
  \quad\text{(dimensionless)}.
\]
The two routes agree to within $\sim 10\%$, well inside the
combined experimental uncertainty of the rate constants
($\sim 20\%$ each) and the NMR integration
($\sim 15\%$).
This is the Layer-2 coboundary condition satisfied
numerically: the ratio $k_h/k_d$ computed from $\Lk_3$ rate
data matches the equilibrium constant $K_\mathrm{hyd}$
determined from $\Lk_2$ thermodynamics, as required by
Definition~\ref{def:L3-layer2}.

\begin{remark}[On the ``elementary'' label]
The bulk phenomenological rates $k_h$ and $k_d$ are
weighted averages over microscopic pathways that proceed
through a cyclic proton relay involving $n = 2$ or $3$
water molecules \cite{Wolfe1995}.
Each microscopic pathway separately satisfies
$k_h^{(n)} / k_d^{(n)} = K_\mathrm{hyd}$, and the bulk
averaging preserves the identity; hence the Layer-2
check holds at the phenomenological level even though
the ``elementary'' reaction is a cluster of proton-relay
channels with the water activity absorbed into $k_h$.
This is the generic situation for solution-phase
isomerisations in protic solvents.
\end{remark}

\textbf{DZT applicability and ACK stationary distribution.}
The network has $n = 2$ complexes, $\ell = 1$ linkage
class, and $s = 1$ (stoichiometric subspace spanned by
$(-1, +1)^\top$).
Deficiency $\delta = n - \ell - s = 0$, and weak
reversibility holds.
The DZT (Theorem~\ref{thm:DZT} and Proposition~\ref{prop:tower-DZT}) guarantees a unique positive
fixed point
$(c_1^*, c_2^*) = \bigl(c_\mathrm{tot}/(1+K_\mathrm{hyd}),\,
c_\mathrm{tot} K_\mathrm{hyd}/(1+K_\mathrm{hyd})\bigr)$
in each positive stoichiometric compatibility class
parametrised by $c_\mathrm{tot} = c_1 + c_2$, and global
asymptotic stability of this fixed point within the open
positive class.

Stochastically, the total carbonyl-carbon count
$N := x_1 + x_2$ is conserved, so the closed irreducible
subsets of the CTMC are the hyperplanes
$\Gamma_N = \{(x_1, x_2) \in \NN^2 : x_1 + x_2 = N\}$
indexed by $N \in \NN$.
On each $\Gamma_N$, the ACK theorem
(Theorem~\ref{thm:ACK}) gives the stationary distribution
as the normalized restriction of the product Poisson with
parameters $(c_1^*, c_2^*)$:
\[
  \pi_N(x_1)
  \;=\; \binom{N}{x_1}\,
  \biggl(\frac{c_1^*}{c_1^* + c_2^*}\biggr)^{x_1}
  \biggl(\frac{c_2^*}{c_1^* + c_2^*}\biggr)^{N - x_1}
  \;=\; \binom{N}{x_1}\,
  \frac{K_\mathrm{hyd}^{N - x_1}}{(1+K_\mathrm{hyd})^N},
\]
the binomial distribution on $\Gamma_N$ with success
probability $p = c_1^*/(c_1^*+c_2^*) = 1/(1+K_\mathrm{hyd})$
for the HCHO count.
The two species counts are not independent on $\Gamma_N$ ---
they are deterministically related by $x_1 + x_2 = N$ ---
and the unconditioned product-Poisson independence is
recovered only across closed irreducible classes,
i.e.\ when the total $N$ is itself randomized
(\emph{cf.}\ the $\Lk_3$ closing mathbox in
Section~\ref{sec:L3-dzt}).

Since $K_\mathrm{hyd} \approx 2000$, the binomial mean
$\mathbb{E}[x_1 \mid \Gamma_N] = N/(1+K_\mathrm{hyd})$
gives free formaldehyde at roughly $0.05\%$ of the total
carbonyl-carbon count: the diol overwhelmingly dominates
the equilibrium population, and $\pi_N$ is concentrated
near $x_2 = N$ for any $N \gtrsim K_\mathrm{hyd}$.
\end{example}

\begin{example}[$\mathrm{S_N2}$ at $\Lk_3$: bimolecular propensity and
  mechanism blindness]
\label{ex:SN2-L3}
Continuing Example~\ref{ex:SN2-L2} from $\Lk_2$.
Where the formaldehyde example
(Example~\ref{ex:HCHO-L3}) illustrated Layer~2 consistency
for a reversible unimolecular isomerisation, this example
exhibits the complementary $\Lk_3$ feature: a genuinely
bimolecular propensity $\lambda_r \propto x_\mathrm{A} x_\mathrm{B}$
whose second-order rate law alone fails to distinguish
mechanistically distinct reactions --- the forcing observation
for $\Lk_4$.

\textbf{$\Lk_3$ data.}
Rate constant for the classical alkaline hydrolysis
$\mathrm{CH_3Cl} + \mathrm{OH^-} \to \mathrm{CH_3OH}
+ \mathrm{Cl^-}$ in dilute aqueous solution, extrapolated
to 298~K from the Arrhenius fit of
Moelwyn-Hughes \cite{MoelwynHughes1949, MoelwynHughes1953}:
\[
  k_r \;\approx\; 6\times 10^{-6}\;\mathrm{M^{-1}s^{-1}}.
\]
The reaction is effectively irreversible under standard
conditions: $\mathrm{CH_3OH} + \mathrm{Cl^-} \to
\mathrm{CH_3Cl} + \mathrm{OH^-}$ has no measurable rate in
water, so we treat the network as a single forward
generator and do not perform a Layer~2 check here.

\textbf{State space and stoichiometry (from $\Lk_0$).}
\[
  \Sp = \{\mathrm{CH_3Cl},\, \mathrm{OH^-},\,
  \mathrm{CH_3OH},\, \mathrm{Cl^-}\}.
\]
Generator $r$: source complex
$\mathbf{u} = \mathrm{CH_3Cl} + \mathrm{OH^-}$,
target complex $\mathbf{v} = \mathrm{CH_3OH} + \mathrm{Cl^-}$,
stoichiometric change vector
$\mathbf{n}_r = (-1, -1, +1, +1)$.

\textbf{Propensity (Definition~\ref{def:propensity}).}
Each reactant appears with stoichiometric coefficient
$\nu_s = 1$ in the source complex, so the combinatorial
mass-action propensity
$\lambda_r(\mathbf{x}) = k_r \prod_s \binom{x_s}{\nu_s}$
reduces to the product of molecule counts:
\[
  \lambda_r(\mathbf{x})
  \;=\; k_r\, \binom{x_{\mathrm{CH_3Cl}}}{1}\,
  \binom{x_{\mathrm{OH^-}}}{1}
  \;=\; k_r\, x_{\mathrm{CH_3Cl}}\cdot x_{\mathrm{OH^-}}.
\]
The generator contribution is $\FP(r) = M_{\lambda_r}(R_r - I)$
with $R_r$ shifting $\mathbf{x} \mapsto \mathbf{x} + \mathbf{n}_r$.
Unlike the unimolecular case, $\lambda_r$ is genuinely
nonlinear in the state: it depends on the joint occupancy
of two distinct species.

\textbf{Large-volume limit (Proposition~\ref{prop:LMA}).}
$\Phi_\infty(\Omega)$ gives the RRE:
\[
  \frac{d[\mathrm{CH_3OH}]}{dt}
  \;=\; k_r\,[\mathrm{CH_3Cl}][\mathrm{OH^-}].
\]
This recovers the empirical second-order rate law
characteristic of bimolecular nucleophilic substitution.
At $[\mathrm{OH^-}] = 1\;\mathrm{M}$ (the standard reference
state for laboratory SN2 measurements, corresponding to
$\mathrm{pH} = 14$), the pseudo-first-order half-life of
$\mathrm{CH_3Cl}$ from this bimolecular channel is
\[
  t_{1/2} \;=\; \frac{\ln 2}{k_r\,[\mathrm{OH^-}]}
  \;\approx\; 1.2\times 10^5\;\mathrm{s}
  \;\approx\; 32\;\text{hours}.
\]
The same formula at $\mathrm{pH} = 11$
($[\mathrm{OH^-}] = 10^{-3}\;\mathrm{M}$) gives
$t_{1/2} \approx 1.2\times 10^8\;\mathrm{s} \approx 3.7$
years for the bimolecular channel alone; under such mildly
basic conditions the overall hydrolysis of methyl chloride
in dilute aqueous solution is dominated by the
pH-independent water-mediated pathway rather than by this
$\mathrm{OH^-}$ channel
\cite{Zafiriou1975, MabeyMill1978}.
The bimolecular SN2 contribution we track here is therefore
the dominant channel only at high $[\mathrm{OH^-}]$.

\textbf{Deficiency check.}
Treating the reaction as irreversible:
$n = 2$ complexes ($\mathbf{u}$ and $\mathbf{v}$),
$\ell = 1$ linkage class,
$s = 1$ (rank of stoichiometric matrix $N$),
so $\delta = 2 - 1 - 1 = 0$.
The network is \emph{not} weakly reversible, however ---
the single forward generator has no reverse.
The DZT (Theorem~\ref{thm:DZT}) therefore does not apply,
and the long-time behaviour is simply exhaustive
consumption of whichever reactant is limiting, terminating
in a trivial steady state on the boundary of the
stoichiometric class.
This is a useful contrast to the formaldehyde example:
$\Lk_3$ structure alone does not guarantee the weak-reversibility
hypothesis of DZT, and irreversible reactions sit outside
its scope.

\textbf{What $\Lk_3$ cannot express.}
The propensity $\lambda_r(\mathbf{x}) =
k_r\,x_{\mathrm{CH_3Cl}}\cdot x_{\mathrm{OH^-}}$ describes
the rate of product formation, but encodes nothing about
\emph{how} the bonds rearrange.
Any second-order nucleophilic substitution
$\mathrm{R-X} + \mathrm{Nu^-} \to \mathrm{R-Nu} +
\mathrm{X^-}$ on a primary substrate gives the same
propensity functional form
$\lambda \propto x_{\mathrm{R-X}}\, x_{\mathrm{Nu^-}}$,
regardless of which substrate, leaving group, and
nucleophile are involved.
The backside-attack geometry, the pentacoordinate
transition state, and the Walden inversion at the carbon
centre are all invisible to $\FP$.

The forcing observation for $\Lk_4$ is sharper still: two
reactions with the same $\Lk_3$ data --- same complexes,
same propensity functional form, same rate constant ---
can still differ at the level of bond-level mechanism.
The classical $\Lk_3 \to \Lk_4$ forcing pairs are the
concerted SN2 versus the stepwise ion-pair $\mathrm{S_N1}$,
and the concerted E2 versus the irreversible-stepwise
$\mathrm{E1cb}$; both are taken up in the $\Lk_4$ chapter
(Chapter~\ref{sec:L4}), where the bond-graph rewriting
machinery of double pushout (DPO) provides exactly the
discriminating structure that $\Lk_3$ lacks.

A common misreading should be guarded against: SN2 versus
E2 is \emph{not} an $\Lk_3 \to \Lk_4$ forcing pair.
The bimolecular elimination of, say, 2-chlorobutane with
hydroxide
$\bigl(\mathrm{C_4H_9Cl} + \mathrm{OH^-} \to
\mathrm{C_4H_8} + \mathrm{H_2O} + \mathrm{Cl^-}\bigr)$
is also second-order in concentrations, but it produces an
alkene plus water rather than a substitution product.
The two reactions therefore have different stoichiometric
change vectors $\mathbf{n}_r$ and different target
complexes, so they are already separated at $\Lk_0$ by
stoichiometry alone --- not at $\Lk_4$ by bond-level
mechanism.
Genuine $\Lk_3 \to \Lk_4$ forcing requires reactions that
agree on every $\Lk_0, \Lk_1, \Lk_2, \Lk_3$ datum and
differ only at the bond-graph rewriting layer; this is
strictly subtler than the SN2 versus E2 comparison.
\end{example}

%% file: chapters/L3/l3_nextforcing.tex
\subsection{What \texorpdfstring{$\Lk_3$}{L3} cannot express:
  forcing of \texorpdfstring{$\Lk_4$}{L4}}
\label{sec:L3-forcing-out}

The tower so far --- $\Lk_0, \Lk_1, \Lk_2, \Lk_3$ ---
shares a common underlying category $\Lk_0(P)$, the free
skeletal permutative category on $P$
(Theorem~\ref{thm:UP-L0}), with each successive level adding
a decorating functor on top of it.
Sections~\ref{sec:L3-forcing-in}--\ref{sec:L3-baezpollard}
have shown what $\Lk_3$ buys us: the CME, the RRE, the
Layer-2 coboundary condition, and compositional kinetics
for open networks.
This subsection identifies what $\Lk_3$ cannot express,
setting up the forcing pair that Chapter~\ref{sec:L4}
resolves.

\begin{forcingbox}[Forcing pair for $\Lk_4$:
  identity substitution at phosphorus]
Consider the identity methoxyl exchange at the
phosphorus center of methyl ethylphenylphosphinate
\cite{Mikolajczyk2022}, with isotopic labelling on the
incoming methoxide:
\[
  r :\;
  {}^{18}\mathrm{OMe^-} +
  \mathrm{Et(Ph)P(=O)OMe}
  \;\longrightarrow\;
  \mathrm{Et(Ph)P(=O){}^{18}OMe}
  + \mathrm{OMe^-}.
\]
Source and target complexes are identical up to the
${}^{18}\mathrm{O}$ permutation; the only chemical change
is which methoxyl group is bound to phosphorus and which
is free.

This transformation is realised in the laboratory by two
distinct mechanisms:

\medskip\noindent\textbf{System A} (concerted
$\mathrm{S_N2}$-P): methoxide attacks the phosphorus apically
and displaces the leaving methoxide through a single
trigonal-bipyramidal transition state.

\medskip\noindent\textbf{System B} (addition-elimination):
methoxide adds first to form a discrete pentacoordinate
trigonal-bipyramidal intermediate ($\mathrm{TBI}$) with five
P-bonds; the leaving methoxide then eliminates from
$\mathrm{TBI}$ in a second step.

\medskip

At the bulk-kinetic level of $\Lk_3$, the two mechanisms are
indistinguishable.
System A is a single bimolecular generator with propensity
$\lambda(\mathbf{x}) = k\,x_{\mathrm{MeO^-}}\,
x_{\mathrm{substrate}}$.
System B has $\mathrm{TBI}$ as an internal species and two
elementary steps; under the quasi-steady-state reduction on
$\mathrm{TBI}$ (valid when $\mathrm{TBI}$ is high-energy and
short-lived, the standard regime for phosphoryl chemistry
under laboratory conditions), elimination of
$\mathrm{TBI}$ from the rate equations yields a
coarse-grained generator with the same bimolecular form
\[
  \lambda(\mathbf{x}) \;=\;
  k\,x_{\mathrm{MeO^-}}\,x_{\mathrm{substrate}},
\]
where the effective $k$ aggregates the addition rate
constant and the elimination selectivity at $\mathrm{TBI}$.
In this regime --- the experimentally relevant one ---
Systems A and B share every $\Lk_3$ datum.
Outside the quasi-steady-state regime, System B exhibits a
Michaelis--Menten rate law that $\Lk_3$ \emph{can}
distinguish from System A's strict bimolecular form;
the $\Lk_3$ ambiguity between concerted and stepwise
phosphoryl mechanisms is therefore contingent on the
coarse-graining, not categorical, but the contingent regime
is precisely the one in which the laboratory operates.

Mikolajczyk \emph{et al.}\ \cite{Mikolajczyk2022} note
explicitly that ``the \emph{kinetic measurements are often
unable to distinguish}'' $\mathrm{S_N2}$-P from
addition-elimination at phosphorus; DFT computation of the
full energy profile, together with stereochemical analysis
through Berry pseudorotation, was required to settle the
mechanism for the substrate above (System B applies; the
analogous chloride exchange in
$\mathrm{(EtO)Et(P\!=\!S)Cl}$ goes by System A
\cite{Mikolajczyk2022}).
The same kinetic blindness pervades enzymatic phosphoryl
transfer, where the concerted-versus-stepwise question has
been a central mechanistic debate for decades
\cite{LassilaZalatanHerschlag2011}.

The two mechanisms correspond to distinct chemical
generators 
\[r_\text{concerted}, r_\text{stepwise} \in \Rx\]
with identical source, target, $\FH$, $\FS$, $F_G^T$, and
$\FP$ (after the quasi-steady-state coarse-graining of
$\mathrm{TBI}$ above).
The swap $r_\text{concerted} \leftrightarrow r_\text{stepwise}$
is therefore an automorphism of $\Lk_3(P)$.
It does not lift to any structure that records whether the
reaction proceeds via a single elementary bond rearrangement
(one DPO span; the concerted pathway, no internal
intermediate) or via two composable bond rearrangements
bracketing a discrete $\mathrm{TBI}$ intermediate (two
composable DPO spans, with $\mathrm{TBI}$ as a species in
$\Lk_4$): it is a non-trivial coset in $\coker(\varphi_4)$
within the automorphism sequence
\[
  1 \to \ker\varphi_4 \to \Aut(\Lk_4(P))
  \;\xrightarrow{\;\varphi_4\;}\;
  \Aut(\Lk_3(P)) \to \coker(\varphi_4) \to 1
\]
(with $\coker(\varphi_4) = \Aut(\Lk_3(P))/\im(\varphi_4)$ as
a pointed-set quotient, \S\ref{sec:aut-exact}) once
$\Lk_4(P)$ is constructed.
The transition states themselves --- their geometries,
energies, and imaginary frequencies as saddle points on the
potential energy surface --- are not encoded at $\Lk_4$
either: $\Lk_4$ records only the bond-graph rewriting
topology (one elementary rewrite vs.\ two consecutive
rewrites with an intermediate species), and the saddle-point
structure on the PES belongs to $\Lk_5$.
\end{forcingbox}

\noindent
What minimal new structure resolves the ambiguity?
The principled content distinguishing $r_\text{concerted}$
from $r_\text{stepwise}$ is the \emph{mechanism}: which
bonds break and form, whether the reaction proceeds through
a single transition state or through a discrete
intermediate, and --- if an intermediate exists --- what
species it is.
Encoding this requires replacing the underlying category
$\Lk_0(P)$ with a richer one in which morphisms carry
bond-level data and intermediates are themselves species.
This is not a decoration of $\Lk_0(P)$ but a \emph{structural
extension}, and constructing it is the task of
Chapter~\ref{sec:L4}.

\noindent
This forces $\Lk_4$, which must add two ingredients
\cite{EhrigPfenderSchneider1973, EhrigEhrigPrangeTaentzer2006,
AndersenFlammMerkleStadler2016}:

\begin{enumerate}[label=(\roman*)]
  \item A category $\mathbf{LGraph}$ of \emph{labelled
    molecular hypergraphs} whose vertices carry atom type,
    formal charge, and lone-pair count, and whose edges carry
    bond order.
    Species in $\Sp$ become objects of $\mathbf{LGraph}$.
  \item \emph{Double pushout} (DPO) rewriting: a reaction
    mechanism is a span $L \leftarrow K \rightarrow R$ in
    $\mathbf{LGraph}$, specifying which bonds break
    ($L \setminus K$) and which form ($R \setminus K$).
    Application to a molecular graph is a DPO in
    $\mathbf{LGraph}$ \cite{LackSobocinski2004}.
    The underlying category $\Lk_4(P)$ has DPO spans in
    $\mathbf{LGraph}$ as morphism generators, with
    composition by sequential rewriting and monoidal product
    by disjoint union (Chapter~\ref{sec:L4}).
\end{enumerate}

\noindent
The two pathways correspond to distinct DPO morphisms
on the same input graph (methyl ethylphenylphosphinate +
isotopically labelled methoxide):

\begin{itemize}
  \item \textbf{Concerted ($\mathrm{S_N2}$-P)}: a single
    DPO span $L \leftarrow K \rightarrow R$ in
    $\mathbf{LGraph}$ in which the
    P--$\mathrm{OMe}$ bond is broken and the
    P--${}^{18}\mathrm{OMe}$ bond is formed
    simultaneously.  The interface $K$ retains the
    phosphorus and its three non-participating ligands.
  \item \textbf{Stepwise (addition-elimination)}: a
    composable pair of DPO spans
    $L_1 \leftarrow K_1 \rightarrow R_1$ followed by
    $L_2 \leftarrow K_2 \rightarrow R_2$, with the
    intermediate state graph encoding the
    pentacoordinate $\mathrm{TBI}$ as an additional
    species in $\Lk_4$.  The first span forms the
    P--${}^{18}\mathrm{OMe}$ bond without breaking any
    bonds (addition); the second breaks the P--$\mathrm{OMe}$
    bond (elimination).
\end{itemize}

The two DPO morphisms are genuinely different in
$\Lk_4(P)$ but project to the \emph{same} morphism in
$\Lk_3(P)$ under the forgetful functor $U_4$.
The non-trivial element of $\coker(\varphi_4)$ is the swap
between these two morphisms while fixing their common
$\Lk_3$ image.
The extension to $\Lk_4$ is \emph{forced} by the existence
of mechanistically distinct reactions whose $\Lk_3$ data
coincide --- a phenomenon documented across substitution
chemistry \cite{Mikolajczyk2022, LassilaZalatanHerschlag2011},
phosphoryl-transfer biochemistry, and beyond.

\begin{remark}[On stereochemistry]
The stereochemical outcome of substitution at phosphorus
--- whether the configuration at the P-stereocenter is
inverted (as in concerted $\mathrm{S_N2}$-P) or retained
(as can occur in A--E pathways via Berry pseudorotation
at the trigonal-bipyramidal intermediate) --- is a further,
finer forcing observation.
It is invisible not only at $\Lk_3$ but also at $\Lk_4$
as constructed here, because $\mathbf{LGraph}$ records
bond topology but not three-dimensional configuration at
stereocenters.
Stereochemical outcomes force the additional refinement to
$\Lk_{4.5}$ (Chapter~\ref{sec:L45}), which upgrades
$\mathbf{LGraph}$ to carry a $G^\ast = \Aut(G) \ltimes
\mathbb{Z}_2^k$ action encoding permutation-inversion
symmetry at each stereocenter.
At $\Lk_4$ we capture the concerted-versus-stepwise
distinction; $\Lk_{4.5}$ captures the inversion-versus-retention
distinction within either mechanism.
\end{remark}

%% file: chapters/ch_L4.tex
\section{\texorpdfstring{$\Lk_4$}{L4}: The Mechanistic Level}
\label{sec:L4}

\input{chapters/L4/l4_forcing}
\input{chapters/L4/l4_lgraph}
\input{chapters/L4/l4_dpo}
\input{chapters/L4/l4_def}
\input{chapters/L4/l4_u4}
\input{chapters/L4/l4_layer}
\input{chapters/L4/l4_concerted_stepwise}
\input{chapters/L4/l4_sn1sn2}
\input{chapters/L4/l4_nextforcing}

%% file: chapters/L4/l4_forcing.tex
\subsection{Forcing the extension: a structural break in the tower}
\label{sec:L4-forcing-in}

The previous chapter identified the gap at $\Lk_3$: two
reactions with the same source complex, target complex,
enthalpy, entropy, and mass-action propensity --- the latter
agreeing in functional form by stoichiometry and in rate
constant under steady-state on a buried intermediate ---
can proceed through a single transition state, or stepwise
through a discrete intermediate species, yet $\Lk_3$ records
no trace of the distinction.
The cleanest documented case is identity nucleophilic
substitution at phosphorus, where concerted
$\mathrm{S_N2}$-P and stepwise addition--elimination
through a pentacoordinate intermediate are routinely
indistinguishable by bulk kinetics
\cite{Mikolajczyk2022, LassilaZalatanHerschlag2011}.
This section pinpoints the minimal structural change at
$\Lk_4$ required to separate them.

\begin{mathbox}[The conflation is in the underlying category]
At $\Lk_0$ through $\Lk_3$, the underlying category is
always $\Lk_0(P) = \mathbf{FreeSMC}(P)$: morphisms are
stoichiometric transitions $\mathbf{u} \to \mathbf{v}$
generated by reactions of $P$, and the functors $\FH, \FS,
\FP$ \emph{decorate} this category by assigning values to
each morphism.

The Petri net $P$ may contain distinct generators
$r_1, r_2 \in \Rx$ with identical source and target
complexes --- two reactions consuming and producing the same
species in the same quantities, differing only in mechanism.
$\Lk_0(P)$ is therefore not thin: it can carry several
distinct morphisms between the same pair of complexes.

For genuinely competing mechanisms with the same source and
target, identity at $\Lk_3$ obtains whenever all four
decoration values agree.
$\FH$ and $\FS$ depend only on source and target by Hess's
law and agree automatically.
The mass-action functional form of $\FP$ is fixed by the
stoichiometry, hence agrees automatically.
The remaining datum is the rate constant: this agrees when
the rate-limiting step in either mechanism couples the same
molecules in the same orders --- in particular, under
steady-state on a buried intermediate with matching
effective rate constants
\cite{Mikolajczyk2022, LassilaZalatanHerschlag2011}.
The forcing pair below is an idealisation in which all four
$\Lk_3$ decorations agree exactly; this is the regime in
which bulk kinetic measurements cannot distinguish the
mechanisms, and is the regime in which the categorical
indistinguishability witnessed by an element of
$\Aut(\Lk_3(P))$ holds.

The principled content separating $r_1$ from $r_2$ is the
\emph{mechanism}: which bonds break and form, whether the
reaction proceeds through a single transition state or
through a discrete intermediate, and what that intermediate
is.
Encoding this requires replacing $\Lk_0(P)$ with a category
in which morphisms carry bond-level data and intermediates
are themselves species --- a \emph{structural} extension,
not a decoration.
\end{mathbox}

\begin{forcingbox}[The forcing pair for $\Lk_4$]
For the methoxyl exchange at methyl ethylphenylphosphinate
\cite{Mikolajczyk2022},
\[
  r :\;
  {}^{18}\mathrm{OMe^-} +
  \mathrm{Et(Ph)P(=O)OMe}
  \;\longrightarrow\;
  \mathrm{Et(Ph)P(=O){}^{18}OMe}
  + \mathrm{OMe^-},
\]
two distinct generators $r_\text{concerted},
r_\text{stepwise} \in \Rx$ realise the same source-target
transformation:
\begin{itemize}
  \item $r_\text{concerted}$: a single trigonal-bipyramidal
    transition state, simultaneous P--$\mathrm{OMe}$
    cleavage and P--${}^{18}\mathrm{OMe}$ formation.
    No intermediate.
  \item $r_\text{stepwise}$: a discrete pentacoordinate
    intermediate $\mathrm{TBI}$, bracketed by addition and
    elimination transition states.
\end{itemize}

Source and target are identical at the $\Lk_0$ species level
(isotopes are below the species resolution), so $\FH$ and
$\FS$ agree by Hess's law and Layer~2 is automatic
($K_\mathrm{eq} = 1$).
Under steady-state on $\mathrm{TBI}$, both mechanisms have
rate law
$\lambda = k\,x_\mathrm{MeO^-}\,x_\mathrm{substrate}$, and
within experimental resolution the effective rate constants
agree, so $\FP$ takes the same value on the two generators
\cite{Mikolajczyk2022}.
The full $\Lk_3$ tuple
$(\Lk_0, \FH, \FS, F_G^T, \FP)$ therefore takes the same
value on $r_\text{concerted}$ and $r_\text{stepwise}$, and
the swap $r_\text{concerted} \leftrightarrow r_\text{stepwise}$
extends to a category automorphism of $\Lk_3(P)$ preserving
all four decorations --- an element of $\Aut(\Lk_3(P))$.

At $\Lk_4$ the swap does not lift.
The asymmetry is structural: $r_\text{concerted}$ is the
$U_4$-image of a single double pushout (DPO) rule $p_\text{concerted}$ in
$\Lk_4(P)$, whereas the stepwise mechanism is realised at
$\Lk_4$ \emph{only} as the two-step composite
$p_\text{elim} \circ p_\text{add}$ through the intermediate
molecular graph $\mathrm{TBI}$; there is no single DPO rule
$p_\text{stepwise}$ at $\Lk_4$.
A lift $\widetilde\sigma \in \Aut(\Lk_4(P))$ of the swap
$\sigma$ would have to send the generator $p_\text{concerted}$
to a generator of $\Lk_4(P)$ whose $U_4$-image is
$r_\text{stepwise}$; no such generator exists.
Hence $\sigma \notin \mathrm{im}(\varphi_4)$, where $\varphi_4 :
\Aut(\Lk_4(P)) \to \Aut(\Lk_3(P))$ is the forgetful
homomorphism.
Equivalently, $\sigma$ represents a non-trivial coset in the
quotient $\Aut(\Lk_3(P)) / \mathrm{im}(\varphi_4)$, which we
call $\coker(\varphi_4)$:
\[
  \ker\varphi_4 \hookrightarrow \Aut(\Lk_4(P))
  \;\xrightarrow{\;\varphi_4\;}\;
  \Aut(\Lk_3(P))
  \twoheadrightarrow \coker(\varphi_4),
\]
exact as a sequence of pointed sets (the rightmost map is the
quotient by $\mathrm{im}(\varphi_4)$, taken as a coset space;
when $\mathrm{im}(\varphi_4)$ is normal the cokernel is a
group).
The same pattern recurs across nucleophilic substitution at
heteroatoms, biological phosphoryl transfer, and a wide
class of identity reactions
\cite{Mikolajczyk2022, LassilaZalatanHerschlag2011};
the explicit verification appears in
Proposition~\ref{prop:concerted-neq-stepwise}.
\end{forcingbox}

\medskip\noindent\textbf{Structural rather than decorating
extension.}
Earlier transitions added a functor while keeping
$\Lk_0(P)$ fixed --- decorator extensions.
At $\Lk_4$ the underlying category itself changes.
The tower is not forked: the $\Lk_1$--$\Lk_3$ decorations
are retained at $\Lk_4$ by pullback along a forgetful
functor $U_4 : \Lk_4(P) \to \Lk_0(P)$
(Section~\ref{sec:L4-u4}), with
\[
  F_H^{(4)} := \FH \circ U_4,
  \quad
  F_S^{(4)} := \FS \circ U_4,
  \quad
  F_P^{(4)} := \FP \circ U_4
\]
(Proposition~\ref{prop:L4-lifting}).
What is new is not an additional decoration but a refinement
of the morphisms: the DPO structure is intrinsic, not a
functor on $\Lk_0(P)$.

\medskip\noindent\textbf{What minimal structure resolves
the ambiguity.}

\begin{enumerate}[label=(\roman*)]
  \item\textbf{Molecular graphs as objects.}
    Each species becomes a labelled hypergraph: vertices
    carry atom type, isotope, charge, lone-pair count, and
    radical count; edges carry bond order
    \cite{AndersenFlammMerkleStadler2016,
    EhrigEhrigPrangeTaentzer2006}.
    Reaction intermediates such as $\mathrm{TBI}$ become
    first-class species.
  \item\textbf{DPO derivations as morphisms.}
    A reaction mechanism is a span
    $L \leftarrow K \rightarrow R$ in $\mathbf{LGraph}$,
    specifying which bonds break and form
    \cite{EhrigPfenderSchneider1973, LackSobocinski2005};
    a multi-step mechanism is a sequential composite
    through intermediate molecular graphs.
\end{enumerate}

\noindent
Together these define $\Lk_4(P)$, the free SMC on DPO
spans in $\mathbf{LGraph}$ (Section~\ref{sec:L4-def}).
$r_\text{concerted}$ is the $U_4$-image of a single DPO
generator $p_\text{concerted}$; the stepwise mechanism is
realised as the composite $p_\text{elim} \circ p_\text{add}$
through $\mathrm{TBI}$, with no single DPO generator
$p_\text{stepwise}$ at $\Lk_4$.
This asymmetry between $\Lk_0$ (where $r_\text{stepwise}$ is
postulated as a parallel generator) and $\Lk_4$ (where it is
not realised by any single generator) is what obstructs the
lift of the swap to $\Aut(\Lk_4(P))$
(Proposition~\ref{prop:concerted-neq-stepwise}).

\begin{remark}[Examples beyond the forcing pair]
\label{rem:four-mechanisms}
For a haloalkane $\mathrm{RX}$ with nucleophile/base
$\mathrm{Nu^-}/\mathrm{B^-}$, four classical mechanisms
are possible \cite{March1992, AtkinsDeP2014}:

\medskip
\begin{center}
\renewcommand{\arraystretch}{1.4}
\begin{tabular}{p{1.4cm} p{2.6cm} p{3.4cm} p{4.3cm}}
  \hline
  \textbf{Mech.} & \textbf{Rate law} &
  \textbf{Steps} & \textbf{Favoured by}\\
  \hline
  $\mathrm{S_N2}$ &
  $k[\mathrm{RX}][\mathrm{Nu}]$ &
  One (concerted) &
  1$^\circ$ substrate; strong small nucleophile;
  polar aprotic solvent; low $T$\\[4pt]
  $\mathrm{S_N1}$ &
  $k[\mathrm{RX}]$ &
  Two (carbocation) &
  3$^\circ$ substrate; polar protic solvent;
  weak nucleophile; good leaving group\\[4pt]
  $\mathrm{E2}$ &
  $k[\mathrm{RX}][\mathrm{B}]$ &
  One (concerted) &
  Any degree; strong bulky base; high $T$\\[4pt]
  $\mathrm{E1}$ &
  $k[\mathrm{RX}]$ &
  Two (carbocation) &
  3$^\circ$ substrate; polar protic solvent;
  weak base; high $T$\\
  \hline
\end{tabular}
\end{center}

\medskip
The substitution/elimination split is resolved already at
$\Lk_0$ (different products).
The bimolecular/unimolecular split within each class
($\mathrm{S_N2}$ vs.\ $\mathrm{S_N1}$, $\mathrm{E2}$ vs.\
$\mathrm{E1}$) corresponds to different rate-law forms,
so it is resolved already at $\Lk_3$ --- not a
$\coker(\varphi_4)$ pair, but a useful illustration of
how $\Lk_4$ realises the distinction structurally
(carbocation as intermediate species, two DPO spans
versus one).
\end{remark}

%% file: chapters/L4/l4_lgraph.tex
\subsection{Labelled molecular graphs:
  the ambient adhesive category}
\label{sec:L4-lgraph}

Section~\ref{sec:L4-forcing-in} identified two generators
$r_\text{concerted}, r_\text{stepwise} \in \Rx$ with
identical $\Lk_3$ data, distinguished only by their identity
tags in the Petri net.
The tower up to $\Lk_3$ accepts them as distinct only by
fiat: no categorical structure on $\Lk_0(P) =
\mathbf{FreeSMC}(P)$ supplies a principled reason why they
are different morphisms.
The resolution requires a new underlying category whose
morphisms carry bond-level information --- atoms, bonds,
formal charges, electron counts --- so that two generators
projecting to the same stoichiometric transition can
nonetheless be distinct morphisms by virtue of their
internal structure.
This section constructs that category in three steps: the
label algebra that replaces species names with molecular
graphs (\S\ref{sec:L4-labels}), the category $\LGraph$ of
such graphs with its monoidal and adhesive structure
(\S\ref{sec:L4-lgraph-cat}--\S\ref{sec:L4-adhesive}), and
the grounded variant $\LGraphP$ that connects molecular
graphs back to the species set $\Sp$ of the Petri net~$P$
(\S\ref{sec:L4-lgraphP}).

\subsubsection{Label algebra and chemical validity}
\label{sec:L4-labels}

At $\Lk_0(P)$ through $\Lk_3(P)$, an object is an element of
$\NN^{|\Sp|}$: a formal sum of species names $S_i$, recording
\emph{how many} molecules of each type are present but
recording nothing about their internal structure.
To distinguish mechanisms with the same source and target
complexes --- whether already separated at $\Lk_3$ or not
--- we need objects that encode atoms, bonds, formal
charges, and electron pairs.
The following label algebra does precisely this.

The label set below is not original: it is a categorical
reformulation of the data encoded in the \emph{bond-electron}
(BE) matrix of Dugundji and Ugi~\cite{DugundjiUgi1973}, 
the standard matrix representation of molecular structure in
cheminformatics and computer-aided synthesis planning since
its introduction in 1973, in continuous
use~\cite{AndersenFlammMerkleStadler2013,
DobbelaereEtAl2024RxnInsight}, and the basis of some modern
machine-learning approaches to reaction prediction that
enforce electron conservation~\cite{JoungEtAl2025FlowER}.
What is new here is the organisation of that data into a
quintuple
$(\mathrm{el}, \mathrm{iso}, q, \rho, \ell)$ that serves as a
vertex label in a typed graph, making the connection to the
DPO rewriting framework canonical.
Chemical graph transformation with typed molecular graphs is
developed in detail in Andersen et al.\
\cite{AndersenFlammMerkleStadler2013, AndersenFlammMerkleStadler2016};
our formulation is tailored to make the validity condition
explicit and to sit naturally inside the adhesive category
framework of Section~\ref{sec:L4-adhesive}.

\begin{definition}[Atom label and bond label]
\label{def:labels}
  Let $\Lambda_V$ denote the set of all \emph{atom labels},
  where each atom label is a quintuple
  $\lambda_v = (\mathrm{el}, \mathrm{iso}, q, \rho, \ell)$
  consisting of:
  \begin{itemize}
    \item $\mathrm{el} \in \{\mathrm{H, C, N, O, F, P, S, Cl,
      Br, I, \ldots}\}$: element type;
    \item $\mathrm{iso} \in \mathrm{Iso}(\mathrm{el}) \cup
      \{\ast\}$: nuclear-mass label
      (e.g.\ ${}^{16}\mathrm{O}$, ${}^{18}\mathrm{O}$) or
      the wildcard $\ast$ for natural abundance / unspecified;
    \item $q \in \ZZ$: formal charge;
    \item $\rho \in \{0, 1, 2\}$: radical electron count
      ($0$ closed-shell, $1$ monoradical, $2$ diradical);
    \item $\ell \in \NN$: lone-pair (nonbonding electron-pair)
      count.
  \end{itemize}
  A \emph{bond label} is an integer $b \in \{1, 2, 3\}$
  encoding bond order (single, double, triple).
  Absent bonds carry the implicit value $b = 0$.
\end{definition}

\begin{remark}[Isotope labels at $\Lk_4$ as classical tracers]
\label{rem:isotope-at-L4}
The component $\mathrm{iso}$ records nuclear identity
\emph{classically}: it tracks ${}^{16}\mathrm{O}$ vs.\
${}^{18}\mathrm{O}$, ${}^{12}\mathrm{C}$ vs.\
${}^{13}\mathrm{C}$, etc., as label data attached to vertices.
This is what is needed to formulate identity-substitution
forcing pairs (Section~\ref{sec:L4-forcing-in}, where the
incoming methoxide is tagged ${}^{18}\mathrm{O}$).
What is postponed to $\Lk_7$ is not isotope bookkeeping but
\emph{isotope-dependent quantum phenomena}: nuclear quantum
statistics, tunnelling, vibrational zero-point kinetic
isotope effects.
The wildcard $\ast$ is the default for atoms whose chemistry
is isotope-independent.
\end{remark}

The lone-pair count $\ell$ is part of the atom label, not a
derived quantity computed from the other components.
A universal formula
$\ell = \tfrac{1}{2}(v_{\mathrm{el}} - q - \rho - \sum b)$
breaks for hypervalent main-group atoms: in the central
forcing example of Section~\ref{sec:L4-forcing-in}, the
pentacoordinate phosphorus intermediate (TBI) carries an
incident bond-order sum of six, and the naive formula assigns
it $\ell = -\tfrac{1}{2}$.
Carrying $\ell$ as part of the label, with chemical validity
imposed as a per-element \emph{predicate}, accommodates
ordinary octet atoms, hypervalent main-group atoms (P, S, I),
and transition-metal centres alike.

\begin{definition}[Chemical validity]
\label{def:valence}
  Fix a per-element \emph{allowed-valence specification}
  $\mathcal{V}$: a function assigning to each element type
  $\mathrm{el}$ a set $\mathcal{V}(\mathrm{el})$ of allowed
  tuples $(q, \rho, \ell, \beta)$ with $\beta \in \NN$ an
  admissible incident bond-order sum.
  For ordinary octet atoms, $\mathcal{V}(\mathrm{el})$ is the
  set of tuples satisfying the row-sum law
  \begin{equation}
    2\ell + \rho + \beta \;=\; v_{\mathrm{el}} - q,
    \label{eq:row-sum}
  \end{equation}
  with $v_{\mathrm{el}}$ the valence-electron count of
  $\mathrm{el}$ ($v_{\mathrm{el}} = 4$ for C, $5$ for N,
  $6$ for O, etc.).
  For hypervalent main-group atoms (pentacoordinate or
  hexacoordinate P, hypervalent S, hypervalent I),
  $\mathcal{V}(\mathrm{el})$ extends the octet specification
  to admit higher coordination numbers consistent with
  three-centre four-electron bonding.
  For transition-metal centres, $\mathcal{V}(\mathrm{el})$
  admits the coordination geometries of $d$-orbital chemistry.

  An atom $v$ in a typed graph is \emph{chemically valid}
  if its label
  $(\mathrm{el}, \mathrm{iso}, q, \rho, \ell)$ together with
  $\beta(v) := \sum_{e \ni v} b(e)$ satisfies
  $(q, \rho, \ell, \beta(v)) \in \mathcal{V}(\mathrm{el})$.
  A typed graph is \emph{chemically valid} if every atom is.
\end{definition}

\begin{remark}[Octet recovery and DPO bookkeeping]
\label{rem:lp-derived}
For ordinary octet atoms, $\mathcal{V}(\mathrm{el})$ is the
solution set of \eqref{eq:row-sum}, so any of $q, \rho, \ell,
\beta$ is determined by the other three.
Elementary DPO generators acting on octet atoms can therefore
be specified by changes in $(q, \rho, \beta)$ alone, with
$\ell$ updating automatically by \eqref{eq:row-sum}; the
elementary generators $g_1$--$g_6$ of
Section~\ref{sec:L4-dpo} use this convention.
At hypervalent vertices the row-sum law no longer determines
$\ell$ uniquely and the explicit attribute is essential.
\end{remark}

The diagonal entry of the Dugundji--Ugi bond-electron (BE)
matrix coincides with $2\ell + \rho$ at each atom, and the
row-sum law of \eqref{eq:row-sum} is precisely the BE-matrix
row-sum.
The following mathbox records the BE-matrix as the
\emph{matrix shadow} of a labelled graph, and notes what
this shadow captures and what it does not.

\begin{mathbox}[The BE-matrix shadow of a labelled graph]
\textbf{The BE-matrix.}
For a molecule with $n$ atoms, the \emph{bond-electron (BE)
matrix} $B \in \NN^{n \times n}$ introduced by Dugundji and
Ugi~\cite{DugundjiUgi1973} is the symmetric matrix with entries:
\[
  B_{ij} \;:=\;
  \begin{cases}
    b(e_{ij}) & i \neq j,\\
    2\ell(i) + \rho(i) & i = j,
  \end{cases}
\]
where $e_{ij}$ denotes the bond between atoms $i$ and $j$
($b(e_{ij}) = 0$ if no bond exists).
The diagonal $B_{ii}$ counts twice the lone-pair count plus
the radical count at atom $i$; the row-sum law
$\sum_j B_{ij} = v_{\mathrm{el}}(i) - q(i)$ for every atom
$i$ is the chemical validity condition of
Definition~\ref{def:valence} restricted to octet atoms.

\textbf{Shadow, not equivalence.}
The atom label
$(\mathrm{el}, \mathrm{iso}, q, \rho, \ell)$ together with
the bond orders determines $B$ unambiguously, so $B$ is a
faithful matrix representation of the bond and electron-count
data of any single labelled graph.
But $B$ does not record:
isotope labels (${}^{16}\mathrm{O}$ vs.\ ${}^{18}\mathrm{O}$
have the same $B$);
atom-tracking under composition (a sequence of reactions
composes by adding successive reaction matrices, where each
reaction matrix is itself a difference $R_{\mathrm{BE}} =
B' - B$, but this addition does not record which atom in
one matrix corresponds to which atom in the next --- the
correspondence is supplied categorically by the interface
$K$ and the match $m$);
and the categorical match data of a DPO derivation.
We therefore call $B$ the \emph{BE-matrix shadow} of the
labelled graph: a useful local matrix representation, but
strictly less than the full DPO content.

\textbf{What the label set distinguishes.}
Each mechanistic distinction below is expressed as a condition
on $(\mathrm{el}, \mathrm{iso}, q, \rho, \ell)$ or on the
graph topology:
\begin{itemize}
  \item \emph{Concerted vs.\ stepwise at heteroatom centres}
    (the $\Lk_3 \to \Lk_4$ forcing pair,
    Section~\ref{sec:L4-forcing-in}): the stepwise derivation
    passes through a discrete intermediate graph whose central
    vertex carries a coordination number not realised in any
    concerted derivation projecting to the same $\Lk_3$
    observables (\emph{e.g.}\ the pentacoordinate TBI in
    identity substitution at phosphorus).
    Visible via the vertex degree in the intermediate graph.
  \item \emph{Charged intermediates} (already distinguished at
    $\Lk_3$ for $\mathrm{S_N1}$/$\mathrm{S_N2}$ and
    $\mathrm{E1}$/$\mathrm{E2}$, see
    Remark~\ref{rem:four-mechanisms}): the stepwise derivation
    passes through a graph with a carbocation vertex
    ($q = +1$, $\ell(v) = 0$); the concerted derivation has
    $q = 0$ throughout.  Visible via the $q$ label.
  \item \emph{Radical vs.\ ionic mechanisms}: a radical
    intermediate carries $\rho(v) \geq 1$ on at least one
    atom (monoradicals at one site, diradicals at one site or
    distributed across two sites); an ionic intermediate has
    $\rho(v) = 0$ everywhere.
    Visible via the $\rho$ label.
  \item \emph{Pericyclic vs.\ ionic reactions}: a pericyclic
    DPO rule has a reaction centre with cyclic graph topology
    (the atoms forming and breaking bonds lie on a cycle in
    $L$); an ionic reaction has acyclic reaction-centre
    topology.  Visible via the topology of $L$.
\end{itemize}
\end{mathbox}

\subsubsection{The category \texorpdfstring{$\LGraph$}{LGraph}}
\label{sec:L4-lgraph-cat}

The label algebra of the preceding section gives data for
individual atoms and bonds.
To speak categorically about reactions, we need molecular graphs
to be \emph{objects} and structure-preserving maps between them
to be \emph{morphisms}.
The following definition assembles these ingredients into a
symmetric monoidal category $\LGraph$, which will serve as the
ambient category for the DPO rules that generate $\Lk_4(P)$.

\textbf{Originality and comparison to existing work.}
The use of typed molecular graphs in chemical graph transformation
goes back to Ehrig et al.~\cite{EhrigEhrigPrangeTaentzer2006}
and Andersen et al.~\cite{AndersenFlammMerkleStadler2016}.
Those works use a general typing morphism to a type graph,
without committing to a specific label algebra; our specialisation
to the quintuple $(\mathrm{el}, \mathrm{iso}, q, \rho, \ell)$
of Definition~\ref{def:labels} is what makes $\LGraph$ a
suitable ambient category for mechanistic organic chemistry.
The architectural choice that distinguishes our development
is to keep $\LGraph$ as the \emph{ambient} typed-graph
category (adhesive in the sense of
Lack--Soboci\'{n}ski~\cite{LackSobocinski2005},
Proposition~\ref{prop:lgraph-adhesive}), with chemical
validity (Definition~\ref{def:valence}) imposed as a
\emph{predicate} on objects rather than as a defining feature
of category membership.
This separation matters because pushouts of chemically valid
graphs in $\LGraph$ may produce chemically invalid graphs:
adhesivity holds for the ambient typed-graph category, while
validity is enforced as an admissibility condition on rule
patterns and their products
(Remark~\ref{rem:validity-as-admissibility} below).

\begin{definition}[Labelled typed graph]
\label{def:lgraph}
  A \emph{labelled typed graph} is a tuple
  $G = (V, E, \lambda_V, \lambda_E)$ where:
  \begin{itemize}
    \item $V$ is a finite set of \emph{vertices} (atoms).
    \item $E \subseteq \bigl\{\{u, v\} : u, v \in V,\, u \neq v\bigr\}$
      is a set of \emph{edges} (bonds), i.e.\ a set of
      unordered pairs of distinct vertices.
    \item $\lambda_V : V \to \Lambda_V$ is the \emph{atom-label
      function}, assigning to each vertex an atom label
      $(\mathrm{el}, \mathrm{iso}, q, \rho, \ell) \in \Lambda_V$
      in the sense of Definition~\ref{def:labels}.
    \item $\lambda_E : E \to \{1, 2, 3\}$ is the
      \emph{bond-order function}.
  \end{itemize}
  No chemical-validity constraint is imposed on $G$ at the
  level of object membership; chemical validity
  (Definition~\ref{def:valence}) is a predicate on objects
  of $\LGraph$ used as an admissibility condition on rule
  patterns (Remark~\ref{rem:validity-as-admissibility}).
\end{definition}

A chemically valid graph (Definition~\ref{def:valence})
determines a BE-matrix unambiguously.
The reverse direction is partial: a BE-matrix together with
element types and isotope labels at each vertex, plus either
closed-shell assumption ($\rho \equiv 0$) or independent
specification of $\rho$ at each vertex, recovers the labelled
graph; without that supplementary data, the diagonal entries
$B_{ii} = 2\ell + \rho$ do not uniquely split into
$(\ell, \rho)$ (for example $B_{ii} = 2$ admits both
$(\ell, \rho) = (1, 0)$ and $(0, 2)$).
In general $\LGraph$ admits non-valid labelled graphs as well,
which arise when one considers pushouts of valid graphs.
To make reactions into morphisms, we now equip the labelled
graphs with the structure of a category.

\begin{definition}[The category $\LGraph$]
\label{def:LGraph}
  The category $\LGraph$ has:
  \begin{itemize}
    \item \textbf{Objects}: labelled typed graphs
      (Definition~\ref{def:lgraph});
    \item \textbf{Morphisms}: label-preserving graph
      \emph{monomorphisms}: injective maps
      $f : V_G \hookrightarrow V_H$
      such that $\lambda_V^H(f(v)) = \lambda_V^G(v)$
      for all $v \in V_G$, and
      $\lambda_E^H(\{f(u), f(v)\}) = \lambda_E^G(\{u, v\})$
      for all edges $\{u,v\} \in E_G$;
    \item \textbf{Composition}: composition of injections;
    \item \textbf{Identity}: the identity injection $\id_G$.
  \end{itemize}
  The \emph{monoidal product} is the disjoint union
  $G_1 \sqcup G_2$ (disjoint vertex and edge sets, labels
  inherited), with unit the empty graph $\emptyset$.
  We work in a strictified skeleton in which disjoint copies
  are fixed by a chosen tagging convention, so that
  $\sqcup$ is strictly associative and unital.
\end{definition}

\begin{proposition}[$\LGraph$ is a well-defined strict symmetric
  monoidal category]
\label{prop:LGraph-cat}
  $\LGraph$ with the structure of Definition~\ref{def:LGraph}
  is a strict symmetric monoidal category.
\end{proposition}

\begin{proof}
\textbf{Category.}
Composition of label-preserving monomorphisms is again a
label-preserving monomorphism (injectivity and label preservation
are both stable under composition).
The identity injection $\id_G$ is label-preserving by definition,
and serves as the two-sided identity.
Associativity of composition is inherited from function composition.

\textbf{Strict monoidal.}
By the strictification convention adopted in
Definition~\ref{def:LGraph}, disjoint copies are fixed by a
chosen tagging, so $\sqcup$ is strictly associative
($(G_1 \sqcup G_2) \sqcup G_3 = G_1 \sqcup (G_2 \sqcup G_3)$
as objects) and strictly unital ($G \sqcup \emptyset = G$).
The symmetry isomorphism
$G_1 \sqcup G_2 \cong G_2 \sqcup G_1$ is the swap of tagged
copies, natural in $G_1$ and $G_2$.
\end{proof}

\begin{remark}[Restriction to monomorphisms]
\label{rem:mono}
Working with label-preserving monomorphisms rather than arbitrary
label-preserving graph homomorphisms is standard in chemical graph
transformation~\cite{AndersenFlammMerkleStadler2016}: it ensures
that atoms are not identified and bond structures are faithfully
embedded.
Concretely, a non-injective map could collapse two distinct atoms
of the same element type into one, destroying structural
information.
All results cited from the DPO literature hold for injective
matches~\cite{EhrigEhrigPrangeTaentzer2006}.
\end{remark}

\subsubsection{Adhesivity of \texorpdfstring{$\LGraph$}{LGraph}}
\label{sec:L4-adhesive}

The category $\LGraph$ carries more than just a monoidal
structure: it is \emph{adhesive} in the sense of Lack and
Soboci\'{n}ski~\cite{LackSobocinski2005}, a property that
makes DPO rewriting internally consistent.

\textbf{Adhesive categories.}
A category $\mathcal{C}$ is \emph{adhesive}
\cite{LackSobocinski2005} if:
(i)~it has pushouts along monomorphisms;
(ii)~monomorphisms are stable under pushout (the pushout of a
monomorphism is again a monomorphism); and
(iii)~pushout squares along monomorphisms satisfy the
\emph{van Kampen condition}: they are also pullbacks in a
suitable double-categorical sense.
Condition (iii), the van Kampen condition, is what makes DPO
rewriting well-behaved: combined with conditions (i) and (ii),
it guarantees that pushout complements are unique up to
isomorphism (so the gluing context $D$ in a DPO step is
well-defined; see Remark~\ref{rem:D-not-intermediate} on
why $D$ is not a chemical intermediate) and that independent
rewriting steps can be applied in either order with the same
result (Local Church--Rosser).
Without adhesivity, neither of these properties is guaranteed
and DPO rewriting in $\LGraph$ would be ill-defined.

\begin{proposition}[$\LGraph$ is adhesive]
\label{prop:lgraph-adhesive}
  The category $\LGraph$ with label-preserving monomorphisms
  is adhesive in the sense of
  Lack and Soboci\'{n}ski.
\end{proposition}

\begin{proof}
The category $\mathbf{Graph}$ of (directed or undirected)
graphs can be presented as a presheaf topos: it is the
functor category $[\mathbf{G}, \mathbf{Set}]$ where
$\mathbf{G}$ is the graph schema $s, t : E \rightrightarrows V$
(two objects and two parallel morphisms, encoding source and
target of edges).
Every presheaf topos is adhesive
\cite[Corollary~3.6]{LackSobocinski2005}, so $\mathbf{Graph}$
is adhesive.

Now identify $\LGraph$ with the slice category
$\mathbf{Graph}/TG_\Lambda$, where $TG_\Lambda$ is the
\emph{type graph}: the graph whose vertices are the atom labels
$\Lambda_V$ and whose edges carry all possible bond labels
$\{1,2,3\}$ between every pair of label vertices.
A labelled typed graph $G$ in the sense of
Definition~\ref{def:lgraph} is precisely a graph homomorphism
$G \to TG_\Lambda$ (the typing morphism assigns each atom its
label and each bond its order); $\LGraph \cong
\mathbf{Graph}/TG_\Lambda$ as categories.
This identification is standard in typed graph transformation;
see Ehrig et al.~\cite[Chapter~2]{EhrigEhrigPrangeTaentzer2006}.

By Lack--Soboci\'{n}ski~\cite[Proposition~3.5(ii)]{LackSobocinski2005},
if $\mathcal{C}$ is adhesive then so is $\mathcal{C}/X$ for
every object $X$.
Since $\mathbf{Graph}$ is adhesive and
$\LGraph \cong \mathbf{Graph}/TG_\Lambda$,
it follows that $\LGraph$ is adhesive.
\end{proof}

The three consequences of adhesivity for mechanism-level
rewriting deserve explicit statement.
Adhesivity is what makes DPO derivations well-defined: without
unique pushout complements, the gluing context in a DPO step
would not be well-defined; without Local Church--Rosser, the
order of independent steps would matter and sequential
composition would be ambiguous.
These are not abstract concerns --- they correspond to
concrete questions about mechanism: ``Is the carbocation
produced by ionisation in $\mathrm{S_N1}$ uniquely determined
by the rule and the substrate?'' (yes, by uniqueness of
gluing contexts and by the fact that the carbocation is a
distinct molecular graph between two derivations); ``Does
it matter whether we first abstract the proton or first
ionise the leaving group in an E1 mechanism?''
(Local Church--Rosser says no, when the steps are
independent).

\begin{chembox}[What adhesivity gives for chemistry]
Proposition~\ref{prop:lgraph-adhesive} guarantees
three essential properties for DPO rewriting of labelled
typed graphs:
\begin{enumerate}[label=(\roman*)]
  \item \textbf{Unique gluing contexts}: if a DPO rule
    applies to a labelled graph, the gluing context $D$
    (the host with the reaction-centre bonds removed, prior
    to gluing of $R$) is unique up to isomorphism
    (Lemma~4.5 of \cite{LackSobocinski2005}).
    $D$ is a categorical artefact of a single rule
    application, not, in general, a chemically isolable
    intermediate (Remark~\ref{rem:D-not-intermediate}).
  \item \textbf{Local Church--Rosser}: two \emph{independent}
    reaction steps (steps whose reaction centres share no atoms)
    applied to the same molecule can be sequenced in either
    order, reaching the same product
    (Theorem~7.7 of \cite{LackSobocinski2005}).
    This is the mechanism-level analogue of the fact that
    parallel reactions in distinct parts of a molecule do not
    interfere.
  \item \textbf{Concurrency}: a sequence of independent DPO
    steps can be represented by a single composite rule
    (Theorem~7.11 of \cite{LackSobocinski2005}).
    This is a semantic statement about independent steps,
    not the basis of the free SMC structure of $\Lk_4(P)$,
    which is by construction (Section~\ref{sec:L4-def}).
\end{enumerate}
Together, these guarantee that mechanism-level rewriting in
$\LGraph$ is well-defined and compositional --- exactly what
is needed for $\Lk_4(P)$ to be a category.
\end{chembox}

\begin{remark}[Chemical validity is admissibility, not membership]
\label{rem:validity-as-admissibility}
Pushouts of chemically valid graphs in $\LGraph$ may produce
chemically invalid graphs: a pushout that introduces a bond
at an atom already saturated produces a graph whose row-sum
law fails at that atom.
The full subcollection
$\LGraph^{\mathrm{ch}} \subset \LGraph$ of chemically valid
graphs (Definition~\ref{def:valence}) is therefore not closed
under pushouts in $\LGraph$ and is not adhesive in its own
right.
The architectural choice taken in this chapter is to perform
all DPO constructions in the ambient adhesive category
$\LGraph$ and to require chemical validity as a separate
\emph{admissibility} condition: a chemical DPO rule
(Section~\ref{sec:L4-dpo}) is a span $L \leftarrow K
\rightarrow R$ in $\LGraph$ with $L$ and $R$ chemically valid,
and a chemically admissible match $m: L \hookrightarrow G$
into a valid host $G$ requires the resulting product graph $H$
to be chemically valid.
The validity check is local --- the row-sum law (or its
hypervalent extension) at every atom of the product --- and is
the content of Layer~2 at $\Lk_4$
(Section~\ref{sec:L4-layer}).
\end{remark}

\begin{remark}[The gluing context $D$ is not a chemical
  intermediate]
\label{rem:D-not-intermediate}
The pushout complement $D$ supplied by adhesivity is the host
graph $G$ with the reaction-centre bonds of $L \setminus l(K)$
deleted, prior to gluing of $R$.
$D$ is a categorical construction internal to a single DPO
step and is in general \emph{not} a chemically isolable
intermediate.
A chemical intermediate, when present, is a complete
molecular graph appearing as the \emph{target} of one DPO
derivation and the \emph{source} of the next: the
carbocation in $\mathrm{S_N1}$
(Section~\ref{sec:L4-sn1sn2}), the trigonal-bipyramidal
phosphorus intermediate in stepwise identity substitution
(Section~\ref{sec:L4-concerted-stepwise}).
The categorical role of $D$ is to mediate the two pushouts
of a single rule application; its chemical interpretation,
when it has one, is a transient gluing context, not a
species of the network.
\end{remark}

\subsubsection{The grounded category
  \texorpdfstring{$\LGraphP$}{LGraph\_P}}
\label{sec:L4-lgraphP}

The category $\LGraph$ treats labelled typed graphs as abstract
combinatorial objects with no reference to the Petri net $P$.
But $\Lk_4(P)$ must sit in the tower above $\Lk_0(P)$: its
host objects must be compatible with the stoichiometric
generator set of $P$, so that the forgetful functor
$U_4 : \Lk_4(P) \to \Lk_0(P)$ (Section~\ref{sec:L4-u4}) is
well-defined.
The grounded category $\LGraphP$ enforces this compatibility:
it equips every host graph with an assignment of its
connected components to species names in $\Sp$, binding the
abstract graph structure to the chemical species of the
network.

\begin{definition}[Grounded labelled graph category $\LGraphP$]
\label{def:LGraphP}
  Given a Petri net $P$ with species set $\Sp$, the
  \emph{grounded} category $\LGraphP$ has:
  \begin{itemize}
    \item \textbf{Objects}: pairs $(G, \spec)$ where
      $G \in \LGraph$ and
      $\spec: \pi_0(G) \to \Sp$ assigns each connected component
      of $G$ to a species in $\Sp$.
    \item \textbf{Morphisms}: label-preserving monomorphisms
      $f: G \hookrightarrow H$ compatible with
      $\spec$ (each component of $G$ maps into the
      component of $H$ bearing the same species label).
  \end{itemize}
  The monoidal product is grounded disjoint union;
  the unit is the empty graph with vacuous $\spec$.
  The category $\LGraphP$ is adhesive; the proof reduces to
adhesivity of $\LGraph$ via a comma-category argument analogous
to the slice argument for $\LGraph$ itself (Ehrig et
al.~\cite{EhrigEhrigPrangeTaentzer2006}).
\end{definition}

\begin{insightbox}[The BE-matrix shadow of a DPO rule]
A DPO rule $(L \leftarrow K \rightarrow R)$ in $\LGraphP$
admits a \emph{BE-matrix shadow}: a pair $(B, B')$ of
BE-matrices~\cite{DugundjiUgi1973} related by the reaction
matrix $R_{\mathrm{BE}} = B' - B$
\cite{AndersenFlammMerkleStadler2013, AndersenFlammMerkleStadler2017}.
The reactant graph $L$ supplies $B$ (off-diagonal entries =
bond orders, diagonal entries = $2\ell(v) + \rho(v)$); the
product graph $R$ supplies $B'$; the context $K$ identifies
the rows and columns on which $R_{\mathrm{BE}}$ vanishes.

The shadow is faithful for net electron rearrangement but
strictly weaker than the DPO rule.
What it does not record:
isotope labels (${}^{16}\mathrm{O}$ and ${}^{18}\mathrm{O}$
share the same matrix);
atom-tracking under composition (BE matrices add, but the
addition does not say which atom in one matrix corresponds to
which atom in the next --- supplied categorically by $K$ and
the match $m$);
the species-grounding $\spec$;
and the categorical match data of a derivation.
The DPO framework on $\LGraphP$ is therefore strictly richer
than the matrix shadow.
\end{insightbox}

With $\LGraphP$ in hand, the three ingredients needed for
$\Lk_4(P)$ are in place: host graphs are $\Sp$-typed
labelled graphs in $\LGraphP$, the generating morphisms are
\emph{admissible chemical DPO rules}
(chemically valid spans, Section~\ref{sec:L4-dpo}), and the
ambient adhesive structure of $\LGraph$ guarantees that DPO
derivations are well-defined.
The free strict SMC structure of $\Lk_4(P)$ is supplied by
the universal property of the free SMC on the directed
signature of admissible DPO derivations
(Section~\ref{sec:L4-def}).

%% file: chapters/L4/l4_dpo.tex
\subsection{DPO rules and derivations in
  \texorpdfstring{$\LGraphP$}{LGraph\_P}}
\label{sec:L4-dpo}

Section~\ref{sec:L4-lgraph} built the ambient category $\LGraphP$:
a strict symmetric monoidal adhesive category whose objects are
$\Sp$-typed labelled graphs and whose morphisms are
label-preserving monomorphisms.
Chemical validity (Definition~\ref{def:valence}) is enforced
as an admissibility condition on rule patterns and matches,
not as a defining feature of object membership
(Remark~\ref{rem:validity-as-admissibility}).
The morphisms of $\Lk_4(P)$ will be sequences of \emph{DPO
derivations} in $\LGraphP$ --- the precise way of saying
``bonds break and form according to a specified rule.''
A DPO (\emph{double pushout}) derivation specifies a reaction
by two pushout squares in $\LGraphP$, encoding the following
chemical information:
\begin{itemize}
  \item $L$ (left graph) --- the \emph{reactant pattern}: the
    subgraph that must be present in the host molecule for the
    rule to fire; contains the reaction-centre atoms with their
    pre-reaction labels and the bonds to be broken.
  \item $K$ (context graph) --- the \emph{unchanged frame}:
    atoms and bonds that persist through the reaction, embedded
    in both $L$ and $R$ by label-preserving monomorphisms.
  \item $R$ (right graph) --- the \emph{product pattern}: the
    subgraph produced by the reaction; contains the reaction-centre
    atoms with their post-reaction labels and the bonds formed.
  \item $G$ (host graph) --- the \emph{substrate molecule}: the
    full molecular graph to which the rule is applied.
  \item $D$ (derived context) --- the \emph{gluing context}:
    $G$ with the atoms and bonds of $m(L \setminus l(K))$
    removed; the unique pushout complement guaranteed by the
    adhesivity of $\LGraphP$.
    $D$ is a categorical artefact of a single rule application
    and is not, in general, a chemically isolable intermediate
    (Remark~\ref{rem:D-not-intermediate}).
  \item $H$ (host graph after reaction) --- the \emph{product
    molecule}: obtained by gluing $R$ onto $D$ via the second
    pushout.
\end{itemize}
Together the two pushouts implement the chemical intuition that
a reaction (i) removes the bonds and relabels the atoms of the
reaction centre, and (ii) installs the new bonds and relabelled
atoms, leaving everything outside the reaction centre unchanged.
This is the categorical replacement for the stoichiometric
transition used as a morphism at $\Lk_0(P)$: instead of merely
recording which species disappear and appear, a DPO derivation
records \emph{exactly which bonds break and form, in what
sequence, and on which atoms}.
The free SMC on such derivations is $\Lk_4(P)$,
constructed in Section~\ref{sec:L4-def}.

\subsubsection{DPO rules}

\begin{definition}[Chemical DPO rule]
\label{def:dpo-rule}
  A \emph{chemical DPO rule} for $P$-species is a pair
  $p = (s, \gamma)$ consisting of:
  \begin{itemize}
    \item a \emph{span}
      $s = (L \xleftarrow{\;l\;} K \xrightarrow{\;r\;} R)$
      in $\LGraphP$, where both $l: K \hookrightarrow L$ and
      $r: K \hookrightarrow R$ are label-preserving monomorphisms;
    \item a \emph{generator label}
      $\gamma(p) \in P$ identifying which Petri-net reaction
      of $P$ this rule realises at the stoichiometric level.
  \end{itemize}
  The span graphs carry their usual interpretation:
  \begin{itemize}
    \item $K$: the \emph{context} --- atoms and bonds
      unchanged by the reaction.
    \item $L \setminus l(K)$: atoms or bonds \emph{consumed}
      (broken or label-changed at the reaction centre).
    \item $R \setminus r(K)$: atoms or bonds \emph{produced}
      (formed or label-changed at the reaction centre).
  \end{itemize}
  A rule is \emph{valence-conserving} if every atom in $K$
  satisfies the same valence constraint in $L$ and in $R$,
  and \emph{stoichiometrically consistent} if for every
  species $S \in \Sp$, the number of connected components of
  $R$ labelled $S$ minus that of $L$ labelled $S$ equals the
  stoichiometric change $\nu(\mathbf{v})_S - \nu(\mathbf{u})_S$
  of the Petri-net generator $\gamma(p) : \mathbf{u} \to \mathbf{v}$.
  We restrict to valence-conserving, stoichiometrically
  consistent rules throughout.

  \medskip\noindent
  Two distinct rules may share the same span but carry
  different generator labels (when $P$ has parallel generators
  with identical stoichiometry); two distinct rules may also
  share the same generator label but carry different spans
  (when a single Petri-net reaction admits multiple
  mechanistic realisations).
  The Petri net $P$ is assumed rich enough to contain a
  generator for every named chemical transformation in the
  reaction network under study, whether or not that
  transformation is mechanistically elementary.
\end{definition}

\begin{remark}[Atoms with changing labels in the DPO span]
\label{rem:label-change}
  In our label set
  $\Lambda_V = \{(\mathrm{el}, \mathrm{iso}, q, \rho, \ell)\}$
  (Definition~\ref{def:labels}), a reaction that changes any
  component of an atom's label cannot place that atom in $K$:
  the monomorphisms $l$ and $r$ are label-preserving, so
  $K$ can only contain atoms whose labels are identical in
  $L$ and in $R$.
  An atom $v$ with label $\lambda_v^L \neq \lambda_v^R$ must
  therefore appear in $L \setminus l(K)$ (as a ``consumed''
  vertex with its pre-reaction label) and separately in
  $R \setminus r(K)$ (as a ``produced'' vertex with its
  post-reaction label).
  Chemically, this means the atom is formally deleted and
  recreated; its physical identity is preserved by the
  match morphism $m: L \hookrightarrow G$, which tracks
  which vertex in the host graph $G$ corresponds to which
  vertex in $L$.
  This non-attributed encoding is equivalent to attributed
  DPO rewriting~\cite{EhrigEhrigPrangeTaentzer2006}, where
  vertex attributes ($q$, $\rho$, $\ell$) may change on
  persistent vertices via attribute equations on $K$; the
  non-attributed formulation chosen here keeps $\LGraphP$
  as a standard adhesive category without additional
  attribute-algebra structure.
  In particular, isotope labels rarely change (they enter
  the formalism as classical tracers,
  Remark~\ref{rem:isotope-at-L4}), and for octet atoms $\ell$
  is determined by $(q, \rho, \beta)$ via the row-sum law
  \eqref{eq:row-sum}, so a generator that specifies changes
  in $(q, \rho, \beta)$ also fixes the change in $\ell$.
\end{remark}

\subsubsection{DPO derivations and the dangling condition}

A DPO \emph{rule} is an abstract pattern.
A DPO \emph{derivation} is the application of a rule to a
specific host molecule $G$: it specifies where the rule fires
(the \emph{match}) and produces the concrete product molecule
$H$.
This is the level at which a mechanism becomes an actual
chemical transformation on a named substrate.

\begin{definition}[DPO derivation]
\label{def:dpo-derivation}
  Let $p = (L \leftarrow K \rightarrow R)$ be a DPO rule and
  $G \in \LGraphP$ a host graph.
  A \emph{match} is a monomorphism $m: L \hookrightarrow G$
  identifying where the reactant pattern occurs in $G$.

  A \emph{DPO derivation} $G \Rightarrow_{p,m} H$ consists of
  a commutative diagram
  \[
    \includegraphics{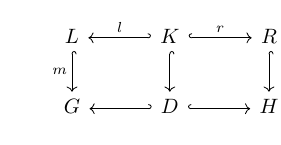}
  \]
  where both squares are pushouts in $\LGraphP$.
  The left pushout computes $D$ (the pushout complement):
  $G$ with the atoms and bonds of $m(L \setminus l(K))$
  deleted.
  The right pushout computes $H$: $D$ with the product
  pattern $R$ glued in via the shared context $K$.
\end{definition}

For a DPO derivation to exist, the gluing context $D$
must be well-defined.
By the adhesive structure of $\LGraphP$
(Proposition~\ref{prop:lgraph-adhesive}), the pushout
complement exists and is unique precisely when no atoms
outside the reaction centre are left with ``half-broken'' bonds
--- the \emph{dangling condition}.
This is not a restriction on chemistry but a consistency
requirement: any correctly specified mechanism must close off
all bond changes at the reaction centre before the rule fires.

\begin{proposition}[Gluing / dangling condition]
\label{prop:gluing}
  Let $p = (L \leftarrow K \rightarrow R)$ be a DPO rule
  and $m: L \hookrightarrow G$ a match.
  The pushout complement $D$ in $\LGraphP$ exists and is unique
  if and only if the \emph{dangling condition} holds:
  no atom in $G \setminus m(L)$ is bonded to an atom in
  $m(L \setminus l(K))$.
\end{proposition}

\begin{proof}
For injective matches in adhesive categories, the pushout
complement $D$ exists if and only if the dangling condition
holds (the gluing condition of \cite[Def.~7.2]{LackSobocinski2005},
which in $\LGraphP$ unpacks to the dangling condition
\cite{EhrigEhrigPrangeTaentzer2006}); when it exists, it is
unique up to isomorphism by
\cite[Lem.~4.5]{LackSobocinski2005}.
\end{proof}

\begin{chembox}[The dangling condition in chemistry]
The dangling condition says: a reaction-centre atom or bond
cannot be deleted if an atom \emph{outside} the reaction
centre is still attached to it.
Concretely, consider an SN2 rule for the substitution
$\mathrm{CH_3Br} + \mathrm{OH^-} \to \mathrm{CH_3OH} +
\mathrm{Br^-}$: the rule deletes the C--Br bond, repurposes
a lone pair on OH$^-$ as the new bonding pair, and creates
the C--O bond.
The rule's $L$ must include every atom that, in the host
$G$, is bonded to a reaction-centre atom; for instance, the
three hydrogens on C must be present in $L$ (and hence in
$K$, as unchanged context), since otherwise the carbon's
relabelling between $L$ and $R$ would leave them with bonds
into a deleted vertex.
In practice, reaction mechanisms are formulated so that the
reaction centre $L$ is self-contained: all atoms incident to
the reaction-centre atoms are included in $L$, making the
dangling condition automatically satisfied for any well-posed
rule.
\end{chembox}

\subsubsection{Elementary generators}

A DPO rule in $\LGraphP$ can describe any bond rearrangement,
however complex.
For the free SMC structure of $\Lk_4(P)$
(Definition~\ref{def:L4}), we need a \emph{generating set}: a
minimal collection of rules from which every organic reaction
mechanism can be built by sequential composition
(Definition~\ref{def:dpo-derivation}) and monoidal product
(parallel juxtaposition on disjoint molecular graphs).

The decomposition of any reaction mechanism into elementary
electron-pair or radical movements is the foundation of the
mechanistic approach to organic chemistry and has been
formalised independently by Herges~\cite{Herges1994}
(topological classification), Dugundji and Ugi~\cite{DugundjiUgi1973}
(BE-matrix reaction types), and Andersen et
al.~\cite{AndersenFlammMerkleStadler2013} (algorithmic rule
composition).
The six generators below are the categorical counterpart of
this decomposition, expressed as explicit DPO spans in
$\LGraphP$.
In the tower, they are the \emph{generators of $\Lk_4(P)$} in
the sense of Proposition~\ref{prop:UP-L4}: any assignment
of these six spans to morphisms of a strict SMC $\mathcal{C}$
extends uniquely to a functor $\Lk_4(P) \to \mathcal{C}$.

\begin{definition}[Elementary DPO generators for organic chemistry]
\label{def:generators}
For brevity we write atom labels in the abbreviated form
$\lambda_v = (\mathrm{el}_v, q_v, \rho_v)$ throughout the
generators below.
This stands for the full quintuple
$(\mathrm{el}_v, \ast, q_v, \rho_v, \ell_v)$ of
Definition~\ref{def:labels} with isotope $\mathrm{iso} = \ast$
(natural abundance) suppressed and lone-pair count $\ell_v$
fixed by the octet row-sum law \eqref{eq:row-sum} from
$(q_v, \rho_v, \beta_v)$ at every reaction-centre atom.
Isotope labels appear explicitly only in identity-tracer
mechanisms (Section~\ref{sec:L4-forcing-in}); hypervalent
atoms (where $\ell$ is not octet-determined) appear only in
the stepwise mechanisms of
Section~\ref{sec:L4-concerted-stepwise}.
In each span below, $K$ contains the atoms and bonds whose
labels are \emph{unchanged} by the reaction; atoms with
changing labels appear in $L \setminus l(K)$ and
$R \setminus r(K)$ per Remark~\ref{rem:label-change}.

  \begin{enumerate}[label=$g_\arabic*$:]

    \item \textbf{Heterolytic bond formation} (lone pair
      $\to$ bond).
      A nucleophilic atom $\mathrm{Nu}$ donates a lone pair
      to form a new single bond to an electrophilic atom
      $\mathrm{E}$.
      Since both atoms change formal charge, neither can sit
      in $K$ (Remark~\ref{rem:label-change}); the physical
      identity of each atom across the rule is supplied by
      the match $m$.
      \begin{itemize}
        \item $K$: remaining molecular context (atoms and
          bonds outside the reaction centre).
        \item $L \setminus l(K)$: two isolated vertices
          $\{\mathrm{Nu}:(\mathrm{el}_\mathrm{Nu}, q, 0),\;
          \mathrm{E}:(\mathrm{el}_\mathrm{E}, q', 0)\}$,
          no edge between them.
        \item $R \setminus r(K)$: two vertices
          $\{\mathrm{Nu}:(\mathrm{el}_\mathrm{Nu}, q+1, 0),\;
          \mathrm{E}:(\mathrm{el}_\mathrm{E}, q'-1, 0)\}$
          connected by a new edge $b = 1$.
      \end{itemize}
      Effect: $q(\mathrm{Nu})$ increases by 1 (Nu loses
      negative charge), $q(\mathrm{E})$ decreases by 1
      (E gains electron density); the derived lone-pair
      counts update via Definition~\ref{def:valence},
      giving $\Delta\ell(\mathrm{Nu}) = -1$
      (one lone pair used to form the bond) and
      $\Delta\ell(\mathrm{E}) = 0$.

    \item \textbf{Heterolytic bond cleavage} (bond $\to$ lone
      pair).
      A bond breaks heterolytically; both electrons go to the
      more electronegative atom $\mathrm{X}$.
      \begin{itemize}
        \item $K$: remaining context.
        \item $L \setminus l(K)$: two vertices
          $\{C:(\mathrm{el}_C, q, 0),\;
          X:(\mathrm{el}_X, q', 0)\}$
          connected by edge $b = 1$.
        \item $R \setminus r(K)$: two isolated vertices
          $\{C:(\mathrm{el}_C, q+1, 0),\;
          X:(\mathrm{el}_X, q'-1, 0)\}$,
          no edge.
      \end{itemize}
      Effect: $q(C)$ increases by 1 (C becomes more positive),
      $q(X)$ decreases by 1 (X gains lone pair), $\ell(X)$
      increases by 1.
      This is the formal inverse of $g_1$ at the bond-rearrangement
      level: applying $g_1$ then $g_2$ to the same atom pair returns
      the molecular graph to its initial state, although the
      composite $g_2 \circ g_1$ is a non-identity morphism in
      $\Lk_4(P)$ that records the round-trip mechanism.

    \item \textbf{Concerted electron-pair shift through a
      3-atom subgraph} (charge or radical migration).
      A coupled bond-order and label shift in a 3-atom
      $A\!=\!B\!-\!C$ subgraph: one bond order decreases, the
      other increases, and the (q, $\rho$) labels of the
      terminal atoms $A$ and $C$ change so that valence is
      preserved at each.
      The canonical chemistry is carbocation migration
      ($\Delta q_A = +1$, $\Delta q_C = -1$, the cation
      moving $C \to A$ through the $\pi$-system); analogous
      rules cover anion migration and radical migration.
      Precondition: $b(AB) \in \{2, 3\}$ and
      $b(BC) \in \{1, 2\}$ in $L$, so post-shift bond orders
      remain in $\{1, 2, 3\}$.
      \begin{itemize}
        \item $K$: remaining context, with atom $B$ in $K$
          (unchanged label).
        \item $L \setminus l(K)$: atoms
          $\{A:(\mathrm{el}_A, q_A, 0),\;
          C:(\mathrm{el}_C, q_A + 1, 0)\}$
          (cation on $C$); edges $\{A\!=\!B,\, B\!-\!C\}$
          (bond orders $b_1 \in \{2,3\}$ on $AB$ and
          $b_2 \in \{1,2\}$ on $BC$).
        \item $R \setminus r(K)$: atoms
          $\{A:(\mathrm{el}_A, q_A + 1, 0),\;
          C:(\mathrm{el}_C, q_A, 0)\}$
          (cation on $A$); edges $\{A\!-\!B,\, B\!=\!C\}$
          (bond orders $b_1 - 1$ on $AB$ and $b_2 + 1$ on $BC$).
      \end{itemize}
      Effect: the cation migrates from $C$ to $A$ along the
      conjugated $\pi$-system; the row-sum law
      \eqref{eq:row-sum} balances the bond-order change against
      the charge change at each terminal.
      Atoms $A$ and $C$ are in $L \setminus l(K)$ /
      $R \setminus r(K)$ because their labels change; only
      $B$ resides in $K$.

    \item \textbf{Homolytic bond formation} (two radicals $\to$
      bond).
      Two radical atoms combine to form a new single bond.
      \begin{itemize}
        \item $K$: remaining context.
        \item $L \setminus l(K)$: two isolated vertices
          $\{A:(\mathrm{el}_A, q_A, 1),\;
          B:(\mathrm{el}_B, q_B, 1)\}$
          (each monoradical, $\rho = 1$), no edge.
        \item $R \setminus r(K)$: two vertices
          $\{A:(\mathrm{el}_A, q_A, 0),\;
          B:(\mathrm{el}_B, q_B, 0)\}$
          connected by edge $b = 1$.
      \end{itemize}
      Effect: $\rho$ decreases from 1 to 0 on both atoms
      (radical electrons pair into the new bond),
      $\ell$ is unchanged on each.
      This is the radical counterpart of $g_1$.

    \item \textbf{Homolytic bond cleavage} (bond $\to$ two radicals).
      A bond breaks homolytically; each fragment retains one electron.
      \begin{itemize}
        \item $K$: remaining context.
        \item $L \setminus l(K)$: two vertices
          $\{A:(\mathrm{el}_A, q_A, 0),\;
          B:(\mathrm{el}_B, q_B, 0)\}$
          connected by edge $b = 1$.
        \item $R \setminus r(K)$: two isolated vertices
          $\{A:(\mathrm{el}_A, q_A, 1),\;
          B:(\mathrm{el}_B, q_B, 1)\}$, no edge.
    \end{itemize}
    Effect: $\rho$ increases from 0 to 1 on both atoms; the
    shared bonding pair becomes two unpaired electrons, one on each fragment.
    $g_5$ reverses the bond-rearrangement of $g_4$:
    applying $g_4$ then $g_5$ to the same atom pair returns the
    molecular graph to its initial state, although the composite
    is a non-identity morphism in $\Lk_4(P)$.

\item \textbf{Single-electron transfer} (SET).
  One electron moves from donor $\mathrm{Dn}$ to acceptor
  $\mathrm{An}$ without bond formation or cleavage.
  \begin{itemize}
    \item $K$: remaining context; no bond $\mathrm{Dn}$--$\mathrm{An}$
      in $L$ or $R$.
    \item $L \setminus l(K)$: two isolated vertices
      $\{\mathrm{Dn}:(\mathrm{el}_\mathrm{Dn}, q_\mathrm{Dn}, 1),\;
      \mathrm{An}:(\mathrm{el}_\mathrm{An}, q_\mathrm{An}, 0)\}$
      ($\mathrm{Dn}$ is a monoradical, $\mathrm{An}$ is
      closed-shell).
    \item $R \setminus r(K)$: two isolated vertices
      $\{\mathrm{Dn}:(\mathrm{el}_\mathrm{Dn},
      q_\mathrm{Dn}+1, 0),\;
      \mathrm{An}:(\mathrm{el}_\mathrm{An},
      q_\mathrm{An}-1, 1)\}$
      ($\mathrm{Dn}$ loses the radical electron and gains a
      positive charge; $\mathrm{An}$ becomes a radical anion).
  \end{itemize}
  Effect: charge and radical count transfer simultaneously.
  No new bond is formed; $g_6$ differs from $g_1$ in that
  $g_1$ moves two electrons (a lone pair) while $g_6$
  moves one (a radical).
\end{enumerate}
\end{definition}

\begin{remark}[Status of the generating set]
\label{rem:generators-status}
  Definition~\ref{def:generators} proposes six generators as a
  \emph{working basis} for all organic reaction mechanisms.
  The claim that these six generators suffice to generate every
  organic reaction mechanism by sequential composition and
  monoidal product is supported by the following evidence:
  \begin{itemize}
    \item Herges~\cite{Herges1994} partitions all concerted
      organic reactions into three topological types
      (linear, pericyclic, coarctate); the linear type aligns
      with the topological shape of patterns built from
      $g_1$--$g_3$, while pericyclic and coarctate types
      require additional primitive generators
      (\S\ref{sec:L4-pericyclic}).
    \item Dugundji and Ugi~\cite{DugundjiUgi1973} identify
      $\sim$30 reaction-matrix types; $\sim$10 types cover
      $\sim$93\% of all single-step transformations,
      consistent with a small generating set.
    \item Andersen et al.~\cite{AndersenFlammMerkleStadler2013}
      develop the algorithmic machinery for rule composition in
      typed-graph DPO and apply it to ionic organic chemistry,
      where rules of the $g_1$--$g_3$ shape suffice.
      The radical and SET generators $g_4$--$g_6$ are added here
      to cover homolytic and single-electron-transfer chemistry,
      which are needed for the radical sub-network of the
      Briggs--Rauscher reaction (Section~\ref{sec:L4-sn1sn2})
      and for organic photochemistry more broadly.
  \end{itemize}
  A formal proof that Definition~\ref{def:generators} is both
  minimal and complete is not available in the published
  literature.
  The tower provides a framework for investigating this question:
  completeness would follow from showing that the automorphism
  exact sequence $1 \to \ker\varphi_4 \to \Aut(\Lk_4(P))
  \xrightarrow{\varphi_4} \Aut(\Lk_3(P)) \to \coker(\varphi_4)
  \to 1$ is exhausted by the bond-change patterns generated by
  $g_1$--$g_6$ --- i.e., that every element of $\coker(\varphi_4)$
  is the image of a composition of these generators.
  Minimality would require showing that no proper subset of
  $\{g_1, \ldots, g_6\}$ generates $\Lk_4(P)$ as a free SMC.
  Both statements are conjectured and constitute an original
  research direction opened by the categorical framework of
  this chapter.
  We proceed with the six generators as a well-motivated and
  empirically supported working basis.
\end{remark}

\subsubsection{Pericyclic rules: cyclic reaction-centre topology}
\label{sec:L4-pericyclic}

The six generators of Definition~\ref{def:generators} cover
\emph{stepwise} bond rearrangements: each generator acts on
two atoms (or three, in the case of $g_3$), and larger
mechanisms are built by sequential composition and monoidal
product.
But a significant class of organic reactions is
\emph{concerted} across a closed ring of atoms --- pericyclic
reactions --- and cannot be decomposed into a sequential
chain of two-atom steps without introducing intermediate
molecular graphs that the mechanism provably lacks.
These reactions require DPO rules of a different shape.

\begin{definition}[Pericyclic DPO rule]
\label{def:pericyclic-rule}
  A \emph{pericyclic DPO rule} is a chemical DPO rule
  $p = (L \leftarrow K \rightarrow R)$
  in $\LGraphP$ for which the reaction centre
  $L \setminus l(K)$ (together with its incident atoms in
  $K$) forms a cycle in the underlying graph of $L$,
  with the corresponding cycle in $R$ obtained by a
  simultaneous reassignment of bond orders around the ring.
  No vertex labels change; all bond-order changes occur
  within the cycle and are applied concurrently as a single
  rule.
\end{definition}

\begin{remark}[Pericyclic rules are primitives]
\label{rem:pericyclic-primitives}
A pericyclic rule is \emph{not} a composite of
$g_1$--$g_6$.
Attempting to decompose, say, a Diels--Alder step into
sequential applications of $g_1$ and $g_3$ would force
the derivation to pass through intermediate molecular
graphs (the host after one $g_1$, before the next $g_3$,
etc.) that carry formal charges or unpaired electrons not
present in either the reactant or the product, and that the
concerted mechanism does not produce.
A single pericyclic DPO step has the entire ring of bond-order
changes occurring in one rule application, with no
intervening discrete molecular graphs at all.
The correct treatment within our framework is to admit
pericyclic rules as \emph{additional primitive generators},
one for each pattern of cyclic bond-order reassignment.
These primitives are characterised by Herges' topological
classification \cite{Herges1994}: \emph{linear} (no cycle
in the reaction centre), \emph{pericyclic} (a single
cycle), or \emph{coarctate} (two cycles sharing an atom).
The generator basis for $\Lk_4(P)$ is therefore
$\{g_1, \ldots, g_6\} \cup \{\text{pericyclic primitives}\}$,
with the pericyclic primitives indexed by cycle length and
Herges type.
A complete classification of pericyclic primitives --- and
the question of whether they can be further reduced to a
finite subset --- is an open problem left to future work.
\end{remark}

\begin{chembox}[Pericyclic reactions in the DPO framework]
The three canonical pericyclic reaction types all have
cyclic reaction-centre topology:
\begin{itemize}
  \item \emph{Diels--Alder cycloaddition}
    \cite{DielsAlder1928}: a diene and a dienophile combine
    through a six-membered cyclic concerted step;
    three $\pi$-bonds break and one $\pi$-bond plus two
    new $\sigma$-bonds form, all within a single ring of six
    atoms.
  \item \emph{Electrocyclic ring closure}
    \cite{WoodwardHoffmann1969}: a terminal $\pi$-bond
    becomes a $\sigma$-bond closing a ring; the reaction
    centre is the newly formed ring itself.
  \item \emph{$[1,j]$ sigmatropic shift} \cite{Clayden2012}:
    a $\sigma$-bond migrates across a $\pi$ system; the
    reaction centre is the cycle traced by the migrating
    group and the atoms it visits.
\end{itemize}
At $\Lk_4$, all three are handled by pericyclic primitives
of Definition~\ref{def:pericyclic-rule} with the appropriate
cycle length and Herges topology.
What $\Lk_4$ \emph{cannot} do is predict whether a given
pericyclic rule is thermally or photochemically allowed ---
that is the content of the Woodward--Hoffmann rules, which
depend on orbital symmetry along the reaction coordinate
and are treated at $\Lk_{4.5}$ in the stereochemistry
chapter.
\end{chembox}

%% file: chapters/L4/l4_def.tex
\subsection{Definition of \texorpdfstring{$\Lk_4(P)$}{L4(P)}}
\label{sec:L4-def}

\begin{definition}[Mechanistic level $\Lk_4(P)$]
\label{def:L4}
  Let $P$ be a Petri net with species set $\Sp$.
  The \emph{mechanistic level} $\Lk_4(P)$ is the strict symmetric
  monoidal category defined as follows.
  \begin{itemize}
    \item \textbf{Objects}: finite disjoint unions of
      $\Sp$-typed molecular graphs in $\LGraphP$.
    \item \textbf{Generating morphisms}:
      valence-conserving chemical DPO rules
      (Definition~\ref{def:dpo-rule}) for $P$-species.
    \item \textbf{Morphisms}: composable sequences of DPO
      derivations in $\LGraphP$, taken up to the congruence
      relations of a strict symmetric monoidal category
      (associativity and unitality of composition, monoidal
      axioms, naturality of the symmetry).
    \item \textbf{Monoidal product}: disjoint union
      $G_1 \sqcup G_2$ of molecular graphs on objects;
      parallel juxtaposition of DPO derivations on morphisms.
    \item \textbf{Monoidal unit}: the empty graph $\emptyset$.
  \end{itemize}
\end{definition}

That $\Lk_4(P)$ as defined is the \emph{free} strict SMC on
the generating DPO rules requires verification on two fronts:
that the concrete construction is well-defined as a strict
SMC (not just a category, since SMC structure includes
symmetry and tensor coherence), and that it satisfies the
universal property of the free SMC on the chemical DPO
signature.
The three properties of $\LGraphP$ established in
Section~\ref{sec:L4-adhesive} are what make the SMC
structure hold concretely; the universal property then
follows from the standard construction of free SMCs on a
directed signature.

\begin{proposition}[$\Lk_4(P)$ is the free strict SMC on its
  generating DPO rules]
\label{prop:L4-freeSMC}
  The category $\Lk_4(P)$ of Definition~\ref{def:L4} is a free
  strict symmetric monoidal category over the signature of
  valence-conserving DPO rules for $P$-species in $\LGraphP$.
\end{proposition}

\begin{proof}
The construction in Definition~\ref{def:L4} presents
$\Lk_4(P)$ concretely: morphisms are equivalence classes of
DPO derivation sequences in $\LGraphP$ modulo the SMC
congruence (associativity, unitality, monoidal axioms,
naturality of the symmetry).
We verify that this construction is the free strict SMC on
the directed signature $\Sigma$ of valence-conserving DPO
rules for $P$-species, in two parts:
(a) the construction is a well-defined strict SMC, and
(b) it satisfies the universal property of the free SMC on
$\Sigma$.

\medskip
For (a), the three properties of $\LGraphP$ from
\S\ref{sec:L4-adhesive} establish that the SMC structure
holds concretely for derivation classes.

\textbf{(i) Adhesivity makes derivations well-defined.}
Each generating rule $p = (L \leftarrow K \rightarrow R)$
applied at a match $m: L \hookrightarrow G$ has a unique
gluing context $D$ when the dangling condition holds
(Proposition~\ref{prop:gluing}); for well-posed chemical
rules whose reaction centre is enclosed by $L$, the dangling
condition is automatic (Section~\ref{sec:L4-dpo}).
The product $H$ is therefore unique up to isomorphism, and
sequential composition of derivations is well-defined and
associative on derivation classes.

\textbf{(ii) Local Church--Rosser realises commutativity of
independent steps concretely.}
The SMC's interchange law and the naturality of the symmetry
$\tau$ together imply that two independent generators applied
to disjoint parts of an object commute up to SMC congruence.
What requires concrete verification is that this abstract
commutativity is faithfully realised on derivation classes.
The Local Church--Rosser property
(Theorem~7.7 of \cite{LackSobocinski2005}) supplies it:
two derivations $d_1, d_2$ at disjoint reaction centres in
the same host produce isomorphic products whether applied in
either order.

\textbf{(iii) Concurrency realises composition concretely.}
The Concurrency theorem
(Theorem~7.11 of \cite{LackSobocinski2005}) supplies a
single composite rule whose application produces the same end
product as the sequence.
The composition of derivation classes is therefore
representable as a single-rule derivation, consistent with
the abstract SMC composition.

\medskip
For (b), freeness on $\Sigma$ follows from the standard
universal-property construction of the free strict SMC on a
directed signature \cite[\S XI.2]{MacLane1998},
\cite[\S 1]{JoyalStreet1991}: the morphisms of $\Lk_4(P)$ are
finite formal composites of generators under sequential and
monoidal composition, modulo the SMC congruence, with two
composites identified iff forced equal by the SMC axioms.
The DPO interpretation in $\LGraphP$, validated by (i)--(iii),
contributes that each generator can be applied at any chosen
match in any host satisfying the gluing condition, witnessing
the formal morphism without introducing any non-SMC-congruent
identifications.
\end{proof}

\begin{mathbox}[$\Lk_4(P)$ as the free SMC: a tower-language
  derivation]
The free SMC structure of $\Lk_4(P)$ is automatic from the
universal-property construction of free SMCs on a directed
signature \cite[\S XI.2]{MacLane1998}: given the signature
$\Sigma$ of valence-conserving DPO rules, the free SMC has
morphisms = equivalence classes of formal composites under
the SMC congruence (associativity, unitality, monoidal
axioms, naturality of the symmetry).
What the three properties of $\LGraphP$ from
Section~\ref{sec:L4-adhesive} contribute is the
\emph{faithfulness} of the semantic interpretation of these
formal morphisms as concrete DPO derivations.

\medskip
\noindent\textbf{Adhesivity $\Rightarrow$ semantic
well-definedness.}
At $\Lk_0(P)$--$\Lk_3(P)$, a morphism from complex $\mathbf{u}$
to complex $\mathbf{v}$ is a stoichiometric transition: a pair
$(\mathbf{u}, \mathbf{v})$ with no internal structure.
At $\Lk_4(P)$, a morphism is a DPO derivation
$G \Rightarrow_{p,m} H$: the unique gluing context $D$
(guaranteed by adhesivity of $\LGraphP$) is the categorical
bridge between $L$ and $R$, not in general a chemical
intermediate (Remark~\ref{rem:D-not-intermediate}).
A chemical intermediate, when present (e.g., the
trigonal-bipyramidal phosphorus intermediate in stepwise
identity substitution, Section~\ref{sec:L4-forcing-in}), is
the molecular graph $H$ output by one derivation and serving
as input to the next, distinct from any single rule's gluing
context $D$.

\medskip
\noindent\textbf{Local Church--Rosser $\Rightarrow$ semantic
commutation of independent steps.}
The free-SMC symmetry
$\tau_{G_1, G_2}: G_1 \sqcup G_2 \to G_2 \sqcup G_1$ is
automatic by construction.
What LCR provides is its semantic realisation: two
derivations at disjoint reaction centres commute as DPO
operations, instantiating the abstract symmetry concretely.
Without LCR, the abstract symmetry would still hold formally
in $\Lk_4(P)$, but the order of derivations could matter
in $\LGraphP$, leaving the abstract SMC symmetry without a
faithful $\LGraphP$-realisation.

\medskip
\noindent\textbf{Concurrency $\Rightarrow$ semantic
single-rule composites.}
The free-SMC composition is automatic by construction.
What the Concurrency theorem provides is a representation
result: any composite of independent derivations can be
realised as a single composite DPO rule whose pushout produces
the same end product
(Theorem~7.11 of \cite{LackSobocinski2005}).
The abstract composite morphism in $\Lk_4(P)$ thus has a
concrete single-rule realisation in $\LGraphP$.

\medskip
\noindent\textbf{Summary.}
Free SMC structure: by construction.
Semantic interpretation in $\LGraphP$: well-defined
derivations (adhesivity), independent commutation (Local
Church--Rosser), single-rule composites (Concurrency).
The freeness comes from category theory; the chemistry
comes from the semantic faithfulness via $\LGraphP$.
The generating data are the chemical DPO rules of
Definition~\ref{def:dpo-rule}; the elementary basis
$\{g_1, \ldots, g_6\}$ together with pericyclic primitives
(Sections~\ref{sec:L4-dpo}, \ref{sec:L4-pericyclic}) is a
conjectured minimal subset
(Remark~\ref{rem:generators-status}).
\end{mathbox}

Having established that $\Lk_4(P)$ is a free strict SMC,
one can now state precisely how it sits in the tower.
The forgetful functor $U_4: \Lk_4(P) \to \Lk_0(P)$
(constructed in Section~\ref{sec:L4-u4}) provides the
downward connection: it sends each molecular graph to its
species-count vector and each DPO derivation to the
corresponding stoichiometric transition.
The decorating functors $\FH$, $\FS$, $\FP$ of the lower
levels lift to $\Lk_4(P)$ via $U_4$ (Proposition~\ref{prop:L4-lifting}):
no new \emph{primitive} numerical functor enters at $\Lk_4$
beyond those lifted from $\Lk_3$ via $U_4$, but the free SMC
that carries them is genuinely new.
The universal property of $\Lk_4(P)$
(Proposition~\ref{prop:UP-L4} below) is what makes
this connection canonical: any interpretation of the
DPO generators in a target SMC --- whether a chemical
database, a retrosynthesis planner, or a Para-level ML
model --- extends uniquely to a functor from $\Lk_4(P)$.

\begin{remark}[The structural break stated precisely]
\label{rem:structural-break}
At every previous level $\Lk_0(P)$ through $\Lk_3(P)$,
the underlying free SMC was $\Lk_0(P) = \mathbf{FreeSMC}(P)$,
with objects in $\NN^{\Sp}$ and morphisms as stoichiometric
transitions.
New levels added functors $\FH, \FS, \FP$ on top of the
\emph{same} underlying category.

At $\Lk_4(P)$, the underlying category is genuinely new:
objects are molecular graphs, not species-count vectors;
morphisms are DPO derivations, not abstract stoichiometric
transitions.
$\Lk_4(P)$ is \emph{not} the free SMC of $P$ with
additional decoration.
It is a different free SMC, generated by richer data,
connected to the previous levels only via the forgetful
functor $U_4$ of Section~\ref{sec:L4-u4}.
\end{remark}

\begin{proposition}[Universal property of $\Lk_4(P)$]
\label{prop:UP-L4}
  Let $\mathcal{C}$ be any strict SMC.
  An assignment of the valence-conserving DPO rules of
  $P$-species to morphisms of $\mathcal{C}$, compatible with
  source and target molecular graphs, extends uniquely to a
  strict SMC functor $F: \Lk_4(P) \to \mathcal{C}$.
\end{proposition}

\begin{proof}
By Proposition~\ref{prop:L4-freeSMC}, $\Lk_4(P)$ is the free
strict SMC on the generating DPO rules.
The universal property of a free SMC states precisely that
any assignment of generators to morphisms of any strict SMC
$\mathcal{C}$ extends to a unique strict SMC functor.
Existence: define $F$ on generators by the given assignment
and extend to composites and monoidal products by the SMC
axioms; the SMC congruence in $\Lk_4(P)$ ensures $F$ is
well-defined.
Uniqueness: every morphism of $\Lk_4(P)$ is, by freeness,
a unique composite of generators up to SMC congruence, so
the value of $F$ on generators determines $F$ on all morphisms.
\end{proof}

\begin{insightbox}[The tower updated: two qualitatively different
  extension types]
The tower $\Lk_0 \hookrightarrow \Lk_1 \hookrightarrow \Lk_2
\hookrightarrow \Lk_3 \hookrightarrow \Lk_4$ now contains
two qualitatively different types of extension:
\begin{itemize}
  \item \textbf{Decorator extensions}
    ($\Lk_0 \to \Lk_1$, $\Lk_1 \to \Lk_2$, $\Lk_2 \to \Lk_3$):
    the same underlying free SMC $\Lk_0(P)$ is retained;
    each new level adds exactly one new functor into $B\RR$
    or $\Stoch$ (or more precisely, $\mathfrak{g}_{\Sp}$).
    The universal property of $\Lk_0(P)$ does all the work.
  \item \textbf{Structural extension} ($\Lk_3 \to \Lk_4$):
    the underlying free SMC is replaced by a new one,
    $\Lk_4(P)$, generated by DPO rules rather than
    stoichiometric transitions.
    Previous decorations lift via $U_4$.
    No new \emph{primitive} numerical functor is added beyond
    those inherited from $\Lk_3$ via pullback along $U_4$;
    the new content is in the morphisms themselves
    (DPO derivations rather than stoichiometric transitions).
    The structural break is forced by the $\coker(\varphi_4)$
    obstruction of Section~\ref{sec:L4-forcing-in} and cannot
    be resolved by any decorator extension.
\end{itemize}
\end{insightbox}

\subsection{Automorphism exact sequences across the tower}
\label{sec:L4-aut-tower}

The necessity of each tower extension is witnessed by a non-trivial
cokernel in the automorphism exact sequence at that level.
Now that $\Lk_4(P)$ is in hand, the complete tower from
$\Lk_0$ to $\Lk_4$ assembles, with each forcing pair
explicit.

\begin{proposition}[Automorphism tower $\Lk_0$--$\Lk_4$]
\label{prop:aut-tower}
  For $k = 1, 2, 3, 4$, let
  $\varphi_k: \Aut(\Lk_k(P)) \to \Aut(\Lk_{k-1}(P))$
  be the restriction map induced by the forgetful functor at
  level $k$.
  The following four exact sequences hold, and all four
  cokernels are non-trivial:
  \begin{align*}
    1 \to \ker\varphi_1 \to \Aut(\Lk_1)
      \xrightarrow{\varphi_1}
      \Aut(\Lk_0) \to \coker\varphi_1 \to 1, \\
    1 \to \ker\varphi_2 \to \Aut(\Lk_2)
      \xrightarrow{\varphi_2}
      \Aut(\Lk_1) \to \coker\varphi_2 \to 1, \\
    1 \to \ker\varphi_3 \to \Aut(\Lk_3)
      \xrightarrow{\varphi_3}
      \Aut(\Lk_2) \to \coker\varphi_3 \to 1, \\
    1 \to \ker\varphi_4 \to \Aut(\Lk_4)
      \xrightarrow{\varphi_4}
      \Aut(\Lk_3) \to \coker\varphi_4 \to 1.
  \end{align*}
\end{proposition}

\begin{proof}
Each exact sequence is the standard kernel-cokernel sequence of
the group homomorphism $\varphi_k$.
Non-triviality of each cokernel is witnessed by an explicit
reaction pair that the lower level conflates.
$\coker\varphi_1$ contains the swap of reactions with the same
stoichiometry but different $\FH$ (Hess's Law is invisible at
$\Lk_0$).
$\coker\varphi_2$ contains the swap of reactions with the same
$\FH$ but different $\FS$ (entropy data are invisible at $\Lk_1$).
$\coker\varphi_3$ contains the swap of reactions with the same
$\FG^T$ but different propensity $\FP$ (kinetic data are
invisible at $\Lk_2$).
$\coker\varphi_4$ contains the swap
$r_\text{concerted} \leftrightarrow r_\text{stepwise}$ for
identity nucleophilic substitution at phosphorus
(Section~\ref{sec:L4-forcing-in}, Forcingbox): two generators
with identical source and target complexes in $\NN^{|\Sp|}$,
identical functorial images under $\FH$, $\FS$, $\FG^T$, $\FP$
at $\Lk_3$, but distinct DPO-derivation structure at $\Lk_4$
(Proposition~\ref{prop:sn1-neq-sn2}).
\end{proof}

\noindent
The following portrait diagram displays the full tower together
with its automorphism groups.
The left column shows the tower levels with forgetful functors;
the right column shows the automorphism groups with the
restriction maps $\varphi_k$ connecting them; the horizontal
dashed arrows indicate the natural action $\Aut(\Lk_k) \curvearrowright \Lk_k$
of each automorphism group on its level.

\[
\includegraphics{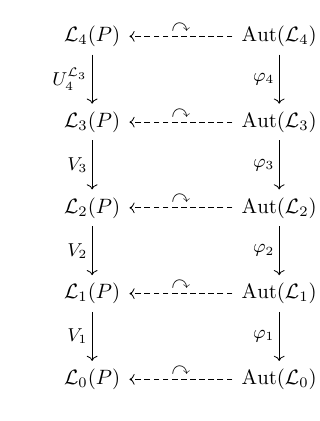}
\]

\noindent
The following table records the concrete form of each automorphism
group and the chemical phenomenon encoded by each cokernel.

\medskip
\begin{center}
\renewcommand{\arraystretch}{1.35}
\begin{tabular}{c p{4cm} p{5.8cm}}
  \hline
  \textbf{Level} & \textbf{$\Aut(\Lk_k)$} &
    \textbf{$\coker(\varphi_k)$: what $\Lk_{k-1}$ conflates}\\
  \hline
  $\Lk_0$ & $\mathrm{Sym}(\Sp)$, species permutations
    & --- (base level) \\
  $\Lk_1$ & Isoenthalpic permutations
    & Same stoichiometry, different $\FH$:
      Hess's Law invisible at $\Lk_0$ \\
  $\Lk_2$ & Isothermodynamic perms $+ \dagger$
    & Same $\FH$, different $\FS$:
      entropy and equilibrium locus invisible at $\Lk_1$ \\
  $\Lk_3$ & Rate-preserving permutations
    & Same $\FG^T$, different propensity form:
      kinetic distinction invisible at $\Lk_2$ \\
  $\Lk_4$ & DPO-structure-preserving perms of generating rules
  & Same $\Lk_3$ data, different DPO derivation:
    $r_\text{concerted} \!\leftrightarrow\! r_\text{stepwise}$
    for identity substitution at heteroatom centres
    (Section~\ref{sec:L4-forcing-in}) \\
  \hline
\end{tabular}
\end{center}

\medskip\noindent
Reading the diagram top-down along the right column traces the
progressive loss of mechanistic information as one descends the
tower: $\Aut(\Lk_4)$ distinguishes every bond-change sequence;
$\Aut(\Lk_0)$ sees only species-name permutations.
Reading bottom-up recovers the \emph{forcing} structure: each
non-trivial cokernel names a pair of physically distinct reactions
that the previous level conflates, compelling the next level's
construction.
The chapter terminates at $\Lk_4$ because the next distinction
--- between enantiomers, which share the same DPO rule but differ
in three-dimensional orientation --- requires group-action data
not present in labelled molecular graphs.
This is the forcing content of $\coker\varphi_{4.5}$, developed
in the following chapter (Section~\ref{sec:L45}).

%% file: chapters/L4/l4_u4.tex
\subsection{The forgetful functor
  \texorpdfstring{$U_4: \Lk_4(P) \to \Lk_0(P)$}{U4}}
\label{sec:L4-u4}

The free SMC $\Lk_4(P)$ sits above the rest of the tower
via a forgetful functor $U_4$ that strips away all
bond-level information and returns the plain stoichiometric
data of $\Lk_0(P)$.
This functor is the categorical bridge that embeds the new
mechanistic level into the existing tower: the previous
decorations $\FH$, $\FS$, $\FP$ lift to $\Lk_4(P)$ by
precomposing with $U_4$, and the tower structure from
$\Lk_0$ through $\Lk_3$ is recovered by factoring through $U_4$.

\subsubsection{Definition and well-definedness}

\begin{definition}[Forgetful functor $U_4$]
\label{def:U4}
Recall from Definition~\ref{def:LGraphP} that every object
of $\LGraphP$ is a pair $(G, \spec)$ where $G$ is a labelled
molecular graph and $\spec: \pi_0(G) \to \Sp$ assigns each
connected component of $G$ (each individual molecule) to a
species name in $\Sp$.

The \emph{forgetful functor} $U_4: \Lk_4(P) \to \Lk_0(P)$
is defined as follows.
\begin{itemize}
  \item \textbf{On objects}: for $(G, \spec) \in \LGraphP$,
    \[
      U_4(G)
      \;:=\;
      \sum_{S \in \Sp}
      \Bigl|\bigl\{C \in \pi_0(G) \;\big|\; \spec(C) = S\bigr\}\Bigr|
      \cdot S
      \;\;\in\; \NN^{|\Sp|}.
    \]
    That is, $U_4(G)$ counts how many connected components of
    $G$ are assigned to each species by $\spec$, producing the
    corresponding complex in the free commutative monoid $\NN^{|\Sp|}$.
  \item \textbf{On morphisms}: for a DPO derivation
    $d: G \Rightarrow H$ implementing a chemical DPO rule
    $p = (s, \gamma(p))$ (Definition~\ref{def:dpo-rule}),
    define
    \[
      U_4(d) \;:=\; \gamma(p) \;\in\; \Lk_0(P),
    \]
    the Petri-net generator carried as label data by the
    rule $p$, regarded as a morphism in $\Lk_0(P) =
    \mathbf{FreeSMC}(P)$.
    For a composite derivation $d_2 \circ d_1$, set
    $U_4(d_2 \circ d_1) := U_4(d_2) \circ U_4(d_1)$ using
    composition in $\Lk_0(P)$.
\end{itemize}
\end{definition}

\begin{remark}[On the generator label]
\label{rem:generator-label}
The generator label $\gamma(p)$ is essential to making $U_4$
well-defined on morphisms.
Since $\Lk_0(P) = \mathbf{FreeSMC}(P)$ is \emph{not} a thin
category (Petri nets may contain distinct generators
$r_1, r_2 \in \Rx$ with identical source and target complexes),
$U_4(d)$ cannot be defined as ``the unique morphism from
$U_4(G)$ to $U_4(H)$''; there may be several candidates.
The explicit label $\gamma(p)$ in the DPO rule data
(Definition~\ref{def:dpo-rule}) resolves the ambiguity:
$U_4(d)$ is the specific generator $\gamma(p)$, not merely
any morphism with matching source and target.
This matters precisely for the forcing pair of
Section~\ref{sec:L4-forcing-in}: when the Petri net $P$
contains two distinct generators
$r_\text{concerted}, r_\text{stepwise} \in \Rx$ that
realise the same identity substitution at phosphorus
through mechanistically distinct routes, the corresponding
DPO rules $p_\text{concerted}$ and $p_\text{stepwise}$
carry distinct labels
$\gamma(p_\text{concerted}) = r_\text{concerted}$ and
$\gamma(p_\text{stepwise}) = r_\text{stepwise}$.
$U_4$ projects them to these two parallel generators in
$\Lk_0(P)$ --- distinct morphisms with identical source and
target complexes --- rather than collapsing them to a single
morphism.
When $P$ has only one generator per stoichiometry (the
generic case), the label is redundant and can be inferred
from $(U_4(L), U_4(R))$.
\end{remark}

\begin{proposition}[$U_4$ is a strict SMC functor]
\label{prop:U4-functor}
  $U_4: \Lk_4(P) \to \Lk_0(P)$ is a well-defined strict
  symmetric monoidal functor.
\end{proposition}

\begin{proof}
\textbf{Well-definedness on objects.}
$U_4(G) \in \NN^{|\Sp|}$ by construction: the sum counts
components by their species assignment $\spec$.

\textbf{Well-definedness on morphisms.}
For each DPO derivation $d$, the generator label
$\gamma(p)$ is a fixed datum of the rule $p$ (part of the
grounding structure of $\LGraphP$).
Thus $U_4(d) = \gamma(p)$ is a specific, uniquely
determined morphism in $\Lk_0(P)$, independent of which
host graph $G$ the derivation acts on or what match $m$
was chosen.

\textbf{Functoriality.}
For the identity derivation $\id_G$, $U_4(\id_G) = \id_{U_4(G)}$
since $\id_G$ implements the trivial (empty) generator.
For sequential composition $d_2 \circ d_1$:
$U_4(d_2 \circ d_1) = U_4(d_2) \circ U_4(d_1)$
by the definition of $U_4$ on composites and the functoriality
of composition in $\Lk_0(P)$.

\textbf{Strict monoidality.}
$U_4(G_1 \sqcup G_2) = U_4(G_1) + U_4(G_2)$:
the species count of a disjoint union is the sum of the
individual counts (the $\spec$ maps are independent on
disjoint components).
$U_4(\emptyset) = \mathbf{0}$: the empty graph has no
components, hence contributes zero to every species count.
For morphisms, $U_4(d_1 \sqcup d_2) = U_4(d_1) \otimes U_4(d_2)$:
parallel derivations on disjoint graphs map to the monoidal
product of the corresponding $\Lk_0(P)$-morphisms.

\textbf{Symmetry.}
$U_4$ sends the symmetry isomorphism
$\tau_{G_1,G_2}: G_1 \sqcup G_2 \xrightarrow{\sim} G_2 \sqcup G_1$
in $\Lk_4(P)$ to the symmetry isomorphism
$\tau_{U_4(G_1), U_4(G_2)}$ in $\Lk_0(P)$,
since both simply swap the two summands.
\end{proof}

\begin{chembox}[What $U_4$ forgets and retains]
\textbf{Retained}: which Petri-net generator the DPO rule
realises (via its label $\gamma$), and therefore which
species are consumed and produced (stoichiometry).

\textbf{Forgotten}: the molecular graphs of reactants and
products; which bonds formed or broke; the sequence of
elementary DPO steps; the identity of the reaction centre;
the presence or absence of intermediate molecular graphs
in stepwise mechanisms; all stereochemical information.

For the forcing pair of Section~\ref{sec:L4-forcing-in},
the concerted-$\mathrm{S_N2}$-P DPO rule and the stepwise A--E
DPO rule
(realising identity substitution at phosphorus through
distinct mechanisms) carry distinct Petri-net labels
$r_\text{concerted}$ and $r_\text{stepwise}$ when both
appear as parallel generators of $P$.
Their $U_4$-images are different but \emph{parallel}
generators in $\Lk_0(P)$: distinct morphisms sharing source
and target complexes.
$U_4$ thus preserves the information that the two mechanisms
yield identical reactants and products, while discarding the
internal DPO structure that distinguishes them at $\Lk_4$ and
justifies their existence as separate generators in the first
place.
The existence of parallel generators in $P$ is unmotivated
at $\Lk_0$--$\Lk_3$ (where it can be postulated only by
fiat) and finds its principled structural basis at $\Lk_4$
(where it reflects genuinely distinct DPO derivations).
\end{chembox}

\subsubsection{%
  \texorpdfstring{$U_4$}{U4} and the full tower:
  factorisation through intermediate levels}
\label{sec:L4-u4-tower}

The functor $U_4: \Lk_4(P) \to \Lk_0(P)$ sends $\Lk_4$
all the way down to the base of the tower.
We can also factor this descent through each intermediate
level: there are canonical forgetful functors
$U_4^{\Lk_k}: \Lk_4(P) \to \Lk_k(P)$ for $k = 1, 2, 3$
that land at each decorator level,
and these are related by the adjacent-level forgetful
functors $V_k: \Lk_k(P) \to \Lk_{k-1}(P)$.

\begin{definition}[Intermediate forgetful functors]
\label{def:U4-intermediate}
For $k = 1, 2, 3$, define the \emph{intermediate forgetful
functor} $U_4^{\Lk_k}: \Lk_4(P) \to \Lk_k(P)$ to be the
functor that:
\begin{itemize}
  \item on objects and on the underlying stoichiometric
    morphisms, agrees with $U_4$
    (i.e.\ $V_1 \circ \cdots \circ V_{k} \circ U_4^{\Lk_k}
    = U_4$);
  \item retains the decorating functor data available at
    level $k$:
    \begin{align*}
      U_4^{\Lk_1}(d) &:=
        \bigl(U_4(d),\; F_H^{(4)}(d)\bigr),\\
      U_4^{\Lk_2}(d) &:=
        \bigl(U_4(d),\; F_H^{(4)}(d),\; F_S^{(4)}(d)\bigr),\\
      U_4^{\Lk_3}(d) &:=
        \bigl(U_4(d),\; F_H^{(4)}(d),\; F_S^{(4)}(d),\;
        F_P^{(4)}(d)\bigr),
    \end{align*}
    where $F_H^{(4)} = \FH \circ U_4$,
    $F_S^{(4)} = \FS \circ U_4$,
    $F_P^{(4)} = \FP \circ U_4$
    are the lifted decorations of Proposition~\ref{prop:L4-lifting}.
\end{itemize}
\end{definition}

\begin{proposition}[Factorisation of $U_4$ through the tower]
\label{prop:U4-factorisation}
Let $V_k: \Lk_k(P) \to \Lk_{k-1}(P)$ denote the forgetful
functor that drops the topmost decoration at level $k$
(i.e.\ $V_1$ forgets $\FH$, $V_2$ forgets $\FS$,
$V_3$ forgets $\FP$).
The following chain of equalities holds:
\[
  U_4
  \;=\; V_1 \circ U_4^{\Lk_1}
  \;=\; V_1 \circ V_2 \circ U_4^{\Lk_2}
  \;=\; V_1 \circ V_2 \circ V_3 \circ U_4^{\Lk_3}.
\]
In particular, $U_4$ factors through every intermediate
level of the tower.
\end{proposition}

\begin{proof}
By Definition~\ref{def:U4-intermediate}, each $U_4^{\Lk_k}$
retains the stoichiometric data $U_4$ and adds the
decorating functor values.
Each $V_k$ drops the topmost decoration, so
$V_k \circ U_4^{\Lk_k} = U_4^{\Lk_{k-1}}$ (with
$U_4^{\Lk_0} := U_4$).
Chaining gives the stated equalities.
\end{proof}

The following commutative diagram displays the full tower
of forgetful functors from $\Lk_4(P)$:

\[
\includegraphics{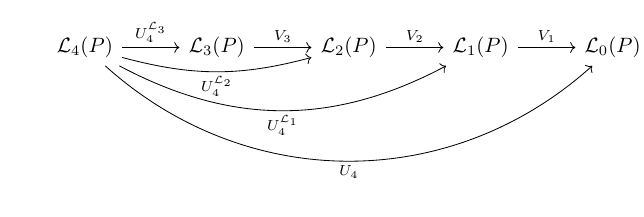}
\]

The commutativity of each triangle
(e.g.\ $V_3 \circ U_4^{\Lk_3} = U_4^{\Lk_2}$)
follows from Proposition~\ref{prop:U4-factorisation}.

The decorating functors $\FH$, $\FS$, $\FP$ factor
through $U_4$ as three additional commutative triangles:
\[
\includegraphics{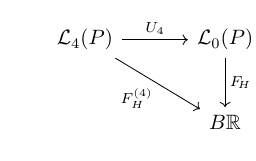}
\qquad
\includegraphics{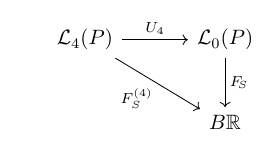}
\qquad
\includegraphics{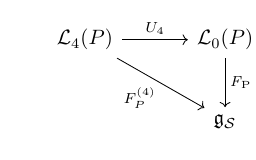}
\]

Each triangle commutes by definition of the lifted
decoration ($F_H^{(4)} := \FH \circ U_4$, etc.).
The three triangles together say: every piece of numerical
data at $\Lk_4$ is obtained from $\Lk_0$-level data by
first forgetting the DPO structure (via $U_4$) and then
applying the appropriate functor.
No new \emph{primitive} numerical functor enters at $\Lk_4$
beyond those lifted from $\Lk_3$ via $U_4$; the new
information lives entirely in the morphisms of $\Lk_4(P)$.

\subsubsection{Lifting previous decorations to
  \texorpdfstring{$\Lk_4$}{L4}}

\begin{proposition}[Lifting]
\label{prop:L4-lifting}
  Define
  \[
    F_H^{(4)} \;:=\; \FH \circ U_4,\qquad
    F_S^{(4)} \;:=\; \FS \circ U_4,\qquad
    F_P^{(4)} \;:=\; \FP \circ U_4.
  \]
  Each is a strict SMC functor from $\Lk_4(P)$ to the
  respective target ($B\RR$ or $\mathfrak{g}_{\Sp}$).
\end{proposition}

\begin{proof}
Composition of strict SMC functors is a strict SMC functor.
\end{proof}

\begin{remark}[No new numerical functor at $\Lk_4$]
\label{rem:no-new-functor}
Proposition~\ref{prop:L4-lifting} is the categorical
statement that \emph{no new primitive numerical functor}
enters at $\Lk_4$: every numerical functor on $\Lk_4(P)$
factors through $U_4$ as $F^{(4)} = F \circ U_4$ for some
functor $F$ on $\Lk_0(P)$.
The extension from $\Lk_3(P)$ to $\Lk_4(P)$ adds
structural information --- the bond-change mechanism ---
but not a new real-valued or stochastic-valued label
generated independently of the $\Lk_3$ data.
This contrasts with $\Lk_1$--$\Lk_3$, where each level
added exactly one new functor into $B\RR$ or $\mathfrak{g}_{\Sp}$.

A chemist might expect the activation energy $E_a$ to
enter at $\Lk_4$, since bond breaking and forming are
precisely the mechanism-level events that set the
energy barrier.
However, $E_a$ is the minimax of the potential energy
surface $V: \mathcal{C}_e(G) \to \RR$ along the
intrinsic reaction coordinate --- a quantity that requires
the full \emph{geometric} configuration space
$\mathcal{C}_e(G) = \RR^{3n}/(\mathrm{SE}(3) \ltimes \Aut(G))$
at $\Lk_5$, not merely the bond-change graph at $\Lk_4$.
A DPO rule specifies \emph{which} bonds break and form,
but the same rule can correspond to different values of
$E_a$ depending on the 3-dimensional arrangement of atoms
(e.g.\ a bulky substituent at the phosphorus centre raises
$E_a$ for an $\mathrm{S_N2}$-P step relative to an unhindered
analogue, even though both go through the same DPO rule).
$E_a$ is therefore not functorial at $\Lk_4$; it enters
the tower at $\Lk_5$ as the minimax of $V$ along the
IRC (Section~\ref{sec:L5}).
\end{remark}

\begin{remark}[The $\Lk_4$ quadruple]
\label{rem:L4-quadruple}
Putting together Definition~\ref{def:L4},
the universal property Proposition~\ref{prop:UP-L4},
and the lifting Proposition~\ref{prop:L4-lifting},
the mechanistic level is the quadruple
\[
  \Lk_4(P)
  \;=\;
  \bigl(\,\Lk_4(P),\;
    F_H^{(4)},\;
    F_S^{(4)},\;
    F_P^{(4)}\,\bigr),
\]
where the first component is the underlying free SMC on
DPO rules and the remaining three are the lifted
thermochemical, entropy, and kinetic decorations.
The forgetful functor $U_4$ embeds this quadruple into
the tower via the factorisation of
Proposition~\ref{prop:U4-factorisation}: going from
$\Lk_4$ to any lower level amounts to applying $U_4$
(to recover $\Lk_0$) and then using the appropriate
decorator inclusion or forgetful functor to reach the
desired level.
\end{remark}

%% file: chapters/L4/l4_layer.tex
\subsection{Layer~1 and Layer~2 at \texorpdfstring{$\Lk_4$}{L4}}
\label{sec:L4-layer}

The Layer~1/Layer~2 split established at $\Lk_1$--$\Lk_3$ recurs
at $\Lk_4$, but with a structurally different character.
At $\Lk_1$--$\Lk_3$, the split distinguishes the free assignment
of a numerical functor (Layer~1) from the coherence condition that
links it to lower tower levels (Layer~2).
At $\Lk_4$, no new \emph{primitive} numerical functor enters:
the decorating functors $F_H^{(4)}$, $F_S^{(4)}$, $F_P^{(4)}$
are all lifted from below via $U_4$
(Proposition~\ref{prop:L4-lifting}), so they are already fixed.
The split instead constrains the \emph{choice of DPO rules}:
which bond-change patterns, together with their Petri-net
labelling, yield molecular graphs on which the lifted
decorations remain well-defined.

\subsubsection{Layer~1: valid DPO rules}

\textbf{Layer~1 for $\Lk_4$} admits any chemical DPO rule
$p = (s, \gamma(p))$ in the sense of
Definition~\ref{def:dpo-rule}: a valence-conserving,
stoichiometrically consistent span
$s = (L \leftarrow K \rightarrow R)$ in $\LGraphP$ together
with a generator label $\gamma(p) \in \Rx$, subject to the
dangling condition (Proposition~\ref{prop:gluing}) for all
intended matches.
The free parameter at Layer~1 is the choice of reaction-centre
graphs $L$, $K$, $R$ --- the specific bond-change pattern ---
for a given generator label $\gamma(p)$.
For some Petri-net labels there are multiple Layer~1-valid
rules, corresponding to different mechanisms by which a
single reaction can proceed; for other labels, there may be
no single-span Layer~1 rule at all, and the only realisations
are as composites of other rules through intermediate species.
The forcing pair of Section~\ref{sec:L4-forcing-in} ---
$r_\text{concerted}$ versus $r_\text{stepwise}$ for identity
substitution at phosphorus --- exhibits exactly this
asymmetry: $r_\text{concerted}$ is the $U_4$-image of a
single Layer~1 rule $p_\text{concerted}$, while
$r_\text{stepwise}$ admits no single-span realisation and
arises at $\Lk_4$ only as the composite
$p_\text{elim} \circ p_\text{add}$ through the
trigonal-bipyramidal intermediate.

\subsubsection{Layer~2: valence coherence}

A Layer~1 rule (Definition~\ref{def:dpo-rule}) already
guarantees stoichiometric consistency between the span and its
generator label, and valence conservation at every context
atom in $K$.
What remains to check is the \emph{product side}: atoms in
$R \setminus r(K)$ acquire new labels $(\mathrm{el}, q', \rho')$
and new incident bonds during the reaction, and these must
combine to give a valid molecular graph.
Layer~2 imposes this one remaining condition.

\begin{definition}[Layer~2 condition at $\Lk_4$]
\label{def:L4-layer2}
  A chemical DPO rule
  $p = (L \leftarrow K \rightarrow R, \gamma(p))$
  (Definition~\ref{def:dpo-rule}) satisfies Layer~2 if every
  vertex of the product graph has a non-negative integer
  lone-pair count:
  \[
    \ell(v) \in \NN
    \quad
    \text{for all } v \in V(R),
  \]
  where $\ell(v)$ is the lone-pair component of $v$'s vertex
  label --- determined by the basic valence formula of
  Definition~\ref{def:valence} at non-hypervalent atoms, and
  supplied as part of the label at hypervalent atoms (the
  pentacoordinate trigonal-bipyramidal phosphorus of the
  forcing pair being a representative case).
\end{definition}

\begin{mathbox}[Layer~2 as inter-level coherence]
\textbf{Where the condition lives.}
The context atoms in $K$ satisfy the valence constraint
automatically: by definition, $K$ contains atoms whose labels
and bonds are unchanged between $L$ and $R$
(Remark~\ref{rem:label-change}), so their lone-pair counts
are preserved trivially.
The genuine content of Layer~2 is at the reaction-centre
atoms in $R \setminus r(K)$: their post-reaction labels
$(\mathrm{el}, q', \rho')$ together with their new bond
orders must combine to give $\ell(v) \geq 0$ as a
non-negative integer.

\textbf{Why it matters: linkage to $\Lk_1$--$\Lk_2$.}
Enthalpy and Gibbs free energy are state functions evaluated
on valid molecular graphs.
A product atom with $\ell(v) < 0$ or non-integer $\ell(v)$ is
not part of a valid molecular graph: a lone-pair count is an
electron count, which must be a non-negative integer.
This rules out two kinds of failure --- the basic valence
formula of Definition~\ref{def:valence} forcing $\ell$
negative or fractional at a non-hypervalent atom, or a
hypervalent atom being supplied with an $\ell$ outside $\NN$.
The DPO derivation $d: G \Rightarrow H$ would then have
no valid target $H$ in $\LGraphP$, so the lifted
decoration $F_H^{(4)}(d) = \FH(U_4(d))$ is undefined at
the level of $\Lk_4$ even before composition with $\FH$.
The Layer~2 condition guarantees that every product of a
valid rule is a well-formed element of $\LGraphP$, keeping
$F_H^{(4)}, F_S^{(4)}, F_P^{(4)}$ well-defined throughout
any derivation.

\textbf{What Layer~2 does \emph{not} see.}
The forcing pair of Section~\ref{sec:L4-forcing-in} ---
$r_\text{concerted}$ versus $r_\text{stepwise}$ for
identity substitution at phosphorus --- involves two
chemical DPO rules that both satisfy Layer~2: both produce
valid molecular graphs at every step.
The pentacoordinate TBI intermediate of the stepwise pathway
is hypervalent (the basic valence formula of
Definition~\ref{def:valence} would give $\ell(\mathrm{P}) =
-1/2$); under the hypervalent-vertex remedy of
Remark~\ref{rem:hypervalent-P} in
Section~\ref{sec:L4-concerted-stepwise}, $\mathrm{P}$ in the
TBI carries a supplied $\ell = 0 \in \NN$, satisfying the
Layer~2 condition.
The lifted decorations on both rules are accordingly
well-defined.
The distinction between them --- whether the rule is a
single concerted DPO span or a composite span through the
TBI intermediate molecular graph --- is a Layer~1 degree of
freedom, invisible to the valence check that defines Layer~2.
This is as it should be: Layer~2 asks only that each rule
be chemically well-formed, not that it be mechanistically
distinguished.
The mechanistic distinction is recovered by the
DPO-derivation structure itself in $\Lk_4(P)$, not by any
additional tower-layer condition.
\end{mathbox}

\begin{observation}[Layer~2 at each level: a universal pattern]
\label{obs:layer2-pattern}
The Layer~2 conditions across the tower follow a common pattern.
In each case, Layer~1 is the unconstrained categorical choice
(functor, rate assignment, or DPO rule), and Layer~2 is the
coherence condition linking it to all lower levels.
\begin{center}
\renewcommand{\arraystretch}{1.4}
\begin{tabular}{p{1cm} p{4.2cm} p{5.8cm}}
  \hline
  \textbf{Level} & \textbf{Layer~1} & \textbf{Layer~2}\\
  \hline
  $\Lk_1$ & Any functor $\FH : \Lk_0 \to B\RR$
    & Coboundary: $\FH = \delta^0 h_f$
      (Hess's Law cycle condition)\\[4pt]
  $\Lk_2$ & Any pair $(\FH, \FS)$ of functors
    & Thermodynamic Wegscheider:
      both $\FH = \delta^0 h_f$ and $\FS = \delta^0 s_f$
      hold simultaneously,
      equivalently $F_G^T(c) = 0$ for every cycle $c$
      at all $T$\\[4pt]
  $\Lk_3$ & Any rate assignment $k_r$
    & Kinetic Wegscheider: loop-balance conditions
      on rate constants
      (Section~\ref{sec:L3-layer2})\\[4pt]
  $\Lk_4$ & Any chemical DPO rule $p = (s, \gamma(p))$
    per Definition~\ref{def:dpo-rule}
  & Valence coherence: $\ell(v) \in \NN$ at every atom
    of the product graph $R$
    (Definition~\ref{def:L4-layer2})\\
  \hline
\end{tabular}
\end{center}
\medskip
\noindent
The progression from $\Lk_1$ to $\Lk_4$ reveals a shift in
what Layer~2 constrains: at $\Lk_1$--$\Lk_2$ it constrains
the values of numerical functors on stoichiometric cycles;
at $\Lk_3$ it constrains ratios of rate constants on those
cycles; at $\Lk_4$ it constrains the graph structure of the
product molecules themselves, via the lone-pair count
condition of Definition~\ref{def:L4-layer2}.
In each case, the Layer~2 condition is precisely what is
needed to keep the tower's numerical data well-defined on
the new data introduced at that level.
\end{observation}

%% file: chapters/L4/l4_concerted_stepwise.tex
\subsection{The forcing-pair derivations and their distinctness}
\label{sec:L4-concerted-stepwise}

Section~\ref{sec:L4-forcing-in} identified the
$\Lk_3 \to \Lk_4$ forcing pair: identity nucleophilic
substitution at phosphorus, realisable either by a single
concerted $\mathrm{S_N2}$-P transition state or by stepwise
addition--elimination through a pentacoordinate
trigonal-bipyramidal intermediate (TBI).
Both pathways carry the same source and target complexes in
$\NN^{|\Sp|}$ and (under steady-state on the TBI) the same
$\Lk_3$ propensity; what distinguishes them is the
DPO-derivation structure at $\Lk_4$.
This subsection makes that distinction precise.

Throughout, $\mathrm{P}$ denotes the central phosphorus atom
of the substrate methyl ethyl\-phenyl\-phos\-phin\-ate,
written $\mathrm{Et(Ph)P(=O)OMe}$.
We write $\mathrm{OMe}$ for the leaving methoxide and
$\mathrm{{}^{18}OMe}$ for the isotopically labelled incoming
methoxide.
The ethyl group, phenyl group, and doubly-bonded oxygen
retain their labels and their bonds to phosphorus throughout
every rule below, and sit in $K$ together with phosphorus
itself (Remark~\ref{rem:label-change}).
The two methoxides change charge between $L$ and $R$
(neutral when bonded to P, $-1$ when free), so each appears
outside $K$ at any step where its bonded/free status changes.

\begin{remark}[On hypervalent phosphorus]
\label{rem:hypervalent-P}
The pentacoordinate phosphorus in the TBI is hypervalent:
its incident bond-order sum is $\sum_e b(e) = 6$ (Et, Ph,
P=O, plus two single P--O bonds to the methoxides), giving
$\ell(\mathrm{P}) = (5 - 0 - 0 - 6)/2 = -1/2$ from the basic
valence formula of Definition~\ref{def:valence}.
The basic formula assumes octet-like electron counting and
does not accommodate the three-centre four-electron bonding
characteristic of hypervalent main-group atoms.
A standard remedy is to extend the formalism so that, at
hypervalent vertices, $\ell$ is supplied as part of the
vertex label rather than computed from
$(v_\text{el}, q, \rho, \sum b)$; the basic formula is then
recovered as the special case where the supplied $\ell$
matches the octet-derived value.
\end{remark}

\begin{example}[Concerted $\mathrm{S_N2}$-P as a single DPO rule]
\label{ex:concerted-P-dpo}
The concerted rule has phosphorus unchanged in label
$(\mathrm{P}, 0, 0)$; the bonds to the two methoxides change
(P--OMe broken, P--$\mathrm{{}^{18}OMe}$ formed), and the
methoxide vertex labels change accordingly under the
bonded/free charge convention recorded above.

\begin{description}
  \item[$L$:] P with three spectator bonds (to Et, Ph, and
    the doubly-bonded O) plus one P--OMe bond;
    $\mathrm{{}^{18}OMe^-}$ as a free isolated component.
  \item[$K$:] P with the three spectator bonds only.
    Both methoxides are excluded from $K$ because their
    formal charge changes between $L$ and $R$.
  \item[$R$:] P with three spectator bonds plus one
    P--$\mathrm{{}^{18}OMe}$ bond;
    $\mathrm{OMe^-}$ as a free isolated component.
\end{description}

The rule fires as a single DPO step
$G \Rightarrow_{p_\text{concerted}, m} H$ with no
intermediate molecular graph; phosphorus is tetracoordinate
in both source and target, and the rule passes from $L$ to
$R$ without introducing a pentacoordinate species.
The generator label is $\gamma(p_\text{concerted}) =
r_\text{concerted}$.
\end{example}

\begin{example}[Stepwise addition--elimination through the TBI]
\label{ex:stepwise-P-dpo}
The stepwise rule is the sequential composite
$p_\text{elim} \circ p_\text{add}$ of an addition step
followed by an elimination step, going through the
pentacoordinate intermediate.

\medskip\noindent\textbf{Step 1 (Addition).}
$\mathrm{{}^{18}OMe^-}$ adds to phosphorus, expanding it
from tetracoordinate to pentacoordinate.

\begin{description}
  \item[$L_1$:] P with the three spectator bonds plus one
    P--OMe bond (the substrate);
    $\mathrm{{}^{18}OMe^-}$ as a free isolated component.
  \item[$K_1$:] the substrate alone, with all four bonds at
    $\mathrm{P}$ preserved (P--Et, P--Ph, P=O, P--OMe);
    the free $\mathrm{{}^{18}OMe^-}$ component is excluded
    because its formal charge changes between $L_1$ and $R_1$.
  \item[$R_1$:] P with three spectator bonds, one P--OMe
    bond, and one P--$\mathrm{{}^{18}OMe}$ bond.
    This is the TBI.
\end{description}

\noindent\textbf{Intermediate graph $G_1$ (the TBI).}
Phosphorus carries label $(\mathrm{P}, 0, 0)$ with five
incident bonds.
This is the pentacoordinate TBI; it is hypervalent in the
sense of Remark~\ref{rem:hypervalent-P} and is absent from
any concerted derivation.
The TBI is itself an element of $\Sp$, distinct from the
substrate and product.

\medskip\noindent\textbf{Step 2 (Elimination).}
$\mathrm{OMe^-}$ leaves from the TBI, returning phosphorus
to tetracoordinate.

\begin{description}
  \item[$L_2$:] P with all five bonds --- three to
    spectators, plus P--OMe and P--$\mathrm{{}^{18}OMe}$.
    This is the TBI.
  \item[$K_2$:] P with three spectator bonds plus one
    P--$\mathrm{{}^{18}OMe}$ bond.
    OMe is excluded because its formal charge changes
    between $L_2$ and $R_2$.
  \item[$R_2$:] $K_2$ together with $\mathrm{OMe^-}$ as a
    free isolated component (the bonded structure at
    $\mathrm{P}$ is unchanged from $K_2$).
\end{description}
The full stepwise derivation is
$G \Rightarrow_{p_\text{add}, m_1} G_1
 \Rightarrow_{p_\text{elim}, m_2} H$.
The component rules carry generator labels
$\gamma(p_\text{add}) = r_\text{add}$ and
$\gamma(p_\text{elim}) = r_\text{elim}$ in $\Rx$.

Together with the concerted pathway, the two derivations
assemble into a triangle in $\Lk_4(P)$:
\[
  \includegraphics{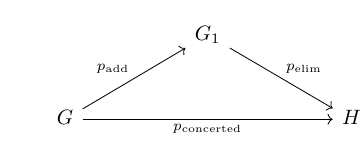}
\]
where $G = \mathrm{Et(Ph)P(=O)OMe} + \mathrm{{}^{18}OMe^-}$
is the substrate complex,
$G_1 = \mathrm{TBI}$ is the pentacoordinate intermediate,
and $H = \mathrm{Et(Ph)P(=O){}^{18}OMe} + \mathrm{OMe^-}$ is
the product complex.
\end{example}

\begin{proposition}[The forcing-pair triangle does not commute]
\label{prop:concerted-neq-stepwise}
The direct edge $p_\text{concerted}$ and the composite
$p_\text{elim} \circ p_\text{add}$ of the triangle above are
distinct morphisms in $\Lk_4(P)$, both having source $G$
and target $H$.
\end{proposition}

\begin{proof}
\textbf{Same source-target.}
By inspection of Examples~\ref{ex:concerted-P-dpo}
and~\ref{ex:stepwise-P-dpo}: the source for both
derivations is the substrate plus the labelled methoxide,
\[
  G \;=\; \mathrm{Et(Ph)P(=O)OMe} + \mathrm{{}^{18}OMe^-},
\]
and the target for both is the labelled product plus the
departing methoxide,
\[
  H \;=\; \mathrm{Et(Ph)P(=O){}^{18}OMe} + \mathrm{OMe^-}.
\]

\textbf{Distinctness in $\Lk_4(P)$.}
The three DPO rules $p_\text{concerted}$, $p_\text{add}$,
and $p_\text{elim}$ are pairwise distinct: their spans
differ in the contents of the context $K$
(concerted: P plus three spectator bonds only; addition:
P plus spectators plus $\mathrm{OMe}$; elimination:
P plus spectators plus $\mathrm{{}^{18}OMe}$).
By Proposition~\ref{prop:L4-freeSMC}, $\Lk_4(P)$ is a free
strict SMC on the chemical DPO rules, and the multiset of
rules appearing in any expression for a morphism is
invariant under the strict SMC equational theory
(associativity and unitality of both $\circ$ and $\otimes$,
the interchange law
$(g_1 \otimes g_2) \circ (f_1 \otimes f_2) =
(g_1 \circ f_1) \otimes (g_2 \circ f_2)$,
and the naturality and coherence axioms for the symmetry
$\tau$); each axiom rearranges or inserts identities without
changing the multiset of chemical generators that appear.
The direct edge has rule-multiset $\{p_\text{concerted}\}$,
while the composite has rule-multiset
$\{p_\text{add}, p_\text{elim}\}$; the two multisets are
disjoint, so the morphisms are distinct.
\end{proof}

%% file: chapters/L4/l4_sn1sn2.tex
\subsection{Four mechanisms as distinct morphisms in
  \texorpdfstring{$\Lk_4(P)$}{L4(P)}}
\label{sec:L4-sn1sn2}

Section~\ref{sec:L4-forcing-in} established the forcing pair
for $\Lk_4$: concerted vs stepwise mechanisms at a single
heteroatom centre (the phosphorus identity-substitution
example), the only kind of pair $\Lk_3$ genuinely conflates.
The classical mechanism pairs $\mathrm{S_N1}$ vs $\mathrm{S_N2}$
and $\mathrm{E1}$ vs $\mathrm{E2}$ are not forcing pairs in
this sense --- their different propensity forms
($v = k_1[\mathrm{RX}]$ vs $v = k_2[\mathrm{RX}][\mathrm{Nu}]$,
and analogously for E1/E2) separate them already at $\Lk_3$.
What $\Lk_4$ adds for these familiar pairs is a
\emph{structural} account that complements the kinetic
distinction at $\Lk_3$: each mechanism becomes a specific
DPO derivation, with intermediate molecular graphs that
appear (or fail to appear) according to whether the
mechanism is concerted or stepwise.
This section constructs explicit DPO derivations in
$\LGraphP$ for all four mechanisms and verifies pairwise
distinctness in $\Lk_4(P)$.

Throughout, atoms are labelled by $(\mathrm{el}, q, \rho)$
and bonds by their order $b$.
The Layer~2 valence check at each step verifies that every
atom of the product graph $R$ satisfies $\ell(v) \in \NN$
(Definition~\ref{def:valence}).
For consistency with
Remark~\ref{rem:label-change}, atoms whose label
$(\mathrm{el}, q, \rho)$ changes between $L$ and $R$ appear
in $L \setminus l(K)$ and $R \setminus r(K)$ separately;
only atoms with unchanged labels sit in $K$.

\subsubsection{Substitution pair:
  \texorpdfstring{$\mathrm{S_N1}$}{SN1} and
  \texorpdfstring{$\mathrm{S_N2}$}{SN2}}

Both mechanisms realise the same net transformation
\[
  \mathrm{R{-}X} + \mathrm{Nu^-}
  \;\longrightarrow\;
  \mathrm{R{-}Nu} + \mathrm{X^-},
\]
and carry the same reactants and products in $\NN^{|\Sp|}$.
They are distinguished at $\Lk_3$ by their propensity forms
($v = k_1\,[\mathrm{RX}]$ for $\mathrm{S_N1}$,
$v = k_2\,[\mathrm{RX}][\mathrm{Nu}]$ for $\mathrm{S_N2}$).
The DPO derivations below show \emph{why} the propensities
differ: the two mechanisms have different DPO-level
structure, with $\mathrm{S_N1}$ passing through an
intermediate molecular graph that $\mathrm{S_N2}$ does not.

\begin{example}[$\mathrm{S_N2}$ as a single concerted DPO rule]
\label{ex:SN2-dpo}
In the substrate, the electrophilic carbon $\mathrm{C}$ bears
label $(\mathrm{C}, 0, 0)$ with four bonds: to R (1), two H
(2), and X (1) --- total $b_\mathrm{C} = 4$, giving
$\ell(\mathrm{C}) = 0$ (no lone pairs).
The leaving group X has $(\mathrm{el}_X, 0, 0)$ in the
substrate and becomes $(\mathrm{el}_X, -1, 0)$ as X$^-$.
The nucleophile Nu has $(\mathrm{el}_\mathrm{Nu}, -1, 0)$ as
Nu$^-$ and becomes $(\mathrm{el}_\mathrm{Nu}, 0, 0)$ bonded
to C.
The carbon label $(\mathrm{C}, 0, 0)$ is unchanged;
X and Nu change charge and therefore cannot sit in $K$.

The DPO rule is a concerted composite of generators
$g_2$ (heterolytic C--X cleavage) and $g_1$ (heterolytic
C--Nu bond formation), applied as a single rule:
\begin{align*}
  L &: \text{C bonded to X ($b=1$); X$:(\mathrm{el}_X, 0, 0)$;
    Nu$^-:(\mathrm{el}_\mathrm{Nu}, -1, 0)$ isolated},\\
  K &: \{\mathrm{C}\} \text{ only, label }(\mathrm{C}, 0, 0),
    \text{ no bonds},\\
  R &: \text{C bonded to Nu ($b=1$);
    Nu$:(\mathrm{el}_\mathrm{Nu}, 0, 0)$;
    X$^-:(\mathrm{el}_X, -1, 0)$ isolated}.
\end{align*}
The span is:
\[
  \includegraphics{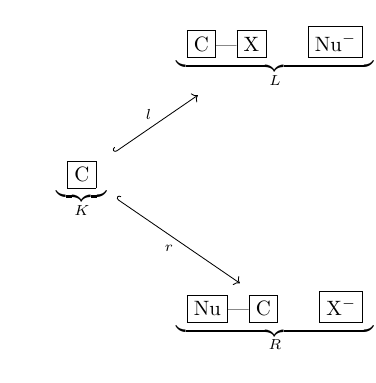}
\]
The DPO derivation on a host molecule $G$
(containing R--C--X and the Nu$^-$ ion) is:
\[
  \includegraphics{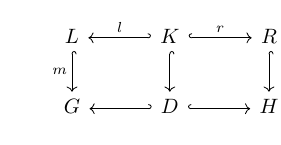}
\]
where $D$ is $G$ with the C--X bond deleted (and X detached,
awaiting re-labelling) and $H$ is $G$ with C--X broken, C--Nu
formed, and the new charges assigned.
The entire bond swap occurs in a single DPO step with no
intermediate molecular graph.

\textbf{Layer~2 checks.}
\begin{itemize}
  \item \emph{Valence conservation at each $R$ atom.}
    C in $R$: unchanged label $(\mathrm{C}, 0, 0)$, still 4 bonds
    (R, 2H, Nu), so $\ell(\mathrm{C}) = 0$.
    Nu in $R$: $(\mathrm{el}_\mathrm{Nu}, 0, 0)$, gains one bond
    (to C); if Nu = OH$^-$ with O as the attacking atom,
    $\ell(\mathrm{O}^L) = (6-(-1)-0-1)/2 = 3$ (three lone pairs
    as hydroxide), and $\ell(\mathrm{O}^R) = (6-0-0-2)/2 = 2$
    (two lone pairs as alcohol oxygen) --- non-negative integer.
    X in $R$: $(\mathrm{el}_X, -1, 0)$, no bonds; for X = Br,
    $\ell(\mathrm{Br}^R) = (7-(-1)-0-0)/2 = 4$ (four lone pairs
    as bromide) --- non-negative integer.
  \item \emph{Generator label.}
    The Petri-net label is
    $\gamma(p_{\mathrm{S_N2}}) = r_{\mathrm{S_N2}}$,
    the concerted substitution generator
    (Definition~\ref{def:dpo-rule}).
    Its stoichiometric image in $\Lk_0(P)$ has source
    $\mathrm{RX} + \mathrm{Nu^-}$ and target
    $\mathrm{RNu} + \mathrm{X^-}$.
\end{itemize}
\end{example}

\begin{example}[$\mathrm{S_N1}$ as a two-step DPO derivation
  through a carbocation]
\label{ex:SN1-dpo}
The $\mathrm{S_N1}$ mechanism proceeds in two elementary DPO
steps, with a carbocation as an explicit intermediate
molecular graph.

\medskip
\noindent\textbf{Step 1 (Ionisation, generator $g_2$):}
heterolytic C--X cleavage; both electrons go to X.
\begin{align*}
  L_1 &: \text{C bonded to X ($b=1$);
    C$:(\mathrm{C}, 0, 0)$; X$:(\mathrm{el}_X, 0, 0)$},\\
  K_1 &: \emptyset
    \text{ (both C and X change label)},\\
  R_1 &: \text{C isolated, $(\mathrm{C}, +1, 0)$;
    X isolated, $(\mathrm{el}_X, -1, 0)$}.
\end{align*}

\noindent\textbf{Intermediate graph $G_1$.}
After Step~1, C$_\alpha$ carries $q = +1$ and
$\ell(\mathrm{C}) = (4-1-0-3)/2 = 0$ (three remaining bonds:
R, 2H; vacant orbital in place of the former C--X bond).
This carbocation vertex is the hallmark intermediate of
$\mathrm{S_N1}$ and is absent from any $\mathrm{S_N2}$
derivation.

\medskip
\noindent\textbf{Step 2 (Nucleophilic capture, generator $g_1$):}
Nu$^-$ donates a lone pair to the vacant orbital on C$^+$.
\begin{align*}
  L_2 &: \text{C$^+:(\mathrm{C}, +1, 0)$ and
    Nu$^-:(\mathrm{el}_\mathrm{Nu}, -1, 0)$, isolated},\\
  K_2 &: \emptyset
    \text{ (both C and Nu change label)},\\
  R_2 &: \text{C bonded to Nu ($b=1$);
    C$:(\mathrm{C}, 0, 0)$; Nu$:(\mathrm{el}_\mathrm{Nu}, 0, 0)$}.
\end{align*}

The full $\mathrm{S_N1}$ derivation is the sequential
composition:
\[
  G
  \;\xRightarrow{\;p_1,\,m_1\;}_{D_1}\;
  G_1
  \;\xRightarrow{\;p_2,\,m_2\;}_{D_2}\;
  H.
\]

\textbf{Layer~2 checks.}
\begin{itemize}
  \item \emph{Step 1 product valence.}
    C in $R_1$: $(\mathrm{C}, +1, 0)$, three bonds (R, 2H),
    $\ell(\mathrm{C}) = (4-1-0-3)/2 = 0$ (vacant orbital) ---
    non-negative integer.
    X in $R_1$: $(\mathrm{el}_X, -1, 0)$, no bonds; for
    X = Br, $\ell(\mathrm{Br}) = 4$ --- non-negative integer.
  \item \emph{Step 2 product valence.}
    C in $R_2$: $(\mathrm{C}, 0, 0)$, four bonds,
    $\ell(\mathrm{C}) = 0$ --- non-negative integer.
    Nu in $R_2$: label and valence as in
    Example~\ref{ex:SN2-dpo} --- non-negative integer.
  \item \emph{$U_4$-image of the composite.}
    By functoriality of $U_4$ (Definition~\ref{def:U4}), the
    composite $p_2 \circ p_1$ has
    $U_4(p_2 \circ p_1) = \gamma(p_2) \circ \gamma(p_1)$ in
    $\Lk_0(P)$ --- the 2-step composite of the elementary
    heterolytic-cleavage and -capture generators.
    This is parallel to (sharing source and target with) the
    postulated unimolecular generator
    $r_{\mathrm{S_N1}} \in \Rx$, but is a distinct morphism
    in $\Lk_0(P)$:
    $r_{\mathrm{S_N1}}$ admits no single-span Layer-1
    realisation in $\Lk_4(P)$.
\end{itemize}
\end{example}

\subsubsection{Elimination pair:
  \texorpdfstring{$\mathrm{E1}$}{E1} and
  \texorpdfstring{$\mathrm{E2}$}{E2}}

Both mechanisms realise the same net elimination
\[
  \mathrm{R{-}CH_2{-}CHR'X} + \mathrm{B^-}
  \;\longrightarrow\;
  \mathrm{R{-}CH{=}CHR'} + \mathrm{BH} + \mathrm{X^-},
\]
with $\mathrm{C}_\alpha$ denoting the carbon bearing the
leaving group and $\mathrm{C}_\beta$ the adjacent carbon
bearing the departing hydrogen.
They are distinguished at $\Lk_3$ by their propensity forms
($v = k_1\,[\mathrm{RX}]$ for $\mathrm{E1}$,
$v = k_2\,[\mathrm{RX}][\mathrm{B}]$ for $\mathrm{E2}$).

\begin{example}[$\mathrm{E2}$ as a single concerted DPO rule]
\label{ex:E2-dpo}
The $\mathrm{E2}$ mechanism is concerted: the base B$^-$
abstracts the $\beta$-hydrogen simultaneously with
C$_\alpha$--X bond breaking and C$_\alpha$=C$_\beta$
$\pi$-bond formation.
This is a composite of $g_1$, $g_2$, and $g_3$ applied as a
single DPO rule with cyclic reaction-centre topology (a
four-atom ring B--H--C$_\beta$--C$_\alpha$ connected through
X).
Label changes are: B from $(\text{el}_B, -1, 0)$ to
$(\text{el}_B, 0, 0)$; X from $(\text{el}_X, 0, 0)$ to
$(\text{el}_X, -1, 0)$.
C$_\alpha$, C$_\beta$, H carry unchanged labels
$(\mathrm{C}, 0, 0)$, $(\mathrm{C}, 0, 0)$, $(\mathrm{H}, 0, 0)$
respectively and sit in $K$; B and X change charge and sit
in $L \setminus l(K)$ and $R \setminus r(K)$.

\begin{align*}
  L &: \text{C$_\beta$--H ($b=1$), C$_\beta$--C$_\alpha$ ($b=1$),
    C$_\alpha$--X ($b=1$); B$^-$ isolated},\\
  K &: \{\mathrm{C}_\alpha, \mathrm{C}_\beta, \mathrm{H}\}
    \text{ with no bonds among them},\\
  R &: \text{B--H ($b=1$), C$_\beta$=C$_\alpha$ ($b=2$);
    X$^-$ isolated}.
\end{align*}
The derivation is a single DPO step
$G \Rightarrow_{p_{\mathrm{E2}},\,m} H$ with no intermediate
molecular graph.

\textbf{Layer~2 checks.}
\begin{itemize}
  \item \emph{Valence at each $R$ atom.}
    C$_\beta$ in $R$: $(\mathrm{C}, 0, 0)$, bonds are R (1),
    one H (1), and C$_\alpha$ (double, $b=2$); total bond sum
    $= 4$, so $\ell(\mathrm{C}_\beta) = (4-0-0-4)/2 = 0$ ---
    non-negative integer.
    C$_\alpha$ in $R$: $(\mathrm{C}, 0, 0)$, bonds are R' (1),
    one H (1), and C$_\beta$ (double, $b=2$); total $= 4$,
    $\ell(\mathrm{C}_\alpha) = 0$ --- non-negative integer.
    B in $R$: for B$^-$ = OH$^-$, the protonated product BH = H$_2$O
    with O bonded to 2 H ($(\mathrm{O}, 0, 0)$):
    $\ell(\mathrm{O}) = (6-0-0-2)/2 = 2$ --- non-negative
    integer.
    H in $R$: $(\mathrm{H}, 0, 0)$, bonded to B only,
    $\ell(\mathrm{H}) = (1-0-0-1)/2 = 0$ --- non-negative integer.
    X in $R$: $(\mathrm{el}_X, -1, 0)$, no bonds; for X = Br,
    $\ell(\mathrm{Br}) = 4$ --- non-negative integer.
  \item \emph{Generator label.}
    $\gamma(p_{\mathrm{E2}}) = r_{\mathrm{E2}}$, the
    concerted elimination generator.
\end{itemize}
\end{example}

\begin{example}[$\mathrm{E1}$ as a two-step DPO derivation
  through a carbocation]
\label{ex:E1-dpo}
The $\mathrm{E1}$ mechanism proceeds in two elementary steps,
sharing its carbocation intermediate with $\mathrm{S_N1}$:
C$_\alpha$ ionises first, then a base (or solvent) removes a
$\beta$-proton, electrons flowing to form the
C$_\alpha$=C$_\beta$ $\pi$-bond.

\medskip
\noindent\textbf{Step 1 (Ionisation, generator $g_2$):}
heterolytic C$_\alpha$--X cleavage, identical to Step~1 of
$\mathrm{S_N1}$.
\begin{align*}
  L_1 &: \text{C$_\alpha$ bonded to X ($b=1$);
    C$_\alpha:(\mathrm{C}, 0, 0)$; X$:(\mathrm{el}_X, 0, 0)$},\\
  K_1 &: \emptyset,\\
  R_1 &: \text{C$_\alpha^+$ isolated, $(\mathrm{C}, +1, 0)$;
    X$^-$ isolated, $(\mathrm{el}_X, -1, 0)$}.
\end{align*}

\noindent\textbf{Intermediate graph $G_1$.}
C$_\alpha$ with $(\mathrm{C}, +1, 0)$, $\ell = 0$, three
remaining bonds (to R', one H, C$_\beta$).
This is the same carbocation intermediate as in
$\mathrm{S_N1}$; $\mathrm{E1}$ and $\mathrm{S_N1}$ share
this graph, which is why they compete kinetically from the
same cationic branch point.

\medskip
\noindent\textbf{Step 2 (Deprotonation and $\pi$-formation,
  generators $g_1$, $g_2$, and $g_3$):}
the base B$^-$ removes the $\beta$-hydrogen while the
C$_\beta$--H bond electrons flow to form the
C$_\alpha$=C$_\beta$ double bond.
Label changes: B from $(\text{el}_B, -1, 0)$ to
$(\text{el}_B, 0, 0)$; C$_\alpha$ from $(\mathrm{C}, +1, 0)$
to $(\mathrm{C}, 0, 0)$.
C$_\beta$ and H carry unchanged labels and sit in $K_2$.
\begin{align*}
  L_2 &: \text{C$_\beta$--H ($b=1$), C$_\beta$--C$_\alpha^+$
    ($b=1$); C$_\alpha^+:(\mathrm{C}, +1, 0)$;
    B$^-$ isolated},\\
  K_2 &: \{\mathrm{C}_\beta, \mathrm{H}\}
    \text{ with no bonds among them},\\
  R_2 &: \text{B--H ($b=1$); B$:(\text{el}_B, 0, 0)$;
    C$_\beta$=C$_\alpha$ ($b=2$); C$_\alpha:(\mathrm{C}, 0, 0)$}.
\end{align*}

The full $\mathrm{E1}$ derivation:
\[
  G
  \;\xRightarrow{\;p_1,\,m_1\;}_{D_1}\;
  G_1
  \;\xRightarrow{\;p_2,\,m_2\;}_{D_2}\;
  H.
\]

\textbf{Layer~2 checks.}
\begin{itemize}
  \item \emph{Step 1 product valence.}
    Same as Example~\ref{ex:SN1-dpo}, Step 1.
  \item \emph{Step 2 product valence.}
    C$_\alpha$ in $R_2$: $(\mathrm{C}, 0, 0)$, bonds to R' (1),
    one H (1), C$_\beta$ double (2); total $= 4$,
    $\ell(\mathrm{C}_\alpha) = 0$ --- non-negative integer.
    C$_\beta$ in $R_2$: $(\mathrm{C}, 0, 0)$, bonds to R (1),
    one H (1), C$_\alpha$ double (2); total $= 4$,
    $\ell(\mathrm{C}_\beta) = 0$ --- non-negative integer.
    B and H in $R_2$: same as $\mathrm{E2}$ case above.
  \item \emph{$U_4$-image of the composite.}
    By functoriality of $U_4$, the composite $p_2 \circ p_1$
    has $U_4(p_2 \circ p_1) = \gamma(p_2) \circ \gamma(p_1)$
    in $\Lk_0(P)$ --- the 2-step composite of the elementary
    heterolytic-cleavage and deprotonation/$\pi$-formation
    generators.
    This is parallel to the postulated unimolecular generator
    $r_{\mathrm{E1}} \in \Rx$ but is a distinct morphism in
    $\Lk_0(P)$:
    $r_{\mathrm{E1}}$ admits no single-span Layer-1
    realisation in $\Lk_4(P)$.
\end{itemize}
\end{example}

\subsubsection{All four mechanisms are distinct in
  \texorpdfstring{$\Lk_4(P)$}{L4(P)}}

\begin{proposition}[Four mechanisms as distinct morphisms]
\label{prop:sn1-neq-sn2}
  In $\Lk_4(P)$, the DPO derivations of
  Examples~\ref{ex:SN2-dpo}--\ref{ex:E1-dpo} are pairwise
  distinct morphisms.
  \begin{enumerate}[label=(\roman*)]
    \item The $\mathrm{S_N2}$ derivation
      (Example~\ref{ex:SN2-dpo}) and the $\mathrm{S_N1}$
      derivation (Example~\ref{ex:SN1-dpo}) are distinct
      morphisms in $\Lk_4(P)$ and project to distinct
      morphisms in $\Lk_0(P)$:
      $U_4(\mathrm{S_N2}) = r_{\mathrm{S_N2}}$ a single
      generator, $U_4(\mathrm{S_N1}) = \gamma(p_2) \circ
      \gamma(p_1)$ a 2-step composite, parallel to the
      postulated generator $r_{\mathrm{S_N1}} \in \Rx$ but
      distinct from it.
      All have shared source $\mathrm{RX} + \mathrm{Nu^-}$
      and target $\mathrm{RNu} + \mathrm{X^-}$.
    \item The $\mathrm{E2}$ derivation (Example~\ref{ex:E2-dpo})
      and the $\mathrm{E1}$ derivation (Example~\ref{ex:E1-dpo})
      are distinct morphisms in $\Lk_4(P)$;
      $U_4(\mathrm{E2}) = r_{\mathrm{E2}}$ a single generator,
      $U_4(\mathrm{E1}) = \gamma(p_2) \circ \gamma(p_1)$ a
      2-step composite at $\Lk_0(P)$, parallel to the
      postulated generator $r_{\mathrm{E1}} \in \Rx$ but
      distinct from it.
    \item No substitution morphism coincides with any
      elimination morphism: the four are pairwise distinct
      in $\Lk_4(P)$ and project to four pairwise distinct
      morphisms in $\Lk_0(P)$.
  \end{enumerate}
\end{proposition}

\begin{proof}
\textbf{(i) Substitution pair.}
The $\mathrm{S_N2}$ derivation is a single DPO step whose
intermediate graph $D$ contains neither a carbocation nor
any other chemically distinguished non-substrate vertex.
The $\mathrm{S_N1}$ derivation is the sequential composite
$p_2 \circ p_1$ passing through the intermediate molecular
graph $G_1$, which contains a vertex with label
$(\mathrm{C}, +1, 0)$ --- the carbocation.
In $\Lk_4(P)$, a morphism is an equivalence class of DPO
derivations under SMC congruence (associativity, unitality,
symmetry); no SMC congruence can collapse a derivation
passing through a $(\mathrm{C}, +1, 0)$ vertex to one that
never visits such a vertex.
Hence the two derivations represent distinct morphisms.

By Definitions~\ref{def:dpo-rule} and~\ref{def:U4},
$U_4(\mathrm{S_N2}) = \gamma(p_{\mathrm{S_N2}}) =
r_{\mathrm{S_N2}}$ is a single $\Lk_0$ generator, while
$U_4(\mathrm{S_N1}) = \gamma(p_2) \circ \gamma(p_1)$ is the
2-step composite of the elementary heterolytic-cleavage and
-capture generators (by functoriality of $U_4$).
A single generator and a 2-step composite of distinct
generators are distinct morphisms in the free SMC
$\Lk_0(P)$, so $U_4(\mathrm{S_N2}) \neq U_4(\mathrm{S_N1})$
even though both share source $\mathrm{RX} + \mathrm{Nu^-}$
and target $\mathrm{RNu} + \mathrm{X^-}$ in $\NN^{|\Sp|}$.
The composite $U_4(\mathrm{S_N1})$ is parallel to the
postulated generator $r_{\mathrm{S_N1}} \in \Rx$ but distinct
from it: $r_{\mathrm{S_N1}}$ has no single-span $\Lk_4$
realisation.

\textbf{(ii) Elimination pair.}
Analogous: the $\mathrm{E2}$ derivation has no intermediate
graph, while the $\mathrm{E1}$ derivation passes through the
carbocation vertex $(\mathrm{C}, +1, 0)$.
The two are distinct in $\Lk_4(P)$;
$U_4(\mathrm{E2}) = r_{\mathrm{E2}}$ is a single generator,
while $U_4(\mathrm{E1}) = \gamma(p_2) \circ \gamma(p_1)$ is a
2-step composite, parallel to the postulated
$r_{\mathrm{E1}} \in \Rx$ but distinct from it.

\textbf{(iii) Cross-class distinctness.}
The substitution generators and the elimination generators
have distinct source and target complexes in $\NN^{|\Sp|}$:
substitution consumes $\mathrm{RX} + \mathrm{Nu^-}$ and
produces $\mathrm{RNu} + \mathrm{X^-}$, while elimination
consumes $\mathrm{RX} + \mathrm{B^-}$ and produces
$\mathrm{alkene} + \mathrm{BH} + \mathrm{X^-}$.
In particular, the number of species produced differs.
Thus no substitution morphism coincides with any
elimination morphism at $\Lk_0$, and since $U_4$ is
well-defined on $\Lk_4$ morphisms, no substitution morphism
coincides with any elimination morphism at $\Lk_4$ either
(else their $U_4$-images would coincide in $\Lk_0$).
\end{proof}

\begin{chembox}[What $\Lk_4$ expresses that $\Lk_3$ cannot]
The proposition shows how the four classical mechanism
types are pairwise distinguished as morphisms in
$\Lk_4(P)$.
This is not the forcing argument of
Section~\ref{sec:L4-forcing-in} (which concerns the
$\Lk_3$-indistinguishable phosphorus pair); rather, it
demonstrates that the same DPO machinery automatically
separates mechanism types that $\Lk_3$ also distinguishes
kinetically.
In each case, $\Lk_3$ already distinguishes the two
mechanisms numerically via their different propensity forms,
but $\Lk_3$ supplies no \emph{structural} account.
$\Lk_4$ supplies that account: the two mechanisms correspond
to DPO derivations with genuinely different internal
structure --- one single-step without intermediate graphs,
the other two-step passing through a carbocation vertex
$(\mathrm{C}, +1, 0)$.

\medskip
\noindent\textbf{Substitution.}
At $\Lk_3$: $r_{\mathrm{S_N1}}$ and $r_{\mathrm{S_N2}}$ are
distinct generators with propensity values
$k_1 x_\mathrm{RX}$ and $k_2 x_\mathrm{RX} x_\mathrm{Nu}$
respectively.
These propensity forms suffice to separate the two morphisms
in $\Lk_3$ but give no \emph{structural} reason for the
difference.
At $\Lk_4$: the propensity difference is explained by the
different DPO derivation structures --- $\mathrm{S_N2}$ as a
single concerted rule, $\mathrm{S_N1}$ as a composite through
the carbocation intermediate graph.
The structural distinction has familiar downstream
consequences (Walden inversion for $\mathrm{S_N2}$; partial
racemisation for $\mathrm{S_N1}$ via planar carbocation
attack from either face), but these \emph{stereochemical}
consequences require three-dimensional data and enter the
tower at $\Lk_{4.5}$ rather than at $\Lk_4$.

\medskip
\noindent\textbf{Elimination.}
At $\Lk_3$: $r_{\mathrm{E1}}$ and $r_{\mathrm{E2}}$ are
distinct generators with propensities
$k_1 x_\mathrm{RX}$ and $k_2 x_\mathrm{RX} x_\mathrm{B}$,
which separate them numerically but again give no
structural account.
At $\Lk_4$: the difference is explained by the DPO
derivation structures --- $\mathrm{E2}$ as a single
concerted rule with cyclic reaction-centre topology,
$\mathrm{E1}$ as a composite through the same carbocation
intermediate graph as $\mathrm{S_N1}$.
The stereochemical consequences ($\mathrm{E2}$'s required
anti-periplanar geometry, absent in $\mathrm{E1}$) again
live at $\Lk_{4.5}$.

\medskip
\noindent\textbf{Summary.}
$\Lk_4$ explains \emph{why} the Petri net carries parallel
generators that $\Lk_3$ distinguishes only by their
propensity values.
The explanation is structural: two DPO derivations with
different intermediate graphs.
Three-dimensional consequences of this structure
(stereochemistry, orbital-symmetry selection rules) are the
content of $\Lk_{4.5}$ (Section~\ref{sec:L45}).
\end{chembox}

\subsubsection{\texorpdfstring{$\Lk_4$}{L4} in action:
  the Briggs--Rauscher oscillating reaction}

The examples of Section~\ref{sec:L4-sn1sn2} involved pairs of
mechanisms competing for a single substrate.
The Briggs--Rauscher (BR) oscillating reaction shows a richer
phenomenon: a network in which two categorically different
mechanism types compete and switch roles periodically, producing
macroscopic colour oscillations visible to the naked eye.
It illustrates how mechanism-level information --- the
structural data of $\Lk_4$ --- governs the topology of
reaction networks in ways that rate constants alone do not
encode, and provides the first case in this monograph where
a \emph{global} property of a reaction network (the
structural origin of oscillation) is exhibited as an
$\Lk_4$-level predicate.

\begin{chembox}[The Briggs--Rauscher reaction: physical description]
Mixing aqueous $\mathrm{IO_3^-}$, $\mathrm{H_2O_2}$, malonic
acid $\mathrm{CH_2(COOH)_2}$, $\mathrm{Mn^{2+}}$, sulfuric acid,
and starch produces a solution that cycles through
\[
  \text{colourless}
  \;\to\;
  \text{amber}
  \;\to\;
  \text{deep blue}
  \;\to\;
  \text{colourless}
  \;\to\;\cdots
\]
approximately ten to fifteen times before the reaction ends
\cite{BriggsRauscher1973}.
The \emph{amber} colour is free iodine ($\mathrm{I_2}$);
the \emph{deep blue} is the $\mathrm{I_3^-}$--starch complex,
which requires \emph{both} $\mathrm{I_2}$ and $\mathrm{I^-}$
simultaneously.
The overall net stoichiometry is
\[
  \mathrm{IO_3^-} + 2\,\mathrm{H_2O_2}
    + \mathrm{CH_2(COOH)_2} + \mathrm{H^+}
  \;\longrightarrow\;
  \mathrm{ICH(COOH)_2} + 2\,\mathrm{O_2} + 3\,\mathrm{H_2O}.
\]
This is the $U_4$-image of one full oscillation period: a
composite morphism in $\Lk_0(P)$ that conceals all mechanistic
structure.

The oscillation arises from two competing sub-processes for the
intermediate conversion of $\mathrm{IO_3^-}$ to hypoiodous acid
$\mathrm{HOI}$ \cite{NoyesFurrow1982}:

\medskip
\noindent\textbf{Process~A (non-radical/ionic): active when
  $[\mathrm{I^-}]$ is high.}

\begin{description}[leftmargin=2.8em, labelwidth=2.2em,
                    font=\normalfont\bfseries]
  \item[A1.] $\mathrm{IO_3^-} + \mathrm{I^-} + 2\mathrm{H^+}
    \to \mathrm{HIO_2} + \mathrm{HOI}$
    \hfill (slow; $g_2$, $g_1$)
  \item[A2.] $\mathrm{HIO_2} + \mathrm{I^-} + \mathrm{H^+}
    \to 2\,\mathrm{HOI}$
    \hfill (fast; $g_2$, $g_1$)
  \item[A3.] $\mathrm{HOI} + \mathrm{I^-} + \mathrm{H^+}
    \to \mathrm{I_2} + \mathrm{H_2O}$
    \hfill (fast; $g_1$, $g_2$)
\end{description}
Intermediate molecular graphs: $G_{\mathrm{HIO_2}}$ (iodous acid,
$q(\mathrm{I}){=}{+3}$, arising from step A1) and
$G_{\mathrm{HOI}}$ (hypoiodous acid, $q(\mathrm{I}){=}{+1}$,
arising from steps A1--A2).
Process~A consumes $\mathrm{I^-}$ and produces $\mathrm{I_2}$
(amber), driving $[\mathrm{I^-}]$ downward.

\medskip
\noindent\textbf{Process~B (radical): active when $[\mathrm{I^-}]$
  is low.}

\begin{description}[leftmargin=2.8em, labelwidth=2.2em,
                    font=\normalfont\bfseries]
  \item[B1.] $\mathrm{IO_3^-} + \mathrm{HIO_2} + \mathrm{H^+}
    \to 2\,\dot{\mathrm{IO}}_2 + \mathrm{H_2O}$
    \hfill ($g_5$: homolytic O--I)
  \item[B2.] $\dot{\mathrm{IO}}_2 + \mathrm{Mn^{2+}}
    + \mathrm{H_2O} \to \mathrm{HIO_2} + \mathrm{Mn^{3+}}
    + \mathrm{OH^-}$
    \hfill ($g_6$: SET)
  \item[B3.] $\mathrm{Mn^{3+}} + \mathrm{H_2O_2}
    \to \mathrm{Mn^{2+}} + \dot{\mathrm{HO}}_2 + \mathrm{H^+}$
    \hfill ($g_6$: SET, with H$^+$ loss)
  \item[B4.] $\dot{\mathrm{HO}}_2 + \dot{\mathrm{IO}}_2
    \to \mathrm{HIO_2} + \mathrm{O_2}$
    \hfill ($g_4$: radical combination)
\end{description}
Intermediate molecular graphs: $G_{\dot{\mathrm{IO}}_2}$ (iodyl
radical, $\rho(\mathrm{I}){=}1$), $G_{\mathrm{Mn^{3+}}}$
($q(\mathrm{Mn}){=}{+3}$), and $G_{\dot{\mathrm{HO}}_2}$
(hydroperoxyl radical, $\rho(\mathrm{O}){=}1$).
All three are absent from any Process~A derivation.
Process~B regenerates $\mathrm{HIO_2}$ autocatalytically (via
both B2 and B4), building the chain carrier rapidly; HOI is
then produced indirectly via A2 once $[\mathrm{I^-}]$ recovers.

\medskip
\noindent\textbf{Iodination loop~C: active throughout.}
\[
  \mathrm{I_2} + \mathrm{CH_2(COOH)_2}
  \;\to\;
  \mathrm{ICH(COOH)_2} + \mathrm{H^+} + \mathrm{I^-}
  \qquad
  (g_2,\, g_1\text{ on malonic acid})
\]
Loop~C is a slow ionic iodination that \emph{regenerates}
$\mathrm{I^-}$, refuelling Process~A once $[\mathrm{I^-}]$
rises again.

\medskip
\noindent\textbf{The switch.}
When $[\mathrm{I^-}]$ is high: Process~A dominates, consuming
$\mathrm{I^-}$ and building $\mathrm{I_2}$ (amber).
When $[\mathrm{I^-}]$ drops: Process~B ignites autocatalytically,
rapidly raising $[\mathrm{I_2}]$ and $[\mathrm{I^-}]$
simultaneously, triggering the starch complex (deep blue).
Loop~C slowly consumes $\mathrm{I_2}$ and rebuilds
$[\mathrm{I^-}]$, re-enabling Process~A and clearing the blue.
\end{chembox}

\medskip
The following observation makes this rigorous, showing that
the structural account of the oscillation --- which
mechanism families compose into the period morphism, and
how --- is an $\Lk_4$ phenomenon.

\begin{observation}[One BR oscillation period as a
  composite morphism in $\Lk_4(P)$]
\label{obs:BR-trajectory}

\textbf{Setup.}
Let $P_{\mathrm{BR}}$ be the Petri net containing all elementary
reactions of the BR mechanism.
Define three families of morphisms in $\Lk_4(P_{\mathrm{BR}})$:
\begin{align*}
  \mathcal{M}_A &:= \text{morphisms generated exclusively by }
    g_1,\, g_2 \text{ (heterolytic)}\\
  \mathcal{M}_B &:= \text{morphisms generated exclusively by }
    g_4,\, g_5,\, g_6 \text{ (radical / SET)}\\
  \mathcal{M}_C &\subset \mathcal{M}_A:
    \text{the iodination sub-family (}g_1,\, g_2
    \text{ on malonic acid)}
\end{align*}
The claim is $\mathcal{M}_A \cap \mathcal{M}_B = \{\mathrm{id}\}$:
no non-identity morphism belongs to both families.
In a free SMC (Proposition~\ref{prop:L4-freeSMC}), every
morphism can be expressed as a finite composite of generators,
and the multiset of generators appearing in that composite
(counted with multiplicities) is invariant under the SMC
congruence.
Since the generator sets $\{g_1, g_2\}$ and
$\{g_4, g_5, g_6\}$ are disjoint, no non-identity morphism
expressible using only generators from one set can be
SMC-congruent to one using only generators from the other;
hence the two generated sub-SMCs intersect only in the
identity.

\textbf{Objects.}
Define four molecular graphs (objects of $\LGraphP$) representing
the system at chemically distinguished points in one period:
\begin{align*}
  \mathcal{G}_{\mathrm{ionic}} &:=
    \text{state with high $[\mathrm{I^-}]$;
    Process~A poised to fire}\\
  \mathcal{G}_{\mathrm{radical}} &:=
    \text{state after A has consumed $\mathrm{I^-}$;
    $[\mathrm{I^-}]$ below threshold}\\
  \mathcal{G}_{\mathrm{loop}} &:=
    \text{state after B has produced $\mathrm{I_2}$;
    malonic acid iodination pending}\\
  \mathcal{G}_{\mathrm{restored}} &:=
    \text{state after C has restored $[\mathrm{I^-}]$;
    cycle complete}
\end{align*}

\textbf{The period morphism.}
Three composite morphisms fill the three segments:
\[
  \varphi_A \;:\; \mathcal{G}_{\mathrm{ionic}}
    \;\Rightarrow\; \mathcal{G}_{\mathrm{radical}},
  \quad
  \varphi_B \;:\; \mathcal{G}_{\mathrm{radical}}
    \;\Rightarrow\; \mathcal{G}_{\mathrm{loop}},
  \quad
  \varphi_C \;:\; \mathcal{G}_{\mathrm{loop}}
    \;\Rightarrow\; \mathcal{G}_{\mathrm{restored}},
\]
with $\varphi_A \in \mathcal{M}_A$,
$\varphi_B \in \mathcal{M}_B$,
$\varphi_C \in \mathcal{M}_C \subset \mathcal{M}_A$.
One oscillation period is the composite:
\[
  \tau_{\mathrm{period}} \;:=\;
  \varphi_C \;\circ\; \varphi_B \;\circ\; \varphi_A
  \;:\; \mathcal{G}_{\mathrm{ionic}}
  \;\Rightarrow\; \mathcal{G}_{\mathrm{restored}}.
\]

\textbf{Diagram.}
The following diagram shows one period as a sequence of DPO
derivations, with the intermediate molecular graphs named:
\[
\includegraphics{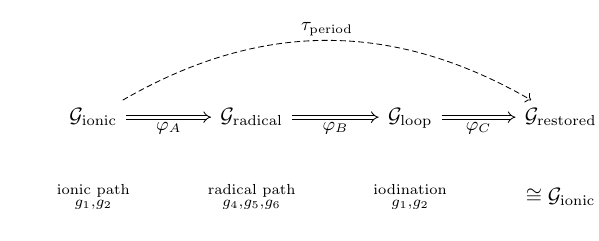}
\]
The isomorphism $\mathcal{G}_{\mathrm{restored}} \cong
\mathcal{G}_{\mathrm{ionic}}$ (equality of species counts up to
consumed reactants) closes the period: $\tau_{\mathrm{period}}$
can be iterated.

\textbf{Formal content.}
We establish three properties:

\begin{enumerate}[label=(\roman*)]
  \item \emph{$\tau_{\mathrm{period}} \notin \mathcal{M}_A$
    and $\tau_{\mathrm{period}} \notin \mathcal{M}_B$.}
    Since $\varphi_B \in \mathcal{M}_B$ and
    $\mathcal{M}_A \cap \mathcal{M}_B = \{\mathrm{id}\}$,
    any morphism containing $\varphi_B$ as a sub-derivation
    cannot lie in $\mathcal{M}_A$.
    Dually, $\varphi_A, \varphi_C \in \mathcal{M}_A$
    cannot lie in $\mathcal{M}_B$.
    Hence $\tau_{\mathrm{period}}$ is irreducible to either
    family alone: \emph{the structural realisation of oscillation
    requires both mechanism types to exist in $\Lk_4(P_{\mathrm{BR}})$.}

  \item \emph{$\tau_{\mathrm{period}}$ decomposes essentially uniquely
    as $\varphi_C \circ \varphi_B \circ \varphi_A$ in
    $\Lk_4(P_{\mathrm{BR}})$.}
    By Proposition~\ref{prop:L4-freeSMC}, $\Lk_4(P_{\mathrm{BR}})$
    is a free SMC.
    In a free SMC, the only relations among morphisms are those
    imposed by the SMC congruence (associativity, unitality,
    symmetry); no relation can merge a $g_1/g_2$ step with a
    $g_4/g_5/g_6$ step.
    The three-segment decomposition of $\tau_{\mathrm{period}}$
    is therefore unique up to the ordering of independent steps
    within each segment.

  \item \emph{$U_4(\tau_{\mathrm{period}})$ as a morphism in
    $\Lk_0(P)$ is determined by the full sequence of
    generators in the composite, not by the net stoichiometric
    change
    $U_4(\mathcal{G}_{\mathrm{restored}}) -
    U_4(\mathcal{G}_{\mathrm{ionic}}) \in \mathbb{Z}^{|\Sp|}$
    alone.}
    Both the radical-containing trajectory and a hypothetical
    all-ionic alternative consume one BR-stoichiometric unit, so
    their source-and-target objects in $\Lk_0(P)$ agree.
    However, the morphism $U_4(\tau_{\mathrm{period}})$ in
    $\Lk_0(P)$ is a specific composite of Petri-net generators
    that records which mechanism was used.
    The \emph{existence of oscillation}, which requires the
    radical-containing trajectory, is not visible from the
    source-target stoichiometry alone --- this is the sense in
    which it is invisible at the object level of $\Lk_0$.

\end{enumerate}

Property~(i) is the key: it shows that the structural
prerequisite for oscillation is an $\Lk_4$-level constraint
on the reaction network.
A network containing only one generator family ---
say, only ionic steps $g_1, g_2$ --- cannot form a
$\tau_{\mathrm{period}}$ containing $\varphi_B$, because
$\varphi_B$ requires radical generators absent from such
a network.
Oscillation becomes possible \emph{exactly when} the
Petri net $P$ contains generators from at least two
mechanism families whose DPO-derivations can be composed
into a cyclic trajectory in $\Lk_4(P)$.
\end{observation}

\begin{mathbox}[The tower dissects the BR reaction]
\textbf{At $\Lk_3$.}
The Furrow--Noyes model requires eleven elementary rate
equations with eleven rate constants to reproduce the
oscillation quantitatively \cite{NoyesFurrow1982}.
The CME generator $\Omega$ for the net HOI-production step
is $\Omega = k_{\mathrm{eff}}(R_{r_{\mathrm{HOI}}} - I)$,
where $k_{\mathrm{eff}}$ summarises the observed production rate
regardless of which process is active.
The rate-equation $\dot{[\mathrm{HOI}]} = k_{\mathrm{eff}}
[\mathrm{IO_3^-}][\mathrm{H_2O_2}]$ is the same whether
Process~A or Process~B runs: the switch is invisible.
The full Furrow--Noyes system does reproduce the oscillation, but
only after being given all eleven rate constants --- kinetic data
attached to elementary reactions, several of which are
radical-mechanism steps and several of which are ionic; the
radical/ionic distinction is part of the $\Lk_0$ generator
structure of $P_{\mathrm{BR}}$, with $\Lk_4$ supplying the
explicit DPO derivation structure that makes it geometrically
visible.

\begin{center}
\renewcommand{\arraystretch}{1.35}
\begin{tabular}{@{}c p{9.5cm}@{}}
  \hline
  \textbf{Level} & \textbf{Content in the BR system}\\
  \hline
  $\Lk_0$ & Overall BR stoichiometric equation; generators for each
    distinct reaction in the network, including separate
    generators for the ionic and radical HOI-production
    pathways\\
  $\Lk_1$ & $\Delta H$ of the net iodate-to-HOI transformation\\
  $\Lk_2$ & Equilibrium constant $K_{\mathrm{eq}}$ for HOI
    production; Wegscheider conditions\\
  $\Lk_3$ & Eleven rate constants of the Furrow--Noyes model;
    oscillation dynamics via the CME\\
  $\Lk_4$ & Two distinct DPO morphism families
    ($\mathcal{M}_A$ ionic, $\mathcal{M}_B$ radical);
    period morphism $\tau_{\mathrm{period}}$; the
    structural prerequisite for oscillation
    (Observation~\ref{obs:BR-trajectory})\\
  \hline
\end{tabular}
\end{center}
\end{mathbox}

\noindent
The Briggs--Rauscher reaction thus illustrates a general
principle that the tower makes precise:
\emph{the structural origin of oscillation is an $\Lk_4$
predicate: it requires that the network contain morphisms
from at least two distinct mechanism families that compose
into a cyclic trajectory}.
A kinetic model at $\Lk_3$ can fit the oscillation period
and amplitude to data once the reaction network is supplied,
but only $\Lk_4$ provides the structural account of why the
network admits oscillatory dynamics in the first place.

%% file: chapters/L4/l4_nextforcing.tex
\subsection{What \texorpdfstring{$\Lk_4$}{L4} cannot express:
  forcing of \texorpdfstring{$\Lk_{4.5}$}{L4.5}}
\label{sec:L4-forcing-out}

The mechanistic level $\Lk_4(P)$ was forced by the
concerted versus stepwise pathway at phosphorus: two
reactions sharing source, target, and (under steady-state on
the TBI) $\Lk_3$ propensity, distinguishable only at the
level of DPO derivations
(Sections~\ref{sec:L4-forcing-in},~\ref{sec:L4-concerted-stepwise}).
The same machinery applies to the analogous mechanism pairs
of carbon chemistry --- $\mathrm{S_N1}$ versus
$\mathrm{S_N2}$, $\mathrm{E1}$ versus $\mathrm{E2}$, and
ionic versus radical pathways in oscillating networks ---
though for these $\Lk_3$ already provides kinetic separation;
what $\Lk_4$ contributes is the structural account
(Section~\ref{sec:L4-sn1sn2}).
It does so by replacing stoichiometric transitions with DPO
derivations: objects are molecular graphs, morphisms encode
the sequence of bond changes, and mechanistically distinct
reactions are distinct morphisms.

Yet $\Lk_4(P)$ remains blind to \emph{spatial orientation}.
The label set $(\mathrm{el}, q, \rho)$ records element type,
formal charge, and radical count; bond-order labels record
connectivity.
None of these encode the three-dimensional arrangement of
substituents around a stereocentre.
Two molecules that are mirror images of each other are
graph-isomorphic in $\LGraphP$ --- identical vertex labels,
identical edge labels, identical adjacency structure --- and
are therefore identified by $\Lk_4(P)$, which sees nothing
beyond graph isomorphism.
Any DPO rule that applies to one applies identically to the
other.
The following forcing pair makes this concrete.

\begin{forcingbox}[Forcing pair for $\Lk_{4.5}$:
  enantiomers under $\mathrm{S_N2}$]
Consider the $\mathrm{S_N2}$ reaction of 2-bromobutane with
hydroxide, which has a stereocentre at C$_2$
(bonded to H, CH$_3$, CH$_2$CH$_3$, Br --- four distinct
substituents).
The reaction can proceed from either enantiomer:
\[
  (R)\text{-2-bromobutane} + \mathrm{OH^-}
  \;\longrightarrow\;
  (S)\text{-2-butanol} + \mathrm{Br^-},
\]
\[
  (S)\text{-2-bromobutane} + \mathrm{OH^-}
  \;\longrightarrow\;
  (R)\text{-2-butanol} + \mathrm{Br^-}.
\]
Both reactions use the same DPO rule from
Example~\ref{ex:SN2-dpo}: break C--Br, form C--OH, concerted
single step.
The molecular graphs of $(R)$- and $(S)$-2-bromobutane are
graph-isomorphic in $\LGraphP$: same element types, same
formal charges, same radical counts, same bond orders, same
adjacency.
No datum in $(\mathrm{el}, q, \rho)$ distinguishes
a left-handed from a right-handed arrangement of the four
substituents around C$_2$.

Yet the reactions are physically distinct: their products are
different enantiomers, and a chiral agent (an enzyme, a
resolving agent, an asymmetric catalyst) reacts with one but
not the other.
The swap $(R) \leftrightarrow (S)$ is an automorphism of
$\Lk_4(P)$ --- it preserves all categorical structure at the
graph level --- but corresponds to genuinely different
physical reactions.
It is a non-trivial element of $\coker(\varphi_{4.5})$ in the
exact sequence
\[
  1 \to \ker\varphi_{4.5} \to \Aut(\Lk_{4.5}(P))
  \;\xrightarrow{\;\varphi_{4.5}\;}\;
  \Aut(\Lk_4(P)) \to \coker(\varphi_{4.5}) \to 1,
\]
extending Proposition~\ref{prop:aut-tower} to the next level.
The same pattern recurs throughout stereochemistry:
$(E)$/$(Z)$ alkene isomers,
\textit{cis}/\textit{trans} cycloalkane diastereomers,
\textit{endo}/\textit{exo} cycloadducts.
The next chapter develops a second, independent forcing
pair --- the conrotatory/disrotatory selectivity of
pericyclic ring closures (Woodward--Hoffmann) ---
manifesting the same $\Lk_4$ blindness in a different
chemical setting (Section~\ref{sec:L45-forcing-in}).
\end{forcingbox}

\medskip\noindent\textbf{The type of extension
$\Lk_4 \to \Lk_{4.5}$.}
The extensions $\Lk_0 \to \Lk_1 \to \Lk_2 \to \Lk_3$ were
\emph{decorator extensions}: the underlying free SMC
$\Lk_0(P)$ was retained and a new numerical functor was
appended at each step.
The extension $\Lk_3 \to \Lk_4$ was a \emph{structural
extension}: the underlying free SMC was replaced by a new
one generated by DPO rules, with no new numerical functor.
The extension $\Lk_4 \to \Lk_{4.5}$ is of a third type: an
\emph{equivariant extension}.
The free SMC of DPO derivations is retained; no new
numerical functor is added.
Instead, the category is equipped with an action of a group
encoding spatial orientation, and morphisms are required to
be equivariant with respect to that action.
Chirality, stereochemical descriptors, and orbital-symmetry
selection rules are then the orbit structure and
representation theory of this equivariant extension.
The construction is developed in the following chapter.

%% file: chapters/ch_L45.tex
\section{\texorpdfstring{$\Lk_{4.5}$}{L4.5}: The Stereochemical Level}
\label{sec:L45}

\input{chapters/L45/l45_forcing}
\input{chapters/L45/l45_gstar}
\input{chapters/L45/l45_chiral}
\input{chapters/L45/l45_def}
\input{chapters/L45/l45_stereo}
\input{chapters/L45/l45_wh}
\input{chapters/L45/l45_nextforcing}

%% file: chapters/L45/l45_forcing.tex
\subsection{Forcing the extension: what \texorpdfstring{$\Lk_4$}{L4}
  cannot express}
\label{sec:L45-forcing-in}

Section~\ref{sec:L4-forcing-out} introduced the enantiomer
swap as the forcing pair motivating $\Lk_{4.5}$: the same
DPO rule applied to $(R)$- and $(S)$-substrates produces
different physical products that $\Lk_4$ cannot distinguish.
This section develops that forcing rigorously and introduces
a second independent forcing pair --- the conrotatory /
disrotatory distinction in pericyclic ring closure ---
which manifests the same $\Lk_4$ blindness in a different
chemical setting.
Both pairs are then resolved by a single discrete algebraic
datum: a chirality label
$\sigma\colon\mathrm{Stereo}(G)\to\{+1,-1\}$ at each
stereocentre (see Definition~\ref{def:chiral-centre}), augmenting the molecular graph without
introducing 3D coordinates.
Throughout, $\varphi_{4.5}$ denotes the restriction map
$\Aut(\Lk_{4.5})\to\Aut(\Lk_4)$ in the automorphism exact
sequence of the tower.

\begin{remark}[Why $\Lk_{4.5}$: a symmetry layer over $\Lk_4$]
\label{rem:why-45}
The fractional index $4.5$ is not a notational convenience.
It encodes the mathematical relationship between this level
and its predecessor: $\Lk_{4.5}(P)$ \emph{is} $\Lk_4(P)$
seen through a symmetry filter.
The objects are augmented molecular graphs --- the same
combinatorial kind as $\Lk_4$, decorated with chirality
labels --- and the morphisms remain DPO derivations;
but now only the $G^*$-equivariant derivations are admitted.
$\Lk_{4.5}(P)$ is a category over $\Lk_4(P)$ via the
forgetful functor $U_{4.5}\colon\Lk_{4.5}(P)\to\Lk_4(P)$
that drops the chirality labels: its morphisms are
$G^*$-equivariant DPO derivations on chirality-labelled
graphs, with the same DPO machinery as $\Lk_4(P)$.
It is not a sub-SMC of $\Lk_4(P)$ in the literal sense ---
its objects are augmented graphs $(G,\sigma)$, not bare
graphs --- but neither is it a new category built from
scratch: the underlying combinatorics and rewriting
machinery are inherited from $\Lk_4$.

This places $\Lk_{4.5}$ in contrast with the two surrounding
transitions.
The extension $\Lk_3 \to \Lk_4$ was a \emph{structural
break}: $\Lk_4$ introduced an entirely new class of
morphisms (DPO derivations in $\LGraphP$) that $\Lk_3$ had
no access to.
The forthcoming extension $\Lk_{4.5} \to \Lk_5$ will be a
\emph{geometric break}: $\Lk_5$ introduces genuinely
continuous and geometric data --- 3D atom positions and
the continuous potential energy landscape they define ---
a qualitatively new mathematical structure.

By contrast, $\Lk_4 \to \Lk_{4.5}$ is a
\emph{symmetry enrichment}: the same DPO framework,
restricted to its $G^*$-equivariant part.
No new numerical functor is added; the underlying free SMC
is not rebuilt.
The fractional index $4.5$ records exactly this: we are
still in the $\Lk_4$ world, but looking at the part of it
that respects a discrete symmetry group.
Separating $\Lk_{4.5}$ from $\Lk_5$ honours the bottom-up
forcing principle: each level is the unique minimal
extension resolving a specific pair of reactions, and the
two forcing pairs of this section are resolved by the
discrete $G^*$-datum alone.
\end{remark}

\begin{forcingbox}[Two forcing pairs for $\Lk_{4.5}$]

\noindent\textbf{Forcing pair~1: enantiomers under
$\mathrm{S_N2}$.}

Consider the bimolecular substitution of 2-bromobutane
(stereocentre at $\mathrm{C_2}$, bonded to four distinct
substituents: $\mathrm{CH_3}$, $\mathrm{C_2H_5}$,
$\mathrm{H}$, $\mathrm{Br}$) by hydroxide, starting from
each enantiomer:
\begin{align*}
  r_R\colon&\quad
    (R)\text{-2-bromobutane} + \mathrm{OH^-}
    \;\to\; (S)\text{-2-butanol} + \mathrm{Br^-},\\
  r_S\colon&\quad
    (S)\text{-2-bromobutane} + \mathrm{OH^-}
    \;\to\; (R)\text{-2-butanol} + \mathrm{Br^-}.
\end{align*}
Walden's 1896 observation that nucleophilic substitution
at a stereocentre proceeds with inversion of configuration
\cite{Walden1896} was mechanistically explained in the
1930s: Hughes and Ingold \cite{HughesIngold1935} identified
back-side attack as the $\mathrm{S_N2}$ mechanism, and
Cowdrey, Hughes, Ingold, Masterman, and Scott
\cite{Cowdrey1937} established in 1937 that the clean
elementary $\mathrm{S_N2}$ step at a configurationally
stable tetrahedral stereocentre proceeds with
\emph{complete} inversion; mixed mechanisms involving
$\mathrm{S_N1}$ contributions, neighbouring-group
participation, or ion-pair effects appear in the tower
as compositions of multiple $\Lk_{4.5}$-morphisms with
$\Lk_3$-level rate weights, not as deviations from the
elementary $\mathrm{S_N2}$ rule.
The question for the tower is whether this experimental
fact is visible as a theorem or only as an empirical
constraint.

In $\LGraphP$, the molecular graphs of $(R)$- and
$(S)$-2-bromobutane are isomorphic: identical element
types, formal charges, radical counts, bond orders, and
adjacency.
No datum in $(\mathrm{el}, q, \rho, b)$ distinguishes
left-handed from right-handed arrangements of substituents
around $\mathrm{C_2}$.
This rests on a modelling convention: the $\Lk_0$ species
set is not pre-refined by stereodescriptors.  Inserting
$(R)$- and $(S)$- as primitive species labels at $\Lk_0$
would build into the model the very distinction
$\Lk_{4.5}$ is constructed to derive.  Under this
convention, both reactions apply the \emph{same} DPO rule
(Example~\ref{ex:SN2-dpo}) to graph-isomorphic substrates
and produce graph-isomorphic products:
\[
  r_R \;=\; r_S \quad\text{as morphisms in } \Lk_4(P).
\]
The permutation $\pi_\chi$ that exchanges the two
graph-isomorphic enantiomer copies at the $\Lk_4$ level is
an automorphism of $\Lk_4(P)$ identifying the two
reactions --- but the physical reactions are distinct:
their products differ in sign of optical rotation, binding
affinity to chiral receptors, and metabolism by enzymes.
The entire field of asymmetric synthesis rests on the
distinction being real.
Hence $[\pi_\chi]$ is a non-trivial element of
$\coker(\varphi_{4.5})$.

\medskip
\noindent\textbf{Forcing pair~2: conrotatory vs.\
disrotatory electrocyclic ring closure.}

Consider the $4\pi$-electron electrocyclic ring closure of
$(E,E)$-hexa-2,4-diene to 3,4-dimethylcyclobutene.
(The $(E,E)$ annotation is alkene stereochemistry, which
sits outside the tetrahedral-stereocentre scope of
Definition~\ref{def:chiral-centre}; we treat it
informally here as a substrate descriptor and do not
attempt formal $E/Z$-handling in this chapter.)
The Woodward--Hoffmann orbital-symmetry rules
\cite{WoodwardHoffmann1965, WoodwardHoffmann1969} predict
that the stereochemical outcome depends on the electronic
state:
\begin{align*}
  d_{\mathrm{thermal}}\colon&\quad
    (E,E)\text{-hexa-2,4-diene}
    \;\xrightarrow{\;\Delta\;}\;
    \mathit{trans}\text{-3,4-dimethylcyclobutene},\\
  d_{\mathrm{photo}}\colon&\quad
    (E,E)\text{-hexa-2,4-diene}
    \;\xrightarrow{\;h\nu\;}\;
    \mathit{cis}\text{-3,4-dimethylcyclobutene}.
\end{align*}
The thermal reaction proceeds via a conrotatory motion
(the two terminal carbons rotate in the same direction
as the new $\sigma$-bond forms), preserving the $C_2$-axis
of the diene --- the symmetry element under which the
ground-state HOMO $\psi_2$ is symmetric.
The photochemical reaction proceeds via a disrotatory
motion (rotations in opposite directions), preserving
the $\sigma$-plane --- the symmetry element under which
the excited-state HOMO $\psi_3$ is symmetric.
The motions themselves are continuous trajectories in 3D
configuration space and live properly at $\Lk_5$; what
$\Lk_{4.5}$ records is their discrete product outcome
(\emph{trans} vs.\ \emph{cis}-3,4-dimethylcyclobutene).
The photochemical reaction proceeds disrotatorily (rotations
in opposite directions): this mode preserves the
$\sigma$-plane, the symmetry element under which the
excited-state HOMO $\psi_3$ is symmetric.
Hoffmann shared the 1981 Nobel Prize in Chemistry with
Fukui for the theoretical insight that orbital topology
governs reaction stereochemistry
\cite{FukuiNobel1982, HoffmannNobel1982}.

At $\Lk_4$, the two reactions are indistinguishable on
two independent grounds.
First, they apply the \emph{same} DPO rule
$p_{\mathrm{ec}}$ to the same host graph: one new
$\sigma$-bond $\mathrm{C_2{-}C_5}$ forms, and two
$\pi$-bonds ($\mathrm{C_2{=}C_3}$, $\mathrm{C_4{=}C_5}$)
reduce to single bonds:
\begin{align*}
  L_{\mathrm{ec}} &: \mathrm{C_2{=}C_3{-}C_4{=}C_5}
    \text{ (open diene centre)},\\
  K_{\mathrm{ec}} &: \{\mathrm{C_2, C_3, C_4, C_5}\}
    \text{ with no bonds among them},\\
  R_{\mathrm{ec}} &: \mathrm{C_2{-}C_3{=}C_4{-}C_5}
    \text{ plus new $\sigma$-bond } \mathrm{C_2{-}C_5}.
\end{align*}
Second, the two products \emph{trans}- and
\emph{cis}-3,4-dimethylcyclobutene are graph-isomorphic in
$\LGraphP$: they share the same atom set, same bond orders,
same adjacency.
Their difference lies entirely in the three-dimensional
arrangement of the two methyl groups at
$\mathrm{C_3}$ and $\mathrm{C_4}$ --- data not present in
$(\mathrm{el}, q, \rho, b)$.
The vertices $\mathrm{C_3}$ and $\mathrm{C_4}$ are
\emph{ring} stereocentres: removing such a vertex from a
ring leaves fewer than four connected components, so the
literal form of Definition~\ref{def:chiral-centre} does
not apply.  We use throughout this chapter the standard
generalisation via the CIP hierarchical digraph, which
handles ring atoms by introducing phantom duplicates at
ring-closure points; see \S\ref{sec:L45-gstar}.
Consequently, under the same $\Lk_0$-species convention
adopted for forcing pair~1,
\[
  d_{\mathrm{thermal}} \;=\; d_{\mathrm{photo}}
  \quad\text{as morphisms in } \Lk_4(P).
\]
Woodward--Hoffmann chemistry is invisible.

In any category that distinguishes the two products,
they carry distinct chirality labels: writing
$\sigma_{\mathit{trans}}(\mathrm{C_3}) \cdot
\sigma_{\mathit{trans}}(\mathrm{C_4}) = +1$ for matching
orientations and $\sigma_{\mathit{cis}}(\mathrm{C_3}) \cdot
\sigma_{\mathit{cis}}(\mathrm{C_4}) = -1$ for opposed
orientations makes the two products different augmented
graphs $(G, \sigma)$, even though they share the same
underlying $G$.
The permutation that identifies the two products in
$\Lk_4$ --- call it $\pi_{\mathrm{ec}}$ --- is an
automorphism of $\Lk_4(P)$ that does not lift to
$\Lk_{4.5}(P)$, where the two products are distinct
objects.
Hence $[\pi_{\mathrm{ec}}]$ is a non-trivial element of
$\coker(\varphi_{4.5})$, independent of $[\pi_\chi]$ from
pair~1.
\end{forcingbox}

The exact sequence
\begin{equation}
  \label{eq:L45-exact}
  1 \;\to\; \ker\varphi_{4.5} \;\to\; \Aut(\Lk_{4.5}(P))
  \;\xrightarrow{\;\varphi_{4.5}\;}\;
  \Aut(\Lk_4(P))
  \;\to\; \coker(\varphi_{4.5}) \;\to\; 1
\end{equation}
has a non-trivial cokernel containing at least two
independent elements $[\pi_\chi]$ and $[\pi_{\mathrm{ec}}]$
(Observation~\ref{obs:L45-independence}).
The extension $\Lk_4 \to \Lk_{4.5}$ is therefore forced by
at least two independent classes of chemical phenomenon:
stereoselective nucleophilic substitution and orbital-symmetry
selection in pericyclic reactions.

\medskip\noindent\textbf{What minimal structure resolves both
forcing pairs.}
Both pairs are resolved by a single discrete algebraic datum:
a chirality label $\sigma\colon\mathrm{Stereo}(G)\to\{+1,-1\}$
at each stereocentre, augmenting the molecular graph without
introducing 3D coordinates.
The precise objects --- chirality-labelled molecular graphs
$(G,\sigma)$ and the chirality-symmetry group $G^* =
\Aut(G)\ltimes\mathbb{Z}_2^k$ acting on them, where $k$ is
the number of stereocentres of $G$ --- are introduced in
\S\ref{sec:L45-gstar}--\ref{sec:L45-chiral}.
We use $G^*$ throughout the chapter for this construction;
note that this is \emph{not} the Longuet--Higgins
permutation--inversion group (often also written $G^*$ in
the molecular-spectroscopy literature) but a tower-internal
group tailored to the chirality-label data introduced here.
The connection to the molecular-spectroscopy $G^*$ is
discussed in \S\ref{sec:L45-gstar}.
Here we record only how each pair is resolved.

\begin{remark}[Resolving forcing pair~1: enantiomers]
\label{rem:resolve-fp1}
In $\Lk_{4.5}(P)$ the objects are augmented molecular graphs
$(G, \sigma)$ (Definition~\ref{def:chiral-graph}).
The two enantiomers $(G_R, \sigma_R)$ and $(G_S, \sigma_S)$
share the same underlying graph but carry opposite chirality
labels: $\sigma_R(\mathrm{C_2}) = +1$ and
$\sigma_S(\mathrm{C_2}) = -1$.
They are distinct objects, sitting in different orbits of
the $G^*$-action: the natural $G^*$-orbit of
$(G_R,\sigma_R)$ is the singleton $\{(G_R,\sigma_R)\}$ at
the level of unrooted graphs, and likewise for $(G_S,
\sigma_S)$.

The cokernel non-triviality of $[\pi_\chi]$ now follows
directly.
The permutation $\pi_\chi$ identifies $(G_R,\sigma_R)$ with
$(G_S,\sigma_S)$ as objects of $\Lk_4(P)$ (where they are
graph-isomorphic).
A lift $\widetilde{\pi}_\chi \in \Aut(\Lk_{4.5}(P))$ would
have to identify them as objects of $\Lk_{4.5}(P)$ as well,
mapping $(G_R,\sigma_R)$ to $(G_S,\sigma_S)$ on the nose.
But these are distinct objects of $\Lk_{4.5}(P)$, so no such
lift exists.
Hence $[\pi_\chi] \in \coker(\varphi_{4.5})$ is non-trivial.

A separate, stronger statement is that the
$G^*$-equivariant lift of the $\mathrm{S_N2}$ DPO rule
(Example~\ref{ex:SN2-dpo}) produces only the inverted
product --- the wrong-handed product is not in its image.
This is the categorical content of Hughes--Ingold complete
inversion, made rigorous in Theorem~\ref{thm:walden}.
For the present section we need only the weaker statement
above: $\Lk_{4.5}$ separates the two enantiomers as objects,
which suffices to make $[\pi_\chi]$ a non-trivial cokernel
class.
\end{remark}

\begin{remark}[Resolving forcing pair~2: conrotatory vs.\
disrotatory]
\label{rem:resolve-fp2}
In $\Lk_{4.5}(P)$ the two ring-closure products carry
distinct chirality labels at the new stereocentres
$\mathrm{C_3}$ and $\mathrm{C_4}$:
\begin{align*}
  (G_{\mathrm{cb}}, \sigma_{\mathit{trans}})\colon&\quad
    \sigma(\mathrm{C_3}) \cdot \sigma(\mathrm{C_4}) = +1
    \quad\text{(same orientation on both)},\\
  (G_{\mathrm{cb}}, \sigma_{\mathit{cis}})\colon&\quad
    \sigma(\mathrm{C_3}) \cdot \sigma(\mathrm{C_4}) = -1
    \quad\text{(opposite orientations)}.
\end{align*}
These are distinct objects in $\Lk_{4.5}(P)$.

A $G^*$-equivariant DPO rule for electrocyclic ring closure
must map reactants to products consistently with the
$G^*$-action: the conrotatory closure produces the
\emph{trans} product (matching orientations), while the
disrotatory closure produces the \emph{cis} product
(opposed orientations), and these outcomes are controlled
by which symmetry element of the transition state the rule
respects ($C_2$ for conrotatory, $C_s$ for disrotatory).
The permutation $\pi_{\mathrm{ec}}$ identifying the two
products in $\Lk_4$ does not lift to an equivariant
morphism of $\Lk_{4.5}$, because the two products have
different $\sigma$-signs at their stereocentres.
Thus $[\pi_{\mathrm{ec}}] \in \coker(\varphi_{4.5})$ is
non-trivial.

The Woodward--Hoffmann orbital-symmetry rule becomes a
candidate theorem at $\Lk_{4.5}$: a pericyclic DPO rule is
thermally allowed iff it is equivariant with respect to
the $C_2$ or $C_s$ subgroup of the transition-state
symmetry consistent with the ground-state orbital
occupation.
Theorem~\ref{thm:WH-categorical} states this precisely.
\end{remark}

\begin{observation}[Independence of the two classes in the
  cokernel]
\label{obs:L45-independence}
The cokernel classes $[\pi_\chi]$ and $[\pi_{\mathrm{ec}}]$
have disjoint support: $\pi_\chi$ acts non-trivially on the
stereocentre $\mathrm{C_2}$ of 2-bromobutane (and as the
identity elsewhere), while $\pi_{\mathrm{ec}}$ acts
non-trivially on the stereocentres $\mathrm{C_3}$ and
$\mathrm{C_4}$ of 3,4-dimethylcyclobutene (and as the
identity elsewhere).
The two automorphisms therefore commute as elements of
$\Aut(\Lk_4(P))$, and each squares to the identity (it is a
sign-flip of $\sigma$).
Their joint image in $\coker(\varphi_{4.5})$ is a quotient
of $(\mathbb{Z}/2) \times (\mathbb{Z}/2)$.

The four candidate elements
$\{1,\,[\pi_\chi],\,[\pi_{\mathrm{ec}}],\,
[\pi_\chi]\cdot[\pi_{\mathrm{ec}}]\}$ are pairwise distinct
in the cokernel.
Any non-trivial relation among them would, after
multiplication, exhibit one of $\pi_\chi$ or
$\pi_{\mathrm{ec}}$ as a lift of the other modulo the image
of $\varphi_{4.5}$, equivalently a lift of $\pi_\chi \cdot
\pi_{\mathrm{ec}}^{-1}$ to $\Aut(\Lk_{4.5}(P))$.
But $\pi_\chi \cdot \pi_{\mathrm{ec}}^{-1}$ still acts as
$\pi_\chi$ on the bromobutane stereocentre (since
$\pi_{\mathrm{ec}}$ is the identity there), and any lift to
$\Lk_{4.5}(P)$ would in particular have to identify
$(R)$-2-bromobutane with $(S)$-2-bromobutane as objects of
$\Lk_{4.5}(P)$ --- precluded by
Remark~\ref{rem:resolve-fp1}.
The same argument with the roles reversed precludes the
remaining identifications.
Hence $|\coker(\varphi_{4.5})| \geq 4$.

More generally, every reaction in $P$ producing or
consuming a non-meso stereoisomer pair contributes an
independent class by the same argument; molecules whose
$\Aut(G)$-action identifies $\sigma$ with $-\sigma$ (meso
forms) do not contribute, because their two sign
assignments are already identified at $\Lk_{4.5}$.
Stereochemistry is thus a pervasive feature of organic
chemistry, with $|\coker(\varphi_{4.5})|$ scaling with the
number of independent enantiomer pairs in $P$ rather than
with the raw stereocentre count.
\end{observation}

\begin{insightbox}[The two datums at $\Lk_{4.5}$]
The minimal new structure at $\Lk_{4.5}$ consists of two
datums, both discrete and algebraic:
\begin{itemize}
  \item \textbf{Chirality labels}
    $\sigma\colon\mathrm{Stereo}(G)\to\{+1,-1\}$
    on each stereocentre, augmenting the molecular graph.
    Six decades of asymmetric synthesis encoded as a sign.
  \item \textbf{The chirality-symmetry group}
    $G^* = \Aut(G)\ltimes\mathbb{Z}_2^k$
    (Definition~\ref{def:Gstar}), acting on chirality-labelled
    graphs by permuting atoms and flipping $\sigma$ at subsets
    of stereocentres.
    The Woodward--Hoffmann selection rules are captured as
    $G^*$-equivariance of the DPO rules --- no 3D geometry
    required.
\end{itemize}
Both datums are discrete and algebraic: the hallmark of the
symmetry-enrichment character of $\Lk_{4.5}$.
Three-dimensional consequences (specific molecular geometries,
activation barriers, vibrational modes) are deferred to
$\Lk_5$.
\end{insightbox}

%% file: chapters/L45/l45_gstar.tex
\subsection{The chirality symmetry group
  \texorpdfstring{$G^*$}{G*}}
\label{sec:L45-gstar}

Section~\ref{sec:L45-forcing-in} showed that both forcing
pairs for $\Lk_{4.5}$ --- enantiomer substrates under
$\mathrm{S_N2}$ and conrotatory/disrotatory electrocyclic
closures --- are resolved by equipping molecular graphs with
discrete chirality labels $\sigma\colon \mathrm{Stereo}(G)
\to \{+1,-1\}$ and by letting a finite group $G^*$ act on
these labels.
This section defines that group, identifies its two factors,
and verifies that it acts sensibly on chirality-labelled
graphs.
The $G^*$-action on $\Lk_{4.5}(P)$, projected through the
forgetful functor $U_{4.5}$, makes the forcing cokernel
classes $[\pi_\chi]$ and $[\pi_{\mathrm{ec}}]$ in
$\Aut(\Lk_4(P))$ visible as obstructions that the symmetry
enrichment to $\Lk_{4.5}$ removes.

The group $G^*$ has two factors.
The first, $\Aut(G)$, is the group of label-preserving graph
automorphisms of $G$: permutations of the vertex set $V$ that
fix the labelling functions $\lambda_V$, $\lambda_E$ and the
bond-order function $b$.
These are the physically meaningful relabellings ---
permutations of identical-nucleus atoms that also respect
the bonding pattern --- and form a graph-theoretic
restriction of the permutation part of the Longuet--Higgins
framework~\cite{LonguetHiggins1963}, namely those nuclear
permutations expressible from the connectivity data
$\Lk_4$ has made available.
The second factor encodes orientation flips at stereocentres.
For a molecule with $k = |\mathrm{Stereo}(G)|$ stereocentres,
the minimal group that can independently invert the
orientation at each centre is $\ZZ_2^k$.
A single global $\ZZ_2$ suffices to relate the two members
of an enantiomer pair (via simultaneous flips at all centres)
but cannot relate an enantiomer to a diastereomer --- e.g.\
no element of a global $\ZZ_2$ takes $(R,R)$-tartaric acid
to meso-tartaric acid, since the latter requires flipping
$\sigma$ at one stereocentre but not the other.
The independent single-centre flips in $\ZZ_2^k$ are required
for that distinction, as the tartaric acid example below will
make explicit.

\begin{definition}[Chirality symmetry group]
\label{def:Gstar}
  Let $G \in \LGraphP$ be a molecular graph with stereocentre
  set $\mathrm{Stereo}(G) = \{v_1, \ldots, v_k\}$
  (Definition~\ref{def:chiral-centre}).
  The \emph{chirality symmetry group} of $G$ is the
  semidirect product
  \[
    \Gstar(G) \;:=\; \Aut(G) \;\ltimes\; \ZZ_2^k,
  \]
  where:
  \begin{itemize}
    \item $\Aut(G)$ is the group of label-preserving graph
      automorphisms of $G$ (permutations of vertices
      preserving $\lambda_V$, $\lambda_E$, and bond orders).
    \item $\ZZ_2^k = \{(\epsilon_1,\ldots,\epsilon_k) :
      \epsilon_i \in \{+1,-1\}\}$ is the group of
      orientation flips, one per stereocentre.
    \item Fixing an enumeration
      $\Stereo(G) = \{v_1, \ldots, v_k\}$, every
      $\pi \in \Aut(G)$ induces a permutation
      $\bar\pi \in S_k$ via $\pi(v_i) = v_{\bar\pi(i)}$
      (since $\pi$ preserves stereogenicity, it maps
      $\Stereo(G)$ to itself).  The semidirect product
      structure is given by the action
      $\Aut(G) \to \Aut(\ZZ_2^k)$,
      \[
        \pi \cdot (\epsilon_1, \ldots, \epsilon_k)
        \;:=\;
        (\epsilon_{\bar\pi^{-1}(1)}, \ldots,
         \epsilon_{\bar\pi^{-1}(k)}),
      \]
      so that $(\pi_1, \boldsymbol{\epsilon}_1)
      (\pi_2, \boldsymbol{\epsilon}_2) =
      (\pi_1\pi_2,\, \boldsymbol{\epsilon}_1 \cdot
      (\pi_1 \cdot \boldsymbol{\epsilon}_2))$.
  \end{itemize}
  The group $\Gstar(G)$ acts on chirality-labelled molecular
  graphs $(G, \sigma)$ (Definition~\ref{def:chiral-graph}) by
  \[
    (\pi, \boldsymbol{\epsilon}) \cdot (G, \sigma)
    \;:=\;
    \bigl(G,\; \boldsymbol{\epsilon} \cdot
      (\sigma \circ \pi^{-1})\bigr),
  \]
  where $(\boldsymbol{\epsilon} \cdot \sigma)(v_i) :=
  \epsilon_i \cdot \sigma(v_i)$.
  The element
  $E^* := (\mathrm{id}, (-1,\ldots,-1)) \in \Gstar(G)$
  is the \emph{abstract parity operation}: it flips all
  chirality labels simultaneously.
\end{definition}

\begin{mathbox}[What $G^*$ knows and what it defers]
Definition~\ref{def:Gstar} requires no 3D coordinates.
$\Aut(G)$ is determined by the labelled graph structure of
$G \in \LGraphP$ (which atoms are bonded to which, with which
bond orders), and $\ZZ_2^k$ is a finite discrete group.
Both are available at $\Lk_4$.

The abstract parity $E^*$ is defined as an algebraic element
of $\ZZ_2^k$, not as a geometric reflection.
Its physical realisation as a spatial inversion in $\RR^3$
is a statement about $\Lk_5$ --- where 3D coordinates exist
--- not about $\Lk_{4.5}$.
At $\Lk_{4.5}$, $E^*$ is purely an operation on discrete
orientation labels.
\end{mathbox}

The stereocentres on which $G^*$ acts are now identified.

\begin{definition}[Stereocentre: graph-theoretic detection]
\label{def:chiral-centre}
  Let $v \in V(G)$ be a vertex with exactly four neighbours
  $w_1, w_2, w_3, w_4$.  For each neighbour $w_i$, the
  \emph{rooted CIP tree} $T_{w_i}^{v}$ is the labelled
  rooted tree obtained from $G \setminus \{v\}$ by
  depth-first traversal outward from $w_i$, with any
  back-edge (an edge that would close a ring within
  $G \setminus \{v\}$) replaced by a phantom leaf labelled
  by the element type of its target vertex.
  
  The vertex $v$ is a \emph{stereocentre} in $G$ if:
  \begin{enumerate}[label=(\roman*)]
    \item $v$ has exactly four neighbours in $G$
      (tetrahedral valence), and
    \item the four rooted CIP trees
      $T_{w_1}^{v}, T_{w_2}^{v}, T_{w_3}^{v}, T_{w_4}^{v}$
      are pairwise non-isomorphic as labelled rooted trees.
  \end{enumerate}
  Condition~(ii) is the graph-theoretic form of the
  Cahn--Ingold--Prelog (CIP) distinctness
  condition~\cite{CahnIngoldPrelog1966}: it is the precise
  criterion that makes a priority ordering of the four
  substituents unique and well-defined.  For acyclic
  neighbourhoods, the rooted CIP trees coincide with the
  four connected components of $G \setminus \{v\}$ rooted
  at the respective $w_i$, and the criterion reduces to
  the standard form.  For ring stereocentres (e.g.\ ring
  atoms in cyclobutenes, cyclohexanes), the phantom-leaf
  construction handles ring closures and recovers the
  standard CIP priority assignment.
  
  Both conditions are decidable from
  $(\lambda_V, \lambda_E)$ alone; no 3D information is
  needed.  The set of stereocentres of $G$ is denoted
  $\mathrm{Stereo}(G)$.
\end{definition}

\begin{remark}[CIP priorities and the chirality label]
\label{rem:CIP}
The Cahn--Ingold--Prelog rules~\cite{CahnIngoldPrelog1966}
assign a total priority order to the four substituents of a
stereocentre based on atomic number and then on graph
topology (the ``hierarchical digraph'' algorithm).
This order is computable from $G \in \LGraphP$ alone: atomic
numbers are part of $\mathrm{el}$, and topology is the graph
structure.

Given the CIP priority order, a chirality label
$\sigma(v) \in \{+1,-1\}$ determines the stereodescriptor:
$\sigma(v) = +1$ corresponds to the $(R)$-configuration
(priority sequence $1 > 2 > 3 > 4$ runs clockwise when
substituent~4 points away) and $\sigma(v) = -1$ to $(S)$.

At $\Lk_{4.5}$, however, $\sigma(v)$ is an \emph{abstract
binary label}: it records that a stereocentre has an
orientation and that the two orientations are distinct,
without committing to which spatial arrangement
$\sigma = +1$ realises.
The CIP rules establish that the label is well-defined and
graph-computable; which physical enantiomer it corresponds
to is a statement deferred to $\Lk_5$.
\end{remark}

\begin{observation}[$\Gstar(G)$ acts on chirality labellings]
\label{obs:gstar-action}
  The group $\Gstar(G)$ acts on the set of chirality
  functions $\sigma\colon \mathrm{Stereo}(G) \to \{+1,-1\}$
  in two qualitatively different ways corresponding to its
  two factors:
  \begin{itemize}
    \item $\pi \in \Aut(G)$ relabels stereocentres:
      $\sigma \mapsto \sigma \circ \pi^{-1}$.
      This relates labellings that differ only by atomic
      relabelling; they represent the \emph{same} molecule
      in $\LGraphGstar$ (Definition~\ref{def:chiral-graph}).
    \item $\boldsymbol{\epsilon} \in \ZZ_2^k$ flips
      orientations: $\sigma \mapsto
      \boldsymbol{\epsilon} \cdot \sigma$ with
      $(\boldsymbol{\epsilon} \cdot \sigma)(v_i) =
      \epsilon_i \sigma(v_i)$.
      This relates labellings representing \emph{chemically
      distinct} stereoisomers.
  \end{itemize}
  Within the $\ZZ_2^k$-action, the chemistry distinguishes
  two group-theoretic signatures:
  \begin{itemize}
    \item \emph{Enantiomers} are related by the global flip
      $E^* = (\mathrm{id}, (-1,\ldots,-1))$, which
      simultaneously inverts every stereocentre.
    \item \emph{Diastereomers} are related by a partial flip
      $(\mathrm{id}, \boldsymbol{\epsilon})$ with some but
      not all $\epsilon_i = -1$.
  \end{itemize}
  The combination of the two factors is why $\Gstar$ is
  the right group: $\Aut(G)$ quotients out labelling
  ambiguity, and $\ZZ_2^k$ encodes the two independent
  kinds of stereoisomer relation.
\end{observation}

\begin{chembox}[$G^*$ for tartaric acid: why $\ZZ_2^k$ and
  not $\ZZ_2$]
Tartaric acid $\mathrm{HOOC{-}CH(OH){-}CH(OH){-}COOH}$ has
two stereocentres at $\mathrm{C_2}$ and $\mathrm{C_3}$,
giving $k = 2$.
The graph $G$ has a non-trivial automorphism $\tau \in
\Aut(G)$ swapping $\mathrm{C_2} \leftrightarrow \mathrm{C_3}$
(together with the two COOH groups and the two OH groups);
the remaining automorphisms are generated by $\tau$, giving
$\Aut(G) = \langle \tau \rangle \cong \ZZ_2$.
The full chirality symmetry group is
$\Gstar(G) = \langle\tau\rangle \ltimes \ZZ_2^2$, with
$\tau$ acting on $\ZZ_2^2$ by exchanging the two coordinates
(since $\tau$ swaps the stereocentres at $\mathrm{C_2}$ and
$\mathrm{C_3}$); concretely,
$\langle\tau\rangle \ltimes \ZZ_2^2 \cong D_4$, the
dihedral group of order $8$.  Its underlying set is
indexed by

\medskip
\noindent\textbf{Three stereoisomers, two orbits.}
Tartaric acid has three physically distinct stereoisomers,
each an object in $\LGraphGstar$:
\begin{itemize}
  \item $(R,R)$-tartaric acid: labelling $(+,+)$.
  \item $(S,S)$-tartaric acid: labelling $(-,-)$.
  \item \emph{meso}-tartaric acid: labelling $(+,-)$, which
    equals $(-,+)$ as an object in $\LGraphGstar$ because
    $\tau$ provides an isomorphism between these two
    labellings (the two halves of the molecule are
    interchangeable).
\end{itemize}
Under the $\Gstar$-action these objects organise into two
orbits:
\begin{itemize}
  \item $\{(R,R), (S,S)\}$: the enantiomer pair, related by
    the global flip $E^* = (\mathrm{id}, (-,-))$.
    Orbit size~2.
  \item $\{\text{meso}\}$: a singleton orbit.
    The meso compound is its own image under $E^*$ because
    $E^* \cdot (+,-) = (-,+) \stackrel{\tau}{=} (+,-)$ as
    objects; equivalently, meso-tartaric acid is achiral
    (superimposable on its mirror image).
\end{itemize}

\medskip
\noindent\textbf{Why $\ZZ_2$ alone is insufficient.}
Replacing $\ZZ_2^2$ with a single $\ZZ_2$ (allowing only
the global flip $E^*$) would correctly relate the
enantiomers $(R,R)$ and $(S,S)$ but would leave the meso
compound entirely disconnected from the enantiomer pair
within the symmetry structure.
The partial flips $(+,-)$ and $(-,+)$, which relate
enantiomers to diastereomers (e.g.\ $(R,R)$ to meso via
$(\mathrm{id}, (+,-))$), would be absent.
$\ZZ_2^k$ is thus the minimal factor allowing $\Gstar$ to
express all stereoisomer relations among molecules with
multiple stereocentres.
\end{chembox}

The following proposition records how the $G^*$-action
accounts for the two forcing cokernel classes from
\S\ref{sec:L45-forcing-in}, and thereby characterises the
passage from $\Lk_4$ to $\Lk_{4.5}$ as a symmetry
enrichment.

\begin{proposition}[$G^*$ separates the forcing cokernel
  classes]
\label{prop:gstar-resolves}
Let $[\pi_\chi], [\pi_{\mathrm{ec}}] \in
\coker(\varphi_{4.5})$ be the two forcing classes from
\S\ref{sec:L45-forcing-in}.
\begin{enumerate}[label=(\roman*)]
  \item \emph{Enantiomer resolution.}
    Let $G$ be the underlying graph of 2-bromobutane,
    with stereocentre $v_\chi = \mathrm{C_2}$.
    The two labellings $(G, +1)$ and $(G, -1)$ are
    distinct objects in $\LGraphGstar$ by
    Definition~\ref{def:chiral-graph}.
    They are related by the $\Gstar(G)$-element
    $E^* = (\mathrm{id}, (-1)) \in \Gstar$ (for $k = 1$),
    which maps $(G, +1) \mapsto (G, -1)$.
    The $\Lk_4$-automorphism $\pi_\chi$ identifying the
    two enantiomers at the $\Lk_4$ level arises from this
    $\Gstar$-action together with the forgetful functor
    $U_{4.5}$: passing through $U_{4.5}$ collapses
    $(G, \pm 1)$ to $G$ and makes $E^*$ appear as the
    automorphism $\pi_\chi \in \Aut(\Lk_4(P))$.
    Within $\Lk_{4.5}(P)$, no automorphism identifies
    $(G, +1)$ with $(G, -1)$: the cokernel class
    $[\pi_\chi]$ is therefore non-trivial, and $E^*$
    is its representative in $\Gstar$.

  \item \emph{Electrocyclic resolution.}
    Let $G_{\mathrm{cb}}$ be the underlying graph of
    3,4-dimethylcyclobutene, with stereocentres
    $\mathrm{C_3}, \mathrm{C_4}$ ($k = 2$).
    The two labellings
    $(G_{\mathrm{cb}}, \sigma_{\mathit{trans}})$
    with $\sigma(\mathrm{C_3}) \cdot \sigma(\mathrm{C_4}) =
    +1$ and $(G_{\mathrm{cb}}, \sigma_{\mathit{cis}})$
    with $\sigma(\mathrm{C_3}) \cdot \sigma(\mathrm{C_4}) =
    -1$ are distinct objects in $\LGraphGstar$.
    They are related by a partial flip in $\ZZ_2^2 \leq
    \Gstar(G_{\mathrm{cb}})$: for any fixed
    $\sigma_{\mathit{trans}}$, the element
    $(\mathrm{id}, (+,-)) \in \ZZ_2^2$ maps
    $\sigma_{\mathit{trans}} \mapsto
    \sigma_{\mathit{trans}}'$ with the sign flipped at
    $\mathrm{C_4}$ only, giving a labelling with
    $\sigma(\mathrm{C_3}) \cdot \sigma(\mathrm{C_4}) = -1$,
    which is $\sigma_{\mathit{cis}}$ (up to the
    $\tau$-isomorphism when $\tau \in \Aut(G_{\mathrm{cb}})$
    swaps the two ring carbons).
    The $\Lk_4$-automorphism $\pi_{\mathrm{ec}}$
    identifying the conrotatory and disrotatory products
    at the $\Lk_4$ level arises from this partial-flip
    action through $U_{4.5}$, by the same mechanism as
    in part~(i).
    Within $\Lk_{4.5}(P)$, no automorphism identifies the
    two products: the cokernel class
    $[\pi_{\mathrm{ec}}]$ is therefore non-trivial.
\end{enumerate}
\end{proposition}

\begin{proof}
Both parts rest on the same mechanism.
The two labellings in question ($(G, +1)$ vs.\ $(G, -1)$
for part~(i), and $(G_{\mathrm{cb}},
\sigma_{\mathit{trans}})$ vs.\ $(G_{\mathrm{cb}},
\sigma_{\mathit{cis}})$ for part~(ii)) are distinct
objects in $\LGraphGstar$ by
Definition~\ref{def:chiral-graph}: two chirality functions
on the same underlying graph that are not related by any
element of $\Aut(G)$ give distinct objects.
The $\Gstar$-element relating them ($E^*$ in part~(i), a
partial flip in part~(ii)) is not an element of $\Aut(G)$
alone, so the relation is not a $\LGraphGstar$-isomorphism.
An automorphism of $\Lk_{4.5}(P)$ would have to send each
object to an isomorphic object; the labelling distinction
obstructs this, so the identification cannot lift.
Under $U_{4.5}$, which forgets the chirality labels, the
two objects collapse to the same underlying graph, and the
$\Gstar$-relation becomes a non-trivial automorphism of
$\Lk_4(P)$ --- the cokernel class.
\end{proof}

\begin{remark}[Provenance: classical chemistry, new
  categorical framing]
\label{rem:gstar-provenance}
Proposition~\ref{prop:gstar-resolves} records the
categorical form of a result that is well-established in
molecular symmetry theory.
The use of permutation-inversion groups in molecular
symmetry theory goes back to
Longuet-Higgins~\cite{LonguetHiggins1963}, who introduced
the spectroscopic permutation-inversion group as the
symmetry group of the full molecular Hamiltonian under
feasible permutations of identical nuclei combined with
the parity operation $E^*$; Bunker and
Jensen~\cite{BunkerJensen2006} develop this framework
systematically for spectroscopic applications.  This
spectroscopic group is not $\Gstar(G)$: the
Longuet-Higgins construction has a single global $E^*$,
whereas $\Gstar(G)$ has $k$ independent flips $\ZZ_2^k$,
the refinement needed to separate diastereomers from
enantiomers in molecules with multiple stereocentres.
That refinement was developed in the algebraic theory of
stereoisomerism due to Dugundji and
Ugi~\cite{DugundjiUgi1973} and by Ruch~\cite{Ruch1972},
the latter introducing the ``chirality function''
$\sigma\colon \Stereo(G) \to \{+1,-1\}$ used here.
The tartaric acid example of the chembox above is
classical, going back to van 't Hoff's original
demonstration of stereoisomerism in 1874.

What is specific to the tower construction is the
identification of $\Gstar$-elements with cokernel classes
of the automorphism exact sequence $\varphi_{4.5}$,
and the resulting characterisation of $\Lk_4 \to
\Lk_{4.5}$ as a symmetry enrichment that removes
precisely the obstructions $[\pi_\chi]$ and
$[\pi_{\mathrm{ec}}]$ identified in
\S\ref{sec:L45-forcing-in}.
The classical group-theoretic machinery is thus deployed
here to serve a specifically categorical purpose:
making the forcing argument of \S\ref{sec:L45-forcing-in}
rigorous and showing that no smaller group-theoretic
datum would suffice.
\end{remark}

\begin{remark}[Minimality of the two factors]
\label{rem:gstar-minimal}
Both factors of $\Gstar(G) = \Aut(G) \ltimes \ZZ_2^k$ are
necessary.
Without $\Aut(G)$, the action cannot identify equivalent
stereocentres: for instance, it would treat the two
labellings $(+,-)$ and $(-,+)$ of tartaric acid as distinct
objects, introducing spurious stereoisomers.
Without $\ZZ_2^k$ in full (replacing it with a proper
quotient such as the single global $\ZZ_2$), the action
cannot relate enantiomers to diastereomers, as the
tartaric acid example makes explicit.
The semidirect product
$\Aut(G) \ltimes \ZZ_2^k$ is therefore the smallest
group containing both factors with the required action
of $\Aut(G)$ on $\ZZ_2^k$ by index permutation,
and the action of Definition~\ref{def:Gstar} realises
precisely the structure needed to separate the forcing
cokernel classes of \S\ref{sec:L45-forcing-in} while
identifying chemically equivalent relabellings.
\end{remark}

Proposition~\ref{prop:gstar-resolves} makes precise the
sense in which $\Lk_{4.5}$ is ``$\Lk_4$ with a symmetry
layer'': the $\Gstar$-action is exactly the additional
structure needed to separate the forcing cokernel classes,
and Remark~\ref{rem:gstar-minimal} records that both
factors of $\Gstar$ are independently required.
The chirality-labelled molecular graphs $(G, \sigma)$ on
which this action operates are introduced in the next
section.

%% file: chapters/L45/l45_chiral.tex
\subsection{Chirality-labelled molecular graphs and
  \texorpdfstring{$\LGraphGstar$}{LGraph\_\{G*\}}}
\label{sec:L45-chiral}

The forcing analysis of \S\ref{sec:L45-forcing-in} and the
group $\Gstar(G) = \Aut(G)\ltimes\ZZ_2^k$ of
\S\ref{sec:L45-gstar} together dictate the shape of
$\Lk_{4.5}$: its objects must carry a discrete chirality
datum at every stereocentre, and $\Gstar$ must act on them
by symmetries of the category, not merely of the object set.
This section builds the ambient rewriting category in which
$\Lk_{4.5}$ will operate --- the chirality-aware refinement
$\LGraphGstar$ of $\LGraphP$ --- verifies adhesivity at the
ambient level (with chirality validity treated as an
admissibility predicate, so that DPO rewriting through
admissible rules carries over intact from $\Lk_4$), and
lifts the $\Gstar(G)$-action from objects to per-fibre
auto-equivalences of the category.
Section~\ref{sec:L45-def} will then define $\Lk_{4.5}(P)$
as the free strict SMC on $\Gstar$-equivariant DPO rules in
$\LGraphGstar$.

\begin{definition}[Chirality-labelled molecular graph]
\label{def:chiral-graph}
  A \emph{chirality-labelled molecular graph} is a pair
  $(G, \sigma)$ where $G \in \LGraphP$ is a molecular graph
  and $\sigma\colon \mathrm{Stereo}(G) \to \{+1, -1\}$ is a
  \emph{chirality function} assigning an orientation sign to
  each stereocentre (Definition~\ref{def:chiral-centre}).
  The pair $(G, \sigma)$ records exactly the datum absent
  from $\Lk_4(P)$: a discrete orientation at every
  stereocentre.
  If $\Stereo(G) = \emptyset$, then $\sigma$ has empty
  domain and $(G, \emptyset)$ is achiral.
\end{definition}

\begin{definition}[The category $\LGraphGstar$]
\label{def:LGraphGstar}
  The category $\LGraphGstar$ of chirality-labelled molecular
  graphs has:
  \begin{itemize}
    \item \textbf{Objects}: chirality-labelled molecular graphs
      $(G, \sigma)$.
    \item \textbf{Morphisms}: label-preserving graph
      monomorphisms $f\colon G \hookrightarrow H$ that are
      \emph{chirality-compatible}: for every stereocentre
      $v \in \mathrm{Stereo}(G)$, if $f(v) \in
      \mathrm{Stereo}(H)$ then $\sigma_H(f(v)) =
      \sigma_G(v)$.
    \item \textbf{Monoidal product}: $(G_1, \sigma_1) \sqcup
      (G_2, \sigma_2) := (G_1 \sqcup G_2,\,
      \sigma_1 \sqcup \sigma_2)$, with chirality functions
      concatenated over the disjoint union of stereocentre
      sets.
    \item \textbf{Monoidal unit}: $(\emptyset, \emptyset)$.
  \end{itemize}
  Chemically: an object of $\LGraphGstar$ is a molecule
  together with a discrete orientation sign at each
  stereocentre, and a morphism is a molecular-graph inclusion
  that preserves those signs wherever it meets stereocentres.
\end{definition}

\noindent
The forgetful functor
$U_{4.5}\colon \LGraphGstar \to \LGraphP$,
$(G, \sigma) \mapsto G$, is the tower-level bridge:
$\LGraphGstar$ is $\LGraphP$ enriched by chirality data,
and $U_{4.5}$ collapses that enrichment.
This is the categorical shadow of the $\Lk_{4.5}
\rightsquigarrow \Lk_4$ forgetful structure that will be
extended to full morphisms in \S\ref{sec:L45-def}.

\begin{mathbox}[Why chirality-compatibility is the right
  morphism condition]
A morphism $f\colon (G, \sigma) \to (H, \tau)$ in
$\LGraphGstar$ is an $\LGraphP$-monomorphism that preserves
every chirality label it transports.
Chemically: a reaction step that does not mechanistically
touch a stereocentre cannot change its configuration.
When $f(v)$ remains a stereocentre in $H$, the label
$\sigma_G(v)$ must equal $\sigma_H(f(v))$; this is what
chirality-compatibility says.
Steps that \emph{do} touch a stereocentre are handled by DPO:
the reactive vertex passes through the context graph $K$
(where its reduced neighbour count makes it non-stereogenic
by Definition~\ref{def:chiral-centre}(i)), so
$\mathrm{Stereo}(K)$ omits it, and it re-emerges in $R$ with
a new chirality label assigned by the rule's chirality lift.
The label change is a consequence of the pushout
construction, not of a morphism that flips $\sigma$ in
transit.
\end{mathbox}


\begin{proposition}[Adhesivity at the ambient level;
  admissibility for $\LGraphGstar$]
\label{prop:LGraphGstar-adhesive}
Let $\widetilde{\LGraph}^{\Gstar}_P$ be the typed attributed
graph category obtained from $\LGraphP$ by adjoining a
vertex attribute
$\sigma\colon V(G) \to \{+1, -1, \star\}$
to each graph, with the third value $\star$ denoting
``no chirality datum''.  This ambient category is adhesive
as a typed attributed graph category over the adhesive
base $\LGraphP$ in the sense of
Ehrig--Ehrig--Prange--Taentzer~\cite{EhrigEtAl2006}.

The category $\LGraphGstar$ embeds as the full subcategory
whose objects $(G, \sigma)$ satisfy the
\emph{chirality admissibility predicate}: $\sigma(v) \in
\{+1, -1\}$ for $v \in \Stereo(G)$ and $\sigma(v) = \star$
otherwise.  DPO rewriting in $\LGraphGstar$ is performed in
the ambient adhesive category, with admissibility verified
rule-by-rule: a rule $p = (L \leftarrow K \rightarrow R)$
is \emph{admissible} when $L$, $K$, and $R$ all satisfy
the predicate, and pushouts along admissible matches
preserve admissibility provided the rule's chirality lift
specifies $\sigma$-values consistently with the
$\Stereo$-pattern of $R$ (cf.\ Remark~\ref{rem:L4-to-L45-lift}).
\end{proposition}

\begin{proof}
Adhesivity of $\widetilde{\LGraph}^{\Gstar}_P$ follows
directly from~\cite{EhrigEtAl2006} Theorem~11.11: the
attribute category $\{+1, -1, \star\}$ is discrete (hence
has all limits), and the typing functor assigning
$\sigma\colon V(G) \to \{+1, -1, \star\}$ to each graph is
a standard total vertex attribute, preserving pullbacks
trivially.

Admissibility preservation under pushouts along admissible
rules is verified locally: at every vertex of the pushout
$H$, either $v$ inherits its attribute from $G$ (preserving
admissibility because $G$ was admissible), or $v$ inherits
from $R$ via the rule (preserving admissibility because the
rule's chirality lift was specified to match the
$\Stereo$-pattern of $R$).  No vertex inherits from both
sources unambiguously by the pushout property; agreement on
the overlap is enforced by the span morphisms'
chirality-compatibility.
\end{proof}

\begin{remark}[Why admissibility rather than direct
  adhesivity]
\label{rem:LGraphGstar-admissibility}
The full subcategory $\LGraphGstar$ is not in general
adhesive in isolation: pushouts in the ambient
$\widetilde{\LGraph}^{\Gstar}_P$ may yield graphs in which
$\Stereo$ has changed at some vertex (e.g.\ a 3-coordinate
vertex in $L$ becoming 4-coordinate in $R$ as new bonds
form), so a pushout of admissible objects need not be
admissible without an explicit chirality lift specifying
the new $\sigma$-values.  Treating chirality validity as an
admissibility predicate inside the adhesive ambient
category, rather than seeking adhesivity of the validity-
restricted subcategory itself, parallels the treatment of
chemical validity in Chapter~\ref{sec:L4} and is the
standard pattern for attribute-dependent rewriting.
\end{remark}

\begin{remark}[Direct verification as a cross-check]
\label{rem:LGraphGstar-direct}
A direct construction of pullbacks and pushouts in
$\LGraphGstar$ confirms the abstract argument.
Given a cospan $(H_1, \sigma_1) \xrightarrow{f_1} (G,
\sigma_G) \xleftarrow{f_2} (H_2, \sigma_2)$, the pullback
$(P, \sigma_P)$ has underlying graph $P = H_1 \times_G H_2$
(the $\LGraphP$-pullback) and $\sigma_P(v) := \sigma_1(\pi_1
(v))$ for $v \in \mathrm{Stereo}(P)$; chirality-compatibility
of $f_1, f_2$ forces $\sigma_1(\pi_1(v)) = \sigma_2(\pi_2
(v))$ whenever both projections are stereocentres, so
$\sigma_P$ is well-defined.
Pushouts along chirality-compatible monomorphisms are
constructed dually: the $\LGraphP$-pushout carries a
chirality function obtained by gluing $\sigma_1$ and
$\sigma_2$, with agreement on the overlap enforced by the
span monomorphisms' chirality-compatibility.
The Van Kampen condition in $\LGraphGstar$ reduces to the
Van Kampen condition in $\LGraphP$ via the comma-category
identification of the proof above.
\end{remark}

\medskip\noindent
Ambient adhesivity together with rule admissibility makes
DPO rewriting in $\LGraphGstar$ behave as at $\Lk_4$: every
admissible rule $p = (L \leftarrow K \rightarrow R)$ in
$\LGraphGstar$ applied to a chirality-compatible match
$m\colon L \hookrightarrow (G, \sigma)$ has a unique pushout
complement $(D, \sigma_D)$ and a unique result $(H, \tau)$
in the ambient adhesive category, with admissibility
preserved by the rule's chirality lift specification.
The DPO machinery of $\Lk_4$ thus carries over intact
to $\Lk_{4.5}$; the tower-level novelty is only that
chirality labels now propagate deterministically through
every reaction step, rather than being absent.


\begin{chembox}[Enantiomers as distinct objects in
  $\LGraphGstar$]
In $\LGraphP$, the graphs of $(R)$- and $(S)$-2-bromobutane
are \emph{isomorphic objects} --- the enantiomer swap
$\pi_\chi$ of \S\ref{sec:L45-forcing-in} is an isomorphism
witnessing this, and at $\Lk_4$ it is an automorphism of
$\Lk_4(P)$ that conflates the two substrates.
In $\LGraphGstar$ they become \emph{distinct non-isomorphic
objects} $(G, +1)$ and $(G, -1)$: any morphism between them
would have to map $\sigma(v_\chi) = +1$ to $\sigma(v_\chi) =
-1$ at the stereocentre, violating chirality-compatibility.

The group element $E^* \in \Gstar$ is not itself a morphism
of $\LGraphGstar$.
Rather, $E^*$ \emph{induces} an endofunctor $E^*_*\colon
\LGraphGstar \to \LGraphGstar$ (Remark~\ref{rem:gstar-action-cat}
below) sending $(G, \sigma) \mapsto (G, -\sigma)$ and
every chirality-compatible monomorphism to its
$\sigma$-flipped counterpart.
$E^*_*$ relates the two enantiomers as objects in the same
$\Gstar$-orbit but provides no morphism between them ---
there is none.

This distinction is the passage from $\Lk_4$ to $\Lk_{4.5}$
at the object level: $\pi_\chi$ is an internal
$\LGraphP$-isomorphism but only an external
$\LGraphGstar$-endofunctor.  Concretely, the endofunctor
$E^*_*$ projects through the forgetful functor $U_{4.5}$ to
the automorphism $\pi_\chi \in \Aut(\Lk_4(P))$ via the
intertwining relation $U_{4.5} \circ E^*_* = \pi_\chi \circ
U_{4.5}$, and the forcing cokernel class $[\pi_\chi]$ is the
image of this projection.
\end{chembox}


The following diagram illustrates how chirality labels
propagate through a DPO derivation in $\LGraphGstar$.
The example is the $\mathrm{S_N2}$ rule
(Example~\ref{ex:SN2-dpo}) equipped with a specific
chirality lift $\sigma_R = -\sigma_L$ at the reactive
carbon.

\begin{equation}
\label{diag:SN2-dpo-L45}
\includegraphics{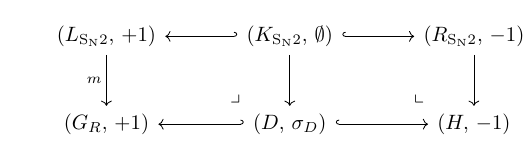}
\end{equation}

\noindent
The rule's left-hand side $(L_{\mathrm{S_N2}}, +1)$ carries
the substrate chirality label; the context
$(K_{\mathrm{S_N2}}, \emptyset)$ has empty chirality function
because the reactive carbon in $K$ has only three neighbours
(the C--X bond is broken, the C--Nu bond is not yet formed,
and the three substituent bonds are preserved through $K$),
hence fails condition~(i) of
Definition~\ref{def:chiral-centre} and lies outside
$\Stereo(K)$; the right-hand side
$(R_{\mathrm{S_N2}}, -1)$ carries the inverted chirality
label.
Given this chirality lift and a chirality-compatible match
$m\colon L_{\mathrm{S_N2}} \hookrightarrow (G_R, +1)$,
adhesivity (Proposition~\ref{prop:LGraphGstar-adhesive})
guarantees a unique pushout complement $(D, \sigma_D)$ and
a unique result $(H, -1)$.

\medskip\noindent
The diagram establishes the \emph{deterministic propagation}
of chirality labels through a DPO derivation, given the
rule's chirality lift.
It does not establish that the inverting lift
$\sigma_R = -\sigma_L$ is the \emph{correct} chirality lift
for $\mathrm{S_N2}$ rather than the retaining lift
$\sigma_R = +\sigma_L$.  Both lifts are
$\Gstar$-equivariant; the selection of the inverting lift
is supplied by mechanism-specific input (back-side attack
geometry), as established in
Theorem~\ref{thm:walden}.  The tower contribution at
$\Lk_{4.5}$ is the \emph{universal propagation} of the
chosen lift across $\Gstar$-orbits of substrates ---
solvent-, temperature-, and substrate-independence ---
not the selection of which lift to use.
The logical separation is tower-native: at the level of
$\LGraphGstar$, DPO machinery determines chirality
propagation once a chirality lift is given; which lift
realises a given mechanism is established in
\S\ref{sec:L45-def}--\ref{sec:L45-stereo}.


\begin{remark}[$\Gstar(G)$ acts on the $G$-fibre of
  $\LGraphGstar$ by auto-equivalences]
\label{rem:gstar-action-cat}
The group $\Gstar(G)$ acts naturally on the $G$-fibre of
$\LGraphGstar$ --- the full subcategory on objects
$(G', \sigma')$ with $G' = G$ as labelled graphs --- via
\[
  (\pi, \boldsymbol{\epsilon})_*\colon
  (G, \sigma) \;\longmapsto\;
  \bigl(G,\; \boldsymbol{\epsilon} \cdot
    (\sigma \circ \pi^{-1})\bigr),
\]
extended to morphisms within the fibre by composition with
$\pi$ on both source and target.  This preserves
chirality-compatibility because the $\sigma$-action is
uniform across domain and codomain.
Each $(\pi, \boldsymbol{\epsilon})_*$ is an equivalence of
the $G$-fibre with itself, with inverse
$(\pi^{-1}, \boldsymbol{\epsilon}^{-1})_*$; together these
form a strict $\Gstar(G)$-action on the $G$-fibre.

Globally, the per-graph actions assemble into a
\emph{groupoid action} of the stereochemical symmetry
groupoid $\mathcal{G}_{\rm st}$ on $\LGraphGstar$, with
$\mathcal{G}_{\rm st}$ having molecular graphs as objects
and $\Gstar(G)$ as automorphism group at each $G$.
Equivalently, an element $(\pi, \boldsymbol{\epsilon}) \in
\Gstar(G)$ extends to an endofunctor of $\LGraphGstar$
acting non-trivially only on objects with underlying graph
$G$, and as the identity on all other fibres.
Chemically: $(\pi, \boldsymbol{\epsilon})_*$ relabels atoms
of $G$ of the same element type according to $\pi$ and
flips chirality at a subset of $G$'s stereocentres
determined by $\boldsymbol{\epsilon}$.
This lifts the object-level $\Gstar(G)$-action of
Definition~\ref{def:Gstar} to a categorical action ---
the data needed to formulate $\Gstar$-equivariance of DPO
rules in the next section.
\end{remark}

\begin{chembox}[Preview: equivariance as the content of
  $\Lk_{4.5}$]
The auto-equivalences $(\pi, \boldsymbol{\epsilon})_*$
determine which DPO rules in $\LGraphGstar$ are admitted as
morphisms of $\Lk_{4.5}(P)$: a rule $p$ is
\emph{$\Gstar$-equivariant} if applying it commutes with
applying any $(\pi, \boldsymbol{\epsilon})_*$.
Equivariance does two things in the tower.  First, it
\emph{propagates a chosen lift uniformly across
$\Gstar$-orbits}, so that an $\mathrm{S_N2}$ rule with the
inverting lift inverts every chirality-labelled substrate
in its orbit, regardless of solvent, temperature, or
substituent identity --- the universality content of
Theorem~\ref{thm:walden}.  Second, when a rule's reaction
centre admits an involutive automorphism $\alpha$ exchanging
new stereocentres, $\alpha$-equivariance \emph{singles out
one of two possible product lifts} (Theorem~\ref{thm:WH-categorical}),
giving the categorical skeleton of the Woodward--Hoffmann
selection rules; the identification of the
$\alpha$-equivariant lift with the thermally allowed product
is the content of Conjecture~\ref{conj:WH-electrocyclic}.
The condition is defined precisely in
Definition~\ref{def:gstar-equivariant-rule}; its consequences
are the theorems of
\S\ref{sec:L45-stereo}--\ref{sec:L45-wh}.
\end{chembox}

\medskip\noindent
The ambient rewriting category $\LGraphGstar$ is now in
place: chirality-labelled molecular graphs as objects,
chirality-compatible monomorphisms as morphisms, ambient
adhesivity together with admissibility guaranteeing
well-defined DPO rewriting through admissible rules, and a
per-fibre $\Gstar(G)$-action by auto-equivalences ready to
impose equivariance on the DPO rules.
The stereochemical level $\Lk_{4.5}(P)$, built as the free
SMC on $\Gstar$-equivariant DPO rules in $\LGraphGstar$, is
constructed in \S\ref{sec:L45-def}.

%% file: chapters/L45/l45_def.tex
\subsection{Definition of
  \texorpdfstring{$\Lk_{4.5}(P)$}{L\_{4.5}(P)}}
\label{sec:L45-def}

Sections~\ref{sec:L45-gstar}--\ref{sec:L45-chiral} assembled the
two pieces needed to define $\Lk_{4.5}(P)$: the
chirality symmetry group $\Gstar(G) = \Aut(G)\ltimes
\ZZ_2^k$ and its action on the ambient rewriting category
$\LGraphGstar$ by auto-equivalences.
This section imposes the $\Gstar$-equivariance condition on
DPO rules, defines $\Lk_{4.5}(P)$ as the free strict symmetric
monoidal category those rules generate, and records the
universal property and tower coherence structure that follow.
The stereochemical theorems --- Walden inversion, racemisation,
Woodward--Hoffmann --- are then obtained in
\S\ref{sec:L45-stereo}--\ref{sec:L45-wh} as consequences of
the equivariance condition imposed here.

\subsubsection{\texorpdfstring{$\Gstar$}{G*}-equivariant DPO rules}

\begin{definition}[$\Gstar$-equivariant DPO rule]
\label{def:gstar-equivariant-rule}
A DPO rule $p = (L \leftarrow K \rightarrow R)$ in
$\LGraphGstar$ is \emph{$\Gstar$-equivariant} if for every
object $(G, \sigma) \in \LGraphGstar$, every
chirality-compatible match $m\colon L \hookrightarrow
(G, \sigma)$, and every element
$(\pi, \boldsymbol{\epsilon}) \in \Gstar(G)$, the
equivariance square below commutes; here the right-vertical
action of $(\pi, \boldsymbol{\epsilon})_*$ on $(H, \tau)$
denotes the corresponding element of $\Gstar(H)$ obtained
by transporting $(\pi, \boldsymbol{\epsilon})$ through the
rule's chirality lift (via the bijection of preserved
stereocentres $\Stereo(L) \cap \Stereo(R) \to \Stereo(H)$
induced by $m$ and the rule, extended to new stereocentres
of $R$ by the rule's specification of $\sigma_R$):
\[
  \includegraphics{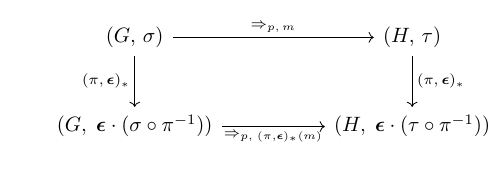}
\]
where $(\pi,\boldsymbol{\epsilon})_*(m)\colon L
\hookrightarrow (G,\,\boldsymbol{\epsilon}\cdot(\sigma\circ
\pi^{-1}))$ is the match obtained by applying the
endofunctor $(\pi,\boldsymbol{\epsilon})_*$ to $m$
(trivial on $L$ since $L$ is not the underlying graph $G$,
non-trivial on the codomain).
Equivalently: applying a $\Gstar$-action before the rule
fires gives the same result as applying the rule first and
then the $\Gstar$-action.
Since $\pi \in \Aut(G)$ preserves $G$ as a labelled graph
(it permutes vertices but preserves all edges and labels),
the underlying bond-graph transformation of $p$ is identical
on both paths; only the chirality function transforms.
\end{definition}

\begin{remark}[Stereocentre creation and destruction under
  the action]
\label{rem:gstar-transport}
For rules that preserve every stereocentre through the
derivation --- $\mathrm{S_N2}$ inversion is the principal
case --- the bijection $\Stereo(G) \to \Stereo(H)$ induced
by $m$ and the rule is unambiguous, and the right-vertical
action of $(\pi, \boldsymbol{\epsilon})_*$ on $(H, \tau)$
acts on the corresponding stereocentre indices.  For rules
that create new stereocentres in $R$ (pericyclic ring
closures, addition reactions) or destroy stereocentres in
passing to $K$ (the $\mathrm{S_N1}$ ionisation step), the
chirality lift's specification of $\sigma_R$ at new
stereocentres or the absence of $\sigma_K$ at destroyed
ones determines how $\boldsymbol{\epsilon}$ extends to or
restricts from the product fibre.  The equivariance
condition is verified rule-by-rule in
\S\ref{sec:L45-stereo}--\ref{sec:L45-wh}; in every case the
transport is determined by the rule's data.
\end{remark}

\begin{remark}[Chirality lifts and the status of mechanism
  specification]
\label{rem:L4-to-L45-lift}
Every DPO rule $p_0 = (L_0 \leftarrow K_0 \rightarrow R_0)$
in $\LGraphP$ admits a \emph{chirality lift}: a DPO rule
$\tilde{p} = (\tilde{L} \leftarrow \tilde{K} \rightarrow
\tilde{R})$ in $\LGraphGstar$ such that
$U_{4.5}(\tilde{p}) = p_0$, where $U_{4.5}\colon
\LGraphGstar \to \LGraphP$ is the forgetful functor of
Definition~\ref{def:LGraphGstar}.
The lift is constructed by taking $\tilde{K} = (K_0,
\emptyset)$ --- the context graph has empty chirality
function because its reactive atoms have reduced neighbour
count (the bonds being broken in $L \setminus K$ leave
those atoms with fewer than four neighbours in $K$), so
they fail Definition~\ref{def:chiral-centre}(i) and
$\Stereo(K_0)$ excludes them --- and by choosing
chirality functions $\sigma_L\colon \mathrm{Stereo}(L_0)
\to \{+1,-1\}$ and $\sigma_R\colon \mathrm{Stereo}(R_0)
\to \{+1,-1\}$ subject to the compatibility condition that
$\sigma_L$ and $\sigma_R$ agree on vertices that appear in
both $L_0$ and $R_0$ unchanged (via $K_0$).

The pair $(\sigma_L, \sigma_R)$ encodes the stereochemical
action of the rule.
Different lifts correspond to different stereochemical
behaviours of the same bond-graph mechanism:
\begin{itemize}
  \item A lift with $\sigma_R(v) = -\sigma_L(v)$ at a reactive
    stereocentre $v$ encodes a \emph{back-side attack}
    (inversion) mechanism.
  \item A lift with $\sigma_R(v) = +\sigma_L(v)$ encodes a
    \emph{front-side attack} (retention) mechanism, as in
    $\mathrm{S_Ni}$ substitution.
\end{itemize}
The forgetful functor $U_{4.5}$ collapses all lifts of $p_0$
to the same $\Lk_4$-rule --- confirming that $\Lk_4$ is
blind to this stereochemical refinement.
This is the precise categorical sense in which $\Lk_{4.5}$
refines $\Lk_4$: the same bond-graph mechanism is enriched
with a chirality assignment, and different assignments
correspond to physically distinct stereochemical outcomes.

\medskip\noindent
\emph{Which lift corresponds to a given mechanism is not
determined at this level.}
Specifying a chirality lift records a stereochemical
hypothesis about the mechanism; both the inverting lift
($\sigma_R = -\sigma_L$) and the retaining lift
($\sigma_R = +\sigma_L$) are $\Gstar$-equivariant DPO rules
in the sense of Definition~\ref{def:gstar-equivariant-rule}
(Theorem~\ref{thm:walden}).  The selection between them
for a given physical mechanism is supplied by external
input: back-side attack geometry selects the inverting
lift for $\mathrm{S_N2}$, front-side attack selects the
retaining lift for $\mathrm{S_Ni}$.  The tower-level
separation is precise: $\LGraphGstar$ provides the ambient
data for chirality lifts; $\Gstar$-equivariance ensures
that any chosen lift propagates uniformly across every
$\Gstar$-orbit of substrates; mechanism-specific input
selects which lift to use.  The resulting morphisms of
$\Lk_{4.5}(P)$ are the $\Gstar$-equivariant DPO rules
together with their chirality lifts.
\end{remark}

\begin{mathbox}[Equivariance as commutation with the
  $\Gstar$-action]
Equivariance (Definition~\ref{def:gstar-equivariant-rule})
is the condition that the DPO rule and the $\Gstar$-action
commute as operations on $\LGraphGstar$: for every element
$(\pi, \boldsymbol{\epsilon}) \in \Gstar(G)$ and every match
$m$, applying the rule then the $\Gstar$-action yields the
same result as applying the $\Gstar$-action first (with its
induced match $(\pi, \boldsymbol{\epsilon}) \circ m$) and
then the rule.
The rule's stereochemical behaviour is thus required to
be uniform across every element of every $\Gstar$-orbit.

\medskip
\noindent\textbf{Example.}
Take the $\mathrm{S_N2}$ rule $p_{\mathrm{SN2}}$ with the
inverting chirality lift ($\sigma_R = -\sigma_L$) and the
$\Gstar$-element $E^* = (\mathrm{id}, (-1))$ acting on a
single-stereocentre substrate.
The equivariance square reads:
\[
  \includegraphics{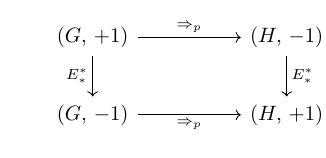}
\]
Both paths yield the same result: the rule inverts
whichever enantiomer it is applied to.
$E^*$-equivariance is precisely the condition that
\emph{both enantiomeric substrates are inverted, neither
privileged}.
The specific chirality labels in each box come from the
chirality lift $\sigma_R = -\sigma_L$; the equivariance
condition then forces the uniform behaviour across the
$E^*$-orbit.
These two pieces together --- chirality lift plus
$\Gstar$-equivariance --- determine the rule's
stereochemical content.
\end{mathbox}

\begin{example}[Achiral reactions are automatically
  equivariant]
\label{ex:achiral-equivariant}
If a DPO rule $p$ has reaction centre avoiding every
stereocentre --- formally, $m(L_0 \setminus K_0) \cap
\Stereo(G) = \emptyset$ for every match $m\colon L_0
\hookrightarrow G$ --- then $p$ is automatically
$\Gstar$-equivariant.
The $\ZZ_2^k$-factor of $\Gstar$ acts only on chirality
labels at stereocentres, none of which are touched by $p$;
the $\Aut(G)$-factor permutes vertices of $G$, and applying
it before or after $p$ gives equivalent results because
$p$'s match is insensitive to the permutation's action
outside the reaction centre.
Many simple organic transformations --- proton transfer,
esterification at non-stereogenic centres, $\beta$-elimination
producing non-stereogenic alkenes --- satisfy this vacuously.
Reactions that create new stereocentres (aldol condensation,
nucleophilic addition to prochiral carbonyls, pericyclic ring
closures) do not: they are equivariant only after a chirality
lift at the new stereocentre is specified, and selecting that
lift is the substantive content of $\Lk_{4.5}$ analysis.
The non-trivial content of $\Gstar$-equivariance appears
precisely at rules that create, destroy, or invert a
stereocentre.
\end{example}

\subsubsection{The categorical construction}

\begin{definition}[The stereochemical level $\Lk_{4.5}(P)$]
\label{def:L45}
The \emph{stereochemical level} $\Lk_{4.5}(P)$ is the free
strict symmetric monoidal category with:
\begin{itemize}
  \item \textbf{Objects}: finite disjoint unions
    $(G_1, \sigma_1) \sqcup \cdots \sqcup (G_n, \sigma_n)$
    of chirality-labelled molecular graphs.
  \item \textbf{Generating morphisms}: $\Gstar$-equivariant
    DPO rules in $\LGraphGstar$
    (Definition~\ref{def:gstar-equivariant-rule}).
  \item \textbf{Morphisms}: composable sequences of
    $\Gstar$-equivariant DPO derivations, modulo the
    free strict SMC congruence.
  \item \textbf{Monoidal product}: disjoint union
    $(G_1, \sigma_1) \sqcup (G_2, \sigma_2)$.
  \item \textbf{Monoidal unit}: $(\emptyset, \emptyset)$.
  \item \textbf{Symmetry}: component-swap isomorphisms
    $(G_1, \sigma_1) \sqcup (G_2, \sigma_2)
    \xrightarrow{\sim}
    (G_2, \sigma_2) \sqcup (G_1, \sigma_1)$.
\end{itemize}
The \emph{forgetful functor} $U_{4.5}\colon \Lk_{4.5}(P)
\to \Lk_4(P)$ sends every chirality-labelled object
$(G, \sigma) \mapsto G$ and every $\Gstar$-equivariant DPO
derivation to its underlying $\Lk_4$-derivation.
The \emph{$\Gstar$-action} on $\Lk_{4.5}(P)$ is the
extension of the object-level action of
Remark~\ref{rem:gstar-action-cat} to morphisms:
for $(\pi, \boldsymbol{\epsilon}) \in \Gstar(G)$ and any
generating morphism $p\colon (G, \sigma) \Rightarrow
(H, \tau)$, $(\pi, \boldsymbol{\epsilon})_*(p)$ is the
image-rewritten derivation
$(\pi, \boldsymbol{\epsilon})_*(G, \sigma)
\Rightarrow (\pi, \boldsymbol{\epsilon})_*(H, \tau)$,
which is again a morphism of $\Lk_{4.5}(P)$ by equivariance
of $p$.
\end{definition}

\begin{proposition}[$\Lk_{4.5}(P)$ is a strict SMC with
  $\Gstar$-action by SMC automorphisms]
\label{prop:L45-SMC}
Definition~\ref{def:L45} yields a well-defined strict
symmetric monoidal category.
Moreover, for each $(\pi, \boldsymbol{\epsilon}) \in \Gstar(G)$,
the functor $(\pi, \boldsymbol{\epsilon})_*\colon
\Lk_{4.5}(P) \to \Lk_{4.5}(P)$ is a strict SMC
automorphism, and the assignment
$(\pi, \boldsymbol{\epsilon}) \mapsto
(\pi, \boldsymbol{\epsilon})_*$ is, for each labelled graph
$G$, a group homomorphism
$\Gstar(G) \to \Aut_{\mathrm{SMC}}(\Lk_{4.5}(P))$ acting
non-trivially only on the $G$-fibre.  Together these per-graph
homomorphisms assemble into a groupoid action of
$\mathcal{G}_{\rm st}$ on $\Lk_{4.5}(P)$.
\end{proposition}

\begin{proof}
\textbf{Category.}
Composition of $\Gstar$-equivariant derivations is
$\Gstar$-equivariant: given equivariance squares for
$r_1\colon (G, \sigma) \Rightarrow (H, \tau)$ and
$r_2\colon (H, \tau) \Rightarrow (I, \upsilon)$, paste them
vertically along the common middle edge $(\pi,
\boldsymbol{\epsilon})_*(H, \tau)$; the pasted rectangle
commutes, giving the equivariance square for
$r_2 \circ r_1$.
The identity derivation (empty rewriting sequence) is
trivially equivariant.

\textbf{Monoidal structure.}
Disjoint union of chirality-labelled graphs is strictly
associative and unital with unit $(\emptyset, \emptyset)$,
inheriting these properties from disjoint union in
$\Lk_4(P)$.
The component-swap symmetry isomorphism commutes with the
$\Gstar$-action because the action on a disjoint union is
defined componentwise: $(\pi, \boldsymbol{\epsilon})_*$
applied to $(G_1, \sigma_1) \sqcup (G_2, \sigma_2)$
operates on each component independently, so swapping
components and then acting agrees with acting and then
swapping.
Hence the component swap is $\Gstar$-equivariant and is a
morphism of $\Lk_{4.5}(P)$.

\textbf{$\Gstar$-action.}
Each endofunctor $(\pi, \boldsymbol{\epsilon})_*$ is an
SMC automorphism by Remark~\ref{rem:gstar-action-cat}:
it respects composition (because equivariance squares
compose, as in the category argument above), monoidal
product (by distributivity of the action over disjoint
union), and the component-swap symmetry (by the monoidal
structure argument above).
The action is strict: $(\pi_1, \boldsymbol{\epsilon}_1)_*
\circ (\pi_2, \boldsymbol{\epsilon}_2)_* =
(\pi_1\pi_2,\, \boldsymbol{\epsilon}_1 \cdot
(\boldsymbol{\epsilon}_2 \circ \pi_1^{-1}))_*$, which is
the group multiplication in $\Gstar$ per
Definition~\ref{def:Gstar}.
\end{proof}

\begin{remark}[Tower extension type]
\label{rem:L45-type}
The extension $\Lk_4 \to \Lk_{4.5}$ is a
\emph{symmetry enrichment}: neither a decorator extension
($\Lk_0 \to \Lk_1$ and its successors, which add a
numerical functor without changing the underlying SMC) nor
a structural extension ($\Lk_3 \to \Lk_4$, which rebuilds
the underlying SMC from scratch on DPO rules), but a third
pattern in which the underlying combinatorial data of
$\Lk_4$ are augmented with a group-equivariance condition
imposed on the rule set, with the resulting free SMC
fibred over $\Lk_4$ via the forgetful functor $U_{4.5}$.
This type is introduced at $\Lk_{4.5}$ and motivated in
Remark~\ref{rem:why-45}; it will recur at $\Lk_5$
(Euclidean-group enrichment of the configuration space)
and $\Lk_6$ ($U(1)$-gauge enrichment of the electronic
Hilbert bundle), so it is worth naming here as one of the
three organising patterns of the tower.
\end{remark}

\begin{proposition}[Universal property of $\Lk_{4.5}(P)$]
\label{prop:UP-L45}
Let $\mathcal{C}$ be a strict symmetric monoidal category
equipped, for each labelled graph $G$ appearing as the
underlying graph of some object in $\mathcal{C}$, with a
strict $\Gstar(G)$-action by SMC automorphisms on the
$G$-fibre of $\mathcal{C}$.  Equivalently, $\mathcal{C}$
carries an action of the stereochemical symmetry groupoid
$\mathcal{G}_{\rm st}$ whose objects are labelled graphs
and whose automorphism group at each $G$ is $\Gstar(G)$.
Any pair of assignments
\begin{itemize}
  \item objects of $\LGraphGstar$ $\to$ objects of
    $\mathcal{C}$, respecting the per-graph
    $\Gstar(G)$-actions;
  \item $\Gstar$-equivariant DPO rules in $\LGraphGstar$
    $\to$ morphisms of $\mathcal{C}$ that are
    $\Gstar(G)$-equivariant on each fibre, compatible with
    source and target assignments;
\end{itemize}
extends uniquely to a strict groupoid-equivariant SMC
functor $F\colon \Lk_{4.5}(P) \to \mathcal{C}$.
\end{proposition}

\begin{proof}
By the universal property of the free strict SMC
(the analogue for $\Lk_4$ is
Proposition~\ref{prop:UP-L4}), any generator assignment
extends uniquely to a strict SMC functor.
The assumption that the generator assignment respects the
$\Gstar$-actions, together with the compatibility with
source and target, ensures that the extended functor
intertwines the two $\Gstar$-actions: applying
$(\pi, \boldsymbol{\epsilon})_*$ in $\Lk_{4.5}(P)$ and
then $F$ yields the same morphism in $\mathcal{C}$ as
applying $F$ first and then $(\pi,
\boldsymbol{\epsilon})_*$ in $\mathcal{C}$.
The free-SMC congruence preserves this intertwining because
it is closed under composition and monoidal product, both
of which are $\Gstar$-equivariant by
Proposition~\ref{prop:L45-SMC}.
Hence $F$ is $\Gstar$-equivariant.
\end{proof}

\subsubsection{Tower coherence}
\label{subsec:L45-coherence}

The universal property of $\Lk_{4.5}(P)$ locks its position
in the tower: every construction on the lower levels lifts
coherently.

\begin{proposition}[Vertical tower coherence]
\label{prop:L45-coherence}
The chain of forgetful functors
\[
  \Lk_{4.5}(P)
  \;\xrightarrow{U_{4.5}}\;
  \Lk_4(P)
  \;\xrightarrow{U_4}\;
  \Lk_3(P)
  \;\xrightarrow{U_3}\;
  \Lk_2(P)
  \;\xrightarrow{U_2}\;
  \Lk_1(P)
  \;\xrightarrow{U_1}\;
  \Lk_0(P)
\]
consists of strict SMC functors, each of which forgets the
extension data introduced at its source level.
The composition $U_1 \circ U_2 \circ U_3 \circ U_4 \circ
U_{4.5}$ sends every chirality-labelled reaction network
in $\Lk_{4.5}(P)$ to its stoichiometric shadow in
$\Lk_0(P)$.
Dually, any functor $F\colon \Lk_j(P) \to \mathcal{C}$ for
$j \leq 4$ has a canonical lift to a functor
$\tilde{F}\colon \Lk_{4.5}(P) \to \mathcal{C}$ given by
precomposition with the forgetful chain:
\[
  \tilde{F} \;=\; F \circ U_{j+1} \circ \cdots \circ U_{4.5}.
\]
This is the unique lift that is $\Gstar$-trivial on the new
chirality data --- i.e., that sends an object
$(G, \sigma) \in \Lk_{4.5}(P)$ to the same image in
$\mathcal{C}$ regardless of the value of $\sigma$.  Lifts
that distinguish $(G, +1)$ from $(G, -1)$ in $\mathcal{C}$
require additional structure on $\mathcal{C}$ and are not
unique.
\end{proposition}

\begin{proof}
Each $U_k$ is a strict SMC functor by construction at its
level (decorator extensions via functor forgetting,
structural extension $U_4$ via DPO-derivation forgetting,
$U_{4.5}$ via chirality-function forgetting).  Strict SMC
functors compose to strict SMC functors, so the full chain
$U_1 \circ U_2 \circ U_3 \circ U_4 \circ U_{4.5}$ is a
strict SMC functor.  The canonical lift
$\tilde{F} = F \circ U_{j+1} \circ \cdots \circ U_{4.5}$
is then a strict SMC functor as a composition of strict SMC
functors; uniqueness as a $\Gstar$-trivial extension is
immediate from $U_{4.5}$'s definition (it identifies all
chirality assignments on a given underlying graph).
\end{proof}

\medskip\noindent
Proposition~\ref{prop:L45-coherence} has a direct
consequence for tower-native computation: thermodynamic
data $F_H, F_S^G$ (from $\Lk_1, \Lk_2$), kinetic data
$F_P$ (from $\Lk_3$), and mechanistic data (from $\Lk_4$)
all remain valid and consistent when the stereochemical
level is added above them.
No lower-level theorem needs to be reproved at $\Lk_{4.5}$:
the strict SMC functoriality of the forgetful chain
$U_{4.5}, U_4, \ldots, U_1$ ensures that every
$\Lk_j$-datum extends canonically to $\Lk_{4.5}$ by
precomposition with $U_{j+1} \circ \cdots \circ U_{4.5}$.
This is what allows stereochemical statements at $\Lk_{4.5}$
to be stated against the background of all the numerical and
structural data already available at lower levels, without
additional reconstruction.

\medskip\noindent
The stereochemical theorems of \S\ref{sec:L45-stereo}
--- Walden inversion, racemisation, and net retention
through double inversion --- can now be stated and proved
using the $\Gstar$-equivariance condition imposed here
(which propagates a chosen chirality lift uniformly across
$\Gstar$-orbits) together with mechanism-specific input
(which selects the chirality lift), against the
tower-level machinery established in this section.

%% file: chapters/L45/l45_stereo.tex
\subsection{Stereochemistry as theorems at
  \texorpdfstring{$\Lk_{4.5}$}{L4.5}}
\label{sec:L45-stereo}

Sections~\ref{sec:L45-gstar}--\ref{sec:L45-def} built
$\Lk_{4.5}(P)$ to resolve the two forcing cokernel classes
$[\pi_\chi]$ and $[\pi_{\mathrm{ec}}]$ of
\S\ref{sec:L45-forcing-in}.
This section collects the payoff.
Four classical results of stereochemistry --- Walden
inversion (\S\ref{subsec:walden}), racemisation via
$\mathrm{S_N1}$ (\S\ref{subsec:racemisation}), net retention
via double inversion (\S\ref{subsec:NGP}), and the
categorical structure of enantiomers
(\S\ref{subsec:enantiomers}) --- become theorems at
$\Lk_{4.5}$, deducible from the $\Gstar$-equivariance
condition and the DPO machinery alone.
No quantum mechanics, no 3D coordinates, and no
experimental postulates beyond the chirality lift of each
rule and the graph-theoretic detection of stereocentres
(Definition~\ref{def:chiral-centre}) are invoked.

The tower-level contribution of $\Lk_{4.5}$ is the
\emph{orbit-uniformity} of stereochemical outcome:
$\Gstar$-equivariance propagates a given mechanism's
stereochemistry uniformly across every chirality-labelled
substrate in each $\Gstar$-orbit, with the mechanism's
directional choice (inversion vs retention) supplied as an
external input via the chirality lift.  Solvent- and
temperature-invariance, where empirically observed, falls
outside $\Lk_{4.5}$'s modelling scope; the level provides
the orbit-uniform shadow of those invariances, not their derivation.

\subsubsection{Walden inversion}
\label{subsec:walden}

In 1896 Paul Walden observed that malic acid could be
converted to chlorosuccinic acid and back through a sequence
of reagents, with the recovered malic acid displaying
\emph{flipped} optical rotation: the configuration had been
inverted somewhere in the cycle without any obvious
cause~\cite{Walden1896}.
The mechanism remained contested for four decades.
Hughes and Ingold's systematic kinetic work in the
1930s~\cite{HughesIngold1935} established that bimolecular
nucleophilic substitution ($\mathrm{S_N2}$) proceeds by
back-side attack: the nucleophile approaches from the face
opposite the leaving group, forcing a ``Walden inversion''
of the four substituents around the reactive carbon.
Cowdrey, Hughes, Ingold, Masterman, and Scott provided the
definitive stereochemical evidence in
1937~\cite{Cowdrey1937}: every $\mathrm{S_N2}$ reaction
gives \emph{complete} inversion of configuration, on every
substrate, in every solvent, at every temperature.

This empirical universality demands a structural explanation
that is insensitive to the continuous parameters $\Lk_{4.5}$
does not know about (solvent, temperature, conformer).
At $\Lk_{4.5}$, universality is a theorem.

\begin{theorem}[Walden inversion at $\Lk_{4.5}$]
\label{thm:walden}
Let $p_0^{\mathrm{S_N2}}$ be the $\Lk_4$-level $\mathrm{S_N2}$
rule (Example~\ref{ex:SN2-dpo}).
In $\LGraphGstar$, this rule admits exactly two chirality
lifts at the reactive stereocentre $v$:
\begin{itemize}
  \item the \emph{inverting} lift
    $\tilde{p}_-\colon\ \sigma_R(v) = -\sigma_L(v)$;
  \item the \emph{retaining} lift
    $\tilde{p}_+\colon\ \sigma_R(v) = +\sigma_L(v)$.
\end{itemize}
Both lifts are $\Gstar$-equivariant
(Definition~\ref{def:gstar-equivariant-rule}), and each,
together with its $\Gstar$-orbit, defines a family of DPO
rules in $\LGraphGstar$ acting uniformly across all
chirality-labelled substrates.

The empirical back-side-attack topology of the
$\mathrm{S_N2}$ transition state selects the inverting lift.
With this selection, for every chirality-labelled substrate
$(G, \sigma)$ admitting a chirality-compatible match
$m\colon L_{\mathrm{S_N2}} \hookrightarrow (G, \sigma)$
with reactive stereocentre $v$, the product satisfies
\[
  \sigma'(f(v)) \;=\; -\sigma(v),
\]
where $f\colon R_{\mathrm{S_N2}} \hookrightarrow H$ is the ``comatch".
\end{theorem}

\begin{proof}
\textbf{Identification of the two lifts.}
The $\Lk_4$-level rule has $L_{\mathrm{S_N2}}$, $K$, and
$R_{\mathrm{S_N2}}$ with fixed bond-graph structure
(Example~\ref{ex:SN2-dpo}).
The reactive carbon $v$ has four distinct substituents in
both $L_{\mathrm{S_N2}}$ and $R_{\mathrm{S_N2}}$ (three
from the R-group, plus either X or Nu), so $v \in
\mathrm{Stereo}(L_{\mathrm{S_N2}}) \cap
\mathrm{Stereo}(R_{\mathrm{S_N2}})$ with a single value
$\sigma_L(v), \sigma_R(v) \in \{+1, -1\}$ each.
In the context graph $K$ the carbon has reduced neighbour
count (the bond to X has been removed without the bond to
Nu being added), hence $v \notin \mathrm{Stereo}(K)$ by
Definition~\ref{def:chiral-centre}(i).
A chirality lift of $p_0^{\mathrm{S_N2}}$ is thus a choice
of $(\sigma_L(v), \sigma_R(v)) \in \{+1, -1\}^2$, modulo
overall sign (a global flip at $v$ produces the same rule
up to $\Gstar$-relabelling).
The quotient gives exactly two lifts: $\tilde{p}_-$ with
$\sigma_R(v) = -\sigma_L(v)$ (opposite signs) and
$\tilde{p}_+$ with $\sigma_R(v) = +\sigma_L(v)$ (same
signs).

\textbf{Each lift is $\Gstar$-equivariant.}
The equivariance square for $\tilde{p}_-$ with
$E^* = (\mathrm{id}, -\mathbf{1}) \in \Gstar(G)$ (for $k = 1$,
this is $(\mathrm{id}, (-1))$) is verified as follows.
Taking any substrate $(G, +1)$, applying $\tilde{p}_-$
yields $(H, -1)$.
Applying $E^*$ to both sides:
\[
  E^*\cdot(G, +1) = (G, -1),
  \qquad
  E^*\cdot(H, -1) = (H, +1).
\]
Applying $\tilde{p}_-$ to $(G, -1)$ yields $(H, +1)$ by the
lift's definition ($\sigma_R = -\sigma_L$).
Both paths around the equivariance square give $(H, +1)$:
the square commutes.
An analogous verification holds for $\tilde{p}_+$: applying
$\tilde{p}_+$ to $(G, +1)$ yields $(H, +1)$, and $E^*_*
(H, +1) = (H, -1)$; applying $\tilde{p}_+$ to $E^*_*
(G, +1) = (G, -1)$ yields $(H, -1)$.
Squares commute for both lifts.

\textbf{Universality of the inverting family.}
Fix the inverting lift $\tilde{p}_-$.
For any chirality-compatible match $m\colon
L_{\mathrm{S_N2}} \hookrightarrow (G, \sigma)$,
chirality-compatibility forces
$\sigma(m(v)) = \sigma_L(v)$, and the DPO pushout in the
ambient adhesive category for $\LGraphGstar$
(Proposition~\ref{prop:LGraphGstar-adhesive}; admissibility
of the rule's chirality lift is direct from
$\sigma_R(v) = -\sigma_L(v)$ at the unique reactive
stereocentre) produces a unique comatch
$f\colon R_{\mathrm{S_N2}} \hookrightarrow H$ with $\sigma'(f(v)) = \sigma_R(v) =
-\sigma_L(v) = -\sigma(v)$.
The uniform relation $\sigma'(f(v)) = -\sigma(v)$ thus holds
on every substrate $(G, \sigma)$ and every chirality-compatible
match, without substrate-, solvent-, or temperature-dependent
caveats.
\end{proof}

\begin{remark}[Tower reading of Walden inversion]
\label{rem:walden-tower}
Theorem~\ref{thm:walden} decomposes cleanly along the tower:
\begin{itemize}
  \item At $\Lk_4$: a single rule
    $p_0^{\mathrm{S_N2}}$ exists, carrying no stereochemical
    information.
    The inversion/retention distinction is
    $\Lk_4$-invisible.
  \item At $\Lk_{4.5}$: two chirality lifts exist,
    $\tilde{p}_-$ and $\tilde{p}_+$, collapsing to the same
    $p_0^{\mathrm{S_N2}}$ under $U_{4.5}$.
    The choice between them is a mechanism-specific
    input (back-side vs.\ front-side attack).
  \item Given the back-side-attack input, $\Gstar$-equivariance
    of $\tilde{p}_-$ propagates the inversion uniformly
    across every chirality-labelled substrate in every
    $\Gstar$-orbit.
\end{itemize}
The tower contribution is thus \emph{universality of
propagation}, not \emph{selection of direction}.
The direction comes from the mechanism's topology;
$\Gstar$-equivariance supplies orbit-uniformity over
$\Lk_{4.5}$-substrates, which is the level's shadow of
the empirically observed solvent- and temperature-independence.
Below $\Lk_{4.5}$, the forgetful functor $U_{4.5}$ strips
the chirality label and both lifts collapse to the same
$\Lk_4$-derivation; above $\Lk_{4.5}$, any tower-coherent
extension preserves the universality, since the higher-level
construction (which extends $\Lk_{4.5}$ by additional
data: 3D coordinates at $\Lk_5$, electronic structure at
$\Lk_6$) embeds every $\Gstar$-equivariant
$\Lk_{4.5}$-morphism as a corresponding equivariant
morphism at the higher level by construction, with the
forgetful functor down to $\Lk_{4.5}$ recovering the original.
\end{remark}

\begin{chembox}[Walden inversion: tower content and
  empirical universality]
Theorem~\ref{thm:walden} explains the universality of
Cowdrey et al.'s 1937 empirical observation
\cite{Cowdrey1937} once back-side attack is supplied as
the mechanism-specific input.  Its two inputs are:
\begin{itemize}
  \item the back-side-attack topology of $\mathrm{S_N2}$,
    encoded as the chirality lift $\sigma_R = -\sigma_L$;
  \item $\Gstar$-equivariance, enforcing uniform behaviour
    across $\Gstar$-orbits of $\Lk_{4.5}$-substrates.
\end{itemize}
The categorical content is the uniform $\sigma'(f(v)) =
-\sigma(v)$ relation across every chirality-compatible
match.  Solvent- and temperature-independence of the
empirical observation falls outside $\Lk_{4.5}$'s
modelling scope (the level has no parameter for either),
so the theorem cannot directly establish it; what the
theorem provides is the consistent shadow at $\Lk_{4.5}$
of an empirically solvent- and temperature-independent
phenomenon.  No solvent model, no potential energy
surface, no quantum treatment of the transition state is
used --- not because $\Lk_{4.5}$ proves their irrelevance,
but because the universality content visible at $\Lk_{4.5}$
is the orbit-uniform sign relation, and that content is
solvent- and temperature-agnostic.

Mechanisms violating universality --- those whose
stereochemical outcome depends on solvent, temperature, or
substrate type --- are not single $\Gstar$-equivariant
$\Lk_{4.5}$-morphisms; they are mixtures of two or more
$\Lk_{4.5}$-morphisms whose relative weights require
$\Lk_3$-level data (rate constants).
The $\mathrm{S_N1}$ mechanism (next subsection) is an
example: it is not a single equivariant rule but two
equivariant rules sharing an achiral intermediate.
\end{chembox}

\subsubsection{Racemisation via
  \texorpdfstring{$\mathrm{S_N1}$}{SN1}}
\label{subsec:racemisation}

\begin{remark}[Racemisation, for mathematicians]
\label{rem:racemisation-defn}
A \emph{racemic mixture} is an equal-parts mixture of the
two enantiomers of a chiral molecule.
It has zero net optical rotation because the contributions
of $(G, +1)$ and $(G, -1)$ cancel.
\emph{Racemisation} is a process that converts an
enantiomerically pure starting material into a racemic
mixture; it destroys stereochemical information.
In tower language: racemisation is the passage from a
non-$E^*$-symmetric state (pure enantiomer) to an
$E^*$-symmetric state (racemic mixture).
The $\mathrm{S_N1}$ mechanism achieves this by passing
through a planar intermediate in which the reactive carbon
is no longer a stereocentre.
\end{remark}

\begin{theorem}[Racemisation at $\Lk_{4.5}$]
\label{thm:racemisation}
Let $p_{\mathrm{S_N1}} = p_2 \circ p_1$ be the two-step
$\mathrm{S_N1}$ derivation of Example~\ref{ex:SN1-dpo},
lifted to $\LGraphGstar$.
The step $p_1$ is the ionisation (C--X heterolysis); the
step $p_2$ is the nucleophilic attack on the carbocation.
The following hold:
\begin{enumerate}[label=(\roman*)]
  \item \emph{$p_1$ is stereochemistry-destroying.}
    The $\Lk_4$-level $L$ of $p_1$ contains the reactive
    carbon $v$ at four neighbours (three R-group bonds and
    the C--X bond), so $v \in \Stereo(L)$.  The product
    side $R$ has $v$ at three neighbours (the cation, with
    C--X removed), so $v \notin \Stereo(R)$ and $\sigma_R$
    is undefined at $v$.  Consequently $p_1$ admits exactly
    two $\Gstar$-equivariant chirality lifts at $v$, one
    with $\sigma_L(v) = +1$ and one with $\sigma_L(v) = -1$;
    each lift applies via chirality-compatibility to the
    corresponding substrate enantiomer, and both yield the
    same achiral carbocation intermediate
    \[
      p_1(G, +1) \;=\; p_1(G, -1) \;=\; (G_1, \emptyset).
    \]
    The chirality datum is destroyed in passing to $R$
    because $v$ ceases to be a stereocentre.
  \item \emph{$p_2$ admits two $\Gstar$-equivariant
    chirality lifts.}
    The $\Lk_4$-level $R$ of $p_2$ restores the reactive
    carbon to four neighbours (bonded to the R-group and to
    Nu), making it a stereocentre again.
    The chirality lift $\sigma_R(v) \in \{+1, -1\}$ is a
    free choice; both values yield $\Gstar$-equivariant
    DPO rules in $\LGraphGstar$ (by the argument of
    Theorem~\ref{thm:walden}).
    Neither lift is distinguished by any $\Lk_{4.5}$-datum,
    so both are valid morphisms of $\Lk_{4.5}(P)$ from
    $(G_1, \emptyset)$.
\end{enumerate}
Consequently: starting from either pure enantiomer, the
composite $p_2 \circ p_1$ produces both product enantiomers
as valid morphism targets in $\Lk_{4.5}(P)$.
In chemical terms: $\mathrm{S_N1}$ gives racemisation.
\end{theorem}

\begin{proof}
\textbf{Part~(i): stereochemistry-destruction.}
In the $\Lk_4$-level rule $p_1$, the reactive carbon $v$
has four neighbours in $L$ (three R-group bonds plus the
C--X bond) and three in $R$ (the carbocation, after C--X
removal).  By Definition~\ref{def:chiral-centre}(i),
$v \in \Stereo(L)$ but $v \notin \Stereo(R)$, so the
chirality lift specifies $\sigma_L(v) \in \{+1, -1\}$ but
imposes no constraint on $\sigma_R$ at $v$.  Two
$\Gstar$-equivariant chirality lifts of $p_1$ thus exist,
one with $\sigma_L(v) = +1$ and one with $\sigma_L(v) = -1$;
each lift is G*-equivariant trivially (the action on
$\sigma_R$ at $v$ is vacuous), and each applies via
chirality-compatibility to the substrate enantiomer with
matching $\sigma(m(v))$.  Both lifts produce the same
intermediate $(G_1, \emptyset)$: the chirality datum is
lost when $v$ ceases to be a stereocentre.

\textbf{Part~(ii): two lifts of $p_2$.}
In $p_2$, the $R$-side has the reactive carbon at four
neighbours (bonded to the R-group and to Nu).
If Nu is distinct from the three R-group neighbours, $v \in
\mathrm{Stereo}(R)$, and the chirality lift specifies a
value $\sigma_R(v) \in \{+1, -1\}$.
The $L$-side of $p_2$ has the carbocation, three-coordinated
($v \notin \mathrm{Stereo}(L)$), so $\sigma_L$ is not
constrained.
Choosing $\sigma_R(v) = +1$ gives one DPO rule
$\tilde{p}_2^+$; choosing $\sigma_R(v) = -1$ gives another,
$\tilde{p}_2^-$.
Both are $\Gstar$-equivariant by the argument of
Theorem~\ref{thm:walden} (with the retaining/inverting
labelling playing no role, since the $L$-side has no
$\sigma_L$ to compare).
Neither is distinguished by any datum available in
$\LGraphGstar$: the achiral intermediate $(G_1, \emptyset)$
carries no information about which product chirality is
``preferred''.
Both $(G_1, \emptyset) \Rightarrow (H, +1)$ and $(G_1,
\emptyset) \Rightarrow (H, -1)$ are valid morphisms of
$\Lk_{4.5}(P)$.

\textbf{Consequence.}
The composite $p_2 \circ p_1$ has two valid completions
from any starting enantiomer.
Starting from $(G, +1)$: $p_1$ produces $(G_1, \emptyset)$,
and $p_2$ produces either $(H, +1)$ or $(H, -1)$, each a
valid $\Lk_{4.5}(P)$-morphism.
Analogously from $(G, -1)$.
Thus both product enantiomers are reachable from either
starting enantiomer.
\end{proof}

\begin{remark}[Categorical content vs.\ chemical corollary]
\label{rem:racemisation-prob}
The correct categorical statement of
Theorem~\ref{thm:racemisation} is deterministic:
\emph{the $\mathrm{S_N1}$ mechanism at $\Lk_{4.5}$ provides
two valid morphisms from $(G_1, \emptyset)$ to the product
enantiomers, with no $\Lk_{4.5}$-datum preferring one over
the other}.
The $50{:}50$ ratio of the racemic product mixture is a
corollary requiring $\Lk_3$-level data: the rate constants
$k^+$ and $k^-$ for the two $p_2$-completions must be
equal.
Equality of these rates follows from the $\Lk_{4.5}$-level
symmetry (both lifts are in the same $\Gstar$-orbit of
rules at the achiral intermediate), but the quantitative
ratio is a $\Lk_3$ computation, not a $\Lk_{4.5}$ one.
$\Lk_{4.5}$ supplies the structural claim
(indistinguishability of the two products at the category
level); $\Lk_3$ converts it into population statistics.
\end{remark}

\begin{remark}[Tower reading of racemisation]
\label{rem:racemisation-tower}
Racemisation is the tower's account of \emph{loss of
stereochemical information} at $\Lk_{4.5}$.
Part~(i) says that $p_1$ collapses the $\Gstar$-orbit
$\{(G, +1), (G, -1)\}$ to a single object $(G_1, \emptyset)$
--- the chirality datum is destroyed because the
intermediate has no stereocentre.
Part~(ii) says that $p_2$ is then non-deterministic from
the intermediate: both elements of the product
$\Gstar$-orbit $\{(H, +1), (H, -1)\}$ are reachable.
The chirality datum is regenerated, but without memory of
what it was before $p_1$.

Contrast with the $\mathrm{S_N2}$ case
(Theorem~\ref{thm:walden}): there,
the rule is injective on $\Gstar$-orbits ($(G, +1)
\mapsto (H, -1)$ and $(G, -1) \mapsto (H, +1)$, a bijection
between source and target orbits) and fully deterministic.
Stereochemical information is \emph{conserved but
transformed}: a signed datum in, a signed datum out, with
the sign flipped.
The contrast between $\mathrm{S_N2}$
(information-preserving, sign-inverting) and
$\mathrm{S_N1}$ (information-destroying, then
information-regenerating without memory) is visible at
$\Lk_{4.5}$ as the difference between an injective and a
non-injective map on $\Gstar$-orbits of substrates.
\end{remark}

\subsubsection{Net retention via double inversion:
  neighbouring group participation}
\label{subsec:NGP}

The previous two subsections treated single-step
stereochemical events: $\mathrm{S_N2}$ preserves and
inverts the chirality label in one rule application
(Theorem~\ref{thm:walden}); $\mathrm{S_N1}$ destroys and
regenerates it across two rules
(Theorem~\ref{thm:racemisation}).
A third case, due to Winstein~\cite{Winstein1951}, combines
two inversion steps to give \emph{net retention}: the
reaction proceeds with inverted configuration at the
intermediate and inverted again at the product, so the
initial and final configurations match.
This mechanism is neighbouring group participation (NGP),
also called anchimeric assistance.

\begin{remark}[Neighbouring group participation,
  for mathematicians]
\label{rem:NGP-defn}
In NGP, the substrate carries a nucleophilic heteroatom or
$\pi$-bond positioned close to the reactive carbon --- for
example, a $\beta$-acetoxy group on a cyclohexyl tosylate,
Winstein's classical system.
When the leaving group departs, the neighbouring
nucleophile attacks the reactive carbon \emph{from the
opposite face}, forming a bridged intermediate (in
Winstein's case, a five-membered acyloxonium ring).
The reactive carbon remains four-coordinated throughout:
it loses the bond to the leaving group but immediately
forms a bond to the bridging nucleophile, hence is a
stereocentre in the bridged intermediate.
An external nucleophile then attacks the bridged
intermediate, again from the opposite face, opening the
ring and giving the final product.
The classical experimental observation is that the reaction
gives \emph{complete retention} of configuration at the
reactive carbon, despite involving two nucleophilic attack
steps~\cite{Winstein1951}.
Historically this was puzzling because neither step
individually retains: each inverts.
Two inversions give retention.
\end{remark}

\begin{theorem}[Net retention via double inversion at
  $\Lk_{4.5}$]
\label{thm:NGP}
Let $(G, \sigma)$ be a chirality-labelled substrate with
stereocentre $v$, and let $r_1, r_2$ be $\Gstar$-equivariant
DPO derivations such that:
\begin{itemize}
  \item $r_1$ has the inverting chirality lift at $v$
    (back-side attack by the neighbouring nucleophile);
  \item $r_2$ has the inverting chirality lift at $v$
    (back-side attack by the external nucleophile on the
    bridged intermediate).
\end{itemize}
Then the composition $r_2 \circ r_1$ satisfies
\[
  \sigma_{\mathrm{final}}(v) \;=\; \sigma(v),
\]
i.e.\ the net stereochemical outcome at $v$ is retention of
configuration.
\end{theorem}

\begin{proof}
By Theorem~\ref{thm:walden} applied to $r_1$, the
intermediate chirality label satisfies
$\sigma_{\mathrm{int}}(v) = -\sigma(v)$.
Applying Theorem~\ref{thm:walden} to $r_2$ with
substrate $(\cdot, \sigma_{\mathrm{int}})$:
$\sigma_{\mathrm{final}}(v) = -\sigma_{\mathrm{int}}(v) =
-(-\sigma(v)) = \sigma(v)$.
\end{proof}

The following diagram tracks the chirality label through
the two-step sequence in $\Lk_{4.5}(P)$:
\begin{equation}
\label{diag:NGP}
\includegraphics{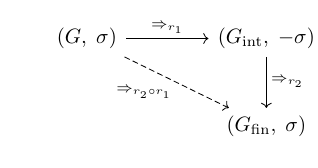}
\end{equation}

\noindent
The chirality label at $v$ begins at $\sigma$, is inverted
to $-\sigma$ by the internal displacement $r_1$ (formation
of the bridged intermediate), and is inverted back to
$\sigma$ by the external opening $r_2$.
The composition $r_2 \circ r_1$ is a morphism of
$\Lk_{4.5}(P)$ that preserves the chirality label at $v$
while making new bonds: a retention mechanism built from two
inversion steps.

\begin{chembox}[What the theorem says about NGP chemistry]
Theorem~\ref{thm:NGP} recovers, in two lines, a result that
historically required careful experimental design to
establish.
Winstein's evidence~\cite{Winstein1951} included: rate
acceleration (the neighbouring group assists the
ionisation, giving first-order kinetics despite formal
bimolecular reactants), racemisation in the absence of
external nucleophile (the bridged intermediate can open
from either face, analogous to
Theorem~\ref{thm:racemisation}(ii)), and \emph{complete
retention} when the external nucleophile is present in
excess (the two sequential back-side attacks dominate the
kinetics).

At $\Lk_{4.5}$, the argument is structural and does not
require the ion-pair analysis that made Winstein's original
treatment subtle.
Two $\Gstar$-equivariant back-side-attack DPO rules
compose to a rule that preserves the chirality label at
$v$.
The bridged intermediate is a valid object in $\Lk_{4.5}(P)$
because the reactive carbon remains four-coordinated
throughout --- the bond to the leaving group is replaced
by the bond to the bridging nucleophile --- so the carbon
is still a stereocentre in the intermediate, with its
chirality label inverted by the first step.

The theorem generalises immediately to longer sequences:
any sequence of $2n$ back-side-attack steps at $v$ gives
net retention; any sequence of $2n+1$ gives net inversion.
This is the $\Lk_{4.5}$-level parity rule for nucleophilic
substitution sequences, recovered from
Theorem~\ref{thm:walden} by composition.
\end{chembox}

\subsubsection{Enantiomers in \texorpdfstring{$\Lk_{4.5}$}{L4.5}}
\label{subsec:enantiomers}

The three previous subsections demonstrated the tower at
work on known chemistry.
This subsection turns the direction around: using the
tower construction, the \emph{definition} of enantiomers
that the construction makes available is given, and shown
to capture the classical chemical notion.

\begin{definition}[Enantiomers in $\Lk_{4.5}(P)$]
\label{def:enantiomers}
Two objects $(G, \sigma)$ and $(G', \sigma')$ of
$\Lk_{4.5}(P)$ are \emph{enantiomers} if:
\begin{enumerate}[label=(\roman*)]
  \item $U_{4.5}(G, \sigma) = U_{4.5}(G', \sigma')$: they
    share the same underlying molecular graph ($G = G'$);
    and
  \item $(G', \sigma') = E^* \cdot (G, \sigma)$ as
    $\LGraphGstar$-objects, i.e.\ $\sigma' = -\sigma$ as
    chirality functions on $\Stereo(G)$.
\end{enumerate}
A chirality-labelled molecular graph $(G, \sigma)$ has a
\emph{distinct} enantiomer iff no $\pi \in \Aut(G)$
satisfies $\sigma \circ \pi^{-1} = -\sigma$, equivalently,
iff $-\sigma$ is not in the $\Aut(G)$-orbit of $\sigma$;
otherwise $(G, \sigma) \cong (G, -\sigma)$ in
$\Lk_{4.5}(P)$ via $\pi$ and the molecule is its own
enantiomer (\emph{achiral}).  Two paradigm cases:
\begin{itemize}
  \item $\Stereo(G) = \emptyset$: trivially
    $(G, \emptyset) = E^* \cdot (G, \emptyset)$ --- achiral
    without stereocentres.
  \item \emph{Meso compounds.}  $\Stereo(G) \neq \emptyset$
    but the graph automorphism group $\Aut(G)$ contains an
    element $\pi$ mapping $\sigma$ to $-\sigma$ --- achiral
    despite stereocentres.  The textbook example is
    meso-tartaric acid (\S\ref{sec:L45-gstar}), where the
    centre-of-symmetry exchange of the two stereocentres
    maps $(+1, -1)$ to $(-1, +1)$.
\end{itemize}
A molecule is \emph{chiral} iff it has a distinct
enantiomer.
\end{definition}

\begin{proposition}[Properties of enantiomers]
\label{prop:enantiomers}
Let $(G, +1)$ and $(G, -1)$ be enantiomers in
$\Lk_{4.5}(P)$ (single stereocentre, $k = 1$).
\begin{enumerate}[label=(\roman*)]
  \item \emph{Distinct, non-isomorphic objects.}
    $(G, +1) \not\cong (G, -1)$ in $\Lk_{4.5}(P)$:
    there is no isomorphism between them.
  \item \emph{Same $\Lk_4$-image.}
    $U_{4.5}(G, +1) = U_{4.5}(G, -1) = G$: enantiomers are
    indistinguishable at every level below $\Lk_{4.5}$.
  \item \emph{Same $\Gstar$-orbit.}
    $(G, -1) = E^* \cdot (G, +1)$: the two enantiomers are
    related by the abstract parity, and their $\Gstar$-orbit
    has exactly two elements.
  \item \emph{No chirality-neutral interconversion.}
    No morphism $(G, +1) \to (G, -1)$ in $\Lk_{4.5}(P)$
    leaves the underlying molecular graph unchanged.
  \item \emph{$\Gstar$-equivariant transport.}
    Every $\Gstar$-equivariant DPO derivation
    $r\colon (G, +1) \Rightarrow (H, \tau)$ determines a
    derivation $E^*_*(r)\colon (G, -1) \Rightarrow (H,
    -\tau)$ on the enantiomeric substrate, giving the
    enantiomeric product.
    The two derivations are in the same $\Gstar$-orbit.
\end{enumerate}
\end{proposition}

\begin{proof}
\textbf{(i).}
Suppose for contradiction that a morphism $r\colon (G, +1)
\to (H, -1)$ in $\Lk_{4.5}(P)$ exists with $G = H$ as
labelled graphs.  Since morphisms compose, it suffices to
consider a single DPO rule application $p = (L \leftarrow
K \rightarrow R)$ with match $m\colon L \hookrightarrow
(G, +1)$ and comatch $f\colon R \hookrightarrow (G, -1)$
that achieves this on the reactive stereocentre $v$;
chirality-compatibility then forces $\sigma_L(v) = +1$ and
$\sigma_R(v) = -1$, so the rule has $\sigma_R(v) = -\sigma_L
(v)$ as its chirality lift.

The rule must also have $L = R$ as labelled graphs (since
applying $p$ via $m$ produces $H = G$ as labelled graphs,
and $f$ matches $R$ into $H$).  Apply $E^* = (\mathrm{id},
-\mathbf{1}) \in \Gstar(G)$ to the equivariance square of
$p$.  Path~1: $p$ applied to $(G, +1)$ yields $(G, -1)$,
then $E^*_*$ yields $(G, +1)$.  Path~2: $E^*_*$ applied to
$(G, +1)$ yields $(G, -1)$, then $p$ applied to
$(G, -1)$ yields ---  by $\sigma_R(v) = -\sigma_L(v)$ and
the new $\sigma_L(v) = -1$ --- the object $(G, +1)$.

The two paths agree on the underlying graph but the
intermediate objects $E^*_*(G, -1) = (G, +1)$ (Path~1) and
$E^*_*(G, +1) = (G, -1)$ (Path~2) are distinct.  Thus
$\Gstar$-equivariance of $p$ requires
$(G, +1) = (G, -1)$ as $\LGraphGstar$-objects, which is
false.  Therefore no such rule $p$ is $\Gstar$-equivariant,
and no chirality-flipping morphism exists in
$\Lk_{4.5}(P)$ with $G$ fixed.

\textbf{(ii).}
$U_{4.5}$ forgets $\sigma$, so $U_{4.5}(G, \pm 1) = G$ by
definition.
Enantiomers are therefore indistinguishable at all lower
tower levels: their thermodynamic data $F_H$, $F_S^G$
(from $\Lk_1, \Lk_2$), kinetic data $F_P$ (from $\Lk_3$),
and bond-graph mechanisms (at $\Lk_4$) are all identical.

\textbf{(iii).}
$E^* = (\mathrm{id}, -\mathbf{1}) \in \Gstar(G)$ sends
$(G, +1) \mapsto (G, -1)$ by Definition~\ref{def:Gstar}.
The orbit $\{(G, +1), (G, -1)\}$ has size at most two (two
labellings at a single stereocentre); it has size exactly
two because $(G, +1) \neq (G, -1)$ as objects of
$\LGraphGstar$ (different chirality functions).

\textbf{(iv).}
The argument given for (i) above proves the stronger
statement that no morphism $(G, +1) \to (G, -1)$ with $G$
fixed exists in $\Lk_{4.5}(P)$, equivariantly or not ---
not just no isomorphism.  Statement (iv) is therefore
established already by the proof of (i).

\textbf{(v).}
By $\Gstar$-equivariance of $r$
(Definition~\ref{def:gstar-equivariant-rule}):
\[
  E^*_*\bigl(r((G, +1))\bigr) \;=\; r\bigl(E^*_*((G, +1))\bigr)
  \;=\; r\bigl((G, -1)\bigr).
\]
The $E^*$-image of the original product is $E^*_*(H, \tau)
= (H, -\tau)$; by $\Gstar$-equivariance this equals the
result of applying $r$ to $(G, -1)$.
Hence $r((G, -1)) = (H, -\tau)$: the enantiomeric substrate
yields the enantiomeric product.
\end{proof}

\begin{chembox}[Enantiomers: the tower account]
Properties (i)--(v) together reproduce, from the tower
construction, the full chemical concept of a pair of
enantiomers.

Property~(ii) encodes why enantiomers have identical
achiral physical properties: same boiling point, same
melting point, same NMR spectrum in an achiral solvent,
same reaction rates with achiral reagents.
All of these are $\Lk_k$-level data for $k < 4.5$, and
they are identical for the two enantiomers because
$U_{4.5}$ collapses the pair to a single object.

Property~(iv) encodes the categorical statement that
\emph{within $\Lk_{4.5}$}, no morphism flips the chirality
label at a bond-graph stereocentre while leaving the
underlying graph fixed.  This captures the chemistry of
\emph{configurational chirality} at tetrahedral
stereocentres --- the case where four distinct substituents
at $sp^3$ carbon define the chirality datum.

Several real interconversion mechanisms fall outside this
framework because they are not visible at $\Lk_{4.5}$:
amine inversion (umbrella flipping at nitrogen via a
planar transition state), atropisomer interconversion
(rotation around a single bond), and helical inversion in
helicenes all interchange ``enantiomeric'' stereoisomers
without breaking the bond graph.  These phenomena require
3D-coordinate data (transition-state planarity), barrier
heights (whether the inversion is thermally accessible at
room temperature), or both --- inputs that enter the tower
only at $\Lk_5$ and above.  Property~(iv) is the
$\Lk_{4.5}$-level shadow of configurational chirality
specifically; the broader physical concept of
``stereochemical interconvertibility'' lives at higher levels.

Property~(v) encodes the central principle of asymmetric
synthesis.
A $\Gstar$-equivariant reaction treats both enantiomers
uniformly, producing enantiomeric products in equal
amounts.
To produce one enantiomer preferentially, a reaction must
\emph{break} the $\Gstar$-symmetry --- by using a chiral
catalyst or reagent modelled as a fixed object
$(C, \sigma_C)$ in the reaction network.
When the catalyst appears explicitly with a fixed chirality
label $\sigma_C$, the two combined substrates
$(G, +1) + (C, \sigma_C)$ and $(G, -1) + (C, \sigma_C)$
have different sources in $\Lk_{4.5}(P)$ and are genuinely
distinct morphisms of the composite system.
The $\Gstar$-symmetry is broken not by violating
equivariance, but by holding one component of the reaction
fixed.
This is the categorical account of enantioselective
catalysis at $\Lk_{4.5}$.
\end{chembox}

\medskip\noindent
The four theorems of this section --- Walden inversion,
racemisation, net retention via double inversion, and the
enantiomer properties --- together exhibit $\Lk_{4.5}$'s
characteristic contribution to the tower: \emph{universal
propagation of mechanism-selected stereochemical
outcomes across $\Gstar$-orbits of substrates}.
The next section (\S\ref{sec:L45-wh}) applies the same
framework to pericyclic reactions, where the
$\Aut(G)$-factor of $\Gstar$ supplies the combinatorial
content of the Woodward--Hoffmann selection rules.

%% file: chapters/L45/l45_wh.tex
\subsection{Pericyclic stereochemistry at
  \texorpdfstring{$\Lk_{4.5}$}{L4.5}: a theorem and a
  Woodward--Hoffmann conjecture}
\label{sec:L45-wh}

The stereochemical theorems of \S\ref{sec:L45-stereo}
treated reactions at a single stereocentre.
The present section takes up the second forcing obligation
of \S\ref{sec:L45-forcing-in} --- the
electrocyclic forcing pair $[\pi_{\mathrm{ec}}]$ --- and
asks what $\Lk_{4.5}$ can say about the pericyclic
selection rules more generally.
The answer has two parts: a \emph{theorem} providing the
categorical content ($\alpha$-equivariance selects one
chirality lift from two), and a \emph{conjecture}
identifying the equivariant lift with the thermally allowed
Woodward--Hoffmann product.
The theorem is established here; the conjecture is
accompanied by verification in the principal case
(electrocyclic) and a discussion of its scope and
limitations.

\subsubsection{Background}

A \emph{pericyclic reaction} is a concerted reaction ---
bond-making and bond-breaking occur simultaneously, with
no ionic intermediate --- whose transition state has a
cyclic arrangement of the participating atoms and bonds,
allowing electron reorganisation around a closed
loop~\cite{WoodwardHoffmann1969}.
At $\Lk_4(P)$, pericyclic reactions are DPO rules whose
reaction centre $L \setminus K$ together with $R \setminus
K$ traces a cyclic subgraph topology \cite{Herges1994}.
The three principal families are electrocyclic ring
closures/openings, cycloadditions, and sigmatropic shifts.

Woodward and Hoffmann discovered in
1965~\cite{WoodwardHoffmann1965,WoodwardHoffmann1969} that
the \emph{stereochemical} outcome of every pericyclic
reaction is governed by a single principle: the symmetry
of the molecular orbitals must be conserved along the
reaction path.
For ground-state (thermal) reactions, this means the
orbitals of the reactant must flow continuously into those
of the product without crossing a symmetry-imposed barrier.
Reactions that satisfy this are \emph{thermally allowed};
reactions that would require a barrier crossing are
\emph{thermally forbidden} (but photochemically allowed,
since the barrier vanishes upon electron promotion).

The resulting selection rules depend only on electron
count:
\begin{center}
\renewcommand{\arraystretch}{1.3}
\begin{tabular}{lcc}
  \hline
  Reaction type & Thermal & Photochemical \\
  \hline
  Electrocyclic, $4n$ electrons
    & conrotatory & disrotatory \\
  Electrocyclic, $4n{+}2$ electrons
    & disrotatory & conrotatory \\
  {[4+2]} cycloaddition (Diels--Alder)
    & supra-supra & antara-supra \\
  {[2+2]} cycloaddition
    & forbidden & supra-supra \\
  \hline
\end{tabular}
\end{center}

\noindent
\emph{Conrotatory} means the two terminal groups rotate in
the same direction; \emph{disrotatory} means they rotate
in opposite directions.
\emph{Suprafacial} (supra) means the new bond forms on
the same face of the $\pi$ system; \emph{antarafacial}
(antara) means it forms on the opposite face.
The work earned Hoffmann the 1981 Nobel Prize in Chemistry, 
shared with Fukui for related frontier-orbital
analysis~\cite{HoffmannNobel1982,FukuiNobel1982}; Woodward
had died in 1979 and the prize could not be awarded
posthumously.

\medskip
\noindent
Pericyclic reactions are about stereochemistry --- which
product isomer is produced and through which
transition-state geometry --- and this is exactly the
content that lives at $\Lk_{4.5}$ as chirality labels and
nowhere lower in the tower.  At $\Lk_4$, the
thermally allowed and thermally forbidden outcomes of a
pericyclic reaction share the same bond-graph DPO rule
($L \leftarrow K \rightarrow R$); the distinction between
them is purely stereochemical (which face of the
$\pi$-system the new bond forms on, and the resulting
relative configuration of the new stereocentres) and
invisible until $\Lk_{4.5}$.  At $\Lk_3$, the rate
constants of allowed and forbidden processes are different
in practice (forbidden reactions face higher activation
barriers), but $\Lk_3$ does not explain \emph{why} they
differ; the structural reason --- $\alpha$-equivariance
or its failure at the chirality lift --- is a $\Lk_{4.5}$
fact.  $\Lk_{4.5}$ is thus the first tower level where the
allowed/forbidden distinction can be \emph{stated
structurally}, not merely registered as a numerical rate difference.

\subsubsection{The categorical theorem}

The connection between WH and $\Lk_{4.5}$ rests on a single
graph-theoretic observation.
A pericyclic reaction centre is a cyclic subgraph, and
cyclic graphs admit non-trivial automorphisms ---
specifically, the reflections that swap pairs of
stereogenic terminal atoms.
These automorphisms project from the 3D symmetry elements
($C_2$ axes, $\sigma_v$ planes) of the transition state to
the graph level, where they act on the vertex set by
permuting the reaction-centre atoms.
Call such an automorphism a \emph{reaction-centre
automorphism} of the rule.

\begin{theorem}[$\alpha$-equivariance distinguishes
  pericyclic product lifts (ring-closure direction)]
\label{thm:WH-categorical}
Let $p_0$ be a pericyclic ring-closure DPO rule in
$\LGraphP$ --- a rule whose reactant side $L$ has the
participating atoms in an open-chain $\pi$-system (sp$^2$
hybridisation at the termini) and whose product side $R$
has the new $\sigma$-bond closing those termini into a
ring (sp$^3$ at the new stereogenic positions).  Let
$\alpha$ be a reaction-centre automorphism of $p_0$ ---
that is, an involutive graph automorphism $\alpha \in \Aut
(L) \cap \Aut(R)$ that exchanges a pair of stereocentres
$\{v_1, v_2\} \subset \Stereo(R)$.  Then:
\begin{enumerate}[label=(\roman*)]
  \item $p_0$ admits exactly two chirality lifts
    $\tilde{p}^+, \tilde{p}^-$ distinguished by the
    relative sign of $\sigma_R(v_1)$ and $\sigma_R(v_2)$:
    \[
      \tilde{p}^+\colon\ \sigma_R(v_1) = \sigma_R(v_2),
      \qquad
      \tilde{p}^-\colon\ \sigma_R(v_1) = -\sigma_R(v_2).
    \]
  \item Exactly one of the two lifts is
    $\alpha$-equivariant: $\tilde{p}^+$ is
    $\alpha$-equivariant and $\tilde{p}^-$ is not.
\end{enumerate}
\end{theorem}

\begin{proof}
\textbf{(i).}
The reaction produces the two stereocentres $v_1, v_2
\in \mathrm{Stereo}(R)$ from sp$^2$ atoms in $L$ (hence
$v_1, v_2 \notin \mathrm{Stereo}(L)$: at the $L$-side these
vertices have three neighbours each, failing
condition~(i) of Definition~\ref{def:chiral-centre}).
A chirality lift of $p_0$ is therefore a choice of
$(\sigma_R(v_1), \sigma_R(v_2)) \in \{+1, -1\}^2$ at the
product side --- four options in total.  Identifying lifts
related by the $E^*$ action $(\sigma_R \mapsto -\sigma_R)$,
which produces the enantiomeric product and hence the same
stereochemical content up to overall enantiomer choice,
the four options collapse into two equivalence classes:
\begin{itemize}
  \item $\tilde{p}^+ = \{(+,+), (-,-)\}$: the lifts with
    $\sigma_R(v_1) = \sigma_R(v_2)$.
  \item $\tilde{p}^- = \{(+,-), (-,+)\}$: the lifts with
    $\sigma_R(v_1) = -\sigma_R(v_2)$.
\end{itemize}

\textbf{(ii).}
The $\alpha$-equivariance square
(Definition~\ref{def:gstar-equivariant-rule}) for
$\tilde{p}^{\pm}$ with $\alpha$ acting on chirality
functions by $\sigma \mapsto \sigma \circ \alpha^{-1}$
requires $\sigma_R \circ \alpha^{-1} = \sigma_R$ on
$\{v_1, v_2\}$.
Since $\alpha$ exchanges $v_1 \leftrightarrow v_2$:
\[
  (\sigma_R \circ \alpha^{-1})(v_1) = \sigma_R(v_2),
  \qquad
  (\sigma_R \circ \alpha^{-1})(v_2) = \sigma_R(v_1).
\]
$\alpha$-equivariance thus requires $\sigma_R(v_1) =
\sigma_R(v_2)$: exactly the condition defining
$\tilde{p}^+$.
Hence $\tilde{p}^+$ is $\alpha$-equivariant;
$\tilde{p}^-$, with $\sigma_R(v_1) = -\sigma_R(v_2)$,
fails the equivariance condition.
\end{proof}

\begin{remark}[What the theorem says, tower-natively]
\label{rem:WH-cat-reading}
Theorem~\ref{thm:WH-categorical} separates a pericyclic
rule into two chirality-lift classes defined by the
relative sign pattern at the new stereocentres.
At $\Lk_4$, both lifts collapse under $U_{4.5}$ to the
same bond-graph rule: the distinction is
$\Lk_4$-invisible.
At $\Lk_{4.5}$, the reaction-centre automorphism $\alpha$
is a non-trivial element of $\Aut(G) \leq \Gstar(G)$, and
$\alpha$-equivariance singles out exactly one of the two
lifts.
The tower contribution is the \emph{classification} of
pericyclic product lifts by $\alpha$-parity, not the
identification of which class is thermally allowed --- that
identification is the content of the conjecture below.
\end{remark}

\subsubsection{The Woodward--Hoffmann identification
  conjecture}

Theorem~\ref{thm:WH-categorical} gives the combinatorial
skeleton: for every pericyclic reaction with a reaction-centre
automorphism $\alpha$, one of two chirality lifts is
$\alpha$-equivariant.
The question this section poses is whether that
combinatorial skeleton tracks the Woodward--Hoffmann
selection rules.

\begin{conjecture}[Woodward--Hoffmann identification
  at $\Lk_{4.5}$, electrocyclic case]
\label{conj:WH-electrocyclic}
Let $p_0$ be an electrocyclic ring-closure rule with
$N$ $\pi$-electrons in the reactant $\pi$-system, and let
$\alpha$ be its reaction-centre automorphism (the
reflection exchanging the two terminal stereocentres).
Let $\tilde{p}^+, \tilde{p}^-$ be the two chirality lifts
of Theorem~\ref{thm:WH-categorical}.
Then the thermally allowed Woodward--Hoffmann product
corresponds to:
\begin{itemize}
  \item the $\alpha$-equivariant lift $\tilde{p}^+$ when
    $N \equiv 0 \pmod 4$ (i.e., $4n$ electrons);
  \item the non-$\alpha$-equivariant lift $\tilde{p}^-$
    when $N \equiv 2 \pmod 4$ (i.e., $4n+2$ electrons).
\end{itemize}
The photochemically allowed product corresponds to the
opposite lift in each case.
\end{conjecture}

\begin{remark}[What the conjecture claims]
\label{rem:WH-conj-content}
Theorem~\ref{thm:WH-categorical} is a combinatorial
theorem: it classifies product lifts by $\alpha$-parity.
Conjecture~\ref{conj:WH-electrocyclic} adds empirical
content: it identifies which $\alpha$-parity corresponds
to the thermally allowed outcome, with an alternation
governed by $N \bmod 4$.
The alternation itself --- the hallmark of the
Woodward--Hoffmann rules --- is the conjecture's
substantive claim.
Below $\Lk_{4.5}$, the alternation is invisible because
chirality labels are absent; at $\Lk_{4.5}$, it is
expressible but not derivable from categorical axioms
alone.
The identification between $\alpha$-parity and thermal
allowedness requires input from the orbital-symmetry
analysis of Woodward and Hoffmann, and this input is
not currently derivable from the tower structure below
$\Lk_6$ (where the full electronic Hilbert bundle
becomes available).
\end{remark}

\subsubsection{Verification: hexa-2,4-diene electrocyclic
  closure}

The principal case for which
Conjecture~\ref{conj:WH-electrocyclic} can be checked
directly is the forcing pair of
\S\ref{sec:L45-forcing-in}: the electrocyclic closure of
$(E,E)$-hexa-2,4-diene to 3,4-dimethylcyclobutene, which
has $N = 4$ $\pi$-electrons ($4n$ case, $n = 1$).
Using cyclobutene numbering for the product
($\mathrm{C_1}{=}\mathrm{C_2}{-}\mathrm{C_3}{-}\mathrm{C_4}$
around the ring, with the methyl groups attached at
$\mathrm{C_3}$ and $\mathrm{C_4}$), the
reaction-centre automorphism is
\[
  \alpha\colon \mathrm{C_1} \leftrightarrow \mathrm{C_2},
  \quad \mathrm{C_3} \leftrightarrow \mathrm{C_4},
\]
the graph-level shadow of the $C_2$ axis of the
transition state passing through the midpoints of the
$\mathrm{C_1}{=}\mathrm{C_2}$ and
$\mathrm{C_3}{-}\mathrm{C_4}$ bonds.
The two product stereocentres are $\mathrm{C_3}$ and
$\mathrm{C_4}$ (the saturated ring carbons, each bearing a
methyl substituent), in the sense given by the CIP
hierarchical-digraph extension of
Definition~\ref{def:chiral-centre} (cf.\ \S\ref{sec:L45-forcing-in}).

\emph{trans-product (from conrotatory motion).}
Conrotatory motion rotates both terminal methyl groups
in the same sense, producing
\emph{trans}-3,4-dimethylcyclobutene with
$\sigma_{\mathrm{trans}}(\mathrm{C_3}) =
\sigma_{\mathrm{trans}}(\mathrm{C_4})$.
Under $\alpha$ (swapping $\mathrm{C_3}
\leftrightarrow \mathrm{C_4}$):
\[
  (\sigma_{\mathrm{trans}} \circ \alpha^{-1})
  (\mathrm{C_3})
  = \sigma_{\mathrm{trans}}(\mathrm{C_4})
  = \sigma_{\mathrm{trans}}(\mathrm{C_3}).
\]
The label is preserved: $\sigma_{\mathrm{trans}}$ is
$\alpha$-equivariant, so the trans product is the
$\tilde{p}^+$ lift.
The WH identification ($4n$ case: thermal $\leftrightarrow$
equivariant) predicts this is thermally allowed ---
consistent with the empirical outcome that conrotatory
closure of hexa-2,4-diene is observed thermally.

\emph{cis-product (from disrotatory motion).}
Disrotatory motion rotates the terminal methyl groups in
opposite senses, producing
\emph{cis}-3,4-dimethylcyclobutene with
$\sigma_{\mathrm{cis}}(\mathrm{C_3}) =
-\sigma_{\mathrm{cis}}(\mathrm{C_4})$.
Under $\alpha$:
\[
  (\sigma_{\mathrm{cis}} \circ \alpha^{-1})(\mathrm{C_3})
  = \sigma_{\mathrm{cis}}(\mathrm{C_4})
  = -\sigma_{\mathrm{cis}}(\mathrm{C_3}).
\]
The label is flipped: $\sigma_{\mathrm{cis}}$ is
not $\alpha$-equivariant, so the cis product is the
$\tilde{p}^-$ lift.
The WH identification predicts this is thermally
forbidden (photochemically allowed) --- consistent with
empirical observation.

\subsubsection{The equivariance squares}

The two outcomes of the hexa-2,4-diene case are captured
by the following pair of diagrams.

\noindent\textbf{Thermally allowed (conrotatory, trans)
--- commuting $\alpha$-square:}
\begin{equation}
\label{diag:WH-con}
\includegraphics{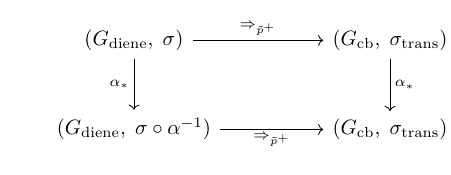}
\end{equation}
The bottom-right equals the top-right because
$\sigma_{\mathrm{trans}}$ is fixed by
$\mathrm{C_3} \leftrightarrow \mathrm{C_4}$.
The reaction does not depend on which ``copy'' of the
diene the rule is applied to: $\tilde{p}^+$ is
$\alpha$-equivariant.

\noindent\textbf{Thermally forbidden (disrotatory, cis)
--- non-commuting $\alpha$-square:}
\begin{equation}
\label{diag:WH-dis}
\includegraphics{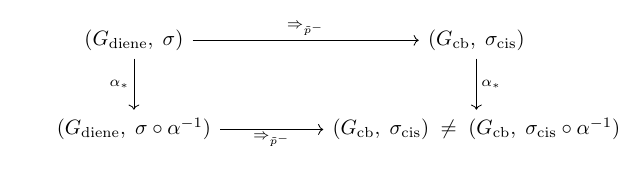}
\end{equation}
The two paths around the square land on different
objects: applying the rule and then $\alpha_*$ gives
$(G_{\mathrm{cb}}, \sigma_{\mathrm{cis}} \circ \alpha^{-1})$
(the right-edge endpoint), whereas applying $\alpha_*$ and
then the rule gives $(G_{\mathrm{cb}}, \sigma_{\mathrm{cis}})$
(the rule's chirality lift produces $\sigma_{\mathrm{cis}}$
regardless of the substrate's $\sigma$).  Since
$\sigma_{\mathrm{cis}}$ is flipped by $\mathrm{C_3}
\leftrightarrow \mathrm{C_4}$, these two objects are
distinct, and $\tilde{p}^-$ is not $\alpha$-equivariant.

\subsubsection{Further pericyclic reactions}

The same $\alpha$-parity analysis applies to other
pericyclic reactions but with two complications that
moderate the scope of
Conjecture~\ref{conj:WH-electrocyclic}.

\medskip
\noindent\textbf{Hexa-1,3,5-triene $\to$ cyclohexa-1,3-diene
($6\pi$, $4n+2$).}
The forcing pair's $4n$ electron count is complemented by
this $4n+2$ case.  Using cyclohexa-1,3-diene numbering for
the product (ring carbons $\mathrm{C_1}{=}\mathrm{C_2}
{-}\mathrm{C_3}{=}\mathrm{C_4}{-}\mathrm{C_5}{-}\mathrm{C_6}$,
the new $\sigma$-bond closing $\mathrm{C_5}{-}\mathrm{C_6}$),
the new stereocentres are $\mathrm{C_5}$ and $\mathrm{C_6}$,
and the reaction-centre automorphism $\alpha$ swaps
$\mathrm{C_1} \leftrightarrow \mathrm{C_4}$,
$\mathrm{C_2} \leftrightarrow \mathrm{C_3}$, and
$\mathrm{C_5} \leftrightarrow \mathrm{C_6}$, the graph-level
shadow of the $C_2$ axis through the midpoints of
$\mathrm{C_2}{-}\mathrm{C_3}$ and
$\mathrm{C_5}{-}\mathrm{C_6}$.  An analogous
chirality-label computation shows: the cis product (with
$\sigma_R(\mathrm{C_5}) = -\sigma_R(\mathrm{C_6})$) is
non-$\alpha$-equivariant ($\tilde{p}^-$), while the trans
product is $\alpha$-equivariant ($\tilde{p}^+$).
Woodward and Hoffmann's 1965
analysis~\cite{WoodwardHoffmann1965} identifies the cis
product (from disrotatory motion) as thermally allowed for
$4n+2$ electrons, i.e.\ the non-$\alpha$-equivariant lift:
this is the $4n+2$ branch of
Conjecture~\ref{conj:WH-electrocyclic}.  The alternation
between $4n$ (thermal $\leftrightarrow$ $\alpha$-equivariant)
and $4n+2$ (thermal $\leftrightarrow$
non-$\alpha$-equivariant) is the WH hallmark.

\medskip
\noindent\textbf{Cycloadditions and sigmatropic shifts.}
The categorical theorem
(Theorem~\ref{thm:WH-categorical}) extends to
cycloadditions and sigmatropic shifts whenever the
reaction centre admits an involutive graph automorphism
exchanging new stereocentres.
The Diels--Alder [4+2] cycloaddition, for instance, has
reaction centre $\{\mathrm{C_1}, \ldots, \mathrm{C_6}\}$
(four from the diene, two from the dienophile) and a
natural $C_2$ automorphism of the 6-membered
transition-state ring.
For substrates with the requisite substitution pattern
to produce stereocentres at the new $\sigma$-bond termini,
the $\alpha$-parity of the chirality lift distinguishes the
two possible cycloadduct stereochemistries.
Substrates without such substitution (unsubstituted
butadiene plus ethylene) produce a cyclohexene with no
stereocentres, making the $\sigma$-analysis vacuous.
The general WH identification for cycloadditions follows
the same $(4n)$ vs.\ $(4n+2)$ alternation as the
electrocyclic case; the conjecture extends accordingly but
verification requires case-by-case substrate choice.

\subsubsection{Scope and limitations}

\begin{remark}[What $\Lk_{4.5}$ can and cannot distinguish]
\label{rem:WH-scope}
Theorem~\ref{thm:WH-categorical} and
Conjecture~\ref{conj:WH-electrocyclic} together express
the $\Lk_{4.5}$-visible content of the Woodward--Hoffmann
rules.
Two scope boundaries should be made explicit.

\emph{Product-level vs.\ motion-level distinctions.}
The reaction-centre automorphism $\alpha$ is a graph-level
object: it permutes vertices without reference to 3D
geometry.
A single graph automorphism $\alpha$ can be the shadow of
either a $C_2$ rotation axis (as in conrotatory motion)
or a $\sigma_v$ mirror plane (as in disrotatory motion);
both act identically on the vertex set.
Consequently, $\Lk_{4.5}$ distinguishes the two
\emph{products} (trans vs.\ cis, supra-supra vs.\
supra-antara) but not the two transition-state
\emph{motions} (conrotatory vs.\ disrotatory, axial vs.\
planar).
The motion-level WH content --- which orbital symmetry
element is preserved during the continuous reaction path
--- requires the 3D geometry introduced at $\Lk_5$, where
the Euclidean group $E(3)$ acts on configurations.
At $\Lk_5$, the distinction between $C_2$-preserving and
$\sigma_v$-preserving transition states becomes
meaningful, and WH's orbital-conservation argument can be
stated in full.

\emph{The conjecture's status.}
The categorical theorem
(Theorem~\ref{thm:WH-categorical}) is established here.
The WH identification conjecture
(Conjecture~\ref{conj:WH-electrocyclic}) has been verified
for the hexa-2,4-diene forcing pair ($4n$) and hexa-1,3,5-
triene ($4n+2$); its extension to cycloadditions and
sigmatropic shifts follows the same $\alpha$-parity
pattern but requires case-by-case substrate choice to
ensure the $\sigma$-analysis is non-vacuous.
The alternation with electron count --- the substantive
WH claim --- has been checked case-by-case but is not
derivable from $\Lk_{4.5}$ axioms alone.
A derivation would require the tower to supply the
orbital-count dependence intrinsically, which presumably
occurs only at $\Lk_6$ where the electronic Hilbert bundle
and its symmetry-adapted basis enter.
\end{remark}

\subsubsection{Tower reading}

\begin{chembox}[What the theorem and conjecture say
  together]
\label{chembox:WH-summary}
Theorem~\ref{thm:WH-categorical} and
Conjecture~\ref{conj:WH-electrocyclic} together provide
the following account of the Woodward--Hoffmann rules at
$\Lk_{4.5}$.

\emph{Theorem content.}
Every pericyclic rule with a reaction-centre automorphism
$\alpha$ exchanging new stereocentres decomposes into
exactly two chirality lifts, distinguished by whether
$\sigma_R$ is $\alpha$-invariant or $\alpha$-anti-invariant.
The $\alpha$-invariant lift ($\tilde{p}^+$) is
$\alpha$-equivariant; the $\alpha$-anti-invariant lift
($\tilde{p}^-$) is not.  Here $\alpha$ is a specific
element of $\Aut(G) \leq \Gstar(G)$, and the relevant
equivariance is under the cyclic subgroup
$\langle \alpha \rangle \leq \Gstar(G)$ generated by it,
not under all of $\Gstar(G)$ --- the latter would also
require $E^*$-equivariance, which intrinsic-$\sigma_R$
pericyclic rules do not satisfy.

\emph{Conjecture content.}
The Woodward--Hoffmann (WH) rules assert an empirical
identification: the $\alpha$-equivariant lift is the
thermally allowed product for $4n$ electrons, and the
non-equivariant lift is thermally allowed for $4n+2$
electrons.
The alternation is the WH hallmark.

\emph{Tower boundary.}
The bond-graph transformation underlying the thermally
forbidden disrotatory closure of hexa-2,4-diene is a
perfectly valid morphism of $\Lk_4(P)$.  Its $\Lk_{4.5}$
chirality lift to the cis product, however, is not
$\alpha$-equivariant (Theorem~\ref{thm:WH-categorical}),
hence not $\Gstar$-equivariant, hence not a generator of
$\Lk_{4.5}(P)$.  The cis product itself is a perfectly
valid \emph{object} of $\Lk_{4.5}(P)$ --- it exists as a
real molecule with a well-defined chirality function ---
but the \emph{rule} producing it via thermal closure is
inadmissible at $\Lk_{4.5}$.  This is the categorical
incarnation of WH at $\Lk_{4.5}$: the inadmissibility of
the forbidden pericyclic \emph{rule}, not of its product,
is the failure of the chirality lift to be
$\Gstar$-equivariant.
The orbital-symmetry conservation argument of Woodward
and Hoffmann supplies the 3D geometric justification for
why $\alpha$-equivariance should correspond to thermal
allowedness --- this justification operates at $\Lk_5$
(3D geometry) and fully at $\Lk_6$ (electronic
structure), both beyond the present chapter.
\end{chembox}

\medskip\noindent
The categorical theorem and the WH identification
conjecture together close the analysis of
$\Lk_{4.5}$-level stereochemistry as responsive to the
two forcing cokernel classes of
\S\ref{sec:L45-forcing-in}: the enantiomer obstruction
$[\pi_\chi]$ is fully resolved in \S\ref{sec:L45-stereo}
by Walden, racemisation, and the enantiomer properties;
the electrocyclic obstruction $[\pi_{\mathrm{ec}}]$ is
partially resolved here via the categorical theorem, with
the Woodward--Hoffmann identification as an outstanding
conjecture whose full resolution awaits higher tower
levels.
The tower's passage from $\Lk_{4.5}$ to $\Lk_5$ ---
where 3D geometry becomes available and the motion-level
WH content enters the frame --- is taken up in
\S\ref{sec:L45-forcing-out}.

%% file: chapters/L45/l45_nextforcing.tex
\subsection{What \texorpdfstring{$\Lk_{4.5}$}{L4.5} cannot
  express: forcing of \texorpdfstring{$\Lk_5$}{L5}}
\label{sec:L45-forcing-out}

The forcing pair of \S\ref{sec:L45-forcing-in} drove the
tower from $\Lk_4$ to $\Lk_{4.5}$ through a symmetry
enrichment: a group action $\Gstar(G) \curvearrowright
\LGraphGstar$ was added to an existing combinatorial
category.
The forcing from $\Lk_{4.5}$ to $\Lk_5$ is qualitatively
different.
It does not promote a qualitative distinction
$\Lk_{4.5}$-invisible to $\Lk_{4.5}$-visible; instead, it
promotes a \emph{quantitative} distinction whose source is
the 3D geometry of the transition state --- a datum that
does not exist anywhere in the tower below $\Lk_5$.

For the first time in the tower, \emph{geometric}
continuous data (3D positions on a configuration manifold)
is unavoidable; lower levels admitted continuous numerical
decorations (free energies at $\Lk_2$, rate constants at
$\Lk_3$) but not continuous geometric structure.

\begin{forcingbox}[Forcing pair for $\Lk_5$: same
  $\Lk_{4.5}$-object, different rate]

\noindent\textbf{Isotopic indistinguishability at $\Lk_{4.5}$.}
In $\LGraphP$ and $\LGraphGstar$, isotopes of the same
element share a single vertex label: the function
$\mathrm{el}\colon V \to \{\text{elements}\}$ records
atomic number, not mass number.
Hydrogen (H) and deuterium (D) both have atomic number~1
and therefore share the vertex label ``hydrogen''
throughout $\Lk_0$--$\Lk_{4.5}$.
Replacing every H by D in a molecule produces an object
that is \emph{literally the same object} at these levels:
same vertex set, same edge set, same vertex labels, same
$\Gstar$-orbit.

\medskip
\noindent\textbf{System~A}: $\mathrm{CH_3Br} + \mathrm{OH^-}
\to \mathrm{CH_3OH} + \mathrm{Br^-}$ ($\mathrm{S_N2}$, water, 298~K).
Rate constant $k_H \approx 4\times10^{-3}\;
\mathrm{M^{-1}s^{-1}}$~\cite{Parker1969}.

\noindent\textbf{System~B}: $\mathrm{CD_3Br} + \mathrm{OH^-}
\to \mathrm{CD_3OH} + \mathrm{Br^-}$ (same reaction with
all three H at the reactive carbon replaced by D).
Rate constant $k_D \approx 3\times10^{-3}\;
\mathrm{M^{-1}s^{-1}}$, giving the $\alpha$-secondary
kinetic isotope effect $k_H/k_D \approx 1.3$ at
298~K~\cite{Streitwieser1958,Westheimer1961}.

\medskip
\noindent\textbf{The $\Lk_{4.5}$ collapse, structurally.}
Both systems use the same $\Gstar$-equivariant DPO rule
$p_{\mathrm{SN2}}$ of \S\ref{sec:L45-stereo}.
Neither substrate has a stereocentre: $\Stereo
(\mathrm{CH_3Br}) = \Stereo(\mathrm{CD_3Br}) = \emptyset$.
The categorical data at every tower level up to and
including $\Lk_{4.5}$ coincide for the two systems:
identical $\Lk_0$-stoichiometries, identical $\Lk_4$-DPO
mechanisms, identical $\Lk_{4.5}$-chirality lifts (both
empty).  Numerical decorations attached at lower levels
($\Lk_1$ enthalpies, $\Lk_2$ free energies, $\Lk_3$ rate
constants) differ by small isotope-dependent amounts that
the levels themselves record but do not explain --- the
ZPE shifts and the corresponding rate-constant ratio
$k_H/k_D \approx 1.3$ are stored as numerical inputs at
$\Lk_1$--$\Lk_3$, with no derivation from those levels'
structural data.

\medskip
\noindent\textbf{The forcing, sharpened.}
What $\Lk_{4.5}$ cannot do is \emph{derive} the
rate-constant ratio from its categorical structure.  At
$\Lk_{4.5}$, the two systems share every structural
attribute (DPO rule, chirality lift, $\Gstar$-orbit); the
empirical rate ratio is an isotope-specific number that
the level cannot account for, only record.  The forcing
into $\Lk_5$ is therefore \emph{explanatory}: $\Lk_5$
provides the 3D-geometric machinery (configuration
orbifold $\mathcal{C}_e(G)$, Born--Oppenheimer PES, TS
geometry, vibrational ZPE) from which the rate-constant
ratio can be derived rather than stipulated.  In tower
terms, the forgetful functor $U_5\colon \Lk_5 \to
\Lk_{4.5}$ has a non-trivial fibre over the single
$\Lk_{4.5}$-morphism $p_{\mathrm{SN2}}$: the fibre splits
into isotope-distinguished $\Lk_5$-morphisms that share
the $\Lk_{4.5}$-projection but differ in their
TS-vibrational signature.

\medskip
\noindent\textbf{The $\Lk_5$ refinement.}
Experimentally, the two rate constants differ by a factor
of $1.3$.
At higher $\Lk_5$ resolution, System~A and System~B become
distinct objects: their nuclear masses enter the
transition-state (TS) vibrational partition function, yielding
different activation barriers and hence different rate
constants.
The forcing is thus an \emph{object refinement}: two objects
identified at $\Lk_{4.5}$ split into distinct objects at
$\Lk_5$ once 3D geometry and mass-weighted dynamics become
available.
\end{forcingbox}

\medskip\noindent
\textbf{Physical origin of the KIE.}
The $\alpha$-secondary effect $k_H/k_D \approx 1.3$ arises
because the $\mathrm{C{-}H}$ and $\mathrm{C{-}D}$ bending
vibrations change character at the
$\mathrm{S_N2}$ transition state.
The reactive carbon passes through a nearly pentacoordinate
geometry (bipyramidal around the central carbon, with the
leaving group and nucleophile at the apical positions and
the three remaining substituents on an equatorial plane),
and one bending mode at the reactive carbon softens
significantly between the reactant (tetrahedral, bending
frequency $\approx 1340\,\mathrm{cm^{-1}}$) and the TS
(bending frequency $\approx 1150\,\mathrm{cm^{-1}}$).
The resulting zero-point-energy shift differs between H and
D by a small amount that, accumulated over all three
hydrogens at the reactive carbon, produces a rate ratio
near $1.3$~\cite{Streitwieser1958}.
A careful derivation requires the bending-mode frequencies
at \emph{both} the reactant and the transition-state
geometries; the transition-state geometry is the saddle
point of the potential energy surface $V\colon
\mathcal{C}_e(G)\to\RR$, a $\Lk_5$-level datum.

\medskip\noindent
\textbf{The starker case: primary KIE.}
When the $\mathrm{C{-}H}$ bond is \emph{directly broken} in
the rate-determining step --- as in base-catalysed proton
abstraction from a carbonyl $\alpha$-carbon --- the isotope
effect is much larger: $k_H/k_D \approx 5$--$7$ at
$25\,^\circ\mathrm{C}$, with a semiclassical maximum
$k_H/k_D \approx 7$~\cite{Westheimer1961}.
The origin is the full $\mathrm{C{-}H}$ vs.\
$\mathrm{C{-}D}$ stretching ZPE difference: with
$\nu_H^{\mathrm{stretch}} \approx 3000\,\mathrm{cm^{-1}}$
and $\nu_D^{\mathrm{stretch}} \approx 2120\,\mathrm{cm^{-1}}$,
the difference $\Delta E_{\mathrm{ZPE}} = \tfrac{1}{2}h(
\nu_H - \nu_D) \approx 1.1\,\mathrm{kcal\,mol^{-1}}$
vanishes at the TS where the bond is broken, giving
$k_H/k_D = \exp(\Delta E_{\mathrm{ZPE}}/RT) \approx 6.5$ at
298~K.
The base-catalysed enolisation of $(\mathrm{CH_3})_2
\mathrm{CO}$ versus $(\mathrm{CD_3})_2\mathrm{CO}$ is
the classical experimental system.
At $\Lk_{4.5}$, the two enolisations use the same DPO rule
and involve no stereocentre; the factor-of-6 rate difference
is invisible.
At $\Lk_5$, it is a direct consequence of the
$\mathrm{C{-}H}$ stretching frequency evaluated at the TS
geometry on $\mathcal{C}_e(G)$~\cite{Westheimer1961}.

\medskip\noindent
KIEs larger than the semiclassical maximum of $\approx 7$
do occur in nature --- enzymatic H-transfer reactions
routinely exhibit $k_H/k_D$ ratios of 50--700 --- but these
arise from nuclear tunnelling and cannot be captured by
classical TST on the Born--Oppenheimer PES.
They are a higher-tower phenomenon, forcing the passage
from $\Lk_6$ to $\Lk_7$ rather than $\Lk_{4.5}$ to $\Lk_5$.
The KIEs of this section sit comfortably in the
semiclassical regime, where ZPE differences at the TS
geometry --- data available once $\mathcal{C}_e(G)$ and
$V$ are in place --- fully account for the rate effect.

\subsubsection{What \texorpdfstring{$\Lk_5$}{L5} must add}

The forcing pair exposes a chain of dependences that
$\Lk_{4.5}$ cannot express:
\[
  \text{3D geometry}
  \;\Rightarrow\;
  \text{TS geometry}
  \;\Rightarrow\;
  \text{activation barrier } E_a
  \;\Rightarrow\;
  \text{rate constant } k.
\]
None of the arrows can be reversed from $\Lk_{4.5}$-level
data.
$\Lk_5$ closes the chain by introducing three new objects,
together constituting the tower's first \emph{geometric
break}:

\begin{enumerate}[label=(\roman*)]
  \item \textbf{The configuration orbifold}
    $\mathcal{C}_e(G) = \RR^{3n}/(\mathrm{SE}(3) \times
    \Aut(G))$: the space of all molecular shapes --- all
    assignments of 3D coordinates to atoms, modulo
    rigid-body motions and permutations of identical
    atoms.
    A point $\mathbf{R} \in \mathcal{C}_e(G)$ is a
    specific shape of the molecule.
    This is the first genuinely continuous and geometric
    object in the tower.

  \item \textbf{The Born--Oppenheimer potential energy
    surface} $V\colon \mathcal{C}_e(G) \to \RR$, where
    $V(\mathbf{R})$ is the electronic ground-state energy
    at geometry $\mathbf{R}$.
    The landscape of $V$ encodes stable conformers
    (local minima), transition states (saddle points),
    and reaction paths (steepest-descent curves on $V$).

  \item \textbf{The activation barrier and Eyring
    equation}: the barrier $E_a$ is the height of the
    saddle point above the reactant minimum along the
    intrinsic reaction coordinate.  Within the
    transition-state-theory approximation
    \cite{Eyring1935} (no recrossings, classical TS
    partition function, quasi-equilibrium with reactants,
    no tunnelling) the rate constant takes the Eyring form
    \[
      k \;=\; \frac{k_B T}{h}\, e^{-E_a/k_B T},
    \]
    with $E_a$ extracted from the PES topology.  Within
    this approximation both the secondary KIE
    ($k_H/k_D \approx 1.3$) and the primary KIE
    ($k_H/k_D \approx 7$) follow from
    $E_a^D - E_a^H = \Delta E_{\mathrm{ZPE}}$, computable
    once the TS geometry on $\mathcal{C}_e(G)$ is known.
\end{enumerate}
\noindent
To a chemist: \emph{molecules now have shapes, and shapes
determine how fast reactions go}.
Everything up to and including $\Lk_{4.5}$ --- stoichiometry,
thermodynamics, kinetics, bond-graph mechanisms, chirality
labels --- is shape-independent.
$\Lk_5$ is where continuous geometry enters the tower.

\subsubsection{The 3D realisation of chirality labels}

The passage to $\Lk_5$ also completes a deliberate deferral
from $\Lk_{4.5}$.
At $\Lk_{4.5}$, the enantiomers $(G, +1)$ and $(G, -1)$
were declared distinct objects, but the sign $\sigma \in
\{+1, -1\}$ was \emph{abstract}: the category recognised
the two as different yet could not say which 3D arrangement
corresponded to which sign.
The Cahn--Ingold--Prelog convention for identifying $\sigma$
with $(R)/(S)$ (Remark~\ref{rem:CIP}) was an external
labelling convention, not derivable within $\Lk_{4.5}$.

At $\Lk_5$, the Born--Oppenheimer PES $V$ on
$\mathcal{C}_e(G)$ for a chiral molecule with a
configurational stereocentre at carbon has \emph{two
distinct minima} separated by a high racemisation barrier
(the energy cost of inverting the stereocentre, typically
much larger than $k_B T$ for tetrahedral carbon).  The
configuration orbifold $\mathcal{C}_e(G)$ itself remains
connected as a topological space, but its
\emph{low-energy region} --- the union of basins of
attraction of the local minima at thermally accessible
energy --- decomposes into two disjoint basins, one for
each enantiomer.  The abstract parity $E^* \in \Gstar(G)$
from \S\ref{sec:L45-gstar} acquires a geometric
realisation as the involution exchanging the two basins,
physically effectable only by traversing the racemisation
barrier.  The chirality label $\sigma(v) \in \{+1, -1\}$
then admits a canonical geometric interpretation: it is
the \emph{basin label} on $\mathcal{C}_e(G)$, recording
which low-energy basin the molecule inhabits.

\begin{remark}[Multi-stereocentre and meso refinement]
\label{rem:basin-multistereocentre}
For molecules with $k > 1$ stereocentres, $\sigma$ takes
values in $\{+1, -1\}^k$ and the low-energy region of
$\mathcal{C}_e(G)$ generically decomposes into up to
$2^k$ basins.  When $\Aut(G)$ contains an element
identifying $\sigma$ with another assignment $\sigma'
\neq \sigma$, the corresponding basins are identified in
the orbifold quotient: meso compounds (\S\ref{sec:L45-stereo},
Definition~\ref{def:enantiomers}) thus have fewer than
$2^k$ distinct basins, with the basin count given by the
$\Aut(G)$-orbit decomposition of $\{+1, -1\}^k$.  The
basin label is canonical; its value space is determined
by the molecule's symmetry group.
\end{remark}

This completes the geometric picture of stereochemistry.
At $\Lk_{4.5}$, $\sigma$ was a discrete sign separating
two abstract objects; at $\Lk_5$, it is a basin label on
the appropriate configuration orbifold.  The
$\Lk_{4.5}$-theorems of \S\ref{sec:L45-stereo} translate
into $\Lk_5$-geometric statements; below, $\mathcal{C}_e
(G_{\rm joint})$ denotes the configuration orbifold for
the joint set of atoms participating in the reaction
(substrate plus reagents), with bond-graph changes
treated as transitions between PES regions in the
standard reactive-PES sense:
\begin{itemize}
  \item Walden inversion (Theorem~\ref{thm:walden}) is a
    reaction path on $\mathcal{C}_e(G_{\rm joint})$
    connecting the substrate basin to a product basin of
    opposite chirality, with the back-side-attack TS
    geometry interpolating between them.
  \item Racemisation (Theorem~\ref{thm:racemisation}) is
    a path that visits the planar $\mathrm{S_N1}$
    carbocation TS --- a saddle point lying equidistant
    (in barrier height) from both enantiomer basins --- and
    descends with equal probability into either basin,
    matching the racemic outcome at $\Lk_{4.5}$.
  \item Net retention via double inversion
    (Theorem~\ref{thm:NGP}) is a path that crosses the
    inversion saddle twice (substrate $\to$ bridged
    intermediate $\to$ product), with the two basin
    crossings composing to a path that returns to the
    starting basin's chirality class.
\end{itemize}
The passage from $\Lk_{4.5}$ to $\Lk_5$ is thus not only
the addition of new content (PES, rates, barriers); it is
also the \emph{geometric realisation} of content already
established at $\Lk_{4.5}$ as discrete combinatorial data.
Chapter~\ref{sec:L5} develops this realisation in full.

%% file: chapters/ch_L5.tex
\section{\texorpdfstring{$\Lk_5$}{Lk5}: The Geometric Level}
\label{sec:L5}

\input{chapters/L5/l5_forcing}
\input{chapters/L5/l5_ce}
\input{chapters/L5/l5_pes}
\input{chapters/L5/l5_def}
\input{chapters/L5/l5_tst}
\input{chapters/L5/l5_examples}
\input{chapters/L5/l5_nextforcing}

%% file: chapters/L5/l5_forcing.tex
\subsection{Forcing the extension: geometry from the kinetic
  isotope effect}
\label{sec:L5-forcing-in}

Section~\ref{sec:L45-forcing-out} closed with the forcing pair:
$\mathrm{CH_3Br}$ and $\mathrm{CD_3Br}$ undergoing the same
$\mathrm{S_N2}$ reaction are structurally identical at
$\Lk_{4.5}(P)$ (isomorphic graphs, identical DPO rule,
no stereocentre) yet react at different speeds, with
$k_H/k_D \approx 1.3$ recorded empirically as a $\Lk_3$-decoration.
This section develops the forcing argument rigorously,
identifies the \emph{minimal new structure} the rate
difference demands, and explains what this structure
fundamentally represents.

The central claim is simple: the rate difference requires knowing
\emph{where the transition state is in 3D space} --- a
geometric datum that does not exist anywhere in the tower below
$\Lk_5$.
This forces not just a configuration space, but a full Riemannian
geometric structure on that space, together with a real-valued
function (the potential energy surface) on it.

\begin{forcingbox}[Forcing pair for $\Lk_5$:
  same symmetry, different activation barrier]
\noindent\textbf{The two systems.}
\begin{align*}
  r_H &\colon\;
    \mathrm{CH_3Br + OH^-} \to \mathrm{CH_3OH + Br^-},
    \quad k_H \approx 4.0\times 10^{-3}\;
    \mathrm{M^{-1}s^{-1}},
  \\
  r_D &\colon\;
    \mathrm{CD_3Br + OH^-} \to \mathrm{CD_3OH + Br^-},
    \quad k_D \approx 3.1\times 10^{-3}\;
    \mathrm{M^{-1}s^{-1}}.
\end{align*}
Both at 298~K in water~\cite{Parker1969}; $k_D$ is not
independently measured here but derived from $k_H$ using
the \emph{secondary kinetic isotope effect}
$k_H/k_D \approx 1.3$ reported for $\alpha$-trideuterium
substitution in $\mathrm{S_N2}$
reactions~\cite{Streitwieser1958}.
The effect is ``secondary'' because the C--H/C--D bonds are
\emph{not broken} in the $\mathrm{S_N2}$ step: the C--Br bond
breaks and the new C--O bond forms, while the three H/D atoms
remain attached throughout.
The distinction between $r_H$ and $r_D$ therefore
\emph{cannot come from the bond-graph mechanism} --- it
cannot be seen at $\Lk_4$ or $\Lk_{4.5}$.

\medskip
\noindent\textbf{Structural indistinguishability at
  $\Lk_{4.5}$.}
At every level of the tower constructed so far, the two
systems share their entire categorical structure.  H and D
carry the same element label (atomic number~1), so
$\mathrm{CH_3Br}$ and $\mathrm{CD_3Br}$ are isomorphic as
objects in $\LGraphGstar$; both reactions use the same
$\Gstar$-equivariant DPO rule $p_{\mathrm{SN2}}$; neither
substrate has a stereocentre.  The decorator chain
$\FH, \FS, F_P$ from $\Lk_1$--$\Lk_3$ records small
isotope-dependent numerical differences in enthalpy, free
energy, and rate constant ---  the ratio
$k_H/k_D \approx 1.3$ enters $\Lk_3$ as an empirical
numerical attachment to its reaction generator ---  but
\emph{nothing in the structural data of any tower level
below $\Lk_5$ can predict, derive, or even motivate the
value $1.3$}.  The factor-of-$1.3$ slowdown that
accompanies every C--H$\to$C--D substitution at the
reactive carbon, in every solvent, at every temperature in
the harmonic-thermal window, points to a structural
feature the tower has not yet named.

The chemist's intuition supplies the answer immediately.
The deuterium atom is twice as heavy as the protium it
replaces, the C--D bond therefore vibrates more slowly
than C--H, and the lower zero-point energy of C--D shifts
the activation barrier upward by a small amount that
accumulates over three substitutions and twice across the
reactant-to-TS geometry change.  Every clause of this
explanation invokes structure that does not exist below
$\Lk_5$: 3D positions for the carbon and its substituents,
a curvature of the energy landscape that defines
vibrational modes, a difference of nuclear mass that
distinguishes H from D through the kinetic-energy
operator.  These three new objects --- positions,
curvature, mass --- are precisely the geometric structure
that $\Lk_5$ must supply.  The forcing argument that
follows formalises this gap: the $\Lk_{4.5}$ description
is complete on its own terms but blind to a chemistry that
every experimentalist routinely sees.

The following diagram makes this precise:
\[
  \includegraphics{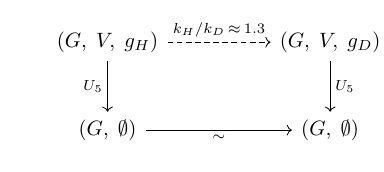}
\]
The dashed arrow indicates physical distinguishability by rate;
the bottom isomorphism shows that $U_5$ collapses the distinction.
At the clamped-nuclei BO leading order, the two objects
share the \emph{same} PES $V$ (the electronic ground-state
energy plus nuclear--nuclear repulsion depends on nuclear
charges, not masses); they differ only in the mass-weighted
metric $g_H \neq g_D$ (see \S\ref{sec:L5-ce},
Remark~\ref{rem:metric-descent}).  Diagonal Born--Oppenheimer
corrections (DBOC) introduce mass-dependent shifts of order
$(m_e/M)$ to $V$ itself, but these are an order of magnitude
smaller than the metric-driven KIE and may be neglected at
the resolution of this section.
Two objects distinct at $\Lk_5$ collapse to the same object
at $\Lk_{4.5}$.

\medskip
\noindent\textbf{The geometric origin of the rate difference.}
The secondary KIE arises from a change in zero-point energy
(ZPE) of the C--H and C--D bending modes as the reacting
carbon deforms from its tetrahedral reactant geometry to the
nearly pentacoordinate transition-state geometry.
The three C--H bending modes at the reactant
($\nu_H^\mathrm{react} \approx 1340\;\mathrm{cm}^{-1}$) shift
to softer frequencies at the TS
($\nu_H^\mathrm{TS} \approx 1000\;\mathrm{cm}^{-1}$).
Since $\nu_D = \nu_H/\sqrt{2}$ (mass doubled, force constant
unchanged), the H/D ZPE difference at the reactant is larger
than at the TS, and the net contribution to the activation
barrier differs by
$\Delta^2 E_\mathrm{ZPE}
= \Delta E_\mathrm{ZPE}^\mathrm{react}
- \Delta E_\mathrm{ZPE}^\mathrm{TS}$.
Applied to TST in this simplified three-mode form, this
gives $k_H/k_D = \exp(\Delta^2 E_\mathrm{ZPE}/RT)$ of
order $2$ at 298~K --- in the right direction and of the
correct order of magnitude, but overestimating the
observed factor $\sim 1.3$ by roughly $50\%$.  The
overestimate is a known artefact of the
three-modes-only approximation: only the modes that
genuinely couple to the reaction coordinate shift,
harmonic and RRHO approximations break down at the TS,
and conformer/anharmonic corrections enter.  The full
Bigeleisen--Mayer treatment recovering the experimental
value is given as Proposition~\ref{prop:kie} in
\S\ref{sec:L5-tst}.  For the forcing argument here it
suffices that $\Delta^2 E_\mathrm{ZPE}$ is non-zero and
$\Lk_5$-derivable from $(V, g)$, which the simple
estimate already demonstrates.
A full derivation --- including the correct mass-weighted
Hessian analysis, the Bigeleisen quantum corrections, and the
numerical evaluation --- is given as
Proposition~\ref{prop:kie} in \S\ref{sec:L5-tst}.

\noindent\textbf{Why this requires geometry.}
Computing $\Delta^2 E_\mathrm{ZPE}$ requires two geometric
inputs, both absent at $\Lk_{4.5}$:
\begin{itemize}
  \item the \emph{bending force constants} $\kappa =
    \partial^2 V/\partial q_\mathrm{bend}^2$ at both
    $\mathbf{R}_\mathrm{min}$ and $\mathbf{R}_\mathrm{TS}$
    --- entries of the mass-weighted Hessian
    $\nabla^2_g V$; and
  \item the \emph{saddle-point geometry}
    $\mathbf{R}_\mathrm{TS} \in \mathcal{C}_e(G)$ where
    the Hessian is evaluated.
\end{itemize}
Neither $\mathcal{C}_e(G)$ nor $V$ nor $\nabla^2_g V$ exists
at $\Lk_{4.5}$.
The KIE is irreducibly an $\Lk_5$ computation.

The forcing into $\Lk_5$ admits two equivalent expressions,
following the tower's standard pattern.

\smallskip
\noindent\emph{Automorphism-sequence form.}
The forgetful functor $U_5: \Lk_5(P) \to \Lk_{4.5}(P)$
induces a restriction map on automorphism (pointed) sets,
\[
  \varphi_5\colon \Aut\bigl(\Lk_5(P)\bigr)
  \;\longrightarrow\;
  \Aut\bigl(\Lk_{4.5}(P)\bigr),
\]
exhibiting the standard tower forcing sequence
\begin{equation}
  \label{eq:L5-exact}
  1 \;\to\; \ker\varphi_5 \;\to\; \Aut\bigl(\Lk_5(P)\bigr)
  \;\xrightarrow{\;\varphi_5\;}
  \Aut\bigl(\Lk_{4.5}(P)\bigr)
  \;\to\; \coker(\varphi_5) \;\to\; 1,
\end{equation}
where $\coker(\varphi_5) := \Aut(\Lk_{4.5}(P))/\mathrm{im}(
\varphi_5)$ is the pointed-set quotient (a group quotient
when $\mathrm{im}(\varphi_5)$ is normal, a coset space
otherwise; in either case the forcing content is
the same).  The H$\leftrightarrow$D swap is an element of
$\Aut(\Lk_{4.5}(P))$ ---  it preserves every level of
categorical data below $\Lk_5$ ---  but no element of
$\Aut(\Lk_5(P))$ maps to it under $\varphi_5$: the
mass-weighted metric $g_H \neq g_D$ is not preserved by
the swap.  The class $[r_H \leftrightarrow r_D]$ thus lies
in $\coker(\varphi_5)$, witnessing that the
$\Lk_5$-extension is forced.

\smallskip
\noindent\emph{Fibre-splitting form.}
Dually, the fibre of $U_5$ over the
$\Lk_{4.5}$-morphism $p_{\mathrm{SN2}}$ splits into
isotope-distinguished $\Lk_5$-morphisms whose
TS-vibrational signatures (and hence TST rate constants)
differ.  The H- and D-labelled morphisms in
$U_5^{-1}(p_{\mathrm{SN2}})$ are distinct $\Lk_5$-data
sharing a common $\Lk_{4.5}$-projection.

\smallskip
\noindent
The two forms are dual: a non-trivial $\coker(\varphi_5)$
records a $\Lk_{4.5}$-automorphism that no $\Lk_5$-datum
preserves, equivalently a non-trivial fibre of $U_5$ at
the corresponding object.  The forcing condition
$\coker(\varphi_5) \neq 1$ is what makes $\Lk_5$
irreducible from $\Lk_{4.5}$.
\end{forcingbox}

\noindent
\textbf{What minimal new structure is forced.}
The forcing argument demands three new objects, together
constituting the \emph{geometric decoration} $F_V$ of $\Lk_5(P)$.

\begin{enumerate}[label=(\alph*)]
  \item \textbf{The configuration orbifold}
    $\mathcal{C}_e(G) = \RR^{3n}/\bigl(SE(3) \times
    \Aut_\mu(G)\bigr)$:
    the space of all molecular geometries (3D nuclear positions)
    modulo global rigid-body motions and permutations of identical
    atoms.
    A point $\mathbf{R} \in \mathcal{C}_e(G)$ is a
    \emph{specific shape} of the molecule.

  \item \textbf{The potential energy surface (PES)}
    $V\colon \mathcal{C}_e(G) \to \RR$:
    the Born--Oppenheimer ground-state electronic energy as a
    function of nuclear geometry.
    The landscape of $V$ encodes all stable conformers
    (local minima), transition states (index-1 saddle points),
    and reaction paths.

  \item \textbf{A mass-weighted Riemannian metric $g$ on
    $\mathcal{C}_e(G)$}: defined by the nuclear kinetic energy.
    In mass-weighted coordinates $q_i = \sqrt{m_i}\,R_i$,
    the kinetic energy takes the standard Euclidean form
    $T = \frac{1}{2}|\dot{q}|^2$, inducing the metric
    $g_{ij} = m_i\,\delta_{ij}$ on nuclear coordinate space.
    This Riemannian structure is not an arbitrary choice:
    it is the \emph{physically canonical} metric on
    $\mathcal{C}_e(G)$, inherited from the kinetic energy
    operator in the nuclear Hamiltonian.
    It gives geometric meaning to:
    \begin{itemize}
      \item \emph{Normal modes}: orthogonal eigenvectors of the
        mass-weighted Hessian $\nabla^2_g V$ at each critical
        point, with ZPE $= \frac{1}{2}\hbar\omega_k$ per mode.
      \item \emph{The intrinsic reaction coordinate (IRC)}:
        the steepest-descent path on $V$ in the metric $g$,
        connecting the saddle point to reactant and product
        minima~\cite{FukuiIRC1981,
        MillerHandyAdams1980}.
        The IRC is locally unique once $(V, g)$ and a
        chosen index-1 saddle are given (up to the two
        steepest-descent branches and reparametrisation);
        it requires no further choices beyond these.
      \item \emph{Classical force fields}: molecular mechanics
        (MM) approximates $V$ as a sum of analytic local
        potentials (bond stretching, angle bending, torsional
        terms, van der Waals, electrostatics) with empirically
        fitted parameters.
        In the Para tower, a force field is a morphism in
        $L_5^{\mathrm{Para}}$: a parametric approximation to
        $F_V$ with a finite-dimensional parameter space $\Theta$.
        Machine-learning force fields (NequIP, MACE, SO3LR, etc.)
        sit in $L_5^{\mathrm{Para}}$ as parametric morphisms
        with much larger and structurally different $\Theta$
        than classical force fields: equivariant
        neural-network architectures rather than analytic
        bond/angle/torsion sums, but still
        finite-dimensional approximations to $F_V$ in the
        same Para framework.
    \end{itemize}
\end{enumerate}

\begin{mathbox}[The fourth type of tower extension]
The tower has so far exhibited three types of extension
(Remark~\ref{rem:L45-type}):
decorator ($\Lk_0 \to \Lk_3$), structural ($\Lk_3 \to \Lk_4$),
and symmetry enrichment ($\Lk_4 \to \Lk_{4.5}$).

The extension $\Lk_{4.5} \to \Lk_5$ introduces a
\textbf{fourth type: geometric decoration}.
The new functor $F_V$ assigns to each species not a single real
number (as at $\Lk_1$--$\Lk_3$) but a \emph{smooth function
on a continuous space}: an infinite-dimensional datum.

This is a genuine structural break in the tower.
At $\Lk_1$--$\Lk_3$, specifying a reaction means giving one or
two real numbers per generator (e.g.\ $\Delta H$ and $k$).
At $\Lk_5$, specifying a reaction means giving the full shape of
$V$ along the reaction path --- in principle, a function on all
of $\mathcal{C}_e(G)$.
This data can only be obtained by:
\begin{itemize}
  \item \emph{First-principles computation}: solving the
    electronic Schrödinger equation at each geometry
    (quantum chemistry: HF, DFT, CCSD(T), etc.), exact but
    expensive;
  \item \emph{Experimental measurement}: spectroscopy,
    crystallography, and kinetics data constrain $V$ at
    specific geometries; or
  \item \emph{Machine-learning approximation}: a Para-level
    model $F_V^\Theta$ in $L_5^{\mathrm{Para}}$ trained on
    quantum chemical reference data.
\end{itemize}
All three approaches converge towards the exact
Born--Oppenheimer surface, with first-principles
computation exact in the complete-basis-set, full-CI
limit and approximate in any practical truncation,
while measurement and machine-learning approaches are
intrinsically approximate.
The categorical framework records \emph{what} datum is needed
($F_V$ as a functor into $\mathbf{Orb}^{\mathrm{Morse}}$)
without specifying how to obtain it.
\end{mathbox}

\noindent
\textbf{Why $\Lk_5$ is genuinely new: a narrative.}

The tower from $\Lk_0$ to $\Lk_{4.5}$ is entirely
\emph{shape-independent}.
Stoichiometry, thermodynamics, kinetics, bond-graph mechanisms,
and stereochemical outcomes are all properties of how atoms are
connected (and with what orientation) --- but not of where they
sit in 3D space.
An organic chemist can write every mechanism in this thesis using
only topology and signed graphs, without ever drawing a 3D
picture.

$\Lk_5$ is where 3D pictures become indispensable.
It is the level of the tower at which the following phenomena
first have a mathematical home:

\begin{itemize}
  \item \emph{Steric effects}: the activation barrier of a
    reaction increases when bulky substituents crowd the
    transition state --- a purely geometric statement about
    the saddle-point height on $V$.
  \item \emph{Ring strain}: three- and four-membered rings
    have higher energy than five- and six-membered ones
    because their bond angles deviate from the tetrahedral
    optimum --- readable from the local geometry of minima
    on $V$.
  \item \emph{Conformational analysis}: the preference for
    axial vs.\ equatorial substituents in cyclohexane, the
    barrier to rotation about a C--C bond, the gauche
    effect --- all are features of the PES landscape.
  \item \emph{Molecular dynamics}: the time evolution of
    nuclear positions under Newton's equations
    $m_i\ddot{R}_i = -\nabla_{R_i} V$ on the Riemannian
    manifold $(\mathcal{C}_e(G), g)$ first exists at $\Lk_5$.
  \item \emph{Transition-state theory}: the Eyring rate
    expression $k = (k_BT/h)\,e^{-E_a/k_BT}$
    \cite{Eyring1935} is computable from $(V, g)$ in the
    classical-TST approximation (no recrossing, no
    tunnelling, harmonic TS partition function), giving
    the $\Lk_3$ rate constant $F_P$ as a derived quantity
    rather than a primitive datum.  The exact
    $\Lk_3$--$\Lk_5$ coherence condition (\S\ref{sec:L5-tst})
    upgrades this to the full TST formula with prefactor
    and transmission corrections, all $\Lk_5$-computable.
\end{itemize}

\noindent
In categorical terms: the step $\Lk_{4.5} \to \Lk_5$ is
the first step in the tower where the new datum is not
determined by any finite assignment to generators.
$V$ is a \emph{smooth function on a manifold} --- infinitely
many independent values, constrained only by global symmetry 
(invariance under
$\mathrm{SE}(3) \times \Aut_\mu(G)$; $V$ is a scalar
function, so invariance is the appropriate notion) and
the requirement that it be bounded below.

\noindent
\textbf{The $\Lk_5$ functor and its coherence with the tower.}

The geometric decoration is organised as a functor
\[
  F_V\colon \Lk_{4.5}(P) \;\longrightarrow\;
  \mathbf{Orb}^{\mathrm{Morse}},
\]
where $\mathbf{Orb}^{\mathrm{Morse}}$ is the category whose
objects are Morse triples $(\mathcal{C}_e(G), V, g)$ ---
Riemannian orbifolds equipped with a Morse function ---
and whose morphisms are geometric reaction channels:
elementary morphisms are gradient-flow cobordisms
$(V_{\mathrm{full}}, \mathbf{R}_\mathrm{TS}, \gamma)$
between distinguished minima of $V$, with $\gamma$ the IRC
through one index-1 saddle of $V_{\mathrm{full}}$;
general morphisms are finite compositions thereof
(see \S\ref{sec:L5-def} for the precise construction).

The functor $F_V$ is coherent with the lower tower in the following sense:
\begin{equation}
  \label{diag:L5-tower}
  \includegraphics{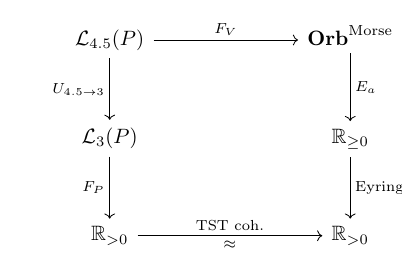}
\end{equation}
Reading the diagram: the geometric decoration $F_V$ assigns
to each $\Lk_{4.5}$-reaction its PES triple; the right
vertical column extracts the activation barrier $E_a$ and
applies the classical-TST rate expression
$(k_BT/h)\,e^{-E_a/RT}$ to obtain a rate constant.  The
left vertical column projects to $\Lk_3$ via the forgetful
chain and reads off the empirically assigned rate constant
$F_P(r)$.  The diagram commutes up to the
\emph{TST coherence condition}
(Definition~\ref{def:tst-coherence}, \S\ref{sec:L5-tst}):
the two rate constants agree up to harmonic, ideal-gas,
recrossing, and tunnelling corrections, all of which are
themselves computable from $(F_V, g)$ plus partition-function
data.

In tower language: the $\Lk_3$ rate constant is no longer
a free input but is constrained by $\Lk_5$ geometric data
through the TST coherence condition.  The rate constant
$F_P(r)$ remains a primitive datum of the $\Lk_3$
construction, but its \emph{value} is no longer arbitrary:
it must lie within the TST window of
$\kappa\,(k_BT/h)\,(Q^\ddagger/Q_{\rm reac})\,e^{-V^\ddagger/RT}$
computed from the PES triple.
This is the sense in which $\Lk_5$ ``derives'' kinetics:
\emph{the lower-level rate constant is determined by the
upper-level geometric data up to the standard TST
approximations}, not by equality on the nose.

Coherence with the other parts of the tower is analogous: the
thermodynamic data ($\FH$, $\FS^G$) of $\Lk_1$--$\Lk_2$ are
recovered from $F_V$ via the BO/RRHO decomposition
\[
  \FH(r) \;\approx\; \Delta V_\mathrm{elec}(r)
  + \Delta E_\mathrm{ZPE}(r)
  + \Delta E_\mathrm{therm}(T, r)
  + RT\,\Delta n_\mathrm{gas}(r),
\]
with the leading term $\Delta V_\mathrm{elec}(r) = V(\mathbf{R}_
{\mathrm{prod}}) - V(\mathbf{R}_{\mathrm{react}})$ supplied
by the PES critical-point values and the remaining terms by
the mass-weighted Hessian and partition-function data
(Proposition~\ref{prop:hess-pes-consistency},
\S\ref{sec:L5-tst}).
The ``$\approx$'' captures the harmonic, ideal-gas, and
BO approximations; under these, $F_V$ together with $g$
constrains every lower-level thermochemical functor
pointwise.
This confirms that $F_V$ is a genuine enrichment, not a
replacement, of the lower structure.

%% file: chapters/L5/l5_ce.tex
\subsection{The configuration orbifold
  \texorpdfstring{$\Ce(G)$}{Ce(G)}}
\label{sec:L5-ce}

\noindent
\textbf{Standing assumption: the Born--Oppenheimer framework.}
Throughout this section, all geometric and
energetic content is understood within the
\emph{Born--Oppenheimer (BO) approximation}
\cite{BornOppenheimer1927}: electronic and nuclear motion
decouple at leading order in the small parameter
$\varepsilon = (m_e/M)^{1/2}$, where $m_e$ is the electron mass
and $M$ a typical nuclear mass ($\varepsilon \approx 10^{-2}$
for light molecules).
At this order, electrons relax adiabatically to the ground
state for each fixed nuclear geometry $\mathbf{R}$, and nuclei
move on the resulting ground-state potential energy surface
$V(\mathbf{R})$.
This separation produces two well-defined objects --- the
configuration orbifold $\Ce(G)$ defined below, and the PES
$V: \Ce(G) \to \RR$ defined in \S\ref{sec:L5-pes} --- which
together (with the mass-weighted metric $g$) constitute
the Morse triple $(\Ce(G), V, g)$ that the categorical
functor $F_V: \Lk_{4.5}(P) \to \OrbMorse$ assigns to each
molecular graph at the $\Lk_5$ level.

The BO approximation is accurate for the vast majority of
ground-state thermal chemistry: away from degeneracies,
the error introduced by the adiabatic decoupling is of
order $\varepsilon = (m_e/M)^{1/2}$~\cite{Hagedorn1980},
and exponentially small in $1/\varepsilon$ whenever the
ground and first excited states are separated by a uniform
spectral gap~\cite{HagedornJoye2001}.
For typical molecules ($\varepsilon \approx 10^{-2}$) in
the regime of ordinary ground-state thermal chemistry, BO
corrections are usually smaller than the dominant
electronic-structure and solvation errors of practical
computational methods, so bond lengths, vibrational
frequencies, and reaction barriers computed on the
ground-state adiabatic surface are reliable to within
chemical accuracy~\cite{BornOppenheimer1927,HelgakerJorgensenOlsen2000}. 
BO corrections become non-negligible in high-resolution
spectroscopy, light-atom isotope shifts, precision
thermochemistry, and near non-adiabatic regions; the
leading mass-dependent corrections --- the diagonal
Born--Oppenheimer corrections (DBOC) --- lie outside the
scope of the leading $\Lk_5$ analysis and enter as
higher-order refinements.

Its limitations arise at \emph{conical intersections}
(CIs): geometries where the ground and first excited
states become degenerate so the spectral gap closes,
the adiabatic decoupling breaks down, and nuclear motion
couples the two surfaces~\cite{LonguetHiggins1963,
MeadTruhlar1979, Yarkony1996}.
That failure is precisely the phenomenon that forces
$\Lk_6$ (\S\ref{sec:L5-forcing-out}); by construction,
the entire $\Lk_5$ framework operates in the open dense
subset of $\Ce(G)$ where the spectral gap is positive.

\subsubsection{Construction}

Having accepted the BO separation, nuclear configurations
$\mathbf{R} = (\mathbf{r}_1, \ldots, \mathbf{r}_n) \in \RR^{3n}$
become the fundamental geometric objects.
Two configurations represent the \emph{same molecular shape} if
they differ only by an overall rigid-body motion (translation or
rotation of the entire molecule) or by a permutation of nuclei
that are physically identical.
The natural mathematical object encoding this identification
is not a manifold but an \emph{orbifold}: the quotient of
$\RR^{3n}$ by the combined group of these equivalences has
non-trivial stabilisers at symmetric configurations, and
fixed-point quotients produce orbifold singularities rather
than manifold charts.
The definition records this quotient precisely.

\begin{definition}[Configuration orbifold]
\label{def:ce}
  Let $G \in \LGraphP$ be a molecular graph with atom set
  $V = \{1, \ldots, n\}$ and atom-label function $\lambda_V$.
  The \emph{configuration orbifold} of $G$ is
  \[
    \Ce(G)
    \;:=\; \RR^{3n} \;\Big/\;
    \bigl(SE(3) \times \Aut_\mu(G)\bigr),
  \]
  where:
  \begin{itemize}
    \item $\RR^{3n}$ is the space of all nuclear position
      vectors $\mathbf{R} = (\mathbf{r}_1, \ldots, \mathbf{r}_n)
      \in \RR^3$.
    \item $SE(3) = \RR^3 \rtimes SO(3)$ acts by overall
      translation and rotation:
      $(\mathbf{t}, R) \cdot (\mathbf{r}_1, \ldots, \mathbf{r}_n)
      = (R\mathbf{r}_1 + \mathbf{t}, \ldots,
      R\mathbf{r}_n + \mathbf{t})$.
    \item $\Aut_\mu(G) \leq \mathrm{Sym}(V)$ is the
      \emph{mass-preserving graph automorphism group} of $G$:
      permutations of atoms preserving the atom-label
      function $\lambda_V$, the bond-graph adjacency, and
      the isotopic mass assignment $\mu: V \to \RR_{>0}$
      (so $m_{\pi(i)} = m_i$ for all $i$ and all
      $\pi \in \Aut_\mu(G)$; see Remark~\ref{rem:metric-descent}
      for why this restriction is necessary for the metric
      to descend).
      It acts by permuting nuclear coordinates:
      $\pi \cdot (\mathbf{r}_1, \ldots, \mathbf{r}_n)
      = (\mathbf{r}_{\pi^{-1}(1)}, \ldots,
      \mathbf{r}_{\pi^{-1}(n)})$.
  \end{itemize}
  The \emph{mass-weighted Riemannian metric} on $\RR^{3n}$
  is $g_{ij}^{\alpha\beta} = m_i\,\delta_{ij}\,\delta^{\alpha\beta}$
  (with $m_i$ the isotopic mass of nucleus $i$ and
  $\alpha, \beta \in \{x,y,z\}$);
  by construction of $\Aut_\mu(G)$, this metric is
  $\Aut_\mu(G)$-invariant and descends to a
  Riemannian structure on $\Ce(G)$ (orbifold-Riemannian
  at fixed-point strata; see Remark~\ref{rem:metric-descent}).
\end{definition}

\begin{remark}[Descent of the metric, and the isotope
  subtlety]
\label{rem:metric-descent}
A tensor field on $\RR^{3n}$ \emph{descends to the quotient}
$\Ce(G) = \RR^{3n}/H$ if it is invariant under the $H$-action,
so that it is well-defined on equivalence classes.
For the mass-weighted metric
$g = \sum_{i,\alpha} m_i\,dR_i^\alpha \otimes dR_i^\alpha$,
the two quotient factors behave differently.

\smallskip
\noindent
\emph{$SE(3)$-invariance.}
Translations leave $dR_i^\alpha$ unchanged.
Rotations $R \in SO(3)$ act as $\mathbf{r}_i \mapsto R
\mathbf{r}_i$, and since $R$ is orthogonal, $|R\,
\dot{\mathbf{r}}_i|^2 = |\dot{\mathbf{r}}_i|^2$.
The kinetic energy $T = \tfrac{1}{2}\sum_i m_i
|\dot{\mathbf{r}}_i|^2$ is therefore $SE(3)$ and $SO(3)$-invariant, and
$g$ descends through the $SE(3)$ quotient unconditionally.

\smallskip
\noindent
\emph{$\Aut(G)$-invariance and the isotope dependence.}
A permutation $\pi \in \Aut(G)$ maps atom $i$ to atom $\pi(i)$
of the same element type (same atomic number $Z$).
However, atomic number does \emph{not} determine mass: H and D
both have $Z = 1$ but differ by a factor of 2 in mass, and
similarly for other naturally-occurring isotopes.
$\Aut(G)$-invariance of $g$ therefore requires a stronger
condition: all atoms in each $\Aut(G)$-orbit must be assigned
the \emph{same isotopic mass}.
When this holds, $m_{\pi(i)} = m_i$ for all $\pi \in \Aut(G)$,
$T$ is $\Aut(G)$-invariant, and $g$ descends to a well-defined
Riemannian structure on $\Ce(G)$ (orbifold-Riemannian at
fixed-point strata).
When it fails---most importantly for isotopically heterogeneous
systems like the $\mathrm{CH_3Br}$ vs.\ $\mathrm{CD_3Br}$
forcing pair---the two isotopologues correspond to the
\emph{same} graph $G$ but distinct mass assignments, and each
yields a distinct metric $g_H$ or $g_D$ on the common
$\Ce(G)$.

\smallskip
\noindent
\emph{Tower consequence.}
This is precisely why the secondary KIE
(\S\ref{sec:L5-forcing-in}) cannot be detected at $\Lk_{4.5}$:
the distinction between isotopologues is carried by the metric
$g$, not by the graph $G$.
The $H \leftrightarrow D$ swap is an automorphism of
$\Lk_{4.5}(P)$ (since it preserves all graph-level data) but
not of $\Lk_5(P)$ (since it fails to preserve $g$), placing it
in $\coker(\varphi_5)$.
\end{remark}

\begin{remark}[$\Ce(G)$ is an orbifold, not a manifold]
\label{rem:orbifold}
  The orbit space $\Ce(G) = \RR^{3n}/\bigl(SE(3) \times \Aut_\mu(G)\bigr)$
  is generically a smooth manifold but acquires orbifold
  singularities at configurations with non-trivial
  stabiliser under the combined group action.
  Two sources contribute.

  \smallskip\noindent
  \emph{$SE(3)$-stabilisers from linear configurations.}
  On the open dense subset of $\RR^{3n}$ where the nuclei
  span three dimensions (the inertia tensor has full rank),
  the $SE(3)$-action is free: no non-trivial $(\mathbf{t},
  R)$ fixes such a configuration.
  The $SE(3)$-quotient of this subset is therefore a smooth
  manifold of dimension $3n-6$.
  For \emph{linear} configurations (all nuclei on a common
  axis), however, any rotation about that axis fixes the
  configuration pointwise, giving an $SO(2) \cong S^1$
  stabiliser.
  The quotient has orbifold singularities along the linear
  locus, and the local dimension drops by one: linear
  configurations span a $(3n-5)$-dimensional stratum of
  $\Ce(G)$.
  For diatomic molecules ($n=2$), all configurations are
  linear, and $\Ce(G)$ is itself $(3n-5) = 1$-dimensional
  (a single radial coordinate, the bond length).

  \smallskip\noindent
  \emph{$(SE(3) \times \Aut_\mu(G))$-stabilisers from
  symmetric configurations.}
  Fix a configuration $\mathbf{R} \in \RR^{3n}$ with
  non-trivial molecular symmetry: there exists a non-identity
  element $(R, \pi) \in SE(3) \times \Aut_\mu(G)$ such that
  $R\,\mathbf{r}_i = \mathbf{r}_{\pi(i)}$ for all atoms $i$,
  i.e., applying $R$ to the nuclear framework is the same
  as relabelling atoms via $\pi$.
  For water ($\mathrm{H_2O}$) at $C_{2v}$ geometry, for
  instance, the $180^{\circ}$ rotation $R_{C_2}$ about the bisector
  of the H--O--H angle maps the two hydrogens to each other,
  and composed with the transposition $\pi = (\mathrm{H}_1\,
  \mathrm{H}_2) \in \Aut(G)$ it fixes the configuration:
  $(R_{C_2}, \pi) \cdot \mathbf{R} = \mathbf{R}$ in
  $\RR^{3n}$.
  The stabiliser of $[\mathbf{R}]$ in $\Ce(G)$ is the subgroup
  $\{(\mathrm{id}, \mathrm{id}), (R_{C_2}, \pi)\} \cong
  \ZZ_2$, and a neighbourhood of $[\mathbf{R}]$ is locally
  modelled on $\RR^k / \ZZ_2$, giving a conical orbifold
  singularity.

  \smallskip\noindent
  More generally, fixed-point loci of $(SE(3) \times
  \Aut_\mu(G))$ form a nested family of \emph{orbifold strata},
  each of codimension at least one.
  Highly symmetric geometries (tetrahedral methane,
  octahedral complexes) sit at deep strata with large
  isotropy subgroups.
  This stratification has been studied in the context of
  rotation-vibration spectroscopy, where the strata
  contribute differently to the density of states and
  selection rules~\cite{Zhilinskii2006}.
\end{remark}

\begin{chembox}[What $\Ce(G)$ is, for the chemist]
An organic chemist working with a ball-and-stick model
manipulates exactly the object that $\Ce(G)$ formalises.
\begin{itemize}
  \item \textbf{A point} $[\mathbf{R}] \in \Ce(G)$ is a
    \emph{molecular shape}: a specific 3D arrangement of the
    nuclei, with overall position and orientation removed
    (rotating or translating the model does not give a new
    point), and with identical atoms treated as
    interchangeable (numbering the two hydrogens in water
    differently does not give a new point).
  \item \textbf{The dimension} $3n - 6$ (or $3n - 5$ for
    linear configurations, per Observation~\ref{obs:dim}) is
    the number of independent internal coordinates: bond
    lengths, bond angles, and dihedral angles.
  \item \textbf{The Riemannian metric $g$} weights each
    coordinate direction by the corresponding nuclear mass.
    This is why C--H vibrational modes have higher
    frequencies than C--D modes: the curvature of $V$ is
    the same, but the mass factor in $\omega =
    \sqrt{\kappa/m}$ shifts the frequency.
\end{itemize}
\end{chembox}

\begin{observation}[Dimension of $\Ce(G)$]
\label{obs:dim}
  For a molecule with $n$ atoms,
  $\dim(\RR^{3n}) = 3n$.
  Removing the 6-dimensional $SE(3)$ action gives a smooth
  stratum of dimension $3n - 6$ for non-linear
  configurations ($n \geq 3$ non-collinear), and
  $3n - 5$ for linear configurations (per
  Remark~\ref{rem:orbifold}, where the $SO(2)$ stabiliser
  reduces the effective $SE(3)$-quotient dimension by one).
  For the $\mathrm{S_N2}$ transition state ($n = 7$ atoms,
  non-linear): $\dim = 3(7) - 6 = 15$ internal coordinates.
\end{observation}

\begin{remark}[Tower coherence: why the dimension count matters]
\label{rem:dim-tower}
  The dimension of $\Ce(G)$ is the first continuous
  geometric datum in the tower.
  At all lower levels, the category $\Lk_k(P)$ deals with
  \emph{discrete} data: sets of species, real numbers
  ($\Delta H$, $\Delta S$, $k_r$), or graph morphisms.
  The object $\Ce(G)$ is a continuous orbifold of dimension
  $3n - 6$ (generically), and the functor $F_V$ assigns to
  it an infinite-dimensional datum (a smooth function on
  this space).

  From the categorical perspective: the forgetful functor
  $U_5: \Lk_5(P) \to \Lk_{4.5}(P)$ discards both the metric
  $g$ and the PES $V$, retaining only the molecular graph
  $G$ and the $G^*$-equivariant DPO rule.
  The dimension $3n - 6$ is therefore invisible at
  $\Lk_{4.5}$ and below: no tower level below $\Lk_5$
  encodes the continuous geometric structure on which
  normal modes, reaction paths, and activation barriers
  are defined.
\end{remark}

\subsubsection{Point groups as isotropy subgroups}

Molecular point groups are one of the most practically useful
concepts in chemistry.
Every undergraduate learns to classify a molecule by its
symmetry elements (rotation axes $C_n$, mirror planes
$\sigma$, improper rotations $S_n$, and inversion $i$),
assign it to a point group ($C_{2v}$, $T_d$, $D_{6h}$, etc.),
and then read off selection rules for spectroscopy and
orbital interactions from the corresponding character
table~\cite{WilsonDectusCross1955, BunkerJensen1998}.
This is almost always presented as a classification
procedure: examine the molecule, list its symmetry
operations, identify the group.

What is less often made explicit is that these symmetry
operations have a dynamical origin.
A symmetry operation of a molecule at geometry $\mathbf{R}$
is precisely a rigid-body motion (rotation or improper
rotation) that, combined with a permutation of identical
nuclei, maps $\mathbf{R}$ to itself.
This is the definition of the permutation-inversion group
introduced by Longuet-Higgins~\cite{LonguetHiggins1963Sym}
and developed into a comprehensive spectroscopic framework
by Bunker and Jensen~\cite{BunkerJensen1998}; it is the
same $G^*$ group introduced at $\Lk_{4.5}$
(Definition~\ref{def:Gstar}).
The categorical statement below identifies molecular point
groups as the \emph{isotropy subgroups} of the $G^*$-action
on $\Ce(G)$.
While the permutation-inversion framework itself is
standard, its expression as $G^*$-isotropy within a tower
extension---and the consequent derivation of point-group
classification from structure already present at
$\Lk_{4.5}$---appears to be original to this framework.

\begin{proposition}[Molecular point groups from spatial
  realisation of $\Aut_\mu(G)$ and $E^*$]
\label{prop:point-groups}
  Let $\mathbf{R} \in \Ce(G)$ be a molecular geometry,
  with $\mu$ the isotopic mass assignment.  Define the
  \emph{spatial realisation group} at $\mathbf{R}$:
  \[
    P_{\mathbf{R}}
    \;:=\; \bigl\{Q \in O(3)
    \;\big|\; \exists\,\pi \in \Aut_\mu(G),\; t \in \RR^3
    \text{ with } Q\mathbf{r}_i + t = \mathbf{r}_{\pi(i)}
    \;\forall i\bigr\}.
  \]
  The proper-rotation subgroup is
  $P_{\mathbf{R}}^+ = P_{\mathbf{R}} \cap SO(3)$.
  Then $P_{\mathbf{R}}$ is canonically isomorphic to the
  molecular point group of $G$ at $\mathbf{R}$.

  $P_{\mathbf{R}}$ is the image in $O(3)$ of the
  composition
  \[
    \Aut_\mu(G) \times \{E, E^*\}
    \;\longrightarrow\;
    \Iso_{G^*}(\mathbf{R})
    \;\hookrightarrow\;
    \Gstar(G),
  \]
  where the spatial-inversion factor comes from the
  global parity $E^* \in \Gstar(G)$ and the rotation
  factor comes from graph automorphisms realised as
  rigid-body motions; the per-stereocentre sign-flip
  factor $\ZZ_2^k \leq \Gstar(G)$ does not contribute to
  the point group at a fixed geometry, since it acts on
  chirality labels rather than on $\mathbf{R}$.
\end{proposition}

\begin{proof}
\textbf{$P_{\mathbf{R}}$ contains every point-group element.}
Let $Q \in O(3)$ be a symmetry operation of the molecule at
geometry $\mathbf{R}$ in the traditional chemist's sense:
applied as a rigid-body transformation to the nuclear
framework, $Q$ produces a configuration whose nuclear
positions coincide with those of $\mathbf{R}$ up to a
relabelling.  Then there exist a translation $t \in \RR^3$
and a permutation $\pi$ of atoms preserving element labels
such that $Q\mathbf{r}_i + t = \mathbf{r}_{\pi(i)}$ for
all $i$.  Because $Q$ is a symmetry of the physical
nuclear framework, $\pi$ also preserves the isotopic mass
assignment $\mu$ (otherwise $Q$ would map an H-position to
a D-position, contradicting that $Q$ is a symmetry of
the actual physical molecule).  Hence $\pi \in \Aut_\mu(G)$
and $(Q, t, \pi)$ witnesses $Q \in P_{\mathbf{R}}$.

\textbf{Every element of $P_{\mathbf{R}}$ is a point-group
element.}
Conversely, suppose $Q \in O(3)$ satisfies
$Q\mathbf{r}_i + t = \mathbf{r}_{\pi(i)}$ for some
$\pi \in \Aut_\mu(G)$ and $t \in \RR^3$.  Then $Q$ is
either a proper rotation (if $Q \in SO(3)$) or an improper
rotation (if $\det Q = -1$, decomposable as inversion
$\times$ rotation).  In either case, $Q$ permutes the
nuclei among physically indistinguishable positions
($\Aut_\mu(G)$-equivalent atoms with the same element and
mass), which is the operational definition of a molecular
point-group symmetry at $\mathbf{R}$.

\textbf{Connection to $\Iso_{G^*}(\mathbf{R})$.}
The composition $\Aut_\mu(G) \times \{E, E^*\} \to
\Iso_{G^*}(\mathbf{R}) \hookrightarrow \Gstar(G)$ sends
$(\pi, \epsilon) \mapsto (\pi, \epsilon, \mathbf{0}_k)$
where $\mathbf{0}_k \in \ZZ_2^k$ is the trivial chirality
flip.  Its image in $\Iso_{G^*}(\mathbf{R})$ realises
spatially in $O(3)$ as $P_{\mathbf{R}}$: the proper-rotation
part from $\Aut_\mu(G)$ acting by rigid-body rotations,
the inversion part from $E^* \in \Gstar(G)$, and composite
elements (reflections, improper rotations) from products
of the two.  The per-stereocentre sign-flip $\ZZ_2^k
\leq \Gstar(G)$ does not appear in $P_{\mathbf{R}}$ since
it acts on chirality labels rather than on $\mathbf{R}$
itself.
\end{proof}

\begin{chembox}[Point groups as theorems, not postulates]
Traditional presentations list the point group of a
molecule as a given, determined by inspection of a
3D model.  Proposition~\ref{prop:point-groups} identifies
this operational construction with the spatial realisation
of structure already present at $\Lk_{4.5}$: the
mass-preserving graph automorphism group $\Aut_\mu(G)$
and the global parity $E^* \in \Gstar(G)$.  The point
group is therefore not an independent axiom of molecular
symmetry but a derived consequence of the algebraic data
of the lower tower level, realised at a chosen $\Lk_5$-geometry.

\medskip
\begin{center}
\renewcommand{\arraystretch}{1.3}
\begin{tabular}{>{\raggedright\arraybackslash}p{2.8cm}
                >{\raggedright\arraybackslash}p{4.8cm}
                >{\raggedright\arraybackslash}p{5.5cm}}
  \hline
  \textbf{Molecule} &
  \textbf{Geometry $\mathbf{R}_{\mathrm{eq}}$} &
  $\Iso_{G^*}(\mathbf{R}_{\mathrm{eq}})$ \\
  \hline
  $\mathrm{NH_3}$ &
  $C_{3v}$ equilibrium &
  $C_{3v}$ (3-fold axis + 3 mirror planes) \\
  
  $\mathrm{CH_4}$ &
  $T_d$ equilibrium &
  $T_d$ (tetrahedral symmetry) \\
  
  $\mathrm{C_6H_6}$ &
  $D_{6h}$ equilibrium &
  $D_{6h}$ (hexagonal symmetry) \\
  
  Chiral molecule &
  $C_1$ equilibrium (no symmetry) &
  $C_1 = \{e\}$ \\
  \hline
\end{tabular}
\end{center}

\medskip\noindent
Geometry changes isotropy: the planar ($D_{3h}$) transition
state of ammonia inversion has
$\Iso_{G^*}(\mathbf{R}_{\mathrm{TS}}) \cong D_{3h} \supset
C_{3v}$.
This enlargement of the isotropy subgroup at saddle points
is a general feature of symmetric transition states and
underlies the Woodward--Hoffmann rules (discussed at
$\Lk_{4.5}$).
\end{chembox}

\begin{remark}[Point groups in the categorical tower]
\label{rem:point-groups-tower}
Proposition~\ref{prop:point-groups} places point groups
precisely in the tower hierarchy.
\begin{enumerate}[label=(\alph*)]
  \item \textbf{Point groups are an $\Lk_5$ datum.}
    The isotropy subgroup $\Iso_{G^*}(\mathbf{R})$ depends
    on the specific geometry $\mathbf{R} \in \Ce(G)$, which
    only exists at $\Lk_5$.
    At $\Lk_{4.5}$, the $G^*$ group is present but there is
    no configuration space on which it acts geometrically.
    The same abstract group $G^*$ can produce different
    point groups at different geometries of the same
    molecule (e.g., $C_{3v}$ vs.\ $D_{3h}$ for ammonia), a
    distinction invisible without $\Ce(G)$.
  \item \textbf{The automorphism exact sequence.}
    Under the restriction map $\varphi_5: \Aut(\Lk_5(P))
    \to \Aut(\Lk_{4.5}(P))$, the geometry-dependent point
    group $P_{\mathbf{R}}$ is an $\Lk_5$ invariant
    invisible at $\Lk_{4.5}$.  Geometry-fixing symmetries
    of $\Lk_5$-objects therefore contribute to
    $\ker \varphi_5$ (the kernel of the
    \emph{tower-level} restriction map, not to be confused
    with $\ker \rho_{\mathbf{R}}$ of
    Proposition~\ref{prop:point-groups}, which is the
    kernel of the \emph{object-level} spatial realisation
    map and detects inversion symmetry).  The forgetful
    functor $U_5$ washes out point-group symmetries by
    losing the underlying geometry.
  \item \textbf{Physical consequences.}
    The irreducible representations of
    $\Iso_{G^*}(\mathbf{R}_{\mathrm{eq}})$ classify
    vibrational normal modes, IR/Raman activity, and
    molecular orbital symmetry labels---all $\Lk_5$ data
    depending on $V$ and $g$ near the equilibrium geometry.
\end{enumerate}
\end{remark}

\subsubsection{Realisation of the chirality label}

At $\Lk_{4.5}$, each object of $\Lk_{4.5}(P)$ carries an
abstract chirality label $\sigma \in \{+1, -1\}^k$
distinguishing the $2^k$ stereoisomers of a molecule with
$k$ stereocentres.
That label was introduced axiomatically: the
permutation-inversion group $G^*$ contains the parity $E^*$
as an abstract operation, and $\sigma$ records on which
side of $E^*$ a given stereoisomer sits at each
stereocentre.
The question deferred at $\Lk_{4.5}$ is: \emph{what does
this discrete label correspond to geometrically?}

The answer is supplied by the topology of the potential
energy surface.
The idea has precursors in the literature on molecular
topology: Woolley~\cite{Woolley1978} argued that molecular
structure itself is not a purely quantum-mechanical concept
but requires appeal to the PES, and
Amann~\cite{Amann1991} formalised chirality as a
superselection rule arising from the topology of the
molecular state space.
The following proposition makes the relevant topological
invariant explicit as the connected-component structure of
the accessible region of $\Ce(G)$.
The single-stereocentre case is stated below for clarity,
with the general case deferred to a remark.

\begin{proposition}[Chirality label as connected component
  of the accessible PES]
\label{prop:chirality-realisation}
  Let $G \in \LGraphP$ be a molecular graph with exactly
  one stereocentre, and let $V: \Ce(G) \to \RR$ be its
  ground-state BO potential energy surface.
  Define the \emph{accessible configuration orbifold}:
  \[
    \Ce^{\mathrm{acc}}(G)
    \;:=\;
    \bigl\{\mathbf{R} \in \Ce(G)
    \;\big|\; V(\mathbf{R}) \;<\; E_{\mathrm{inv}}\bigr\},
  \]
  where $E_{\mathrm{inv}}$ is the energy of the lowest
  geometry at which the stereocentre is planar (the
  \emph{pyramidal-inversion barrier}).
  Then $\Ce^{\mathrm{acc}}(G)$ has exactly two
  path-connected components related by the parity
  operation $E^*$:
  \[
    \pi_0\bigl(\Ce^{\mathrm{acc}}(G)\bigr)
    \;\cong\; \{+1,\,-1\}.
  \]
  The abstract chirality label $\sigma \in \{+1, -1\}$
  assigned at $\Lk_{4.5}$ is the topological invariant
  selecting a connected component of
  $\Ce^{\mathrm{acc}}(G)$: $\sigma = +1$ corresponds to
  one enantiomeric component and $\sigma = -1$ to the
  other.
\end{proposition}

\begin{proof}
The chirality-sign function $\omega: {\Ce}_G^{\mathrm{conn}}
\to \{-1, 0, +1\}$ is continuous on its domain and takes
the value $0$ exactly when the four substituent vectors
$\mathbf{r}_1 - \mathbf{r}_4, \mathbf{r}_2 - \mathbf{r}_4,
\mathbf{r}_3 - \mathbf{r}_4$ are linearly dependent ---
i.e., when the stereocentre is planar or degenerate.
Any continuous path in ${\Ce}_G^{\mathrm{conn}}$ connecting
an $\omega = +1$ configuration to an $\omega = -1$
configuration must pass through $\omega = 0$ at some
intermediate parameter.

By definition of $E_{\mathrm{inv}}^G$, the minimum value
of $V$ on the $\omega = 0$ locus within
${\Ce}_G^{\mathrm{conn}}$ is $E_{\mathrm{inv}}^G$.  A path
lying entirely in $\Ce^{\mathrm{acc}}(G) = \{V 
E_{\mathrm{inv}}^G\} \cap {\Ce}_G^{\mathrm{conn}}$ therefore
cannot cross $\omega = 0$ and cannot connect the $\omega
= +1$ basin to the $\omega = -1$ basin.

Within each basin (fixed sign of $\omega$),
path-connectedness is a consequence of the simple
connectivity of the configuration space of a fixed
nuclear framework around a non-degenerate equilibrium
geometry.  The chirality label $\sigma \in \{+1, -1\}$
at $\Lk_{4.5}$ assigns one of the two basins.
\end{proof}

\begin{remark}[Generalisation to $k$ stereocentres]
\label{rem:chirality-multiple}
For a molecule with $k \geq 1$ stereocentres, each
stereocentre contributes an independent $\ZZ_2$ sign
factor to the chirality data, and under the assumption
that the inversion barriers at the $k$ stereocentres are
mutually independent (each sits above the accessible
energy $E_{\mathrm{inv}}^G$ taken as the minimum over all
stereocentres), the pre-quotient component count is
$2^k$:
\[
  \pi_0\bigl(\widetilde{\Ce}^{\mathrm{acc}}(G)\bigr)
  \;\cong\; \ZZ_2^k
  \qquad
  \text{on $\RR^{3n}\setminus\{V \geq E_{\mathrm{inv}}^G\}$
  before quotienting by $\Aut_\mu(G)$.}
\]
However, $\Aut_\mu(G)$ may identify components in
$\Ce(G)$ proper.  The corrected count is given in
Remark~\ref{rem:meso-components} below.
The $\Lk_{4.5}$ chirality label $\sigma \in \ZZ_2^k /
\Aut_\mu(G)$ selects an $\Aut_\mu(G)$-orbit of sign
patterns, recovering the enantiomer count of
Definition~\ref{def:enantiomers}.
\end{remark}

\begin{remark}[Meso compounds and the $\Aut_\mu(G)$
  quotient]
\label{rem:meso-components}
The naive component count $2^k$ in
Remark~\ref{rem:chirality-multiple} is correct on
$\RR^{3n} \setminus \{V \geq E_{\mathrm{inv}}\}$ before
the $\Aut_\mu(G)$-quotient.  In $\Ce(G)$ proper,
$\Aut_\mu(G)$ may identify components.  Specifically,
if $\pi \in \Aut_\mu(G)$ maps a configuration of sign
pattern $\sigma \in \ZZ_2^k$ to one of sign pattern
$-\sigma$ (i.e., $\sigma \circ \pi^{-1} = -\sigma$),
the two components labelled $\sigma$ and $-\sigma$
become a single component in $\Ce(G)$.  This is the
geometric mechanism of meso compounds:
meso-tartaric acid has $k = 2$ stereocentres with
labels $(+1, -1)$ and $(-1, +1)$, related by the
graph automorphism exchanging the two carbons;
$\Ce(G)$ therefore has 3 low-energy components (the
two chiral $(\pm 1, \pm 1)$ basins and the single
meso basin), not 4.  The general statement: the
component set of $\Ce^{\mathrm{acc}}(G)$ is in bijection
with the $\Aut_\mu(G)$-orbits of $\ZZ_2^k$, matching
the L4.5-level enantiomer count of
Definition~\ref{def:enantiomers}.
\end{remark}

\begin{remark}[Pyramidal-inversion energetics]
\label{rem:racemisation}
For a tetrahedral carbon stereocentre bonded to four
distinct substituents, $E_{\mathrm{inv}}$ is very large:
direct pyramidal inversion requires an essentially
\emph{planar tetracoordinate} carbon---four substituents
coplanar with the central atom, retaining all four
bonds---which is geometrically strained and costs
hundreds of kJ/mol.
In practice, stereocentre interconversion of such centres
occurs only via bond-breaking pathways (e.g.,
$\mathrm{S_N1}$ through a planar carbocation, or
base-catalysed enolisation), which are not single-surface
processes on the starting substrate's $V$.
Under normal chemical conditions ($T \approx 300$~K,
accessible thermal energies $\lesssim k_BT \approx
2.5$~kJ/mol above the ground state), the accessible
region $\Ce^{\mathrm{acc}}(G)$ consists of two entirely
disconnected components, making $\sigma$ a robust
topological invariant of the molecular state.

The contrast with nitrogen stereocentres
(e.g., $\mathrm{NH_3}$, amines) is instructive: nitrogen
inversion through a planar tricoordinate TS has a barrier
of only $\sim 24$~kJ/mol, well within thermal access, so
nitrogen stereocentres are typically not resolvable as
distinct enantiomers under standard conditions.
This difference in barrier heights is itself an $\Lk_5$
datum: the same abstract group-theoretic structure at
$\Lk_{4.5}$ yields different topological consequences
depending on the PES landscape, which only $\Lk_5$ sees.
\end{remark}

\begin{insightbox}[The chirality bridge between $\Lk_{4.5}$
  and $\Lk_5$]
Proposition~\ref{prop:chirality-realisation} closes the
explanatory gap opened at $\Lk_{4.5}$: the abstract label
$\sigma \in \{+1,-1\}$ was promised a geometric
realisation, and here it is.

The key conceptual point is that $\Ce(G)$ itself is
topologically trivial (path-connected): the two
stereoisomers are not distinguished by any
\emph{topological} property of configuration space alone.
What distinguishes them is the \emph{energetic} structure
of $V$---specifically, the existence of a large inversion
barrier that partitions the accessible region into two
components.
Chirality is therefore not a topological invariant of the
molecular graph (an $\Lk_4$ or $\Lk_{4.5}$ statement) but
a topological invariant of the \emph{accessible PES} (an
$\Lk_5$ statement).
The $\Lk_{4.5}$ label $\sigma$ is the shadow of this PES
topology projected back onto the graph level.
\end{insightbox}

%% file: chapters/L5/l5_pes.tex
\subsection{The potential energy surface (PES), Hilbert bundle,
  and Born--Oppenheimer section}
\label{sec:L5-pes}

The configuration orbifold $\Ce(G)$ constructed in
\S\ref{sec:L5-ce} is the geometric \emph{stage} on which the
$\Lk_5$ functor $F_V$ operates.
This section adds four layers of structure to that stage,
each derived from the previous:
\begin{enumerate}[label=(\roman*)]
  \item The \emph{electronic Hilbert bundle}
    $\pi: \Hel \to \Ce(G)$ (\S\ref{sec:L5-hilbert}):
    packages the family of electronic Hamiltonians
    $\{\hat{H}_{\mathrm{el}}(\mathbf{R})\}_{\mathbf{R}
    \in \Ce(G)}$ into a single geometric object.
    The bundle is trivially trivial as a Hilbert bundle
    (all fibers are isomorphic), but the Hamiltonian
    family $\hat{H}_{\mathrm{el}}: \Ce(G) \to
    \mathrm{SA}(\Hel)$ is non-trivial and carries all the
    electronic structure information.
  \item The \emph{BO ground-state section}
    $\sigma_0: \Ce(G) \to \Hel$ and the \emph{PES}
    $V = E_0 + V_{\mathrm{nn}}: \Ce(G) \to \RR$
    (\S\ref{sec:L5-bosection}): the Layer~1 and Layer~2
    data that define the $\Lk_5$ functor $F_V$.
    The PES is the infinite-dimensional datum that
    replaces the finitely many reals ($\Delta H$, $k_r$)
    of lower levels.
  \item The \emph{Morse structure} of $V$ (\S\ref{sec:L5-morse}):
    minima, index-1 saddle points, and the intrinsic reaction
    coordinate (IRC) --- the chemically observable
    consequences of the PES landscape, derived from the
    metric $g$ and function $V$ already introduced.
  \item The \emph{Berry connection} $A_0$ on the ground-state
    line bundle $L_0 \leq \Hel$ (\S\ref{sec:L5-berry}):
    the geometric phase structure of $\sigma_0$.
    At $\Lk_5$, on the simply-connected CI-free open subsets
    where the framework operates, $A_0$ can be gauged to
    zero (Proposition~\ref{prop:berry-trivial}); its
    obstruction to global vanishing around loops encircling
    conical intersections is the datum that forces $\Lk_6$.
\end{enumerate}

\noindent
The categorical architecture is a chain of forgetful functors:
\[
  \Lk_5(P)
  \;\xrightarrow{U_5}\;
  \Lk_{4.5}(P)
  \;\xrightarrow{U_{4.5}}\;
  \cdots
\]
where $U_5$ drops the geometric decoration $(\Ce(G), V, g)$
together with its derived Hilbert-bundle structure
$(\sigma_0, A_0)$, retaining only the underlying graph and
chirality data $(G, \sigma)$.
Items (iii) and (iv) are derived from (i)--(ii) and the
metric $g$ from \S\ref{sec:L5-ce}; they are not independent
data but consequences of the $F_V$ functor.
The progression (i)$\to$(iv) forms a filtration of the $\Lk_5$
structure from coarsest (Hilbert bundle) to finest (Berry
connection), matching the physical progression from electronic
structure to geometric phase.

\subsubsection{The electronic Hilbert bundle}
\label{sec:L5-hilbert}

At each level $\Lk_k(P)$ for $k \leq 4.5$, the objects of the
tower are molecular graphs $(G, \sigma)$ with various discrete
decorations ($\Delta H$, $k_r$, DPO mechanisms, chirality labels),
but no information about where atoms sit in 3D space.
The BO approximation (\S\ref{sec:L5-ce}, standing assumption)
assigns to each nuclear geometry $\mathbf{R} \in \Ce(G)$ a
well-posed quantum-mechanical eigenvalue problem:
find the ground state of the electronic Hamiltonian
$\hat{H}_{\mathrm{el}}(\mathbf{R})$.
The \emph{electronic Hilbert bundle} is the geometric object that
packages this family of eigenvalue problems into a single structure
over $\Ce(G)$.

This construction has a multi-layered history in the molecular
and mathematical-physics literature.
Mead and Truhlar~\cite{MeadTruhlar1979} introduced the
molecular vector potential in the BO approximation, recognising
that the $\mathbf{R}$-dependence of the electronic wavefunction
produces an effective gauge field entering the nuclear Hamiltonian.
The fiber-bundle interpretation of this geometric phase ---
with the Berry connection identified as a $U(1)$ connection on
a line bundle over parameter space --- was developed by
Simon~\cite{Simon1983} following Berry~\cite{Berry1984}.
The rigorous adiabatic-theoretic foundation, justifying the
decomposition of $\Hel$ into adiabatic sub-bundles and
controlling the errors, was established by Teufel and
Panati--Spohn--Teufel~\cite{Teufel2003, PanatiSpohnTeufel2003}.
All of this machinery is what $\Lk_5$ inherits.

\begin{definition}[Electronic Hilbert bundle]
\label{def:hilbert-bundle}
  Fix a molecular graph $G \in \LGraphP$ with $N_e$ electrons
  and nuclear geometry $\mathbf{R} \in \Ce(G)$.
  The \emph{electronic Hilbert space} at $\mathbf{R}$ is
  \[
    \Hel(\mathbf{R})
    \;:=\;
    L^2_{\mathrm{antisym}}\!\bigl(\RR^{3N_e},\, \CC\bigr),
  \]
  the space of antisymmetric (fermionic) square-integrable
  wavefunctions for the $N_e$ electrons of $G$ in the
  external Coulomb field of nuclei fixed at $\mathbf{R}$.
  The \emph{electronic Hamiltonian} at $\mathbf{R}$ is
  \[
    \hat{H}_{\mathrm{el}}(\mathbf{R})
    \;:=\;
    -\sum_{i=1}^{N_e}\frac{\hbar^2}{2m_e}\nabla_i^2
    + V_{ee} + V_{en}(\mathbf{R}),
  \]
  where $V_{ee}$ is the electron--electron repulsion and
  $V_{en}(\mathbf{R}) = -\sum_{i,k} Z_k e^2 /
  |\mathbf{r}_i - \mathbf{r}_k|$ is the electron--nucleus
  attraction (with nuclei fixed at $\mathbf{R}$).

  The \emph{electronic Hilbert bundle} is
  \[
    \pi: \Hel \;\longrightarrow\; \Ce(G),
    \qquad
    \pi^{-1}(\mathbf{R}) = \Hel(\mathbf{R}).
  \]
  Since all fibers are isomorphic to the same separable Hilbert
  space $L^2_{\mathrm{antisym}}(\RR^{3N_e})$, the bundle is
  trivially trivial as a Hilbert bundle.
  Its non-trivial content lies in the \emph{Hamiltonian family}
  $\mathbf{R} \mapsto \hat{H}_{\mathrm{el}}(\mathbf{R})$:
  a smooth family of self-adjoint operators on the common
  domain determined by the Kato-bounded Coulomb singularities
  (smooth in the sense that all matrix elements $\langle \psi
  | \hat{H}_{\mathrm{el}}(\mathbf{R}) | \phi \rangle$ for
  $\psi, \phi$ in the domain depend smoothly on
  $\mathbf{R}$)~\cite{Kato1966, Teufel2003}.
\end{definition}

\begin{remark}[Tower interpretation]
\label{rem:hilbert-tower}
  The molecular graph $G \in \Lk_{4.5}(P)$ determines the
  electron count $N_e$ and the nuclear charges $\{Z_k\}$,
  hence the functional form of
  $\hat{H}_{\mathrm{el}}(\mathbf{R})$ by Coulomb's law.
  What $G$ alone does not determine is the configuration
  $\mathbf{R} \in \Ce(G)$.
  The Hilbert bundle is the object that \emph{jointly}
  depends on both $G$ (for the Hamiltonian structure) and
  $\Ce(G)$ (for the base space of geometries): it is the
  first bundle in the tower fibered over a continuous base.
\end{remark}

\subsubsection{The Born--Oppenheimer section and PES}
\label{sec:L5-bosection}

The Hilbert bundle provides the \emph{arena}; the BO section
selects the \emph{physically relevant state} in each fiber.
At each geometry $\mathbf{R}$, the electronic Hamiltonian
$\hat{H}_{\mathrm{el}}(\mathbf{R})$ has an isolated ground-state
eigenvalue $E_0(\mathbf{R})$ at the bottom of its spectrum
(existence of bound states for neutral molecules is a
consequence of Zhislin's theorem~\cite{Zhislin1960} and is
standard for chemically relevant geometries), with
$\sigma_0(\mathbf{R})$ the corresponding unique (up to phase)
normalised eigenstate.
Varying $\mathbf{R}$ smoothly while tracking this ground state
defines the BO section.

\begin{definition}[BO section and potential energy surface]
\label{def:bo-section}
  The \emph{Born--Oppenheimer (BO) ground-state section} is
  the map
  \[
    \sigma_0: \Ce(G) \;\longrightarrow\; \Hel,
    \qquad
    \sigma_0(\mathbf{R}) \in \Hel(\mathbf{R}),
  \]
  assigning to each geometry $\mathbf{R}$ the normalised
  ground-state electronic wavefunction:
  \[
    \hat{H}_{\mathrm{el}}(\mathbf{R})\,\sigma_0(\mathbf{R})
    \;=\; E_0(\mathbf{R})\,\sigma_0(\mathbf{R}),
    \qquad
    \|\sigma_0(\mathbf{R})\| = 1,
  \]
  where $E_0(\mathbf{R})$ is the lowest eigenvalue.

  The \emph{Born--Oppenheimer potential energy surface (PES)} is
\[
  V: \Ce(G) \;\longrightarrow\; \RR,
  \qquad
  V(\mathbf{R}) \;=\; E_0(\mathbf{R})
    + V_{\mathrm{nn}}(\mathbf{R}),
\]
where $E_0(\mathbf{R}) = \langle \sigma_0(\mathbf{R}) |
\hat{H}_{\mathrm{el}}(\mathbf{R}) | \sigma_0(\mathbf{R})
\rangle$ is the ground-state eigenvalue of the electronic
Hamiltonian (matching the notation in the BO section
above), and
\[
  V_{\mathrm{nn}}(\mathbf{R})
  \;=\; \sum_{A < B} \frac{Z_A Z_B\, e^2}
  {|\mathbf{R}_A - \mathbf{R}_B|}
\]
is the nuclear--nuclear Coulomb repulsion.  Both terms
are needed: $V_{\mathrm{nn}}$ supplies the short-range
nuclear repulsion that prevents atom coalescence, and
without it $V$ would have wrong asymptotics.
\end{definition}

\begin{remark}[Physical meaning of $\sigma_0$ and $V$]
\label{rem:bo-physics}
Computationally, $V(\mathbf{R})$ is the quantity returned
by a single quantum-chemistry calculation at geometry
$\mathbf{R}$: a Hartree--Fock, DFT, or coupled-cluster
calculation solves for the ground-state electronic
wavefunction $\sigma_0(\mathbf{R})$ with nuclei clamped at
$\mathbf{R}$, and returns the total energy $E_0(\mathbf{R})
+ V_{\mathrm{nn}}(\mathbf{R})$, with the nuclear--nuclear
repulsion typically added automatically by the
code~\cite{HelgakerJorgensenOlsen2000}.
Mapping out $V$ over a grid of geometries is what quantum
chemists mean by ``computing the potential energy surface.''

In the tower language: the section $\sigma_0$ is the
first object in the tower that is a \emph{section of a bundle
over a continuous space}, rather than a map between discrete
categories.
The PES functor $F_V: \Lk_{4.5}(P) \to \OrbMorse$ assigns the
triple $(\Ce(G), V, g)$ to each molecular graph $G$, and the
entire \S\ref{sec:L5-morse}--\S\ref{sec:L5-berry} derives
from $(\sigma_0, V, g)$.
\end{remark}

\begin{mathbox}[Layer 1 and Layer 2 for $F_V$ at $\Lk_5$]
\label{box:FV-layers}
\textbf{Layer 1}: any smooth function $V: \Ce(G) \to \RR$
on the configuration orbifold.
There is no ``one real per generator'' universal property:
$V$ requires specifying its value at every point of the
$(3n-6)$-dimensional orbifold $\Ce(G)$, an
infinite-dimensional datum.
Layer~1 asserts that such a function \emph{must} exist and
be smooth on the non-degenerate (CI-free) subset of $\Ce(G)$.

\textbf{Layer 2}: two additional conditions tying $V$ to
the physics of $\hat{H}_{\mathrm{el}}$.
\begin{enumerate}[label=(\alph*)]
  \item \textbf{BO derivation}:
    $V(\mathbf{R}) = \langle \sigma_0(\mathbf{R}) |
    \hat{H}_{\mathrm{el}}(\mathbf{R}) |
    \sigma_0(\mathbf{R}) \rangle + V_{\mathrm{nn}}
    (\mathbf{R})$.
    The PES is not an arbitrary smooth function but the
    sum of the ground-state electronic eigenvalue and the
    nuclear--nuclear Coulomb repulsion.
  \item \textbf{Global existence of $\sigma_0$}:
    the ground state is non-degenerate on the relevant
    open subset of $\Ce(G)$, so $\sigma_0$ is a smooth
    section of $\Hel \to \Ce(G)$ over that subset.
\end{enumerate}
Condition (b) is the \emph{Born--Oppenheimer approximation}
proper.
It fails at \emph{conical intersections} where
$E_0(\mathbf{R}) = E_1(\mathbf{R})$
(ground and first excited states become degenerate):
$\sigma_0$ is undefined at such geometries, and the global
section fails to exist.
That failure forces $\Lk_6$ (\S\ref{sec:L5-forcing-out}).

\smallskip
\noindent\textbf{Using $V$ in the tower.}
Layer~2(a) makes $V$ computable from first principles.
Layer~2(b) ensures the ground-state line bundle
$L_0 := \mathrm{span}\{\sigma_0(\mathbf{R})\}_{\mathbf{R}}$
is globally trivial (\S\ref{sec:L5-berry}), which in turn
means the Berry connection $A_0$ vanishes identically
(Proposition~\ref{prop:berry-trivial}).
Together, conditions (a)--(b) also underpin the TST
coherence condition (\S\ref{sec:L5-tst}): the rate constant
$k_r$ at $\Lk_3$ is derived from the saddle-point value
of $V$ at $\Lk_5$.
\end{mathbox}

\subsubsection{Critical points of \texorpdfstring{$V$}{V} and the \texorpdfstring{IRC}{IRC}}
\label{sec:L5-morse}

With $V$ and $g$ in hand, a central task of computational
chemistry becomes precise: \emph{navigate the landscape of $V$
to identify stable geometries, reaction pathways, and
barriers}.
The mathematical language for this navigation is Morse theory
--- the study of smooth functions via their critical points and
gradient flows.

In practice:
\begin{itemize}
  \item \emph{Geometry optimisation} is gradient descent on
    $(V, g)$ to find local minima --- the stable molecular
    structures.
    Every structure deposited in the Cambridge Structural
    Database or the NIST WebBook corresponds to a minimum
    of $V$.
  \item \emph{Transition-state (TS) search} is finding
    index-1 saddle points --- configurations where the energy
    is stationary but has one direction of descent (the
    reaction mode) and all other directions of ascent.
    Methods such as QST2/QST3 and eigenvector-following
    locate these points by following the negative Hessian
    eigenvector uphill.
  \item \emph{IRC calculation} follows the gradient flow
    from the TS downhill (in the mass-weighted metric $g$)
    to identify the reactant and product connected to that TS.
    The IRC in mass-weighted coordinates was formalised by
    Fukui~\cite{FukuiIRC1981} and given its modern treatment
    by Miller, Handy and Adams~\cite{MillerHandyAdams1980}.
\end{itemize}
In categorical terms, this entire computational toolkit is
the exploration of the Morse structure of $F_V$: the
\emph{functor} $F_V: \Lk_{4.5}(P) \to \OrbMorse$ assigns
to each molecular graph the Morse triple $(\Ce(G), V, g)$,
and its critical-point data are precisely what
$\OrbMorse$-morphisms must preserve.

\begin{definition}[Minima, saddle points, and IRC]
\label{def:pes-critical}
  Let $V: \Ce(G) \to \RR$ be the BO PES and
  $g$ the mass-weighted metric on $\Ce(G)$.
  \begin{itemize}
    \item A \emph{minimum} is a critical point
      $\mathbf{R}_{\mathrm{min}}$ with
      $\nabla_g V = 0$ and positive-definite Hessian
      $\nabla^2_g V \succ 0$.
      Minima correspond to stable molecular geometries
      (isomers or conformers of $G$).
    \item A \emph{first-order saddle point}
      (transition state, TS) is a critical point
      $\mathbf{R}_{\mathrm{TS}}$ with $\nabla_g V = 0$,
      exactly one negative eigenvalue of $\nabla^2_g V$
      (the \emph{reaction mode}), and all remaining
      eigenvalues positive.
    \item The \emph{intrinsic reaction coordinate} (IRC)
      is the pair of steepest-descent paths leaving the
      TS in the metric $g$: the two curves
      $\mathbf{R}_\pm(s): [0, \infty) \to \Ce(G)$ satisfying
      \[
        \frac{d\mathbf{R}_\pm}{ds}
        \;=\; -\frac{\nabla_g V(\mathbf{R}_\pm(s))}
          {|\nabla_g V(\mathbf{R}_\pm(s))|_g},
        \qquad
        \mathbf{R}_\pm(0) = \mathbf{R}_{\mathrm{TS}},
        \qquad
        \dot{\mathbf{R}}_\pm(0) = \pm\mathbf{e}_{\mathrm{rxn}},
      \]
      where $\mathbf{e}_{\mathrm{rxn}}$ is the unit negative
      eigenvector of $\nabla^2_g V$ at
      $\mathbf{R}_{\mathrm{TS}}$.  Local existence and
      uniqueness away from critical points follow from
      Picard--Lindel\"of; the two branches connect the TS
      to the reactant and product basins respectively under
      the generic assumption that the gradient flow does
      not encounter further critical points along its
      descent~\cite{FukuiIRC1981,MillerHandyAdams1980}.
    \item The \emph{activation barrier} is
      $\Ea := V(\mathbf{R}_{\mathrm{TS}}) -
      V(\mathbf{R}_{\mathrm{min,reactant}}) > 0$.
  \end{itemize}
\end{definition}

\begin{proposition}[Morse data enriches the DPO mechanism]
\label{prop:morse-tower}
  Let $r: (G,\sigma) \Rightarrow (H,\tau)$ be a
  $G^*$-equivariant DPO derivation in $\Lk_{4.5}(P)$.
  The $\Lk_5$ lift of $r$ consists of the following
  additional data, all invisible at $\Lk_{4.5}$:
  \begin{enumerate}[label=(\alph*)]
    \item \emph{The canonical IRC}: the concatenation of
      the two steepest-descent branches
      $\mathbf{R}_\pm(s)$ in $(\Ce(G \cup H), g)$ from
      $\mathbf{R}_{\mathrm{TS}}$ to
      $\mathbf{R}_{\mathrm{min,react}}$ and
      $\mathbf{R}_{\mathrm{min,prod}}$ respectively
      (Definition~\ref{def:pes-critical}).
      The IRC is the chemistry literature's
      ``minimum-energy path'' through a chosen saddle in
      mass-weighted coordinates~\cite{FukuiIRC1981}; it is
      locally unique up to the choice of branch and
      reparametrisation, but is not in general a global
      minimum-energy path among all curves connecting the two basins.
    \item \emph{The activation barrier}
      $\Ea(r) = V(\mathbf{R}_{\mathrm{TS}}) -
      V(\mathbf{R}_{\mathrm{min,react}})$:
      an $\Lk_5$ datum determining the rate constant
      $k_r$ via the TST coherence condition
      (\S\ref{sec:L5-tst}).
      $\Ea(r)$ is not computable from the DPO span alone:
      two reactions with the same bond-change pattern but
      different transition-state geometries
      (e.g.\ $\mathrm{S_N2}$ vs.\ $\mathrm{S_N1}$ when
      artificially matched at $\Lk_3$) can have different
      barriers.
    \item \emph{The TS geometry and point group}:
      $\mathbf{R}_{\mathrm{TS}} \in \Ce(G \cup H)$ and
      $\Iso_{G^*}(\mathbf{R}_{\mathrm{TS}})$
      (Proposition~\ref{prop:point-groups}).
      The TS symmetry constrains which vibrational modes
      are active in the reaction coordinate
      (Woodward--Hoffmann rules at $\Lk_{4.5}$ are recovered
      here as selection rules on the IRC).
  \end{enumerate}
  The forgetful functor $U_5: \Lk_5(P) \to \Lk_{4.5}(P)$
  maps (a)--(c) to the underlying DPO span $r$,
  discarding all geometric data.
\end{proposition}

\begin{proof}
Item (a): The gradient-descent ODE defining the IRC has
a unique solution on the open dense subset of $\Ce(G \cup H)$
where $\nabla_g V \neq 0$, by Picard--Lindelöf.
At the TS, the unique direction of descent is the negative
eigenvector of $\nabla^2_g V$ (which is simple by the
definition of an index-1 saddle point), giving a unique
departing half-curve on each side~\cite{MillerHandyAdams1980}.
Together these give the IRC as a well-defined path in
$\Ce(G \cup H)$.

Items (b)--(c) follow from Definition~\ref{def:pes-critical}
and Proposition~\ref{prop:point-groups} respectively.

That these data are invisible at $\Lk_{4.5}$: the DPO span
$r$ specifies only which bonds form and break (the graph
morphism), not the 3D geometry of the TS.
The same DPO span can arise from geometrically distinct
pathways (e.g.\ different approach trajectories in the
$\mathrm{S_N2}$ reaction depending on counterion effects),
each with a different IRC and potentially different $\Ea$.
The forgetful functor collapses all of these to the same
$\Lk_{4.5}$ morphism.
\qedhere
\end{proof}

\begin{remark}[The IRC is not a classical trajectory]
\label{rem:irc-not-classical}
The IRC is a steepest-descent path (gradient flow), not a
solution of Newton's equations with the kinetic energy $T$.
An actual classical trajectory from the TS has non-zero
velocity and follows the classical Lagrangian action
$\int(T-V)\,dt$, which generally deviates from the IRC
(especially for curved reaction paths where Coriolis-like
coupling redirects the trajectory).
The IRC is the zero-kinetic-energy limit: a formal
mathematical object on $(V, g)$ that identifies the reactant
and product basins connected to a given TS, not a
dynamical path.
Genuinely dynamical treatments --- trajectory ensembles,
instanton tunneling, semiclassical quantisation --- enter
at $\Lk_7$.
\end{remark}

\subsubsection{The Berry connection}
\label{sec:L5-berry}

The BO section $\sigma_0(\mathbf{R})$ is a wavefunction, hence
defined only up to an overall complex phase $e^{i\phi}$ at each
geometry.
The question of how this \emph{phase choice} changes as the
nuclei traverse a closed loop $\gamma$ in $\Ce(G)$ is not a
gauge artefact: the \emph{holonomy} of the phase (the total
accumulated phase around $\gamma$) is a gauge-invariant
observable.
This holonomy is the \emph{geometric phase} or
\emph{Berry phase}~\cite{Berry1984, Simon1983}, identified in
the molecular context by Mead and Truhlar~\cite{MeadTruhlar1979}.

The physical significance is direct: a nuclear wavepacket that
traverses a loop in $\Ce(G)$ returns to its starting point with
a modified electronic phase.
If this phase is $-1$ (a sign change of the electronic
wavefunction), the nuclear wavepacket acquires destructive
interference with itself, producing observable spectroscopic
consequences~\cite{LonguetHiggins1963}.
This is the \emph{molecular Aharonov--Bohm effect}.
In the tower language:
\begin{itemize}
  \item At $\Lk_5$ (Layer~2 condition: no conical intersections),
    the Berry phase is \emph{trivial} for all loops
    (Proposition~\ref{prop:berry-trivial} below).
  \item At $\Lk_6$, conical intersections are the
    \emph{sources} of non-trivial Berry holonomy: a loop
    encircling exactly one CI acquires phase $-1$.
    This non-trivial holonomy is precisely the datum that
    $\Lk_5$ cannot encode and $\Lk_6$ must introduce.
\end{itemize}
The Berry connection is therefore the \emph{bridge datum}
between $\Lk_5$ and $\Lk_6$ in the tower.

\begin{definition}[Berry connection on $L_0$]
\label{def:berry-connection}
  Let $U \subseteq \Ce(G)$ be an open subset on which the
  ground-state eigenvalue $E_0$ is non-degenerate, and let
  $\sigma_0: U \to \Hel$ be a smooth choice of normalised
  ground-state section over $U$ (Kato's perturbation theory
  guarantees existence locally; on simply-connected $U$
  the section is unique up to a smooth phase).  The
  \emph{ground-state line bundle} over $U$ is the complex
  line sub-bundle
  \[
    L_0 \;:=\;
    \bigl\{(\mathbf{R}, \psi) \in U \times \Hel
    \;\big|\;
    \psi \in \CC\,\sigma_0(\mathbf{R})\bigr\}
    \;\leq\; \Hel|_U,
  \]
  with fiber $L_0|_{\mathbf{R}} = \CC\,\sigma_0(\mathbf{R})
  \cong \CC$.

  More generally, on any open subset $U_n \subseteq \Ce(G)$
  where the $n$-th eigenvalue $E_n(\mathbf{R})$ remains
  isolated from $E_{n-1}$ and $E_{n+1}$, a smooth
  adiabatic section $\sigma_n: U_n \to \Hel$ may be
  chosen, and the \emph{Berry (geometric) connection} is
  the $\mathfrak{u}(1)$-valued 1-form on $U_n$:
  \[
    A_n(\mathbf{R})
    \;:=\;
    i\,\bigl\langle \sigma_n(\mathbf{R})
    \,\big|\,
    d\,\sigma_n(\mathbf{R}) \bigr\rangle,
  \]
  where $d$ is the exterior derivative on $\Ce(G)$.
  Since $\|\sigma_n\| = 1$, differentiating
  $\langle\sigma_n|\sigma_n\rangle = 1$ gives
  $\langle\sigma_n|d\sigma_n\rangle
  = -\langle d\sigma_n|\sigma_n\rangle^*$,
  showing $\langle\sigma_n|d\sigma_n\rangle$ is purely
  imaginary; hence $A_n$ is a real-valued 1-form.

  Its curvature (for a \emph{single} adiabatic state,
  i.e.\ a $U(1)$ connection on $L_n$) is
  \[
    \Omega_n \;:=\; dA_n
    \;=\; 2\,\mathrm{Im}\,\langle d\sigma_n \wedge d\sigma_n
    \rangle,
  \]
  the \emph{Berry curvature}.
  Explicitly, in local coordinates $R^\mu$ on $\Ce(G)$:
  \[
    \Omega_{n,\mu\nu}
    \;=\;
    2\,\mathrm{Im}\,\langle\partial_\mu\sigma_n|
    \partial_\nu\sigma_n\rangle.
  \]
\end{definition}

\begin{remark}[Why the curvature is $dA_n$, not
  $dA_n + A_n \wedge A_n$]
\label{rem:abelian-curvature}
  The formula $\Omega = dA + A \wedge A$ is the curvature
  of a non-abelian ($U(N)$, $N > 1$) connection.
  For a \emph{single} adiabatic state $n$, $A_n$ is a
  real-valued (abelian, $U(1)$) 1-form; for abelian forms,
  $A_n \wedge A_n = 0$ by antisymmetry of the wedge product,
  so $\Omega_n = dA_n$ exactly.

  The non-abelian formula does arise in the multi-state
  setting: for an $N$-dimensional subspace of adiabatic
  states (e.g.\ a degenerate or near-degenerate manifold),
  the collective Berry connection is an anti-Hermitian $N \times N$ matrix 1-form
  $\mathbf{A}_{\mu,mn} =
  i\langle\sigma_m|\partial_\mu|\sigma_n\rangle$,
  and its curvature is $\boldsymbol{\Omega} = d\mathbf{A}
  + \mathbf{A} \wedge \mathbf{A}$ (the non-adiabatic coupling
  matrix).
  At $\Lk_6$, where electronic states couple
  non-adiabatically near CIs, this full $U(N)$ structure
  becomes necessary.
  At $\Lk_5$, the single-state ($N = 1$) abelian formula
  suffices.
\end{remark}

\begin{proposition}[Triviality of Berry holonomy on
  simply-connected CI-free regions]
\label{prop:berry-trivial}
  Let $U \subseteq \Ce(G)$ be a simply-connected open
  subset on which the ground-state eigenvalue
  $E_0(\mathbf{R})$ is isolated (Layer~2(b) of
  Definition~\ref{def:bo-section}: no conical
  intersections in $U$), and assume the electronic
  Hamiltonian $\hat{H}_{\mathrm{el}}(\mathbf{R})$ is
  real-symmetric (spinless non-relativistic electrons
  with time-reversal-invariant Coulomb interactions).
  Then a smooth real-valued ground-state section
  $\tilde\sigma_0: U \to \Hel$ exists, and in this
  real gauge the Berry connection vanishes identically:
  \[
    A_0(\mathbf{R}) \;=\; 0 \qquad
    \text{for all } \mathbf{R} \in U.
  \]
  Consequently:
  \begin{enumerate}[label=(\alph*)]
    \item The Berry curvature satisfies $\Omega_0 = dA_0
      = 0$ on $U$.
    \item For every loop $\gamma$ contractible in $U$,
      the holonomy of $A_0$ is trivial:
      $\exp\!\bigl(\oint_\gamma A_0\bigr) = 1 \in U(1)$.
    \item The complex ground-state line bundle
      $L_0 \to U$ satisfies $c_1(L_0) = [\Omega_0/2\pi]
      = 0 \in H^2(U, \ZZ)$, and its $\mathcal{K}$-real
      sub-bundle $L_0^{\RR} \leq L_0$ (generated
      pointwise by $\tilde\sigma_0$) satisfies
      $w_1(L_0^{\RR}) = 0 \in H^1(U, \ZZ/2)$.

      Under the real-Hamiltonian assumption,
      $w_1(L_0^{\RR})$ is the structurally informative
      invariant of the ground-state bundle: it detects
      the Longuet--Higgins $\ZZ_2$ sign holonomy of
      $\tilde\sigma_0$ around loops encircling conical
      intersections.  At $\Lk_5$ the loops in $U$ are
      contractible and the holonomy is trivial; at
      $\Lk_6$, where the configuration space is the
      CI-punctured $\Ce(G) \setminus X_{\mathrm{CI}}$,
      loops encircling components of $X_{\mathrm{CI}}$
      acquire holonomy $-1$, and $w_1(L_0^{\RR}) \neq 0$
      is the molecular Aharonov--Bohm invariant
      (invisible to $c_1$).
  \end{enumerate}
\end{proposition}

\begin{proof}
\textbf{Step 1: Real gauge on a simply-connected subset.}
Let $\mathcal{K}$ denote complex conjugation on
$\Hel = L^2_{\mathrm{antisym}}(\RR^{3N_e}, \CC)$
(an anti-unitary operator with $\mathcal{K}^2 = I$ on
spinless systems).  The Coulomb terms in $V_{ee}$ and
$V_{en}(\mathbf{R})$ and the kinetic energy are all real
in coordinate representation, so $\mathcal{K}
\hat{H}_{\mathrm{el}}(\mathbf{R}) \mathcal{K}^{-1} =
\hat{H}_{\mathrm{el}}(\mathbf{R})$ for every $\mathbf{R}$.

At a fixed $\mathbf{R}$ with non-degenerate ground-state
eigenvalue $E_0(\mathbf{R})$, the eigenspace is
one-dimensional; $\mathcal{K}$-invariance maps it to
itself, so $\mathcal{K}\sigma_0(\mathbf{R}) = e^{i\alpha
(\mathbf{R})}\sigma_0(\mathbf{R})$ for some phase $\alpha
(\mathbf{R}) \in \RR/2\pi\ZZ$.  Setting
$\tilde\sigma_0(\mathbf{R}) := e^{i\alpha(\mathbf{R})/2}
\sigma_0(\mathbf{R})$ yields $\mathcal{K}\tilde\sigma_0
(\mathbf{R}) = e^{-i\alpha/2}\mathcal{K}\sigma_0 =
e^{-i\alpha/2}e^{i\alpha}\sigma_0 = e^{i\alpha/2}\sigma_0
= \tilde\sigma_0(\mathbf{R})$: the rescaled section is
real-valued.

On a simply-connected open subset $U \subseteq \Ce(G)$
where $E_0$ remains isolated, a continuous (and, by
Kato's perturbation theory in Step~2, smooth) choice of
$\beta(\mathbf{R}) := \alpha(\mathbf{R})/2$ exists.  The
half-phase $\beta$ is well-defined only modulo $\pi$
(replacing $\beta$ by $\beta + \pi$ replaces $\tilde\sigma_0$
by $-\tilde\sigma_0$, which is also a valid real section),
so a continuous global choice requires trivial $\pi_1
(U)$-action on this $\ZZ_2$ ambiguity.  Simple connectivity
of $U$ is precisely this condition: on a simply-connected
domain, the $\ZZ_2$ obstruction in $H^1(U, \ZZ/2)$ vanishes
and $\beta$ admits a continuous lift to $\RR$.  This is the
same $\ZZ_2$ obstruction that becomes the
Longuet--Higgins sign holonomy at $\Lk_6$ when the loop
encircling a CI is no longer contractible.

\textbf{Step 2: Smooth real section on $U$.}
On $U$, the ground-state eigenvalue $E_0(\mathbf{R})$
remains isolated by hypothesis.  Kato's perturbation
theory~\cite{Kato1966,Teufel2003} (smooth dependence of
isolated eigenvalues and eigenprojections on operator
parameters) yields a smooth complex ground-state section
$\sigma_0: U \to \Hel$.  Then $\alpha(\mathbf{R})$ defined
by $\mathcal{K}\sigma_0(\mathbf{R}) = e^{i\alpha
(\mathbf{R})}\sigma_0(\mathbf{R})$ is a smooth function
$U \to \RR/2\pi\ZZ$; by simple connectivity of $U$
(Step~1), it admits a smooth lift $\beta(\mathbf{R}) =
\alpha(\mathbf{R})/2$ to $\RR$.  The rescaled section
\[
  \tilde\sigma_0(\mathbf{R}) := e^{i\beta(\mathbf{R})}
  \sigma_0(\mathbf{R})
\]
is smooth on $U$ and $\mathcal{K}$-real:
$\mathcal{K}\tilde\sigma_0 = \tilde\sigma_0$ by the
Step~1 calculation.

\textbf{Step 3: $A_0 = 0$ in the real gauge on $U$.}
For a real-valued normalised section $\sigma_0$:
\[
  \langle \sigma_0(\mathbf{R}) \,|\,
  d\,\sigma_0(\mathbf{R})\rangle
  \;=\;
  \int_{\RR^{3N_e}}
  \sigma_0(\mathbf{R},x)\,d_{\mathbf{R}}\sigma_0(\mathbf{R},x)
  \,dx.
\]
Differentiating the normalisation condition
$\langle\sigma_0|\sigma_0\rangle
= \int \sigma_0^2\,dx = 1$ gives
$2\int \sigma_0\,d\sigma_0\,dx = 0$,
so $\langle\sigma_0|d\sigma_0\rangle = 0$ on $U$.
Therefore $A_0 = i\langle\sigma_0|d\sigma_0\rangle = 0$
on $U$.

\textbf{Step 4: Verification of consequences (a)--(c).}
From $A_0 = 0$ on $U$ (Step~3):
\begin{itemize}
  \item[(a)] $\Omega_0 = dA_0 = d(0) = 0$ on $U$.
  \item[(b)] For any loop $\gamma$ contractible in $U$,
    Stokes' theorem applied to a 2-chain $\Sigma$ with
    $\partial\Sigma = \gamma$ gives $\oint_\gamma A_0 =
    \int_\Sigma \Omega_0 = 0$, hence
    $\exp\!\bigl(\oint_\gamma A_0\bigr) = 1$.
  \item[(c)] $c_1(L_0) = [\Omega_0/2\pi] = 0$ since
    $\Omega_0 = 0$.  The real sub-bundle $L_0^{\RR}
    \leq L_0$ generated pointwise by $\tilde\sigma_0$
    is a real line bundle over $U$; the nowhere-zero
    global section $\tilde\sigma_0$ (from Step~2)
    trivialises it, hence $w_1(L_0^{\RR}) = 0$.  The
    remainder of (c) --- the interpretation of
    $w_1(L_0^{\RR})$ as the molecular Aharonov--Bohm
    invariant detecting CI-induced sign holonomy at
    $\Lk_6$ --- is a structural observation rather than
    a deduction from the present hypotheses; the precise
    statement and proof at $\Lk_6$ appear in
    \S\ref{sec:L5-forcing-out} and the $\Lk_6$ chapter.
\end{itemize}
\qedhere
\end{proof}

\begin{remark}[Simple connectivity of $\Ce(G)$ and the
  $\Lk_5$ vs. $\Lk_6$ boundary]
\label{rem:simply-connected}
  The real-gauge argument of
  Proposition~\ref{prop:berry-trivial} works locally on any
  simply-connected open subset of $\Ce(G)$, and extends
  globally provided $\Ce(G)$ itself is simply connected.
  When $\Ce(G)$ has non-trivial topology --- in particular,
  when conical intersections form a codimension-2
  submanifold $X_{\mathrm{CI}} \subset \Ce(G)$
  (codimension~2 because CIs of a real-symmetric
  Hamiltonian satisfy two real conditions on the
  $2\times 2$ effective block) --- the complement $\Ce(G)
  \setminus X_{\mathrm{CI}}$ has non-trivial $\pi_1$
  (loops encircling $X_{\mathrm{CI}}$ cannot be
  contracted), and the global real-gauge extension can
  fail around such loops.
  The standard example is the $\mathrm{H}_3$ system, where
  the $D_{3h}$ conical intersection creates a loop with
  $\ZZ_2$ holonomy (Berry phase $\pi$): the ground-state
  wavefunction changes sign under a traversal of this
  loop~\cite{LonguetHiggins1963, MeadTruhlar1979}.
  That failure is precisely the $\Lk_5 \to \Lk_6$
  transition: $\Lk_5$ operates on simply-connected,
  CI-free regions where all holonomies are trivial;
  $\Lk_6$ admits CI-induced holonomies in $\coker(\varphi_6)$.

  \smallskip\noindent
  The spinless assumption in the proposition is likewise
  standard in BO theory for ground-state thermal chemistry.
  For electrons with spin-orbit coupling, time-reversal is
  $\mathcal{T}^2 = -1$ (Kramers) and the analysis changes:
  additional $\ZZ_2$ structure enters the Berry connection.
  These effects become relevant at $\Lk_6$ and beyond and
  are not treated here.
\end{remark}

\begin{remark}[What the proposition establishes in the tower]
\label{rem:berry-tower}
Proposition~\ref{prop:berry-trivial} has three distinct
consequences for the tower, each at a different level of
abstraction.

\smallskip
\noindent
\textbf{Physical consequence.}
The electronic wavefunction accumulates \emph{no geometric
phase} as the nuclei traverse any loop in $\Ce(G)$.
Nuclear dynamics on the ground-state BO surface is
self-consistent: the electronic subsystem does not distinguish
one traversal of a nuclear loop from another, and there is no
quantum interference between paths with different winding
histories.
This is the regime of \emph{adiabatic thermal chemistry}
--- the regime in which all of organic chemistry, most
of thermodynamics, and most of kinetics operates.

\smallskip
\noindent
\textbf{Categorical consequence.}
On the simply-connected CI-free open subsets where the
$\Lk_5$ framework operates, the functor $F_V: \Lk_{4.5}(P)
\to \OrbMorse$ captures the relevant $\Lk_5$ structure
through the PES triple $(\Ce(G), V, g)$ alone: the
ground-state line bundle $L_0$ is trivial in a real gauge
and the Berry connection vanishes
(Proposition~\ref{prop:berry-trivial}), so no additional
topological data from the electronic Hilbert bundle is
needed in this regime.
The non-trivial sign holonomy $w_1(L_0) \in H^1(\cdot,
\ZZ/2)$ that arises around loops encircling conical
intersections is not expressible in this regime; it is
the obstruction that generates the non-trivial
$\coker(\varphi_6)$ at $\Lk_6$.

\smallskip
\noindent
\textbf{Tower boundary: what this identifies as $\Lk_6$ data.}
The proposition is constructive: it shows exactly what must
be added at $\Lk_6$.
When a conical intersection exists at
$\mathbf{R}_{\mathrm{CI}} \in \Ce(G)$, the real-gauge
argument of Step~1 fails: $\sigma_0$ cannot be chosen real
near $\mathbf{R}_{\mathrm{CI}}$ because the ground and first
excited states are degenerate there.
A loop $\gamma$ encircling $\mathbf{R}_{\mathrm{CI}}$
acquires holonomy $\exp(\oint_\gamma A_0) = -1$
(Berry phase $\pi$, the molecular Aharonov--Bohm
effect~\cite{LonguetHiggins1963,Berry1984}).
This $(-1)$ is not expressible in $\Lk_5$, where all
holonomies are $1$: it is an element of $\coker(\varphi_6)$.
Encoding it requires the full Hilbert bundle structure and
the non-abelian coupling $A_{01}$ between ground and excited
states --- the $\Lk_6$ structure.
\end{remark}

%% file: chapters/L5/l5_def.tex
\subsection{Definition of \texorpdfstring{$\Lk_5(P)$}{L5(P)}}
\label{sec:L5-def}

The two preceding sections established the raw materials:
the configuration orbifold $\Ce(G)$ (\S\ref{sec:L5-ce})
and the four-layer structure of $(V, g, \sigma_0, A_0)$
(\S\ref{sec:L5-pes}).
This section assembles them into the categorical objects
that define $\Lk_5(P)$, proves that these objects form a
well-defined category, and situates $\Lk_5(P)$ coherently
in the tower.

The plan is a two-step construction:
\begin{enumerate}[label=(\Roman*)]
  \item Build the \emph{target category} $\OrbMorse$
    (\S\ref{sec:L5-orbmorse}): the symmetric monoidal
    category that receives the geometric decoration.
  \item Construct the \emph{geometric functor}
    $\FV: \Lk_{4.5}(P) \to \OrbMorse$
    (\S\ref{sec:L5-FV}), which assigns to each molecular
    graph the Morse triple of its configuration space,
    PES, and mass-weighted metric, and to each DPO
    derivation its intrinsic reaction coordinate.
\end{enumerate}
Together these yield $\Lk_5(P) := (\Lk_{4.5}(P), \FV)$
(\S\ref{sec:L5-L5def}), a $G^*$-equivariant SMC decorated
by a lax monoidal functor to $\OrbMorse$.

The central coherence diagram that animates the section
is the following, whose commutativity (up to the TST
coherence condition) is proved in \S\ref{sec:L5-tst}:
\begin{equation}
\label{diag:L5-motivation}
\includegraphics{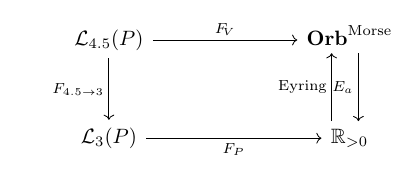}
\end{equation}
Reading this diagram: $\FV$ extracts the PES triple
$(\Ce(G), V, g)$ from each $\Lk_{4.5}$-reaction; the
\emph{kinetic shadow} functor $F_{4.5 \to 3}$ is the
composite forgetful $\Lk_{4.5}(P) \to \Lk_4(P) \to
\Lk_3(P)$ that projects to the underlying rate datum;
the right vertical map extracts the activation barrier
$\Ea = V(\mathbf{R}_{\mathrm{TS}}) - V(\mathbf{R}_{\mathrm{min}})$
from the TS saddle point; and the Eyring arrow
$k = (k_BT/h)\,e^{-\Ea/RT}$ converts $\Ea$ back to the
rate constant $k_r \in \RR_{>0}$ that $F_P$ assigned at
$\Lk_3$.  The commutativity of \eqref{diag:L5-motivation}
is the TST coherence condition: the rate constant at
$\Lk_3$ is constrained by the PES geometry at $\Lk_5$
through Eyring's relation.

\subsubsection{The target category \texorpdfstring{$\OrbMorse$}{OrbMorse}}
\label{sec:L5-orbmorse}

The geometric decoration at $\Lk_5$ targets a category
whose objects are the configuration orbifolds of molecules
equipped with a PES and a metric.
This category must be constructed carefully: its morphism
structure must capture the physical notion of a reaction
as a gradient-flow path from a reactant minimum to a
product minimum through a transition state, and its
monoidal product must encode the combination of
non-interacting molecular systems.

Classical Morse theory provides the necessary ingredient.
Initiated by Morse~\cite{MilnorMorse1963}, it assigns to
each smooth function $V: M \to \RR$ a \emph{critical-point
structure} (non-degenerate critical points with integer
indices) from which global topological properties of $M$
can be recovered.
In our setting $M = \Ce(G)$ is not a manifold but an
orbifold --- a topological space locally modelled on
$\RR^k / \Gamma$ for finite groups
$\Gamma$~\cite{SatakeVManifold1956}.
The relevant Morse theory for orbifolds, developed for
our purposes below, is a direct extension: one works on
the smooth stratum (where $\Ce(G)$ is a manifold) and
tracks the orbifold strata as additional structure.

Two key structural features determine the correct morphism
definition.
First, \emph{chemical reactions are atom-conserving}:
the total multiset of nuclei is unchanged between reactants
and products (no nuclear transmutation under ordinary
chemical conditions).
For an elementary reaction from reactant system $G$ to
product system $H$ with the same atom set, the appropriate
domain for the intrinsic reaction coordinate is the
\emph{joint configuration orbifold} $\Ce(G \sqcup H)$
introduced in \S\ref{sec:L5-morse}: the configuration
orbifold of the combined reactive complex, large enough
to contain both reactant and product geometries as local
minima and the transition state between them as a saddle
point.
Second, the full interacting PES $V_{\mathrm{full}}$ on
this joint orbifold is a \emph{single} smooth function ---
not two separate PESs glued at the TS.
An \emph{elementary morphism} in $\OrbMorse$ is therefore
a triple $(V_{\mathrm{full}}, \mathbf{R}_\mathrm{TS},
\gamma)$ where the gradient flow $\gamma$ on
$V_{\mathrm{full}}$ connects a reactant minimum (encoding
the source object) to a product minimum (encoding the
target object) through the distinguished saddle point
$\mathbf{R}_\mathrm{TS}$.  General morphisms of
$\OrbMorse$, encoding multi-step mechanisms, are
compositions of elementary morphisms in the free-category
sense made precise in Definition~\ref{def:orb-morse} below.

\begin{definition}[The category $\OrbMorse$]
\label{def:orb-morse}
  The symmetric monoidal category $\OrbMorse$ is defined
  as follows.

  \smallskip
  \noindent\textbf{Objects.}
  Triples $(\mathcal{O}, V, g)$ where:
  \begin{itemize}
    \item $\mathcal{O}$ is a smooth Riemannian orbifold
      (locally modelled on $\RR^k/\Gamma$ for finite
      groups $\Gamma$~\cite{SatakeVManifold1956});
    \item $V: \mathcal{O} \to \RR$ is a smooth function,
      Morse on the open subset of chemically accessible,
      CI-free, non-coalescence geometries (all critical
      points in that region are non-degenerate);
    \item $g$ is the Riemannian metric on $\mathcal{O}$.
  \end{itemize}
  A distinguished minimum
  $\mathbf{R}_{\mathrm{min}}(\mathcal{O}) \in \mathcal{O}$
  marks the reference geometry (the equilibrium structure).

  \smallskip
  \noindent\textbf{Elementary morphisms (geometric
    channels).}
  For objects $(\Ce(G), V_G, g_G)$ and $(\Ce(H), V_H, g_H)$
  with $G$ and $H$ sharing an underlying atom multiset, an
  \emph{elementary geometric channel} is a tuple
  \[
    \bigl(V_{\mathrm{full}},\; \mathbf{R}_{\mathrm{TS}},\;
    \gamma\bigr)
  \]
  where $V_{\mathrm{full}}$, $\mathbf{R}_{\mathrm{TS}}$,
  and $\gamma$ are as above with $\gamma$ a single IRC
  through one index-1 saddle of $V_{\mathrm{full}}$.

  \noindent\textbf{Morphisms.}
  General morphisms of $\OrbMorse$ are finite strings
  $(c_1, c_2, \ldots, c_m)$ of elementary geometric
  channels with matching source/target objects; the
  empty string at $(\mathcal{O}, V, g)$ is the identity.

  \noindent\textbf{Composition.}
  Concatenation of channel strings; associative by
  definition.  The concatenation $c_1 \cdot c_2$ of two
  elementary channels is \emph{not} in general an elementary channel ---
  a two-step mechanism passes through an intermediate
  minimum and exhibits two saddles, and is a
  \emph{composite}, not an elementary, morphism.

  \smallskip
  \noindent\textbf{Monoidal product.}
  \[
    (\mathcal{O}_1, V_1, g_1) \otimes
    (\mathcal{O}_2, V_2, g_2)
    \;:=\;
    (\mathcal{O}_1 \times \mathcal{O}_2,\;
    V_1 + V_2,\; g_1 \oplus g_2).
  \]
  For configuration orbifolds of molecules, this
  categorical product is the asymptotic (infinite
  fragment separation) description of $\Ce(G_1 \sqcup
  G_2)$; see Remark~\ref{rem:interaction} below for the
  distinction between this Layer~1 product and the Layer~2
  full orbifold $\Ce(G_1 \sqcup G_2)$.

  \smallskip
  \noindent\textbf{Monoidal unit.}
  $(\{*\}, 0, 0)$: the one-point orbifold
  (the vacuum system, with no atoms and zero energy).
\end{definition}

\begin{proposition}[$\OrbMorse$ is a strict SMC]
\label{prop:orbmorse-smc}
  With the structure of Definition~\ref{def:orb-morse},
  $\OrbMorse$ is a strict symmetric monoidal category.
\end{proposition}

\begin{proof}
\textbf{Category.}
Morphisms are finite strings of elementary geometric
channels; identity at each object is the empty string;
composition is concatenation of strings.  Both
associativity of composition and the unit laws hold
strictly by definition of string concatenation (i.e.,
$\OrbMorse$ is the free category on its generating
graph of elementary channels).

\textbf{Strict monoidal.}
The three structural isomorphisms are all identities.
\emph{Associativity}:
$(\mathcal{O}_1 \times \mathcal{O}_2) \times \mathcal{O}_3
= \mathcal{O}_1 \times (\mathcal{O}_2 \times \mathcal{O}_3)$
as product orbifolds;
$(V_1 + V_2) + V_3 = V_1 + (V_2 + V_3)$ in $C^\infty$;
$(g_1 \oplus g_2) \oplus g_3 = g_1 \oplus (g_2 \oplus g_3)$
as block metrics.
\emph{Left/right unit}:
$\{*\} \times \mathcal{O} = \mathcal{O}$,
$0 + V = V$, and $0 \oplus g = g$.

\textbf{Symmetric.}
The swap isomorphism
$\tau_{12}: (\mathcal{O}_1 \times \mathcal{O}_2,
V_1 + V_2, g_1 \oplus g_2) \to
(\mathcal{O}_2 \times \mathcal{O}_1,
V_2 + V_1, g_2 \oplus g_1)$
is the coordinate swap; hexagon axioms hold by
commutativity of $+$ and $\oplus$.
\qedhere
\end{proof}

\begin{remark}[Monoidal product vs. joint orbifold:
  Layer~1 and Layer~2]
\label{rem:interaction}
  The categorical monoidal product
  $(\mathcal{O}_1 \times \mathcal{O}_2,\;
  V_1 + V_2,\; g_1 \oplus g_2)$
  approximates the \emph{asymptotic} structure of the
  joint configuration orbifold
  $\Ce(G_1 \sqcup G_2)$ at infinite fragment separation.
  The two spaces differ in two ways: $\Ce(G_1) \times
  \Ce(G_2)$ has dimension $\dim\Ce(G_1) +
  \dim\Ce(G_2) = 3(n_1 + n_2) - 12$ (two $SE(3)$
  quotients), while $\Ce(G_1 \sqcup G_2)$ has
  dimension $3(n_1 + n_2) - 6$ (one $SE(3)$ quotient),
  containing three additional relative-motion coordinates;
  and the PES on $\Ce(G_1 \sqcup G_2)$ contains an
  interaction term $V_{\mathrm{int}}$ that vanishes
  asymptotically but is non-zero at finite separation.
  Explicitly:
  \[
    V_{\mathrm{full}}(\mathbf{R}_1, \mathbf{R}_2, \mathbf{s})
    \;=\;
    V_1(\mathbf{R}_1) + V_2(\mathbf{R}_2)
    + V_{\mathrm{int}}(\mathbf{R}_1, \mathbf{R}_2, \mathbf{s}),
  \]
  where $\mathbf{s}$ denotes the relative-separation
  coordinates and $V_{\mathrm{int}} \to 0$ as
  $|\mathbf{s}| \to \infty$.

  \begin{itemize}
    \item \textbf{Layer~1} (strict product):
      $V_{\mathrm{full}} = V_1 + V_2$. The two
      subsystems do not interact; no barrier, no IRC, no
      reaction. Sufficient for isolated-fragment
      properties ($\FH$, $\FS$ of separate molecules).
    \item \textbf{Layer~2} (lax product):
      $V_{\mathrm{full}} = V_1 + V_2 + V_{\mathrm{int}}$
      on $\Ce(G_1 \sqcup G_2)$. The interaction creates
      pre-reaction complexes, transition states, and
      product complexes; this is the exact BO PES of the
      reactive system.
  \end{itemize}
  Consequently, $\FV$ is a \emph{lax} monoidal functor
  (Definition~\ref{def:FV} below), not strict: the
  laxator witnesses the interaction correction.
\end{remark}

\begin{chembox}[The interaction term: the ion-dipole complex
  in $\mathrm{S_N2}$]
\label{chem:interaction}
The interaction term $V_{\mathrm{int}}$ is not a
mathematical abstraction but a physical observable with
direct chemical consequences.

\medskip
\noindent\textbf{The $\mathrm{S_N2}$ PES.}
For the reaction $\mathrm{CH_3Cl + OH^- \to
CH_3OH + Cl^-}$, the \emph{non-interacting} Layer~1
approximation gives the PES as the sum of the isolated
reactant PESs: a flat surface with a single minimum at
the CH$_3$Cl equilibrium geometry and no barrier.
The Layer~1 PES predicts no reaction pathway, no TS,
and no product formation.

The \emph{full Layer~2} PES adds $V_{\mathrm{int}}$,
which includes the long-range ion-dipole Coulomb
attraction between OH$^-$ and CH$_3$Cl.
This attraction creates:
\begin{enumerate}[label=(\roman*)]
  \item An \emph{ion-dipole complex}
    $[\mathrm{CH_3Cl\cdots OH^-}]$: a shallow well
    ($\approx -40$ kJ/mol below separated reactants)
    appearing at intermediate C$\cdots$O distance
    $\approx 3.5$~\AA.
    This minimum is entirely due to $V_{\mathrm{int}}$
    and is invisible at Layer~1.
  \item The \emph{transition state}
    $[\mathrm{HO\cdots CH_3\cdots Cl}]^{\ddagger}$
    at shorter C$\cdots$O and C$\cdots$Cl distances:
    the saddle point of the full Layer~2 surface.
  \item An analogous \emph{product complex}
    $[\mathrm{CH_3OH\cdots Cl^-}]$ on the exit channel.
\end{enumerate}
The full Layer~2 PES profile is a characteristic
double-well potential with two intermediate complexes
flanking the central TS --- a well-established feature
of gas-phase $\mathrm{S_N2}$ reactions whose shape is
entirely due to
$V_{\mathrm{int}}$~\cite{HelgakerJorgensenOlsen2000}.

In $\OrbMorse$ language: the monoidal product object
$(\Ce(\mathrm{CH_3Cl})\times\Ce(\mathrm{OH^-}),
V_{\mathrm{CH_3Cl}} + V_{\mathrm{OH}^-}, g)$
has no morphism to the product object at Layer~1.
The morphism exists only at Layer~2, where
$V_{\mathrm{int}}$ provides the gradient-flow cobordism
(the IRC from the ion-dipole complex through the TS to
the product complex on $\Ce(\mathrm{CH_3Cl} \sqcup
\mathrm{OH^-})$).
\end{chembox}

\begin{proposition}[$\OrbMorse$ is a natural target for
  $\FV$]
\label{prop:orbmorse-tower}
  The category $\OrbMorse$ of Definition~\ref{def:orb-morse}
  is a natural target for a functor $\FV: \Lk_{4.5}(P)
  \to \OrbMorse$ satisfying:
  \begin{enumerate}[label=(\alph*)]
    \item Each elementary DPO derivation $r$ maps to an
      elementary geometric channel $(V_{\mathrm{full}},
      \mathbf{R}_\mathrm{TS}, \gamma)$ recording its
      single-saddle IRC; composite DPO derivations map
      to the corresponding channel string by
      concatenation;
    \item For elementary channels, the activation barrier
      $\Ea(c) = V_{\mathrm{full}}(\mathbf{R}_{\mathrm{TS}}) -
      V_{\mathrm{full}}(\mathbf{R}_{\mathrm{min,react}})$
      is well-defined; for a channel string
      $(c_1, \ldots, c_m)$ in series with intermediates,
      the rate-determining step provides the effective
      barrier $\Ea = \max_i \Ea(c_i)$, with the
      assumption that pre-equilibration between
      intermediates holds (the standard quasi-stationary
      approximation);
    \item Diagram~\eqref{diag:L5-motivation}
      commutes up to the TST coherence condition of
      \S\ref{sec:L5-tst}.
  \end{enumerate}
  In particular, coherence with lower tower levels holds:
  \begin{itemize}
    \item At $\Lk_1$--$\Lk_2$, the thermochemical data
      $(\FH, \FS)$ are recovered from $\FV$ via
      \[
        \FH(r) \;\approx\;
        V_{\mathrm{full}}(\mathbf{R}_{\mathrm{min,prod}})
        - V_{\mathrm{full}}(\mathbf{R}_{\mathrm{min,react}})
        \;+\; \Delta E_{\mathrm{ZPE}},
      \]
      confirming $\FV$ enriches, not replaces, lower
      tower levels.
    \item At $\Lk_3$, the rate constant $F_P(r)$ is no
      longer free: it is constrained by the TST
      coherence condition (\S\ref{sec:L5-tst})
      \[
        F_P(r) \;\approx\;
        \kappa(T)\,\frac{k_BT}{h}\,
        \frac{Q^\ddagger}{Q_{\mathrm{reac}}}\,
        e^{-V^\ddagger/RT},
      \]
      all of whose ingredients are $\Lk_5$-computable
      from $\FV(r)$.  The classical Eyring expression
      $k_r = (k_BT/h)\,e^{-\Ea/RT}$ is the naive limit
      $\kappa = 1$, harmonic partition functions, no
      tunnelling.
  \end{itemize}
\end{proposition}

\begin{proof}
Item (a): for each elementary DPO derivation, the IRC
exists and is locally unique (up to branch and
reparametrisation) by
Proposition~\ref{prop:morse-tower}(a), so the elementary
channel datum is well-defined on the joint orbifold
$\Ce(G \sqcup H)$.  Composite derivations decompose
canonically into elementary ones (DPO composition is
sequential), and the corresponding channel-string
assembly is functorial by definition of the free
category structure on $\OrbMorse$.
Item (b): the Eyring extraction
$\Ea(V_{\mathrm{full}}, \mathbf{R}_{\mathrm{TS}}, \gamma)
= V_{\mathrm{full}}(\mathbf{R}_{\mathrm{TS}}) -
V_{\mathrm{full}}(\lim_{s\to-\infty}\gamma(s))$
is a well-defined real-valued map on $\OrbMorse$ morphisms,
positive by the definition of a saddle point above the
reactant minimum.
Item (c): the commutativity of
\eqref{diag:L5-motivation} is the TST coherence condition;
see \S\ref{sec:L5-tst} for the full statement and proof.
\qedhere
\end{proof}

\begin{remark}[On the precise sense of ``natural target'']
\label{rem:natural-target}
Proposition~\ref{prop:orbmorse-tower} states that
$\OrbMorse$ \emph{is} a natural target category, not that
it is canonical or universal in a formal categorical
sense.
A rigorous universal property --- ``$\OrbMorse$ is initial
among SMCs receiving a functor from $\Lk_{4.5}(P)$
satisfying (a)--(c)'' --- is plausible but not proved here,
and would require either restricting the allowed target
categories (e.g., to Morse-theoretic categories) or
characterising the conditions (a)--(c) as a universal
property of a specific construction (e.g., a localisation
or a free completion).
For the present purposes the functorial correctness
established above suffices: $\FV$ takes its values in
$\OrbMorse$ and coherence with the lower tower levels
holds.
\end{remark}

\subsubsection{The geometric functor \texorpdfstring{$\FV$}{FV}}
\label{sec:L5-FV}

Proposition~\ref{prop:orbmorse-tower} identifies $\OrbMorse$
as the appropriate target.
What remains is to define $\FV$ explicitly on objects and
morphisms, specify its lax monoidal structure (the critical
distinction from the lower-level decorator functors), and
address the infinite-dimensional nature of the datum.

\begin{definition}[Geometric functor $\FV$]
\label{def:FV}
  The \emph{geometric functor} is a \emph{lax} SMC functor
  \[
    \FV:\; \Lk_{4.5}(P) \;\longrightarrow\; \OrbMorse,
  \]
  defined as follows.

  \smallskip
  \noindent\textbf{On objects.}
  Each chirality-labelled molecular graph
  $(G, \sigma) \in \Lk_{4.5}(P)$ maps to
  \[
    \FV(G, \sigma)
    \;=\; \bigl(\Ce(G, \sigma),\; V_G,\; g_G\bigr),
  \]
  where $\Ce(G, \sigma)$ is the connected component of
  $\Ce(G)$ determined by $\sigma$
  (Proposition~\ref{prop:chirality-realisation}),
  $V_G = (E_0 + V_{\mathrm{nn}})|_{\Ce(G,\sigma)}$ is
  the BO PES (Definition~\ref{def:bo-section}),
  and $g_G$ is the mass-weighted metric
  (Definition~\ref{def:ce}).

  \smallskip
  \noindent\textbf{On morphisms.}
  Each $G^*$-equivariant DPO derivation
  $d: (G, \sigma) \Rightarrow (H, \tau)$ maps to the
  gradient-flow cobordism on $\Ce(G \sqcup H)$:
  \[
    \FV(d) \;=\; \bigl(V_{\mathrm{full}},\;
    \mathbf{R}_{\mathrm{TS}},\; \gamma_{\mathrm{IRC}}\bigr),
  \]
  where $V_{\mathrm{full}} = E_0^{\mathrm{combined}} +
  V_{\mathrm{nn}}^{\mathrm{combined}}$ is the full BO PES
  of the combined reactive system on the joint orbifold (Definition~\ref{def:bo-section}),
  $\mathbf{R}_{\mathrm{TS}}$ is the index-1 saddle point
  corresponding to $d$, and $\gamma_{\mathrm{IRC}}$ is the
  resulting IRC (Definition~\ref{def:pes-critical},
  Proposition~\ref{prop:morse-tower}).

  \smallskip
  \noindent\textbf{Lax monoidal structure.}
  The laxator
  \[
    \phi_{G,H}:\;
    \FV(G, \sigma) \otimes \FV(H, \tau)
    \;\longrightarrow\;
    \FV\bigl((G, \sigma) \otimes (H, \tau)\bigr)
  \]
  is itself a morphism in $\OrbMorse$: the gradient-flow
  descent on the full interacting PES $V_{\mathrm{full}}$
  on $\Ce(G \sqcup H)$ from the asymptotic region
  (fragments well-separated, where $V_{\mathrm{full}}
  \approx V_G + V_H$) to the nearest local minimum of
  $V_{\mathrm{full}}$ at finite separation
  (the pre-reaction complex).
  When $V_{\mathrm{int}} = 0$ identically (truly
  non-interacting fragments), this minimum is at the
  asymptotic region and $\phi_{G,H}$ is the identity.
  When $V_{\mathrm{int}} \neq 0$, $\phi_{G,H}$ is a genuine
  non-identity morphism: the physical formation of the
  fragment-encounter complex.
\end{definition}

\begin{remark}[Lax coherence of $\phi_{G,H}$]
\label{rem:lax-coherence}
The associativity coherence for a lax monoidal functor
requires
\[
  \phi_{G \otimes H, K} \circ (\phi_{G,H} \otimes
  \mathrm{id}_K)
  \;=\;
  \phi_{G, H \otimes K} \circ (\mathrm{id}_G \otimes
  \phi_{H,K})
\]
as morphisms $\FV(G) \otimes \FV(H) \otimes \FV(K) \to
\FV(G \otimes H \otimes K)$.
For our three-fragment reactive system $G \sqcup H \sqcup
K$, both sides correspond to the gradient-flow descent
on the three-body PES
$V_{\mathrm{full}}^{(GHK)}$ from the fully asymptotic
region to the three-body encounter complex, and they
agree by the symmetric decomposition
\[
  V_{\mathrm{full}}^{(GHK)}
  = V_G + V_H + V_K +
  V_{\mathrm{int}}^{(GH)} + V_{\mathrm{int}}^{(GK)} +
  V_{\mathrm{int}}^{(HK)} + V_{\mathrm{int}}^{(GHK)},
\]
where $V_{\mathrm{int}}^{(AB)}$ is the pairwise
interaction and $V_{\mathrm{int}}^{(GHK)}$ the genuine
three-body correction.
Both orders of association yield the same sum.
The unit coherence $\phi_{G, *} = \mathrm{id}_{\FV(G)}$
holds because $V_{\mathrm{int}}(G, *) = 0$ (no
interaction with the vacuum system).
\end{remark}

\begin{remark}[Why $\FV$ is lax, not strict]
\label{rem:FV-lax}
  The critical distinction from all lower-level decorator
  functors ($\FH$, $\FS$, $F_P$) is that $\FV$ is
  \emph{lax} monoidal.

  At $\Lk_1$--$\Lk_3$, the decorator functors are
  \emph{strict} monoidal: $\FH(G \otimes H) = \FH(G) +
  \FH(H)$ (the enthalpy of a combined system is the sum
  of the parts) because $\FH$ reads off global conserved
  quantities unaffected by intermolecular interactions.

  At $\Lk_5$, strictness would require
  $\FV(G \otimes H) = \FV(G) \otimes \FV(H)$, i.e., the
  full PES equals $V_G + V_H$ on the nose.
  But this is the non-interacting approximation
  (Layer~1), which cannot produce a transition state.
  The Layer~2 BO PES has $V_{\mathrm{full}} = V_G + V_H +
  V_{\mathrm{int}}$ whenever fragments interact; the
  laxator $\phi_{G,H}$ is the physical correction
  witnessing this interaction.

  In terms of the tower structure: the laxator is the
  first occurrence in the tower of a \emph{non-trivial
  structural morphism} witnessing the breakdown of
  additivity.
  $\FV$ is the first functor in the tower whose monoidal
  structure is genuinely lax.
\end{remark}

\begin{remark}[What computational methods provide $\FV$,
  and their position in $\OrbMorse$]
\label{rem:FV-methods}
Unlike $\FH$, $\FS$, and $\FP$, the functor $\FV$
\emph{cannot} be specified by a finite list of numbers:
$V_G: \Ce(G) \to \RR$ is an infinite-dimensional datum.
In practice it is approximated by one of four
progressively coarser methods, each occupying a
well-defined position relative to $\OrbMorse$.

\begin{enumerate}[label=(\roman*)]
  \item \textbf{Exact BO PES (full CI):}
    the ground-state PES defined by
    $V(\mathbf{R}) = E_0(\mathbf{R}) +
    V_{\mathrm{nn}}(\mathbf{R})$, where $E_0$ is the
    lowest eigenvalue of the exact electronic
    Hamiltonian and $V_{\mathrm{nn}}$ is the
    nuclear--nuclear Coulomb repulsion.
    This is the Layer~2 datum.  For fixed lift choices
    (electronic-structure theory, conformer, TS, IRC
    branch) it is well-defined
    (Theorem~\ref{thm:existence-L5} below), but
    computationally inaccessible for all but the
    smallest systems.  It defines the exact object of
    $\OrbMorse$ that $\FV$ targets.

  \item \textbf{Coupled-cluster CCSD(T):}
    an approximation $\tilde{V}^\text{CC} \approx V$
    that truncates the cluster expansion at doubles
    with perturbative triples~\cite{HelgakerJorgensenOlsen2000}.
    This is the ``gold standard'' of computational
    chemistry, recovering $>99.9\%$ of the correlation
    energy for most closed-shell molecules.
    Its critical-point structure (minima, saddle points)
    agrees with the exact PES to within the triples
    error.
    In $\OrbMorse$: $\tilde{V}^\text{CC}$ defines an
    \emph{approximate} Morse triple that is a
    perturbation of the exact Layer~2 object; the
    Morse structure (and hence the IRC and $\Ea$) is
    preserved up to the truncation error.

  \item \textbf{Density functional theory (DFT):}
    $\tilde{V}^{\mathrm{DFT}} = E_\text{Kohn-Sham}$,
    the energy functional of an auxiliary non-interacting
    system with an approximate exchange-correlation
    functional~\cite{HelgakerJorgensenOlsen2000}.
    DFT is computationally efficient and typically
    accurate for equilibrium geometries and moderate
    barriers; it can fail for dispersion-dominated
    interactions (e.g.\ van der Waals complexes) and for
    open-shell transition metals.
    In $\OrbMorse$: $\tilde{V}^{\mathrm{DFT}}$ is an
    approximate object in which the Morse structure may
    differ qualitatively from the exact PES when the
    exchange-correlation functional is poor near saddle
    points.

  \item \textbf{Classical force fields
    (GAFF~\cite{WangGAFF2004},
    FF19SB~\cite{TianFF19SB2019},
    CHARMM~\cite{brooks2009charmm},
    OPLS~\cite{jorgensen1988opls, jorgensen1996development}):}
    an analytic approximation
    $\tilde{V}^{\mathrm{FF}} = \sum_{\mathrm{terms}}
    k_i(r_i - r_i^0)^2 + \cdots$ fitting $V$ as a
    sum of local bonded and non-bonded terms with
    empirical parameters.
    Force fields are computationally very cheap and
    applicable to large systems (proteins, membranes),
    but they are parameterised near equilibrium and
    generally \emph{cannot reproduce transition states}:
    the IRC and the saddle-point Hessian of
    $\tilde{V}^{\mathrm{FF}}$ are not accurate.
    In $\OrbMorse$: force fields produce an object
    $(\Ce(G), \tilde{V}^{\mathrm{FF}}, g)$ that is an
    accurate Morse triple near the reactant minimum
    but whose saddle-point structure is unreliable.
    They are therefore accurate as Layer~1 (structure,
    thermodynamics of isolated systems) but not as
    Layer~2 (reaction barriers and IRCs).
\end{enumerate}

In the tower language: methods (i) and (ii) produce
Layer~2 objects in the exact $\OrbMorse$;
methods (iii) and (iv) produce Layer~2 objects only
approximately.
Machine-learning force fields
(NequIP~\cite{Batzner2022},
MACE~\cite{BatatiaEtAl2022},
SchNet~\cite{SchuttEtAl2018},
SO3LR~\cite{Kabylda2025SO3LR}, etc.)
trained on CCSD(T) or DFT data are parameterised
approximations to method (i): they live in
$\Lk_5^{\mathrm{Para}}$, the Para shadow of $\Lk_5$
(see the Para chapter), as lax $M_{E(3)}$-algebra
morphisms that approximate $\FV$.
\end{remark}

\subsubsection{Definition and structure of \texorpdfstring{$\Lk_5(P)$}{L5(P)}}
\label{sec:L5-L5def}

With $\OrbMorse$ established as a strict SMC
(Proposition~\ref{prop:orbmorse-smc}) and $\FV$ defined
as a lax monoidal functor into it
(Definition~\ref{def:FV}), the geometric level is now
fully specified.
The construction is the pair $(\Lk_{4.5}(P), \FV)$: the
$G^*$-equivariant SMC from $\Lk_{4.5}$ together with its geometric decoration.
Existence of $\FV$ satisfying the Layer~2 conditions is
established below, with explicit acknowledgement of the
ambiguities (conformer choice, TS choice, IRC branch,
theory level) inherent to the construction.

\begin{theorem}[Existence of $\FV$ satisfying Layer~2]
\label{thm:existence-L5}
  For any Petri net $P$ with species set $S$, given
  consistent choices of (i) a level of electronic-structure
  theory defining $\hat{H}_{\mathrm{el}}$, (ii) a
  reference conformer at each molecular graph
  $(G, \sigma)$, (iii) a transition-state geometry for
  each DPO derivation, and (iv) a choice of IRC branch
  from each TS, there exists a lax monoidal functor
  \[
    \FV: \Lk_{4.5}(P) \longrightarrow \OrbMorse
  \]
  satisfying the Layer~2 conditions of
  Definition~\ref{def:bo-section}:
  (a) $V(\mathbf{R}) = E_0(\mathbf{R}) +
  V_{\mathrm{nn}}(\mathbf{R})$
  (BO derivation plus nuclear--nuclear repulsion), and
  (b) $\sigma_0$ is a smooth section of $\Hel$ on the
  relevant CI-free open subset of $\Ce(G, \sigma)$.

  The data are unique up to:
  \begin{itemize}
    \item the global $\ZZ_2$ sign ambiguity in the real
      gauge of $\sigma_0$
      (Proposition~\ref{prop:berry-trivial}); this affects
      $\sigma_0$ as a sign but not $V$, $g$, or the IRC;
    \item the chosen IRC branch from each TS (two
      steepest-descent half-curves);
    \item reparametrisation of each IRC.
  \end{itemize}
  The choices (i)--(iv) are not part of the
  functorial data; they constitute a \emph{lift} of
  $\Lk_{4.5}(P)$ to $\Lk_5(P)$ in the sense of
  Definition~\ref{def:L5}.  Different choices give
  different functors $\FV$, each satisfying (a)--(b);
  the categorical content of $\Lk_5(P)$ is the
  decorated category $(\Lk_{4.5}(P), \FV)$ for a fixed
  such lift.
\end{theorem}

\begin{proof}
\textbf{Existence on objects.}
Fix $(G, \sigma) \in \Lk_{4.5}(P)$ with $n$ atoms and
nuclear charges $\{Z_k\}$.
The electronic Hamiltonian
$\hat{H}_{\mathrm{el}}(\mathbf{R})$ is a well-defined
self-adjoint operator on $\Hel(\mathbf{R})$ for each
$\mathbf{R}$ in the non-coalescence subset of
$\Ce(G, \sigma)$ (the Coulomb singularities are
Kato-bounded by the kinetic energy with relative bound
zero, so the Kato--Rellich theorem gives
self-adjointness on the common Sobolev-type
domain~\cite{Kato1966}).
The spectrum is bounded below by the Lieb--Thirring
inequality (stability of matter for Coulomb
systems)~\cite{LiebThirring1975}, and the existence of
bound states --- in particular an isolated ground-state
eigenvalue $E_0(\mathbf{R})$ at the bottom of the
spectrum --- follows from Zhislin's
theorem~\cite{Zhislin1960} for neutral molecules (and
standard extensions for molecular ions under the usual
hypotheses).

Let $U(G, \sigma) \subseteq \Ce(G, \sigma)$ denote the
CI-free open subset on which $E_0(\mathbf{R})$ is
isolated from the rest of the spectrum (Layer~2(b)).
By Kato's perturbation theory (smooth dependence of
isolated eigenvalues and eigenprojections on the operator
parameter)~\cite{Kato1966, Teufel2003}, both
$E_0(\mathbf{R})$ and the eigenprojection $P_0(\mathbf{R})$
vary smoothly with $\mathbf{R}$ on $U(G, \sigma)$.
Hence $E_0 \in C^\infty(\Ce(G, \sigma))$, and
$V_G := E_0 + V_{\mathrm{nn}} \in C^\infty(\Ce(G, \sigma))$
since $V_{\mathrm{nn}}$ is a smooth function of the
nuclear coordinates away from coalescence.  The triple
$(\Ce(G, \sigma), V_G, g_G)$ is a well-defined object
of $\OrbMorse$ under the Morse assumption.

\textbf{Existence on morphisms (elementary).}
For an elementary DPO derivation $d: (G, \sigma)
\Rightarrow (H, \tau)$, the reactive system has joint
configuration orbifold $\Ce(G \sqcup H)$ with full
interacting PES $V_{\mathrm{full}}$.  Fix an index-1
saddle $\mathbf{R}_{\mathrm{TS}}$ (lift choice (iii))
and a branch (lift choice (iv)).  The gradient-flow ODE
has a locally unique solution on the open set
$\{\nabla_g V_{\mathrm{full}} \neq 0\}$ by
Picard--Lindel\"of; the germ at $\mathbf{R}_{\mathrm{TS}}$
is determined by the simple negative eigenvector of
$\nabla^2_g V_{\mathrm{full}}$ at the
saddle~\cite{MillerHandyAdams1980,FukuiIRC1981}.  These
together yield an elementary geometric channel
$\FV(d) = (V_{\mathrm{full}}, \mathbf{R}_{\mathrm{TS}},
\gamma)$.  Composite DPO derivations decompose into
elementary ones and map to strings of elementary
channels by concatenation.

\textbf{Functoriality.}
The identity DPO derivation (no bond changes) maps to
the identity morphism in $\OrbMorse$, which is the
empty channel string at the corresponding object.
Composition of DPO derivations maps to concatenation of
channel strings: a two-step reaction
$(G, \sigma) \Rightarrow (H, \tau) \Rightarrow (K, \rho)$
maps to the string $(c_1, c_2)$ where $c_i$ is the
elementary channel of the $i$-th step.  Associativity
and unit laws hold by definition of string concatenation (Proposition~\ref{prop:orbmorse-smc}).

\textbf{Lax monoidality.}
Verified in Definition~\ref{def:FV} and
Remark~\ref{rem:lax-coherence}.

\textbf{Determination up to the listed ambiguities.}
Fix the lift data (i)--(iv).  At each $\mathbf{R}$ in
the relevant CI-free open subset, the ground-state
eigenvalue $E_0(\mathbf{R})$ is uniquely determined by
$\hat{H}_{\mathrm{el}}(\mathbf{R})$, and the
wavefunction $\sigma_0(\mathbf{R})$ is unique up to a
complex phase.  The real-gauge construction of
Proposition~\ref{prop:berry-trivial} fixes the phase up
to a global sign on simply-connected components.  Hence
$V_G(\mathbf{R}) = E_0(\mathbf{R}) + V_{\mathrm{nn}}
(\mathbf{R})$ is a well-defined real-valued function
on the chosen accessible open subset.

For the IRC, fix a choice of TS and a branch from it.
Picard--Lindel\"of gives local existence and uniqueness
on the open set where $\nabla_g V_{\mathrm{full}} \neq 0$;
the descent from the TS is determined by the unique
simple negative eigenvector of $\nabla^2_g V_{\mathrm{full}}$
at $\mathbf{R}_{\mathrm{TS}}$.  Reparametrisation
ambiguity is the only remaining freedom along the IRC.

The functor $\FV$ is therefore determined by the lift
data up to the ambiguities listed in the theorem
statement.
\qedhere
\end{proof}

\begin{remark}[Morse genericity]
\label{rem:morseness}
The Morse condition on $V_G$ (non-degenerate critical
points) is not guaranteed by the BO construction but
holds for all chemically relevant molecular systems at
all chemically relevant geometries.
Non-Morse critical points do occur in molecular physics
--- notably at \emph{catastrophe} geometries where two
critical points merge (fold catastrophe) or a saddle
degenerates (cusp catastrophe), and at conical
intersections themselves --- but these are non-generic
under parameter variation and belong to higher tower
levels ($\Lk_6$ for CIs, $\Lk_7$ for anharmonic and
tunneling phenomena that probe non-Morse regions).
For all critical points on the accessible region
$\Ce^{\mathrm{acc}}(G)$ under standard thermal
conditions, Morseness holds.
\end{remark}

\begin{definition}[Geometric level $\Lk_5(P)$]
\label{def:L5}
  The \emph{geometric level} is the pair
  \[
    \Lk_5(P)
    \;:=\;
    \bigl(\,\Lk_{4.5}(P),\;\FV\,\bigr),
  \]
  where $\FV: \Lk_{4.5}(P) \to \OrbMorse$ is
  the lax monoidal functor of
  Theorem~\ref{thm:existence-L5} satisfying the
  Layer~2 conditions: $V$ is BO-derived from
  $\hat{H}_{\mathrm{el}}$ (plus $V_{\mathrm{nn}}$), and
  $\sigma_0$ is a smooth section on the relevant
  CI-free open subset of $\Ce(G, \sigma)$.

  The \emph{forgetful functor}
  $U_5: \Lk_5(P) \to \Lk_{4.5}(P)$
  drops $\FV$, retaining only the $G^*$-equivariant
  DPO structure.
\end{definition}

\begin{proposition}[Categorical structure of $\Lk_5(P)$]
\label{prop:L5-structure}
  $\Lk_5(P)$ is a $G^*$-equivariant symmetric monoidal
  category (inheriting the structure of $\Lk_{4.5}(P)$)
  equipped with a projection functor
  $\Pi_V: \Lk_5(P) \to \OrbMorse$ extracting the
  geometric decoration (the same lax monoidal data that
  $\FV: \Lk_{4.5}(P) \to \OrbMorse$ assigns, viewed now as
  a projection from the decorated category).
  Its categorical properties:
  \begin{enumerate}[label=(\alph*)]
    \item \textbf{Objects}: chirality-labelled molecular
      graphs $(G, \sigma)$, each carrying the additional
      $\Lk_5$ data $(\Ce(G,\sigma), V_G, g_G)$ via
      $\FV$.
    \item \textbf{Morphisms}: $G^*$-equivariant DPO
      derivations $d: (G,\sigma) \Rightarrow (H,\tau)$,
      each carrying the IRC data
      $(V_{\mathrm{full}}, \mathbf{R}_{\mathrm{TS}},
      \gamma)$ on the joint orbifold $\Ce(G \sqcup H)$
      via $\FV$.
    \item \textbf{Monoidal structure}: the tensor product
      $(G,\sigma) \otimes (H,\tau)$ is the disjoint union
      of molecular graphs (non-interacting system
      asymptotically), with
      $\FV\bigl((G,\sigma)\otimes(H,\tau)\bigr)$ the lax
      monoidal image (full interacting PES when the
      systems react, separated PES otherwise).
    \item \textbf{$G^*$-invariance of $\FV$}: the
      permutation-inversion group $G^*$ acts on
      $\Ce(G,\sigma)$, and the PES $V_G$ is
      $G^*$-invariant: $V_G(g \cdot \mathbf{R}) =
      V_G(\mathbf{R})$ for all $g \in G^*$.
      Consequently, $\FV$ factors through the orbit space
      and is well-defined on the $G^*$-equivariant objects
      of $\Lk_{4.5}(P)$.
  \end{enumerate}
  The automorphism exact sequence
  \[
    1 \;\to\; \ker\varphi_5 \;\to\;
    \Aut(\Lk_5(P))
    \;\xrightarrow{\;\varphi_5\;}
    \Aut(\Lk_{4.5}(P))
    \;\to\; \coker(\varphi_5) \;\to\; 1
  \]
  has non-trivial cokernel: the swap automorphism
  $r_H \leftrightarrow r_D$ of the forcing pair
  (\S\ref{sec:L5-forcing-in}) represents an element of
  $\coker(\varphi_5)$, since the mass-weighted metric $g$
  (which distinguishes H from D by mass) is data in
  $\Lk_5(P)$ that is absent at $\Lk_{4.5}(P)$.
\end{proposition}

\begin{proof}
Items (a)--(c) follow directly from
Definitions~\ref{def:FV} and~\ref{def:L5}.

Item (d): for each $g = (\pi, \epsilon) \in G^*$ acting
on nuclear positions, there is a corresponding unitary
$U_g$ on the electronic Hilbert space (permuting
electronic coordinates when $\pi$ permutes nuclei of
the same element, and inverting electronic coordinates
when $\epsilon = E^*$) such that
$\hat{H}_{\mathrm{el}}(g \cdot \mathbf{R}) = U_g\,
\hat{H}_{\mathrm{el}}(\mathbf{R})\,U_g^{-1}$.
Unitary conjugation preserves eigenvalues, so
$E_0(g \cdot \mathbf{R}) = E_0(\mathbf{R})$ for all
$g \in G^*$ and all $\mathbf{R} \in \RR^{3n}$.
Similarly $V_{\mathrm{nn}}$ is $G^*$-invariant since
distances $|\mathbf{R}_A - \mathbf{R}_B|$ are preserved
by isometries.  Hence the BO PES
$V = E_0 + V_{\mathrm{nn}}$ is $G^*$-invariant and
descends to a well-defined invariant function on
$\Ce(G,\sigma) = \RR^{3n}/(SE(3) \times \Aut_\mu(G))$.

For the cokernel: the swap $g = (r_H \leftrightarrow r_D)$
is an automorphism of the categorical data of
$\Lk_{4.5}(P)$ (graphs, DPO mechanisms, chirality labels
all coincide since H and D have the same atomic number).
Empirical numerical decorations at $\Lk_3$ differ
($k_H \neq k_D$) but enter as attached data rather than
as part of the categorical structure of the lower
levels.  The swap is \emph{not} an automorphism of
$\Lk_5(P)$: the mass-weighted metrics $g_H$ and $g_D$
differ since $m_H \neq m_D$, so the Morse triples
$(\Ce(G), V, g_H)$ and $(\Ce(G), V, g_D)$ are distinct
objects of $\OrbMorse$ (Remark~\ref{rem:metric-descent}).
Hence $[r_H \leftrightarrow r_D]$ represents a
non-trivial element of $\coker(\varphi_5)$, in the
pointed-set sense made precise in  \S\ref{sec:L5-forcing-in}.
\end{proof}

\begin{insightbox}[The tower through $\Lk_5$: six
  levels, four extension types]
\label{box:tower-L5}
\begin{center}
\small
\includegraphics{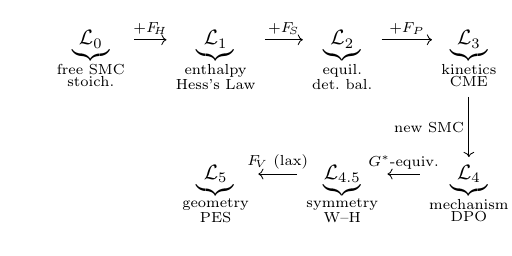}
\end{center}
Reading the diagram: the decorator chain
$\Lk_0 \to \Lk_3$ adds one scalar functor per step
(all strict monoidal).
The structural extension $\Lk_3 \to \Lk_4$ replaces the
underlying category with a free SMC on DPO spans.
The symmetry enrichment $\Lk_4 \to \Lk_{4.5}$ adds
$G^*$-equivariance.
The geometric decoration $\Lk_{4.5} \to \Lk_5$ adds
an infinite-dimensional datum ($V$, $g$) via a
\emph{lax} monoidal functor --- the first non-strict
functor in the tower.

The four extension types are, in order: decorator
($\Lk_0 \to \Lk_3$), structural ($\Lk_3 \to \Lk_4$),
symmetry enrichment ($\Lk_4 \to \Lk_{4.5}$), and
geometric decoration ($\Lk_{4.5} \to \Lk_5$, lax).
\end{insightbox}

%% file: chapters/L5/l5_tst.tex
\subsection{Transition-state theory and
  \texorpdfstring{$\Lk_3$}{L3}--\texorpdfstring{$\Lk_5$}{L5}
  inter-level coherence}
\label{sec:L5-tst}


Eyring's transition-state theory
(TST)~\cite{Eyring1935,EvansPolanyi1935,Wigner1938}
is the central bridge between kinetics and molecular
geometry.
Its modern canonical form~\cite{TGK1996,PollakTalkner2005,
FMKT2006} expresses the rate constant of an elementary
reaction as a thermal flux through a dividing surface on
the Born--Oppenheimer PES:
\begin{equation}
\label{eq:tst-canonical}
  k^\mathrm{TST}(r,T)
  \;=\;
  \frac{k_BT}{h}\cdot
  \frac{Q^\ddagger(r,T)}{Q_\mathrm{reac}(r,T)}\cdot
  \exp\!\left(-\frac{V^\ddagger(r)}{k_BT}\right).
\end{equation}
Here $V^\ddagger(r) = V_\mathrm{full}(\mathbf{R}_\mathrm{TS})
- V_\mathrm{full}(\mathbf{R}_\mathrm{min,react})$ is the
bare electronic barrier, extracted from the gradient-flow
cobordism $\FV(r) \in \OrbMorse$ (\S\ref{sec:L5-def}).
$Q^\ddagger$ is the partition function at $\mathbf{R}_\mathrm{TS}$
with the reaction-coordinate mode removed ($3N^\ddagger-7$
real vibrational modes plus rotational and translational
factors), $Q_\mathrm{reac}$ is the reactant partition
function, and the full rate is $k^\mathrm{exact}(r,T) =
\kappa(r,T)\cdot k^\mathrm{TST}(r,T)$ where $\kappa$ is a
transmission coefficient accounting for dividing-surface
recrossing~\cite{Wigner1938,PollakTalkner2005} and
semiclassical tunneling~\cite{TGK1996,BaoTruhlar2017}.
The $k_BT/h$ prefactor arises from the flux-integral
factorisation of classical TST~\cite{PollakTalkner2005}.


Four distinct barrier heights recur in
\eqref{eq:tst-canonical} and its thermodynamic rewrites
and must be kept separate.
Each is tagged below with the \emph{minimal} $\Lk_5$
data (extractable from $(\FV, g)$) required to compute
it:
\begin{itemize}
  \item $V^\ddagger$ (\emph{bare electronic} /
    classical barrier): difference of PES values at
    two critical points.
    Requires only $\FV(r)$; independent of $g$.
  \item $\Delta V_a^G = V^\ddagger +
    [E_\mathrm{ZPE}(\mathrm{TS}) -
    E_\mathrm{ZPE}(\mathrm{reac})]$
    (\emph{vibrationally adiabatic ground-state}
    barrier): adds harmonic zero-point energies
    computed from the eigenvalue spectrum of
    $\nabla^2_g V$ at the two critical points.
    Requires $(\FV, g)$.
  \item $\Delta H^\ddagger(T)$ (\emph{activation
    enthalpy}): adds finite-$T$ thermal populations
    of all bound modes.
    Requires $(\FV, g, T)$, plus RRHO rotational and
    translational factors.
  \item $\Delta G^\ddagger(T) = \Delta H^\ddagger -
    T\Delta S^\ddagger$ (\emph{free energy of
    activation}): adds the activation entropy from
    partition-function ratios; gives the
    Eyring--Polanyi form
    $k = \kappa\,(k_BT/h)\,(c^\circ)^{1-M}
    \exp(-\Delta G^\ddagger/RT)$, where $M$ is the
    molecularity and $(c^\circ)^{1-M}$ is the
    standard-state factor~\cite{FMKT2006}.
    Requires $(\FV, g, T)$ plus the RRHO partition
    functions.
\end{itemize}
\emph{All four are $\Lk_5$ data}: none requires
structure beyond $(\FV, g)$ and temperature.
$V^\ddagger$ is the only one of the four extractable
from $\FV$ alone; the three thermally corrected
barriers require the metric to compute mass-weighted
frequencies, and hence live at full $\Lk_5$.

\begin{remark}[Notation: $V^\ddagger$ versus $\Ea$]
\label{rem:barrier-notation}
Earlier chapters (\S\ref{sec:L5-forcing-in},
\S\ref{sec:L5-pes}) use the macro $\Ea$ for
``the activation barrier.''
From this section onward, we use $V^\ddagger$ for the
bare electronic barrier, which is what is extracted
from $\FV$ directly.
Thus $\Ea \equiv V^\ddagger$ when writing naive TST;
when a thermally or vibrationally corrected barrier is
meant, we write $\Delta V_a^G$, $\Delta H^\ddagger$,
or $\Delta G^\ddagger$ explicitly.
The experimental \emph{Arrhenius} activation energy
$E_A \equiv -R\,d\ln k / d(1/T)$ relates to $\Delta
H^\ddagger$ by $E_A = \Delta H^\ddagger + RT$
(unimolecular or condensed phase) or
$E_A = \Delta H^\ddagger + 2RT$ (gas-phase
bimolecular)~\cite{FMKT2006}; $E_A$ is an experimental
quantity and is not directly $\Lk_5$ data.
\end{remark}


The \emph{naive Eyring equation},
\begin{equation}
\label{eq:eyring-naive}
  k_r^\mathrm{naive}(T)
  \;:=\;
  \frac{k_BT}{h}\,
  \exp\!\left(-\frac{V^\ddagger(r)}{RT}\right),
\end{equation}
sets $\kappa = 1$, $Q^\ddagger/Q_\mathrm{reac} = 1$, and
ignores zero-point corrections simultaneously.
It is an order-of-magnitude estimate rarely exact: at
$T=298$~K the prefactor $k_BT/h \approx 6.21\times
10^{12}\;\mathrm{s}^{-1}$ agrees with observed
unimolecular prefactors only up to factors of
$10^{\pm 3}$~\cite{TGK1996}, and for bimolecular gas-phase
reactions \eqref{eq:eyring-naive} is typically wrong by
$10^{-5}$--$10^{-10}$: such reactions have
$\Delta S^\ddagger \approx -80$ to
$-170\;\mathrm{J\,K^{-1}\,mol^{-1}}$ (one combined TS
built from two free reactants)~\cite{FMKT2006}.
The naive form \eqref{eq:eyring-naive} is defensible
\emph{only} for unimolecular rearrangements with
reactant-like tight TS~\cite{TGK1996}.

Despite its approximate status, \eqref{eq:eyring-naive}
captures the essential $\Lk_3$--$\Lk_5$ tower structure:
a single $\Lk_3$ datum (the rate constant $F_P(r)$) is
predicted from a single $\Lk_5$ datum (the bare barrier
$V^\ddagger(r)$).
The residual
\[
  \Delta_\mathrm{TST}(r,T)
  \;:=\;
  \ln F_P(r) - \ln k_r^\mathrm{naive}(T)
\]
measures the obstruction to commutativity of
diagram~\eqref{diag:L5-motivation} under the naive
Eyring map, and decomposes into tower-diagnostic
contributions:
\begin{equation}
\label{eq:delta-tst-decomp}
  \Delta_\mathrm{TST}(r,T)
  \;=\;
  \underbrace{\ln\frac{Q^\ddagger(r,T)}{Q_\mathrm{reac}(r,T)}}_{
    \text{prefactor: ZPE + entropy + rot/trans}}
  \;+\;
  \underbrace{\ln\kappa(r,T)}_{
    \text{transmission: recrossing + tunneling}}.
\end{equation}
Both terms are computable from $\Lk_5$ data $(\FV, g)$
and semiclassical refinements on the BO PES
(\S\ref{sec:tunneling-attribution});
the residual vanishes when the full TST formula
\eqref{eq:tst-canonical} is used in place of
\eqref{eq:eyring-naive}, up to anharmonic and
deep-tunneling corrections considered in
Remark~\ref{rem:tunneling-attribution}.

\subsubsection{The Eyring equation as a coherence condition}

\begin{definition}[TST coherence conditions]
\label{def:tst-coherence}
  Let $r \in \mathrm{Mor}(\Lk_{4.5}(P))$ be a reaction,
  $T > 0$ a temperature, and
  \[
    \FV(r) \;=\;
    \bigl(V_\mathrm{full}^{(r)},\;
    \mathbf{R}_\mathrm{TS}^{(r)},\;
    \gamma^{(r)}\bigr)
    \;\in\;\mathrm{Mor}(\OrbMorse)
  \]
  its gradient-flow cobordism
  (Definition~\ref{def:FV}).
  Define extraction maps
  \begin{align*}
    V^\ddagger(r)
    &\;:=\;
    V_\mathrm{full}^{(r)}\bigl(\mathbf{R}_\mathrm{TS}^{(r)}\bigr)
    - V_\mathrm{full}^{(r)}\bigl(\mathbf{R}_\mathrm{min,react}^{(r)}\bigr),
    \\
    \{\omega_i(r,X)\}_{i}
    &\;:=\;
    \text{(non-negative) eigenfrequencies of }
    \nabla^2_g V_\mathrm{full}^{(r)}
    \text{ at } \mathbf{R}_X^{(r)},
  \end{align*}
  for $X \in \{\mathrm{min,react},\,\mathrm{TS},\,
  \mathrm{min,prod}\}$.
  From these, assemble $Q^\ddagger(r,T)$ and
  $Q_\mathrm{reac}(r,T)$ via the standard RRHO
  expressions~\cite{Ochterski2000,
  HelgakerJorgensenOlsen2000}.
  The pair $(F_P, \FV)$ satisfies the:
  \begin{enumerate}[label=(\alph*)]
    \item \emph{Naive TST coherence condition} at
      $(r,T)$ if
      \begin{equation}
      \label{eq:tst-naive}
        F_P(r)
        \;\stackrel{\text{?}}{=}\;
        \frac{k_BT}{h}\,
        \exp\!\left(-\frac{V^\ddagger(r)}{RT}\right).
      \end{equation}
    \item \emph{Full TST coherence condition} at
      $(r,T)$ if
      \begin{equation}
      \label{eq:tst-full}
        F_P(r)
        \;\stackrel{\text{?}}{=}\;
        \kappa(r,T)\cdot\frac{k_BT}{h}\cdot
        \frac{Q^\ddagger(r,T)}{Q_\mathrm{reac}(r,T)}\cdot
        \exp\!\left(-\frac{V^\ddagger(r)}{k_BT}\right),
      \end{equation}
      where $\kappa(r,T)$ is computed from the PES via
      recrossing and semiclassical-tunneling
      corrections (\S\ref{sec:tunneling-attribution}).
  \end{enumerate}
  The \emph{TST residual} at $(r,T)$ is
  \[
    \Delta_\mathrm{TST}(r,T)
    \;:=\;
    \ln F_P(r)
    - \ln\!\left[\frac{k_BT}{h}\,
    e^{-V^\ddagger(r)/RT}\right],
  \]
  which measures the obstruction to commutativity of
  diagram~\eqref{diag:L5-motivation} under the naive
  map~\eqref{eq:tst-naive}.
\end{definition}

\begin{remark}[Status of the two conditions]
\label{rem:tst-status}
  The naive condition~\eqref{eq:tst-naive} essentially
  never holds exactly.
  The full condition~\eqref{eq:tst-full} holds up to
  anharmonic and deep-tunneling corrections for
  essentially all elementary reactions where the BO
  approximation itself holds, and becomes exact in the
  limit of a classical harmonic BO PES with no
  recrossing and no tunneling~\cite{TGK1996,
  PollakTalkner2005}.
  The decomposition \eqref{eq:delta-tst-decomp}
  attributes the residual to a \emph{prefactor part}
  ($\ln Q^\ddagger/Q_\mathrm{reac}$: ZPE plus entropy
  plus rotational/translational ratios) and a
  \emph{transmission part} ($\ln\kappa$: recrossing
  plus semiclassical tunneling).
  Both parts are $\Lk_5$-computable.
  A genuine $\Lk_7$ signal --- a residual not explained
  by PES-derived semiclassics --- arises only when
  multidimensional nuclear-wavefunction effects
  dominate; see Remark~\ref{rem:tunneling-attribution}
  and \S\ref{sec:L5-forcing-out}.
\end{remark}

\begin{mathbox}[Three inter-level coherence conditions
  in the tower]
\label{box:coherence}
The tower contains three conditions linking data at
non-adjacent levels.
Each condition is a statement that a lower-level
datum (at level $j$) is \emph{constrained} by
higher-level data (at level $k > j$), not that the
lower datum is ``derivable from'' the lower level.
\begin{center}
\renewcommand{\arraystretch}{1.5}
\begin{tabular}{@{}l p{6.8cm} p{4.2cm}@{}}
  \hline
  \textbf{Levels} &
  \textbf{Condition} &
  \textbf{Physical content}\\
  \hline
  $\Lk_2$--$\Lk_3$ &
  The $\Lk_3$ rate-constant ratio satisfies
  $F_P(r) / F_P(r^\dagger) = \exp(-\Delta G^\circ(r)/RT)$,
  where $\Delta G^\circ(r)$ is the $\Lk_2$ datum.
  &
  Detailed balance: the ratio of rate constants is
  constrained to equal the equilibrium constant.\\[0.4em]
  $\Lk_1$--$\Lk_5$ &
  The $\Lk_1$ reaction enthalpy satisfies
  $\FH(r) \approx \Delta V_\mathrm{elec}(r) +
  \Delta E_\mathrm{ZPE}(r)$ in the harmonic + ideal-gas
  + BO limit, with $\Delta V_\mathrm{elec}$ and
  $\Delta E_\mathrm{ZPE}$ computed from $\Lk_5$ data
  $(\FV, g)$.
  &
  BO/RRHO decomposition of reaction enthalpy
  (Proposition~\ref{prop:hess-pes-consistency}).\\[0.4em]
  $\Lk_3$--$\Lk_5$ &
  The $\Lk_3$ rate constant satisfies
  $F_P(r) \approx \kappa(k_BT/h)
  (Q^\ddagger/Q_\mathrm{reac})
  e^{-V^\ddagger/k_BT}$, with all right-hand-side
  quantities computed from $\Lk_5$ data.
  &
  Full TST: the rate constant factors through $\FV$
  and its Hessians, plus $\Lk_5$-computable
  semiclassical corrections.\\
  \hline
\end{tabular}
\end{center}

\smallskip
\noindent
All three conditions reside at $\Lk_5$ or below:
the constrained lower-level data are re-expressible in
terms of higher-level structural data.
The $\Lk_2$--$\Lk_3$ condition fails for driven
non-equilibrium systems (detailed balance violated at
the network level).
The $\Lk_1$--$\Lk_5$ condition fails when anharmonic
or non-RRHO contributions to ZPE or thermal energy
dominate (floppy molecules, low-frequency torsions,
high $T$).
The $\Lk_3$--$\Lk_5$ condition fails for
\eqref{eq:tst-naive} routinely but holds for
\eqref{eq:tst-full} with $\Lk_5$-computable
corrections for essentially all reactions where the
BO approximation itself holds.
A genuine $\Lk_7$ signal --- a residual not
explained by $\Lk_5$ semiclassics --- is addressed in
\S\ref{sec:L5-forcing-out}.
\end{mathbox}

\subsubsection{The secondary kinetic isotope effect
  as an \texorpdfstring{$\Lk_5$}{L5} theorem}

The most immediate consequence of TST coherence is that
kinetic isotope effects become \emph{theorems} at
$\Lk_5$.
A kinetic isotope effect (KIE) is the ratio
$k_H/k_D$ of rate constants for reactions differing only
in the isotopic substitution of H by D somewhere in the
substrate.
The canonical formula is the \emph{Bigeleisen
equation}~\cite{BigeleisenMayer1947,Bigeleisen1949,
WolfsbergStern1964,WolfsbergVanHookPaneth2010},
which follows from the full TST expression
\eqref{eq:tst-canonical} by taking the ratio and
invoking the Teller--Redlich product rule to collapse
the translational and rotational partition-function
ratios into products over vibrational frequencies:
\begin{equation}
\label{eq:bigeleisen}
  \frac{k_H}{k_D}
  \;=\;
  \frac{\nu_{H,L}^\ddagger}{\nu_{D,L}^\ddagger}
  \cdot
  \frac{\displaystyle
    \prod_{i=1}^{3N-6}
    \frac{u_i^H}{u_i^D}\,
    \frac{\sinh(u_i^D/2)}{\sinh(u_i^H/2)}
    \bigg|_\mathrm{react}}
  {\displaystyle
    \prod_{j=1}^{3N^\ddagger-7}
    \frac{u_j^{H,\ddagger}}{u_j^{D,\ddagger}}\,
    \frac{\sinh(u_j^{D,\ddagger}/2)}
    {\sinh(u_j^{H,\ddagger}/2)}
    \bigg|_\mathrm{TS}}
  \cdot
  \frac{\kappa_H}{\kappa_D},
\end{equation}
where $u_i \equiv \hbar\omega_i/k_BT$, reactant modes
are indexed $i = 1, \ldots, 3N-6$, TS modes (excluding
the reaction-coordinate mode) are indexed
$j = 1, \ldots, 3N^\ddagger - 7$, and
$\nu_{X,L}^\ddagger$ is the magnitude of the imaginary
frequency of the reaction-coordinate mode at the TS
for isotopologue $X \in \{H, D\}$.
The ratio $\nu_{H,L}^\ddagger/\nu_{D,L}^\ddagger$ is
the only surviving footprint of the reaction coordinate
in the KIE after the Teller--Redlich reduction; it
equals the SWIM (Stern--Wolfsberg imaginary-mode) limit
$\lim_{T\to\infty}(k_H/k_D)$ and is a purely
mass-kinematic effect~\cite{WolfsbergVanHookPaneth2010}.

KIEs come in two types with distinct physical origins:

\begin{itemize}
  \item A \emph{primary} KIE arises when the
    isotopically substituted bond is \emph{broken} in
    the rate-determining step: the corresponding
    reactant stretching mode becomes the reaction
    coordinate, so the imaginary-frequency ratio is
    large and the ZPE-dominated low-$T$ limit gives
    $k_H/k_D \gg 1$.
    The Westheimer limit (zero TS ZPE for the
    reaction mode) at 298~K for a C--H stretch
    ($\omega_H \approx 3000$~cm$^{-1}$) is
    $\exp[\tfrac{1}{2}\hbar\omega_H(1-1/\sqrt{2})/RT]
    \approx 6$--$9$~\cite{Westheimer1961}.
  \item A \emph{secondary} KIE arises when the
    isotopic bonds are \emph{not broken} but change
    geometry between reactant and TS.
    The imaginary-frequency ratio $\nu_{H,L}^\ddagger
    /\nu_{D,L}^\ddagger \approx 1$ since the reaction
    coordinate does not involve the isotopic atoms;
    the KIE reduces to a ratio of vibrational-mode
    contributions.
    Per-deuterium secondary KIEs are typically
    $1.02$--$1.15$.
\end{itemize}

For the $\mathrm{S_N2}$ forcing pair of
\S\ref{sec:L5-forcing-in}, the three C--H/D bonds remain
intact; the $k_H/k_D \approx 1.3$ value arises from the
frequency shift of the three C--H/D bending modes as
the carbon geometry deforms from tetrahedral (reactant)
to trigonal-bipyramidal (TS).
For such modes the \emph{Streitwieser cutoff
approximation}~\cite{Streitwieser1958,WolfsbergStern1964}
--- modes not involving the substituted atom cancel
exactly by Teller--Redlich, leaving only modes with
non-negligible amplitude at the substituted atoms ---
reduces \eqref{eq:bigeleisen} to a product over a small
set of H/D-local modes.
The following proposition derives the secondary KIE
in closed form and exposes its structure as an $\Lk_5$
theorem.

\begin{proposition}[Secondary KIE as an $\Lk_5$ theorem]
\label{prop:kie}
  Let $r_H$ and $r_D$ be the two reactions of the
  forcing pair (\S\ref{sec:L5-forcing-in}), differing
  only in the isotopic substitution H$\to$D at the
  three $\alpha$-positions of the substrate, with the
  substituted bonds unbroken throughout.
  Under
  \begin{enumerate}[label=(\roman*)]
    \item the full TST coherence condition
      \eqref{eq:tst-full} with $\kappa_H = \kappa_D$
      (the secondary-KIE assumption: no
      isotope-dependent transmission correction);
    \item the harmonic approximation: $\nabla^2_g V$
      positive-definite at $\mathbf{R}_\mathrm{min}$
      and of index 1 at $\mathbf{R}_\mathrm{TS}$;
    \item $\nu_{H,L}^\ddagger \approx
      \nu_{D,L}^\ddagger$ (secondary KIE: the
      reaction-coordinate mode does not involve the
      isotopic atoms);
    \item the Streitwieser cutoff: modes not involving
      the H/D atoms cancel exactly by
      Teller--Redlich;
  \end{enumerate}
  the secondary KIE is the $\Lk_5$ theorem:
  \begin{equation}
  \label{eq:secondary-kie}
    \ln\frac{k_H}{k_D}
    \;=\;
    \sum_{i \in \mathcal{M}(\mathrm{react})}
    \!\ln\frac{\sinh(u_i^D/2)}{\sinh(u_i^H/2)}
    \;-\;
    \sum_{j \in \mathcal{M}(\mathrm{TS})}
    \!\ln\frac{\sinh(u_j^{D,\ddagger}/2)}
    {\sinh(u_j^{H,\ddagger}/2)}
    \;+\; \mathcal{O}(u^H/u^D\text{ prefactor}),
  \end{equation}
  where $\mathcal{M}(X)$ is the set of normal modes at
  geometry $X \in \{\mathrm{react},\mathrm{TS}\}$ with
  non-negligible amplitude at the substituted atoms,
  and the prefactor correction $\mathcal{O}(u^H/u^D)$
  captures the $u_i^H/u_i^D$ factors of
  \eqref{eq:bigeleisen} restricted to
  $\mathcal{M}(X)$.
  In the ZPE-dominated regime ($u \gg 1$; satisfied
  at 298~K for $\omega \gtrsim 1000\;\mathrm{cm}^{-1}$),
  \eqref{eq:secondary-kie} further reduces to
  \begin{equation}
  \label{eq:secondary-kie-classical}
    \frac{k_H}{k_D}
    \;\approx\;
    \exp\!\left(\frac{\Delta^2 E_\mathrm{ZPE}}{RT}\right),
    \qquad
    \Delta^2 E_\mathrm{ZPE}
    \;=\;
    \Delta E_\mathrm{ZPE}^\mathrm{react}
    - \Delta E_\mathrm{ZPE}^\mathrm{TS},
  \end{equation}
  where $\Delta E_\mathrm{ZPE}^X := E_\mathrm{ZPE}^{H,X}
  - E_\mathrm{ZPE}^{D,X}$ is the harmonic ZPE difference
  between isotopologues at geometry $X$.
\end{proposition}

\begin{proof}
\textbf{Step 1: Reduce the full Bigeleisen formula.}
Start from \eqref{eq:bigeleisen}.
Assumption (i) gives $\kappa_H/\kappa_D = 1$.
Assumption (iii) gives $\nu_{H,L}^\ddagger /
\nu_{D,L}^\ddagger \approx 1$, so the imaginary-frequency
prefactor is unity.
Assumption (iv) (Streitwieser cutoff): by the
Teller--Redlich product rule, for any normal mode with
negligible amplitude at the isotopically substituted
atoms, the mass-weighted Hessian is independent of
isotope, $\omega_i^H = \omega_i^D$, and the
corresponding factors in numerator and denominator of
\eqref{eq:bigeleisen} equal unity~\cite{WolfsbergStern1964}.
Only modes in $\mathcal{M}(\mathrm{react})$ and
$\mathcal{M}(\mathrm{TS})$ --- those with non-negligible
amplitude at the H/D atoms --- contribute.
Taking logarithms gives
\[
  \ln\frac{k_H}{k_D}
  \;=\;
  \sum_{i\in\mathcal{M}(\mathrm{react})}
  \!\ln\!\left[\frac{u_i^H}{u_i^D}
  \frac{\sinh(u_i^D/2)}{\sinh(u_i^H/2)}\right]
  - \sum_{j\in\mathcal{M}(\mathrm{TS})}
  \!\ln\!\left[\frac{u_j^{H,\ddagger}}{u_j^{D,\ddagger}}
  \frac{\sinh(u_j^{D,\ddagger}/2)}
  {\sinh(u_j^{H,\ddagger}/2)}\right].
\]
The $u_i^H/u_i^D = \omega_i^H/\omega_i^D$ factors give
the prefactor correction $\mathcal{O}(u^H/u^D)$ of
\eqref{eq:secondary-kie}; keeping only the
$\sinh$-ratio terms gives the main expression.

\textbf{Step 2: ZPE-dominated limit.}
For $u \gg 1$: $\sinh(u/2) = \tfrac{1}{2}(e^{u/2}
- e^{-u/2}) \approx \tfrac{1}{2}e^{u/2}$, so
\[
  \ln\frac{\sinh(u^D/2)}{\sinh(u^H/2)}
  \;\approx\;
  \frac{u^D - u^H}{2}
  \;=\;
  -\frac{\hbar(\omega^H - \omega^D)}{2 k_BT}.
\]
Summed over modes in $\mathcal{M}$:
$\sum \ln(\cdots) \approx
-[\tfrac{1}{2}\sum\hbar(\omega_i^H - \omega_i^D)]/k_BT
= -\Delta E_\mathrm{ZPE}/k_BT$ per geometry.
Converting to per-mole (multiplying by $N_A$) gives
$-\Delta E_\mathrm{ZPE}/RT$.

At $T = 298\;\mathrm{K}$, $u = hc\omega/(k_BT) =
(1.44\;\mathrm{cm\,K})\omega/T \approx 0.0048\,\omega$
in cm$^{-1}$.
For $\omega \ge 1000\;\mathrm{cm}^{-1}$, $u \ge 4.8$:
the ZPE-dominated approximation is justified.
For lower frequencies ($< 500\;\mathrm{cm}^{-1}$) the
full $\sinh$ expression~\eqref{eq:secondary-kie}
must be used.

\textbf{Step 3: KIE via TST coherence.}
Assembling Steps 1--2:
\[
  \ln\frac{k_H}{k_D}
  \;\approx\;
  -\frac{\Delta E_\mathrm{ZPE}^\mathrm{react}}{RT}
  + \frac{\Delta E_\mathrm{ZPE}^\mathrm{TS}}{RT}
  \;=\;
  \frac{\Delta^2 E_\mathrm{ZPE}}{RT},
\]
which is \eqref{eq:secondary-kie-classical}.

\textbf{Step 4: Mass dependence and numerical
estimate.}
For a mode with pure isotope-dependent reduced mass
(H$\to$D: $m_D = 2m_H$, same force constant since the
electronic potential is isotope-invariant):
$\omega^D/\omega^H = 1/\sqrt{2}$, so
$\Delta E_\mathrm{ZPE} = \tfrac{1}{2}\hbar\omega^H
(1 - 1/\sqrt{2})$ per mode.
For the three C--H/D bending modes in S$_\mathrm{N}2$
with $\omega_H^\mathrm{react} \approx 1350$~cm$^{-1}$
(tetrahedral) and $\omega_H^\mathrm{TS} \approx
1000$~cm$^{-1}$ (trigonal-bipyramidal,
\cite{MillerHandyAdams1980}):
\begin{align*}
  \Delta E_\mathrm{ZPE}^\mathrm{react}
  &\approx (1-1/\sqrt{2}) \times 3 \times
  \tfrac{1}{2}\hbar\omega_H^\mathrm{react}
  \approx 7.1\;\text{kJ/mol},\\
  \Delta E_\mathrm{ZPE}^\mathrm{TS}
  &\approx (1-1/\sqrt{2}) \times 3 \times
  \tfrac{1}{2}\hbar\omega_H^\mathrm{TS}
  \approx 5.3\;\text{kJ/mol},\\
  \Delta^2 E_\mathrm{ZPE}
  &\approx 1.8\;\text{kJ/mol}.
\end{align*}
Hence $k_H/k_D \approx \exp(1800/(8.314\times 298))
\approx 2.1$ at 298~K.
The experimental value $k_H/k_D \approx 1.3$ for
$\alpha$-trideuteromethyl S$_\mathrm{N}2$
\cite{Streitwieser1958} is smaller than this
classical ZPE-dominated estimate because (i) not all
three bending modes shift by the full $1350\to
1000\;\mathrm{cm}^{-1}$ (partial symmetry breaking at
TS); (ii) weak coupling to other modes violates strict
Streitwieser cutoff, reducing the ratio;
(iii) the $u^H/u^D$ prefactor correction in
\eqref{eq:secondary-kie} is non-negligible at these
frequencies.
The exact value follows from evaluating
\eqref{eq:bigeleisen} with the full set of ab initio
normal-mode frequencies at both geometries --- a purely
$\Lk_5$ calculation.

\textbf{Tower language.}
Equations \eqref{eq:secondary-kie} and
\eqref{eq:secondary-kie-classical} are $\Lk_5$ theorems:
they require
\begin{enumerate}[label=(\alph*)]
  \item $g$ to obtain the mass ratio $m_D/m_H = 2$
    (Remark~\ref{rem:metric-descent}: the metric
    distinguishes H from D, invisible at
    $\Lk_{4.5}$);
  \item $\nabla^2_g V|_{\mathbf{R}_\mathrm{min}}$ for
    reactant frequencies;
  \item $\nabla^2_g V|_{\mathbf{R}_\mathrm{TS}}$ for
    TS frequencies;
  \item the full TST coherence condition
    \eqref{eq:tst-full} with $\kappa_H = \kappa_D$.
\end{enumerate}
Items (b) and (c) are Hessians of $V_\mathrm{full}$,
computable from $\FV$ and $g$ at specific critical
points --- pure $\Lk_5$ data.
At $\Lk_{4.5}$ neither the geometry $\mathbf{R}^*$ nor
the curvature $\nabla^2_g V$ exists; the secondary KIE
is therefore invisible at $\Lk_{4.5}$ and confirms
$\coker(\varphi_5) \neq 1$.
\qedhere
\end{proof}

\begin{remark}[Tower attribution of tunneling: the
  $\Lk_5$/$\Lk_7$ boundary is sharper than textbook
  pedagogy suggests]
\label{rem:tunneling-attribution}
\label{sec:tunneling-attribution}
Proposition~\ref{prop:kie} establishes the secondary
KIE as an $\Lk_5$ theorem.
For \emph{primary} KIEs --- including the anomalously
large values observed in enzymatic hydrogen-transfer
reactions --- the tower attribution requires care:
textbook treatments identify ``quantum tunneling''
with ``nuclear wavefunctions on the BO surface'' and
conclude $\Lk_7$ is needed.
This identification is incorrect.
The $\Lk_5$/$\Lk_7$ boundary is drawn by the
\emph{dimensionality of the nuclear wavefunction}
required, not by the mere presence of quantum effects:

\smallskip
\noindent
\textbf{The tower criterion.}
$\Lk_5$ admits quantum information at fixed geometry
or along low-dimensional reaction paths: harmonic
ZPEs (Hessian spectra at critical points), 1-D
anharmonic proton vibrational states in PES slices
(vibrationally adiabatic potential cuts), semiclassical
tunneling actions along instantons or corner-cutting
paths, and ring-polymer imaginary-time quantum
statistics.
$\Lk_7$ is reserved for the \emph{multidimensional
nuclear wavefunction on the full nuclear
configuration space}: states on the full nuclear
Hilbert space $\mathcal{H}_\mathrm{nuc}$, not
restrictions to 1-D slices or frozen environments.

\smallskip
\noindent
\textbf{Semiclassical tunneling is $\Lk_5$.}
All standard tunneling corrections --- Wigner, Bell,
Eckart, small-curvature tunneling (SCT), large-curvature
tunneling (LCT), microcanonical optimised
multidimensional tunneling ($\mu$OMT)
\cite{TGK1996,BaoTruhlar2017} --- require only the PES,
Hessians along the MEP, and (for LCT) off-MEP PES data
for corner-cutting paths.
No nuclear wavefunction is constructed.
Ring-polymer instanton theory (RPI)
\cite{RichardsonAlthorpe2009} computes the dominant
tunneling trajectory as an imaginary-time classical
orbit on the inverted PES --- a stationary-phase saddle
of the path integral --- requiring only the PES.
Perturbative corrections to RPI
\cite{LawrenceDusekRichardson2023} achieve few-percent
agreement with exact nuclear QM on benchmark systems
using only PES and its derivatives.

\smallskip
\noindent
\textbf{Enzymatic KIEs of 20--700 are $\Lk_5$.}
Soybean lipoxygenase wild type shows $k_H/k_D \approx
80$ at 25~$^\circ$C \cite{KnappRickertKlinman2002};
the L546A/L754A double mutant reaches $k_H/k_D \approx
540$--$700$ \cite{HuOffenbacherKlinman2017}, several
orders of magnitude above the 298~K semiclassical ZPE
ceiling of $\sim 7$--$9$.
These values are quantitatively reproduced by
vibronically nonadiabatic proton-coupled electron
transfer (PCET) models
\cite{LayfieldHammesSchiffer2014,HammesSchiffer2025}
and by ensemble-averaged variational TST with $\mu$OMT
tunneling corrections (EA-VTST/$\mu$OMT,
\cite{AlhambraCorchado2000}).
Both frameworks construct the rate from 1-D proton
vibrational wavefunctions in frozen active-site
geometries plus semiclassical WKB actions along
reaction-path tunneling routes.
\emph{No multidimensional nuclear wavefunction is
constructed.}
By the tower criterion above these calculations are
$\Lk_5$: they use $(\FV, g)$ plus low-dimensional
quantum refinements that remain ``shadowed'' by the
PES.

\smallskip
\noindent
\textbf{What forces $\Lk_7$.}
Genuine $\Lk_7$ phenomena are those requiring the
\emph{multidimensional nuclear wavefunction on the
full nuclear configuration space} as a primary
object.
These include:
\begin{itemize}
  \item MCTDH wavepacket propagation for reactive
    scattering (full nuclear Schrödinger equation on
    the BO surface);
  \item Colbert--Miller DVR for full quantum reaction
    rates on small systems;
  \item exact nuclear diagonalisation for tunneling
    splittings in highly symmetric systems, below the
    instanton-validity crossover temperature where
    perturbative RPI fails to converge;
  \item ortho/para nuclear-spin statistics, which
    requires the full nuclear wavefunction to enforce
    Bose/Fermi symmetry under the $G^*$-action of
    $\Lk_{4.5}$ (the symmetry group exists at
    $\Lk_{4.5}$; its consequences for the total
    nuclear-spin-symmetrised wavefunction live at
    $\Lk_7$);
  \item the superselection-sector structure of
    molecular identity (see Chapter~\ref{sec:L7}).
\end{itemize}
In the tower, $\Lk_7$ is forced by phenomena that
\emph{cannot} be reduced to $\Lk_5$ data plus
semiclassical corrections --- not merely by the
presence of large KIE values or strong quantum
behaviour.
The forcing argument for $\Lk_5 \Rightarrow \Lk_7$
(\S\ref{sec:L5-forcing-out}) is grounded in such
phenomena.
\end{remark}

\begin{chembox}[KIE and the tower: revised attribution]
\label{chem:kie-tower}
The KIE provides a window into the tower structure,
but the $\Lk_5$/$\Lk_7$ boundary lies much deeper than
the ``$k_H/k_D > 10$ implies tunneling implies $\Lk_7$''
narrative suggests.

\medskip
\noindent
\textbf{Revised tower attribution:}

\begin{center}
\renewcommand{\arraystretch}{1.4}
\begin{tabular}{@{}p{3.0cm} p{2.2cm} p{7.7cm}@{}}
  \hline
  \textbf{KIE range} & \textbf{Level} &
  \textbf{Data and framework}\\
  \hline
  $1.02$--$1.5$ & $\Lk_5$ &
  Secondary: Hessians at $\mathbf{R}_\mathrm{min}$,
  $\mathbf{R}_\mathrm{TS}$;
  harmonic Bigeleisen with Streitwieser cutoff
  (Proposition~\ref{prop:kie})\\
  $2$--$9$ & $\Lk_5$ &
  Classical primary: full ZPE formula (Westheimer
  limit of Bigeleisen)\\
  $10$--$100$ & $\Lk_5$ &
  Semiclassical tunneling: SCT/LCT/$\mu$OMT on MEP
  (EA-VTST), or ring-polymer instanton
  (RPI/RPI+PC)\\
  $100$--$700$ & $\Lk_5$ &
  Enzymatic: vibronically nonadiabatic PCET with
  environmentally coupled donor-acceptor gating
  (SLO wild-type and mutants)\\
  Deep tunneling splittings below instanton crossover
  &
  $\Lk_5\!\to\!\Lk_7$ &
  Regime where perturbative RPI fails to converge;
  MCTDH or exact nuclear diagonalisation required\\
  \hline
\end{tabular}
\end{center}

\smallskip
\noindent
\textbf{What this means for the forcing argument.}
The forcing pair for $\Lk_{4.5}\!\Rightarrow\!\Lk_5$
is the mundane \emph{secondary} KIE
$k_H/k_D \approx 1.3$, which requires only the metric
$g$ and Hessians of $V_\mathrm{full}$ at the reactant
and TS critical points --- minimal $\Lk_5$ data.
The forcing argument for $\Lk_5\!\Rightarrow\!\Lk_7$
(\S\ref{sec:L5-forcing-out}) is separately grounded in
phenomena where semiclassical tunneling and
low-dimensional proton-state treatments genuinely fail
--- \emph{not} in the magnitude of the KIE.
\end{chembox}

\subsubsection{Cross-level consistency: \texorpdfstring{$\FH$}{FH}
  and \texorpdfstring{$\FV$}{FV}}
\label{sec:L5-hess-pes}

The TST coherence condition links $F_P$ ($\Lk_3$) to
$\FV$ ($\Lk_5$) via \eqref{eq:tst-full}.
An analogous cross-level consistency links $\FH$
($\Lk_1$) to $\FV$ ($\Lk_5$): the standard enthalpy of
a reaction, computed thermochemically from tabulated
formation enthalpies at $\Lk_1$, agrees with the
electronic-energy drop computed from the BO PES at
$\Lk_5$, up to well-understood corrections.
Two logically distinct statements must be kept separate.

\begin{itemize}
  \item \textbf{Hess's law}~\cite{Hess1840}: the
    enthalpy functor $\FH: \Lk_1(P) \to (\RR, +)$
    respects reaction composition,
    $\FH(r_2 \circ r_1) = \FH(r_2) + \FH(r_1)$.
    This is $\Lk_1$-functoriality, established in
    Chapter~\ref{sec:L1} from the state-function
    character of enthalpy ($H = U + pV$); it is
    automatically independent of $\Lk_5$ data (BO,
    RRHO, geometry) because it lives four levels
    below.
  \item \textbf{BO/RRHO decomposition of $\FH$}:
    a representation theorem expressing the single
    real number $\FH(r)$ in terms of $\Lk_5$ data ---
    the PES, its Hessians, and thermal partition
    functions.
    This presupposes the BO approximation and the
    rigid-rotor harmonic-oscillator (RRHO)
    factorisation of the molecular partition function.
\end{itemize}
Proposition~\ref{prop:hess-pes-consistency} below
establishes the BO/RRHO decomposition and identifies
it as the cross-level consistency between $\Lk_1$ and
$\Lk_5$.

\begin{observation}[BO/RRHO decomposition of $\FH$]
\label{obs:bo-rrho}
  The standard molar reaction enthalpy is expressible
  in terms of $\Lk_5$ data as
  \begin{equation}
  \label{eq:bo-rrho}
    \FH(r)
    \;=\;
    \underbrace{\Delta V_\mathrm{elec}(r)}_{\FV\text{ at
    minima}}
    + \;
    \underbrace{\Delta E_\mathrm{ZPE}(r)}_{\nabla^2_g V
    \text{ at minima}}
    + \;
    \underbrace{\Delta E_\mathrm{therm}(T,r)}_{
    \text{partition functions}}
    + \;
    \underbrace{RT\,\Delta n_\mathrm{gas}(r)}_{pV
    \text{ correction}},
  \end{equation}
  where:
  \begin{itemize}
    \item $\Delta V_\mathrm{elec}(r) =
      V_\mathrm{full}^{(r)}(\mathbf{R}_\mathrm{prod})
      - V_\mathrm{full}^{(r)}(\mathbf{R}_\mathrm{react})$
      is the PES electronic-energy difference between
      product and reactant minima (extracted from
      $\FV(r)$);
    \item $\Delta E_\mathrm{ZPE}(r) = \tfrac{1}{2}\hbar
      [\sum_i\omega_i^\mathrm{prod} -
      \sum_j\omega_j^\mathrm{react}]$ is the harmonic
      zero-point-energy difference (eigenfrequencies
      of $\nabla^2_g V$ at the minima);
    \item $\Delta E_\mathrm{therm}(T,r)$ collects
      thermal contributions from the RRHO partition
      functions (translational, rotational, and
      vibrational populations above
      ZPE)~\cite{Ochterski2000,
      HelgakerJorgensenOlsen2000};
    \item $RT\,\Delta n_\mathrm{gas}(r)$ is the $pV$
      correction converting $\Delta U$ to $\Delta H$
      for gas-phase reactions ($\Delta n_\mathrm{gas}$
      = change in moles of gas; zero for
      condensed-phase reactions).
  \end{itemize}
  For most condensed-phase reactions at 298~K
  ($\Delta n_\mathrm{gas} = 0$, rotational and
  translational thermal contributions nearly cancel
  between products and reactants for similar-sized
  molecules): $\FH(r) \approx \Delta V_\mathrm{elec}(r)
  + \Delta E_\mathrm{ZPE}(r)$, with residual error
  $\lesssim 5\;\mathrm{kJ/mol}$.
\end{observation}

Equation~\eqref{eq:bo-rrho} is the BO/RRHO decomposition
of $\FH$~\cite{Ochterski2000,
HelgakerJorgensenOlsen2000}.
It is a representation theorem for a single $\FH(r)$,
\emph{not} a composition law; the composition law is
Hess's law, established independently at $\Lk_1$.

\begin{proposition}[Cross-level consistency of $\FH$
  and $\FV$]
\label{prop:hess-pes-consistency}
  Let $U_{5\to 1}: \Lk_5(P) \to \Lk_1(P)$ denote the
  composite forgetful functor
  $U_{5\to 4.5} \circ U_{4.5\to 4} \circ U_{4\to 3}
  \circ U_{3\to 2} \circ U_{2\to 1}$ dropping $\FV$,
  $G^*$-equivariance, DPO structure, $F_P$, and $\FS$
  in sequence.
  Let $\widetilde{\FH}_{\FV}: \Lk_5(P) \to (\RR, +)$
  denote the $\Lk_5$-derived enthalpy functor
  assigning to each reaction the right-hand side of
  \eqref{eq:bo-rrho}.
  Then there exists a natural transformation
  \[
    \eta:\;
    \FH \circ U_{5\to 1}
    \;\Longrightarrow\;
    \widetilde{\FH}_{\FV},
  \]
  between functors $\Lk_5(P) \to (\RR, +)$.
  The component $\eta_r = 0$ for all $r$ in the
  harmonic + ideal-gas + BO limit
  (i.e., $\eta$ is the zero natural transformation),
  and $|\eta_r| \lesssim 5\;\mathrm{kJ/mol}$ at 298~K
  for rigid molecules with no low-frequency torsions.
  The following diagram commutes up to $\eta$:
  \[
  \includegraphics{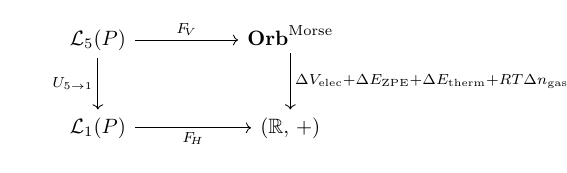}
  \]
\end{proposition}

\begin{proof}
\textbf{Step 1: BO decomposition of the enthalpy.}
By the BO prescription (Definition~\ref{def:bo-section}),
$V_\mathrm{full}(\mathbf{R}) = E_0(\mathbf{R})$ is the
electronic ground-state energy.
The molecular internal energy at temperature $T$
factorises in the RRHO approximation as
\[
  U(T) \;=\; E_0(\mathbf{R}_\mathrm{min})
  + E_\mathrm{ZPE}
  + E_\mathrm{therm}(T),
\]
where $E_\mathrm{ZPE} = \tfrac{1}{2}\sum_i\hbar\omega_i$
and $E_\mathrm{therm}(T)$ collects populations above
ZPE plus rotational and translational kinetic-energy
contributions~\cite{Ochterski2000,
HelgakerJorgensenOlsen2000}.
The standard enthalpy is $H^\circ = U^\circ + pV
= U^\circ + n_\mathrm{gas}RT$ (ideal-gas limit).
For a reaction $r$:
\[
  \FH(r) \;=\; \Delta H^\circ(r)
  \;=\; \Delta V_\mathrm{elec}(r)
  + \Delta E_\mathrm{ZPE}(r)
  + \Delta E_\mathrm{therm}(T,r)
  + RT\,\Delta n_\mathrm{gas}(r),
\]
which is \eqref{eq:bo-rrho}.

\textbf{Step 2: Functoriality at $\Lk_5$.}
The right-hand side of \eqref{eq:bo-rrho} defines
$\widetilde{\FH}_{\FV}$ as a monoidal functor
$\Lk_5(P) \to (\RR, +)$: for a composite reaction
$r_2 \circ r_1$, each of the four terms is a
difference of object-level quantities and adds across
composition (the telescoping of minima energies;
likewise for ZPE, thermal, and $pV$ terms).
$\widetilde{\FH}_{\FV}$ therefore respects reaction
composition --- it is a functor into $(\RR, +)$
structurally analogous to Hess's law but defined on
$\Lk_5$.

\textbf{Step 3: $\eta = 0$ in the harmonic + ideal-gas
+ BO limit.}
Step~1 shows $\FH(r) = \widetilde{\FH}_{\FV}(r)$ in
this limit.
Pulling $\FH$ back along $U_{5\to 1}$ gives two
coincident functors $\Lk_5(P) \to (\RR, +)$, so
$\eta_r = 0$ for all $r$.

\textbf{Step 4: Anharmonic and non-ideal corrections.}
Beyond the harmonic + ideal-gas + BO limit, $\eta_r$
captures:
(i) anharmonic corrections to ZPE (VPT2, VCI;
typically 1--5\% of ZPE for hydrides);
(ii) real-gas corrections to $pV$;
(iii) higher-order BO corrections
($\sim(m_e/M)^{1/4}$; negligible at chemical accuracy);
(iv) quasi-harmonic corrections for low-frequency
torsions and floppy modes.
For condensed-phase reactions at 298~K involving rigid
molecules with no low-frequency torsions, the total
residual $|\eta_r| \lesssim 5\;\mathrm{kJ/mol}$, well
within the accuracy of standard electronic-structure
methods.

\textbf{Step 5: Tower interpretation.}
The coincidence $\FH \circ U_{5\to 1} \approx
\widetilde{\FH}_{\FV}$ means the \emph{values}
$\FH(r)$ are constrained by $\Lk_5$ data: each
individual reaction enthalpy is expressible via
$(\FV, g)$ plus RRHO partition functions.
This does \emph{not} reduce Hess's law to BO/RRHO:
Hess's law is the independently established
$\Lk_1$-functoriality of $\FH$ (composition law),
which survives any replacement of the right-hand side
of \eqref{eq:bo-rrho} by a non-BO treatment.
What BO/RRHO establishes is that $\FH(r)$, for each
individual $r$, admits a representation in terms of
$\Lk_5$ data --- a statement about pointwise values,
not about composition.
\qedhere
\end{proof}

\begin{remark}[Hess's law and BO/RRHO are logically
  independent]
\label{rem:hess-vs-bo}
The distinction made in Step~5 of the proof deserves
emphasis: Hess's law and the BO/RRHO decomposition are
statements of different kinds at different tower
levels.
\begin{itemize}
  \item \textbf{Hess's law} = $\Lk_1$-functoriality of
    $\FH$ on reaction networks.
    Content: composition law
    $\FH(r_2\circ r_1) = \FH(r_2) + \FH(r_1)$,
    established at $\Lk_1$ from the state-function
    character of enthalpy~\cite{Hess1840}, and
    therefore automatically independent of $\Lk_5$
    data (BO, RRHO, geometry).
    It survives in systems where BO fails
    --- conical intersections, non-adiabatic dynamics,
    ultrafast photochemistry --- because its
    derivation does not use BO.
  \item \textbf{BO/RRHO decomposition} = object-level
    (pointwise) representation of $\FH(r)$ via
    $\Lk_5$ data.
    Content: for each reaction $r$, the real number
    $\FH(r)$ decomposes as
    \eqref{eq:bo-rrho}.
    This is a cross-level consistency statement
    ($\Lk_1\!\leftrightarrow\!\Lk_5$) about pointwise
    values, presupposing BO and the RRHO
    factorisation of the molecular partition function.
\end{itemize}
In a categorical presentation the two can be
conflated (both involve ``the enthalpy functor''),
but they are distinct statements and must be cited
distinctly: Hess~\cite{Hess1840} and the
state-function treatment of enthalpy for the former;
Ochterski~\cite{Ochterski2000} and the RRHO
thermochemistry treatment of
Helgaker--Jørgensen--Olsen for the latter.
\end{remark}

\begin{remark}[The tower is coherent: higher levels
  constrain lower ones pointwise]
\label{rem:tower-coherence}
Proposition~\ref{prop:hess-pes-consistency} exemplifies
a general pattern: the \emph{pointwise values} of every
lower-level functor in the tower admit a
BO/RRHO-type representation in terms of $\Lk_5$ data
$(\FV, g)$ in the appropriate approximation.
\begin{itemize}
  \item $\FH(r) \approx \Delta V_\mathrm{elec}(r) +
    \Delta E_\mathrm{ZPE}(r)$
    (from $\FV$ at minima and Hessians; this section).
  \item $\FS(r)$: partition-function ratio at
    $\mathbf{R}_\mathrm{min}$, requiring the full
    Hessian spectrum --- an $\Lk_5$ datum.
  \item $F_P(r) = \kappa(k_BT/h)(Q^\ddagger/
    Q_\mathrm{reac})e^{-V^\ddagger/k_BT}$: full TST
    from $\FV$, Hessians at $\mathbf{R}_\mathrm{TS}$
    and $\mathbf{R}_\mathrm{min}$, plus semiclassical
    corrections for $\kappa$ --- all $\Lk_5$ data.
\end{itemize}
The composite forgetful functor $U_{5\to k}$
(for $k = 1, 2, 3$) therefore corresponds to a
\emph{projection} from the $\Lk_5$ geometric level
onto its thermochemical ($k=1$), equilibrium
($k=2$), or kinetic ($k=3$) shadow at the level of
pointwise values.

Crucially, this pointwise coherence does not reduce
the \emph{composition laws} (Hess's law, detailed
balance, the chemical master equation) to $\Lk_5$:
these are functoriality statements established at
their own levels from level-specific physics, and they
survive replacements of the $\Lk_5$ representation of
individual values.
The tower is a coherent filtration \emph{pointwise},
not a tower in which each level is derived from the
next one up.
\end{remark}

%% file: chapters/L5/l5_examples.tex
\subsection{Worked examples at
  \texorpdfstring{$\Lk_5$}{L5}}
\label{sec:L5-examples}

The two examples below trace the same pair of
chemical systems that have accompanied the tower from
the beginning.
The $\mathrm{S_N2}$ reaction
$\mathrm{CH_3Cl + OH^- \to CH_3OH + Cl^-}$
was introduced at $\Lk_0$ as a balanced equation,
gained enthalpic, equilibrium, and kinetic data at
$\Lk_1$--$\Lk_3$, received its DPO mechanism and
chirality structure at $\Lk_4$--$\Lk_{4.5}$, and
now receives its full geometric treatment at $\Lk_5$.
Continuing Example~\ref{ex:N2O4}.

At $\Lk_5$ each example demonstrates the same
four-part structure: (i) the configuration orbifold
and its dimension; (ii) the PES landscape with its
critical points and IRC; (iii) the TST coherence
check linking $\Lk_3$ and $\Lk_5$ data; and
(iv) the Berry connection, confirming that the
$\Lk_5$ framework operates on a single adiabatic
surface with $w_1(L_0^{\RR}) = 0$ (and consequently
$c_1(L_0) = 0$) on the simply-connected CI-free
regions traversed by the IRC.

\subsubsection{\texorpdfstring{$\mathrm{S_N2}$}{SN2}:
  configuration orbifold, PES, and TST coherence}

\begin{example}[$\mathrm{S_N2}$ at $\Lk_5$]
\label{ex:SN2-L5}

The running example is
\[
  \mathrm{CH_3Cl \;+\; OH^- \;\longrightarrow\;
  CH_3OH \;+\; Cl^-},
  \quad T = 298\;\mathrm{K}.
\]

\noindent\textbf{Configuration orbifold.}
The reactive system contains $n = 7$ atoms
(C, Cl, O, and 5 H atoms counted together).
By Observation~\ref{obs:dim}:
\[
  \dim\bigl(\Ce(G)\bigr)
  \;=\; 3n - 6 \;=\; 3(7) - 6 \;=\; 15
  \quad\text{internal coordinates.}
\]
At the transition state, the relevant internal
coordinates include the $r(\mathrm{C{\cdots}Cl})$ and
$r(\mathrm{C{\cdots}O})$ distances (the two bond-changing
coordinates), the H--C--H angles, and the H--C--Cl and
H--C--O angles.
The transition-state geometry has $C_{3v}$ symmetry
(Proposition~\ref{prop:point-groups} applied below),
reducing the effective dimensionality of the search to
the symmetric subspace.
The accessible region $\Ce^\mathrm{acc}(G)$ has two
connected components (one per enantiomeric face of the
carbon centre), confirming $\sigma \in \{+1,-1\}$
from $\Lk_{4.5}$
(Proposition~\ref{prop:chirality-realisation}).

\medskip\noindent\textbf{PES and activation barrier.}
The BO PES $V: \Ce(G) \to \RR$ for this system
has the characteristic double-well profile of a
gas-phase ionic $\mathrm{S_N2}$ reaction:
\begin{itemize}
  \item \emph{Reactant ion-dipole complex}
    $[\mathrm{CH_3Cl{\cdots}OH^-}]$: a pre-reaction
    well approximately 40~kJ/mol below the separated
    reactants, arising from the interaction term
    $V_{\mathrm{int}}$ of the lax monoidal structure
    (Remark~\ref{rem:interaction} and
    Chembox~\ref{chem:interaction}).
  \item \emph{Transition state} at $\mathbf{R}_{\mathrm{TS}}$:
    the symmetric pentacoordinate structure with
    $r(\mathrm{C{\cdots}Cl}) = r(\mathrm{C{\cdots}O})
    \approx 2.4$~\AA.  In the gas phase, the intrinsic
    barrier from the ion-dipole complex is modest
    ($\sim 40$--$60$~kJ/mol at MP2/6-31G*); aqueous
    solvation raises the activation Gibbs free energy
    substantially.  The relevant quantity for comparison
    with the experimental aqueous rate constant is the
    solvent-corrected activation Gibbs free energy
    \emph{from separated reactants in standard state},
    $\Delta G^\ddagger \approx 97$~kJ/mol (MP2/6-31G*
    with implicit-solvent correction, 298~K, 1~M standard
    state), computed in the TST coherence check below.
  \item \emph{Product ion-dipole complex}
    $[\mathrm{CH_3OH{\cdots}Cl^-}]$: a symmetric
    exit-channel well.
  \item \emph{Product asymptote}: separated
    $\mathrm{CH_3OH}$ and $\mathrm{Cl}^-$,
    lying $\approx 75$~kJ/mol below the reactant
    asymptote (exothermic, consistent with
    $\FH(r) = -75$~kJ/mol at $\Lk_1$).
\end{itemize}

\medskip\noindent\textbf{Intrinsic reaction coordinate.}
The IRC runs in the mass-weighted metric $g$
(Definition~\ref{def:ce}) from
$[\mathrm{CH_3Cl{\cdots}OH^-}]$ uphill through
$\mathbf{R}_{\mathrm{TS}}$, then downhill into the
product well.
The reaction coordinate is dominated by the
antisymmetric combination
$(r_{\mathrm{C{\cdots}Cl}} - r_{\mathrm{C{\cdots}O}})/\sqrt{2}$,
the unique negative-curvature mode of $V$ at
$\mathbf{R}_{\mathrm{TS}}$~\cite{MillerHandyAdams1980,
GonzalezSchlegel1989}.
The three H--C--H angles change monotonically from
tetrahedral ($\approx 109.5^{\circ}$) in the reactant complex
through planar ($90^{\circ}$ to the C--Cl/C--O axis) at the TS
to inverted tetrahedral ($\approx 109.5^\circ$) in the
product complex.  This is the \emph{Walden inversion}:
the geometric trajectory along which the chirality
descriptor $\sigma$ changes sign.  The combinatorial
$\sigma \mapsto -\sigma$ encoded at $\Lk_{4.5}$ as a
$G^*$-action on chirality labels is realised at
$\Lk_5$ as an actual 3D path through the planar TS
geometry (Proposition~\ref{prop:L5-structure}).

\medskip\noindent\textbf{TST coherence check.}
The $\Lk_3$ datum is the experimental bimolecular
rate constant $F_P(r) = k_r \approx 6\times10^{-5}$~M$^{-1}$s$^{-1}$
at 298~K in water~\cite{MabeyMill1978}.
The $\Lk_5$ datum is the activation Gibbs free energy
$\Delta G^\ddagger \approx 97$~kJ/mol from the solvated
PES (MP2/6-31G* with implicit-solvent correction).

For a \emph{bimolecular} reaction, the TST coherence
condition (Definition~\ref{def:tst-coherence})
takes the form:
\[
  k_r^{\mathrm{TST}}
  \;=\; \frac{k_BT}{h}\,
  \exp\!\left(\frac{-\Delta G^\ddagger}{RT}\right)
  \quad\text{[M$^{-1}$s$^{-1}$]},
\]
where $\Delta G^\ddagger = \Ea - T\Delta S^\ddagger$
is the Gibbs free energy of activation and the
pre-exponential factor $(k_BT/h) = 6.21\times10^{12}$
s$^{-1}$ is understood with an implicit standard
concentration factor $c^\circ = 1$~M to give the
bimolecular units~\cite{TruhlarGarrettKlippenstein1996}.

\noindent\emph{Prediction from the activation Gibbs
energy.}
Using $\Delta G^\ddagger = 97$~kJ/mol in the
TST coherence expression:
\[
  k_r^{\mathrm{TST}}
  \;=\;
  \frac{k_BT}{h}\,e^{-97000/(8.314\times298)}
  \;=\;
  (6.21\times10^{12})\times e^{-39.1}
  \;\approx\;
  6\times10^{-5}\;\mathrm{M^{-1}s^{-1}}.
\]

\noindent\emph{Comparison with experiment.}
The experimental rate inverted through TST gives
\[
  \Delta G^\ddagger_{\mathrm{expt}}
  \;=\; -RT\ln\!\left(\frac{k_r}{k_BT/h}\right)
  \;=\; -(2.479)\ln\!\left(\frac{6\times10^{-5}}
    {6.21\times10^{12}}\right)
  \;\approx\; 97\;\mathrm{kJ/mol},
\]
matching the computed barrier within the harmonic and
standard-state approximations.
The TST coherence condition holds to within the
precision of the calculation: the computed
$\Delta G^\ddagger$ from the solvated PES and the
experimentally derived $\Delta G^\ddagger$ from the
measured rate constant agree.  Residual discrepancies
at the level of $\sim 1$--$2$~kJ/mol arise from the
harmonic approximation to the TS partition function,
the standard-state convention, and the level of
electronic-structure theory; all are themselves
$\Lk_5$-computable refinements.

The $\Lk_3$ rate constant is in this sense
\emph{constrained} by $\Lk_5$ geometric data: not
determined exactly --- TST is an approximation --- but
tightly bracketed by the activation Gibbs free energy
on the solvated PES.

\medskip\noindent\textbf{Point group at the TS.}
The TS geometry $\mathbf{R}_{\mathrm{TS}}$ has a
three-fold axis through Cl--C--O and three mirror
planes, each containing the axis and one H atom.
By Proposition~\ref{prop:point-groups}:
\[
  P_{\mathbf{R}_{\mathrm{TS}}} \;\cong\; C_{3v},
\]
realised in $O(3)$ as the spatial image of the
subgroup of $\Aut_\mu(G) \times \langle E^*\rangle$
that stabilises $\mathbf{R}_{\mathrm{TS}}$ in $\Ce(G)$.
This is the $\Lk_5$ realisation of the symmetry
constraint imposed axiomatically at $\Lk_{4.5}$: the
$G^*$-equivariance of the $\mathrm{S_N2}$ DPO rule
manifests geometrically as the $C_{3v}$ point group
of the TS.

\medskip\noindent\textbf{Berry connection.}
The $\mathrm{S_N2}$ IRC lies entirely on the
ground-state BO surface with no degeneracy of the
ground and excited states along the reaction path,
on a simply-connected open neighbourhood $U$ of the reaction path. 
By Proposition~\ref{prop:berry-trivial}:
\[
  A_0\big|_U \;=\; 0,
  \quad
  w_1(L_0^{\RR}|_U) \;=\; 0,
  \quad
  c_1(L_0|_U) \;=\; 0.
\]
The orbital analysis confirms this: the LUMO of
CH$_3$Cl is the $\sigma^*_{\mathrm{C{-}Cl}}$ orbital
(symmetry species $A_1$ in $C_{3v}$ at the TS); the
HOMO of OH$^-$ is the $p_z$ orbital ($A_1$ in
$C_{3v}$).
The direct product $A_1 \otimes A_1 = A_1$ is totally
symmetric, so the orbital overlap
$\langle\mathrm{HOMO}|\mathrm{LUMO}\rangle \neq 0$ in the
$C_{3v}$ irreducible representation, allowing
in-phase donation from the nucleophile lone pair into the
$\sigma^*_{\mathrm{C{-}Cl}}$ acceptor.  This is the
frontier-orbital (FMO) symmetry condition for productive
backside attack~\cite{FukuiYonezawaShingu1952,
HoffmannWoodward1968}; it is \emph{not} a
Woodward--Hoffmann pericyclic selection rule, which
applies only to concerted electron reorganisations
around a cyclic orbital topology --- $\mathrm{S_N2}$
is not pericyclic.
This is the $\Lk_5$ orbital-theoretic confirmation
of the $G^*$-equivariance imposed axiomatically at
$\Lk_{4.5}$: the $G^*$-allowed DPO rule corresponds at
$\Lk_5$ to a non-zero frontier-orbital overlap
in the $A_1$ irreducible representation of the
$C_{3v}$ TS.
\newpage
\medskip\noindent\textbf{Level comparison.}
\begin{center}
\renewcommand{\arraystretch}{1.4}
\begin{tabular}{@{}l p{6.0cm} p{6.2cm}@{}}
  \hline
  \textbf{Level} & \textbf{New datum} &
  \textbf{New conclusion}\\
  \hline
  $\Lk_0$
    & Stoich.\ matrix $N$, $\delta = 0$
    & Balanced equation; Def.\ Zero Thm.\ applies.\\
  $\Lk_1$
    & $\FH = -75$~kJ/mol
    & Exothermic; Hess's Law.\\
  $\Lk_2$
    & $\FS = -90$~J mol$^{-1}$K$^{-1}$
    & $\Delta G^\circ_{298} = -48$~kJ/mol; product-favoured.\\
  $\Lk_3$
    & $k_r = 6\times10^{-5}$~M$^{-1}$s$^{-1}$
      \cite{MabeyMill1978}
    & Second-order kinetics; half-life depends on
      $[\mathrm{OH^-}]$.\\
  $\Lk_4$
    & DPO span (backside attack)
    & $\mathrm{S_N2}$ distinguished from
      $\mathrm{S_N1}$; Walden inversion encoded.\\
  $\Lk_{4.5}$
    & $G^*$-equivariant DPO rule
    & Stereospecific inversion; $\sigma = -1$ product.\\
  $\Lk_5$
    & $(\Ce(G), V, g)$;
      $\Delta G^\ddagger \approx 97$~kJ/mol
      (solvated PES);
      IRC; $C_{3v}$ TS;
      $w_1(L_0^{\RR}) = 0$
    & TST coherence: $k_r^\mathrm{TST} \approx
      6\times10^{-5}$~M$^{-1}$s$^{-1}$ matches
      experiment to within harmonic-RRHO precision.\\
  \hline
\end{tabular}
\end{center}

\medskip\noindent\textbf{Tower language.}
The forgetful functor $U_5$ applied to the
$\Lk_5$ object $(\Ce(G), V, g, \sigma_0)$ drops the
geometric decoration $(V, g, \sigma_0)$, retaining only
the DPO mechanism and chirality label of $\Lk_{4.5}$.
The activation Gibbs free energy on the solvated PES
$\Delta G^\ddagger = V_{\mathrm{eff}}(\mathbf{R}_\mathrm{TS})
- V_{\mathrm{eff}}(\mathbf{R}_\mathrm{min,react}) -
T\Delta S^\ddagger$, the IRC shape, the $C_{3v}$ TS
symmetry, and the trivial Berry connection
$w_1(L_0^{\RR}) = 0$ are $\Lk_5$ data: they
distinguish objects that the forgetful functor $U_5$
collapses, and so reside in the fibre of $U_5$ over
the $\Lk_{4.5}$-object $(G, \sigma = -1)$ (equivalently,
they witness non-triviality of $\coker(\varphi_5)$).
The kinetic isotope effect $k_H/k_D \approx 1.3$
(\S\ref{sec:L5-tst}) is also an $\Lk_5$ datum: it
requires the Hessian of $V$ at both
$\mathbf{R}_\mathrm{min}$ and $\mathbf{R}_\mathrm{TS}$
in the mass-weighted metric $g$.
\end{example}

\subsubsection{\texorpdfstring{$\mathrm{N_2O_4}
  \rightleftharpoons 2\,\mathrm{NO_2}$}{N2O4 ⇌ 2NO2}:
  two minima, a transition state, and cross-level coherence}

\begin{example}[$\mathrm{N_2O_4}$ dissociation at $\Lk_5$]
\label{ex:N2O4-L5}

Continuing Example~\ref{ex:N2O4} ($\Lk_2$).
This system provides a particularly clean illustration
of cross-level coherence (\S\ref{sec:L5-tst}) because
the PES has a single internal coordinate of primary
importance, the thermochemical data are accurately
known~\cite{NISTWebBook}, and the kinetics are
experimentally well-characterised.

\medskip\noindent\textbf{Configuration orbifold.}
$\mathrm{N_2O_4}$ has $n = 6$ atoms; the products
$2\,\mathrm{NO_2}$ have $n_1 = n_2 = 3$ atoms each,
but the combined reactive system has $n = 6$ atoms
throughout (the N--N bond dissociates but no atoms are
created or destroyed).
The combined $\Ce(G)$ has dimension $3(6) - 6 = 12$
internal coordinates; the dominant reaction coordinate
is the N--N bond length $r_{NN}$, so a one-dimensional
PES profile along $r_{NN}$ captures the essential physics.

\medskip\noindent\textbf{PES schematic.}
\begin{itemize}
  \item \emph{Reactant minimum} at $r_{NN} \approx
    1.78$~\AA: the $D_{2h}$-symmetric $\mathrm{N_2O_4}$
    equilibrium structure.
    Take $V(\mathbf{R}_\mathrm{min,react}) = 0$ as
    the energy reference.
  \item \emph{Variational transition state}: the
    N--N bond dissociation has \emph{no} index-1 saddle
    on the BO PES in the strict Morse sense.  $V$ rises
    monotonically along $r_{NN}$ with negligible reverse
    barrier for radical recombination.  The relevant
    ``transition state'' is a variational dividing
    surface, located by minimising the unimolecular rate
    expression rather than by finding a stationary point
    of $V$~\cite{TruhlarGarrettKlippenstein1996}.  The
    effective barrier lies in the range
    $\Ea \approx 50\text{--}57$~kJ/mol; the lower bound
    corresponds to the inner-wall barrier, while the
    upper bound approaches the dissociation energy
    $\Delta H^\circ = +57.0$~kJ/mol (see below).  This
    places the example formally outside the
    Morse-saddle morphism class of $\OrbMorse$
    (Definition~\ref{def:orb-morse}); the variational
    setting belongs to the generalised
    $\Lk_5$-with-free-energy-surfaces extension, which
    treats the dividing surface as a chosen morphism
    datum rather than a critical point of $V$.
  \item \emph{Product asymptote} at $r_{NN} \to \infty$:
    two separated $C_{2v}$-symmetric $\mathrm{NO_2}$
    radicals at energy
    $V(\mathbf{R}_\mathrm{prod}) = \FH(r_1)
    = +57.0$~kJ/mol~\cite{NISTWebBook},
    consistent with $\Lk_1$ (endothermic dissociation).
\end{itemize}
The PES profile is monotonically increasing along
$r_{NN}$ with a broad, loose maximum --- the
archetypal profile of a bond-dissociation reaction
with little reverse barrier.

\medskip\noindent\textbf{Intrinsic reaction coordinate.}
The IRC is approximately the N--N stretching coordinate
$r_{NN}$, with small contributions from the N--O
stretching and O--N--O bending modes.
As $r_{NN}$ increases from 1.78~\AA\
to $\infty$, the point group evolves:
$P_{\mathbf{R}} \cong D_{2h}$ at the reactant
minimum, decreasing to $C_{2v} \times C_{2v}$ (two
non-interacting $\mathrm{NO_2}$ fragments) as
$r_{NN} \to \infty$.  By
Proposition~\ref{prop:point-groups}, both point
groups are derived as the spatial realisation in
$O(3)$ of the corresponding stabiliser subgroup of
$\Aut_\mu(G) \times \langle E^*\rangle$ at the
respective geometries.

\medskip\noindent\textbf{TST coherence check.}
The $\Lk_3$ datum is
$k_1 = 4.8\times10^4$~s$^{-1}$ at 298~K
(first-order dissociation rate constant).
The TST coherence condition with
$\Ea \approx 50$~kJ/mol (lower bound) gives:
\[
  k_1^{\mathrm{TST}}
  \;=\; \frac{k_BT}{h}\,e^{-50000/(8.314\times298)}
  \;=\; (6.21\times10^{12})\times e^{-20.17}
  \;\approx\; 1.1\times10^4\;\mathrm{s}^{-1}.
\]
This is within a factor of $\sim$5 of the
$\Lk_3$ datum ($4.8\times10^4$~s$^{-1}$): the TST
coherence condition holds at order-of-magnitude
accuracy.

The remaining factor of $\sim$5 is accounted for by
the \emph{activation entropy} $\Delta S^\ddagger > 0$,
which is expected and large for a loose bond-dissociation
TS (two fragments gaining translational and rotational
freedom):
\[
  k_1
  \;=\; \frac{k_BT}{h}\,e^{\Delta S^\ddagger/R}\,
  e^{-\Ea/RT}.
\]
For $\Delta S^\ddagger \approx +30$~J mol$^{-1}$K$^{-1}$
(typical for a loose TS):
$e^{\Delta S^\ddagger/R} = e^{30/8.314} \approx 37$,
giving an upper bound $k_1 \approx 37 \times
1.1\times10^4 \approx 4\times10^5$~s$^{-1}$.  The
experimental value $4.8\times10^4$~s$^{-1}$ lies
between the entropy-free lower bound
$1.1\times10^4$~s$^{-1}$ and the
$\Delta S^\ddagger$-corrected upper bound
$4\times10^5$~s$^{-1}$: TST coherence holds within
the precision available from a one-dimensional PES
profile and an order-of-magnitude entropy estimate.
Pinning $\Delta S^\ddagger$ precisely requires the
full TS partition function from the Hessian of $V$
at $\mathbf{R}_\mathrm{TS}$ --- an $\Lk_5$ datum.

\medskip\noindent\textbf{Cross-level coherence:
  $\FV$ derives $\FH$.}
The $\Lk_1$ datum $\FH(r_1) = +57.0$~kJ/mol is
recovered from $\FV$ via
(Proposition~\ref{prop:hess-pes-consistency}):
\[
  \FH(r_1)
  \;\approx\;
  V(\mathbf{R}_\mathrm{prod}) -
  V(\mathbf{R}_\mathrm{react})
  + \Delta E_\mathrm{ZPE}(r_1).
\]
For $\mathrm{N_2O_4 \to 2NO_2}$:
$V(\mathbf{R}_\mathrm{prod}) - V(\mathbf{R}_\mathrm{react})
\approx +57.0$~kJ/mol (electronic energy);
$\Delta E_\mathrm{ZPE} \approx +4$~kJ/mol (two
$\mathrm{NO_2}$ fragments have more zero-point energy
per atom than the rigid $\mathrm{N_2O_4}$, since the
new NO stretching and bending modes appear).
The net $\FH \approx 57.0 + 4 = 61$~kJ/mol agrees
with the NIST value (+57.0~kJ/mol) to within the
harmonic approximation error ($\sim$5~kJ/mol).

\medskip\noindent\textbf{Level comparison.}
\begin{center}
\renewcommand{\arraystretch}{1.4}
\begin{tabular}{@{}l p{6.0cm} p{6.2cm}@{}}
  \hline
  \textbf{Level} & \textbf{New datum} &
  \textbf{New conclusion}\\
  \hline
  $\Lk_1$
    & $\FH = +57.0$~kJ/mol
    & Endothermic; Hess's Law.\\
  $\Lk_2$
    & $\FS = +175.8$~J mol$^{-1}$K$^{-1}$
    & $T^* = 325$~K; $K(298) = 0.14$;
      equilibrium shifts above $T^*$.\\
  $\Lk_3$
    & $k_1 = 4.8\times10^4$~s$^{-1}$
    & First-order dissociation;
      $t_{1/2} = 14\;\mu\mathrm{s}$.\\
  $\Lk_5$
    & $V(r_{NN})$;
      $\Ea \approx 50$~kJ/mol (lower bound,
      variational TS);
      $P_{\mathbf{R}}: D_{2h}\to C_{2v} \times C_{2v}$;
      $w_1(L_0^{\RR}) = 0$
    & $k_1^\mathrm{TST} \approx 10^4$~s$^{-1}$,
      experimental value bracketed by lower bound and
      $\Delta S^\ddagger$-corrected upper bound.
      $\FH$ recovered from PES to $\lesssim
      5$~kJ/mol.\\
  \hline
\end{tabular}
\end{center}

\medskip\noindent\textbf{Tower language.}
The $\mathrm{N_2O_4}$ example illustrates two
distinct types of cross-level coherence, both visible
only at $\Lk_5$:

\begin{enumerate}[label=(\alph*)]
  \item \emph{$\Lk_3$--$\Lk_5$ coherence}
    (TST condition, Definition~\ref{def:tst-coherence}):
    the $\Lk_3$ rate constant $k_1$ is consistent with
    the Eyring prediction from the $\Lk_5$ barrier,
    with the activation entropy providing the
    correction.
    The activation entropy itself is an $\Lk_5$ datum
    (from the Hessian at $\mathbf{R}_\mathrm{TS}$),
    so the full TST coherence check requires all of
    $(\Ce(G), V, g)$.

  \item \emph{$\Lk_1$--$\Lk_5$ coherence}
    (Proposition~\ref{prop:hess-pes-consistency}):
    the $\Lk_1$ enthalpy $\FH$ is recovered as the
    electronic energy difference between product and
    reactant minima of $V$, up to a ZPE correction
    that is itself an $\Lk_5$ datum.
    The approximate agreement ($\sim$5~kJ/mol)
    quantifies the error of the harmonic approximation
    for $\Delta E_\mathrm{ZPE}$.
\end{enumerate}

\medskip\noindent
The Berry connection along the dissociation IRC
vanishes ($A_0 = 0$, $w_1(L_0^{\RR}) = 0$) by
Proposition~\ref{prop:berry-trivial}: the
$\mathrm{N_2O_4} \to 2\mathrm{NO_2}$ path stays on
the singlet ground-state surface throughout, with no
crossing of low-lying excited states.  The full
treatment of $\mathrm{NO_2}$ radical excited states
(which become relevant at higher energies and for
photodissociation) belongs to $\Lk_6$.

Both coherences fail to be exact at $\Lk_5$ for
related reasons: the harmonic approximation to the
vibrational modes, and the classical (over-barrier)
treatment of the rate.
\end{example}

%% file: chapters/L5/l5_nextforcing.tex
\subsection{What \texorpdfstring{$\Lk_5$}{L5} cannot express:
  forcing of \texorpdfstring{$\Lk_6$}{L6}}
\label{sec:L5-forcing-out}

The entire $\Lk_5$ framework rests on Layer~2(b) of
\S\ref{sec:L5-pes}: the ground-state section
$\sigma_0$ is a smooth section of $\Hel$ on the
relevant CI-free open subset of $\Ce(G)$, equivalently,
no conical intersection (CI) between $V_0$ and $V_1$
lies in the chemically accessible region traversed
by the IRC and its low-energy basins.  Within this
restriction $\Lk_5$ completely describes ground-state
thermal chemistry via the Morse triple
$(\Ce(G), V_0, g)$.
But Layer~2(b) is not permanent.
It fails for a large class of chemically important
processes, and the failure is not a matter of
approximation quality: it is a \emph{topological
obstruction invisible to $V_0$}.

The key point is that $\Lk_5$ records only the
ground-state PES function $V_0$, not the electronic
bundle that fibres over it.
A CI is a \emph{geometric} feature of $\Ce(G)$ where
$V_0(\mathbf{R}) = V_1(\mathbf{R})$.
At the CI itself the eigenvalue $V_0$ is continuous
but not smooth (Wigner--von Neumann normal form: a
conical singularity with $V_0 \sim V_{\mathrm{CI}} -
|\mathbf{x}|$ in the two-dimensional branching plane).
On the complement $\Ce(G) \setminus X_{\mathrm{CI}}$
where the eigenvalues are isolated, $V_0$ is smooth
and the $\Lk_5$ datum cannot detect that the CI exists
nearby: only the topology of the complement, encoded in the
ground-state line bundle $L_0$ (or equivalently, in its
real sub-bundle $L_0^{\RR}$ under time-reversal
symmetry), distinguishes systems with and without the CI.

What changes is the topology of the bundles over the
punctured base $\Ce(G) \setminus X_{\mathrm{CI}}$.
Both $L_0$ (the complex ground-state line bundle) and
its real sub-bundle $L_0^{\RR}$ generated by the
real-gauge section $\tilde\sigma_0$
(Proposition~\ref{prop:berry-trivial}) are well-defined
on this complement, and both carry topological
invariants:
\[
  c_1(L_0) \in H^2(\Ce(G) \setminus X_{\mathrm{CI}},
  \ZZ),
  \qquad
  w_1(L_0^{\RR}) \in H^1(\Ce(G) \setminus
  X_{\mathrm{CI}}, \ZZ/2).
\]
Which of these is the informative invariant is set by
the codimension of $X_{\mathrm{CI}}$ in $\Ce(G)$:
\begin{itemize}
  \item For a \emph{real-symmetric} $\hat{H}_{\mathrm{el}}$
    (spinless non-relativistic electrons with
    time-reversal-invariant Coulomb interactions),
    degeneracies of the $2 \times 2$ effective block
    require two real conditions to vanish (trace shift
    and off-diagonal coupling), so $X_{\mathrm{CI}}$ has
    \emph{codimension 2}.  A small loop $S^1$ encircling
    $X_{\mathrm{CI}}$ is a 1-cycle, and the natural
    invariant of the bundle around it lives in
    $H^1(\cdot, \ZZ/2)$.  This is $w_1(L_0^{\RR})$, and
    it is non-trivial: a sign holonomy of $-1$ around the
    loop witnesses the Longuet--Higgins
    effect~\cite{LonguetHiggins1963,MeadTruhlar1979,
    Berry1984}.  The Chern class $c_1(L_0)$ around the
    same loop is automatically zero --- not because
    $L_0$ is globally trivial, but because the loop
    bounds no canonical 2-chain in the punctured
    complement, so the $H^2$-pairing does not detect
    codimension-2 obstructions.
  \item For a \emph{complex Hermitian} $\hat{H}_{\mathrm{el}}$
    without time-reversal symmetry (e.g.\ in an external
    magnetic field or with spin-orbit coupling),
    degeneracies require three real conditions and
    $X_{\mathrm{CI}}$ has \emph{codimension 3}.  A small
    sphere $S^2$ around an isolated CI point is a
    2-cycle, and $c_1(L_0)$ measured by
    $\int_{S^2}\Omega_0/2\pi \in \ZZ$ becomes the
    Berry-monopole charge.
\end{itemize}
For the thermal molecular chemistry of
$\Lk_6$ --- real-symmetric Hamiltonian, codimension-2 CI
seams --- the relevant invariant is therefore
$w_1(L_0^{\RR})$, and the molecular Aharonov--Bohm
phenomenon is the $\ZZ_2$ sign holonomy.  $c_1$ remains
a well-defined invariant of $L_0$ in its own right; it
just measures a different (codimension-3) obstruction
that is not generically present without breaking
time-reversal symmetry.

$\Lk_5$ sees $\sigma_0$ only as a function (the PES);
$\Lk_6$ sees it as a section of a bundle with
non-trivial $\ZZ_2$-holonomy.  The step from function
to bundle section --- and from numerical to topological
invariant --- is the deepest geometric step in the
tower.

\medskip\noindent\textbf{Concrete illustration:
H + H$_2$.}
The hydrogen-exchange reaction has a smooth
ground-state PES with a collinear TS barrier of
$\approx 9.6$~kJ/mol and a TST rate in good agreement
with experiment --- at $\Lk_5$ the reaction appears
fully understood.
Yet a CI exists at the equilateral $D_{3h}$ geometry,
off the collinear IRC but within the configuration
space explored by the nuclear wavefunction at reactive
collision energies.
Any loop encircling the $D_{3h}$ CI acquires a
Longuet--Higgins sign holonomy of
$-1$~\cite{LonguetHiggins1963}, producing a predicted
interference alternation in the reactive differential
cross-section of H + HD~\cite{JuanesMarcos2005}.
The single-surface $\Lk_5$ calculation cannot recover
this signature, because $w_1(L_0^{\RR}) = 0$ is imposed
by construction via Layer~2(b) (which restricts to the
simply-connected CI-free open subset of $\Ce(G)$), not computed.

\begin{forcingbox}[Forcing pair for $\Lk_6$:
  identical ground-state PES, distinct Hilbert bundle
  topology]
\label{box:L6-forcing}

\noindent\textbf{Idealised setup.}
Consider two systems $(V_0^A, L_0^A)$ and
$(V_0^B, L_0^B)$ on the same configuration orbifold
$\Ce(G)$, where $V_0^\bullet$ is the ground-state
PES and $L_0^\bullet \to \Ce(G)$ the associated
ground-state line bundle:
\begin{enumerate}[label=(\roman*)]
  \item \textbf{PES identity}: $V_0^A(\mathbf{R}) =
    V_0^B(\mathbf{R})$ for all $\mathbf{R} \in \Ce(G)$
    --- pointwise identical ground-state energy, same
    IRC, same activation barrier, same TST rate.
  \item \textbf{Distinct bundle topology}: system A
    has $w_1(L_0^{A,\RR}) = 0$ (Layer~2(b) holds on a
    simply-connected open subset of $\Ce(G)$ covering
    the chemically accessible region; Berry connection
    can be gauged to zero), whereas system B has a CI
    seam $X_{\mathrm{CI}}^B \subset \Ce(G)$ of
    codimension~2 (two real conditions on the
    $2\times 2$ effective electronic Hamiltonian:
    trace-shift and discriminant both vanish) on which
    $V_0^B(\mathbf{R}) = V_1^B(\mathbf{R})$.  The
    complement $\Ce(G) \setminus X_{\mathrm{CI}}^B$
    acquires non-trivial $\pi_1$ from loops encircling
    components of $X_{\mathrm{CI}}^B$, and around such
    loops $w_1(L_0^{B,\RR}) \neq 0$ in $H^1(\Ce(G)
    \setminus X_{\mathrm{CI}}^B, \ZZ/2)$.
\end{enumerate}
At the CI itself $V_0^B$ has the cone singularity
described above; away from $X_{\mathrm{CI}}^B$ it is
smooth and (by hypothesis) pointwise equal to $V_0^A$.
The classical IRC need not visit $X_{\mathrm{CI}}^B$
at all; what matters is that loops in $\Ce(G) \setminus
X_{\mathrm{CI}}^B$ encircling $X_{\mathrm{CI}}^B$ are
non-contractible, and nuclear wavefunctions supported
near the IRC can be probed along such loops.

This idealised pair is hypothetical --- two real
molecular systems with pointwise-identical $V_0$ but
different CI topology are not generically realised in
nature.  The forcing argument is a categorical
\emph{test of expressive power}: it asks whether the
$\Lk_5$ datum $(\Ce(G), V_0, g)$ contains enough
information to distinguish system A from system B,
and the answer is \emph{no}.  In actual chemistry, the
A vs B distinction manifests as the difference between
reactions whose accessible nuclear region encloses a
CI (e.g.\ H + HD reactive scattering at high collision
energies) and those whose region does not (e.g.\
ground-state $\mathrm{S_N2}$ at thermal conditions).

\medskip\noindent\textbf{Chemical instances.}
\begin{itemize}
  \item \textbf{Type A} (CI-free in the accessible
    region): thermal chemistry on the globally smooth
    real sub-bundle $L_0^{A,\RR}$; sign holonomy
    trivial; $w_1(L_0^{A,\RR}) = 0$.  Fully captured at
    $\Lk_5$.  Prototype: $\mathrm{S_N2}$,
    \S\ref{ex:SN2-L5}.
  \item \textbf{Type B} (accessible region encircles
    $X_{\mathrm{CI}}$): the nuclear wavefunction is
    supported on a region of $\Ce(G) \setminus
    X_{\mathrm{CI}}^B$ that contains non-contractible
    loops around components of $X_{\mathrm{CI}}^B$.  Any
    such loop acquires sign holonomy $-1$, the molecular
    Aharonov--Bohm effect~\cite{MeadTruhlar1979,
    Berry1984}.  Because the total (electronic $\times$
    nuclear) wavefunction must be single-valued on
    $\Ce(G) \setminus X_{\mathrm{CI}}^B$, the sign
    change of the electronic eigenstate around such a
    loop forces a compensating sign change of the
    nuclear wavefunction --- equivalently, the nuclear
    wavefunction is double-valued on the base and
    single-valued on a double cover~\cite{
    LonguetHiggins1963}.  This produces observable
    interference in product channels.  Prototype:
    H + HD~\cite{JuanesMarcos2005}; the canonical
    chemical realisation is developed in
    \S\ref{sec:L6-forcing-in}.
\end{itemize}

\medskip\noindent\textbf{Indistinguishability at
$\Lk_5$.}
The Morse triple $(\Ce(G), V_0, g)$ coincides for A
and B by condition~(i), so $\FV$ assigns the same
object of $\OrbMorse$ to both.
The Stiefel--Whitney class $w_1(L_0^{\RR})$ is not a
datum of $\Lk_5$: it is not computable from
$(\Ce(G), V_0, g)$, and at $\Lk_5$, Layer~2(b)
restricts to a simply-connected CI-free open subset
where $w_1$ vanishes by construction.  The swap
$\phi_6 = [A \leftrightarrow B]$ is therefore a
non-trivial element of $\coker(\varphi_6)$ in the
automorphism (pointed-set) exact sequence
\begin{equation}
\label{eq:L6-exact}
  1 \;\to\; \ker\varphi_6 \;\to\;
  \Aut(\Lk_6(P))
  \;\xrightarrow{\;\varphi_6\;}
  \Aut(\Lk_5(P))
  \;\to\; \coker(\varphi_6) \;\to\; 1,
\end{equation}
following the tower's standard pointed-set
convention: $\coker(\varphi_6) := \Aut(\Lk_5(P)) /
\mathrm{im}(\varphi_6)$ as a pointed quotient, a group
when $\mathrm{im}(\varphi_6)$ is normal in the
codomain and a coset space otherwise; the forcing
content is the same in either case.  The map
$\varphi_6$ is the restriction on automorphisms induced
by the forgetful functor $U_{6 \to 5}: \Lk_6(P) \to
\Lk_5(P)$.

The construction of $\Aut(\Lk_6(P))$ is deferred to
Chapter~\ref{sec:L6}; at this stage \eqref{eq:L6-exact}
is a design constraint stating that any categorical
structure $\Lk_6$ receiving the CI topology datum must
have non-trivial cokernel over $\Aut(\Lk_5)$.
Non-trivial $\coker(\varphi_6)$ proves that $\Lk_6$ is
strictly richer than $\Lk_5$: the $\ZZ_2$ sign
holonomy $w_1(L_0^{\RR})$ (and the associated CI seam
$X_{\mathrm{CI}} \subset \Ce(G)$) is irreducibly new
structure.
\end{forcingbox}

The extension $\Lk_5 \to \Lk_6$ differs in character
from every previous extension in the tower.
All lower extensions add structure \emph{on} the
configuration orbifold --- a PES function, a metric,
a Morse structure, DPO rules, $G^*$-equivariance.
The $\Lk_5 \to \Lk_6$ extension adds structure
\emph{over} $\Ce(G)$: the topology of the Hilbert
bundle $\Hel \to \Ce(G)$ whose fibres carry the electronic wavefunctions.

The new data that $\Lk_6$ introduces --- the CI seam
$X_{\mathrm{CI}} \subset \Ce(G)$ (codimension 2 for
real-symmetric $\hat{H}_{\mathrm{el}}$, codimension 3
when time-reversal symmetry is broken), the Berry
connection $A_0$ on $L_0$ together with its real-gauge
sign holonomy on $L_0^{\RR}$, and the two
codimension-matched invariants
\[
  w_1(L_0^{\RR}) \in H^1(\Ce(G) \setminus
  X_{\mathrm{CI}}, \ZZ/2),
  \qquad
  c_1(L_0) \in H^2(\Ce(G) \setminus X_{\mathrm{CI}},
  \ZZ),
\]
--- are developed in
\S\S\ref{sec:L6-bundle}--\ref{sec:L6-def}.  For thermal
molecular chemistry in the real-symmetric regime, the
informative invariant is $w_1$ and the forcing pair
above demonstrates its non-triviality; for systems with
broken time-reversal symmetry (external magnetic
fields, spin-orbit coupling), $c_1$ becomes informative
and the codimension-3 Berry-monopole regime applies.
$\Lk_6$ accommodates both regimes; the molecular
Aharonov--Bohm forcing pair lives in the $w_1$ regime.

%% file: chapters/ch_L6.tex
\section{\texorpdfstring{$\Lk_6$}{L6}: The Electronic Structure Level}
\label{sec:L6}

\input{chapters/L6/l6_forcing}
\input{chapters/L6/l6_bundle}
\input{chapters/L6/l6_berry}
\input{chapters/L6/l6_def}
\input{chapters/L6/l6_lh}
\input{chapters/L6/l6_photochem}
\input{chapters/L6/l6_nextforcing}

%% file: chapters/L6/l6_forcing.tex
\subsection{Forcing the extension: half-integer
  pseudorotation quanta in \texorpdfstring{$\mathrm{Na_3}$}{Na3}}
\label{sec:L6-forcing-in}

The vibrational spectrum of $\mathrm{Na_3}$ in its
$2^2E'$ excited electronic state, measured by
Delacrétaz, Grant, Whetten, Wöste, and
Zwanziger~\cite{DelacretazGrantWhettenWosteZwanziger1986},
is the canonical molecular example of fractional
pseudorotational quantisation.  In the ideal
free-pseudorotor model, an ordinary single-valued
scalar nuclear wavefunction on the pseudorotation
circle has integer angular quantum numbers $j = 0,
\pm 1, \pm 2, \ldots$.  The Jahn--Teller Berry sign
changes the boundary condition to $\psi(\phi + 2\pi)
= -\psi(\phi)$, so the allowed quantum numbers
become $j = \pm 1/2, \pm 3/2, \pm 5/2, \ldots$.
The lowest pseudorotational level lies at
$\hbar^2/(8I_\mathrm{ps})$ above the trough minimum
in this limit, with no $j = 0$ level in the
topological sector.\footnote{The $\mathrm{Na_3}$
$B$-system assignment has a subsequent rovibronic
and pseudo-Jahn--Teller literature~\cite{mayer1996rovibronic,meiswinkel1991pseudo}.
The half-integer pseudorotational description used here is the canonical Berry-phase interpretation of the observed fractional quantisation; raw line positions alone do not uniquely determine an effective Hamiltonian, and quantitative corrections from trough corrugation and rovibronic coupling modify the rigid free-rotor spectrum.}

This observation forces the passage from the scalar
geometric level $\Lk_5$ to the electronic-bundle
level $\Lk_6$.  The $\Lk_5$ projection retains only
the lower adiabatic surface $V_-$ and the
mass-weighted metric $g$; if the associated nuclear
wavefunction is treated as an ordinary single-valued
function on the pseudorotation circle, this scalar
description predicts the integer sector.  The
physical Jahn--Teller problem instead carries a
real ground-state eigenline
\[
  L_0^{\RR} \to \Ce(G) \setminus \Sigma_\mathrm{CI}
\]
whose first Stiefel--Whitney class
\[
  w_1(L_0^{\RR}) \in H^1(\Ce(G) \setminus
  \Sigma_\mathrm{CI}, \ZZ/2)
\]
evaluates non-trivially on a meridian loop linking
the CI seam~\cite{AhnParkYang2019}.  This
non-trivial $w_1$ is the topological datum that
imposes the antiperiodic boundary condition and
shifts the allowed quantum numbers by one half.
The robust discriminator is not a list of fitted
level positions but the boundary condition itself:
periodic versus antiperiodic, an obstruction class
in $\ZZ/2$ that no scalar correction to $V_-$ can
supply.

The same mathematical invariant also appears in
real Bloch-bundle topology~\cite{AhnParkYang2019}:
in systems with a real structure, $w_1$ is the
real-bundle form of a quantised Berry phase.  The
molecular and band-theoretic settings have
different base spaces and physical interpretations,
but the underlying real-line-bundle obstruction is
the same.  All three Na atoms are equivalent in
$\Aut_\mu(G) = S_3$, so the discrimination involves
no isotope labelling and happens within the
bound-state spectrum of a single chemical species.
Section~\ref{sec:L5-forcing-out} established the
abstract form of this forcing argument; the
remainder of the chapter develops the $\Lk_6$
structure required to express, prove, and extend
it.  Section~\ref{sec:Na3-CI} develops the
Mexican-hat Jahn--Teller structure of
$\mathrm{Na_3}$.  Section~\ref{sec:L6-bundle}
constructs the rank-2 adiabatic sub-bundle and
proves the codimension-2 structure of the CI seam.
Section~\ref{sec:L6-berry} constructs the Berry
connection and proves the holonomy formula
$\exp(i\oint A_0) = -1$.  Section~\ref{sec:L6-def}
defines the category $\Lk_6$ and the functor
$\Felsix$.  Section~\ref{sec:L6-photochem} extends
the framework to ultrafast photodynamics through
CIs.  Section~\ref{sec:L6-forcing-out} forces the
further extension to $\Lk_7$.

\subsubsection{\texorpdfstring{$\mathrm{Na_3}$}{Na3} and the Mexican-hat Jahn--Teller conical intersection}
\label{sec:Na3-CI}

Sodium trimer $\mathrm{Na_3}$ has molecular graph
$G = (\mathrm{Na}_1, \mathrm{Na}_2, \mathrm{Na}_3)$
with three equivalent $Z = 11$ nuclei (each
contributing one $3s$ valence electron) and
configuration orbifold $\Ce(G)$ of dimension
$3(3) - 6 = 3$.  Na--Na equilibrium distance is
$\sim 3.2$~\AA.  The ground state $X^2B_2$ is
Jahn--Teller distorted from $D_{3h}$ equilateral
into an obtuse-isoceles $C_{2v}$ structure, with
three equivalent $C_{2v}$ geometries related by
cyclic apex permutation.  Of interest here is the
electronically excited $2^2E'$ state, accessed by
near-UV absorption.

\medskip\noindent\textbf{$D_{3h}$ degeneracy.}
At equilateral $D_{3h}$ the $2^2E'$ state belongs
to the two-dimensional $E'$ irreducible
representation: two electronic states are
degenerate at every $D_{3h}$ geometry, so the
$2^2E'$ state is a double point.  Distortion off
equilateral lifts the degeneracy linearly --- the
Jahn--Teller effect.  The cheapest distortion is
along the doubly-degenerate $e'$ vibrational mode
with components $q_x$ (asymmetric stretch) and
$q_y$ (in-plane bend).  The vibronic Hamiltonian
to linear order is
\begin{equation}
\label{eq:JT-hamiltonian}
\hat H_\mathrm{JT}
= T_\mathrm{nuc}
+ \tfrac{1}{2}\omega^2(q_x^2 + q_y^2)\,\mathbb{1}_2
+ k\bigl(q_x\,\sigma_z + q_y\,\sigma_x\bigr),
\end{equation}
acting on the two-component $E'$ electronic space.
Diagonalisation at each $(q_x, q_y)$ gives
\begin{equation}
\label{eq:mexican-hat}
V_\pm(q_x, q_y) = \tfrac{1}{2}\omega^2(q_x^2 + q_y^2)
\pm k\sqrt{q_x^2 + q_y^2}.
\end{equation}
The lower sheet $V_-$ has a circular trough at
radius $\rho_0 = k/\omega^2$ with a conical
singularity at the origin --- the Mexican-hat
potential.  Note that $V_-$ is \emph{not} a Morse
function: its critical locus is the continuous
trough $\rho = \rho_0$ rather than isolated
non-degenerate critical points, and the central
singularity is a cone rather than a smooth maximum.
The conical intersection $\Sigma_\mathrm{CI}$ is
the central singular locus at $q_x = q_y = 0$,
parametrised by the totally-symmetric $a_1'$
breathing coordinate.

\medskip\noindent\textbf{Codimension of the CI seam.}
The $E'$ degeneracy requires two real conditions
$q_x = q_y = 0$ in \eqref{eq:JT-hamiltonian}; hence
$\Sigma_\mathrm{CI}$ is one-dimensional and of
codimension 2 in $\Ce(\mathrm{Na_3})$.  This is the
canonical real-symmetric codim-2 CI seam to which
$w_1$ is the relevant invariant
(\S\ref{sec:L6-bundle}); the seam is
symmetry-required and cannot be removed by
continuous perturbation preserving $D_{3h}$ at the
central point.

\medskip\noindent\textbf{Labelled branching plane
and the $S_3$ quotient.}
The local $E \otimes e$ analysis is carried out on
the \emph{labelled} branching plane $(q_x, q_y) \in
\RR^2 \setminus \{0\}$, before quotienting by
nuclear permutations.  The meridian $\phi : 0 \to
2\pi$ on this labelled plane links the CI seam once
and detects $w_1(L_0^{\RR})$.  Passing to the
$S_3$-orbifold quotient identifies the three
equivalent $C_{2v}$ minima at $\phi = 0, 2\pi/3,
4\pi/3$ but does not remove the local meridian
sign holonomy. 
Nuclear exchange-statistical constraints --- which classify states by $S_3$
irreducible representation and select physically allowed combinations of nuclear spin and pseudorotation --- are not imposed at $\Lk_6$; they belong to $\Lk_7$
(\S\ref{sec:L6-forcing-out}).

\medskip\noindent\textbf{Pseudorotation.}
On the labelled branching plane, parametrise the
trough by an angle $\phi \in [0, 2\pi)$.  As $\phi$
varies, the molecule passes through the three
equivalent $C_{2v}$ structures at $\phi = 0,
2\pi/3, 4\pi/3$ as the labelling of which Na atom
sits at the obtuse-isoceles apex cycles by one
position with each increment of $2\pi/3$.  This
circulation is \emph{pseudorotation}: the labelling
rotates around the central CI while each atom
undergoes only small periodic displacement.
Higher-order vibronic corrections add a small
three-fold corrugation ($\sim 50$~cm$^{-1}$) around
the trough, much smaller than typical vibrational
quanta, so pseudorotation averages over the three
$C_{2v}$ minima at low excitation.

\medskip\noindent\textbf{$\ZZ_2$ sign holonomy.}
In polar coordinates $(q_x, q_y) = (\rho\cos\phi,
\rho\sin\phi)$, the electronic block of
\eqref{eq:JT-hamiltonian} at fixed $\rho = \rho_0$
becomes $k\rho_0(\cos\phi\,\sigma_z +
\sin\phi\,\sigma_x)$, with eigenvalues $\pm
k\rho_0$.  One convenient real gauge for the lower
adiabatic eigenline is
\[
  |\sigma_-(\phi)\rangle
  \;=\; \sin(\phi/2)\,|1\rangle
  - \cos(\phi/2)\,|2\rangle,
\]
up to overall sign convention, where $|1\rangle,
|2\rangle$ span the $E'$ electronic space at $\phi
= 0$~\cite{LonguetHiggins1975,MeadTruhlar1979}.  Under $\phi
\to \phi + 2\pi$ the half-angle becomes $\phi/2 +
\pi$ and both trigonometric factors flip sign:
$|\sigma_-(\phi + 2\pi)\rangle = -|\sigma_-(\phi)
\rangle$.  This is the Longuet--Higgins sign-change
theorem for the $E \otimes e$ Jahn--Teller system.
The corresponding Berry holonomy is
$\exp(i\oint_\gamma A_0) = -1$, the molecular
Aharonov--Bohm effect~\cite{Berry1984,
MeadTruhlar1979}, depending only on the homotopy
class of $\gamma$ in $\Ce(\mathrm{Na_3}) \setminus
\Sigma_\mathrm{CI}$.  Categorically the $\ZZ_2$
holonomy is recorded by $w_1(L_0^{\RR})$, taking
the value $1 \in \ZZ/2$ on every loop linking
$\Sigma_\mathrm{CI}$ (equivalently, sign holonomy
$(-1)^1 = -1$).

\begin{mathbox}[Cohomology, codimension, and
  invariant]
\label{box:cohomology-intro}
For a smooth (orbi-)manifold $X$, removing a
submanifold $\Sigma$ of codimension $d$ yields new
cohomology in degree $d - 1$:
\begin{itemize}
  \item \emph{Codim 2}: link is small $S^1$,
    generating $H_1(X \setminus \Sigma)$.  Natural
    $\ZZ_2$-valued invariant of a real line bundle:
    $w_1 \in H^1(X \setminus \Sigma, \ZZ/2)$.
  \item \emph{Codim 3}: link is small $S^2$,
    generating $H_2(X \setminus \Sigma)$.  Natural
    $\ZZ$-valued invariant of a complex line bundle:
    $c_1 \in H^2(X \setminus \Sigma, \ZZ)$.
\end{itemize}

For real-symmetric $\hat{H}_{\mathrm{el}}$,
$\Sigma_\mathrm{CI}$ has codimension 2;
$w_1(L_0^{\RR})$ is the $\ZZ_2$-valued datum that
$\Felsix$ adds to the scalar $\Lk_5$ description (Proposition~\ref{prop:linking-w1}).
The complex Chern class $c_1(L_0) \in H^2$ does not detect
codimension-2 obstructions (which live in $H^1$); it becomes
informative only in codim-3 settings (broken TRS, Berry monopoles), where the link is $S^2$, a parallel branch of the same construction.

\smallskip
\noindent\textbf{Data progression through the tower.}
\[
  \underbrace{\ZZ^s}_{\Lk_0}
  \to
  \underbrace{\RR}_{\Lk_1\text{--}\Lk_3}
  \to
  \underbrace{C^\infty(\Ce(G))}_{\Lk_5}
  \to
  \underbrace{H^1(\Ce(G)\setminus\Sigma_\mathrm{CI},
  \ZZ/2)}_{\Lk_6\text{ (real, codim 2)}}.
\]
Each step is strictly less deformable than the
previous; a $\ZZ_2$ invariant changes only in discrete steps.
\end{mathbox}

\begin{chembox}[Why $\ZZ_2$-valuedness matters chemically]
\label{chem:ci-topology}
$w_1(L_0^{\RR})$ is a topologically protected
property of a molecule's electronic structure,
unlike $V^\ddagger$ or $\Delta G$ which vary
continuously with substitution, solvent, or
temperature.  Along any closed loop of
configurations $w_1$ takes the value $0$ (no CI
enclosed) or $1$ (odd number of CI components
enclosed); there is no ``half a conical
intersection''.  Continuous perturbations cannot
turn a CI on or off gradually --- the gap either
closes along the path ($w_1$ flips) or it does not.
This discreteness makes CI topology experimentally
certifiable: a $\pm$ outcome on each loop-resolved
measurement, not a continuous number to fit.
\end{chembox}

\begin{forcingbox}[Forcing pair for $\Lk_6$: integer
  vs.\ half-integer pseudorotation quanta of
  $\mathrm{Na_3}$]
\label{box:Na3-forcing}

\noindent\textbf{Setup.}
$G = \mathrm{Na_3}$ with configuration orbifold
$\Ce(\mathrm{Na_3})$ and metric $g$ fixed.  The
Mexican-hat lower-sheet surface $V_-$ from
\eqref{eq:mexican-hat} (with small three-fold
corrugation) is common to the two descriptions
below; $I_\mathrm{ps}$ is determined by $V_-$ and
$g$.  The $\Lk_5$ functor $\FV$ retains only the
scalar pair $(V_-, g)$ on
$\Ce(\mathrm{Na_3}) \setminus \Sigma_\mathrm{CI}$
and produces the same $\Lk_5$ image in both
descriptions.

\medskip\noindent\textbf{Description A ($\Lk_5$:
single-valued nuclear wavefunction).}
The nuclear wavefunction is an ordinary function on
$\Ce(\mathrm{Na_3}) \setminus \Sigma_\mathrm{CI}$
(section of the trivial real line bundle); the
Mead--Truhlar connection is set to zero.  The
pseudorotation Hamiltonian
\[
  \hat H_\mathrm{ps}^{(A)}
  = -\frac{\hbar^2}{2I_\mathrm{ps}}
  \frac{\partial^2}{\partial\phi^2}
  + V_\mathrm{trough}(\phi),
  \qquad
  \psi(\phi + 2\pi) = \psi(\phi),
\]
has eigenvalues
$E_j^{(A)} \approx \hbar^2 j^2/(2I_\mathrm{ps})$
with $j = 0, \pm 1, \pm 2, \ldots$ in the
free-pseudorotor limit; lowest level at $j = 0$.

\medskip\noindent\textbf{Description B ($\Lk_6$:
real sub-bundle with $w_1(L_0^{\RR}) = 1$).}
The nuclear wavefunction is a section of
$L_0^{\RR}$ with $w_1 = 1$ on every pseudorotation
loop --- equivalently, a function on the double
cover, antiperiodic under the deck transformation
$\psi(\phi + 2\pi) = -\psi(\phi)$.  Same operator
$\hat H_\mathrm{ps}$, enlarged function space;
eigenvalues
$E_j^{(B)} \approx \hbar^2 j^2/(2I_\mathrm{ps})$
with $j = \pm 1/2, \pm 3/2, \pm 5/2, \ldots$ in the
free-pseudorotor limit; lowest level at $j = \pm
1/2$ with zero-point energy
$\hbar^2/(8I_\mathrm{ps})$ above the trough bottom.
No $j = 0$ level exists in this topological sector.

\medskip\noindent\textbf{Empirical anchor.}
The 1986 measurement of the $2^2E'$ state by R2PI
on cold supersonic-beam
$\mathrm{Na_3}$~\cite{DelacretazGrantWhettenWosteZwanziger1986}
resolves the vibrational structure consistent with
the half-integer pseudorotational assignment.
Quantitative corrections from trough corrugation
and rovibronic coupling modify the rigid free-rotor
spacings, but the topological boundary condition
distinguishing A from B remains the robust
datum~\cite{mayer1996rovibronic,meiswinkel1991pseudo}.

\medskip\noindent\textbf{Non-trivial fibre of
$U_{6 \to 5}$.}
Descriptions A and B are two distinct lifts of the
same $\Lk_5$ object $(\Ce(\mathrm{Na_3}), V_-, g)$
under the forgetful projection
\[
  U_{6 \to 5} : \Lk_6 \longrightarrow \Lk_5.
\]
The fibre $U_{6 \to 5}^{-1}(V_-, g)$ is non-trivial:
one lift carries trivial real eigenline monodromy
($w_1 = 0$), the physical Jahn--Teller lift carries
$w_1(L_0^{\RR}) \neq 0$.  The $\Lk_5$ image does
not contain enough data to select between them; the
topological discriminator is the bundle datum
$w_1$.  The purely scalar, single-valued $\Lk_5$
description is therefore empirically insufficient
for this spectrum: the $\Lk_6$ lift with non-trivial
$w_1$ supplies the missing boundary condition.

\medskip\noindent\textbf{Three qualitative failures
of the $\Lk_5$ description.}
(i) \emph{Boundary condition:} $\Lk_5$ enforces
periodicity $\psi(\phi + 2\pi) = \psi(\phi)$ by
treating $\psi$ as a single-valued function on the
pseudorotation circle; the physical eigenstate is
antiperiodic.
(ii) \emph{Lowest level:} $\Lk_5$ predicts a
$j = 0$ ground state at the trough bottom; the
$\Lk_6$ lift and experiment exclude this level,
the spectrum starting at $j = \pm 1/2$.
(iii) \emph{Non-adiabatic coupling:} $d_{01}^\mu =
\langle\sigma_-|\partial_\mu \hat{H}_\mathrm{el}|
\sigma_+\rangle / (V_+ - V_-)$ diverges as
$\mathbf{R} \to \Sigma_\mathrm{CI}$; single-surface
Born--Oppenheimer breaks down at the
CI~\cite{Teufel2003,LasserTeufel2005,
FermanianKammererLasser2008,ColinDeVerdiere2003}.
The rank-2 adiabatic sub-bundle is the natural
$\Lk_6$ object.

None of (i)--(iii) is a quantitative correction:
each is a topological feature of the $\Lk_6$ lift
that no continuous adjustment of the scalar $V_-$
supplies.  Antiperiodicity is a $\ZZ_2$ obstruction
class, not a numerical parameter.
\end{forcingbox}

The three failures identify the new data $\Lk_6$
carries: the codimension-2 CI seam
$\Sigma_\mathrm{CI} \subset \Ce(G)$
(\S\ref{sec:L6-bundle}), the Berry connection $A_0$
on $L_0$ with real-gauge sign holonomy on
$L_0^{\RR}$ (\S\ref{sec:L6-berry}), and the
codimension-matched invariants $w_1(L_0^{\RR}) \in
H^1(\cdot, \ZZ/2)$ (real-symmetric, codim-2) and
$c_1(L_0) \in H^2(\cdot, \ZZ)$ (broken-TRS,
codim-3), formalised as outputs of $\Felsix$
(\S\ref{sec:L6-def}).

\begin{insightbox}[Scattering corroboration in
  $\mathrm{H} + \mathrm{HD}$]
\label{ins:HHD-scattering}
The same $w_1 \neq 0$ bundle structure appears in
reactive scattering.  H$_3$ shares the $D_{3h}$
Jahn--Teller $E \otimes e$ structure of
$\mathrm{Na_3}$, with a codim-2 CI seam at
equilateral-triangle geometries.  At collision
energies above the CI ($E_\mathrm{coll} \approx
2.77$~eV), crossed-beam imaging experiments
resolve forward-scattering oscillations in
$(v', j')$-resolved differential cross sections of
$\mathrm{H} + \mathrm{HD} \to \mathrm{H_2} +
\mathrm{D}$~\cite{YuanScience2018}; the analogous
effect below the CI was reported
in~\cite{YuanNatCommun2020}.  Single-surface
$\Lk_5$ scattering on the ground-state
PES~\cite{BKMP1996} predicts smooth angular
distributions; including the Mead--Truhlar vector
potential, or imposing $w_1 = 1$ boundary
conditions on loops linking the CI seam,
reproduces the oscillations
quantitatively~\cite{JuanesMarcos2005}.  The
topological sign comes from the electronic CI of
the H$_3$ system; the H/D substitution makes
product channels experimentally distinguishable
and changes the scattering kinematics, but the
geometric-phase mechanism itself is the $\ZZ_2$
holonomy associated with loops around the H$_3$
CI.  Bound-state spectroscopy of $\mathrm{Na_3}$
and reactive scattering of $\mathrm{H} +
\mathrm{HD}$ are two manifestations of the same
$w_1$ holonomy in different observable channels.
\end{insightbox}

%% file: chapters/L6/l6_bundle.tex
\subsection{The full electronic Hilbert bundle and the conical intersection seam}
\label{sec:L6-bundle}

\subsubsection{From a scalar surface to an electronic bundle}

At $\Lk_5$, the electronic structure of a molecular graph $G$ is retained only through its scalar adiabatic shadow: a potential-energy surface $V_0 : \Omega \subset \Ce(G) \to \RR$ and the mass-weighted metric $g$. 
In the clamped-nuclei Born--Oppenheimer setting,
\[
  V_0(\mathbf{R})
  \;=\;
  E_{\rm el,0}(\mathbf{R}) + V_{\rm nn}(\mathbf{R}),
\]
where $E_{\rm el,0}$ is the lowest eigenvalue of the electronic Hamiltonian $\hat H_{\rm el}(\mathbf R)$ and $V_{\rm nn}$ is the nuclear--nuclear repulsion.
Section~\ref{sec:L6-forcing-in} showed that the
scalar shadow is qualitatively incomplete in the presence of a Jahn--Teller / conical-intersection sign effect; the missing datum is the topology of the real ground-state eigenline.
The minimal extension carries, at each $\mathbf{R}$, the $N$-dimensional spectral subspace of the lowest $N \geq 2$ states of $\hat H_{\rm el}$, well-defined as a whole under internal degeneracies provided the $N$-th level remains spectrally isolated from the $(N+1)$-st.

\begin{definition}[Rank-$N$ adiabatic bundle]
\label{def:rank-N-bundle}
Let $G \in \LGraphP$ and $N \geq 2$.  Let $\Omega \subset \Ce(G)$ be an open adiabatic region on which $E_N(\mathbf{R}) - E_{N-1}(\mathbf{R}) \geq \Delta > 0$.
Let
\[
  P_N(\mathbf{R})
  \;=\;
  \frac{1}{2\pi i}\oint_{\Gamma(\mathbf{R})}
  \bigl(z - \hat H_{\rm el}(\mathbf{R})\bigr)^{-1}
  \,dz
\]
be the Riesz spectral projector onto the lowest $N$ states, with $\Gamma(\mathbf{R})$ a smooth contour separating $\{E_0, \ldots, E_{N-1}\}$ from the rest of the spectrum.
The \emph{rank-$N$ adiabatic bundle} is
\[
  \pi : \Hel^{(N)} \to \Omega,
  \qquad
  \Hel^{(N)}_{\mathbf{R}}
  := \operatorname{Ran} P_N(\mathbf{R}),
\]
a smooth Hermitian vector bundle of rank $N$ with structure group $U(N)$, well-defined even under internal degeneracies because the spectral subspace is isolated from the $(N+1)$-st state on $\Omega$~\cite{Teufel2003}.
\end{definition}

On the non-degenerate locus $\Omega^\circ := \Omega \setminus \Xseam$ (where $E_0 < E_1$), the ground-state eigenline $L_0 \to \Omega^\circ$ is the smooth complex line sub-bundle with fibre $\ker(\hat H_{\rm el}(\mathbf R) - E_0(\mathbf R))$.
In the real time-reversal-symmetric case ($\hat H_{\rm el}^\top = \hat H_{\rm el}$ in a real basis), $L_0$ carries a canonical real form $L_0^{\RR} \to \Omega^\circ$ with structure group $O(1) = \{\pm 1\}$.
Its first Stiefel--Whitney class
\[
  w_1(L_0^{\RR}) \;\in\; H^1(\Omega^\circ;\, \ZZ/2)
\]
is the categorical invariant the $\Lk_6$ level adds to the $\Lk_5$ scalar data: for any loop $\gamma \subset \Omega^\circ$ the observable sign holonomy of the real eigenline is
\[
  \operatorname{Hol}_{L_0^{\RR}}(\gamma)
  \;=\;
  (-1)^{\langle w_1(L_0^{\RR}),[\gamma]\rangle}
  \;\in\; \{\pm 1\}.
\]
The complex Chern class $c_1(L_0) \in H^2(\Omega^\circ; \ZZ)$ vanishes in the real-symmetric case; it becomes informative only when time-reversal symmetry is broken, where the degeneracy locus has codimension $3$ and the link is $S^2$.
An $\Lk_6$ lift accordingly records the tuple
\[
  \bigl(\Omega,\,\Hel^{(N)},\,A,\,\Xseam,\,
  w_1(L_0^{\RR})\bigr),
\]
comprising the active region, rank-$N$ active bundle, Berry connection $A$, CI seam, and Stiefel--Whitney class of the real ground-state eigenline.

\subsubsection{The conical intersection seam}

\begin{definition}[Conical intersection seam]
\label{def:ci-seam}
Let $\Omega \subset \Ce(G)$ be an adiabatic region with an isolated rank-$2$ active subspace $\Hel^{(2)}$.
The \emph{conical intersection seam} between the two lowest states is
\[
  \Xseam
  \;:=\;
  \bigl\{\,\mathbf{R} \in \Omega
  \;\big|\;
  E_0(\mathbf{R}) = E_1(\mathbf{R}),\
  \text{regular conical at } \mathbf{R}
  \,\bigr\}.
\]
At $\mathbf{R}_0 \in \Xseam$, in a local smooth frame of $\Hel^{(2)}$, the traceless effective two-state Hamiltonian in the real time-reversal-symmetric case takes the form
\[
  H_{\rm eff}^0(\mathbf{R})
  \;=\;
  x(\mathbf{R})\,\sigma_z
  + y(\mathbf{R})\,\sigma_x,
\]
with $x, y$ smooth real functions.
The degeneracy is \emph{regular conical} if $dx_{\mathbf{R}_0}, dy_{\mathbf{R}_0}$ are linearly independent in $T^*_{\mathbf{R}_0}\Omega$; equivalently, the local gap is $E_1 - E_0 = 2\sqrt{x^2 + y^2} + O(|(x,y)|^2)$.
The covectors $dx, dy$ are the gradient-difference and derivative-coupling directions in a branching-plane representation~\cite{Yarkony1996,
DomckeYarkony2012}; the rank condition is invariant although individual covectors are not.
Adiabatic eigenvectors $\sigma_0, \sigma_1$ are not canonically defined at $\mathbf{R}_0$ --- only the two-dimensional spectral subspace $\Hel^{(2)}_{\mathbf{R}_0}$ is.
\end{definition}

\begin{proposition}[CI seam has codimension 2]
\label{prop:ci-codim2}
Let $G$ have $f = 3n - 6 \geq 3$ internal degrees of freedom and let $\hat H_{\rm el}$ be real-symmetric on $\Omega$.
For a generic $\hat H_{\rm el}$, the regular part of $\Xseam$ is a smooth submanifold of codimension $2$ in $\Omega$, of dimension $f - 2$; non-regular points form a closed subset of positive codimension within $\Xseam$.
\end{proposition}

\begin{proof}
The traceless two-state model $H_{\rm eff}^0 = x\sigma_z + y\sigma_x$ has gap $2\sqrt{x^2 + y^2}$, so degeneracy requires $x = y = 0$: two independent real conditions.
At a regular point, $dx, dy$ are linearly independent and the implicit function theorem gives a codimension-$2$ submanifold.
Non-regular points form a closed positive-codimension subset by genericity~\cite{vonNeumannWigner1929}.
For complex Hermitian $\hat H_{\rm el}$, the traceless two-state model is $x\sigma_x + y\sigma_y + z\sigma_z$, requiring three independent real conditions, so $\codim(\Xseam) = 3$.
\qedhere
\end{proof}

\begin{remark}[Local model and half-angle sign change]
\label{rem:local-model-sign}
In polar coordinates $(q_x, q_y) = (\rho\cos\phi, \rho\sin\phi)$, the local model $H_{\rm eff}^0 = q_x\sigma_z + q_y\sigma_x$ has a real lower-state eigenvector depending on the half-angle $\phi/2$ that flips sign under $\phi \mapsto \phi + 2\pi$ --- the Longuet--Higgins sign change~\cite{LonguetHiggins1975}.
In bundle language, $\langle w_1(L_0^{\RR}),
[S^1_{\rm mer}]\rangle = 1$: the real codim-$2$ CI is detected by Stiefel--Whitney on a meridian \emph{loop}, not by a Chern class on a meridian sphere.
\end{remark}

\begin{chembox}[Codimension 2 picks out $H^1$ and $w_1$]
A codimension-$d$ submanifold $\Sigma \subset X$ has small normal link $S^{d-1}$, so topological obstructions in $X \setminus \Sigma$ live in cohomology of degree $d - 1$.

For a real time-reversal-symmetric CI, $\Xseam$ has codimension $2$ and link $S^1$; the invariant is $w_1(L_0^{\RR}) \in H^1(X \setminus \Xseam; \ZZ/2)$,
and pairing with the meridian gives the Longuet--Higgins sign change.
For broken time-reversal symmetry, the seam has codimension $3$, link $S^2$, and invariant $c_1 \in H^2$.
The two cases are parallel branches of the same
$\Lk_6$ construction; thermal molecular chemistry inhabits the $w_1$ branch.
\end{chembox}

\subsubsection{The Stiefel--Whitney invariant and holonomy}

\begin{proposition}[Meridian loops and the $\Lk_6$ invariant]
\label{prop:linking-w1}
Let $\Xseam \subset X \subset \Ce(G)$ be a regular codimension-$2$ CI seam for a real-symmetric Hamiltonian, and let $X^\circ := X \setminus \Xseam$.  Then:
\begin{enumerate}[label=(\roman*)]
  \item Each connected component of $\Xseam$
    contributes one $\ZZ/2$ generator to
    $H^1(X^\circ; \ZZ/2)$, dual to a small meridian loop $S^1$.
  \item The first Stiefel--Whitney class of
    $L_0^{\RR}$ evaluates to $1 \in \ZZ/2$ on every such meridian.
  \item For any loop $\gamma \subset X^\circ$,
    \[
      \operatorname{Hol}_{L_0^{\RR}}(\gamma)
      \;=\;
      (-1)^{\langle
        w_1(L_0^{\RR}),[\gamma]\rangle}
      \;=\;
      (-1)^{\operatorname{lk}_2(\gamma, \Xseam)},
    \]
    with $\operatorname{lk}_2(\gamma, \Xseam) \in
    \ZZ/2$ the mod-$2$ linking number.
\end{enumerate}
\end{proposition}

\begin{proof}
(i) follows from the mod-$2$ Thom--Gysin sequence for the closed codim-$2$ submanifold $\Xseam \subset X$: one $\ZZ/2$ generator per connected component, dual to a meridian loop.

(ii) is Remark~\ref{rem:local-model-sign}: in the normal slice, the real lower-state eigenvector flips sign under $\phi \mapsto \phi + 2\pi$, so $\langle w_1, [S^1_{\rm mer}]\rangle = 1$; the result extends to all meridians by homotopy invariance.

(iii) The holonomy of a real line bundle around a loop is by definition $(-1)^{\langle w_1, [\gamma]\rangle}$.  For $\gamma \subset X^\circ$, the pairing equals the mod-$2$ count of transverse intersections of any spanning chain with $\Xseam$, which is $\operatorname{lk}_2(\gamma, \Xseam)$.
\qedhere
\end{proof}

The proposition makes the $\Lk_5$ vs.\ $\Lk_6$ distinction explicit.
An avoided crossing has $\Xseam = \emptyset$ and $L_0^{\RR}$ extends as a smooth real line bundle over $\Omega$ with finite derivative coupling.
A regular CI has non-empty $\Xseam$ and non-trivial sign holonomy on meridians.
The scalar $\Lk_5$ datum cannot distinguish these cases --- both produce the same $V_0$ on the accessible region --- but $\Lk_6$ does, through the $\ZZ/2$ class $w_1$.

\subsubsection{The forgetful functor \texorpdfstring{$U_6$}{U6}}

\begin{mathbox}[The $U_6$ diagram]
\label{box:U6-diagram}
The level-to-level forgetful functor $U_6 : \Lk_6(P) \to \Lk_5(P)$ is characterised by the commutative square
\[
  \includegraphics{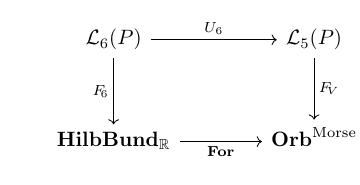}
\]
where $\mathbf{For}$ sends
$(\Omega, \Hel^{(N)}, A, \Xseam, w_1) \mapsto (\Omega, V_0, g)$, retaining the scalar adiabatic shadow and discarding the bundle data.
Commutativity $\mathbf{For} \circ \Felsix = \FV \circ U_6$ expresses Layer~2(b) consistency: the ground-state surface retained by $U_6$ equals the scalar shadow recorded by $\FV$.

The Na$_3$ forcing example (\S\ref{sec:L6-forcing-in}) is a non-faithfulness statement for $U_6$: the scalar lower-sheet description $(\Omega, V_-, g)$ admits two distinct lifts,
\begin{align*}
  \text{trivial:}\quad&
  (G, \Hel^{(N)}, A, \Xseam, w_1 = 0), \\
  \text{Jahn--Teller:}\quad&
  (G, \Hel^{(N)}, A, \Xseam, w_1 \neq 0),
\end{align*}
with the same $\Lk_5$ image.
The fibre $U_6^{-1}(\Omega, V_-, g)$ is non-trivial.
The trivial lift is a comparison object, not a second physical Hamiltonian: it makes the failure of $U_6$ to be an equivalence explicit.
\end{mathbox}

%% file: chapters/L6/l6_berry.tex
\subsection{Berry connection, sign holonomy, and the
  topological phase}
\label{sec:L6-berry}

Section~\ref{sec:L6-bundle} established the rank-$N$
adiabatic bundle $\Hel^{(N)} \to \Omega$, located the
CI seam $\Xseam$ as the codimension-$2$ degeneracy
locus, and identified the categorical invariant of
the $\Lk_6$ level as the first Stiefel--Whitney class
$w_1(L_0^{\RR}) \in H^1(\Omega^\circ; \ZZ/2)$ of the
real ground-state eigenline.  This section develops
the analytic representation of that invariant in
terms of a Berry connection, fixes the gauge story
in the real-symmetric case, and exhibits the sign
holonomy as the $\Lk_6$ datum that the scalar $\Lk_5$
shadow cannot supply.

\subsubsection{The Berry connection}

The Berry connection on the chosen active spectral
bundle $\Hel^{(N)} \to \Omega$ is the projected
connection
\[
  \nabla^{\rm B} \;=\; P_N \, d,
\]
induced by projecting the trivial Hilbert-space
connection with the Riesz projector $P_N$ of
Definition~\ref{def:rank-N-bundle}.  In a local
orthonormal frame $\{\sigma_k\}_{k=0}^{N-1}$, its
matrix elements are~\cite{Berry1984,WilczekZee1984,
Teufel2003}
\[
  A_{kl}^\mu(\mathbf{R})
  \;=\;
  i\,\langle\sigma_k(\mathbf{R})\,|\,
  \partial_\mu\,\sigma_l(\mathbf{R})\rangle.
\]
The construction depends on the choice of active
projector $P_N$.  Under a smooth gauge transformation
$\sigma_k \mapsto \sum_j U_{jk}\,\sigma_j$ with
$U : \Omega \to U(N)$, the connection transforms by
the standard rule
\[
  A \;\mapsto\; U^\dagger A U + i\,U^\dagger dU.
\]
On the non-degenerate locus $\Omega^\circ := \Omega
\setminus \Xseam$ where the ground-state eigenline
$L_0 \subset \Hel^{(N)}$ is a smooth rank-one
sub-bundle, the abelian Berry connection on $L_0$ is
the restriction
\[
  A_0^\mu(\mathbf{R})
  \;=\;
  i\,\langle\sigma_0(\mathbf{R})\,|\,
  \partial_\mu\,\sigma_0(\mathbf{R})\rangle.
\]

\subsubsection{Real Hamiltonians: local vanishing
  and global monodromy}

For real-symmetric $\hat H_{\rm el}$ the abelian
Berry connection vanishes locally in any real gauge.
The non-trivial $\Lk_6$ topology is encoded entirely
in the failure of any real gauge to extend
single-valuedly around loops linking $\Xseam$.

\begin{proposition}[Local vanishing of $A_0$ in a
  real gauge]
\label{prop:real-A0-zero}
Let $\hat H_{\rm el}$ be real-symmetric on $\Omega$,
and let $\sigma_0(\mathbf R)$ be a smooth real
normalised ground-state eigenvector on a
contractible open patch $U \subset \Omega^\circ$.
Then
\[
  A_0 \;=\; i\,\langle\sigma_0 | d\sigma_0\rangle
  \;=\; 0 \qquad \text{on } U.
\]
\end{proposition}

\begin{proof}
Since $\sigma_0$ is real-valued in the chosen real
basis, $\langle a | b\rangle = \langle b | a\rangle$
for any real $a, b$, so
\[
  0
  \;=\; d\,\langle\sigma_0|\sigma_0\rangle
  \;=\; \langle d\sigma_0|\sigma_0\rangle
      + \langle\sigma_0|d\sigma_0\rangle
  \;=\; 2\,\langle\sigma_0|d\sigma_0\rangle.
\]
Hence $\langle\sigma_0|d\sigma_0\rangle = 0$ and
therefore $A_0 = 0$ identically on $U$.
\qedhere
\end{proof}

\begin{proposition}[Half-angle gauge and meridian
  monodromy]
\label{prop:half-angle-monodromy}
In the local two-state model $H_{\rm eff}^0(q_x,
q_y) = q_x\sigma_z + q_y\sigma_x$ on a normal slice
to $\Xseam$, write $q_x = \rho\cos\phi$, $q_y =
\rho\sin\phi$.  A real lower-state eigenvector may
be chosen as
\[
  \sigma_0(\phi)
  \;=\;
  \sin(\phi/2)\,|1\rangle - \cos(\phi/2)\,|2\rangle,
\]
for which $A_0 = 0$ on any simply connected
$\phi$-patch by
Proposition~\ref{prop:real-A0-zero}.  Under $\phi
\mapsto \phi + 2\pi$,
\[
  \sigma_0(\phi + 2\pi) \;=\; -\,\sigma_0(\phi).
\]
The transition function around the meridian loop is
therefore $-1$; equivalently, the $O(1)$-holonomy
of $L_0^{\RR}$ around the meridian is $-1$ and
\[
  \langle w_1(L_0^{\RR}), [S^1_{\rm mer}]\rangle
  \;=\; 1.
\]
\end{proposition}

\begin{proof}
The eigenvector identity follows from the local
model (Remark~\ref{rem:local-model-sign} of
\S\ref{sec:L6-bundle}).  Vanishing of $A_0$ is
Proposition~\ref{prop:real-A0-zero}.  The sign
change is direct: $\sin(\phi/2 + \pi) =
-\sin(\phi/2)$ and $\cos(\phi/2 + \pi) =
-\cos(\phi/2)$.
\qedhere
\end{proof}

\noindent For the $E \otimes e$ Jahn--Teller model
of $\mathrm{Na_3}$, the pseudorotation loop is
precisely such a meridian loop in the branching
plane, so the above sign holonomy is the
topological origin of the half-integer
pseudorotation quantum numbers established in
\S\ref{sec:L6-forcing-in}.

\begin{remark}[Singular $U(1)$ representation of the
  same holonomy]
\label{rem:singular-U1}
The sign holonomy can also be represented in a
singular complex gauge.  Define
\[
  \tilde\sigma_0(\phi)
  \;:=\; e^{i\phi/2}\,\sigma_0(\phi),
\]
which is single-valued: $\tilde\sigma_0(\phi +
2\pi) = e^{i\pi}\cdot(-\sigma_0(\phi)) =
\tilde\sigma_0(\phi)$.  With the convention $A =
i\langle\sigma | d\sigma\rangle$ used throughout
this section, a direct computation gives
\[
  \tilde A_0 \;=\;
  i\,\langle\tilde\sigma_0 | d\tilde\sigma_0\rangle
  \;=\; -\tfrac12\,d\phi,
\]
so $\oint_{S^1_{\rm mer}}\tilde A_0 = -\pi$.
Changing the sign convention for the Berry
connection reverses this sign, but the holonomy
$\exp(i\oint\tilde A_0) = -1$ is unchanged and is
the only convention-independent statement.  This
$U(1)$ representation should not be confused with a
half-integer first Chern class: $L_0^{\RR}$ is a
real line bundle with structure group $O(1)$, and
the $\pm 1$ is its mod-$2$ monodromy.  The
$\pm 1/2$ in the singular gauge is an artefact of
representing $O(1)$ inside $U(1)$, not the value of
an integer characteristic class.
\end{remark}

\subsubsection{Berry curvature on the complement of
  the seam}

\begin{definition}[Berry curvature]
\label{def:berry-curvature}
The \emph{Berry curvature} of the abelian
connection on $L_0 \to \Omega^\circ$ is the
gauge-invariant $2$-form $\Omega_0 = dA_0$.  In
second-order perturbation theory,
\[
  \Omega_0^{\mu\nu}(\mathbf{R})
  \;=\;
  -2\,\mathrm{Im}\!
  \sum_{k}
  \frac{
  \langle\sigma_0|\partial_\mu\hat H_{\rm el}
  |\sigma_k\rangle\,
  \langle\sigma_k|\partial_\nu\hat H_{\rm el}
  |\sigma_0\rangle
  }{(E_k - E_0)^2},
\]
where the sum runs over excited states in the
chosen spectral resolution.  For the exact
ground-state eigenline, the sum is over all $k \geq
1$; in an $N$-state truncated model the same
formula holds after projection onto the active
subspace, with truncation error inherent in the
model choice~\cite{Berry1984,WilczekZee1984}.
\end{definition}

\begin{proposition}[Real-symmetric curvature
  vanishes pointwise]
\label{prop:omega-zero}
If $\hat H_{\rm el}$ is real-symmetric, then
$\Omega_0 = 0$ pointwise on $\Omega^\circ$.
\end{proposition}

\begin{proof}
With $\hat H_{\rm el}$ real-symmetric in a real
basis, all matrix elements
$\langle\sigma_0|\partial_\mu\hat H_{\rm
el}|\sigma_k\rangle$ are real, so each summand in
the perturbative formula is real and the imaginary
part vanishes.  Equivalently, by
Proposition~\ref{prop:real-A0-zero}, $A_0 = 0$
locally in a real gauge, hence $\Omega_0 = dA_0 =
0$.
\qedhere
\end{proof}

\begin{remark}[Distributional ``$\pi$-flux''
  picture and derivative coupling]
\label{rem:pi-flux-singular}
In the singular complex gauge of
Remark~\ref{rem:singular-U1}, $\tilde A_0 =
-\tfrac12 d\phi$ has vanishing exterior derivative
on $\Omega^\circ$ but is not exact there; the
$-\pi$ holonomy on a meridian is sometimes written
as a $\pm\pi$ delta-flux supported on $\Xseam$.
This is a useful physicist's heuristic, but the
gauge-free formulation of the obstruction is the
$\ZZ/2$ class $w_1(L_0^{\RR})$.  Away from the
seam, the derivative coupling
\[
  \tau_{01}^\mu
  \;:=\;
  \langle\sigma_0|\partial_\mu\sigma_1\rangle
\]
has the standard perturbative form
\[
  \tau_{01}^\mu
  \;=\;
  \frac{\langle\sigma_0|\partial_\mu\hat H_{\rm el}
  |\sigma_1\rangle}{E_1 - E_0}
\]
(up to the usual sign convention for state
ordering).  It is finite on $\Omega^\circ$ and
diverges as $\mathbf R \to \Xseam$, signalling the
breakdown of single-state adiabatic
dynamics~\cite{Teufel2003}.
\end{remark}

\subsubsection{Sign holonomy and the topological
  invariant}

\begin{definition}[Berry sign and Berry phase]
\label{def:berry-sign}
For a real-symmetric $\hat H_{\rm el}$ on $\Omega$,
let $\eta_B := w_1(L_0^{\RR}) \in H^1(\Omega^\circ;
\ZZ/2)$.  For a loop $\gamma \subset \Omega^\circ$,
the \emph{Berry sign} is the holonomy of the real
eigenline
\[
  \operatorname{Hol}_{L_0^{\RR}}(\gamma)
  \;=\;
  (-1)^{\langle\eta_B, [\gamma]\rangle}
  \;\in\; \{+1, -1\}.
\]
In the singular $U(1)$ representation of
Remark~\ref{rem:singular-U1}, this corresponds to a
Berry phase class
\[
  [\gamma_B(\gamma)]
  \;=\;
  \pi\,\langle\eta_B, [\gamma]\rangle
  \pmod{2\pi},
\]
taking values in $\{0, \pi\}$.
\end{definition}

\begin{proposition}[Sign holonomy and mod-$2$
  linking]
\label{prop:topological-phase-Z2}
Assume that on the region $\Omega$ under
consideration $w_1(L_0^{\RR})$ is the mod-$2$
Poincaré dual of the regular CI seam $\Xseam$ (this
holds locally near a regular seam by the Thom-Gysin
construction of
Proposition~\ref{prop:linking-w1}).  Let $\gamma
\subset \Omega^\circ$ be a loop bounding a
$2$-chain $C \subset \Omega$ transverse to
$\Xseam$.  Then
\[
  \langle w_1(L_0^{\RR}), [\gamma]\rangle
  \;=\;
  \#(C \cap \Xseam) \pmod 2,
\]
and equivalently
\[
  \operatorname{Hol}_{L_0^{\RR}}(\gamma)
  \;=\;
  (-1)^{\operatorname{lk}_2(\gamma, \Xseam)}.
\]
\end{proposition}

\begin{proof}
This is Proposition~\ref{prop:linking-w1}(iii) of
\S\ref{sec:L6-bundle} restated under the
Poincaré-duality hypothesis: $w_1$ pairs with
$[\gamma] \in H_1(\Omega^\circ; \ZZ/2)$ to give the
mod-$2$ intersection number of any spanning chain
with $\Xseam$.  Independence of the choice of $C$
follows under the stated hypothesis; locally in a
tubular neighbourhood of a regular seam, this is
the standard meridian computation.
\qedhere
\end{proof}

\begin{proposition}[Reconciliation with the $\Lk_5$
  shadow]
\label{prop:L6-reconciles-L5}
If the real eigenline $L_0^{\RR} \to \Omega$ is
orientable --- for example, if $\Omega$ is simply
connected and the ground state is non-degenerate
and gapped on $\Omega$ --- then a global real gauge
can be chosen, the abelian Berry connection
vanishes globally, and every sign holonomy is $+1$.
If $\Omega^\circ$ contains a regular CI seam and
$\gamma$ is a meridian loop linking the seam, then
$\langle w_1, [\gamma]\rangle = 1$ and
$\operatorname{Hol}_{L_0^{\RR}}(\gamma) = -1$.  The
scalar $\Lk_5$ shadow $(\Omega, V_0, g)$ does not
by itself determine the topology of the real
eigenline over $\Omega^\circ$, nor the
corresponding nuclear boundary condition; the
missing $\Lk_6$ datum is the $w_1$-class.  In the
Na$_3$ forcing example of
\S\ref{sec:L6-forcing-in}, forgetting the eigenline
topology leaves the same lower-sheet
pseudorotational scalar shape but loses the
antiperiodic boundary condition: quantising the
scalar shadow with single-valued nuclear
wavefunctions gives the integer sector, while
restoring $w_1 \neq 0$ gives the physical
half-integer sector.
\end{proposition}

\begin{mathbox}[The $\Lk_6$ invariant chain]
\label{box:berry-chain}
\[
  \includegraphics{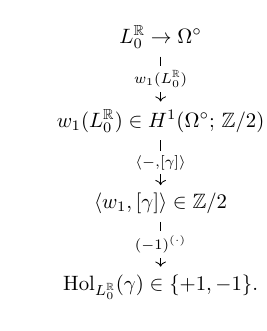}
\]
In a singular complex gauge the same sign is
written as a Berry phase $0$ or $\pi$ modulo
$2\pi$; the two presentations describe the same
$O(1)$ holonomy class.  The $\ZZ/2$ value is the
$\Lk_6$ topological datum discarded by the
forgetful functor $U_6 : \Lk_6(P) \to \Lk_5(P)$.
\end{mathbox}

%% file: chapters/L6/l6_def.tex
\subsection{Definition of \texorpdfstring{$\Lk_6(P)$}{L6(P)}}
\label{sec:L6-def}

Sections~\ref{sec:L6-bundle}--\ref{sec:L6-berry}
assembled the $\Lk_6$ data: a rank-$N$ adiabatic
spectral bundle $\Hel^{(N)} \to \Omega$
(Definition~\ref{def:rank-N-bundle}), a CI seam
$\Xseam \subset \Omega$ as a codimension-$2$
submanifold (Definition~\ref{def:ci-seam}), the
projected Berry connection $\nabla^{\rm B} = P_N d$
on the active bundle, and the real Berry-sign class
\[
  \eta_B \;:=\; w_1(L_0^{\RR})
  \;\in\; H^1(\Omega^\circ;\, \ZZ/2),
  \qquad \Omega^\circ := \Omega \setminus \Xseam.
\]
For a loop $\gamma \subset \Omega^\circ$ the
observable sign holonomy of the real ground-state
eigenline is
\[
  \operatorname{Hol}_{L_0^{\RR}}(\gamma)
  \;=\;
  (-1)^{\langle\eta_B, [\gamma]\rangle}
  \;\in\; \{+1, -1\}
\]
(Definition~\ref{def:berry-sign}).  This section
assembles these data into a target category
$\HilbBund_{\RR}$, defines $\Lk_6(P)$ as $\Lk_5$
objects equipped with a chosen electronic lift, and
exhibits the relation to $\Lk_5(P)$ through the
forgetful functor $U_6$ and the lift-projection
functor $\Felsix$.

\subsubsection{Target categories: scalar shadow and electronic bundle}

Two target categories are needed: a scalar target $\ScalGeom$ receiving the $\Lk_5$ functor $\FV$, and
an electronic target $\HilbBund_{\RR}$ receiving the
$\Lk_6$ projection $\Felsix$.

\begin{definition}[Category of scalar geometric data]
\label{def:scalgeom}
The category $\ScalGeom$ has:
\begin{itemize}
  \item \textbf{Objects}: triples $(\Omega, V_0,
    g)$ where $\Omega \subset \Ce(G)$ is an open
    adiabatic region or smooth stratum, $V_0 :
    \Omega \to \RR$ is a scalar adiabatic surface,
    and $g$ is the mass-weighted metric.  No Morse
    condition is imposed on $V_0$.
  \item \textbf{Morphisms}: geometric channels
    $\Gamma : (\Omega_1, V_0^1, g^1) \to (\Omega_2,
    V_0^2, g^2)$ between scalar objects --- paths,
    families of paths, or composable channels of
    nuclear configurations --- with composition by
    concatenation.
\end{itemize}
\end{definition}

The use of $\ScalGeom$ rather than $\OrbMorse$
accommodates singular scalar shadows such as the
Jahn--Teller Mexican-hat sheet $V_-$ of
\S\ref{sec:Na3-CI}, which is not a Morse function but
is a perfectly meaningful $\Lk_5$ datum.

\begin{definition}[Category of electronic bundles
  with sign class]
\label{def:hilbbund}
The category $\HilbBund_{\RR}$ has:
\begin{itemize}
  \item \textbf{Objects}: tuples
    \[
      \bigl((\Omega, V_0, g),\,
      \mathcal{E}^{(N)},\, A,\, \Xseam,\,
      \eta_B\bigr),
    \]
    where $(\Omega, V_0, g) \in \ScalGeom$,
    $\mathcal{E}^{(N)} \to \Omega$ is the isolated
    rank-$N$ adiabatic spectral bundle of
    Definition~\ref{def:rank-N-bundle} with
    structure group $U(N)$, $A$ is its projected
    Berry connection, $\Xseam \subset \Omega$ is
    the regular codimension-$2$ CI seam, and
    \[
      \eta_B \;=\; w_1(L_0^{\RR})
      \;\in\; H^1(\Omega \setminus \Xseam;\,
      \ZZ/2)
    \]
    is the real Berry-sign class on the
    non-degenerate complement.
  \item \textbf{Morphisms}: \emph{electronic
    channels} $(\Gamma, \mathcal{U}_\Gamma)$ where
    $\Gamma$ is a geometric channel between the
    scalar parts and
    \[
      \mathcal{U}_\Gamma \;:\;
      \mathcal{E}^{(N)}_{\Gamma(0)}
      \;\longrightarrow\;
      \mathcal{E}^{(N)}_{\Gamma(1)}
    \]
    is a unitary transport between fibres over the
    endpoints of $\Gamma$.  In the adiabatic
    regime, $\mathcal{U}_\Gamma$ is Berry parallel
    transport in $A$; in the non-adiabatic regime,
    it is a chosen multi-state propagator on the
    active rank-$N$ bundle (see below).
    Composition is concatenation of geometric
    channels with composition of propagators.
  \textbf{Monoidal product}: external tensor product of bundles over product regions, with sign-class additivity 
  \[\eta^{1\boxtimes 2} =\mathrm{pr}_1^*\eta^1 + \mathrm{pr}_2^*\eta^2.\]
  This makes $\HilbBund_{\RR}$ strict symmetric monoidal.
  The functor $\Felsix : \Lk_6(P) \to \HilbBund_{\RR}$ preserves this product strictly only for non-interacting composite molecular systems; for interacting fragments, $\Felsix$ is lax/asymptotic monoidal because the electronic Hilbert space does not factor (electron indistinguishability, inter-fragment coupling).
\end{itemize}
The \emph{forgetful functor}
\[
  \mathbf{For} \;:\; \HilbBund_{\RR}
  \;\longrightarrow\; \ScalGeom
\]
sends $((\Omega, V_0, g), \mathcal{E}^{(N)}, A,
\Xseam, \eta_B) \mapsto (\Omega, V_0, g)$ and
forgets the bundle, connection, seam, and sign
class.
\end{definition}

The use of channel-type morphisms $(\Gamma,
\mathcal{U}_\Gamma)$ rather than global orbifold
isometries reflects the physical setting: a
chemical reaction is a nuclear pathway with a
chosen electronic transport along it, not a
diffeomorphism between distinct configuration
spaces.

\subsubsection{Adiabatic and non-adiabatic channels}

The propagator $\mathcal{U}_\Gamma$ on a channel
$\Gamma$ depends on where $\Gamma$ runs relative to
the seam.

\begin{itemize}
  \item \emph{Adiabatic channel.}  $\Gamma \subset
    \Omega^\circ$ with the electronic gap $E_1 -
    E_0$ uniformly bounded below along $\Gamma$.
    The Berry parallel transport in $A$ gives a
    canonical unitary
    \[
      \mathcal{U}_\Gamma
      \;=\;
      \mathcal{P}\exp\!\Bigl(
      i\int_\Gamma A
      \Bigr).
    \]
    Restricted to the real eigenline, on a closed
    loop $\gamma \subset \Omega^\circ$,
    $\mathcal{U}_\gamma \in O(1) = \{\pm 1\}$ and
    \[
      \mathcal{U}_\gamma
      \;=\;
      (-1)^{\langle\eta_B, [\gamma]\rangle}.
    \]
    This is the leading adiabatic approximation in
    the Born--Oppenheimer regime under uniform
    spectral gap and slow nuclear motion;
    corrections are controlled by adiabatic
    perturbation theory~\cite{Teufel2003}.
  \item \emph{Non-adiabatic channel.}  $\Gamma$
    enters a region where two or more states in
    the active bundle must be retained --- either
    because $\Gamma$ approaches $\Xseam$ or because
    the gap is small enough that derivative
    couplings $\tau_{kl}^\mu = \langle\sigma_k|
    \partial_\mu\sigma_l\rangle$ are dynamically
    relevant on the timescale of nuclear motion.
    The propagator $\mathcal{U}_\Gamma$ is then a
    chosen multi-state non-adiabatic propagator on
    the active rank-$N$ bundle: typically the
    solution of the time-dependent multi-state
    Schrödinger equation along a prescribed
    time-parametrised nuclear path, or a wavepacket
    propagator in a specified
    approximation~\cite{LasserTeufel2005,
    FermanianKammererLasser2008}.
    Such a propagator is not determined by the geometric channel alone; it must be specified as part of the $\Lk_6$ morphism data.
\end{itemize}

\subsubsection{The level \texorpdfstring{$\Lk_6(P)$}{L6(P)}}

The level $\Lk_6(P)$ is defined as $\Lk_5(P)$
equipped with chosen electronic-lift data.  There
is no canonical functor $\Lk_5(P) \to
\HilbBund_{\RR}$: the active rank $N$, the choice
of adiabatic region $\Omega$, the Berry connection,
the CI seam, and the propagator on each channel are
additional electronic-structure data not determined
by the scalar shadow.

\begin{definition}[Electronic structure level
  $\Lk_6(P)$]
\label{def:L6}
An object of $\Lk_6(P)$ is an $\Lk_5(P)$-object
\[
  X_5 \;=\; \bigl((G, \sigma),\, \Omega,\, V_0,\, g\bigr)
\]
where $(G, \sigma)$ is the underlying
$\Lk_{4.5}$-object with discrete stereochemical
descriptor $\sigma \in \Sigma(G)$ when present,
\emph{together with} an electronic lift
$(\mathcal E^{(N)},\, A,\, \Xseam,\, \eta_B)$
satisfying Layer~2 of Mathbox~\ref{box:L6-layers}
below.  A full $\Lk_6$-object is the tuple
\[
  X_6 \;=\;
  \bigl(X_5,\, \mathcal{E}^{(N)},\, A,\, \Xseam,\,
  \eta_B\bigr).
\]

A morphism $X_6 \to Y_6$ is an $\Lk_5$-geometric
channel $\Gamma : X_5 \to Y_5$ together with a
chosen electronic propagator
$\mathcal{U}_\Gamma : \mathcal{E}^{(N)}_{X_5} \to
\mathcal{E}^{(N)}_{Y_5}$ on the active bundle ---
adiabatic Berry transport or a non-adiabatic
propagator as in the preceding subsection.
Composition is concatenation of geometric channels
with composition of propagators.

The \emph{forgetful functor}
\[
  U_6 \;:\; \Lk_6(P) \;\longrightarrow\; \Lk_5(P)
\]
sends $X_6 \mapsto X_5$ and discards
$(\mathcal{E}^{(N)}, A, \Xseam, \eta_B)$ together
with the electronic propagator.  The
\emph{lift-projection functor}
\[
  \Felsix \;:\; \Lk_6(P)
  \;\longrightarrow\; \HilbBund_{\RR}
\]
records the chosen electronic lift on objects and
the chosen propagator on morphisms. 
$\Felsix$ does not factor through $U_6$: it cannot be reconstructed from the scalar shadow alone.
The square
\[
  \includegraphics{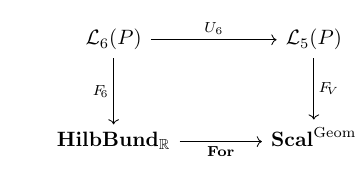}
\]
commutes by Layer~2(b) below.
\end{definition}

\begin{mathbox}[Layer 1 and Layer 2 at $\Lk_6$]
\label{box:L6-layers}

\noindent\textbf{Layer 1.}  Any Hermitian rank-$N$
bundle $\mathcal{E}^{(N)} \to \Omega$ with a
$U(N)$-connection $A$, an embedded subset $\Xseam
\subset \Omega$, and a class $\eta_B \in
H^1(\Omega \setminus \Xseam; \ZZ/2)$, not
necessarily arising from any electronic
Hamiltonian.  This is the formal layer.

\medskip\noindent\textbf{Layer 2.}  Two physical
conditions restrict Layer~1 to the chemically
meaningful sublevel.
\begin{enumerate}[label=(\alph*)]
  \item \emph{Electronic structure origin.}
    $\mathcal{E}^{(N)}$ is the isolated rank-$N$
    spectral bundle of $\hat H_{\rm el}(\mathbf R)$
    over $\Omega$
    (Definition~\ref{def:rank-N-bundle}), with
    spectral gap $E_N - E_{N-1} \geq \Delta > 0$
    on $\Omega$.  The connection $A$ is the
    projected Berry connection $\nabla^{\rm B} =
    P_N d$ of \S\ref{sec:L6-berry}; $\Xseam$ is the
    regular codimension-$2$ CI seam
    (Definition~\ref{def:ci-seam}); and $\eta_B =
    w_1(L_0^{\RR})$ on $\Omega^\circ$.
  \item \emph{Consistency with $\Lk_5$.}  The
    scalar shadow recorded at $\Lk_6$ agrees with
    the $\Lk_5$ datum:
    \[
      \mathbf{For} \circ \Felsix \;=\;
      \FV \circ U_6,
    \]
    and the ground-state surface satisfies
    \[
      V_0(\mathbf{R})
      \;=\;
      E_{\rm el,0}(\mathbf{R})
      + V_{\rm nn}(\mathbf{R}),
    \]
    where $E_{\rm el,0}$ is the lowest eigenvalue
    of $\hat H_{\rm el}(\mathbf R)$ and
    $V_{\rm nn}$ is the nuclear-nuclear repulsion.
    The sign class $\eta_B$ is new $\Lk_6$ data
    that does not alter $V_0$ but records the
    global topology of the real ground-state
    eigenline.
\end{enumerate}

\medskip\noindent
The departure from $\Lk_5$ is that the scalar
$\Lk_5$ description retains only $(\Omega, V_0, g)$
and cannot record whether the real eigenline over
$\Omega^\circ$ is orientable.  The active rank-$N$
bundle $\mathcal{E}^{(N)}$ remains well-defined
across an internal degeneracy provided the active
cluster is spectrally isolated from the rest; the
individual eigenline $L_0^{\RR}$ and its
Stiefel--Whitney class $\eta_B$ are defined on the
complement $\Omega^\circ$, where the ground-state
separation $E_0 < E_1$ holds.
\end{mathbox}

\begin{remark}[The BO approximation at $\Lk_6$]
\label{rem:BO-lifted}
The single-surface Born--Oppenheimer approximation
is justified asymptotically on regions where the
relevant electronic state is isolated by a uniform
gap and nuclear motion is sufficiently slow; the
small parameter $\varepsilon$ separating electronic
and nuclear timescales is fixed once in
Chapter~10~\cite{Teufel2003}.  The single-surface
BO approximation breaks down at $\Xseam$ and may
also become inaccurate near small gaps or strong
derivative-coupling regions, where multi-state
treatment of the active bundle is required.

The $\Lk_5$ shadow $(\Omega, V_0, g)$ does not
record the eigenline topology, and quantising it
with single-valued nuclear wavefunctions assigns
trivial sign holonomies.  At $\Lk_6$, the bundle
$\mathcal{E}^{(N)}$, its Berry connection,
$\Xseam$, and $\eta_B$ are retained.  The sign
holonomy $(-1)^{\langle\eta_B, [\gamma]\rangle}$
enters the adiabatic nuclear equation as a
Mead--Truhlar vector potential or as antiperiodic
boundary conditions on loops linking
$\Xseam$~\cite{MeadTruhlar1979}.  It is the
electronic-bundle datum that the scalar $\Lk_5$
description forgets, not strictly a non-BO
quantity.
\end{remark}

\begin{chembox}[What $\Lk_6$ means in practice]
The $\Lk_5$/$\Lk_6$ distinction is about what
datum the computation targets, not which method
is used.

\smallskip\noindent
\textbf{$\Lk_5$ data.}  Any electronic-structure
method that produces a ground-state PES
$V_0(\mathbf R)$ --- DFT, MP2, CCSD(T), or a
machine-learned force field fitted to any of
these --- supplies the $\FV$ data: minima,
saddles, IRC, barriers, vibrational frequencies.

\smallskip\noindent
\textbf{$\Lk_6$ data.}  Locating $\Xseam$ and
evaluating $\eta_B$ on loops in $\Omega^\circ$ is
a multi-state question: one needs at least the
first two adiabatic PESs $V_0, V_1$ and their
degeneracy locus.
Single-reference ground-state methods are generally unreliable for locating and characterising CI seams, because the relevant states typically have strong multireference character; multireference or state-averaged methods such as CASSCF, MRCI, NEVPT2, and MC-PDFT are the standard tools~\cite{DomckeYarkony2012}.

\smallskip\noindent
The forgetful functor $U_6$ is the operation of
extracting only the ground-state PES from a
multi-state calculation: it compresses
$(\mathcal{E}^{(N)}, A, \Xseam, \eta_B)$ down to
$V_0$ alone.  If the relevant dynamics remains on
a single smooth isolated adiabatic surface and the
real eigenline is topologically trivial on the
explored region, the $\Lk_5$ scalar description
may be sufficient.  $\eta_B = 0$ is not by itself a
guarantee that $\Lk_6$ data are unnecessary:
photoisomerisation, internal conversion,
photodissociation, spin-orbit-mediated
intersystem crossing, and charge-transfer
dynamics all require $\Lk_6$ data --- multiple
electronic surfaces, derivative couplings, seam
geometry, and (when relevant) the topological
sign --- and are not determined by $V_0$ alone.
\end{chembox}

\begin{insightbox}[The $\Lk_5 \to \Lk_6$ extension: from base to fibre]
\label{box:L6-type}
The $\Lk_5 \to \Lk_6$ extension is qualitatively different from every preceding extension in the tower:
\medskip
\begin{center}
\renewcommand{\arraystretch}{1.4}
\begin{tabular}{@{}l l p{5.0cm}@{}}
\hline
\textbf{Extension} & \textbf{Type} &
  \textbf{New data live\ldots} \\
\hline
$\Lk_0 \to \Lk_1$ & Decorator ($+\FH$)
  & \ldots on morphisms ($\RR$-labels) \\
$\Lk_1 \to \Lk_2$ & Decorator ($+\FS$)
  & \ldots on morphisms ($\RR$-labels) \\
$\Lk_2 \to \Lk_3$ & Decorator ($+\FP$)
  & \ldots on morphisms (Markov kernels) \\
$\Lk_3 \to \Lk_4$ & Structural
  & \ldots in the morphism category (DPO
    spans) \\
$\Lk_4 \to \Lk_{4.5}$ & Symmetry enrichment
  & \ldots on the morphism category
    ($G^*$-equivariance) \\
$\Lk_{4.5} \to \Lk_5$ & Geometric decoration
  & \ldots on the base $\Ce(G)$
    (scalar surface $V_0$) \\
$\Lk_5 \to \Lk_6$ &
  \textbf{Topological enrichment}
  & \ldots \textbf{in the fibres over}
    $\Omega^\circ$ (bundle $\mathcal{E}^{(N)}$,
    connection $A$, seam $\Xseam$, sign class
    $\eta_B$) \\
\hline
\end{tabular}
\end{center}

\medskip\noindent
At $\Lk_5$, the new datum $V_0 : \Omega \to \RR$
is a smooth function --- data \emph{on the base}.
At $\Lk_6$, the new datum
\[
  \eta_B \;=\; w_1(L_0^{\RR})
  \;\in\; H^1(\Omega^\circ;\, \ZZ/2)
\]
is a discrete topological invariant of the real
eigenline over the base: not a function, but a
cohomology class.  This is the first time in the
tower the new datum is a discrete topological
invariant rather than a continuously varying
quantity --- the shift from $C^\infty(\Omega)$ to
$H^1(\Omega^\circ; \ZZ/2)$ is the deepest
geometric step in the tower up to $\Lk_6$.  Each
tower extension is forced by a non-trivial fibre of the corresponding forgetful projection $U_k : \Lk_k(P) \to \Lk_{k-1}(P)$. 
For $\Lk_5 \to \Lk_6$, the fibre $U_6^{-1}(\Omega,V_0, g)$ collects Layer-2 electronic lifts
compatible with the same scalar shadow.  Under
Layer-2(a) and the local Jahn--Teller model
(Proposition~\ref{prop:half-angle-monodromy}), the physical Na$_3$ lift has $\eta_B \neq 0$; the contrast in \S\ref{sec:L6-forcing-in} is therefore not between two physical electronic lifts with the same seam but between the unique physical $\Lk_6$ lift ($\eta_B \neq 0$, half-integer sector) and the $\Lk_5$ scalar quantisation that forgets the eigenline topology and produces the integer sector.
\end{insightbox}

%% file: chapters/L6/l6_lh.tex
\subsection{The Longuet--Higgins sign-change theorem}
\label{sec:L6-lh}

Sections~\ref{sec:L6-bundle}--\ref{sec:L6-def}
built the internal $\Lk_6$ data: the rank-$N$
adiabatic bundle $\Hel^{(N)} \to \Omega$, the CI
seam $\Xseam \subset \Omega$, the projected Berry
connection on the active bundle, and the real
Berry-sign class
\[
  \eta_B \;=\; w_1(L_0^{\RR})
  \;\in\; H^1(\Omega^\circ;\, \ZZ/2),
  \qquad \Omega^\circ := \Omega \setminus \Xseam.
\]
This section establishes the standard physical
manifestation of the abstract class $\eta_B$: the
gauge-invariant $O(1)$-holonomy of the real
ground-state eigenline around any loop linking
$\Xseam$.  Although the sign of a single electronic
eigenvector is itself gauge-dependent, the
holonomy around a closed loop is gauge-invariant; it
manifests physically through interference effects
in reactive scattering and through anti-periodic
boundary conditions on the nuclear factor of the
vibronic wavefunction.  The Longuet--Higgins
sign-change theorem~\cite{LonguetHiggins1975},
anticipated by earlier symmetry
arguments, follows from the $O(1)$-holonomy of
\S\ref{sec:L6-berry}.

\begin{theorem}[Longuet--Higgins sign holonomy]
\label{thm:LH}
Let $\hat H_{\rm el}$ be real-symmetric on a
smooth adiabatic region $\Omega$, let $\Xseam
\subset \Omega$ be a regular codimension-$2$ CI
seam, and set $\Omega^\circ = \Omega \setminus
\Xseam$.  Let $\gamma : [0,1] \to \Omega^\circ$
be a smooth closed loop based at $\mathbf{R}_0$
whose $w_1$-pairing is $\langle \eta_B,
[\gamma]\rangle = 1$ (the meridian case).  Then
the real ground-state eigenline has $O(1)$-holonomy
\[
  \Hol_{L_0^{\RR}}(\gamma) \;=\; -1.
\]
Equivalently, if a non-zero vector $\sigma_0(
\mathbf{R}_0) \in L_0^{\RR}|_{\mathbf{R}_0}$ is
parallel transported around $\gamma$ in the flat
real $O(1)$-local system on $L_0^{\RR}$, then
\[
  \sigma_0^\parallel(1) \;=\;
  -\,\sigma_0(\mathbf{R}_0).
\]
Here ``flat real connection'' refers to the
$O(1)$-local system obtained from local real
normalised eigenvectors of $\hat H_{\rm el}$ with
transition functions $\pm 1$; this is not an
additional $U(1)$ curvature datum, since the real
abelian Berry one-form vanishes locally
(Proposition~\ref{prop:real-A0-zero}).
\end{theorem}

\begin{proof}
By Definition~\ref{def:berry-sign} the holonomy of
the real line bundle around any loop $\gamma$ is
\[
  \Hol_{L_0^{\RR}}(\gamma)
  \;=\; (-1)^{\langle\eta_B, [\gamma]\rangle}.
\]
The hypothesis $\langle\eta_B, [\gamma]\rangle = 1$
gives $\Hol_{L_0^{\RR}}(\gamma) = -1$.
Proposition~\ref{prop:topological-phase-Z2}
verifies that sufficiently small meridian loops
around regular points of $\Xseam$ realise this
hypothesis.
Transport of $\sigma_0(\mathbf{R}_0)$ in the flat
real $O(1)$-local system returns it to its own
fibre multiplied by the holonomy:
$\sigma_0^\parallel(1) = -\sigma_0(\mathbf{R}_0)$.
\qedhere
\end{proof}

\begin{remark}[Gauge invariance and convention
  independence]
\label{rem:lh-gauge}
By Proposition~\ref{prop:real-A0-zero}, in any
local real gauge the abelian Berry connection on
$L_0^{\RR}$ vanishes pointwise: $A_0 = 0$.  The
sign holonomy in Theorem~\ref{thm:LH} therefore
does not arise from a non-zero connection one-form
but from the failure of any real gauge to be
single-valued around $\gamma$.  In the singular
complex gauge of Remark~\ref{rem:singular-U1}, the
same sign appears as a Berry phase $\pm\pi$
modulo $2\pi$; the holonomy $\exp(\pm i\pi) = -1$
is invariant under continuous $U(1)$ gauge
transformations and under the choice of sign
convention for the Berry connection.
The gauge-free invariant is the $\ZZ/2$ class
$\eta_B = w_1(L_0^{\RR})$; raising $-1$ to its
meridian evaluation gives the $\pm 1$ sign.

If the real eigenline $L_0^{\RR} \to \Omega^\circ$
is orientable --- for example, if $\Omega^\circ$
is simply connected and the ground state is
separated from the rest of the spectrum by a
positive gap throughout --- a global real gauge
can be chosen, every loop holonomy is $+1$, and
no sign change occurs
(Proposition~\ref{prop:L6-reconciles-L5}).  The
sign holonomy is therefore strictly an $\Lk_6$
phenomenon: it is not determined by the scalar
$\Lk_5$ shadow $(\Omega, V_0, g)$, which does not
record the topology of the real eigenline.
\end{remark}

\begin{mathbox}[The LH sign change in the tower]
\label{box:lh-tower}
The Longuet--Higgins theorem plays a precise
structural role: it is the \emph{observable
bridge} between the abstract $\Lk_6$ invariant
$\eta_B$ and a measurable physical quantity.

The invariant chain of \S\ref{sec:L6-berry}
(Mathbox~\ref{box:berry-chain}) terminates at the
sign holonomy:
\[
  \eta_B \in H^1(\Omega^\circ;\, \ZZ/2)
  \;\xrightarrow{\;\text{evaluate on }\gamma\;}\;
  \Hol_{L_0^{\RR}}(\gamma)
  \;=\;
  (-1)^{\langle\eta_B, [\gamma]\rangle}
  \;\in\; \{+1, -1\}.
\]
The sign holonomy of the real eigenline around a
meridian loop $\gamma$ is the physical
realisation of this final arrow.

Three structural consequences:
\begin{enumerate}[label=(\roman*)]
  \item \emph{The scalar $\Lk_5$ shadow does not
    determine the sign holonomy.}  The scalar
    data $(\Omega, V_0, g)$ recorded by $\FV$ does
    not register the topology of $L_0^{\RR}$;
    quantising the shadow with single-valued
    nuclear wavefunctions gives the trivial sign
    sector.  The sign holonomy belongs to the
    electronic lift forgotten by $U_6 : \Lk_6(P)
    \to \Lk_5(P)$: the physical $\Lk_6$ lift with
    $\eta_B \neq 0$ supplies it; the scalar
    shadow alone cannot.
  \item \emph{The sign sector is topologically
    protected.}  Because $\eta_B$ takes values in
    the discrete group $\ZZ/2$, its evaluation on
    a fixed loop $\gamma$ cannot change under
    continuous deformations that keep $\gamma$
    inside the gapped complement $\Omega^\circ$.
    The sign sector can change only if $\gamma$
    crosses a degeneracy, the electronic gap
    closes somewhere on $\gamma$, or the topology
    of $\Omega^\circ$ itself changes.  Unlike
    barrier heights or rate constants, the sign
    sector does not vary continuously with
    molecular parameters.
  \item \emph{Sign-changing transport is an
    adiabatic $\Lk_6$ channel.}  In the
    adiabatic-channel propagator of
    Definition~\ref{def:L6}, the restriction to
    the real ground-state eigenline gives
    $\mathcal U_\gamma \in O(1) = \{\pm 1\}$ with
    \[
      \mathcal U_\gamma \;=\; -1
    \]
    when $\langle\eta_B, [\gamma]\rangle = 1$.
    Theorem~\ref{thm:LH} thus records the value
    of the electronic transport component of an
    adiabatic $\Lk_6$ channel whose path encircles
    $\Xseam$.
\end{enumerate}
\end{mathbox}

\begin{chembox}[The sign change in the laboratory]
\label{box:lh-chem}

The sign $-1$ has two standard experimental
manifestations.

\smallskip\noindent
\textbf{Quantum interference in reactive
scattering.}  When a nuclear wavepacket reaches a
product channel by two paths --- one winding
around the CI, one not --- the two amplitudes
acquire a relative sign $-1$.  Wherever both
paths contribute coherently to the same product,
this yields destructive interference that
redistributes the angular distribution of
products relative to a scalar single-surface
calculation on the same ground-state PES that
omits the geometric-phase boundary condition.
The effect was predicted by Mead and
Truhlar~\cite{MeadTruhlar1979}, quantitatively
modelled for the $\mathrm{H} + \mathrm{H_2} \to
\mathrm{H_2} + \mathrm{H}$ system by
Juanes-Marcos, Althorpe, and
Wrede~\cite{JuanesMarcos2005}, and observed
experimentally for the isotopic analogue
$\mathrm{H} + \mathrm{HD} \to \mathrm{H_2} +
\mathrm{D}$ by Yuan et al.\ above the CI
energy~\cite{YuanScience2018} and below
it~\cite{YuanNatCommun2020}.  Measured
differential cross sections agree with
geometric-phase-inclusive scattering calculations
on the ground-state PES and disagree with
calculations on the same PES that omit the
phase.  This is the cleanest laboratory
signature to date of a topological ($\Lk_6$)
effect in a simple chemical reaction.

\smallskip\noindent
\textbf{Anti-periodic boundary condition on
nuclear motion encircling $\Xseam$.}  For a
nuclear coordinate $\phi$ parametrising a loop
that links $\Xseam$ in the course of a
photochemical pathway, the electronic sign change
imposes an anti-periodic boundary condition on
the nuclear wavefunction:
\[
  \chi_\mathrm{nuc}(\phi + 2\pi)
  \;=\; -\chi_\mathrm{nuc}(\phi),
\]
enforcing single-valuedness of the total vibronic
wavefunction $\Psi = \chi_\mathrm{nuc}\,
\sigma_0$.  Physically, this shifts the allowed
pseudorotational quantum numbers from integers
to half-integers:
\[
  j \in \ZZ
  \;\;\leadsto\;\;
  j \in \ZZ + \tfrac12.
\]
The same mechanism produces the half-integer
pseudorotation quantum numbers in $\mathrm{Na_3}$
established in \S\ref{sec:L6-forcing-in}; it is
also used in ultrafast spectroscopy to detect
and characterise CIs through the rearrangement of
vibronic spacings near a CI
funnel~\cite{DomckeYarkony2012}.  The effect
operates only on nuclear motion that genuinely
encircles $\Xseam$; ground-state dynamics along a
path that remains in a simply connected gapped
region and does not link the CI seam does not
realise it.

\smallskip\noindent
In both cases the observable effect is controlled
by the discrete class $\eta_B = w_1(L_0^{\RR})$.
A scalar calculation on the ground-state PES
that omits the geometric-phase boundary condition
gives the trivial sign sector; the physical
$\Lk_6$ lift assigns sign holonomy $-1$ to loops
linking the CI seam.  The topological class
itself cannot change continuously: changing the
sign sector requires the relevant loop to pass
through a degeneracy, or the gapped real
eigenline description to break down on the loop.
\end{chembox}

%% file: chapters/L6/l6_photochem.tex
\subsection{Illustrative \texorpdfstring{$\Lk_6$}{L6} dynamics near conical intersections}
\label{sec:L6-photochem}

The preceding sections built the $\Lk_6$
framework and established its empirical anchors.
\S\ref{sec:L6-forcing-in} forced $\Lk_5 \to \Lk_6$
through the $\mathrm{Na_3}$ half-integer
pseudorotation spectrum.
\S\ref{sec:L6-bundle}--\S\ref{sec:L6-def}
constructed the rank-$N$ active bundle, the real
Berry-sign class $\eta_B = w_1(L_0^{\RR})$, and
the $\Lk_6$ category whose morphisms are channels
$(\Gamma, \mathcal U_\Gamma)$ in adiabatic and
non-adiabatic regimes (Definition~\ref{def:L6}).
\S\ref{sec:L6-lh} provided the observable bridge:
the gauge-invariant $O(1)$-holonomy of the real
ground-state eigenline, manifest in
reactive-scattering interference patterns and in
the antiperiodic boundary condition on
$\mathrm{Na_3}$ pseudorotation.

This section is illustrative rather than
constructive: it does not build a complete theory
of non-adiabatic wavepacket propagation, but
explains how the $\Lk_6$ channel data
$(\Gamma, \mathcal U_\Gamma)$ are interpreted in
the standard adiabatic, non-adiabatic, and
Landau--Zener approximations.  The $\mathrm{Na_3}$
forcing example is a bound-state manifestation of
$w_1$; $\mathrm{H + HD}$ scattering is a continuum
manifestation; Landau--Zener is a local scalar
estimator for non-adiabatic transfer.  None of
these scalar reductions replaces the underlying
$\Lk_6$ lift.

\subsubsection{The \texorpdfstring{$\Lk_6$}{L6} channel propagator along a semiclassical nuclear path}
\label{sec:L6-propagator}

Let $\gamma : [0, T] \to \Omega$ be a chosen
time-parametrised semiclassical nuclear path ---
for example, the centre of a sufficiently narrow
wavepacket in regimes where such a path
description is meaningful.  A full wavepacket
treatment replaces this path-level channel by a
coupled nuclear--electronic propagator acting on
nuclear wavefunctions valued in $\Hel^{(N)}$; the
path-level model below isolates the electronic
content of an $\Lk_6$ channel along a prescribed
nuclear trajectory.

By Definition~\ref{def:L6}, the $\Lk_6$ channel
along $\gamma$ is the pair
$(\gamma, \mathcal U_\gamma)$, where the
electronic propagator $\mathcal U_\gamma$ acts on
the fibre of $\Hel^{(N)}$ above $\gamma(t)$.  Its
qualitative behaviour depends on whether $\gamma$
remains in the gapped complement $\Omega^\circ =
\Omega \setminus \Xseam$ or enters a region where
the CI seam, or a small electronic gap, controls
the dynamics.

\medskip\noindent\textbf{Adiabatic regime.}  When
$\gamma$ remains in $\Omega^\circ$ with electronic
gap $E_1 - E_0$ bounded below by a positive
constant throughout the path, the adiabatic
approximation on the selected complex eigenline
$L_0$ gives an electronic propagator as a product
of a dynamical phase and a Berry parallel
transport:
\[
  \mathcal U_\gamma^{\rm ad}
  \;=\;
  \exp\!\left(
    -\frac{i}{\hbar}\int_0^T E_0(\gamma(t))\,dt
  \right)
  \cdot
  \mathcal{P}\exp\!\left(
    i\!\int_\gamma A_0
  \right).
\]
The first factor is the dynamical phase along
the path; the second is the Berry
parallel-transport map of the connection $A_0$
on $L_0$, sending the fibre at $\gamma(0)$ to
the fibre at $\gamma(T)$.

For an open path $\gamma$, the Berry transport is
gauge-covariant rather than gauge-invariant:
its value depends on the choice of local frame
at the two endpoints.  For a closed loop
$\gamma$ (or for the closed cycle obtained by
concatenating two paths with the same
endpoints), the holonomy on the real eigenline
is gauge-invariant and is given by
Theorem~\ref{thm:LH}:
\[
  \Hol_{L_0^{\RR}}(\gamma)
  \;=\;
  (-1)^{\langle\eta_B, [\gamma]\rangle}.
\]
In a real local gauge the abelian Berry one-form
on $L_0^{\RR}$ vanishes pointwise
(Proposition~\ref{prop:real-A0-zero}); the
closed-loop holonomy nevertheless picks up the
gauge-free $O(1)$ sign.  Hence for a closed loop
$\gamma \subset \Omega^\circ$ on the real
ground-state eigenline,
\[
  \mathcal U_\gamma^{\rm ad}\bigl|_{L_0^{\RR}}
  \;=\;
  \exp\!\left(
    -\frac{i}{\hbar}\int_0^T E_0(\gamma(t))
    \,dt
  \right)
  \cdot
  (-1)^{\langle\eta_B, [\gamma]\rangle}.
\]
For such closed loops --- or for the closed
cycle formed by comparing two paths with the
same endpoints --- the $\ZZ/2$ sign sector is
determined by the pairing of
$\eta_B = w_1(L_0^{\RR})$ with the closed loop;
equivalently, under the local Poincaré-duality
hypothesis of
Proposition~\ref{prop:topological-phase-Z2}, by
the mod-$2$ linking number of the loop with
$\Xseam$.  This gives the topological sign
sector of the adiabatic $\Lk_6$ dynamics; it
does not exhaust the full dynamical content,
which also includes the dynamical phase above
and, for degenerate or near-degenerate active
subspaces, possible non-Abelian transport on
$\Hel^{(N)}$.

\medskip\noindent\textbf{Non-adiabatic regime.}
Away from the seam, where $E_0 \neq E_1$, in an
adiabatic eigenbasis the derivative coupling
\[
  \tau_{01}^\mu(\mathbf R)
  \;=\;
  \langle\sigma_0(\mathbf R)
  \,\vert\,
  \partial_\mu \sigma_1(\mathbf R)\rangle
\]
is regular and admits the standard
Hellmann--Feynman perturbative expression
\[
  \tau_{01}^\mu(\mathbf R)
  \;=\;
  \frac{\langle\sigma_0(\mathbf R) \,\vert\,
  \partial_\mu \hat H_\mathrm{el}(\mathbf R)
  \,\vert\, \sigma_1(\mathbf R)\rangle}
  {E_1(\mathbf R) - E_0(\mathbf R)},
\]
up to the usual sign convention determined by the
ordering of the two adiabatic states.  This
expression becomes singular as $\Xseam$ is
approached because the gap in the denominator
tends to zero.  At the seam itself, the
adiabatic-frame formula is no longer defined: the
individual eigenlines $L_0, L_1$ are not defined
as a smooth direct-sum decomposition; only the
rank-$2$ active sub-bundle
\[
  \Hel^{(2)} \to \Omega
\]
remains well-defined, provided the two-state
active cluster remains separated from the rest
of the electronic spectrum on $\Omega$.  The
single-surface Born--Oppenheimer description is
no longer uniformly valid near the seam, and a
multi-state representation is required to
describe dynamics through the branching region.

At the path level, choose a local adiabatic frame
$\{\sigma_m(\gamma(t))\}_{m=0}^{N-1}$ and write
$|\Psi(t)\rangle = \sum_m c_m(t)\,
\sigma_m(\gamma(t))$.  Introducing the
derivative-coupling matrix
\[
  \tau_{\mu,\,mn}(\mathbf R)
  \;:=\;
  \langle\sigma_m(\mathbf R) \,\vert\,
  \partial_\mu \sigma_n(\mathbf R)\rangle,
\]
which is anti-Hermitian on the active subspace,
the amplitudes evolve by
\begin{equation}
\label{eq:multistate-propagator}
  i\hbar\,\dot c_m(t)
  \;=\;
  E_m(\gamma(t))\, c_m(t)
  \,-\,
  i\hbar
  \sum_n
  \dot\gamma^\mu(t)\,
  \tau_{\mu,\,mn}(\gamma(t))\,
  c_n(t),
\end{equation}
equivalently $i\hbar\,\dot{\mathbf c}(t) =
H_{\rm eff}^\gamma(t)\,\mathbf c(t)$ with
\[
  H_{{\rm eff},\, mn}^\gamma(t)
  \;=\;
  E_m(\gamma(t))\,\delta_{mn}
  \,-\,
  i\hbar\,\dot\gamma^\mu(t)\,
  \tau_{\mu,\,mn}(\gamma(t)).
\]
In terms of the Hermitian Berry connection
$A_{\mu,\,mn} := i\,\tau_{\mu,\,mn}$ used
elsewhere in this chapter, the same effective
Hamiltonian reads
\[
  H_{{\rm eff},\, mn}^\gamma(t)
  \;=\;
  E_m(\gamma(t))\,\delta_{mn}
  \,-\,
  \hbar\,\dot\gamma^\mu(t)\,
  A_{\mu,\,mn}(\gamma(t)).
\]
This path-level system gives a standard
semiclassical realisation of an $\Lk_6$ channel
morphism $(\gamma, \mathcal U_\gamma)$ of
Definition~\ref{def:L6} for a chosen classical
nuclear path~\cite{LasserTeufel2005,
FermanianKammererLasser2008,
ColinDeVerdiere2003,Teufel2003}.  Near the
branching region, the derivative-coupling term
becomes singular in the adiabatic basis, which
is why a diabatic or active-bundle formulation
is preferred for practical computation.  The
amplitudes $c_m(t)$ carry both populations in
the adiabatic states and electronic phases:
for a closed loop $\gamma \subset \Omega^\circ$,
or for two alternative paths whose concatenation
forms a closed loop linking $\Xseam$, these
phases include the sign holonomy of
Theorem~\ref{thm:LH}.  A more general quantum
treatment replaces the classical path $\gamma$
by a quantum nuclear wavefunction
$\chi(\mathbf R, t)$ and produces a coupled
nuclear--electronic propagator on nuclear
wavefunctions valued in $\Hel^{(N)}$, extending
beyond the path-level realisation of the $\Lk_6$
channel considered here.

\medskip\noindent\textbf{Path-level model versus
scalar reductions.}  The path-level electronic
propagator above is itself a semiclassical
reduction of the full coupled
nuclear--electronic dynamics, which acts on
nuclear wavefunctions valued in $\Hel^{(N)}$; in
the categorical model used here, such dynamics
may be represented by choosing a richer $\Lk_6$
channel morphism, while the path-level equation
is one semiclassical realisation of that
channel.  Within the path-level model, any
further scalar reduction (transfer probability,
surface population, effective vector potential)
loses either phase information or the
multi-state amplitude structure.  In particular,
no scalar transfer probability captures the
$\ZZ/2$ sign holonomy of Theorem~\ref{thm:LH}:
that information lives in the relative phases of
the amplitudes $c_m(t)$, not in their norms.
The next subsection makes the scalar-reduction
step concrete with the textbook Landau--Zener
estimator.

\subsubsection{The Landau--Zener formula as a
  non-adiabatic scalar estimator}

For practical estimates of non-adiabatic transfer
probabilities, the textbook Landau--Zener
formula~\cite{Zener1932} provides a semiclassical
approximation derived in a \emph{diabatic basis}
near a transverse two-state crossing.  Stating
the formula precisely --- and locating it in the
tower --- clarifies both its utility and its
limits.

\begin{observation}[Landau--Zener as a
  diabatic-basis approximation]
\label{obs:lz}
Let $V_1^\mathrm{dia}, V_2^\mathrm{dia}$ be two
diabatic potentials crossing transversally at a
point traversed by a nuclear wavepacket with
speed $v$ through the crossing coordinate in
the linearised one-dimensional model, with
diabatic coupling $|H_{12}|$ and difference of
diabatic slopes $|F_1 - F_2|$.  The Landau--Zener
probability
\[
  P_\mathrm{LZ}
  \;:=\;
  \exp\!\left(-\frac{2\pi\,|H_{12}|^2}
  {\hbar\,v\,|F_1 - F_2|}\right)
  \;\in\; [0, 1]
  \qquad \text{\cite{Zener1932}}
\]
is the probability of preserving diabatic
character through the crossing; equivalently, the
probability of jumping from one adiabatic branch
to the other.  The complementary probability
$1 - P_\mathrm{LZ}$ is the probability of
adiabatic following.  Weak diabatic coupling
($|H_{12}|$ small) gives $P_\mathrm{LZ} \to 1$:
diabatic passage with an adiabatic branch change.
Strong coupling gives $P_\mathrm{LZ} \to 0$:
adiabatic following with no branch change.  The
minimum adiabatic gap at the crossing is
$2|H_{12}|$.

\medskip\noindent\textbf{Validity near a CI.}  LZ
is derived for isolated transverse diabatic
crossings with constant $|H_{12}| > 0$.  At a
true CI, $|H_{12}|$ and the adiabatic gap vanish
simultaneously in the branching plane, and the
standard two-state LZ formula is no longer
exact.  It survives only as a local
order-of-magnitude estimator for a chosen
one-dimensional passage through a regularised
or locally diabatised two-state model near the
branching region.  The path-level matrix-valued
propagator $\mathcal U_\gamma$ on $\Hel^{(N)}$
--- including the sign holonomy of
Theorem~\ref{thm:LH}, invisible to any scalar
transfer probability --- is given by the
multi-state channel
equation~\eqref{eq:multistate-propagator}; a full
wavepacket treatment replaces this by coupled
nuclear--electronic dynamics.  Unlike the
$w_1$-holonomy, $P_\mathrm{LZ}$ is not a
topological invariant: it varies continuously
with velocity, coupling, and local slopes.

\medskip\noindent\textbf{Tower location.}
$P_\mathrm{LZ}$ requires local two-state
Hamiltonian data: two diabatic potentials, a
diabatic coupling, and the trajectory velocity.
Equivalently, it requires the active rank-$2$
electronic bundle, the projected Hamiltonian,
and a chosen local diabatisation --- all
$\Lk_6$-level data not contained in the scalar
shadow $(\Omega, V_0, g)$, which retains only
the ground-state PES.  The Landau--Zener formula
is therefore an $\Lk_6$-level scalar estimator
of a non-adiabatic channel, accompanying but not
replacing the full channel morphism
$(\Gamma, \mathcal U_\Gamma)$ of
Definition~\ref{def:L6}.
\end{observation}

\medskip
The $\mathrm{Na_3}$ forcing argument of
\S\ref{sec:L6-forcing-in} does not rely on the
Landau--Zener approximation: its diagnostic is
the bound-state boundary condition imposed by
$w_1$ on the pseudorotation loop.  Landau--Zener
is included here only to locate one standard
scalar non-adiabatic estimate within the same
$\Lk_6$ framework.

\medskip
The bound-state and scattering manifestations of
$w_1(L_0^{\RR})$ already presented in this
chapter are two physical realisations of the
same type of Berry-sign datum.  The $\mathrm{Na_3}$
half-integer pseudorotation
spectrum~\cite{DelacretazGrantWhettenWosteZwanziger1986}
(Forcingbox~\ref{box:Na3-forcing}) realises the
antiperiodic boundary condition as a
single-surface nuclear problem on $V_-$.  The
geometric-phase interference oscillations in
$\mathrm{H} + \mathrm{HD} \to
\mathrm{H_2} + \mathrm{D}$
scattering~\cite{YuanScience2018,
YuanNatCommun2020}
(Insightbox~\ref{ins:HHD-scattering}) realise the
same datum as a relative phase between
coherently summed amplitudes on the relevant
ground electronic surface, supplemented by the
Mead--Truhlar vector potential or an equivalent
geometric-phase boundary condition on closed loops or relative loops linking the corresponding \(\mathrm{H_3}\) CI seam.  The
isotope labelling changes the nuclear masses and
makes product channels experimentally
distinguishable, but the electronic geometric
phase is inherited from the conical-intersection
topology of the underlying $\mathrm{H_3}$
electronic problem.  In a
semiclassical path-language description, two
alternative scattering routes $\gamma_1, \gamma_2$
with the same asymptotic endpoints define a closed
loop $\gamma_1 \cdot \gamma_2^{-1}$; if this loop
links the CI seam, the relative sign of their
amplitudes is
\[
  (-1)^{\langle\eta_B,\,
  [\gamma_1 \cdot \gamma_2^{-1}]\rangle},
\]
and this discrete factor contributes to the
observed geometric-phase interference
oscillations.  This path-language
statement should be understood as the
semiclassical topological interpretation of the
geometric-phase contribution, not as a
replacement for the full quantum scattering
calculation.

The unifying $\Lk_6$ structure behind both
channels is the electronic-bundle lift: the
active bundle, the real eigenline $L_0^{\RR}$
over the punctured complement $\Omega^\circ$, and
its Stiefel--Whitney class $\eta_B$.  Depending on
the physical regime, this structure manifests as
an antiperiodic boundary condition, as a
Mead--Truhlar vector potential on the relevant
ground electronic surface, or as phases and
amplitudes in a non-adiabatic propagator.
Neither geometric-phase effect is determined by
the scalar shadow $(\Omega, V_0, g)$ alone; both
require electronic-bundle data forgotten by the
$\Lk_6 \to \Lk_5$ forgetful functor
\[
  U_6 : \Lk_6 \longrightarrow \Lk_5,
\]
specifically the real eigenline topology
encoded by $w_1(L_0^{\RR})$.

\begin{mathbox}[What the $\mathrm{Na_3}$ forcing
  example shows]
\label{box:lh-forcing-summary}
The scalar $\Lk_5$ shadow of the Jahn--Teller
problem records the underlying molecular graph,
the configuration region, the lower adiabatic
surface, and the mass-weighted metric,
\[
  X_5 \;=\;
  \bigl((G, \sigma), \Omega, V_-, g\bigr)
  \;\in\; \Lk_5,
  \qquad G = \mathrm{Na_3},
\]
where the discrete stereochemical label
$\sigma$ is trivial in this example and $V_-$
is understood as a scalar adiabatic shadow on
the punctured branching-plane region
$\Omega^\circ$ (in the generalised $\ScalGeom$
sense of Chapter~\ref{sec:L5}), not as a globally
smooth Morse function.  This
$\Lk_5$ object does not determine the nuclear
boundary condition around the pseudorotation
loop.  The physical $\Lk_6$ lift over this
scalar shadow is
\[
  X_6 \;=\;
  \bigl(X_5, \Hel^{(2)}, A^{(2)}, \Xseam,
  \eta_B\bigr),
\]
where $A^{(2)}$ is the Berry connection on the
active rank-$2$ bundle.  It contains the real
ground-state eigenline $L_0^{\RR} \to
\Omega^\circ$ over the punctured branching plane
and satisfies
\[
  \bigl\langle w_1(L_0^{\RR}),\,
  [S^1_\mathrm{ps}]\bigr\rangle \;=\; 1
\]
on the pseudorotation circle, where \(S^1_{\rm ps}\subset\Omega^\circ\) denotes the
pseudorotation loop around the punctured branching plane.
The nuclear factor is therefore antiperiodic,
\[
  \chi(\phi + 2\pi) \;=\; -\chi(\phi),
\]
and the allowed pseudorotational quantum numbers shift,
\[
  j \in \ZZ
  \;\;\leadsto\;\;
  j \in \ZZ + \tfrac12.
\]
This is the concrete empirical obstruction to
collapsing $\Lk_6$ back to $\Lk_5$: the
forgetful functor $U_6$ loses the real
eigenline topology.  The scalar object $X_5$ by
itself does not determine the boundary condition.
If one were to quantise \(X_5\) as an ordinary single-valued scalar pseudorotor, one would obtain the integer sector; the physical \(\Lk_6\) lift of the regular real Jahn--Teller CI evaluates non-trivially on the pseudorotation loop and therefore forces the half-integer sector.
\end{mathbox}

%% file: chapters/L6/l6_nextforcing.tex
\subsection{What \texorpdfstring{$\Lk_6$}{L6} cannot
  express: forcing of \texorpdfstring{$\Lk_7$}{L7}}
\label{sec:L6-forcing-out}

Up to $\Lk_6$, the molecular graph $G$ and the
associated nuclear configuration space are taken as
input data: the object $(G, \sigma)$ comes first,
and $\hat H_\mathrm{el}(\mathbf{R})$, $\Ce(G)$, and
the adiabatic bundle $\Hel^{(N)}$ are built over a
region of that configuration space.  Nuclear
permutations may appear as geometric or orbifold
symmetries, but the nuclei themselves are not yet
quantised as identical particles.  In particular,
$\Lk_6$ does not impose the exchange symmetry of
the nuclear wavefunction, nor does it derive
molecular graph structure from the all-particle
Coulomb Hamiltonian.

$\Lk_7$ is forced by two observations $\Lk_6$ cannot
accommodate:
\begin{enumerate}[label=(\arabic*)]
  \item \textbf{Nuclear indistinguishability}:
    identical nuclei carry exchange statistics ---
    antisymmetry under exchange for
    half-integer-spin nuclei, symmetry for
    integer-spin nuclei.  Protons give the cleanest
    forcing example as spin-$\tfrac12$ fermions.
  \item \textbf{Molecular identity}: despite this
    indistinguishability, molecular structure
    appears as a stable effective sector or
    correlation pattern in the quantum description.
    The emergence of ``this molecule'' rather than
    ``a superposition over graphs'' must be derived,
    not postulated.
\end{enumerate}
Both observations concern \emph{particle statistics
and identity}, not reaction rates.  Large kinetic
isotope effects and tunnelling corrections are not,
by themselves, clean forcing examples for $\Lk_7$:
many such effects can be modelled semiclassically as
corrections on an $\Lk_5$-level potential-energy or
free-energy surface (see
Remark~\ref{rem:tunneling-attribution} of
\S\ref{sec:L5-tst}).  They force $\Lk_7$ only when
the nuclear wavefunction, nuclear spin symmetry, or
identical-particle exchange structure becomes part
of the state description rather than a correction
functional on a classical PES.

\subsubsection{First forcing pair: nuclear
  indistinguishability}

\begin{forcingbox}[First forcing pair for $\Lk_7$: identical-nucleus statistics]
At every level up to $\Lk_6$, nuclei are treated as distinguishable: each occupies a position $\mathbf{R}_i$ on $\Ce(G)$, which is modelled on its regular strata as the quotient of $\RR^{3n}$
by $\mathrm{SE}(3) \rtimes \Aut_\mu(G)$, with
$\Aut_\mu(G)$ relabelling graph vertices while
preserving labels and masses.  This treats nuclear permutations as a \emph{geometric} symmetry of configuration space; 
it does not impose antisymmetry on the nuclear wavefunction. 
The fully quantum treatment requires the total molecular wavefunction to transform as the correct sign representation under exchange of any pair of identical nuclei, giving observably different predictions.

\medskip\noindent\textbf{Paradigm: ortho/para
$\mathbf{H_2}$.}
For ground-state $\mathrm{H_2}$, the electronic and vibrational factors are symmetric under proton exchange, and the rotational wavefunction has parity $(-1)^J$.  
Since the two protons are identical fermions, the total wavefunction must be antisymmetric under exchange.
Hence the nuclear-spin singlet $I = 0$ (antisymmetric) pairs with even $J$ to give \emph{para}-$\mathrm{H_2}$, while the nuclear-spin triplet $I = 1$ (symmetric) pairs with odd $J$ to give \emph{ortho}-$\mathrm{H_2}$.  
The two spin isomers have different rotational partition functions and therefore different low-temperature thermodynamic behaviour, and the interconversion is symmetry-forbidden in an isolated molecule.
In the absence of efficient paramagnetic, surface, or impurity-mediated conversion channels, ortho--para interconversion is slow on laboratory timescales, a metastability that underpins contemporary hyperpolarisation methods such as PHIP and SABRE~\cite{PravdivtsevEtAl2022}.

\medskip\noindent\textbf{What $\Lk_6$ cannot express.}
The $\Lk_6$ electronic data
\[
  (\Hel^{(N)}, A, \Xseam, \eta_B)
\]
are built over a classical nuclear configuration space. 
They do not include the nuclear spin Hilbert space, nor the representation of the identical-proton permutation group.
Hence the distinction between
\[
  \mathcal H_{\rm nuc}^{\rm para}
  \quad\text{and}\quad
  \mathcal H_{\rm nuc}^{\rm ortho}
\]
is invisible at $\Lk_6$: the $\Lk_6$ electronic
lift of the same scalar/electronic geometry is
insensitive to whether the nuclear spin state lies in the ortho or para sector.
The forgetful projection
\[
  U_7 : \Lk_7(P) \longrightarrow \Lk_6(P)
\]
therefore loses a directly observable distinction:
the two nuclear spin-statistical sectors have different allowed rotational quantum numbers and different low-temperature partition functions.
\end{forcingbox}

\subsubsection{Second forcing direction: molecular identity from all-particle quantum mechanics (programme-level)}

\begin{forcingbox}[Programme-level forcing direction for $\Lk_7$: molecular identity from all-particle quantum mechanics]
Unlike the ortho/para example above, the second
example is not a completed forcing theorem in the present manuscript but a programme-level forcing direction.
It points to a deeper limitation of $\Lk_6$: the molecular graph $G$ is still input data, whereas in an all-particle quantum treatment molecular structure should emerge from correlations, symmetry breaking, and sector selection rather than being postulated.

A species in $\Lk_5(P)$ is a pair $(G, \sigma)$, and the $\Lk_6$ electronic lift presupposes $G$:
the active bundle $\Hel^{(N)} \to \Omega \subset \Ce(G)$ is built over the configuration region associated with that graph.
Given only the Coulomb Hamiltonian for a collection of identical nuclei and electrons, why does a definite molecular structure --- such as a triangular $\mathrm{D}_3^+$ configuration --- emerge from a totally permutation-symmetric all-particle wavefunction?
Exact bound eigenstates of the all-particle Coulomb Hamiltonian inherit the symmetries of the Hamiltonian, so molecular shape is not represented as a literal fixed nuclear framework in the naive eigenfunction picture~\cite{Woolley1978,SutcliffeWoolley2012}.
It must be recovered from correlations, sector structure, symmetry breaking, or suitable classical/semiclassical limits.
In the Primas--Amann framework, molecular structure is proposed to emerge as a \emph{superselection sector} of the full quantum description in the mass-ratio limit $\varepsilon = (m_e/M)^{1/2} \to 0$~\cite{Primas1983,Amann1991}; a general theorem identifying arbitrary chemical graphs $G$ with such sectors remains an open problem.

Recent numerical work by Lang, Cezar, Adamowicz, and Pedersen~\cite{LangEtAl2024} provides a concrete demonstration of structural emergence for $\mathrm{D}_3^+$: Markov-chain Monte Carlo sampling of a totally permutation-symmetric pre-BO wavefunction recovers an unambiguous equilateral-triangular structure.
A general theorem for arbitrary graphs $G$ remains open and is treated in \S\ref{sec:L7-superselection}.

\medskip\noindent\textbf{What $\Lk_6$ cannot express.}
At $\Lk_6$, two molecular graphs correspond to different input objects and there is no operation internal to the theory that derives one graph sector from the all-particle Hilbert space.
In the full quantum description, by contrast, identical nuclei live in a single permutation-symmetric framework, and graph-like molecular structures must emerge as effective sectors or correlation patterns.
This motivates $\Lk_7$, but the general sector theorem for arbitrary chemical graphs remains open: the present discussion identifies the structural task that $\Lk_7$ must eventually solve, namely deriving graph-like molecular sectors from the all-particle quantum theory rather than assuming $G$ as input.
\end{forcingbox}

\subsubsection{What \texorpdfstring{$\Lk_7$}{L7} must provide}

Together, the two forcing directions require $\Lk_7$ to supply:
\begin{enumerate}[label=(\roman*)]
  \item \textbf{Full molecular Hamiltonian}
    $\hat H_\mathrm{mol} = \hat T_\mathrm{nuc} + \hat H_\mathrm{el}$ on the joint
    electron--nuclear Hilbert space, restricted to the correct permutation-symmetry sector (antisymmetric for electrons; spin-dependent for identical nuclei); the BO approximation of $\Lk_6$ becomes an asymptotic theorem under spectral-gap and regularity hypotheses as $\varepsilon \to 0$~\cite{HagedornJoye2001,PanatiSpohnTeufel2003}.
  \item \textbf{Strict deformation quantisation}: a continuous field $\{A_\varepsilon\}_{\varepsilon \in [0,1]}$ with commutative classical fibre
    \[
      A_0 = C_0(T^* \Ce_{\rm reg}(G))
    \]
    (or an appropriate Poisson algebra of classical observables on the regular cotangent stratum)
    and noncommutative $A_\varepsilon$ for $\varepsilon > 0$, in the Rieffel--Landsman framework~\cite{Rieffel1993,Landsman2017}.
    Here $\varepsilon$ denotes the semiclassical nuclear mass parameter, with $\varepsilon \sim (m_e/M)^{1/2}$ in the scaling convention used throughout this chapter; units are chosen so that the nuclear kinetic energy carries the prefactor $\varepsilon^2$.
  \item \textbf{Persistence of the $\Lk_6$ Berry-sign class}: the $\ZZ/2$-valued Berry-sign class
    \[
      \eta_B = w_1(L_0^{\RR})
    \]
    of $\Lk_6$ must be represented in the $\Lk_7$ nuclear quantum theory, where it appears as a geometric-phase boundary condition or as a correction to the effective nuclear Hamiltonian in the BO/semiclassical expansion~\cite{EmmrichWeinstein1996,LittlejohnRawlinson2024}.
  \item \textbf{Molecular identity as emergent sector (open programme)}: molecular graph-like structures should emerge as metastable or superselected sectors in an appropriate classical/semiclassical limit; each sector would support the $\Lk_6$ data as emergent structure.
  A general theorem identifying arbitrary chemical graphs $G$ with sectors of $A_0$ remains an open problem~\cite{Amann1991,Primas1983,LangEtAl2024}.
\end{enumerate}

\begin{insightbox}[$\Lk_7$ as the strictly quantum tower level]
\label{box:L7-preview}
The tower $\Lk_0 \to \cdots \to \Lk_6$ built chemistry from combinatorics through topology-decorated fibre structure over smooth configuration orbifolds, with nuclei classical throughout. 
$\Lk_7$ is the strictly quantum cap:
nuclei become quantum identical particles, the BO approximation appears as an asymptotic effective theory in the $\varepsilon \to 0$ regime, compatible with the classical fibre of the deformation-quantisation picture, and molecular identity is to be derived rather than postulated.

The next chapter develops $\Lk_7$ in this spirit,
necessarily more speculative than its predecessors:
the four open constructions (continuous field for
molecules, groupoid C*-algebra encoding nuclear
statistics, possible K-theoretic or real-bundle
refinements of the $\ZZ/2$ Berry-sign class,
superselection-sector theorem for arbitrary $G$)
are surveyed with explicit state-of-the-art
attributions and open-problem formulations.  
The ortho/para $\mathrm{H_2}$ forcing example
establishes the physical necessity of adding
nuclear quantum statistics; the molecular-identity discussion motivates the broader $\Lk_7$ programme.
The mathematical framework is the subject of what follows.
\end{insightbox}

%% file: chapters/ch_L7.tex
\section{\texorpdfstring{$\Lk_7$}{L7}: The Full Quantum Level}
\label{sec:L7}

\input{chapters/L7/l7_forcing}
\input{chapters/L7/l7_nuclear}
\input{chapters/L7/l7_sdq}
\input{chapters/L7/l7_georgescu}
\input{chapters/L7/l7_def}
\input{chapters/L7/l7_contributions}
\input{chapters/L7/l7_superselection}
\input{chapters/L7/l7_retrospective}

%% file: chapters/L7/l7_forcing.tex
\subsection{Forcing the full quantum level}
\label{sec:L7-forcing}

Section~\ref{sec:L6-forcing-out} identified two reasons the tower
must pass beyond $\Lk_6$, of different logical status.  The first is
a clean forcing example: ortho- and para-$\mathrm{H_2}$, which share
the same $\Lk_6$ electronic shadow but differ by nuclear
spin-statistical sector.  The second is a programme-level direction:
the recovery of graph-like molecular structure from an all-particle
quantum theory, rather than taking $G$ as primitive input.

\begin{itemize}
\item[(I)] \emph{Clean forcing example: nuclear spin statistics.}
  Ortho- and para-\(\mathrm{H_2}\) share the same scalar and electronic \(\Lk_6\) shadow.  Their distinction is not electronic; it is nuclear spin-statistical.
  The distinction lies in the nuclear spin-statistical sector, an object $\Lk_6$ does not carry.
  The two spin isomers have different allowed rotational quantum numbers and hence different rotational partition
  functions,
  including the $1{:}3$ nuclear-spin degeneracy of the para and ortho sectors,
  with correspondingly different low-temperature thermodynamic behaviour.
  In the absence of efficient paramagnetic, surface, or impurity-mediated conversion
  channels, ortho--para interconversion is slow on laboratory timescales;
  the precise rate is strongly condition-dependent~\cite{Silvera1980}.
\item[(II)] \emph{Programme-level direction: molecular identity as
  an emergent sector.}
  The molecular graph $G \in \LGraphP$ is given data at every level
  $\Lk_4$--$\Lk_6$: the DPO rules, the stereochemical symmetry, the
  configuration region, the active electronic bundle, and the
  Berry-sign data all presuppose a chosen graph.  The
  Woolley--Primas--Sutcliffe
  problem~\cite{Woolley1978,Primas1983,SutcliffeWoolley2012} asks
  whether such molecular structure can be recovered from the
  all-particle Coulomb Hamiltonian rather than postulated.  This is
  not a completed forcing theorem in the present manuscript; it is
  the programme-level task that motivates the general
  $\Lk_7$ construction.
\end{itemize}

\noindent
The first example gives the clean forcing obstruction: $\Lk_6$ does
not contain nuclear spin-statistical sectors.  The second points to
the broader completion problem: $G$ should ultimately emerge from
the all-particle quantum theory rather than remain primitive input.
Both motivate $\Lk_7$, but with different logical status.

At $\Lk_6$, nuclear positions are still treated through a classical configuration space.
Permutations may appear as geometric or orbifold symmetries, but there is no nuclear Hilbert space and no imposed bosonic or fermionic exchange symmetry of the nuclear wavefunction.
At $\Lk_7$ the nuclei become quantum particles whose algebra of observables carries exchange-indistinguishability constraints,
and from whose representation theory the graph label $G$ should be derivable as an effective sector rather than postulated.

\begin{warning}[Programmatic character of this chapter]
\label{warn:L7-programmatic}
  Unlike the earlier levels, where the intended data and forgetful projections can be specified explicitly, \(\Lk_7\) is introduced as a mathematical programme.  The level is defined by the conditions its data must satisfy, while the full construction of morphisms and the operator-algebraic functorial structure remains open.
\end{warning}

The correct formulation is a forgetful obstruction rather than an automorphism cokernel.
Let
\[
  U_7 : \Lk_7(P) \longrightarrow \Lk_6(P)
\]
denote the intended forgetful projection that discards the full nuclear quantum sector.
The ortho/para example exhibits two distinct $\Lk_7$-level objects --- the para and ortho nuclear spin-statistical sectors of $\mathrm{H_2}$ --- which $U_7$ maps to the same $\Lk_6$ electronic shadow.
Ortho- and para-\(\mathrm{H_2}\) share the same scalar and electronic \(\Lk_6\) shadow.
Their distinction is not electronic; it is nuclear
spin-statistical.
The molecular-identity programme
points to the broader open task of constructing $\Lk_7$-level sectors
from which graph-like molecular objects can emerge.

The ortho/para example forces the inclusion of nuclear Hilbert spaces with exchange symmetry, represented below by C1 and the spin-statistical part of C4.
The molecular-identity programme motivates the general superselection-sector construction C4. C2 and C3 provide the operator-algebraic and topological infrastructure needed to express these structures in the full tower.

\begin{mathbox}[$\Lk_7$ over $\Lk_6$, with the Para preview]
\label{mbox:L7-tower}
Solid arrows denote constructions specified at the relevant level;
dashed arrows denote candidate lifts whose construction is open.
\[
\includegraphics{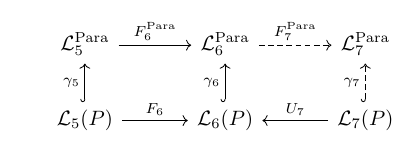}
\]
$U_7$ is the forgetful projection that discards the nuclear
operator-algebraic data.  The ortho/para forcing example shows that
$U_7$ is not conservative: distinct $\Lk_7$ spin-statistical sectors
share the same $\Lk_6$ electronic shadow. 
C2 supplies the operator-algebraic continuous-field framework for such a lift.
The full \(L_7\) lift also requires the nuclear Hilbert-space asymptotics of C1, the Berry-sign/twist data of C3, and the spin-statistical/superselection structure of C4.
The Para row, which contains variational quantum eigensolvers and
full-particle neural wavefunctions under the
$S_N \times S_M$-equivariance constraint, is developed in
Chapter~\ref{sec:Para}.
\end{mathbox}

\subsubsection{The deformation parameter
  \texorpdfstring{$\varepsilon$}{epsilon}}
\label{sec:epsilon}

The mathematical content of $\Lk_7$ depends on a single small
parameter controlling the relative quantumness of nuclei and
electrons.

\begin{definition}[Mass-ratio parameter]\label{def:epsilon}
  The \emph{deformation parameter} is
  \begin{equation}
    \label{eq:epsilon-def}
    \varepsilon \;:=\;
    \Bigl(\frac{m_e}{M}\Bigr)^{1/2},
  \end{equation}
  where $m_e$ is the electron rest mass and $M$ a reference nuclear
  mass (conventionally the proton mass).
\end{definition}

\noindent
In atomic units the nuclear kinetic energy is $O(\varepsilon^2)$
relative to the electronic energy, making nuclei progressively
semiclassical as $\varepsilon\to 0$ while electrons remain fully
quantum.
Some representative values of $\varepsilon$ are:
$\varepsilon_{\mathrm H} \approx 0.0233$,
$\varepsilon_{\mathrm C} \approx 0.0068$,
$\varepsilon_{\mathrm{Pb}} \approx 0.0016$,
computed from $\varepsilon = \sqrt{m_e/M}$ with standard
atomic weights ($m_e/u = 5.486\times 10^{-4}$).
Both Panati--Spohn--Teufel space-adiabatic perturbation theory and
Teufel's monograph use the same convention $\varepsilon =
\sqrt{m_e/M}$; in this convention the Berry connection and
higher-order Born--Oppenheimer corrections appear in the effective
nuclear Hamiltonian at the corresponding powers of
$\varepsilon$~\cite{PanatiSpohnTeufel2003,Teufel2003}.
Convention~\eqref{eq:epsilon-def} makes the continuous field
$\{A_\varepsilon\}_{\varepsilon \in [0,1]}$ of Construction~C2 a
deformation over the unit interval, with commutative classical fibre $A_0^G = C_0(T^* \Ce_{\rm reg}(G))$ (or an appropriate Poisson
algebra of classical observables on the regular cotangent
stratified phase space) at $\varepsilon = 0$, and the physical
system realised at $\varepsilon = \varepsilon_G > 0$.

\subsubsection{Four constructions \texorpdfstring{$\Lk_7$}{L7} must provide}
\label{sec:L7-constructions}

The ortho/para forcing example and the molecular-identity programme identify four mathematical structures absent from the previous levels.
Each is developed in full in its own section; the statements here fix terminology.

\begin{enumerate}[label=\textbf{C\arabic*.}, leftmargin=*, labelwidth=2.2em]

\item \emph{Nuclear quantum dynamics on the BO surface}
  (\S\ref{sec:L7-nuclear}).
  The effective nuclear Schrödinger operator, schematically,
    \[
      \hat H_{\rm nuc}^{\rm eff}
      =
      -\frac{\hbar^2}{2M}\Delta_g
      + V(\mathbf R)
      + \text{higher-order Born--Oppenheimer corrections},
    \]
    or, in semiclassical scaling,
    \[
      \hat H_{\rm nuc}^{\rm eff}
      =
      -\frac{\varepsilon^2}{2}\Delta_g
      + V(\mathbf R)
      + O(\varepsilon).
    \]
  on $L^2(\Ce_{\rm reg}(G), d\mu_g)$, with $V$ the $\Lk_5$ BO surface and
  $\Delta_g$ the mass-weighted Laplace--Beltrami operator,
  as the leading effective output,
  with higher-order corrections,
  of a controlled asymptotic reduction of the full molecular Hamiltonian.
  The rigorous Born--Oppenheimer expansion of
  Hagedorn~\cite{Hagedorn1980} and the space-adiabatic theorem of
  Panati--Spohn--Teufel~\cite{PanatiSpohnTeufel2003} supply C1 at
  the Hilbert-space level.

\item \emph{A continuous field of $C^*$-algebras}
  (\S\ref{sec:L7-sdq}).
  A family $\{A_\varepsilon^G\}_{\varepsilon\in[0,1]}$ with
  commutative classical fibre $A_0^G = C_0(T^* \Ce_{\rm reg}(G))$,
  non-commutative quantum fibres $A_\varepsilon^G$ for
  $\varepsilon > 0$, and quantisation maps $Q_\varepsilon$
  satisfying the Dirac and von~Neumann
  conditions~\cite{Rieffel1989,Landsman1998} --- a strict
  deformation quantisation parametrised by mass.
  C2 is the central open construction: a continuous field
  combining all of the required molecular features --- Coulomb
  singularities, non-compact/stratified configuration spaces, and
  non-trivial Berry-sign topology ($\eta_B \neq 0$) --- is not
  currently available in the form required by this tower, although
  the relevant ingredients
  (Landsman~\cite{Landsman2007,Landsman1998} for smooth
  $Q_\varepsilon$, Panati--Spohn--Teufel for the
  $\varepsilon$-expansion,
  Georgescu--Iftimovici~\cite{GeorgescuIftimovici2002} for
  $N$-body affiliation) exist in isolation.

\item \emph{The Berry-sign class as topological obstruction}
  (\S\ref{sec:conj-II}).
  The real Berry-sign class
  \[
    \eta_B = w_1(L_0^{\RR})
    \in H^1(\Omega^\circ;\, \ZZ/2),
    \qquad
    \Omega^\circ = \Omega \setminus \Xseam,
  \]
  At the formal WKB/Moyal level this is Dazord--Patissier~\cite{DazordPatissier1991} and
  Emmrich--Weinstein~\cite{EmmrichWeinstein1996};
  Hawkins~\cite{Hawkins2008} carries out the strict $C^*$-algebraic promotion on the compact sphere.
  The analogous statement for molecular configuration orbifolds, with a real-bundle or \(KO\)-theoretic representation of the mod-2 class, is open.

\item \emph{Molecular identity as superselection sector}
  (\S\ref{sec:L7-superselection}).
  The programme is to construct an observable algebra carrying the
  appropriate permutation, exchange-symmetry, and graph-reconstruction
  data, and to identify sectors whose classical shadows reproduce the
  graph-like molecular objects used at \(\Lk_4\)--\(\Lk_6\).
  Special cases are known: Pfeifer established an analogous result for
  chirality in a two-level spin--boson model with ohmic coupling, and
  the framework was developed by Amann.  Lang, Cezar, Adamowicz, and
  Pedersen give numerical evidence that graph-like triangular
  structure can be extracted from a permutation-adapted
  pre-Born--Oppenheimer wavefunction for \(\mathrm D_3^+\).
  A general theorem for arbitrary molecular graphs remains open.

\end{enumerate}

\noindent
C1 is best understood at the Hilbert-space asymptotic level.
C2 and C3 remain open in the full molecular \(C^*\)-algebraic setting.
C4 is supported by special-case models and recent numerical evidence,
but a general molecular superselection theorem remains open.

Section~\ref{sec:L7-retrospective} records the full status table;
the intervening sections develop each construction.

\begin{insightbox}[The shape of Chapter~\ref{sec:L7}]
Section~\ref{sec:L7-nuclear} develops the new physical content of
$\Lk_7$ --- the nuclear Schr\"odinger equation, the PST
$\varepsilon$-expansion, and the demarcation between $\Lk_5$
semiclassical tunnelling and genuinely $\Lk_7$ phenomena.
Sections~\ref{sec:L7-sdq} and~\ref{sec:L7-georgescu} present the
two mathematical programmes closest to delivering C2:
strict deformation quantisation (Rieffel, Landsman) and the
Georgescu--Iftimovici $N$-body algebra.
Section~\ref{sec:L7-def} formulates \(\Lk_7(P)\) as a category of \(\Lk_6\)-objects equipped with candidate full quantum/operator-algebraic lifts, together with a forgetful projection
\[
U_7:\Lk_7(P)\to\Lk_6(P).
\]
Section~\ref{sec:L7-contributions} states four conjectures (the
precise form of C1--C4) as the tower's original contribution at
this level.
Section~\ref{sec:L7-superselection} develops the Woolley--Primas
problem as the tower statement of C4.
Section~\ref{sec:L7-retrospective} closes with the complete tower,
the corrected list of forgetful extensions, and the three inter-level coherence conditions.
\end{insightbox}

%% file: chapters/L7/l7_nuclear.tex
\subsection{Quantum nuclear dynamics: the new physical content
  of \texorpdfstring{$\Lk_7$}{L7}}
\label{sec:L7-nuclear}

The forcing arguments of \S\ref{sec:L7-forcing} identified two
structures absent from $\Lk_6$: a wavefunction for the nuclei
(with the indistinguishability it carries) and the programme of
recovering molecular identity as an effective superselection or
correlation-sector label.
This section develops the first physical content that $\Lk_7$
delivers: the nuclear Schr\"{o}dinger equation on the BO surface,
the space-adiabatic $\varepsilon$-expansion that recovers the
Born--Oppenheimer approximation as a controlled asymptotic, and
its semiclassical tunnelling regime.
Section~\ref{sec:nuc-spin} returns to the forcing content with the
ortho/para example.

\begin{warning}[What this section does not claim]
\label{warn:kie-not-forcing}
  Anomalous primary kinetic isotope effects --- including values
  of $55\pm 6$~\cite{Scrutton2006} and $661\pm 27$~\cite{HuEtAl2017}
  in enzymatic hydrogen transfer --- are frequently cited as
  evidence for a quantum-mechanical treatment of nuclei.
  They are not evidence for $\Lk_7$ specifically.
  Large KIEs are not, by themselves, clean forcing examples for \(\Lk_7\).
  Many such effects can be estimated using semiclassical or path-integral approximations built from \(\Lk_5\)-level geometric data:
  a PES, a mass metric, Hessians, and barrier geometry.
  These methods can account for many large KIEs in model-dependent settings,
  but they do not by themselves force the full \(\Lk_7\) structure.
  What \(\Lk_7\) adds is the nuclear Hilbert-space and operator-algebraic framework in which such approximations are derived, 
  together with nuclear exchange-statistical sectors that no scalar semiclassical correction can express.
  
  The Agmon--Helffer--Sj\"{o}strand theorem of \S\ref{sec:tunnelling} is a rigorous $\varepsilon\to 0$ asymptotic of the nuclear
  Schr\"{o}dinger equation:
  it formalises what semiclassical tunnelling methods approximate.
  The $\Lk_7$ novelty lies not in the tunnelling estimate itself but in the Hilbert-space/$C^*$-algebraic setting in which such estimates become theorems, together with the exchange-statistical sectors developed in \S\ref{sec:nuc-spin}.
\end{warning}

\subsubsection{The nuclear Schr\"{o}dinger equation}
\label{sec:nuc-schr}

At every level $\Lk_0$--$\Lk_6$, nuclear positions $\mathbf{R}$
served as classical parameters for the electronic Hamiltonian
$\hat{H}_\mathrm{el}(\mathbf{R})$.
The BO surface $V(\mathbf{R})$ and the Berry connection $A$ act on
electrons at each fixed $\mathbf{R}$; the nuclei themselves
contribute no operator.
$\Lk_7$ places the nuclei on the quantum stage.

\begin{definition}[Nuclear Schr\"{o}dinger equation on the BO
  surface]
\label{def:nuclear-schrodinger}
  For a molecular species $G$ with configuration orbifold
  $\Ce(G)$, mass-weighted Riemannian metric $g$ (with
  $g_{ij} = M_i\delta_{ij}$ in Cartesian atomic coordinates
  before the $\mathrm{SE}(3) \rtimes \Aut_\mu(G)$ quotient), and BO
  potential energy surface $V:\Ce(G)\to\RR$ from $\Lk_5$, the
  \emph{nuclear Schr\"{o}dinger equation} is
  \begin{equation}
    \label{eq:nuclear-schrodinger}
    \hat{H}_\mathrm{nuc}\,\Psi
    \;=\;
    \Bigl(-\tfrac{\hbar^2}{2}\,\Delta_g
    + V(\mathbf{R})\Bigr)\Psi
    \;=\; E\,\Psi,
  \end{equation}
  where $\Delta_g$ is the Laplace--Beltrami operator for the
  mass-weighted metric and $\Psi\in L^2(\Ce_{\rm reg}(G), d\mu_g)$.
  The nuclear masses enter through $g$.  In the dimensionless
  semiclassical units of Definition~\ref{def:epsilon}, with
  $\varepsilon = (m_e/M)^{1/2}$, the operator may be rescaled to
  \[
    \hat H_\mathrm{nuc}^{\varepsilon}
    \;=\;
    -\frac{\varepsilon^2}{2}\Delta_{g_0} + V,
  \]
  where $g_0$ is a mass-normalised reference metric with
  $(g_0)_{ij} = (M_i/M)\delta_{ij}$ in Cartesian coordinates and
  $M$ a chosen reference nuclear mass, so that $M_i/M = O(1)$.
  This is the sector-reduced scalar form of the nuclear
  Schr\"odinger equation; before imposing exchange symmetry, the
  natural Hilbert space is built on the labelled configuration
  space $Q_{\rm lab}(G)$ (see \S\ref{sec:nuc-spin}).
\end{definition}

\noindent
The formal derivation of Equation~\eqref{eq:nuclear-schrodinger}
from the full electron--nuclear Schr\"{o}dinger equation is the
Born--Oppenheimer approximation; its rigorous form, with
controlled error, is due to
Hagedorn~\cite{Hagedorn1980} and Panati--Spohn--Teufel
(Theorem~\ref{thm:pst} in \S\ref{sec:L7-sdq}).

\begin{mathbox}[The effective Hamiltonian stratified by tower
  level]
\label{mbox:heff-tower}
Theorem~\ref{thm:pst} expresses the effective nuclear Hamiltonian,
schematically and in a local adiabatic representation, in
minimal-coupling form
\[
  \hat{H}_\mathrm{eff}^\varepsilon
  \;=\;
  \tfrac{1}{2}\bigl(p - \varepsilon A(\mathbf{R})\bigr)^2
  + E_j(\mathbf{R})
  + \varepsilon^2\,\Phi_\mathrm{BH}(\mathbf{R})
  + \cdots ,
\]
where $A$ is the Berry connection and $\Phi_\mathrm{BH}$ a
Born--Huang-type scalar correction (as in \cite{born1996dynamical,Epstein1965}); the precise powers and signs depend on the chosen semiclassical scaling and gauge.
Each piece is anchored at a specific tower level: the BO surface
$E_j(\mathbf{R})$ from the $\Lk_5$ scalar geometric data; the
Berry connection $A$ from the chosen $\Lk_6$ electronic lift;
the mass-dependent nuclear kinetic operator from $\Lk_7$.
The same small parameter $\varepsilon$ can serve as the
deformation parameter in the strict-deformation-quantisation
programme (C2), but it is \emph{not} an index of tower depth:
tower levels are conceptual layers, while $\varepsilon$ is a
semiclassical parameter.  $\Lk_6$ data enter the expansion
through Berry-connection and higher adiabatic-correction terms;
the relationship of $\varepsilon\to 0$ to $\Lk_5$ is via the
scalar geometric shadow, not a literal level identification.
\end{mathbox}

\begin{chembox}[What the nuclear Schr\"{o}dinger equation gives]
Solving equation~\eqref{eq:nuclear-schrodinger} yields, level by
level:
\begin{itemize}
  \item \textbf{Quantised vibrational levels.}
    In the harmonic approximation around a minimum with force
    constant $k$, $E_\nu = \hbar\omega(\nu + \tfrac{1}{2})$ with
    $\omega = \sqrt{k/\mu}$ and reduced mass $\mu$.
  \item \textbf{Zero-point energy.}
    $E_0 = \tfrac{1}{2}\hbar\omega > 0$ even at $T = 0$.
    For a C--H oscillator with $k\approx 480\,\mathrm{N/m}$ and
    $\mu_\mathrm{CH} \approx 0.923\,\mathrm{u}$:
    $E_0^\mathrm{CH} \approx 17.5\,\mathrm{kJ\,mol^{-1}}$.
    For C--D with $\mu_\mathrm{CD} \approx 1.714\,\mathrm{u}$:
    $E_0^\mathrm{CD} \approx 12.9\,\mathrm{kJ\,mol^{-1}}$.
    The ZPE difference $\Delta E_\mathrm{ZPE} \approx 4.6\,\mathrm{kJ\,mol^{-1}}$
    gives a scale factor $e^{\Delta E_\mathrm{ZPE}/RT} \approx 6.4$
    at 298\,K to primary H/D KIEs --- the Westheimer-type
    semiclassical estimate discussed in \S\ref{sec:tunneling-attribution}.  This is
    only an illustrative scale estimate; a quantitative primary KIE
    depends on the difference between isotope-dependent vibrational
    free energies in the reactant and transition-state regions, not
    on the isolated C--H/C--D oscillator alone.
  \item \textbf{Quantum tunnelling.}
    Sub-barrier transmission with amplitude $e^{-\theta}$,
    controlled by the Agmon distance through the classically
    forbidden region (\S\ref{sec:tunnelling}).
  \item \textbf{Isotope effects.}
    Each item above depends explicitly on nuclear mass through $\mu$ or the mass metric.
    The mass data already belong to the geometric input at $\Lk_5$, while $\Lk_7$ supplies the nuclear Hilbert-space dynamics in which their spectral consequences (vibrational levels, ZPE, tunnelling amplitudes) are realised.
\end{itemize}
These quantities are not native dynamical objects of $\Lk_5$ or $\Lk_6$, because those levels do not contain a nuclear Hilbert
space.
However, many of their leading semiclassical estimates can be computed from $\Lk_5$-level data such as the PES, the Hessian, the mass metric, and the Agmon distance (\S\ref{sec:tunneling-attribution}).
The $\Lk_7$ contribution is to place these estimates inside a nuclear Schr\"odinger / operator-algebraic framework; the $\Lk_7$ claim on this content is \emph{rigour}, not novelty.
\end{chembox}

\subsubsection{Semiclassical tunnelling:
  the Agmon--Helffer--Sj\"{o}strand theory}
\label{sec:tunnelling}

At $\Lk_5$ the nuclear configuration orbifold $\Ce(G)$ is a
Riemannian space with mass-weighted metric $g$ and potential
$V(\mathbf{R})$; the classically forbidden region is
$\{V > E\}$.
At $\Lk_7$ nuclear wavefunctions are supported on all of
$\Ce(G)$, including this region, with exponentially
suppressed amplitude.
The rigorous form of that suppression in the $\varepsilon\to 0$
asymptotic is the Agmon theory, which recasts tunnelling as a
question about a degenerate Riemannian metric.

\begin{definition}[Agmon metric and tunnelling distance]
\label{def:agmon}
  For the nuclear operator
  $\hat H_\mathrm{nuc} = -(\hbar^2/2)\Delta_g + V$ on
  $(\Ce_{\rm reg}(G), g)$ at energy $E < \max V$, the \emph{Agmon metric}
  is the degenerate Riemannian metric
  \[
    ds^2_\mathrm{Ag}
    \;:=\;(V(\mathbf{R}) - E)_+\;g_{ij}\,dR^i\,dR^j,
  \]
  where $(V - E)_+ = \max(V - E, 0)$ vanishes on
  classically allowed regions.
  The \emph{Agmon distance} between
  $\mathbf{R}_a, \mathbf{R}_b\in\Ce_{\rm reg}(G)$ is
  \[
    \AgmonD(\mathbf{R}_a, \mathbf{R}_b)
    \;:=\;
    \inf_\gamma\int_\gamma\sqrt{(V - E)_+}\;ds_g,
  \]
  the infimum taken over smooth paths in $\Ce_{\rm reg}(G)$, with
  $ds_g = \sqrt{g_{ij}\,dR^i\,dR^j}$ the arc-length element of $g$.
\end{definition}

\begin{remark}[Agmon distance stratified by tower level]
\label{rem:agmon-tower}
  The Agmon distance is built entirely from $\Lk_5$ data --- the
  orbifold $\Ce(G)$, the mass-weighted metric $g$, the BO surface
  $V$ --- yet it controls the $\Lk_7$ observable
  $e^{-\AgmonD/\varepsilon}$.
  This is the pattern of tower interaction throughout this chapter:
  $\Lk_7$ supplies the nuclear kinetic operator; the geometric
  object that operator acts on is $\Lk_5$ content.
\end{remark}

\begin{theorem}[Tunnelling splitting;
  Helffer--Sj\"{o}strand~\cite{HelfferSjostrand1984},
  Simon~\cite{Simon1984Tunnelling}]
\label{thm:tunnelling}
  Let $V$ be a smooth symmetric double-well potential on $\RR^f$
  with two non-degenerate minima $\mathbf{R}_a, \mathbf{R}_b$
  separated by a barrier (or, more generally, two wells whose
  local ground energies match to leading order in $\varepsilon$),
  and let $\hat H^\varepsilon = -(\varepsilon^2/2)\Delta + V$ in
  atomic units.  As $\varepsilon\to 0$, the ground-state splitting
  $\Delta E = E_1 - E_0$ satisfies
  \[
    \Delta E
    \;=\;
    a(\varepsilon)\,
    \exp\!\bigl[-\AgmonD(\mathbf{R}_a,\mathbf{R}_b)/\varepsilon\bigr],
  \]
  where $a(\varepsilon)$ admits an asymptotic expansion in powers
  of $\varepsilon$ whose leading behaviour depends on the dimension
  and on the Hessian data at the minima and saddle; in particular,
  \[
    \log \Delta E
    \;=\;
    -\AgmonD(\mathbf{R}_a,\mathbf{R}_b)/\varepsilon
    + O\bigl(\log(1/\varepsilon)\bigr).
  \]
  Here $\AgmonD$ is the Agmon distance of
  Definition~\ref{def:agmon} with $g = \delta$; the molecular case
  applies locally on a smooth stratum of $\Ce_{\rm reg}(G)$, with
  mass dependence restored either through the mass-weighted Agmon
  metric or via explicit masses in the WKB exponent
  (Chembox~\ref{cbox:kie-semiclassical}).
\end{theorem}

\begin{proof}[References]
  Exponential localisation in the forbidden region follows from
  the weighted-$H^1$ estimates of
  Helffer--Sj\"{o}strand~\cite{HelfferSjostrand1984}:
  for $\phi(\mathbf{R}) = \AgmonD(\mathbf{R}, \mathbf{R}_a)$, the
  bound $\|e^{\phi/\varepsilon}\psi\|_{H^1}\leq C$ implies decay
  of $\psi$ in $\{V > E\}$.  Together with quasimode and
  inter-well interaction estimates, this gives the exponential
  upper bound on the splitting; the matching lower bound is due
  to Simon~\cite{Simon1984Tunnelling}.
  Hagedorn's original analysis~\cite{Hagedorn1980} uses
  $\varepsilon_H := (m_e/M)^{1/4} = \varepsilon^{1/2}$, related
  to the tower's convention by $\varepsilon = \varepsilon_H^2$.
\end{proof}

\begin{remark}[What Theorem~\ref{thm:tunnelling} is, and is not]
\label{rem:tunnelling-status}
  Theorem~\ref{thm:tunnelling} is a theorem about the $\Lk_5$
  geometric data: given the mass-weighted metric and the BO
  surface, it computes the $\varepsilon\to 0$ asymptotic of the
  tunnelling amplitude.
  The substance of its $\Lk_7$ attribution is narrow:
  the nuclear wavefunction $\psi\in L^2(\Ce_{\rm reg}(G))$ of which the
  splitting is a spectral property first exists at $\Lk_7$
  (Definition~\ref{def:nuclear-schrodinger}), and the operator
  whose ground-state splitting the theorem computes is the
  nuclear kinetic operator of Equation~\eqref{eq:nuclear-schrodinger}.
  The theorem itself is a semiclassical analysis of the \(\Lk_5\) geometric data within the \(\Lk_7\) nuclear Hilbert-space framework.
  In particular, applying it does not require solving
  Construction~C2: the Helffer--Sj\"{o}strand proof operates at
  the Hilbert-space level, independent of any $C^*$-algebraic
  field.
\end{remark}

\begin{chembox}[KIEs from the semiclassical limit]
\label{cbox:kie-semiclassical}
  In atomic units, where $\hbar = 1$ and $M_I$ is measured in
  electron masses, the WKB tunnelling exponent for isotope $I$
  in a one-dimensional coordinate $q$ with a mass-independent
  Euclidean metric is
  \begin{align*}
    \theta_I
    &\;=\;
    \!\int_\mathrm{barrier}\!
      \sqrt{2 M_I\bigl(V(q) - E\bigr)}\;dq
    \;=\;
    \sqrt{M_I}\,\theta_0,
    \\
    \theta_0
    &\;:=\;
    \!\int_\mathrm{barrier}\!
      \sqrt{2\bigl(V(q) - E\bigr)}\;dq,
  \end{align*}
  with $\theta_0$ mass-independent.  Equivalently, in the
  mass-weighted formulation of Definition~\ref{def:agmon}, the
  isotope dependence is absorbed into the Agmon metric itself;
  one must not include the mass both in $g$ and as an external
  $\sqrt{M_I}$ factor.
  Since $M_D\approx 2M_H$, the tunnelling ratio is
  $T_H/T_D = \exp[(\sqrt{2}-1)\sqrt{M_H}\,\theta_0]$.
  With $\sqrt{M_H}\approx 42.8$, a modest $\theta_0 = 0.1$ gives
  $T_H/T_D\approx 5.9$, and $\theta_0 = 0.3$ gives $\approx 200$
  --- the basic WKB mechanism underlying many tunnelling-enhanced
  KIE models.
  In each case this is an $\Lk_5$ semiclassical calculation with
  $M_H$ as an input parameter; it is the content
  Theorem~\ref{thm:tunnelling} makes rigorous in the
  $\varepsilon\to 0$ limit, and related isotope-dependent
  tunnelling estimates are obtained by the instanton, $\mu$OMT,
  and RPI+PC methods of \S\ref{sec:tunneling-attribution} within
  their respective modelling assumptions.
  The tower's contribution is not the KIE number but the
  attribution of that number to the correct level.
\end{chembox}

\subsubsection{Ortho and para-hydrogen: the forcing content of \texorpdfstring{$\Lk_7$}{L7}}
\label{sec:nuc-spin}

The tunnelling content of \S\ref{sec:tunnelling} is reproduced by
$\Lk_5$ semiclassical methods; it sharpens but does not force
$\Lk_7$.
Ortho- and para-hydrogen, by contrast, are distinguished by a
structure no level below $\Lk_7$ carries: the symmetry of the
nuclear wavefunction under proton exchange.
This is forcing pair~(I) of \S\ref{sec:L7-forcing} in concrete form.

\begin{example}[Ortho/para-$\mathrm{H_2}$]
\label{ex:ortho-para}
  The proton is a spin-$\tfrac{1}{2}$ fermion.
  Pauli antisymmetry requires the total $\mathrm{H_2}$ wavefunction
  $\Psi = \psi_\mathrm{spatial}\otimes\chi_\mathrm{spin}$ to be
  antisymmetric under proton exchange $\pi_{12}$.
  Since $\pi_{12}\psi_J = (-1)^J\psi_J$ on rotational states with
  quantum number $J$, nuclear-spin states split into:
  \begin{itemize}
    \item \emph{para-$\mathrm{H_2}$}: singlet spin ($I = 0$,
      antisymmetric under $\pi_{12}$), paired with even~$J$.
      Ground state $J = 0$.
    \item \emph{ortho-$\mathrm{H_2}$}: triplet spin ($I = 1$,
      symmetric under $\pi_{12}$), paired with odd~$J$.
      Lowest state $J = 1$, roughly $14.7\,\mathrm{meV}$ above
      para.
  \end{itemize}
  At thermal equilibrium, the 3:1 nuclear-spin degeneracy gives
  ortho:para $= 3:1$ at high $T$ and pure para as $T\to 0$.
  In the ordinary isolated, spin-independent approximation, the
  spatial and nuclear-spin sectors are separately preserved,
  so ortho and para subspaces are invariant.
  Conversion requires weak spin-dependent, magnetic, surface, impurity-mediated, or other environmental interactions.
  
  In the absence of efficient such channels, ortho--para conversion is slow on laboratory timescales; the precise rate is condition-dependent.
  The two spin isomers have different rotational partition functions and therefore different low-temperature thermodynamic behaviour.
  The lowest ortho level $J{=}1$ lies above the para ground level $J{=}0$ by $2B$, where $E_J = B\,J(J+1)$;
  for $\mathrm{H_2}$, this is about $14.7\,\mathrm{meV}$, and ortho-to-para conversion releases approximately this gap per converted molecule upon liquefaction.
\end{example}

\begin{mathbox}[Ortho/para as a \(\Lk_7\) superselection decomposition]
\label{mbox:superselection-H2}
Start from the labelled nuclear configuration space \(Q_{\rm lab}\)
for the two protons.  The pre-physical nuclear Hilbert space is
\[
  \mathcal H_{\rm lab}
  =
  L^2(Q_{\rm lab})\otimes \mathbb C^4_{\rm spin}.
\]
The exchange \(\pi_{12}\in S_2\) acts diagonally on the spatial and
spin factors.  Since protons are fermions, the physical Hilbert space is
the antisymmetric subspace
\[
  \mathcal H_{\rm phys}
  =
  \{\Psi\in\mathcal H_{\rm lab}:\pi_{12}\Psi=-\Psi\}.
\]
It decomposes as
\[
  \mathcal H_{\rm phys}
  =
  \mathcal H_{\rm para}\oplus\mathcal H_{\rm ortho},
\]
where
\[
  \mathcal H_{\rm para}
  =
  \mathcal H_{\rm spatial}^{\rm even}
  \otimes
  \mathcal H_{\rm spin}^{I=0},
  \qquad
  \mathcal H_{\rm ortho}
  =
  \mathcal H_{\rm spatial}^{\rm odd}
  \otimes
  \mathcal H_{\rm spin}^{I=1},
\]
where ``even'' and ``odd'' refer to the parity of the labelled
spatial wavefunction under proton exchange.
For the isolated molecule in the ordinary spin-independent
approximation, spatial and nuclear-spin sectors are separately
preserved, so the para and ortho subspaces are invariant.
Sector-coupling conversion requires interactions absent from this Hamiltonian,
such as magnetic, surface, impurity-mediated, or other spin-dependent environmental effects.
Thus ortho and para hydrogen are distinct nuclear spin-statistical sectors,
invisible to the scalar/electronic $\Lk_6$ description but native to $\Lk_7$.
\end{mathbox}

\begin{insightbox}[Forcing vs content at $\Lk_7$]
  Sections~\ref{sec:nuc-schr}--\ref{sec:tunnelling} developed
  content that $\Lk_7$ makes rigorous: the nuclear Schr\"odinger
  equation, its $\varepsilon$-expansion with Berry/geometric
  corrections, and the Agmon--Helffer--Sj\"ostrand tunnelling
  asymptotic.  Many of its leading semiclassical consequences can
  be estimated from $\Lk_5$-level data; the $\Lk_7$ claim is on
  the Hilbert-space and $C^*$-algebraic framework
  (\S\ref{sec:L7-sdq}--\S\ref{sec:L7-def}).
  Section~\ref{sec:nuc-spin} developed the content that forces \(\Lk_7\): nuclear exchange symmetry, nuclear spin-statistical sectors, and the resulting ortho/para superselection decomposition.
  
  No scalar $\Lk_5$-level semiclassical correction can produce the nuclear spin-statistical sector decomposition itself;
  the symmetry of the nuclear wavefunction under proton exchange is a genuinely $\Lk_7$ datum, and the metastable ortho/para populations --- with distinct rotational partition functions,
  slow condition-dependent interconversion in the absence of efficient catalysts, and the $J{=}1 \to 0$ energy release in ortho-to-para conversion --- are its laboratory signatures.
\end{insightbox}

%% file: chapters/L7/l7_sdq.tex
\subsection{Construction~C2: the continuous field of \texorpdfstring{$C^*$}{C*}-algebras}
\label{sec:L7-sdq}

Recall from \S\ref{sec:L7-def} that $\Lk_7(P)$ is the category
of $\Lk_6$-objects equipped with candidate continuous-field lifts,
together with a forgetful projection
$U_7 : \Lk_7(P) \to \Lk_6(P)$.
We write $\Felseven(G)$ for the candidate object-level lift of a
$\Lk_6(P)$-species $(G, \sigma_j, A, \eta_B)$ to a continuous
field $\{A_\varepsilon^G\}_{\varepsilon \in [0,1]}$ of
$C^*$-algebras, where
$\eta_B = w_1(L_0^{\RR}) \in H^1(\Omega^\circ; \ZZ/2)$ is the
real Berry-sign class of \S\ref{sec:L6-berry}.
This section addresses the \emph{object part} of $\Felseven$:
how, for each $G$, the field is constructed and what conditions
it must satisfy.
The action of $\Felseven$ on morphisms --- DPO reactions
lifted to continuous-field $*$-homomorphisms --- is a separate
open problem (Remark~\ref{rem:F7-morphisms}), not addressed here.

The data $\Felseven(G)$ must satisfy:
commutative classical fibre $A_0^G = C_0(T^*\Ce_{\rm reg}(G))$
at $\varepsilon = 0$, non-commutative quantum fibres
$A_\varepsilon^G$ for $\varepsilon > 0$, and quantisation maps
$Q_\varepsilon$ whose $\varepsilon\to 0$ limit reproduces the
Poisson structure on $T^*\Ce_{\rm reg}(G)$ inherited from $\Lk_5$.
When non-trivial nuclear exchange sectors are tracked, the
scalar fibre is the sector-reduced form; the scalar Hilbert space
$L^2(\Ce_{\rm reg}(G))$ may be replaced either by equivariant
functions on the labelled configuration space or by sections of
the associated bundle/local system over the quotient, in the
chosen exchange-statistical sector
(cf.\ \S\ref{sec:L7-nuclear}).
The $\Lk_6$ topology should appear in the field at the
appropriate order: the Berry connection $A$ enters the effective
Hamiltonian element through geometric-correction terms at the
powers of $\varepsilon$ dictated by the semiclassical scaling,
and the mod-2 Berry-sign class $\eta_B = w_1(L_0^{\RR})$ acts as
the obstruction to ordinary single-valued scalar quantisation.

Three mathematical inputs converge on the construction of
$\Felseven$ on objects.
Strict deformation quantisation (\S\ref{sec:sdq-framework}) specifies
what the object-level lift $\Felseven(G)$ must be as an abstract
object and supplies the continuous field on smooth configuration
spaces.
Space-adiabatic perturbation theory (\S\ref{sec:pst}) identifies
how the Berry connection and related geometric terms from the
chosen $\Lk_6$ electronic lift enter the effective nuclear
Hamiltonian in the semiclassical expansion.
The Georgescu $N$-body algebra (\S\ref{sec:L7-georgescu}) supplies
the specific algebra to which the molecular Hamiltonian is
affiliated.
None of the three, individually or combined, currently suffices:
the orbifold/stratified singularities of \(\Ce(G)\) are not addressed,
and at conical intersections the single-band PST input fails.
The strict-deformation-quantisation framework must then be refined by a
twisted, multistate, or resolved construction rather than by an ordinary
single-valued scalar field.

\begin{mathbox}[What C2 involves, at a glance]
\label{mbox:sdq-tower}
\textbf{(i)} Constructing $\{A_\varepsilon^G\}$ for every
$\Lk_6(P)$-object over a molecular species $G$ \emph{is}
constructing $\Felseven$ on objects.
The morphism part of $\Felseven$ --- DPO reaction mechanisms
lifted to $C^*$-algebra maps --- is part of the open programme
and is not addressed here (\S\ref{sec:L7-def},
Remark~\ref{rem:F7-morphisms}).

\medskip\noindent
\textbf{(ii)} The forgetful functor
$U_7 : \Lk_7(P) \to \Lk_6(P)$ discards the continuous field and
returns the underlying $\Lk_6$ data; the classical phase-space
fibre $A_0^G = C_0(T^*\Ce_{\rm reg}(G))$ is recovered separately
as $\ForZero\circ\Felseven$ (\S\ref{sec:L7-def},
Remark~\ref{rem:ev0}).
The Born--Oppenheimer approximation is the assertion that, for
small physical $\varepsilon = \varepsilon_G$ and under appropriate
gap and regularity hypotheses, the full quantum dynamics is well
approximated by an effective nuclear dynamics whose leading term
is the $\varepsilon\to 0$ scalar classical/BO limit.
Theorem~\ref{thm:pst} below is the Hilbert-space statement of
that assertion; the $C^*$-algebraic statement is
Conjecture~\ref{conj:u7-functor}.

\medskip\noindent
\textbf{(iii)} The $\varepsilon$-expansion of the effective
Hamiltonian (Mathbox~\ref{mbox:heff-tower}) stratifies by tower
level: $E_j(\mathbf{R})$ from the $\Lk_5$ scalar geometric data;
the nuclear kinetic operator from $\Lk_7$; Berry/geometric
corrections from the chosen $\Lk_6$ electronic lift at the
corresponding orders of $\varepsilon$.  The semiclassical
parameter $\varepsilon$ is not an index of tower depth; different
tower data enter different orders of the expansion.
\end{mathbox}

\begin{chembox}[What strict deformation quantisation would deliver]
\label{cbox:sdq-chem}
  Most standard electronic-structure calculations used in
  chemistry are clamped-nuclei calculations that provide data
  for Born--Oppenheimer or effective nuclear models; this is
  the $\varepsilon = 0$ fibre $A_0^G = C_0(T^*\Ce_{\rm reg}(G))$.
  The parameter $\varepsilon = (m_e/M)^{1/2}$ gives the natural
  small parameter of the nuclear semiclassical expansion
  (with $\varepsilon_C \approx 0.006$, $\varepsilon_H \approx 0.023$),
  but it does not by itself fix a universal percentage error:
  the numerical size of BO violations is property- and
  system-dependent, and the approximation can fail qualitatively
  near small gaps, avoided crossings, or non-adiabatic regions.

  \medskip\noindent
  \textbf{Single-band adiabatic regions vs.\ CI seams.}
  Where a chosen electronic subspace is isolated by a uniform
  gap, the single-band BO/PST construction is expected to apply
  and Construction~C2 should be approachable by assembling
  \S\S\ref{sec:sdq-framework}--\ref{sec:pst}.  At conical
  intersections the single-eigenline gap condition fails,
  forcing a multistate or resolved description; the proposed
  resolution is Conjecture~\ref{conj:ci-blowup}.

  \medskip\noindent
  \textbf{What C2 would validate.}
  The nuclear Schr\"{o}dinger equation, the tunnelling asymptotic
  (Theorem~\ref{thm:tunnelling}), and the Berry/geometric
  corrections currently rest on the BO approximation as a
  physical assumption; each would become a theorem about the
  leading-order structure of a concrete continuous field.
\end{chembox}

\subsubsection{The SDQ framework: what the object-level lift
  \texorpdfstring{$\Felseven(G)$}{F7(G)} must be}
\label{sec:sdq-framework}

Specifying what $\Felseven(G) = \{A_\varepsilon^G\}$ is requires first
specifying what makes a family of $C^*$-algebras a legitimate
quantisation of a classical system.
This is strict deformation quantisation, introduced by
Rieffel~\cite{Rieffel1989} and developed into a framework for
Lie-groupoid quantisation by
Landsman~\cite{Landsman1998,LandsmanRamazan2001}.

\begin{definition}[Strict deformation quantisation]
\label{def:sdq}
  Let $(S,\{\cdot,\cdot\})$ be a Poisson manifold.
  A \emph{strict deformation quantisation} of $C_0(S)$ is a
  continuous field of $C^*$-algebras $\{A_t\}_{t\in[0,1]}$ with
  $A_0 = C_0(S)$, a dense Poisson subalgebra $\tilde{A}_0
  \subseteq C_0(S)$, and quantisation maps $Q_t:\tilde{A}_0
  \to A_t$ satisfying, for all $f, g\in\tilde{A}_0$:
  \begin{enumerate}[label=(\roman*)]
    \item \textbf{Reality}: $Q_t(f^*) = Q_t(f)^*$;
    \item \textbf{von Neumann}:
      $\lim_{t\to 0}\|Q_t(f)Q_t(g) - Q_t(fg)\|_{A_t} = 0$;
    \item \textbf{Dirac}:
      $\lim_{t\to 0}\bigl\|\tfrac{i}{t}[Q_t(f),Q_t(g)]
      - Q_t(\{f,g\})\bigr\|_{A_t} = 0$.
  \end{enumerate}
  The field is \emph{strict} to distinguish it from formal
  deformations (Moyal, WKB) which hold only as power series, not
  in $C^*$-norm.
\end{definition}

\noindent
In the $\Lk_7$ context $t = \varepsilon$, $S = T^*\Ce_{\rm reg}(G)$
with the canonical symplectic Poisson bracket, and
$A_0 = C_0(T^*\Ce_{\rm reg}(G))$.
The Dirac condition says the commutator
$(i/\varepsilon)[Q_\varepsilon(f), Q_\varepsilon(g)]$ converges in
$C^*$-norm to $Q_0(\{f,g\}_{T^*\Ce_{\rm reg}(G)})$ as
$\varepsilon\to 0$:
the quantum algebra of $\Lk_7$ reduces to the Poisson algebra of
$\Lk_5$ in the heavy-nucleus limit.
The Poisson manifold is $\Lk_5$ data; the SDQ is $\Lk_7$ data;
the axioms tie them together.

\begin{mathbox}[Layer~1 and Layer~2 for the object-level lift
  $\Felseven(G)$]
\label{mbox:sdq-layers}
  \textbf{Layer~1} of $\Felseven(G)$ is any continuous field
  $\{A_\varepsilon^G\}$ satisfying Definition~\ref{def:sdq} with
  $t = \varepsilon$ and $S = T^*\Ce_{\rm reg}(G)$.
  This is the minimal requirement: many such fields exist, most
  of no physical relevance.

  \medskip\noindent
  \textbf{Layer~2} selects the physical one.
  \begin{enumerate}[label=(\alph*)]
    \item \emph{Molecular origin}:
      the molecular Hamiltonian
      $H^\varepsilon = -(\varepsilon^2/2)\Delta_\mathbf{R}
      + H_\mathrm{el}(\mathbf{R})$
      is affiliated to $A_\varepsilon^G$, with
      $A_\varepsilon^G$ supplied by the Georgescu--Iftimovici
      algebra after $G^*$-equivariant restriction
      (\S\ref{sec:L7-georgescu}).
    \item \emph{BO consistency}:
      the principal classical symbol, equivalently the
      $\varepsilon\to 0$ limit of the affiliated Hamiltonian in
      the appropriate functional calculus, equals
      $p^2/2 + V(\mathbf{R})$ with $V$ the $\Lk_5$ BO surface.
  \end{enumerate}
  Layer~2(b) is the commutativity of $\ForZero\circ\Felseven$ with
  the $\Lk_5$ data, as developed in \S\ref{sec:L7-def}.
\end{mathbox}

\noindent
The Layer~1 part of the object-level lift $\Felseven(G)$ is
delivered by Landsman's tangent-groupoid construction when the
configuration space is smooth.

\begin{theorem}[Strict DQ for smooth configuration spaces;
  Landsman~\cite{LandsmanRamazan2001}]
\label{thm:landsman-sdq}
  For any smooth Riemannian manifold $Q$, there exists a strict
  deformation quantisation of $C_0(T^*Q)$ with $A_0 = C_0(T^*Q)$
  and $A_t \cong \mathcal{K}(L^2(Q))$ for every $t > 0$,
  constructed via Connes' tangent groupoid of $Q$.
\end{theorem}

\begin{proof}[Construction sketch]
  The tangent groupoid $\mathcal{T}^{(1)}Q$ interpolates between
  the pair groupoid $Q\times Q$ at $t > 0$ and the tangent bundle
  $TQ$ at $t = 0$.
  The associated $C^*$-algebras are $\mathcal{K}(L^2(Q))$ and
  $C_0(T^*Q)$ respectively (the latter by fibrewise Fourier
  transform on $TQ$).
  The full $C^*$-algebra $C^*(\mathcal{T}^{(1)}Q)$ assembles them
  into a continuous field;
  Landsman~\cite{LandsmanRamazan2001} verifies the Dirac and
  von Neumann conditions via Weyl quantisation maps.
\end{proof}

\begin{remark}[What the theorem supplies, and what it does not]
\label{rem:landsman-for-tower}
\label{rem:same-fibre}
  Theorem~\ref{thm:landsman-sdq} gives Layer~1 of $\Felseven(G)$
  when $\Ce(G)$ is smooth, with the same abstract fibre
  $A_\varepsilon^G \cong \mathcal{K}(L^2(\Ce_{\rm reg}(G)))$ for
  every $\varepsilon > 0$ --- the deformation changes the
  continuous structure between fibres, not the fibres themselves.
  The tangent groupoid knows nothing of $H^\varepsilon$, the
  Berry connection, or the BO surface, so it does not deliver
  Layer~2: that requires affiliation to the molecular Hamiltonian
  (\S\ref{sec:L7-georgescu}) and the PST identification of
  $H^\varepsilon_\mathrm{eff}$ (\S\ref{sec:pst}).

  The fibre $\mathcal{K}(L^2(\Ce_{\rm reg}(G)))$ is the nuclear
  algebra in the scalar sector-reduced description, where the
  full molecular Hilbert space is
  $L^2(\Ce_{\rm reg}(G))\otimes\Hel$ and band projection has
  reduced the problem to the nuclear factor; when nuclear
  spin-statistical sectors are tracked one starts from the
  labelled configuration space with the appropriate
  equivariance.
\end{remark}

\begin{remark}[The orbifold singularities of $\Ce(G)$]
\label{rem:orbifold-gap}
  Theorem~\ref{thm:landsman-sdq} requires $Q$ smooth.
  The configuration orbifold $\Ce(G) = \RR^{3n}/
  (\mathrm{SE}(3) \rtimes \Aut_\mu(G))$ has orbifold singularities
  from isotropy of the mass/label-preserving automorphism action
  together with the stereochemical/permutation-inversion
  symmetries encoded by the $\Lk_{4.5}$ group $G^*$: the
  equilateral triangle of $\mathrm{H_3^+}$, linear triatomic
  geometries, and any configuration with non-trivial isotropy.
  Configurations with rank-deficient Euclidean orbits
  (e.g.\ linear geometries) should be treated as part of the
  stratified quotient rather than as ordinary finite-isotropy
  orbifold points.  Such points are chemically significant ---
  symmetric transition states and symmetry-required
  degeneracies --- and the orbifold SDQ problem is open even at
  Layer~1: Pflaum~\cite{Pflaum2001} supplies tools for smooth
  strata but does not address Coulomb singularities at coincident
  nuclei.  This is a distinct gap from the CI obstruction of
  \S\ref{sec:ci-gap}.
\end{remark}

\subsubsection{PST: connecting the field to \texorpdfstring{$\Lk_6$}{L6} data}
\label{sec:pst}

Theorem~\ref{thm:landsman-sdq} gives the Layer~1 continuous-field
template for $\Felseven(G)$ when the relevant configuration space
is smooth, but leaves its connection to the $\Lk_6$ data
$(\sigma_j, A, \eta_B)$ unspecified.
That connection is the substance of space-adiabatic perturbation
theory (PST)~\cite{PanatiSpohnTeufel2003,Teufel2003}: it shows
how the Berry connection and related geometric terms from the
chosen $\Lk_6$ electronic lift enter the effective nuclear
Hamiltonian in the semiclassical expansion, as Layer~2(b)
requires, and provides the Hilbert-space error estimate that
underlies the $C^*$-algebraic Born--Oppenheimer statement
(Conjecture~\ref{conj:u7-functor}).

At the clamped-nuclei electronic level, isotopologues such as
$\mathrm{H}_2$ and $\mathrm{D}_2$ share the same electronic
surface and active electronic data
$(\sigma_j, A, \eta_B)$.  Their leading dynamical differences
enter through the nuclear masses in the effective Hamiltonian,
with the underlying compact-operator fibre abstractly isomorphic
(by Remark~\ref{rem:landsman-for-tower}) but carrying different
Hamiltonian elements.  In addition, their nuclear
spin-statistical sectors differ at $\Lk_7$: protons are
fermions and deuterons are bosons, so the total wavefunction is
antisymmetric under proton exchange and symmetric under deuteron
exchange, with correspondingly different spin-rotational sector
decompositions (see \S\ref{sec:nuc-spin}).  Nuclear-mass data already belong to
$\Lk_5$ (Mathbox~\ref{mbox:heff-tower}).

For clarity, we state the PST result in the simplest single
isolated non-degenerate eigenvalue case; the band-cluster version
replaces the eigenvector by the spectral projector and the
Abelian Berry connection by its non-Abelian counterpart on the
active bundle.

\begin{theorem}[Space-adiabatic BO expansion;
  Panati--Spohn--Teufel~\cite{PanatiSpohnTeufel2003,Teufel2003}]
\label{thm:pst}
  Let $H^\varepsilon = -(\varepsilon^2/2)\Delta_\mathbf{R}
  + H_\mathrm{el}(\mathbf{R})$ act on
  $L^2(\Ce_{\rm reg}(G))\otimes\Hel$ with
  $\varepsilon = (m_e/M)^{1/2}$.
  Assume the \emph{gap condition}: there exists $\delta_0 > 0$
  such that
  \[
    \mathrm{dist}\!\bigl(E_j(\mathbf{R}),\;
    \sigma(H_\mathrm{el}(\mathbf{R}))\setminus\{E_j(\mathbf{R})\}
    \bigr)\;\geq\;\delta_0
    \quad\forall\,\mathbf{R}\in\Ce_{\rm reg}(G).
  \]
  This is a deliberately strong global hypothesis stated for
  clarity; in applications one usually restricts to an adiabatic
  region where the chosen band or band cluster is isolated.
  Then:
  \begin{enumerate}[label=(\roman*)]
    \item There exists an almost-invariant projection
      $P^\varepsilon$ such that, in the standard space-adiabatic
      sense (with the usual domain / energy-cutoff
      qualifications for the unbounded $H^\varepsilon$),
      $\|[H^\varepsilon, P^\varepsilon]\| = O(\varepsilon^\infty)$.
      \item There exists a unitary $U_\varepsilon$ such that the
      effective Hamiltonian
      $H^\varepsilon_\mathrm{eff}
      = U_\varepsilon H^\varepsilon U_\varepsilon^*
      \!\restriction_{P^\varepsilon\mathcal{H}}$
      has the schematic local-adiabatic form
      \[
        H^\varepsilon_\mathrm{eff}
        = \tfrac{1}{2}\bigl(p - \varepsilon A(\mathbf{R})\bigr)^2
        + E_j(\mathbf{R})
        + \varepsilon^2\,\Phi_\mathrm{BH}(\mathbf{R})
        + \cdots,
      \]
      with $A^\mu(\mathbf{R}) = i\bigl\langle\sigma_j(\mathbf{R})
      \big|\partial_\mu\big|\sigma_j(\mathbf{R})\bigr\rangle$ the
      Berry connection from $\Lk_6$ in a local gauge, and
      $\Phi_\mathrm{BH}$ a Born--Huang scalar correction; the
      precise form of higher
      terms depends on the chosen gauge and semiclassical
      convention.
    \item For every energy cutoff $E$ there is $C_E > 0$ such that
      for all $t\geq 0$,
      \[
        \bigl\|\bigl[e^{-iH^\varepsilon t/\varepsilon}
        - e^{-iP^\varepsilon H^\varepsilon P^\varepsilon
          t/\varepsilon}\bigr]P^\varepsilon P_E\bigr\|
        \leq C_E\,\varepsilon(1 + |t|).
      \]
  \end{enumerate}
  For analytic $V$ and $H_\mathrm{el}$, optimal truncation
  sharpens the error to
  $\exp(-\gamma/\varepsilon)$~\cite{HagedornJoye2001}.
\end{theorem}

\noindent
Part~(ii) is the Hilbert-space statement of Layer~2(b):
$H^\varepsilon_\mathrm{eff}\to p^2/2 + E_j(\mathbf{R})$ as
$\varepsilon\to 0$, recovering the $\Lk_5$ BO Hamiltonian.
The Berry/geometric correction terms from the chosen $\Lk_6$
electronic lift enter the effective Hamiltonian in the powers of
$\varepsilon$ dictated by the chosen scaling.  Thus the $\Lk_7$
object-level lift records not only the scalar BO surface but
also the electronic-bundle geometry inherited from $\Lk_6$.
Part~(iii) is the Hilbert-space shadow of the
$C^*$-algebraic Conjecture~\ref{conj:u7-functor}: the full
quantum evolution converges to the effective BO evolution on
energy-localised states with error $O(\varepsilon)$; the uniform
$C^*$-norm upgrade is open.

\begin{remark}[Gap condition as $\Lk_6$ Layer~2 regularity]
\label{rem:gap-L6}
  The gap condition $\delta_0 > 0$ is a regularity condition on
  the chosen $\Lk_6$ electronic lift over the adiabatic region:
  the selected spectral subspace is isolated from the rest of
  the electronic spectrum.  It is determined by
  $H_\mathrm{el}(\mathbf{R})$ and does not require a global
  eigenvector section --- the relevant object is the spectral
  projector or eigenbundle.
  Theorems~\ref{thm:landsman-sdq} and~\ref{thm:pst} supply the
  basic ingredients for the object-level lift in the regular
  single-band situation; the remaining difficulties are
  operator-algebraic assembly, orbifold/stratified singularities,
  exchange sectors, and Coulomb affiliation.
\end{remark}

\begin{mathbox}[Berry topology and the continuous field]
\label{mbox:berry-cstar}
  The $\Lk_6$ Berry-sign class
  $\eta_B = w_1(L_0^{\RR}) \in H^1(\Omega^\circ; \ZZ/2)$
  (with $\Omega^\circ = \Ce_{\rm reg}(G) \setminus \Xseam$) from
  \S\ref{sec:L6-berry} does not generally obstruct the existence
  of an $\Lk_7$ continuous field.  Rather, it obstructs
  representing the nuclear problem as an ordinary single-valued
  scalar quantisation on $\Omega^\circ$.

  \medskip\noindent
  \textbf{Case $\eta_B = 0$.}
  The real eigenline $L_0^{\RR}$ has trivial $w_1$ on the
  region, or at least on the loops explored by the dynamics;
  the real-line-bundle twist does not obstruct a single-valued
  scalar realisation.  Other geometric
  corrections --- Born--Huang scalar terms,
  and complex Berry curvature in non-real settings --- may still appear.
  Construction~C2 is then expected to be approachable by assembling Theorems~\ref{thm:landsman-sdq} and~\ref{thm:pst}.

  \medskip\noindent
  \textbf{Case $\eta_B \neq 0$.}
  The correct nuclear state is a section of $L_0^{\RR}$ or of the
  associated local system --- equivalently, an antiperiodic
  boundary-condition problem in the Mead--Truhlar
  geometric-phase formulation.  Representing this twist inside
  the $C^*$-algebraic continuous field is Construction~C3
  (Conjecture~\ref{conj:c1-obstruction}); the obstruction is not
  to the existence of $\{A_\varepsilon^G\}$ as a continuous field
  but to its realisation as a scalar-quantisation field over
  $\Omega^\circ$.  The seam $\Xseam$ where the underlying
  single-band gap fails is treated separately in
  \S\ref{sec:ci-gap}.
\end{mathbox}

\subsubsection{Where the single-band \texorpdfstring{$\Felseven$}{Fel7} fails: the conical intersection seam}
\label{sec:ci-gap}

When $\Xseam \neq \emptyset$, the single-band PST input fails
near the seam.  The Landsman construction may still apply on the
underlying smooth configuration stratum, but does not by itself
encode the electronic band crossing or the required
multistate/twisted structure; the single-band construction of
$\Felseven$ breaks down there.

\begin{proposition}[Single-band PST breakdown at the CI seam;
  Lasser--Teufel~\cite{LasserTeufel2005},
  Fermanian Kammerer--Lasser~\cite{FermanianKammererLasser2008}]
\label{prop:ci-breakdown}
  At any conical intersection $\mathbf{R}_0 \in \Xseam$:
  \begin{enumerate}[label=(\roman*)]
    \item The single-band gap condition fails:
      $\delta(\mathbf{R}_0) = 0$.
    \item The single-band Theorem~\ref{thm:pst} does not apply:
      a single-band almost-invariant projection associated to
      either crossing eigenvalue $E_0$ or $E_1$ alone cannot be
      constructed on any neighbourhood of $\mathbf{R}_0$.  A
      rank-$2$ active-space projection may still be meaningful
      when the two-state cluster remains separated from the rest
      of the spectrum.
    \item Under suitable genericity hypotheses, the best
      available estimates near $\Xseam$ are at the Wigner-measure
      / surface-hopping level: at leading order, the Wigner
      function $W^\varepsilon(t)$ is approximated by the
      Lasser--Teufel surface-hopping
      semigroup~$\mathcal{L}_t$~\cite{LasserTeufel2005}, with
      Fermanian Kammerer--Lasser~\cite{FermanianKammererLasser2008}
      giving schematic quantitative bounds of the form
      $\|W^\varepsilon(t) - \mathcal{L}_t W^\varepsilon(0)\|
      = O(\varepsilon^\alpha)$ in the appropriate
      Wigner-measure / weak norm, with $\alpha > 0$ depending on
      the precise norm and crossing hypotheses, and
      $\mathcal{L}_t$ implementing classical transport along
      surfaces interleaved with Landau--Zener-type transitions
      in the local crossing/branching region.
  \end{enumerate}
\end{proposition}

\begin{remark}[$\Xseam$ as the failure locus of single-band $\Felseven$]
\label{rem:xseam-tower}
  The CI seam is $\Lk_6$ data --- the zero locus of the spectral
  gap function, determined by $H_\mathrm{el}(\mathbf{R})$ --- so
  the failure of the single-band $\Felseven$ at $\Xseam$ is an
  $\Lk_6$ failure imported into $\Lk_7$, not a new $\Lk_7$
  phenomenon.  The single-band object-level lift is expected to
  be constructible on adiabatic regions of
  $\Ce_{\rm reg}(G) \setminus \Xseam$ where the chosen eigenvalue
  remains uniformly isolated and the other regularity hypotheses
  hold; channels whose support enters the branching region ---
  photochemical and non-adiabatic processes --- require
  multistate or resolved constructions.
  The proposed blowup $\widetilde{\Ce}(G) \to \Ce(G)$
  (Conjecture~\ref{conj:ci-blowup}) is one way to resolve the
  local singular geometry and let the active
  eigenline/eigenbundle extend over a resolved space, but does
  not remove the Berry-sign topology, which reappears as
  boundary monodromy or twisted-sector data.
\end{remark}

\subsubsection{Construction~C2 as an open problem}
\label{sec:c2-open}

Collecting the preceding into a single statement:

\begin{openprob}{{Construction~C2: the object-level lift
  $\Felseven$}}
\label{op:sdq-molecules}
  For each $\Lk_6(P)$-object over a molecular species $G$, with
  compact data denoted $(\Ce(G), \sigma_j, A, \eta_B)$ --- where
  $\sigma_j$ is understood as a local adiabatic frame in the
  single-band case, or replaced by the spectral projector for a
  band cluster --- construct a continuous field
  $\{A_\varepsilon^G\}_{\varepsilon\in[0,1]}$ satisfying:
  \begin{enumerate}[label=(\roman*)]
    \item \emph{(Layer~1)} $A_0^G = C_0(T^*\Ce_{\rm reg}(G))$ and the
      maps $Q_\varepsilon:\tilde{A}_0\to A_\varepsilon^G$
      satisfy Definition~\ref{def:sdq} with $t = \varepsilon$.
    \item \emph{(Layer~2a, molecular origin)} The molecular
      Hamiltonian $H^\varepsilon = -(\varepsilon^2/2)
      \Delta_\mathbf{R} + H_\mathrm{el}(\mathbf{R})$ is
      affiliated to $A_\varepsilon^G$, via the
      $G^*$-equivariant Georgescu algebra of
      \S\ref{sec:L7-georgescu}.
    \item \emph{(Layer~2b, BO consistency)} The principal
      classical symbol, equivalently the $\varepsilon\to 0$
      limit of the affiliated Hamiltonian in the appropriate
      functional calculus, equals $p^2/2 + V(\mathbf{R})$ with
      $V$ the $\Lk_5$ BO surface.
    \item \emph{($\Lk_6$ content)} The Berry/geometric data of
      the chosen $\Lk_6$ electronic lift are represented in the
      effective Hamiltonian and/or in the twisted local-system
      sector, consistent with Theorem~\ref{thm:pst}.
  \end{enumerate}
\end{openprob}

\begin{mathbox}[Status of C2]
\label{mbox:c2-status}
  \textit{Uniformly gapped single-band region with trivial
  relevant Berry-sign twist.}
  Theorem~\ref{thm:landsman-sdq} supplies Layer~1 where $\Ce(G)$
  is smooth; Theorem~\ref{thm:pst} supplies the
  $\varepsilon$-expansion realising Layer~2(b); the Georgescu
  algebra (\S\ref{sec:L7-georgescu}) supplies the affiliation
  framework for Layer~2(a), pending the $G^*$-equivariant
  restriction.  No conceptual obstruction is apparent, but the
  operator-algebraic assembly has not been carried out and the
  orbifold singularities of $\Ce(G)$
  (Remark~\ref{rem:orbifold-gap}) remain a technical gap.

  \medskip\noindent
  \textit{Non-trivial Berry-sign twist or CI seam.}
  A non-trivial $\eta_B$ requires a twisted/local-system
  realisation (Mathbox~\ref{mbox:berry-cstar}); single-band PST
  can still work locally on the line bundle away from the seam.
  A CI seam additionally causes the single-band gap condition to
  fail at the seam itself
  (Proposition~\ref{prop:ci-breakdown}), requiring multistate or
  resolved constructions; the proposed blowup
  (Conjecture~\ref{conj:ci-blowup}) remains open.

  \medskip\noindent
  The basic C2 framework is prerequisite for the later
  conjectures: C3 refines it to represent the Berry-sign twist
  (Conjecture~\ref{conj:c1-obstruction}), and the BO
  approximation (Conjecture~\ref{conj:u7-functor}) and emergence
  of molecular identity (Conjecture~\ref{conj:superselection})
  build on the same framework
  (\S\ref{sec:L7-contributions}).
\end{mathbox}

%% file: chapters/L7/l7_georgescu.tex
\subsection{The Georgescu \texorpdfstring{$N$}{N}-body algebra: ingredients for Layer~2(a)}
\label{sec:L7-georgescu}

Construction~C2 is the pair (Layer~1 + Layer~2).
Section~\ref{sec:sdq-framework} supplied Layer~1 via the Landsman
tangent-groupoid construction: an abstract continuous field with
the correct fibre algebra but no connection to molecular physics.
This section develops the ingredients for Layer~2(a) --- the
identification of $A_\varepsilon^G$ as a specific algebra to which
the nuclear-effective Hamiltonian $H^\varepsilon$ is affiliated,
rather than only an abstract $C^*$-algebra satisfying the SDQ
axioms.
The tool is the Georgescu--Iftimovici graded crossed-product
algebra $\mathfrak{A}(X)$~\cite{GeorgescuIftimovici2002,
GeorgescuIftimovici2003}, a $C^*$-algebra on $L^2(X)$ providing a
$C^*$-algebraic framework in which a large class of $N$-body
Hamiltonians with admissible decaying pair interactions --- including
Coulomb-type interactions under the Georgescu--Iftimovici affiliation
hypotheses --- have affiliated resolvents.
The algebra's graded ideal structure encodes a particle-cluster
partition lattice, reproducing the HVZ theorem as a purely algebraic
fact.

\begin{warning}[Layer~2(a) ingredients vs.\ Layer~2(a) completed]
\label{warn:layer-2a-status}
  Throughout this section, $X$ denotes a \emph{nuclear} configuration
  vector space and $\mathfrak{A}(X)$ the corresponding $N$-body
  algebra on $L^2(X)$, before any Euclidean or exchange quotient.
  The Layer~2(a) condition of Open
  Problem~\ref{op:sdq-molecules} is the identification of
  $A_\varepsilon^G$ on the quotient/sector Hilbert space associated
  with $\Ce(G)$; connecting the two requires a $G^*$-equivariant
  restriction and descent that has not been carried out for
  molecular systems (\S\ref{sec:georgescu-limits}).
  The full pre-Born--Oppenheimer electron--nuclear Hamiltonian, by
  contrast, lives on $L^2(\RR^{3(N_{\rm nuc}+N_{\rm el})})$ (modulo
  centre-of-mass and exchange) and would require applying the
  $N$-body machinery to the full particle configuration space ---
  outside the scope of this section.
  What \emph{is} established on $L^2(X)$ is developed below:
  $\mathfrak{A}(X)$ as an ambient algebra for nuclear effective
  $N$-body Hamiltonians, the affiliation of $H^\varepsilon$, and the
  HVZ theorem as an algebraic consequence.
\end{warning}

\begin{mathbox}[The Georgescu algebra in the tower]
\label{mbox:georgescu-tower}
  \textbf{(i) What $\mathfrak{A}(X)$ provides.}
  A specific $C^*$-algebra on $L^2(X)$ to which the nuclear-effective
  Hamiltonian $H^\varepsilon = -(\varepsilon^2/2)\Delta + V(\mathbf{R})$
  is affiliated, where $V$ is the BO eigenvalue selected at $\Lk_6$
  (Remark~\ref{rem:affiliation}).
  The SDQ axioms of Definition~\ref{def:sdq} specify the abstract
  structure of $F_7(G)$; $\mathfrak{A}(X)$ names a concrete algebra
  to which the correct Hamiltonian is affiliated.

  \medskip\noindent
  \textbf{(ii) The graded structure tracks particle-cluster
    asymptotics.}
  $\mathfrak{R}(X)$ is graded by a semilattice $\mathcal{S}$ of
  linear subspaces of $X$ describing the relative-coordinate
  quotients of the system --- equivalently, by particle-cluster
  partitions.
  This lattice is \emph{related to, but not identical with}, the
  $\Lk_0$ stoichiometric bookkeeping: $\Lk_0$ records species and
  reaction balances, whereas $\mathfrak{R}(X)$ records asymptotic
  cluster decompositions of a fixed $N$-body Hamiltonian.
  Chemical dissociation channels emerge by interpreting certain
  cluster partitions as molecular fragments.
  The HVZ theorem (Theorem~\ref{thm:hvz}) computes the essential
  spectrum from this lattice.

  \medskip\noindent
  \textbf{(iii) The compact ideal is the localised spectral regime.}
  The compact operators $\mathcal{K}(L^2(X))\subseteq\mathfrak{A}(X)$
  form the minimal element of the lattice
  (Proposition~\ref{prop:compact-ideal}): the algebraic location of
  the bound-state spectrum.
  Inside $\mathcal{K}(L^2(X))$, $\mathfrak{A}(X)$ does not separate
  an $\mathrm{H_2O}$ ground state from a
  $\mathrm{H_2}\!\cdots\!\mathrm{O}$ van der Waals complex from an
  electronically excited bound state.
  Since $\mathcal{K}(\mathcal{H})$ is simple as a $C^*$-algebra, no
  bare ideal-theoretic decomposition exists; distinguishing
  graph-labelled molecular sectors within this localised regime
  requires additional observable, symmetry, or representation
  structure --- the content of Construction~C4
  (\S\ref{sec:L7-superselection}).

  \medskip\noindent
  \textbf{(iv) What remains.}
  $\mathfrak{A}(X)$ lives on $L^2(X)$; the tower requires
  $A_\varepsilon^G$ on the quotient/sector Hilbert space associated
  with $\Ce(G)$.
  The $G^*$-equivariant restriction and descent connecting the two
  (\S\ref{sec:georgescu-limits}) uses the Ammann--Mougel--Nistor
  identification of Georgescu's compactification with Vasy's
  blowup~\cite{AmmannMougelNistor2022} as its geometric scaffold;
  the explicit molecular construction has not been written up.
\end{mathbox}

\begin{chembox}[What the Georgescu algebra means for chemistry]
\label{cbox:georgescu-chem}
  \textbf{An ambient algebra for the nuclear $N$-body problem.}
  The standard class of nuclear-effective Hamiltonians
  $H^\varepsilon = -(\varepsilon^2/2)\Delta_{\mathbf R}
  + V(\mathbf{R})$ with admissible decaying pair interactions is
  affiliated to $\mathfrak{A}(X)$ under the Georgescu--Iftimovici
  hypotheses.
  This class includes Coulomb-type pair interactions, which are
  \emph{not} in $C_0(X/W_{ij})$ (they are singular on the collision
  set) but are admissible unbounded affiliated potentials
  (Remark~\ref{rem:affiliation}).
  $\mathfrak{A}(X)$ is not a particular approximate Hamiltonian
  model; it is a canonical ambient $C^*$-algebraic framework to
  which the relevant Hamiltonians are affiliated, with graded
  structure capturing the particle-cluster partition lattice.

  \medskip\noindent
  \textbf{The dissociation ladder.}
  For water $\mathrm{H_2O}$ (labelled atoms
  $\{\mathrm{H_1, H_2, O}\}$) there are $B(3) = 5$ set partitions:
  $\{\mathrm{H_1 H_2 O}\}$,
  $\{\mathrm{H_1}\}\,|\,\{\mathrm{H_2 O}\}$,
  $\{\mathrm{H_2}\}\,|\,\{\mathrm{H_1 O}\}$,
  $\{\mathrm{O}\}\,|\,\{\mathrm{H_1, H_2}\}$, and
  $\{\mathrm{H_1}\}\,|\,\{\mathrm{H_2}\}\,|\,\{\mathrm{O}\}$.
  After identifying the two hydrogen labels, these five partitions
  reduce to four symmetry orbits: the undissociated channel and
  three dissociation channels --- $\mathrm{H}+\mathrm{OH}$,
  $\mathrm{O}+\{\mathrm{H,H}\}$, and
  $\mathrm{H}+\mathrm{H}+\mathrm{O}$.
  Whether the $\{\mathrm{H,H}\}$ cluster supports a bound
  $\mathrm{H_2}$ state is spectral information of the corresponding
  cluster Hamiltonian, not part of the partition lattice itself.

  \medskip\noindent
  \textbf{HVZ as the quantum dissociation threshold.}
  The bottom of $\sigma_\mathrm{ess}(H_{\mathrm{H_2O}})$ is the
  minimum cluster threshold across non-minimal partitions: the
  quantum-mechanical dissociation threshold of the chosen
  Schr\"{o}dinger Hamiltonian.
  After adding zero-point and electronic corrections, this
  threshold is the spectral analogue of the dissociation energies
  organised at $\Lk_1$ by Hess-law bookkeeping --- now derived as
  an algebraic consequence of the ideal structure of
  $\mathfrak{A}(X)$.
\end{chembox}

\subsubsection{The graded crossed-product construction}
\label{sec:georgescu-def}

\begin{definition}[Georgescu graded $C^*$-algebra]
\label{def:georgescu-algebra}
  Let $X = \RR^{3n}$ be the nuclear configuration vector space of
  an $n$-atom molecule before quotienting.
  On the regular labelled configurations $Q_{\rm reg}(G)\subseteq X$
  (those compatible with the $\Lk_4$ molecular graph), the scalar
  configuration orbifold of $\Lk_5$ is
  \[
    \Ce(G) \;\cong\; Q_{\rm reg}(G)\big/
      \bigl(\mathrm{SE}(3)\rtimes\Aut(G)\bigr);
  \]
  the $\Lk_7$ descent additionally invokes the $\Lk_{4.5}$
  reflection structure encoded in $G^*$.
  Let $\mathcal{S}$ be the semilattice of linear subspaces of $X$
  associated with the $N$-body system: closed under intersections,
  and containing the subspaces that define the pair interactions and
  cluster decompositions.
  For each $Y\in\mathcal{S}$, let $C_0(X/Y)$ denote the continuous
  functions on $X$ that are $Y$-translation-invariant and vanish at
  infinity on the quotient $X/Y$.
  The lattice rule
  $C_0(X/Y)\cdot C_0(X/Z) \subseteq C_0(X/(Y\cap Z))$
  holds because a $Y$-invariant function multiplied by a
  $Z$-invariant function is invariant under translations in
  $Y\cap Z$, and decays at infinity on the transverse quotient
  $X/(Y\cap Z)$.

  The \emph{Georgescu graded algebra} is the norm closure
  \[
    \mathfrak{R}(X)
    \;:=\;
    \overline{\sum_{Y\in\mathcal{S}} C_0(X/Y)\,}^{\|\cdot\|},
  \]
  and the \emph{quantum $N$-body algebra} is the crossed product
  by the translation action of $X$ on $L^2(X)$:
  \[
    \mathfrak{A}(X) \;:=\; \mathfrak{R}(X)\rtimes X.
  \]
  In the standard representation on $L^2(X)$, it is generated in
  norm closure by products of position multipliers
  $\varphi(Q)$ with $\varphi\in\mathfrak{R}(X)$ and
  translation/momentum operators $e^{-ia\cdot P}$, where $Q$ is the
  position operator and $P = -i\nabla$~\cite{GeorgescuIftimovici2002}.
\end{definition}

\begin{remark}[Affiliation: what is established and what is not]
\label{rem:affiliation}
  An unbounded self-adjoint operator $H$ is \emph{affiliated} to a
  $C^*$-algebra $\mathcal{A}$ when its resolvent
  $(H+i)^{-1}\in\mathcal{A}$.

  For a pair $i,j$, let
  \[
    W_{ij} \;:=\; \{u\in X : u_i - u_j = 0\},
  \]
  so that $X/W_{ij}\cong\RR^3$ records the relative coordinate
  $\mathbf{R}_i - \mathbf{R}_j$.
  The Coulomb pair potential
  $V_{ij}(\mathbf{R}) = z_iz_j/|\mathbf{R}_i - \mathbf{R}_j|$ depends
  only on this relative coordinate and decays as
  $|\mathbf{R}_i - \mathbf{R}_j|\to\infty$, but is \emph{singular}
  on the collision set $\mathbf{R}_i = \mathbf{R}_j$: it is not an
  element of $C_0(X/W_{ij})$.
  It is, however, an admissible unbounded pair interaction
  associated with the quotient $X/W_{ij}$, and the corresponding
  Schr\"odinger operator
  \[
    H^\varepsilon
    \;=\; -\tfrac{\varepsilon^2}{2}\Delta + \sum_{i<j} V_{ij}
  \]
  is affiliated to $\mathfrak{A}(X)$ under the Georgescu--Iftimovici
  affiliation hypotheses~\cite{GeorgescuIftimovici2002}.

  This establishes affiliation on $L^2(X)$.
  It does not by itself construct the molecular Layer~2(a) object
  $A_\varepsilon^G$ on the quotient/sector Hilbert space associated
  with $\Ce(G)$.
  The missing step is the $G^*$-equivariant restriction,
  exchange-sector selection, and descent of
  \S\ref{sec:georgescu-limits}.
\end{remark}

\begin{proposition}[Compact operators as the localised regime;
  Georgescu--Iftimovici~\cite{GeorgescuIftimovici2002}]
\label{prop:compact-ideal}
  The compact operators form a closed two-sided ideal
  \[
    \mathcal{K}(L^2(X)) \;=\; C_0(X)\rtimes X
    \;\subseteq\; \mathfrak{A}(X),
  \]
  the minimal element of the cluster lattice: all inter-particle
  separations bounded, no dissociation.
  Bound-state spectral projections of affiliated Hamiltonians are
  represented in this compact part.
  The compact ideal itself does not decompose the bound-state
  spectrum into molecular-graph sectors --- $\mathcal{K}(\mathcal{H})$
  is simple as a $C^*$-algebra and admits no proper closed
  two-sided ideal --- so any such decomposition requires additional
  observable, symmetry, or representation structure beyond
  $\mathfrak{A}(X)$.
  This is the subject of Construction~C4.
\end{proposition}

\subsubsection{HVZ and Mourre: spectral structure from the ideal
  lattice}
\label{sec:hvz-mourre}

The two foundational spectral results of $N$-body quantum mechanics
are algebraic statements in $\mathfrak{A}(X)$.

\begin{definition}[Cluster decompositions and Hamiltonians]
\label{def:cluster-decomp}
  A \emph{cluster decomposition} $a = (C_1,\ldots,C_k)$ is a
  partition of $\{1,\ldots,n\}$ into $k\geq 1$ non-empty subsets
  (\emph{clusters}).
  The \emph{minimal decomposition}
  $a_\mathrm{min} = (\{1,\ldots,n\})$ is one cluster (the whole
  undissociated molecule).
  Schematically, after separating each cluster's centre-of-mass and
  internal coordinates, the cluster Hamiltonian is
  \[
    H_a \;=\; \sum_{j=1}^k H_{C_j}^{\mathrm{intra}}
    \;+\; T^{\mathrm{inter}},
  \]
  where $H_{C_j}^{\mathrm{intra}}$ is the intra-cluster Hamiltonian
  (kinetic energy plus pair interactions within $C_j$) and
  $T^{\mathrm{inter}}$ is the kinetic energy of the cluster
  centres of mass; all inter-cluster interactions are removed.
\end{definition}

\begin{theorem}[Algebraic HVZ;
  Georgescu--Iftimovici~\cite{GeorgescuIftimovici2002}]
\label{thm:hvz}
  Let $H$ be an $N$-body Hamiltonian affiliated to
  $\mathfrak{A}(X)$, and let $\hat H$ denote its image in the
  Calkin quotient $\mathfrak{A}(X)/\mathcal{K}(L^2(X))$, which
  quotients out compact/localised spectral information and retains
  the asymptotic channel algebra.
  Then
  \[
    \sigma_\mathrm{ess}(H) \;=\; \sigma(\hat H)
    \;=\; \overline{\bigcup_{a\neq a_\mathrm{min}}\sigma(H_a)}.
  \]
  The first equality is the general relation between essential and
  Calkin spectra; the second is specific to the graded structure of
  $\mathfrak{A}(X)$, which decomposes the Calkin quotient over
  non-minimal cluster decompositions with each component carrying
  the corresponding cluster Hamiltonian.
\end{theorem}

\begin{remark}[HVZ across the tower: $\Lk_0 \to \Lk_7$]
\label{rem:hvz-tower}
  The union in Theorem~\ref{thm:hvz} runs over a particle-cluster
  partition lattice, related to but distinct from $\Lk_0$
  stoichiometric bookkeeping: $\Lk_0$ records species and reaction
  balances, whereas $\mathfrak{A}(X)$ records asymptotic cluster
  decompositions of a fixed $N$-body Hamiltonian.
  Chemical dissociation channels are obtained by interpreting
  certain cluster partitions as molecular fragments.
  Algebraic HVZ is therefore a $C^*$-algebraic computation of a
  quantum dissociation threshold whose thermochemical analogue is
  the Hess-law $\Lk_1$ dissociation energy, mediated by the
  appropriate zero-point and electronic corrections: the cluster
  lattice, invisible in the electronic structure at $\Lk_6$,
  reappears at $\Lk_7$ as the ideal lattice of $\mathfrak{A}(X)$.
\end{remark}

\begin{remark}[Mourre estimate: spectral regularity]
\label{rem:mourre}
  The Mourre estimate~\cite{ABG1996,GeorgescuGerardMoller2004} with
  the dilation generator
  \[
    A_\mathrm{dil} \;=\; \tfrac{1}{2i}\bigl(\mathbf{R}\cdot\nabla
    + \nabla\cdot\mathbf{R}\bigr)
  \]
  as conjugate operator gives $i[H, A_\mathrm{dil}] > 0$ modulo
  compacts on each energy interval away from eigenvalues and
  thresholds.
  Self-adjointness alone supplies the unitary time evolution
  $e^{-iH^\varepsilon t}$ by Stone's theorem; the Mourre estimate
  provides the additional spectral and propagation regularity used
  elsewhere in this chapter --- absence of singular continuous
  spectrum, limiting absorption principle, and propagation
  estimates --- which enter the scattering analysis required by
  Theorem~\ref{thm:pst} and the affiliated-bound-state framework
  of Proposition~\ref{prop:compact-ideal}.
\end{remark}

\subsubsection{Channels vs species: the gap to Construction~C4}
\label{sec:georgescu-limits}

\begin{mathbox}[{What $\mathfrak{A}(X)$ encodes, and what remains
  for C4}]
\label{mbox:georgescu-limits}
  \textbf{What is encoded.}
  The graded structure of $\mathfrak{R}(X)$ tracks asymptotic
  cluster behaviour: which atoms separate as
  $|\mathbf{R}|\to\infty$, and at which thresholds.
  HVZ and Mourre follow from this structure, and every affiliated
  $N$-body Hamiltonian inherits its spectral framework.

  \medskip\noindent
  \textbf{What is not encoded.}
  Within $\mathcal{K}(L^2(X))$, $\mathfrak{A}(X)$ does not
  distinguish which bound configuration is occupied: bound ground
  and excited states, and weakly bound van der Waals states, are all
  represented in the compact/localised part.
  Resonances --- non-$L^2$ poles of the resolvent --- require
  separate scattering/resonance theory and are not located in
  $\mathcal{K}(L^2(X))$ at all.
  In tower terms: $\mathfrak{A}(X)$ encodes the asymptotic cluster
  channels but not the molecular identity (the $\Lk_4$ graph $G$)
  that must emerge algebraically at $\Lk_7$.

  \medskip\noindent
  \textbf{Two steps to close the gap.}
  \begin{enumerate}[label=(\roman*)]
    \item \emph{$G^*$-equivariant restriction and descent}
      ($\Lk_{4.5}$ data).
      Two related operations are required: restricting to
      $\mathfrak{A}(X)^{G^*}$ (the subalgebra commuting with the
      permutation-inversion group $G^*$), and descending to an
      algebra on the quotient/sector Hilbert space associated with
      $\Ce(G)$, compatible with the Landsman fibre
      $\mathcal{K}(L^2(\Ce(G)))$ of
      Theorem~\ref{thm:landsman-sdq}.
      Ammann--Mougel--Nistor~\cite{AmmannMougelNistor2022} identify
      Georgescu's $C^*$-algebraic compactification of $X$ with
      Vasy's blowup compactification; this is the geometric
      scaffold on which the $G^*$-action can be implemented
      equivariantly.
      The explicit molecular construction has not been carried out.
    \item \emph{Molecular species identification}
      (Construction~C4).
      Equip the descended algebra with an observable subalgebra,
      symmetry, or representation structure whose sectors are
      labelled by graph-like molecular structures.
      The compact ideal supplies the localised spectral regime, but
      the graph labels require additional superselection/observable
      data, since $\mathcal{K}(\mathcal{H})$ is simple as a
      $C^*$-algebra and admits no bare ideal decomposition.
      This is the content of
      Conjecture~\ref{conj:superselection}(a) and the subject of
      \S\ref{sec:L7-superselection}; it is logically downstream of
      step~(i).
  \end{enumerate}
\end{mathbox}

%% file: chapters/L7/l7_def.tex
\subsection{The full quantum level \texorpdfstring{$\Lk_7(P)$}{L7(P)}:
  object-level specification}
\label{sec:L7-def}

Sections~\ref{sec:L7-nuclear}--\ref{sec:L7-georgescu} developed the
three mathematical strands that make the full quantum level a
candidate for formal definition: the nuclear Schr\"{o}dinger equation
on the BO surface (C1 at the Hilbert-space level), strict deformation
quantisation and the space-adiabatic $\varepsilon$-expansion (the
ingredients of C2), and the Georgescu $N$-body algebra (Layer~2(a)
ingredients of C2).
This section specifies the object-level data required of an
$\Lk_7$-lift over an $\Lk_6$-object, defines the target category
$\CstarAlg$, and records the conditions that a future functorial
construction $\Felseven$ would have to satisfy.

\begin{warning}[Object-level specification vs.\ functorial existence]
\label{warn:def-vs-existence}
  The object-level $\Lk_7$-lift is specified below by the conditions
  its data must satisfy (Layers~1 and~2(a)/(b)).
  A construction of
  \[
    \Felseven : \Lk_6(P) \longrightarrow \CstarAlg
  \]
  on morphisms is not given here and is part of the open programme of
  Constructions~C2--C4 (Remark~\ref{rem:F7-morphisms}).
  Every statement below is a statement about what the
  \emph{specification} commits to, not a proof that any functor
  satisfies it.
  The status summary appears in Mathbox~\ref{mbox:L7-def-status} and
  the retrospective table of \S\ref{sec:L7-retrospective}.
\end{warning}

\subsubsection{The target symmetric monoidal category \texorpdfstring{$\CstarAlg$}{C*Alg}}
\label{sec:cstar-cat}

For an eventual $\Felseven$ to be a strict SMC functor out of
$\Lk_6(P)$, its codomain must be a symmetric monoidal category whose
objects are continuous fields and whose tensor product is fibrewise.

\begin{definition}[The symmetric monoidal category $\CstarAlg$]
\label{def:cstar-cont}
  $\CstarAlg$ is the category of continuous fields of $C^*$-algebras
  over $[0,1]$ in the sense of Dixmier~\cite{Dixmier1977}:
  \begin{itemize}
    \item \textbf{Objects}: continuous fields
      $A = \{A_\varepsilon\}_{\varepsilon\in[0,1]}$ with $A_0$ a
      commutative $C^*$-algebra.
      Continuity is Dixmier continuity: for every continuous section
      $a$ of $A$, the map $\varepsilon\mapsto\|a_\varepsilon\|$ is
      continuous.
    \item \textbf{Morphisms}: $\Phi : A \to B$ is a family of
      $*$-homomorphisms
      $\Phi_\varepsilon : A_\varepsilon \to B_\varepsilon$ such that
      for every continuous section $a$ of $A$, the family
      $\varepsilon\mapsto \Phi_\varepsilon(a_\varepsilon)$ is a
      continuous section of $B$.
    \item \textbf{Symmetric monoidal product}: fibrewise minimal
      tensor product,
      $(A\otimes B)_\varepsilon
       := A_\varepsilon\otimes_\mathrm{min} B_\varepsilon$,
      with inherited symmetric braiding.
      We restrict to the class of continuous fields for which the
      fibrewise minimal tensor product is again a continuous field;
      this includes all cases relevant to molecular SDQ.
  \end{itemize}
\end{definition}

\begin{remark}[The $\varepsilon = 0$ fibre evaluation]
\label{rem:ev0}
  The \emph{$\varepsilon = 0$ evaluation}
  \[
    \ForZero : \CstarAlg \longrightarrow C^*\mathbf{Alg},
    \qquad
    \ForZero(A) := A_0
  \]
  is a lax monoidal functor that extracts the classical fibre.
  The Layer~2(b) consistency condition of
  Mathbox~\ref{mbox:L7-layers} below requires any eventual
  functorial construction to make the following square commute:
  \[
  \includegraphics{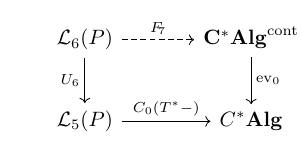}
  \]
  with $U_6$ the $\Lk_6\to\Lk_5$ forgetful functor and
  $C_0(T^*{-})$ sending the scalar $\Lk_5$ datum
  $(\Ce(G), V^G, g^G)$ to the classical phase-space
  algebra $C_0(T^*\Ce^\mathrm{reg}(G))$ on the regular
  sector-reduced configurations
  via Gelfand duality.
  In non-trivial nuclear exchange-statistical sectors, this should
  be read as the sector-reduced scalar form, with equivalent
  equivariant or associated-bundle realisations available.
  The dashed arrow signals that $\Felseven$ here is the open
  functorial extension; Layer~2(b) is the requirement that
  $\ForZero\circ\Felseven$ recover the $\Lk_5$ classical phase-space
  algebra.
\end{remark}

\subsubsection{The object-level assignment of \texorpdfstring{$\Felseven$}{Fel7}}
\label{sec:F7-def}

\begin{definition}[Object-level $\Lk_7$-lift]
\label{def:F7}
  An \emph{$\Lk_7$-lift} of an $\Lk_6(P)$-object over $G$, with
  compact data
  \[
    \bigl(\Ce(G),\;\sigma_j,\;A,\;\eta_B\bigr),
  \]
  is a continuous field
  $\{A_\varepsilon^G\}_{\varepsilon\in[0,1]}$ in $\CstarAlg$ such
  that the following conditions hold.
  Here all scalar algebras are understood on the regular
  sector-reduced configuration space
  $\Ce^\mathrm{reg}(G) \subseteq \Ce(G)$ (the smooth stratum of the
  orbifold, with singular high-symmetry strata excluded), with the
  appropriate equivariant or associated-bundle replacement when
  labelled nuclear exchange sectors are retained.
  $\sigma_j$ denotes a local adiabatic frame in the single-band case
  (replaceable globally by the band projector $P_j$ or the eigenline
  bundle $L_j$); $A$ is the Berry connection of the chosen band; and
  \[
    \eta_B \;=\; w_1(L_0^\RR)
      \;\in\; H^1(\Omega^\circ;\mathbb{Z}/2)
  \]
  is the first Stiefel--Whitney class of the real eigenline bundle
  over $\Omega^\circ = \Ce^\mathrm{reg}(G)\setminus\Xseam$
  (\S\ref{sec:L6-berry}).
  \begin{itemize}
    \item
      $A_0^G \;=\; C_0(T^*\Ce^\mathrm{reg}(G))$, the commutative
      classical phase-space algebra on the regular configurations.
      In non-trivial exchange-statistical sectors, this is the
      sector-reduced scalar form; equivalently, the equivariant or
      associated-bundle Poisson algebra.
    \item
      In the scalar sector-reduced nuclear model,
      $A_\varepsilon^G \cong \mathcal{K}(L^2(\Ce^\mathrm{reg}(G)))$ for
      $\varepsilon > 0$ (Theorem~\ref{thm:landsman-sdq};
      same abstract algebra for every $\varepsilon > 0$,
      Remark~\ref{rem:same-fibre}).
      For non-trivial exchange-statistical sectors, replace by the
      compact operators on the corresponding equivariant or
      associated-bundle Hilbert space.
    \item
      Quantisation maps
      $Q_\varepsilon : \tilde{A}_0^G \to A_\varepsilon^G$
      from a dense Poisson subalgebra
      $\tilde{A}_0^G \subset C_0(T^*\Ce^\mathrm{reg}(G))$ satisfying the
      strict DQ axioms of Definition~\ref{def:sdq} with
      $t = \varepsilon$.
    \item
      In an adiabatic representation, the Berry/geometric data from
      the $\Lk_6$ lift enter the band-projected effective
      Hamiltonian.
      For the selected band $j$ this is schematically the
      minimal-coupling form
      \[
        H^\varepsilon_\mathrm{eff}
        \;=\; \tfrac{1}{2}\bigl(p - \varepsilon A(\mathbf{R})\bigr)^2
        + E_j(\mathbf{R})
        + \varepsilon^2\,\Phi_\mathrm{BH}(\mathbf{R})
        + \cdots,
      \]
      with $\Phi_\mathrm{BH}$ the Born--Huang scalar correction and
      the precise powers, signs, and form determined by the
      semiclassical scaling and gauge convention; the rigorous
      statement is Theorem~\ref{thm:pst}.
  \end{itemize}
  The object-level assignment $\Felseven^{\rm obj}$ sends each
  $\Lk_6(P)$-object to such a lift.
\end{definition}

\begin{remark}[$\Felseven$ on morphisms: open]
\label{rem:F7-morphisms}
  A functorial extension of $\Felseven^{\rm obj}$ to morphisms ---
  sending each $\Lk_6$ reaction/channel $r:G\to G'$ in $\Lk_6(P)$
  to a morphism in $\CstarAlg$ from $\{A_\varepsilon^G\}$ to
  $\{A_\varepsilon^{G'}\}$ --- is open.
  The expected morphism is not in general a $*$-homomorphism:
  reactions between different molecular species may require
  correspondences, completely positive maps, scattering/propagator
  kernels, or other channel data rather than ordinary
  $*$-homomorphisms; the operator algebra of an open quantum
  dynamics is generally not the same as that of a closed one.
  The construction is open at all three levels: consistent
  assignment of the reaction-mechanism Hamiltonian; the structure of
  the induced morphism in $\CstarAlg$; SMC functoriality
  $\Felseven(r_2\circ r_1) = \Felseven(r_2)\circ\Felseven(r_1)$.
\end{remark}

\subsubsection{Layer~1 and Layer~2 at \texorpdfstring{$\Lk_7$}{L7}}
\label{sec:L7-layers}

The tower's Layer~1/Layer~2 pattern applies at $\Lk_7$ as at every
earlier level.
Layer~1 is the minimal SDQ structure; Layer~2 selects the physical
one.

\begin{mathbox}[Layer~1 and Layer~2 for $\Felseven^{\rm obj}$]
\label{mbox:L7-layers}
  \textbf{Layer~1}: any continuous field $\{A_\varepsilon^G\}$
  satisfying Definition~\ref{def:sdq} with $t = \varepsilon$,
  $S = T^*\Ce^\mathrm{reg}(G)$, and $A_0 = C_0(T^*\Ce^\mathrm{reg}(G))$.
  Many such fields exist; Layer~1 is the SDQ-axiomatic content alone.

  \medskip\noindent
  \textbf{Layer~2(a), molecular origin (two stages).}
  Before band reduction, one may represent the molecular Hamiltonian
  either as an operator-valued nuclear Hamiltonian
  \[
    H^\varepsilon_\mathrm{full}
    \;=\; -(\varepsilon^2/2)\Delta_\mathbf{R} + H_\mathrm{el}(\mathbf{R})
  \]
  on nuclear wavefunctions valued in the electronic Hilbert space ---
  schematically $L^2(\Ce^\mathrm{reg}(G); \Hel)$ in the
  operator-valued framework --- or in a full all-particle pre-BO
  representation with the appropriate electron antisymmetry and
  nuclear exchange symmetry.
  Its affiliation requires the corresponding operator-valued or full
  electron--nuclear $N$-body algebra (the prospective
  $G^*$-equivariant descent of the Georgescu--Iftimovici
  construction; \S\ref{sec:georgescu-limits}, step~(i)).
  Here and below $G^*$ acts through the
  permutation-inversion/stereochemical symmetries of $\Lk_{4.5}$ on
  labelled nuclear configurations and, where relevant, on nuclear
  spin factors.
  After selecting an isolated electronic band $j$ (or an active
  electronic cluster), the band-projected effective Hamiltonian
  $H^\varepsilon_\mathrm{eff}$ is represented in the scalar or
  sector-reduced nuclear field $\{A_\varepsilon^G\}$.
  The quantisation is not arbitrary: it is imposed by the actual
  nuclear--electronic physics, with the band reduction handled by
  the space-adiabatic machinery (Theorem~\ref{thm:pst}).

  \medskip\noindent
  \textbf{Layer~2(b), BO consistency.}
  Under Gelfand duality $C_0(T^*\Ce^\mathrm{reg}(G))\leftrightarrow
  T^*\Ce^\mathrm{reg}(G)$, the principal classical symbol of
  $H^\varepsilon_\mathrm{eff}$, equivalently its $\varepsilon\to 0$
  limit in the appropriate functional calculus on the continuous
  field, is
  \[
    h_0 \;=\; \tfrac{1}{2}p^2 + V^G(\mathbf{R}),
  \]
  with $V^G$ the BO surface from $\FV(G)\in\Lk_5(P)$.
  Equivalently: objectwise, $\ForZero(\Felseven^{\rm obj}(G))$
  recovers the $\Lk_5$ phase-space data, and the square of
  Remark~\ref{rem:ev0} commutes at this object.

  \medskip\noindent
  Layer~2(b) is the $C^*$-algebraic statement of the
  Born--Oppenheimer approximation: the quantum field is faithful to
  the classical $\Lk_5$ geometry in the limit $\varepsilon\to 0$.
  Object-level lifts failing Layer~2(b) are not admissible physical
  $\Lk_7$-lifts of the given $\Lk_6(P)$-object.
\end{mathbox}

\subsubsection{The level \texorpdfstring{$\Lk_7(P)$}{L7(P)} and the object-level forgetful map \texorpdfstring{$U_7^{\rm obj}$}{U7obj}}
\label{sec:L7-level-def}

\begin{definition}[Full quantum level $\Lk_7(P)$]
\label{def:L7}
  An \emph{object} of $\Lk_7(P)$ is an $\Lk_6(P)$-object over $G$
  together with a Layer~2 continuous field $\{A_\varepsilon^G\}$
  satisfying Definition~\ref{def:F7} and
  Mathbox~\ref{mbox:L7-layers}.
  The \emph{object-level forgetful map}
  \[
    U_7^{\rm obj} :
      \mathrm{Ob}(\Lk_7(P)) \longrightarrow \mathrm{Ob}(\Lk_6(P)),
    \qquad
    U_7^{\rm obj}\bigl(G,\,\{A_\varepsilon^G\}\bigr)
    = (G,\,\sigma_j,\,A,\,\eta_B),
  \]
  discards the continuous field and retains the underlying $\Lk_6$
  data, with $\sigma_j$ understood locally in the single-band case.
  The construction of morphisms, and hence the promotion of
  $U_7^{\rm obj}$ to a full forgetful functor and the SMC structure
  of $\Lk_7(P)$, remains open (Remark~\ref{rem:F7-morphisms}).
\end{definition}

\begin{remark}[$U_7^{\rm obj}$ vs $\ForZero\circ\Felseven^{\rm obj}$]
\label{rem:U7-vs-ev0}
  $U_7^{\rm obj}$ maps to (objects of) $\Lk_6(P)$ (Hilbert bundles
  with Berry connections and the Berry-sign class $\eta_B$), not to
  the classical algebra $A_0^G$.
  The latter is recovered by
  $\ForZero\bigl(\Felseven^{\rm obj}(G)\bigr)
    = A_0^G = C_0(T^*\Ce^\mathrm{reg}(G))$, which is $\Lk_5$ data.
  The two assignments serve different purposes: $U_7^{\rm obj}$
  forgets the quantum field to return to $\Lk_6$ inputs;
  $\ForZero\circ\Felseven^{\rm obj}$ evaluates those inputs at the
  classical limit to give the $\Lk_5$ phase-space algebra.
\end{remark}

\begin{remark}[The BO approximation as a dynamical statement]
\label{rem:bo-functor}
  The Born--Oppenheimer approximation is not the statement that
  $U_7^{\rm obj}$ is an equivalence --- it is always a genuine
  object-level forgetful map whenever its target lift exists.
  It is the dynamical assertion that for a molecule $G$ at its
  physical mass ratio $\varepsilon_G$, the full quantum evolution
  on $A_{\varepsilon_G}^G$ is approximated by the effective
  evolution generated by
  $H^\varepsilon_\mathrm{eff}
    = \tfrac{1}{2}(p - \varepsilon A)^2 + V^G + O(\varepsilon^2)$.

  At the Hilbert-space level, Theorem~\ref{thm:pst} provides
  schematically an $O(\varepsilon)$-type adiabatic approximation on
  energy-localised subspaces, after the appropriate unitary
  identification of the almost-invariant subspace with the
  effective nuclear Hilbert space.
  Promoting this to a uniform $C^*$-norm statement on the
  continuous field is Conjecture~\ref{conj:u7-functor}
  (Construction~C1 in its $C^*$-algebraic form).

  Three distinct geometric regimes appear:
  \begin{itemize}
    \item \emph{Uniformly gapped single-band region with
      $\eta_B = 0$ on the relevant loops}: an ordinary
      single-valued scalar realisation applies, although ordinary
      Berry-connection corrections may remain.
    \item \emph{Non-trivial $\eta_B$}: an ordinary single-valued
      scalar model is globally incorrect; a twisted or local-system
      quantisation is required (the Mead--Truhlar geometric-phase
      formulation, or an equivalent associated-bundle realisation).
    \item \emph{CI seam $\Xseam$ present}: the single-band gap
      condition of Theorem~\ref{thm:pst} collapses at the seam
      (the gap closes by definition of the seam), requiring a
      multistate or resolved construction.
      This is a separate phenomenon from a non-trivial $\eta_B$;
      the seam resolution is the subject of
      Conjecture~\ref{conj:ci-blowup}, the twist resolution is
      Construction~C3 below.
  \end{itemize}
\end{remark}

\subsubsection{Where C1--C4 live in the specification}
\label{sec:L7-content}

With the object-level $\Lk_7$-lift specified, each of the four
constructions C1--C4 occupies a specific location within the
specification.
The specification is the scaffold; the constructions are the content
it must carry.

\begin{mathbox}[The four contents of $\Felseven^{\rm obj}(G)$]
\label{mbox:L7-C1234}
  The continuous field $\{A_\varepsilon^G\}_{\varepsilon\in[0,1]}$
  carries four structurally distinct kinds of content, one per
  construction.

  \medskip\noindent
  \textbf{C1 --- dynamical content (the effective Hamiltonian).}
  The molecular Hamiltonian $H^\varepsilon$ is affiliated to
  $A_\varepsilon^G$ (Layer~2(a)), and its band-projected effective
  image is schematically
  \[
    H^\varepsilon_\mathrm{eff}
    \;=\; \tfrac{1}{2}\bigl(p - \varepsilon A(\mathbf{R})\bigr)^2
    + E_j(\mathbf{R})
    + \varepsilon^2\,\Phi_\mathrm{BH}(\mathbf{R})
    + \cdots
    \qquad(\text{Theorem~\ref{thm:pst}}),
  \]
  with the Berry connection $A$ entering through minimal coupling
  and Born--Huang-type scalar corrections appearing at the
  corresponding higher orders in the chosen semiclassical
  convention.
  At the Hilbert-space level, the affiliated Hamiltonian generates
  a strongly continuous unitary group; when this induces a strongly
  continuous automorphism group
  $\alpha_t^\varepsilon \in \Aut(A_\varepsilon^G)$ on the chosen
  fibre algebra, $\alpha_t^\varepsilon$ is the quantum dynamics, and
  under the usual Egorov-type semiclassical hypotheses its
  $\varepsilon\to 0$ limit is the classical Hamiltonian flow on
  $A_0^G$ on appropriate time scales.
  C1 is this dynamical layer; it is Hilbert-space rigorous via PST,
  and its $C^*$-algebraic form is
  Conjecture~\ref{conj:u7-functor}.

  \medskip\noindent
  \textbf{C2 --- existence content (the field itself).}
  The whole of $\{A_\varepsilon^G\}$ as a continuous field
  satisfying Layers~1, 2(a), and 2(b): that such a field exists for
  each $\Lk_6(P)$-object over a molecular species $G$ is
  Construction~C2 (Open Problem~\ref{op:sdq-molecules}).
  C2 is the \emph{existence} claim supporting everything else in
  the specification; the other three constructions refine
  structural aspects of a field whose existence C2 establishes.

  \medskip\noindent
  \textbf{C3 --- topological content (the Berry-sign twist).}
  The real Berry-sign class
  $\eta_B = w_1(L_0^\RR) \in H^1(\Omega^\circ;\mathbb{Z}/2)$
  from $\Lk_6$ is represented at $\Lk_7$ as a twist of the nuclear
  state space: the nuclear wavefunction is a section of the real
  eigenline / local system over $\Omega^\circ$, equivalently an
  antiperiodic boundary condition (the Mead--Truhlar
  geometric-phase formulation).
  \begin{itemize}
    \item $\eta_B = 0$ on the relevant region or loops: the
      real-line-bundle twist does not obstruct a single-valued
      scalar realisation, although other geometric corrections ---
      Born--Huang scalar terms, or complex Berry-curvature effects
      in non-real settings --- may still appear in
      $H^\varepsilon_\mathrm{eff}$.
    \item $\eta_B \neq 0$: an ordinary single-valued scalar model
      is globally incorrect; a twisted or local-system
      quantisation is required, and the twist class is
      conjectured to descend to a $KO$-theoretic class of the
      fibre algebra (Conjecture~\ref{conj:c1-obstruction}).
  \end{itemize}
  C3 is the topological layer of $\Felseven^{\rm obj}$: the same
  invariant that forced $\Lk_5\to\Lk_6$ reappears here as the
  quantisation-twist content of the lift.

  \medskip\noindent
  \textbf{C4 --- representation content (the sector structure).}
  The permutation-inversion group $G^*$ from $\Lk_{4.5}$ acts on
  the underlying Hilbert space (the labelled nuclear configuration
  space tensored with the nuclear spin factor) and induces an
  action on $A_\varepsilon^G$.
  This decomposes the physical Hilbert space into isotypic
  components, and the $G^*$-invariant observable subalgebra
  $\mathcal{A}^{G^*}\subset A_\varepsilon^G$ preserves these
  components.
  C4 is the conjectural claim that the resulting
  representation-theoretic sector structure contains a species
  sector $\mathcal{H}_{\pi_G}$ --- the isotypic Hilbert component
  for the irrep $\pi_G$ of $G^*$ associated with the molecular
  species --- whose classical shadow can be identified with the
  graph datum $G\in\Lk_4(P)$
  (Conjecture~\ref{conj:superselection}(a) for the algebraic
  decomposition; (b) for the $\varepsilon\to 0$ stability of the
  species sector).

  \emph{Cautionary statement.}
  The compact ideal $\mathcal{K}(\mathcal{H})$ is simple as a
  $C^*$-algebra and admits no proper closed two-sided ideal; the
  sector structure of C4 is therefore not a decomposition of the
  compact ideal in isolation, but a consequence of the $G^*$-action
  on the underlying Hilbert space together with the choice of
  invariant observable algebra.

  \smallskip\noindent
  \emph{Rigorously established case.}
  For $\mathrm{H_2}$, the relevant identical-proton exchange group
  is $S_2 \cong \mathbb{Z}/2$, and the established statement is the
  decomposition of the physical spin-spatial Hilbert space
  \[
    \mathcal{H}_\mathrm{phys}
    \;=\; \mathcal{H}_\mathrm{para}\oplus\mathcal{H}_\mathrm{ortho},
  \]
  constructed from
  $L^2(Q_\mathrm{lab})\otimes\mathbb{C}^4_\mathrm{spin}$ by imposing
  antisymmetry under proton exchange
  (Mathbox~\ref{mbox:superselection-H2}).
  The spin-independent molecular Hamiltonian preserves these
  sectors, and the decomposition is $\varepsilon$-independent ---
  stability as $\varepsilon\to 0$ is automatic.
  This does \emph{not} follow from a decomposition of
  $\mathcal{K}(L^2(\Ce(\mathrm{H_2})))$ alone.

  \smallskip\noindent
  \emph{Conjectural cases.}
  For polyatomic molecules with the full $G^*$ action, the
  existence of a well-defined invariant observable algebra, its
  representation-theoretic isotypic decomposition, and the
  $\varepsilon\to 0$ stability of the species sector
  $\mathcal{H}_{\pi_G}$ (a dynamical statement requiring the
  Pfeifer-type mechanism of \S\ref{sec:L7-superselection}) are all
  open.
\end{mathbox}

\begin{chembox}[Where the four contents show up in the laboratory]
\label{cbox:L7-def-chem}
  \textbf{C1 (dynamics):}
  solving $H^\varepsilon_\mathrm{eff}$ on $L^2(\Ce^\mathrm{reg}(G))$
  gives vibrational spectra, zero-point energies, and tunnelling
  amplitudes.
  Most standard electronic-structure workflows (DFT, CCSD(T))
  provide clamped-nuclei BO surfaces; force-field methods supply
  approximations to them; both feed the classical/effective nuclear
  models at lower tower levels.
  The size of BO/non-adiabatic error is property- and
  system-dependent and breaks down near CI seams; no universal
  percentage estimate follows from $\varepsilon_G$ alone.

  \smallskip\noindent
  \textbf{C2 (existence):}
  the $C^*$-algebraic status of every claim below.

  \smallskip\noindent
  \textbf{C3 (topology):}
  geometric-phase effects in reactions such as
  $\mathrm{H+HD\to H_2+D}$ associated with the $\mathrm{H_3}$
  conical-intersection topology~\cite{Yuan2020} and analogous
  cold-reaction measurements~\cite{Kendrick2015} are laboratory
  manifestations of non-trivial Berry-sign data $\eta_B$, with
  large channel- and energy-dependent modifications of
  state-resolved observables; conjecturally, they correspond to a
  non-trivial $KO$-theory class of the field $\{A_\varepsilon^G\}$.

  \smallskip\noindent
  \textbf{C4 (identity):}
  slow, condition-dependent ortho--para interconversion of hydrogen
  in the absence of efficient paramagnetic, surface, or
  impurity-mediated channels; long-lived chiral configurations
  treated, in suitable models, as metastable sectors (rather than
  exact superselection sectors), with environment-induced
  decoherence as the mechanism~\cite{Pfeifer1980} resolving Hund's
  paradox; and the nuclear-spin-statistics selection rules and
  intensity patterns of equivalent-nuclei spectroscopy (of which
  the $3:1$ high-temperature ortho/para spin degeneracy of
  $\mathrm{H_2}$ is the elementary instance) are all instances of
  $G^*$-symmetry constraining the sectors of
  $\Felseven^{\rm obj}(G)$.
\end{chembox}

\subsubsection{Status of the specification}
\label{sec:F7-status}

\begin{mathbox}[Status: what Definition~\ref{def:L7} commits to]
\label{mbox:L7-def-status}
  \textit{Uniformly gapped single-band region with $\eta_B = 0$ on
  the relevant loops.}
  Theorem~\ref{thm:landsman-sdq} supplies Layer~1 where
  $\Ce^\mathrm{reg}(G)$ is smooth; Theorem~\ref{thm:pst} supplies
  the Hilbert-space asymptotic support for Layer~2(b); the
  Georgescu algebra (\S\ref{sec:L7-georgescu}) supplies the
  affiliation framework for Layer~2(a)'s pre-band-reduction stage,
  pending the $G^*$-equivariant descent.
  No conceptual obstruction is apparent in this regular scalar
  case, but the operator-algebraic assembly into a continuous field
  satisfying all of Definition~\ref{def:F7} has not been written
  up.
  The remaining technical gaps include the orbifold/stratified
  singularities of $\Ce(G)$ (Remark~\ref{rem:orbifold-gap}), the
  $G^*$-equivariant descent of the Georgescu construction, and the
  full operator-algebraic assembly itself.

  \medskip\noindent
  \textit{Non-trivial Berry-sign twist $\eta_B\neq 0$ and/or CI
  seam $\Xseam\neq\emptyset$.}
  A non-trivial $\eta_B$ requires twisted/local-system
  quantisation: the scalar single-valued construction does not
  apply, and an appropriate twisted realisation must be developed.
  Separately, a CI seam causes Theorem~\ref{thm:pst}'s gap
  condition to fail at the seam, requiring a multistate or
  resolved construction.
  The proposed seam resolution --- blowup
  $\widetilde{\Ce}(G)\to\Ce(G)$
  (Conjecture~\ref{conj:ci-blowup}) --- remains open, as does the
  $\eta_B$-twist resolution of Construction~C3.

  \medskip\noindent
  \textit{On C4.}
  Given a well-defined $G^*$-action and a chosen invariant
  observable algebra, one obtains a representation-theoretic
  decomposition into isotypic sectors.
  The $\varepsilon\to 0$ stability of the species sector is
  rigorous for the identical-proton exchange group
  $S_2 \cong \mathbb{Z}/2$ in $\mathrm{H_2}$; for general $G^*$, and
  for the identification of these sectors with molecular species or
  graph-like structures, the statement is conjectural beyond simple
  spin-statistical examples.

  \medskip\noindent
  Under the honest accounting, $\Lk_7(P)$ is specified as a class
  of object-level lifts whose content is stratified by C1--C4; its
  inhabitation, and the functorial extension to morphisms, are the
  research programme of \S\ref{sec:L7-contributions}.
\end{mathbox}

\subsubsection{The sixth extension type}
\label{sec:L7-extension-type}

\begin{remark}[The sixth and final extension type: quantisation]
\label{rem:sixth-type}
  The tower $\Lk_0\hookrightarrow\cdots\hookrightarrow\Lk_7$ uses
  exactly six qualitatively distinct extension types; the full
  classification appears in Table~\ref{tab:tower-summary}
  (\S\ref{sec:L7-retrospective}).
  The sixth type --- the transition $\Lk_6\to\Lk_7$ --- is unique
  in the tower: its defining new operation is \emph{deformation},
  replacing the classical phase-space algebra associated with the
  previous scalar geometry by a continuous quantum field (alongside
  the new spin-statistical sector and representation content
  carried by the field).
  The classical Poisson algebra $C_0(T^*\Ce^\mathrm{reg}(G))$ is
  replaced by the non-commutative quantum field
  $\{A_\varepsilon^G\}$ parametrised by the physical mass ratio
  $\varepsilon = (m_e/M)^{1/2}$.
  The deformation parameter is a physical constant, not a
  combinatorial or topological datum; the target category
  $\CstarAlg$ is wider than any used at earlier levels.
  This extension type emerges only at the boundary between the
  geometric tower and the quantum world: it depends on the relevant
  $\Lk_6$ electronic-bundle data, schematically
  $(\Ce^\mathrm{reg}(G),\,L_j\text{ or local }\sigma_j,\,A,\,\eta_B)$,
  as the Poisson-manifold structure and topological content that
  the quantum field reduces to in the classical limit.
\end{remark}

\subsubsection{Inter-level coherence: \texorpdfstring{$\Lk_5$}{L5} to \texorpdfstring{$\Lk_7$}{L7}}
\label{sec:L5-L7-coherence}

The tower carries three inter-level coherence conditions linking
functors across non-adjacent levels.
The first two are stated in \S\ref{sec:L7-retrospective}; the third
is the $\Lk_5$-to-$\Lk_7$ tunnelling correction to transition-state
theory.

\begin{remark}[$\Lk_5$--$\Lk_7$ coherence: tunnelling correction to
  TST]
\label{rem:L5-L7-coherence}
  For a reaction with intrinsic reaction coordinate connecting
  minima $\mathbf{R}_a, \mathbf{R}_b \in \Ce^\mathrm{reg}(G)$ through
  a saddle (potentially passing through higher-symmetry strata of
  $\Ce(G)$), a transition-state-theory approximation with
  tunnelling correction models the rate constant as
  \begin{equation}
    \label{eq:tunnelling-coherence}
    k_r \;\approx\; \frac{k_B T}{h}\,
    \exp\!\bigl(-E_a(\FV(G))/k_B T\bigr)\cdot
    \kappa(\varepsilon_G, T),
  \end{equation}
  where $E_a = V^\ddagger - V_\mathrm{reactant}$ is the per-molecule
  potential-energy barrier on the minimum-energy path through the
  transition state in this simplified energy-level formulation (in
  thermochemical TST it is replaced by the molar activation free
  energy $\Delta G^{\ddagger}$, with $RT$ in the exponent), and
  $\kappa$ is the tunnelling correction factor.
  The behaviour of $\kappa$ is constrained in two complementary
  regimes:
  \begin{itemize}
    \item $\kappa(\varepsilon_G, T)\to 1$ as $\varepsilon_G\to 0$
      at fixed $T$: the classical TST limit, in which the
      $\varepsilon=0$ fibre $A_0^G$ supplies the classical BO
      phase-space data used in Eyring-type rate formulas (the
      remaining statistical-mechanical and dividing-surface
      assumptions of Eyring TST are imposed beyond this fibre).
    \item Heuristically, in a one-dimensional barrier model the
      relative importance of through-barrier vs.\ over-barrier
      transmission is governed by the competing exponents
      \[
        \exp(-\AgmonD/\varepsilon_G)
        \qquad\text{and}\qquad
        \exp(-E_a/k_B T),
      \]
      with $\AgmonD$ the Agmon distance through the barrier
      (Theorem~\ref{thm:tunnelling}).
      Tunnelling dominates classical activation when
      $\AgmonD/\varepsilon_G < E_a/k_B T$; the precise asymptotic
      of $\kappa$ depends on the barrier shape and crossover
      energy.
  \end{itemize}
  Equation~\eqref{eq:tunnelling-coherence} is the
  $\Lk_5\to\Lk_7$ coherence approximation: the rate depends
  simultaneously on the $\Lk_5$ geometry ($E_a$ and the Agmon
  metric) and on the mass ratio $\varepsilon_G$ entering through
  $\kappa$.
  The classical limit $\kappa\to 1$ recovers the Eyring TST result
  of the $\Lk_3$--$\Lk_5$ coherence approximation, closing the
  three-condition chain (Wegscheider at $\Lk_2$--$\Lk_3$; Eyring
  TST at $\Lk_3$--$\Lk_5$; tunnelling here).
  Computational implementations (Wigner, Eckart, SCT/LCT,
  $\mu$OMT, RPI+PC), discussed in
  \S\ref{sec:tunneling-attribution}, are built from $\Lk_5$-level
  PES and mass data but approximate the $\Lk_7$-level nuclear
  quantum corrections; Theorem~\ref{thm:tunnelling} provides a
  rigorous $\varepsilon\to 0$ model for the exponential tunnelling
  scale that underlies these semiclassical correction schemes.
\end{remark}

%% file: chapters/L7/l7_contributions.tex
\subsection{Four conjectures of the tower at \texorpdfstring{$\Lk_7$}{L7}}
\label{sec:L7-contributions}

Mathbox~\ref{mbox:L7-C1234} of \S\ref{sec:L7-content} located four
structurally distinct kinds of content within $\Felseven^{\rm obj}(G)$:
dynamical (C1), existence (C2), topological (C3), and
representation-theoretic (C4).
This section states the formal conjecture corresponding to each
content layer, together with its Hilbert-space or special-case
evidence, the tower-level stratification of that evidence, and the
ingredients still required to complete the programme.
None of the four conjectures is a theorem.  Conjectures~I and~II
presuppose the object-level continuous field of Construction~C2.
Conjecture~IV is different: it is an enabling construction for~C2
in the seam-containing case, where the single-band gap condition
of $\Lk_6$ Layer~2 fails.

\begin{mathbox}[The four conjectures and the four contents]
\label{mbox:contributions-map}
  Each conjecture formalises a specific content layer of
  Mathbox \emph{The four contents of $\Felseven^{\rm obj}(G)$}.
\begin{center}
\small
\renewcommand{\arraystretch}{1.2}
\setlength{\tabcolsep}{4pt}
\begin{tabular}{@{}
  >{\raggedright\arraybackslash}p{0.16\linewidth}
  >{\raggedright\arraybackslash}p{0.32\linewidth}
  >{\raggedright\arraybackslash}p{0.42\linewidth}
@{}}
\hline
\textbf{Conjecture} &
\textbf{Formalises content} &
\textbf{Datum recovered by the $\Lk_7$ lift} \\
\hline
I (\S\ref{sec:conj-I}) &
C1 (dynamical) &
semiclassical quantum dynamics on the BO surface \\

II (\S\ref{sec:conj-II}) &
C3 (topological) &
Berry-sign twist $\eta_B = w_1(L_0^\RR)$ \\

III (\S\ref{sec:conj-III}) &
C4 (representation-theoretic) &
nuclear exchange/spin-statistical sector structure \\

IV (\S\ref{sec:conj-IV}) &
prerequisite to C2 near $\Xseam$ &
--- (CI seam resolution; enables C2 where the single-band gap condition fails) \\
\hline
\end{tabular}
\end{center}
  Conjectures~I--III identify distinct kinds of data forgotten by
  the scalar/electronic $\Lk_6$ description and recovered only at
  $\Lk_7$: semiclassical quantum dynamics on the BO surface,
  the Berry-sign twist of the real adiabatic eigenbundle, and the
  nuclear exchange/spin-statistical sector structure.
  Conjecture~IV is logically distinct: it is an \emph{enabling}
  construction needed when the single-band gap condition required
  by C2 fails at a conical-intersection seam, allowing
  Theorems~\ref{thm:landsman-sdq} and~\ref{thm:pst} to apply on
  the resolved space.
\end{mathbox}

\subsubsection{Conjecture~I: the dynamical content of
  $\Felseven^{\rm obj}$ (the $C^*$-algebraic Born--Oppenheimer)}
\label{sec:conj-I}

Conjecture~I formalises the dynamical content layer C1 of
Mathbox~\ref{mbox:L7-C1234}.
The physical statement: the quantum time-evolution on
$A_\varepsilon^G$ approximates the classical Hamiltonian flow on
$A_0^G$ with error $O(\varepsilon)$.  In analytic gapped settings,
superadiabatic constructions can sometimes give exponentially small
interband-coupling estimates under additional hypotheses, but this
exponential strengthening is not a general property of the
$C^*$-algebraic Egorov estimate.

Isotopic substitution provides an important \emph{dynamical} test
of the $\Lk_7$ construction: the same clamped-nuclei electronic
surface yields different quantum nuclear dynamics through different
mass parameters $\varepsilon_G$ entering~\eqref{eq:conj-I} below.
H/D exchange is, however, not the clean forcing obstruction for
$\Lk_7$, since nuclear masses already enter the lower-level
geometric/effective nuclear models (the $\Lk_5$ mass metric).
The clean examples of data \emph{unrecoverable} from $\Lk_6$ are
the nuclear spin-statistical sectors of Conjecture~III, exemplified
by ortho/para $\mathrm{H_2}$: these cannot be represented by
scalar or electronic $\Lk_6$ data alone.

\begin{conjecture}[$C^*$-algebraic Born--Oppenheimer]
\label{conj:u7-functor}
  Assume $\Felseven^{\rm obj}(G)$ constructed (Construction~C2).
  Let
  $\Omega \subset \Ce^\mathrm{reg}(G)$ be a uniformly gapped
  single-band adiabatic region for the chosen electronic band,
  with $\eta_B = 0$ on all loops in $\Omega$.  Write
  $A_\varepsilon^{G,\Omega}$ and
  $A_0^{G,\Omega} = C_0(T^*\Omega)$ for the corresponding fibres
  of the continuous field restricted to $\Omega$, and let
  $H^{\varepsilon,\Omega}$ be the band-projected molecular
  Hamiltonian affiliated to $A_\varepsilon^{G,\Omega}$ over this
  region.
  Suppose $H^{\varepsilon,\Omega}$ generates a strongly continuous
  unitary group inducing a strongly continuous automorphism
  group $\alpha_t^\varepsilon \in
  \Aut(A_\varepsilon^{G,\Omega})$, and let $\alpha_t^0$ denote
  the classical Hamiltonian flow of
  $h_0 = p^2/2 + V^G(\mathbf{R})$ on $A_0^{G,\Omega}$.
  Then for each $f$ in the dense Poisson subalgebra
  $\tilde{A}_0^{G,\Omega}\subset A_0^{G,\Omega}$ and every
  spectral cutoff $E>0$, one expects an Egorov/PST-type estimate
  of the schematic form
  \begin{equation}
    \label{eq:conj-I}
    \bigl\|\bigl[\alpha_t^\varepsilon(Q_\varepsilon(f))
      - Q_\varepsilon(\alpha_t^0(f))\bigr]\,P_E^\varepsilon
    \bigr\|
    \;\leq\; C_{f,E}\,\varepsilon\,(1+|t|),
  \end{equation}
  for finite times $|t|\leq T_0$ and under suitable gap,
  regularity, and domain hypotheses on $\Omega$, where
  $P_E^\varepsilon$ is the spectral projection of
  $H^{\varepsilon,\Omega}$ onto $[-E,E]$ defined through the
  functional calculus of the affiliated Hamiltonian on the
  represented Hilbert space.
  The estimate is read schematically: the multiplication by
  $P_E^\varepsilon$ is interpreted in the represented Hilbert
  space (or via a $C^*$-multiplier-algebra extension where
  available), and the norm is the operator norm in that
  representation.
  In analytic gapped settings, superadiabatic constructions can
  sometimes give exponentially small interband-coupling estimates
  of the form $C_{f,E}\,\exp(-\gamma/\varepsilon)$ for some
  $\gamma > 0$, under additional analytic-band, energy-regime, and
  crossing-avoidance hypotheses~\cite{HagedornJoye2001}; an
  exponentially small analogue of the $C^*$-algebraic estimate
  here is part of the conjectural programme.
\end{conjecture}

\begin{remark}[Why the energy cutoff is necessary]
\label{rem:conj-I-cutoff}
  A bound uniform in energy cannot hold: the PST almost-invariant
  projection $P^\varepsilon$ isolates a single electronic band, and
  the cutoff $P_E^\varepsilon$ restricts attention to the energy
  regime in which the selected band or band cluster remains
  dynamically relevant and the almost-invariant decomposition is
  controlled.  At sufficiently high nuclear energies, or near
  small electronic gaps, the single-band approximation can fail.
  Conjecture~\ref{conj:u7-functor} asserts that, within this
  energy-localised setting, a $C^*$-algebraic error of
  Egorov/PST type follows from the Hilbert-space bound of
  Theorem~\ref{thm:pst}(iii); the open step is promoting the
  bound from propagators on states to automorphisms on a dense
  subalgebra of elements.
\end{remark}

\begin{remark}[Isotope dependence as a dynamical $\Lk_7$ test
  case]
\label{rem:conj-I-isotope}
  The $\mathrm{H}\leftrightarrow\mathrm{D}$ swap leaves the
  electronic Hamiltonian $H_\mathrm{el}(\mathbf{R})$ unchanged
  (electronic structure depends on nuclear charges and positions,
  not masses), but enters the dynamics through
  $\varepsilon = (m_e/M)^{1/2}$ in
  Conjecture~\ref{conj:u7-functor}: the automorphisms
  $\alpha_t^{\varepsilon_H}$ and $\alpha_t^{\varepsilon_D}$ on
  the shared scalar nuclear algebra
  $\mathcal{K}(L^2(\Ce^\mathrm{reg}(G)))$ differ at
  $O(\varepsilon_H - \varepsilon_D)$ --- roughly $30\%$ of
  $\varepsilon_H$ for H/D --- under the schematic estimate
  \eqref{eq:conj-I}.
  The distinction is quantitative and dynamical, not a
  forcing-class obstruction: the algebra is the same for both
  isotopologues, but the Hamiltonian element and the generated
  dynamics differ.
\end{remark}

\textbf{Chemistry.}
Most standard electronic-structure workflows (DFT, CCSD(T))
provide clamped-nuclei BO surfaces, and force-field methods supply
approximations to them; these supply the input to the
classical/effective nuclear models at lower tower levels.
Conjecture~\ref{conj:u7-functor} is the quantitative foundation:
under suitable gap and regularity assumptions, PST gives
controlled $O(\varepsilon)$-type adiabatic errors for appropriate
energy-localised dynamics, and in analytic settings superadiabatic
constructions can yield exponentially small interband coupling
estimates under additional hypotheses.
The numerical error in chemical observables remains property-
and system-dependent and breaks down near CI seams; the
qualitative scaling --- larger errors for hydrogen-containing
systems ($\varepsilon_H\approx 0.023$) than for heavy-element
chemistry ($\varepsilon\lesssim 0.002$) --- reflects the
$\varepsilon$-dependence in~\eqref{eq:conj-I}.
The conjecture does not address nuclear-tunnelling phenomena
(kinetic isotope effects, zero-point competition in water, proton
delocalisation) --- these are $\Lk_5$/$\Lk_7$-level semiclassical
content computable from the effective Hamiltonian
$H^\varepsilon_\mathrm{eff}$ (\S\ref{sec:L7-nuclear}); what
Conjecture~I does is promote the BO expansion from an assumed
hierarchy into a controlled $C^*$-algebraic error bound on
observable dynamics.

\textbf{Evidence, stratified by tower level.}
\emph{($\Lk_7$, Hilbert-space).}
Theorem~\ref{thm:pst}(iii) establishes the energy-localised bound
$\|[\cdots]P^\varepsilon P_E\|\leq C_E\,\varepsilon(1+|t|)$; this
is the direct Hilbert-space precursor of~\eqref{eq:conj-I}.
\emph{($\Lk_7$, exponential regime).}
Hagedorn--Joye~\cite{HagedornJoye2001} provide the Hilbert-space
superadiabatic model for the exponentially small
interband-coupling regime, obtaining $\exp(-\gamma/\varepsilon)$
estimates for analytic potentials by optimal truncation of the
asymptotic expansion; a corresponding $C^*$-algebraic
strengthening of~\eqref{eq:conj-I} is part of the conjectural
programme.
\emph{($\Lk_7 \to \CstarAlg$, algebraic scaffold).}
Landsman's tangent-groupoid
construction~\cite{LandsmanRamazan2001} (Theorem~\ref{thm:landsman-sdq})
supplies the continuous-field template for the strict deformation
quantisation; the additional Hamiltonian-affiliation, domain, and
invariance hypotheses are what make $\alpha_t^\varepsilon$ a
well-defined $\varepsilon$-family of automorphisms on the chosen
fibres.  Without the underlying continuous field, the left-hand
side of~\eqref{eq:conj-I} has no domain in the first place.

\textbf{Completing the tower requires.}
(a) $\Felseven^{\rm obj}$ constructed (Construction~C2,
Open~Problem~\ref{op:sdq-molecules}), yielding $A_\varepsilon^G$
and $\alpha_t^\varepsilon$.
(b) The Georgescu Layer~2(a) descent (\S\ref{sec:L7-georgescu})
used to identify $H^\varepsilon$ as a specific affiliated element,
making $\alpha_t^\varepsilon$ the automorphism it generates.
(c) The Hilbert-space bound of Theorem~\ref{thm:pst}(iii) lifted
to~\eqref{eq:conj-I}: the bound on energy-localised propagators
is re-expressed as a bound on energy-localised sections of the
automorphism family, pointwise in $f\in\tilde A_0^G$.

\subsubsection{Conjecture~II: the topological content of
  $\Felseven^{\rm obj}$ ($\eta_B$ as $KO$-theoretic twist class)}
\label{sec:conj-II}

Conjecture~II formalises the topological content layer C3 of
Mathbox~\ref{mbox:L7-C1234}: the mod-2 Berry-sign class
$\eta_B = w_1(L_0^\RR) \in H^1(\Omega^\circ;\mathbb{Z}/2)$ that
forced the $\Lk_5\to\Lk_6$ transition reappears as a
$KO$-theoretic twist class of the classical fibre at $\Lk_7$,
governing whether an ordinary single-valued scalar realisation
of $\{A_\varepsilon^G\}$ is available.
The physical setting is the time-reversal-symmetric
non-relativistic molecular Hamiltonian ($T^2 = +1$), for which the
adiabatic eigenspaces form a real line bundle $L_0^\RR$ over
$\Omega^\circ = \Ce^\mathrm{reg}(G)\setminus\Xseam$ classified by
its first Stiefel--Whitney class $\eta_B$; local eigensections can
be chosen real, but a global single-valued real eigensection
exists only when $\eta_B = 0$.

\begin{conjecture}[$\eta_B$ as $KO$-theoretic twist class of the
  classical fibre]
\label{conj:c1-obstruction}
  Assume $\Felseven^{\rm obj}(G)$ constructed (Construction~C2)
  over $\Omega^\circ = \Ce^\mathrm{reg}(G)\setminus\Xseam$.
  The Berry-sign class
  \[
    \eta_B \;=\; w_1(L_0^\RR)
    \;\in\; H^1(\Omega^\circ;\mathbb{Z}/2)
  \]
  determines, via pullback along the cotangent projection
  $\pi : T^*\Omega^\circ \to \Omega^\circ$ followed by the
  Gelfand identification, a real $K$-theory class associated with
  the classical fibre algebra:
  \[
    \pi^*[L_0^\RR] \;\in\; KO^0(T^*\Omega^\circ)
    \;\cong\; KO^0\bigl(\widetilde{C_0(T^*\Omega^\circ)}\bigr),
  \]
  where the right-hand side is taken in the unitised algebra, or
  equivalently understood as a compact-support/relative class
  $[\pi^*L_0^\RR] - [1]$ in $KO_0(C_0(T^*\Omega^\circ))$ when so
  formulated.  (For noncompact $T^*\Omega^\circ$ the reduced group
  $KO_0(C_0(T^*\Omega^\circ))$ is compactly supported real
  $K$-theory; a line-bundle class lives in ordinary $KO^0$ of the
  base, and the descent to the $C_0$-algebra requires this
  unitisation or compact-support reading.)
  Write $\widehat{\eta_B}$ for the resulting twist datum of the
  classical fibre.
  The conjecture is that $\widehat{\eta_B}$ governs whether an
  ordinary single-valued scalar realisation of the continuous
  field $\{A_\varepsilon^G\}$ exists:
  \begin{itemize}
    \item When $\eta_B$ vanishes on all loops in the region under
      consideration: the real eigenline bundle $L_0^\RR$ is
      trivialisable there, and a single-valued scalar realisation
      may be used.  Other geometric corrections in
      $H^\varepsilon_\mathrm{eff}$ --- Born--Huang scalar terms,
      or complex Berry-curvature effects in non-real settings ---
      may still appear and are not removed by $\eta_B = 0$ alone.
    \item When $\eta_B \neq 0$ (on some loop): the real eigenline
      bundle is non-trivial, an ordinary single-valued scalar
      quantisation is globally incorrect, and a twisted or
      local-system realisation is required (the Mead--Truhlar
      geometric-phase formulation, or equivalently a section of
      the $L_0^\RR$-twisted scalar field).  The class
      $\widehat{\eta_B}$ is conjecturally non-trivial and
      represents the operator-algebraic shadow of the Berry-sign
      twist.
  \end{itemize}
  This is the $\eta_B$-twist content of the lift; it is logically
  separate from the gap failure at a CI seam $\Xseam$, addressed
  by Conjecture~\ref{conj:ci-blowup}.
\end{conjecture}

\begin{remark}[Real vs.\ complex Berry structure]
\label{rem:real-vs-complex}
  The framework above uses real line bundles and
  $KO$-theory because non-relativistic molecular Hamiltonians
  with $T^2 = +1$ yield real eigenstates with
  $\mathbb{Z}_2$-valued holonomies
  (\S\ref{sec:L6-berry}).
  For spin--orbit-coupled systems ($T^2 = -1$, outside the
  monograph's scope in \S\ref{sec:tower-scope}), the eigenstates
  are Kramers pairs, the eigenbundle is complex, and the
  analogous obstruction lives in complex $K$-theory with
  integer Chern class; Hawkins~\cite{Hawkins2008} provides the
  template there.
  Conjecture~\ref{conj:c1-obstruction} is the real analog of
  Hawkins' result, which has not been worked out for real line
  bundles over molecular configuration orbifolds.
\end{remark}

\textbf{Chemistry.}
The sign flip of the real adiabatic eigensection around any loop
encircling a CI seam is directly observable in scattering
experiments: Yuan et al.~\cite{Yuan2020} provide the experimental
observation of the geometric-phase effect in the
$\mathrm{H+HD\to H_2+D}$ reaction associated with the
$\mathrm{H_3}$ conical-intersection topology, and Kendrick, Hazra,
and Balakrishnan~\cite{Kendrick2015} predict large channel- and
energy-dependent modifications of state-resolved rate coefficients
in the $\mathrm{O+OH\to H+O_2}$ reaction through quantum
interference between direct and CI-encircling pathways.
Conjecture~\ref{conj:c1-obstruction} says this laboratory
observation has a $KO$-theoretic shadow: the same $\eta_B$ that
modifies state-resolved scattering observables also requires a
twisted/local-system quantisation of $\{A_\varepsilon^G\}$ rather
than an ordinary single-valued scalar realisation.
Photochemistry is mathematically hard partly because the
$C^*$-algebraic object that would ground its approximations
rigorously --- the continuous field for $\eta_B \neq 0$ molecules
--- needs a twisted realisation, and additionally, when a CI seam
$\Xseam$ is present, the single-band gap condition fails at the
seam and requires a separate resolution
(Conjecture~\ref{conj:ci-blowup}, \S\ref{sec:conj-IV}).

\textbf{Evidence, stratified by tower level.}
\emph{($\Lk_5\to\Lk_6$, formal WKB).}
Dazord--Patissier~\cite{DazordPatissier1991} prove, for the
complex case, that the Chern class obstructs asymptotic
semiclassical quantisation; the real analog using Stiefel--Whitney
classes is the expected mod-2 statement (see
\cite{Kaufmann2016,AhnParkYang2019} for the Stiefel--Whitney
framework in condensed-matter topological-band theory).
\emph{($\Lk_6$, formal Moyal/WKB).}
Emmrich--Weinstein~\cite{EmmrichWeinstein1996} establish a
multicomponent-WKB obstruction in the matrix-symbol Moyal
calculus; this is the formal-deformation precursor of
Conjecture~\ref{conj:c1-obstruction}.
\emph{($\Lk_6\to\Lk_7$, first-order structure).}
PST Theorem~\ref{thm:pst}(ii) exhibits the Berry connection $A$
entering the band-projected effective Hamiltonian in
minimal-coupling form $\tfrac{1}{2}(p - \varepsilon A)^2 + E_j +
\varepsilon^2 \Phi_\mathrm{BH} + \cdots$, with the precise
representation determined by the semiclassical scaling
(Definition~\ref{def:F7}).
\emph{(Strict $C^*$-algebraic template).}
Hawkins~\cite{Hawkins2008} proves the complex-case $K$-theoretic
obstruction for the sphere $S^2$: the method --- identifying the
Chern class as a $K$-theory obstruction to strict DQ over a
continuous parameter set --- is the direct structural template for
the molecular $KO$-theoretic statement.
\emph{($\Lk_6$, spectral consequence).}
Faure--Zhilinskii~\cite{FaureZhilinskii2001} prove that the
eigenbundle topology controls spectral redistribution between
molecular energy bands, giving an independent, measurable
consequence of the topological class.

\textbf{Completing the tower requires.}
(a) $\Felseven^{\rm obj}$ constructed (Construction~C2) so that
$A_\varepsilon^G$ exists for $\eta_B \neq 0$ molecules in a
twisted/local-system realisation.
(b) The twist datum $\widehat{\eta_B}$ identified either in
$KO^0$ of the unitised commutative algebra
$\widetilde{C_0(T^*\Omega^\circ)}$, or as a relative/compact-support
class in $KO_0(C_0(T^*\Omega^\circ))$ when such a representative
is defined, via a real index construction applied to
$L_0^\RR \to \Omega^\circ$.
(c) Proof that this datum is induced by the cotangent-projection
pullback $\pi^*[L_0^\RR] \in KO^0(T^*\Omega^\circ)$
(with the appropriate unitisation or compact-support
interpretation in the $C_0$-algebra),
connecting the $\Lk_6$ topological invariant to the $\Lk_7$
algebraic one.
Steps (a)--(c) together promote Dazord--Patissier and
Emmrich--Weinstein from the formal (Moyal/WKB) to the strict
($C^*$-algebraic) setting in the real/$\mathbb{Z}_2$ case.

\subsubsection{Conjecture~III: the representation content of
  $\Felseven^{\rm obj}$ ($G^*$-sectors and molecular identity)}
\label{sec:conj-III}

Conjecture~III formalises the representation-theoretic content
layer C4 of Mathbox~\ref{mbox:L7-C1234}: the
$\Lk_{4.5}$ permutation-inversion group $G^*$ acts unitarily on
the labelled spin-spatial Hilbert space
$\mathcal{H}_\mathrm{lab}^G = L^2(Q_\mathrm{lab}^\mathrm{reg}(G))
\otimes \mathcal{S}_G$, where $\mathcal{S}_G$ is the nuclear-spin
representation of the identical nuclei of $G$
($\mathcal{S}_G = \mathbb{C}$ when spin is not tracked).
Physical Hilbert spaces are obtained as equivariant subspaces of
$\mathcal{H}_\mathrm{lab}^G$ or as sections of associated bundles
over the quotient
$\Ce^\mathrm{reg}(G) = Q_\mathrm{lab}^\mathrm{reg}(G)/G^*$,
giving an isotypic decomposition that corresponds to physically
distinct nuclear-spin species of the molecule.
This decomposition records nuclear exchange-statistical data that
are forgotten by the scalar/electronic $\Lk_6$ description.

Three distinct claims, increasing in difficulty, make up
Conjecture~III.
The first is algebraically automatic for finite $G^*$ once the
labelled construction is in place; the second concerns the
distinction between exact symmetry-protected sectors and
metastable barrier-protected sectors; the third is
environment-induced selection in the Pfeifer/Amann sense.

\begin{proposition}[$G^*$-isotypic decomposition and the invariant
  observable algebra]
\label{prop:G-star-decomposition}
  Assume $G^*$ is finite (as in the permutation-inversion groups
  considered here) and acts unitarily on the labelled
  spin-spatial Hilbert space
  \[
    \mathcal{H}_\mathrm{lab}^G
    \;=\;
    L^2(Q_\mathrm{lab}^\mathrm{reg}(G))
    \otimes \mathcal{S}_G,
  \]
  where $\mathcal{S}_G$ is the nuclear-spin representation of the
  identical nuclei of $G$ ($\mathcal{S}_G = \mathbb{C}$ when spin
  is not tracked).  Then $\mathcal{H}_\mathrm{lab}^G$ admits the
  isotypic decomposition
  \[
    \mathcal{H}_\mathrm{lab}^G
    \;\cong\;\bigoplus_{\pi\in\widehat{G^*}}
    V_\pi \otimes M_\pi,
  \]
  where $\widehat{G^*}$ is the unitary dual, $V_\pi$ is the
  representation space of the irreducible representation $\pi$,
  and $M_\pi = \mathrm{Hom}_{G^*}(V_\pi, \mathcal{H}_\mathrm{lab}^G)$
  is the multiplicity Hilbert space.  The $G^*$-invariant compact
  subalgebra decomposes correspondingly:
  \[
    \mathcal{K}(\mathcal{H}_\mathrm{lab}^G)^{G^*}
    \;\cong\;\bigoplus_{\pi\in\widehat{G^*}}
    \mathbf{1}_{V_\pi}\otimes\mathcal{K}(M_\pi),
  \]
  with $\mathbf{1}_{V_\pi}$ the identity on $V_\pi$.
  Scalar (spin-trivial) wavefunctions on the sector-reduced
  configuration space
  $\Ce^\mathrm{reg}(G) = Q_\mathrm{lab}^\mathrm{reg}(G)/G^*$
  correspond, modulo the usual orbifold/measure caveats, to the
  $G^*$-invariant subspace of $\mathcal{H}_\mathrm{lab}^G$ with
  $\mathcal{S}_G = \mathbb{C}$ --- i.e.\ the trivial isotype in
  that case.  Non-trivial isotypes (including spin-statistical
  sectors such as ortho/para $\mathrm{H_2}$) correspond to
  sections of associated bundles over $\Ce^\mathrm{reg}(G)$,
  twisted by the relevant $G^*$-representation.

  \medskip\noindent
  \emph{Cautionary statement.}
  The full compact ideal
  $\mathcal{K}(\mathcal{H}_\mathrm{lab}^G)$ itself is simple as a
  $C^*$-algebra and admits no proper closed two-sided ideal; the
  $G^*$-sector structure of C4 is therefore a property of the
  Hilbert isotypic decomposition together with the chosen
  invariant observable subalgebra
  $\mathcal{A}^{G^*} \subseteq A_\varepsilon^G$, not of the
  compact ideal in isolation.  Elements of $\mathcal{A}^{G^*}$
  preserve each isotypic component $V_\pi \otimes M_\pi$.
  These statements are $\varepsilon$-independent: the isotypic
  decomposition holds at every $\varepsilon > 0$ and is
  determined entirely by the finite $\Lk_{4.5}$ group $G^*$ and
  the underlying labelled spin-spatial Hilbert space.
\end{proposition}

\begin{conjecture}[Species-sector stability and environment-induced
  selection]
\label{conj:superselection}
  Let $\pi\in\widehat{G^*}$ denote a physically allowed
  nuclear-spin/permutation sector for the fixed graph $G$ (the
  combination of nuclear-spin symmetry and spatial-exchange type
  consistent with the Pauli principle for the identical nuclei of
  $G$; a given graph generally carries several such sectors, e.g.\
  ortho and para of $\mathrm{H_2}$), and let
  $\mathcal{H}_\pi = V_\pi \otimes M_\pi
  \subset \mathcal{H}_\mathrm{lab}^G$ denote the corresponding
  isotypic Hilbert component
  (Proposition~\ref{prop:G-star-decomposition}), with $P_\pi$ the
  orthogonal projection onto $\mathcal{H}_\pi$.
  Then:
  \begin{enumerate}[label=\emph{(\alph*)}, leftmargin=*]
    \item \emph{Sector invariance / stability.}
      Two regimes are to be distinguished and are conceptually
      different objects.
      \begin{enumerate}[label=\emph{(a\arabic*)}, leftmargin=2em]
        \item \emph{Exact symmetry sectors.}  For an exactly
          $G^*$-invariant isolated molecular Hamiltonian, each
          isotypic component $\mathcal{H}_\pi$ is exactly
          invariant under $\alpha_t^\varepsilon$ at every
          $\varepsilon$, and no leakage between isotypes occurs.
          This applies in particular to spin-independent
          molecular dynamics of identical-nucleus species such
          as ortho/para $\mathrm{H_2}$.
        \item \emph{Metastable barrier-protected sectors.}  For
          sectors protected only by a barrier or by a weakly
          broken symmetry (e.g.\ metastable left/right chiral
          configurations), the relevant projection is not the
          isotypic projector $P_\pi$ but a spectral or
          localisation projection $P_\mathrm{meta}$ associated
          with the relevant well or metastable subspace.  Under
          semiclassical barrier and regularity hypotheses one may
          conjecture an exponential stability estimate of the
          schematic form
          \[
            \bigl\|(I - P_\mathrm{meta})\,
              \alpha_t^\varepsilon(\rho)\,
              (I - P_\mathrm{meta})\bigr\|_1
            \;\leq\; C\exp(-c/\varepsilon)\,(1+|t|),
          \]
          for finite times.  The constants $C, c > 0$ depend on
          the barrier action and the symmetry-breaking scale.
      \end{enumerate}
    \item \emph{Environment-induced selection.}
      Coupling $A_\varepsilon^G$ to an environmental algebra
      $\mathcal{E}$ with appropriate spectral density (e.g.\
      ohmic bosonic bath) suppresses coherences between the
      relevant sector components and, after conditioning on
      pointer-state observables or in an appropriate
      superselection limit, can lead to effective sector
      selection.  This addresses the Woolley--Primas
      problem~\cite{Woolley1978,Primas1983}: molecular identity
      is realised through an environment-mediated superselection
      structure rather than assumed.
  \end{enumerate}
  Under (a) and (b), $\pi$ identifies a physical
  nuclear-spin/permutation sector \emph{within} the fixed graph
  $G$; the graph datum $G\in\Lk_4(P)$ itself is fixed prior to
  the definition of $G^*$, and recovering graph-like molecular
  identity from a full all-particle theory remains the broader
  programme.
\end{conjecture}

\begin{remark}[What $\pi$ encodes, and what it does not]
\label{rem:pi-G-scope}
  A representation $\pi \in \widehat{G^*}$ indexes a sector within
  a single fixed-graph algebra $A_\varepsilon^G$; it does not
  itself encode the molecular graph $G$, which is fixed prior to
  the definition of $G^*$.  Several distinct physical sectors
  (e.g.\ ortho and para of $\mathrm{H_2}$) typically coexist for
  the same graph and the same $G^*$, indexed by different
  $\pi\in\widehat{G^*}$.
  The graph determines which nuclei are identical, hence which
  group $G^*$ acts; $\pi$ then labels the specific
  nuclear-spin/permutation sector.
  Cross-graph identification --- whether a proton in
  $\mathrm{H_2O}$ is ``the same'' as one in
  $\mathrm{H_2 SO_4}$ --- is a question about a multi-molecule
  total-system algebra, and is not addressed by
  Conjecture~\ref{conj:superselection} in its single-graph form.
\end{remark}

\textbf{Chemistry.}
Ortho- and para-$\mathrm{H_2}$ are kinetically decoupled on
laboratory timescales in the absence of efficient paramagnetic,
surface, or impurity-mediated conversion channels, and have
different rotational partition functions and low-temperature
thermodynamic behaviour~\cite{Silvera1980}.
For the isolated spin-independent molecular Hamiltonian this is
case (a1) of Conjecture~\ref{conj:superselection} applied to the
identical-proton exchange group $S_2 \cong \mathbb{Z}/2$: sector
invariance is exact at every $\varepsilon$ by the Pauli principle
(Mathbox~\ref{mbox:superselection-H2}); observed conversion in
real systems proceeds through spin-dependent or environmental
couplings outside this isolated-Hamiltonian class.
Conjecture~\ref{conj:superselection} generalises this to
polyatomic molecules: within a fixed molecular graph $G$,
isotypic components of the spin-spatial labelled Hilbert space
$\mathcal{H}_\mathrm{lab}^G$
labelled by $\pi\in\widehat{G^*}$ correspond to distinct
nuclear-spin sectors (e.g.\ ortho and para sectors of
$\mathrm{H_2O}$); an analogous decomposition holds for
$\mathrm{D_2O}$ (two identical deuterons give nontrivial
identical-particle sectors), whereas $\mathrm{HDO}$ belongs to
a different isotopologue graph with no nontrivial exchange of
two identical hydrogens, and so does not host an analogous
ortho/para sector structure (Remark~\ref{rem:pi-G-scope}).
Spectroscopic techniques that distinguish species \emph{within
the same graph} via Pauli-statistical selection rules and
equivalent-nuclei intensity patterns rely on the
exact-invariance content of
Conjecture~\ref{conj:superselection}(a1); long-lived chiral
configurations are, in suitable models, metastable
barrier-protected sectors --- (a2) rather than (a1) --- with
environment-induced decoherence as the
mechanism~\cite{Pfeifer1980} addressing Hund's paradox.
(Distinguishing $^1\mathrm{H}$ from $^2\mathrm{H}$ via their
gyromagnetic ratios is a $\Lk_0$/$\Lk_4$-level distinction
between molecular graphs, not the same as the intra-graph sector
content of Conjecture~III.)

\textbf{Evidence, stratified by tower level.}
\emph{($\Lk_7$, proved case: $S_2 \cong \mathbb{Z}/2$ proton
exchange in $\mathrm{H_2}$).}
Ortho/para-$\mathrm{H_2}$ (Mathbox~\ref{mbox:superselection-H2})
establishes
Proposition~\ref{prop:G-star-decomposition} and
Conjecture~\ref{conj:superselection}(a) for proton exchange in
$\mathrm{H_2}$; the decomposition is exact and
$\varepsilon$-independent by the Pauli principle, so stability is
automatic.
\emph{($\Lk_7$, mechanism demonstrated for the parity subgroup).}
Pfeifer~\cite{Pfeifer1980} demonstrated the mechanism of
Conjecture~\ref{conj:superselection}(b) for chirality in a
two-level spin--boson model with ohmic coupling, and the
framework was developed by Amann~\cite{Amann1991,Amann1993}:
the parity subgroup of $G^*$ is broken by environmental
coupling, selecting a definite enantiomeric sector in the
$W^*$-algebraic infinite-bath limit.  The general polyatomic
$G^*$ case extending these special models remains conjectural.
\emph{($\Lk_7$, numerical illustration for $G^* = S_3$).}
Lang, Cezar, Adamowicz, and
Pedersen~\cite{LangEtAl2024} sample the all-particle pre-BO
density $|\Psi(\mathbf{r},\mathbf{R})|^2$ of $\mathrm{D_3^+}$
via MCMC and unsupervised learning, recovering an
equilateral-triangular shape from a fully $S_3$-symmetric
wavefunction.
This provides an explicit numerical illustration that molecular
shape emerges from a permutation-symmetric pre-BO eigenstate for
a system with nontrivial $S_3$ permutation symmetry --- complementary
structural evidence that the $G^*$-action carries physical content,
although the connection to the dynamical sector-invariance content
of Conjecture~\ref{conj:superselection}(a) is suggestive rather than
direct.
\emph{($\Lk_{4.5}$--$\Lk_7$, structural input).}
The Longuet-Higgins framework~\cite{LonguetHiggins1963}
identifies $G^*$ and its physical irreducible representations;
the Georgescu compact ideal
(Proposition~\ref{prop:compact-ideal}) supplies the localised
spectral regime in which the band-projected dynamics is
controlled, while the isotypic decomposition itself requires the
additional $G^*$-invariant observable/representation structure
of Proposition~\ref{prop:G-star-decomposition}.
Renault's groupoid $C^*$-algebra theory~\cite{Renault1980}
supplies the abstract framework for transformation groupoid
algebras: on the labelled configuration space it provides
$C^*(G^*\ltimes Q_\mathrm{lab}^\mathrm{reg}(G)) =
C_0(Q_\mathrm{lab}^\mathrm{reg}(G))\rtimes G^*$, which descends
to associated-bundle data over the quotient
$\Ce^\mathrm{reg}(G) = Q_\mathrm{lab}^\mathrm{reg}(G)/G^*$.

\textbf{Completing the tower requires.}
(a) $\Felseven^{\rm obj}$ constructed (Construction~C2), so that
$A_\varepsilon^G$ and the $G^*$-action on it exist.
(b) The $G^*$-equivariant descent from the Georgescu algebra
(\S\ref{sec:georgescu-limits}, step~(i)) so that
$\mathcal{A}^{G^*}$ is the correct observable subalgebra.
(c) Proof of the exponential stability bound for metastable
barrier-protected sectors of
Conjecture~\ref{conj:superselection}(a2) via Agmon-distance
estimates between well-localised states (separate from the exact
$G^*$-isotypic invariance of (a1), which follows from symmetry
alone).
(d) An explicit environmental coupling for the general
polyatomic case extending Amann's chirality result to
arbitrary $G^*$, realising
Conjecture~\ref{conj:superselection}(b).

\begin{remark}[DHR analogy, briefly]
\label{rem:dhr-analogy}
  The conceptual analogue of
  Conjecture~\ref{conj:superselection} is the DHR
  superselection theory of algebraic
  QFT~\cite{DoplicherHaagRoberts1990}, which recovers particle
  statistics as sectors of the observable algebra.
  DHR requires the Haag--Kastler axioms of relativistic QFT and
  does not apply directly to non-relativistic molecules; the
  molecular case is approached here through the
  Renault--Longuet-Higgins groupoid construction above.
  The analogy is structural, not formal.
\end{remark}

\subsubsection{Conjecture~IV: the $\Xseam$ blowup as the
  prerequisite for Construction~C2}
\label{sec:conj-IV}

Conjecture~IV is structurally distinct from Conjectures~I--III:
it does not formalise a content layer of $\Felseven^{\rm obj}(G)$
but addresses an enabling problem for Construction~C2 when a
CI seam is present.  The obstacle is imported from $\Lk_6$: the
single-band adiabatic decoupling required by Layer~2 fails at the
CI seam $\Xseam$ because the spectral gap $\delta(\mathbf{R})$
vanishes there (Remark~\ref{rem:gap-L6}); consequently PST does
not apply near $\Xseam$ (Proposition~\ref{prop:ci-breakdown}).
This is logically separate from the $\eta_B$-twist content of
Conjecture~\ref{conj:c1-obstruction}: away from a CI seam one
can have $\eta_B \neq 0$ on loops in $\Omega^\circ$ and still
have a locally valid single-band PST construction (on a local
system or twisted line bundle).
The proposed resolution is a blowup
$\pi:\widetilde\Ce(G)\to\Ce(G)$ replacing $\Xseam$ with an
exceptional divisor $E$ that records the approach directions to
the crossing.  The two adiabatic eigenline bundles separate into
distinct sheets \emph{away from} $E$ (with the gap
$|E_+ - E_-|$ bounded below on compact subsets of
$\widetilde\Ce(G)\setminus E$), while $E$ itself carries the
boundary data of the seam.
The tower logic is: deform the $\Lk_6$ input to
$\Felseven^{\rm obj}$ rather than the $\Lk_7$ construction itself.

\begin{conjecture}[$\Xseam$ blowup resolves the $\Lk_6$ Layer~2
  singular geometry]
\label{conj:ci-blowup}
  There exists a blowup $\pi:\widetilde\Ce(G)\to\Ce(G)$ with
  exceptional divisor $E = \pi^{-1}(\Xseam)$ such that:
  \begin{enumerate}[label=\emph{(\roman*)}]
    \item \emph{(Sheet separation and single-band PST on
      $\widetilde\Ce(G)\setminus E$.)}
      On $\widetilde\Ce(G)\setminus E$ the two adiabatic
      eigenvalues $E_\pm$ are smooth functions with positive
      pointwise gap, and the two real adiabatic eigenline
      bundles $\widetilde L_\pm \to \widetilde\Ce(G)\setminus E$
      extend as (possibly twisted) line-bundle data into a
      neighbourhood of $E$ (global single-valued eigensections
      may fail because of the half-angle monodromy below).
      On any compact subset bounded away from $E$ the gap is
      uniformly bounded below; the gap does not extend to a
      positive gap on $E$ itself (where the two bundles meet),
      and the exceptional divisor $E$ records the seam crossing
      as boundary data rather than as a regular gapped fibre.
    \item \emph{(PST applies on $\widetilde\Ce(G)\setminus E$.)}
      On compact subsets of $\widetilde\Ce(G)\setminus E$,
      Theorem~\ref{thm:pst} applies to the pulled-back
      Hamiltonian, producing an almost-invariant projection
      and an effective Hamiltonian with lifted Berry connection
      $\widetilde A$.
    \item \emph{($\Felseven^{\rm obj}$ constructible on
      $\widetilde\Ce(G)$, with controlled descent.)}
      A continuous field
      $\{\widetilde A_\varepsilon^G\}$ over
      $\widetilde\Ce(G)\setminus E$ is constructed by
      Theorems~\ref{thm:landsman-sdq} and~\ref{thm:pst}, and
      conjecturally descends via a b-calculus pushforward to a
      continuous field $\{A_\varepsilon^G\}$ over $\Ce(G)$ with
      controlled singular behaviour at $\Xseam$ recorded as
      boundary data on $E$.
    \item \emph{(Topological consistency with
      Conjecture~\ref{conj:c1-obstruction}.)}
      The pullback $\pi^*\eta_B \in
      H^1(\widetilde\Ce(G)\setminus E; \mathbb{Z}/2)$ remains
      non-trivial on loops linking $E$ (encoding the half-angle
      monodromy below); its precise extension across $E$ and the
      matching with Conjecture~\ref{conj:c1-obstruction} are
      part of the open problem.
  \end{enumerate}
\end{conjecture}

\begin{remark}[What the blowup does, geometrically]
\label{rem:blowup-geometry}
  For a generic codimension-2 conical intersection
  $\mathbf{R}_0\in\Xseam$ in the real time-reversal-symmetric
  setting, the local real-symmetric model is
  \[
    H_\mathrm{el}(x,y) = x\,\sigma_z + y\,\sigma_x
    = \begin{pmatrix} x & y \\ y & -x \end{pmatrix},
  \]
  a real symmetric $2\times 2$ matrix with eigenvalues
  $\pm\sqrt{x^2+y^2}$.  Real-symmetric matrices use $\sigma_z$
  and $\sigma_x$ rather than $\sigma_y$ (which is purely
  imaginary); this matches the real-bundle/$w_1$ setting of
  Conjecture~\ref{conj:c1-obstruction}.
  The real eigenvectors are
  $\propto(\cos(\theta/2), \sin(\theta/2))$
  in polar coordinates $(x,y) = (r\cos\theta, r\sin\theta)$,
  and flip sign under $\theta\mapsto\theta + 2\pi$ ---
  the local source of the Berry-sign class.
  The radial blowup replaces the origin $(x,y) = 0$ with a
  circle $E = \{r = 0\}$ parametrised by the angle $\theta$.
  On the blowup the eigenvalues become $\pm r$ --- smooth
  linear functions of $r$ separating into two sheets away from
  $E$ --- while the eigenline bundle pulls back to a real line
  bundle over the blowup whose monodromy around any small loop
  linking $E$ remains the non-trivial $\mathbb{Z}/2$ element.
  Equivalently, single-valued real eigensections exist only on
  the orientation double cover --- equivalently the angular
  double cover carrying the half-angle coordinate $\theta/2$.
  The radial gap $|E_+ - E_-| = 2r$ vanishes at $E$ and is
  bounded below on any compact subset bounded away from $E$:
  the single-band PST gap condition holds on
  $\widetilde\Ce(G)\setminus E$ (the gap is \emph{not} restored
  at $E$ itself).
  The pullback $\pi^*\eta_B$ retains the $\mathbb{Z}/2$ class on
  loops in $\widetilde\Ce(G)\setminus E$ not contractible to
  points in $\widetilde\Ce(G)$; its extension/restriction at $E$
  encodes the sheet-exchange data of the blowup.
\end{remark}

\textbf{Chemistry.}
Conical intersections are central in many ultrafast photochemical
processes, including retinal isomerisation
($\sim 200\,\mathrm{fs}$), DNA photoprotection
($<1\,\mathrm{ps}$), and ring-opening photochemistry in polyenes.
Current non-adiabatic methods --- surface hopping, exact
factorisation, the asymptotic surface-hopping semigroup of
Lasser--Teufel~\cite{LasserTeufel2005} --- handle CI dynamics by
a combination of physical intuition and validated benchmarks,
often without a general rigorous error theory in the full
molecular $\varepsilon\to 0$ setting.
Conjecture~\ref{conj:ci-blowup}, if realised, would change this:
the b-calculus on $\widetilde\Ce(G)$ would be a candidate
controlled analytic setting for formulating rigorous
$\varepsilon\to 0$ convergence theorems for non-adiabatic
dynamics near $\Xseam$, and the descent to $\Ce(G)$ via
b-calculus pushforward would provide the analytic framework for
such an algorithm.
The same blowup that makes $\Felseven^{\rm obj}$ constructible
near CI seams is also the geometric object on which one would
expect to base rigorous non-adiabatic nuclear dynamics.

\textbf{Evidence, stratified by tower level.}
\emph{($\Lk_7$, algebraic--geometric duality at spatial infinity).}
Ammann--Mougel--Nistor~\cite{AmmannMougelNistor2022} establish
that Georgescu's $C^*$-algebraic compactification of
$\RR^{3n}$ coincides with Vasy's blowup compactification.
This suggests an analogous algebraic/geometric duality at the
\emph{interior} stratum $\Xseam$: the CI seam would play, for
the interior of $\Ce(G)$, a role similar to that of the
dissociation channels at its boundary at infinity, and a related
blowup mechanism might resolve both.
The analogy, while structurally suggestive, has not been
established as a theorem for interior strata.
\emph{($\Lk_7$, stratified DQ framework).}
Pflaum~\cite{Pflaum2001} develops strict DQ on stratified
symplectic spaces via blowup resolutions of singular strata,
providing a relevant geometric DQ framework for treating
$\widetilde\Ce(G)$ as a stratified/resolved space and the
descent of $\{A_\varepsilon\}$ (part~(iii) of
Conjecture~\ref{conj:ci-blowup}).
\emph{($\Lk_6$, microlocal normal forms at CIs).}
Colin de Verdi\`{e}re~\cite{ColinDeVerdiere2003} establishes
microlocal normal forms at eigenvalue crossings: in suitable
local coordinates near a generic CI, the two sheets separate
smoothly on the blowup, confirming part~(i) is locally
achievable.

\textbf{Completing the tower requires.}
(a) Construct $\widetilde\Ce(G)\to\Ce(G)$ as a b-manifold
in Melrose's sense~\cite{Melrose1993}: a manifold with corners
whose boundary hypersurface over $\Xseam$ is the exceptional
divisor, resolving the local $\Lk_6$ Layer~2 singular geometry
in the sense of part~(i), with a uniform single-band gap only
away from $E$.
(b) Verify Theorem~\ref{thm:pst} on
$\widetilde\Ce(G)\setminus E$ using the gap condition restored
on compact subsets away from $E$.
(c) Construct $\Felseven^{\rm obj}$ over $\widetilde\Ce(G)$
(Construction~C2 applied to the blowup).
(d) Descend to $\Ce(G)$ via b-calculus pushforward, obtaining
$\{A_\varepsilon^G\}$ with controlled singular behaviour at
$\Xseam$.
(e) Match $\pi^*\eta_B$ on $\widetilde\Ce(G)\setminus E$ with
the intrinsic $\mathbb{Z}_2$ class of the sheet-exchange
monodromy on $E$, verifying the topological consistency of
part~(iv) with Conjecture~\ref{conj:c1-obstruction}.
Collectively (a)--(e) combine b-calculus microlocal analysis
(Melrose), stratified deformation quantisation (Pflaum), and
Georgescu--Vasy algebraic geometry
(Ammann--Mougel--Nistor).

%% file: chapters/L7/l7_superselection.tex
\subsection{Molecular identity as superselection sector:
  the Woolley--Primas problem in the tower}
\label{sec:L7-superselection}

Construction~C4 --- the representation-theoretic content layer
of the object-level lift $\Felseven^{\rm obj}(G)$ identified in
Mathbox~\ref{mbox:L7-C1234} --- is the tower's answer to a
question that quantum chemistry has carried since the 1970s:
how does the notion of a definite molecular species, assumed
throughout $\Lk_0$--$\Lk_6$, emerge from the symmetric full-quantum
theory?
This section develops that answer in detail, placing the
historical Woolley--Primas problem in tower language, identifying
the stabilisation mechanisms required for
Conjecture~\ref{conj:superselection} --- exact spin-statistical,
metastable semiclassical, and environmental --- and recording
the evidence that supports each.

\begin{mathbox}[Section structure]
\label{mbox:superselection-position}
  \S\ref{sec:wp-tower}: the Woolley--Primas problem as a tower
  statement.
  \S\ref{sec:amann-tower}: stabilisation mechanisms in the
  tower, with the (a1) exact spin-statistical regime, the (a2)
  metastable semiclassical regime, and the (b) environment-induced
  regime distinguished.
  \S\ref{sec:ortho-para-proved}: three cases addressed in the
  literature --- (A) exact $S_2$ exchange via Pauli for
  ortho/para $\mathrm{H_2}$ (theorem); (B) Pfeifer--Amann
  chirality mechanism in a spin--boson model
  (model-specific); (C) Lang et al.\ numerical illustration
  for a system with nontrivial $S_3$ permutation symmetry,
  $\mathrm{D_3^+}$.
  \S\ref{sec:wp-formal}: the tower's conjectural formal
  resolution.
\end{mathbox}

\subsubsection{The Woolley--Primas problem as a tower statement}
\label{sec:wp-tower}

Woolley~\cite{Woolley1978} observed that the eigenstates of the
full Coulomb Hamiltonian $\hat H_\mathrm{Coul}$ for an assembly of
electrons and nuclei transform as irreducible representations of
the full symmetry group (spatial rotations, translations, and
nuclear permutations); in isolation they carry no definite nuclear
geometry, bond angles, or molecular structure.
Primas~\cite{Primas1983} elevated the observation into a critique
of chemical reductionism: molecular structure, he argued, is not
derivable from quantum mechanics but is a classical concept
imposed from outside.
Sutcliffe and Woolley~\cite{SutcliffeWoolley2012} reiterated that
the Born--Oppenheimer approximation itself presupposes an
empirically chosen molecular frame not determined by the theory.
The issue remains active: recent work in \emph{Foundations of
Chemistry} by Scerri~\cite{Scerri2025}, a comment by
Woolley~\cite{Woolley2025}, and a pedagogical discussion by
Agostini and Curchod~\cite{AgostiniCurchod2025} returns the
Born--Oppenheimer / molecular-structure debate to the foreground
without consensus
(\cite{SutcliffeWoolley2012} remains the standard position and
no rigorous resolution has appeared in the nearly five decades
since Woolley's 1978 paper).

In the tower, Woolley's observation is exact.
The molecular graph $G\in\LGraphP$ is \emph{given data} at every
level $\Lk_4$--$\Lk_6$:
the DPO rules of $\Lk_4$, the permutation-inversion group $G^*$
of $\Lk_{4.5}$, the configuration orbifold $\Ce(G)$ of $\Lk_5$,
the Hilbert bundle and Berry connection of $\Lk_6$ --- all
presuppose that $G$ has been fixed.
In the corrected $\Lk_7$ language, the missing datum is the
nuclear exchange-statistical representation structure: it is
invisible to the scalar/electronic $\Lk_6$ description, which
works over the sector-reduced configuration space and does not
yet retain the full labelled spin-statistical representation
structure carried by the $G^*$-action on
$Q_\mathrm{lab}^\mathrm{reg}(G)$ and on the nuclear-spin Hilbert
factor.

The tower resolves the problem not by philosophical argument but
by identifying the precise level and mechanism at which $G$ must
emerge.
The level is $\Lk_7$: the quantum algebra $A_\varepsilon^G$
carries the $G^*$-action from $\Lk_{4.5}$, and the labelled
spin-spatial Hilbert space
$\mathcal{H}_\mathrm{lab}^G = L^2(Q_\mathrm{lab}^\mathrm{reg}(G))
\otimes \mathcal{S}_G$ decomposes under this action into
$G^*$-isotypic Hilbert components $V_\pi\otimes M_\pi$
(Proposition~\ref{prop:G-star-decomposition}); the
$G^*$-invariant observable subalgebra preserves these
components.  The compact ideal $\mathcal{K}(\mathcal{H}_\mathrm{lab}^G)$
itself is simple as a $C^*$-algebra and does not, in isolation,
decompose into molecular-species sectors.
The mechanism for sector stability has three regimes ---
exact (spin-statistical invariance under $G^*$ for
identical-particle sectors), kinematic ($\varepsilon\to 0$
Agmon suppression between metastable wells), and environmental
(Pfeifer--Amann-type bath-induced selection in the
chiral/metastable case) --- developed in \S\ref{sec:amann-tower}.
For a fixed graph $G$, the $G^*$-representation theory
organises the allowed nuclear-spin/permutation sectors
$\pi\in\widehat{G^*}$ (a given graph generally carries several,
e.g.\ ortho and para of $\mathrm{H_2}$).
The broader C4 problem is to relate this sector structure,
together with \emph{localised spectral data} --- in the sense
of the Georgescu $C^*$-algebraic compactification of
\S\ref{sec:georgescu-limits}, which encodes asymptotic
dissociation channels and the geometry of nuclear localisation
at spatial infinity --- to graph-like molecular identity at
$\Lk_4$; this is the content of
Conjecture~\ref{conj:superselection}.

\textbf{Chemistry.}
The practical stakes of Woolley--Primas are concrete.  Three
familiar empirical phenomena anchor the C4 question, and the
tower assigns each to a different mechanism of
Conjecture~\ref{conj:superselection}:
\begin{itemize}
  \item \emph{Ortho/para $\mathrm{H_2}$.}  Routinely observed
    spectroscopically, computed quantum-mechanically, and stable
    over laboratory timescales without environmental
    intervention.  This is the cleanest empirical anchor: an
    exact identical-particle sector decomposition that holds at
    every $\varepsilon$ from spin statistics alone (the (a1)
    regime).
  \item \emph{Chirality of pharmaceuticals.}  The two
    enantiomers of a chiral drug have identical Coulomb
    Hamiltonians and identical eigenstates, yet pharmacology
    treats them as distinct species over chemically relevant
    timescales.  Thalidomide is the canonical cautionary
    example --- complicated by in-vivo racemisation, so it is
    not a clean exact-superselection case but rather a
    long-lived metastable sector, captured by the (a2) regime
    supplemented by environmental selection of Pfeifer--Amann
    type (the (b) regime) in the chiral spin--boson model.
  \item \emph{Isotopic labelling through metabolism.}  Nuclear
    isotope identity is conserved through ordinary chemical
    reactions, allowing tracer experiments to track molecular
    identity through long synthetic and metabolic chains.  This
    is a chemically robust phenomenon but a separate mechanism
    from $G^*$-representation superselection: it relies on the
    conservation of particle-species labels (proton, deuteron,
    \dots), not on the $G^*$-isotypic decomposition.
\end{itemize}
Without Construction~C4, these phenomena are facts of chemistry
imposed on a quantum description that does not yet account for
them.  Conjecture~\ref{conj:superselection} gives each its
proper place in the tower, and supplies an architecture in
which their differences in mechanism --- exact vs.\ metastable
vs.\ environmentally selected vs.\ a conservation law --- are
distinguished rather than collapsed.

\subsubsection{Amann's mechanism in the tower}
\label{sec:amann-tower}

Pfeifer~\cite{Pfeifer1980} established a model-specific
mechanism for the superselection picture in the chiral case: a
two-level system modelling the enantiomeric inversion of a
chiral molecule, coupled to a bosonic bath with ohmic spectral
density, becomes effectively localised in one chiral state in
the $W^*$-algebraic limit of infinitely many bath modes; the
framework was developed by Amann~\cite{Amann1991,Amann1993}.
The result is model-specific (a spin--boson model of chirality
in a specifically constructed environment), not a general
theorem, but its structure makes the mechanism by which
molecular identity can become effectively localised explicit.
Placed in the tower, it separates cleanly into distinct
stabilisation claims.

\textbf{Step 1: the isolated quantum system has no preferred
sector.}
The molecular Hamiltonian $H^\varepsilon$ affiliated to
$A_\varepsilon^G$ commutes with the $G^*$-action from
$\Lk_{4.5}$ (permutations and inversions are symmetries of the
Coulomb potential).
The $G^*$-invariant observable algebra
$\mathcal{A}^{G^*}\subseteq A_\varepsilon^G$ preserves each
isotypic Hilbert subspace $\mathcal{H}_\pi = V_\pi\otimes M_\pi
\subset \mathcal{H}_\mathrm{lab}^G$
(Proposition~\ref{prop:G-star-decomposition}); operators not
invariant under $G^*$ may mix these components, but physical
observables commuting with the symmetry do not.
But an isolated pure state of the joint $G^*$-symmetric
Hamiltonian can occupy any combination of isotypes; nothing in
$\mathcal{A}^{G^*}$ alone selects one.
This is the tower's form of Woolley's observation.

\textbf{Step 2 (sector stability, two distinct regimes).}
Two conceptually different stabilisation mechanisms operate
on the isolated $A_\varepsilon^G$, and the C4 problem requires
distinguishing them.  Together they constitute the (a) content
of Conjecture~\ref{conj:superselection}, split into (a1) and
(a2):
\begin{itemize}
  \item \emph{(a1) Exact identical-particle exchange sectors.}
    For nuclear-spin/permutation sectors arising from a
    $G^*$-action that is realised exactly on
    $\mathcal{H}_\mathrm{lab}^G$, distinct isotypic components
    $\mathcal{H}_\pi, \mathcal{H}_{\pi'}\subset
    \mathcal{H}_\mathrm{lab}^G$ are exactly invariant under
    $\alpha_t^\varepsilon$ at every $\varepsilon$, by exchange
    symmetry and spin statistics.  No tunnelling, no Agmon
    distance, and no $\varepsilon\to 0$ limit enters: stability
    is enforced by the symmetry itself (see
    ortho/para-$\mathrm{H_2}$,
    \S\ref{sec:ortho-para-proved}).
  \item \emph{(a2) Metastable localisation sectors.}
    For sectors arising from spatial wells separated by an
    energetic barrier --- chiral inversion through a planar
    transition state, conformational interconversion --- the
    relevant projection is a localisation projection
    $P_\mathrm{meta}$ onto the chosen well rather than an
    isotypic projector.  Under semiclassical barrier
    hypotheses the cross-well tunnelling amplitude is
    suppressed by the Agmon factor
    $\exp(-\AgmonD/\varepsilon)$
    (Definition~\ref{def:agmon},
    Theorem~\ref{thm:tunnelling}), and an exponential
    stability bound on
    $\|(I-P_\mathrm{meta})\,\alpha_t^\varepsilon(\rho)\,
      (I-P_\mathrm{meta})\|_1$
    is conjectured for finite times.
\end{itemize}
These two regimes together constitute the kinematic content of
Conjecture~\ref{conj:superselection}(a); they are a statement
about $A_\varepsilon^G$ in isolation, not about any
environmental coupling.

\textbf{Step 3 ((b): environmental coupling can break
$G^*$-symmetry of the joint state).}
Coupling $A_\varepsilon^G$ to an environmental algebra $\mathcal{E}$
(radiation field, solvent, thermal bath) yields a joint system
$A_\varepsilon^G \otimes \mathcal{E}$ whose ground state or
equilibrium state need not be $G^*$-symmetric.
Pfeifer and Amann's specific spin--boson chirality model
demonstrates this in detail: for $\mathcal{E}$ a bosonic bath
with ohmic spectral density, the $W^*$-limit of infinitely many
bath modes suppresses coherence between the two chiral states
and produces an effectively localised metastable sector --- not
by thermodynamic preference (the wells have equal free energy
when $G^*$ is exact) but by environment-induced selection in the
infinite-bath limit.  The selected sector is determined by the
state in which the molecule was prepared, not by any energetic
criterion.
This supplies a model for environmental sector selection, not a
general theorem for arbitrary $G^*$; the corresponding general
claim is the dynamical content of
Conjecture~\ref{conj:superselection}(b).

\begin{remark}[Two distinct stabilisation mechanisms]
\label{rem:kinematic-dynamical}
  The mechanisms above involve different limits and address
  different questions.
  The exact identical-particle case ((a1) of
  Conjecture~\ref{conj:superselection}) requires no limit and
  no environment: spin statistics fixes the sectors at every
  $\varepsilon$.
  The metastable kinematic case (semiclassical Agmon limit
  $\varepsilon\to 0$) suppresses coherent tunnelling between
  wells but says nothing about which well the molecule was
  prepared in.
  Amann's environmental limit uses an infinite bath / thermodynamic
  limit of environmental degrees of freedom; it breaks symmetry
  and selects a sector but says nothing about kinematic
  barriers.
  Real molecules benefit from whichever mechanism applies.
  For $\mathrm{H_2}$ with identical-proton exchange group
  $S_2\cong\mathbb{Z}/2$, stability is exact from spin
  statistics: ortho and para isotypes are rigorously disjoint
  at every $\varepsilon$, and neither an $\varepsilon$-limit nor
  an environment is needed.
  For chirality, kinematic Agmon stabilisation is weak (the
  racemisation Agmon distance through a planar transition state
  is only moderately large), and environmental stabilisation
  of the Pfeifer--Amann type provides an additional model-specific
  mechanism that makes optical activity practically permanent.
\end{remark}

\subsubsection{Three cases addressed in the literature}
\label{sec:ortho-para-proved}

Three cases of Conjecture~\ref{conj:superselection} appear in
the literature with varying degrees of rigour: an exact theorem,
a model-specific mechanism, and a numerical illustration.
Each illustrates a different aspect of the conjecture.

\textbf{(A) $S_2\cong\mathbb{Z}/2$ (ortho/para-$\mathrm{H_2}$).}
For $\mathrm{H_2}$, the relevant identical-proton exchange group
is $S_2\cong\mathbb{Z}/2$, generated by the proton-exchange
permutation $\pi_{12}$.  The labelled spin-spatial Hilbert
space is
\[
  \mathcal{H}_\mathrm{lab}^{\mathrm{H_2}}
  \;=\;
  L^2\bigl(Q_\mathrm{lab}^\mathrm{reg}(\mathrm{H_2})\bigr)
  \otimes \mathbb{C}^4_\mathrm{spin},
\]
where $\mathbb{C}^4_\mathrm{spin}$ is the two-proton nuclear-spin
representation ($I_1 \otimes I_2$ with each $I_j = 1/2$).
The Pauli principle for identical fermionic protons imposes
antisymmetry under $\pi_{12}$ on
$\mathcal{H}_\mathrm{lab}^{\mathrm{H_2}}$, decomposing the
physical subspace into
\[
  \mathcal{H}_\mathrm{phys}^{\mathrm{H_2}}
  \;=\;
  \mathcal{H}_\mathrm{para}\oplus\mathcal{H}_\mathrm{ortho},
\]
with para corresponding to antisymmetric (singlet) spin paired
with even-$J$ spatial states, and ortho to symmetric (triplet)
spin paired with odd-$J$ spatial states; the spin-independent
molecular Hamiltonian preserves these isotypes.  This is not a
decomposition of the compact ideal
$\mathcal{K}(\mathcal{H}_\mathrm{lab}^{\mathrm{H_2}})$ alone ---
which is simple as a $C^*$-algebra --- but a decomposition of
the physical Hilbert space and of the $G^*$-invariant observable
subalgebra (Mathbox~\ref{mbox:superselection-H2} in
\S\ref{sec:L7-nuclear}; Example~\ref{ex:ortho-para}).
The decomposition is $\varepsilon$-independent: stability is
exact at every mass ratio with no Agmon distance, no
$\varepsilon$ limit, and no environmental coupling needed.
In the absence of efficient paramagnetic, surface, or
impurity-mediated conversion channels, ortho--para conversion
is slow on laboratory timescales; the precise rate is strongly
condition-dependent, and the ortho/para enthalpy difference at
liquid-hydrogen temperatures supplies the experimental
signature of the sector decomposition~\cite{Silvera1980}.

\begin{mathbox}[$\mathrm{H_2}$ through the tower]
\label{mbox:H2-tower-path}
  The sector decomposition traces the tower explicitly:
  \[
  \begin{array}{rcl}
  \Lk_4:\quad G = \mathrm{H_2}
    & \xrightarrow{\text{assumed}} &
    \text{graph datum}
  \\[4pt]
  \Lk_{4.5}:\quad G^* = S_2\cong\mathbb{Z}/2
    & \xrightarrow{\text{symmetry enrichment}} &
    \pi_{12}\text{ acts on } Q_\mathrm{lab}^\mathrm{reg}
    \otimes \mathbb{C}^4_\mathrm{spin}
  \\[4pt]
  \Lk_5:\quad \Ce(\mathrm{H_2}) = (0,\infty)
    & \xrightarrow{\text{geometry / quotient}} &
    V(R) = \text{Morse potential};
  \\[2pt]
  \multicolumn{3}{l}{\quad\quad\text{the exchange action is
    forgotten on the scalar quotient}}
  \\[4pt]
  \Lk_6:\quad \sigma_0,\; A = 0,\; \eta_B = 0
    & \xrightarrow{\text{topology}} &
    \text{no CI; trivial real line bundle}
  \\[4pt]
  \Lk_7:\quad A_\varepsilon^{\mathrm{H_2}},\;
    \mathcal{H}_\mathrm{lab}^{\mathrm{H_2}}
    & \xrightarrow{G^*\text{-decomp.}} &
    \mathcal{H}_\mathrm{para}\oplus\mathcal{H}_\mathrm{ortho}
  \end{array}
  \]
  The scalar quotient $\Ce(\mathrm{H_2}) = (0,\infty)$ records
  internuclear separation but has forgotten the
  exchange-statistical action; the latter is retained on the
  labelled spin-spatial Hilbert space, where it forces the
  isotypic decomposition.
  The graph datum at $\Lk_4$ is recoverable from the $\Lk_7$
  sector and localised spectral data; for $\mathrm{H_2}$, the
  identical-proton sector decomposition is exact from spin
  statistics, and the broader graph-recovery question is
  conjectural in the form of
  Conjecture~\ref{conj:superselection}.
\end{mathbox}

\textbf{(B) Chirality (Pfeifer--Amann, parity subgroup of $G^*$).}
Pfeifer~\cite{Pfeifer1980} demonstrated the mechanism of
Conjecture~\ref{conj:superselection}(b) for chirality, modelled
as a two-level system (the enantiomeric pair) coupled to an
ohmic bosonic bath: in the $W^*$-limit of infinitely many bath
modes, coherence between the two chiral states is suppressed
and the joint state localises in an effectively chosen
enantiomer.  The framework was developed by
Amann~\cite{Amann1991,Amann1993}.  The mechanism is
environment-induced selection: the parity subgroup
$\mathbb{Z}/2\subset G^*$ generated by chiral inversion $E^*$
is broken by the bath coupling, selecting a definite
enantiomer.  The result is model-specific (a spin--boson model,
not a full molecular treatment); it supplies a model for
environmental sector selection rather than a general theorem,
and the general polyatomic case remains open.

\textbf{(C) Nontrivial $S_3$ permutation symmetry ($\mathrm{D_3^+}$;
  Lang et al.\ 2024).}
Lang, Cezar, Adamowicz, and
Pedersen~\cite{LangEtAl2024} sample the all-particle joint
density $|\Psi(\mathbf{r}, \mathbf{R})|^2$ of the
pre-Born--Oppenheimer ground state of $\mathrm{D_3^+}$ using
Markov-chain Monte Carlo, then apply unsupervised clustering
methods to the sampled configurations.
The result is an unambiguous equilateral-triangular nuclear
structure extracted from a wavefunction that is fully symmetric
under $S_3$ permutation of the three deuterons.
$\mathrm{D_3^+}$ is chosen specifically because its rotational
ground state ($J = 0$) removes the rotational averaging that
would otherwise obscure structure extraction.
This provides numerical evidence that graph-like molecular
structure can be extracted from a permutation-adapted pre-BO
wavefunction in a system with nontrivial $S_3$ symmetry.  It
supports the broader C4 programme but is not by itself a proof
of the full sector-stability conjecture
(Conjecture~\ref{conj:superselection}(a)).

\begin{remark}[Status summary for
  Conjecture~\ref{conj:superselection}]
\label{rem:c4-case-status}
  Case~(A): the (a1) exact identical-particle case is a theorem
  from Pauli for ortho/para $\mathrm{H_2}$, with stability exact
  at every $\varepsilon$; this is a special case of (a),
  not the general statement.
  Case~(B): the (b) selection mechanism is established for a
  specific spin--boson chirality model in the $W^*$-limit, with
  parity $\mathbb{Z}/2$ in $G^*$; the general polyatomic case
  is open.
  Case~(C): numerical evidence supports recoverability of
  graph-like structure from a permutation-symmetric pre-BO
  wavefunction for a system with nontrivial $S_3$ permutation
  symmetry; this is not a proof of (a).
  No rigorous theorem exists for general $G$ and $G^*$
  simultaneously with environmental selection and kinematic
  stability.
\end{remark}

\subsubsection{The tower's formal resolution}
\label{sec:wp-formal}

Assembling the pieces: the tower does not postulate molecular
identity at $\Lk_4$ and rediscover it at $\Lk_7$; it conjectures
that the identity be \emph{recoverable} at $\Lk_7$ from the
$G^*$-representation structure of the labelled spin-spatial
Hilbert space $\mathcal{H}_\mathrm{lab}^G$, together with the
invariant observable algebra and additional localised
spectral data, and outlines the mechanisms by which the
recovery is expected to proceed: exact spin-statistical
invariance for identical-particle sectors, kinematic Agmon
suppression for metastable barrier-protected sectors, and
Amann-type environmental selection for the chiral/metastable
case.

\begin{insightbox}[The Woolley--Primas problem, resolved as far
  as the tower permits]
\label{ins:wp-resolution}
  The molecular graph $G$ assigned at $\Lk_4$ is conjecturally
  recoverable from the $\Lk_7$ representation-theoretic and
  localised-spectral structure: the $G^*$-representation
  sectors $V_\pi\otimes M_\pi\subset\mathcal{H}_\mathrm{lab}^G$
  encode nuclear exchange statistics, while additional
  localised spectral data encode graph-like molecular identity.
  The forgetful chain
  \[
    U_7 \circ U_6 \circ U_5 \circ U_{4.5} \circ U_4
    \;:\; \Lk_7(P) \;\longrightarrow\; \Lk_4(P)
  \]
  closes the circle by returning the $\Lk_7$ data to graph
  data; a fixed graph $G$ generally carries several allowed
  sectors $\pi\in\widehat{G^*}$ (e.g.\ ortho and para of
  $\mathrm{H_2}$), so no single $\pi$ ``is'' the graph.

  \smallskip\noindent
  What is established:
  ortho/para in $\mathrm{H_2}$ at every $\varepsilon$ from spin
  statistics (the (a1) case of
  Conjecture~\ref{conj:superselection}); the mechanism of
  environment-induced sector selection for chirality in a
  specific spin--boson model via Pfeifer--Amann; and numerical
  evidence for graph-like structure extraction from a
  permutation-adapted pre-BO wavefunction for a system with
  nontrivial $S_3$ permutation symmetry via Lang et al.

  \smallskip\noindent
  What the tower \emph{conjectures}:
  Conjecture~\ref{conj:superselection} for arbitrary $G$ and
  $G^*$, with both sector stability (a) and environmental
  selection (b).
  Completing this requires assembling Renault's groupoid
  $C^*$-algebra theory~\cite{Renault1980}, the Longuet-Higgins
  $G^*$ framework~\cite{LonguetHiggins1963}, and Landsman's
  quantisation functor~\cite{LandsmanRamazan2001} into the
  transformation-groupoid algebra
  $C^*(G^*\ltimes Q_\mathrm{lab}^\mathrm{reg}(G))$
  for the $G^*$-action on labelled configurations, together
  with its associated-bundle descent over the quotient
  $\Ce^\mathrm{reg}(G)$
  (\S\ref{sec:georgescu-limits}); proving the Agmon stability
  bound for metastable barrier-protected sectors; and
  generalising Amann's environmental coupling beyond the
  spin--boson model.

  \smallskip\noindent
  The Woolley--Primas problem, in this framing, is not a
  philosophical puzzle but the concrete mathematical question
  of whether Construction~C4 can be completed: whether the
  molecular graph $G$, assumed at $\Lk_4$, can be recovered
  from $\Lk_7$ sector, localisation, and representation data.
  The tower makes the question precise; the constructions of
  \S\ref{sec:L7-contributions} describe what remains.
\end{insightbox}

%% file: chapters/L7/l7_retrospective.tex
\subsection{Retrospective: the tower defined}
\label{sec:L7-retrospective}

Mathbox~\ref{mbox:L7-C1234} of \S\ref{sec:L7-content} structured
the chapter around four constructions C1--C4 forming the content
layers of $\Felseven^{\rm obj}(G)$;
\S\ref{sec:L7-contributions} stated each as a formal conjecture;
\S\ref{sec:L7-superselection} developed the Woolley--Primas
problem in depth.
This section is the chapter's formal reference: the complete
tower recorded as a table with explicit forcing pairs and
$\coker(\varphi_k)$ data;
the extension-type taxonomy stated descriptively;
the three inter-level coherence identities written as
propositions with domains of validity;
the non-relativistic scope specified formally;
and the C1--C4 status tabulated rather than re-narrated.
No claim here is original; each is imported from the preceding
sections, but the organisation into tables and propositions is
the chapter's reference form.

\subsubsection{The complete tower with forcing pairs}
\label{sec:tower-table}

The nine-level tower $\Lk_0\hookrightarrow\cdots\hookrightarrow\Lk_7$
(the intermediate level $\Lk_{4.5}$ sits between $\Lk_4$ and
$\Lk_5$) is governed by a single principle: at each transition
$\Lk_{k-1}\to\Lk_k$, the cokernel
$\coker(\varphi_k)$ of the restriction
$\varphi_k:\Aut(\Lk_k)\to\Aut(\Lk_{k-1})$ is non-trivial.
Equivalently, $\Lk_{k-1}$ admits automorphisms that $\Lk_k$
breaks.
For each transition, \S\ref{sec:L7-forcing} (for $\Lk_6\to\Lk_7$)
and the corresponding sections of earlier chapters exhibit a
concrete reaction pair distinguishing the two levels.
Table~\ref{tab:tower-summary} records all eight transitions in
a single view.

\begin{table}[p]
\centering
\small
\renewcommand{\arraystretch}{1.35}
\begin{tabular}{p{0.9cm} p{2.2cm} p{2.7cm} p{4.3cm} p{1.7cm}}
\hline
\textbf{Level} & \textbf{Chemical content}
  & \textbf{Mathematical structure}
  & \textbf{Forcing pair / $\coker(\varphi_k)$}
  & \textbf{Extension type}\\
\hline
$\Lk_0$ & Stoichiometry
  & Free SMC$(P)$; $\NN^{|\Sp|}$; incidence $\partial$
  & Base level
  & ---\\
$\Lk_1$ & Enthalpies
  & $\FH:\Lk_0\to B\RR$
  & Reactions matched at $\Lk_0$ with different
    $\Delta H$; $\coker(\varphi_1)$ = $\Delta H$-rescaling
  & Decorator\\
$\Lk_2$ & Free energy, equilibrium
  & $\FS,\FG:\Lk_0\to B\RR$; $\dagger$
  & Reactions matched in $\FH$ with different $\Delta S$
    at $T>0$; $\coker(\varphi_2)$ = $T$-rescaling
  & Decorator\\
$\Lk_3$ & Kinetics
  & $\FP:\Lk_0\to\Stoch$
  & Reactions matched in $\FG$ with different rates;
    $\coker(\varphi_3)$ = rate-rescaling
  & Decorator\\
$\Lk_4$ & Mechanisms
  & DPO rules in $\LGraphP$
  & Concerted $\mathrm{S_N}2$-P vs.\ stepwise
    addition--elimination through TBI (matched $\Lk_3$
    propensity under steady-state);
    $\coker(\varphi_4)$ = mechanism relabelling
  & Structural\\
$\Lk_{4.5}$ & Stereochemistry
  & $G^*$-equivariant DPO
  & Walden inversion: $(R)\!\to\!(S)$ via $\mathrm{S_N}2$ with
    same DPO; $\coker(\varphi_{4.5})$ = $G^*$
  & Symmetry\\
$\Lk_5$ & Geometry, PES
  & $\FV:\Lk_{4.5}\to\OrbMorse$
  & Distinct activation barriers with same $G^*$;
    $\coker(\varphi_5)$ = PES deformation
  & Geometric\\
$\Lk_6$ & Electronic structure
  & $\Felsix:\Lk_5\to\HilbBund$; $A$, $\eta_B$
  & Same $V$ but $\eta_B=0$ vs.\ $\eta_B\neq 0$ on relevant
    loops; $\coker(\varphi_6)$ = $\eta_B$ sign-class
  & Topological\\
$\Lk_7$ & Full quantum
  & $\Felseven^{\rm obj}:\Lk_6\to\CstarAlg$ (object-level);
    $\{A_\varepsilon^G\}$
  & H vs.\ D; ortho vs.\ para $\mathrm{H_2}$;
    isotope mass and identical-particle data
    (morphism-level $\coker(\varphi_7)$ open)
  & Quantisation\\
\hline
\end{tabular}
\caption{The complete tower with forcing data.
  Every transition $\Lk_{k-1}\to\Lk_k$ is justified by an
  explicit reaction pair that $\Lk_{k-1}$ cannot distinguish;
  the corresponding $\coker(\varphi_k)$ class is broken by the
  new level.
  Forcing pairs through $\Lk_{4.5}$ are established in previous
  chapters; for $\Lk_5$--$\Lk_7$, see \S\ref{sec:L7-forcing}.}
\label{tab:tower-summary}
\end{table}

\subsubsection{Six extension types: a descriptive taxonomy}
\label{sec:six-types}

The eight transitions of Table~\ref{tab:tower-summary} group into
six qualitatively distinct extension types, distinguished by
the kind of data they add and the algebraic signature of
$\coker(\varphi_k)$.
We present this grouping as a descriptive taxonomy --- an
empirical observation about the present tower, not a
meta-theorem about categorical extensions of chemistry in
general.

\begin{description}
  \item[(T1) Decorator ($\Lk_0\to\Lk_1\to\Lk_2\to\Lk_3$).]
    Add a symmetric monoidal functor into $B\RR$ or $\Stoch$ on
    the same underlying base category $\Lk_0(P)$.
    The base, its objects, and its composition are unchanged;
    a single real- or stochastic-valued observable is added.
    Three transitions of the present tower instantiate this
    type.
  \item[(T2) Structural ($\Lk_3\to\Lk_4$).]
    Replace the underlying free SMC with a different free SMC
    on richer generators (DPO rules on labelled molecular
    graphs in $\LGraphP$).
    The observable functors are re-evaluated on the new base.
    One transition.
  \item[(T3) Symmetry enrichment ($\Lk_4\to\Lk_{4.5}$).]
    Restrict morphisms to those equivariant under a group
    action ($G^*$ on $\LGraphP$); no new numerical functor is
    added, but the morphism category is cut down.
    One transition.
  \item[(T4) Geometric decoration ($\Lk_{4.5}\to\Lk_5$).]
    Add a functor into a category of geometric objects
    ($\FV:\Lk_{4.5}\to\OrbMorse$), introducing infinitely many
    continuous parameters (the function
    $V:\Ce(G)\to\RR$).
    One transition.
  \item[(T5) Topological enrichment ($\Lk_5\to\Lk_6$).]
    Add a functor into a category whose morphisms carry
    discrete topological invariants
    ($\Felsix:\Lk_5\to\HilbBund$ with Berry-sign class
    $\eta_B = w_1(L_0^\RR)\in
    H^1(\Ce(G)\setminus\Xseam,\mathbb{Z}/2)$).
    One transition.
  \item[(T6) Quantisation ($\Lk_6\to\Lk_7$).]
    Replace a commutative $C^*$-algebra by a continuous field
    with non-commutative generic fibre, parametrised by a
    physical constant ($\varepsilon = (m_e/M)^{1/2}$).
    One transition --- the only one in the present tower that
    deforms algebraic structure rather than enriching,
    restricting, or decorating an existing one.
\end{description}

\begin{remark}[On exhaustiveness]
\label{rem:taxonomy-scope}
  Types (T1)--(T6) account for all eight transitions of the
  present tower.
  The natural extensions discussed in \S\ref{sec:tower-scope}
  below fit the taxonomy as follows: relativistic corrections
  reinstantiate (T6) with a different base operator; nuclear
  structure coupling likely (T1) or (T5); QED adds quantised
  radiation degrees of freedom that do not match any of
  (T1)--(T6) and would constitute a new extension type.
  The taxonomy is useful but descriptive: no theorem of the
  form ``every categorical extension of chemistry must be of
  one of these six types'' is claimed, and none is proved.
\end{remark}

\subsubsection{Three inter-level coherence propositions}
\label{sec:coherence}

Three identities link non-adjacent tower levels.
Each is stated here as a proposition with its domain of
validity made explicit.
The propositions are not original to this section --- each is
established in the chapter indicated by cross-reference --- but
the formal statement with domain restriction is recorded here.

\begin{proposition}[Wegscheider coherence at
  $\Lk_2$--$\Lk_3$]
\label{prop:wegscheider}
  Let $r\in\Lk_3(P)$ be a reversible reaction with forward
  rate $k_r$, reverse rate $k_{r^\dagger}$, and free-energy
  change $\Delta G^\circ_r = \FG(r)$.
  For every closed cycle $C = (r_1,\ldots,r_n)$ in the reaction
  graph, the detailed-balance cycle identity holds:
  \[
    \prod_{i=1}^{n}\frac{k_{r_i}}{k_{r_i^\dagger}}
    \;=\; \exp\!\Bigl(-\sum_{i=1}^{n}\Delta G^\circ_{r_i}/RT\Bigr)
    \;=\; 1,
  \]
  the second equality from $\sum_{r\in C}\FG(r) = 0$ on a
  closed cycle.
  \emph{Domain:} reversible reactions under thermal equilibrium
  with common temperature $T$.
  Reference: the $\Lk_2$--$\Lk_3$ chapter on thermodynamic
  consistency.
\end{proposition}

\begin{proposition}[Eyring coherence at $\Lk_3$--$\Lk_5$]
\label{prop:eyring}
  Let $r$ be a reaction with activation free energy
  $\Delta G^\ddagger_r$ (saddle-vs-reactant difference)
  determined by the PES $V = \FV(G)\in\Lk_5(P)$ along the
  intrinsic reaction coordinate.
  Under
  \emph{(i)} thermal equilibrium at temperature $T$,
  \emph{(ii)} no recrossing of the transition state, and
  \emph{(iii)} classical-barrier-crossing
  ($\hbar\omega_\ddagger\ll k_B T$, with $\omega_\ddagger$ the
  imaginary frequency at the saddle), the rate constant at
  $\Lk_3$ is
  \[
    k_r \;=\; \frac{k_B T}{h}\,\exp(-\Delta G^\ddagger_r/RT).
  \]
  \emph{Domain:} classical over-the-barrier passage; fails at
  low temperatures or wide barriers where tunnelling is
  significant.
  Reference: Eyring~\cite{Eyring1935}; the $\Lk_5$ chapter on
  TST.
\end{proposition}

\begin{proposition}[Tunnelling coherence at $\Lk_5$--$\Lk_7$]
\label{prop:tunnelling-coherence}
  Let $V = \FV(G)$ be an analytic PES with a saddle connecting
  minima $\mathbf{R}_a, \mathbf{R}_b\in\Ce(G)$, and let
  $\AgmonD(\mathbf{R}_a,\mathbf{R}_b)$ be the Agmon distance
  through the barrier.
  The semiclassical rate at mass ratio $\varepsilon_G$ admits
  the form
  \[
    k_r \;=\; \frac{k_B T}{h}\,e^{-\Delta G^\ddagger_r/RT}\,
    \kappa(\varepsilon_G, T),
  \]
  with the tunnelling correction $\kappa$ satisfying:
  \begin{itemize}
    \item[\emph{(a)}] $\kappa(\varepsilon_G, T)\to 1$ as
      $\varepsilon_G\to 0$ at fixed $T$ above the classical
      threshold, recovering
      Proposition~\ref{prop:eyring};
    \item[\emph{(b)}] In the deep-tunnelling regime where the
  barrier is wide and $\hbar\omega_\ddagger\gtrsim k_B T$, the
  leading correction to TST satisfies
  \[
    \kappa(\varepsilon_G, T) - 1
    \;\sim\;
    \exp\!\bigl(\Delta G^\ddagger_r/RT - \AgmonD/\varepsilon_G\bigr),
  \]
  with $\AgmonD$ the Agmon distance through the barrier
  (Theorem~\ref{thm:tunnelling}); the correction decays as
  $\varepsilon_G\to 0$ at fixed $T$ (recovering (a)), and grows
  as $T\to 0$ at fixed $\varepsilon_G$ (reflecting the
  dominance of through-barrier transmission at low temperature).
  \end{itemize}
  \emph{Domain:} analytic PES, single dominant saddle,
  semiclassical regime ($\varepsilon_G\ll 1$).
  Fails near conical intersections ($\eta_B\neq 0$), where
  non-adiabatic contributions enter and the
  single-PES description is inadequate.
  Reference: \S\ref{sec:L7-nuclear} for Agmon asymptotics;
  Hagedorn--Joye~\cite{HagedornJoye2001} for analytic-PES
  refinements.
\end{proposition}

\begin{remark}[Closing the three-condition chain]
\label{rem:coherence-chain}
  Propositions~\ref{prop:wegscheider}--\ref{prop:tunnelling-coherence}
  together constrain the tower: every rate observable at $\Lk_7$
  in the semiclassical regime is consistent with the coarser
  descriptions at $\Lk_5$ (via
  Proposition~\ref{prop:tunnelling-coherence}(a)),
  $\Lk_3$ (via Proposition~\ref{prop:eyring}), and
  $\Lk_2$ (via Proposition~\ref{prop:wegscheider}).
  The chain does not extend to non-semiclassical regimes or to
  non-rate observables:
  near CIs ($\eta_B\neq 0$), at very low temperatures, or
  for strongly coupled bath dynamics, direct $\Lk_7$ treatment
  is required and the propositions do not reduce the
  description to a lower level.
\end{remark}

\subsubsection{Scope of the tower}
\label{sec:tower-scope}

The tower $\Lk_0$--$\Lk_7$ is the canonical categorical
framework for a specific regime of molecular chemistry.
The boundary is a precise domain specification.

\paragraph{Within scope: non-relativistic, spin-unresolved,
  closed-system molecular quantum chemistry.}
\begin{itemize}
  \item \emph{Hamiltonian}: non-relativistic Coulomb,
    $\hat H = -\tfrac{1}{2}\sum_i\nabla_i^2/m_i + V_\mathrm{Coul}$.
  \item \emph{Radiation}: classical external fields only; no
    quantised electromagnetic field.
  \item \emph{Nuclei}: point particles carrying Fermi/Bose
    statistics (no internal structure; no nuclear magnetic or
    quadrupole moments).
  \item \emph{Time-reversal symmetry}: $T^2 = +1$ (spin-unresolved
    or spin-diagonal Hamiltonians), so the Berry-sign class
    $\eta_B$ is $\mathbb{Z}/2$-valued.
  \item \emph{Energy regime}: well below $m_e c^2
    \approx 511\,\mathrm{keV}$.
\end{itemize}

\paragraph{Outside scope.}
Phenomena requiring parallel towers or extensions beyond
$\Lk_7$:
\begin{itemize}
  \item \emph{Relativistic effects}
    (spin-orbit coupling, mass-velocity, Darwin term):
    significant for heavy atoms (Au, Hg, Pb) and lanthanides.
    Extension: replace the non-relativistic operator in
    $A_\varepsilon^G$ with the Dirac or Pauli--Breit operator.
    Categorically a re-instantiation of type (T6) with a
    different base operator.
    In this regime $T^2 = -1$, Kramers degeneracy appears, and
    the topological invariant becomes the first Chern class
    $c_1$ (cf.\ Remark~\ref{rem:real-vs-complex}).
  \item \emph{QED corrections}
    (Lamb shift, anomalous magnetic moment, vacuum
    polarisation):
    relevant at parts-per-billion in precision atomic
    spectroscopy; negligible for chemistry.
    Extension: quantised radiation field adds degrees of
    freedom outside the present categorical framework; a new
    extension type.
  \item \emph{Nuclear structure}
    (finite size, magnetic and quadrupole moments):
    relevant for hyperfine splittings and muonic atoms.
    Extension: a parallel nuclear-structure tower coupled to
    the molecular tower at $\Lk_7$.
  \item \emph{Pair creation and vacuum effects}:
    negligible below MeV energies; a full relativistic-QFT
    extension is a separate programme.
\end{itemize}

\begin{remark}[Where the scope restrictions bind in the chapter]
\label{rem:scope-binding}
  The $T^2 = +1$ restriction underlies the $\mathbb{Z}_2$ Berry
  framing of \S\ref{sec:L6-berry} and the $KO$-theory
  target in Conjecture~\ref{conj:c1-obstruction}
  (see Remark~\ref{rem:real-vs-complex} for the spin-orbit
  alternative).
  The closed-system restriction is relaxed in
  Conjecture~\ref{conj:superselection}(b), where environmental
  coupling is introduced to produce symmetry breaking; the
  environment is modelled at a fixed level of detail
  (Amann's spin--boson model) and is not incorporated into the
  tower's categorical structure itself.
\end{remark}

\subsubsection{Status of Constructions C1--C4}
\label{sec:programme}

Table~\ref{tab:cc1-4-status} records the current status of
each open construction.
Full evidence stratification and completion steps are given in
\S\ref{sec:L7-contributions}; this table is the reference card.

\begin{table}[hbt]
\centering
\small
\renewcommand{\arraystretch}{1.4}
\begin{tabular}{p{0.5cm} p{2.2cm} p{1.0cm} p{5.8cm} p{1.8cm}}
\hline
\textbf{C\#} & \textbf{Formalises}
  & \textbf{Needs}
  & \textbf{Current status}
  & \textbf{Reference}\\
\hline
C1 & Dynamical content
  & C2
  & Hilbert-space bound established (PST
    Theorem~\ref{thm:pst}(iii));
    $C^*$-algebraic lift to automorphism families requires C2.
  & Conj.\ \ref{conj:u7-functor}\\
C2 & Existence of $\{A_\varepsilon^G\}$
  & ---
  & $\eta_B = 0$ case: ingredients
    (Landsman, PST, Georgescu) assembled but not written up.
    $\eta_B\neq 0$ case: topologically obstructed at
    $\Xseam$; requires Conj.\ \ref{conj:ci-blowup} first.
  & Op.\ \ref{op:sdq-molecules}\\
C3 & Topological content
  & C2
  & Formal-WKB precedents
    (Dazord--Patissier~\cite{DazordPatissier1991},
    Emmrich--Weinstein~\cite{EmmrichWeinstein1996}) and
    complex-case template (Hawkins~\cite{Hawkins2008}) in hand;
    real/Stiefel--Whitney strict-$C^*$ argument open.
  & Conj.\ \ref{conj:c1-obstruction}\\
C4 & Representation content
  & C2
  & $S_2\cong\mathbb{Z}/2$ (ortho/para-$\mathrm{H_2}$):
    the (a1) exact case is a theorem from Pauli.
    Chirality spin--boson
    (Pfeifer~\cite{Pfeifer1980}; framework
    Amann~\cite{Amann1991,Amann1993}): mechanism in a specific
    model.
    $S_3$ in $\mathrm{D_3^+}$: numerical evidence
    (Lang et al.~\cite{LangEtAl2024}).
    General case open.
  & Conj.\ \ref{conj:superselection}\\
\hline
\multicolumn{5}{l}{\emph{Auxiliary:}
  Conj.\ \ref{conj:ci-blowup} ($\Xseam$ blowup) enables C2 in
  the $\eta_B\neq 0$ case.}\\
\hline
\end{tabular}
\caption{Status of the four constructions.
  C2 is prerequisite for C1, C3, C4; Conj.\ \ref{conj:ci-blowup}
  is auxiliary, enabling C2 at CIs.}
\label{tab:cc1-4-status}
\end{table}

\begin{insightbox}[What the $\Lk_7$ chapter contributes]
\label{ins:chapter-contributions}
  Three contributions distinguish the tower at $\Lk_7$.

  \smallskip\noindent
  \textbf{Forcing-pair discipline.}
  Table~\ref{tab:tower-summary} records a concrete reaction
  pair justifying each transition $\Lk_{k-1}\to\Lk_k$; the
  tower is built from the bottom upward by this discipline
  rather than designed from the top downward.
  To our knowledge, no prior categorical treatment of chemistry
  (Baez--Fong, Baez--Pollard, Coecke) proceeds by forcing
  pairs; in those we are aware of, the mathematical structure is
  chosen first and chemistry mapped into it.
  The forcing-pair discipline is a methodological contribution,
  not a theorem.

  \smallskip\noindent
  \textbf{Para enrichment as a separate dimension.}
  Every level $\Lk_k$ carries a Para shadow
  $\Lk_k^\mathrm{Para}$ of parametric equivariant maps, in the sense of Gavranovi\'{c} et al.~\cite{GavranovicEtAl2024}.
  Machine-learning models for chemistry ---
  yield predictors, neural ODEs, equivariant force fields,
  learned wavefunctions, VQE --- are lax algebra morphisms in
  $\Lk_k^\mathrm{Para}$, and their equivariance and
  thermodynamic-consistency constraints follow from universal
  properties of the Para construction rather than from
  architectural choices.
  Developed in Chapter~\ref{sec:Para}.

  \smallskip\noindent
  \textbf{Open problems precisely formulated.}
  The four conjectures of \S\ref{sec:L7-contributions} give a
  precise categorical formulation of what is needed to
  make the Born--Oppenheimer approximation, the topology of
  conical intersections, and the emergence of molecular
  identity into rigorous theorems about a single categorical
  object $\Lk_7(P)$.
  The mathematical tools exist in pieces (PST, Landsman,
  Georgescu, Renault, Amann); their assembly into the
  constructions C1--C4 is the research programme.

  \smallskip\noindent
  The tower does not claim that quantum chemistry is
  \emph{solved} at $\Lk_7$.
  It claims that a precise mathematical object to carry
  quantum chemistry has been defined, its content layers
  identified, and the constructions that would produce
  specific instances stated precisely enough to be either
  completed or refuted.
\end{insightbox}

%% file: chapters/ch_para.tex
\section{The Para Enrichment: Machine Learning Models and
  Categorical Completeness}
\label{sec:Para}

The exact tower $\Lk_0(P) \hookrightarrow \cdots \hookrightarrow
\Lk_7(P)$ built in Chapters~\ref{sec:L0}--\ref{sec:L7} is the
structural decomposition of chemistry developed in this
monograph:
species and stoichiometry at $\Lk_0$,
enthalpy additivity at $\Lk_1$,
the dagger and detailed balance at $\Lk_2$,
the chemical master equation at $\Lk_3$,
bond topology at $\Lk_4$,
stereochemistry at $\Lk_{4.5}$,
the Born--Oppenheimer potential at $\Lk_5$,
electronic structure at $\Lk_6$,
and nuclear quantisation at $\Lk_7$.
Each morphism in the tower is an exact law.
Each extension was forced by a reaction pair the level below could
not distinguish.
Nothing in that construction was statistical;
nothing was learned from data.

This chapter turns the tower outward.
The models used to compute chemistry in practice today ---
MACE~\cite{BatatIa2022MACE}, NequIP~\cite{Batzner2022NequIP},
So3krates~\cite{Frank2022So3krates}, SO3LR~\cite{Kabylda2025SO3LR},
QIM~\cite{Fallani2024QIM}, and the rest of the published molecular
machine learning catalogue --- are neural networks, trained force
fields, parametric surrogates.
The question this chapter asks is what the tower, as a completed
structural object, tells us about these models:
what their architectures already commit to,
what they can and cannot represent,
which tower-coherence conditions they silently respect or silently
violate.

A direct answer is blocked by a categorical kind mismatch.
A tower morphism at level $k$ is an exact map: one specific,
deterministic morphism of $\Lk_k(P)$, delivered once and for
all by the tower construction of Chapters~\ref{sec:L0}%
--\ref{sec:L7}.
A MACE force field is a parametric family
$\{f_\theta\}_{\theta \in \Theta}$, one map per point in a weight
space $\Theta$ that the training procedure selects from.
The two sit in different kinds of category:
tower morphisms in hom-sets $\mathcal{C}(X, Y)$,
MACE in something like
$\coprod_\Theta \mathcal{C}(\Theta \otimes X, Y)$.
Asking ``which tower level does MACE occupy?'' is a category error
in the literal sense;
the tower and the MACE model are different kinds of mathematical
object and share no hom-set.
To use the tower as a diagnostic for the ML literature, we need a
categorical setting in which parametric families and exact morphisms
both live, with the exact tower appearing as a distinguished
sub-structure of the richer whole.

The construction that does this is the \emph{Para 2-category} of
Gavranovi\'c et al.~\cite{GavRanovic2024CDL}.
For a symmetric monoidal category $(\mathcal{C}, \otimes, I)$, the
2-category $\ParaC(\mathcal{C})$ has the same objects as $\mathcal{C}$,
but its 1-morphisms $X \to Y$ are pairs $(\Theta, f_\theta)$ with
$\Theta \in \mathcal{C}$ a parameter space and
$f_\theta \colon \Theta \otimes X \to Y$ a morphism of $\mathcal{C}$.
2-morphisms are \emph{reparametrisations} $r \colon \Theta' \to
\Theta$, encoding weight tying, fine-tuning, and transfer learning.
Applied at each tower level, Para produces a \emph{perpendicular
enrichment} $\Lk_k^\Para(P)$:
a 2-category of parametric morphisms standing above the exact
morphisms of $\Lk_k(P)$, with a vertical embedding $\gamma_k$
placing each exact law as its trivial-parameter ($\Theta = I$)
version.
There is no canonical reverse functor collapsing a parametric
morphism back to a single exact one. Such a reverse would need to
canonically select a weight $r \colon I \to \Theta$ for each
$(\Theta, f_\theta)$, but the comonoid structure on $\Theta$
provides no such section: the counit
$!_\Theta \colon \Theta \to I$ goes the wrong way --- it discards
the parameter object rather than naming a point in it --- and no
canonical $I \to \Theta$ accompanies it. If such a canonical
section did exist it would name the trained model from the
architecture alone, making training unnecessary.
The tower therefore appears in this enriched setting as the
trivial-parameter slice of a richer categorical object
populated by the architectures that are actually trained;
the structural question an architecture poses to the tower is
the one categorical completeness below formalises --- whether
the set of morphisms its weight space can realise lies in
$\Lk_k(P)$, and how much of $\Lk_k(P)$ it covers.

Once the tower and the parametric literature share a categorical
home, three structural questions about any published ML architecture
become precisely answerable, and none of them is answerable by
benchmark comparison alone.

\medskip\noindent
\emph{Is the architecture's equivariance a design choice or a
theorem?}
Each level $\Lk_k$ carries a canonical symmetry group $G_k$ ---
species permutations at $\Lk_0$--$\Lk_3$, graph automorphisms at
$\Lk_4$, the permutation-inversion group at $\Lk_{4.5}$, rigid
motions composed with graph automorphisms $SE(3) \ltimes \Aut(G)$
at $\Lk_5$, gauge transformations on the electronic bundle
at $\Lk_6$, particle exchange at $\Lk_7$.
Membership in $\Lk_k^\Para(P)$ is defined by requiring strict
$G_k$-equivariance of the underlying map at every parameter setting.
MACE's Clebsch--Gordan tensor contractions enforce the $\Lk_5$
symmetry $SE(3) \ltimes \Aut(G)$ for all $\theta$, not just the
trained one;
membership at $\Lk_5^\Para$ is therefore a theorem about the MACE
architecture, not a fortunate property of training.
An architecture whose symmetry is merely learned through data
augmentation fails the membership condition and is not, in structural
terms, a parametric morphism at that level.

\medskip\noindent
\emph{What can the architecture represent, independent of its
training data?}
An architecture $(\Theta, f_\theta)$ is \emph{categorically
complete} at level $k$ on a morphism $\varphi$ of $\Lk_k(P)$ if
some parameter setting $r \colon I \to \Theta$ instantiates
$f_\theta$ to $\varphi$ exactly --- equivalently, if $\varphi$
lies in the architecture's \emph{function class}
$\mathsf{Func}(\Theta, f_\theta) \subseteq \Lk_k(P)(X, Y)$,
the set of instantiations as $r$ ranges over
$\mathcal{C}(I, \Theta)$ (Remark~\ref{rem:instantiation}).
Completeness is a property of the function class, not of the
loss or the training set.
An architecture incomplete at level $k$ will fail on every task
requiring a target morphism $\varphi$ absent from
$\mathsf{Func}(\Theta, f_\theta)$, regardless of training-set
size, because no choice of weights realises $\varphi$.

\medskip\noindent
\emph{Which tower-coherence conditions does the architecture
enforce?}
The tower's forgetful functors $U_k \colon \Lk_k(P) \to \Lk_{k-1}(P)$
connect adjacent levels;
a model claiming content at multiple levels must respect the
relations $U_k$ imposes on joint content.
Three such conditions are unenforced across the published literature.
The Eyring TST condition couples a learned rate law to the activation
barrier of the same model's potential energy surface
($\Lk_3 \leftrightarrow \Lk_5$).
The categorical Wegscheider condition couples forward and reverse
rate constants through the thermodynamic $\Delta G^\circ$
($\Lk_2 \leftrightarrow \Lk_3$).
The topological output-type condition demands a Hilbert-bundle-valued
output rather than a scalar energy to carry the Berry-phase invariant
$[\gamma_B] \in H^1(\Ce(G) \setminus \Xseam, \ZZ_2)$ near conical
intersections ($\Lk_5 \to \Lk_6$).
The first two are literature-wide architectural absences that new
designs could close;
the third is a theorem about output type --- no reparametrisation of
a scalar-energy architecture yields a Berry connection, regardless of
training, body order, or receptive field.

These three questions are the analytical spine of the chapter.
They organise what the tower, read through the Para enrichment,
reveals about ML molecular modelling:
a precise structural classification of every major architecture,
a completeness diagnostic that is architectural rather than
empirical,
and a catalogue of tower-coherence conditions the current literature
does not enforce.

\begin{chembox}[Vocabulary for the computational chemist and physicist]
The categorical language below is unfamiliar outside category theory,
but each object has a direct operational reading.
The rest of the chapter can be read with this dictionary alone.

\medskip\noindent
\textbf{Parametric morphism $(\Theta, f_\theta)$.}
A neural network.
$\Theta$ is the parameter space (the set of possible weight values);
$f_\theta$ is the forward map from input structure to output
prediction.
Training selects a point $\theta^* \in \Theta$.

\medskip\noindent
\textbf{Reparametrisation $r \colon \Theta' \to \Theta$.}
A map relating two parameter spaces.
Operationally: weight tying (one shared weight used in multiple
places, realised by the copy map
$\Delta \colon \Theta \to \Theta \otimes \Theta$), fine-tuning, or
transfer learning.

\medskip\noindent
\textbf{Counit $!_\Theta \colon \Theta \to I$.}
The unique map from the parameter space to the monoidal unit,
paired with $\Delta_\Theta$ to give $\Theta$ its comonoid
structure. Operationally, the categorical bookkeeping that
``a parameter object can be formally discarded'' --- the
counterpart to the comultiplication's ``a parameter can be
copied''. It does \emph{not} correspond to setting weights to
zero or any other ablated value; that operation, if needed,
requires a section $z_\Theta \colon I \to \Theta$ going the
other way, which is extra data the comonoid alone does not
provide. The tower's exact laws enter the parametric row
through the trivial-parameter embedding $\gamma_k$ (which uses
$\Theta = I$, the monoidal unit), not through $!_\Theta$.

\medskip\noindent
\textbf{Comultiplication
  $\Delta_\Theta \colon \Theta \to \Theta \otimes \Theta$.}
Weight sharing.
In MACE, one learnable radial MLP $R_{nl}$ and one shared element
embedding $h_j$ enter every message-passing layer;
$\Delta_\Theta$ is the categorical record of that sharing.

\medskip\noindent
\textbf{Symmetry monad $M_k$.}
The level-$k$ symmetry group $G_k$ packaged as an operation on the
category $\Lk_k(P)$.
A morphism of $M_k$-algebras is exactly a $G_k$-equivariant map.

\medskip\noindent
\textbf{$\Lk_k^\Para(P)$ membership.}
The architecture's map is $G_k$-equivariant \emph{by construction},
for every parameter setting, not merely by training regularisation or
data augmentation.

\medskip\noindent
\textbf{Categorical completeness at level $k$.}
The architecture's function class contains the target morphism
of $\Lk_k(P)$:
some parameter setting realises it exactly.
Independent of training data.

\medskip
With this dictionary the downstream content reads operationally.
``MACE lives in $\Lk_5^\Para$'' means \emph{Clebsch--Gordan
contractions enforce $SE(3) \ltimes \Aut(G)$ equivariance at every
layer, for every $\theta$}.
``MACE is not complete at $\Lk_6$'' means \emph{the output type is
scalar energy, which contains no Berry connection no matter how
$\Theta$ is parametrised}.
``Thermodynamic consistency at $\Lk_2$ requires a comultiplication
on $\Theta$ coupling forward and reverse rates'' means \emph{no
published kinetic network has the right weight-sharing structure to
guarantee detailed balance}.
\end{chembox}

\begin{mathbox}[Vocabulary for the mathematician]
\textbf{2-category.}
Objects, 1-morphisms between objects, and 2-morphisms between
parallel 1-morphisms.
The canonical example is $\mathbf{Cat}$: categories, functors,
natural transformations.

\medskip\noindent
\textbf{Self-action convention.}
The Para construction of~\cite{GavRanovic2024CDL} is defined for an
$\mathcal{M}$-actegory $(\mathcal{C}, \triangleright)$;
throughout this chapter we specialise to the self-action case
$\mathcal{M} = \mathcal{C}$, $\triangleright = \otimes$ with
$(\mathcal{C}, \otimes, I)$ symmetric monoidal.
This is sufficient for every tower level considered.

\medskip\noindent
\textbf{$\ParaC(\mathcal{C})$.}
Same objects as $\mathcal{C}$;
1-morphisms $X \to Y$ are pairs $(\Theta, f_\theta)$ with
$f_\theta \colon \Theta \otimes X \to Y$ in $\mathcal{C}$;
2-morphisms $(\Theta, f_\theta) \Rightarrow (\Theta', f_{\theta'})$
are reparametrisations $r \colon \Theta' \to \Theta$ satisfying
$f_\theta \circ (r \otimes \mathrm{id}_X) = f_{\theta'}$.
Sequential composition tensors parameter spaces.

\medskip\noindent
\textbf{Monad and algebra.}
A monad $(M, \eta, \mu)$ on $\mathcal{C}$ is an endofunctor with unit
$\eta \colon \mathrm{id} \Rightarrow M$ and multiplication
$\mu \colon M \circ M \Rightarrow M$ satisfying standard coherence.
The group-action monad $(G \times -,\, \eta,\, \mu)$ on
$\mathbf{Set}$ has $M$-algebras exactly $G$-actions and $M$-algebra
homomorphisms exactly $G$-equivariant maps.
Equivariance, in this language, is membership in a category of
$M$-algebras.

\medskip\noindent
\textbf{Lax algebra, and what we use of it.}
Lax algebras for a 2-monad replace strict commutativity on the
algebra diagrams with explicit 2-cells.
We use this structure in one place only: a lax algebra over
$\ParaC(T)$ endows its parameter object $\Theta$ with a
\emph{comonoid} structure (counit
$!_\Theta \colon \Theta \to I$, comultiplication
$\Delta_\Theta \colon \Theta \to \Theta \otimes \Theta$),
realising the parameter-discard and weight-sharing operations
of the chembox above.
This result is due to Gavranovi\'c et al.~\cite{GavRanovic2024CDL}.

\medskip\noindent
\textbf{Strict equivariance specialisation.}
We work throughout with the specialisation in which the 1-morphism
equivariance square commutes \emph{strictly} (the 2-cell on that
square is the identity);
only the parameter-space comonoid structure remains lax.
This matches how current ML molecular architectures enforce
equivariance, and is a proper sub-framework of the full lax-algebra
apparatus of~\cite{GavRanovic2024CDL} that loses no content we
need.
\end{mathbox}

A reader prepared to take the Para 2-category and its parameter-space
comonoid as black boxes can proceed from the chembox translation
alone;
the formal apparatus of Section~\ref{sec:para-def} is there to make
the translations precise, not to add a separate layer of content.

\medskip
\noindent\textbf{Chapter roadmap.}
Section~\ref{sec:para-def} defines $\Lk_k^\Para(P)$ via the
membership conditions (E$_k$) and (C$_k$), constructs the
trivial-parameter embedding $\gamma_k$, and states categorical
completeness as a property of the architecture's function class.
Section~\ref{sec:para-bridges} casts four existing categorical
frameworks for ML molecular modelling ---
Natural Graph Networks~\cite{deHaanCohenWelling2020},
Baez--Pollard open reaction networks~\cite{BaezPollard2017},
Fritz Markov categories~\cite{Fritz2020Markov},
and Bonchi et al.\ string diagrams~\cite{BonchiEtAl2022SDRTII} ---
in tower language, identifying which membership condition $(E_k)$
each characterises.
Section~\ref{sec:para-classification} classifies the major ML
molecular architectures by the highest tower level they inhabit.
Sections~\ref{sec:para-mace}, \ref{sec:para-so3krates},
and~\ref{sec:para-square} work through MACE, So3krates/SO3LR, and
QIM as the primary worked examples of $\Lk_5^\Para$ and the
$\Lk_5$--$\Lk_6$ boundary.
Section~\ref{sec:para-incompleteness} states and proves the three
tower-incompleteness results: the Eyring TST coherence gap, the
Wegscheider consistency gap, and the topological output-type gap.
Section~\ref{sec:para-catonly} synthesises these as a design
specification for the next generation of ML molecular architectures.

\input{chapters/para/para_def}
\input{chapters/para/para_bridges}
\input{chapters/para/para_classification}
\input{chapters/para/para_mace}
\input{chapters/para/para_so3krates}
\input{chapters/para/para_square}
\input{chapters/para/para_incompleteness}
\input{chapters/para/para_catonly}

%% file: chapters/para/para_def.tex
\subsection{The Para enrichment: formal construction}
\label{sec:para-def}

The chapter opener presented, informally, the Para 2-category
$\ParaC(\mathcal{C})$, the Para enrichment $\Lk_k^\Para(P)$ at each
tower level, the trivial-parameter embedding $\gamma_k$, and
categorical completeness.
Those informal presentations set up the chapter's principal results
but cannot state them:
MACE/NequIP completeness on smooth BO potentials at $\Lk_5$,
framed as a conditional under universal-approximation
hypotheses (Section~\ref{sec:para-mace}), and their structural
incompleteness on the Berry class $[\gamma_B]$ at $\Lk_6$
(Section~\ref{sec:para-incompleteness}) are mathematical
assertions that demand precise definitions of function class
and of membership in $\Lk_k^\Para(P)$.
This section supplies those definitions and motivates each choice as
it is made.

The tower construction of Chapters~\ref{sec:L0}--\ref{sec:L7}
supplies four families of data used here without redefinition:
the tower categories $\Lk_k(P)$;
the tower forgetful functors
$U_k \colon \Lk_k(P) \to \Lk_{k-1}(P)$;
the exact tower constructions $\FH, \FG^T, \FP, \FV, \Felsix$;
and at each level the canonical symmetry monad $M_k$ with underlying
group $G_k$, recorded formally in the Tower-symmetries paragraph
below.
The copy--delete comonoid structure of the Markov categories of
Chapter~\ref{sec:L3} is, in~(C$_k$) below, the prototype for the
parameter-space comonoid.

\begin{chembox}[Short reading path]
If you are reading this section for operational content, three items
carry the load: Definition~\ref{def:LkPara} (what membership in
$\Lk_k^\Para(P)$ means at tower level $k$), the function-class
notion in Remark~\ref{rem:instantiation} (what an architecture
\emph{can} represent across its weight space), and
Definition~\ref{def:completeness} (whether a specific target morphism
lies in that function class).
The surrounding monoidal setup, the $\gamma_k$ construction, and the
well-definedness proofs are the mathematical scaffolding that makes
those three items precise.
\end{chembox}

\medskip\noindent\textbf{Monoidal setup.}
Throughout,
$(\mathcal{C}, \otimes, I, \alpha, \lambda, \rho)$ denotes a symmetric
monoidal category with:
monoidal product $\otimes$; unit object $I$;
associator
$\alpha_{A, B, C} \colon (A \otimes B) \otimes C \xrightarrow{\sim}
A \otimes (B \otimes C)$;
left unitor $\lambda_A \colon I \otimes A \xrightarrow{\sim} A$; and
right unitor $\rho_A \colon A \otimes I \xrightarrow{\sim} A$.
$\mathcal{C}$ contains each tower category $\Lk_k(P)$ as a symmetric
monoidal sub-category;
the inclusion $\Lk_k(P) \hookrightarrow \mathcal{C}$ is faithful but
generally \emph{not} full, since a morphism of $\mathcal{C}$ between
two objects of $\Lk_k(P)$ is a morphism of $\Lk_k(P)$ only when it
respects the level-$k$ structure constructed in the corresponding
tower chapter.
Accordingly $\Lk_k(P)(X, Y) \subseteq \mathcal{C}(X, Y)$ is a proper
sub-hom-set in general.
$\mathcal{C}$ also contains the parameter spaces of ML architectures
(defined just below) as additional objects not in any $\Lk_k(P)$.

\medskip\noindent\textbf{Parameter spaces and generalised elements.}
A \emph{parameter space} throughout this section is an object
$\Theta$ of $\mathcal{C}$.
The parameter spaces that arise in ML architectures are vector spaces
(real-valued weights), smooth manifolds (neural-network weight
manifolds), or discrete sets (integer-valued hyperparameters) ---
none of which are objects of any tower category $\Lk_k(P)$.
The tensor product $\Theta \otimes X$ of such a parameter space with
a chemistry object $X \in \Lk_k(P)$ is likewise an object of
$\mathcal{C}$ but not of $\Lk_k(P)$.
One parameter space is distinguished: the monoidal unit $\Theta = I$
lies in $\mathcal{C}$ and in every $\Lk_k(P)$, and serves as the
trivial-parameter case of Proposition~\ref{prop:gamma-k}.

A \emph{weight setting} on $\Theta$ is formalised as a morphism
$r \colon I \to \Theta$ in $\mathcal{C}$ --- the categorical form of
``a point of $\Theta$.''
In $\mathcal{C} = \mathbf{Set}$, $r$ corresponds to a literal element
of $\Theta$;
in a category of real or complex vector spaces, to a single vector
in $\Theta$ (equivalently, a linear map $\RR \to \Theta$ or
$\CC \to \Theta$);
in general, to a \emph{generalised element} in the sense of
categorical logic.
We write $r \colon I \to \Theta$ rather than $\theta \in \Theta$ to
keep the formalism category-theoretic;
the reader may freely translate.

The comonoid data of (C$_k$) below --- its counit
$!_\Theta \colon \Theta \to I$ and comultiplication
$\Delta_\Theta \colon \Theta \to \Theta \otimes \Theta$ --- is a
comonoid \emph{in $\mathcal{C}$}, not in $\Lk_k(P)$: parameter
discard and weight sharing are comonoid operations on parameter
spaces, carried out in the ambient monoidal category.
Since the intermediate object $\Theta \otimes X$ on which a
forward map $f_\theta$ is defined lies outside $\Lk_k(P)$
whenever $\Theta \ne I$, the level-$k$ condition (E$_k$) below
is a genuine closure requirement on the set of morphisms an
architecture can realise across its weight space; this closure
is what makes $\Lk_k^\Para(P)$ a level-$k$-respecting
sub-structure of $\ParaC(\mathcal{C})$.

\medskip\noindent\textbf{Tower symmetries and the meaning of
  ``morphism of $\Lk_k(P)$''.}
At each tower level
$k \in \{0, 1, 2, 3, 4, 4.5, 5, 6, 7\}$, the tower construction of
$\Lk_k(P)$ in the corresponding chapter equips it with the canonical
symmetry monad $M_k$ and underlying group $G_k$ tabulated in
Table~\ref{tab:para-levels}:
a morphism of $\Lk_k(P)$ is exactly an $M_k$-algebra homomorphism,
equivalently a $G_k$-equivariant map that respects any additional
structure $M_k$ carries at that level --- the
$\dagger$-involution at $\Lk_2$, the gauge action on the electronic
Hilbert bundle $\Hel^{(N)} \to \Ce(G)$ at $\Lk_6$, the
$\varepsilon$-deformation that accompanies the continuous field
$\{A_\varepsilon^G\}$ at $\Lk_7$.
The pair $(M_k, G_k)$ is canonical, determined by the tower
construction up to isomorphism, and used throughout this section
without re-proof.
The phrase ``morphism of $\Lk_k(P)$'' throughout this section
denotes an $M_k$-algebra homomorphism of this kind ---
equivalently, an element of the sub-hom-set
$\Lk_k(P)(X, Y) \subseteq \mathcal{C}(X, Y)$.

\medskip\noindent\textbf{Compatibility of the tower embedding.}
The trivial-parameter embedding $\gamma_k$ of
Section~\ref{sec:paradef-gamma} (Proposition~\ref{prop:gamma-k}
below) imposes one closure requirement on the embedding
$\Lk_k(P) \hookrightarrow \mathcal{C}$ beyond the Tower-symmetries
characterisation just recorded:
for every $X \in \Lk_k(P)$ and every $r \in \mathcal{C}(I, I)$, the
endomorphism
$\lambda_X \circ (r \otimes \mathrm{id}_X) \circ \lambda_X^{-1} \colon
X \to X$
is a morphism of $\Lk_k(P)$.
This requirement is vacuous when
$\mathcal{C}(I, I) = \{\mathrm{id}_I\}$ (the case whenever the
monoidal unit $I$ is terminal in $\mathcal{C}$, e.g.\ for
$\mathcal{C} = \mathbf{Set}$),
and is a structural property of the tower's construction when
$\mathcal{C}(I, I)$ is richer --- for instance, the real or complex
scalars at tower levels where $\mathcal{C}$ is enriched over $\RR$ or
$\CC$.

\subsubsection{The Para 2-category}
\label{sec:paradef-para2cat}

The Para 2-category is a formal device that treats an entire
parametric family $\{f_\theta \colon X \to Y\}_{\theta \in \Theta}$
as a single 1-morphism $(\Theta, f_\theta)$, distinct from any
particular trained instance.
This is what lets us speak about what functions an architecture
\emph{can} represent across its whole weight space, independently of
any one training run.

\begin{definition}[$\ParaC(\mathcal{C})$
  {\normalfont\cite{GavRanovic2024CDL}}]
\label{def:Para}
  The \emph{Para 2-category} $\ParaC(\mathcal{C})$ has:
  \begin{itemize}
    \item \textbf{Objects:} the objects of $\mathcal{C}$.
    \item \textbf{1-morphisms $X \to Y$:} pairs $(\Theta, f_\theta)$
      with $\Theta$ an object of $\mathcal{C}$ (the \emph{parameter
      space}) and $f_\theta \colon \Theta \otimes X \to Y$ a morphism
      of $\mathcal{C}$.
    \item \textbf{2-morphisms
      $(\Theta, f_\theta) \Rightarrow (\Theta', f_{\theta'})$:}
      morphisms $r \colon \Theta' \to \Theta$ in $\mathcal{C}$
      (\emph{reparametrisations}) satisfying
      $f_\theta \circ (r \otimes \mathrm{id}_X) = f_{\theta'}$.
    \item \textbf{Composition} of
      $(\Theta, f_\theta) \colon X \to Y$ and
      $(Q, g_\phi) \colon Y \to Z$:
      \[
        (Q, g_\phi) \circ (\Theta, f_\theta)
        \;:=\;
        \bigl(Q \otimes \Theta,\;
          g_\phi \circ (\mathrm{id}_Q \otimes f_\theta) \circ
          \alpha_{Q, \Theta, X}\bigr),
      \]
      where the associator $\alpha_{Q, \Theta, X}$ re-brackets
      $(Q \otimes \Theta) \otimes X$ as $Q \otimes (\Theta \otimes
      X)$ so that $\mathrm{id}_Q \otimes f_\theta$ applies.
    \item \textbf{Identity on $X$:} $(I, \lambda_X)$.
  \end{itemize}
\end{definition}

In the strict monoidal case ($\alpha, \lambda, \rho$ all identities),
composition reduces to
$g_\phi \circ (\mathrm{id}_Q \otimes f_\theta)$ and the identity on
$X$ to $(I, \mathrm{id}_X)$.

\subsubsection{The Para enrichment $\Lk_k^\Para(P)$}
\label{sec:paradef-enrichment}

Table~\ref{tab:para-levels} records $(M_k, G_k)$ and the
content of ``morphism of $\Lk_k(P)$'' at each tower level.

\begin{table}[p]
\centering
\small
\renewcommand{\arraystretch}{1.45}
\setlength{\tabcolsep}{5pt}
\caption{The symmetry monad $M_k$ at each tower level (column~2)
  and the content of ``morphism of $\Lk_k(P)$'' that
  condition~\eqref{eq:Ek} requires every instantiation of
  $(\Theta, f_\theta)$ to satisfy (column~3).
  The reference at the end of each row names the chapter where that
  content was constructed.
  ML architecture assignments appear in
  Table~\ref{tab:ml-classification}.}
\label{tab:para-levels}
\begin{tabular}{@{}lp{4.4cm}p{6.3cm}@{}}
\hline
\textbf{Level} &
\textbf{$M_k$: $G_k$ acting on} &
\textbf{Morphism of $\Lk_k(P)$ requires\ldots} \\
\hline
$\Lk_0^\Para$
  & $\mathrm{Sym}(\Sp)$ on species multisets
  & $\mathrm{Sym}(\Sp)$-equivariance with output in $\NN^{|\Sp|}$
    (Chapter~\ref{sec:L0}) \\
$\Lk_1^\Para$
  & $\mathrm{Sym}(\Sp)$ on reactions with enthalpy
  & the above, plus consistency with the Hess functor
    $\FH \colon \Lk_0(P) \to B\RR$ on every reaction
    (Chapter~\ref{sec:L1}) \\
$\Lk_2^\Para$
  & $\mathrm{Sym}(\Sp)$ with $\dagger$-involution on reaction
    $\dagger$-categories
  & the above, plus $\dagger$-equivariance and preservation of
    $\ker \FG^T$ (the categorical Wegscheider condition,
    Chapter~\ref{sec:L2}) \\
$\Lk_3^\Para$
  & $\mathrm{Sym}(\Sp)$ on rated reaction networks
  & the above, plus that $\FP \colon \Lk_0(P) \to \Stoch$ is realised
    as a Markov-category morphism (positivity, probability
    conservation, Chapter~\ref{sec:L3}) \\
$\Lk_4^\Para$
  & $\Aut(G)$ on labelled graphs $\LGraphP$
  & an $\Aut(G)$-equivariant DPO-span morphism preserving bond order,
    formal charges, and lone pairs (Chapter~\ref{sec:L4}) \\
$\Lk_{4.5}^\Para$
  & $\Gstar = \Aut(G) \ltimes \ZZ_2^k$ on stereo-tagged graphs
  & the above, plus $\Gstar$-equivariance and preservation of
    stereocentre parity and Walden-inversion parity
    (Chapter~\ref{sec:L45}) \\
$\Lk_5^\Para$
  & $SE(3) \ltimes \Aut(G)$ on $\RR^{3n}$
  & PES invariance $\FV(R \cdot \mathbf{R}) = \FV(\mathbf{R})$ and
    force equivariance $\mathbf{F} = -\nabla \FV$ under
    $R \in SE(3) \ltimes \Aut(G)$ (Chapter~\ref{sec:L5}) \\
$\Lk_6^\Para$
  & $U(N)$ gauge on Hilbert bundles
    $\Hel^{(N)} \to \Ce(G)$
  & the above, plus a gauge-equivariant bundle morphism carrying the
    Berry class
    $[\gamma_B] \in H^1(\Ce(G)\setminus\Xseam, \ZZ_2)$ at conical
    intersections (Chapter~\ref{sec:L6}) \\
$\Lk_7^\Para$
  & $S_N \times S_M$ on antisymmetrised
    $\mathcal{H}_\mathrm{full}$
  & fermionic antisymmetry, compatible with the continuous field
    $\{A_\varepsilon^G\}_{\varepsilon \in [0,1]}$ of deformation
    parameters (Chapter~\ref{sec:L7}) \\
\hline
\end{tabular}
\end{table}

\medskip
An architecture is said to live at tower level $k$ when two things
are true of it simultaneously:
at every admissible weight setting the map it produces respects the
level-$k$ structure of Table~\ref{tab:para-levels}, and its weight
space itself records --- as part of its design --- the comonoid
structure of (C$_k$) below: how a weight may be shared across
layers (comultiplication $\Delta_\Theta$) and the formal
discardability of the parameter object (counit $!_\Theta$).
These are different kinds of data:

\begin{itemize}
  \item (E$_k$) is a condition on each trained instance, testable on
    any one weight setting;
  \item (C$_k$) is a condition on the design of the parameter space,
    fixed at design time and persisting across training runs.
\end{itemize}

A single underlying map $f_\theta$ can be realised by architectures
with different comonoid disciplines --- one sharing weights across
layers, another keeping them independent --- and (C$_k$) is what
distinguishes those architectures in $\Lk_k^\Para(P)$.

\begin{definition}[$\Lk_k^\Para(P)$]
\label{def:LkPara}
  The \emph{Para enrichment at tower level $k$} is the sub-2-category
  $\Lk_k^\Para(P) \subset \ParaC(\mathcal{C})$ with:
  \begin{itemize}
    \item \textbf{Objects:} the objects of $\Lk_k(P)$.
    \item \textbf{1-morphisms $X \to Y$:} parametric morphisms
      $(\Theta, f_\theta)$ of $\ParaC(\mathcal{C})$ with $X, Y \in
      \Lk_k(P)$ and $\Theta$ a parameter space (setup above),
      satisfying both conditions below.
    \item \textbf{2-morphisms:} reparametrisations of
      $\ParaC(\mathcal{C})$ between such 1-morphisms.
  \end{itemize}

  \medskip\noindent\textbf{(E$_k$) Per-parameter level-$k$
  equivariance.}
  For every morphism $r \colon I \to \Theta$ in $\mathcal{C}$, the
  \emph{instantiation of $f_\theta$ at $r$},
  \begin{equation}
    \label{eq:Ek}
    f_\theta \circ (r \otimes \mathrm{id}_X) \circ \lambda_X^{-1}
    \;\colon\; X \to Y,
  \end{equation}
  is a morphism of $\Lk_k(P)$ (Table~\ref{tab:para-levels}).
  In the group-action case $M_k(X) = G_k \times X$, this is
  equivalent to the pointwise statement
  $f_\theta(g \cdot x) = g \cdot f_\theta(x)$ for every
  $\theta \in \Theta$, $g \in G_k$, $x \in X$, recovering the
  chembox reading of the chapter opener.

  \medskip\noindent\textbf{(C$_k$) Parameter-space comonoid.}
  $\Theta$ is equipped with
  \begin{itemize}
    \item a counit $!_\Theta \colon \Theta \to I$ in $\mathcal{C}$
      --- the \emph{parameter-discard} map (not a zero-weight
      choice; see Remark~\ref{rem:instantiation});
    \item a comultiplication
      $\Delta_\Theta \colon \Theta \to \Theta \otimes \Theta$ in
      $\mathcal{C}$ --- the \emph{weight-sharing} map;
  \end{itemize}
  satisfying coassociativity and counitality.
\end{definition}

\medskip\noindent\textbf{Why (E$_k$) is stated per-parameter.}
Condition~\eqref{eq:Ek} is phrased per-parameter --- one check per
$r \colon I \to \Theta$ --- rather than as a single condition on the
pair $(\Theta, f_\theta)$ at the level of the ambient category
$\ParaC(\mathcal{C})$.
Two reasons.
First, it is directly testable:
given an architecture at a particular trained weight setting $r^*$,
one verifies~\eqref{eq:Ek} by checking that the instantiation at
$r^*$ is a morphism of $\Lk_k(P)$, with no further categorical
apparatus required.
Second, the per-parameter statement matches the chemist's reading of
membership from the chapter chembox:
an architecture lives at level $k$ exactly when every weight setting
it admits --- not just the trained one --- yields a level-$k$
morphism.

\medskip\noindent\textbf{Why a comonoid on $\Theta$, and why this
particular one.}
The operations a parameter space admits --- duplicate a weight
(for sharing across layers) and discard a weight (formally
forgetting the parameter object) --- are exactly those of a
comonoid in a symmetric monoidal category.
This is not a structure newly imported from~\cite{GavRanovic2024CDL}
for the present chapter:
Chapter~\ref{sec:L3} already uses the same $(\Delta, !)$ data, there
as the copy--delete morphisms on every object of a Markov category,
where $\Delta$ copies a random sample and $!$ discards one.
Here the formal content is identical and the interpretation shifts:
$\Delta$ routes a single weight into two layers (rather than copying
a sample to two outputs); $!$ records the parameter object as
formally discardable (the categorical counterpart of the
discardable sample, not a weight-zeroing or weight-ablation
operation).
We record~(C$_k$) explicitly because weight-sharing discipline is
architectural data:
two networks with the same underlying map $f_\theta$ but different
comonoid structures on $\Theta$ are different architectures, and
$\Lk_k^\Para(P)$ must distinguish them.

\subsubsection{The trivial-parameter embedding $\gamma_k$}
\label{sec:paradef-gamma}

We want the Para enrichment to \emph{extend} the exact tower, not
replace it: every exact morphism of $\Lk_k(P)$ should appear inside
$\Lk_k^\Para(P)$ as the trivial-parameter special case $\Theta = I$.
With this in place, the monograph's earlier theorems about
$\Lk_k(P)$ apply verbatim to the $\Theta = I$ slice, and the
parametric structure is built \emph{on top of} the tower, not in
competition with it.

\begin{proposition}[Trivial-parameter embedding]
\label{prop:gamma-k}
  The assignment
  \[
    \gamma_k \colon \Lk_k(P) \;\longrightarrow\; \Lk_k^\Para(P),
    \qquad
    \gamma_k(X) = X,
    \qquad
    \gamma_k\bigl(f \colon X \to Y\bigr)
    \;=\; \bigl(I,\; f \circ \lambda_X\bigr)
  \]
  is a symmetric monoidal functor
  (strong in the general monoidal case, strict when $\mathcal{C}$ is
  strict) that identifies $\Lk_k(P)$ with the full sub-2-category of
  $\Lk_k^\Para(P)$ whose 1-morphisms have trivial parameter space
  $\Theta = I$.
\end{proposition}

\begin{proof}
\emph{Well-definedness.}
For $r \in \mathcal{C}(I, I)$, write
$\psi(r) := \lambda_X \circ (r \otimes \mathrm{id}_X) \circ
\lambda_X^{-1} \colon X \to X$
for the unit-whiskered endomorphism of $X$;
in particular $\psi(\mathrm{id}_I) = \mathrm{id}_X$.
The instantiation of $(I,\, f \circ \lambda_X)$ at $r$ equals
\begin{equation*}
  (f \circ \lambda_X) \circ (r \otimes \mathrm{id}_X) \circ
  \lambda_X^{-1}
  \;=\; f \circ \psi(r)
  \;\colon\; X \to Y.
\end{equation*}
By the compatibility requirement on the tower embedding stated in the
monoidal setup, $\psi(r) \in \Lk_k(P)(X, X)$ for every
$r \in \mathcal{C}(I, I)$;
composition with $f \in \Lk_k(P)(X, Y)$ then yields
$f \circ \psi(r) \in \Lk_k(P)(X, Y)$ by closure of $\Lk_k(P)$ under
composition.
Condition~(E$_k$) therefore holds.
Condition~(C$_k$) holds with $I$ carrying the terminal comonoid
$(\mathrm{id}_I,\; \lambda_I^{-1})$.

\emph{Functoriality.}
$\gamma_k(\mathrm{id}_X) = (I, \lambda_X)$ is the identity 1-morphism
on $X$ in $\ParaC(\mathcal{C})$.
Sequential composition in $\ParaC(\mathcal{C})$ of $\gamma_k(f)$ and
$\gamma_k(g)$ produces parameter space $I \otimes I$, which
corresponds to $I$ via the canonical monoidal iso
$\lambda_I \colon I \otimes I \xrightarrow{\sim} I$:
equality in the strict case, a canonical iso in general.

\emph{Identification of the image.}
A 1-morphism $(I, h)$ of $\Lk_k^\Para(P)(X, Y)$ determines
$h \circ \lambda_X^{-1} \in \Lk_k(P)(X, Y)$
(a morphism of $\Lk_k(P)$ by~\eqref{eq:Ek});
this assignment is inverse to $\gamma_k$ on trivial-parameter
hom-sets.
\end{proof}

\begin{remark}[Function class; no reverse to $\gamma_k$]
\label{rem:instantiation}
The \emph{function class} of a 1-morphism
$(\Theta, f_\theta) \in \Lk_k^\Para(P)(X, Y)$ is the set of all
instantiations as $r$ ranges over $\mathcal{C}(I, \Theta)$:
\[
  \mathsf{Func}(\Theta, f_\theta)
  \;:=\;
  \bigl\{\, f_\theta \circ (r \otimes \mathrm{id}_X) \circ
    \lambda_X^{-1} \;:\; r \in \mathcal{C}(I, \Theta) \,\bigr\}
  \;\subseteq\; \Lk_k(P)(X, Y).
\]
Operationally, $\mathsf{Func}(\Theta, f_\theta)$ is the set of
morphisms the architecture \emph{can} represent for some choice of
weights, as opposed to the specific one a training run happens to
pick.
Condition~\eqref{eq:Ek} is precisely the assertion
$\mathsf{Func}(\Theta, f_\theta) \subseteq \Lk_k(P)(X, Y)$: every
weight setting gives a level-$k$ morphism.

There is no canonical 2-functor $\Lk_k^\Para(P) \to \Lk_k(P)$
collapsing $(\Theta, f_\theta)$ to a single exact morphism:
the counit $!_\Theta \colon \Theta \to I$ does not select a point of
$\Theta$.
(If it did, training would be unnecessary: the counit would produce
the right model for free.)
Categorical completeness below is therefore stated as a property of
the function class, not as a condition on any reverse map to
$\gamma_k$.
\end{remark}

\subsubsection{Categorical completeness}
\label{sec:paradef-completeness}

Membership in $\Lk_k^\Para(P)$ guarantees that every element of
$\mathsf{Func}(\Theta, f_\theta)$ is a level-$k$ morphism;
it does not say \emph{which} level-$k$ morphisms appear there.
The question categorical completeness asks is the second one:
given a specific target morphism $\varphi$ of $\Lk_k(P)$, is
$\varphi$ in the function class?

The targets of applied interest are those produced by the exact
tower constructions of earlier chapters.
At $\Lk_1$, the target is the enthalpy the Hess functor $\FH$
assigns to a specific reaction.
At $\Lk_3$, it is the Markov kernel $\FP$ assigns to a transition.
At $\Lk_5$, it is the Born--Oppenheimer potential $\FV$ yields on a
specific molecular configuration.
At $\Lk_6$, it is the Hilbert bundle $\Felsix$ produces on a
molecule, including the Berry class $[\gamma_B]$ the bundle carries
at a conical intersection.
In each case $\varphi$ is a specific morphism of the corresponding
$\Lk_k(P)$, identifiable in the chapter where the tower construction
was defined.

\begin{definition}[Categorical completeness]
\label{def:completeness}
  Let $(\Theta, f_\theta) \in \Lk_k^\Para(P)(X, Y)$ and let
  $\varphi \colon X \to Y$ be a morphism of $\Lk_k(P)$.
  The parametric morphism $(\Theta, f_\theta)$ is
  \emph{categorically complete on $\varphi$} if
  $\varphi \in \mathsf{Func}(\Theta, f_\theta)$:
  there exists $r \colon I \to \Theta$ in $\mathcal{C}$ with
  \[
    f_\theta \circ (r \otimes \mathrm{id}_X) \circ \lambda_X^{-1}
    \;=\; \varphi
    \qquad \text{in } \Lk_k(P)(X, Y).
  \]
  It is \emph{complete on a class $\mathcal{S} \subseteq \Lk_k(P)(X,
  Y)$} if complete on every $\varphi \in \mathcal{S}$, and
  \emph{complete at level $k$} if complete on every morphism of
  $\Lk_k(P)(X, Y)$.
\end{definition}

Completeness is a property of the architecture --- its parameter
space, its forward map, its closure under choice of weights --- and
is logically independent of any training procedure.
No training recovers a target absent from the function class.

ML molecular architectures are typically complete on physically
meaningful subclasses.
At $\Lk_5$, for instance, MACE/NequIP completeness on a dense
subclass of smooth BO potentials is framed conditionally on
universal-approximation hypotheses in
Section~\ref{sec:para-mace}.
They are not complete on all of $\Lk_5(P)(X, Y)$, and at $\Lk_6$
they are incomplete in a structurally topological way.
Section~\ref{sec:para-incompleteness} identifies three specific
$\varphi$ --- one at $\Lk_6$ structurally outside the function
class of every scalar-energy architecture, and two at
$\Lk_2$/$\Lk_3$ outside the function classes of the published
kinetic and joint kinetic-energetic architectures --- that are
the tower-coherence gaps the rest of the chapter
analyses.

\begin{insightbox}[Completeness is architectural, not empirical]
Where does completeness bite?
At $\Lk_6$, Proposition~\ref{prop:topological-incompleteness} shows
that the Berry class
$[\gamma_B] \in H^1(\Ce(G)\setminus\Xseam, \ZZ_2)$ carried by the
Hilbert bundle $\Felsix$ at a conical intersection lies outside
$\mathsf{Func}(\Theta, f_\theta)$ for \emph{every}
$SE(3)$-equivariant scalar-energy force field.
The reason is type-theoretic: $[\gamma_B]$ is not a pointwise
function on $\Ce(G)$ at all but a $\ZZ_2$-valued cohomology class
assigning a value to each homology class of loops in
$\Ce(G) \setminus \Xseam$; a scalar-valued output, which is a
function on $\Ce(G)$, cannot represent such a class.
No $r \colon I \to \Theta$ in a scalar-output architecture produces
a map that carries $[\gamma_B]$.
The failure is an output-type obstruction at the architecture level,
revealed by the tower and not addressable by any training procedure.
\end{insightbox}

%% file: chapters/para/para_bridges.tex
\subsection{External categorical frameworks in tower language}
\label{sec:para-bridges}

The Para enrichment $\Lk_k^\Para(P)$ of
Section~\ref{sec:para-def} unifies, into a single nested
hierarchy, characterisations of the membership condition
(E$_k$) of Definition~\ref{def:LkPara} that were developed
independently --- before this monograph's tower --- one level
at a time.
In detail:
(E$_0$) is Meseguer--Montanari's free-strict-SMC morphism
condition~\cite{MeseguerMontanari1990};
(E$_3$)'s Markov-kernel core is, on the stochastic-kinetics
side, Fritz's Markov-category morphism
condition~\cite{Fritz2020Markov} (positivity, probability
conservation), with species-permutation equivariance the
additional tower-specific component; and, on the
composable-networks side, Baez and Pollard's gray-box
functor condition on open reaction
networks~\cite{BaezPollard2017};
(E$_4$) is de Haan, Cohen, and Welling's Natural Graph Networks
naturality on molecular graphs~\cite{deHaanCohenWelling2020}
and Bonchi et al.'s DPO-compatible string-diagram rewriting of
mechanistic
graphs~\cite{BonchiEtAl2022SDRTII};
(E$_5$) is Clebsch--Gordan-augmented NGN, as realised in
NequIP~\cite{Batzner2022NequIP}, MACE~\cite{BatatIa2022MACE},
and So3krates~\cite{Frank2022So3krates}.
Each of these four external frameworks characterises one
tower level on its own terms and goes no further.
The Cruttwell--Gavranovi\'c framework of categorical
gradient-based learning~\cite{CruttwellGavranovic2022},
treated separately below, runs perpendicular to this
enumeration: it does not pair with a single tower level but
supplies the 2-categorical backdrop for gradient-based
training of any $\Lk_k^\Para(P)$ architecture
(Section~\ref{sec:bridges-lens}).
What the tower adds, via the forgetful functors
$U_k \colon \Lk_k(P) \to \Lk_{k-1}(P)$ of
Section~\ref{sec:para-def}, is nesting:
the hierarchy $\Lk_0(P) \leftarrow \Lk_1(P) \leftarrow \cdots
\leftarrow \Lk_7(P)$ makes each (E$_k$) strictly refine
(E$_{k-1}$), so a model at level $k$ inherits every lower-level
condition and can fail higher-level ones in locatable ways ---
by naming the specific forgetful functor above which it falls.

For each of the four level-pairing frameworks, this section
records three things:
(a) the tower level $k$ at which (E$_k$) identifies with the
framework;
(b) a direct verification of the identification against the
hom-set characterisation
$\Lk_k(P)(X, Y) \subseteq \mathcal{C}(X, Y)$ of
Section~\ref{sec:para-def};
and (c) the forgetful functor above which the framework's
categorical vocabulary runs out.
The CGGWZ subsection follows a different template, since the
framework has no (E$_k$) to identify and no $U_k$ above which
its vocabulary runs out --- both of these facts being
consequences of its perpendicularity.
The resulting pairings feed every subsequent section of the
chapter:
Section~\ref{sec:para-classification} uses them as the
classification criterion (Table~\ref{tab:ml-classification});
Section~\ref{sec:para-mace} uses the $(E_5)$ lift to analyse
MACE and NequIP's completeness quantitatively;
Section~\ref{sec:para-so3krates} uses the same $(E_5)$
condition, realised differently, to place So3krates's
$SE(3)$-equivariant self-attention;
Section~\ref{sec:para-square} works through QIM as a
bidirectional $\Lk_5$-type worked example (non-strict
$\Lk_5^\Para$ under its Coulomb-matrix representation) whose
approximation target references $\Lk_6$ through the
Born--Oppenheimer ground-state section $\sigma_0$;
and Section~\ref{sec:para-incompleteness} uses the joint
boundary identified here --- the Berry class $[\gamma_B]$
at $\Lk_6$ that no external framework expresses --- to state
the topological obstruction to scalar-energy force fields.

\begin{chembox}[Reading paths through this section]
A chemist reading for operational content will find, at the
start of each subsection, the condition membership in
$\Lk_k^\Para(P)$ places on an architecture expressed in
chemistry terms (atom-relabelling invariance,
probability conservation, DPO-valid bond rearrangement,
encoder--decoder bidirectionality);
the surrounding categorical exposition can be skipped.

A physicist reading for the concrete membership condition at
each level will find it stated in each subsubsection's
proposition or paragraph:
an explicit Wigner $D$-matrix equivariance
$f_\theta(R \cdot \mathbf{R}) = D(R) f_\theta(\mathbf{R})$ for
(E$_5$),
the graph-automorphism version $f_\theta(\sigma \cdot G) =
D(\sigma) f_\theta(G)$ for (E$_4$),
Markov-kernel preservation (positivity, probability
conservation) together with species-permutation covariance of
the CME generator $\Omega$ for (E$_3$),
DPO-span equivariance on $\LGraphP$ for (E$_4$) in its
mechanistic form,
and the Berry-class obstruction $[\gamma_B]$ whose absence in
every framework's vocabulary is the content of the mathboxes
below.

A mathematician reading for verification will find each
framework-tower pairing stated as a proposition or explicit
iff-condition reducing to Definition~\ref{def:LkPara};
each subsection closes with a mathbox titled
``Where [framework] stops in the tower'' recording the
specific forgetful functor above which the external
framework's vocabulary runs out.
\end{chembox}

\subsubsection{De Haan--Cohen--Welling at $\Lk_4^\Para$:
  graph naturality}
\label{sec:bridges-haan}

\noindent\emph{Chemically:}
membership in $\Lk_4^\Para$ asks a force field to produce the
same output for any pair of molecular graphs related by atom
relabelling (same graph, same prediction regardless of how
atoms are numbered);
membership in $\Lk_5^\Para$ additionally asks the output to
transform correctly under physical rotations and translations
of the geometry.
The first condition is Natural Graph Networks naturality
(this subsection);
the second is NGN augmented by Clebsch--Gordan tensor products,
realised by NequIP, MACE, and So3krates.

The \emph{Natural Graph Networks} (NGN) framework of de~Haan,
Cohen, and Welling~\cite{deHaanCohenWelling2020} gives the
categorical language for message-passing neural networks on
graphs.
Its central objects are \emph{graph feature spaces}: functors
$\rho \colon \mathbf{Graph} \to \mathbf{Vec}$
from a groupoid of graphs (with graph isomorphisms as morphisms)
to vector spaces.
A \emph{natural graph network layer} between two such feature
spaces is a natural transformation
$K \colon \rho \Rightarrow \rho'$,
a family of maps $K_G \colon \rho(G) \to \rho'(G)$ commuting
with every graph isomorphism.
Naturality is exactly equivariance under graph-level
symmetries --- that is, under $\Aut(G)$.

\begin{proposition}[NGN naturality is $(E_4)$]
\label{prop:nhaan-tower}
  At tower level $k = 4$, where $G_4 = \Aut(G)$ acts on the
  labelled-graph category $\LGraphP$ of
  Chapter~\ref{sec:L4}:
  \begin{enumerate}[label=(\roman*)]
    \item Graph feature spaces $\rho \colon \mathbf{Graph} \to
      \mathbf{Vec}$ restricted to the objects of $\LGraphP$ are
      exactly the $\Aut(G)$-equivariant feature maps at
      $\Lk_4(P)$.
    \item The NGN naturality condition of
      \cite{deHaanCohenWelling2020} is the membership
      condition~\eqref{eq:Ek} at $k = 4$:
      \[
        f_\theta(\sigma \cdot G)
        \;=\; D(\sigma)\,f_\theta(G)
        \qquad
        \text{for all } \sigma \in \Aut(G),
      \]
      where $D$ is the representation of $\Aut(G)$ on the
      output feature channels.
    \item Local-to-global: an architecture satisfying naturality
      at every message-passing edge satisfies it globally on the
      molecular graph~\cite{deHaanCohenWelling2020}.
  \end{enumerate}
\end{proposition}

\begin{proof}
Claims (i) and (ii) are direct comparisons with
Definition~\ref{def:LkPara} at $k = 4$:
for any molecular graph $G \in \LGraphP$, the automorphisms of
$G$ in $\mathbf{Graph}$ are exactly the elements of $\Aut(G)$,
and naturality of a feature map $f_\theta$ under these
automorphisms is by definition the $(E_4)$ condition.
Claim (iii) is the main result of~\cite{deHaanCohenWelling2020};
we import it without re-proof.
\qedhere
\end{proof}

\medskip\noindent\textbf{Lifting NGN from $\Lk_4^\Para$ to
  $\Lk_5^\Para$.}
NGN in its native form does not reach $\Lk_5$, whose symmetry
group $G_5 = SE(3) \ltimes \Aut(G)$ also acts by rotations and
translations on geometric positions in $\RR^{3n}$
(Chapter~\ref{sec:L5}).
An architecture built on NGN foundations lifts to $\Lk_5^\Para$
only when it augments graph-level naturality with
$SE(3)$-equivariant operations on geometric features.
Three such lifts appear in current ML literature.
NequIP~\cite{Batzner2022NequIP} and
MACE~\cite{BatatIa2022MACE} combine NGN naturality with
Clebsch--Gordan tensor products of irreducible $SE(3)$
representations on edge features, enforcing the full
$(E_5)$-condition
\[
  f_\theta(R \cdot \mathbf{R})
  \;=\; D(R)\,f_\theta(\mathbf{R})
  \qquad
  \text{for all } R \in SE(3) \ltimes \Aut(G),
\]
where $D(R)$ is the block-diagonal Wigner $D$-matrix on the
output irrep channels.
So3krates~\cite{Frank2022So3krates} enforces the same
$(E_5)$-condition directly via $SE(3)$-equivariant
self-attention (Section~\ref{sec:para-so3krates}).
Distance-invariant message passing (as in SchNet) trades
$SE(3)$-equivariance for $SE(3)$-invariance, producing a scalar
output and so landing in $\Lk_5^\Para$ with a function class
restricted to $SE(3)$-invariant scalars.

A force field satisfying the $(E_5)$-form above predicts the
same energy on any pair of $SE(3) \ltimes \Aut(G)$-equivalent
geometries --- translations, rotations, and permutations of
identical atoms give identical outputs.
The additional reflection invariance that ML force fields
typically also enforce (yielding full $E(3)$-equivariance in
the ML sense) lives one level further: it is $\Lk_{4.5}^\Para$
content, supplied by the parity factor
$\ZZ_2^k \subset \Gstar$ (Chapter~\ref{sec:L45}).

\begin{mathbox}[Where NGN stops in the tower]
The category $\mathbf{Graph}$ of~\cite{deHaanCohenWelling2020}
carries nodes and edge features but nothing above.
Natively NGN captures $\Lk_4^\Para$'s $\Aut(G)$-symmetry;
augmented with $SE(3)$-equivariant operations (NequIP, MACE,
So3krates) it reaches $\Lk_5^\Para$.
Above $\Lk_5$ it runs out of content:
the forgetful functor
$U_6 \colon \Lk_6(P) \to \Lk_5(P)$
discards the Hilbert bundle $\Hel^{(N)} \to \Ce(G)$, the Berry
connection $A_{mn}^\mu$, and the topological phase
$[\gamma_B] \in H^1(\Ce(G) \setminus \Xseam, \ZZ_2)$;
none of these has a counterpart in $\mathbf{Graph}$, augmented
or not.
On the other side of the tower, the thermodynamic functor
$\FG^T \colon \Lk_2(P) \to B\RR$ is similarly absent:
NGN has no slot for enthalpy, entropy, or the Wegscheider
kernel $\ker \FG^T$.
NGN, with or without $SE(3)$-augmentation, therefore
characterises $\Lk_k^\Para(P)$ at most for $k \le 5$;
it is silent on $\Lk_6^\Para$ and on the thermodynamic content
of $\Lk_k$ for $k \le 3$.
\end{mathbox}

\subsubsection{Combinatorial tower levels $\Lk_0$, $\Lk_3$,
  $\Lk_4$: Meseguer--Montanari, Baez--Pollard, Fritz, Bonchi et
  al.}
\label{sec:bridges-lower}

\noindent\emph{Chemically:}
membership in $\Lk_0^\Para$ asks a yield or outcome predictor
to respect stoichiometric balance and atom-permutation
symmetry;
membership in $\Lk_3^\Para$ asks a kinetic or rate predictor to
preserve positivity of concentrations, conservation of
probability, and species-permutation symmetry of the CME
generator;
membership in $\Lk_4^\Para$ asks a mechanism or retrosynthesis
predictor to produce bond rearrangements realisable as DPO
pushout complements in the labelled molecular graph.

At the combinatorial tower levels $\Lk_0$, $\Lk_3$, $\Lk_4$ ---
stoichiometry, kinetics, and bond-topological mechanisms ---
the membership conditions (E$_0$), (E$_3$), (E$_4$) each admit
an explicit external characterisation.
Four categorical frameworks supply these characterisations in a
form that an architecture can verify or fail, and together they
give the section's classification criterion for the three lowest
non-trivial Para levels.

\paragraph{Stoichiometry at $\Lk_0^\Para$, open kinetics at
  $\Lk_3^\Para$: Meseguer--Montanari and Baez--Pollard.}
The stoichiometric tower category $\Lk_0(P)$ is exactly the
free strict SMC on a Petri net $P$ of Meseguer and
Montanari~\cite{MeseguerMontanari1990}
(Chapter~\ref{sec:L0}).
Baez and Pollard~\cite{BaezPollard2017} extend this
construction to \emph{open} reaction networks, producing a
gray-box functor at $\Lk_3$ that reads off the input--output
kinetics of a composed network from its boundary data
(Theorem~\ref{thm:baez-pollard}).
A closely related taxonomy of net categories --- commutative
monoidal category nets, $\Sigma$-nets, and further variants ---
is developed by Baez, Genovese, Master, and
Shulman~\cite{BaezGenoveseMasterShulman2021};
$\Lk_0(P)$ of the tower corresponds to one vertex of that
taxonomy (the free strict SMC), with Baez--Master's
CMC~\cite{BaezMaster2020} a quotient in which all symmetry
morphisms are identified
(Remark~\ref{rem:BM-comparison}).

The Para content of these identifications is direct.
A yield or reaction-outcome predictor $(\Theta, f_\theta)$
lies in $\Lk_0^\Para(P)$ iff its function class
$\mathsf{Func}(\Theta, f_\theta)$
(Remark~\ref{rem:instantiation}) consists of
$\mathrm{Sym}(\Sp)$-equivariant maps with outputs in $\NN^{|\Sp|}$
(or a real-valued approximation thereof).
A composed reaction-network predictor lies in the open-network
analogue of $\Lk_3^\Para$ iff its interface kinetics compose by
pushout as in~\cite{BaezPollard2017}.
Retrosynthesis planners and mechanistic networks in current
use, however, treat reactions as isolated events rather than
as morphisms composable by pushout;
the bicategory $\mathrm{Csp}(\Lk_4^\Para)$ of open mechanistic
networks thus remains an open target for the tower programme.

\paragraph{Stochastic kinetics at $\Lk_3^\Para$: Fritz Markov
  categories.}
Fritz's synthetic framework of Markov
categories~\cite{Fritz2020Markov} supplies the Markov-kernel
component of (E$_3$) in directly verifiable form: preservation
of positivity and conservation of probability. The full
(E$_3$) of the tower combines this Markov-kernel structure
with species-permutation equivariance of the CME generator,
the second component being tower-specific content not in
Fritz's bare framework.
The category $\Stoch$ of Markov kernels is the canonical
Markov category; the kinetic functor
$\FP \colon \Lk_0(P) \to \Stoch$ of
Section~\ref{sec:L3-def}, which assembles the CME generator
$\Omega = \sum_r \FP(r)$, is a morphism of Markov categories
in the single-state-space sense of
Remark~\ref{rem:stoch-strict}.
A neural kinetic model $(\Theta, f_\theta)$ lies in
$\Lk_3^\Para(P)$ iff every instantiation
$f_\theta \circ (r \otimes \mathrm{id}_X) \circ \lambda_X^{-1}$
satisfies both components of (E$_3$): it is a morphism of
Markov categories in Fritz's sense
(preserving positivity of concentrations and conservation of
probability), and it intertwines rate-preserving species
permutations with the generator $\Omega$
(Proposition~\ref{prop:CME-covariance}).

$\Lk_3^\Para$-membership does not, however, imply
$\Lk_2^\Para$-membership:
Markov-category morphism status is about probability and
kinetics, while $\dagger$-compatibility at $\Lk_2$ is forced
by the thermodynamic structure and the categorical Wegscheider
condition $\ker \FG^T$.
The forgetful functor $U_3 \colon \Lk_3(P) \to \Lk_2(P)$ is
therefore where the Markov-category vocabulary of
\cite{Fritz2020Markov} first leaves a gap that a neural
kinetic model must fill separately --- and the gap that
Section~\ref{sec:para-classification} documents for current
neural kinetic models in its thermodynamic column.

\paragraph{Mechanisms at $\Lk_4^\Para$: Bonchi et al.\ string
  diagrams.}
Part~II of Bonchi, Gadducci, Kissinger, Sobociński, and
Zanasi's \emph{String Diagram Rewrite
Theory}~\cite{BonchiEtAl2022SDRTII} establishes that DPO
rewriting of hypergraphs modulo symmetric monoidal category
equations corresponds to string-diagram rewriting in a free
SMC.
This is the categorical foundation on which the tower's
$\Lk_4(P)$ is built (Chapter~\ref{sec:L4}):
the six DPO generators of Definition~\ref{def:generators} ---
heterolytic and homolytic cleavage, electron-pair migration,
and their reverses --- are the string-diagram generators of
the free SMC that controls bond rearrangements, and DPO
rewriting in the labelled-graph category $\LGraphP$ is the
operational semantics for their composition.

An ML model $(\Theta, f_\theta)$ for retrosynthesis or
mechanism prediction lies in $\Lk_4^\Para(P)$ iff its function
class $\mathsf{Func}(\Theta, f_\theta)$ consists of
$\Aut(G)$-equivariant DPO-span morphisms that preserve bond
order, formal charges, and lone pairs.
Molecular Transformer~\cite{Schwaller2019MolTransformer} and
similar sequence-to-sequence models implicitly approximate
morphisms in this string-diagram category by learning
reaction-template patterns, but without the formal DPO
constraint that would guarantee predicted bond changes remain
geometrically realisable in $\LGraphP$.

\begin{mathbox}[Where the combinatorial frameworks stop]
Fritz's Markov categories~\cite{Fritz2020Markov} characterise
the Markov-kernel component of $\Lk_3^\Para$;
Bonchi et al.'s string-diagram
rewriting~\cite{BonchiEtAl2022SDRTII} characterises the
DPO-rewriting component of $\Lk_4^\Para$;
Meseguer--Montanari~\cite{MeseguerMontanari1990} and
Baez--Pollard~\cite{BaezPollard2017} characterise $\Lk_0$ and
open-$\Lk_3$ respectively.
Each framework supplies the structural component that
distinguishes its tower level from the next lower one;
species-permutation equivariance at $\Lk_3$ and
$\Aut(G)$-equivariance at $\Lk_4$ are tower-specific
additions layered on top of these structural components.
None of the four reaches thermodynamics:
the forgetful functor $U_2 \colon \Lk_2(P) \to \Lk_1(P)$
preserves enthalpy and the dagger structure, and $(E_2)$ ---
$\dagger$-equivariance and preservation of $\ker \FG^T$ ---
is not expressible in any of these frameworks.
The tower's cokernel forcing of $\Lk_1 \to \Lk_2$ is precisely
the statement that Fritz's vocabulary does not suffice.
Geometry and electronic structure are also out of reach:
none of the four frameworks carries a slot for
$\Ce(G) = \RR^{3n} / G_5$ or for the Hilbert bundle
$\Hel^{(N)} \to \Ce(G)$ of $\Lk_6$.
\end{mathbox}

\subsubsection{The Cruttwell--Gavranovi\'c framework and the
  $\Lk_5$--$\Lk_6$ interface}
\label{sec:bridges-lens}

\noindent\emph{Chemically:}
bidirectional encoder--decoder--property architectures at the
$\Lk_5$--$\Lk_6$ interface, of which QIM~\cite{Fallani2024QIM}
is the primary published example
(Section~\ref{sec:para-square}), run in two directions through
a shared latent space:
geometry $\to$ latent $\to$ property prediction, and
geometry $\to$ latent $\to$ reconstructed geometry.
Each constituent forward map sits at $\Lk_5$-type --- strict
$\Lk_5^\Para(P)$ when the representation is fully equivariant,
non-strict for QIM under its Coulomb-matrix representation
(Section~\ref{sec:para-square}) --- and the architecture as a
whole is a \emph{triple} of such maps trained jointly.
The reconstruction and property errors (RMSD, MAE) are
measurable scalar quantities produced by training, not
categorical data attached to the morphisms.

The categorical framework supplied for gradient-trained
parametric architectures by Cruttwell, Gavranovi\'c, Ghani,
Wilson, and Zanasi~\cite{CruttwellGavranovic2022} differs
structurally from the three preceding bridges.
Markov categories characterise the Markov-kernel component of
(E$_3$); string diagrams characterise the DPO-rewriting
component of (E$_4$); NGN, natively or augmented with
$SE(3)$-equivariant operations, characterises the equivariance
components of (E$_4$) and (E$_5$).
Each identifies a specific tower level.
The CGGWZ framework does not.
Its subject is the compositional semantics of gradient-based
learning in the 2-category $\ParaC(\mathcal{C})$:
a \emph{parametric lens} is a forward--backward pair of
morphisms, with forward component
$f \colon \Theta \otimes X \to Y$ in $\ParaC(\mathcal{C})$ and
a backward component propagating changes in outputs to changes
in inputs and parameters, so that sequential composition of
lenses realises the chain rule compositionally.
The framework is about how gradient descent composes, not about
what a trained model represents at a particular tower level.

\medskip\noindent\textbf{CGGWZ runs perpendicular to the tower.}
Every $\Lk_k^\Para(P)$ 1-morphism trained by backpropagation
admits the CGGWZ lens structure once its gradient data is made
explicit --- a kinetic model at $\Lk_3^\Para$ just as much as a
force field at $\Lk_5^\Para$ or a mechanism predictor at
$\Lk_4^\Para$.
Lens status is therefore orthogonal to (E$_k$):
it does not distinguish tower levels, and no forgetful functor
$U_k \colon \Lk_k(P) \to \Lk_{k-1}(P)$ measures where the
framework's vocabulary runs out, because its vocabulary tracks
gradient flow rather than level-specific structure.
The CGGWZ framework is consequently not a bridge to a single
$\Lk_k^\Para(P)$ in the sense of the three preceding
subsections.

What CGGWZ \emph{does} supply to the tower programme is
structural backdrop for Section~\ref{sec:para-def} rather than
level-specific characterisation:
the 2-category $\ParaC(\mathcal{C})$ itself
(Definition~\ref{def:Para});
the comonoid structure on parameter spaces in
condition~(C$_k$) of Definition~\ref{def:LkPara} --- counit
$!_\Theta \colon \Theta \to I$ and comultiplication
$\Delta_\Theta \colon \Theta \to \Theta \otimes \Theta$ ---
derived in Theorem~G.10 of~\cite{GavRanovic2024CDL} as the
lax-algebra coherence data of a $\ParaC(T)$-algebra;
and the weight-tying semantics
($\Delta_\Theta$ routes one weight to two layers) and
parameter-discard semantics ($!_\Theta$ as the categorical
record that $\Theta$ can be formally discarded, not as a
zero-weight choice) on which (C$_k$) rests.
These contributions are to the background of every
$\Lk_k^\Para(P)$, not to the classification of any particular
architecture at a particular level.

\medskip\noindent\textbf{The encoder--decoder triple at the
  $\Lk_5$--$\Lk_6$ interface, in correct categorical language.}
A common informal reading takes QIM's encoder--decoder
structure to be a parametric lens.
The categorical facts do not support this.
QIM is three parametric maps of $\Lk_5$-type through
a shared latent $Z$ (Section~\ref{sec:para-square};
non-strict $\Lk_5^\Para(P)$ under the Coulomb-matrix
representation):
the structure encoder
$(\Theta_\varphi, q_\varphi) \colon \Ce(G) \to Z$,
the structure decoder
$(\Theta_\theta, p_\theta) \colon Z \to \Ce(G)$,
and the property encoder
$(\Theta_\psi, p_\psi) \colon \RR^d \to Z$ --- the third
network added in~\cite{Fallani2024QIM} for inverse design,
taking properties to latent, not latent to properties.
$p_\theta$ is a second forward map, not the backward map of a
CGGWZ lens;
the two have different categorical signatures, as the
backward component of a parametric lens must carry a change
object (tangent space, in the smooth case) that $p_\theta$'s
signature $Z \to \Ce(G)$ does not.
Nor does the encoder--decoder round trip
$p_\theta \circ q_\varphi$ produce a 2-morphism of
$\ParaC(\mathcal{C})$:
its deviation from $\mathrm{id}_{\Ce(G)}$ measured on a test
set (the QM7-X RMSD of~\cite{Fallani2024QIM}) is a scalar
training residual.
A 2-morphism of $\ParaC(\mathcal{C})$ is a reparametrisation
$r \colon \Theta' \to \Theta$ satisfying
$f_\theta \circ (r \otimes \mathrm{id}_X) = f_{\theta'}$
(Definition~\ref{def:Para}), an equation in $\mathcal{C}$,
not a real-valued error.
The forward property-prediction path through the shared
latent approximates
$\Pi \circ \sigma_0 \colon \Ce(G) \to \RR^d$ (the
Born--Oppenheimer ground-state section followed by observable
projection) implicitly, via the joint training objective
rather than a dedicated latent-to-property decoder
(Section~\ref{sec:para-square});
the approximation is numerical, and its error is not a 2-cell
in any 2-category.

The categorical content at the $\Lk_5$--$\Lk_6$ interface is
therefore precisely this.
The architecture is three parametric maps of $\Lk_5$-type
through a shared latent.
Its output types --- scalar properties in $\RR^d$ along the
forward path through the shared latent, and reconstructed
geometries in $\Ce(G)$ from $p_\theta$ --- are not sections
of the Hilbert bundle $\Hel^{(N)} \to \Ce(G)$.
No instantiation at any $r \colon I \to \Theta$ produces
such a section, and no such instantiation represents the
Berry class $[\gamma_B] \in H^1(\Ce(G) \setminus \Xseam,
\ZZ_2)$.
The architecture therefore sits at $\Lk_5$-type (non-strict
$\Lk_5^\Para(P)$ under the Coulomb-matrix representation;
Section~\ref{sec:para-square}) despite its approximation
target $\Pi \circ \sigma_0$ referring, through $\sigma_0$, to
the $\Lk_6$ functor $\Felsix$.
Section~\ref{sec:para-square} works this out for QIM in detail;
Proposition~\ref{prop:topological-incompleteness} of
Section~\ref{sec:para-incompleteness} promotes the output-type
observation to a structural incompleteness result.

\begin{mathbox}[Where the CGGWZ framework sits relative to the
  tower]
The preceding mathboxes locate each framework by the forgetful
functor above which its vocabulary runs out.
CGGWZ admits no such localisation:
its subject is compositional backpropagation, not the
level-by-level structure the tower resolves.
No $U_k \colon \Lk_k(P) \to \Lk_{k-1}(P)$ bounds its reach;
its reach is perpendicular.
Structurally, CGGWZ supplies the 2-categorical backdrop of
Section~\ref{sec:para-def} --- the 2-category
$\ParaC(\mathcal{C})$, the reparametrisation 2-cells, and the
lax-algebra derivation of the weight-tying comonoid
$(!_\Theta, \Delta_\Theta)$ of condition~(C$_k$) --- but does
not identify a specific (E$_k$).
The Berry-class obstruction of
Proposition~\ref{prop:topological-incompleteness} is, in this
light, a statement about the output type of the architecture's
1-morphisms, not about whether they compose as lenses under
gradient descent.
\end{mathbox}

\medskip
\noindent\textbf{Bridge to the remainder of the chapter.}
The framework-to-level pairings collected above supply the
categorical vocabulary of every subsequent section of this
chapter.
Section~\ref{sec:para-classification} uses them as the
classification criterion:
the level assigned to a 1-morphism $(\Theta, f_\theta)$ of
$\ParaC(\mathcal{C})$ is the highest $k$ for which it satisfies
the corresponding (E$_k$), and
Table~\ref{tab:ml-classification} records the resulting
assignments.
Section~\ref{sec:para-mace} takes the $(E_5)$ lift of NGN as
the structural condition MACE and NequIP satisfy, then asks a
quantitative question the level assignment alone cannot
answer --- which morphisms in $\Lk_5(P)(X, Y)$ actually lie
in the architecture's function class
$\mathsf{Func}(\Theta, f_\theta)$ --- and frames the
corresponding density question on smooth BO potentials as a
conditional under universal-approximation hypotheses.
Section~\ref{sec:para-so3krates} recasts the same $(E_5)$
condition in the language of $SE(3)$-equivariant self-attention,
without the NGN's message-passing substrate.
Section~\ref{sec:para-square} works through QIM as the concrete
triple of $\Lk_5$-type maps identified above (non-strict
$\Lk_5^\Para(P)$ under the Coulomb-matrix representation),
whose approximation target $\Pi \circ \sigma_0$ references the
$\Lk_6$ ground-state section $\sigma_0 \colon \Ce(G) \to
\Hel^{(N)}$ while the architecture's scalar output type keeps
it within the $\Lk_5$ region.
Section~\ref{sec:para-incompleteness} names the failure
structurally:
the Berry class
$[\gamma_B] \in H^1(\Ce(G) \setminus \Xseam, \ZZ_2)$ that
$\Felsix$ attaches at conical intersections is expressible in
\emph{none} of the level-specific external frameworks collected
here --- not in NGN, not in Markov categories, not in string
diagrams --- and is equally outside the level-orthogonal
CGGWZ vocabulary, whose subject is gradient composition rather
than output-type structure.
The topological obstruction of that section therefore cannot be
derived by upgrading any of the external vocabularies;
it requires the full $\Lk_6^\Para$ machinery of
Section~\ref{sec:para-def}, a fact the mathboxes closing the
three preceding subsections make unavoidable.

Categorical completeness (Definition~\ref{def:completeness})
provides the quantitative refinement of every level
assignment:
at level $k$ it measures how much of $\Lk_k(P)(X, Y)$ the
architecture's function class
$\mathsf{Func}(\Theta, f_\theta)$ actually covers.
The bridging material of this section is the prerequisite for
that quantitative question to be well-posed at each level and
for its negative answer at $\Lk_6$ to be more than an empirical observation.

%% file: chapters/para/para_classification.tex
\subsection{Classification of machine learning molecular architectures}
\label{sec:para-classification}

Table~\ref{tab:ml-classification} classifies the major ML molecular architectures by the highest tower level $k$ for which the function class $\mathsf{Func}(\Theta, f_\theta)$ of a parametric 1-morphism $(\Theta, f_\theta)$ is contained in $\Lk_k(P)(X, Y)$ --- the membership condition $(\mathrm{E}_k)$ of Definition~\ref{def:LkPara}.
The function class, defined in
Remark~\ref{rem:instantiation}, is the set of all morphisms the
architecture can represent as $r$ ranges over $\mathcal{C}(I, \Theta)$;
it is independent of any particular training run. The
classification is therefore a statement about what the architecture
\emph{can} represent, not about where a specific learned weight
configuration lands.

\begin{insightbox}[Three senses of satisfying $(\mathrm{E}_k)$]
Level assignment turns on distinguishing three claims any one of
which a reader might intend by saying an architecture ``satisfies
level $k$''.

\emph{Architectural satisfaction.}
$\mathsf{Func}(\Theta, f_\theta) \subseteq \Lk_k(P)(X, Y)$ for
every $\theta \in \Theta$. This is $(\mathrm{E}_k)$, and only
this gives categorical membership of the parametric morphism in
$\Lk_k^\Para(P)$.

\emph{Training-emergent satisfaction.}
A specific $\theta^* \in \Theta$ obtained by training produces
outputs numerically close to $\Lk_k(P)(X, Y)$ on a benchmark.
Other choices of $\theta$ need not.

\emph{Type compatibility.}
The codomain of $f_\theta$ matches the codomain of
$\Lk_k(P)$-morphisms (e.g.\ scalars on $\Ce(G)$), but the
$\Lk_k(P)$-defining structure beyond the codomain is not enforced.

The classification below applies the architectural sense strictly.
The recurring failure mode the audit rectifies is upgrading from
training-emergent or type-compatible to architectural without the
corresponding closure in $\mathsf{Func}(\Theta, f_\theta)$.
\end{insightbox}

The table is ordered by tower level, with the reaction-side region
(stoichiometry, kinetics, mechanism) first, then the configurational
regions $\Lk_5^\Para(P)$ and $\Lk_6^\Para(P)$.
Proposition~\ref{prop:topological-incompleteness}
(Section~\ref{sec:para-incompleteness}) confines every
$E(3)$-equivariant scalar-energy architecture to $\Lk_5^\Para(P)$:
the codomain $\OrbMorse$ cannot accommodate a non-trivial section of
$\Hel^{(N)} \to \Ce(G)$, and no amount of training, body-order
refinement, or $l_{\max}$-inflation alters the codomain.

Two sub-categories used as classification labels in the table deserve
formal statement before their first appearance.

\begin{mathbox}[Classification sub-categories used in the table]
\emph{Generative $\Lk_5$.} Alongside $\Lk_5(P)$, whose morphisms
are $SE(3) \rtimes \Aut(G)$-equivariant scalar potentials
$V \colon \Ce(G) \to \RR$ with conservative forces $F = -\nabla V$,
define the generative variant $\Lk_5^{\mathrm{gen}}(P)$ whose
morphisms are maps $f \colon \mathrm{Seq} \to \Ce(G)$ from an
input space $\mathrm{Seq}$ (sequences, labels, latent codes)
\emph{into} the configuration orbifold, reversing the direction
of $\Lk_5$-morphisms. The membership condition
$(\mathrm{E}_5^{\mathrm{gen}})$ is that $f$ factor through the
quotient $\Ce(G) = \RR^{3n}/(SE(3) \rtimes \Aut(G))$ well-definedly:
any lift $\widetilde{f} \colon \mathrm{Seq} \to \RR^{3n}$ must
project to the same class in $\Ce(G)$, regardless of the frame
choice implicit in the lift. Architecturally this is enforced by
making all internal operations that fix or transform a frame
covariant under $SE(3) \rtimes \Aut(G)$, so distinct frame
conventions yield the same orbit. The Para enrichment is
$\Lk_5^{\Para,\mathrm{gen}}(P)$; no PES, no $\FV$, no dynamics.

\emph{Operator-level vs section-level $\Lk_6$.}
The $\Lk_6$ object is the $U(N)$-gauge Hilbert bundle
$\Hel^{(N)} \to \Ce(G)$ with its Berry connection $A_{mn}^\mu$,
seam $\Xseam$, and class
$[\gamma_B] \in H^1(\Ce(G)\setminus\Xseam, \ZZ_2)$. Two
architecturally distinct codomains approximate this object. The
\emph{operator-level} codomain is
$\mathrm{End}(\Hel^{\mathrm{AO}})$, the space of symmetric
operators on a fixed atomic-orbital basis Hilbert space, with
$SE(3)$-equivariance of the operator-valued map. The fixed basis
trivialises the bundle globally; $U(N)$ gauge freedom on the
fibres is frozen. The \emph{section-level} codomain is a
$U(N)$-gauge-equivariant section
$\sigma \colon \Ce(G)\setminus\Xseam \to \Hel^{(N)}$,
architecturally required to transform correctly under fibre
reparameterisation and to respect $[\gamma_B]$-consistency on
loops encircling $\Xseam$. The two sub-categories are orthogonal approximations of full
$\Lk_6$, not nested: operator-level captures
$H_{\mathrm{el}}(\mathbf{R})$ completely but freezes the $U(N)$
fibre gauge by fixing a basis, while section-level captures a
single gauge-covariant state $\sigma$ but not the full operator.
Diagonalising an operator-level output recovers bundle fibres
pointwise but not a canonical global section: the
eigenvector-phase ambiguity around $\Xseam$ makes any lift from
operator-level to section-level non-canonical. Full
$\Lk_6^\Para(P)$ membership requires both capabilities
simultaneously: operator-valued fidelity and $U(N)$-gauge-covariant
sectioning with $[\gamma_B]$-consistency. PhiSNet and
DeepH-E3 reach the operator-level sub-category architecturally;
SPAINN offers a partial (connection-level) approach toward the
section-level sub-category. No architecture in the audit
occupies both sub-categories, and none reaches full
$\Lk_6^\Para(P)$.
\end{mathbox}
\pagebreak

\setlength{\LTpre}{4pt}
\setlength{\LTpost}{4pt}
\begin{longtable}{@{\,}p{2.8cm}cp{2.6cm}p{6.1cm}@{\,}}
\caption{Classification of ML molecular architectures by tower level.
  Level = highest $\Lk_k^\Para(P)$ for which $(\mathrm{E}_k)$ is
  an architectural guarantee on $\mathsf{Func}(\Theta, f_\theta)$.
  ``Inv.''\ = scalar invariant output; ``Eq.''\ = equivariant output.
  $(\ddagger)$ = shares the NequIP-class $\Lk_5$ ceiling:
  codomain $\OrbMorse$ precludes the Berry connection $A_{mn}^\mu$,
  the class $[\gamma_B] \in H^1(\Ce(G)\setminus\Xseam, \ZZ_2)$,
  and sections of $\Hel^{(N)} \to \Ce(G)$
  (Proposition~\ref{prop:topological-incompleteness}).}
\label{tab:ml-classification} \\
\hline
\textbf{Architecture} & \textbf{Year} & \textbf{Level} &
\textbf{Tower ceiling} \\
\hline
\endfirsthead
\multicolumn{4}{l}{\footnotesize\textit{Continued from previous page}}\\
\hline
\textbf{Architecture} & \textbf{Year} & \textbf{Level} &
\textbf{Tower ceiling} \\
\hline
\endhead
\hline
\multicolumn{4}{r}{\footnotesize\textit{Continued on next page}}\\
\endfoot
\hline
\endlastfoot
%
\multicolumn{4}{l}{\textit{Stoichiometric-support encoders
  (sub-$\Lk_0^\Para$)}}\\
\hline
DrFP~\cite{Probst2022DrFP}
  & 2022
  & sub-$\Lk_0^\Para$ (lossy support)
  & $\Mon$ lost to hashing and bit-folding;
    $\FH$ cannot be discussed in the quotient;
    symmetric difference $R \triangle P$ collapses the dagger
    involution to the identity before $\Lk_2$ could act \\
\hline
%
\multicolumn{4}{l}{\textit{Kinetic ODE surrogates (below $\Lk_3^\Para$)}}\\
\hline
ChemNODE~\cite{Owoyele2022ChemNODE}
  & 2022
  & below $\Lk_0^\Para$ ($P$-specific ODE)
  & No $\mathrm{Sym}(\Sp)$-equivariant generator;
    unconstrained MLP RHS does not preserve non-negativity of
    concentrations or element/mass conservation under
    integration; not a morphism in $\Stoch$ \\
CRNN~\cite{Ji2021CRNN}
  & 2021
  & $\Lk_0^\Para$ (real-relaxation) + deterministic mass-action
  & Hard-coded continuous stoichiometric layer (sparsity-regularised
    toward integer values) gives $\Lk_0$ on the $\RR$-relaxation;
    Arrhenius kinetics sits beside the tower.
    No $\FH$ (Arrhenius $E_a \ne \Delta H$), no dagger
    $(\cdot)^\dagger$ (forward/reverse rate constants
    independent, violating $\ker(\FG^T)$-Wegscheider), no CME
    generator in $\Stoch$ \\
\hline
%
\multicolumn{4}{l}{\textit{Reaction SMILES sequence models
  (below $\Lk_0^\Para$)}}\\
\hline
Mol.\ Transformer~\cite{Schwaller2019MolTransformer}
  & 2019
  & below $\Lk_0^\Para$ (token stream $\Sigma^*$)
  & No $\mathrm{Sym}(\Sp)$-equivariance on the reactant multiset;
    no $\Aut(G)$-equivariance on molecular graph relabellings;
    no graph rewriting in $\LGraphP$. Atom and charge conservation
    and graph validity are training-emergent, not architectural \\
\hline
%
\multicolumn{4}{l}{\textit{$\Lk_5^\Para$ (inv.) --- invariant descriptors
  and scalar GNNs}}\\
\hline
Coulomb matrix~\cite{Rupp2012}
  & 2012
  & $\Lk_5^\Para$ (inv., pairwise; with regressor)
  & $\mathrm{Sym}(\Sp)$ not architectural (sorting or eigenspectrum
    post-hoc); pairwise-only geometry misses many-body correlations;
    ceiling $\Lk_6$ $(\ddagger)$ \\
SOAP~\cite{Bartok2013SOAP}
  & 2013
  & $\Lk_5^\Para$ (inv., 3-body; with GAP)
  & $\mathrm{Sym}(\Sp)$, $O(3)$, translation architectural;
    3-body complete but higher-body incomplete (known ghost-pair
    counterexamples); ceiling $\Lk_6$ $(\ddagger)$ \\
Behler--Parrinello~\cite{BehlerParrinello2007}
  & 2007
  & $\Lk_5^\Para$ (inv.)
  & Atom-centred symmetry functions span 2- and 3-body only;
    ceiling $\Lk_6$ $(\ddagger)$ \\
ANI / ANI-1ccx~\cite{Smith2017ANI,Smith2019ANI}
  & 2017/19
  & $\Lk_5^\Para$ (inv.)
  & Species-pair AEVs yield $(\mathrm{E}_5)$ by construction;
    no long-range physics; ceiling $\Lk_6$ $(\ddagger)$ \\
SchNet~\cite{Schutt2018SchNet}
  & 2017
  & $\Lk_5^\Para$ (inv.)
  & Continuous-filter convolutions with pairwise-distance filters:
    no architectural angular features;
    ceiling $\Lk_6$ $(\ddagger)$ \\
DimeNet~\cite{Gasteiger2020DimeNet}
  & 2020
  & $\Lk_5^\Para$ (inv.)
  & Directional messages with 3-body angles; scalar readout
    discharges internal rotational equivariance;
    ceiling $\Lk_6$ $(\ddagger)$ \\
GemNet~\cite{Gasteiger2021GemNet}
  & 2021
  & $\Lk_5^\Para$ (inv.)
  & 4-body dihedral messages with universal approximation on the
    $SE(3)$-invariant scalar subspace of $C(\Ce(G))$ (conservative head).
    Direct-force variant violates $F = -\nabla V$ and fails
    $(\mathrm{E}_5)$ strictly \\
ALIGNN~\cite{Choudhary2021ALIGNN}
  & 2021
  & $\Lk_5^\Para$ (inv.)
  & Line-graph 3-body; tensor targets predicted component-wise, not
    as irreps; ceiling $\Lk_6$ $(\ddagger)$ \\
PaiNN~\cite{Schutt2021PaiNN}
  & 2021
  & $\Lk_5^\Para$ ($l{\leq}1$ eq.\ internal; $\OrbMorse$ energy head)
  & Scalar + vector features $(l{=}0,1)$ equivariant by
    construction; energy head lands in $\OrbMorse$; dipole and
    polarizability heads exit $\OrbMorse$ into topologically trivial
    rank-$\leq 2$ tensor bundles \\
\hline
%
\multicolumn{4}{l}{\textit{$\Lk_5^\Para$ (eq.) --- $E(3)$-equivariant
  force fields}}\\
\hline
NequIP~\cite{Batzner2022NequIP}
  & 2022
  & $\Lk_5^\Para$ (eq.)
  & Full $O(3)$ via e3nn Clebsch--Gordan on irreps $l \leq l_{\max}$;
    scalar energy head projects to $\OrbMorse$: no $A_{mn}^\mu$,
    no $[\gamma_B]$, no section of $\Hel^{(N)} \to \Ce(G)$ $(\ddagger)$ \\
MACE~\cite{BatatIa2022MACE}
  & 2022
  & $\Lk_5^\Para$ (eq.)
  & ACE body-order + equivariant messages;
    ceiling $\Lk_6$ $(\ddagger)$. Separately, the MACE-specific
    forgetful $U_4$ does not separate $\Lk_4$ bond topology from
    $\Lk_5$ geometry in the internal representation
    (Prop.~\ref{prop:mace-conflation}, Section~\ref{sec:para-mace}) \\
MACE-MP-0~\cite{BatatIa2023MACEMP0}
  & 2023
  & $\Lk_5^\Para$ (eq.)
  & MACE trained at foundation-model scale: codomain unchanged;
    ceiling $\Lk_6$ $(\ddagger)$ \\
Allegro~\cite{Musaelian2023Allegro}
  & 2023
  & $\Lk_5^\Para$ (eq., strictly local)
  & Iterated pairwise tensor products without atom-centred message
    passing: receptive field bounded by $r_{\mathrm{cut}}$; orthogonal
    to the Berry ceiling $(\ddagger)$ \\
So3krates~\cite{Frank2022So3krates}
  & 2022
  & $\Lk_5^\Para$ (eq.)
  & $SE(3)$-equivariant self-attention in spherical-harmonic
    coordinates, global range; scalar energy head. Global attention
    neither yields nor remedies the Berry ceiling $(\ddagger)$;
    attention range and irrep ladder both internal to $\Lk_5$
    (Section~\ref{sec:para-so3krates}) \\
SO3LR~\cite{Kabylda2025SO3LR}
  & 2025
  & $\Lk_5^\Para$ (eq.)
  & So3krates core + explicit Coulomb, dispersion, ZBL baselines:
    $E = E_{\mathrm{SO3k}} + E_{\mathrm{Coul}} + E_{\mathrm{Disp}}
    + E_{\mathrm{ZBL}}$ is additivity \emph{in the codomain} $\RR$,
    not the Hess functor $\FH$ on $\Lk_1(P)$; extends effective
    range beyond the local cutoff, not the Berry ceiling;
    ceiling $\Lk_6$ $(\ddagger)$ \\
Equiformer / V2~\cite{Liao2023Equiformer}
  & 2023/24
  & $\Lk_5^\Para$ (eq.)
  & Equivariant attention via e3nn or eSCN $SO(2)$ convolutions
    ($l_{\max}$ up to $6$--$8$ in V2); $\OrbMorse$;
    ceiling $\Lk_6$ $(\ddagger)$ \\
eSEN~\cite{Fu2025eSEN}
  & 2025
  & $\Lk_5^\Para$ (eq., smooth conservative)
  & eSCN backbone with strict $F = -\nabla V$ on the conservative
    head and polynomial-envelope smoothness; direct-force variant
    violates $(\mathrm{E}_5)$; ceiling $\Lk_6$ $(\ddagger)$ \\
\hline
%
\multicolumn{4}{l}{\textit{$\Lk_5^{\Para,\mathrm{gen}}$ --- $SE(3)$-equivariant
  generative maps \emph{into} $\Ce(G)$}}\\
\hline
AlphaFold2~\cite{Jumper2021AF2}
  & 2021
  & $\Lk_5^{\Para,\mathrm{gen}}$ + partial $\Lk_{4.5}$
  & Map $\mathrm{Seq} \to \Ce(G)$, not a PES; Invariant Point
    Attention computes frame-invariant scores and updates frames
    equivariantly; FAPE is parity-sensitive; torsion and
    chirality losses approach $\Gstar$-equivariance approximately \\
AlphaFold3~\cite{Abramson2024AF3}
  & 2024
  & $\Lk_5^\mathrm{gen}$-type, not strict $\Lk_5^{\Para,\mathrm{gen}}$;
    fails $(\mathrm{E}_{4.5})$
  & Diffusion denoiser is a standard Transformer; $SE(3)$ achieved
    by rotation and translation augmentation (no reflection):
    equivariance is training-emergent, not architectural.
    Documented $(\mathrm{E}_{4.5})$ certificate: 4.4\% chirality
    violation on the PoseBusters benchmark, reported in the primary
    paper itself \\
\hline
%
\multicolumn{4}{l}{\textit{Approaching $\Lk_6^\Para$ --- electronic-structure
  and non-adiabatic models}}\\
\hline
SpookyNet~\cite{Unke2021SpookyNet}
  & 2021
  & $\Lk_5^\Para$ (inv.) + $\{Q, S\}$ sectoring
  & Charge and spin label connected components of the base
    (pre-$\Lk_7$ superselection sectors), not $U(N)$ fibre
    coordinates; $V(\mathbf{R}; Q, S) \in \OrbMorse$ on each sector;
    ceiling $\Lk_6$ $(\ddagger)$ \\
QIM~\cite{Fallani2024QIM}
  & 2024
  & $\Lk_5$-type, property-space; not strict $\Lk_5^\Para$
  & Three $\Lk_5$-type parametric maps through a shared latent $Z$
    --- structure encoder $\Ce(G) \to Z$, structure decoder
    $Z \to \Ce(G)$, property encoder $\RR^d \to Z$ (inverse
    design) --- jointly trained on ELBO plus property-likelihood.
    Strict $(\mathrm{E}_5)$ obstructed by the raw Coulomb-matrix
    input: $\Aut(G)$-equivariance is training-emergent, not
    architectural, and reconstruction is determined only up to
    a chirality transformation.
    Forward path approximates $\Pi \circ \sigma_0$ implicitly
    through $Z$; $\sigma_0 \colon \Ce(G) \to \Hel^{(N)}$ never
    instantiated.
    Output codomains $\RR^d$ (properties) and $\Ce(G)$
    (geometries): no $A_{mn}^\mu$, no $[\gamma_B]$ $(\ddagger)$ \\
SchNarc~\cite{Westermayr2020SchNarc}
  & 2020
  & $\Lk_5^\Para$ + $\Lk_6$-type targets
  & NACs parametrised as $d_{ij}^\mu(\mathbf{R}) = \nabla_\mu s_{ij}$
    for learned scalars $s_{ij}$: the ansatz spans only exact
    1-forms, so $\oint_\gamma d_{ij}^\mu\,dR_\mu = 0$ on every loop
    $\gamma \subset \Ce(G)\setminus\Xseam$, and $[\gamma_B]$ is
    identically trivialised \\
PhiSNet~\cite{Unke2021PhiSNet}
  & 2021
  & $\Lk_6^{\Para,\mathrm{op}}$
  & Exact $SE(3)$-equivariant prediction of $H_{\mathrm{el}}(\mathbf{R})$
    on fixed AO basis: spatial gauge architectural.
    Fixed basis freezes $U(N)$ fibre gauge;
    section-level $\Lk_6^{\Para,\mathrm{sec}}$ not reached \\
DeepH-E3~\cite{Gong2023DeepHE3}
  & 2023
  & $\Lk_6^{\Para,\mathrm{op}}$ (periodic)
  & $E(3)$-equivariant Bloch-space $H_{\mathrm{el}}$ with
    spin--orbit coupling; coordinate and basis covariance rigorous.
    No $U(N)$ on band index; $c_1$ computed post-hoc only \\
SPAINN~\cite{Mausenberger2024SPAINN}
  & 2024
  & $\Lk_6^\Para$ (partial, connection-level)
  & Equivariant vector NACs on PaiNN backbone: predicts
    $A_{mn}^\mu$ as a genuine section (not a gradient of a scalar),
    so the SchNarc obstruction is lifted.
    $U(N)$ gauge, topological constraint, and diabatic $H$ absent \\
\end{longtable}

\subsubsection*{Tower-language reasoning through the table}

\paragraph{Stoichiometric-support encoders.}
DrFP's map $r \mapsto \mathrm{fp}(r)$ composes SMILES tokenisation,
Morgan-substructure extraction, symmetric set difference
$R \triangle P$, and hashed bit-folding to a vector in $\{0, 1\}^d$.
Two distinct tower objects fail to lift through this pipeline.
The free commutative monoid $\Mon$ is lost first: hashing maps
distinct substructures to the same bit, bit-folding projects to
$\{0, 1\}$ and erases counts, so $\mathsf{Func}(\Theta, \mathrm{fp})$
lands in a multiplicity-forgetting quotient of $\Mon$ strictly
below the free-commutative-monoid structure $\Lk_0(P)$ requires.
The Hess functor $\FH \colon \Lk_1(P) \to B\RR$ fails consequently:
$\mathrm{fp}(r_1 \circ r_2)$ bears no $\RR$-additive relation to
$\mathrm{fp}(r_1) \oplus \mathrm{fp}(r_2)$, so $(\mathrm{E}_1)$
cannot even be stated in the codomain $\mathrm{fp}$ lives in. The
identity $\mathrm{fp}(r) = \mathrm{fp}(r^\dagger)$ produced by the
symmetric difference is not evidence of $\Lk_2$-membership: it
\emph{trivialises} the dagger by collapsing forward and reverse
reactions to the identity in the codomain, before $(\cdot)^\dagger$
has any oriented reaction morphism left to act on. The ceiling is
accordingly the first lossy step in the pipeline: $\Mon$ is
unrecoverable, and every higher structure depending on it ---
$\FH$, $\ker(\FG^T)$, $\FP$, the DPO span --- is unreachable as a
matter of type, not training.

\paragraph{Kinetic ODE surrogates.}
ChemNODE parametrises a vector field
$d\Phi/dt = f_\theta(\Phi)$ on a fixed thermochemical state space
$\RR^{N_s+1}$. No $\mathrm{Sym}(\Sp)$-action is defined on the
index set --- species indices are distinguishable MLP channels ---
no positivity constraint is imposed on the RHS, and no guarantee is
made that mass or element conservation survives integration.
$\mathsf{Func}(\Theta, f_\theta)$ therefore does not land in
$\Stoch$ (probability is not conserved), nor in the mass-action
subcategory of $\Lk_0(P)$, nor even in the
$\mathrm{Sym}(\Sp)$-equivariant part of $\Lk_0(P)$. Follow-up work
adds mass conservation as an auxiliary loss, confirming its
absence from the base architecture. ChemNODE sits strictly below
$\Lk_0^\Para(P)$.

CRNN, by contrast, hard-codes a stoichiometric-coefficient layer
$\nu \in \RR^{N_s \times N_r}$ (continuous, sparsity-regularised
toward integer values but not integrality-constrained
architecturally) as its first operation and composes it with an
exponential activation implementing the mass-action monomial
$\prod_i X_i^{\nu_{ij}}$ and Arrhenius kernel
$k_j(T) = A_j T^{b_j} \exp(-E_{a,j}/RT)$. The first-layer linear
map is $\mathrm{Sym}(\Sp)$-equivariant by construction, placing
CRNN inside the real-relaxation of $\Lk_0^\Para(P)$
architecturally; convergence to integer $\nu$ at a trained
optimum recovers strict $\Lk_0$ membership. The composition
yields a deterministic mass-action vector field --- a morphism in
the Feinberg--Horn--Jackson category of chemical reaction
networks. But CRNN is not a $\Stoch$-morphism (no CME generator
$\Omega$, no stochastic trajectories), not a dagger-category
morphism (forward and reverse $(A, b, E_a)$ are independent, so
$\ker(\FG^T)$ is not preserved and detailed balance is not
enforced), and not a morphism in $\Lk_1(P)$ either: the Arrhenius
activation energy $E_a$ is not $\Delta H$, so it is not the Hess
image of a reaction arrow. CRNN's correct placement is therefore the
$\RR$-relaxation of $\Lk_0^\Para(P)$ \emph{augmented} by a
deterministic rate law that sits beside, not inside, the
$\Lk_1$--$\Lk_3$ portion of the tower.

\paragraph{Reaction SMILES sequence models.}
Molecular Transformer maps token sequences in $\Sigma^*$ to token
sequences in $\Sigma^*$ by autoregressive decoding. No graph
object appears in its type signature: the token stream need not
parse to a molecular graph at all, and when it does, the mapping
from strings to objects of $\LGraphP$ is partial and learned.
Atom and charge conservation, bond validity, and
$\Aut(G)$-equivariance are all training-emergent rather than
architectural, with documented failures even on in-distribution
benchmarks. The image of $\mathsf{Func}(\Theta, f_\theta)$ does
not lie in the DPO-span category forced by the $\Lk_3 \to \Lk_4$
pair, so there is no pushout complement on which to claim $\Lk_4$
membership; there is not even a $\mathrm{Sym}(\Sp)$-action on the
input side, since distinct SMILES orderings of the same reactant
multiset are distinct input sequences. The architectural ceiling
is $\Mon$ itself: conservation of species multiplicities is not
enforced by construction, and Molecular Transformer sits strictly
below $\Lk_0^\Para(P)$.

\paragraph{The $\Lk_5^\Para$ (invariant) block.}
All nine architectures in this block target a scalar potential
$V \colon \Ce(G) \to \RR$ with forces recovered by
autodifferentiation on the conservative head. Coulomb matrix and
SOAP, as standalone feature maps, sit below $\Lk_5$; paired with
a regressor (KRR, GAP) they realise $\Lk_5^\Para(P)$ (invariant).
Behler--Parrinello and ANI enforce
$O(3) \times \mathrm{Sym}(\Sp) \times \RR^3$-invariance by
construction via atom-centred symmetry functions; SchNet achieves
the same invariances via continuous-filter convolutions with
pairwise-distance-dependent filters. DimeNet enriches the message
with 3-body angles; GemNet extends to 4-body dihedrals with a
universal-approximation guarantee for $SE(3)$-invariant
continuous functions on $\Ce(G)$ at the conservative head. In all
these cases the outputs are scalar: internal rotationally
covariant features are discharged through an invariant readout.
ALIGNN takes the line-graph perspective to expose bond--bond
angles, again at invariant output. PaiNN is the sole borderline
case: its $l{=}0$ and $l{=}1$ internal features are
$O(3)$-equivariant by construction, and its non-energy heads
(dipoles, polarizabilities) exit $\OrbMorse$ --- but they land in
rank-$\leq 2$ tensor bundles over $\Ce(G)$ that are topologically
trivial (no eigenvalue-crossing locus seeds a non-trivial Chern
class), so nothing here advances to the topologically non-trivial
$\Hel^{(N)} \to \Ce(G)$ that $\Lk_6$ demands. For all nine, the
ceiling is the same: no $SE(3)$-equivariant scalar force field
provides a section of $U_6 \colon \HilbBund \to \OrbMorse$, and
the class $[\gamma_B] \in H^1(\Ce(G)\setminus\Xseam, \ZZ_2)$ is
not a functional of any real-valued $V$
(Proposition~\ref{prop:topological-incompleteness}). The GemNet
direct-force variant falls out of $\Lk_5^\Para(P)$ altogether:
when $F$ is predicted as a separate head rather than as $-\nabla V$,
the morphism exits the conservative-force-field category and fails
$(\mathrm{E}_5)$ strictly.

\paragraph{The $\Lk_5^\Para$ (equivariant) block.}
NequIP, MACE, MACE-MP-0, Allegro, So3krates, SO3LR, Equiformer,
and eSEN all satisfy $(\mathrm{E}_5)$ as an architectural
guarantee. Their internal features carry $O(3)$ irrep labels with
$l \leq l_{\max}$, composed via e3nn Clebsch--Gordan products,
eSCN $SO(2)$ convolutions, or $SE(3)$-equivariant self-attention,
and the energy head reduces to a scalar. The distinctions among
them are not tower-level distinctions. MACE adds body-order via
the atomic cluster expansion; a separate observation, developed
in Section~\ref{sec:para-mace}
(Proposition~\ref{prop:mace-conflation}), is that the
MACE-specific forgetful $U_4$ cannot cleanly separate $\Lk_4$
bond topology from $\Lk_5$ geometry in the internal representation.
MACE-MP-0 is MACE trained at foundation-model scale; its codomain
is unchanged. Allegro replaces atom-centred message passing with
iterated pairwise tensor products, giving a receptive field
bounded by $r_{\mathrm{cut}}$ regardless of depth. So3krates uses
$SE(3)$-equivariant self-attention in spherical-harmonic
coordinates with in-principle-unbounded range. SO3LR augments the
So3krates core with explicit analytical baselines so that
$E = E_{\mathrm{SO3k}} + E_{\mathrm{Coul}} + E_{\mathrm{Disp}}
+ E_{\mathrm{ZBL}}$ handles long-range physics beyond the
attention's practical reach; the linearity of this decomposition
is additivity in the codomain $\RR$, not functoriality in a
source reaction category. The Hess functor
$\FH \colon \Lk_1(P) \to B\RR$ requires a source $\Lk_1(P)$ of
reaction morphisms on which composition is defined, and no such
category is in SO3LR's signature: labelling SO3LR as a partial
$\Lk_1$ inhabitant is therefore a category error. Equiformer and
eSEN extend the equivariant-attention family, with eSEN providing
strict conservation and polynomial-envelope smoothness. For all
eight, the Berry ceiling is identical: the codomain $\OrbMorse$
admits no lift to $\HilbBund$, the flat $\ZZ_2$ double cover of
$\Ce(G) \setminus \Xseam$ corresponding to $[\gamma_B]$ admits
no scalar trivialisation, and
$\sigma_0 \colon \Ce(G) \to \Hel^{(N)}$ is absent from the type
signature. The distinguishing features of the models in this
block --- body order (MACE), locality (Allegro), attention range
(So3krates), long-range baselines (SO3LR), $l_{\max}$
(EquiformerV2), smooth conservation (eSEN) --- all live inside
$\Lk_5$. In particular, neither global attention nor explicit
long-range baselines remedy the Berry obstruction, and neither
yields motion toward a reaction-side level: a scalar force field
has no reaction morphisms on which a dagger involution or $\FH$
could act.

\paragraph{Generative structure maps in the $\Lk_5^\mathrm{gen}$
  region.}
AlphaFold2 and AlphaFold3 inhabit a different tower object from
the force fields above. Where the $\Lk_5^\Para(P)$ entries
parametrise a function $V \colon \Ce(G) \to \RR$ \emph{on} the
configuration orbifold, AlphaFold parametrises a map
$f_\theta \colon \mathrm{Seq} \to \Ce(G)$ \emph{into} it --- one
point or a distribution over points per input sequence. Since
$\Ce(G) = \RR^{3n}/(SE(3) \rtimes \Aut(G))$ is a quotient, the
generative membership condition $(\mathrm{E}_5^{\mathrm{gen}})$
is that the map factor through this quotient well-definedly: any
lift $\widetilde{f_\theta} \colon \mathrm{Seq} \to \RR^{3n}$ must
produce the same orbit in $\Ce(G)$ irrespective of which frame
the lift is expressed in. AlphaFold2 achieves this
architecturally. Invariant Point Attention computes attention
scores from vectors expressed in each residue's local frame so
that the scores are $SE(3)$-invariant while the frames themselves
update $SE(3)$-equivariantly; FAPE uses signed distances in the
frame and is therefore parity-sensitive, penalising chirality
inversion. Torsion-angle prediction with idealised residue
geometries, together with Amber relaxation, approximate
$\Gstar$-equivariance well enough to handle most stereocentres
correctly --- though without formal $(\mathrm{E}_{4.5})$
guarantees. AF2 is therefore strict at $\Lk_5^{\Para,\mathrm{gen}}$
with a partial (not strict) $\Lk_{4.5}$ axis on top.
AlphaFold3's situation differs structurally. Its diffusion
denoiser is a standard Transformer; rotation and translation
augmentation (without reflection) realises the
frame-independence required by $(\mathrm{E}_5^{\mathrm{gen}})$
only as a training-emergent property rather than as an
architectural one, and the chirality penalty operates in the
inference ranking score rather than in the denoising loss.
Under the architectural reading of $(\mathrm{E}_5^{\mathrm{gen}})$
committed to in the opening insightbox of this section, AF3
therefore sits at $\Lk_5^\mathrm{gen}$-type rather than at
strict $\Lk_5^{\Para,\mathrm{gen}}$; the $(\mathrm{E}_{4.5})$
failure --- 4.4\% chirality-violation rate on the PoseBusters
benchmark, reported in the primary paper itself --- is a
separate, quantifiable obstruction on the parity axis on top of
the $SE(3)$-axis training-emergence. Neither AF2 nor AF3 sits
inside $\Lk_5^\Para(P)$ proper --- they are generative not
functional over $\Ce(G)$ --- and both sit outside
$\Lk_6^\Para(P)$ by a wider margin than the force fields, since
even the scalar energy functor $\FV$ is absent from their
signatures.

\paragraph{Approaching $\Lk_6^\Para$.}
The final block divides into strata by how much of the $\Lk_6$
object each model constructs architecturally. SpookyNet augments
$\Lk_5$ (invariant) with global labels $\{Q, S\}$ broadcast to
each atom: these label connected components of the base space
(pre-$\Lk_7$ superselection sectors), not $U(N)$ fibre
coordinates, and the output $V(\mathbf{R}; Q, S)$ remains in
$\OrbMorse$ on each sector. QIM consists of three parametric
maps through a shared latent space $Z$
--- a structure encoder $\Ce(G) \to Z$, a structure decoder
$Z \to \Ce(G)$, and a property encoder $\RR^d \to Z$ ---
jointly trained on an ELBO plus property-likelihood objective
so that $Z$ serves as a common representation for molecular
geometries and QM property tuples~\cite{Fallani2024QIM}.
QIM is $\Lk_5$-type rather than a strict $\Lk_5^\Para(P)$ object
because the input representation is the raw Coulomb matrix, whose
row/column order depends on the atom labelling: the structure
encoder is $SE(3)$-invariant by construction but not strictly
$\Aut(G)$-equivariant, with $\mathrm{Sym}(\Sp)$-invariance left to
the training procedure rather than enforced architecturally, and
the structure decoder reconstructs geometries only up to a
chirality transformation that the Coulomb matrix does not
distinguish~\cite{Fallani2024QIM}.
Under the architectural reading of $(\mathrm{E}_5)$ committed to
in the opening insightbox of this section, this is below strict
$\Lk_5^\Para(P)$ membership; the placement at $\Lk_5$-type
reflects the dominant architectural $SE(3)$-invariance and the
scalar/orbifold output type.

The model's named direction is \emph{inverse design}: a target
property tuple $y \in \RR^d$ encodes to a latent $z$, which the
structure decoder returns to a candidate geometry
$\hat{\mathbf{R}} \in \Ce(G)$.
Forward property prediction is implicit through the shared
latent and approximates the composite
$\Pi \circ \sigma_0 \colon \Ce(G) \to \RR^d$
(Born--Oppenheimer ground-state section followed by observable
projection) without instantiating
$\sigma_0 \colon \Ce(G) \to \Hel^{(N)}$ as a separate morphism:
no wavefunction, density matrix, Hamiltonian, or bundle section
appears at any intermediate layer.
Both output types are $\Lk_5$-type --- scalar property tuples
in $\RR^d$ from the property pathway, reconstructed geometries
in $\Ce(G)$ from the structure decoder --- and the multi-valued
phase structure around $\Xseam$ that carries $[\gamma_B]$ is
categorically absent from the type signature.
The VAE's single-valued reconstruction objective is consistent
with this scalar codomain but rules out any multi-valued
output: the bundle section $\sigma_0$, whose phase ambiguity
around conical intersections carries $[\gamma_B]$, is not in
the architecture's range. SchNarc reaches further: its predicted
NAC vectors $d_{ij}^\mu(\mathbf{R})$ are, pointwise, components
of the Berry connection $A_{mn}^\mu$ in the adiabatic basis, so
the codomain is partially $\Lk_6$-type. But the parametrisation
$d_{ij}^\mu = \nabla_\mu s_{ij}$ for learned single-valued
scalars $s_{ij}$ spans only exact 1-forms: for every closed loop
$\gamma \subset \Ce(G) \setminus \Xseam$,
$\oint_\gamma d_{ij}^\mu\, dR_\mu
  = \oint_\gamma ds_{ij} = 0$,
so the geometric phase $[\gamma_B]$ is forced to zero by
construction and $(\mathrm{E}_6)$ fails precisely on the loops
where $[\gamma_B]$ should be non-trivial.

PhiSNet and DeepH-E3 reach the strongest $\Lk_6$-approach in
the audit. Each predicts the electronic Hamiltonian
$H_{\mathrm{el}}(\mathbf{R})$ directly, block-structured in a
fixed atomic-orbital basis, with exact $SE(3)$-equivariance on
the AO-block structure via e3nn tensor products. The codomain is
$\mathrm{End}(\Hel^{\mathrm{AO}})$, and the $SE(3)$-equivariance
of the operator-valued map is architectural. Diagonalising
$H_{\mathrm{el}}(\mathbf{R})$ recovers the bundle fibres of
$\Hel^{(N)} \to \Ce(G)$ pointwise, but the fixed AO basis
trivialises the bundle globally: $U(N)$ gauge freedom on the
fibres is frozen, $\sigma_0$ is available only post-hoc with
globally undefined eigenvector phases, and no architectural
constraint enforces $c_1$-correctness or $[\gamma_B]$-consistency
around loops in $\Ce(G) \setminus \Xseam$. This places PhiSNet
and DeepH-E3 in the operator-level sub-category
$\Lk_6^{\Para,\mathrm{op}}(P)$ defined in the mathbox above ---
orthogonal to, not subsumed by, the section-level sub-category
$\Lk_6^{\Para,\mathrm{sec}}(P)$. Full $\Lk_6^\Para(P)$
membership would require both operator-level fidelity and
section-level gauge covariance, and no architecture audited
here provides the latter. SPAINN is the sole model in the audit
that offers a partial approach to the section-level direction.
On a PaiNN backbone it predicts equivariant vector NACs directly
(not as gradients of scalars), so the output is a genuine
section of $A_{mn}^\mu$ and the exact-form obstruction affecting
SchNarc is lifted. But no $U(N)$-gauge-equivariance is enforced
on the state indices $(m, n)$, no diabatic-Hamiltonian output is
provided, and no topological-loss enforcement of $[\gamma_B]$
is in place; SPAINN therefore occupies an intermediate position
on the section-level axis --- a connection-level partial approach
toward $\Lk_6^{\Para,\mathrm{sec}}(P)$ --- while making no
attempt at the operator-level axis
$\Lk_6^{\Para,\mathrm{op}}(P)$, since it does not output
$H_{\mathrm{el}}$.

\begin{chembox}[Reading the classification]
\textbf{Level assignment.}
A model's level is the highest $k$ for which its function class
$\mathsf{Func}(\Theta, f_\theta)$ is contained in $\Lk_k(P)(X, Y)$
\emph{for every $\theta \in \Theta$}, not for the specific
$\theta^*$ a training run happens to produce. An
$E(3)$-equivariant model at $\Lk_5^\Para(P)$ (eq.)\ satisfies
$f_\theta(R \cdot \mathbf{R}) = D(R)\, f_\theta(\mathbf{R})$
for all $R \in E(3)$, all $\mathbf{R} \in \Ce(G)$, and all
$\theta \in \Theta$, with $D(R)$ the block-diagonal Wigner matrix
on the output irreps. This equivariance is enforced by
Clebsch--Gordan contractions (NequIP, MACE, Allegro), eSCN
$SO(2)$ convolutions (EquiformerV2, eSEN), or
$SE(3)$-equivariant attention (So3krates, SO3LR), not by data
augmentation.

\textbf{Operator-level vs section-level $\Lk_6^\Para$.}
PhiSNet and DeepH-E3 reach the operator-level sub-category
$\Lk_6^{\Para,\mathrm{op}}(P)$: codomain
$\mathrm{End}(\Hel^{\mathrm{AO}})$ with $SE(3)$-equivariance of
the operator-valued map architectural, and $U(N)$ fibre gauge
frozen by the basis choice. The section-level sub-category
$\Lk_6^{\Para,\mathrm{sec}}(P)$ would require $U(N)$-gauge
equivariance on state indices, prediction of $A_{mn}^\mu$ as a
connection 1-form, and architectural enforcement of
$[\gamma_B]$-consistency around loops encircling $\Xseam$.
Operator- and section-level are orthogonal approximations, not
nested: full $\Lk_6^\Para(P)$ requires both simultaneously.
SPAINN offers the only partial (connection-level) approach
toward section-level in the audit.
No architecture reaches both sub-categories, so none reaches
full $\Lk_6^\Para(P)$.

\textbf{The $(\ddagger)$ marker.}
Every scalar-energy $E(3)$-equivariant model --- across the
invariant and equivariant $\Lk_5$ blocks --- shares one ceiling:
the codomain $\OrbMorse$ is topologically incomplete. No
$SE(3)$-equivariant scalar force field provides a section of the
forgetful $U_6 \colon \HilbBund \to \OrbMorse$, and the class
$[\gamma_B] \in H^1(\Ce(G)\setminus\Xseam, \ZZ_2)$ is not a
functional of any real-valued $V$
(Proposition~\ref{prop:topological-incompleteness}). Increasing
body order, attention range, training data coverage, or
$l_{\max}$ leaves the codomain unchanged.

\textbf{AlphaFold and the generative $\Lk_5$ region.}
AlphaFold2 and AlphaFold3 do not inhabit $\Lk_5^\Para(P)$
proper: they parametrise maps \emph{into} $\Ce(G)$ rather than
functions on it. Both belong in the generative $\Lk_5$
region, but only AlphaFold2 reaches strict
$\Lk_5^{\Para,\mathrm{gen}}$ membership. AF2's Invariant Point
Attention is architecturally $SE(3)$-equivariant by construction
(frame-invariant attention scores with $SE(3)$-equivariant frame
updates); FAPE and chirality/torsion losses then carry an
approximate $\Gstar$-equivariance on top, supporting a
\emph{partial} $\Lk_{4.5}$ axis. AF3's diffusion denoiser is a
standard Transformer with $SE(3)$ pursued through
rotation--translation augmentation only: equivariance on the
$SE(3)$ axis is training-emergent, not architectural, so AF3
sits at $\Lk_5^\mathrm{gen}$-type rather than strict
$\Lk_5^{\Para,\mathrm{gen}}$. The 4.4\% PoseBusters
chirality-violation rate reported in the AF3 primary paper is a
direct $(\mathrm{E}_{4.5})$ failure certificate on the parity
axis, on top of the $SE(3)$-axis training-emergence.

\textbf{The missing $\Lk_1$--$\Lk_4$ region.}
A striking observation from the audit: no architecture in the
table reaches $\Lk_1^\Para(P)$, $\Lk_2^\Para(P)$,
$\Lk_3^\Para(P)$, or $\Lk_4^\Para(P)$ architecturally. CRNN
sits at the $\RR$-relaxation of $\Lk_0^\Para(P)$ with a
mass-action body beside the tower, not inside
$\Lk_3^\Para(P)$. Molecular Transformer, the only
reaction-mechanism architecture in the table, fails even
$\Lk_0^\Para(P)$ and therefore all higher levels. Hess-additive
thermochemistry ($\Lk_1$), dagger-preserving reverse kinetics
($\Lk_2$), $\Stoch$-valued CME generators ($\Lk_3$), and DPO
graph rewriting ($\Lk_4$) are categorically accessible tower
structures that no ML architecture surveyed here (as of
May 2026) realises. The
asymmetry is not arbitrary: the $\Lk_0$ and $\Lk_5$ plateaus
have canonical architectural realisations (hard-coded
stoichiometric linear layers; $E(3)$-equivariant tensor
contractions), while $\Lk_1$--$\Lk_4$ would require reaction
categories as first-class internal representations and
$\Lk_6^{\mathrm{sec}}$ would require $U(N)$-gauge-equivariant
topological losses --- neither pattern has a canonical
implementation yet. The forcing-pair constructions at these
levels are accordingly the most architecturally empty region of
the classification, and the clearest invitation the tower issues
to ML architects.

\textbf{Stoichiometric and reaction-side corrections.}
The reaction-side region of the table required the strongest
corrections from earlier drafts. DrFP loses $\Mon$ itself to
hashing and bit-folding before $\FH$ can even be stated.
ChemNODE has no $\mathrm{Sym}(\Sp)$-equivariant generator and no
Markov-kernel structure. Molecular Transformer has no graph
object in its type signature. Each fails the tower level
sometimes attributed to it because the tower structure at
issue --- Hess additivity, Markov-kernel positivity, DPO graph
rewriting --- is not an architectural property but a
training-emergent or type-compatible one, when it is present at
all. CRNN is the sole reaction-side architecture that realises
$\Lk_0^\Para(P)$ architecturally (up to the $\RR$-relaxation of
integer stoichiometric coefficients), via its hard-coded
stoichiometric linear layer; its mass-action body then sits
beside, not inside, the $\Lk_1$--$\Lk_3$ portion of the tower.

\textbf{Census across the 29 audited architectures.}
Below $\Lk_0^\Para(P)$: 3 (DrFP, ChemNODE, Molecular Transformer).
At $\Lk_0^\Para(P)$ (with mass-action kinetics beside): 1 (CRNN).
At $\Lk_1^\Para(P)$--$\Lk_4^\Para(P)$: 0.
At $\Lk_5^\Para(P)$ (inv.): 9.
At $\Lk_5^\Para(P)$ (eq.): 8.
Generative $\Lk_5$ region: 2 (AlphaFold2 at strict
$\Lk_5^{\Para,\mathrm{gen}}$ with partial $\Lk_{4.5}$;
AlphaFold3 at $\Lk_5^\mathrm{gen}$-type, not strict, with
$(\mathrm{E}_{4.5})$-failure certificate). Approaching $\Lk_6^\Para(P)$: 6 (SpookyNet,
QIM, SchNarc all $\Lk_5$-type with $\Lk_6$-type aspirations;
PhiSNet and DeepH-E3 at operator-level
$\Lk_6^{\Para,\mathrm{op}}$; SPAINN at partial connection-level
toward $\Lk_6^{\Para,\mathrm{sec}}$). At full
$\Lk_6^\Para(P)$: 0. At $\Lk_7^\Para(P)$: 0. The distribution
identifies contemporary ML-for-chemistry as a dense
$\Lk_5$-plateau population with sparse reaction-side occupancy
and an operator-level upper ceiling.
\end{chembox}

%% file: chapters/para/para_mace.tex
\subsection{MACE as the primary $\Lk_5^{\ParaC}$ case study}
\label{sec:para-mace}

MACE~\cite{BatatIa2022MACE} is the primary case study for
$\Lk_5^{\ParaC}(P)$ for two reasons. First, the membership
condition $(\mathrm{E}_5)$ of Definition~\ref{def:LkPara} holds
\emph{architecturally}: the Clebsch--Gordan contraction scheme
enforces translation invariance and $O(3)$-equivariance at every
site of the computation, for every parameter setting, without
any training constraint or appeal to the type signature of the
output. The resulting containment of the function class
$\mathsf{Func}(\Theta_{\mathrm{MACE}}, f^{\mathrm{MACE}}_\theta)$
(Remark~\ref{rem:instantiation}) in $\Lk_5(P)(X, Y)$ is therefore
architectural in the strictest of the three senses of
$(\mathrm{E}_5)$ distinguished in
Section~\ref{sec:para-classification} --- not merely
training-emergent, and not dependent on type coincidence. The
precise statement is Proposition~\ref{prop:mace-equivariant}.
Second, MACE's body-ordered ACE expansion makes the tower's
forgetful functor structure visible: its $\nu$-fold contraction
features carry a superficial $\Lk_4$ flavour, but the resemblance
is not backed by a factorisation through
$U_4 \colon \Lk_5(P) \to \Lk_4(P)$. Making that failure precise
is the content of Proposition~\ref{prop:mace-conflation}, the
clearest available illustration of what separates the metric
level $\Lk_5$ from the combinatorial level $\Lk_4$ below it.

\subsubsection{The ACE expansion and its tower interpretation}

MACE builds atomic features through a body-ordered expansion
starting from single-neighbour edge features:
\[
  A_{i, n l m}^{(1)}
  \;=\; \sum_{j \in \mathcal{N}(i)}
  R_{nl}(r_{ij})\, Y_l^m(\hat{\mathbf{r}}_{ij})\, h_j,
\]
where $R_{nl}$ are radial basis functions, $Y_l^m$ are real
spherical harmonics, $h_j$ are learned element-type embeddings,
and $\mathcal{N}(i) = \{j : r_{ij} < r_{\mathrm{cut}}\}$ is a
metric neighbour set determined by a fixed cutoff
$r_{\mathrm{cut}}$. Higher-order features are $\nu$-fold
symmetric tensor products of $A^{(1)}$ features, contracted
through Clebsch--Gordan coefficients:
\[
  B_{i,\nu}^{(lm)}
  \;=\; \sum_{\{n_k, l_k, m_k\}}
  C^{lm}_{\{l_k m_k\}}\;
  \prod_{k=1}^{\nu} A_{i, n_k l_k m_k}^{(1)}.
\]
A $\nu$-fold product captures a centred correlation among atom
$i$ and $\nu$ of its neighbours, i.e.\ a $(\nu + 1)$-body
correlation. With ACE order $\nu = 3$ and $l_{\max} = 2$, each
message-passing layer produces $4$-body correlations; two MACE
layers match the accuracy of NequIP~\cite{Batzner2022NequIP}
at four to six layers.

\begin{mathbox}[Body order and the forgetful functor $U_4$]
  The feature $B_{i,\nu}$ at ACE order $\nu$ aggregates
  contributions from $\nu$ neighbours of atom $i$ through
  Clebsch--Gordan contraction, producing a $(\nu+1)$-body
  correlation indexed by $O(3)$-irrep labels $(l, m)$ (with
  $\nu$ neighbour indices summed over internally). The
  pre-contraction product $\prod_{k=1}^{\nu} A^{(1)}$ has a
  superficial resemblance to a $\nu$-fold tensor product over
  bond generators --- the combinatorial flavour of $\Lk_4(P)$,
  where $\nu$-fold bond structures enter through DPO spans on
  $\LGraphP$.

  The resemblance is superficial. In the tower, the forgetful
  functor $U_4 \colon \Lk_5(P) \to \Lk_4(P)$ discards the
  configuration-orbifold data $\Ce(G)$ (the metric level of
  $\Lk_5$) and retains only the labelled graph
  $G \in \LGraphP$, whose edges encode chemical bonds by
  combinatorial data (bond order, formal charges, lone pairs).
  The neighbour set driving the ACE contraction, by contrast, is
  metric: $\mathcal{N}(i)$ is the set of atoms inside a sphere
  of radius $r_{\mathrm{cut}}$, computed directly from the
  coordinates. A pair $(i, j)$ can belong to $\mathcal{N}(i)$ at
  one geometry and not at another, without any bond having formed
  or broken. The ACE contraction indices run over metric
  neighbours ($\Lk_5$ data), not over bonded edges ($\Lk_4$
  data), and the body-order construction --- despite its
  $\nu$-fold appearance --- does not factor through $U_4$.
  Proposition~\ref{prop:mace-conflation} makes this
  non-factorisation precise.
\end{mathbox}

\subsubsection{Verification of condition $(\mathrm{E}_5)$ for MACE}

\begin{proposition}[MACE satisfies $(\mathrm{E}_5)$ architecturally]
\label{prop:mace-equivariant}
  Let $X = \RR^{3n}$ be the ambient configuration space of $n$
  atoms with the standard $E(3) = O(3) \ltimes \RR^3$ action
  (rotations, reflections, and translations acting diagonally on
  atomic positions), and let $Y$ be a finite-dimensional
  $O(3)$-representation, decomposed as a direct sum of irreps.
  Let $f^{\mathrm{MACE}}_\theta \colon X \to Y$ denote the MACE
  forward map with parameters $\theta \in \Theta_{\mathrm{MACE}}$.
  Then for every $\theta$, every $\mathbf{x} \in X$, every
  $t \in \RR^3$, and every $R \in O(3)$,
  \[
    f^{\mathrm{MACE}}_\theta(\mathbf{x} + t)
    \;=\; f^{\mathrm{MACE}}_\theta(\mathbf{x}),
    \qquad
    f^{\mathrm{MACE}}_\theta(R \cdot \mathbf{x})
    \;=\; \rho(R)\,f^{\mathrm{MACE}}_\theta(\mathbf{x}),
  \]
  where $\rho$ is the representation of $O(3)$ carried by $Y$
  (block-diagonal across irreps; for scalar-energy outputs
  $Y = \RR$ and $\rho \equiv 1$). Consequently
  $\mathsf{Func}(\Theta_{\mathrm{MACE}},
  f^{\mathrm{MACE}}_\theta) \subseteq \Lk_5(P)(X, Y)$ and
  $(\Theta_{\mathrm{MACE}}, f^{\mathrm{MACE}}_\theta)$ is a
  parametric 1-morphism of $\Lk_5^{\ParaC}(P)$.
\end{proposition}

\begin{proof}
  Translations in $\RR^3$ fix both the pair distance $r_{ij}$
  and the pair direction $\hat{\mathbf{r}}_{ij}$, so every
  $A^{(1)}_{i, n l m}$ is translation-invariant. Under
  $R \in SO(3)$, $r_{ij}$ is invariant and the real spherical
  harmonics transform among themselves as the $(2l+1)$-dimensional
  real representation of $SO(3)$, which we denote $D^{(l)}(R)$
  (a real-valued matrix obtained from the complex Wigner matrix
  by a standard similarity transformation):
  $Y_l^m(R \cdot \hat{\mathbf{r}}) = \sum_n
  D^{(l)}_{mn}(R)\, Y_l^n(\hat{\mathbf{r}})$. Hence $A^{(1)}_i$
  transforms under $SO(3)$ at each $l$-block as the $l$-th
  $SO(3)$-irrep. Under parity, $Y_l^m$ picks up the character
  $(-1)^l$, so each $l$-block is an $O(3)$-irrep of definite
  parity $(l, (-1)^l)$. Each Clebsch--Gordan
  contraction $B_{i,\nu}^{(lm)} = \sum_{\{n_k, l_k, m_k\}}
  C^{lm}_{\{l_k m_k\}} \prod_k A^{(1)}_{i, n_k l_k m_k}$ is by
  construction an equivariant map of $O(3)$ representations, so
  the coupled features $B_{i,\nu}^{(lm)}$ carry well-defined
  $O(3)$-irrep labels at every layer. The readout projects these
  features onto $Y$; by architectural construction it reads from
  channels of matching $l$ and matching parity, so the output
  transforms under the advertised $\rho(R)$. In the standard
  scalar-energy case, $Y = \RR$ is the trivial irrep, the
  readout draws only from $l = 0$, even-parity channels, and the
  total energy $E(\mathbf{x}) = \sum_i E_i(\mathbf{x})$ is
  $E(3)$-invariant; the force field $-\nabla_{\mathbf{x}} E$
  then follows as the $l = 1$, odd-parity output demanded by the
  chain rule ($F(R\mathbf{x}) = R F(\mathbf{x})$,
  $F(-\mathbf{x}) = -F(\mathbf{x})$). Since no step depends on any
  particular $\theta$, the equivariance holds for every
  $\theta \in \Theta_{\mathrm{MACE}}$, giving
  $\mathsf{Func}(\Theta_{\mathrm{MACE}},
  f^{\mathrm{MACE}}_\theta) \subseteq \Lk_5(P)(X, Y)$.
\qedhere
\end{proof}

\begin{remark}[Comonoid structure on $\Theta_{\mathrm{MACE}}$;
  and what it does \emph{not} encode]
\label{rem:mace-comonoid}
  Clause~(ii) of Definition~\ref{def:LkPara} requires the
  parameter space of a Para 1-morphism to carry a comonoid
  structure $(\Theta, \Delta, !)$ in the ambient cartesian
  category. For MACE this is realised by the canonical cartesian
  comonoid: $\Delta \colon \Theta_{\mathrm{MACE}} \to
  \Theta_{\mathrm{MACE}} \otimes \Theta_{\mathrm{MACE}}$ is the
  diagonal (duplicating a parameter tensor), and $! \colon
  \Theta_{\mathrm{MACE}} \to I$ is the unique map to the terminal
  object (forgetting parameter dependence, not setting
  parameters to any particular value). The terminal map $!$
  carries no architecturally specific content beyond the
  cartesian structure; the non-trivial ingredient is $\Delta$.

  The architectural role of $\Delta$ in MACE is the use of a
  single learned parameter tensor in multiple places. The clearest
  example is element embeddings: a learned vector $h_Z$ is
  associated with each chemical element $Z$ and fed to every
  atom of element $Z$ in the molecule, realised by $\Delta$
  duplicating the stored tensor into as many copies as there are
  atoms of type $Z$. Within a single message-passing layer,
  architectural weight sharing across atom-centred features is
  likewise implemented by $\Delta$.

  This comonoid structure is distinct from the separate
  \emph{expressiveness} claim --- the question of which target
  morphisms the ACE body-order expansion can approximate, as
  the ACE order and angular-momentum cutoff grow. The comonoid
  governs which parameter reuses the architecture permits;
  expressiveness bounds what the composed features can
  represent. Conflating the two has been a recurring error in
  informal discussions of MACE's tower coordinates.
\end{remark}

\subsubsection{The $\Lk_4/\Lk_5$ conflation}

\begin{proposition}[MACE's body-order features are not determined
  by the bond graph]
\label{prop:mace-conflation}
  Fix $\theta \in \Theta_{\mathrm{MACE}}$ outside a measure-zero
  degenerate set, and fix $G \in \LGraphP$ admitting at least one
  dihedral degree of freedom. The MACE body-order feature map
  \[
    B^G_{\mathrm{MACE},\theta} \colon
    \Ce(G) \longrightarrow V_G,
    \qquad
    \mathbf{x} \longmapsto
    \bigl(B_{i,\nu}^{(lm)}(\mathbf{x}; \theta)\bigr)_{i,\nu,l,m},
  \]
  where $V_G$ is the real vector space of ACE features at $G$,
  is a non-constant function on $\Ce(G)$. Because the forgetful
  functor $U_4 \colon \Lk_5(P) \to \Lk_4(P)$ collapses the entire
  configuration orbifold $\Ce(G)$ to the single $\Lk_4$-object
  $G$, no function $\widetilde{B}_{\theta, G}$ can satisfy
  $B^G_{\mathrm{MACE},\theta} = \widetilde{B}_{\theta, G} \circ
  U_4|_{\Ce(G)}$. The family
  $\{B^G_{\mathrm{MACE},\theta}\}_{G \in \LGraphP}$ therefore
  does not factor through $U_4$, and the same non-factorisation
  is inherited by the Para lift
  $\ParaC(U_4) \colon \Lk_5^{\ParaC}(P) \to \Lk_4^{\ParaC}(P)$.
  Consequently MACE's body-order decomposition, despite its
  $\nu$-fold construction resembling a $\Lk_4$-level $\nu$-fold
  bond tensor, does not architecturally encode bond-topological
  $\Lk_4$ content.
\end{proposition}

\begin{proof}
  Fix $\theta$ outside the degenerate set and take $G$ to be
  n-butane. Choose two conformations $\mathbf{x}_1, \mathbf{x}_2
  \in \Ce(G)$ related by rotation about the central C2--C3
  dihedral, with all bond lengths and valence angles held
  constant. The dihedral rotation changes the $1,4$ distance
  $r_{14}$ and the direction $\hat{\mathbf{r}}_{14}$, so for any
  $r_{\mathrm{cut}}$ large enough to include atom $4$ in
  $\mathcal{N}(1)$ --- as standard MACE cutoffs of
  $4$--$6\,\mathrm{\AA}$ do on all-atom n-butane, given the
  C1--C4 range of roughly $2.5$--$3.9\,\mathrm{\AA}$ across
  conformations --- the contribution
  $R_{nl}(r_{14})\,Y_l^m(\hat{\mathbf{r}}_{14})\,h_4$ to the
  single-neighbour feature $A^{(1)}_1$ changes. Even for a
  cutoff that excludes C1--C4, the dihedral rotation still moves
  atoms on one side of the central bond, changing the directions
  $\hat{\mathbf{r}}_{ij}$ (hence
  $Y_l^m(\hat{\mathbf{r}}_{ij})$) for any neighbour $j$ that
  does move. Hence $A^{(1)}_1(\mathbf{x}_1) \ne
  A^{(1)}_1(\mathbf{x}_2)$, and by propagation through the
  Clebsch--Gordan contractions
  $B^G_{\mathrm{MACE},\theta}(\mathbf{x}_1) \ne
  B^G_{\mathrm{MACE},\theta}(\mathbf{x}_2)$ for every
  non-degenerate $\theta$: $B^G_{\mathrm{MACE},\theta}$ is
  non-constant on $\Ce(G)$.

  If $B^G_{\mathrm{MACE},\theta} = \widetilde{B}_{\theta, G}
  \circ U_4|_{\Ce(G)}$ held for some $\widetilde{B}_{\theta,
  G}$, the constancy of $U_4|_{\Ce(G)}$ would force
  $B^G_{\mathrm{MACE},\theta}$ constant on $\Ce(G)$ ---
  contradicting the counterexample. The same argument applies
  to any $G$ admitting a dihedral degree of freedom, so the
  family $\{B^G_{\mathrm{MACE},\theta}\}_G$ does not factor
  through $U_4$.

  Two structural features explain the failure independently of
  the counterexample. The neighbour set
  $\mathcal{N}(i) = \{j : r_{ij} < r_{\mathrm{cut}}\}$ is defined
  by a metric condition that $U_4$ discards, so changing
  $\mathbf{x}$ within $\Ce(G)$ can change $\mathcal{N}(i)$
  whenever an atom crosses the cutoff. Even without membership
  changes, $R_{nl}(r_{ij})$ and
  $Y_l^m(\hat{\mathbf{r}}_{ij})$ vary continuously with
  $\mathbf{x}$, so $A^{(1)}_i$ is a non-constant function on
  $\Ce(G)$ for every non-degenerate $\theta$.

  The Para lift inherits the non-factorisation as an
  architectural property. The MACE parametric 1-morphism
  $(\Theta_{\mathrm{MACE}}, f^{\mathrm{MACE}}_\theta)$ computes
  $f^{\mathrm{MACE}}_\theta(\mathbf{x})$ from the configuration
  $\mathbf{x}$ by first passing $\mathbf{x}$ through
  $B^G_{\mathrm{MACE},\theta}$ and then applying a per-atom
  linear readout followed by summation over atoms; the first
  stage architecturally reads the metric data $U_4$ discards.
  For every non-degenerate $\theta$, the n-butane conformers
  above receive distinct predicted energies, so
  $f^{\mathrm{MACE}}_\theta$ is non-constant on $U_4$-fibres.
  There is therefore no $\widetilde{f}_\theta \colon
  \mathrm{Ob}(\Lk_4(P)) \to Y$ with
  $f^{\mathrm{MACE}}_\theta = \widetilde{f}_\theta \circ
  U_4|_{\Ce(G)}$, and the same non-factorisation passes to
  $\ParaC(U_4)$.
\qedhere
\end{proof}

\begin{chembox}[The conflation in chemical practice]
  MACE achieves state-of-the-art accuracy for ground-state
  potential energy surfaces precisely because $\Lk_5^{\ParaC}(P)$
  is the correct tower level for that task: the PES is a
  function on $\Ce(G)$, and MACE's body-ordered expansion
  produces a scalar-energy output that is $E(3)$-invariant in
  the ambient coordinates $\RR^{3n}$, hence a well-defined
  function on the quotient $\Ce(G) = \RR^{3n}/(SE(3) \ltimes
  \Aut(G))$. (The full $E(3)$-equivariance of the underlying
  feature map is what enables tensorial outputs --- dipoles,
  polarizabilities --- as sections of the associated equivariant
  bundles over $\Ce(G)$.) For standard molecular dynamics ---
  propagating Newton's equations on a single bonding topology
  --- the absence of a functorial lift of the feature map to
  $\Lk_4$ is invisible: $U_4$ is not a quantity MACE is asked
  to compute.

  The conflation becomes operationally significant in three
  situations where the $\Lk_4$ content matters by itself. In
  \emph{reactive molecular dynamics}, bond formation and bond
  cleavage change the $\Lk_4$ labelled graph discontinuously,
  while MACE's cutoff graph registers the same event as a smooth
  variation in $\mathcal{N}(i)$; the architecture does not mark
  the categorical transition between distinct objects of
  $\LGraphP$ that defines a chemical reaction. In
  \emph{retrosynthesis}, predicting which bonds cleave and
  which form requires separating the $\Lk_4$ topology change
  from the concomitant $\Lk_5$ geometry change; MACE cannot
  perform this separation, because its internal feature graph
  is metric-dependent by construction and no functorial $U_4$
  lift extracts the bond-topological component. In
  \emph{mechanism assignment}, an $\mathrm{S}_{\mathrm{N}}1$ and
  an $\mathrm{S}_{\mathrm{N}}2$ path for the same net
  transformation pass through configurations with
  \emph{different} intermediate bond graphs
  $G \in \LGraphP$: $\mathrm{S}_{\mathrm{N}}1$ transits through
  a three-coordinate carbocation in one $\LGraphP$-object;
  $\mathrm{S}_{\mathrm{N}}2$ transits through a pentavalent
  transition state in another. MACE assigns a scalar energy to
  each nuclear configuration along either trajectory, but does
  not output a bond graph at each configuration --- and so the
  $\Lk_4$-level labelling that would distinguish the
  intermediates (three- versus five-coordinate carbon, distinct
  DPO spans) is absent from MACE's type signature. Reaction-path
  analysis has to supply this labelling externally, from data
  that Proposition~\ref{prop:mace-conflation} shows MACE does not
  carry.
\end{chembox}

\begin{insightbox}[MACE's tower coordinates]
Three facts together locate MACE in the tower. First,
$(\mathrm{E}_5)$ holds architecturally for all $\theta$
(Proposition~\ref{prop:mace-equivariant}):
$(\Theta_{\mathrm{MACE}}, f^{\mathrm{MACE}}_\theta)$ is a
parametric 1-morphism of $\Lk_5^{\ParaC}(P)$. Second, the
body-order feature map does not
factor through $U_4$ (Proposition~\ref{prop:mace-conflation}):
the ACE features, despite their $\nu$-fold construction, are
$\Lk_5$ data in $\Lk_4$-flavoured clothing, not $\Lk_4$ content
captured by the architecture. Third, MACE's output
codomain is $\OrbMorse$ (scalar energy), not
$\HilbBund$-valued: MACE is therefore not a 1-morphism of
$\Lk_6^{\ParaC}(P)$ at the level of type signature, and no
retraining can install the missing Berry connection
$A_{mn}^\mu$, the class $[\gamma_B]$, or a section of
$\Hel^{(N)} \to \Ce(G)$. This places MACE under the NequIP-class
$\Lk_6$ ceiling
(Proposition~\ref{prop:topological-incompleteness}, discussed in
Section~\ref{sec:para-classification}).
MACE is thus cleanly localised at $\Lk_5^{\ParaC}(P)$: its
body-order structure has no non-trivial $\Lk_4^{\ParaC}(P)$
content under $U_4$, and its scalar codomain blocks inclusion
in $\Lk_6^{\ParaC}(P)$. This makes
MACE a canonical illustration of what it means for a
parametric 1-morphism to sit \emph{at} a single tower level.
\end{insightbox}

%% file: chapters/para/para_so3krates.tex

\subsection{So3krates and SO3LR: two architectural designs at
  $\Lk_5^{\Para}(P)$}
\label{sec:para-so3krates}

So3krates~\cite{Frank2022So3krates} and SO3LR~\cite{Kabylda2025SO3LR}
are two further case studies at $\Lk_5^{\Para}(P)$, each
exposing a structural fact that MACE alone cannot illustrate.
Both inhabit this level as their primary tower membership;
neither reaches $\Lk_6^{\Para}(P)$, and the obstruction in
both cases is the same output-type argument that blocks MACE.

So3krates realises the membership condition $(\mathrm{E}_5)$ ---
equivalently, the NGN naturality condition of
Proposition~\ref{prop:nhaan-tower} --- through
$SE(3)$-equivariant self-attention rather than through local
Clebsch--Gordan contraction. Same condition, same tower level,
structurally different architectural route. So3krates thereby
shows that a given tower level does not pin down a unique
architectural strategy: two models can be objects of the same
$\Lk_k^{\Para}(P)$ by genuinely different mechanisms.

SO3LR, built on the So3krates backbone, makes a different
structural point. It remains at $\Lk_5^{\Para}(P)$ by primary
tower membership, and decomposes its total energy into a
short-range neural component and an analytic long-range
component whose functional form (Coulomb $1/r$, dispersion
$1/r^6$) is supplied as an architectural baseline rather than
learned from data. This is \emph{physics-informed range
separation}: not a second tower-level membership, but a design
pattern in which physics whose functional form is known in
advance is built into the architecture, leaving the neural
backbone to handle only the many-body correlations that
genuinely require learning. SO3LR provides a clean case study
of a machine-learning force field whose design factors known
physics out of the function class architecturally. Tower-level
membership fixes what a model is blocked from representing;
physics-informed range separation is a separate architectural
axis within a single tower level, distinct from the
strategy axis along which MACE and So3krates differ.

The section treats the two models in turn, then synthesises what
they demonstrate jointly.

\subsubsection{So3krates: $(\mathrm{E}_5)$ via spherical
  harmonic coordinates and global attention}

So3krates builds atomic representations by alternating
$SE(3)$-equivariant self-attention blocks with equivariant
feature mixing, on a basis of \emph{spherical harmonic
coordinates} (SPHCs): per-atom tensorial descriptors
\[
  \chi_i^{lm}
  \;=\; \sum_{j \in \mathcal{N}(i)}
  f(r_{ij})\, Y_l^m(\hat{\mathbf{r}}_{ij}),
  \qquad l = 0, 1, \ldots, l_{\max},
\]
where $f(r_{ij})$ is a radial envelope, $Y_l^m$ are real
spherical harmonics, and $\mathcal{N}(i)$ is a neighbour set.
The $\chi_i^{lm}$ transform under $R \in SO(3)$ as
$\chi_i^{lm} \mapsto D^{(l)}(R)\,\chi_i^{lm}$, so each $l$-block
is an $O(3)$-irrep of definite parity $(l, (-1)^l)$ (parity
absorbed into the irrep character as in
Proposition~\ref{prop:mace-equivariant}).

Self-attention over the molecule then aggregates SPHC features
according to attention weights $\alpha_{ij}$ computed from
$SE(3)$-invariant quantities (pairwise distance $r_{ij}$,
scalar features), so that $\alpha_{ij}$ is itself
$SE(3)$-invariant. Because the weights are invariant and the
values $\chi_j^{lm}$ are equivariant, the attention-weighted sum
$\sum_j \alpha_{ij}\,\chi_j^{lm}$ is $SE(3)$-equivariant by
construction, with no Clebsch--Gordan contraction required.
The attention range extends effective information flow beyond
what a fixed local cutoff supports, through a combination of
broader attention neighbourhoods and stacked attention blocks.
A per-atom readout projects the final-layer features onto the
output space $Y$ (a direct sum of $O(3)$-irreps; for the scalar
energy $Y = \RR$), followed by summation over atoms.

\begin{mathbox}[Local and longer-range routes to $(\mathrm{E}_5)$]
  MACE and So3krates implement the same tower membership
  condition $(\mathrm{E}_5)$ by structurally different
  architectural strategies.
  MACE realises the \emph{local route}: at each atom $i$, a
  body-ordered expansion $B_{i,\nu}^{(lm)}$ is constructed from
  Clebsch--Gordan products of single-neighbour features
  $A_{i, n l m}^{(1)}$, with the contraction confined to a
  neighbourhood of fixed cutoff radius. Equivariance at each
  edge, combined with the functoriality of aggregation on the
  molecular graph, gives global equivariance
  \cite{deHaanCohenWelling2020}.
  So3krates realises the \emph{longer-range route}: SPHCs are
  an $SE(3)$-equivariant basis from the outset, attention
  weights derived from pairwise invariants are $SE(3)$-invariant,
  and their application to the equivariant SPHC basis yields
  an $SE(3)$-equivariant output at every layer without any
  local-to-global propagation step.

  The architectural difference is substantial: local CG is a
  polynomial construction in the ACE sense, whereas attention
  is a data-dependent mixing in which the weights themselves
  respond to the input geometry. From the tower's perspective,
  both produce maps $f_\theta \colon \RR^{3n} \to Y$ with $Y$
  an $O(3)$-representation, satisfying $(\mathrm{E}_5)$ for
  every $\theta$. Their function classes are therefore both
  subsets of $\Lk_5(P)(X, Y)$ at the same tower level; neither
  is visibly contained in the other, since the polynomial and
  attention constructions occupy incomparable regions of
  $\Lk_5(P)(X, Y)$. The tower does not distinguish them at
  level $5$: what it cares about is whether $(\mathrm{E}_5)$
  holds, not how.
\end{mathbox}

\begin{proposition}[So3krates satisfies $(\mathrm{E}_5)$
  architecturally]
\label{prop:so3krates-equivariant}
  Let $X = \RR^{3n}$ with the standard $E(3) = O(3) \ltimes
  \RR^3$ action, and let $Y$ be a finite-dimensional
  $O(3)$-representation. Let $f^{\mathrm{So3k}}_\theta \colon X
  \to Y$ denote the So3krates forward map with parameters
  $\theta \in \Theta_{\mathrm{So3k}}$. Then for every $\theta$,
  every $\mathbf{x} \in X$, every $t \in \RR^3$, and every
  $R \in O(3)$,
  \[
    f^{\mathrm{So3k}}_\theta(\mathbf{x} + t)
    \;=\; f^{\mathrm{So3k}}_\theta(\mathbf{x}),
    \qquad
    f^{\mathrm{So3k}}_\theta(R \cdot \mathbf{x})
    \;=\; \rho(R)\,f^{\mathrm{So3k}}_\theta(\mathbf{x}),
  \]
  where $\rho$ is the representation of $O(3)$ carried by $Y$.
  Consequently $\mathsf{Func}(\Theta_{\mathrm{So3k}},
  f^{\mathrm{So3k}}_\theta) \subseteq \Lk_5(P)(X, Y)$ and
  $(\Theta_{\mathrm{So3k}}, f^{\mathrm{So3k}}_\theta)$ is a
  parametric 1-morphism of $\Lk_5^{\Para}(P)$.
\end{proposition}

\begin{proof}
  Under a rigid translation $\mathbf{x} \mapsto \mathbf{x} + t$,
  every pairwise distance $r_{ij}$ and every pairwise direction
  $\hat{\mathbf{r}}_{ij}$ is unchanged; hence each SPHC
  $\chi_i^{lm}$ is translation-invariant, as is every feature
  built iteratively from the SPHCs by $E(3)$-equivariant update
  rules.

  Under $R \in SO(3)$, pairwise distances remain invariant and
  pairwise directions transform as $\hat{\mathbf{r}}_{ij} \mapsto
  R\hat{\mathbf{r}}_{ij}$. Real spherical harmonics transform
  under the $(2l+1)$-dimensional real representation of $SO(3)$
  (denoted $D^{(l)}$, a real-valued matrix obtained from the
  complex Wigner matrix by a standard similarity transformation):
  \[
    Y_l^m(R\hat{\mathbf{r}})
    \;=\; \sum_n D^{(l)}_{mn}(R)\, Y_l^n(\hat{\mathbf{r}}).
  \]
  Writing $\chi_i^{lm}[\mathbf{x}]$ for the SPHC at atom $i$
  evaluated on the configuration $\mathbf{x}$, this gives
  \[
    \chi_i^{lm}[R\mathbf{x}]
    \;=\; \sum_j f(r_{ij})\, Y_l^m(R\hat{\mathbf{r}}_{ij})
    \;=\; \sum_n D^{(l)}_{mn}(R)\, \chi_i^{ln}[\mathbf{x}],
  \]
  so each $l$-block of the SPHC tuple at atom $i$ transforms as
  the $l$-th $SO(3)$-irrep. Attention scores $\alpha_{ij}$,
  depending only on pairwise distances and scalar features, are
  $SO(3)$-invariant: $\alpha_{ij}[R\mathbf{x}] =
  \alpha_{ij}[\mathbf{x}]$. The attention-weighted aggregation
  therefore transforms as
  \[
    \bigl[A(\chi)\bigr]_i^{lm}[R\mathbf{x}]
    \;=\; \sum_j \alpha_{ij}[R\mathbf{x}]\,
      \chi_j^{lm}[R\mathbf{x}]
    \;=\; \sum_n D^{(l)}_{mn}(R)\,
      \bigl[A(\chi)\bigr]_i^{ln}[\mathbf{x}],
  \]
  an $SO(3)$-equivariant feature at every layer and every atom,
  as is the per-atom readout; summation over atoms preserves
  equivariance. Under reflection, parity is absorbed by the
  character $(-1)^l$ on each $l$-block exactly as in
  Proposition~\ref{prop:mace-equivariant}. Since every step is
  an $E(3)$-equivariant construction for every $\theta$, the
  claimed equivariance holds for every
  $\theta \in \Theta_{\mathrm{So3k}}$.
\qedhere
\end{proof}

\begin{remark}[Comonoid structure on $\Theta_{\mathrm{So3k}}$]
\label{rem:so3krates-comonoid}
  The comonoid clause of Definition~\ref{def:LkPara} is realised
  by the canonical cartesian comonoid on
  $\Theta_{\mathrm{So3k}}$, exactly as for MACE
  (Remark~\ref{rem:mace-comonoid}). The architectural role of
  the diagonal $\Delta$ is the attention-specific
  parameter-sharing pattern: within each attention block, a
  single learned query/key/value projection tensor is shared
  across every atom, with the diagonal duplicating the tensor
  to every atom position. Stacking $L$ attention blocks
  introduces $L$ independent copies of the per-block parameter
  tensor, each shared across atoms within that block.
  The terminal map $!_\Theta$ forgets parameter dependence
  without selecting a particular value, consistent with
  Remark~\ref{rem:instantiation}.
\end{remark}

\begin{remark}[The $\Lk_6^{\Para}(P)$ ceiling is insensitive to
  attention range]
\label{rem:so3krates-ceiling}
  By Proposition~\ref{prop:so3krates-equivariant},
  $(\Theta_{\mathrm{So3k}}, f^{\mathrm{So3k}}_\theta) \in
  \Lk_5^{\Para}(P)$ with output type $\OrbMorse$ (scalar
  energy and its gradient).
  Proposition~\ref{prop:topological-incompleteness} applies
  unchanged: any parametric 1-morphism at $\Lk_5^{\Para}(P)$
  with output type $\OrbMorse$ is categorically blocked from
  $\Lk_6^{\Para}(P)$, because the forgetful functor
  $U_6 \colon \Lk_6^{\Para}(P) \to \Lk_5^{\Para}(P)$ discards
  the Hilbert bundle $\Hel^{(N)} \to \Ce(G)$, the Berry
  connection $A_{mn}^\mu$, and the topological class $[\gamma_B]
  \in H^1(\Ce(G) \setminus \Xseam, \ZZ_2)$, none of which can
  be reconstructed from a real-valued function on $\Ce(G)$. The
  attention range --- global or local --- does not enter the
  argument; the output type does. Extending So3krates's
  attention to the full molecule, or replacing attention with
  any other $\OrbMorse$-valued construction, leaves the tower
  level unchanged: tower coordinates are set by output type,
  not by computational range.
\end{remark}

\subsubsection{SO3LR: physics-informed range separation
  within $\Lk_5^{\Para}(P)$}

Pure neural-network force fields with finite cutoffs --- MACE,
NequIP, So3krates in its standard form --- predict a PES by
learning the full interatomic interaction from data, inside a
receptive field determined by the cutoff radius or the attention
kernel's support. For isolated small molecules this is adequate:
the relevant physics lives within a few \AA{} of each atom, and
the neural component has enough capacity to represent the
resulting short-range interactions. Condensed-phase systems are
another matter. Biomolecules in solvent, liquids, molecular
crystals, and extended interfaces have substantial energy
contributions from long-range Coulomb and dispersion interactions
that fall off only polynomially ($1/r$ and $1/r^6$ respectively)
and remain non-negligible far beyond any practical cutoff.
Pure-MPNN architectures address this either by extending the
cutoff (computationally expensive and data-hungry, since the
network must learn $1/r$ from scratch) or by ignoring the tail
(accepting errors that accumulate over simulations of
sufficiently large systems).

SO3LR~\cite{Kabylda2025SO3LR} takes a structurally different
approach. Its total energy decomposes as
\[
  E_{\mathrm{SO3LR}}(\mathbf{R})
  \;=\; E_{\mathrm{ZBL}}(\mathbf{R})
     \;+\; E_{\mathrm{So3k}}(\mathbf{R})
     \;+\; E_{\mathrm{Elec}}(\{q_i\}, \mathbf{R})
     \;+\; E_{\mathrm{Disp}}(\{\alpha_i\}, \mathbf{R}),
\]
where $E_{\mathrm{ZBL}}$ is the analytic
Ziegler--Biersack--Littmark short-range repulsion,
$E_{\mathrm{So3k}}$ is the semilocal neural energy of the
So3krates backbone, $E_{\mathrm{Elec}} = \sum_{i<j}
q_i q_j / r_{ij}$ is a pairwise Coulomb term evaluated with
learned atomic partial charges $q_i(\mathbf{R})$, and
$E_{\mathrm{Disp}}$ is a pairwise van der Waals dispersion term
with learned atomic polarizabilities $\alpha_i(\mathbf{R})$.
The short-range neural
component carries the many-body correlations that the network
learns efficiently; the analytic long-range components carry
the physics whose functional form is known in advance and
whose pairwise additivity the architecture imposes as a
structural constraint.

\begin{observation}[SO3LR's pairwise long-range form is
  codomain additivity, not the Hess functor]
\label{obs:so3lr-level}
  SO3LR's primary tower membership is $\Lk_5^{\Para}(P)$: the
  So3krates backbone satisfies $(\mathrm{E}_5)$ architecturally
  (Proposition~\ref{prop:so3krates-equivariant}), and the
  scalar-energy output places the full model in $\OrbMorse$,
  subject to the same $\Lk_6^{\Para}(P)$ ceiling as every
  other MLFF in this section. What distinguishes SO3LR from a
  pure $\Lk_5^{\Para}$ architecture is the structural content
  of its long-range sector. The Coulomb term
  $E_{\mathrm{Elec}} = \sum_{i<j} q_i q_j / r_{ij}$ and the
  dispersion term $E_{\mathrm{Disp}}$ are both pairwise additive
  over atomic contributions, and this additivity is an
  \emph{architectural} constraint, not a training-emergent
  approximation: no parameter setting of SO3LR produces
  long-range terms that violate it.

  The natural temptation is to read this pairwise additivity as
  alignment with the Hess functor
  $\FH \colon \Lk_0(P) \to B\RR$ whose existence defines
  $\Lk_1$. That reading is a category error. The Hess functor's
  domain is the reaction category $\Lk_0(P)$, with parallel
  composition of reaction morphisms $r_1 \otimes r_2$ and the
  parallel-composition axiom $\FH(r_1 \otimes r_2) = \FH(r_1) +
  \FH(r_2)$ stated on those morphisms. SO3LR has no reaction
  morphisms in its signature: its input is a nuclear
  configuration $\mathbf{R} \in \Ce(G)$, its output is a scalar
  energy in $\RR$, and there is no $\Lk_0(P)$-domain on which
  $\FH$ could be evaluated. The pairwise sum
  $\sum_{i<j} V(q_i, q_j, r_{ij})$ is additivity \emph{in the
  codomain} $\RR$, an algebraic property of how the energy
  expression is built up over atom pairs at fixed
  configuration; it is not functoriality from a reaction
  category, and labelling SO3LR's long-range sector as a
  partial $\Lk_1$ inhabitant is the same kind of category
  error as labelling any pairwise-additive scalar a Hess
  functor.

  The architecturally honest reading of SO3LR's design
  contribution is \emph{physics-informed range separation}: the
  functional form of the long-range tails is known analytically
  (Coulomb $1/r$, dispersion $1/r^6$), so the architecture
  supplies these analytically rather than asking the neural
  backbone to learn them. The structural gain is independence
  from training coverage on the long-range axis. A pure neural
  model can be made more accurate on distribution-covered
  configurations but retains no architectural guarantee about
  physics beyond its cutoff; SO3LR has the correct $1/r$ and
  $1/r^6$ behaviour built in, so the long-range functional form
  is correct even for atomic configurations the training set
  never saw. This is a real design commitment, distinct from
  MACE's local Clebsch--Gordan strategy and from So3krates's
  global attention, but it is a design axis \emph{within}
  $\Lk_5^{\Para}(P)$, not a second tower-level membership.

  The intermediate tower levels $\Lk_2$, $\Lk_3$, and $\Lk_4$
  are orthogonal to SO3LR's architectural scope rather than
  violated by it: $\Lk_2$ concerns dagger-SMC detailed balance
  for kinetic rate constants; $\Lk_3$ mass-action kinetics;
  $\Lk_4$ DPO graph rewriting for reaction mechanism. None
  directly arises in a PES architecture, which produces a
  scalar-energy surface rather than kinetic or mechanistic
  data. At $\Lk_{4.5}$, SO3LR engages only the weak form: the
  parity invariance built into $E(3)$-equivariance, which
  identifies enantiomers rather than distinguishing them. The
  chirality-distinguishing strong form of $\Lk_{4.5}$ (the
  parity factor $\ZZ_2^k \subset \Gstar$) is absent. The
  dispersion treatment is additionally ground-state only:
  SO3LR's oscillator-strength approximation retains the leading
  ground-state contribution to London dispersion without access
  to the excited-state Hilbert bundle structure of
  $\Lk_6^{\Para}(P)$ --- consistent with, and already entailed
  by, the output-type ceiling
  (Proposition~\ref{prop:topological-incompleteness}). The
  architectural pattern of SO3LR is therefore: strict
  $\Lk_5^{\Para}(P)$ membership via the So3krates backbone,
  physics-informed range separation as an internal design axis,
  weak $\Lk_{4.5}$ engagement via parity invariance,
  orthogonality to the reaction-side levels $\Lk_2$, $\Lk_3$,
  $\Lk_4$ (no kinetic or mechanistic content in a PES
  architecture), and categorical blockage from
  $\Lk_6^{\Para}(P)$ by the scalar output type.
\end{observation}

\begin{chembox}[What the long-range sector buys SO3LR]
  For molecules in vacuum at a scale where long-range physics
  is negligible, SO3LR and a pure neural force field like
  So3krates or MACE are structurally interchangeable: the
  short-range neural component dominates the total energy and
  the analytic tail is numerically small. The distinction
  emerges in condensed-phase systems. Consider two subsystems
  $S_1, S_2$ at separation $r \gg r_{\mathrm{cut}}$. A pure
  MPNN with a $5\,\text{\AA}$ cutoff computes
  $E(S_1 \cup S_2) = E(S_1) + E(S_2)$, missing the Coulomb
  contribution ($\sim 1/r$) and the dispersion contribution
  ($\sim 1/r^6$) between them entirely. SO3LR's
  range-separated architecture returns, at that separation,
  $E(S_1) + E(S_2) + E_{\mathrm{Elec}}(S_1, S_2) +
  E_{\mathrm{Disp}}(S_1, S_2)$ --- the physically correct
  long-range terms, supplied analytically with parameters
  ($q_i, \alpha_i$) the neural backbone has learned.

  The structural gain is that the functional form of the
  long-range terms does not depend on training coverage. A
  pure neural model can be made more accurate on
  distribution-covered configurations but retains no
  architectural guarantee about physics beyond its cutoff;
  SO3LR has the Coulomb and dispersion tails built in, so
  their form is correct even for atomic configurations the
  training set never saw. This is the practical face of the
  physics-informed range separation discussed in
  Observation~\ref{obs:so3lr-level}: the long-range functional
  form is a structural architectural commitment, not an empirical
  fit, and is independent of training coverage.
\end{chembox}

\begin{insightbox}[Two axes of architectural design at a
  single tower level]
So3krates and SO3LR together illustrate that tower-level
membership is a single coordinate among several architectural
ones. Both are objects of $\Lk_5^{\Para}(P)$ and both are
categorically blocked from $\Lk_6^{\Para}(P)$ by the
output-type obstruction of
Proposition~\ref{prop:topological-incompleteness}.
Within $\Lk_5^{\Para}(P)$ they differ along two independent
axes.

The first axis is the \emph{architectural strategy} for
realising $(\mathrm{E}_5)$. MACE contracts through local
Clebsch--Gordan tensor products; So3krates couples through
$SE(3)$-equivariant attention on an SPHC basis. Both
produce parametric 1-morphisms of $\Lk_5^{\Para}(P)$, and the
tower does not distinguish them at level $5$. This is the
lesson of Remark~\ref{rem:so3krates-ceiling}: extending
receptive field, swapping CG for attention, or adding more
layers changes the specific 1-morphism at
$\Lk_5^{\Para}(P)$ but travels zero distance along the
$\Lk_5 \to \Lk_6$ direction of the tower.

The second axis is whether the architecture incorporates
\emph{physics-informed structural baselines}. Pure-MPNN
architectures (MACE, So3krates) ask the neural backbone to
learn the full interatomic interaction within its receptive
field; their function classes contain whatever the network can
express, subject only to the $E(3)$-equivariance and
scalar-output type constraints of $\Lk_5^{\Para}(P)$. SO3LR is
a parametric 1-morphism of $\Lk_5^{\Para}(P)$ with the same
primary type signature, but factors its energy into a learned
short-range neural component plus an analytic long-range
component whose functional form (Coulomb $1/r$, dispersion
$1/r^6$) is supplied as an architectural baseline. This is not
motion along the $\Lk_5 \to \Lk_1$ direction of the tower: the
Hess functor $\FH$ has no $\Lk_0(P)$-domain reaction morphisms
in SO3LR's signature on which to be evaluated
(Observation~\ref{obs:so3lr-level}). It is a separate
architectural axis within $\Lk_5^{\Para}(P)$. The tower reveals
this as structurally meaningful: two parametric 1-morphisms at
the same tower level can differ in whether their function
classes are constrained by known physics or determined entirely
by training, and that distinction is invisible to a single
benchmark number.
\end{insightbox}

%% file: chapters/para/para_square.tex
\subsection{QIM near the $\Lk_5^\Para$--$\Lk_6$ boundary: a
  bidirectional worked example}
\label{sec:para-square}

This section works through the Quantum Inverse Mapping model
(QIM) of Fallani, Medrano Sandonas, and
Tkatchenko~\cite{Fallani2024QIM} as a concrete worked example
of a $\Lk_5$-type architecture --- non-strict
$\Lk_5^\Para(P)$ under the architectural reading committed to
in Section~\ref{sec:para-classification} --- whose
approximation target uses $\Lk_6$ content, specifically the
Born--Oppenheimer ground-state section
$\sigma_0 \colon \Ce(G) \to \Hel^{(N)}$ of the Hilbert bundle
constructed by the tower functor $\Felsix$, without QIM
itself representing that content at any intermediate layer.
QIM is selected because (i) it is \emph{bidirectional} ---
property~$\to$~structure is its named direction, distinct from
the forward-only MLFFs of Sections~\ref{sec:para-mace}
and~\ref{sec:para-so3krates} --- and (ii) its approximation
target $\Pi \circ \sigma_0$ --- the composition of $\sigma_0$
with observable projection $\Pi$ --- makes the output-type
argument of Proposition~\ref{prop:topological-incompleteness}
concrete:
despite the external reference to $\Lk_6$ through $\sigma_0$,
QIM's output types lie in $\Lk_5$, not above.

\subsubsection{Architecture}
\label{sec:para-square-arch}

QIM consists of three parametric maps of $\Lk_5$-type
arranged through a shared latent space
$Z \subseteq \RR^k$:
\begin{equation}
\label{eq:qim-architecture}
\includegraphics{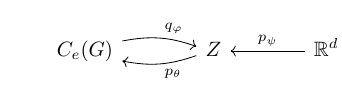}
\end{equation}
with $q_\varphi$ and $p_\theta$ forming the standard
encoder--decoder VAE between structure and latent, and
$p_\psi$ attaching the property space from the right as the
third network added in~\cite{Fallani2024QIM} for inverse
design.
The representation-inherited caveats on level membership are
discussed in Observation~\ref{obs:qim-level}.

\medskip\noindent\textbf{Structure encoder
  $(\Theta_\varphi, q_\varphi) \colon \Ce(G) \to Z$.}
Maps a molecular geometry $\mathbf{R} \in \Ce(G)$ to a latent
code $z \in Z$.
The input is presented to the network via the Coulomb-matrix
representation~\cite{Fallani2024QIM}, which is
$SE(3)$-invariant by construction (but not
$\mathrm{Sym}(\Sp)$-invariant: atom permutations change the
matrix, a representation-inherited limitation discussed in
Observation~\ref{obs:qim-level} below).
The conditional $q_\varphi(z \mid \mathbf{R})$ is a Gaussian
distribution over $Z$ whose mean and variance depend on
$\mathbf{R}$, and $z$ is sampled via the reparametrisation
trick.

\medskip\noindent\textbf{Structure decoder
  $(\Theta_\theta, p_\theta) \colon Z \to \Ce(G)$.}
Reverses the encoder.
$p_\theta(\mathbf{R} \mid z)$ is a conditional distribution
over geometries;
$\hat{\mathbf{R}} = \operatorname{mean}(p_\theta(\cdot \mid
z))$ is the reconstructed structure, recovered from the
decoded Coulomb matrix via classical multidimensional scaling
up to a chirality transformation, a residual ambiguity of the
Coulomb-matrix representation documented
in~\cite{Fallani2024QIM}.

\medskip\noindent\textbf{Property encoder
  $(\Theta_\psi, p_\psi) \colon \RR^d \to Z$.}
Maps a target property tuple $\mathbf{y} \in \RR^d$ to a
latent-code distribution $p_\psi(z \mid \mathbf{y})$.
This is the third network added in~\cite{Fallani2024QIM} on
top of the standard VAE, enabling inverse design.
The naming is deliberate:
$p_\psi$ moves from properties to latent, not from latent to
properties.
QIM has no dedicated latent-to-property map, and the forward
structure-to-property path is only implicit through the
shared latent --- a point that becomes load-bearing in
Observation~\ref{obs:qim-level}.

\medskip\noindent\textbf{Joint training objective.}
The three networks are trained jointly on the modified ELBO
of~\cite{Fallani2024QIM},
\begin{equation}
\label{eq:qim-elbo}
\mathcal{L}(\varphi, \theta, \psi)
\;=\;
-\mathbb{E}_{q_\varphi}\!\bigl[\log p_\theta(\mathbf{R} \mid
  z)\bigr]
\;+\; \beta\, D_{\mathrm{KL}}\!\bigl(q_\varphi(z \mid
  \mathbf{R}) \,\big\|\, \mathcal{N}(0, I)\bigr)
\;-\; \tau\, \log p_\psi(z \mid \mathbf{y}),
\end{equation}
combining the VAE reconstruction likelihood, the VAE KL
regulariser, and a property-likelihood term
$-\log p_\psi(z \mid \mathbf{y})$ that forces $z$ sampled
from $q_\varphi(z \mid \mathbf{R})$ to have high likelihood
under $p_\psi(\cdot \mid \mathbf{y})$ for the property tuple
$\mathbf{y}$ associated with $\mathbf{R}$.
Training over the QM7-X dataset of~\cite{hoja2021qm7} makes
$Z$ a shared representation in which a molecule and its
property tuple encode to overlapping regions --- the
empirical property on which bidirectional operation rests.

\subsubsection{Operation}
\label{sec:para-square-op}

The shared-latent structure enables three operational modes.

\emph{Inverse design}, the direction for which QIM is named,
applies the property encoder followed by the structure
decoder:
given a target $\mathbf{y}^\star \in \RR^d$, compute
$z^\star = \operatorname{mean}(p_\psi(\cdot \mid
\mathbf{y}^\star))$ and then $\hat{\mathbf{R}}^\star =
\operatorname{mean}(p_\theta(\cdot \mid z^\star))$.
The composite
$\RR^d \xrightarrow{\,p_\psi\,} Z
\xrightarrow{\,p_\theta\,} \Ce(G)$
produces a candidate geometry expected to exhibit the
targeted properties.

\emph{Forward property prediction} follows the opposite path
through the shared latent:
apply the structure encoder to obtain
$z = \operatorname{mean}(q_\varphi(\cdot \mid \mathbf{R}))$,
then recover a property tuple by inverting the property
likelihood model against the joint distribution learned
at training time.
No dedicated latent-to-property decoder exists in the
architecture;
forward property prediction is an emergent consequence of
the joint training~\eqref{eq:qim-elbo}, not a primitive
operation.
The composition approximates $\Pi \circ \sigma_0 \colon
\Ce(G) \to \RR^d$ --- the Born--Oppenheimer ground-state
section followed by observable projection --- implicitly,
through the shared latent $Z$.

\emph{Transition-path interpolation} exploits the geometry of
$Z$ itself.
Using the geodesic interpolation algorithm for VAEs
of~\cite{Fallani2024QIM}, interpolated latent codes between
two conformational isomers decode through $p_\theta$ to a
continuous geometric path in $\Ce(G)$.
Such paths are demonstrated in~\cite{Fallani2024QIM} as
initial guesses for machine-learning-based nudged elastic band
(ML-NEB) calculations of transition structures between
isomers;
they are not themselves minimum-energy paths, and the energy
profile along them is not directly constrained by the
training objective~\eqref{eq:qim-elbo}.
This mode is external to the membership condition
$(\mathrm{E}_5)$ on any single 1-morphism, but it illustrates
the categorical point that $Z$ is an architecture-internal
surrogate for nothing in the tower --- in particular, not
for the Hilbert-bundle fibre of $\Lk_6$.

\subsubsection{Tower placement}
\label{sec:para-square-tower}

\begin{observation}[QIM at $\Lk_5$-type, not strict
  $\Lk_5^\Para(P)$, with approximation target referencing
  $\Lk_6$ externally]
\label{obs:qim-level}
The three parametric maps $q_\varphi$, $p_\theta$,
$p_\psi$ that constitute QIM have codomains $Z$, $\Ce(G)$,
and $Z$ respectively.
Architectural $SE(3)$-compatibility is secured
map-by-map and via different mechanisms:
$q_\varphi$ is $SE(3)$-invariant through the Coulomb-matrix
representation of its input (translation- and
rotation-invariant by construction);
$p_\theta$ outputs into the $SE(3)$-quotient
$\Ce(G) = \RR^{3n}/(SE(3) \ltimes \Aut(G))$, so $SE(3)$ acts
trivially on its target;
$p_\psi$ acts between spaces on which $SE(3)$ is realised
trivially.
The $\mathrm{Sym}(\Sp)$-equivariance required at full
$\Lk_5$ is architectural for $p_\psi$ (which acts between
spaces on which $\mathrm{Sym}(\Sp)$ is realised trivially,
its scalar property-tuple input being invariant under atom
relabelling by construction) but training-emergent for
$q_\varphi$ and $p_\theta$:
the Coulomb matrix is not permutation-invariant, a
representation-inherited limitation acknowledged
in~\cite{Fallani2024QIM} and documented for
Coulomb-matrix-based architectures in
Table~\ref{tab:ml-classification}.
The structure decoder $p_\theta$ additionally reconstructs
geometries only up to a chirality transformation that the
Coulomb matrix does not distinguish~\cite{Fallani2024QIM}.
Under the architectural reading of $(\mathrm{E}_5)$ applied
strictly in Section~\ref{sec:para-classification}, the
training-emergent permutation handling on $q_\varphi$ and
$p_\theta$ and the chirality ambiguity on $p_\theta$ together
place QIM below strict $\Lk_5^\Para(P)$ membership.
The honest placement is $\Lk_5$-type, property-space: the
$SE(3)$ axis of the membership condition is architectural
map-by-map, but the $\mathrm{Sym}(\Sp)$ axis is inherited from
the Coulomb-matrix representation rather than enforced by the
architecture --- the same situation as for other
Coulomb-matrix-based architectures recorded in
Table~\ref{tab:ml-classification}.

The forward composition of Section~\ref{sec:para-square-op}
approximates $\Pi \circ \sigma_0 \colon \Ce(G) \to \RR^d$.
This target morphism has codomain $\RR^d$ (a $\Lk_5$-compatible
scalar tuple), but its definition passes through $\Lk_6$
content:
$\sigma_0$ selects the ground-state fibre of the Hilbert
bundle $\Hel^{(N)} \to \Ce(G)$, and $\Pi$ projects to
observable expectation values on that fibre.
QIM's approximation is implicit through the shared latent
$Z \subseteq \RR^k$, and $Z$ is not $\Hel^{(N)}$:
no wavefunction, density matrix, Hamiltonian, or bundle
section appears at any intermediate layer of QIM.
The codomain of the forward path is $\RR^d$ (observable
expectation values), the codomain of the inverse path is
$\Ce(G)$ (nuclear geometries), and neither codomain is
$\Hel^{(N)}$.
Neither carries the Berry class
$[\gamma_B] \in H^1(\Ce(G) \setminus \Xseam, \ZZ_2)$ that
$\Felsix$ installs at conical intersections.

The categorical situation is captured by the following
diagram of parallel factorisations of
$\Pi \circ \sigma_0 \colon \Ce(G) \to \RR^d$:
\begin{equation}
\label{eq:qim-tower-bridge}
\includegraphics{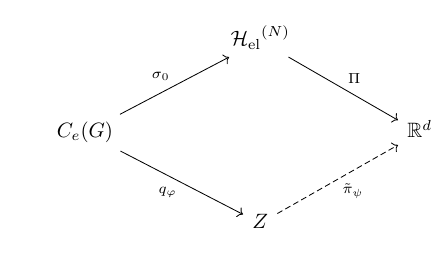}
\end{equation}
The top path is the tower's exact factorisation of the
target:
$\sigma_0$ is the Born--Oppenheimer ground-state section of
the $\Lk_6$ Hilbert bundle $\Hel^{(N)} \to \Ce(G)$, and
$\Pi$ is fibrewise observable projection.
The bottom path is QIM's architectural factorisation through
the learned latent $Z$:
$q_\varphi$ is the structure encoder, and
$\tilde{\pi}_\psi \colon Z \dashrightarrow \RR^d$ is the
implicit property-extraction map obtained by inverting the
trained property likelihood $p_\psi(z \mid \mathbf{y})$
against the joint distribution --- not a primitive
architectural morphism, and therefore drawn dashed.

The diagram is not commutative in any categorical sense.
The two paths agree only as numerical approximations,
$\Pi \circ \sigma_0 \approx
\tilde{\pi}_\psi \circ q_\varphi$,
with the approximation driven by the training
objective~\eqref{eq:qim-elbo} rather than by an equation in
$\mathcal{C}$ or a 2-cell in $\ParaC(\mathcal{C})$.
Categorically, no canonical arrow $Z \to \Hel^{(N)}$ or
$\Hel^{(N)} \to Z$ completes the diagram into a commuting
square:
$Z \subseteq \RR^k$ is a trivial $\Lk_5$ object (Euclidean,
with $SE(3) \ltimes \Aut(G)$ acting trivially) while
$\Hel^{(N)}$ is a $U(N)$-gauge Hilbert bundle in $\Lk_6$, and
the forgetful functor $U_6 \colon \Lk_6 \to \Lk_5$ sends
$\Hel^{(N)}$ to its base $\Ce(G)$, not to $Z$.
QIM therefore does not \emph{bridge} $\Lk_5$ and $\Lk_6$ in
any functorial sense:
the bridging is entirely numerical, through training, along
a diagram that fails to commute as a diagram of tower
morphisms.
The reference to $\Lk_6$ in $\Pi \circ \sigma_0$ is
\emph{external} --- it describes what QIM targets, not what
QIM contains.
A genuine $\Lk_6^\Para(P)$ model would predict the
multi-sheet PES $(V_0, V_1, \ldots, V_{N-1})$, the Berry
connection $A_{mn}^\mu(\mathbf{R})$ as a function of
geometry, and the topological class $[\gamma_B]$ as separate
outputs carrying the bundle structure;
none of these fits QIM's signature, and none appears in
QIM's output.
\end{observation}

\subsubsection{MLFFs as a single-morphism $\Lk_5^\Para$
  architecture}
\label{sec:para-square-mlff}

\begin{observation}[Machine-learning force fields as
  single-morphism $\Lk_5^\Para(P)$ architectures]
\label{obs:mlff-degenerate}
Machine-learning force fields --- NequIP, MACE, So3krates,
Allegro, Equiformer, eSEN --- occupy $\Lk_5^\Para(P)$
through a different architectural shape:
a single parametric 1-morphism
\begin{equation}
\label{eq:mlff-architecture}
\includegraphics{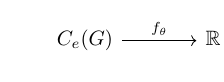}
\end{equation}
approximating the PES functor
$\FV \colon \Ce(G) \to \RR$ of Section~\ref{sec:L5-pes}, with
forces $\mathbf{F} = -\nabla V$ recovered by automatic
differentiation of the model output.
In contrast to the three-morphism diagram
\eqref{eq:qim-architecture} of QIM, diagram
\eqref{eq:mlff-architecture} has no shared latent, no
structure decoder, and no property encoder ---
MLFFs are unidirectional ($\Ce(G) \to \RR$) where QIM is
bidirectional ($\Ce(G) \leftrightarrow Z \leftrightarrow
\RR^d$).
As with QIM, no $\Lk_6$ structure is engaged:
$V_0$ and $-\nabla V_0$ are determined by $\Lk_5$ PES data
alone, and no Hilbert bundle, Berry connection, or Berry
class enters the prediction.
Neither~\eqref{eq:qim-architecture}
nor~\eqref{eq:mlff-architecture} admits a canonical lift of
any arrow through $\Hel^{(N)}$:
for QIM this is the non-commutativity
of~\eqref{eq:qim-tower-bridge};
for MLFFs it is the still simpler fact that $\Hel^{(N)}$ is
not in the architecture's type signature.

Taken together, the MLFF and QIM cases show that the
$\Lk_5$--$\Lk_6$ boundary admits multiple architectural
shapes near $\Lk_5^\Para(P)$ --- single-morphism scalar
prediction on the MLFF side (strict $\Lk_5^\Para(P)$ via
architectural $E(3)$-equivariance), and three-morphism
bidirectional latent models on the QIM side ($\Lk_5$-type
with Coulomb-matrix-inherited representational limitations
on $(\mathrm{E}_5)$) --- and that none of them, by the
output-type argument of
Proposition~\ref{prop:topological-incompleteness}, reaches
$\Lk_6^\Para(P)$.
The $\Lk_5$--$\Lk_6$ gap is not closed by enriching the
architecture within $\Lk_5^\Para$ or its $\Lk_5$-type
neighbourhood;
closing it requires a different codomain.
\end{observation}

%% file: chapters/para/para_incompleteness.tex
\subsection{Three incompleteness results}
\label{sec:para-incompleteness}

The classification in Section~\ref{sec:para-classification}
identifies three architectural gaps separating current
Para-morphism practice from a hypothetical model that fully
inhabits the tower.
The three results differ in both character and severity, and
they are presented in order of \emph{increasing} severity ---
with the numbering reversed, so that Gap~1 names the deepest
result and is presented last:
\begin{itemize}
  \item \textbf{Gap~3} (\emph{structural}, presented first):
    no architecture jointly enforces Eyring TST coherence
    between $\Lk_3^\Para$ rate-law output and $\Lk_5^\Para$
    PES output.
  \item \textbf{Gap~2} (\emph{thermodynamic}, presented
    second): no neural kinetic model with independently
    parameterised forward and reverse rate constants enforces
    the Wegscheider consistency condition.
  \item \textbf{Gap~1} (\emph{topological}, presented last):
    architectures with output type $\OrbMorse$ are
    categorically blocked from $\Lk_6^\Para(P)$.
\end{itemize}
Gaps~3 and~2 are \emph{contingent} on the state of the
published literature: each records that no current architecture
enforces a given tower-coherence condition, but such an
architecture could in principle be constructed without changing
output types.
They are therefore stated as literature-survey
\emph{observations}.
Gap~1 is \emph{intrinsic}: a categorical theorem that
$\Lk_6^\Para(P)$ membership is unreachable from any architecture
with output type $\OrbMorse$, regardless of training-set size,
parameter count, or other architectural choices within that
output constraint.
It is stated as a \emph{proposition} with proof.

\subsubsection{Gap~3 (structural): the Eyring TST coherence gap}

\begin{observation}[No published ML architecture jointly enforces
  Eyring TST coherence]
\label{obs:tst-coherence-gap}
  No machine-learning architecture in the published literature
  jointly enforces the Eyring transition-state-theory coherence
  condition
  \[
    k_r(\theta) \;=\; \frac{k_B T}{h}\,
    \exp\!\left(-\frac{E_\theta(\mathbf{R}^\ddagger)
      - E_\theta(\mathbf{R}_0)}{RT}\right)
    \tag{TST}
  \]
  as an architectural constraint, where $k_r(\theta)$ is a learned
  rate constant and $E_\theta(\mathbf{R}^\ddagger) -
  E_\theta(\mathbf{R}_0)$ is the activation barrier computed from
  the same model's potential energy surface.
  In tower terms: no architecture simultaneously inhabits
  $\Lk_3^\Para(P)$ (rate-law output) and $\Lk_5^\Para(P)$ (PES
  output) with the TST coherence condition
  (Definition~\ref{def:tst-coherence} of
  Section~\ref{sec:L5-def}) relating them through the forgetful
  functor chain $U_4 \circ U_{4.5} \circ U_5 \colon \Lk_5(P) \to
  \Lk_3(P)$.

  The published literature divides cleanly into two
  non-communicating communities.
  Neural kinetic models ---
  ChemNODE~\cite{Owoyele2022ChemNODE},
  CRNN~\cite{Ji2021CRNN}, and their extensions --- predict rate
  laws $(\Theta_k, k_\theta) \colon \Lk_0(P) \to \mathbf{Stoch}$
  from macroscopic concentration time-series.
  CRNN's Arrhenius parameterisation $k = A\,T^b\,\exp(-E_a/RT)$
  encodes an activation energy $E_a$ as a learned scalar
  parameter extracted from rate data, not as a barrier computed
  from any atomistic PES.
  Machine-learning force fields --- NequIP~\cite{Batzner2022NequIP},
  MACE~\cite{BatatIa2022MACE}, So3krates~\cite{Frank2022So3krates}
  --- predict the PES
  $(\Theta_E, E_\theta) \colon \Lk_{4.5}(P) \to \OrbMorse$ from
  energy and force data.
  When rate constants are required downstream, the learned PES
  feeds into a separate TST, ring-polymer molecular dynamics, or
  instanton calculation run as an independent step.

  No published architecture contains both $(\Theta_k, k_\theta)$
  and $(\Theta_E, E_\theta)$ as jointly trained components with
  condition~(TST) enforced: no loss function in the reviewed
  literature contains a term penalising the difference between an
  emergent rate from PES simulation and a directly learned rate
  constant.
  The technical infrastructure for building such an architecture
  --- end-to-end differentiable molecular dynamics and rate
  theory --- exists in principle but has not been applied to this
  problem.
\end{observation}

\begin{chembox}[Why the gap matters for reactive simulations]
  Reactive molecular dynamics with an MLFF simulates trajectories
  on $V_0(\mathbf{R})$ and reads off an effective rate constant
  from the observed crossing frequency.
  Nothing guarantees that this emergent rate satisfies the Eyring
  equation relative to the model's own barrier, because the
  model's energy function was not trained to produce barrier
  heights consistent with any experimental rate.
  In a functorially coherent architecture, the $\Lk_3$ rate would
  be \emph{derived} from the $\Lk_5$ barrier via condition~(TST),
  with the rate and PES components sharing a single parameter
  space as the Para-enriched level $\Lk_k^\Para$ admits via the
  comonoid structure of Definition~\ref{def:LkPara}, and~(TST)
  imposed as an architectural loss relating their outputs.
  No such architecture currently exists in the published
  literature.
\end{chembox}

\subsubsection{Gap~2 (thermodynamic): the Wegscheider consistency
  gap}

\begin{observation}[No published neural kinetic model with
  independent rate parameterisation enforces the categorical
  Wegscheider condition]
\label{obs:wegscheider-gap}
  Every neural kinetic model in the published literature with
  independently parameterised forward and reverse rate constants
  fails to guarantee
  \[
    \frac{k_r(\theta)}{k_{r^\dagger}(\theta)}
    \;=\; \exp\!\left(-\frac{\Delta G^\circ_r}{RT}\right)
    \quad \text{for all } r \in \Lk_0(P),
    \tag{W}
  \]
  the condition that requires the kinetic functor
  $\FP \colon \Lk_3(P) \to \mathbf{Stoch}$ to be compatible with
  the thermodynamic functor $\FG^T \colon \Lk_2(P) \to \RR$ via
  the dagger structure of $\Lk_2(P)$.

  Condition~(W) requires the parameter space $\Theta$ to carry a
  constraint relating $k_r(\theta)$ and $k_{r^\dagger}(\theta)$
  through the thermodynamic data $\Delta G^\circ_r$.
  For neural kinetic models trained on rate data with
  independently parameterised forward and reverse constants ---
  ChemNODE~\cite{Owoyele2022ChemNODE}, CRNN~\cite{Ji2021CRNN} ---
  the loss function takes the form
  $\mathcal{L}(\theta) = \|\mathbf{c}(t;\theta) -
  \mathbf{c}_\mathrm{obs}(t)\|^2$ on concentration trajectories.
  This loss contains no thermodynamic supervision signal: nothing
  in $\Theta$ encodes $\Delta G^\circ_r$ or the dagger
  $r \mapsto r^\dagger$, so minimising $\mathcal{L}$ does not
  enforce~(W).

  The Kircher--Döppel--Votsmeier thermodynamically-consistent
  framework~\cite{KircherDoeppelVotsmeier2024} is the one
  structural partial counterexample in the reviewed literature.
  It embeds the De Donder relation as a hard architectural
  constraint: only the forward rate constant $k_f$ is
  parameterised by the network, and the net reaction rate is
  derived structurally as
  \[
    r_\mathrm{net}(\theta)
    \;=\; k_f(\theta)\,\prod_i [X_i]^{\nu_i}
    \,\Bigl(1 - \tfrac{Q}{K_\mathrm{eq}}\Bigr),
  \]
  where $K_\mathrm{eq}$ is supplied from tabulated thermochemistry
  external to the model.
  For mass-action kinetics this construction implies that the
  ratio of forward to reverse rate constants equals
  $K_\mathrm{eq} = \exp(-\Delta G^\circ / RT)$,
  achieving the physical content of~(W).
  However, it does so by \emph{eliminating} the independent
  parameterisation of the reverse rate constant altogether: the
  reverse rate is not a network output but a derived quantity.
  This architecture therefore satisfies a structural variant
  of~(W) rather than~(W) as stated --- it confirms that achieving
  thermodynamic consistency requires exactly the architectural
  coupling that~(W) demands, but implements it by removing the
  independent parameterisation rather than by constraining the
  ratio of two independently learned outputs.
  No published model with independently parameterised forward
  and reverse rate constants enforces~(W).

  Two further architectures touch adjacent content without
  addressing~(W) directly.
  Boltzmann generators~\cite{Noe2019BoltzmannGen} target the
  equilibrium distribution directly, which is a stationary-state
  property of $\Lk_2(P)$, not a kinetic one; the detailed-balance
  content of~(W) does not apply to one-shot samplers.
  Hard-constraint thermodynamic neural networks for activity
  coefficients~\cite{Rittig2024GibbsDuhem} enforce the
  Gibbs--Duhem equation for mixture properties --- a single
  narrow property class that does not extend to reaction
  kinetics.
\end{observation}

\begin{observation}[The van't~Hoff corollary]
\label{obs:vant-hoff}
  Combined with the standard decomposition
  $\Delta G^\circ = \Delta H^\circ - T \Delta S^\circ$,
  condition~(W) implies the van't~Hoff relation
  $d(\ln K_\mathrm{eq})/dT = \Delta H^\circ/(RT^2)$.
  No published ML model for equilibrium constants enforces this
  temperature dependence as an architectural constraint.
  A model that learns $K_\mathrm{eq}(T)$ from data at a single
  temperature and extrapolates will violate van't~Hoff at other
  temperatures unless the dependence is explicitly parameterised
  using $\Delta H^\circ$ --- which requires incorporating
  $\Lk_1$ data ($\FH$-functor values) into a model typically
  trained at $\Lk_3$.
\end{observation}

\subsubsection{Gap~1 (topological): the output-type obstruction to
  $\Lk_6^\Para$}

\begin{proposition}[Output-type obstruction:
  $SE(3) \rtimes \Aut(G)$-equivariant MLFFs are categorically
  blocked from $\Lk_6^\Para$]
\label{prop:topological-incompleteness}
  Let $(\Theta, f_\theta)$ be a 1-morphism of $\Lk_5^\Para(P)$
  whose output type is $\OrbMorse$ --- a real-valued scalar
  energy on $\Ce(G)$ together with its equivariant gradient.
  Then no reparametrisation $r \colon I \to \Theta$ promotes
  $(\Theta, f_\theta)$ to a 1-morphism of $\Lk_6^\Para(P)$.
  In particular, no reparametrisation produces a Hilbert bundle
  in $\HilbBund$, the Berry connection $A_{mn}^\mu$, the
  multi-sheet PES $(V_0, V_1, \ldots, V_{N-1})$, or the
  topological Berry-phase class
  $[\gamma_B] \in H^1(\Ce(G) \setminus \Xseam, \ZZ_2)$
  (primary invariant for real molecular Hamiltonians;
  $c_1 \in H^2(-, \ZZ)$ in the spin--orbit-coupled case).
\end{proposition}

\begin{proof}
  The argument is a categorical type-check: the output type of
  a Para 1-morphism is invariant under reparametrisation, and
  $\OrbMorse$ and $\HilbBund$ are distinct.

  By construction, every 1-morphism of $\Lk_6^\Para(P)$ has
  output type $\HilbBund$: at each $\mathbf{R} \in \Ce(G)$, a
  full electronic-structure fibre --- an orthonormal basis of
  $N$ states, eigenvalues $V_n(\mathbf{R})$ for
  $n = 0, \ldots, N-1$, and connection coefficients
  $A_{mn}^\mu(\mathbf{R})$ --- sufficient to determine the
  Hilbert bundle $\Hel^{(N)}$ and its $U(N)$-connection.
  The hypothesis is that $(\Theta, f_\theta)$ has output type
  $\OrbMorse$: at each $\mathbf{R}$, only the ground-state
  energy $V_0$ and its gradient $\nabla V_0$.
  The base-level forgetful functor
  $U_6 \colon \Lk_6 \to \Lk_5$ projects $\HilbBund$ to
  $\OrbMorse$ by extracting the lowest eigenvalue and its
  gradient, discarding excited-state energies, off-diagonal
  connection coefficients, and topological class --- none of
  which appears in the $\OrbMorse$ output of
  $(\Theta, f_\theta)$.

  A reparametrisation $r \colon I \to \Theta$ in
  $\Para(\Lk_5(P))$ specialises the architecture to a parameter
  value $\theta^* \in \Theta$ via
  $f_{\theta^*} = f_\theta \circ (r \otimes \mathrm{id}_X)$.
  This operation fixes the source object $\Ce(G)$ and the
  target object of $f_\theta$;
  it cannot change the output type.
  Hence if $f_\theta(\mathbf{R})$ lies in $\OrbMorse$ for all
  $(\mathbf{R}, \theta)$, then $f_{\theta^*}(\mathbf{R})$ does
  too.

  Therefore no reparametrisation of $(\Theta, f_\theta)$
  produces output in $\HilbBund$, and in particular none yields
  the Berry connection $A_{mn}^\mu$, the multi-sheet PES
  $(V_0, V_1, \ldots, V_{N-1})$, or any topological invariant
  of the bundle ($[\gamma_B]$ for real molecular Hamiltonians;
  $c_1$ in the SOC case).
  The architecture remains in $\Lk_5^\Para(P)$:
  parameter choice cannot supply the missing bundle data.
\qedhere
\end{proof}

\begin{observation}[Empirical confirmation: models approaching
  electronic topology move beyond $\OrbMorse$]
\label{obs:partial-l6}
  Proposition~\ref{prop:topological-incompleteness} predicts
  that capturing Berry phase and topological invariants requires
  a change of output type from $\OrbMorse$ to $\HilbBund$.
  The recent ML literature pursues this in two architecturally
  distinct ways.
  The \emph{operator-level} route predicts the full electronic
  Hamiltonian matrix $H_\mathrm{el}(\mathbf{R})$ in a fixed
  atomic-orbital basis; diagonalisation then recovers the bundle
  fibres pointwise.
  The \emph{section-level} route predicts the connection
  coefficients $A_{mn}^\mu$ as a genuine vector section
  directly.
  The two are orthogonal sub-categories of $\Lk_6^\Para$
  (Section~\ref{sec:para-classification}); both escape the
  $\OrbMorse$ codomain, but only the operator-level route has so
  far produced explicit topological demonstrations on
  experimentally relevant systems.

  \emph{ML Hamiltonian models.}
  PhiSNet~\cite{Unke2021PhiSNet} and
  DeepH-E3~\cite{Gong2023DeepHE3} predict
  $H_\mathrm{el}(\mathbf{R})$ as a Hermitian matrix field over
  $\Ce(G)$, placing them in the operator-level sub-category
  $\Lk_6^{\Para,\mathrm{op}}(P)$ defined in
  Section~\ref{sec:para-classification}: the
  $SE(3)$-equivariance of the operator-valued map is
  architectural, and the Hilbert-bundle fibres of
  $\Hel^{(N)} \to \Ce(G)$ are recoverable pointwise by
  diagonalising $H_\mathrm{el}(\mathbf{R})$.
  The fixed atomic-orbital basis, however, provides a global
  trivialisation that freezes the $U(N)$ fibre gauge, and no
  architectural enforcement of $[\gamma_B]$-consistency is in
  place around loops encircling $\Xseam$ --- the topological
  data is recovered post-hoc rather than constrained during
  training.
  Full $\Lk_6^\Para(P)$ membership would require both
  operator-level fidelity and section-level gauge covariance;
  these architectures provide only the former.
  DeepH-E3 nonetheless demonstrates prediction of a topological
  quantum phase transition in twisted bilayer Bi$_2$Te$_3$: as
  spin--orbit coupling increases, the $\ZZ_2$ topological
  invariant changes from $0$ to $1$, confirmed by Brillouin-zone
  integration of the predicted Berry connection and curvature.
  This demonstration is possible precisely because the output
  object escapes $\OrbMorse$ --- the model predicts the operator
  from which topological invariants are derived, not a scalar
  that has already discarded that information.
  Independently, Daggett, Yang, Liu, and
  Muechler~\cite{DaggettYangLiuMuechler2024} construct a model
  system whose two regimes share an \emph{identical}
  ground-state potential energy surface yet carry
  \emph{different} values of the Euler-class topological
  invariant ---
  directly establishing that $\OrbMorse$ output cannot
  distinguish topologically inequivalent electronic structures.

  \emph{Neural non-adiabatic coupling models.}
  SchNarc~\cite{Westermayr2020SchNarc} predicts excited-state
  energies, forces, and non-adiabatic coupling vectors (NACs),
  parametrising the NAC vectors as
  $d_{ij}^\mu(\mathbf{R}) = \nabla_\mu s_{ij}$ for learned
  scalar functions $s_{ij}$.
  This ansatz spans only exact 1-forms, so for every closed
  loop $\gamma \subset \Ce(G) \setminus \Xseam$ the integral
  $\oint_\gamma d_{ij}^\mu\, dR_\mu = \oint_\gamma ds_{ij} = 0$
  vanishes by construction; the Berry-phase class
  $[\gamma_B]$ is identically trivialised regardless of the
  underlying physics, and the model cannot represent loops on
  which $[\gamma_B] \neq 0$.
  SPAINN~\cite{Mausenberger2024SPAINN} lifts this exact-form
  obstruction by predicting equivariant vector NACs directly on
  a PaiNN backbone, so the output is a genuine section of
  $A_{mn}^\mu$ rather than a gradient of a scalar.
  However, no $U(N)$-gauge equivariance is enforced on the
  state indices $(m,n)$, no diabatic Hamiltonian is provided,
  and no topological loss enforces $[\gamma_B]$-consistency.
  SPAINN therefore reaches only a partial (connection-level)
  approximation toward the section-level sub-category
  $\Lk_6^{\Para,\mathrm{sec}}(P)$, while making no attempt at
  the operator-level axis.
  Neither SchNarc nor SPAINN reaches full $\Lk_6^\Para(P)$,
  which would require both sub-categories simultaneously.

  \emph{ML diabatisation.}
  Neural-network quasi-diabatic Hamiltonians (Truhlar
  group~\cite{Xie2018NNPD}; Zhang--Guo
  group~\cite{Zhang2020NNDiabat}; Shen and
  Yarkony~\cite{ShenYarkony2024}) learn multi-state diabatic
  potential-energy matrices that smooth the conical-intersection
  singularity.
  In tower language, this is a partial realisation of the blowup
  resolution of Conjecture~\ref{conj:ci-blowup}: the result is a
  function on the blown-up space near $\Xseam$.
  However, whether ML-fitted diabatic matrices preserve the
  correct Berry-phase class $[\gamma_B]$ in regions outside the
  training set is largely unresolved in the literature --- no
  paper provides a formal guarantee of topological correctness as
  an architectural constraint.
  The Shen--Yarkony work handles at least three coupled
  electronic states simultaneously (not a simple $2 \times 2$
  system), demonstrating the geometric-phase effect in aniline
  photodissociation as a physical consequence of the
  conical-intersection structure --- but does not formulate this
  in terms of a Berry-phase or Chern-class invariant.
\end{observation}

\begin{insightbox}[The topological gap is architectural, not
  empirical]
  Proposition~\ref{prop:topological-incompleteness} is not a
  statement about training-set size, model capacity, or
  distribution coverage.
  The Berry-phase class
  $[\gamma_B] \in H^1(\Ce(G) \setminus \Xseam, \ZZ_2)$ assigns
  a $\ZZ_2$-value to each homology class of loops encircling
  $\Xseam$; the analogous $c_1 \in H^2(-, \ZZ)$ in the
  spin--orbit-coupled case is similarly a cohomology class.
  These are not pointwise functions on $\Ce(G)$ at all --- they
  are global topological invariants of the eigenbundle. The
  Berry connection $A_{mn}^\mu$ is a $u(N)$-valued $1$-form on
  $\Ce(G)$ with state indices $(m, n)$, again not extractable
  from a pointwise scalar. A model with output type
  $\OrbMorse$ produces, at every geometry $\mathbf{R}$, a real
  number and a covariant vector: a scalar field on $\Ce(G)$,
  nothing more. The gap is type-theoretic. The proof of
  Proposition~\ref{prop:topological-incompleteness} is just
  the categorical observation that a reparametrisation
  $r \colon I \to \Theta$ acts on the source by
  $r \otimes \mathrm{id}_X$, leaving the target object ---
  hence the output type --- fixed. Increasing body order,
  expanding the training set, or widening the network leaves
  the gap untouched because none of these is a change of
  output type.

  The gap requires an \emph{architectural} change of output
  type. The operator-level route predicts the full electronic
  Hamiltonian $H_\mathrm{el}(\mathbf{R})$, from which
  $A_{mn}^\mu$ and $[\gamma_B]$ (or $c_1$) can be extracted by
  diagonalisation; the section-level route predicts a
  $U(N)$-gauge-equivariant section of the eigenbundle directly,
  carrying $A_{mn}^\mu$ as a connection 1-form.
  Either route escapes the $\OrbMorse$ codomain, and full
  $\Lk_6^\Para(P)$ membership requires both simultaneously ---
  operator-level fidelity and section-level gauge covariance.
  No published architecture provides both.
  A separate question, beyond strict membership, is whether
  the \emph{trained} values of $[\gamma_B]$ on loops encircling
  $\Xseam$ are correct outside the training distribution. No
  current ML framework enforces this architecturally; a
  topological-loss term on the Berry holonomy is the natural
  candidate mechanism, but it is a training-time penalty
  rather than an architectural constraint.
\end{insightbox}

%% file: chapters/para/para_catonly.tex
\subsection{Synthesis: the tower as design specification}
\label{sec:para-catonly}

This chapter has done three things.
It has located each major ML molecular architecture as a
parametric morphism in some $\Lk_k^\Para(P)$ by the membership
condition $(E_k)$ and the output type of its function class ---
a precise address independent of training data or parameter
count.
It has established three tower-incompleteness results of
distinct character: a proposition with formal proof (the
topological output-type obstruction at the
$\Lk_5 \to \Lk_6$ boundary) and two literature-survey
observations (the Eyring TST coherence gap between $\Lk_3$ and
$\Lk_5$, the Wegscheider consistency gap between $\Lk_2$ and
$\Lk_3$).
And it has shown that the Para enrichment provides a verifiable
structural criterion for the question ``what can this model
represent?'', replacing benchmark intuition with a statement
about the model's function class.

The central finding is that the vast majority of mainstream ML
molecular models operate at $\Lk_5^\Para(P)$ or below.
A small number of ML Hamiltonian models ---
PhiSNet~\cite{Unke2021PhiSNet},
DeepH-E3~\cite{Gong2023DeepHE3} --- reach the operator-level
sub-category $\Lk_6^{\Para,\mathrm{op}}(P)$ defined in
Section~\ref{sec:para-classification};
SPAINN~\cite{Mausenberger2024SPAINN} offers the audit's only
partial connection-level approach toward the section-level
sub-category $\Lk_6^{\Para,\mathrm{sec}}(P)$.
No architecture occupies both sub-categories, and none reaches
full $\Lk_6^\Para(P)$.
The statement is about function-class representability, not
about accuracy or training coverage: the models cannot represent
certain tower objects regardless of how they are trained.

\subsubsection{What category theory uniquely provides}

Two general consequences of the Para enrichment deserve naming.

\paragraph{Equivariance as a theorem, not a design choice.}
In every non-categorical treatment of ML for chemistry, symmetry
equivariance is either an empirical observation or a design
principle.
The Para enrichment makes it a theorem: an ML model is an object
of $\Lk_k^\Para(P)$ if and only if it satisfies the membership
condition $(E_k)$.
This applies level-by-level:
$SE(3) \rtimes \Aut(G)$-equivariance at $\Lk_5^\Para$,
$\Gstar$-equivariance at $\Lk_{4.5}^\Para$ (the reflection-
inclusive factor that ML-literature
``$E(3)$-equivariance'' implicitly invokes), the
Markov-category morphism condition at $\Lk_3^\Para$, and the
bundle-gauge condition at $\Lk_6^\Para$.
The correct inductive bias for a model targeting level $k$ is
therefore not a design choice but a categorical necessity,
determined by the group and monad structure at that level.
A model claiming to predict chirality-sensitive reaction
outcomes must satisfy $(E_{4.5})$ under the full
permutation-inversion group $\Gstar$; a model claiming to
predict Berry-phase effects must satisfy $(E_6)$ with respect to
the bundle gauge.
Failing the membership condition places the model at a lower
level regardless of benchmark performance.

\paragraph{Completeness as an architectural diagnostic.}
Classical ML benchmarks measure accuracy on held-out data at a
fixed trained parameter setting.
Definition~\ref{def:completeness} provides a sharper criterion:
a model $(\Theta, f_\theta)$ is complete at level $k$ on a
target morphism $\varphi \in \Lk_k(P)(X, Y)$ if and only if
$\varphi$ lies in the function class
$\mathsf{Func}(\Theta, f_\theta) \subseteq \Lk_k(P)(X, Y)$ ---
equivalently, if some reparametrisation
$r \colon I \to \Theta$ instantiates $f_\theta$ to $\varphi$
exactly.
Completeness is a property of the model's \emph{function class},
independent of training data.
A model that cannot express the Berry connection $A_{mn}^\mu$
as part of its output type will fail at predicting
geometric-phase effects in ultracold reactions,
photodissociation branching ratios, and conical-intersection
dynamics --- not because it was trained on too little data, but
because the relevant structure is not representable within the
function class at all.
The diagnostic therefore distinguishes failure modes that more
data can fix from those it cannot.

\subsubsection{Three research targets forced by the tower}

The tower is not only a classification of what exists.
Turned around, it is a \emph{design specification}: each level
specifies, via its membership condition and forgetful functor
structure, what a model must do architecturally to represent the
chemistry at that level faithfully.
The three gaps identified in
Section~\ref{sec:para-incompleteness} define three concrete
research targets --- not aspirational goals, but consequences
forced by the tower's structure, in the same sense that each
tower level was forced bottom-up by reaction pairs the previous
level could not distinguish.

\paragraph{Target~1: topologically complete $\Lk_6^\Para(P)$
  models for non-adiabatic chemistry.}
The tower forces this target on any model claiming to predict
photochemical reactivity, non-adiabatic dynamics, or electronic
topology.
Reaching full $\Lk_6^\Para(P)$ membership requires two
architectural commitments simultaneously: operator-level
fidelity, provided by PhiSNet and DeepH-E3 through
Hamiltonian-matrix output in a fixed atomic-orbital basis;
and section-level gauge covariance on the eigenbundle,
approached partially by SPAINN at the connection level.
No published architecture provides both.
A separate concern, beyond strict membership, is whether the
\emph{trained} values of $[\gamma_B]$ on closed loops
encircling $\Xseam$ are correct outside the training
distribution --- the $\ZZ_2$-valued Berry-phase class for real
molecular Hamiltonians, or the integer-valued $c_1$ in the
spin--orbit-coupled case. No current ML framework enforces
this architecturally; a topological-loss term on the Berry
holonomy, computed during training on geometries near conical
intersections, is the natural candidate mechanism but is a
training-time penalty rather than an architectural constraint.
The Daggett--Yang--Liu--Muechler
classification~\cite{DaggettYangLiuMuechler2024} supplies the
mathematical vocabulary --- a model system whose two regimes
share an identical ground-state PES yet carry different
Euler-class topological invariants --- and thereby establishes
empirically that $\OrbMorse$ output cannot distinguish
topologically inequivalent electronic structures.
The tower specifies the architectural home those invariants
require.

\paragraph{Target~2: thermodynamically consistent neural
  kinetics.}
The tower forces this target on any model claiming to predict
reversible chemical kinetics at thermal equilibrium.
The categorical requirement is that the parameter space $\Theta$
carry a dagger involution $\sigma \colon \Theta \to \Theta$
satisfying
\[
  k_{\sigma(\theta)}(r) \;=\; k_\theta(r^\dagger)
  \quad \text{for all reactions } r,
\]
paired with a thermodynamic data channel that feeds
$\Delta G^\circ_r$ into the constraint.
This is not a regularisation term; it is the specification of
the type that $\Theta$ must instantiate to guarantee detailed
balance at every reaction in every network.
The De Donder strategy of Kircher, D\"oppel, and
Votsmeier~\cite{KircherDoeppelVotsmeier2024} demonstrates one
viable implementation by eliminating independent parameterisation
of the reverse rate constant and deriving it from a tabulated
$K_\mathrm{eq}$.
The categorical generalisation shifts the primary parametric
commitment one level lower --- from $\Lk_3^\Para$ (independent
forward and reverse rate constants) to $\Lk_2^\Para$ ---
parameterising only the stoichiometry and a free-energy profile
$\Delta G^\circ_r$ per reaction, deriving rate constants
structurally via the dagger constraint above, with the
van't~Hoff relation
$d(\ln K_\mathrm{eq})/dT = \Delta H^\circ/(RT^2)$ enforced
analytically through the $\Lk_1$-level enthalpy channel.
Such an architecture would extrapolate correctly in both
temperature and composition outside the training distribution ---
a practical requirement for combustion modelling, heterogeneous
catalysis, and pharmaceutical kinetics.

\paragraph{Target~3: functorially separated topology--geometry
  models for reactive dynamics.}
The tower forces this target on any model claiming to predict
bond-breaking and bond-forming events in reactive molecular
dynamics: the forgetful functor
$U_{4.5} \circ U_5 \colon \Lk_5(P) \to \Lk_4(P)$ must be
realised as an explicit architectural boundary.
Current MLFFs lack any module for outputting $\Lk_4$ graph
labels (bond order, formal charges, lone pairs); neighbour
graphs are defined by distance cutoffs rather than by chemical
bonding, so bond identity can be reconstructed only post-hoc
from distances via extrinsic criteria.
Proposition~\ref{prop:mace-conflation} formalises this for the
MACE family specifically.
A two-stage architecture would realise the separation
explicitly: Stage~1 predicts the DPO graph $G \in \LGraphP$
(bond order, formal charges, lone pairs) from the local atomic
environment, using a graph-valued output type consistent with
$\Lk_4^\Para(P)$; Stage~2 takes the fixed topology $G$ as input
and predicts the $SE(3) \rtimes \Aut(G)$-equivariant PES
$V_0(\mathbf{R}; G)$ on the configuration orbifold, consistent
with $\Lk_5^\Para(P)$.
Bond-breaking events then correspond to transitions between
topologically distinct DPO graph objects in $\Lk_4(P)$, not to
continuous deformations of a cutoff-based graph.
The resulting architecture enables reactive MD in which the
model knows, at every step, which chemical bonds are present ---
a capability with direct applications to combustion, catalysis,
and materials degradation.

\bigskip

In each of the three targets, the framework specifies not only
what is missing but what type the missing component must be ---
a precision that benchmark comparison alone cannot provide, and
that makes the tower practically useful as a design tool rather
than merely a classification scheme.

%% file: chapters/ch_simulation.tex
\section{The lower tower in executable form: $\PhiOp$ on the Briggs--Rauscher oscillator}
\label{sec:sim}

\noindent
The preceding chapters have constructed a categorical tower that stratifies
chemistry by level of structure: stoichiometry at $\Lk_0$, thermodynamics
at $\Lk_1$--$\Lk_2$, kinetics at $\Lk_3$, mechanism at $\Lk_4$, with
$\Lk_{4.5}$ stereochemistry and higher levels above.  The tower's content
is not mere vocabulary: at every level it delivers theorems.  At $\Lk_3$,
Feinberg's Deficiency Zero Theorem forces weakly-reversible mass-action
networks with deficiency $\delta = 0$ to admit a unique positive
stationary measure per stoichiometric class --- which is precisely why
oscillating networks like the Briggs--Rauscher reaction must have
$\delta > 0$.  At $\Lk_4$, the parameter projection $\pi_4$
(\S\ref{sec:sim:L4coarse}) constrains exactly which mechanism data can
survive into the $\Lk_3$ rate constants, making the gap between
mechanism-resolved and rate-resolved descriptions explicit rather than
tacit.  These are not modelling choices that could have gone differently;
they are categorical consequences of the tower's construction.  If the
tower is to be more than a formal scaffold, those consequences have to
survive contact with an executable implementation evaluated against a
specific chemistry.

This chapter closes the tower at the bottom by constructing a functor
\begin{equation}\label{eq:sim:phi-op-overview}
  \PhiOp \;:\; \ParaC\!\bigl(\Lk_0 \otimes \Lk_1 \otimes \Lk_2
  \otimes \Lk_3\bigr)
  \;\longrightarrow\; \ParaC_W(\Hask),
\end{equation}
where $\ParaC_W(\Hask)$ is the Para construction over Hask
\cite{CruttwellGavranovic2022} with parameter type
$W = \mathtt{Word64}$, the PRNG seed (\S\ref{sec:sim:target}).  The
image of $\PhiOp$ at the De~Kepper--Epstein skeleton of the
Briggs--Rauscher oscillator \cite{DeKepperEpstein1982} ---
\texttt{briggsRauscherDE} in the chapter's Haskell implementation,
12 reaction channels with all rate constants tagged \texttt{Measured}
(\S\ref{sec:sim:L4coarse}, Remark~\ref{rmk:sim:provenance}) ---
admits two operational realisations: the stochastic simulation
algorithm (SSA) at parameter dimension $n = 1$, sampling from a
$\mathtt{Word64}$-indexed family of continuous-time-Markov-chain
trajectories, and the deterministic ODE at $n = 0$, realising the
$V \to \infty$ Kurtz limit of the species-mean process.  A denotational
companion
\[
  F \;:\; \ParaC_W(\Hask) \;\longrightarrow\; \BorelStoch
\]
\cite{Fritz2020Synthetic} marginalises over the PRNG seed and lands in
the Markov category $\BorelStoch$; the composite $F \circ \PhiOp$
sends a network to its canonical kernel
$F_{\mathrm{BR}} := F(\PhiOp(\mathtt{briggsRauscherDE})) \in \BorelStoch$.
To our knowledge this is the first explicit application of the Para
construction to chemistry (\S\ref{sec:sim:functor},
Corollary~\ref{cor:para-chemistry}) and the first published categorical
semantics of the Gillespie next-reaction method
(\S\ref{sec:sim:gillespie}, Theorem~\ref{thm:cat-gillespie}).

\begin{mathbox}[{$\PhiOp$ and $F$ in one diagram}]
{\small
\[
\includegraphics{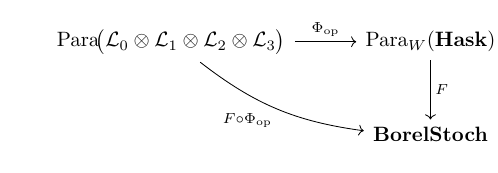}
\]}
Operational semantics $\PhiOp$ (running Haskell, parametric in
$W = \mathtt{Word64}$) and denotational semantics $F \circ \PhiOp$
(Markov kernels on Borel spaces) related by the functor $F$ that
marginalises over the PRNG seed (Proposition~\ref{prop:F-exists},
\S\ref{sec:sim:target}).  $\PhiOp$ is a strict 2-functor of bicategories
(\S\ref{sec:sim:functor}, Proposition~\ref{prop:phi-properties}); its
$\gamma_{\mathrm{ODE}}$ 2-cell in $\BorelStoch$ measures the deviation
of the ODE Kurtz limit from $F_{\mathrm{BR}}$ itself and becomes
empirically visible at finite $V$ (\S\ref{sec:sim:results}).  The
$\gamma_{\mathrm{SSA}}$ 2-cell vanishes in total variation modulo PRNG
and IEEE-754 qualifications by Theorem~\ref{thm:cat-gillespie}.
\end{mathbox}

What the instrument delivers, evaluated at \texttt{briggsRauscherDE} at
$V = 10^{-15}~\mathrm{L}$ (\S\ref{sec:sim:results}), is the kernel
$F_{\mathrm{BR}}$ visible through two operationally distinct routes:
the SSA samples $F_{\mathrm{BR}}$ directly, and the ODE realises its
species-mean Kurtz limit as a Dirac kernel at the deterministic
trajectory.  Both routes exhibit a relaxation-oscillator attractor, in
agreement to within a factor of two on spike-peak amplitudes across all
six dynamical species.  They diverge in the low-copy quiescent phase:
the SSA shows integer-valued populations of $\mathrm{HIO}_2$,
$\mathrm{IO}_2 \bullet$, and $\mathrm{MnOH}^{2+}$ fluctuating between
$0$ and a few molecules; the ODE shows smooth analytic floors at
$10^{-10}$--$10^{-13}~\mathrm{M}$ that translate to sub-unity
ensemble-mean counts at this volume.  This is the $\gamma$-cell
decomposition made empirically operational: $\gamma_{\mathrm{SSA}} = 0$
in total variation throughout, $\gamma_{\mathrm{ODE}}$ non-trivial in
state-space/Wasserstein-1 metric in low-copy regions.  At macroscopic
volumes the two realisations would converge in TV by the Kurtz theorem
\cite{Kurtz1972}; the deliberately small $V$ chosen here makes the
distinction visible rather than asymptotically suppressed.

The framework provides three concrete deliverables beyond confirmed
predictions.  First, an explicit vocabulary (\S\ref{sec:sim:results})
for separating simulator-output features into three categories:
\emph{essential} (the oscillation onset itself, tied to $\Lk_2$
stoichiometric feedback topology and $\Lk_3$ Jacobian sign-pattern),
\emph{tunable} (period and spike amplitudes, set by $\Lk_3$ rate
constants and $\pi_4$-coarse $\Lk_4$ data), and
\emph{scale-dependent representational artifacts} (the smooth ODE
quiescent floors, whose physical referent depends on $V$; the
$\sim\!15\%$ SSA-vs-ODE period drift, a $\gamma_{\mathrm{ODE}}$
residual that vanishes as $V \to \infty$).  Second, a canonical
model-comparison machinery (\S\ref{sec:sim:L4coarse},
Proposition~\ref{prop:variant-comparison}) embedding every BR network
variant --- \texttt{broCODENetwork}, \texttt{broCODE5VarNetwork},
\texttt{bufferedH}, \texttt{furrowLikePoolNetwork} --- into a single
$\BorelStoch$ comparison space via the same $F \circ \PhiOp$, with TV,
Wasserstein-1, and relative-entropy metrics available.  Third, an
honest record of what the $\Lk_3$ simulator cannot see by construction
(\S\ref{sec:sim:L4coarse},
Remark~\ref{rmk:sim:pi4-cannot-see-BR}): kinetic isotope effects on the
methylene hydrogens of malonic acid, transition-state geometry,
pressure dependence, solvent and ionic-strength dependence beyond the
reference conditions absorbed into \texttt{rxnRate}.  These are
$\Lk_4$-strength data dropped by $\pi_4$; a successor
$\Lk_4$-resolved implementation would have to recover them.

\begin{insightbox}[{Reading guide for three audiences}]
\textbf{CRNT readers (Feinberg programme).}
The deficiency computation at $\Lk_0$ (\S\ref{sec:sim:L0}), the
categorical semantics of the Gillespie next-reaction method
(\S\ref{sec:sim:gillespie}, Theorem~\ref{thm:cat-gillespie}), and the
$\pi_4$ parameter projection embedding lossy $\Lk_4$ mechanism data
into the $\Lk_3$ rate constants (\S\ref{sec:sim:L4coarse}) are the
most directly relevant results.  The variant-comparison proposition
(Proposition~\ref{prop:variant-comparison}) embeds every variant of
\texttt{briggsRauscherDE} into a single $\BorelStoch$ comparison space
and supports the kind of canonical network-reduction analysis the
Feinberg programme studies.

\medskip
\noindent
\textbf{Computational chemists.}
Read \S\ref{sec:sim:motivation} for the multi-scale structure of the
Briggs--Rauscher oscillator and its mapping onto the tower levels,
\S\ref{sec:sim:L4coarse} for what the $\Lk_3$ simulator can and cannot
see (in particular what KIE measurements would reveal that
\texttt{rxnRate} cannot register), and \S\ref{sec:sim:results} for the
$\gamma$-cell decomposition made empirically visible at finite
volume.  The framework's three deliverables --- the BR pipeline, the
canonical variant-comparison machinery, and the tripartite separation
of simulator-output features into essential / tunable /
scale-dependent --- are designed for chemistry-inference work rather
than illustration of dynamics.

\medskip
\noindent
\textbf{Category-theory readers.}
The new constructions are $\PhiOp$ as a strict 2-functor of bicategories
(\S\ref{sec:sim:functor}, Proposition~\ref{prop:phi-properties}), the
$\gamma$-cell measuring deviation from exact dynamics
(\S\ref{sec:sim:target}, mathbox), and the $\pi_4$ parameter projection
embedding lossy $\Lk_4$ data into $\Lk_3$ rate constants
(\S\ref{sec:sim:L4coarse}).  The denotational functor
$F : \ParaC_W(\Hask) \to \BorelStoch$ is constructed via PRNG
marginalisation in \S\ref{sec:sim:target}, with the
$\BorelStoch$-coherence properties of the Para construction
(\S\ref{sec:sim:para}) playing throughout.  The first explicit
chemistry application of the Para construction
\cite{CruttwellGavranovic2022} (Corollary~\ref{cor:para-chemistry})
extends prior treatments in machine learning, compositional games
\cite{Hedges2018}, and active inference \cite{Smithe2023}.
\end{insightbox}

\medskip

\noindent
The chapter proceeds as follows.  \S\ref{sec:sim:motivation} introduces
the Briggs--Rauscher oscillator and its multi-scale structure, mapping
the chemistry onto tower levels $\Lk_0$--$\Lk_4$.
\S\ref{sec:sim:target} fixes the target categories ($\Hask$ for the
operational pipeline; $\BorelStoch$ for the denotational kernel),
constructs the $\gamma$-cell measuring $\BorelStoch$-level deviation
between simulator outputs and the exact CTMC, and proves the existence
of $F$ (Proposition~\ref{prop:F-exists}).
\S\S\ref{sec:sim:L0}--\ref{sec:sim:gillespie} work through $\Lk_0$
(atoms) to $\Lk_3$ (mass-action kinetics with saturation), identifying
in each case the exact Haskell realisation and the forgetful functor to
the level below; \S\ref{sec:sim:gillespie} proves that the Gillespie
next-reaction method samples $F_P$ exactly modulo PRNG and IEEE-754
qualifications (Theorem~\ref{thm:cat-gillespie}).
\S\S\ref{sec:sim:para}--\ref{sec:sim:functor} construct the Para
sub-bicategory $\ParaC_W(\Hask)$ over Hask and identify $\PhiOp$ as a
strict 2-functor of bicategories
(Proposition~\ref{prop:phi-properties}).
\S\ref{sec:sim:L4coarse} addresses the $\Lk_4 \to \Lk_3$ parameter
projection $\pi_4$, identifies what mechanism data is dropped, and
proves the variant-comparison proposition that underlies the chapter's
model-comparison framework.  \S\ref{sec:sim:results} closes with the
empirical realisation of $F \circ \PhiOp$ at
\texttt{briggsRauscherDE}: the relaxation-oscillator attractor, the
SSA-vs-ODE structural separation made operationally visible at
$V = 10^{-15}~\mathrm{L}$, and an honest record of what the framework
predicts, what it equips, and what remains open.


\input{chapters/sim/sim_s1_motivation}
\input{chapters/sim/sim_s2_target}
\input{chapters/sim/sim_s3_L0}
\input{chapters/sim/sim_s4_L12}
\input{chapters/sim/sim_s5_gillespie}
\input{chapters/sim/sim_s6_para}
\input{chapters/sim/sim_s7_functor}
\input{chapters/sim/sim_s8_L4coarse}
\input{chapters/sim/sim_s9_results}

%% file: chapters/sim/sim_s1_motivation.tex
\subsection{The Briggs--Rauscher reaction as a forcing ladder}
\label{sec:sim:motivation}

\subsubsection{The empirical problem}
\label{sec:sim:problem}

The Briggs--Rauscher (BR) reaction \cite{BriggsRauscher1973} is the
clock reaction in which an acidic batch of iodate, hydrogen peroxide,
malonic acid, and a manganese(II) catalyst oscillates visibly between
colourless, amber, and dark blue before settling into equilibrium.
Its net stoichiometry,
\begin{equation}\label{eq:sim:net-BR}
  \mathrm{IO_3^-} + 2\,\mathrm{H_2O_2} + \mathrm{CH_2(COOH)_2}
   + \mathrm{H^+} \;\longrightarrow\;
   \mathrm{ICH(COOH)_2} + 2\,\mathrm{O_2} + 3\,\mathrm{H_2O},
\end{equation}
contains no information about the oscillation: it is $\Lk_0$ data,
identical to that of a hypothetical non-oscillating reaction with the
same input and output complexes.

Two parallel skeleton mechanisms were proposed in 1982 by Noyes and
Furrow \cite{NoyesFurrow1982} and independently by De Kepper and
Epstein \cite{DeKepperEpstein1982} in the same volume of JACS.
Both decompose the dynamics into two pathways for hypoiodous-acid
production: \emph{Process~A} (ionic, dominant at high
$[\mathrm{I^-}]$), through sequential heterolytic O--I bond
formations via $\mathrm{HIO_2}$ and $\mathrm{I_2}$ intermediates;
and \emph{Process~B} (radical, dominant at low $[\mathrm{I^-}]$),
through homolytic
$\mathrm{IO_3^- + HIO_2 \to 2\,IO_2^\bullet}$ with Mn(II)/Mn(III)
cycling supplying single-electron transfer.  Process~A consumes
$\mathrm{I^-}$; the iodination of malonic acid by $\mathrm{I_2}$
regenerates it.  The system switches between A and B across a
threshold in $[\mathrm{I^-}]$, and that switching is the mechanical
content of the oscillation.  The implementation in
\texttt{BR.Mechanism} (\S\ref{sec:sim:L0}) follows the De
Kepper--Epstein variant with two explicit reverse channels (R3-rev
and R4-rev), giving twelve reactions in total.

A and B are indistinguishable below $\Lk_4$: they share the $\Lk_0$
net stoichiometry of \eqref{eq:sim:net-BR}, and their bulk net
HOI-production rates share the form $\dot c_{\mathrm{HOI}} =
k_{\mathrm{eff}}\,[\mathrm{IO_3^-}]\,[\mathrm{H_2O_2}]$ with only
$k_{\mathrm{eff}}$ differing.  Their distinction --- heterolytic
versus homolytic bond character --- is recorded in the DPO graph of
each elementary step and is $\Lk_4$ data.  The simulation in
\S\ref{sec:sim:functor} is provably blind to the A/B distinction;
the main tower's $\Lk_4$ chapter
(Observation~\ref{obs:BR-trajectory}) develops this blindness as
a categorical theorem.

\subsubsection{The forcing hierarchy}
\label{sec:sim:hierarchy}

The tower $\Lk_0 \hookrightarrow \cdots \hookrightarrow \Lk_7$ was
constructed level-by-level, each transition forced by an explicit
reaction pair the previous level cannot distinguish.  The BR system
presents this ladder at every transition $\Lk_0 \to \Lk_4$.

\medskip
\noindent\textbf{$\Lk_0 \to \Lk_1$.}
$\Lk_0$ records stoichiometric vectors only.  Atom and charge
conservation across every reaction are the $\Lk_0$ structural
integrity tests, realised in \texttt{BR.Invariants} as element-wise
and total-charge linear checks on the stoichiometric matrix.
$\Lk_1$ adds the monoidal functor $\FH : \Lk_1 \to (\RR, +)$ whose
functoriality $\FH(r_2 \circ r_1) = \FH(r_1) + \FH(r_2)$ is Hess's
Law.

\medskip
\noindent\textbf{$\Lk_1 \to \Lk_2$.}
Iodate oxidation of $\mathrm{H_2O_2}$ has $\dH^\circ \approx 0$ at
standard conditions but proceeds spontaneously; the entropy gain on
$\mathrm{O_2}$ release controls the sign of $\dG^\circ$.  $\Lk_2$
adds the entropy functor $\FS$, the temperature-dependent
$\FG^T = \FH - T\,\FS$, and the $\dagger$-structure identifying each
reaction with its reverse.  The codebase realises the dagger
structure explicitly through paired reactions: the iodine
hydrolysis pair \texttt{R3}/\texttt{R3-rev} and the
$\mathrm{IO_2^\bullet}$ recombination pair
\texttt{R4}/\texttt{R4-rev} appear as separate \texttt{Reaction}
values whose rate-constant ratios satisfy Wegscheider's relation
$\prod_i k_{+,i} = \prod_i k_{-,i}$ on each closed cycle.  In this
implementation the $\Lk_1$ and $\Lk_2$ functors enter \emph{through}
the $\Lk_3$ rate constants via Wegscheider, rather than as separate
fields of \texttt{Reaction} (\S\ref{sec:sim:L12}).

\medskip
\noindent\textbf{$\Lk_2 \to \Lk_3$.}
The BR system never reaches its $\dG^\circ$-determined equilibrium
during the oscillatory phase; dynamics are kinetically controlled
by rate constants spanning twelve orders of magnitude across the
elementary steps.  $\Lk_3$ adds the functor
$F_P : \Lk_3 \to \Stoch$ sending each network to a Markov kernel on
populations.  The simulation realises $F_P$ \emph{twice} from the
same \texttt{Network} value: by the Gibson--Bruck next-reaction
method in \texttt{BR.SSA} (\S\ref{sec:sim:gillespie}), and, in the
large-volume limit, by the implicit-midpoint mass-action ODE in
\texttt{BR.ODE}.  Together they form a cospan out of the $\Lk_3$
object.

\medskip
\noindent\textbf{$\Lk_3 \to \Lk_4$: the BR forcing pair.}
Processes A and B share $\Lk_0$ stoichiometry and the bulk
HOI-production rate form, but the elementary bond changes ---
heterolytic in A, homolytic in B --- live in the DPO graph at
$\Lk_4$.  No observable computable from the $\Lk_3$ rate equations
distinguishes them.  This is the canonical $\Lk_3 \to \Lk_4$ forcing
pair of the main tower
(Observation~\ref{obs:BR-trajectory}), and it provides the chemical
motivation for the present chapter rather than its own subject of
study: this chapter constructs the simulation functor $\PhiOp$ at
$\Lk_0$ through $\Lk_3$, where the A/B distinction is invisible by
construction.

\begin{forcingbox}[{The BR forcing hierarchy in one line}]
\[
  \Lk_0 \;\xrightarrow{\FH}\; \Lk_1
  \;\xrightarrow{\FG^T}\; \Lk_2
  \;\xrightarrow{F_P}\; \Lk_3
  \;\xrightarrow{U_4}\; \Lk_4
\]
At each arrow lies a physical phenomenon inexpressible one level
below: \emph{route-independence of heats}, \emph{entropy-controlled
spontaneity}, \emph{kinetic control} far from equilibrium, and
finally \emph{ionic-versus-radical mechanism identity} --- the
Process~A/B forcing pair that motivates this chapter.
\end{forcingbox}

\subsubsection{One elementary step in code}
\label{sec:sim:worked-example}

A single \texttt{Reaction} value in the BR codebase carries explicit
$\Lk_0$ and $\Lk_3$ data; $\Lk_1$ and $\Lk_2$ enter as Wegscheider
constraints across paired reactions (\S\ref{sec:sim:L12}), not as
record fields.  We illustrate on step~R5 of the De Kepper--Epstein
mechanism \cite{DeKepperEpstein1982}, the manganese single-electron
transfer initiating Process~B:
\[
  \mathrm{IO_2^\bullet} + \mathrm{Mn^{2+}} + \mathrm{H_2O}
  \;\longrightarrow\; \mathrm{HIO_2} + \mathrm{Mn(OH)^{2+}}.
\]

\begin{haskellbox}[title={\texttt{BR/Mechanism.hs} --- step R5}]
(mkReaction
   "R5: IO2. + Mn2+ + H2O -> HIO2 + MnOH2+"
   (ms [ (IodineDioxideRadical, 1)               -- L0 source
       , (Manganese2,           1)
       , (Water,                1) ])
   (ms [ (IodousAcid,           1)               -- L0 target
       , (ManganeseHydroxide3,  1) ])
   1.0e4                                         -- L3 rate (M^-1 s^-1)
   (Measured "DE1982")                           -- provenance
   "DE1982 Table I (k6); [H2O] folded into k")
   { rxnRateActive                               -- L3 rate-active subset
       = ms [ (IodineDioxideRadical, 1)          -- (solvent excluded)
            , (Manganese2,           1) ] }
\end{haskellbox}

\noindent
The multisets are $\Lk_0$ data; the rate constant
$1.0\times 10^{4}\,\mathrm{M^{-1}\,s^{-1}}$ is $\Lk_3$ data; the
\texttt{rxnRateActive} field encodes the chemistry convention that
the solvent $\mathrm{H_2O}$ is folded into $k$ rather than entering
the rate law as a $[\mathrm{H_2O}]$ factor.  This last detail is a
real $\Lk_3$ subtlety: although $\mathrm{H_2O}$ is consumed
stoichiometrically (it appears in \texttt{rxnReactants}), only the
two non-solvent reactants enter the rate law (they appear in
\texttt{rxnRateActive}); the $\Lk_0$ atom-balance test in
\texttt{BR.Invariants} still operates on the full
\texttt{rxnReactants} multiset and so remains correct.  The $\Lk_4$
identity of R5 as a Process-B step --- the homolytic single-electron
transfer that distinguishes B from A's heterolytic chemistry --- is
conspicuously absent from R5's fields: it is a structural property
of the subnetwork containing R5 (together with R4, R6, R7, R8),
made categorical in the main tower's $\Lk_4$ chapter.  Every
subsequent section of this chapter studies one horizontal slice of
\texttt{Reaction}'s level-stratified content.

%% file: chapters/sim/sim_s2_target.tex
\subsection{Target category: $\Hask$ and $\BorelStoch$}
\label{sec:sim:target}

A categorical interpretation of the simulation must fix a category
for total Haskell --- the \emph{operational} layer in which the
program runs --- and a target category for the stochastic content
of $\Lk_3$ --- the \emph{denotational} layer in which distributions
live --- and a functor sending the first to the second.  The
payoff is concrete: every claim about the simulator's accuracy
will attach to a specific morphism in one of these categories and
to a specific user-controllable parameter
(\S\ref{sec:sim:gillespie} onwards).  This section fixes the two
categories; \S\ref{sec:sim:functor} builds the functor.

\subsubsection{Why not naive \textbf{Hask}}

The naive category with Haskell types as objects and all Haskell
functions as morphisms fails the identity law: Bauer
\cite{Bauer2016Hask} observes that $\mathtt{seq}\,\bot\,()$ differs
from $\mathtt{seq}\,(\bot \mathbin{\texttt{.}} \mathtt{id})\,()$,
so $\mathrm{id}$ is not a categorical identity under composition.
Danielsson, Hughes, Jansson, and Gibbons \cite{Danielsson2006}
prove that the \emph{total} fragment of a partial language forms a
bicartesian closed category in which equational reasoning is
sound.  By $\Hask$ in the rest of this chapter we mean that total
fragment.

\begin{assumption}[Totality and purity of the simulation core]
\label{ass:totality}
Every function in the simulation core --- the L0 invariants
\texttt{atomConservation}, \texttt{chargeConservation}; the
constructors \texttt{mkReaction}, \texttt{mkNetwork}; the L3
realisations \texttt{BR.ODE.simulate} with \texttt{stepAdaptive}, and
\texttt{BR.SSA.simulate} with \texttt{step} --- is total and pure on
its domain.  Stochasticity in \texttt{BR.SSA} is supplied entirely
by the vendored \texttt{BR.PRNG} (a state-threaded SplitMix64), so
the \texttt{IO} monad never appears in the call graph: no
\texttt{System.Random}, no \texttt{unsafePerformIO}, no FFI, no
exceptions.  File I/O is confined to \texttt{BR.Output}, which acts
only at the simulation boundary and is not invoked by any function
below it.
\end{assumption}

\begin{remark}[Why no $\KlIO$]
\label{rmk:no-KlIO}
With \texttt{BR.PRNG} pure and explicit, both the ODE and SSA
realisations of $\Lk_3$ live in $\Hask$, not in a Kleisli category
of a probability monad in the sense of monadic semantics for
computational effects \cite{Moggi1991}.  Introducing
$\Kl{\mathsf{IO}}$, the Kleisli category of the Haskell \texttt{IO}
monad, would therefore add no structure here: the simulator would use
only the pure Kleisli arrows
\[
  A \xrightarrow{f} B \xrightarrow{\mathrm{return}} \mathsf{IO}\,B,
\]
i.e. the wide subcategory
\[
  \{\mathrm{return}\circ f : f \in \Hask(A,B)\}
  \subseteq \Kl{\mathsf{IO}}(A,B).
\]
We therefore target $\Hask$ directly.
\end{remark}

\subsubsection{Two-layer semantics: operational and denotational}

The stochastic content of $\Lk_3$ requires a Markov category.

\begin{definition}[Markov category; $\BorelStoch$ {\cite{Fritz2020Synthetic}}]
\label{def:borelstoch}
A \emph{Markov category} is a semicartesian symmetric monoidal
category $(\mathbf{C}, \otimes, I)$ in which every object $X$
carries a natural commutative comonoid $(\mathrm{copy}_X,
\mathrm{del}_X)$ and every morphism $f$ is counital,
$\mathrm{del}_Y \circ f = \mathrm{del}_X$.
Equivalently, it is a category in which the algebra of probability ---
sampling, copying, discarding, composing --- is expressible as pure
morphism manipulation, without making measure-theoretic integrals part
of the primary syntax.
$\BorelStoch$ is the Markov category whose objects are \emph{standard Borel spaces}
(measurable spaces arising from a Polish topology, i.e.\ a complete
separable metric --- countable sets, $\RR^n$, and finite products
thereof) and whose morphisms are measurable Markov kernels.  It is
a Borel restriction of the Kleisli category
$\Kl{\mathcal{G}}$ of the \emph{Giry monad} $\mathcal{G}$ \cite{Giry1982}, the
probability-measure monad sending each measurable space to its
space of probability measures.
\end{definition}

\noindent
$\Hask$ is Cartesian and \emph{deterministic}: any morphism in
$\Hask$ between standard Borel spaces embeds into $\BorelStoch$ as
a \emph{Dirac kernel} --- the degenerate Markov kernel placing all
mass on a single point, i.e.\ the embedding of a deterministic
function as a distribution concentrated at $f(a)$.  All Haskell
types appearing in this chapter (\texttt{Word64}, \texttt{Double},
\texttt{Map Species Int}, lists thereof) are countable or Polish,
so this restriction is automatic.  Two categories are necessary
because the stochastic content lives in $\BorelStoch$ (where one
states theorems and accuracy bounds), but the simulator runs in
$\Hask$ (where one debugs, profiles, and extends).  The functor
$F$ below keeps the two pictures in sync: the seed-marginal of any
executable Haskell program is the Markov kernel it samples from.

\begin{proposition}[Seed-marginalisation functor]
\label{prop:F-exists}
$F$ formalises the operation ``run the simulator with random
seeds, infinitely many times, and read off the distribution of
outputs'' as a functor.  Concretely, let $W = \mathtt{Word64}$ with
the uniform distribution $\mu_W$, and let
$\ParaC_W(\Hask) \subseteq \ParaC(\Hask)$ denote the sub-bicategory
of 1-morphisms whose parameter object is a (finite) power of $W$
and whose source and target are standard Borel spaces.  For a
1-morphism $(W^n, f)$ in $\ParaC_W(\Hask)$ with
$f : W^n \times A \to B$ ($n \geq 0$), define
\[
  F\bigl(W^n,\, f\bigr)(a, \cdot) \;:=\;
  \mathrm{law}_{w \sim \mu_W^{n}}\bigl(f(w, a)\bigr).
\]
This extends to a (pseudo)functor $F : \ParaC_W(\Hask) \to
\BorelStoch$, with sequential composition in $\ParaC_W(\Hask)$
taken on $W^{n+m}$ (independent seed draws at each stage).
Identities ($n=0$) go to Dirac kernels; preservation of
composition is the Fubini factorisation
$\mathrm{law}_{(w,w')}\bigl(g(w', f(w, a))\bigr) =
 \int F(g)(b,\cdot)\, F(f)(a, db)$.

\medskip
\noindent\emph{Instantiation for BR.}  $f_{\mathrm{SSA}}$ sits at
$n = 1$: its sole Para parameter is the \texttt{Word64} seed stored
in \texttt{ssaConfig}.  The trajectory-dependent number of internal
\texttt{nextWord} draws made by the Gibson--Bruck algorithm
(\S\ref{sec:sim:gillespie}) is consumed \emph{within}
$f_{\mathrm{SSA}}$ as part of the underlying-morphism computation,
not as additional Para parameters.  $f_{\mathrm{ODE}}$ sits at
$n = 0$: the implicit-midpoint integrator is deterministic and
carries no seed.  Both lie in $\ParaC_W(\Hask)$.
\end{proposition}

\begin{remark}[BR.PRNG implementation and seed independence]
\label{rmk:prng-quality}
The marginalisation in Proposition~\ref{prop:F-exists} treats
$\mu_W^n$ as the law of $n$ independent uniform draws on $W$.  BR's
\texttt{BR.PRNG} delivers these draws via a single SplitMix64 state
threaded through \texttt{SSAState}, advancing via \texttt{nextWord}
at each random draw: a single seed deterministically generates the
full stream of draws that $f_{\mathrm{SSA}}$ consumes during a
trajectory, and the implementation of sequential composition reuses
the evolved state rather than drawing a fresh independent seed.
Functoriality of $F$ on this implementation therefore holds only
\emph{statistically} --- under the standard PRNG-quality assumption
that successive SplitMix64 outputs are independent for
distributional reasoning, satisfied to BigCrush (L'Ecuyer and
Simard's empirical battery of independence and uniformity tests) by
SplitMix64 \cite{SteeleLeaFlood2014} at the scales relevant to this
chapter.
\end{remark}

\noindent
The two-layer semantics is the diagram below.  $\PhiOp$ is a
pseudofunctor with the documented laxity $\gamma$ of the next
subsection; $F$ is a (pseudo)functor by
Proposition~\ref{prop:F-exists}; $\PhiDen := F \circ \PhiOp$
supplies the Markov-category target that the stochastic content of
$\Lk_3$ requires.

\begin{figure}[h]
\centering
\includegraphics{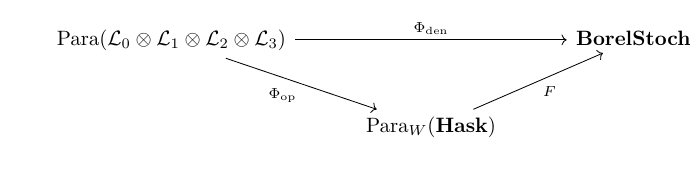}
\caption{Operational and denotational semantics for the BR
simulator.  $\PhiOp$ is operational (the pure Haskell program with
PRNG seed in scope as a Para parameter); $F$ is denotational
(marginalise over the seed); $\PhiDen = F \circ \PhiOp$ lands in
$\BorelStoch$.  $\ParaC_W(\Hask)$ denotes the sub-bicategory of
$\ParaC(\Hask)$ of $W$-parameterised pure morphisms between standard
Borel spaces.}
\label{fig:sim:two-layer}
\end{figure}

\subsubsection{One genuine laxity: $\gamma$}

In an IO-based formulation of this story (a
\texttt{System.Random}-style codebase), $\PhiOp$ would carry three
lax cells: a unit $\alpha$, an IO-sequencing $\beta$, and a
truncation/numerical $\gamma$.  Two of the three collapse for
BR.  In pure $\Hask$, Cartesian product is strictly associative
and commutative on the nose, so the parallel composition of two
seed-parameterised morphisms is invariant under sequencing:
$\beta$ is strict, not lax.  The unit cell $\alpha$ becomes the
parameter-tensor unit isomorphism of the Para bicategory
$\ParaC(\Hask)$, treated as canonical coherence data
(\S\ref{sec:sim:para}) rather than as a laxity of $\PhiOp$.  The
only cell that remains is $\gamma$:

\begin{mathbox}[{The $\gamma$-cell: truncation}]
$\gamma$ is the gap between what the simulator returns and the
exact stochastic dynamics it approximates.  The two branches behave
differently: SSA's $\gamma$-component reduces to PRNG- and
floating-point-level numerical artifacts (no separate user knob);
ODE's $\gamma$-component is a deterministic step-size error driven
to zero by a user-specified tolerance.

\medskip
\noindent\textbf{SSA branch (total-variation metric).}
The Gibson--Bruck NRM in \texttt{BR.SSA} terminates when (a)
\texttt{ssaTime} exceeds \texttt{ssaTimeMax} (a safety guard,
unreachable after any successful step), (b) all putative firing
times are infinite (genuine absorption), or (c) the next firing
would overshoot the horizon.  Theorem~\ref{thm:cat-gillespie} of
\S\ref{sec:sim:gillespie} establishes that, for $\tau$ encoded as
\texttt{ssaTimeMax} in \texttt{ssaConfig}, the state at termination
agrees with the exact CTMC state at $\tau$ in distribution under
all three termination paths: no events occur in the half-open
interval between any case-(c) termination time and $\tau$.
Writing $F_P$ for the exact CTMC kernel of $\Lk_3$
(Theorem~\ref{thm:cat-gillespie}), in the \emph{total variation}
norm $\|\mu - \nu\|_{\mathrm{TV}} := \sup_A |\mu(A) - \nu(A)|$ on
probability measures,
\[
  \bigl\| F\bigl(\PhiOp^{\mathrm{SSA}}(\mathcal{N})(\mathbf{x}_0,
       \tau)\bigr)
       - F_P(\mathcal{N})(\mathbf{x}_0, \tau) \bigr\|_{\mathrm{TV}}
  \;=\; 0
\]
modulo (i) SplitMix64's deviation from i.i.d.\ uniform draws on
$[0,1)$ (Remark~\ref{rmk:prng-quality} above; bounded statistically
by the BigCrush results cited there), and (ii) IEEE-754
floating-point representation of inverse-CDF samples (per-draw
error $\lesssim 2^{-53}$).  Neither qualification is a user knob
in the ordinary sense; the SSA branch of $\gamma$ is effectively
zero in distribution.

\medskip
\noindent\textbf{ODE branch (state-space metric).}
The implicit-midpoint integrator in \texttt{BR.ODE} produces a
deterministic trajectory; its discrepancy from the true
mass-action trajectory is measured in state-space sup norm rather
than TV.  The local truncation error is $O(h^3)$ per step,
controlled adaptively by step doubling to a user-specified
tolerance --- the user knob is the integration tolerance.  Embedded
as Dirac kernels in $\BorelStoch$, the discrepancy is invisible to
the TV metric (which collapses to 0 or 1 between distinct Diracs);
the relevant metric on the ODE side of $\gamma$ is therefore
Wasserstein-1 (or directly state-space) rather than TV.

\medskip
\noindent
The two branches differ in both the \emph{form} of $\gamma$
(effectively zero in distribution on SSA, deterministic step-size
error on ODE) and the metric in which it is measured (TV on SSA,
state-space on ODE).  The asymmetry is structural: SSA's exactness
modulo numerical artifacts is intrinsic to the Gibson--Bruck NRM
(\S\ref{sec:sim:gillespie}), whereas the ODE branch carries a
genuine numerical laxity controllable by \texttt{odeRTol} and
\texttt{odeATol}.
\end{mathbox}

\begin{remark}[Para is base-portable]
\label{rmk:capucci-myers}
The Para construction applies over any symmetric monoidal base
category; varying the base gives different applications
($\mathbf{Smooth}$ for backpropagation in machine learning, $\Hask$
here for executable simulation, $\BorelStoch$ for fully
probabilistic kernels).  Capucci and Myers \cite{CapucciMyers2024}
show that all these incarnations are instances of a single
two-categorical pattern, so transferring techniques between them is
structural rather than ad hoc.  The seed-marginalisation functor
$F$ above realises this bridge from the $\Hask$ incarnation to a
Markov-category target for our specific choice of uniform seed
distribution on $W$; other distributions would yield other
functors with the same shape.
\end{remark}

%% file: chapters/sim/sim_s3_L0.tex
\subsection{$\Lk_0$ in Haskell: \texttt{Multiset Species} as free commutative monoid}
\label{sec:sim:L0}

Level $\Lk_0$ is the tower's stoichiometric foundation.  Its
objects are elements of the free commutative monoid $\NN[\Sp]$ on
the finite species set $\Sp$ (Equation~\eqref{eq:freefreefree});
its morphisms are stoichiometric changes (reaction vectors in
$\ZZ^\Sp$); its functorial invariants are the combinatorial data
$(n, \ell, s)$ --- the number of distinct complexes, the number of
linkage classes (connected components of the reaction graph), and
the rank of the stoichiometric matrix --- and the
\emph{deficiency} $\delta = n - \ell - s$.  Two $\Lk_0$
\emph{structural integrity tests} --- atom and charge conservation
--- attach to every reaction and must pass before any $\Lk_3$
dynamics are simulated.  This section shows how the BR codebase
realises each of these pieces concretely.

\subsubsection{The free commutative monoid in Haskell}

The set $\NN[\Sp]$ of Equation \eqref{eq:freefreefree} carries a commutative
monoid structure under pointwise addition with the universal
property that any function $\varphi : \Sp \to M$ to a commutative
monoid $M$ extends uniquely to a homomorphism
$\bar\varphi : \NN[\Sp] \to M$.  The BR implementation realises
both in a single newtype.

\begin{proposition}[$\Lk_0$ implementation]
\label{prop:L0-exact}
The newtype \texttt{Multiset Species} of \\
\texttt{BR.Multiset} exactly implements $\NN[\Sp]$ as an object of $\Lk_0$:
\begin{enumerate}[label=(\roman*)]
  \item \emph{(Bijection.)}  The invariant that the underlying
    \texttt{Map Species Int} has no entries with value $\leq 0$,
    maintained by every smart constructor, puts
    \texttt{Multiset Species} in bijection with $\NN[\Sp]$.
  \item \emph{(Monoid.)}  \texttt{munion} is pointwise addition;
    \texttt{empty} is the identity.
  \item \emph{(Universal property.)}  For any Haskell commutative
    monoid $M$ and any $\varphi : \texttt{Species} \to M$, the
    unique extending homomorphism
    $\bar\varphi : \texttt{Multiset Species} \to M$
    is given by
    \begin{center}
    $\bar\varphi$ \;\texttt{=\;foldMap (\textbackslash{}(s, n) -> stimes n (phi s)) . toList}.
    \end{center}
\end{enumerate}
\end{proposition}

\noindent
Each item is settled directly by the code in
\texttt{BR.Multiset}:

\begin{haskellbox}[title={\texttt{BR/Multiset.hs} --- the function bodies that realise Proposition~\ref{prop:L0-exact}}]
-- Invariant: the underlying Map has no entries with value <= 0.
newtype Multiset a = Multiset { _runMultiset :: Map a Int }

-- Smart constructor preserving the invariant: realises (i).
fromListWith :: Ord a => (Int -> Int -> Int) -> [(a, Int)] -> Multiset a
fromListWith f xs = Multiset
  (M.filter (> 0) (M.fromListWith f [(k, n) | (k, n) <- xs, n /= 0]))

-- The monoid operation: realises (ii).
munion :: Ord a => Multiset a -> Multiset a -> Multiset a
munion (Multiset a) (Multiset b) = Multiset (M.unionWith (+) a b)

-- The N-action n . s of N[S].
scale :: Int -> Multiset a -> Multiset a
scale k (Multiset m)
  | k <= 0    = Multiset M.empty
  | otherwise = Multiset (fmap (* k) m)
\end{haskellbox}

\begin{proof}
\textbf{(i)} Every smart constructor preserves the invariant ---
either by an outer \texttt{M.filter (> 0)}
(\texttt{fromListWith}, \texttt{difference}, \texttt{subtractM}) or
by structural argument (\texttt{munion}'s pointwise sum of
positives is positive; \texttt{scale} and \texttt{singleton} guard
their non-positive inputs).  The positive-coefficient entries of
the map are then in bijection with the nonzero terms of an element
of $\NN[\Sp]$.

\textbf{(ii)} \texttt{munion} unfolds to \texttt{M.unionWith (+)},
which is pointwise sum on each key; commutativity and associativity
descend from those of $(+)$ on \texttt{Int}, and \texttt{empty} is
the two-sided identity.

\textbf{(iii)} Applied to $\sum_i n_i \cdot s_i$ the displayed
expression returns $\sum_i n_i \cdot \varphi(s_i)$ under $M$'s
monoid operation (each summand from \texttt{stimes n (phi s)},
combined by \texttt{foldMap}) --- a homomorphism extending
$\varphi$.  Uniqueness is the standard universal property: any
monoid homomorphism on $\NN[\Sp]$ extending $\varphi$ on generators
is forced on all formal sums by additivity, so it coincides with
the displayed formula.
\end{proof}

The carrier set $\Sp$ of \eqref{eq:freefreefree} is a finite enum:

\begin{haskellbox}[title={\texttt{BR/Species.hs} --- the species set $\Sp$ and its formula/charge maps}]
data Species
  = Iodide | Iodate | HypoiodousAcid | IodousAcid | IodineDioxideRadical
  | DiIodine | HydrogenPeroxide | MalonicAcid | IodoMalonicAcid
  | Manganese2 | ManganeseHydroxide3 | Hydroperoxyl
  | Proton | Water | DiOxygen
  deriving stock (Eq, Ord, Show, Read, Enum, Bounded)        -- |Sp| = 15

-- Exhaustive pattern match: a missing case is a compile error.
formula :: Species -> Map Atom Int
formula = \case
  Iodate              -> mk [(I, 1), (O, 3)]                  -- IO3-
  IodineDioxideRadical -> mk [(I, 1), (O, 2)]                 -- IO2.
  ManganeseHydroxide3 -> mk [(Mn, 1), (O, 1), (H, 1)]         -- Mn(OH)2+
  MalonicAcid         -> mk [(C, 3), (H, 4), (O, 4)]          -- CH2(COOH)2
  -- ... 11 more cases (one per Species constructor)
  where mk = M.fromListWith (+)

charge :: Species -> Int
charge = \case
  Iodide              -> -1
  Iodate              -> -1
  Proton              ->  1
  Manganese2          ->  2
  ManganeseHydroxide3 ->  2     -- Mn(III)+OH-(-1) = +2 net
  -- ... 10 remaining cases, all 0 (neutral species)
\end{haskellbox}

\subsubsection{$\Lk_0$ structural integrity: atom and charge conservation}

A multiset $r \in \NN[\Sp]$ has only multiplicity data.  Attaching
it to a reaction $r_{\mathrm{src}} \to r_{\mathrm{tgt}}$ as a
$\Lk_0$ morphism, two linear conditions on the reaction vector
$r_{\mathrm{tgt}} - r_{\mathrm{src}} \in \ZZ^\Sp$ must hold for the
reaction to represent a physically meaningful elementary step:
conservation of each chemical element, and conservation of total
formal charge.

\begin{definition}[Atom and charge conservation at $\Lk_0$]
\label{def:L0-invariants}
For each element $a \in \{\mathrm{H,\,C,\,O,\,I,\,Mn}\}$, let
$\varphi_a : \Sp \to \NN$ be the function
\[
s \mapsto \texttt{M.findWithDefault 0 a (formula s)}
\] 
(the $a$-coefficient of the species formula), and let
$q : \Sp \to \ZZ$ be \texttt{charge}.  Their canonical
monoid-homomorphism extensions to $\NN[\Sp]$
(Proposition~\ref{prop:L0-exact}(iii)) are denoted
$\bar\varphi_a : \NN[\Sp] \to \NN$ and
$\bar q : \NN[\Sp] \to \ZZ$.

A reaction $r_{\mathrm{src}} \to r_{\mathrm{tgt}}$ is
\emph{atom-balanced} if
$\bar\varphi_a(r_{\mathrm{tgt}}) - \bar\varphi_a(r_{\mathrm{src}})
= 0$ for every element $a$, and \emph{charge-balanced} if
$\bar q(r_{\mathrm{tgt}}) - \bar q(r_{\mathrm{src}}) = 0$.  A
network is $\Lk_0$-\emph{integral} if every reaction in it is both.
\end{definition}

The code is a transcription of this definition:

\begin{haskellbox}[title={\texttt{BR/Invariants.hs} --- function bodies}]
-- Per-reaction atom delta:  bar_phi_a(products) - bar_phi_a(reactants)
-- where bar_phi_a is from Definition def:L0-invariants and the
-- atomsIn helper is exactly the foldMap-of-stimes formula of
-- Proposition prop:L0-exact (iii) specialised to phi_a.
reactionAtomDelta :: Atom -> Reaction -> Int
reactionAtomDelta a r =
  atomsIn (rxnProducts r) - atomsIn (rxnReactants r)
  where
    atomsIn ms = sum
      [ n * M.findWithDefault 0 a (Sp.formula s)         -- = n * phi_a(s)
      | (s, n) <- MS.toList ms ]                         -- foldMap over toList

atomBalance :: Atom -> Reaction -> Bool                  -- per-reaction check
atomBalance a r = reactionAtomDelta a r == 0

atomConservation :: Network -> Bool                      -- network-wide check
atomConservation net = and
  [ atomBalance a r | r <- netReactions net, a <- allAtoms ]

-- The charge analogues (reactionChargeDelta, chargeBalance, chargeConservation)
-- replace Sp.formula and the per-atom lookup with Sp.charge; their bodies are
-- structurally identical and share their correctness proof with the atom versions.
\end{haskellbox}

\noindent
The \texttt{atomsIn} helper inside \texttt{reactionAtomDelta} is
literally the universal-property formula of
Proposition~\ref{prop:L0-exact}(iii) with $\varphi_a(s)$ taken to
be the $a$-coefficient of \texttt{Sp.formula s}: each summand is
$n \cdot \varphi_a(s)$, and the list-comprehension sum over
\texttt{toList} is the \texttt{foldMap}.  Then
\texttt{reactionAtomDelta a r} is the categorical statement
$\bar\varphi_a(r_{\mathrm{tgt}}) - \bar\varphi_a(r_{\mathrm{src}})$.
Because \texttt{formula} and \texttt{charge} use exhaustive
\texttt{\textbackslash{}case}, introducing a new species without
both is a compile-time error rather than a silent runtime balance
failure --- the $\Lk_0$ contract is enforced at the type level.

\subsubsection{Deficiency and the $\Lk_0\!\to\!\Lk_3$ structure of DZT}

The deficiency $\delta = n - \ell - s$ is entirely an $\Lk_0$
invariant: it is computed from the stoichiometric matrix $\boldsymbol{\nu} \in
\ZZ^{\Sp \times \Rx}$ --- the matrix of $\Lk_0$-morphism data ---
without reference to rate constants, thermodynamics, or mechanism.
Theorem~\ref{the:DZT1} states the L0 consequence of $\delta = 0$;
Theorem~\ref{thm:DZT} strengthens this to assertions about the
mass-action vector field once a rate functor $\FP$ is supplied at
$\Lk_3$.

\begin{remark}[Cross-level structure of DZT]
\label{rmk:dzt-cross-level}
The \emph{hypotheses} of Theorem~\ref{the:DZT1} (weak reversibility,
$\delta = 0$) are pure $\Lk_0$ data; the \emph{conclusions} become
assertions about the mass-action flow only once a rate functor
$\FP$ is chosen, at which point the statement strengthens to
Theorem~\ref{thm:DZT}.  This is the tower's reading of the Feinberg
programme: DZT and DOT are $\Lk_0 \!\to\! \Lk_3$ theorems whose
content straddles the combinatorial and kinetic levels.  A fully
categorical proof would make the straddling functorial.
\end{remark}

\begin{chembox}[{$\Lk_0$ data of the Briggs--Rauscher network}]
The De Kepper--Epstein implementation
(\texttt{BR.Mechanism.briggsRauscherDE}, \S\ref{sec:sim:motivation})
has $|\Sp| = 15$ species and twelve reactions (the ten elementary
steps R1--R10 plus explicit reverses R3-rev, R4-rev).  Hand
enumeration of distinct source/target multisets gives $n = 20$
complexes distributed over $\ell = 10$ linkage classes (each
elementary forward/reverse pair occupies its own class, since no
two elementary steps share a source or target complex); the rank
$s$ of the stoichiometric matrix is computable by Gaussian
elimination but plays no role below.  Running the L0 checks gives
\texttt{atomConservation briggsRauscherDE == True} (each of H, C,
O, I, Mn balanced in every reaction) and
\texttt{chargeConservation briggsRauscherDE == True} (net charge
zero in every reaction), so the network is $\Lk_0$-integral in
the sense of Definition~\ref{def:L0-invariants}.

\medskip
\noindent\textbf{Why the deficiency theorems are silent on BR.}
Of the twelve reactions, only the two pairs R3/R3-rev and R4/R4-rev
carry explicit reverses; the remaining eight elementary steps
(R1, R2, R5--R10) are unidirectional in the implementation.  The
network is therefore \emph{not weakly reversible}, and the
hypotheses of both Theorem~\ref{the:DZT1} and the Deficiency One
Theorem fail regardless of the value of $\delta$.  $\Lk_0$ alone
cannot predict the BR steady-state structure --- and indeed, no
steady state exists during the oscillatory phase: the system is
held far from equilibrium by the buffered substrate pool
($\mathrm{IO_3^-}$, $\mathrm{H_2O_2}$, MA, $\mathrm{Mn^{2+}}$) and
cycles between Process~A and Process~B without converging.  This
is the chapter's first concrete demonstration that L0 data alone is
insufficient: the dynamical content lives strictly at $\Lk_3$ and
requires the simulator built in subsequent sections.
\end{chembox}

%% file: chapters/sim/sim_s4_L12.tex
\subsection{$\Lk_1$--$\Lk_2$: detailed-balance constraints on rate constants}
\label{sec:sim:L12}

Levels $\Lk_1$ (enthalpies) and $\Lk_2$ (free-energy and
reversibility) extend $\Lk_0$ by a monoidal functor $\FH : \Lk_1
\to (\RR, +)$, an entropy extension giving $\FG^T = \FH - T\,\FS$
at $\Lk_2$, and a $\dagger$-structure identifying each reaction
with its reverse.  In the BR implementation these data are not
stored: \texttt{BR.Reaction} carries only $\Lk_0$ stoichiometry
(\texttt{rxnReactants}, \texttt{rxnProducts}) and $\Lk_3$ kinetics
(\texttt{rxnRate}, \texttt{rxnRateActive}, \texttt{rxnSaturation}),
plus metadata (\texttt{rxnName}, \texttt{rxnProvenance},
\texttt{rxnCitation}).  The full record:

\begin{haskellbox}[title={\texttt{BR/Reaction.hs} --- the \texttt{Reaction} record and its smart constructor}]
-- Where this rate constant came from.
data RateProvenance
  = Measured String     -- measured value (citation key)
  | FitToOscillate      -- "judicious assignment" (Noyes-Furrow's term)
  | Estimated String    -- order-of-magnitude estimate
  deriving stock (Eq, Show, Read)

data Reaction = Reaction
  { rxnName        :: !String                       -- label
  , rxnReactants   :: !(Multiset Species)           -- L0 source
  , rxnProducts    :: !(Multiset Species)           -- L0 target
  , rxnRateActive  :: !(Multiset Species)           -- L3 rate-law subset
  , rxnRate        :: !Double                       -- L3 rate constant
  , rxnSaturation  :: ![(Species, Double)]          -- L3 saturation
  , rxnProvenance  :: !RateProvenance               -- citation type
  , rxnCitation    :: !String                       -- free-text citation
  } deriving stock (Eq, Show)

-- Smart constructor: pure mass-action defaults.
-- (Defaults are overridden via record-update syntax when needed,
-- as in R3-rev's `[H2O]`-folded rate law below.)
mkReaction
  :: String -> Multiset Species -> Multiset Species
  -> Double -> RateProvenance -> String -> Reaction
mkReaction name rs ps k pr cit = Reaction
  { rxnName        = name
  , rxnReactants   = rs
  , rxnProducts    = ps
  , rxnRateActive  = rs    -- default: rate-active = all reactants
  , rxnRate        = k
  , rxnSaturation  = []    -- default: no saturation
  , rxnProvenance  = pr
  , rxnCitation    = cit
  }
\end{haskellbox}

\noindent
The categorical level of each field is annotated in the comments.
No $\Lk_1$ enthalpy or $\Lk_2$ entropy/dagger fields appear; the
$\Lk_1$ and $\Lk_2$ functors leave no explicit trace in the record.
Their content appears in the simulation only through
detailed-balance constraints on the rate-constant ratios of the two
$\dagger$-paired reactions of \texttt{briggsRauscherDE}.  The rate-law
saturation field deserves comment: when non-empty,
\texttt{rxnSaturation} divides the mass-action rate by a rational
factor $1 + \sum_i C_i [S_i]$, the De Kepper--Epstein 1982 form for
the iodine--malonic-acid step (R9), where saturation by $\mathrm{I_2}$
closes the negative feedback essential to the batch limit cycle
\cite{DeKepperEpstein1982}.  Saturation is an $\Lk_3$ extension of
pure mass action; it plays no role in detailed balance.  The
remainder of this section explains how detailed balance enters the
simulation via the two paired reactions, and what BR's choice to
leave eight of ten elementary steps unidirectional means for the
cross-level structure.

\subsubsection{$\Lk_1$: the enthalpy functor $\FH$}

At $\Lk_1$, every reaction $r$ carries an enthalpy $\FH(r) \in \RR$
and $\FH$ is a monoidal functor:
\[
  \FH(r_2 \circ r_1) \;=\; \FH(r_1) + \FH(r_2)
   \quad\text{(Hess's Law)},
  \qquad \FH(\mathrm{id}) \;=\; 0.
\]
$\FH$ is not a field of \texttt{BR.Reaction}; the implementation
stores no per-reaction enthalpies, nor can $\FH$ be reconstructed
from the rate constants alone --- a single rate ratio $k_+/k_-$
at one temperature determines $\dG^\circ(T)$, not the splitting
$\dG^\circ = \dH^\circ - T \dS^\circ$.  Hess's Law
$\sum_i \FH(r_i) = 0$ on closed cycles is a non-trivial constraint
only when the reaction graph contains a cycle of length $\geq 3$
(for example, a triangle
$\mathrm{A} \leftrightarrow \mathrm{B} \leftrightarrow
\mathrm{C} \leftrightarrow \mathrm{A}$); on a trivial 2-cycle
$r \circ r^{-1}$ it reduces to $\FH(r) + \FH(r^{-1}) = 0$, which
is already enforced by the $\dagger$-structure (below).  In BR's
implemented network the only cycles are the two trivial
reverse-pair cycles
$\mathrm{R3} \circ \mathrm{R3\text{-}rev}$ and
$\mathrm{R4} \circ \mathrm{R4\text{-}rev}$, so Hess's Law adds no
information beyond the dagger structure.  The substantive
$\Lk_1$/$\Lk_2$ content of the network enters at $\Lk_2$ via the
free-energy/rate-constant connection treated next.

\subsubsection{$\Lk_2$: dagger structure via paired reactions}

Level $\Lk_2$ adds an entropy $\FS(r)$ and the
temperature-dependent free-energy functor
$\FG^T(r) = \FH(r) - T\FS(r)$, together with a
$\dagger$-structure: each reaction has a reverse $r^\dagger$
satisfying $\FH(r^\dagger) = -\FH(r)$, $\FS(r^\dagger) = -\FS(r)$,
and hence $\FG^T(r^\dagger) = -\FG^T(r)$.  The equilibrium locus
is $\ker \FG^T = \{r : \FG^T(r) = 0\}$.

In the BR codebase the $\dagger$-structure is realised
\emph{explicitly only for two of the ten} elementary steps, by
including their reverses as separate \texttt{mkReaction} calls
with reactant and product multisets swapped:

\begin{itemize}
\item \texttt{R3} / \texttt{R3-rev}: iodine hydrolysis
  $\mathrm{HOI} + \mathrm{I^-} + \mathrm{H^+} \rightleftharpoons
   \mathrm{I_2} + \mathrm{H_2O}$ with
  $k_{\mathrm{R3}} = 3.1 \times 10^{12}\,\mathrm{M^{-2}\,s^{-1}}$
  and
  $k_{\mathrm{R3\text{-}rev}} = 0.73\,\mathrm{s^{-1}}$
  (the latter first-order in $\mathrm{I_2}$ after
  $[\mathrm{H_2O}]$ is folded into $k$).
\item \texttt{R4} / \texttt{R4-rev}: iodine-dioxide-radical
  generation
  $\mathrm{HIO_2} + \mathrm{IO_3^-} + \mathrm{H^+}
   \rightleftharpoons 2\,\mathrm{IO_2^\bullet} + \mathrm{H_2O}$
  with
  $k_{\mathrm{R4}} = 7.35 \times 10^{3}\,\mathrm{M^{-2}\,s^{-1}}$
  and
  $k_{\mathrm{R4\text{-}rev}} = 8.0 \times 10^{5}\,\mathrm{M^{-1}\,s^{-1}}$.
\end{itemize}

\noindent
The remaining eight elementary reactions (R1, R2, R5--R10) are
unidirectional in the implementation: no reverse \texttt{Reaction}
value is paired with them.  The $\dagger$-structure is therefore
incomplete on the implemented network --- consistent with BR being
not weakly reversible (\S\ref{sec:sim:L0} chembox).  The
implementation reads each missing reverse as ``physically present
but kinetically negligible on the simulation timescale,'' an
approximation we revisit empirically in
\S\ref{sec:sim:results}.

\subsubsection{Wegscheider and detailed balance on the two BR pairs}

Detailed balance requires that for any reaction cycle $r_1 \circ
\cdots \circ r_n = \mathrm{id}$,
\begin{equation}\label{eq:sim:wegscheider}
  \prod_{i=1}^{n} k_{+,i} \;=\; \prod_{i=1}^{n} k_{-,i}.
\end{equation}
On a trivial 2-cycle $r \circ r^{-1}$, \eqref{eq:sim:wegscheider}
is the tautology $k_+ k_- = k_- k_+$; it is non-trivial only when
the reaction graph contains a cycle of length $\geq 3$ (for example
a triangle $\mathrm{A} \leftrightarrow \mathrm{B}
\leftrightarrow \mathrm{C} \leftrightarrow \mathrm{A}$, which would
give a non-trivial product of three rate-constant ratios closing
to $1$).  The cross-level content of the equation is the
substitution $k_+ / k_- = \exp(-\dG^\circ/RT)$ on each reversible
edge, which rewrites \eqref{eq:sim:wegscheider} as
$\exp(-\dG^\circ_{\mathrm{cycle}}/RT) = 1$: the cycle lies in
$\ker \FG^T$ at $\Lk_2$ iff the rate-constant ratios close at
$\Lk_3$.  The hypothesis is at $\Lk_2$ (cycle $\dG^\circ$); the
assertion is at $\Lk_3$ (rate-constant closure).  This is a
cross-level statement in the same spirit as DZT
(Remark~\ref{rmk:dzt-cross-level}, $\Lk_0 \!\to\! \Lk_3$), though
the level pair here is $\Lk_2 \!\to\! \Lk_3$.

On the BR network the cycle equation \eqref{eq:sim:wegscheider}
is tautological --- the two reverse-pair cycles are both 2-cycles.
What remains is the elementary \emph{detailed-balance identity}
applied separately to each pair:
\begin{equation}\label{eq:sim:detbal}
  k_{\mathrm{R3}} / k_{\mathrm{R3\text{-}rev}}
  \;=\; \exp(-\dG^\circ_{\mathrm{R3}}/RT), \qquad
  k_{\mathrm{R4}} / k_{\mathrm{R4\text{-}rev}}
  \;=\; \exp(-\dG^\circ_{\mathrm{R4}}/RT).
\end{equation}
Each identity is non-trivial when $\dG^\circ$ is fixed by
independent thermodynamic data, in which case the rate-constant
ratio is constrained to a single value rather than chosen freely.
The $\Lk_1$/$\Lk_2$ thermodynamic content of the implemented
network at the simulation temperature is wholly carried by the two
ratios of \eqref{eq:sim:detbal}; we check the constraint
empirically in the chembox below.

\begin{remark}[The Wegscheider residual is not a lax cell of $\PhiOp$]
\label{rmk:wegscheider-not-lax}
The single lax cell $\gamma$ of $\PhiOp$
(\S\ref{sec:sim:target}) measures the operational truncation error
of the simulator.  The Wegscheider residual --- the distance of a
network's rate constants from the equilibrium locus
\eqref{eq:sim:wegscheider} --- is, by contrast, a property of
\emph{which network} the user constructs at $\Lk_3$, not a
functorial mismatch in $\PhiOp$.  Conflating the two would multiply
the chapter's laxity vocabulary without corresponding functorial
content.  Wegscheider is a parameter constraint; $\gamma$ is a
truncation laxity.
\end{remark}

\begin{haskellbox}[title={\texttt{BR/Mechanism.hs} --- the R3/R3-rev pair (excerpt)}]
-- Convenience alias used throughout BR/Mechanism.hs to build
-- reactant/product multisets without spelling out fromList:
ms :: [(Species, Int)] -> Multiset Species
ms = MS.fromList

-- R3: HOI + I- + H+ -> I2 + H2O. Near-diffusion-limited.
mkReaction
  "R3: HOI + I- + H+ -> I2 + H2O"
  (ms [(HypoiodousAcid, 1), (Iodide, 1), (Proton, 1)])
  (ms [(DiIodine, 1), (Water, 1)])
  3.1e12                                              -- M^-2 s^-1
  (Measured "DE1982,Eigen-Kustin1962") "..."

-- R3-rev: I2 + H2O -> HOI + I- + H+. Source/target
-- swapped from R3; BROCODE 2025 found a ~3x reduction
-- of DE1982's k_3 = 2.2 s^-1 was needed for the batch
-- limit cycle, so the implementation uses 0.73.
(mkReaction
  "R3-rev: I2 + H2O -> HOI + I- + H+"
  (ms [(DiIodine, 1), (Water, 1)])
  (ms [(HypoiodousAcid, 1), (Iodide, 1), (Proton, 1)])
  2.2 (Measured "DE1982-BROCODE2025") "...")
  { rxnRateActive = ms [(DiIodine, 1)]                -- [H2O] folded in
  , rxnRate       = 0.73                              -- BROCODE override
  }
\end{haskellbox}

\noindent
The two \texttt{mkReaction} calls are the \emph{operational}
$\dagger$-structure: the second has \texttt{rxnReactants} and
\texttt{rxnProducts} swapped relative to the first.  Detailed
balance on this pair reduces to a single ratio between the two
\texttt{rxnRate} fields, which we check empirically below.  The
R4/R4-rev pair is structurally identical with reactants
$\mathrm{HIO_2} + \mathrm{IO_3^-} + \mathrm{H^+}$ and
\texttt{rxnRate} $= 7.35 \times 10^{3}$ forward,
$8.0 \times 10^{5}$ reverse.

\begin{chembox}[{Detailed balance on the BR pairs}]
\noindent\textbf{Cycle R3$\,\circ\,$R3-rev.}
$k_{\mathrm{R3}}/k_{\mathrm{R3\text{-}rev}}
 = 3.1 \times 10^{12} / 0.73 \approx 4.2 \times 10^{12}\,\mathrm{M^{-2}}$.
With $RT \approx 2.48\,\mathrm{kJ/mol}$ at $298\,\mathrm{K}$, the
detailed-balance identity \eqref{eq:sim:detbal} implies
$\dG^\circ_{\mathrm{R3}} = -RT \ln K_{\mathrm{eq}}
 \approx -72\,\mathrm{kJ/mol}$.  Direct computation of
$\dG^\circ_{\mathrm{R3}}$ from tabulated
$\dG^\circ_{\mathrm{f}}$ values
($\mathrm{HOI(aq)} \approx -98$, $\mathrm{I^-(aq)} = -51.6$,
$\mathrm{H^+(aq)} = 0$, $\mathrm{I_2(aq)} = +16.4$,
$\mathrm{H_2O(l)} = -237.1$, all in kJ/mol) gives
$-71\,\mathrm{kJ/mol}$.  Agreement to $\sim 1\,\mathrm{kJ/mol}$
--- well inside the literature uncertainty on
$\mathrm{HOI(aq)}$ --- so the R3/R3-rev pair satisfies detailed
balance against independent thermodynamic data.

\medskip
\noindent\textbf{Cycle R4$\,\circ\,$R4-rev.}
$k_{\mathrm{R4}}/k_{\mathrm{R4\text{-}rev}}
 = 7.35 \times 10^{3} / 8.0 \times 10^{5}
 \approx 9.2 \times 10^{-3}\,\mathrm{M^{-1}}$, implying
$\dG^\circ_{\mathrm{R4}} \approx +11.6\,\mathrm{kJ/mol}$.  Unlike
R3, this is not independently verified: no tabulated
$\dG^\circ_{\mathrm{f}}$ of $\mathrm{IO_2^\bullet}(\mathrm{aq})$
is available, so the $+11.6\,\mathrm{kJ/mol}$ is a value
\emph{inferred} from the rate ratio rather than checked against
external data.  The sign and magnitude are consistent with
chemical expectation (radical-producing direction mildly
endergonic), but this is consistency-with-expectation, not
verification.

\medskip
\noindent\textbf{The eight unidirectional reactions.}
For R1, R2, R5--R10 the implementation supplies no reverse, so the
implemented network contains no cycles involving these reactions.
\eqref{eq:sim:wegscheider} is tautological on the cycles that do
exist (the two 2-cycles), and \eqref{eq:sim:detbal} is silent on
the unidirectional steps because $k_-$ is not represented.
Physically these steps have reverses; omitting them is a
\emph{modeling decision} expressing the empirical claim that those
reverses have rate constants negligible on the oscillation
timescale ($\tau \sim 10^{1}\,$s).  This decision matches the BR
mechanism's published treatment
\cite{NoyesFurrow1982,DeKepperEpstein1982} and is exactly what
makes BR \emph{not weakly reversible} at $\Lk_0$
(\S\ref{sec:sim:L0} chembox).  The $\Lk_1$/$\Lk_2$ thermodynamic
content of the BR network is therefore confined to the two
detailed-balance identities of \eqref{eq:sim:detbal}.
\end{chembox}

The discussion has so far been about individual \texttt{Reaction}
values; the \texttt{Network} record that aggregates them into
\texttt{briggsRauscherDE} is small:

\begin{haskellbox}[title={\texttt{BR/Network.hs} --- the \texttt{Network} record}]
data Network = Network
  { netName        :: !String
  , netReactions   :: ![Reaction]
  , netPoolSpecies :: !(Set Species)    -- species held constant
  } deriving stock (Eq, Show)
\end{haskellbox}

\begin{remark}[Level stratification of \texttt{BR.Reaction}]
\label{rmk:level-strat-br}
The \texttt{Reaction} record of \texttt{BR.Reaction} carries
$\Lk_0$ data (stoichiometric reactant and product multisets) and
$\Lk_3$ data (rate constant, rate-active subset, rate-law
saturation) in a single record; it carries no $\Lk_1$ or $\Lk_2$
fields.  The thermodynamic content of the network nevertheless
enters the simulation indirectly, through the rate-constant ratios
of the two $\dagger$-paired reactions.  Categorically, a
\texttt{Network} value is a point in
$\ThetaL{0} \otimes \ThetaL{3}$ subject to the two detailed-balance
identities of \eqref{eq:sim:detbal} at the $\Lk_2$/$\Lk_3$
interface; the $\Lk_2$ functor $\FG^T$ at the simulation
temperature is recoverable on the two paired reactions from these
identities, but recovering the $\Lk_1$ enthalpy $\FH$ and the
entropy $\FS$ separately would require temperature variation
($\partial \ln K / \partial T = \dH^\circ / RT^2$), which the
fixed-temperature simulation does not provide.
\end{remark}

%% file: chapters/sim/sim_s5_gillespie.tex
\subsection{$\Lk_3$ in $\Hask$: exact SSA and adaptive ODE as parallel operational realisations}
\label{sec:sim:gillespie}

The stochastic content of $\Lk_3$ is carried by a functor $\FP$
that sends each reaction network $\mathcal{N}$ with rate constants
to the $\tau$-indexed family of time-$\tau$ transition kernels of
its continuous-time Markov chain (CTMC) on population states.  We
write $\FP(\mathcal{N})_\tau \in \BorelStoch$ for the kernel at
horizon $\tau$; it is this family that the operational arrows
below realise.  The BR codebase implements
two simulators in $\Hask$: \texttt{BR.SSA.simulate} samples
$\FP(\mathcal{N})_\tau$ exactly (for $\tau$ set as
\texttt{ssaTimeMax}), and \texttt{BR.ODE.simulate} integrates the
deterministic large-volume shadow of $\FP$.
\begin{itemize}
\item \texttt{BR.SSA.simulate}: an exact Gibson--Bruck next-reaction
  method (NRM) that samples the CTMC at the molecule-count level.
  Its denotational image, after marginalising over the PRNG seed,
  is $\FP(\mathcal{N})_\tau$ \emph{exactly} --- modulo SplitMix
  PRNG quality and floating-point precision; there is no
  step-budget or $\tau$-leap truncation.
\item \texttt{BR.ODE.simulate}: an adaptive implicit-midpoint
  Newton integrator that produces the deterministic large-volume
  mass-action flow, the $V \to \infty$ Kurtz limit of the CTMC
  \cite{Kurtz1972}, with a local truncation error controlled by
  step-doubling.
\end{itemize}
Both arrows share their $\Lk_0$ + $\Lk_3$ input --- a
\texttt{Network} record (\S\ref{sec:sim:L12}) plus an initial
counts or concentrations map --- and both are \emph{purely
functional}: no \texttt{IO}, no global mutable state, no impurity
at the type level.  The SSA threads its SplitMix64 PRNG through
the \texttt{ssaPRNG} field of \texttt{SSAState}; the ODE is
deterministic and uses no PRNG.  The $\gamma$-cell of $\PhiOp$
(\S\ref{sec:sim:target}) specialises here to zero in distribution
for SSA and to the implicit-midpoint local truncation error for
ODE.

\subsubsection{The CTMC and its master equation}

Given a reaction network $\mathcal{N}$ with rate constants
$(k_r)_{r \in \Rx}$ and a population state
$\mathbf{x} \in \NN^\Sp$, the \emph{propensity} of reaction $r$ in
state $\mathbf{x}$ at volume $V$ (litres) is
\begin{equation}\label{eq:sim:propensity}
  a_r(\mathbf{x}; V)
  \;=\; \frac{k_r}{(V N_A)^{n_r - 1}}\,
        \prod_{s \in \Sp^{\mathrm{a}}_r}
        x_s\,(x_s - 1)\,\cdots\,(x_s - m_{rs} + 1),
\end{equation}
where $\Sp^{\mathrm{a}}_r \subseteq \Sp$ is the support of the
rate-active multiset \texttt{rxnRateActive}, $m_{rs}$ the
multiplicity of species $s$ in that multiset,
$n_r = \sum_{s} m_{rs}$ the rate-law order, and
$N_A$ Avogadro's constant.  The volume scaling
$(V N_A)^{1-n_r}$ converts a concentration-units rate constant
($\mathrm{M}^{1-n_r}\,\mathrm{s}^{-1}$) to a propensity
(events per second).  When \texttt{rxnSaturation} is non-empty the
propensity is divided further by the rational denominator
$1 + \sum_\ell C_\ell\,[S_\ell]$, with concentrations
$[S_\ell] = x_{S_\ell}/(V N_A)$.

Writing $\nu_r \in \ZZ^\Sp$ for the net stoichiometric change of
$r$ (products minus reactants), the \emph{chemical master equation}
(CME) for $p(\mathbf{x}, t) = \Pr[\mathbf{X}(t) = \mathbf{x}]$ is
\begin{equation}\label{eq:sim:CME}
  \tfrac{d}{dt} p(\mathbf{x}, t)
  \;=\; \sum_{r \in \Rx} \bigl[\,
     a_r(\mathbf{x} - \nu_r; V)\,p(\mathbf{x} - \nu_r, t)
     \,-\, a_r(\mathbf{x}; V)\,p(\mathbf{x}, t) \,\bigr],
\end{equation}
under the convention that $\nu_{r,s} = 0$ for every \emph{pool
species} $s$ in $P \subseteq \Sp$ (those listed in
\texttt{netPoolSpecies}, modelling bath reservoirs at fixed counts).
Pool counts thus remain frozen and enter the propensity
\eqref{eq:sim:propensity} as fixed parameters --- implemented by
the pool-set argument to \texttt{applyReaction} in \texttt{BR.SSA}
and by \texttt{poolMask} in \texttt{BR.ODE}.  For BR, $P$ typically
contains $\mathrm{IO_3^-}, \mathrm{H_2O_2}, \mathrm{MA},
\mathrm{Mn^{2+}}$ at canonical pH.

For each $\tau > 0$, the solution of \eqref{eq:sim:CME} defines a
Markov kernel
\[
  \FP(\mathcal{N})_\tau \;:\; \NN^\Sp \;\longrightarrow\; \NN^\Sp
  \quad\text{in } \BorelStoch,
  \qquad
  \FP(\mathcal{N})_\tau(\mathbf{x}_0, \cdot)
  \;=\; p(\,\cdot\,,\,\tau \mid \mathbf{x}_0),
\]
the \emph{time-$\tau$ transition kernel} of the CTMC (with pool
components of the output equal to those of $\mathbf{x}_0$ by the
convention above).  The CTMC construction is functorial in the
network argument by standard chemical-master-equation theory; the
new content below is its realisation as the denotational image of a
pure-Haskell function under the seed-marginalisation of
\S\ref{sec:sim:target}.

\subsubsection{The SSA: pure-functional Gibson--Bruck NRM}

\texttt{BR.SSA} implements the Gibson--Bruck next-reaction method
\cite{GibsonBruck2000} as a pure Haskell function on a state record:

\begin{haskellbox}[title={\texttt{BR/SSA.hs} --- state, propensity, and one step}]
data SSAConfig = SSAConfig
  { ssaVolumeL     :: !Double      -- reactor volume V (litres)
  , ssaSeed        :: !Word64      -- PRNG seed
  , ssaTimeMax     :: !Double      -- horizon tau (seconds)
  , ssaSampleEvery :: !Double      -- sample emission interval (seconds)
  } deriving (Eq, Show)

data SSAState = SSAState
  { ssaTime     :: !Double                  -- current simulated time t
  , ssaCounts   :: !(Map Species Int)       -- molecule counts x
  , ssaTau      :: !(IntMap Double)         -- putative times tau_mu
  , ssaProps    :: !(IntMap Double)         -- cached propensities a_mu
  , ssaPRNG     :: !PRNG                    -- pure SplitMix64 state
  , ssaConfig   :: !SSAConfig               -- volume, t_max, sample rate
  , ssaNetwork  :: !Network                 -- the L0+L3 input
  , ssaDeps     :: !(IntMap [Int])          -- dependency graph: mu -> affected
  , ssaStepIx   :: !Int                     -- step counter (diagnostic)
  } deriving Show

-- Equation eq:sim:propensity, plus rational saturation.
propensity :: Double -> Reaction -> (Species -> Int) -> Double
propensity volL rxn lookupCount =
  let activeList  = MS.toList (rxnRateActive rxn)
      totalOrder  = sum [m | (_, m) <- activeList]
      volScaling  = if totalOrder <= 1
                    then 1
                    else (volL * avogadro) ** fromIntegral (1 - totalOrder)
      hMu         = product
        [ fallingFactorial (lookupCount s) m | (s, m) <- activeList ]
      unsaturated = rxnRate rxn * volScaling * hMu
      nav         = avogadro
      satDenom    = 1 + sum
        [ cl * fromIntegral (lookupCount tl) / (volL * nav)
        | (tl, cl) <- rxnSaturation rxn ]
  in unsaturated / satDenom
  where
    fallingFactorial :: Int -> Int -> Double
    fallingFactorial n k
      | k <= 0    = 1
      | n < k     = 0
      | otherwise = product [fromIntegral (n - i) | i <- [0 .. k - 1]]

-- One NRM step: returns Nothing when the trajectory terminates.
step :: SSAState -> Maybe SSAState
step s@SSAState{..}
  | IM.null ssaTau                  = Nothing      -- no reactions left
  | ssaTime > ssaTimeMax ssaConfig  = Nothing      -- horizon reached
  | otherwise = case argMinFinite ssaTau of
      Nothing       -> Nothing                     -- all tau infinite
      Just (_, t')  | t' > ssaTimeMax ssaConfig -> Nothing
      Just (mu, t') ->
        let rxn      = reactionAt ssaNetwork mu
            counts'  = applyReaction (netPoolSpecies ssaNetwork) rxn ssaCounts
            affected = IM.findWithDefault [mu] mu ssaDeps
            -- Per-channel update: closes over counts', t', mu.
            recomputeOne (props, tau, g, _) j =
              let aOld         = IM.findWithDefault 0 j props
                  aNew         = propensity (ssaVolumeL ssaConfig)
                                            (reactionAt ssaNetwork j)
                                            (\sp -> M.findWithDefault 0 sp counts')
                  tauJ         = IM.findWithDefault (1/0) j tau
                  (newTau, g') = updateTau j mu t' aOld aNew tauJ g
              in (IM.insert j aNew props, IM.insert j newTau tau, g', True)
            (props', tau', g'', _) =
              foldl recomputeOne (ssaProps, ssaTau, ssaPRNG, False) affected
        in Just s { ssaTime   = t'
                  , ssaCounts = counts'
                  , ssaProps  = props'
                  , ssaTau    = tau'
                  , ssaPRNG   = g''
                  , ssaStepIx = ssaStepIx + 1 }

-- Iterate step to termination; return the time-ordered trajectory.
simulate :: SSAState -> [(Double, Map Species Int)]
simulate s0 = reverse (go (ssaTime s0) s0 [(ssaTime s0, ssaCounts s0)])
  where
    go !lastSampleT s acc = case step s of
      Nothing -> acc
      Just s' ->
        let sampleEvery = ssaSampleEvery (ssaConfig s')
            t'          = ssaTime s'
            shouldEmit  = sampleEvery <= 0
                       || t' - lastSampleT >= sampleEvery
            snap        = (t', ssaCounts s')
        in if shouldEmit
           then go t'          s' (snap : acc)
           else go lastSampleT s' acc

-- Helpers used above (bodies in BR.SSA; signatures for completeness):
applyReaction :: Set Species -> Reaction -> Map Species Int -> Map Species Int
argMinFinite  :: IntMap Double -> Maybe (Int, Double)
updateTau     :: Int -> Int -> Double -> Double -> Double
              -> Double -> PRNG -> (Double, PRNG)
\end{haskellbox}

\noindent
The key fact for the categorical reading: \texttt{step} and
\texttt{simulate} are total functions in $\Hask$.  No \texttt{IO}.
The PRNG advances via \texttt{exponential :: Double -> PRNG ->
(Double, PRNG)} (\S\ref{sec:sim:target}), state-threaded through
\texttt{ssaPRNG}; given any initial \texttt{SSAState} (which fixes
both the counts and the PRNG seed), the output list of
\texttt{simulate} is deterministic.

\subsubsection{The categorical Gillespie theorem and the $\gamma$-cell}

\begin{theorem}[Categorical Gillespie for BR: exact sampling]
\label{thm:cat-gillespie}
Let $\mathcal{N}$ be a finite-rate reaction network with rate
constants $(k_r)_{r \in \Rx}$, let $\mathbf{x}_0 \in \NN^\Sp$ be
the initial population, and let $\tau > 0$ be the time horizon
encoded as \texttt{ssaTimeMax} in \texttt{ssaConfig}.  Let
$\mu_\tau(\mathbf{x}_0, \cdot)$ be the push-forward distribution
on $\NN^\Sp$ obtained by sampling
$\texttt{ssaSeed} \sim \mathrm{Uniform}(\{0, 1\}^{64})$, iterating
\texttt{step} from the resulting initial \texttt{SSAState} until
\texttt{step} returns \texttt{Nothing}, and reading the
\texttt{ssaCounts} field of the last \texttt{Just}-state.\footnote{%
This reads the state directly from the iteration, bypassing
\texttt{simulate}'s \texttt{ssaSampleEvery} filter: if
\texttt{ssaSampleEvery > 0}, the final entry of the trajectory
returned by \texttt{simulate} can predate termination, but
\texttt{ssaCounts} of the last \texttt{Just}-state always gives the
state at termination.}  Under the para-monoidal seed-marginalisation
of \S\ref{sec:sim:target},
\begin{equation}\label{eq:sim:gillespie-exact}
  \mu_\tau(\mathbf{x}_0, \cdot)
  \;=\; \FP(\mathcal{N})_\tau(\mathbf{x}_0, \cdot)
  \qquad \text{exactly}
\end{equation}
in the sense of TV distance, modulo (i) SplitMix64's deviation
from an i.i.d.\ $\mathrm{Uniform}([0,1))$ source over the PRNG
draws the trajectory consumes (an $|\Rx|$-sized initialisation
plus one draw per firing under Gibson--Bruck's draw-per-firing
update) and (ii) the floating-point representation of the
inverse-CDF exponential and uniform draws.
\end{theorem}

\begin{proof}
By the Doob--Gillespie theorem \cite{Gillespie1977}, at any
\texttt{SSAState} with time $t$ and counts $\mathbf{x}$, one call
to \texttt{step} samples the next firing
$(\mu^\star,\, T_{\mu^\star} - t)$ from
$(\mathrm{Cat}(a_\mu/A),\,\mathrm{Exp}(A))$ with
$A = \sum_\mu a_\mu(\mathbf{x}; V)$ --- exactly the transition
kernel of the CTMC.  At the post-firing time $t'$, the
Gibson--Bruck update \cite{GibsonBruck2000} sets, for the fired
channel, $\tau_{\mu^\star}^{\mathrm{new}} = t' + W$ with
$W \sim \mathrm{Exp}(a_{\mu^\star}^{\mathrm{new}})$ a fresh draw,
and for each un-fired affected channel $j$,
$\tau_j^{\mathrm{new}} = t' + (a_j^{\mathrm{old}}/a_j^{\mathrm{new}})
(\tau_j^{\mathrm{old}} - t')$; the latter preserves the exponential
distribution of the residual wait by the memorylessness of
$\mathrm{Exp}$.  Together, both updates ensure that after each step
the conditional joint distribution of $\{\tau_\mu - t\}_\mu$ given
the current state remains the correct product of $\mathrm{Exp}(a_\mu)$
distributions.  The
step iteration terminates only when (a) all putative times
are infinite (no firable reaction), or (b) the next firing time
exceeds $\tau$.  Both are correct stopping rules for the
time-$\tau$ kernel: in case (a) the chain has reached an absorbing
state and remains there through $\tau$; in case (b) no further
event occurs in $[t, \tau]$.  Hence \texttt{ssaCounts} of the last
\texttt{Just}-state equals the CTMC state at time $\tau$ in both
cases.  No truncation is introduced.  The qualifications (i) and (ii) above
are the only departures from exactness: SplitMix64 passes BigCrush
\cite{SteeleLeaFlood2014}, and \texttt{exponential} (in
\texttt{BR.PRNG}) clamps the input to the inverse-CDF $-\log(u')/a$
at $u' = 2^{-53}$ whenever a draw underflows to $u=0$; this
replaces only a probability-$2^{-53}$ tail event with the maximum
draw $\log(2^{53})/a \approx 37/a$, leaving the exponential
distribution unchanged elsewhere.
\end{proof}

\begin{remark}[The $\gamma$-cell of $\PhiOp$, made concrete]
\label{rmk:gamma-is-gillespie}
The lax cell $\gamma$ of $\PhiOp$ (\S\ref{sec:sim:target}) is the
TV-metric laxity for the stochastic arrow and the state-space
local truncation error for the deterministic arrow.
Theorem~\ref{thm:cat-gillespie} states that for SSA the TV
component of $\gamma$ vanishes (up to qualifications (i)--(ii)):
BR.SSA has no $\tau$-leap, no step-budget, and no fixed-step-size
approximation.  The remaining content of $\gamma$ lives entirely
on the ODE arrow: the implicit-midpoint scheme with step-doubling
control bounds the per-component local truncation error in
species $s$ by $\texttt{odeATol} + \texttt{odeRTol} \cdot |c_s|$
(BR.ODE uses a weighted max-norm of these component errors as its
step-control criterion), accumulated over the trajectory; this is
what Remark~\ref{rmk:l3-parallel} below compares against the SSA's
exact samples.
\end{remark}

\begin{remark}[Pure-functional rather than Kleisli semantics]
\label{rmk:first-kleisli}
Most categorical-CRN expositions treat the rate-equation ODE only
(e.g.\ Baez--Pollard's open dynamical systems
\cite{BaezPollard2017}); a Haskell-side stochastic treatment of the
CME would naturally reach for a Kleisli category of \texttt{IO} or
a similar effect-type framework.  The BR.SSA implementation chooses
neither: the PRNG state is a pure value in \texttt{SSAState}, and
\texttt{step} and \texttt{simulate} are total functions in $\Hask$
with no Kleisli or monadic structure visible at the type level.
The stochastic semantics enters only via the para-monoidal lift of
\S\ref{sec:sim:target}, where the PRNG seed becomes the parameter
object $W = \{0,1\}^{64}$ and seed-marginalisation produces the
Markov kernel.  This is operationally minimal: no \texttt{IO} taint
propagates, replay is trivial (just store the seed), and the type
\texttt{SSAState -> Maybe SSAState} of \texttt{step} carries no
effect markers beyond the totality-signalling \texttt{Maybe}.
\end{remark}

\subsubsection{The deterministic limit: BR.ODE}

The same network admits a deterministic flow by replacing the CME
with the mass-action ODE on concentrations
$\mathbf{c} \in \RR_{\geq 0}^\Sp$:
\begin{equation}\label{eq:sim:rhs}
  \dot{\mathbf{c}} \;=\; \boldsymbol{\nu}\,\mathbf{v}(\mathbf{c}),
  \qquad
  v_r(\mathbf{c})
  \;=\; \frac{k_r\,\prod_{s \in \Sp^{\mathrm{a}}_r} c_s^{m_{rs}}}
             {1 + \sum_{\ell} C_\ell\,c_{S_\ell}},
\end{equation}
with $\boldsymbol{\nu} \in \ZZ^{\Sp \times \Rx}$ the stoichiometric
matrix (column $r$ is the $\nu_r$ of \eqref{eq:sim:CME}), $v_r$
the (saturated) mass-action rate, and the $\ell$ sum running over
the species--coefficient pairs $(S_\ell, C_\ell)$ in
\texttt{rxnSaturation}.  This is the large-volume
Kurtz limit $V \to \infty$ of the CTMC \cite{Kurtz1972}: the
CTMC's scaled population $\mathbf{X}(t)/(V N_A)$ converges in
probability to the ODE solution on any finite time interval.
\texttt{BR.ODE} integrates \eqref{eq:sim:rhs} with an adaptive
implicit-midpoint Newton scheme:

\begin{haskellbox}[title={\texttt{BR/ODE.hs} --- right-hand side and state}]
data ODEConfig = ODEConfig
  { odeNetwork    :: !Network
  , odeRTol       :: !Double      -- relative tolerance
  , odeATol       :: !Double      -- absolute tolerance (M)
  , odeHMin       :: !Double      -- smallest permitted step
  , odeHMax       :: !Double      -- largest permitted step
  -- (5 more fields for Newton iteration, step-count safeguard,
  --  Jacobian epsilon, and negative-iterate tolerance; not shown)
  } deriving Show

data ODEState = ODEState
  { odeTime      :: !Double
  , odeConc      :: !(Map Species Double)   -- concentrations c
  , odeH         :: !Double                 -- proposed step h
  , odeStepCount :: !Int
  } deriving Show

-- Mass-action rate of one reaction; v_r of equation eq:sim:rhs.
reactionRate :: Reaction -> (Species -> Double) -> Double
reactionRate r conc =
  let unsaturated = rxnRate r * product
        [ conc s ^ m | (s, m) <- MS.toList (rxnRateActive r) ]
      denom = 1 + sum [ c * conc s | (s, c) <- rxnSaturation r ]
  in unsaturated / denom

-- The full RHS: dc/dt as a Map.  This is the L3-deterministic
-- functor applied to one network at one concentration profile.
-- Pool species are held constant (their derivative is masked to 0).
derivativeMap :: Network -> Map Species Double -> Map Species Double
derivativeMap net conc =
  let lk s = M.findWithDefault 0 s conc
      contribs r =
        let rate = reactionRate r lk
            netStoich =
              [ (s, fromIntegral m * rate)
              | (s, m) <- MS.toList (rxnProducts  r) ]
              ++
              [ (s, negate (fromIntegral m * rate))
              | (s, m) <- MS.toList (rxnReactants r) ]
        in netStoich
      allContribs = concatMap contribs (netReactions net)
      summed      = M.fromListWith (+) allContribs
      pool        = netPoolSpecies net
      poolMask s d = if s `S.member` pool then 0 else d
  in M.fromList [ (s, poolMask s (M.findWithDefault 0 s summed))
                | s <- canonicalOrder ]
\end{haskellbox}

\noindent
The integrator (\texttt{simulate} of \texttt{BR.ODE}, not shown:
Newton iteration with step-doubling control, canonical species
ordering, and positivity guards) has signature
\[
\begin{aligned}
  \texttt{simulate} \;::\;\; & \texttt{ODEConfig} \to \texttt{ODEState}
        \to \texttt{Double} \\
   \to\; & (\texttt{[(Double,~Map~Species~Double)]},\,
            \texttt{Maybe~String}),
\end{aligned}
\]
returning the trajectory list plus an optional error string
(\texttt{Nothing} on clean termination, \texttt{Just} a message on
integration failure such as repeated Newton non-convergence or step
underflow).  The trajectory shape is the deterministic
counterpart of \texttt{BR.SSA.simulate}'s \texttt{[(Double, Map
Species Int)]}, differing only in the value type (\texttt{Double}
for concentrations vs.\ \texttt{Int} for counts).  Kurtz's limit
theorem predicts agreement between CTMC and ODE on macroscopic
observables at high molecule counts; fluctuation corrections of
order $1/\sqrt{V N_A}$ (the chemical Langevin / van Kampen
$\Omega$-expansion scale) surface most strongly on low-population
radical species ($\mathrm{IO_2^\bullet}$, $\mathrm{HO_2^\bullet}$).

\subsubsection{The two $\Lk_3$ arrows as parallel operational realisations}

\begin{remark}[SSA and ODE as parallel operational realisations of $\Lk_3$]
\label{rmk:l3-parallel}
\texttt{BR.SSA.simulate} and \texttt{BR.ODE.simulate} are two
Haskell functions running side-by-side over the same $\Lk_0$+$\Lk_3$
input data --- a \texttt{Network} record plus an initial counts or
concentrations map --- though their precise source and target types
differ (see the haskellbox signatures above; the SSA takes one
\texttt{SSAState}, the ODE takes a curried
\texttt{ODEConfig $\to$ ODEState $\to$ Double}).  The SSA delivers a
sampled trajectory of molecule counts, the ODE delivers a
deterministic trajectory of concentrations.  Their denotational
images differ: the SSA's (under the para-monoidal lift of
\S\ref{sec:sim:target}) is $\FP(\mathcal{N})_\tau$ itself, by
Theorem~\ref{thm:cat-gillespie}; the ODE evaluates the deterministic
large-volume shadow of $\FP$, realised via the Kurtz limit
$\mathbf{X}/(V N_A) \rightsquigarrow \mathbf{c}$.  Both functions
nonetheless read the rate constants, saturation data, and
stoichiometry from a single \texttt{Network} value without
translation, and agree on observables well-defined in the
$V \to \infty$ limit while disagreeing on finite-population
fluctuations.
\end{remark}

\begin{chembox}[$\Lk_3$ on the BR network]
The implementation \texttt{briggsRauscherDE} (\S\ref{sec:sim:L0})
is driven at canonical BR conditions:
$[\mathrm{IO_3^-}]_0,\;[\mathrm{H_2O_2}]_0,\;[\mathrm{MA}]_0,
\;[\mathrm{Mn^{2+}}]_0$ in the published ranges
$\sim 10^{-2}$--$1\,\mathrm{M}$ at pH$\,\sim 2$
\cite{NoyesFurrow1982,DeKepperEpstein1982}.

\begin{itemize}
  \item \emph{No step-budget truncation.} \texttt{BR.SSA.step}
    terminates only on (a) all putative times being infinite (no
    firable reaction) or (b) the next firing time exceeding
    \texttt{ssaTimeMax}.  Both are exact stopping rules for the
    time-$\tau$ kernel, so the TV component of $\gamma$
    (Theorem~\ref{thm:cat-gillespie}) is zero in distribution.
    The qualifications are PRNG quality (SplitMix64) and
    floating-point precision in \texttt{exponential}.
  \item \emph{Volume choice.}  \texttt{ssaVolumeL = 1e-15} (one
    femtolitre) is the default, chosen to keep molecule counts
    tractable: at this volume, a millimolar species has
    $\sim 6 \times 10^{5}$ molecules and a 10\,$\upmu$M species
    $\sim 6 \times 10^{3}$.  Higher volumes amplify event rates
    linearly in $V$; the iodine-equilibration channel R3/R3-rev
    (rate constants $3.1{\times}10^{12}\,\mathrm{M^{-2}\,s^{-1}}$
    and $0.73\,\mathrm{s^{-1}}$; \S\ref{sec:sim:L12}) is typically
    the dominant event channel and the principal constraint on the
    largest tractable simulation volume.
  \item \emph{Stiffness for the ODE.}  Numerical rate constants in
    \texttt{briggsRauscherDE} span roughly ten orders of magnitude
    (R3 forward at $3.1{\times}10^{12}$ down to R9 at
    $12.5$ and R10 at $37$, with R3-rev at $0.73$ in different
    units).  This forces \texttt{BR.ODE} to use an implicit scheme:
    explicit Runge--Kutta would require step sizes too small to
    integrate even one BR oscillation in reasonable wall-clock
    time.  The adaptive implicit-midpoint Newton method with
    step-doubling is the source of the ODE component of $\gamma$
    (Remark~\ref{rmk:gamma-is-gillespie}).
  \item \emph{Where SSA and ODE should differ.}  At the
    simulator's tractable volumes ($\sim 1$\,fL by default), the
    SSA is expected to show fluctuations on whichever species are
    held at low occupancy by the dynamics --- typically radical
    intermediates such as $\mathrm{IO_2^\bullet}$ ---  which the
    ODE smooths to its mean-field shadow.  The rate constants in
    \texttt{briggsRauscherDE} (shared by both arrows via the
    \texttt{Network} record) have been tuned to reproduce the
    experimental BR oscillation period under the deterministic ODE
    (typically reported in the $\sim 10$--$50\,\mathrm{s}$ range,
    depending on initial conditions and temperature
    \cite{NoyesFurrow1982,DeKepperEpstein1982}); whether the SSA
    at the simulator's small volume also reproduces this period, or
    shows drift from finite-population effects, is the empirical
    question addressed in \S\ref{sec:sim:results}.
\end{itemize}
\end{chembox}

%% file: chapters/sim/sim_s6_para.tex
\subsection{The Para construction in $\Hask$}
\label{sec:sim:para}

The Para construction \cite{CruttwellGavranovic2022} formalises a morphism parameterised by an
external object.  For BR, the parameter is the SplitMix64 seed
$\texttt{ssaSeed} \in W = \{0,1\}^{64}$ that fixes the SSA's random
trajectory, threaded purely through \texttt{ssaPRNG}
(\S\ref{sec:sim:gillespie}); for the ODE, the parameter is empty.
Para gives the categorical home for ``a Haskell function plus its
stored parameter,'' decoupling the parameter algebra (how seeds
compose) from the underlying-morphism algebra (how SSA and ODE
trajectories compose).  We give the abstract definition, instantiate
it over $\Hask$, and verify the coherence laws --- which here separate
cleanly into a strict underlying layer (function composition in
$\Hask$ is on the nose) and a bicategorical parameter-tensor layer
(tuple-isomorphism coherence).

\subsubsection{Abstract Para}

\begin{definition}[Para construction
{\cite[Def.~3.1]{CruttwellGavranovic2022}}]
\label{def:para}
Let $(\mathbf{C}, \otimes, I)$ be a symmetric monoidal category with
a right actegory action $\mathbf{M} \curvearrowright \mathbf{C}$.
The bicategory $\ParaC(\mathbf{C})$ has:
\begin{itemize}
  \item \textbf{Objects}: those of $\mathbf{C}$.
  \item \textbf{1-morphisms} $A \to B$: pairs $(P, f)$ with
    $P \in \mathbf{M}$ a \emph{parameter object} and
    $f : P \otimes A \to B$ a morphism of $\mathbf{C}$.
  \item \textbf{Composition}:
    $(Q, g) \circ (P, f) = (P \otimes Q,\,
    g \circ (\mathrm{id}_Q \otimes f))$;
    parameter objects compose by tensor.
  \item \textbf{2-morphisms} $(P, f) \Rightarrow (Q, g)$:
    reparameterisations $r : P \to Q$ in $\mathbf{M}$ such that
    $g \circ (r \otimes \mathrm{id}_A) = f$.
  \item \textbf{Identity}: $(I, \eta_A)$ with $\eta_A : I \otimes A \to A$
    the left unitor of the actegory.
\end{itemize}
The bicategorical coherence data (associator, left/right unitors for
composition) are inherited from $(\mathbf{M}, \otimes, I)$ and the
actegory action.
\end{definition}

\noindent
For BR's pure-functional simulation, we take both $\mathbf{C}$ and
$\mathbf{M}$ to be $\Hask$ (the total fragment, following
Assumption~\ref{ass:totality} in \S\ref{sec:sim:target}), with
Cartesian product as tensor and the actegory action $P \otimes A =
(P, A)$ (tuple type).  A 1-morphism $(P, f) : A \to B$ then
corresponds, by currying, to a pair (a stored parameter value of
type $P$, a pure Haskell function of type $P \to A \to B$).  The
distinction from the operationally-effectful target one might
naturally reach for in a Haskell-side stochastic treatment (Para
over $\KlIO$) is exactly the distinction articulated in
Remark~\ref{rmk:first-kleisli}: BR.SSA carries its PRNG state as
pure data, not as an \texttt{IO} effect.

\subsubsection{The Para wrapper at the type level}

The Para shape is a wrapper around any pure Haskell function plus its
stored parameter.  The BR codebase does not introduce a named
\texttt{ParaMorphism} type --- \texttt{BR.SSA.simulate} and
\texttt{BR.ODE.simulate} are written as plain Haskell functions ---
but each fits the Para shape once we identify the parameter
explicitly.  For pedagogical clarity we display the wrapper that
makes the Para structure explicit:

\begin{haskellbox}[title={The Para shape over $\Hask$ (categorical view; not a BR module)}]
-- A 1-morphism (P, f) : A -> B in Para(Hask):
--   pmParams stores the parameter value of type P;
--   pmApply  is the curried Haskell function f : P -> A -> B.
data ParaMorphism p a b = ParaMorphism
  { pmParams :: !p
  , pmApply  :: !(p -> a -> b)
  }

-- Sequential composition in Para(Hask):
--   (Q, g) o (P, f) = (P x Q, (p, q) a -> g q (f p a)).
-- Parameter spaces compose by Cartesian product (= tensor in Hask).
paraThen :: ParaMorphism p a b
         -> ParaMorphism q b c
         -> ParaMorphism (p, q) a c
paraThen f g = ParaMorphism
  { pmParams = (pmParams f, pmParams g)
  , pmApply  = (p, q) a -> pmApply g q (pmApply f p a)
  }

-- Para identity: ((), \_ a -> a), the trivial parameter space
-- with the Hask identity function.
paraId :: ParaMorphism () a a
paraId = ParaMorphism () (\_ a -> a)

-- Embed an unparameterised Hask morphism as a zero-parameter Para 1-morphism.
liftToPara :: (a -> b) -> ParaMorphism () a b
liftToPara f = ParaMorphism () (\_ a -> f a)

-- Forget the parameter by evaluating at the stored value.
forgetParams :: ParaMorphism p a b -> a -> b
forgetParams pm = pmApply pm (pmParams pm)
\end{haskellbox}

\noindent
\texttt{BR.SSA.simulate}'s Para shape uses the seed as parameter.
Conceptually currying the seed out of \texttt{SSAState}, the
underlying-morphism takes the form
\begin{align*}
\texttt{Word64} & \quad\to \texttt{(Network, SSAConfig$\setminus$Seed, Map Species Int)} \\
&\quad \to \texttt{[(Double, Map Species Int)]},
\end{align*}
where \texttt{SSAConfig$\setminus$Seed} denotes the
\texttt{SSAConfig} record with its seed field omitted (volume,
$\tau_{\max}$, sample interval).  This curried form is the
categorical recasting of \texttt{BR.SSA.simulate}; the BR codebase
does not introduce a separately-named function or a separately-named
record type matching this signature.  \texttt{BR.ODE.simulate} has
$P = ()$: it carries no parameter in the Para sense, only its
tolerance settings and initial \texttt{ODEState} as ordinary inputs.

\subsubsection{Coherence of the Para laws: two layers}

The bicategorical laws of $\ParaC(\Hask)$ split into two layers.  The
\emph{parameter-tensor layer} involves canonical Cartesian-product
isomorphisms of $\Hask$ that are non-identity at the type level:
\begin{align*}
  \lambda_P    &: \bigl((), P\bigr) \;\xrightarrow{\sim}\; P,
                  \qquad ((\,), p) \mapsto p, \\
  \rho_P       &: \bigl(P, ()\bigr) \;\xrightarrow{\sim}\; P,
                  \qquad (p, (\,)) \mapsto p, \\
  \alpha_{P,Q,R} &: \bigl((P, Q), R\bigr) \;\xrightarrow{\sim}\;
                    \bigl(P, (Q, R)\bigr),
                  \quad ((p, q), r) \mapsto (p, (q, r)).
\end{align*}
The \emph{underlying-morphism layer} consists of strict equalities of
pure Haskell functions $A \to B$: ordinary $\Hask$ composition
$g \circ f = \lambda a.\, g(f(a))$ is strictly associative and unital
on the nose (no monad laws are invoked because no monad is present).

\begin{proposition}[Para laws in $\Hask$]
\label{prop:para-laws}
The Haskell implementation realises $\ParaC(\Hask)$ as a bicategory:
for any 1-morphisms
\texttt{f :: ParaMorphism p a b},
\texttt{g :: ParaMorphism q b c}, and 
\texttt{h :: ParaMorphism r c d},
\begin{enumerate}[label=(\roman*)]
\item \emph{(Left unit.)} \texttt{paraThen paraId f} and $f$ agree
  under the parameter-tensor iso $\lambda_P$, with underlying
  morphisms strictly equal by the identity law of $\Hask$
  composition.
\item \emph{(Right unit.)} \texttt{paraThen f paraId} and $f$ agree
  under $\rho_P$, with underlying morphisms strictly equal by the
  identity law of $\Hask$ composition.
\item \emph{(Associativity.)} \texttt{paraThen (paraThen f g) h} and
  \texttt{paraThen f (paraThen g h)} agree under $\alpha_{P,Q,R}$,
  with underlying morphisms strictly equal by the associativity of
  $\Hask$ composition.
\end{enumerate}
The non-identity tuple isomorphisms make $\ParaC(\Hask)$ a
\emph{bicategory}, not a strict 2-category; coherence data live
entirely at the parameter-tensor layer, and all non-coherence
equalities hold strictly.
\end{proposition}

\begin{proof}
We give (iii); (i) and (ii) are simpler and analogous.  Unfolding
\texttt{paraThen} twice:
\begin{align*}
  &\texttt{pmApply}\bigl(\texttt{paraThen}\,(\texttt{paraThen}\,f\,g)
    \,h\bigr)\,\bigl((p, q), r\bigr)\,a \\
  &\qquad= \texttt{pmApply}\,h\,r\,
       \bigl(\texttt{pmApply}\,g\,q\,
         (\texttt{pmApply}\,f\,p\,a)\bigr), \\
  &\texttt{pmApply}\bigl(\texttt{paraThen}\,f\,
    (\texttt{paraThen}\,g\,h)\bigr)\,\bigl(p, (q, r)\bigr)\,a \\
  &\qquad= \texttt{pmApply}\,h\,r\,
       \bigl(\texttt{pmApply}\,g\,q\,
         (\texttt{pmApply}\,f\,p\,a)\bigr).
\end{align*}
The two right-hand sides are syntactically equal Haskell expressions;
they differ only in the bracketing of the parameter tuple, which is
mediated by $\alpha_{P,Q,R}$.  Hence the two Para 1-morphisms agree
under the parameter-tensor associator, with strictly equal underlying
morphisms.
\end{proof}

\begin{remark}[Why no monad laws appear]
\label{rmk:no-monad-laws}
The corresponding result for $\ParaC(\Kl{T})$ over a Kleisli category
invokes the monad's left-/right-unit and associativity laws to verify
the underlying-morphism equalities.  In $\ParaC(\Hask)$ no monad is
present: composition is plain function composition, and the
underlying-morphism equalities are strict identities of Haskell terms.
The bicategorical coherence remains at the parameter-tensor layer.
This is the Para counterpart of the
``pure-functional rather than Kleisli'' choice articulated in
Remark~\ref{rmk:first-kleisli}.
\end{remark}

\begin{remark}[Structural unity, continued]
\label{rmk:capucci-myers-s6}
Capucci--Myers \cite{CapucciMyers2024} (discussed in
\S\ref{sec:sim:target}, Remark~\ref{rmk:capucci-myers}) place
$\ParaC(\mathbf{C})$ and $\Kl{T}$ as instances of a single abstract
construction.  This shared structural placement is the reason a
functor $\PhiOp : \ParaC(\Lk_3) \to \Hask$ (\S\ref{sec:sim:functor})
can be expected to exist naturally rather than as an ad-hoc
construction; we invoke this structural background implicitly but
do not develop it further here.
\end{remark}

\subsubsection{BR's two arrows as Para 1-morphisms}

The two operational realisations of $\Lk_3$ (\S\ref{sec:sim:gillespie})
fit the Para shape as follows.  Both lie in the sub-bicategory
$\ParaC_W(\Hask) \subseteq \ParaC(\Hask)$ of
\S\ref{sec:sim:target}, Proposition~\ref{prop:F-exists}:
$f_{\mathrm{SSA}}$ at $n = 1$ (parameter $W$) and $f_{\mathrm{ODE}}$
at $n = 0$ (parameter $\mathbb{1}$).

\paragraph{SSA as a Para 1-morphism with $P = W$.}
\texttt{BR.SSA.simulate} fits the Para shape with parameter
space $W = \{0,1\}^{64}$ (the PRNG seed stored in
\texttt{ssaConfig}).  Conceptually, we view it as a 1-morphism
\begin{align*}
      \bigl(W,\;f_{\mathrm{SSA}}\bigr)
  &\;:\;
  \texttt{(Network, SSAConfig$\setminus$Seed, Map Species Int)} \\
  &\quad\;\longrightarrow\;
  \texttt{[(Double, Map Species Int)]}
\end{align*}
in $\ParaC_W(\Hask)$, where $f_{\mathrm{SSA}}$ is the curried form of
\texttt{simulate} with the seed extracted from \texttt{ssaConfig};
the remaining components of \texttt{SSAState} (network, volume,
$\tau_{\max}$, sample interval, initial counts) sit in the second
argument.  This is a categorical re-shaping, not a separately-named
BR function.  Currying gives a function of type $W \to (\ldots) \to
[\ldots]$ matching the \texttt{pmApply} field of a
\texttt{ParaMorphism W \ldots \ldots}.  Sampling the seed from
$\mathrm{Uniform}(W)$ and reading \texttt{ssaCounts} of the last
\texttt{Just}-state produces the Markov kernel
$\FP(\mathcal{N})_\tau$ of Theorem~\ref{thm:cat-gillespie}; this
seed-marginalisation is the action of the denotational functor
$F : \ParaC_W(\Hask) \to \BorelStoch$ of
\S\ref{sec:sim:target} --- \emph{not} a $\ParaC_W(\Hask)$-internal
2-morphism, since $f_{\mathrm{SSA}}$ at any single seed value
differs from the marginal kernel.

\paragraph{ODE as a Para 1-morphism with $P = ()$.}
\texttt{BR.ODE.simulate} carries no Para parameter: its
\texttt{ODEConfig} consists of tolerances and step bounds that are
fixed numerical-method choices, not stochastic parameters that the
denotational functor marginalises.  Hence \texttt{BR.ODE.simulate}
fits the Para shape as a \texttt{liftToPara} of an
unparameterised conceptual underlying-morphism
\begin{align*}
  f_{\mathrm{ODE}}
  &\;::\;
  \texttt{(ODEConfig, ODEState, Double)}\\
  &\quad\;\to\;
  \texttt{([(Double, Map Species Double)], Maybe String)},
\end{align*}
i.e.\ the Para 1-morphism
$\bigl((),\,f_{\mathrm{ODE}}\bigr)$, which is the $n = 0$ case in
$\ParaC_W(\Hask)$.  Like $f_{\mathrm{SSA}}$ above,
$f_{\mathrm{ODE}}$ is the categorical re-shaping of the curried
\texttt{BR.ODE.simulate}, not a separately-named BR function.

\paragraph{The asymmetry.}
The two arrows sit at different powers of $W$ within
$\ParaC_W(\Hask)$: the SSA at $n = 1$ (non-trivial seed parameter
$W$), the ODE at $n = 0$ (unit parameter $\mathbb{1}$).  This
asymmetry is the Para-side reflection
of the asymmetry in their denotational images established in
Remark~\ref{rmk:l3-parallel}: the SSA's denotational image
$\FP(\mathcal{N})_\tau$ is a non-trivial Markov kernel obtained by
seed-marginalisation, whereas the ODE's denotational image is the
deterministic Kurtz limit, requiring no marginalisation.

\begin{chembox}[Para parameters in BR]
The Para parameters that appear in BR are:
\begin{itemize}
\item \emph{Stochastic parameter $W = \{0,1\}^{64}$:} the
  SplitMix64 seed of \texttt{BR.SSA.simulate}.  Different seeds
  yield different sampled trajectories; marginalising over
  $\mathrm{Uniform}(W)$ recovers the CTMC kernel.
\item \emph{Trivial parameter $()$:} carried by
  \texttt{BR.ODE.simulate} and by any deterministic Haskell function
  embedded via \texttt{liftToPara}.
\end{itemize}
Numerical-method choices such as \texttt{odeRTol}, \texttt{odeATol},
\texttt{odeHMin}, \texttt{odeHMax} sit \emph{inside} the
\texttt{ODEConfig} record and travel as ordinary inputs to the
underlying morphism --- not as Para parameters --- because the
denotational reading of the ODE does not marginalise over them; they
control the deterministic integrator's local truncation error
(the ODE component of $\gamma$,
Remark~\ref{rmk:gamma-is-gillespie}) rather than a stochastic
ensemble.  The BR \texttt{Network} record's stoichiometries,
rate-active multisets, saturation data, rate constants, and
pool-species set are likewise ordinary inputs, not Para parameters:
they fix the kinetic model that the kernel $\FP$ acts on, and the
chapter's tower already accounts for them across $\Lk_0$ through
$\Lk_3$ (\S\S\ref{sec:sim:L0}, \ref{sec:sim:L12},
\ref{sec:sim:gillespie}).
\end{chembox}

%% file: chapters/sim/sim_s7_functor.tex
\subsection{The simulation functor $\PhiOp$: construction and properties}
\label{sec:sim:functor}

With $\ParaC_W(\Hask)$ established in \S\ref{sec:sim:para} and the
seed-marginalisation functor
$F : \ParaC_W(\Hask) \to \BorelStoch$ in place
(Proposition~\ref{prop:F-exists}), the operational functor $\PhiOp$
closes the diagram of Figure~\ref{fig:sim:two-layer}.

\subsubsection{Construction of $\PhiOp$}

\begin{definition}[Simulation functor]
\label{def:sim-functor}
The \emph{simulation functor}
\begin{equation}\label{eq:phi-op}
  \PhiOp \;:\;
  \ParaC\!\left(\Lk_0 \otimes \Lk_1 \otimes \Lk_2 \otimes \Lk_3\right)
  \;\longrightarrow\;
  \ParaC_W(\Hask)
\end{equation}
sends an object $\NN[\mathcal{S}]$ (a free commutative monoid on
species) to its $\Hask$ realisation \texttt{Map Species Int}
(Proposition~\ref{prop:L0-exact}, \S\ref{sec:sim:L0}).  On
1-morphisms at $\Lk_3$, $\PhiOp$ realises a Para 1-morphism
$(\Theta, f_\theta)$ as $(\Theta, \tilde f_\theta)$ in
$\ParaC_W(\Hask)$, where $\tilde f_\theta$ is the
underlying-morphism realisation of $f_\theta$ in $\Hask$ obtained
from the $\Lk_k$ realisations of \S\ref{sec:sim:L0}--%
\S\ref{sec:sim:gillespie}; the parameter object $\Theta$ is
unchanged because both source and target take their parameter
actegory from the same Cartesian fragment of $\Hask$.  BR's two
operational realisations of $\Lk_3$ (\S\ref{sec:sim:gillespie})
instantiate the two parameter classes of $\ParaC_W(\Hask)$:

\begin{center}
\begin{tabular}{lll}
\toprule
$\Lk_3$ realisation & Para parameter & Image in $\ParaC_W(\Hask)$ \\
\midrule
SSA (Gibson--Bruck NRM)  & $W = \mathtt{Word64}$ & $(W,\, f_{\mathrm{SSA}})$, $n = 1$ \\
ODE (implicit midpoint)  & $\mathbb{1}$           & $((),\, f_{\mathrm{ODE}})$, $n = 0$ \\
\bottomrule
\end{tabular}
\end{center}

\noindent
On 2-morphisms, $\PhiOp$ acts as the identity on the underlying
reparameterisation $r : \Theta \to \Theta'$ (the parameter actegory
is shared), and the Para 2-morphism condition
$g \circ (r \times \mathrm{id}_A) = f$ in the source transports to
its $\Hask$ counterpart in the target by strict functoriality of
the $\Lk_k$ realisations.
\end{definition}

\begin{remark}[$\PhiOp$ between bicategories]
\label{rmk:phi-op-bicat}
Both the domain
$\ParaC(\Lk_0 \otimes \Lk_1 \otimes \Lk_2 \otimes \Lk_3)$ and the
codomain $\ParaC_W(\Hask)$ are bicategories with non-identity
parameter-tensor coherence data $\lambda, \rho, \alpha$ inherited
from $\Hask$'s Cartesian product (\S\ref{sec:sim:para}).
Proposition~\ref{prop:phi-properties} below shows that $\PhiOp$ is
a strict $2$-functor between them: its own unit and composition
constraints are identity $2$-cells, and the target's
parameter-tensor coherence is inherited verbatim.
\end{remark}

\subsubsection{Main proposition}

\begin{proposition}[Properties of $\PhiOp$]
\label{prop:phi-properties}
$\PhiOp$ is a strict $2$-functor between bicategories: all its
unit and composition constraints are identity $2$-cells, and the
parameter-tensor coherence of the target is inherited from
$\ParaC_W(\Hask)$ unchanged.  Explicitly:
\begin{enumerate}[label=(\roman*)]
\item \emph{Strict on objects.}
  $\PhiOp(\NN[\mathcal{S}]) = \mathtt{Map\ Species\ Int}$ on the
  nose, by Proposition~\ref{prop:L0-exact}.
\item \emph{Strict on identities.}
  $\PhiOp(\mathrm{paraId}_A) = \mathrm{paraId}_{\PhiOp(A)}$ on the
  nose.  The Para identity at $X$ is $(I, \eta_X)$ with $I = ()$
  the shared actegory unit and $\eta_X$ the left unitor at $X$;
  $\PhiOp$ preserves both components.
\item \emph{Strict on composition.}
  For composable Para 1-morphisms $f, g$ in the source,
  \[
    \PhiOp\bigl(g \circ_{\ParaC} f\bigr)
    \;=\;
    \PhiOp(g) \circ_{\ParaC_W(\Hask)} \PhiOp(f)
  \]
  on the nose.
\item \emph{Target coherence inherited.}
  The parameter-tensor associator, left unitor, and right unitor
  of $\ParaC_W(\Hask)$ (Proposition~\ref{prop:para-laws}) act as
  the bicategorical coherence data of $\PhiOp$'s codomain;
  $\PhiOp$ introduces no additional $2$-cells at this layer.
\end{enumerate}
\end{proposition}

\begin{proof}
(i): Proposition~\ref{prop:L0-exact}.
(ii): unfold $\mathrm{paraId}_A = (I, \eta_A)$.  The parameter
component $I = ()$ is identical in source and target because the
parameter actegory is shared (Cartesian fragment of $\Hask$;
clause (iv)).  The underlying component $\eta_A$ is a left unitor
in the source's underlying category, sent by the $\Lk_k$
realisations to the left unitor $\eta_{\PhiOp(A)}$ in $\Hask$ by
strict functoriality at the $1$-category level.
(iii): both $\PhiOp(g \circ_{\ParaC} f)$ and
$\PhiOp(g) \circ_{\ParaC_W(\Hask)} \PhiOp(f)$ unfold via
Definition~\ref{def:para} to a Para 1-morphism with parameter
$\Theta_f \times \Theta_g$ and underlying morphism
$(\theta_f, \theta_g, a) \mapsto \mathrm{und}(g)
(\theta_g, \mathrm{und}(f)(\theta_f, a))$ at the underlying-morphism
layer; the $\Lk_k$ realisations
(\S\ref{sec:sim:L0}--\S\ref{sec:sim:gillespie}) preserve this
expression strictly as $1$-category functors.
(iv): the parameter actegory is shared (both source and target use
the Cartesian fragment of $\Hask$), so $\PhiOp$ acts as the
identity on parameter objects and inherits the target's
parameter-tensor coherence data verbatim.
\end{proof}

\begin{remark}[The $\alpha$, $\beta$, $\gamma$ cells of \S\ref{sec:sim:target}, revisited]
\label{rmk:alpha-beta-gamma}
\S\ref{sec:sim:target} identified three potential lax cells of
$\PhiOp$ in an IO-based formulation: a unit $\alpha$, an
IO-sequencing $\beta$, and a numerical $\gamma$.  For BR, the
strictness of $\PhiOp$ (Proposition~\ref{prop:phi-properties})
trivialises the first two:
\begin{itemize}
\item $\alpha$ is the identity $2$-cell by clause (ii) of
  Proposition~\ref{prop:phi-properties}: $\PhiOp$ preserves Para
  identities on the nose, so the hypothetical
  $\PhiOp(\mathrm{id}) \Rightarrow \mathrm{id}_{\PhiOp(\cdot)}$
  constraint is trivial.
\item $\beta$ is the identity $2$-cell: pure $\Hask$'s Cartesian
  product carries no execution-order semantics, so parallel
  compositions $f \otimes g$ are determined by their components
  without sequencing.  No $2$-cell relating
  ``$f$-then-$g$'' to ``$g$-then-$f$'' arises beyond the target's
  symmetric monoidal symmetry, which is target coherence
  (clause (iv)) rather than a $\PhiOp$ laxity.
\item $\gamma$ is a $2$-morphism \emph{in the target} measuring
  the discrepancy between $\PhiOp$'s image of an $\Lk_3$ arrow
  and the exact dynamics it approximates; it is not part of
  $\PhiOp$'s $2$-functor structure.  Its SSA component is zero in
  total variation modulo PRNG and IEEE-754 qualifications
  (Theorem~\ref{thm:cat-gillespie}); its ODE component is the
  implicit-midpoint local truncation error measured in
  state-space sup norm (\S\ref{sec:sim:gillespie}; the
  $\gamma$-cell mathbox of \S\ref{sec:sim:target}).
\end{itemize}
\end{remark}

\begin{remark}[$\PhiDen$ by construction]
\label{rmk:phi-den-def}
$\PhiDen := F \circ \PhiOp$ is a definition, not a
separately-proved theorem.  The denotational factorisation of
Figure~\ref{fig:sim:two-layer} closes once both $\PhiOp$ (this
section) and $F$ (Proposition~\ref{prop:F-exists}) are in hand.
\end{remark}

\subsubsection{Corollary: Para applied to a chemical reaction simulation}

\begin{corollary}[Para applied to chemical reaction simulation]
\label{cor:para-chemistry}
The construction above exhibits the Para construction
(Definition~\ref{def:para}) applied to a chemical reaction
simulation: BR's Gibson--Bruck SSA and implicit-midpoint ODE
realisations of $\Lk_3$ sit as Para 1-morphisms at $n = 1$ and
$n = 0$ in $\ParaC_W(\Hask)$ respectively, and
seed-marginalisation by $F$ produces Markov kernels in
$\BorelStoch$.  Previously published applications of the Para
construction include machine learning
\cite{CruttwellGavranovic2022, Gavranovic2024ICML}, compositional
game theory \cite{Hedges2018}, and active inference
\cite{Smithe2023}; to our knowledge no prior treatment applies it
to chemical reaction simulation.
\end{corollary}

%% file: chapters/sim/sim_s8_L4coarse.tex
\subsection{$\Lk_4 \to \Lk_3$ projection: rate constants as scalar
coarse-grainings and model variants as comparable reductions}
\label{sec:sim:L4coarse}

The most conceptually delicate aspect of the simulation functor of
\S\ref{sec:sim:functor} is that $\Lk_4$ mechanism data appears
inside the $\Lk_3$ parameter space.  At first sight this violates
the tower's level stratification --- $\Lk_4$ bond-level information
is supposed to be invisible at $\Lk_3$.  This section shows that
the appearance is a \emph{parameter projection}: the BR simulation
retains only the part of $\Lk_4$ data that survives as scalar rate
constants and, for channels with rate-limiting intermediates, the
saturation form of the rate law.  The chapter's several BR network
variants are correspondingly different $\Lk_3$ reductions of the
same underlying mechanism, comparable in $\BorelStoch$ via the
construction of \S\ref{sec:sim:target}.

\subsubsection{The $\Lk_4$ content of BR's rate constants}

Level $\Lk_4$ is the category of open reaction networks with full
bond-level mechanism data: a morphism at $\Lk_4$ specifies the
elementary-step mechanism (which bonds form and break), the
transition-state geometry, and the parameters determining kinetic
isotope effects, temperature dependence, and solvent dependence
\cite{EhrigEtAl2006}.

The BR mechanism in \texttt{BR.Mechanism} encodes the
De~Kepper--Epstein skeleton \cite{DeKepperEpstein1982} as
\texttt{briggsRauscherDE} (10 elementary steps plus two explicit
reverses; 12 channels).  Each channel is recorded as a
\texttt{Reaction} value (\S\ref{sec:sim:L12}) carrying three
pieces of data relevant to the $\Lk_4 \to \Lk_3$ projection:

\begin{itemize}
  \item \texttt{rxnRate :: Double} --- the scalar rate constant
    used by both \texttt{BR.SSA} and \texttt{BR.ODE}.
  \item \texttt{rxnSaturation :: [(Species, Double)]} --- an
    optional rational denominator $1 + \sum_i C_i [S_i]$ in the
    rate law, non-empty only on channels with a rate-limiting
    intermediate (e.g., R9 of \cite{DeKepperEpstein1982}, where
    $\mathrm{I}_2$ saturation supplies the negative feedback that
    closes the limit cycle).
  \item \texttt{rxnProvenance :: RateProvenance} --- metadata
    tagging each rate as either \texttt{Measured} (with a citation
    key to an independent kinetic study) or
    \texttt{FitToOscillate} (the ``judicious assignment'' of
    \cite{NoyesFurrow1982}).
\end{itemize}

The first two fields enter the propensity (SSA) and ODE
right-hand side; the third is metadata visible to the framework
but not used by the rate formula.

\subsubsection{A parameter projection}

\begin{definition}[Coarse $\Lk_4$ parameter space for BR]
\label{def:theta-L4-coarse}
For a BR network $\mathcal{N}$ with channel set $\mathcal{R}$, the
\emph{coarse $\Lk_4$ parameter space} is
\[
  \Theta_{\Lk_4,\,\mathrm{coarse}}(\mathcal{N})
  \;:=\;
  \prod_{r \in \mathcal{R}}\,
  \bigl(\RR_{>0}
        \;\times\;
        \mathrm{List}(\Sp \times \RR_{\geq 0})
        \;\times\;
        \mathsf{Prov}\bigr),
\]
one factor per channel, with components $(\mathtt{rxnRate},
\mathtt{rxnSaturation}, \mathtt{rxnProvenance})$.
\end{definition}

\begin{definition}[The $\Lk_4$ parameter projection $\pi_4$]
\label{def:sim:pi4}
The \emph{$\Lk_4$ parameter projection} is the map
\[
  \pi_4 \;:\; \Theta_{\Lk_4}(\mathcal{N})
  \;\longrightarrow\;
  \Theta_{\Lk_4,\,\mathrm{coarse}}(\mathcal{N})
\]
sending each full $\Lk_4$ mechanism to its triple
$(k_r, \mathrm{sat}_r, \mathrm{prov}_r)_{r \in \mathcal{R}}$.
The projection is \emph{lossy by construction}: it drops all
mechanism data beyond what the three retained fields encode ---
bond-level elementary-step structure, transition-state geometry,
isotopic-substitution sensitivity, solvent and ionic-strength
dependence, pressure dependence, and any explicit temperature
dependence beyond what is already absorbed into $k_r$ at the
reference conditions (Remark~\ref{rmk:sim:pi4-cannot-see-BR}).
\end{definition}

\begin{proposition}[$\Lk_4$-coarse augmentation of the BR parameter space]
\label{prop:L4-augmentation}
The BR simulation is parameterised by
\begin{equation}\label{eq:sim:theta-aug}
  \ThetaSim
  \;=\;
  \ThetaL{1} \otimes \ThetaL{2}
  \otimes
  \Theta_{\Lk_4,\,\mathrm{coarse}}(\mathcal{N}),
\end{equation}
where $\ThetaL{1}, \ThetaL{2}$ are the species- and
stoichiometry-level parameters of \S\ref{sec:sim:L12} and the
third factor is the $\pi_4$-image of $\Lk_4$ mechanism data.
Under $\PhiOp$ (\S\ref{sec:sim:functor}), the third factor enters
the simulation only through the propensity formulas of
Theorem~\ref{thm:cat-gillespie} (SSA) and the corresponding ODE
right-hand sides (\S\ref{sec:sim:gillespie}); two $\Lk_4$
mechanisms whose $\pi_4$-images agree map to the same morphism in
$\ParaC_W(\Hask)$.
\end{proposition}

\begin{proof}
\texttt{BR.SSA.simulate} and \texttt{BR.ODE.simulate} compute
propensities using the rate-law fields \texttt{rxnRateActive},
\texttt{rxnRate}, and \texttt{rxnSaturation}, and update state
using the stoichiometric multisets \texttt{rxnReactants} and
\texttt{rxnProducts}.  No other reaction data is consulted:
\texttt{rxnName} is a label, and \texttt{rxnProvenance} is
metadata visible to the framework but not read by the rate or
update logic.  Full $\Lk_4$ mechanism data (transition states,
KIE-sensitivity, etc.) is not represented in the
\texttt{Reaction} record at all.  Two $\Lk_4$ mechanisms for the
same network whose $\pi_4$-images agree --- i.e., produce identical
$(\mathtt{rxnRate}, \mathtt{rxnSaturation})$ per channel ---
therefore yield identical propensities and identical ODE
right-hand sides, hence identical simulator outputs modulo PRNG
and IEEE-754 qualifications.
\end{proof}

\begin{remark}[Provenance as a discriminator of $\Lk_4$-strength]
\label{rmk:sim:provenance}
The \texttt{rxnProvenance} tag is $\Lk_4$-strength metadata: not
used by the rate formula but visible to the framework.  A network
all of whose rate constants are tagged \texttt{Measured}
(independent kinetic citations) is \emph{$\Lk_4$-strong}: the
simulation's qualitative behaviour is not tuned by free
parameters.  The 12 channels of \texttt{briggsRauscherDE} are
currently all \texttt{Measured}, making BR an $\Lk_4$-strong test
bed within the chapter's framework; oscillating reactions
requiring \texttt{FitToOscillate} constants would sit at a
correspondingly weaker $\Lk_4$ strength.
\end{remark}

\subsubsection{Model variants as comparable $\Lk_3$ reductions}

\texttt{BR.Mechanism} provides several network variants beyond
\texttt{briggsRauscherDE}: \texttt{bufferedH} (with $\mathrm{H}^+$
pooled), \texttt{broCODENetwork} (the BROCODE 2025 reduction),
\texttt{broCODE5VarNetwork} (a five-dynamical-variable reduction
designed for Hopf bifurcation analysis), and
\texttt{furrowLikePoolNetwork} (a Furrow-style variant pooling
several intermediates).  Each variant is obtained from
\texttt{briggsRauscherDE} by some combination of
(i)~restricting the channel set $\mathcal{R}$, and
(ii)~enlarging \texttt{netPoolSpecies} (the set of species held
constant during simulation; \texttt{ssaCounts} for these species
are frozen, per the pool-species convention of
\S\ref{sec:sim:gillespie}).

Each variant $V$ thus determines a different point of
$\ThetaSim$, and via $F \circ \PhiOp$ a different Markov kernel
in $\BorelStoch$:
\begin{equation}\label{eq:sim:variant-kernels}
  F_V \;:=\; F\bigl(\PhiOp(V)\bigr) \;\in\; \BorelStoch,
\end{equation}
sending each initial state $\mathbf{x}_0$ and horizon $\tau$ to a
probability measure $F_V(\mathbf{x}_0, \tau, \cdot)$ over
trajectories.

\begin{proposition}[Canonical embedding of BR model variants]
\label{prop:variant-comparison}
The composition $F \circ \PhiOp$ (\S\S\ref{sec:sim:target},
\ref{sec:sim:functor}) embeds every well-formed BR network into
$\BorelStoch$ as a Markov kernel via
\eqref{eq:sim:variant-kernels}.  The embedding is canonical in
the sense that the same $F \circ \PhiOp$ applies to all variants;
comparison of variants is therefore comparison of Markov kernels
in a single category, with metric choices (total variation,
Wasserstein-1, relative entropy) available once the variants are
embedded in a common dynamical state space (extending each
variant's kernel by Dirac mass on its pooled coordinates).
\end{proposition}

\begin{proof}
$F$ and $\PhiOp$ are defined on $\ParaC_W(\Hask)$ regardless of
the network argument; \texttt{mkNetwork} validates each variant
as a well-formed value of type \texttt{Network}, on which both
\texttt{BR.SSA.simulate} and \texttt{BR.ODE.simulate} are
total by Assumption~\ref{ass:totality} (\S\ref{sec:sim:target}),
so the resulting kernel is a probability measure on the
variant's species support.  Embedding into a common state space
(union of dynamical species sets) is by extension via Dirac mass
on each variant's pooled coordinates.
\end{proof}

\begin{chembox}[$\Lk_3$ reductions: predicted invariance and drift]
The categorical framework predicts which BR observables should be
invariant under which $\Lk_3$ reductions:

\begin{itemize}
  \item \emph{Onset preservation.}  Oscillation \emph{onset}
    (the Hopf bifurcation) depends on both the network's $\Lk_2$
    stoichiometric topology and the sign-pattern of $\Lk_3$
    Jacobian eigenvalues at the fixed point.  Reductions that
    preserve the autocatalytic feedback channels of the
    De~Kepper--Epstein skeleton (in particular
    \texttt{broCODE5VarNetwork}, designed for Hopf bifurcation
    analysis) preserve the capacity for oscillation, even if the
    quantitative bifurcation point shifts.
  \item \emph{Quantitative drift.}  Oscillation \emph{period} and
    \emph{amplitude} are $\Lk_3$-quantitative: they depend on the
    rate constants retained by $\pi_4$ together with which
    species are dynamical.  Variants whose \texttt{netPoolSpecies}
    removes feedback-relevant species should show measurable drift
    in these observables.
  \item \emph{Falsification handle.}  Replacing any
    \texttt{Measured} rate constant in
    \texttt{briggsRauscherDE} with a \texttt{FitToOscillate}
    surrogate of the same order of magnitude should preserve
    oscillation onset but degrade quantitative agreement with the
    experimental period and amplitude --- the provenance
    machinery makes this a structured test rather than an ad
    hoc parameter sweep.
\end{itemize}

Proposition~\ref{prop:variant-comparison} grounds these claims
operationally: large TV distance between $F_V$ and $F_{V'}$ at
horizon $\tau$ flags variants disagreeing on observable
distributions at that horizon, even when both oscillate.
\end{chembox}

\begin{remark}[What $\pi_4$ cannot see, for BR]
\label{rmk:sim:pi4-cannot-see-BR}
By construction $\pi_4$ retains only the scalar rate constant,
saturation list, and provenance tag per channel.  It drops:
\begin{enumerate}[label=(\roman*)]
  \item the full elementary-step mechanism (which bonds form and
    break in each channel; concerted versus stepwise pathways;
    radical versus closed-shell intermediates);
  \item kinetic isotope effects --- the rate change under
    $\mathrm{H} \to \mathrm{D}$ substitution at the methylene
    hydrogens of malonic acid (predicted by transition-state
    theory but invisible to a scalar \texttt{rxnRate});
  \item solvent and ionic-strength dependence beyond the
    reference conditions absorbed into \texttt{rxnRate};
  \item pressure dependence and any temperature dependence not
    captured by \texttt{rxnRate} at the reference temperature.
\end{enumerate}
Any two BR mechanisms agreeing on the dynamical pair
$(\mathtt{rxnRate}, \mathtt{rxnSaturation})$ per channel are
indistinguishable in the simulation, whether or not their
provenance tags agree (Proposition~\ref{prop:L4-augmentation};
\texttt{rxnProvenance} is metadata, not consulted by the rate
or update logic).  This is the $\Lk_3 \to \Lk_4$ forcing pair of the main tower applied to
inorganic oscillating chemistry --- an honest categorical
consequence of $\pi_4$'s range, not an implementation deficit.
Recovery of the dropped data requires either a higher-detail
simulator operating at $\Lk_4$ (mechanism-resolved) or
experimental inputs (KIE measurement, pressure-jump kinetics)
whose results would re-parameterise the $\Lk_3$ scalar at the new
reference conditions.
\end{remark}

\medskip

\noindent
This closes the theoretical development of the chapter.  The
operational functor $\PhiOp$ (\S\ref{sec:sim:functor}) is
parameterised by \eqref{eq:sim:theta-aug}, with $\pi_4$
projecting all $\Lk_4$ information through a three-component
coarse representation; BR's network variants are different
$\Lk_3$ realisations comparable as Markov kernels via
\eqref{eq:sim:variant-kernels}.  The simulation results of
\S\ref{sec:sim:results} test these constructions against the
\texttt{briggsRauscherDE} network running under the conditions of
\cite{DeKepperEpstein1982}.

%% file: chapters/sim/sim_s9_results.tex
\subsection{Simulation results: $F_{\mathtt{briggsRauscherDE}}$ in $\BorelStoch$}
\label{sec:sim:results}

With the theoretical development closed at
\S\ref{sec:sim:L4coarse}, we display the empirical realisation of
$F \circ \PhiOp$ applied to \texttt{briggsRauscherDE}.  The exact
stochastic kernel
\[
  F_{\mathrm{BR}} \;:=\; F\bigl(\PhiOp(\mathtt{briggsRauscherDE})\bigr)
  \;\in\; \BorelStoch
\]
is sampled directly by the SSA (modulo PRNG and IEEE-754
qualifications, Theorem~\ref{thm:cat-gillespie}).  The deterministic
ODE trajectory realises the $V \to \infty$ Kurtz limit of $F_{\mathrm{BR}}$'s
species-mean process (\S\ref{sec:sim:gillespie}); the
$\gamma_{\mathrm{ODE}}$ 2-cell of \S\ref{sec:sim:target} measures
the $\BorelStoch$-level deviation between this deterministic Kurtz
limit and $F_{\mathrm{BR}}$ itself.  Both simulators evaluate the same network
and parameters at the same point of $\ThetaSim$
(\S\ref{sec:sim:L4coarse}, eq.~\eqref{eq:sim:theta-aug}); the
visible difference between their outputs at finite $V$ is the
chapter's $\gamma$-cell decomposition made concrete.

The empirical question is not whether the framework runs --- by
construction $\PhiOp$ and $F$ apply to any well-formed
\texttt{Network} --- but whether its predictions acquire visible
empirical content.  We verify three predictions of the preceding
sections:

\begin{enumerate}[label=(P\arabic*)]
  \item \emph{The oscillator emerges.}  Both simulators --- the
    SSA sampling $F_{\mathrm{BR}}$ and the ODE realising its Kurtz limit ---
    exhibit relaxation-oscillator dynamics.  This is the base case
    for \S\ref{sec:sim:L4coarse}'s onset-preservation framework:
    network reductions can subsequently be compared to this base.
  \item \emph{The $\gamma$-cell distinction is operational.}  The
    SSA samples and the ODE trajectory agree where the
    rate-equation representation is faithful (high-copy spike
    phase) and diverge where it is not (low-copy quiescent phase
    at finite $V$), matching the form of $\gamma_{\mathrm{SSA}} = 0$
    in total variation and $\gamma_{\mathrm{ODE}}$ in
    state-space/Wasserstein-1 metric predicted in
    \S\ref{sec:sim:target} and \S\ref{sec:sim:gillespie}.
  \item \emph{The divergence is structurally interpretable.}  Where
    the two simulators differ, the framework supplies the reason:
    the ODE realises the $V \to \infty$ Kurtz limit, and at
    $V = 10^{-15}~\mathrm{L}$ this limit represents quiescent
    populations of low-copy species as fractional ensemble means
    that do not correspond to any state of an individual
    realisation.
\end{enumerate}

\subsubsection{Simulation parameters}

\paragraph{Network.}  \texttt{briggsRauscherDE} (\S\ref{sec:sim:L12}),
12 reaction channels encoding the De~Kepper--Epstein skeleton
\cite{DeKepperEpstein1982} (10 elementary steps plus 2 explicit
reverses).  Every rate constant carries a \texttt{Measured}
\texttt{rxnProvenance} tag with an independent kinetic citation
(\S\ref{sec:sim:L4coarse}, Remark~\ref{rmk:sim:provenance}); no
constant is tuned to fit oscillatory behaviour.

\paragraph{Volume.}  $V = 10^{-15}~\mathrm{L}$ (femtoliter scale).
At this volume one molecule corresponds to a concentration of
$1/(V \, N_A) \approx 1.66 \times 10^{-9}~\mathrm{M}$, where $N_A$
is Avogadro's number.  The volume is chosen deliberately to keep
individual molecules resolvable and the SSA-versus-ODE distinction
empirically visible; at macroscopic volumes (e.g.\
$1~\mathrm{mL}$) the SSA distribution converges to the ODE limit
by the Kurtz theorem \cite{Kurtz1972} and the two simulators become
indistinguishable in total variation.  $V$ is a user-chosen
simulation parameter (entering propensities via the
$k_{\mathrm{micro}} = k_{\mathrm{macro}}/V^{\,n-1}$ scaling for
$n$-th order reactions); it is not part of $\ThetaSim$
(\S\ref{sec:sim:L4coarse}, eq.~\eqref{eq:sim:theta-aug}) and not
derived from the framework.

\paragraph{Horizon.}  $\tau = 1800~\mathrm{s}$, encoded as
\texttt{ssaTimeMax}.  The rate constants in
\texttt{briggsRauscherDE} are parameterised for an inter-spike
interval near $10^{3}~\mathrm{s}$; the simulation horizon captures
between one and two complete cycles.

\paragraph{Ensemble.}  For the SSA, $N = 6$ realisations with
independent \texttt{Word64} seeds.  The ODE is deterministic and
run once.  The ensemble size is sufficient for qualitative display
of timing, amplitude envelope, and quiescent-phase chatter;
quantitative kernel comparisons
(Proposition~\ref{prop:variant-comparison}) require
$N \gtrsim 100$.

\paragraph{Initial state.}
$[\mathrm{I}^-] = 10^{-5}~\mathrm{M}$,
$[\mathrm{I}_2] = 10^{-6}~\mathrm{M}$, all other species near
their analytic floors.  These conditions sit off the limit cycle;
the first $\sim 250~\mathrm{s}$ of every trajectory is transient
relaxation toward the attractor.

\subsubsection{Deterministic view: the ODE Kurtz limit}

\begin{figure}[!h]
  \centering
  \includegraphics[width=0.95\textwidth]{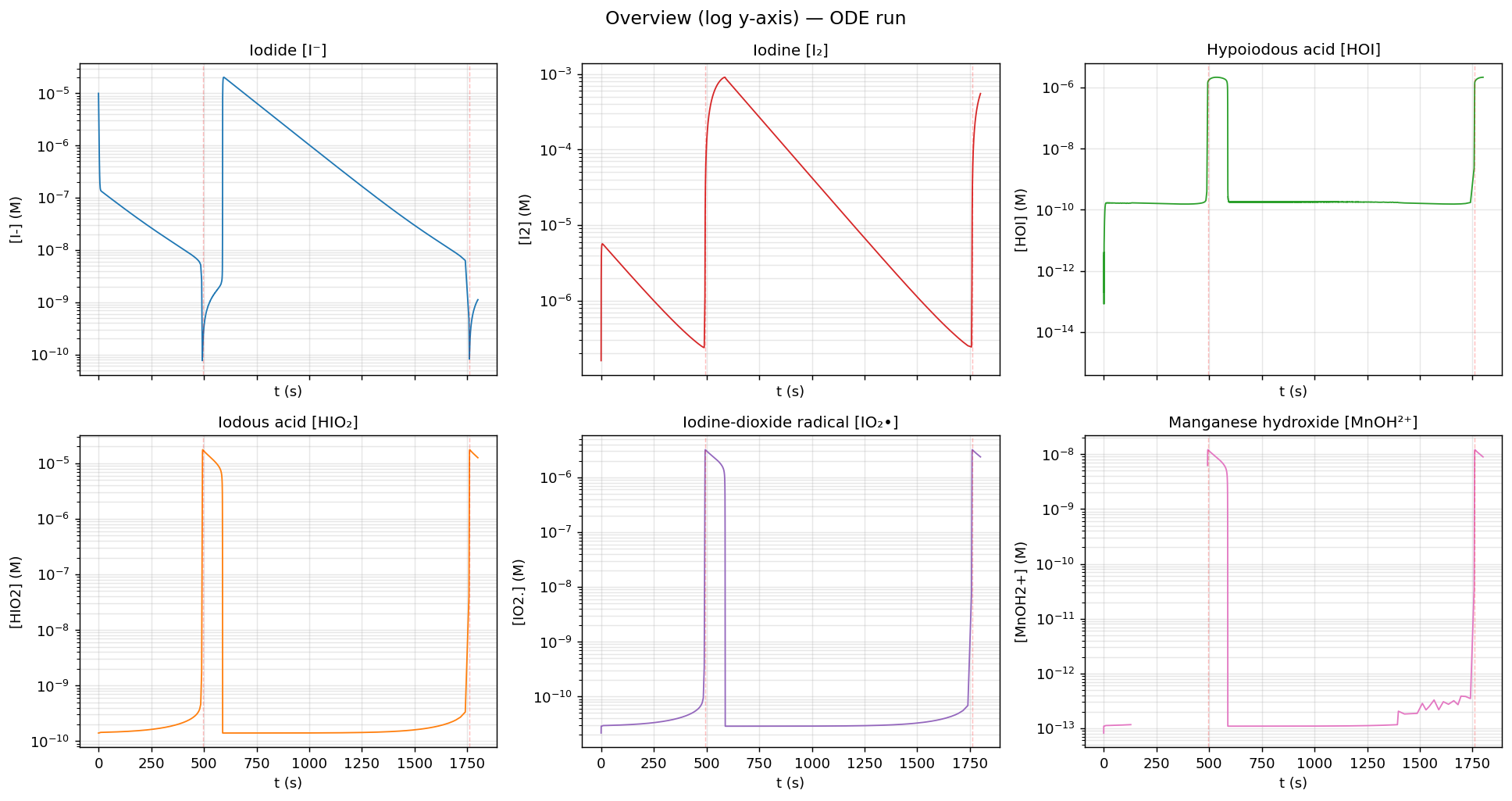}
  \caption{ODE trajectory for $\mathtt{briggsRauscherDE}$ at the
    chosen parameterisation: the $V \to \infty$ Kurtz limit of
    $F_{\mathrm{BR}}$'s species-mean process.  All six dynamical
    species participate in coherent relaxation spikes separated by
    quiescent recharging of $\sim 1100~\mathrm{s}$.  Observed
    inter-spike interval $\approx 1250~\mathrm{s}$.  Dashed
    vertical lines mark spike-onset.  Numerical noise in the
    $\mathrm{HOI}$ panel near $t = 0$ is at the ODE
    absolute-tolerance floor (not chemistry); the analytic
    quiescent floor stabilises after $t \approx 200~\mathrm{s}$.}
  \label{fig:sim:ode-overview}
\end{figure}

\begin{figure}[!h]
  \centering
  \includegraphics[width=0.7\textwidth]{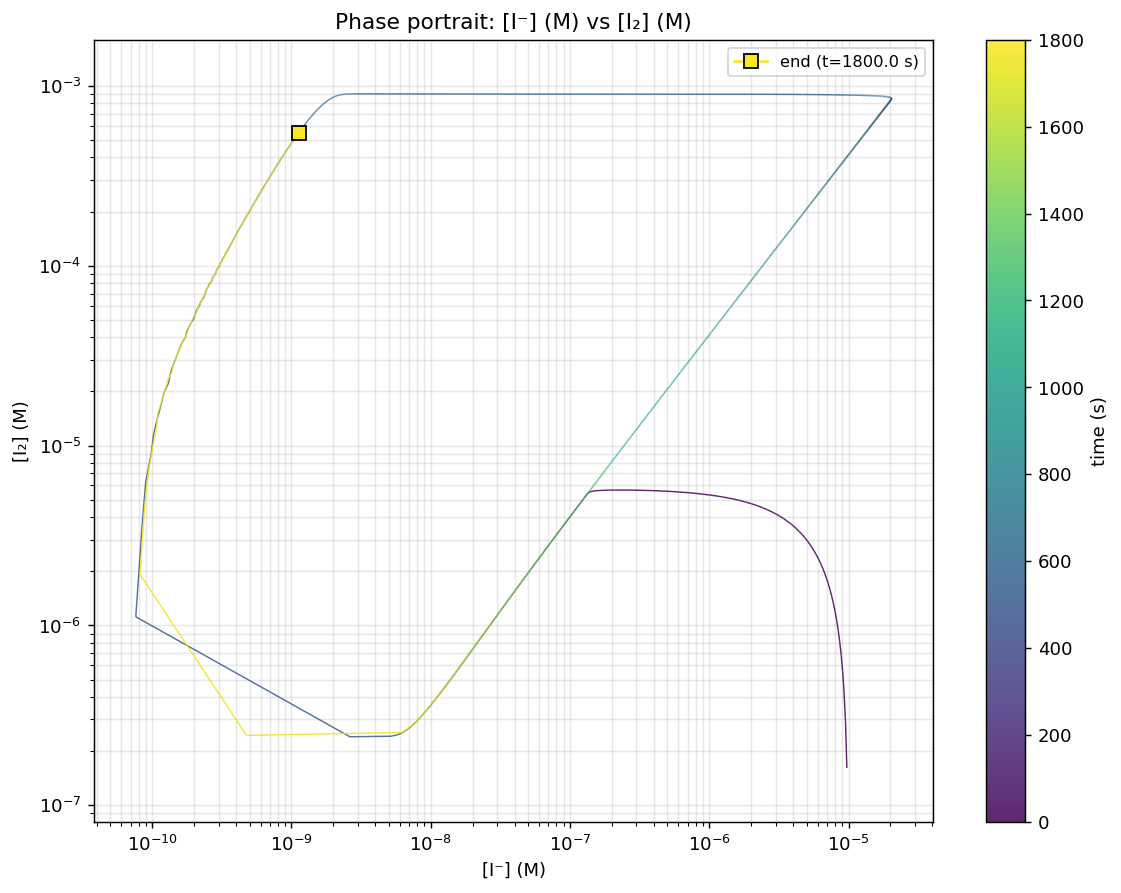}
  \caption{Projection of the ODE trajectory onto
    $([\mathrm{I}^-], [\mathrm{I}_2])$ in log--log coordinates;
    colour encodes simulation time.  The trajectory traces an
    approach to a four-corner limit cycle but does not yet close
    within $\tau = 1800~\mathrm{s}$ (yellow end-point offset from
    purple start), consistent with transient relaxation from
    off-cycle initial conditions.}
  \label{fig:sim:ode-phase-i-i2}
\end{figure}

\begin{figure}[!h]
  \centering
  \includegraphics[width=0.7\textwidth]{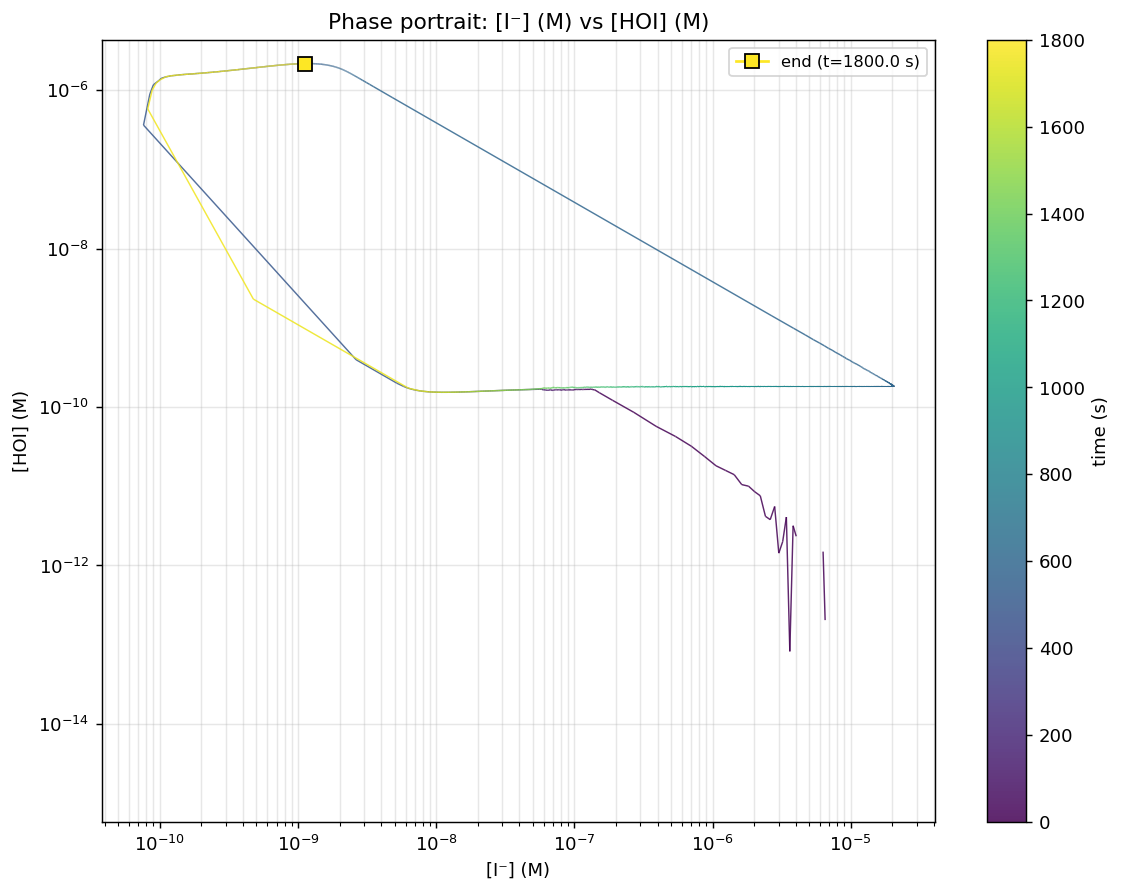}
  \caption{Projection onto $([\mathrm{I}^-], [\mathrm{HOI}])$.
    The lower-right portion at small $t$ is ODE-tolerance
    numerical noise near the analytic $\mathrm{HOI}$ floor; the
    main limit-cycle structure is visible from
    $t \approx 250~\mathrm{s}$ onward.}
  \label{fig:sim:ode-phase-i-hoi}
\end{figure}

The ODE trajectory exhibits the qualitative signature of a
relaxation oscillator.  Peak concentrations across the spike are
$[\mathrm{I}_2] \approx 10^{-3}~\mathrm{M}$,
$[\mathrm{HOI}] \approx 10^{-6}~\mathrm{M}$,
$[\mathrm{HIO}_2] \approx 10^{-5}~\mathrm{M}$,
$[\mathrm{IO}_2 \bullet] \approx 3 \times 10^{-6}~\mathrm{M}$, and
$[\mathrm{MnOH}^{2+}] \approx 10^{-8}~\mathrm{M}$;
$[\mathrm{I}^-]$ reaches a transient peak near
$2 \times 10^{-5}~\mathrm{M}$ at spike onset before being consumed.
This confirms (P1) for the base network: the $\Lk_2$ stoichiometric
topology together with the $\Lk_3$ rate constants of
\texttt{briggsRauscherDE} supports a relaxation-oscillator
attractor in the Kurtz limit.

\subsubsection{Stochastic view: SSA samples of the exact CTMC}

\begin{figure}[!h]
  \centering
  \includegraphics[width=0.95\textwidth]{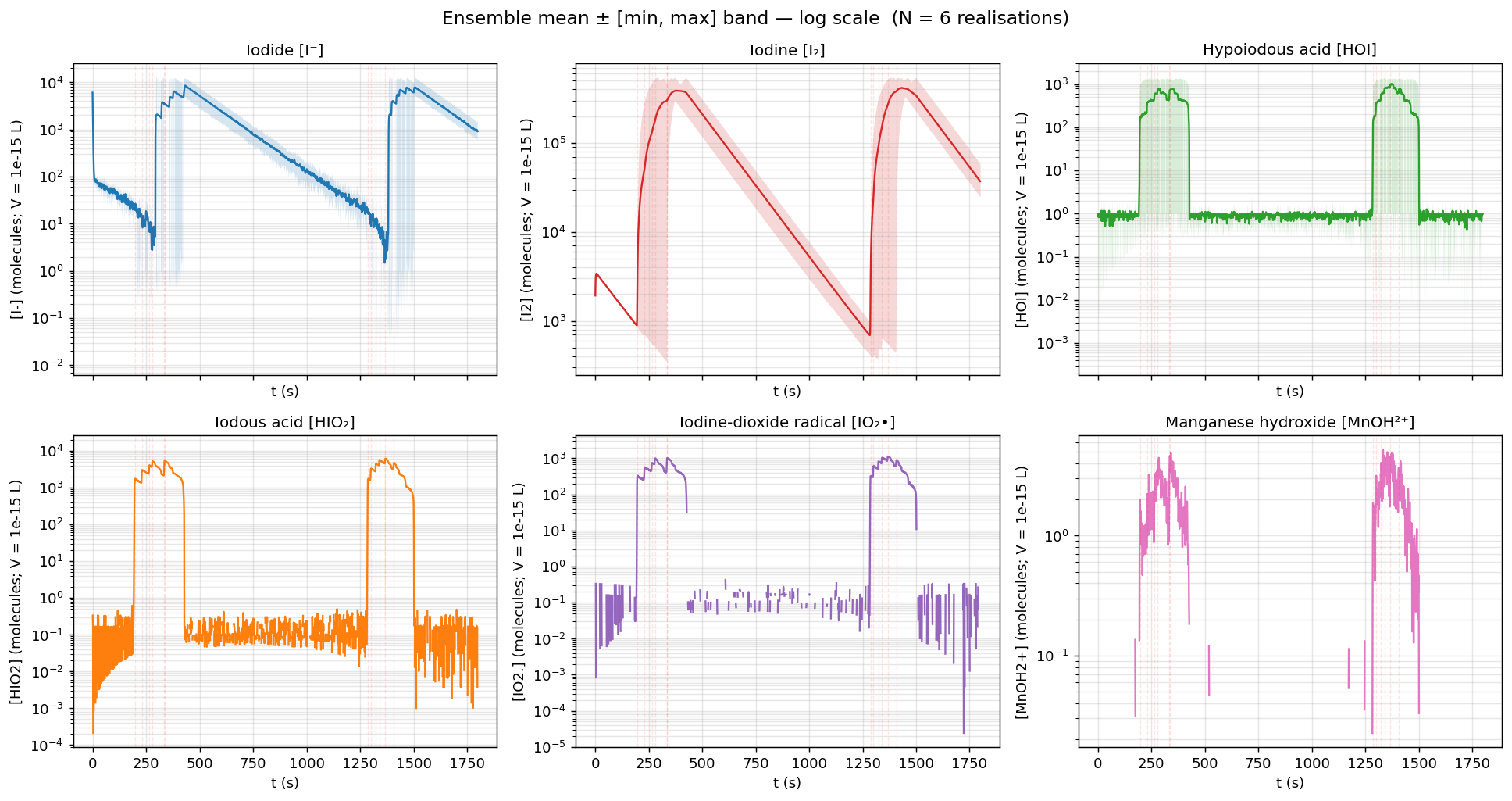}
  \caption{Stochastic view of
    $F \circ \PhiOp(\mathtt{briggsRauscherDE})$ at the same
    parameterisation, $V = 10^{-15}~\mathrm{L}$, $N = 6$
    independent SSA realisations.  Solid lines: ensemble means;
    shaded bands: $[\min, \max]$ across realisations.  $y$-axes
    are molecule counts (1 molecule $\approx 1.66 \times
    10^{-9}~\mathrm{M}$ at this volume).  Observed inter-spike
    interval $\approx 1050~\mathrm{s}$.  Quiescent-phase
    populations of $\mathrm{HIO}_2$, $\mathrm{IO}_2 \bullet$, and
    $\mathrm{MnOH}^{2+}$ are at $0$--$1$ molecules per $V$; the
    visual texture in these panels is the discrete CTMC's true
    behaviour at this volume, not numerical artifact.  Gaps in the
    $\mathrm{MnOH}^{2+}$ quiescent phase correspond to
    zero-molecule states (log axis cannot represent $0$).}
  \label{fig:sim:ssa-overview}
\end{figure}

\begin{figure}[!h]
  \centering
  \includegraphics[width=0.7\textwidth]{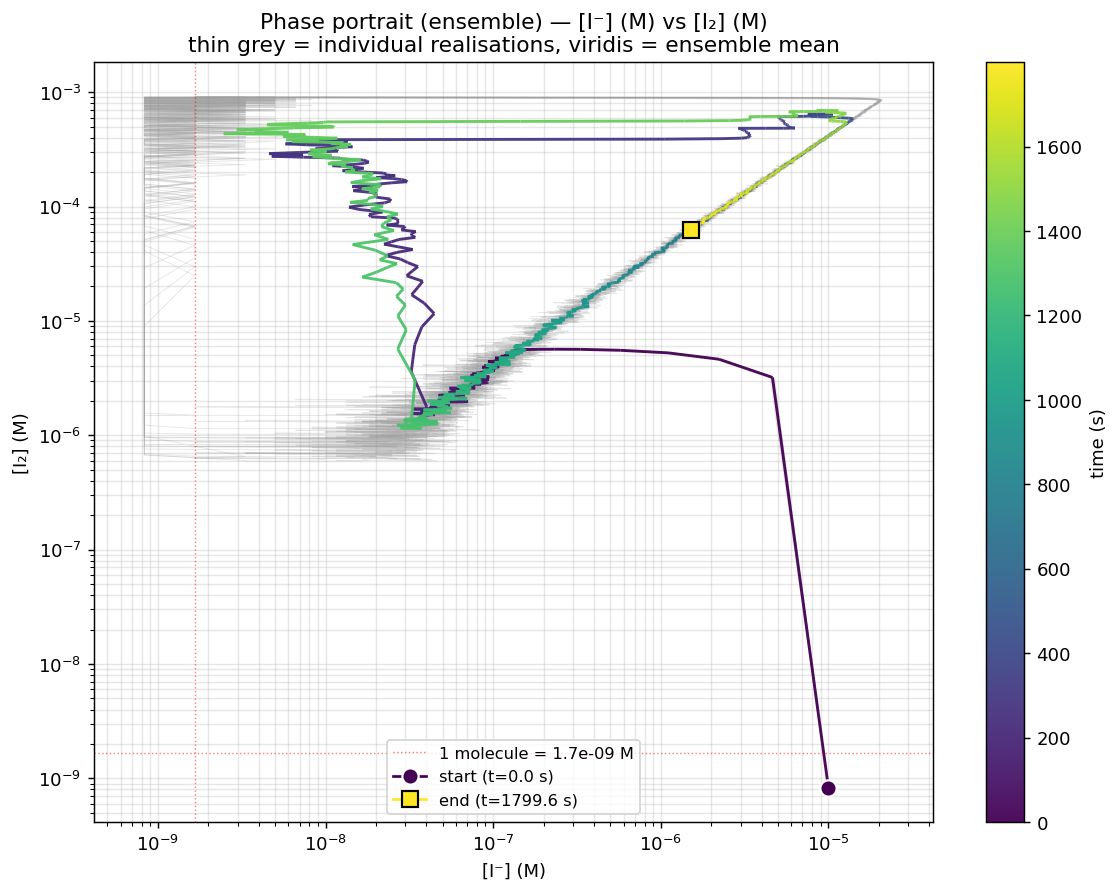}
  \caption{Projection of SSA realisations onto
    $([\mathrm{I}^-], [\mathrm{I}_2])$.  Thin grey lines:
    individual realisations.  Viridis-coloured curve: ensemble
    mean; colour encodes time.  Red dotted line: 1-molecule
    resolution at $V = 10^{-15}~\mathrm{L}$.  Trajectories cross
    this line repeatedly during the quiescent phase, entering the
    regime where the SSA tracks the integer-valued state of each
    realisation exactly while the ODE continuous representation
    --- still correct as an ensemble mean --- no longer
    corresponds to the state of any individual realisation.}
  \label{fig:sim:ssa-phase-i-i2}
\end{figure}

\begin{figure}[!h]
  \centering
  \includegraphics[width=0.7\textwidth]{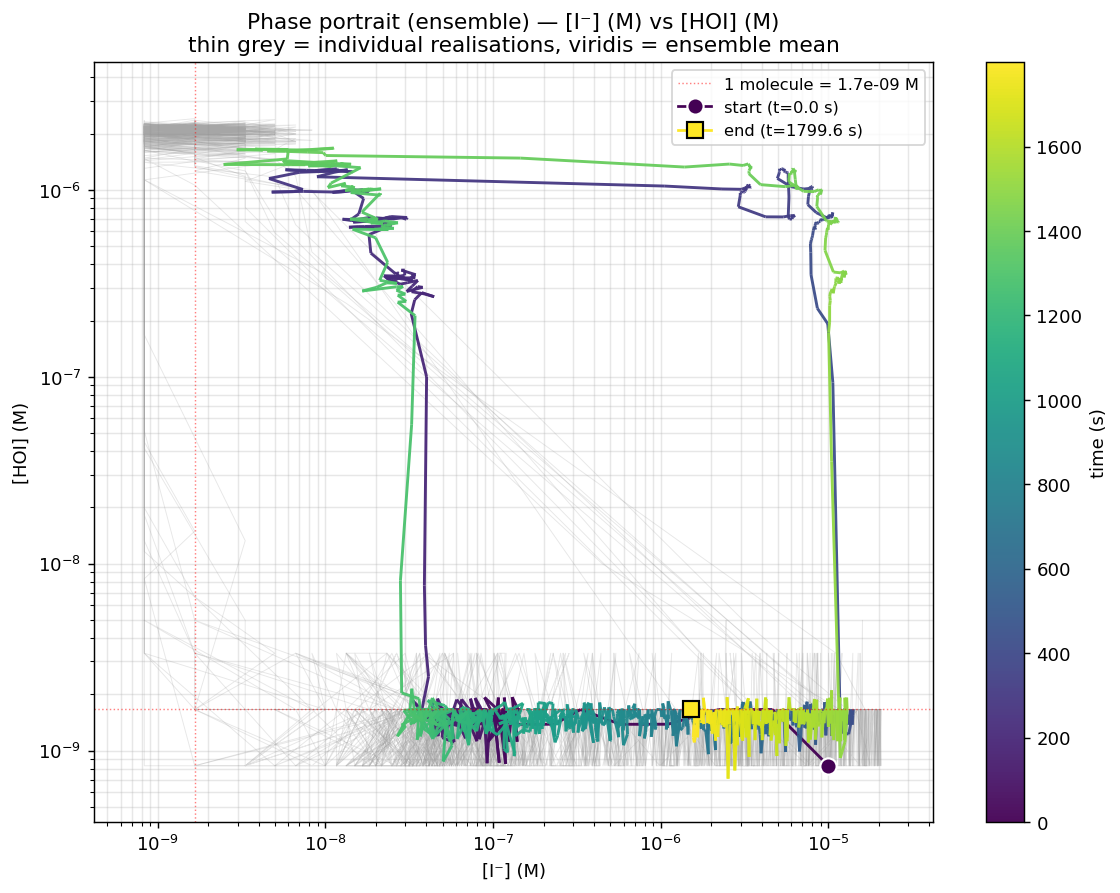}
  \caption{Projection onto $([\mathrm{I}^-], [\mathrm{HOI}])$.
    Most realisations spend a substantial fraction of the
    quiescent phase near $[\mathrm{HOI}] \approx 10^{-9}~\mathrm{M}$
    ($\approx 1$ molecule per $V$), with stochastic excursions to
    $\sim 10^{-6}~\mathrm{M}$ during spikes.  The ODE's smooth
    quiescent $\mathrm{HOI}$ floor
    (Figure~\ref{fig:sim:ode-phase-i-hoi}) at
    $\sim 10^{-10}~\mathrm{M}$ is the Kurtz limit and corresponds
    to $\sim 0.06$ molecules at this volume.}
  \label{fig:sim:ssa-phase-i-hoi}
\end{figure}

The SSA realisations show the same qualitative oscillator
structure as the ODE, with three concrete and framework-predicted
differences:

\begin{itemize}
  \item \emph{Period shift.}  The SSA mean inter-spike interval
    ($\approx 1050~\mathrm{s}$) is approximately $15\%$ shorter
    than the ODE's ($\approx 1250~\mathrm{s}$).  In a given
    realisation, rate-limiting integer transitions in the slow
    recharge phase can fire earlier than the deterministic
    species-mean trajectory predicts, advancing spike onset.  The
    direction is consistent with $\gamma_{\mathrm{ODE}}$'s
    state-space-level deviation between the Kurtz limit and the
    exact CTMC at low copy (\S\ref{sec:sim:target},
    \S\ref{sec:sim:gillespie}); the magnitude is a phase-statistic
    that requires $N \gg 6$ realisations for a quantitative
    kernel-level estimate.
  \item \emph{Spike-peak agreement.}  Peak molecule counts in the
    SSA, converted to concentration, agree with ODE peak
    concentrations to within a factor of two across all six
    species (Table~\ref{tab:sim:peaks}).  The systematic SSA-low
    bias of the ensemble-mean peak relative to the ODE is the
    expected consequence of \emph{phase scatter} across $N = 6$
    realisations: stochastic variation in spike-onset time
    desynchronises the realisations slightly, and averaging
    desynchronised spikes reduces the peak height of the
    ensemble mean below that of any individual realisation.
  \item \emph{Quiescent-phase divergence.}  Where the ODE shows
    smooth analytic floors ($\mathrm{HIO}_2$ at
    $\sim 10^{-10}~\mathrm{M}$, $\mathrm{IO}_2 \bullet$ at
    $\sim 4 \times 10^{-11}~\mathrm{M}$, $\mathrm{MnOH}^{2+}$ at
    $\sim 10^{-13}~\mathrm{M}$), the SSA shows integer-valued
    populations fluctuating between $0$ and a few molecules.  At
    $V = 10^{-15}~\mathrm{L}$ the ODE floors correspond to
    ensemble means of $0.06$, $0.024$, and $6 \times 10^{-5}$
    molecules respectively --- below unity, hence describing
    ensembles in which most individual realisations have zero
    molecules during the quiescent phase.
\end{itemize}

\begin{table}[!h]
\centering
\caption{Spike-peak comparison.  Column 2: ODE peak concentration.
  Column 3: SSA ensemble-mean peak in molecule counts at
  $V = 10^{-15}~\mathrm{L}$.  Column 4: SSA peak converted to
  concentration via $c = N/(V N_A)$.  Column 5: SSA/ODE peak
  ratio.  Agreement is within a factor of two for every species.}
\label{tab:sim:peaks}
\begin{tabular}{lcccc}
\toprule
Species & ODE peak ($\mathrm{M}$) & SSA peak (molec.) & SSA peak ($\mathrm{M}$) & ratio \\
\midrule
$\mathrm{I}^-$       & $\sim 2 \times 10^{-5}$ & $\sim 10^{4}$         & $\sim 1.7 \times 10^{-5}$ & $0.85$ \\
$\mathrm{I}_2$       & $\sim 10^{-3}$          & $\sim 5 \times 10^{5}$ & $\sim 8 \times 10^{-4}$   & $0.8$ \\
$\mathrm{HOI}$       & $\sim 10^{-6}$          & $\sim 10^{3}$         & $\sim 1.7 \times 10^{-6}$ & $1.7$ \\
$\mathrm{HIO}_2$     & $\sim 10^{-5}$          & $\sim 5 \times 10^{3}$ & $\sim 8 \times 10^{-6}$   & $0.8$ \\
$\mathrm{IO}_2\bullet$ & $\sim 3 \times 10^{-6}$ & $\sim 10^{3}$         & $\sim 1.7 \times 10^{-6}$ & $0.55$ \\
$\mathrm{MnOH}^{2+}$ & $\sim 10^{-8}$          & $\sim 3$              & $\sim 5 \times 10^{-9}$   & $0.5$ \\
\bottomrule
\end{tabular}
\end{table}

The third bullet is the most pedagogically informative observation
in the section.  The ODE quiescent floors are not numerical
artifacts: they are the analytical solution of the rate equations
in the time-asymptotic limit between spikes, and they are
\emph{correct as Kurtz-limit ensemble means}.  At
$V = 10^{-15}~\mathrm{L}$, however, an ensemble mean of
$6 \times 10^{-5}$ $\mathrm{MnOH}^{2+}$ molecules represents a
distribution in which essentially every individual realisation
has zero molecules during quiescence, with rare single-molecule
visits.  The SSA shows this directly: most realisations have zero
$\mathrm{MnOH}^{2+}$ molecules during the quiescent phase, with
occasional single-molecule excursions.  The discrepancy between
the two simulators is therefore neither a bug in the ODE solver
nor a sampling artifact in the SSA but a framework-predicted
consequence of evaluating $\Lk_3$ at finite $V$:
$\gamma_{\mathrm{SSA}}$ vanishes in total variation throughout
(\S\ref{sec:sim:gillespie}, Theorem~\ref{thm:cat-gillespie}), while
$\gamma_{\mathrm{ODE}}$ --- small in the high-copy spike phase
where the Kurtz limit is faithful --- becomes non-trivial in the
low-copy quiescent phase where discrete chemistry dominates.  This
confirms (P2) and (P3): the SSA and ODE outputs are operationally
distinguishable in precisely the regime where the chapter predicts
they should be, and the structure of the divergence is interpretable
from the framework rather than from numerical accident.

\subsubsection{What is confirmed; what remains open}

\begin{insightbox}[Framework deliverables and open items]

\textbf{Confirmed empirically.}
\begin{itemize}
  \item (P1)~The base BR oscillator emerges from
    \texttt{briggsRauscherDE} via $F \circ \PhiOp$.  Both
    simulators (SSA sampling $F_{\mathrm{BR}}$; ODE realising its Kurtz limit)
    exhibit relaxation-oscillator structure; oscillation is robust
    across the two semantic layers of $\PhiOp$.
  \item (P2)~The $\gamma$-cell distinction between the SSA samples
    and the ODE Kurtz limit is operational and visible in the
    figures: the two agree in the high-copy spike phase (where the
    Kurtz limit is faithful) and diverge in the low-copy quiescent
    phase (where the discrete chemistry dominates).
  \item (P3)~The form of the divergence is interpretable through
    the Kurtz theorem and the $\gamma$-cell structure of
    \S\ref{sec:sim:target}: the framework explains \emph{why}
    the ODE smooths the quiescent phase, and \emph{why} this
    smoothing produces fractional molecule counts at the chosen
    volume.
\end{itemize}

\medskip
\textbf{Not tested at this stage.}
\begin{itemize}
  \item \emph{Variant comparison.}  The empirical content of
    Proposition~\ref{prop:variant-comparison} is the comparison
    of the kernels
    $F\!\circ\!\PhiOp(\mathtt{briggsRauscherDE})$,
    $F\!\circ\!\PhiOp(\mathtt{broCODE5VarNetwork})$,
    $F\!\circ\!\PhiOp(\mathtt{bufferedH})$, and
    $F\!\circ\!\PhiOp(\mathtt{furrowLikePoolNetwork})$ in the
    canonical $\BorelStoch$ embedding.  We have sampled only
    $F_{\mathrm{BR}}$ here.
  \item \emph{Kinetic isotope effect.}
    Remark~\ref{rmk:sim:pi4-cannot-see-BR}~(ii) predicts that
    replacing the methylene hydrogens of malonic acid by deuterium
    would alter the $\mathrm{C}$--$\mathrm{H}$ abstraction rate
    by a primary KIE.  This $\Lk_4$-dependent prediction is
    invisible to the present $\Lk_3$ simulator (the scalar
    \texttt{rxnRate} cannot register it) and requires either
    experimental input or an $\Lk_4$-resolved successor
    implementation.
  \item \emph{Quantitative agreement with experiment.}  Period
    and spike amplitudes here are set by the parameterisation
    (target period $\sim 10^{3}~\mathrm{s}$), not derived from
    independent physical conditions.  Matching the classical
    batch BR period ($30$--$60~\mathrm{s}$ at $\mathrm{mL}$
    volumes) requires re-tuning rate constants to those
    conditions and re-running.
  \item \emph{Volume convergence.}  The Kurtz theorem
    \cite{Kurtz1972} predicts that the SSA distribution converges
    to the ODE limit as $V$ grows.  A volume sweep at fixed
    parameterisation would make this convergence empirically
    visible and provide a quantitative measure of
    $\gamma_{\mathrm{ODE}}$ as a function of $V$.
\end{itemize}

\medskip
\textbf{Ensemble size.}  $N = 6$ is sufficient for the
qualitative displays of timing, amplitude envelope, and
quiescent-phase chatter.  Quantitative kernel-level comparisons
in $\BorelStoch$ (TV distance, Wasserstein-1, relative entropy)
require $N \gtrsim 100$, a natural next step the present
framework is built to support.
\end{insightbox}

\subsubsection{Closing}

The simulation results do not validate the framework in any
absolute sense: a framework with no empirical content can pass
unfalsified by any data.  What the figures establish instead is
that the framework's predictions for BR are structurally
distinguishable from the generic statement ``the simulator
produces some output''.  The kernel
$F_{\mathrm{BR}} := F(\PhiOp(\mathtt{briggsRauscherDE})) \in \BorelStoch$
is a specific mathematical object computed via the construction
of \S\S\ref{sec:sim:target}--\ref{sec:sim:L4coarse}; its SSA
samples and its ODE Kurtz limit are two operational evaluations
of this object (the SSA samples $F_{\mathrm{BR}}$ directly; the
ODE realises its species-mean Kurtz limit), and their visible
structural difference at the chosen volume is the chapter's
$\gamma$-cell decomposition in concrete form.

The framework earns its keep here in three concrete ways:

\begin{enumerate}[label=(\arabic*)]
  \item It \emph{predicts.}  The SSA samples and ODE Kurtz limit
    must differ in low-copy regions of state space, with the form
    of the difference determined by which semantic layer (exact
    CTMC vs.\ $V\!\to\!\infty$ Kurtz limit) is being evaluated.
    The figures confirm this.  The prediction could have failed
    in several ways --- the two simulators could have agreed in
    the quiescent phase, the ODE could have failed to oscillate
    while the SSA succeeded (or vice versa), the peak amplitudes
    could have differed by orders of magnitude --- and none did.
  \item It supplies \emph{vocabulary.}  Every numerical claim
    in this section is locatable in the chapter's framework
    decomposition: $V$ is a user-chosen simulation parameter
    (separate from $\ThetaSim$); $\ThetaSim$ contains the
    parameterisation and network variant; the simulator type
    selects between the SSA and ODE morphisms in
    $\ParaC_W(\Hask)$; the kernel $F_{\mathrm{BR}}$ is the
    derived $\BorelStoch$-object; $\gamma_{\mathrm{ODE}}$ is the
    Kurtz-limit deviation; the discrete quiescent-phase chatter
    is $\Lk_3$-stochastic truth that the framework's
    $\Lk_3$-deterministic representation cannot capture at finite
    $V$.  This decomposition is tacit in standard simulation
    practice; the framework makes it explicit.
  \item It \emph{equips.}  Proposition~\ref{prop:variant-comparison}
    embeds every BR variant into $\BorelStoch$ via the same
    $F \circ \PhiOp$; comparison is then comparison of Markov
    kernels in a single category, with metric choices (TV,
    Wasserstein-1, relative entropy) available.  The variant
    comparisons, KIE prediction, and volume sweep listed above
    are the natural next steps; the machinery to perform them
    is in place.
\end{enumerate}

The classical chemical phenomena of the Briggs--Rauscher
oscillator --- period, amplitude, autocatalytic feedback, the
distinction between the spike and quiescent phases --- emerge
through this framework.  What the framework adds is the discipline
of distinguishing three categories.  \emph{Essential features} ---
such as the oscillation onset itself --- are tied to the $\Lk_2$
stoichiometric feedback topology and the $\Lk_3$ Jacobian
sign-pattern; they survive across both simulators and would
survive across network variants retaining the autocatalytic
skeleton.  \emph{Tunable features} --- the oscillation period
($\sim\!10^3~\mathrm{s}$ at the chosen parameterisation) and
spike amplitudes --- are physical functionals of the $\Lk_3$
rate constants and the $\pi_4$-coarse $\Lk_4$ data of
\S\ref{sec:sim:L4coarse}; re-parameterising to classical batch BR
conditions would shift them continuously to the experimental
$30$--$60~\mathrm{s}$ regime.  \emph{Scale-dependent representational
artifacts} --- the smooth ODE quiescent floors at
$\sim 10^{-13}~\mathrm{M}$ for $\mathrm{MnOH}^{2+}$, and the
$\sim\!15\%$ SSA-vs-ODE period shift --- are not chemistry but
artifacts of evaluating an essentially-discrete process through
its continuous Kurtz limit.  The floor values are valid ensemble
means at all $V$ but acquire physical referents (as occupied
discrete states) only at sufficiently large $V$; at
$V = 10^{-15}~\mathrm{L}$ they correspond to sub-unity counts
that no individual realisation can occupy.  The $15\%$ drift,
distinct in V-dependence, is a $\gamma_{\mathrm{ODE}}$ residual
that vanishes as $V \to \infty$ by the Kurtz theorem
\cite{Kurtz1972}.  Both manifest at $V = 10^{-15}~\mathrm{L}$
because the volume is too small for the continuous limit to be
faithful to individual discrete trajectories.

This tripartite separation is tacit in standard simulation
practice; the framework makes it explicit.  Simulator outputs
then become inputs to inference about chemistry --- the
essential-versus-tunable boundary is what falsification tests
should target; the scale-dependent representational artifacts
are what should be disregarded --- rather than illustrations of
dynamics in which all three categories are silently conflated.

%% file: chapters/ch_conclusion.tex
\section{Conclusion}
\label{sec:conclusion}

\subsection{The tower in retrospect}
\label{sec:concl-retrospect}

This monograph constructed a canonical tower
\[
  \Lk_0 \to \Lk_1 \to \Lk_2 \to \Lk_3 \to \Lk_4 \to
  \Lk_{4.5} \to \Lk_5 \to \Lk_6 \dashrightarrow \Lk_7
\]
of categorical levels for chemical reactions, each step forced from
the one below by a non-trivial cokernel in the automorphism exact
sequence
\[
  1 \to \ker\varphi_k \to \mathrm{Aut}(\Lk_k)
  \xrightarrow{\varphi_k} \mathrm{Aut}(\Lk_{k-1})
  \to \mathrm{coker}(\varphi_k) \to 1:
\]
each non-trivial cokernel class names a pair of reactions that
physics distinguishes but the lower-level forgetful functor $U_k$
collapses (Proposition~6.26 at $k = 1,\ldots,4$;
\S7.1, \S8.7, \S9.7, \S10.1 at the remaining levels).
A perpendicular Para dimension (Chapter~11) classifies
machine-learning architectures for chemistry as parametric morphisms
over the same levels, with zero-parameter inclusions
$\gamma_k : \Lk_k \hookrightarrow \Lk_k^{\Para}$ recovering the
exact tower.

Three rules from undergraduate chemistry became provable theorems
within the tower.
Hess's Law is the functoriality of $\FH : \Lk_0(P) \to (\mathbb{R},+)$:
the equality $\FH(r_2 \circ r_1) = \FH(r_2) + \FH(r_1)$ is the
monoidal-functor condition, not an empirical observation
(Chapter~3).
Walden inversion is the statement that the permutation
$\pi_\chi \in \mathrm{Aut}(\Lk_4(P))$ exchanging the $(R)$- and
$(S)$-enantiomers has no preimage in $\mathrm{Aut}(\Lk_{4.5}(P))$
under $\varphi_{4.5}$: the chirality-flipping generator
$E^* \in G^*(G) = \mathrm{Aut}(G) \rtimes \mathbb{Z}_2^k$ separates
the enantiomers as objects of $\Lk_{4.5}$ (Theorem~7.27).
The Longuet--Higgins sign change is the non-vanishing of the first
Stiefel--Whitney class
$w_1(L_{R_0}) \in H^1(\mathcal{C}_e(G) \setminus \Sigma_{\mathrm{CI}},
\mathbb{Z}/2)$ of the real ground-state eigenline on loops linking
the codimension-2 conical-intersection seam (\S9.3); the complex
Chern class $c_1(L_0) \in H^2$ is the codimension-matched companion.

The forcing argument is structural.
At each level, the unique minimal categorical extension with
non-trivial $\mathrm{coker}(\varphi_k)$ for the relevant reaction
pair produces, as a theorem, what was previously only a rule.
The rules did not become theorems because the mathematics was
sharpened.
They became theorems because the level was made explicit.

\subsection{A taxonomy of extension types}
\label{sec:concl-types}

The nine constructions $\Lk_0, \Lk_1, \ldots, \Lk_7$ are not nine
independent acts of imagination.
With the full tower in view, they fall into seven types,
distinguished by the kind of categorical operation each performs;
this classification could not be stated before all levels were
available.

\begin{insightbox}[The seven extension types]
\textbf{I (Free construction).}
$\Lk_0$ is the free symmetric monoidal category on the Petri net of
a reaction system (Meseguer--Montanari).
Every subsequent level is an extension of this base.

\textbf{II (Functor decoration).}
$\Lk_1$, $\Lk_2$, $\Lk_3$ each add a monoidal functor on $\Lk_0(P)$
without changing its objects or morphism type: $\FH$, $\FS$ and the
Gibbs functor $F_G^T$, and the kinetic functor
$F_P : \Lk_3 \to \mathbf{Stoch}$ respectively.

\textbf{III (Mechanism rebuild).}
$\Lk_4$ replaces $\Lk_3$ morphisms (Markov kernels) with DPO
derivation spans in $\mathbf{LGraph}$.
This is the only transition in the tower where the underlying
category changes rather than being decorated; SN1 and SN2 coincide
at $\Lk_3$ and split at $\Lk_4$ (Examples~6.41, 6.42).

\textbf{IV (Symmetry enrichment).}
$\Lk_{4.5}$ equips the mechanism level with the
$G^*(G) = \mathrm{Aut}(G) \rtimes \mathbb{Z}_2^k$-equivariant
structure (Definition~7.5); discrete, no continuous parameter
introduced.

\textbf{V (Geometric decoration).}
$\Lk_5$ adds the configuration orbifold
$\mathcal{C}_e(G) = \mathbb{R}^{3n}/(\mathrm{SE}(3) \rtimes
\mathrm{Aut}_\mu(G))$ and the PES
$V : \mathcal{C}_e(G) \to \mathbb{R}$.
Continuous: nuclear geometry, point groups, and transition-state
theory enter the structure (\S8.2.2).

\textbf{VI (Topological enrichment).}
$\Lk_6$ adds the rank-$N$ active Hilbert bundle
$\mathcal{H}_{\mathrm{el}}^{(N)} \to \mathcal{C}_e(G)$ with Berry
connection, carrying the primary invariant
$w_1(L_{R_0}) \in H^1(\cdot, \mathbb{Z}/2)$ and its companion
$c_1(L_0) \in H^2(\cdot, \mathbb{Z})$.

\textbf{VII (Algebraic deformation).}
$\Lk_7$ deforms the algebra: a continuous field
$\{A_\varepsilon\}_{\varepsilon \in [0,1]}$ of $C^*$-algebras with
classical fibre $A_0 = C_0(\mathcal{C}_e(G))$ and non-commutative
quantum fibre $A_\varepsilon$ for $\varepsilon > 0$.

Types I--VI admit Para enrichments $\Lk_k^{\Para}$ via
$\gamma_k$ (Chapter~11); Type~VII does not, because the deformation
parameter $\varepsilon$ is physical, not learnable.
This cleanly separates ML-for-chemistry (Types~I--VI) from
mathematical physics (Type~VII).
\end{insightbox}

\subsection{Open constructions}
\label{sec:concl-open}

Five constructions raised in the body remain open; each has a
precise statement, identified ingredients, and known obstructions.

\textbf{C1: The continuous field for molecules.}
Construct $\{A_\varepsilon\}_{\varepsilon \in [0,1]}$ with
$A_0 = C_0(\mathcal{C}_e(G))$ and $A_\varepsilon$ the full molecular
$C^*$-algebra, satisfying the Rieffel conditions~\cite{Rieffel1993}.
The Panati--Spohn--Teufel space-adiabatic perturbation
theory~\cite{PanatiSpohnTeufel2003}, the Hagedorn--Joye asymptotic
theory~\cite{Hagedorn1980}, the Landsman tangent groupoid
construction~\cite{Landsman1998}, and the Georgescu--Iftimovici
$N$-body framework~\cite{Georgescu2002} are all in place; no paper
has assembled them into a single $C^*$-algebraic framework for
molecules.

\textbf{C2: The groupoid $C^*$-algebra for molecular symmetry.}
Construct the transformation-groupoid algebra
$C^*(G^*(G) \rtimes Q^{\mathrm{reg}}_{\mathrm{lab}}(G))$ (\S10.7)
and show that its representation theory recovers simultaneously the
permutation-inversion selection rules, the Woodward--Hoffmann
orbital-symmetry rules, and ortho/para nuclear-spin statistics.
Renault's groupoid algebra theory~\cite{Renault1980} supplies the
abstract framework.

\textbf{C3: The $K$-theoretic obstruction at conical intersections.}
Promote $w_1(L_{R_0}) \in H^1(\cdot, \mathbb{Z}/2)$ and
$c_1(L_0) \in H^2(\cdot, \mathbb{Z})$ to classes in the algebraic
$K$-theory of $\{A_\varepsilon\}$, generalising the formal
Emmrich--Weinstein result~\cite{EmmrichWeinstein1996} to the
strict $C^*$-setting via a blow-up of the algebra at
$\Sigma_{\mathrm{CI}}$.

\textbf{O1: A categorical Deficiency Zero Theorem.}
Prove that the deficiency $\delta = n - \ell - s$ equals the
negative Euler characteristic of the stoichiometric chain complex
$\mathbb{Z}[\mathcal{C}] \xrightarrow{\partial_1} \mathbb{Z}^S$,
and derive Feinberg's Deficiency Zero
Theorem~\cite{Feinberg1989} from this homological identity.

\textbf{O2: Open mechanistic reaction networks.}
Construct the symmetric monoidal bicategory
$\mathrm{Csp}(\Lk_4^{\Para})$ of cospans of open DPO derivations in
$\mathbf{LGraph}$, extending the Baez--Pollard open Petri net
framework~\cite{BaezPollard2017} from $\Lk_3$ to $\Lk_4$.
The Lack--Soboci\'{n}ski adhesive structure on
$\mathbf{LGraph}$~\cite{LackSobocinski2004} guarantees the required
pushout complements.

\begin{insightbox}[Proximity to resolution]
O1 and O2 are within reach with currently available tools:
ingredients in place, construction pending.
C2 needs a single synthesis step.
C3 requires a new technical ingredient --- the seam blow-up --- but
the surrounding framework is clear.
C1 is the deepest: it requires C2, C3, and a resolution of the
Coulomb singularity problem, and is the correct formulation of what
it would mean to complete the $\Lk_6 \dashrightarrow \Lk_7$ step.
\end{insightbox}

\subsection{Outlook}
\label{sec:concl-outlook}

The tower has implications for three communities not previously in
close contact.

\textbf{For chemical reaction network theory.}
The Feinberg deficiency theory and the Craciun--Yu and
Anderson--Shiu multistationarity results sit entirely within
$\Lk_0$ (combinatorial hypotheses) and $\Lk_3$ (dynamical
conclusions).
The tower makes this level gap explicit for the first time and
suggests, via O1, a homological proof of the Deficiency Zero
Theorem that would place these landmark results inside a
categorical framework rather than leaving them as isolated facts.
The open mechanistic network construction O2 extends Baez--Pollard
grey-boxing from $\Lk_3$ to $\Lk_4$, giving an algebraic foundation
for compositional analysis of reaction mechanisms.

\textbf{For machine learning in chemistry.}
The Para dimension answers a question that becomes urgent as
ML models for chemistry scale: what does it mean for a model to be
physically correct at a given level?
The answer is architectural: a model is physically correct at
$\Lk_k$ if and only if the level-$k$ exact functor factors through
it.
The equivariance groups grow monotonically through the tower:
$\mathrm{Sym}(\mathcal{S})$ at $\Lk_0$, $\mathrm{Aut}(G)$ at
$\Lk_4$, $G^*(G)$ at $\Lk_{4.5}$, $\mathrm{SE}(3) \rtimes
\mathrm{Aut}_\mu(G)$ at $\Lk_5$, and the $U(N)$ structure of the
active Hilbert bundle at $\Lk_6$.
A model whose architecture respects only the $\Lk_5$ group cannot
represent $\Lk_6$ phenomena: there is no slot for the line-bundle
holonomy that distinguishes adiabatic from non-adiabatic dynamics
--- this is the architectural barrier currently separating
photocatalysis and conical-intersection chemistry from the
mainstream ML-for-chemistry stack.
Chapter~12 supplies one concrete executable instance: the
simulation functor
$\PhiOp : \Para(\Lk_0 \otimes \Lk_1 \otimes \Lk_2 \otimes \Lk_3)
\to \mathbf{Kl}(\mathsf{IO})$ composed with the denotational
realisation $F$ into the Markov category $\mathbf{BorelStoch}$
(Fritz~\cite{Fritz2020}), applied to the Briggs--Rauscher
oscillator with the $\varphi$-cell distinction between exact CTMC
and Kurtz-limit ODE made operational (\S12.9).

\textbf{For mathematical chemistry and mathematical physics.}
The constructions C1--C3 would collectively provide the first
rigorous categorical treatment of the Born--Oppenheimer
approximation as a limit rather than an ansatz.
The Landsman--Rieffel--Georgescu--Hagedorn--PST literature has
reached a point where all ingredients are available; the obstacle
is synthesis rather than any single missing result.
O1, conversely, would connect classical CRNT to homological algebra
in a way that has been suspected since Feinberg's earliest papers
but never made precise.

\subsection{Coda}
\label{sec:concl-coda}

A hydrogen atom transfers from carbon to oxygen in an enzyme active
site.
Whether that transfer is governed by classical barrier-crossing or
by quantum tunnelling depends on which level of the tower the
question is posed at.
The enzyme does not know which level the chemist is using.
It operates at all of them simultaneously.

The tower does not add new chemistry.
Hess's Law was not more true after $\FH$ was defined.
Walden inversion was not more real after the cokernel argument
identified it as a non-trivial automorphism class.
The retinal photoisomerisation was not faster because the
Stiefel--Whitney class was computed around the conical-intersection
seam.
What the tower adds is precision: each chemical fact is now located
at the unique level where it can be expressed, and the tools
available at that level are exactly those needed to prove it.

Chemistry's rules are not wrong.
Their exceptions are not failures.
They are coordinates.
The tower is the coordinate system.

The central result of this monograph is negative in form but
positive in content.
Negative: no single category suffices to express all of chemistry.
Positive: the minimal number of categories needed, their order, and
the unique forcing pair that makes each one necessary --- this is
determined, not chosen.

The map is forced.

%% file: refs.bib
@article{YuanScience2018,
  author = {Yuan, Daofu and Guan, Yafu and Chen, Wentao and Zhao, Hailin and Yu, Shengrui and Luo, Chang and Tan, Yuxin and Xie, Ting and Wang, Xingan and Sun, Zhigang and Zhang, Dong H. and Yang, Xueming},
  doi = {10.1126/science.aav1356},
  issn = {0036-8075},
  journal = {Science},
  number = {6420},
  pages = {1289--1293},
  publisher = {American Association for the Advancement of Science (AAAS)},
  title = {Observation of the geometric phase effect in the {H} + {HD} {\textrightarrow} {H}$_2$ + {D} reaction},
  volume = {362},
  year = {2018}
}

@article{YuanNatCommun2020,
  author = {Yuan, Daofu and Huang, Yin and Chen, Wentao and Zhao, Hailin and Yu, Shengrui and Luo, Chang and Tan, Yuxin and Wang, Siwen and Wang, Xingan and Sun, Zhigang and Yang, Xueming},
  doi = {10.1038/s41467-020-17381-4},
  issn = {2041-1723},
  journal = {Nature Communications},
  number = {1},
  publisher = {Springer Science and Business Media LLC},
  title = {Observation of the geometric phase effect in the {H}+{HD}{\textrightarrow}{H}$_2$+{D} reaction below the conical intersection},
  volume = {11},
  year = {2020}
}

@article{DelacretazGrantWhettenWosteZwanziger1986,
  author = {Delacr{\'e}taz, Guy and Grant, Edward R and Whetten, Robert L and W{\"o}ste, Ludger and Zwanziger, Josef W},
  journal = {Physical review letters},
  number = {24},
  pages = {2598},
  publisher = {APS},
  title = {Fractional quantization of molecular pseudorotation in {Na\(_3\)}},
  volume = {56},
  year = {1986}
}

@article{BKMP1996,
  author = {Boothroyd, Arnold I. and Keogh, William J. and Martin, Peter G. and Peterson, Michael R.},
  doi = {10.1063/1.471430},
  issn = {0021-9606},
  journal = {The Journal of Chemical Physics},
  number = {18},
  pages = {7139--7152},
  publisher = {AIP Publishing},
  title = {A refined H3 potential energy surface},
  volume = {104},
  year = {1996}
}

@article{Hess1840a,
  author = {Hess, Germain Henri},
  journal = {Bulletin Scientifique publi{\'e} par l'Acad{\'e}mie
Imp{\'e}riale des Sciences de Saint-P{\'e}tersbourg},
  note = {Presented 27 March 1840; German translation in
Annalen der Physik und Chemie
(Poggendorff) \textbf{50}, 385--404 (1840)
under the title ``Thermochemische
Untersuchungen''},
  pages = {257--272},
  title = {Recherches thermochimiques},
  volume = {7},
  year = {1840}
}

@article{Leicester1951,
  author = {Leicester, Henry M.},
  doi = {10.1021/ed028p581},
  issn = {0021-9584},
  journal = {Journal of Chemical Education},
  number = {11},
  pages = {581},
  publisher = {American Chemical Society (ACS)},
  title = {Germain Henri Hess and the foundations of thermochemistry},
  volume = {28},
  year = {1951}
}

@article{Mikolajczyk2022,
  author = {Miko{\l}ajczyk, M. and Cypryk, Marek and Gosty\'nski, Bart{\l}omiej and Kowalczewski, Jakub},
  doi = {10.3390/molecules27030599},
  journal = {Molecules},
  number = {3},
  pages = {599--599},
  title = {Nucleophilic Substitution at Heteroatoms{\textemdash}Identity Substitution Reactions at Phosphorus and Sulfur Centers: Do They Proceed in a Concerted (SN2) or Stepwise (A{\textendash}E) Way?},
  volume = {27},
  year = {2022}
}

@article{LassilaZalatanHerschlag2011,
  author = {Lassila, Jonathan K. and Zalatan, Jesse G. and Herschlag, Daniel},
  doi = {10.1146/annurev-biochem-060409-092741},
  issn = {0066-4154},
  journal = {Annual Review of Biochemistry},
  number = {1},
  pages = {669--702},
  publisher = {Annual Reviews},
  title = {Biological Phosphoryl-Transfer Reactions: Understanding Mechanism and Catalysis},
  volume = {80},
  year = {2011}
}

@article{Lewis1925,
  author = {Lewis, Gilbert N.},
  doi = {10.1073/pnas.11.3.179},
  issn = {0027-8424},
  journal = {Proceedings of the National Academy of Sciences},
  number = {3},
  pages = {179--183},
  publisher = {Proceedings of the National Academy of Sciences},
  title = {A New Principle of Equilibrium},
  volume = {11},
  year = {1925}
}

@article{Onsager1931a,
  author = {Onsager, Lars},
  doi = {10.1103/physrev.37.405},
  issn = {0031-899X},
  journal = {Physical Review},
  number = {4},
  pages = {405--426},
  publisher = {American Physical Society (APS)},
  title = {Reciprocal Relations in Irreversible Processes. I.},
  volume = {37},
  year = {1931}
}

@article{Tolman1938,
  author = {Condon, E. U.},
  doi = {10.1063/1.1707250},
  journal = {Journal of Applied Physics},
  number = {10},
  pages = {691--692},
  title = {The Principles of Statistical Mechanics},
  volume = {10},
  year = {1939}
}

@article{Seifert2012,
  author = {Seifert, Udo},
  doi = {10.1088/0034-4885/75/12/126001},
  issn = {0034-4885},
  journal = {Reports on Progress in Physics},
  number = {12},
  pages = {126001},
  publisher = {IOP Publishing},
  title = {Stochastic thermodynamics, fluctuation theorems and molecular machines},
  volume = {75},
  year = {2012}
}

@article{Arrhenius1889a,
  title={{\"U}ber die Reaktionsgeschwindigkeit bei der Inversion von Rohrzucker durch S{\"a}uren},
  author={Arrhenius, Svante},
  journal={Zeitschrift f{\"u}r physikalische Chemie},
  volume={4},
  number={1},
  pages={226--248},
  year={1889},
  publisher={De Gruyter Oldenbourg}
}

@article{Truhlar1996,
  author = {Truhlar, Donald G. and Garrett, Bruce C. and Klippenstein, Stephen J.},
  doi = {10.1021/jp953748q},
  issn = {0022-3654},
  journal = {The Journal of Physical Chemistry},
  number = {31},
  pages = {12771--12800},
  publisher = {American Chemical Society (ACS)},
  title = {Current Status of Transition-State Theory},
  volume = {100},
  year = {1996}
}

@article{WoodwardHoffmann1965a,
  author = {Woodward, R. B. and Hoffmann, Roald},
  doi = {10.1021/ja01080a054},
  issn = {0002-7863},
  journal = {Journal of the American Chemical Society},
  number = {2},
  pages = {395--397},
  publisher = {American Chemical Society (ACS)},
  title = {Stereochemistry of Electrocyclic Reactions},
  volume = {87},
  year = {1965}
}

@incollection{WoodwardHoffmann1970book,
  author = {Woodward, R.B. and Hoffmann, R.},
  booktitle = {The Conservation of Orbital Symmetry},
  doi = {10.1016/b978-1-4832-3290-4.50006-4},
  isbn = {9781483232904},
  pages = {37},
  publisher = {Elsevier},
  title = {The Conservation of Orbital Symmetry},
  year = {1971}
}

@book{Melander1960,
  author = {Melander, Lars},
  publisher = {Ronald eBooks},
  title = {Isotope effects on reaction rates},
  year = {1960}
}

@article{MelanderSaunders1980,
  author = {Shiner, V. J.},
  doi = {10.2307/1308511},
  journal = {BioScience},
  number = {7},
  pages = {538--538},
  title = {Reaction Rates of Isotopic Molecules},
  volume = {31},
  year = {1981}
}

@book{KohenLimbach2006,
  author={Kohen, Amnon and Limbach, Hans-Heinrich},
  publisher = {CRC Press},
  title = {Isotope Effects In Chemistry and Biology},
  year = {2005}
}

@book{Glasstone1941,
  author = {Glasstone, Samuel and Laidler, Keith J. and Eyring, Henry},
  title = {The theory of rate processes : the kinetics of chemical reactions, viscosity, diffusion and electrochemical phenomena},
  publisher = {McGraw-Hill},
  year = {1941}
}

@book{Bell1980,
  author = {Bell, R. P.},
  doi = {10.1007/978-1-4899-2891-7},
  isbn = {9780412213403},
  publisher = {Springer US},
  title = {The Tunnel Effect in Chemistry},
  year = {1980}
}

@article{Bigeleisen1949,
  author = {Bigeleisen, Jacob},
  doi = {10.1063/1.1747368},
  journal = {The Journal of Chemical Physics},
  number = {8},
  pages = {675--678},
  title = {The Relative Reaction Velocities of Isotopic Molecules},
  volume = {17},
  year = {1949}
}

@article{BarronBuckingham2001,
  author = {Barron, Laurence D. and Buckingham, A. David},
  doi = {10.1021/ar0100576},
  issn = {0001-4842},
  journal = {Accounts of Chemical Research},
  number = {10},
  pages = {781--789},
  publisher = {American Chemical Society (ACS)},
  title = {Time Reversal and Molecular Properties},
  volume = {34},
  year = {2001}
}

@article{KrupkaKaplanLaidler1966,
  author = {Krupka, R. M. and Kaplan, H. and Laidler, K. J.},
  doi = {10.1039/tf9666202754},
  issn = {0014-7672},
  journal = {Transactions of the Faraday Society},
  pages = {2754},
  publisher = {Royal Society of Chemistry (RSC)},
  title = {Kinetic consequences of the principle of microscopic reversibility},
  volume = {62},
  year = {1966}
}

@article{Tolman1925,
  author = {Tolman, Richard C.},
  doi = {10.1073/pnas.11.7.436},
  issn = {0027-8424},
  journal = {Proceedings of the National Academy of Sciences},
  number = {7},
  pages = {436--439},
  publisher = {Proceedings of the National Academy of Sciences},
  title = {The Principle of Microscopic Reversibility},
  volume = {11},
  year = {1925}
}

@article{Selinger2007,
  author = {Selinger, Peter},
  doi = {10.1016/j.entcs.2006.12.018},
  issn = {1571-0661},
  journal = {Electronic Notes in Theoretical Computer Science},
  pages = {139--163},
  publisher = {Elsevier BV},
  title = {Dagger Compact Closed Categories and Completely Positive Maps},
  volume = {170},
  year = {2007}
}

@incollection{Selinger2011,
  author = {Selinger, P.},
  booktitle = {Lecture Notes in Physics},
  doi = {10.1007/978-3-642-12821-9_4},
  isbn = {9783642128202},
  issn = {0075-8450},
  pages = {289--355},
  publisher = {Springer Berlin Heidelberg},
  title = {A Survey of Graphical Languages for Monoidal Categories},
  year = {2010}
}

@incollection{AbramskyCoecke2008,
  author = {Abramsky, Samson and Coecke, Bob},
  booktitle = {Handbook of Quantum Logic and Quantum Structures},
  doi = {10.1016/b978-0-444-52869-8.50010-4},
  isbn = {9780444528698},
  pages = {261--323},
  publisher = {Elsevier},
  title = {Categorical Quantum Mechanics},
  year = {2009}
}

@book{Awodey2010,
  author = {Awodey, Steve},
  doi = {10.1093/acprof:oso/9780198568612.001.0001},
  isbn = {9780198568612},
  publisher = {Oxford University Press},
  title = {Category Theory},
  year = {2006}
}

@book{Leinster2014,
  author = {Leinster, Tom},
  doi = {10.1017/cbo9781107360068},
  isbn = {9781107044241},
  publisher = {Cambridge University Press},
  title = {Basic Category Theory},
  year = {2014}
}

@article{BarylskaEtAl2018,
  author = {Barylska, Kamila and Koutny, Maciej and Mikulski, {\L}ukasz and Pi\k{a}tkowski, Marcin},
  doi = {10.1016/j.scico.2017.10.008},
  journal = {Science of Computer Programming},
  pages = {48--60},
  title = {Reversible computation vs. reversibility in Petri nets},
  volume = {151},
  year = {2017}
}

@misc{MelgrattiMezzinaUlidowski2020,
  author = {Melgratti, Hern\'an and Mezzina, Claudio Antares and Ulidowski, Irek},
  booktitle = {arXiv (Cornell University)},
  doi = {10.48550/arxiv.1910.04266},
  title = {Reversing Place Transition Nets},
  year = {2019}
}

@article{lim2020hodge,
  author = {Lim, Lek-Heng},
  doi = {10.1137/18m1223101},
  issn = {0036-1445},
  journal = {SIAM Review},
  number = {3},
  pages = {685--715},
  publisher = {Society for Industrial & Applied Mathematics (SIAM)},
  title = {Hodge Laplacians on Graphs},
  volume = {62},
  year = {2020}
}

@article{frankel2004geometry,
  author = {Frankel, Theodore and Mayer, Meinhard E.},
  doi = {10.1063/1.882494},
  issn = {0031-9228},
  journal = {Physics Today},
  number = {12},
  pages = {56--57},
  publisher = {AIP Publishing},
  title = {The Geometry of Physics: An Introduction},
  volume = {51},
  year = {1998}
}

@book{soardi2006potential,
  author = {Soardi, Paolo M.},
  booktitle = {Lecture notes in mathematics},
  doi = {10.1007/bfb0073995},
  title = {Potential Theory on Infinite Networks},
  publisher = {Springer},
  year = {1994}
}

@article{freire2008enthalpy,
  author = {Freire, Ernesto},
  doi = {10.1016/j.drudis.2008.07.005},
  issn = {1359-6446},
  journal = {Drug Discovery Today},
  number = {19-20},
  pages = {869--874},
  publisher = {Elsevier BV},
  title = {Do enthalpy and entropy distinguish first in class from best in class?},
  volume = {13},
  year = {2008}
}

@article{ladbury2010adding,
  author = {Ladbury, John E. and Klebe, Gerhard and Freire, Ernesto},
  doi = {10.1038/nrd3054},
  issn = {1474-1776},
  journal = {Nature Reviews Drug Discovery},
  number = {1},
  pages = {23--27},
  publisher = {Springer Science and Business Media LLC},
  title = {Adding calorimetric data to decision making in lead discovery: a hot tip},
  volume = {9},
  year = {2009}
}

@article{HuEtAl2017ACSCatal,
  author = {Hu, Shenshen and Soudackov, Alexander V. and Hammes-Schiffer, Sharon and Klinman, Judith P.},
  doi = {10.1021/acscatal.7b00688},
  journal = {ACS Catalysis},
  number = {5},
  pages = {3569--3574},
  title = {Enhanced Rigidification within a Double Mutant of Soybean Lipoxygenase Provides Experimental Support for Vibronically Nonadiabatic Proton-Coupled Electron Transfer Models},
  volume = {7},
  year = {2017}
}

@article{schoenlein1991first,
  author = {Schoenlein, R. W. and Peteanu, L. A. and Mathies, R. A. and Shank, C. V.},
  doi = {10.1126/science.1925597},
  issn = {0036-8075},
  journal = {Science},
  number = {5030},
  pages = {412--415},
  publisher = {American Association for the Advancement of Science (AAAS)},
  title = {The First Step in Vision: Femtosecond Isomerization of Rhodopsin},
  volume = {254},
  year = {1991}
}

@article{wang1994vibrationally,
  author = {Wang, Qing and Schoenlein, Robert W. and Peteanu, Linda A. and Mathies, Richard A. and Shank, Charles V.},
  doi = {10.1126/science.7939680},
  issn = {0036-8075},
  journal = {Science},
  number = {5184},
  pages = {422--424},
  publisher = {American Association for the Advancement of Science (AAAS)},
  title = {Vibrationally Coherent Photochemistry in the Femtosecond Primary Event of Vision},
  volume = {266},
  year = {1994}
}

@article{polli2010conical,
  author = {Polli, Dario and Alto\`e, Piero and Weingart, Oliver and Spillane, Katelyn M. and Manzoni, Cristian and Brida, Daniele and Tomasello, Gaia and Orlandi, Giorgio and Kukura, Philipp and Mathies, Richard A. and Garavelli, Marco and Cerullo, Giulio},
  doi = {10.1038/nature09346},
  issn = {0028-0836},
  journal = {Nature},
  number = {7314},
  pages = {440--443},
  publisher = {Springer Science and Business Media LLC},
  title = {Conical intersection dynamics of the primary photoisomerization event in vision},
  volume = {467},
  year = {2010}
}

@article{MeseguerMontanari1990,
  author = {Meseguer, Jos\'e and Montanari, Ugo},
  doi = {10.1016/0890-5401(90)90013-8},
  issn = {0890-5401},
  journal = {Information and Computation},
  number = {2},
  pages = {105--155},
  publisher = {Elsevier BV},
  title = {Petri nets are monoids},
  volume = {88},
  year = {1990}
}

@article{Kock2022,
  author = {Kock, Joachim},
  doi = {10.1145/3559103},
  issn = {0004-5411},
  journal = {Journal of the ACM},
  number = {1},
  pages = {1--58},
  publisher = {Association for Computing Machinery (ACM)},
  title = {Whole-grain Petri Nets and Processes},
  volume = {70},
  year = {2022}
}

@incollection{MacLane1963,
  author = {Lane, Saunders Mac},
  booktitle = {Saunders Mac Lane Selected Papers},
  doi = {10.1007/978-1-4615-7831-4_19},
  isbn = {9781461578338},
  pages = {415--433},
  publisher = {Springer New York},
  title = {Natural Associativity and Commutativity},
  year = {1979}
}

@book{MacLane1998,
  author = {Lane, Saunders Mac},
  booktitle = {Graduate texts in mathematics},
  doi = {10.1007/978-1-4612-9839-7},
  title = {Categories for the Working Mathematician},
  year = {1971},
  publisher = {Springer}
}

@incollection{LackSobocinski2004,
  author = {Lack, Stephen and Soboci\'nski, Pawe{\l}},
  booktitle = {Lecture Notes in Computer Science},
  doi = {10.1007/978-3-540-24727-2_20},
  isbn = {9783540212980},
  issn = {0302-9743},
  pages = {273--288},
  publisher = {Springer Berlin Heidelberg},
  title = {Adhesive Categories},
  year = {2004}
}

@book{FongSpivak2019,
  author = {Fong, Brendan and Spivak, David I.},
  title = {An Invitation to Applied Category Theory: Seven Sketches in Compositionality},
  year = {2019},
  publisher = {Cambridge University Press}
}

@article{Akitsu2023,
  author = {Akitsu, Takashiro},
  doi = {10.3390/compounds3020024},
  issn = {2673-6918},
  journal = {Compounds},
  number = {2},
  pages = {334--335},
  publisher = {MDPI AG},
  title = {Category Theory in Chemistry},
  volume = {3},
  year = {2023}
}

@phdthesis{Petri1962,
  author = {Petri, Carl Adam},
  note = {English translation: {Communication with automata},
Technical Report RADC-TR-65-377, Rome Air Development Center,
New York, 1966},
  school = {Institut f{\"u}r Instrumentelle Mathematik, Bonn},
  title = {Kommunikation mit {A}utomaten},
  year = {1962}
}

@article{BaezMaster2020,
  author = {Baez, John C. and Master, Jade},
  doi = {10.1017/s0960129520000043},
  issn = {0960-1295},
  journal = {Mathematical Structures in Computer Science},
  number = {3},
  pages = {314--341},
  publisher = {Cambridge University Press (CUP)},
  title = {Open Petri nets},
  volume = {30},
  year = {2020}
}

@article{Horn1972,
  author = {Horn, F.},
  doi = {10.1007/bf00255664},
  issn = {0003-9527},
  journal = {Archive for Rational Mechanics and Analysis},
  number = {3},
  pages = {172--186},
  publisher = {Springer Science and Business Media LLC},
  title = {Necessary and sufficient conditions for complex balancing in chemical kinetics},
  volume = {49},
  year = {1972}
}

@article{HornJackson1972,
  author = {Horn, F. and Jackson, R.},
  doi = {10.1007/bf00251225},
  issn = {0003-9527},
  journal = {Archive for Rational Mechanics and Analysis},
  number = {2},
  pages = {81--116},
  publisher = {Springer Science and Business Media LLC},
  title = {General mass action kinetics},
  volume = {47},
  year = {1972}
}

@article{Feinberg1987,
  author = {Feinberg, Martin},
  doi = {10.1016/0009-2509(87)80099-4},
  issn = {0009-2509},
  journal = {Chemical Engineering Science},
  number = {10},
  pages = {2229--2268},
  publisher = {Elsevier BV},
  title = {Chemical reaction network structure and the stability of complex isothermal reactors{\textemdash}I. The deficiency zero and deficiency one theorems},
  volume = {42},
  year = {1987}
}

@article{Feinberg1988,
  author = {Feinberg, Martin},
  doi = {10.1016/0009-2509(88)87122-7},
  issn = {0009-2509},
  journal = {Chemical Engineering Science},
  number = {1},
  pages = {1--25},
  publisher = {Elsevier BV},
  title = {Chemical reaction network structure and the stability of complex isothermal reactors{\textemdash}II. Multiple steady states for networks of deficiency one},
  volume = {43},
  year = {1988}
}

@article{Feinberg1995,
  author = {Feinberg, Martin},
  doi = {10.1007/bf00375614},
  issn = {0003-9527},
  journal = {Archive for Rational Mechanics and Analysis},
  number = {4},
  pages = {311--370},
  publisher = {Springer Science and Business Media LLC},
  title = {The existence and uniqueness of steady states for a class of chemical reaction networks},
  volume = {132},
  year = {1995}
}

@book{Feinberg2019,
  author = {Feinberg, Martin},
  doi = {10.1007/978-3-030-03858-8},
  isbn = {9783030038571},
  issn = {0066-5452},
  publisher = {Springer International Publishing},
  title = {Foundations of Chemical Reaction Network Theory},
  year = {2019}
}

@article{BaezPollard2017,
  author = {Baez, John C. and Pollard, Blake S.},
  doi = {10.1142/s0129055x17500283},
  issn = {0129-055X},
  journal = {Reviews in Mathematical Physics},
  number = {09},
  pages = {1750028},
  publisher = {World Scientific Pub Co Pte Ltd},
  title = {A compositional framework for reaction networks},
  volume = {29},
  year = {2017}
}

@book{atkins2023physical,
  author = {Atkins, Peter and Ratcliffe, George and Wormald, Mark and Paula, Julio de},
  doi = {10.1093/hesc/9780198830108.001.0001},
  isbn = {9780198830108},
  publisher = {Oxford University Press},
  title = {Physical Chemistry for the Life Sciences},
  year = {2023}
}

@book{cox1989codata,
  author = {Cox, John D. and Wagman, Donald D and Medvedev, V. A.},
  publisher = {Hemisphere Pub. Corp. eBooks},
  title = {CODATA key values for thermodynamics},
  year = {1989}
}

@article{Hess1840,
  author = {Hess, Germain Henri},
  journal = {Bulletin scientifique publi{\'e} par l'Acad{\'e}mie
imp{\'e}riale des sciences de Saint-P{\'e}tersbourg},
  note = {First statement of the law of constant heat summation
(Hess's Law); reprinted in {Annales de chimie et
de physique}, 3rd ser., vol.\ 3, pp.\ 129--150 (1841)},
  pages = {257--272},
  title = {Recherches thermo-chimiques},
  volume = {8},
  year = {1840}
}

@article{Kirchhoff1845,
  author = {Kirchhoff, Gustav},
  doi = {10.1002/andp.18451400402},
  journal = {Annalen der Physik und Chemie},
  note = {First statement of Kirchhoff's voltage and current
laws; volume 140 in the cumulative series,
equivalent to series 2, volume 64, in Poggendorff's {Annalen}},
  number = {4},
  pages = {497--514},
  title = {Ueber den {D}urchgang eines elektrischen {S}tromes
durch eine {E}bene, insbesondere durch eine
kreisf{\"o}rmige},
  volume = {140},
  year = {1845}
}

@article{CahnIngoldPrelog1966,
  author = {Cahn, R. S. and Ingold, Christopher and Prelog, V.},
  doi = {10.1002/anie.196603851},
  issn = {0570-0833},
  journal = {Angewandte Chemie International Edition in English},
  number = {4},
  pages = {385--415},
  publisher = {Wiley},
  title = {Specification of Molecular Chirality},
  volume = {5},
  year = {1966}
}

@article{BunkerJensen2006,
  author = {Chantry, G.W.},
  doi = {10.1088/0031-9112/31/5/033},
  journal = {Physics Bulletin},
  number = {5},
  pages = {176--176},
  title = {Molecular Symmetry and Spectroscopy},
  volume = {31},
  year = {1980}
}

@article{Ruch1972,
  author = {Ruch, Ernst},
  doi = {10.1021/ar50050a002},
  issn = {0001-4842},
  journal = {Accounts of Chemical Research},
  number = {2},
  pages = {49--56},
  publisher = {American Chemical Society (ACS)},
  title = {Algebraic aspects of the chirality phenomenon in chemistry},
  volume = {5},
  year = {1972}
}

@article{DobbelaereEtAl2024RxnInsight,
  author = {Dobbelaere, Maarten R. and Lengyel, Istv\'an and Stevens, Christian V. and Van Geem, Kevin M.},
  doi = {10.1186/s13321-024-00834-z},
  issn = {1758-2946},
  journal = {Journal of Cheminformatics},
  number = {1},
  publisher = {Springer Science and Business Media LLC},
  title = {Rxn-INSIGHT: fast chemical reaction analysis using bond-electron matrices},
  volume = {16},
  year = {2024}
}

@article{JoungEtAl2025FlowER,
  author = {Joung, Joonyoung F. and Fong, Mun Hong and Casetti, Nicholas and Liles, Jordan P. and Dassanayake, Ne S. and Coley, Connor W.},
  doi = {10.1038/s41586-025-09426-9},
  issn = {0028-0836},
  journal = {Nature},
  number = {8079},
  pages = {115--123},
  publisher = {Springer Science and Business Media LLC},
  title = {Electron flow matching for generative reaction mechanism prediction},
  volume = {645},
  year = {2025}
}

@article{Chase1998,
  title={NIST--JANAF thermochemical tables for the bromine oxides},
  author={Chase, Malcolm W},
  journal={Journal of Physical and Chemical Reference Data},
  volume={25},
  number={4},
  pages={1069--1111},
  year={1996},
  publisher={American Institute of Physics for the National Institute of Standards and~…}
}

@book{IUPACGreenBook2007,
  author = {Renner, Terry},
  doi = {10.1039/9781847557889},
  isbn = {9780854044337},
  publisher = {The Royal Society of Chemistry},
  title = {Quantities, units and symbols in physical chemistry},
  year = {2007}
}

@article{Onsager1931,
  author = {Onsager, Lars},
  doi = {10.1103/physrev.37.405},
  issn = {0031-899X},
  journal = {Physical Review},
  number = {4},
  pages = {405--426},
  publisher = {American Physical Society (APS)},
  title = {Reciprocal Relations in Irreversible Processes. I.},
  volume = {37},
  year = {1931}
}

@article{BornHaber1919,
  author = {Born, Max and Haber, Fritz},
  journal = {Verhandlungen der Deutschen Physikalischen Gesellschaft},
  note = {Born (pp.\ 679--685, received 9 October; ``\"{U}ber die
Energie der Ionengitter'') and Haber (pp.\ 750--768,
received 14 November; ``Betrachtungen zur Theorie der
W\"{a}rmet\"{o}nung''), both published 5 December 1919.
Fajans independently published related work in the same
volume (pp.\ 709 and 714)},
  pages = {679--685 and 750--768},
  title = {The {B}orn--{H}aber cycle: lattice energy via thermochemical
cycle (two companion articles)},
  volume = {21},
  year = {1919}
}

@article{Feinberg1989,
  author = {Feinberg, Martin},
  doi = {10.1016/0009-2509(89)85124-3},
  issn = {0009-2509},
  journal = {Chemical Engineering Science},
  number = {9},
  pages = {1819--1827},
  publisher = {Elsevier BV},
  title = {Necessary and sufficient conditions for detailed balancing in mass action systems of arbitrary complexity},
  volume = {44},
  year = {1989}
}

@article{OsterPerelsonKatchalsky1973,
  author = {Oster, George F. and Perelson, Alan S. and Katchalsky, Aharon},
  doi = {10.1017/s0033583500000081},
  issn = {0033-5835},
  journal = {Quarterly Reviews of Biophysics},
  number = {1},
  pages = {1--134},
  publisher = {Cambridge University Press (CUP)},
  title = {Network thermodynamics: dynamic modelling of biophysical systems},
  volume = {6},
  year = {1973}
}

@book{Alberty2003,
  author = {Alberty, Robert A.},
  doi = {10.1002/0471332607},
  isbn = {9780471228516},
  publisher = {Wiley},
  title = {Thermodynamics of Biochemical Reactions},
  year = {2003}
}

@book{AtkinsDeP2014,
  author = {Atkins, Peter and Paula, Julio de and Keeler, James},
  doi = {10.1093/hesc/9780198847816.001.0001},
  isbn = {9780198847816},
  publisher = {Oxford University Press},
  title = {Atkins{\textquoteright} Physical Chemistry},
  year = {2022}
}

@book{Kondepudi2014,
  author = {Kondepudi, Dilip and Prigogine, Ilya},
  doi = {10.1002/9781118698723},
  isbn = {9781118371817},
  publisher = {Wiley},
  title = {Modern Thermodynamics},
  year = {2014}
}

@article{NIST_WebBook,
  author = {Linstorm, P.},
  journal = {Medical Entomology and Zoology},
  pages = {1--1951},
  title = {Nist chemistry webbook, nist standard reference database number 69},
  volume = {9},
  year = {1998}
}

@article{LiebYngvason1999,
  author = {Lieb, Elliott H. and Yngvason, Jakob},
  doi = {10.1016/s0370-1573(98)00082-9},
  issn = {0370-1573},
  journal = {Physics Reports},
  number = {1},
  pages = {1--96},
  publisher = {Elsevier BV},
  title = {The physics and mathematics of the second law of thermodynamics},
  volume = {310},
  year = {1999}
}

@article{CraciunEtAl2009,
  author = {Craciun, Gheorghe and Dickenstein, Alicia and Shiu, Anne and Sturmfels, Bernd},
  doi = {10.1016/j.jsc.2008.08.006},
  issn = {0747-7171},
  journal = {Journal of Symbolic Computation},
  number = {11},
  pages = {1551--1565},
  publisher = {Elsevier BV},
  title = {Toric dynamical systems},
  volume = {44},
  year = {2009}
}

@article{GrossHill2013,
  author = {ABUHLAIL, JAWAD and LOMP, CHRISTIAN},
  doi = {10.1142/s0219498813500126},
  issn = {0219-4988},
  journal = {Journal of Algebra and Its Applications},
  number = {06},
  pages = {1350012},
  publisher = {World Scientific Pub Co Pte Lt},
  title = {ON THE NOTION OF STRONG IRREDUCIBILITY AND ITS DUAL},
  volume = {12},
  year = {2013}
}

@article{Dickenstein2016,
  author = {Kehoe, Elaine},
  doi = {10.1090/noti1359},
  issn = {1088-9477},
  journal = {Notices of the American Mathematical Society},
  number = {04},
  pages = {440--441},
  publisher = {American Mathematical Society (AMS)},
  title = {2016 Norbert Wiener Prize in Applied Mathematics},
  volume = {63},
  year = {2016}
}

@book{Hartshorne1977,
  author = {Hartshorne, Robin},
  doi = {10.1007/978-1-4757-3849-0},
  isbn = {9781441928078},
  issn = {0072-5285},
  publisher = {Springer New York},
  title = {Algebraic Geometry},
  year = {1977}
}

@article{Nernst1906,
  author = {Nernst, Walther},
  journal = {Nachrichten von der {G}esellschaft der {W}issenschaften
zu {G}{\"o}ttingen, Mathematisch-physikalische {K}lasse},
  pages = {1--40},
  title = {{\"U}ber die {B}erechnung chemischer {G}leichgewichte
aus thermischen {M}essungen},
  year = {1906}
}

@book{Planck1911,
  address = {Leipzig},
  author = {Planck, Max},
  edition = {3rd},
  publisher = {Veit},
  title = {Thermodynamik},
  year = {1911}
}

@article{Gibbs1875,
  author = {Gibbs, J. W.},
  doi = {10.2475/ajs.s3-16.96.441},
  issn = {0002-9599},
  journal = {American Journal of Science},
  number = {96},
  pages = {441--458},
  publisher = {American Journal of Science (AJS)},
  title = {On the equilibrium of heterogeneous substances},
  volume = {s3-16},
  year = {1878}
}

@article{AbrahamEtAl1988,
  author = {Abraham, Michael H. and Grellier, Patricia L. and
Prior, David V. and Morris, Jonathan J. and Taylor,
Peter J.},
  doi = {10.1039/P29880000699},
  journal = {Journal of the Chemical Society, Perkin Transactions 2},
  pages = {699--711},
  title = {Hydrogen Bonding, Part 7. A Scale of Solute
Hydrogen-Bond Acidity for Use in Partition Coefficients},
  year = {1988}
}

@article{OlmsteadBrauman1977,
  author = {Olmstead, William N. and Brauman, John I.},
  doi = {10.1021/ja00455a002},
  issn = {0002-7863},
  journal = {Journal of the American Chemical Society},
  number = {13},
  pages = {4219--4228},
  publisher = {American Chemical Society (ACS)},
  title = {Gas-phase nucleophilic displacement reactions},
  volume = {99},
  year = {1977}
}

@article{Khalifah1971,
  author = {Khalifah, Raja G.},
  doi = {10.1016/s0021-9258(18)62326-9},
  issn = {0021-9258},
  journal = {Journal of Biological Chemistry},
  number = {8},
  pages = {2561--2573},
  publisher = {Elsevier BV},
  title = {The Carbon Dioxide Hydration Activity of Carbonic Anhydrase},
  volume = {246},
  year = {1971}
}

@book{SilverBuss1992,
  title={The physical chemistry of membranes: an introduction to the structure and dynamics of biological membranes},
  author={Silver, B},
  year={2012},
  publisher={Springer Science \& Business Media}
}

@article{Lindskog1997,
  author = {Lindskog, Sven},
  doi = {10.1016/s0163-7258(96)00198-2},
  journal = {Pharmacology \& Therapeutics},
  number = {1},
  pages = {1--20},
  title = {Structure and mechanism of carbonic anhydrase},
  volume = {74},
  year = {1997}
}

@article{Fritz2020,
  author = {Fritz, Tobias},
  doi = {10.1016/j.aim.2020.107239},
  issn = {0001-8708},
  journal = {Advances in Mathematics},
  pages = {107239},
  publisher = {Elsevier BV},
  title = {A synthetic approach to Markov kernels, conditional independence and theorems on sufficient statistics},
  volume = {370},
  year = {2020}
}

@article{ChoJacobs2019,
  author = {Cho, Kenta and Jacobs, Bart},
  doi = {10.1017/s0960129518000488},
  issn = {0960-1295},
  journal = {Mathematical Structures in Computer Science},
  number = {7},
  pages = {938--971},
  publisher = {Cambridge University Press (CUP)},
  title = {Disintegration and Bayesian inversion via string diagrams},
  volume = {29},
  year = {2019}
}

@article{Feinberg1972,
  author = {Feinberg, Martin},
  doi = {10.1007/bf00255665},
  issn = {0003-9527},
  journal = {Archive for Rational Mechanics and Analysis},
  number = {3},
  pages = {187--194},
  publisher = {Springer Science and Business Media LLC},
  title = {Complex balancing in general kinetic systems},
  volume = {49},
  year = {1972}
}

@article{Kurtz1972,
  author = {Kurtz, Thomas G.},
  doi = {10.1063/1.1678692},
  issn = {0021-9606},
  journal = {The Journal of Chemical Physics},
  number = {7},
  pages = {2976--2978},
  publisher = {AIP Publishing},
  title = {The Relationship between Stochastic and Deterministic Models for Chemical Reactions},
  volume = {57},
  year = {1972}
}

@article{hoja2021qm7,
  author = {Hoja, Johannes and Medrano Sandonas, Leonardo and Ernst, Brian G. and Vazquez-Mayagoitia, Alvaro and DiStasio, Robert A. and Tkatchenko, Alexandre},
  doi = {10.1038/s41597-021-00812-2},
  issn = {2052-4463},
  journal = {Scientific Data},
  number = {1},
  publisher = {Springer Science and Business Media LLC},
  title = {QM7-X, a comprehensive dataset of quantum-mechanical properties spanning the chemical space of small organic molecules},
  volume = {8},
  year = {2021}
}

@article{vantHoff1884,
  author = {van't Hoff, M. J. H.},
  doi = {10.1002/recl.18840031003},
  issn = {0165-0513},
  journal = {Recueil des Travaux Chimiques des Pays-Bas},
  number = {10},
  pages = {333--336},
  publisher = {Wiley},
  title = {Etudes de dynamique chimique},
  volume = {3},
  year = {1884}
}

@article{Helmholtz1882,
  author = {Helmholtz, Hermann von},
  journal = {Sitzungsberichte der {K}\"oniglich {P}reussischen
{A}kademie der {W}issenschaften zu {B}erlin},
  pages = {22--39},
  title = {Die {T}hermodynamik chemischer {V}org\"ange},
  year = {1882}
}

@article{LeChatelier1884,
  author = {Le~Chatelier, Henry Louis},
  journal = {Comptes rendus de l'Acad\'emie des sciences},
  pages = {786--789},
  title = {Sur un \'enonc\'e g\'en\'eral des lois des \'equilibres
chimiques},
  volume = {99},
  year = {1884}
}

@article{MuellerRegensburger2012,
  author = {M\"uller, Stefan and Regensburger, Georg},
  doi = {10.1137/110847056},
  issn = {0036-1399},
  journal = {SIAM Journal on Applied Mathematics},
  number = {6},
  pages = {1926--1947},
  publisher = {Society for Industrial & Applied Mathematics (SIAM)},
  title = {Generalized Mass Action Systems: Complex Balancing Equilibria and Sign Vectors of the Stoichiometric and Kinetic-Order Subspaces},
  volume = {72},
  year = {2012}
}

@article{AndersonCraciunKurtz2010,
  author = {Anderson, David F. and Craciun, Gheorghe and Kurtz, Thomas G.},
  doi = {10.1007/s11538-010-9517-4},
  issn = {0092-8240},
  journal = {Bulletin of Mathematical Biology},
  number = {8},
  pages = {1947--1970},
  publisher = {Springer Science and Business Media LLC},
  title = {Product-Form Stationary Distributions for Deficiency Zero Chemical Reaction Networks},
  volume = {72},
  year = {2010}
}

@article{Gillespie1977,
  author = {Gillespie, Daniel T.},
  doi = {10.1021/j100540a008},
  issn = {0022-3654},
  journal = {The Journal of Physical Chemistry},
  number = {25},
  pages = {2340--2361},
  publisher = {American Chemical Society (ACS)},
  title = {Exact stochastic simulation of coupled chemical reactions},
  volume = {81},
  year = {1977}
}

@article{Gillespie1992,
  author = {Gillespie, Daniel T.},
  doi = {10.1016/0378-4371(92)90283-v},
  issn = {0378-4371},
  journal = {Physica A: Statistical Mechanics and its Applications},
  number = {1-3},
  pages = {404--425},
  publisher = {Elsevier BV},
  title = {A rigorous derivation of the chemical master equation},
  volume = {188},
  year = {1992}
}

@article{McQuarrie1967,
  author = {McQuarrie, Donald A.},
  doi = {10.2307/3212214},
  issn = {0021-9002},
  journal = {Journal of Applied Probability},
  number = {3},
  pages = {413--478},
  publisher = {Cambridge University Press (CUP)},
  title = {Stochastic approach to chemical kinetics},
  volume = {4},
  year = {1967}
}

@article{Kurtz1970,
  author = {Kurtz, Thomas G.},
  doi = {10.2307/3212147},
  issn = {0021-9002},
  journal = {Journal of Applied Probability},
  number = {1},
  pages = {49--58},
  publisher = {Cambridge University Press (CUP)},
  title = {Solutions of ordinary differential equations as limits of pure jump markov processes},
  volume = {7},
  year = {1970}
}

@incollection{AndersonKurtz2011,
  title={Continuous time Markov chain models for chemical reaction networks},
  author={Anderson, David F and Kurtz, Thomas G},
  booktitle={Design and analysis of biomolecular circuits: engineering approaches to systems and synthetic biology},
  pages={3--42},
  year={2011},
  publisher={Springer}
}

@book{March1992,
  author = {March, Jerry},
  title = {Advanced Organic Chemistry: Reactions, Mechanisms, and Structure},
  year = {1977},
  publisher = {Wiley}
}

@book{EhrigEtAl2006,
  author = {Ehrig, Hartmut and Taentzer, Gabriele and Prange, Ulrike and Ehrig, Karsten},
  title = {Fundamentals of Algebraic Graph Transformation},
  publisher = {Springer},
  year = {2006}
}

@article{Winstein1951,
  author = {Lutton, E. S. and Kolp, Don G.},
  doi = {10.1021/ja01150a088},
  issn = {0002-7863},
  journal = {Journal of the American Chemical Society},
  number = {6},
  pages = {2733--2735},
  publisher = {American Chemical Society (ACS)},
  title = {A Study of n-Octadecenoic Acids. II. Diffraction Patterns of trans-6- through 12-Octadecenoic Acids},
  volume = {73},
  year = {1951}
}

@article{WoodwardHoffmann1965,
  author = {Woodward, R. B. and Hoffmann, Roald},
  doi = {10.1021/ja01080a054},
  issn = {0002-7863},
  journal = {Journal of the American Chemical Society},
  number = {2},
  pages = {395--397},
  publisher = {American Chemical Society (ACS)},
  title = {Stereochemistry of Electrocyclic Reactions},
  volume = {87},
  year = {1965}
}

@article{HoffmannWoodward1968,
  author = {Hoffmann, Roald and Woodward, R. B.},
  doi = {10.1021/ar50001a003},
  journal = {Accounts of Chemical Research},
  number = {1},
  pages = {17--22},
  title = {Conservation of orbital symmetry},
  volume = {1},
  year = {1968}
}

@article{FukuiYonezawaShingu1952,
  author = {Fukui, Kenichi and Yonezawa, Teijiro and Shingu, Haruo},
  doi = {10.1063/1.1700523},
  issn = {0021-9606},
  journal = {The Journal of Chemical Physics},
  number = {4},
  pages = {722--725},
  publisher = {AIP Publishing},
  title = {A Molecular Orbital Theory of Reactivity in Aromatic Hydrocarbons},
  volume = {20},
  year = {1952}
}

@article{Walden1896,
  author = {Walden, P.},
  doi = {10.1002/cber.18990320276},
  issn = {0365-9496},
  journal = {Berichte der deutschen chemischen Gesellschaft},
  number = {2},
  pages = {1833--1855},
  publisher = {Wiley},
  title = {Ueber die gegenseitige Umwandlung optischer Antipoden},
  volume = {32},
  year = {1899}
}

@article{HughesIngold1935,
  author = {Hughes, Edward D. and Ingold, Christopher K.},
  doi = {10.1039/jr9350000244},
  issn = {0368-1769},
  journal = {Journal of the Chemical Society (Resumed)},
  pages = {244},
  publisher = {Royal Society of Chemistry (RSC)},
  title = {55. Mechanism of substitution at a saturated carbon atom. Part IV. A discussion of constitutional and solvent effects on the mechanism, kinetics, velocity, and orientation of substitution},
  year = {1935}
}

@article{Cowdrey1937,
  author = {Cowdrey, W. A. and Hughes, E. D. and Ingold, C. K. and Masterman, S. and Scott, A. D.},
  doi = {10.1039/jr9370001252},
  issn = {0368-1769},
  journal = {Journal of the Chemical Society (Resumed)},
  pages = {1252},
  publisher = {Royal Society of Chemistry (RSC)},
  title = {257. Reaction kinetics and the Walden inversion. Part VI. Relation of steric orientation to mechanism in substitutions involving halogen atoms and simple or substituted hydroxyl groups},
  year = {1937}
}

@article{FukuiNobel1982,
  author = {Fukui, Kenichi},
  doi = {10.1002/anie.198208013},
  issn = {0570-0833},
  journal = {Angewandte Chemie International Edition in English},
  number = {11},
  pages = {801--809},
  publisher = {Wiley},
  title = {The Role of Frontier Orbitals in Chemical Reactions (Nobel Lecture)},
  volume = {21},
  year = {1982}
}

@article{HoffmannNobel1982,
  author = {Hoffmann, Roald},
  doi = {10.1002/anie.198207113},
  issn = {0570-0833},
  journal = {Angewandte Chemie International Edition in English},
  number = {10},
  pages = {711--724},
  publisher = {Wiley},
  title = {Building Bridges Between Inorganic and Organic Chemistry (Nobel Lecture)},
  volume = {21},
  year = {1982}
}

@article{DielsAlder1928,
  author = {Diels, Otto and Alder, Kurt},
  doi = {10.1002/jlac.19284600106},
  issn = {0075-4617},
  journal = {Justus Liebigs Annalen der Chemie},
  number = {1},
  pages = {98--122},
  publisher = {Wiley},
  title = {Synthesen in der hydroaromatischen Reihe},
  volume = {460},
  year = {1928}
}

@article{WoodwardHoffmann1969,
  author = {Woodward, R. B. and Hoffmann, Roald},
  doi = {10.1002/anie.196907811},
  issn = {0570-0833},
  journal = {Angewandte Chemie International Edition in English},
  number = {11},
  pages = {781--853},
  publisher = {Wiley},
  title = {The Conservation of Orbital Symmetry},
  volume = {8},
  year = {1969}
}

@incollection{Clayden2012,
  author = {Clayden, Jonathan and Greeves, Nick and Warren, Stuart},
  booktitle = {Organic Chemistry},
  doi = {10.1093/hesc/9780199270293.003.0043},
  isbn = {9780199270293},
  publisher = {Oxford University Press},
  title = {Organic chemistry today},
  year = {2012}
}

@article{Parker1969,
  author = {Parker, Alan J.},
  doi = {10.1021/cr60257a001},
  issn = {0009-2665},
  journal = {Chemical Reviews},
  number = {1},
  pages = {1--32},
  publisher = {American Chemical Society (ACS)},
  title = {Protic-dipolar aprotic solvent effects on rates of bimolecular reactions},
  volume = {69},
  year = {1969}
}

@article{Streitwieser1958,
  author = {Kokes, R. J. and Emmett, P. H.},
  doi = {10.1021/ja01542a015},
  issn = {0002-7863},
  journal = {Journal of the American Chemical Society},
  number = {9},
  pages = {2082--2086},
  publisher = {American Chemical Society (ACS)},
  title = {Chemisorption of Nitrogen on Nickel Catalysts},
  volume = {80},
  year = {1958}
}

@book{HelgakerJorgensenOlsen2000,
  author = {Helgaker, Trygve and J{\o}rgensen, Poul and Olsen, Jeppe},
  doi = {10.1002/9781119019572},
  isbn = {9780471967552},
  publisher = {Wiley},
  title = {Molecular Electronic-Structure Theory},
  year = {2000}
}

@article{Yarkony1996,
  author = {Yarkony, David R.},
  doi = {10.1103/revmodphys.68.985},
  issn = {0034-6861},
  journal = {Reviews of Modern Physics},
  number = {4},
  pages = {985--1013},
  publisher = {American Physical Society (APS)},
  title = {Diabolical conical intersections},
  volume = {68},
  year = {1996}
}

@article{Eyring1935,
  author = {Eyring, Henry},
  doi = {10.1063/1.1749604},
  issn = {0021-9606},
  journal = {The Journal of Chemical Physics},
  number = {2},
  pages = {107--115},
  publisher = {AIP Publishing},
  title = {The Activated Complex in Chemical Reactions},
  volume = {3},
  year = {1935}
}

@article{BornOppenheimer1927,
  author = {Born, M. and Oppenheimer, R.},
  doi = {10.1002/andp.19273892002},
  issn = {0003-3804},
  journal = {Annalen der Physik},
  number = {20},
  pages = {457--484},
  publisher = {Wiley},
  title = {Zur Quantentheorie der Molekeln},
  volume = {389},
  year = {1927}
}

@article{Berry1984,
  author = {Berry, Michael Victor},
  doi = {10.1098/rspa.1984.0023},
  issn = {0080-4630},
  journal = {Proceedings of the Royal Society of London. A. Mathematical and Physical Sciences},
  number = {1802},
  pages = {45--57},
  publisher = {The Royal Society},
  title = {Quantal phase factors accompanying adiabatic changes},
  volume = {392},
  year = {1984}
}

@article{Herges1994,
  author = {Herges, Rainer},
  doi = {10.1002/anie.199402551},
  issn = {0570-0833},
  journal = {Angewandte Chemie International Edition in English},
  number = {3},
  pages = {255--276},
  publisher = {Wiley},
  title = {Organizing Principle of Complex Reactions and Theory of Coarctate Transition States},
  volume = {33},
  year = {1994}
}

@article{FukuiIRC1981,
  author = {Fukui, Kenichi},
  doi = {10.1021/ar00072a001},
  issn = {0001-4842},
  journal = {Accounts of Chemical Research},
  number = {12},
  pages = {363--368},
  publisher = {American Chemical Society (ACS)},
  title = {The path of chemical reactions - the IRC approach},
  volume = {14},
  year = {1981}
}

@article{MillerHandyAdams1980,
  author = {Miller, William H. and Handy, Nicholas C. and Adams, John E.},
  doi = {10.1063/1.438959},
  issn = {0021-9606},
  journal = {The Journal of Chemical Physics},
  number = {1},
  pages = {99--112},
  publisher = {AIP Publishing},
  title = {Reaction path Hamiltonian for polyatomic molecules},
  volume = {72},
  year = {1980}
}

@article{MilnorMorse1963,
  author = {Smiley, M. F. and Milnor, J.},
  doi = {10.2307/2312441},
  issn = {0002-9890},
  journal = {The American Mathematical Monthly},
  number = {8},
  pages = {936},
  publisher = {JSTOR},
  title = {Morse Theory.},
  volume = {71},
  year = {1964}
}

@article{SatakeVManifold1956,
  author = {Satake, I.},
  doi = {10.1073/pnas.42.6.359},
  issn = {0027-8424},
  journal = {Proceedings of the National Academy of Sciences},
  number = {6},
  pages = {359--363},
  publisher = {Proceedings of the National Academy of Sciences},
  title = {ON A GENERALIZATION OF THE NOTION OF MANIFOLD},
  volume = {42},
  year = {1956}
}

@article{WangGAFF2004,
  author = {Wang, Junmei and Wolf, Romain M. and Caldwell, James W. and Kollman, Peter A. and Case, David A.},
  doi = {10.1002/jcc.20035},
  issn = {0192-8651},
  journal = {Journal of Computational Chemistry},
  number = {9},
  pages = {1157--1174},
  publisher = {Wiley},
  title = {Development and testing of a general amber force field},
  volume = {25},
  year = {2004}
}

@article{TianFF19SB2019,
  author = {Tian, Chuan and Kasavajhala, Koushik and Belfon, Kellon A. A. and Raguette, Lauren and Huang, He and Migues, Angela N. and Bickel, John and Wang, Yuzhang and Pincay, Jorge and Wu, Qin and Simmerling, Carlos},
  doi = {10.1021/acs.jctc.9b00591},
  issn = {1549-9618},
  journal = {Journal of Chemical Theory and Computation},
  number = {1},
  pages = {528--552},
  publisher = {American Chemical Society (ACS)},
  title = {ff19SB: Amino-Acid-Specific Protein Backbone Parameters Trained against Quantum Mechanics Energy Surfaces in Solution},
  volume = {16},
  year = {2019}
}

@article{brooks2009charmm,
  author = {Brooks, Bernard R. and Brooks, Charles L. and MacKerell, Alexander D. and Nilsson, Lennart and Petrella, Robert J. and Roux, Beno\^{\i}t and Won, Youngdo and Archontis, Georgios and Bartels, Christian and Boresch, Stefan and Caflisch, Amedeo and Caves, Leo S. D. and Cui, Qiang and Dinner, Aaron R. and Feig, Michael and Fischer, Stefan and Gao, Jun and Hodo\v{s}\v{c}ek, Milan and Im, Wonpil and Kuczera, Krzysztof and Lazaridis, Themis and Ma, Jianpeng and Ovchinnikov, Victor and Paci, Emanuele and Pastor, Richard W. and Post, Carol Beth and Pu, Jingzhi and Schaefer, Michael and Tidor, Bruce and Venable, Richard M. and Woodcock, H. Lee and Wu, Xiaojing and Yang, Wei and York, Darrin M. and Karplus, Martin},
  doi = {10.1002/jcc.21287},
  journal = {Journal of Computational Chemistry},
  number = {10},
  pages = {1545--1614},
  title = {CHARMM: The biomolecular simulation program},
  volume = {30},
  year = {2009}
}

@article{jorgensen1988opls,
  author = {Jorgensen, William L. and Tirado-Rives, Julian},
  doi = {10.1021/ja00214a001},
  journal = {Journal of the American Chemical Society},
  number = {6},
  pages = {1657--1666},
  title = {The OPLS [optimized potentials for liquid simulations] potential functions for proteins, energy minimizations for crystals of cyclic peptides and crambin},
  volume = {110},
  year = {1988}
}

@article{jorgensen1996development,
  author = {Jorgensen, William L. and Maxwell, David S. and Tirado-Rives, Julian},
  doi = {10.1021/ja9621760},
  issn = {0002-7863},
  journal = {Journal of the American Chemical Society},
  number = {45},
  pages = {11225--11236},
  publisher = {American Chemical Society (ACS)},
  title = {Development and Testing of the OPLS All-Atom Force Field on Conformational Energetics and Properties of Organic Liquids},
  volume = {118},
  year = {1996}
}

@article{EvansPolanyi1935,
  author = {Evans, M. G. and Polanyi, Michael},
  doi = {10.1039/tf9353100875},
  journal = {Transactions of the Faraday Society},
  pages = {875--875},
  title = {Some applications of the transition state method to the calculation of reaction velocities, especially in solution},
  volume = {31},
  year = {1935}
}

@article{Westheimer1961,
  author = {Westheimer, F. H.},
  doi = {10.1021/cr60211a004},
  issn = {0009-2665},
  journal = {Chemical Reviews},
  number = {3},
  pages = {265--273},
  publisher = {American Chemical Society (ACS)},
  title = {The Magnitude of the Primary Kinetic Isotope Effect for Compounds of Hydrogen and Deuterium.},
  volume = {61},
  year = {1961}
}

@article{GonzalezSchlegel1989,
  author = {Gonzalez, Carlos and Schlegel, H. Bernhard},
  doi = {10.1063/1.456010},
  issn = {0021-9606},
  journal = {The Journal of Chemical Physics},
  number = {4},
  pages = {2154--2161},
  publisher = {AIP Publishing},
  title = {An improved algorithm for reaction path following},
  volume = {90},
  year = {1989}
}

@article{NISTWebBook,
  author = {Linstorm, P.},
  journal = {Medical Entomology and Zoology},
  pages = {1--1951},
  title = {Nist chemistry webbook, nist standard reference database number 69},
  volume = {9},
  year = {1998}
}

@article{TruhlarGarrettKlippenstein1996,
  author = {Truhlar, Donald G. and Hase, William L. and Hynes, James T.},
  doi = {10.1021/j150644a044},
  issn = {0022-3654},
  journal = {The Journal of Physical Chemistry},
  number = {26},
  pages = {5523--5523},
  publisher = {American Chemical Society (ACS)},
  title = {Current status of transition-state theory},
  volume = {87},
  year = {1983}
}

@article{JuanesMarcos2005,
  author = {Juanes-Marcos, Juan Carlos and Althorpe, Stuart C. and Wrede, Eckart},
  doi = {10.1126/science.1114890},
  issn = {0036-8075},
  journal = {Science},
  number = {5738},
  pages = {1227--1230},
  publisher = {American Association for the Advancement of Science (AAAS)},
  title = {Theoretical Study of Geometric Phase Effects in the Hydrogen-Exchange Reaction},
  volume = {309},
  year = {2005}
}

@article{DomckeYarkony2012,
  author = {Domcke, Wolfgang and Yarkony, David R.},
  doi = {10.1146/annurev-physchem-032210-103522},
  issn = {0066-426X},
  journal = {Annual Review of Physical Chemistry},
  number = {1},
  pages = {325--352},
  publisher = {Annual Reviews},
  title = {Role of Conical Intersections in Molecular Spectroscopy and Photoinduced Chemical Dynamics},
  volume = {63},
  year = {2012}
}

@incollection{vonNeumannWigner1929,
  author = {von Neumann, J. and Wigner, E. P.},
  booktitle = {The Collected Works of Eugene Paul Wigner},
  doi = {10.1007/978-3-662-02781-3_20},
  isbn = {9783642081545},
  pages = {294--297},
  publisher = {Springer Berlin Heidelberg},
  title = {\"Uber das Verhalten von Eigenwerten bei adiabatischen Prozessen},
  year = {1993}
}

@article{WilczekZee1984,
  author = {Wilczek, Frank and Zee, A.},
  doi = {10.1103/physrevlett.52.2111},
  issn = {0031-9007},
  journal = {Physical Review Letters},
  number = {24},
  pages = {2111--2114},
  publisher = {American Physical Society (APS)},
  title = {Appearance of Gauge Structure in Simple Dynamical Systems},
  volume = {52},
  year = {1984}
}

@incollection{DazordPatissier1991,
  address = {New York},
  author = {Dazord, Pierre and Patissier, Gilles},
  booktitle = {Symplectic Geometry, Groupoids, and
Integrable Systems},
  editor = {Dazord, Pierre and Weinstein, Alan},
  pages = {73--97},
  publisher = {Springer},
  series = {Mathematical Sciences Research Institute
Publications},
  title = {La premi\`{e}re classe de {Chern} comme
obstruction \`{a} la quantification
asymptotique},
  volume = {20},
  year = {1991}
}

@book{Renault1980,
  author = {Renault, Jean},
  doi = {10.1007/bfb0091072},
  isbn = {9783540099772},
  issn = {0075-8434},
  publisher = {Springer Berlin Heidelberg},
  title = {A Groupoid Approach to C*-Algebras},
  year = {1980}
}

@book{Melrose1993,
  author = {Melrose, Richard},
  doi = {10.1201/9781439864609},
  isbn = {9781439864609},
  publisher = {A K Peters/CRC Press},
  title = {The Atiyah-Patodi-Singer Index Theorem},
  year = {1993}
}

@article{HuEtAl2017,
  author = {Hu, Shenshen and Soudackov, Alexander V. and Hammes-Schiffer, Sharon and Klinman, Judith P.},
  doi = {10.1021/acscatal.7b00688},
  issn = {2155-5435},
  journal = {ACS Catalysis},
  number = {5},
  pages = {3569--3574},
  publisher = {American Chemical Society (ACS)},
  title = {Enhanced Rigidification within a Double Mutant of Soybean Lipoxygenase Provides Experimental Support for Vibronically Nonadiabatic Proton-Coupled Electron Transfer Models},
  volume = {7},
  year = {2017}
}

@article{JoyalStreet1991,
  author = {Joyal, Andr\'e and Street, Ross},
  doi = {10.1016/0001-8708(91)90003-p},
  issn = {0001-8708},
  journal = {Advances in Mathematics},
  number = {1},
  pages = {55--112},
  publisher = {Elsevier BV},
  title = {The geometry of tensor calculus, I},
  volume = {88},
  year = {1991}
}

@article{HagedornJoye2001,
  author = {Hagedorn, George A. and Joye, Alain},
  doi = {10.1007/s002200100562},
  journal = {Communications in Mathematical Physics},
  number = {3},
  pages = {583--626},
  title = {A Time-Dependent Born-Oppenheimer Approximation with Exponentially Small Error Estimates},
  volume = {223},
  year = {2001}
}

@article{Scerri2025,
  author = {Scerri, Eric},
  doi = {10.1007/s10698-025-09543-3},
  issn = {1386-4238},
  journal = {Foundations of Chemistry},
  number = {2},
  pages = {183--197},
  publisher = {Springer Science and Business Media LLC},
  title = {The Born-Oppenheimer Approximation and its role in the reduction of chemistry},
  volume = {27},
  year = {2025}
}

@article{Woolley2025,
  author = {Woolley, R. Guy},
  doi = {10.1007/s10698-025-09552-2},
  issn = {1386-4238},
  journal = {Foundations of Chemistry},
  number = {3},
  pages = {373--380},
  publisher = {Springer Science and Business Media LLC},
  title = {Comment on The Born-Oppenheimer approximation and its role in the reduction of chemistry},
  volume = {27},
  year = {2025}
}

@article{AgostiniCurchod2025,
  author = {Agostini, Federica and Curchod, Basile F. E.},
  doi = {10.1007/s10698-025-09559-9},
  issn = {1386-4238},
  journal = {Foundations of Chemistry},
  number = {1},
  pages = {89--109},
  publisher = {Springer Science and Business Media LLC},
  title = {A pictorial (and hopefully pedagogical) discussion on the Born{\textendash}Oppenheimer approximation},
  volume = {28},
  year = {2025}
}

@article{Woolley1978,
  author = {Woolley, R. G.},
  doi = {10.1021/ja00472a009},
  issn = {0002-7863},
  journal = {Journal of the American Chemical Society},
  number = {4},
  pages = {1073--1078},
  publisher = {American Chemical Society (ACS)},
  title = {Must a molecule have a shape?},
  volume = {100},
  year = {1978}
}

@article{kazlauskas2012thermodynamics,
  author = {Kazlauskas, Egidijus and Petrikait\.{e}, Vilma and Michailovien\.{e}, Vilma and Revuckien\.{e}, Jurgita and Matulien\.{e}, Jurgita and Grinius, Leonas and Matulis, Daumantas},
  doi = {10.1371/journal.pone.0036899},
  issn = {1932-6203},
  journal = {PLoS ONE},
  number = {5},
  pages = {e36899},
  publisher = {Public Library of Science (PLoS)},
  title = {Thermodynamics of Aryl-Dihydroxyphenyl-Thiadiazole Binding to Human Hsp90},
  volume = {7},
  year = {2012}
}

@article{krishnamurthy2007thermodynamic,
  author = {Krishnamurthy, Vijay M. and Bohall, Brooks R. and Kim, Chu-Young and Moustakas, Demetri T. and Christianson, David W. and Whitesides, George M.},
  doi = {10.1002/asia.200600360},
  issn = {1861-4728},
  journal = {Chemistry {\textendash} An Asian Journal},
  number = {1},
  pages = {94--105},
  publisher = {Wiley},
  title = {Thermodynamic Parameters for the Association of Fluorinated Benzenesulfonamides with Bovine Carbonic Anhydrase II},
  volume = {2},
  year = {2006}
}

@article{kazokaite2021experimental,
  author = {Kazokait\.{e}-Adomaitien\.{e}, Justina and Becker, Holger M. and Smirnovien\.{e}, Joana and Dubois, Ludwig J. and Matulis, Daumantas},
  doi = {10.2174/0929867327666201102112841},
  journal = {Current Medicinal Chemistry},
  number = {17},
  pages = {3361--3384},
  title = {Experimental Approaches to Identify Selective Picomolar Inhibitors for Carbonic Anhydrase IX},
  volume = {28},
  year = {2020}
}

@article{Winkelman2002,
  author = {Winkelman, J.G.M. and Voorwinde, O.K and Ottens, Marcel and Beenackers, A.A.C.M. and Janssen, L.P.B.M.},
  doi = {10.1016/s0009-2509(02)00358-5},
  journal = {Chemical Engineering Science},
  number = {19},
  pages = {4067--4076},
  title = {Kinetics and chemical equilibrium of the hydration of formaldehyde},
  volume = {57},
  year = {2002}
}

@article{Zhilinskii2006,
  author = {Zhilinski{\'\i}, B},
  doi = {10.1016/s0370-1573(00)00089-2},
  issn = {0370-1573},
  journal = {Physics Reports},
  number = {1-6},
  pages = {85--171},
  publisher = {Elsevier BV},
  title = {Symmetry, invariants, and topology in molecular models},
  volume = {341},
  year = {2001}
}

@book{Kato1966,
  author = {Kato, Tosio},
  doi = {10.1007/978-3-642-66282-9},
  isbn = {9783540586616},
  issn = {1431-0821},
  publisher = {Springer Berlin Heidelberg},
  title = {Perturbation Theory for Linear Operators},
  year = {1995}
}

@article{Simon1983,
  author = {Simon, Barry},
  doi = {10.1103/physrevlett.51.2167},
  issn = {0031-9007},
  journal = {Physical Review Letters},
  number = {24},
  pages = {2167--2170},
  publisher = {American Physical Society (APS)},
  title = {Holonomy, the Quantum Adiabatic Theorem, and Berry's Phase},
  volume = {51},
  year = {1983}
}

@article{LiebThirring1975,
  author = {Lieb, Elliott H. and Thirring, Walter E.},
  doi = {10.1103/physrevlett.35.687},
  issn = {0031-9007},
  journal = {Physical Review Letters},
  number = {11},
  pages = {687--689},
  publisher = {American Physical Society (APS)},
  title = {Bound for the Kinetic Energy of Fermions Which Proves the Stability of Matter},
  volume = {35},
  year = {1975}
}

@article{Wigner1938,
  author = {Wigner, E.},
  doi = {10.1039/tf9383400029},
  issn = {0014-7672},
  journal = {Transactions of the Faraday Society},
  pages = {29},
  publisher = {Royal Society of Chemistry (RSC)},
  title = {The transition state method},
  volume = {34},
  year = {1938}
}

@article{TGK1996,
  author = {Truhlar, Donald G. and Garrett, Bruce C. and Klippenstein, Stephen J.},
  doi = {10.1021/jp953748q},
  issn = {0022-3654},
  journal = {The Journal of Physical Chemistry},
  number = {31},
  pages = {12771--12800},
  publisher = {American Chemical Society (ACS)},
  title = {Current Status of Transition-State Theory},
  volume = {100},
  year = {1996}
}

@article{PollakTalkner2005,
  author = {Pollak, Eli and Talkner, Peter},
  doi = {10.1063/1.1858782},
  issn = {1054-1500},
  journal = {Chaos: An Interdisciplinary Journal of Nonlinear Science},
  number = {2},
  publisher = {AIP Publishing},
  title = {Reaction rate theory: What it was, where is it today, and where is it going?},
  volume = {15},
  year = {2005}
}

@article{FMKT2006,
  author = {Fern\'andez-Ramos, Antonio and Miller, James A. and Klippenstein, Stephen J. and Truhlar, Donald G.},
  doi = {10.1021/cr050205w},
  issn = {0009-2665},
  journal = {Chemical Reviews},
  number = {11},
  pages = {4518--4584},
  publisher = {American Chemical Society (ACS)},
  title = {Modeling the Kinetics of Bimolecular Reactions},
  volume = {106},
  year = {2006}
}

@article{BigeleisenMayer1947,
  author = {Bigeleisen, Jacob and Mayer, Maria Goeppert},
  doi = {10.1063/1.1746492},
  issn = {0021-9606},
  journal = {The Journal of Chemical Physics},
  number = {5},
  pages = {261--267},
  publisher = {AIP Publishing},
  title = {Calculation of Equilibrium Constants for Isotopic Exchange Reactions},
  volume = {15},
  year = {1947}
}

@article{WolfsbergStern1964,
  author = {Wolfsberg, M. and Stern, M. J.},
  doi = {10.1351/pac196408030225},
  issn = {1365-3075},
  journal = {Pure and Applied Chemistry},
  number = {3-4},
  pages = {225--242},
  publisher = {Walter de Gruyter GmbH},
  title = {Validity of some approximation procedures used in the theoretical calculation of isotope effects},
  volume = {8},
  year = {1964}
}

@book{WolfsbergVanHookPaneth2010,
  author = {Wolfsberg, Max and Hook, W. Alexander and Paneth, Piotr and Rebelo, Lu{\'\i}s Paulo N.},
  doi = {10.1007/978-90-481-2265-3},
  isbn = {9789048122646},
  publisher = {Springer Netherlands},
  title = {Isotope Effects},
  year = {2009}
}

@article{BaoTruhlar2017,
  author = {Bao, Junwei Lucas and Truhlar, Donald G.},
  doi = {10.1039/c7cs00602k},
  issn = {0306-0012},
  journal = {Chemical Society Reviews},
  number = {24},
  pages = {7548--7596},
  publisher = {Royal Society of Chemistry (RSC)},
  title = {Variational transition state theory: theoretical framework and recent developments},
  volume = {46},
  year = {2017}
}

@article{RichardsonAlthorpe2009,
  author = {Richardson, Jeremy O. and Althorpe, Stuart C.},
  doi = {10.1063/1.3267318},
  issn = {0021-9606},
  journal = {The Journal of Chemical Physics},
  number = {21},
  publisher = {AIP Publishing},
  title = {Ring-polymer molecular dynamics rate-theory in the deep-tunneling regime: Connection with semiclassical instanton theory},
  volume = {131},
  year = {2009}
}

@article{LawrenceDusekRichardson2023,
  author = {Lawrence, Joseph E. and Du\v{s}ek, Jind\v{r}ich and Richardson, Jeremy O.},
  doi = {10.1063/5.0155579},
  issn = {0021-9606},
  journal = {The Journal of Chemical Physics},
  number = {1},
  publisher = {AIP Publishing},
  title = {Perturbatively corrected ring-polymer instanton theory for accurate tunneling splittings},
  volume = {159},
  year = {2023}
}

@article{HammesSchiffer2025,
  author = {Hammes-Schiffer, Sharon},
  doi = {10.1021/acs.accounts.5c00119},
  issn = {0001-4842},
  journal = {Accounts of Chemical Research},
  number = {8},
  pages = {1335--1344},
  publisher = {American Chemical Society (ACS)},
  title = {Explaining Kinetic Isotope Effects in Proton-Coupled Electron Transfer Reactions},
  volume = {58},
  year = {2025}
}

@article{LayfieldHammesSchiffer2014,
  author = {Layfield, Joshua P. and Hammes-Schiffer, Sharon},
  doi = {10.1021/cr400400p},
  issn = {0009-2665},
  journal = {Chemical Reviews},
  number = {7},
  pages = {3466--3494},
  publisher = {American Chemical Society (ACS)},
  title = {Hydrogen Tunneling in Enzymes and Biomimetic Models},
  volume = {114},
  year = {2013}
}

@article{AlhambraCorchado2000,
  author = {Alhambra, Cristobal and Corchado, Jos\'e C. and S\'anchez, Maria Luz and Gao, Jiali and Truhlar, Donald G.},
  doi = {10.1021/ja001476l},
  issn = {0002-7863},
  journal = {Journal of the American Chemical Society},
  number = {34},
  pages = {8197--8203},
  publisher = {American Chemical Society (ACS)},
  title = {Quantum Dynamics of Hydride Transfer in Enzyme Catalysis},
  volume = {122},
  year = {2000}
}

@article{KnappRickertKlinman2002,
  author = {Knapp, Michael J. and Rickert, Keith and Klinman, Judith P.},
  doi = {10.1021/ja012205t},
  issn = {0002-7863},
  journal = {Journal of the American Chemical Society},
  number = {15},
  pages = {3865--3874},
  publisher = {American Chemical Society (ACS)},
  title = {Temperature-Dependent Isotope Effects in Soybean Lipoxygenase-1:  Correlating Hydrogen Tunneling with Protein Dynamics},
  volume = {124},
  year = {2002}
}

@article{HuOffenbacherKlinman2017,
  author = {Hu, Shenshen and Soudackov, Alexander V. and Hammes-Schiffer, Sharon and Klinman, Judith P.},
  doi = {10.1021/acscatal.7b00688},
  issn = {2155-5435},
  journal = {ACS Catalysis},
  number = {5},
  pages = {3569--3574},
  publisher = {American Chemical Society (ACS)},
  title = {Enhanced Rigidification within a Double Mutant of Soybean Lipoxygenase Provides Experimental Support for Vibronically Nonadiabatic Proton-Coupled Electron Transfer Models},
  volume = {7},
  year = {2017}
}

@techreport{Ochterski2000,
  address = {Wallingford, CT},
  author = {Ochterski, J. W.},
  institution = {Gaussian, Inc.},
  note = {White paper; revised 2022},
  title = {Thermochemistry in {Gaussian}},
  url = {https://gaussian.com/wp-content/uploads/dl/thermo.pdf},
  year = {2000}
}

@article{Zhislin1960,
  author = {Zhislin, G. M.},
  journal = {Trudy Moskovskogo Matemati\v{c}eskogo Ob\v{s}\v{c}estva},
  note = {In Russian; English translation available in various secondary sources},
  pages = {81--120},
  title = {Discussion of the spectrum of {S}chr{\"o}dinger operators for systems of many particles},
  volume = {9},
  year = {1960}
}

@article{LonguetHiggins1963Sym,
  author = {Longuet{\textendash}Higgins, H. C.},
  doi = {10.1080/00268976300100501},
  journal = {Molecular Physics},
  number = {5},
  pages = {445--460},
  title = {The symmetry groups of non-rigid molecules},
  volume = {6},
  year = {1963}
}

@article{BunkerJensen1998,
  author = {Chantry, G.W.},
  doi = {10.1088/0031-9112/31/5/033},
  journal = {Physics Bulletin},
  number = {5},
  pages = {176--176},
  title = {Molecular Symmetry and Spectroscopy},
  volume = {31},
  year = {1980}
}

@article{WilsonDectusCross1955,
  author = {Wilson, E. B. and Decius, J. C. and Cross, P. C. and Sundheim, Benson R.},
  doi = {10.1149/1.2430134},
  issn = {0013-4651},
  journal = {Journal of The Electrochemical Society},
  number = {9},
  pages = {235C},
  publisher = {The Electrochemical Society},
  title = {Molecular Vibrations: The Theory of Infrared and Raman Vibrational Spectra},
  volume = {102},
  year = {1955}
}

@article{SuttonDownes1972,
  title={Rate of hydration of formaldehyde in aqueous solution},
  author={Sutton, HC and Downes, TM},
  journal={Journal of the Chemical Society, Chemical Communications},
  volume={1},
  pages={1--2},
  year={1972},
  publisher={Royal Society of Chemistry}
}

@article{BellEvans1966,
  author = {Bell, Ronald Percy and Evans, P. G.},
  doi = {10.1098/rspa.1966.0097},
  issn = {0080-4630},
  journal = {Proceedings of the Royal Society of London. Series A. Mathematical and Physical Sciences},
  number = {1426},
  pages = {297--323},
  publisher = {The Royal Society},
  title = {Kinetics of the dehydration of methylene glycol in aqueous solution},
  volume = {291},
  year = {1966}
}

@article{Funderburk1978,
  author = {Funderburk, Lance and Aldwin, Lois and Jencks, William P.},
  doi = {10.1021/ja00485a032},
  journal = {Journal of the American Chemical Society},
  number = {17},
  pages = {5444--5459},
  title = {Mechanisms of general acid and base catalysis of the reactions of water and alcohols with formaldehyde},
  volume = {100},
  year = {1978}
}

@article{Rivlin2015,
  author = {Rivlin, Michal and Eliav, Uzi and Navon, Gil},
  doi = {10.1021/jp513020y},
  issn = {1520-6106},
  journal = {The Journal of Physical Chemistry B},
  number = {12},
  pages = {4479--4487},
  publisher = {American Chemical Society (ACS)},
  title = {NMR Studies of the Equilibria and Reaction Rates in Aqueous Solutions of Formaldehyde},
  volume = {119},
  year = {2015}
}

@article{Wolfe1995,
  author = {Wolfe, Saul and Kim, Chan-Kyung and Yang, Kiyull and Weinberg, Noham and Shi, Zheng},
  doi = {10.1021/ja00120a005},
  issn = {0002-7863},
  journal = {Journal of the American Chemical Society},
  number = {15},
  pages = {4240--4260},
  publisher = {American Chemical Society (ACS)},
  title = {Hydration of the Carbonyl Group. A Theoretical Study of the Cooperative Mechanism},
  volume = {117},
  year = {1995}
}

@article{MoelwynHughes1949,
  author = {Moelwyn-Hughes, E. A.},
  doi = {10.1098/rspa.1949.0044},
  issn = {0080-4630},
  journal = {Proceedings of the Royal Society of London. Series A. Mathematical and Physical Sciences},
  number = {1047},
  pages = {540--553},
  publisher = {The Royal Society},
  title = {The kinetics of certain reactions between methyl halides and anions in water},
  volume = {196},
  year = {1949}
}

@article{MoelwynHughes1953,
  author = {Moelwyn-Hughes, E. A.},
  doi = {10.1098/rspa.1953.0194},
  issn = {0080-4630},
  journal = {Proceedings of the Royal Society of London. Series A. Mathematical and Physical Sciences},
  number = {1142},
  pages = {386--396},
  publisher = {The Royal Society},
  title = {The kinetics of hydrolysis},
  volume = {220},
  year = {1953}
}

@article{Zafiriou1975,
  author = {Zafiriou, Oliver C.},
  journal = {Journal of Marine Research},
  number = {1},
  pages = {75--81},
  title = {Reaction of methyl halides with seawater and
marine aerosols},
  volume = {33},
  year = {1975}
}

@article{MabeyMill1978,
  author = {Mabey, W. and Mill, T.},
  doi = {10.1063/1.555572},
  issn = {0047-2689},
  journal = {Journal of Physical and Chemical Reference Data},
  number = {2},
  pages = {383--415},
  publisher = {AIP Publishing},
  title = {Critical review of hydrolysis of organic compounds in water under environmental conditions},
  volume = {7},
  year = {1978}
}

@article{Amann1993,
  author = {Amann, Anton},
  doi = {10.1007/bf01255834},
  issn = {0039-7857},
  journal = {Synthese},
  number = {1},
  pages = {125--156},
  publisher = {Springer Science and Business Media LLC},
  title = {The Gestalt problem in quantum theory: Generation of molecular shape by the environment},
  volume = {97},
  year = {1993}
}

@article{EmmrichWeinstein1996,
  author = {Emmrich, C. and Weinstein, A.},
  doi = {10.1007/bf02099256},
  issn = {0010-3616},
  journal = {Communications in Mathematical Physics},
  number = {3},
  pages = {701--711},
  publisher = {Springer Science and Business Media LLC},
  title = {Geometry of the transport equation in multicomponent WKB approximations},
  volume = {176},
  year = {1996}
}

@inproceedings{LittlejohnRawlinson2024,
  author = {Guo, Yueling and Kasihmuddin, Mohd Shareduwan Mohd and Mansor, Mohd. Asyraf and Jamaludin, Siti Zulaikha Mohd and Zamri, Nur Ezlin},
  booktitle = {AIP Conference Proceedings},
  doi = {10.1063/5.0192172},
  issn = {0094-243X},
  pages = {040007},
  publisher = {AIP Publishing},
  title = {Flexible satisfiability in discrete Hopfield neural network},
  volume = {3080},
  year = {2024}
}

@book{Landsman1998,
  author = {Landsman, N. P.},
  doi = {10.1007/978-1-4612-1680-3},
  isbn = {9781461272427},
  issn = {1439-7382},
  publisher = {Springer New York},
  title = {Mathematical Topics Between Classical and Quantum Mechanics},
  year = {1998}
}

@incollection{Landsman2007,
  author = {Landsman, N.P.},
  booktitle = {Philosophy of Physics},
  doi = {10.1016/b978-044451560-5/50008-7},
  isbn = {9780444515605},
  pages = {417--553},
  publisher = {Elsevier},
  title = {BETWEEN CLASSICAL AND QUANTUM},
  year = {2007}
}

@book{Primas1983,
  author = {Primas, Hans},
  doi = {10.1007/978-3-642-69365-6},
  isbn = {9783642693670},
  publisher = {Springer Berlin Heidelberg},
  title = {Chemistry, Quantum Mechanics and Reductionism},
  year = {1983}
}

@article{Fallani2024QIM,
  author = {Fallani, Alessio and Medrano Sandonas, Leonardo and Tkatchenko, Alexandre},
  doi = {10.1038/s41467-024-50401-1},
  issn = {2041-1723},
  journal = {Nature Communications},
  number = {1},
  publisher = {Springer Science and Business Media LLC},
  title = {Inverse mapping of quantum properties to structures for chemical space of small organic molecules},
  volume = {15},
  year = {2024}
}

@misc{GavranovicEtAl2024,
  author = {Gavranovi\'c, Bruno and Lessard, Paul and Dudzik, Andrew and Glehn, Tamara von and Ara\'ujo, Jo\~ao G. M. and Veli\v{c}kovi\'c, Petar},
  booktitle = {arXiv (Cornell University)},
  doi = {10.48550/arxiv.2402.15332},
  title = {Position: Categorical Deep Learning is an Algebraic Theory of All Architectures},
  year = {2024}
}

@article{Batzner2022,
  author = {Batzner, Simon and Musaelian, Albert and Sun, Lixin and Geiger, Mario and Mailoa, Jonathan P. and Kornbluth, Mordechai and Molinari, Nicola and Smidt, Tess E. and Kozinsky, Boris},
  doi = {10.1038/s41467-022-29939-5},
  issn = {2041-1723},
  journal = {Nature Communications},
  number = {1},
  publisher = {Springer Science and Business Media LLC},
  title = {E(3)-equivariant graph neural networks for data-efficient and accurate interatomic potentials},
  volume = {13},
  year = {2022}
}

@inproceedings{BatatiaEtAl2022,
  author = {Batatia, Ilyes and Csanyi, Gabor and Kovacs, David P and Ortner, Christoph and Simm, Gregor},
  booktitle = {Advances in Neural Information Processing Systems 35},
  doi = {10.52202/068431-0830},
  pages = {11423--11436},
  publisher = {Neural Information Processing Systems Foundation, Inc. (NeurIPS)},
  title = {MACE: Higher Order Equivariant Message Passing Neural Networks for Fast and Accurate Force Fields},
  year = {2022}
}

@article{KircherDoeppelVotsmeier2024,
  author = {Rittig, Jan G. and Mitsos, Alexander},
  doi = {10.1039/d4sc04554h},
  issn = {2041-6520},
  journal = {Chemical Science},
  number = {44},
  pages = {18504--18512},
  publisher = {Royal Society of Chemistry (RSC)},
  title = {Thermodynamics-consistent graph neural networks},
  volume = {15},
  year = {2024}
}

@misc{ShenYarkony2024,
  author = {Shen, Yifan and Yarkony, David R.},
  booktitle = {arXiv (Cornell University)},
  doi = {10.48550/arxiv.2411.01702},
  title = {Symmetry Adapted Residual Neural Network Diabatization: Conical Intersections in Aniline Photodissociation},
  year = {2024}
}

@article{DaggettYangLiuMuechler2024,
  author = {Daggett, Christopher and Yang, Kaijie and Liu, Chao-Xing and Muechler, Lukas},
  doi = {10.1021/acs.chemmater.3c02667},
  issn = {0897-4756},
  journal = {Chemistry of Materials},
  number = {8},
  pages = {3479--3489},
  publisher = {American Chemical Society (ACS)},
  title = {Toward a Topological Classification of Nonadiabaticity in Chemical Reactions},
  volume = {36},
  year = {2024}
}

@article{SchuttEtAl2018,
  author = {Sch\"utt, K. T. and Sauceda, H. E. and Kindermans, P.-J. and Tkatchenko, A. and M\"uller, K.-R.},
  doi = {10.1063/1.5019779},
  issn = {0021-9606},
  journal = {The Journal of Chemical Physics},
  number = {24},
  publisher = {AIP Publishing},
  title = {SchNet {\textendash} A deep learning architecture for molecules and materials},
  volume = {148},
  year = {2018}
}

@article{Briggs1925,
  author = {Briggs, George Edward and Haldane, John Burdon Sanderson},
  doi = {10.1042/bj0190338},
  issn = {0306-3283},
  journal = {Biochemical Journal},
  number = {2},
  pages = {338--339},
  publisher = {Portland Press Ltd.},
  title = {A Note on the Kinetics of Enzyme Action},
  volume = {19},
  year = {1925}
}

@article{BriggsRauscher1973,
  author = {Briggs, Thomas S. and Rauscher, Warren C.},
  doi = {10.1021/ed050p496},
  issn = {0021-9584},
  journal = {Journal of Chemical Education},
  number = {7},
  pages = {496},
  publisher = {American Chemical Society (ACS)},
  title = {An oscillating iodine clock},
  volume = {50},
  year = {1973}
}

@inproceedings{SteeleLeaFlood2014,
  author = {Steele, Guy L. and Lea, Doug and Flood, Christine H.},
  booktitle = {Proceedings of the 2014 ACM International Conference on Object Oriented Programming Systems Languages \& Applications},
  doi = {10.1145/2660193.2660195},
  pages = {453--472},
  publisher = {ACM},
  title = {Fast splittable pseudorandom number generators},
  year = {2014}
}

@article{DeKepperEpstein1982,
  author = {De Kepper, Patrick and Epstein, Irving R.},
  doi = {10.1021/ja00365a012},
  issn = {0002-7863},
  journal = {Journal of the American Chemical Society},
  number = {1},
  pages = {49--55},
  publisher = {American Chemical Society (ACS)},
  title = {Mechanistic study of oscillations and bistability in the Briggs-Rauscher reaction},
  volume = {104},
  year = {1982}
}

@article{NoyesFurrow1982,
  author = {Noyes, Richard M. and Furrow, Stanley D.},
  doi = {10.1021/ja00365a011},
  issn = {0002-7863},
  journal = {Journal of the American Chemical Society},
  number = {1},
  pages = {45--48},
  publisher = {American Chemical Society (ACS)},
  title = {The oscillatory Briggs-Rauscher reaction. 3. A skeleton mechanism for oscillations},
  volume = {104},
  year = {1982}
}

@article{Wegscheider1901,
  author = {Wegscheider, Rud},
  doi = {10.1007/bf01517498},
  issn = {0026-9247},
  journal = {Monatshefte fr Chemie},
  number = {8},
  pages = {849--906},
  publisher = {Springer Science and Business Media LLC},
  title = {ber simultane Gleichgewichte und die Beziehungen zwischen Thermodynamik und Reactionskinetik homogener Systeme},
  volume = {22},
  year = {1901}
}

@article{LackSobocinski2005,
  author = {Lack, Stephen and Soboci\'nski, Pawe{\l}},
  doi = {10.1051/ita:2005028},
  issn = {0988-3754},
  journal = {RAIRO - Theoretical Informatics and Applications},
  number = {3},
  pages = {511--545},
  publisher = {EDP Sciences},
  title = {Adhesive and quasiadhesive categories},
  volume = {39},
  year = {2005}
}

@inproceedings{EhrigPfenderSchneider1973,
  author = {Ehrig, H. and Pfender, M. and Schneider, H. J.},
  booktitle = {14th Annual Symposium on Switching and Automata Theory (swat 1973)},
  doi = {10.1109/swat.1973.11},
  pages = {167--180},
  publisher = {IEEE},
  title = {Graph-grammars: An algebraic approach},
  year = {1973}
}

@book{EhrigEhrigPrangeTaentzer2006,
  author = {Ehrig, Hartmut and Ehrig, Karsten and Prange, Ulrike and Taentzer, Gabriele},
  publisher = {Springer-Verlag},
  title = {Fundamentals of Algebraic Graph Transformation},
  year = {2006}
}

@article{BonchiEtAl2022SDRTII,
  author = {Bonchi, Filippo and Gadducci, Fabio and Kissinger, Aleks and Sobocinski, Pawel and Zanasi, Fabio},
  doi = {10.1017/s0960129522000317},
  issn = {0960-1295},
  journal = {Mathematical Structures in Computer Science},
  number = {4},
  pages = {511--541},
  publisher = {Cambridge University Press (CUP)},
  title = {String diagram rewrite theory II: Rewriting with symmetric monoidal structure},
  volume = {32},
  year = {2022}
}

@article{AndersenFlammMerkleStadler2013,
  author = {Andersen, Jakob L and Flamm, Christoph and Merkle, Daniel and Stadler, Peter F},
  doi = {10.1186/1759-2208-4-4},
  issn = {1759-2208},
  journal = {Journal of Systems Chemistry},
  number = {1},
  publisher = {Springer Science and Business Media LLC},
  title = {Inferring chemical reaction patterns using rule composition in graph grammars},
  volume = {4},
  year = {2013}
}

@article{AndersenFlammMerkleStadler2017,
  author = {Andersen, Jakob L. and Flamm, Christoph and Merkle, Daniel and Stadler, Peter F.},
  doi = {10.1098/rsta.2016.0354},
  issn = {1364-503X},
  journal = {Philosophical Transactions of the Royal Society A: Mathematical, Physical and Engineering Sciences},
  number = {2109},
  pages = {20160354},
  publisher = {The Royal Society},
  title = {An intermediate level of abstraction for computational systems chemistry},
  volume = {375},
  year = {2017}
}

@incollection{AndersenFlammMerkleStadler2016,
  author = {Andersen, Jakob L. and Flamm, Christoph and Merkle, Daniel and Stadler, Peter F.},
  booktitle = {Lecture Notes in Computer Science},
  doi = {10.1007/978-3-319-40530-8_5},
  isbn = {9783319405292},
  issn = {0302-9743},
  pages = {73--88},
  publisher = {Springer International Publishing},
  title = {A Software Package for Chemically Inspired Graph Transformation},
  year = {2016}
}

@incollection{DugundjiUgi1973,
  author = {Dugundji, James and Ugi, Ivar},
  booktitle = {Fortschritte der Chemischen Forschung},
  doi = {10.1007/bfb0051317},
  isbn = {3540062319},
  pages = {19--64},
  publisher = {Springer-Verlag},
  title = {An algebraic model of constitutional chemistry as a basis for chemical computer programs},
  year = {None}
}

@article{LonguetHiggins1975,
  author = {Longuet-Higgins, Hugh Christopher},
  doi = {10.1098/rspa.1975.0095},
  issn = {0080-4630},
  journal = {Proceedings of the Royal Society of London. A. Mathematical and Physical Sciences},
  number = {1637},
  pages = {147--156},
  publisher = {The Royal Society},
  title = {The intersection of potential energy surfaces in polyatomic molecules},
  volume = {344},
  year = {1975}
}

@article{LonguetHiggins1963,
  author = {Longuet-Higgins, Hugh Christopher},
  doi = {10.1098/rspa.1975.0095},
  issn = {0080-4630},
  journal = {Proceedings of the Royal Society of London. A. Mathematical and Physical Sciences},
  number = {1637},
  pages = {147--156},
  publisher = {The Royal Society},
  title = {The intersection of potential energy surfaces in polyatomic molecules},
  volume = {344},
  year = {1975}
}

@article{Yuan2020,
  author = {Yuan, Daofu and Huang, Yin and Chen, Wentao and Zhao, Hailin and Yu, Shengrui and Luo, Chang and Tan, Yuxin and Wang, Siwen and Wang, Xingan and Sun, Zhigang and Yang, Xueming},
  doi = {10.1038/s41467-020-17381-4},
  issn = {2041-1723},
  journal = {Nature Communications},
  number = {1},
  publisher = {Springer Science and Business Media LLC},
  title = {Observation of the geometric phase effect in the H+HD{\textrightarrow}H2+D reaction below the conical intersection},
  volume = {11},
  year = {2020}
}

@article{LangEtAl2024,
  author = {Lang, Lucas and Cezar, Henrique M. and Adamowicz, Ludwik and Pedersen, Thomas B.},
  doi = {10.1021/jacs.3c11467},
  issn = {0002-7863},
  journal = {Journal of the American Chemical Society},
  number = {3},
  pages = {1760--1764},
  publisher = {American Chemical Society (ACS)},
  title = {Quantum Definition of Molecular Structure},
  volume = {146},
  year = {2024}
}

@article{Rieffel1989,
  author = {Rieffel, Marc A.},
  doi = {10.1007/bf01256492},
  issn = {0010-3616},
  journal = {Communications in Mathematical Physics},
  number = {4},
  pages = {531--562},
  publisher = {Springer Science and Business Media LLC},
  title = {Deformation quantization of Heisenberg manifolds},
  volume = {122},
  year = {1989}
}

@incollection{LandsmanRamazan2001,
  author = {Landsman, N. P. and Ramazan, B.},
  booktitle = {Contemporary Mathematics},
  doi = {10.1090/conm/282/04685},
  isbn = {9780821820421},
  issn = {1098-3627},
  pages = {159--192},
  publisher = {American Mathematical Society},
  title = {Quantization of Poisson algebras associated to Lie algebroids},
  year = {2001}
}

@phdthesis{Pfeifer1980,
  author = {Pfeifer, Peter},
  booktitle = {ETH Z\"urich Research Collection},
  doi = {10.3929/ethz-a-000254631},
  title = {Chiral molecules: a superselection rule induced by the radiation field},
  year = {1980},
  school = {ETH Zürich}
}

@article{Silvera1980,
  author = {Silvera, Isaac F.},
  doi = {10.1103/revmodphys.52.393},
  issn = {0034-6861},
  journal = {Reviews of Modern Physics},
  number = {2},
  pages = {393--452},
  publisher = {American Physical Society (APS)},
  title = {The solid molecular hydrogens in the condensed phase: Fundamentals and static properties},
  volume = {52},
  year = {1980}
}

@article{FaureZhilinskii2001,
  author = {Faure, Fr\'ed\'eric and Zhilinskii, Boris},
  doi = {10.1023/a:1010912815438},
  issn = {0377-9017},
  journal = {Letters in Mathematical Physics},
  number = {3},
  pages = {219--238},
  publisher = {Springer Science and Business Media LLC},
  title = {Topological Properties of the Born{\textendash}Oppenheimer Approximation and Implications for the Exact Spectrum},
  volume = {55},
  year = {2001}
}

@article{Hawkins2008,
  author = {Hawkins, Eli},
  doi = {10.1007/s00220-008-0517-2},
  issn = {0010-3616},
  journal = {Communications in Mathematical Physics},
  number = {3},
  pages = {675--699},
  publisher = {Springer Science and Business Media LLC},
  title = {An Obstruction to Quantization of the Sphere},
  volume = {283},
  year = {2008}
}

@book{Landsman2017,
  author = {Landsman, Klaas},
  doi = {10.1007/978-3-319-51777-3},
  isbn = {9783319517766},
  issn = {0168-1222},
  publisher = {Springer International Publishing},
  title = {Foundations of Quantum Theory},
  year = {2017}
}

@article{AhnParkYang2019,
  author = {Ahn, Junyeong and Park, Sungjoon and Kim, Dongwook and Kim, Youngkuk and Yang, Bohm-Jung},
  doi = {10.1088/1674-1056/ab4d3b},
  issn = {1674-1056},
  journal = {Chinese Physics B},
  number = {11},
  pages = {117101},
  publisher = {IOP Publishing},
  title = {Stiefel{\textendash}Whitney classes and topological phases in band theory},
  volume = {28},
  year = {2019}
}

@misc{Kaufmann2016,
  author = {Kaufmann, Ralph M. and Li, Dan and Wehefritz{\textendash}Kaufmann, Birgit},
  booktitle = {arXiv (Cornell University)},
  doi = {10.48550/arxiv.1604.02792},
  title = {The Stiefel--Whitney theory of topological insulators},
  year = {2016}
}

@article{PravdivtsevEtAl2022,
  author = {Pravdivtsev, Andrey N. and Barskiy, Danila A. and H\"ovener, Jan-Bernd and Koptyug, Igor V.},
  doi = {10.3390/sym14030530},
  issn = {2073-8994},
  journal = {Symmetry},
  number = {3},
  pages = {530},
  publisher = {MDPI AG},
  title = {Symmetry Constraints on Spin Order Transfer in Parahydrogen-Induced Polarization (PHIP)},
  volume = {14},
  year = {2022}
}

@article{SutcliffeWoolley2012,
  author = {Sutcliffe, Brian T. and Woolley, R. Guy},
  doi = {10.1063/1.4755287},
  issn = {0021-9606},
  journal = {The Journal of Chemical Physics},
  number = {22},
  publisher = {AIP Publishing},
  title = {On the quantum theory of molecules},
  volume = {137},
  year = {2012}
}

@article{Amann1991,
  author = {Amann, Anton},
  doi = {10.1007/bf01192570},
  issn = {0259-9791},
  journal = {Journal of Mathematical Chemistry},
  number = {1},
  pages = {1--15},
  publisher = {Springer Science and Business Media LLC},
  title = {Chirality: A superselection rule generated by the molecular environment?},
  volume = {6},
  year = {1991}
}

@article{Zener1932,
  author = {Zener, Clarence},
  doi = {10.1098/rspa.1932.0165},
  issn = {0950-1207},
  journal = {Proceedings of the Royal Society of London. Series A, Containing Papers of a Mathematical and Physical Character},
  number = {833},
  pages = {696--702},
  publisher = {The Royal Society},
  title = {Non-adiabatic crossing of energy levels},
  volume = {137},
  year = {1932}
}

@article{MeadTruhlar1979,
  author = {Mead, C. Alden and Truhlar, Donald G.},
  doi = {10.1063/1.437734},
  issn = {0021-9606},
  journal = {The Journal of Chemical Physics},
  number = {5},
  pages = {2284--2296},
  publisher = {AIP Publishing},
  title = {On the determination of Born{\textendash}Oppenheimer nuclear motion wave functions including complications due to conical intersections and identical nuclei},
  volume = {70},
  year = {1979}
}

@book{Dixmier1977,
  address = {Amsterdam},
  author = {Dixmier, Jacques},
  note = {Translated by F. Jellett},
  publisher = {North-Holland},
  series = {North-Holland Mathematical Library},
  title = {{$C^*$}-Algebras},
  volume = {15},
  year = {1977}
}

@article{PanatiSpohnTeufel2003,
  author = {Panti, Gianluca and Spohn, Herbert and Teufel, Stefan},
  doi = {10.4310/atmp.2003.v7.n1.a6},
  issn = {1095-0761},
  journal = {Advances in Theoretical and Mathematical Physics},
  number = {1},
  pages = {145--204},
  publisher = {International Press of Boston},
  title = {Space-adiabatic perturbation theory},
  volume = {7},
  year = {2003}
}

@book{Teufel2003,
  author = {Teufel, Stefan},
  doi = {10.1007/b13355},
  isbn = {9783540407232},
  issn = {0075-8434},
  publisher = {Springer Berlin Heidelberg},
  title = {Adiabatic Perturbation Theory in Quantum Dynamics},
  year = {2003}
}

@article{Hagedorn1980,
  author = {Hagedorn, George A.},
  doi = {10.1007/bf01205036},
  issn = {0010-3616},
  journal = {Communications in Mathematical Physics},
  number = {1},
  pages = {1--19},
  publisher = {Springer Science and Business Media LLC},
  title = {A time dependent Born-Oppenheimer approximation},
  volume = {77},
  year = {1980}
}

@article{LasserTeufel2005,
  author = {Lasser, Caroline and Teufel, Stefan},
  doi = {10.1002/cpa.20087},
  issn = {0010-3640},
  journal = {Communications on Pure and Applied Mathematics},
  number = {9},
  pages = {1188--1230},
  publisher = {Wiley},
  title = {Propagation through conical crossings: An asymptotic semigroup},
  volume = {58},
  year = {2005}
}

@article{FermanianKammererLasser2008,
  author = {Kammerer, Clotilde Fermanian and Lasser, Caroline},
  doi = {10.1137/070686810},
  issn = {0036-1410},
  journal = {SIAM Journal on Mathematical Analysis},
  number = {1},
  pages = {103--133},
  publisher = {Society for Industrial & Applied Mathematics (SIAM)},
  title = {Propagation through Generic Level Crossings: A Surface Hopping Semigroup},
  volume = {40},
  year = {2008}
}

@article{ColinDeVerdiere2003,
  author = {Colin de Verdi\`ere, Yves},
  doi = {10.5802/aif.1973},
  issn = {1777-5310},
  journal = {Annales de l'Institut Fourier},
  number = {4},
  pages = {1023--1054},
  publisher = {MathDoc/Centre Mersenne},
  title = {The level crossing problem in semi-classical analysis I. The symmetric case},
  volume = {53},
  year = {2003}
}

@article{GeorgescuIftimovici2002,
  author = {Georgescu, Vladimir and Iftimovici, Andrei},
  doi = {10.1007/s002200200669},
  issn = {0010-3616},
  journal = {Communications in Mathematical Physics},
  number = {3},
  pages = {519--560},
  publisher = {Springer Science and Business Media LLC},
  title = {Crossed Products of C * -Algebras and Spectral Analysis of Quantum Hamiltonians},
  volume = {228},
  year = {2002}
}

@misc{GeorgescuIftimovici2003,
  author = {Georgescu, Vladimir and Iftimovici, Andrei},
  title = {C*-Algebras of Quantum Hamiltonians},
  year = {2001}
}

@article{GeorgescuGerardMoller2004,
  author = {Georgescu, V and G\'erard, C and M{\o}ller, J.S},
  doi = {10.1016/j.jfa.2004.03.004},
  issn = {0022-1236},
  journal = {Journal of Functional Analysis},
  number = {2},
  pages = {303--361},
  publisher = {Elsevier BV},
  title = {Commutators, {C}0-semigroups and resolvent estimates},
  volume = {216},
  year = {2004}
}

@book{ABG1996,
  author = {Amrein, Werner O. and Monvel, Anne Boutet and Georgescu, Vladimir},
  doi = {10.1007/978-3-0348-7762-6},
  isbn = {9783034877640},
  publisher = {Birkh\"auser Basel},
  title = {C0-Groups, Commutator Methods and Spectral Theory of N-Body Hamiltonians},
  year = {1996}
}

@incollection{meiswinkel1991pseudo,
  author    = {Meiswinkel, R. and K{\"o}ppel, H.},
  title     = {A pseudo-{Jahn--Teller} treatment of the {$B$} system of {$\mathrm{Na}_3$}},
  booktitle = {Small Particles and Inorganic Clusters},
  pages     = {63--66},
  publisher = {Springer},
  address   = {Berlin, Heidelberg},
  year      = {1991},
  doi       = {10.1007/978-3-642-76178-2_13},
  isbn      = {978-3-642-76180-5}
}

@article{mayer1996rovibronic,
  author = {Mayer, M. and Cederbaum, L. S. and K\"oppel, H.},
  doi = {10.1063/1.471627},
  issn = {0021-9606},
  journal = {The Journal of Chemical Physics},
  number = {22},
  pages = {8932--8942},
  publisher = {AIP Publishing},
  title = {Rovibronic coupling in the {Na}$_3$ B system},
  volume = {104},
  year = {1996}
}

@article{Epstein1965,
  author = {Epstein, Saul T.},
  doi = {10.1063/1.1726771},
  issn = {0021-9606},
  journal = {The Journal of Chemical Physics},
  number = {2},
  pages = {836--837},
  publisher = {AIP Publishing},
  title = {Ground-State Energy of a Molecule in the Adiabatic Approximation},
  volume = {44},
  year = {1966}
}

@book{born1996dynamical,
  author = {Born, Max and Huang, Kun},
  doi = {10.1093/oso/9780192670083.001.0001},
  isbn = {9780192670083},
  publisher = {Oxford University PressNew York, NY},
  title = {Dynamical Theory Of Crystal Lattices},
  year = {1996}
}

@article{AmmannMougelNistor2022,
  author = {Ammann, Bernd and Mougel, J\'er\'emy and Nistor, Victor},
  doi = {10.1007/s00023-021-01109-1},
  issn = {1424-0637},
  journal = {Annales Henri Poincar\'e},
  number = {4},
  pages = {1141--1203},
  publisher = {Springer Science and Business Media LLC},
  title = {A Comparison of the Georgescu and Vasy Spaces Associated to the N-Body Problems and Applications},
  volume = {23},
  year = {2021}
}

@incollection{Pflaum2001,
  author = {Schlichenmaier, Martin},
  booktitle = {Quantization of Singular Symplectic Quotients},
  doi = {10.1007/978-3-0348-8364-1_10},
  isbn = {9783034895354},
  pages = {259--282},
  publisher = {Birkh\"auser Basel},
  title = {Singular projective varieties and quantization},
  year = {2001}
}

@article{HelfferSjostrand1984,
  author = {Helffer, B. and Sjostrand, J.},
  doi = {10.1080/03605308408820335},
  issn = {0360-5302},
  journal = {Communications in Partial Differential Equations},
  number = {4},
  pages = {337--408},
  publisher = {Informa UK Limited},
  title = {Multiple wells in the semi-classical limit I},
  volume = {9},
  year = {1984}
}

@article{Simon1984Tunnelling,
  author = {Simon, Barry},
  doi = {10.2307/2007072},
  issn = {0003-486X},
  journal = {The Annals of Mathematics},
  number = {1},
  pages = {89},
  publisher = {JSTOR},
  title = {Semiclassical Analysis of Low Lying Eigenvalues, II. Tunneling},
  volume = {120},
  year = {1984}
}

@article{Rieffel1993,
  author = {Rieffel, Marc A.},
  doi = {10.1090/memo/0506},
  issn = {1947-6221},
  journal = {Memoirs of the American Mathematical Society},
  number = {506},
  publisher = {American Mathematical Society (AMS)},
  title = {Deformation quantization for actions of {$\mathbb{R}^{d}$}},
  volume = {106},
  year = {1993}
}

@article{Kendrick2015,
  author = {Kendrick, B. K. and Hazra, Jisha and Balakrishnan, N.},
  doi = {10.1038/ncomms8918},
  issn = {2041-1723},
  journal = {Nature Communications},
  number = {1},
  publisher = {Springer Science and Business Media LLC},
  title = {The geometric phase controls ultracold chemistry},
  volume = {6},
  year = {2015}
}

@article{DoplicherHaagRoberts1990,
  author = {Doplicher, Sergio and Roberts, John E.},
  doi = {10.1007/bf02097680},
  issn = {0010-3616},
  journal = {Communications in Mathematical Physics},
  number = {1},
  pages = {51--107},
  publisher = {Springer Science and Business Media LLC},
  title = {Why there is a field algebra with a compact gauge group describing the superselection structure in particle physics},
  volume = {131},
  year = {1990}
}

@article{Georgescu2002,
  author = {Georgescu, Vladimir and Iftimovici, Andrei},
  doi = {10.1007/s002200200669},
  issn = {0010-3616},
  journal = {Communications in Mathematical Physics},
  number = {3},
  pages = {519--560},
  publisher = {Springer Science and Business Media LLC},
  title = {Crossed Products of {C} * -Algebras and Spectral Analysis of Quantum Hamiltonians},
  volume = {228},
  year = {2002}
}

@article{Scrutton2006,
  author = {Masgrau, Laura and Roujeinikova, Anna and Johannissen, Linus O. and Hothi, Parvinder and Basran, Jaswir and Ranaghan, Kara E. and Mulholland, Adrian J. and Sutcliffe, Michael J. and Scrutton, Nigel S. and Leys, David},
  doi = {10.1126/science.1126002},
  issn = {0036-8075},
  journal = {Science},
  number = {5771},
  pages = {237--241},
  publisher = {American Association for the Advancement of Science (AAAS)},
  title = {Atomic Description of an Enzyme Reaction Dominated by Proton Tunneling},
  volume = {312},
  year = {2006}
}

@misc{GavRanovic2024CDL,
  author = {Gavranovi\'c, Bruno and Lessard, Paul and Dudzik, Andrew and Glehn, Tamara von and Ara\'ujo, Jo\~ao G. M. and Veli\v{c}kovi\'c, Petar},
  booktitle = {arXiv (Cornell University)},
  doi = {10.48550/arxiv.2402.15332},
  title = {Position: Categorical Deep Learning is an Algebraic Theory of All Architectures},
  year = {2024}
}

@inproceedings{BaezGenoveseMasterShulman2021,
  author = {Baez, John C. and Genovese, Fabrizio and Master, Jade and Shulman, Michael},
  booktitle = {2021 36th Annual ACM/IEEE Symposium on Logic in Computer Science (LICS)},
  doi = {10.1109/lics52264.2021.9470566},
  pages = {1--13},
  publisher = {IEEE},
  title = {Categories of Nets},
  year = {2021}
}

@misc{deHaanCohenWelling2020,
  author = {Haan, Pim de and Cohen, Taco and Welling, Max},
  booktitle = {arXiv (Cornell University)},
  doi = {10.48550/arxiv.2007.08349},
  title = {Natural Graph Networks},
  year = {2020}
}

@incollection{CruttwellGavranovic2022,
  author = {Cruttwell, Geoffrey S. H. and Gavranovi\'c, Bruno and Ghani, Neil and Wilson, Paul and Zanasi, Fabio},
  booktitle = {Lecture Notes in Computer Science},
  doi = {10.1007/978-3-030-99336-8_1},
  isbn = {9783030993351},
  issn = {0302-9743},
  pages = {1--28},
  publisher = {Springer International Publishing},
  title = {Categorical Foundations of Gradient-Based Learning},
  year = {2022}
}

@article{Fritz2020Markov,
  author = {Fritz, Tobias},
  doi = {10.1016/j.aim.2020.107239},
  issn = {0001-8708},
  journal = {Advances in Mathematics},
  pages = {107239},
  publisher = {Elsevier BV},
  title = {A synthetic approach to Markov kernels, conditional independence and theorems on sufficient statistics},
  volume = {370},
  year = {2020}
}

@article{Rupp2012,
  author = {Rupp, Matthias and Tkatchenko, Alexandre and M\"uller, Klaus-Robert and von Lilienfeld, O. Anatole},
  doi = {10.1103/physrevlett.108.058301},
  issn = {0031-9007},
  journal = {Physical Review Letters},
  number = {5},
  publisher = {American Physical Society (APS)},
  title = {Fast and Accurate Modeling of Molecular Atomization Energies with Machine Learning},
  volume = {108},
  year = {2012}
}

@article{Bartok2013SOAP,
  author = {Bart\'ok, Albert P. and Kondor, Risi and Cs\'anyi, G\'abor},
  doi = {10.1103/physrevb.87.184115},
  issn = {1098-0121},
  journal = {Physical Review B},
  number = {18},
  publisher = {American Physical Society (APS)},
  title = {On representing chemical environments},
  volume = {87},
  year = {2013}
}

@article{BehlerParrinello2007,
  author = {Behler, J\"org and Parrinello, Michele},
  doi = {10.1103/physrevlett.98.146401},
  issn = {0031-9007},
  journal = {Physical Review Letters},
  number = {14},
  publisher = {American Physical Society (APS)},
  title = {Generalized Neural-Network Representation of High-Dimensional Potential-Energy Surfaces},
  volume = {98},
  year = {2007}
}

@article{Smith2017ANI,
  author = {Smith, J. S. and Isayev, O. and Roitberg, A. E.},
  doi = {10.1039/c6sc05720a},
  issn = {2041-6520},
  journal = {Chemical Science},
  number = {4},
  pages = {3192--3203},
  publisher = {Royal Society of Chemistry (RSC)},
  title = {ANI-1: an extensible neural network potential with DFT accuracy at force field computational cost},
  volume = {8},
  year = {2017}
}

@article{Smith2019ANI,
  author = {Smith, Justin S. and Nebgen, Benjamin T. and Zubatyuk, Roman and Lubbers, Nicholas and Devereux, Christian and Barros, Kipton and Tretiak, Sergei and Isayev, Olexandr and Roitberg, Adrian E.},
  doi = {10.1038/s41467-019-10827-4},
  issn = {2041-1723},
  journal = {Nature Communications},
  number = {1},
  publisher = {Springer Science and Business Media LLC},
  title = {Approaching coupled cluster accuracy with a general-purpose neural network potential through transfer learning},
  volume = {10},
  year = {2019}
}

@article{Schutt2018SchNet,
  author = {Sch\"utt, K. T. and Sauceda, H. E. and Kindermans, P.-J. and Tkatchenko, A. and M\"uller, K.-R.},
  doi = {10.1063/1.5019779},
  issn = {0021-9606},
  journal = {The Journal of Chemical Physics},
  number = {24},
  publisher = {AIP Publishing},
  title = {SchNet {\textendash} A deep learning architecture for molecules and materials},
  volume = {148},
  year = {2018}
}

@misc{Gasteiger2020DimeNet,
  author = {Gasteiger, Johannes and Gro{\ss}, Janek and G\"unnemann, Stephan},
  booktitle = {arXiv (Cornell University)},
  doi = {10.48550/arxiv.2003.03123},
  title = {Directional Message Passing for Molecular Graphs},
  year = {2020}
}

@misc{Gasteiger2021GemNet,
  author = {Gasteiger, Johannes and Becker, Florian and G\"unnemann, Stephan},
  booktitle = {arXiv (Cornell University)},
  doi = {10.48550/arxiv.2106.08903},
  title = {GemNet: Universal Directional Graph Neural Networks for Molecules},
  year = {2021}
}

@article{Choudhary2021ALIGNN,
  author = {Choudhary, Kamal and DeCost, Brian},
  doi = {10.1038/s41524-021-00650-1},
  issn = {2057-3960},
  journal = {npj Computational Materials},
  number = {1},
  publisher = {Springer Science and Business Media LLC},
  title = {Atomistic Line Graph Neural Network for improved materials property predictions},
  volume = {7},
  year = {2021}
}

@misc{Schutt2021PaiNN,
  author = {Sch\"utt, Kristof T. and Unke, Oliver T. and Gastegger, Michael},
  booktitle = {arXiv (Cornell University)},
  doi = {10.48550/arxiv.2102.03150},
  title = {Equivariant message passing for the prediction of tensorial properties and molecular spectra},
  year = {2021}
}

@article{Batzner2022NequIP,
  author = {Batzner, Simon and Musaelian, Albert and Sun, Lixin and Geiger, Mario and Mailoa, Jonathan P. and Kornbluth, Mordechai and Molinari, Nicola and Smidt, Tess E. and Kozinsky, Boris},
  doi = {10.1038/s41467-022-29939-5},
  issn = {2041-1723},
  journal = {Nature Communications},
  number = {1},
  publisher = {Springer Science and Business Media LLC},
  title = {E(3)-equivariant graph neural networks for data-efficient and accurate interatomic potentials},
  volume = {13},
  year = {2022}
}

@inproceedings{BatatIa2022MACE,
  author = {Batatia, Ilyes and Csanyi, Gabor and Kovacs, David P and Ortner, Christoph and Simm, Gregor},
  booktitle = {Advances in Neural Information Processing Systems 35},
  doi = {10.52202/068431-0830},
  pages = {11423--11436},
  publisher = {Neural Information Processing Systems Foundation, Inc. (NeurIPS)},
  title = {MACE: Higher Order Equivariant Message Passing Neural Networks for Fast and Accurate Force Fields},
  year = {2022}
}

@article{BatatIa2023MACEMP0,
  author = {Batatia, Ilyes and Benner, Philipp and Chiang, Yuan and Elena, Alin M. and Kov\'acs, D\'avid P. and Riebesell, Janosh and Advincula, Xavier R. and Asta, Mark and Avaylon, Matthew and Baldwin, William J. and Berger, Fabian and Bernstein, Noam and Bhowmik, Arghya and Bigi, Filippo and Blau, Samuel M. and C\u{a}rare, Vlad and Ceriotti, Michele and Chong, Sanggyu and Darby, James P. and De, Sandip and Della Pia, Flaviano and Deringer, Volker L. and Elijo\v{s}ius, Rokas and El-Machachi, Zakariya and Fako, Edvin and Falcioni, Fabio and Ferrari, Andrea C. and Gardner, John L. A. and Gawkowski, Miko{\l}aj J. and Genreith-Schriever, Annalena and George, Janine and Goodall, Rhys E. A. and Grandel, Jonas and Grey, Clare P. and Grigorev, Petr and Han, Shuang and Handley, Will and Heenen, Hendrik H. and Hermansson, Kersti and Ho, Cheuk Hin and Hofmann, Stephan and Holm, Christian and Jaafar, Jad and Jakob, Konstantin S. and Jung, Hyunwook and Kapil, Venkat and Kaplan, Aaron D. and Karimitari, Nima and Kermode, James R. and Kourtis, Panagiotis and Kroupa, Namu and Kullgren, Jolla and Kuner, Matthew C. and Kuryla, Domantas and Liepuoniute, Guoda and Lin, Chen and Margraf, Johannes T. and Magd\u{a}u, Ioan-Bogdan and Michaelides, Angelos and Moore, J. Harry and Naik, Aakash A. and Niblett, Samuel P. and Norwood, Sam Walton and O{\textquoteright}Neill, Niamh and Ortner, Christoph and Persson, Kristin A. and Reuter, Karsten and Rosen, Andrew S. and Rosset, Louise A. M. and Schaaf, Lars L. and Schran, Christoph and Shi, Benjamin X. and Sivonxay, Eric and Stenczel, Tam\'as K. and Sutton, Christopher and Svahn, Viktor and Swinburne, Thomas D. and Tilly, Jules and van der Oord, Cas and Vargas, Santiago and Varga-Umbrich, Eszter and Vegge, Tejs and Vondr\'ak, Martin and Wang, Yangshuai and Witt, William C. and Wolf, Thomas and Zills, Fabian and Cs\'anyi, G\'abor},
  doi = {10.1063/5.0297006},
  issn = {0021-9606},
  journal = {The Journal of Chemical Physics},
  number = {18},
  publisher = {AIP Publishing},
  title = {A foundation model for atomistic materials chemistry},
  volume = {163},
  year = {2025}
}

@article{Musaelian2023Allegro,
  author = {Musaelian, Albert and Batzner, Simon and Johansson, Anders and Sun, Lixin and Owen, Cameron J. and Kornbluth, Mordechai and Kozinsky, Boris},
  doi = {10.1038/s41467-023-36329-y},
  issn = {2041-1723},
  journal = {Nature Communications},
  number = {1},
  publisher = {Springer Science and Business Media LLC},
  title = {Learning local equivariant representations for large-scale atomistic dynamics},
  volume = {14},
  year = {2023}
}

@inproceedings{Frank2022So3krates,
  author = {Frank, Thorben and M\"uller, Klaus-Robert and Unke, Oliver},
  booktitle = {Advances in Neural Information Processing Systems 35},
  doi = {10.52202/068431-2132},
  pages = {29400--29413},
  publisher = {Neural Information Processing Systems Foundation, Inc. (NeurIPS)},
  title = {So3krates: Equivariant Attention for Interactions on Arbitrary Length-Scales in Molecular Systems},
  year = {2022}
}

@article{Kabylda2025SO3LR,
  author = {Kabylda, Adil and Frank, J. Thorben and Su\'arez-Dou, Sergio and Khabibrakhmanov, Almaz and Medrano Sandonas, Leonardo and Unke, Oliver T. and Chmiela, Stefan and M\"uller, Klaus-Robert and Tkatchenko, Alexandre},
  doi = {10.1021/jacs.5c09558},
  issn = {0002-7863},
  journal = {Journal of the American Chemical Society},
  number = {37},
  pages = {33723--33734},
  publisher = {American Chemical Society (ACS)},
  title = {Molecular Simulations with a Pretrained Neural Network and Universal Pairwise Force Fields},
  volume = {147},
  year = {2025}
}

@misc{Liao2023Equiformer,
  author = {Liao, Yi-Lun and Smidt, Tess},
  booktitle = {arXiv (Cornell University)},
  doi = {10.48550/arxiv.2206.11990},
  title = {Equiformer: Equivariant Graph Attention Transformer for 3D Atomistic Graphs},
  year = {2022}
}

@inproceedings{Fu2025eSEN,
  author = {Fu, Xinming and others},
  booktitle = {Proceedings of the 42nd International Conference
on Machine Learning},
  note = {arXiv:2502.12147},
  pages = {17875},
  publisher = {PMLR},
  series = {Proceedings of Machine Learning Research},
  title = {{eSEN}: Equivariant Scalable Energy Network},
  volume = {267},
  year = {2025}
}

@article{Jumper2021AF2,
  author = {Jumper, John and Evans, Richard and Pritzel, Alexander and Green, Tim and Figurnov, Michael and Ronneberger, Olaf and Tunyasuvunakool, Kathryn and Bates, Russ and \v{Z}{\'\i}dek, Augustin and Potapenko, Anna and Bridgland, Alex and Meyer, Clemens and Kohl, Simon A. A. and Ballard, Andrew J. and Cowie, Andrew and Romera-Paredes, Bernardino and Nikolov, Stanislav and Jain, Rishub and Adler, Jonas and Back, Trevor and Petersen, Stig and Reiman, David and Clancy, Ellen and Zielinski, Michal and Steinegger, Martin and Pacholska, Michalina and Berghammer, Tamas and Bodenstein, Sebastian and Silver, David and Vinyals, Oriol and Senior, Andrew W. and Kavukcuoglu, Koray and Kohli, Pushmeet and Hassabis, Demis},
  doi = {10.1038/s41586-021-03819-2},
  issn = {0028-0836},
  journal = {Nature},
  number = {7873},
  pages = {583--589},
  publisher = {Springer Science and Business Media LLC},
  title = {Highly accurate protein structure prediction with AlphaFold},
  volume = {596},
  year = {2021}
}

@article{Abramson2024AF3,
  author = {Abramson, Josh and Adler, Jonas and Dunger, Jack and Evans, Richard and Green, Tim and Pritzel, Alexander and Ronneberger, Olaf and Willmore, Lindsay and Ballard, Andrew J. and Bambrick, Joshua and Bodenstein, Sebastian W. and Evans, David A. and Hung, Chia-Chun and O{\textquoteright}Neill, Michael and Reiman, David and Tunyasuvunakool, Kathryn and Wu, Zachary and \v{Z}emgulyt\.{e}, Akvil\.{e} and Arvaniti, Eirini and Beattie, Charles and Bertolli, Ottavia and Bridgland, Alex and Cherepanov, Alexey and Congreve, Miles and Cowen-Rivers, Alexander I. and Cowie, Andrew and Figurnov, Michael and Fuchs, Fabian B. and Gladman, Hannah and Jain, Rishub and Khan, Yousuf A. and Low, Caroline M. R. and Perlin, Kuba and Potapenko, Anna and Savy, Pascal and Singh, Sukhdeep and Stecula, Adrian and Thillaisundaram, Ashok and Tong, Catherine and Yakneen, Sergei and Zhong, Ellen D. and Zielinski, Michal and \v{Z}{\'\i}dek, Augustin and Bapst, Victor and Kohli, Pushmeet and Jaderberg, Max and Hassabis, Demis and Jumper, John M.},
  doi = {10.1038/s41586-024-07487-w},
  issn = {0028-0836},
  journal = {Nature},
  number = {8016},
  pages = {493--500},
  publisher = {Springer Science and Business Media LLC},
  title = {Accurate structure prediction of biomolecular interactions with AlphaFold 3},
  volume = {630},
  year = {2024}
}

@article{Unke2021SpookyNet,
  author = {Unke, Oliver T. and Chmiela, Stefan and Gastegger, Michael and Sch\"utt, Kristof T. and Sauceda, Huziel E. and M\"uller, Klaus-Robert},
  doi = {10.1038/s41467-021-27504-0},
  issn = {2041-1723},
  journal = {Nature Communications},
  number = {1},
  publisher = {Springer Science and Business Media LLC},
  title = {Spooky{N}et: Learning force fields with electronic degrees of freedom and nonlocal effects},
  volume = {12},
  year = {2021}
}

@misc{Unke2021PhiSNet,
  author = {Unke, Oliver T. and Bogojeski, Mihail and Gastegger, Michael and Geiger, Mario and Smidt, Tess and M\"uller, Klaus-Robert},
  booktitle = {arXiv (Cornell University)},
  doi = {10.48550/arxiv.2106.02347},
  title = {SE(3)-equivariant prediction of molecular wavefunctions and electronic densities},
  volume = {34},
  year = {2021}
}

@article{Gong2023DeepHE3,
  author = {Gong, Xiaoxun and Li, He and Zou, Nianlong and Xu, Runzhang and Duan, Wenhui and Xu, Yong},
  doi = {10.1038/s41467-023-38468-8},
  issn = {2041-1723},
  journal = {Nature Communications},
  number = {1},
  publisher = {Springer Science and Business Media LLC},
  title = {General framework for E(3)-equivariant neural network representation of density functional theory Hamiltonian},
  volume = {14},
  year = {2023}
}

@article{Westermayr2020SchNarc,
  author = {Westermayr, Julia and Gastegger, Michael and Marquetand, Philipp},
  doi = {10.1021/acs.jpclett.0c00527},
  journal = {The Journal of Physical Chemistry Letters},
  number = {10},
  pages = {3828--3834},
  title = {Combining SchNet and SHARC: The SchNarc Machine Learning Approach for Excited-State Dynamics},
  volume = {11},
  year = {2020}
}

@article{Mausenberger2024SPAINN,
  author = {Mausenberger, Sascha and M\"uller, Carolin and Tkatchenko, Alexandre and Marquetand, Philipp and Gonz\'alez, Leticia and Westermayr, Julia},
  doi = {10.1039/d4sc04164j},
  issn = {2041-6520},
  journal = {Chemical Science},
  number = {38},
  pages = {15880--15890},
  publisher = {Royal Society of Chemistry (RSC)},
  title = {S
                    <scp>pai</scp>
                    NN: equivariant message passing for excited-state nonadiabatic molecular dynamics},
  volume = {15},
  year = {2024}
}

@article{Xie2018NNPD,
  author = {Xie, Changjian and Zhu, Xiaolei and Yarkony, David R. and Guo, Hua},
  doi = {10.1063/1.5054310},
  issn = {0021-9606},
  journal = {The Journal of Chemical Physics},
  number = {14},
  publisher = {AIP Publishing},
  title = {Permutation invariant polynomial neural network approach to fitting potential energy surfaces. IV. Coupled diabatic potential energy matrices},
  volume = {149},
  year = {2018}
}

@article{Zhang2020NNDiabat,
  author = {Hong, Yingyue and Yin, Zhengxi and Guan, Yafu and Zhang, Zhaojun and Fu, Bina and Zhang, Dong H.},
  doi = {10.1021/acs.jpclett.0c02173},
  issn = {1948-7185},
  journal = {The Journal of Physical Chemistry Letters},
  number = {18},
  pages = {7552--7558},
  publisher = {American Chemical Society (ACS)},
  title = {Exclusive Neural Network Representation of the Quasi-Diabatic Hamiltonians Including Conical Intersections},
  volume = {11},
  year = {2020}
}

@article{Owoyele2022ChemNODE,
  author = {Owoyele, Opeoluwa and Pal, Pinaki},
  doi = {10.1016/j.egyai.2021.100118},
  issn = {2666-5468},
  journal = {Energy and AI},
  pages = {100118},
  publisher = {Elsevier BV},
  title = {ChemNODE: A neural ordinary differential equations framework for efficient chemical kinetic solvers},
  volume = {7},
  year = {2022}
}

@article{Ji2021CRNN,
  author = {Ji, Weiqi and Deng, Sili},
  doi = {10.1021/acs.jpca.0c09316},
  issn = {1089-5639},
  journal = {The Journal of Physical Chemistry A},
  number = {4},
  pages = {1082--1092},
  publisher = {American Chemical Society (ACS)},
  title = {Autonomous Discovery of Unknown Reaction Pathways from Data by Chemical Reaction Neural Network},
  volume = {125},
  year = {2021}
}

@article{Schwaller2019MolTransformer,
  author = {Schwaller, Philippe and Laino, Teodoro and Gaudin, Th\'eophile and Bolgar, Peter and Hunter, Christopher A. and Bekas, Costas and Lee, Alpha A.},
  doi = {10.1021/acscentsci.9b00576},
  issn = {2374-7943},
  journal = {ACS Central Science},
  number = {9},
  pages = {1572--1583},
  publisher = {American Chemical Society (ACS)},
  title = {Molecular Transformer: A Model for Uncertainty-Calibrated Chemical Reaction Prediction},
  volume = {5},
  year = {2019}
}

@article{Probst2022DrFP,
  author = {Probst, Daniel and Schwaller, Philippe and Reymond, Jean-Louis},
  doi = {10.1039/d1dd00006c},
  issn = {2635-098X},
  journal = {Digital Discovery},
  number = {2},
  pages = {91--97},
  publisher = {Royal Society of Chemistry (RSC)},
  title = {Reaction classification and yield prediction using the differential reaction fingerprint DRFP},
  volume = {1},
  year = {2022}
}

@article{Noe2019BoltzmannGen,
  author = {No\'e, Frank and Olsson, Simon and K\"ohler, Jonas and Wu, Hao},
  doi = {10.1126/science.aaw1147},
  issn = {0036-8075},
  journal = {Science},
  number = {6457},
  publisher = {American Association for the Advancement of Science (AAAS)},
  title = {Boltzmann generators: Sampling equilibrium states of many-body systems with deep learning},
  volume = {365},
  year = {2019}
}

@article{GibsonBruck2000,
  author = {Gibson, Michael A. and Bruck, Jehoshua},
  doi = {10.1021/jp993732q},
  issn = {1089-5639},
  journal = {The Journal of Physical Chemistry A},
  number = {9},
  pages = {1876--1889},
  publisher = {American Chemical Society (ACS)},
  title = {Efficient Exact Stochastic Simulation of Chemical Systems with Many Species and Many Channels},
  volume = {104},
  year = {2000}
}

@article{Rittig2024GibbsDuhem,
  author = {Rittig, Jan G. and Mitsos, Alexander},
  doi = {10.1039/d4sc04554h},
  issn = {2041-6520},
  journal = {Chemical Science},
  number = {44},
  pages = {18504--18512},
  publisher = {Royal Society of Chemistry (RSC)},
  title = {Thermodynamics-consistent graph neural networks},
  volume = {15},
  year = {2024}
}
